\AddToHook{package/hyperref/before}{\RequirePackage[original]{imakeidx}}
\documentclass{dis}

\usepackage[T1]{fontenc}

\usepackage{amsmath}
\usepackage{amsfonts}
\usepackage{amssymb}
\usepackage{amsthm}
\usepackage{mathtools}

\usepackage{booktabs}
\usepackage{longtable}
\usepackage{tabularx}
\usepackage{graphicx}
\usepackage{xurl}
\makeindex
\usepackage{placeins}
\usepackage{needspace}
\usepackage[shortlabels]{enumitem}

\usepackage{xcolor}

\usepackage{cleveref}

\usepackage{tikz}
\usetikzlibrary{arrows.meta,positioning,calc,fit,backgrounds,
  decorations.pathreplacing}
\definecolor{figaccent}{RGB}{22,66,112}
\colorlet{figink}{black!88}
\colorlet{figrule}{black!55}
\colorlet{figgrey}{black!8}
\colorlet{figtint}{figaccent!7}
\newcommand{\figfont}{\fontsize{8}{9.6}\selectfont}
\newcommand{\figsmall}{\fontsize{7}{8.4}\selectfont}
\newcommand{\figref}[1]{{\figsmall\color{figaccent}#1}}
\tikzset{
  figdiagram/.style={
    >={Stealth[length=4.4pt,width=3.6pt]},
    line cap=round, line join=round,
    font=\figfont, text=black,
    node distance=5mm and 5mm,
    execute at begin picture={\hypersetup{linkcolor=.}}},
  fbox/.style={rectangle, rounded corners=1.8pt, align=center,
    inner xsep=3.5pt, inner ysep=3pt, outer sep=0pt, font=\figfont},
  fres/.style={fbox, draw=figaccent, line width=0.75pt, fill=figtint},
  fmech/.style={fbox, draw=none, fill=figgrey},
  finp/.style={fbox, draw=figrule, line width=0.4pt, fill=white},
  fopen/.style={fbox, draw=figrule, line width=0.4pt, fill=white,
    dash pattern=on 2.2pt off 1.6pt},
  fnote/.style={align=left, inner sep=1.5pt, outer sep=0pt,
    font=\figsmall, text=black!80},
  flab/.style={align=center, inner sep=1.3pt, outer sep=0.5pt,
    font=\figsmall\itshape, text=black, fill=white},
  fhyp/.style={align=center, inner sep=1.3pt, outer sep=0.5pt,
    font=\figsmall, text=black, fill=white},
  fuse/.style={->, draw=figink, line width=0.6pt, rounded corners=2pt},
  ffeed/.style={-, draw=figink, line width=0.6pt, rounded corners=2pt},
  fmap/.style={->, draw=figink, line width=0.5pt, rounded corners=2pt,
    dash pattern=on 2.6pt off 1.7pt},
  flink/.style={-, draw=figink, line width=0.7pt, line cap=round,
    dash pattern=on 0pt off 2.1pt},
  fptr/.style={->, draw=figrule, line width=0.4pt, rounded corners=2pt,
    >={Straight Barb[length=2.4pt,width=3.4pt]}},
  fjoin/.style={circle, fill=figink, draw=none, inner sep=0pt,
    minimum size=3.2pt, outer sep=0pt},
  fgroup/.style={rectangle, rounded corners=3pt, draw=figrule,
    line width=0.35pt, dash pattern=on 1.2pt off 1.4pt,
    inner xsep=5pt, inner ysep=7pt},
  fhead/.style={font=\figfont\scshape, text=black, inner sep=1pt,
    outer sep=0pt, align=left},
  ftag/.style={font=\figsmall\scshape, text=figaccent, inner sep=1.5pt,
    outer sep=0pt, fill=white, align=left},
  fband/.style={rectangle, rounded corners=2pt, draw=none, fill=black!4,
    inner sep=3pt},
  frule/.style={draw=figrule, line width=0.35pt},
}
\counterwithout{figure}{chapter}

\hypersetup{%
  unicode=true,
  colorlinks=true,
  citecolor=black,
  linkcolor=black,
  urlcolor=black,
  filecolor=black,
  pdfborder={0 0 0},
  bookmarksnumbered=true,
  bookmarksopen=true,
  pdftitle={Arithmetic limiting laws over shifted primes: Hausdorff dimension zero,
            Fourier decay, and absolute continuity},
  pdfauthor={Vipin Singh Sehrawat},
  pdfsubject={Number theory; harmonic analysis of singular measures;
              Bernoulli convolutions; cyclotomic lattices},
  pdfkeywords={Shifted primes, Erdős--Wintner laws,
               totient and divisor-sum distributions,
               Hausdorff and packing dimensions, Shannon information dimension,
               quantization dimensions, absolutely normal Liouville numbers,
               concentration functions, Fourier decay, Fourier amplitudes,
               absolute continuity, learning with errors,
               cyclotomic codifferents, lattice smoothing parameter}%
}

\crefname{chapter}{Chapter}{Chapters}
\Crefname{chapter}{Chapter}{Chapters}
\crefname{appendix}{Appendix}{Appendices}
\Crefname{appendix}{Appendix}{Appendices}
\crefname{subappendix}{Appendix}{Appendices}
\Crefname{subappendix}{Appendix}{Appendices}
\crefname{subsubappendix}{Appendix}{Appendices}
\Crefname{subsubappendix}{Appendix}{Appendices}
\crefname{section}{\S}{\S\S}
\Crefname{section}{Section}{Sections}
\crefname{subsection}{\S}{\S\S}
\Crefname{subsection}{Section}{Sections}
\crefname{equation}{}{}
\crefname{figure}{Fig.}{Figs.}
\Crefname{figure}{Fig.}{Figs.}
\crefname{theorem}{Theorem}{Theorems}
\Crefname{theorem}{Theorem}{Theorems}
\crefname{proposition}{Proposition}{Propositions}
\Crefname{proposition}{Proposition}{Propositions}
\crefname{lemma}{Lemma}{Lemmas}
\Crefname{lemma}{Lemma}{Lemmas}
\crefname{corollary}{Corollary}{Corollaries}
\Crefname{corollary}{Corollary}{Corollaries}
\crefname{definition}{Definition}{Definitions}
\Crefname{definition}{Definition}{Definitions}
\crefname{remark}{Remark}{Remarks}
\Crefname{remark}{Remark}{Remarks}

\newcommand{\R}{\mathbb{R}}
\newcommand{\Z}{\mathbb{Z}}
\newcommand{\Q}{\mathbb{Q}}
\newcommand{\N}{\mathbb{N}}
\newcommand{\C}{\mathbb{C}}
\newcommand{\Fp}{\mathbb{F}_p}
\newcommand{\GL}{\mathrm{GL}}
\newcommand{\Mn}{M_n(\Fp)}
\newcommand{\muF}{\mu_f}
\newcommand{\FT}{\widehat{\mu_f}}
\newcommand{\bigast}{\mathop{\raisebox{-1pt}{\scalebox{1.5}{$\ast$}}}}
\providecommand{\Conv}{\bigast}
\providecommand{\csig}{c_{\sigma}}
\providecommand{\gsig}{g_{\sigma}}
\providecommand{\wsig}{\omega^{\sigma}}
\providecommand{\nusig}{\nu^{\sigma}}
\providecommand{\muH}{\mu_{H_{\sigma}}}
\providecommand{\mug}{\mu_{g}}
\providecommand{\mugodd}{\mu_{g}^{\mathrm{odd}}}
\providecommand{\Smer}{\mathfrak{S}_{2}}
\DeclareMathOperator{\Var}{Var}
\DeclareMathOperator{\supp}{supp}

\numberwithin{equation}{chapter}

\theoremstyle{plain}
\newtheorem{theorem}{Theorem}[chapter]
\newtheorem{proposition}[theorem]{Proposition}
\newtheorem{lemma}[theorem]{Lemma}
\newtheorem{corollary}[theorem]{Corollary}

\theoremstyle{definition}
\newtheorem{definition}[theorem]{Definition}
\newtheorem{remark}[theorem]{Remark}

\AddToHook{env/proposition/begin}{\crefalias{theorem}{proposition}}
\AddToHook{env/lemma/begin}{\crefalias{theorem}{lemma}}
\AddToHook{env/corollary/begin}{\crefalias{theorem}{corollary}}
\AddToHook{env/definition/begin}{\crefalias{theorem}{definition}}
\AddToHook{env/remark/begin}{\crefalias{theorem}{remark}}

\AddToHook{cmd/appendix/before}{%
  \crefalias{chapter}{appendix}%
  \crefalias{section}{subappendix}%
  \crefalias{subsection}{subsubappendix}%
}

\makeatletter
\newcommand{\KeepHeadingSpace}[1]{%
  \par
  \begingroup
    \skip@=\lastskip
    \vskip-\skip@
    \Needspace{#1}%
    \vskip\skip@
  \endgroup
}
\AddToHook{env/theindex/begin}{%
  \renewcommand\subitem{\par\nobreak\@idxitem\hspace*{15\p@}}%
}
\makeatother
\AddToHook{cmd/chapter/before}{\KeepHeadingSpace{8\baselineskip}}
\AddToHook{cmd/section/before}{\KeepHeadingSpace{1\baselineskip}}
\AddToHook{env/theorem/before}{\KeepHeadingSpace{1\baselineskip}}
\AddToHook{env/proposition/before}{\KeepHeadingSpace{1\baselineskip}}
\AddToHook{env/lemma/before}{\KeepHeadingSpace{1\baselineskip}}
\AddToHook{env/corollary/before}{\KeepHeadingSpace{1\baselineskip}}
\AddToHook{env/definition/before}{\KeepHeadingSpace{1\baselineskip}}
\AddToHook{env/remark/before}{\KeepHeadingSpace{1\baselineskip}}

\begin{document}

\mathclass{Primary 11N60; Secondary 11H06, 11J82, 11K16, 11K65,
  11L20, 11N13, 11N56, 11R18, 28A78, 42A38, 60E05, 94A17, 94A34, 94A60.}

\abbrevauthors{V. S. Sehrawat}
\abbrevtitle{Arithmetic limiting laws over shifted primes}
\title{Arithmetic limiting laws over shifted primes: Hausdorff dimension zero, Fourier decay, and absolute continuity}

\author{Vipin Singh Sehrawat}
\address{Circle Internet\\
New York, NY, USA\\
E-mail: vipin.sehrawat.cs@gmail.com}

\maketitledis

\setcounter{tocdepth}{3}
\tableofcontents

\begin{abstract}
We study the classical limiting laws of $\varphi(n)/n$ and $\sigma(n)/n$ when
$n$ ranges over the positive integers or the shifted primes $p-1$.
Independent prime-coordinate models connect their geometry, concentration
and Fourier behavior.
Let $\mu_G$ be the limiting law of $\varphi(p-1)/(p-1)$ over the primes,
supported on $[0,\tfrac12]$. We prove that it has Hausdorff dimension
zero and that every Borel set of positive $\mu_G$-measure has packing
dimension one. The lower and upper Shannon information dimensions of all four
ratio laws and their logarithmic images satisfy
$0.55043653<\underline d_1\le\overline d_1<0.570489495$.

The law $\mu_G$ gives full mass to absolutely normal Liouville numbers.
Both $\mu_G$ and its negative-logarithmic image $\muF$ have quantitative
Fourier decay. For $\muF$, we determine sharp concentration asymptotics
and localize maximizing intervals. For the integer divisor-sum law and the
shifted-prime divisor-sum laws with and without the factor at $2$, every
sufficiently short maximizing interval starts at the left endpoint of the
support, in both raw and logarithmic coordinates. Let $U_T$ be uniform on $[0,T]$.
As $T\to\infty$, $(\log T)\FT(U_T)$ converges in distribution to a
random Euler product with an independent uniform phase.

The limiting primitive-root determinant density for uniform square
matrices over $\Fp$ has the same prime-sampling law as the totient ratio.
In every fixed reduced progression, the empirical laws of these two
statistics over primes $p\le x$ converge to their common progression law
with sharp Kolmogorov order $1/\log x$.
We also obtain quantitative transport estimates in growing progressions.

Strongly additive functions with prime values $(\log\ell)^{-A}$ have
absolutely continuous integer and shifted-prime laws for every $A>0$,
disproving the Erd\H{o}s--K\'atai singularity conjecture for $A>2$.
Applications include extremal determinant bounds and certified prime
selection without complete factorization of $p-1$.
The results stated above are unconditional; further
quantitative refinements assume the generalized Riemann hypothesis for
Dirichlet $L$-functions.
\end{abstract}

\makeabstract

\phantomsection
\chapter*{Principal notation}
\addcontentsline{toc}{chapter}{Principal notation}
\noindent Unless a base is indicated, $\log$ denotes the natural logarithm.\par
\noindent In decompositions of the positive integers, a dyadic scale is
$2^k$ with $k\in\Z_{\ge0}$. A \emph{dyadic block} is $\{1\}$, of scale
$1$, or $(H,2H]\cap\mathbb N$, of scale $H$ with $H$ dyadic.\par\smallskip
\begingroup\renewcommand{\arraystretch}{1.04}\scriptsize
\let\vss\vfil
\begin{longtable}{@{}p{0.17\textwidth}p{0.55\textwidth}>{\raggedright\arraybackslash}p{0.20\textwidth}@{}}
\toprule
Symbol & Meaning & \shortstack[l]{Definition / main\\reference}\\
\midrule
\endfirsthead
\toprule
Symbol & Meaning & \shortstack[l]{Definition / main\\reference}\\
\midrule
\endhead
\midrule
\multicolumn{3}{r@{}}{\scriptsize Continued on the next page}\\
\endfoot
\bottomrule
\endlastfoot
$p$, $\ell$ & primes; \(\ell\) is restricted to odd primes only where stated & \Cref{sec:intro}\\
$\varphi$, $\sigma$ & Euler totient; sum of divisors & \Cref{sec:intro}, \Cref{sec:dist-setup}\\
$\omega(n)$ & number of distinct prime factors & \eqref{eq:arithmetic-notation}\\
$\Omega(n)$ & $\sum_{\ell\mid n}v_\ell(n)$, number of prime factors counted with multiplicity; $\Omega(1)=0$ & \Cref{lem:prime-large-smooth-window}\\
$\tau(n)$ & number of positive divisors of $n$, with $\tau=\mathbf d_2$ & \Cref{lem:divisor-energy}\\
$\mathbf d_k(n)$ & number of ordered factorizations of $n$ into $k$ positive integers & \Cref{sec:raj-annulus}\\
$\vartheta(y)$ & Chebyshev sum $\sum_{\ell\le y}\log\ell$ over primes & \eqref{eq:arithmetic-notation}\\
$P^+(n)$, $P^-(n)$ & largest and least prime factors of $n$; $P^+(1)=1$, $P^-(1)=\infty$ & \Cref{sec:ap-law,ssec:friable-unconditional}; \eqref{eq:prime-model-sieve}\\
$s_z(n)$ & $z$-smooth part $\prod_{q\le z}q^{v_q(n)}$, over primes $q$ & \Cref{ssec:sharp-prime-convergence}\\
$\mathsf{E}_t(p)$ & $\prod_{\substack{\ell\le t\\\ell\mid p-1}}(1-1/\ell)^{-1}$ over primes $\ell$, for prime $p$ and real $t\ge3$ & \Cref{sec:explicit-certificates,ssec:friable-unconditional}\\
$\pi(x;m,a)$, $\operatorname{Li}(x)$ & prime-counting function in the class $a\bmod m$; logarithmic integral & \Cref{sec:prime-counting-inputs}\\
$\pi_S(x;m,a)$, $c_S$ & number of primes $p\le x$ in $a\bmod m$ with prescribed divisibility of $p-1$; limiting proportion within that progression & \Cref{prop:ap-bivariate}\\
$p_k$, $N_k$ & $k$th prime; primorial $N_k=p_1\cdots p_k$ & \Cref{sec:det-density}\\
$c(p)$ & limiting primitive-root determinant density $\frac{\varphi(p-1)}{p-1}P(p)$ & \Cref{sec:intro}\\
$c_n(p)$ & finite-\(n\) primitive-root determinant density & \Cref{sec:intro}, \Cref{sec:det-density}\\
$P(p)$ & matrix factor $\prod_{j\ge1}(1-p^{-j})$ & \eqref{eq:def-P}\\
$G$, $\mu_G$ & limiting distribution function and measure of $c(p)$ over the primes & \Cref{thm:phi-dist}\\
$X$, $B_\ell$ & limiting variable $X\overset{d}{=}\tfrac12\prod_{\ell\ge3}(1-1/\ell)^{B_\ell}$; independent Bernoulli indicators with $\mathbb P(B_\ell=1)=1/(\ell-1)$ & \Cref{sec:intro,lem:factor}\\
$f(n)$ & strongly additive logarithmic totient function \(\log(n/\varphi(n))\) & \eqref{eq:f-def}\\
$\mu_f$, $\FT$ & logarithmic law (image of $\mu_G$ under $u\mapsto-\log u$); its Fourier transform & \Cref{ssec:four-scalar-laws,lem:factor}; \eqref{eq:measure-Fourier-dimension}\\
\shortstack[l]{$\mu_F^{\mathrm{int}}$\\$\mu_F^{\mathrm{int},\log}$} & integer totient law of $\varphi(n)/n$; its image under $-\log$ & \Cref{ssec:four-scalar-laws}; \eqref{eq:integer-totient-factorization}\\
\shortstack[l]{$\mu_\sigma^{\mathrm{int}}$\\$\mu_\sigma^{\mathrm{int},\log}$} & integer divisor-sum law of $\sigma(n)/n$; its image under $+\log$ & \Cref{ssec:four-scalar-laws}; \eqref{eq:integer-sigma-factorization}\\
$\mug$, $\muH$ & logarithmic and raw shifted-prime divisor-sum laws, related by $\muH=\exp_*\mug$ & \Cref{ssec:four-scalar-laws,thm:dist-sigma}\\
$\mathsf I$, $\mathsf A$, $\mathsf B$ & integer valuation law, its conditioning on $J_2\ge1$, and shifted-prime valuation law & \Cref{ssec:four-scalar-laws}; \eqref{eq:valuation-laws}\\
$\beta_2$, $\mugodd$ & prime-$2$ head and odd-prime convolution for the logarithmic shifted-prime divisor-sum law & \Cref{ssec:four-scalar-laws}; \eqref{eq:nusig}\\
$\nu_\ell$, $Z_\ell$ & law of $Z_\ell=B_\ell\omega_\ell$; the corresponding summand in $-\log X$, for odd primes $\ell$ & \Cref{lem:factor}; \eqref{eq:local-log-jump}\\
$\omega_\ell$ & logarithmic jump $\log(\ell/(\ell-1))$ & \Cref{sec:intro}, \eqref{eq:def-omega}\\
$d_\ell$, $S(\tau)$ & coefficient $2(\ell-2)/(\ell-1)^2$ and cosine sum $S(\tau)=\sum_{\ell\ge3}d_\ell(1-\cos(\tau\omega_\ell))$, satisfying $|\FT(\tau)|^2\le e^{-S(\tau)}$ & \eqref{eq:def-dell}; \eqref{eq:forward}\\
$M(s)$ & Mellin transform $\mathbb{E}[X^s]$ & \Cref{thm:moments,prop:hadamard}\\
$\operatorname{Law}(Y)$ & probability distribution of $Y$ & \Cref{lem:supp_conv}\\
$\mathcal L$ & set of Liouville numbers & \Cref{sec:liouville-carriers,thm:liouville-carriers}\\
$W$, $W_S$ & odd-prime Bernoulli sum $\sum_{\ell\ge3}B_\ell\omega_\ell$ and the sum with a finite set $S$ deleted & \Cref{prop:endpoint-half}; \eqref{eq:deleted-bernoulli-sum}\\
$W_F$, $M_F(s)$ & limiting normalized Fourier modulus of the shifted-prime logarithmic totient law and its real Mellin transform, $s>-1$ & \Cref{thm:fourier-amplitude-law}\\
$\Smer{}$ & Hardy--Littlewood twin-prime singular series & \eqref{eq:intro-endpoint-constant}\\
$\kappa_{\mathrm{end}}$ & right-endpoint constant \(\Smer{}e^{-\gamma}\) & \eqref{eq:intro-endpoint-constant}; \Cref{prop:endpoint-half}\\
$x_S$, $\kappa_S$ & point $x_S=\tfrac12\prod_{\ell\in S}(1-1/\ell)$ for a finite set $S$ of odd primes; constant in $G(x_S)-G(x_S-\varepsilon)\sim\kappa_S/\log(1/\varepsilon)$ & \Cref{def:anchor-weights}\\
$Q_G(\varepsilon)$, $Q_{\mu_f}(h)$ & relative concentration function of $G$; concentration function of $\mu_f$ & \Cref{sec:intro,thm:sharp-global-modulus}\\
$H_{\mathrm{end}}(\varepsilon)$ & right-endpoint mass $1-G(\tfrac12(1-\varepsilon))$ & \eqref{eq:intro-concentration}; \Cref{ssec:global-modulus-deletion}\\
$C(h)$ & concentration function of $\mu_G$: $\sup_t\{G(t)-G(t-h)\}$ & \Cref{cor:sharp-additive-modulus}\\
$\gamma$ & Euler's constant & \Cref{thm:mertens}\\
$L_X$ & Xylouris bound $5$ on the Linnik constant & \cite[Theorem~1]{Xyl18}\\
$p^*(k)$ & least prime congruent to $1\pmod{N_k}$ & \Cref{sec:det-density}\\
$D_x^U$ & Kolmogorov distance from the empirical prime law of $U$ to $\mu_G$ & \Cref{thm:sharp-prime-convergence}\\
$d_K(\mu,\nu)$ & Kolmogorov distance $\sup_x|\mu(({-\infty},x])-\nu(({-\infty},x])|$ & \Cref{cor:sharp-truncation-inverse}\\
$d_{\mathrm{TV}}(\alpha,\beta)$ & total variation distance $\sup_{A\textup{ Borel}}|\alpha(A)-\beta(A)|$ & \Cref{cor:summable-local-coupling}\\
$\rho(\tau)$ & $\rho(\tau)=(\log\log|\tau|/\log|\tau|)^{1/3}$ for $|\tau|>e$, Fourier bound for the logarithmic and four ratio laws & \eqref{eq:rho-def}\\
$\mathfrak N$ & absolutely normal real numbers & \Cref{lem:iterlog-normality,thm:raw-ratio-normality}\\
$\dim_H$, $\dim_F$ & Hausdorff and Fourier dimension of a measure & \eqref{eq:measure-Hausdorff-dimension}, \eqref{eq:measure-Fourier-dimension}\\
$\underline d_1$, $\overline d_1$ & lower and upper Shannon information dimensions & \Cref{thm:shannon-dimension}\\
$c_L$, $c_L^+$, $d_*$ & constructed lower functional, strengthened scalar lower bound and coding upper bound & \eqref{eq:scalar-functional}, \eqref{eq:scalar-strengthened-lower}\\
$\underline d_\lambda$, $\overline d_\lambda$ & lower and upper local dimensions & \eqref{eq:local-dimensions}\\
$P_{\mathrm{pos}}(\lambda)$, $P_{\mathrm{full}}(\lambda)$ & infimal packing dimensions of positive-mass and full-mass Borel carriers & \eqref{eq:packing-carrier-dimensions}\\
$\mathcal W_r(\alpha,\beta)$ & $r$-Wasserstein distance, $1\le r<\infty$ & \Cref{prop:prime-wasserstein}\\
$G_{Q,b}$, $X_{Q,b}$ & limiting distribution function of $c(p)$ for primes $p\equiv b\pmod Q$; a random variable with that distribution & \Cref{sec:intro,thm:ap-law}\\
$\mathcal P_Q$ & the prime \(2\) and the prime divisors of \(Q\) & \Cref{thm:ap-law}\\
$D(Q)$, $D_{\le y}(Q)$ & progression deletion factor and its product restricted to primes $\ell\le y$ & \Cref{ssec:sharp-prime-convergence}, \eqref{eq:moving-modulus-endpoint}\\
$f_A$, $\mu_A$ & strongly additive function $f_A(n)=\sum_{\ell\mid n,\,\ell\text{ prime}}(\log\ell)^{-A}$; limiting law of $f_A(p-1)$ over primes & \Cref{ssec:slow-jump-ac}; \eqref{eq:slow-mu}\\
$\Lambda_m$, $\Lambda_m^*$ & canonical embedding of $\Z[\zeta_m]$ and its trace dual (the embedded codifferent), for $m\ge3$ & \Cref{sec:cyclotomic-gap}\\
$\eta_\epsilon(\Lambda)$ & smoothing parameter of $\Lambda$ at error $\epsilon$ & \Cref{def:lattice-smoothing}\\
\end{longtable}\endgroup
\clearpage

\chapter{Introduction}
\label{sec:intro}

For a prime $p$ sampled uniformly among the primes at most $x$, consider
Euler's totient ratio
\[
 \frac{\varphi(p-1)}{p-1}
 =\prod_{\substack{\ell\mid p-1\\\ell\textup{ prime}}}
       \left(1-\frac1\ell\right).
\]
We study its limiting distribution as $x\to\infty$.

For odd $p$, the factor at $2$ is always $1/2$. A fixed odd prime
$\ell$ contributes its factor when $p\equiv1\pmod\ell$, one of the
$\ell-1$ nonzero residue classes. Its limiting activation probability
is therefore $1/(\ell-1)$, and the joint limit for any fixed finite
list of such primes has independent coordinates
(\Cref{prop:ap-bivariate}, with $m=1$). 
For uniform integer sampling the activation probability is instead
$1/\ell$. The two-prime calculation in \Cref{rem:prime35-example}
shows the difference explicitly, even after conditioning the integer
to be even.

The resulting totient law has support $[0,1/2]$ and gives full mass to
a set of Hausdorff dimension zero. Quantized entropy measures the
information needed to locate a sample at fine resolution; rational
approximation measures its proximity to rational numbers; concentration
measures the mass of short intervals. The totient and divisor-sum laws
studied here combine dimension-zero carriers with quantitative Fourier
decay. Their logarithmic images are sums of independent prime terms.
Replacing the totient jumps by $(\log\ell)^{-A}$ gives absolutely
continuous integer and shifted-prime laws for every $A>0$.

The monograph is intended for graduate students and researchers in analytic
number theory, probability and harmonic analysis. The common models
come first; the matrix and lattice material is optional. Prerequisites
and reading routes appear at the end of this introduction.

A matrix experiment gives a second interpretation of the same totient
law and motivates several of the applications.

For a prime $p$ and an integer $n\ge1$, let $c_n(p)$ be the probability
that a uniformly chosen $n\times n$ matrix over $\Fp$ has primitive-root
determinant. Its limit is
\begin{equation}\label{eq:def-P}
 c(p):=\lim_{n\to\infty}c_n(p)=\frac{\varphi(p-1)}{p-1}P(p),\qquad
 P(p):=\prod_{j\ge1}(1-p^{-j}).
\end{equation}
Here $\varphi$ and $\sigma$ denote Euler's totient and the sum-of-divisors
function. For integers $n\ge1$ and real $y\ge2$, write
\begin{equation}\label{eq:arithmetic-notation}
 \omega(n):=\#\{\ell:\ell\mid n\},\qquad
 \vartheta(y):=\sum_{\ell\le y}\log\ell,
\end{equation}
where $\ell$ ranges over primes. We write $(a,b)=\gcd(a,b)$ and
$[a,b]=\operatorname{lcm}(a,b)$ for integers $a,b$.
Since $P(p)\to1$, $c(p)$ has the classical shifted-prime totient law.

\section*{Principal results}

We first state the principal results for the shifted-prime totient law
and for additive functions with logarithmic prime values. The Context
section compares them with earlier work.

Throughout, $p$ and $\ell$ denote primes, and sums and products indexed by
$\ell\ge3$ run over odd primes.
Let $(B_\ell)_{\ell\ge3}$ be independent
Bernoulli variables with $\mathbb P(B_\ell=1)=1/(\ell-1)$, and put
\[
 X:=\tfrac12\prod_{\ell\ge3}(1-1/\ell)^{B_\ell},\qquad
 \omega_\ell:=\log\frac\ell{\ell-1}.
\]
Write $G$ for the distribution function of $X$, $\mu_G$ for its law,
$\muF$ for its image under $u\mapsto-\log u$, $\FT$ for the Fourier
transform of $\muF$, and $\mathcal L$ for the set of Liouville numbers.
For integers $Q\ge1$ and $b$ with $(b,Q)=1$, let $G_{Q,b}$ be the
distribution function obtained from $X$ by fixing
$B_\ell=\mathbf1_{b\equiv1\pmod\ell}$ for odd primes $\ell\mid Q$
and leaving the remaining coordinates unchanged. Define
\begin{equation}\label{eq:intro-endpoint-constant}
 \Smer{}:=2\prod_{\ell\ge3}\left(1-\frac1{(\ell-1)^2}\right),
 \qquad \kappa_{\mathrm{end}}:=\Smer{}e^{-\gamma},
\end{equation}
where $\gamma$ is Euler's constant and $\Smer{}$ is the twin-prime
singular series~\cite{HL23}. For $0<\varepsilon<1$ and $h>0$, set
\begin{equation}\label{eq:intro-concentration}
\begin{aligned}
 Q_G(\varepsilon)&:=\sup_t\{G(t)-G(t(1-\varepsilon))\},&
 H_{\mathrm{end}}(\varepsilon)&:=1-G(\tfrac12(1-\varepsilon)),\\
 Q_{\muF}(h)&:=\sup_a\muF([a,a+h]),&
 C(h)&:=\sup_t\{G(t)-G(t-h)\}.
\end{aligned}
\end{equation}
Unless another model is specified, $(B_\ell)_{\ell\ge3}$ denotes this independent family.
Unqualified results are unconditional.

\Needspace{8\baselineskip}
\begin{theorem}[Limiting law, geometry and information]
\label{thm:intro-main}
The values $c(p)$ over the primes have the continuous limiting
distribution function $G$, with $\supp\mu_G=[0,\tfrac12]$.
Moreover, $\dim_H\mu_G=\dim_H\muF=0$ and $\mu_G(\mathcal L)=1$.
For either measure, the lower local dimension is zero and the upper
local dimension is one almost everywhere. Every Borel set of positive
measure has packing dimension one.

Their lower and upper Shannon information dimensions satisfy
\[
0.55043653<\underline d_1\le\overline d_1<0.570489495.
\]
\end{theorem}

\begin{proof}
\Cref{thm:phi-dist} gives the limiting law and support;
\Cref{thm:hausdorff-zero,thm:liouville-carriers} give the carrier
conclusions; \Cref{thm:local-dimension-oscillation} gives the local
and packing dimensions; and \Cref{thm:shannon-dimension} gives the
information bounds.
\end{proof}

\begin{theorem}[Sharp concentration and maximizing intervals]
\label{thm:intro-concentration}
For the shifted-prime totient law $G$ and its negative-logarithmic
image $\muF$, use the constant $\kappa_{\mathrm{end}}$ of
\eqref{eq:intro-endpoint-constant} and the concentration functions of
\eqref{eq:intro-concentration}.
Put $z=\log(1/\varepsilon)$ and $L=\log(1/h)$. As
$\varepsilon,h\downarrow0$,
\[
\begin{aligned}
 Q_G(\varepsilon)&=\frac{\kappa_{\mathrm{end}}}{z}
       -\frac{\kappa_{\mathrm{end}}\pi^2}{12z^3}+O(z^{-4}),\\
 Q_{\muF}(h)&=\frac{\kappa_{\mathrm{end}}}{L}
       -\frac{\kappa_{\mathrm{end}}\pi^2}{12L^3}+O(L^{-4}),\\
 C(h)&=\frac{\kappa_{\mathrm{end}}}{L}
       +\frac{\kappa_{\mathrm{end}}\log3}{L^2}
       +\frac{\kappa_{\mathrm{end}}((\log3)^2-\pi^2/12)}{L^3}
       +O(L^{-4}).
\end{aligned}
\]
The suprema are attained for every $0<\varepsilon<1$ and $h>0$.
For sufficiently small $\varepsilon,h$, every maximizing interval
for $Q_G(\varepsilon)$ or $C(h)$ contains $1/3$, and every maximizing
interval for $Q_{\muF}(h)$ contains $\log3$.
\end{theorem}

\begin{proof}
\Cref{thm:sharp-global-modulus} gives the relative raw and additive
logarithmic assertions; \Cref{cor:sharp-additive-modulus} gives the
additive raw assertions.
\end{proof}

\begin{theorem}[Effective Fourier decay and normality]
\label{thm:intro-fourier}
Let $\mu_G$ be the shifted-prime totient law and $\muF$ its image
under $u\mapsto-\log u$, with Fourier transform $\FT$.
For all sufficiently large $|\tau|$,
\[
 \max\{|\FT(\tau)|,|\widehat{\mu_G}(\tau)|\}\ll
 (\log|\tau|)^{-1/3}(\log\log|\tau|)^{1/3},
\]
with effective constants and threshold. Both laws give full mass to
the absolutely normal numbers.
\end{theorem}

\begin{proof}
\Cref{thm:vk-fourier,thm:raw-ratio-normality} give the two decay
bounds, and \Cref{lem:iterlog-normality} gives normality from them.
\end{proof}

\begin{proposition}[Sharp prime comparison in a fixed progression]
\label{prop:intro-prime-comparison}
Let $Q\ge1$ be fixed, let $(b,Q)=1$, and let $G_{Q,b}$ be the
distribution function defined above by fixing the prime coordinates
dividing $Q$.
For $U(p)=\varphi(p-1)/(p-1)$ or $U(p)=c(p)$,
\[
 \sup_t\left|\frac{\#\{p\le x:p\equiv b\pmod Q,\ U(p)\le t\}}
 {\pi(x;Q,b)}-G_{Q,b}(t)\right|
 \asymp_Q\frac1{\log x}\qquad(x\to\infty).
\]
The upper constants and thresholds may be ineffective.
\end{proposition}

\begin{proof}
\Cref{thm:ap-law} identifies $G_{Q,b}$, and
\Cref{thm:sharp-prime-convergence} gives the two-sided order.
\end{proof}

The second-order term for $C(h)$ arises from the shift
$\log(1/(3h))=\log(1/h)-\log3$; the expansions for $Q_G$ and
$Q_{\muF}$ have no such term. Quantitative endpoint comparisons and
localization rates appear in
\Cref{thm:cofinite-window-lower,thm:sharp-global-modulus,cor:sharp-additive-modulus}.

To describe the transform at a typical frequency, let
$(\Theta_\ell)_{\ell\ge3}$ be independent uniform phases on $[0,2\pi]$,
and put
\[
 W_F:=\kappa_{\mathrm{end}}\exp\!\left\{
 \sum_{\ell\ge3}\log\left|1+\frac{e^{i\Theta_\ell}}{\ell-2}\right|
 \right\},\qquad M_F(s):=\mathbb EW_F^s\quad(s>-1).
\]
The log series converges almost surely. Let $U_T$ be uniform on $[0,T]$,
and let $\Theta_0$ be uniform on $[0,2\pi]$ and independent of $W_F$.

\begin{theorem}[Fourier amplitude law and real moments]
\label{thm:intro-amplitude}
As $T\to\infty$,
\[
 (\log T)\FT(U_T)\Rightarrow e^{i\Theta_0}W_F,\qquad
 (\log T)|\FT(U_T)|\Rightarrow W_F.
\]
For every fixed real $s>-1$, as $T\to\infty$,
\[
 \frac1T\int_0^T|\FT(\tau)|^s\,d\tau
 \sim\frac{M_F(s)}{(\log T)^s}.
\]
For $s\le-1$ the integral is infinite when $T\ge\pi/\omega_3$.
\end{theorem}

\begin{proof}
\Cref{thm:fourier-amplitude-law} gives the modulus limit, the real
moment asymptotics and the divergence threshold.
\Cref{cor:complex-amplitude-law,cor:totient-amplitude-circular-moments}
give the complex limit with its independent uniform angle.
\end{proof}

\begin{corollary}[Density of the totient amplitude]
\label{cor:intro-amplitude-density-tail}
The law of $W_F$ has support $[0,\infty)$ and a strictly positive smooth
density on $[0,\infty)$.
\end{corollary}

\begin{proof}
See \Cref{cor:amplitude-density,cor:amplitude-schwartz-density}.
\end{proof}

Finer density properties and tail asymptotics appear in
\Cref{cor:amplitude-schwartz-density,cor:amplitude-upper-tail}.

The modulus limit in \Cref{thm:intro-amplitude} gives the typical
amplitude scale $1/\log T$ and implies
$\log|\FT(U_T)|/\log\log T\to-1$ in probability.
By \Cref{prop:fourier-resonance}, suitable frequencies have amplitudes
at least of order $(\log\log\tau)/\log\tau$:
\[
 \limsup_{\tau\to\infty}\frac{\log\tau}{\log\log\tau}
 |\FT(\tau)|\ge1.
\]
\begin{theorem}[Absolute continuity for logarithmic prime values]
\label{thm:intro-all-A}
Let $A>0$, and let $f_A$ be the strongly additive function with prime
values $f_A(\ell)=(\log\ell)^{-A}$. The values $f_A(n)$ over the positive
integers and $f_A(p-1)$ over the primes have absolutely continuous
limiting laws.
\end{theorem}

\begin{proof}
Since $\sum_\ell f_A(\ell)/\ell<\infty$, the Erd\H{o}s--Wintner
theorem~\cite{EW39} and \Cref{thm:pure-type} give the limiting laws, with independent odd-prime activation
probabilities $1/\ell$ and $1/(\ell-1)$, respectively.
\Cref{lem:slow-jump-all-ac} applies with $c_\ast=1$, $\beta=0$; convolution with
the respective factor at $2$ preserves absolute continuity.
\end{proof}

For $A>2$, the integer assertion disproves the singularity conjecture of
Erd\H{o}s and K\'atai~\cite[Remark~2 after Theorem~3, p.~297]{EK79}.
Regularity, independent sums and joint models are treated in
\Cref{ssec:slow-jump-ac}; varying the activation probabilities gives
further absolute-continuity criteria (\Cref{ssec:annulus}).

\paragraph*{Geometry and information}
The Hausdorff, local and packing dimension conclusions extend to the
integer and shifted-prime totient and divisor-sum laws and their
logarithmic images. For each arithmetic function, the lower and upper
Shannon information dimensions agree between the two sampling models;
whether the lower and upper dimensions coincide remains open
(\Cref{thm:shannon-dimension}). Their dyadic $L^q$ dimensions are one
for $0<q<1$ and zero for $q>1$, and their positive-order quantization
dimensions are one
(\Cref{cor:renyi-spectrum,thm:quantization}).

Fix independent basic logarithmic samples $Z_i$ and nonzero real weights
$a_i$. Every finite sum $S_n=\sum_{i=1}^n a_iZ_i$ has Hausdorff dimension
zero, while its lower and upper information dimensions tend to one as
$n\to\infty$. The infinite series has information dimension one when
$\sum_i|a_i|<\infty$, or when $\sum_i a_i^2<\infty$ after centering each
sample. A coefficient-count condition gives absolute continuity even for
$a_i=\exp\{-\exp(\exp(ci))\}$ with any fixed $c>0$. A stronger
condition gives smooth densities (\Cref{sec:information-convolution}).

\paragraph*{Rational approximation and joint laws}
The prime-coordinate models yield simultaneous rational approximation
and Liouville carriers (\Cref{sec:liouville-carriers}). Independent
sequences from the raw laws almost surely generate fields of infinite
transcendence degree whose irrational elements are absolutely normal
and Liouville (\Cref{cor:normal-liouville-field}). Existence of
absolutely normal Liouville numbers is classical~\cite{Bu02}.
Joint totient laws at fixed nonzero polynomial prime arguments are
constructed in \Cref{thm:polynomial-root-law}, omitting primes where a
polynomial vanishes and retaining exceptional local factors. The logarithmic
vectors converge in every finite Wasserstein order, and the ratios converge
in every mixed real Mellin moment. Their affine specializations are developed
in \Cref{sec:difference-law}.

\paragraph*{Further Fourier laws}
The Fourier bound extends to all four ratio laws and their fixed
nonzero real powers. For the shifted-prime laws it also holds in every
fixed reduced progression (\Cref{thm:raw-power-progressions-mas16}).
Images of the four laws under fixed nonconstant real-analytic maps on
$(0,\infty)$ are Rajchman (their Fourier transforms tend to zero at
infinity) and give full mass to the absolutely normal numbers
(\Cref{cor:raw-ratio-analytic-images}). The integer logarithmic totient
and the logarithmic divisor-sum laws also have normalized Fourier
amplitude limits, with different symmetry and dependence structures
(\Cref{cor:integer-totient-amplitude,thm:sigma-amplitude}).
For each basic logarithmic law $\nu$, the normalized Fourier modulus
$V_T=(\log T)|\widehat\nu(U_T)|$ satisfies
$\sup_{T\ge3}\|V_T\|_r\asymp_\nu r$ for $r\ge1$ and
$\sup_{T\ge3}\mathbb P(V_T>K)\ll_\nu e^{-c_\nu K}$ for $K\ge1$,
with $c_\nu>0$ (\Cref{thm:valuation-fourier-moments}).

\paragraph*{Concentration and arithmetic applications}
Mellin transforms and endpoint expansions yield sharp estimates for
truncation errors and shortest intervals of prescribed probability
(\Cref{sec:moments,sec:global-modulus}). Quantitative prime sampling is
treated in Kolmogorov and Wasserstein distance, with distinct fixed-
and growing-modulus regimes and further conditional ranges under GRH
(\Cref{ssec:sharp-prime-convergence}).

The arithmetic applications concern explicit determinant bounds,
certified prime selection without complete factorization, and sampling
prime bases without replacement (\Cref{sec:explicit,ssec:certgen}).
The finite-field models distinguish a fixed extension degree with
varying prime base from a fixed field with sampled extension degrees
(\Cref{sec:composite-modulus}). Determinant-constrained learning with
errors (LWE) and an independent appendix on lattice shells and smoothing
provide further applications
(\Cref{sec:syd21-app,app:cyclotomic-smoothing}).

\section*{Context}

\paragraph*{Classical limiting laws}
Schur obtained the complex moments of the integer totient ratio in
1923/24; see~\cite[no.~18, p.~194]{Sch28} and
\cite[p.~713, footnote~1]{EW39}. Schoenberg~\cite{Sch28} proved its
continuous limiting distribution; Erd\H{o}s~\cite{Erd39} proved its
singularity, within the Jessen--Wintner dichotomy~\cite{JW35}.
The Erd\H{o}s--Wintner theorem~\cite{EW39} characterizes real additive
functions with limiting laws (see also~\cite[Theorem~III.4.1]{Ten15});
Davenport~\cite{Dav33} treated the integer divisor-sum ratio.
For shifted primes, Barban, Vinogradov and
Levin~\cite{BVL65} obtained normalized laws for strongly additive
functions of Kubilius class $H$; K\'atai treated $p+1$~\cite{Kat68} and
$p-1$~\cite{Kat69}. Hildebrand's general theorem~\cite{Hil89} uses
Bombieri--Vinogradov~\cite{Bom65,Vin65}; see also
\cite[Theorem~II.8.34]{Ten15}.
K\'atai's pair laws~\cite[Theorems~2 and~3]{Kat69} and their symmetric
series form~\cite[\S~3]{DK19} provide the shifted-pair comparison.
K\'atai's mean-value theorem also treats coprime linear forms at prime
arguments~\cite[Theorem~1]{Kat69}; De Koninck and
K\'atai~\cite[Theorem~B]{DK19} state an affine-form version.
The local-factor correction and the further geometric and concentration
conclusions are recorded in \Cref{sec:difference-law}.
For polynomial prime arguments, the scalar existence theorem is stated
and attributed to Toleuov--Fa\u{\i}nle\u{\i}b by
Aivazidis--Sofos~\cite[Theorem~6.3]{AS15}; their proof of
Theorem~1.3(i) applies it to totient ratios at $p^m-1$.
Deshouillers proved strict increase for a multiplicative class containing
the totient and reciprocal divisor-sum ratios~\cite[Th\'eor\`eme~2, \S~4.1]{Des70}.
He conjectured the support criterion for positive multiplicative
functions~\cite[\S~4.6]{Des70} and recorded the $0<g\le1$ case as
proved~\cite{Des76}. For arbitrary positive multiplicative $g$ with a
continuous limiting probability law along $p-1$,
\Cref{cor:deshouillers-criterion} extends this criterion: the limiting
support is $\overline{\{g(p-1):p\ge3\textup{ prime}\}}$.
He proved the $p+1$ analogue for strongly multiplicative $0<g\le1$
with continuous limiting law~\cite[Th\'eor\`eme~(0)]{Des75},
then dropped strong multiplicativity~\cite[p.~283]{Des75}.

\paragraph*{Carriers and quantitative approximation}
Deshouillers and Hassani~\cite{DH12} recorded continuity on
$p\equiv1\pmod m$ and left nondifferentiability at arithmetic anchors.
Their moment estimate~\cite[(3.1)]{DH12} implies
$G(x_S)-G(x_S-\varepsilon)\gg_S1/\log(1/\varepsilon)$;
\Cref{thm:anchor-sharp} determines the leading constant.
Tenenbaum~\cite[Theorem~1 and the following remark]{Ten12} gives local
power and global concentration lower bounds. Singularity is asserted
in~\cite{DH12,Ten12}, citing Exercise~256 of Tenenbaum's 2008 French
edition (p.~423); see also~\cite[Chapter~III.2, Exercise~256, pp.~440--441]{Ten15}.
Integer-case proofs via Jessen--Wintner appear in
\cite[Chapter~III.2, Exercises~2.4--2.5, pp.~147--149]{TW96}.
Our dimension-zero carriers refine Erd\H{o}s's integer
coverings~\cite[\S~3]{Erd39}; see also~\cite{TW96}.
The joint approximation in \Cref{lem:joint-rational-carrier}(b) is a
form of rational-function closure on a common set of Liouville
numbers~\cite[Theorem~1]{KTW15} and applies to the limit variables.
For dependent coordinates of independent prime factors,
\Cref{prop:prime-product-polynomial-nonconcentration} supplies the
additional polynomial nonconcentration estimate.
Effective integer Erd\H{o}s--Wintner estimates appear in~\cite{TV21};
for the totient and divisor-sum ratios, see
\cite[Remark~(ii) after Theorem~1.2]{TV21} and \cite[Corollaire~1.5]{dlBT12}.
Ford~\cite{For25} approximates truncated shifted-prime factor data.
Weighted Bombieri--Vinogradov and truncated inclusion--exclusion give
the progression Wasserstein rate in \Cref{prop:prime-wasserstein}.
A fixed-progression sieve below the Bombieri--Vinogradov level and
localized bounds for large smooth parts give the sharp Kolmogorov order $1/\log x$ in
every fixed reduced class (\Cref{thm:sharp-prime-convergence});
\Cref{cor:prime-shrinking-anchor} gives counts over unrestricted primes
in shrinking windows at fixed anchors and at moving anchors in the
range of \eqref{eq:prime-moving-anchor}.

\paragraph*{Concentration and arithmetic anchors}
\begingroup\widowpenalty=10000
For the integer totient law, Tjan~\cite[p.~118]{Tja66} obtained relative
concentration $O(1/\log\log(1/h))$.
Erd\H{o}s~\cite{Erd74} proved the sharp divisor-sum order
$1/\log(1/h)$ and stated the totient analogue; Diamond and
Rhoads~\cite{DR84} are credited with its bounded-mean-oscillation upper
bound~\cite[p.~368]{Tou06}.
Toulmonde~\cite[Th\'eor\`eme~2]{Tou06} localized integer totient
maximizers near $1/2$ and proved that the integer analogue of
$Q_G(\varepsilon)$ equals
the mass of $(1-\varepsilon,1]$ up to
$O(\exp(-\tfrac15\sqrt{z\log z}))$, $z=\log(1/\varepsilon)$;
see~\cite[Corollaire~2]{Tou09} for its expansion.
The integer logarithmic totient concentration has order
$1/\log(1/h)$ as $h\downarrow0$ by~\cite[Theorem~1]{EK79};
comparison principles appear in~\cite[\S~2.4]{DK19}.
For $\log(\sigma(p-1)/(p-1))$, this order is stated without proof as a
consequence of~\cite{IK92} in~\cite[remark after Theorem~3, p.~25]{KS05}.
Our Fourier-moment argument proves this order directly for the
shifted-prime totient and the integer and shifted-prime logarithmic
divisor-sum laws
(\Cref{thm:uniform-fourier-moments,thm:sigma-moments}).
For the integer divisor-sum law, Toulmonde's results already give
the leading concentration constant $e^{-\gamma}$ for both raw and
logarithmic windows \cite[\S\S~10.1--10.2, Th\'eor\`eme~4]{Tou06}.
For the odd and full shifted-prime divisor-sum laws, the respective
leading constants are $\kappa_{\mathrm{end}}$ and
$\kappa_{\mathrm{end}}/2$ for both raw and logarithmic windows
(\Cref{thm:sigma-concentration-constants-mas05}). For each of the three laws
the endpoint interval is the unique maximizer for every sufficiently small
width, in either coordinate. The shifted endpoint expansions therefore
give global concentration to all orders. The fixed-progression criterion
is given in \Cref{cor:ap-sigma-saturation,prop:ap-sigma-endpoint-class}.
For the shifted-prime totient law, the Erd\H{o}s--Toulmonde method
gives the sharp constant and places the anchor in every sufficiently
short maximizing interval (\Cref{thm:sharp-global-modulus}).
Cofinite residual laws and convolution modes are treated in
\Cref{thm:cofinite-window-lower,thm:convolution-modes};
compare~\cite[Th\'eor\`emes~1--2]{Tou06b} for the integer law.
\par\endgroup

\paragraph*{Logarithmic prime increments}
For integer laws with prime increments $(\log\ell)^{-A}$, Babu proved an
$L^2$ density for $0<A<2$~\cite{Bab75}. His broader additive and paired
results~\cite{Bab95,Bab92}
are discussed in \Cref{par:slow-jump-integer}.
\Cref{thm:intro-all-A} extends absolute continuity to every $A>0$,
disproving the singularity conjecture of Erd\H{o}s and K\'atai for
$A>2$~\cite[Remark~2 after Theorem~3, p.~297]{EK79}.
For $A>1$, Koukoulopoulos~\cite[Theorem~1.3]{Kou14} proved the integer
concentration order $h^{1/A}$, following the large-$A$
result~\cite[Corollaire~1.4]{dlBT12}. Bernoulli-product comparison gives
the same order for the shifted-prime law (\Cref{par:slow-jump-integer}).
\Cref{lem:slow-jump-all-ac} adapts the conditioning method
of~\cite[Theorem~4.1]{PPPT15} for increments $n^{-s}$ over all integers.
For common prime activations with probabilities summably close to $c/p$,
$c>0$, and finitely many distinct positive exponents,
\Cref{thm:shared-prime-anisotropic,prop:shared-prime-global-regularity}
give the absolute-continuity and regularity results;
\Cref{prop:shared-prime-sobolev} gives the exact isotropic Sobolev threshold.

\paragraph*{Fourier decay and prime cancellation}
The coefficient-counting argument for infinite convolutions goes back to
Wintner~\cite{Win35a,Win35b}.
For stationary Bernoulli convolutions, Erd\H{o}s and Salem studied the
Pisot obstruction~\cite{Erd39b,Sal63}; Erd\H{o}s and Kahane established
almost-everywhere Fourier decay~\cite{Erd40,Kah71};
see~\cite{Sol95,PS96,PSS00,Hoc14,Shm19,Var19} for absolute continuity and
dimension, and~\cite{LS22,Bre21,ARHW22,Rap22,HK38,GMSZ20} for varying
contractions and inhomogeneous two-point convolutions.
Algom, Rodriguez Hertz and Wang~\cite[\S~1.1 and Theorem~1.1]{ARHW23}
prove polynomial Fourier decay for non-atomic self-conformal measures
of uniformly contracting $C^2$ iterated function systems (IFS) that are
not $C^2$-conjugate to a linear IFS, with linearity understood in their sense.
Garsia~\cite{Gar62} treats square-summable scales.
Our prime-indexed jumps are nonlacunary and $\Q$-linearly independent,
with $\omega_\ell\asymp1/\ell$ (\Cref{thm:linind}).

Dimension-zero Rajchman measures with Liouville carriers occur
in~\cite{Blu00}; see also~\cite{Lyo95}. Bugeaud~\cite[p.~120]{Bu02}
constructs one satisfying
$|\widehat\mu(t)|\ll\exp(-c\sqrt{\log(1+|t|)})$ for some $c>0$.
More generally, Polasek and Rela~\cite[Theorem~1.4]{PR26} prove that
for every bounded $f:[0,\infty)\to[0,\infty)$ which is decreasing or
continuous and satisfies $t^{-a}=o(f(t))$ as $t\to\infty$ for every
$a>0$, there is a Liouville-carried probability measure with
$|\widehat\mu(t)|=O(f(|t|))$ as $|t|\to\infty$.
For prescribed Hausdorff and
Fourier dimensions $1\ge\alpha>\beta>0$, with a measure realizing both,
see~\cite[Theorem~1.8]{LL24}; the proof of~\cite[Theorem~1.4]{Kor11}
applies only to $\alpha=\beta$~\cite[\S1.3]{LL24}.

Our prime-cancellation estimates use different prime-size ranges.
A Vaughan decomposition and exponent pairs give cancellation for
$\log(p/(p-1))$, $\log(1+1/p)$ and $1/p$ for sufficiently large $|\tau|$, on dyadic prime intervals
$N<p\le2N$ with $|\tau|^{1/(r+1)+\delta}\le N\le|\tau|^{1-\delta}$,
for the fixed parameters $r,\delta$ of \Cref{thm:annulus},
with applications to activation probabilities $K(\log\ell)^\alpha/\ell$
(\Cref{cor:heavier-activation}).
The quantitative Rajchman bound uses \Cref{prop:vk-prime-prefix},
whose estimate applies for $64\le N\le b|\tau|^{1/5}$ with an effective
absolute constant $b>0$. Its fixed-character extension
(\Cref{prop:vk-prime-prefix-character}) retains the Vinogradov prime-size
range and effective cancellation constants. Combining it with fixed-progression prime mass
gives the periodic-weight refinement of
\Cref{thm:periodic-weight-dissipation}, which is ineffective as proved.

\paragraph*{Extremal orders and determinant densities}
\looseness=-1
The extremal estimates use Mertens' product theorem~\cite{Mer74} and
Linnik's theorem~\cite{Lin44}, including its quantitative
form~\cite[Corollary~18.8]{IK04}; for earlier work,
see~\cite[p.~113]{Gro13} and~\cite{Ram15}.
Balakrishnan's least-prime construction~\cite[Theorem~2]{Bal94} implies
the lower bound for the reciprocal limsup; see also
\cite[proof of Theorem~1(i)]{LP02}, the priority
acknowledgment~\cite{LP12}, and~\cite[\S~3]{ST08}.
Sehrawat, Yeo and Desmedt~\cite[Remark~3(ii)]{SYD21} deduced
$\inf_pc(p)=0$ assuming infinitely many primorial primes $N_k+1$;
Dirichlet's theorem~\cite{Dir37} and $\sum_\ell1/\ell=\infty$ give this
unconditionally (see \Cref{cor:ap-sharp-extrema} for the sharp order).
Singer cycles in $\GL_n(q)$~\cite{AS15} have primitive-root determinant,
but the converse fails for $n\ge2$: the densities are
$\varphi(q^n-1)/(n(q^n-1))$ and $\varphi(q-1)/(q-1)$, respectively.
For $n=1$, proper prime powers have relative density zero among prime
powers, so the latter density has limiting law $G$;
\Cref{prop:endpoint-half} answers the endpoint question
in~\cite[\S~1, p.~4]{AS15}.

\paragraph*{Lattice shells and smoothing}
Micciancio and Regev introduced the smoothing parameter~\cite{MR07};
see~\cite{CDLP13} for its approximation problem. Espitau, Wallet and
Yu use the first dual shell and theta-series reversion~\cite[\S~6.1, Lemma~14]{EWY23};
the revision~\cite[\S~6.1, Proposition~18]{EWY26} adds the second dual
shell for lattices similar to rational lattices.
For every fixed full-rank lattice,
\Cref{prop:fixed-lattice-second-shell} gives a two-term expansion of
the squared smoothing parameter multiplied by the shortest nonzero
dual squared norm, as the smoothing error tends to zero.
The correction coefficient is invariant under dilation; the ratios of
the first three distinct nonzero dual squared norms determine the
remainder exponent.
For cyclotomic tensor decomposition and trace duality,
see~\cite[\S~2.3, pp.~44--45]{LPR13}; for the type-$A$
description, see~\cite[\S~1]{DvW18}.
Kitaoka's $E$-type theorem gives the minimum and minimal
vectors~\cite[Proposition~1 and Theorem~2]{Kit77}, recorded in
\Cref{lem:A-tensor-minimum}.
For two factors of rank at least two, the second distinct nonzero length
is implicit in~\cite[Theorem~2]{DvW18};
\Cref{thm:A-tensor-second-shell} treats $k$-fold products for every $k\ge1$,
including all nonzero squared norms below twice the squared minimum and their
multiplicities.

\section*{Prerequisites and reading guide}

We assume familiarity with elementary analytic number theory,
measure-theoretic probability, Fourier transforms and basic complex
analysis. Dimension and concentration conventions are specified where
they first occur. The prime-counting inputs are collected in
\Cref{sec:prime-counting-inputs}; their divisibility-pattern forms,
including the conditional GRH estimates, are given in
\Cref{app:bv-grh}.

\paragraph*{Online companion}
An interactive visual companion is available at
\begin{center}
\url{https://crypto-vipin.github.io/arithmetic-limiting-laws/}
\end{center}
The companion illustrates prime-coordinate models, dimension, entropy,
concentration and Fourier regularity, and provides a guide to selected
results and open questions. It is supplementary to the mathematical development.

\paragraph*{Models, carriers, rational approximation and information}
Begin with \Cref{sec:distribution}, which supplies the four basic laws,
their supports and their progression variants.
\Cref{sec:dimension-liouville} develops carrier geometry and local
dimensions; \Cref{sec:liouville-carriers} then treats simultaneous
rational approximation, Liouville carriers and rational statistics.
Its continued-fraction results use the packing dimension established
in the geometry chapter. \Cref{sec:information-dimensions} treats
information dimensions, quantization and independent sums. The
infinite-sum absolute-continuity and smoothing results also use the
Fourier bounds of \Cref{sec:raj-annulus}.
\Cref{sec:difference-law} studies joint laws, beginning with polynomial
root models and then treating affine specializations.

\paragraph*{Concentration and arithmetic sampling}
The Mellin transforms and endpoint asymptotics of \Cref{sec:moments}
lead to sharp concentration and localization in \Cref{sec:global-modulus}.
\Cref{ssec:sharp-prime-convergence} compares finite prime samples with
the limiting laws in Kolmogorov and Wasserstein distance;
\Cref{sec:distribution} treats random-product truncations.

\paragraph*{Fourier decay and regularity}
\Cref{ssec:slow-jump-ac} studies densities and regularity when the prime
jumps are logarithmically decreasing, including joint laws with common
prime activations. For the totient and divisor-sum laws,
\Cref{sec:raj-annulus} develops prime-prefix estimates to prove
pointwise Fourier decay and normality.
\Cref{ssec:typical-decay} then studies frequency-averaged amplitudes of
the logarithmic laws, using the Fourier-moment bounds from
\Cref{sec:global-modulus}.
\Cref{ssec:annulus} develops cancellation over a complementary range
of dyadic prime intervals and applies it to varying activation
probabilities.

\paragraph*{Optional arithmetic and matrix applications}
\Cref{sec:explicit} returns to determinant densities at a fixed prime
and develops extremal and explicit bounds. \Cref{ssec:certgen} then
treats certified prime selection and sampling distinct primes for
residue number systems. Finite-ring identities and finite-field degree
models form \Cref{sec:composite-modulus}; determinant-constrained LWE
and matrices with independent entries are treated in
\Cref{sec:syd21-app}. 

\paragraph*{Research questions and supporting calculations}
\Cref{sec:conclusion} records the remaining questions. Power-scale
endpoint comparisons and two equivalent formulations of the conjectural
Buchstab transition are developed in \Cref{ssec:prime-cofactor-detectors}. Certified CDF
enclosures and supporting Ces\`aro-moment arguments are in
\Cref{app:gencert-numerics,app:cesaro-moments}.

\paragraph*{Independent lattice appendix}
\Cref{app:cyclotomic-smoothing} assumes basic Euclidean lattice theory
and cyclotomic fields and may be read separately. It passes from
type-$A$ tensor shells to cyclotomic codifferents and smoothing
estimates. The power-of-two negacyclic case supplies the uniform
parameter connection with prime selection; the arithmetic search also
allows \hyperref[par:gencert-fixed-length]{every fixed transform length}.

\paragraph*{Diagrams}
Twelve diagrams record how constructions and estimates recur across
chapters. The four scalar laws and their valuation models are displayed in
\Cref{fig:four-laws}. The inputs to the rational-statistics and dimension
results are separated in
\Cref{fig:rational-nonconcentration,fig:dimension-mechanisms}; the reuse of
the information bounds and the routes to regularity for independent sums
are followed in \Cref{fig:information-reuse,fig:sum-regularity}.
Dependent arithmetic tuples are compared with their independent reference
models in \Cref{fig:reference-models}. The three kinds of Fourier statement
and the activation-scale criteria are set out in
\Cref{fig:fourier-questions,fig:activation-scale}.
\Cref{fig:certified-search,fig:sampling-regimes,fig:field-ensembles}
accompany the applications, and \Cref{fig:tensor-smoothing} the lattice
appendix. In these diagrams a solid arrow records a use in a proof or a
stated implication, and lines meeting at a dot are inputs needed jointly;
a dashed connector is a labelled map, conditioning, specialization or
comparison, and a dotted connector indicates a shared setting, not a
dependency. Thin grey arrows point to other figures. Boxes with a heavy frame are results, grey boxes
without a frame are arguments or constructions, and boxes with a thin
frame are models, data or hypotheses. The diagrams select proof inputs; they are not complete
accounts of dependencies.

\chapter{Totient and divisor-sum limiting laws}
\label{sec:distribution}
Choose a prime $p$ uniformly from the odd primes up to $x$, and record
which primes divide $p-1$. The totient ratio always contains the factor
$1/2$ from the prime $2$. The factors $2/3$ and $4/5$ occur when
$3\mid p-1$ and $5\mid p-1$, with limiting frequencies $1/2$ and $1/4$.
For any fixed finite set of odd primes, the Chinese remainder theorem
and prime counting in reduced residue classes make the limiting joint
frequencies factorize, as proved in \Cref{prop:ap-bivariate}. This gives
independent coordinates in the limiting model. The four possibilities at
$3$ and $5$ are displayed in \Cref{rem:prime35-example}.

Finite-coordinate limits are only part of the passage to an infinite
random product. One must also control the total contribution of the
omitted primes. This chapter establishes the limiting measures and
their supports, providing the prime-coordinate description used in the
later geometric and analytic arguments. The law of
$\varphi(p-1)/(p-1)$ is classical~\cite{Kat69,Des70};
\Cref{thm:ap-law} extends the form for $p\equiv1\pmod m$
\cite[Proposition~2.1]{DH12} to every fixed reduced class.
Throughout, we write
\begin{equation}\label{eq:f-def}
f(n)\;:=\;\log\frac{n}{\varphi(n)}
\;=\;\sum_{\substack{\ell\mid n\\ \ell\textup{ prime}}}\log\frac{\ell}{\ell-1}\,.
\end{equation}

\section{Setup: \texorpdfstring{$\varphi$}{phi}- and \texorpdfstring{$\sigma$}{sigma}-cases}\label{sec:dist-setup}

The additive function associated with $\sigma/\mathrm{id}$ is
\[
\gsig(n)\;:=\;\log\frac{\sigma(n)}{n}
\;=\;\sum_{\ell^{k}\,\|\,n}\gsig(\ell^{k}),
\qquad
\gsig(\ell^{k})\;=\;\log\!\Bigl(\sum_{j=0}^{k}\ell^{-j}\Bigr).
\]

The logarithmic laws $\mu_f$ and $\mug$ are the limiting laws of
$f(p-1)$ and $\gsig(p-1)$, respectively. Their images under
$u\mapsto e^{-u}$ and $u\mapsto e^u$ are $\mu_G$ and $\muH$, the laws
of $\varphi(p-1)/(p-1)$ and $\sigma(p-1)/(p-1)$.

\subsection{The four scalar laws}\label{ssec:four-scalar-laws}
\index{Totient laws!valuation models}
\index{Divisor-sum laws!valuation models}
\index{Valuation models}

The four scalar laws describe $\varphi(n)/n$ and $\sigma(n)/n$ under
integer sampling by natural density and shifted-prime sampling
$n=p-1$ by relative frequency among primes. The totient ratio records
which primes divide $n$; the divisor-sum ratio also records their
exponents. The coordinate distributions below define the models.
The cited classical results and the proofs in this chapter establish
their arithmetic identifications, atomlessness and supports.
For every prime $\ell$, write
$\omega_\ell:=\log(\ell/(\ell-1))=f(\ell)$.

The integer and shifted-prime local models use independent prime
valuations $J_\ell$ and $V_\ell$, respectively, with
\begin{equation}\label{eq:valuation-laws}
\begin{aligned}
 \mathbb P(J_\ell=k)&=(1-1/\ell)\ell^{-k} &&(k\ge0),\\
 \mathbb P(V_\ell=k)&=
 \begin{cases}
 (\ell-2)/(\ell-1),&k=0,\\
 \ell^{-k},&k\ge1.
 \end{cases}
\end{aligned}
\end{equation}
The arithmetic origin of these probabilities is visible in the tails.
For integer sampling, $\ell^k\mid n$ has density $\ell^{-k}$.
For shifted-prime sampling and fixed $k\ge1$, the condition
$\ell^k\mid p-1$ selects the single reduced class $1\pmod{\ell^k}$,
whose limiting relative frequency is $1/\varphi(\ell^k)$.
Subtracting successive tails gives the masses for $k\ge1$; the mass
at $0$ is the complement of the tail at $1$. The Chinese remainder
theorem gives the corresponding factorization
for any fixed finite collection of prime-power conditions at distinct primes.

These formulas include $\ell=2$: $V_2\ge1$ almost surely and
$\mathbb P(V_2=k)=2^{-k}$ for $k\ge1$.
Let $\mathsf I$ and $\mathsf B$ denote the corresponding product measures
on $\N_0^{\{\ell\textup{ prime}\}}$, and put
$\mathsf A:=\mathsf I(\,\cdot\mid J_2\ge1)$.

Let $\mu_F^{\mathrm{int}}$ and $\mu_\sigma^{\mathrm{int}}$ be the
natural-density limiting laws of $\varphi(n)/n$ and $\sigma(n)/n$,
respectively. These laws give full mass to $(0,1]$ and $[1,\infty)$,
respectively. Their logarithmic factorizations follow from the
Erd\H{o}s--Wintner theorem~\cite{EW39}:
\begin{equation}\label{eq:integer-totient-factorization}
 \begin{gathered}
 \mu_F^{\mathrm{int},\log}:=(-\log)_*\mu_F^{\mathrm{int}}
   =\Conv_{\ell\ge2,\,\ell\textup{ prime}}\nu_\ell^{\mathrm{int}},\\
 \nu_\ell^{\mathrm{int}}
   =\left(1-\frac1\ell\right)\delta_0
       +\frac1\ell\delta_{\omega_\ell}.
 \end{gathered}
\end{equation}
Write $\mu_\sigma^{\mathrm{int},\log}:=\log_*\mu_\sigma^{\mathrm{int}}$;
thus the integer totient law uses $-\log$, and the integer divisor-sum
law uses $+\log$. Davenport's divisor-sum law~\cite{Dav33} satisfies
\begin{equation}\label{eq:integer-sigma-factorization}
 \log_*\mu_\sigma^{\mathrm{int}}
 =\Conv_{\ell\textup{ prime}}
 \left(\left(1-\frac1\ell\right)\delta_0
       +\sum_{k\ge1}\frac{\ell-1}{\ell^{k+1}}
                         \delta_{\gsig(\ell^k)}\right).
\end{equation}
In both factorizations the expected absolute prime increment is
$O(\ell^{-2})$, since the activation probability is $1/\ell$ and the
increment is at most $\omega_\ell\ll1/\ell$. The series converge
almost surely. Their atomlessness is a consequence of
\Cref{lem:levy-atomless}, proved below.
For the shifted-prime divisor-sum model, define, for every odd prime $\ell$,
\begin{equation}\label{eq:nusig}
 \nusig_\ell:=\frac{\ell-2}{\ell-1}\delta_0
       +\sum_{k\ge1}\ell^{-k}\delta_{\gsig(\ell^k)},
\end{equation}
and put
\[
 \beta_2:=\sum_{k\ge1}2^{-k}\delta_{\gsig(2^k)},\qquad
 \mugodd:=\Conv_{\substack{\ell\ge3\\\ell\textup{ prime}}}\nusig_\ell.
\]
The identification
$\mug=\beta_2*\mugodd$, and hence $\muH=\exp_*\mug$,
is proved in \Cref{thm:dist-sigma}.

The prime-$2$ logarithmic totient head under $\mathsf I$ is
$\tfrac12(\delta_0+\delta_{\log2})$; under either $\mathsf A$ or
$\mathsf B$ it is $\delta_{\log2}$. For the divisor-sum model the
head under $\mathsf I$ is $\tfrac12\delta_0+\tfrac12\beta_2$, and
under either $\mathsf A$ or $\mathsf B$ it is $\beta_2$.
Thus even conditioning makes the prime-$2$ heads agree, but leaves
the odd-prime integer activation probabilities at $1/\ell$, whereas
the shifted-prime probabilities are $1/(\ell-1)$. The resulting
finite-head probabilities can be compared directly in
\Cref{rem:prime35-example}. In particular, the comparison of valuation
measures in \Cref{lem:valuation-likelihood} uses the even-conditioned
integer law $\mathsf A$, not the unconditioned law $\mathsf I$.

The definitions above specify the one-dimensional marginals. The same-input pair in
\Cref{cor:joint-phi-sigma} shares the valuation at each prime, whereas
\Cref{thm:independent-sums} explicitly takes independent copies.
\Cref{fig:four-laws} displays how the two valuation models, the two arithmetic maps and
the exponential maps give the four laws, together with the valuations shared within
each same-input pair and the conditioning at the prime~$2$.

\begin{figure}[t]
\centering
% Grid (mm): two mirrored columns of width 49.8, gutter 35.0.
%   lanes x = 11.95, 37.85 (integer) and 96.75, 122.65 (shifted prime);
%   column centres 24.9 and 109.7; gutter centre 67.3.
%   rows (top y): headings 0, measures -9, logarithmic laws -34, raw laws -50.8.
% The \vphantom's give the boxes of one row identical heights and baselines.
\begin{tikzpicture}[figdiagram, x=1mm, y=1mm,
  topbox/.style={finp, inner xsep=3pt, minimum width=21.9mm},
  lawbox/.style={minimum width=23.9mm},
  brlab/.style={fnote, align=center}]
\def\FphL{\vphantom{\sum_\ell\mathbf 1_{\{J_\ell\ge1\}}\gsig(\ell^{J_\ell})}}%
\def\FphM{\vphantom{\sim\mu_F^{\mathrm{int},\log}\mu_g}}%
\def\FphR{\vphantom{\mu_F^{\mathrm{int}}\muH(}}%
\def\FphS{\vphantom{\varphi(p-1)/(p-1)}}%
\def\FphT{\vphantom{(T}}%
% ---- column headings -----------------------------------------------------
\node[fhead, anchor=north] (hI) at (24.9,0) {Integer sampling};
\node[fhead, anchor=north] (hB) at (109.7,0) {Shifted-prime sampling};
\draw[frule] (0,-3.35) -- (49.8,-3.35);
\draw[frule] (84.8,-3.35) -- (134.6,-3.35);
\node[fnote, anchor=north] (hI2) at (24.9,-3.95) {$n$ by natural density};
\node[fnote, anchor=north] (hB2) at (109.7,-3.95) {$n=p-1$, frequency among primes};
% ---- top row: the three valuation measures --------------------------------
\node[topbox, anchor=north] (I) at (24.9,-9.0)
  {$\mathsf I$: independent\\ valuations $J_\ell$\\ \figref{\eqref{eq:valuation-laws}}};
\node[topbox, anchor=north] (B) at (109.7,-9.0)
  {$\mathsf B$: independent\\ valuations $V_\ell$\\ \figref{\eqref{eq:valuation-laws}}};
\node[finp, inner xsep=2.6pt] (A) at (67.3,0 |- I.center)
  {$\mathsf A=\mathsf I(\,\cdot\mid J_2\ge1)$\\ even-conditioned};
\draw[fmap] (I) -- node[flab, above] {conditioning} node[fhyp, below] {on $J_2\ge1$} (A);
\draw[fmap, -] (A) -- node[flab, above] {comparison} node[fhyp, below] {$\mathsf A\neq\mathsf B$} (B);
% note in the gutter, below the top row and clear of its connectors
\node[fnote, anchor=north, align=center] (note) at (67.3,-22.6)
  {under $\mathsf A$ and $\mathsf B$:\\ \mbox{prime-$2$} heads agree;\\ at odd $\ell$,\\
   $\mathbb P(J_\ell\ge1)=1/\ell$,\\ $\mathbb P(V_\ell\ge1)=1/(\ell-1)$};
% ---- forks: one trunk per column, two labelled maps ------------------------
\draw[fmap, -] (I.south) -- (24.9,-23.4) coordinate (fI);
\draw[fmap] (fI) -| (11.95,-34.0);
\draw[fmap] (fI) -| (37.85,-34.0);
\draw[fmap, -] (B.south) -- (109.7,-23.4) coordinate (fB);
\draw[fmap] (fB) -| (96.75,-34.0);
\draw[fmap] (fB) -| (122.65,-34.0);
\node[flab, anchor=base] (lT)  at (11.95,-27.85)  {totient: $\mathbf 1_{\{J_\ell\ge1\}}$};
\node[flab, anchor=base] (lD)  at (37.85,-27.85)  {divisor sum: $J_\ell$};
\node[flab, anchor=base] (lTp) at (96.75,-27.85)  {totient: $\mathbf 1_{\{V_\ell\ge1\}}$};
\node[flab, anchor=base] (lDp) at (122.65,-27.85) {divisor sum: $V_\ell$};
% ---- logarithmic variables --------------------------------------------------
\node[fmech, lawbox, anchor=north] (T)  at (11.95,-34.0)
  {$\FphL\sum_\ell\mathbf 1_{\{J_\ell\ge1\}}\,\omega_\ell$\\[1pt] $\FphM\sim\mu_F^{\mathrm{int},\log}$};
\node[fmech, lawbox, anchor=north] (D)  at (37.85,-34.0)
  {$\FphL\sum_\ell\gsig(\ell^{J_\ell})$\\[1pt] $\FphM\sim\mu_\sigma^{\mathrm{int},\log}$};
\node[fmech, lawbox, anchor=north] (Tp) at (96.75,-34.0)
  {$\FphL\sum_\ell\mathbf 1_{\{V_\ell\ge1\}}\,\omega_\ell$\\[1pt] $\FphM\sim\mu_f$};
\node[fmech, lawbox, anchor=north] (Dp) at (122.65,-34.0)
  {$\FphL\sum_\ell\gsig(\ell^{V_\ell})$\\[1pt] $\FphM\sim\mug$};
% ---- raw laws -----------------------------------------------------------------
\node[fres, lawbox, anchor=north] (R1) at (11.95,-50.8)
  {$\FphR\mu_F^{\mathrm{int}}$: law of\\ $\FphS\varphi(n)/n$\\ \figref{\FphT\eqref{eq:integer-totient-factorization}}};
\node[fres, lawbox, anchor=north] (R2) at (37.85,-50.8)
  {$\FphR\mu_\sigma^{\mathrm{int}}$: law of\\ $\FphS\sigma(n)/n$\\ \figref{\FphT\eqref{eq:integer-sigma-factorization}}};
\node[fres, lawbox, anchor=north] (R3) at (96.75,-50.8)
  {$\FphR\mu_G$: law of\\ $\FphS\varphi(p-1)/(p-1)$\\ \figref{\FphT\Cref{thm:phi-dist}}};
\node[fres, lawbox, anchor=north] (R4) at (122.65,-50.8)
  {$\FphR\muH$: law of\\ $\FphS\sigma(p-1)/(p-1)$\\ \figref{\FphT\Cref{thm:dist-sigma}}};
% ---- exponential images ---------------------------------------------------------
\draw[fmap] (T)  -- node[fhyp, right=1pt] {$u\mapsto e^{-u}$} (R1);
\draw[fmap] (D)  -- node[fhyp, right=1pt] {$u\mapsto e^{u}$}  (R2);
\draw[fmap] (Tp) -- node[fhyp, right=1pt] {$u\mapsto e^{-u}$} (R3);
\draw[fmap] (Dp) -- node[fhyp, right=1pt] {$u\mapsto e^{u}$}  (R4);
% ---- same-input braces -------------------------------------------------------
\draw[frule, decorate, decoration={brace, mirror, amplitude=3pt}]
  ($(R1.south west)+(0.4,-0.7)$) -- ($(R2.south east)+(-0.4,-0.7)$)
  node[midway, below=2.8pt, brlab] (brI) {same input, shared $J_\ell$: \figref{\Cref{cor:joint-phi-sigma}}};
\draw[frule, decorate, decoration={brace, mirror, amplitude=3pt}]
  ($(R3.south west)+(0.4,-0.7)$) -- ($(R4.south east)+(-0.4,-0.7)$)
  node[midway, below=2.8pt, brlab] (brB) {same input, shared $V_\ell$: \figref{\Cref{cor:joint-phi-sigma}}};
\end{tikzpicture}
\caption{Four scalar laws from two valuation models and two arithmetic maps;
$\gsig(m)=\log(\sigma(m)/m)$. Conditioning aligns the \mbox{prime-$2$} heads, but
$\mathsf A\neq\mathsf B$. Under either arithmetic map, the images of $\mathsf A$
and $\mathsf B$ have equal lower and equal upper local dimensions
(likelihood comparison, \Cref{lem:valuation-likelihood,cor:valuation-local-dimensions});
the integer and shifted-prime laws have equal lower and equal upper information
dimensions (finite-entropy coupling, \Cref{cor:arithmetic-entropy-invariance}).}
\label{fig:four-laws}
\end{figure}

\paragraph*{Models and hypotheses}\phantomsection\label{par:models-hypotheses}
The logarithmic laws are independent prime-coordinate sums. Totient
uses Bernoulli activations; divisor-sum retains the valuations of
\eqref{eq:valuation-laws}. Their finite absolute first moments allow
\Cref{lem:transfers}\textup{(i)} to identify the lower and upper
information dimensions of each logarithmic law with those of its raw
exponential image. In the vector models of
\Cref{sec:affine-tuple-models}, different primes are independent, but
components at one prime share a residue variable.

\Needspace{12\baselineskip}
\emph{Intensity notation.}
After omission of finitely many primes, the coefficient of
$\log\log y$ in the cumulative hazard has the following notation.
\begin{center}
\begin{tabularx}{\linewidth}{@{}p{.17\linewidth}X@{}}
\toprule
Notation & Use \\
\midrule
$c_\ast$ & The covering criterion \eqref{eq:covering-zero-hazard},
  the cumulative-hazard clause of \Cref{prop:renyi-collapse}\textup{(b)},
  and the summable perturbation in \Cref{cor:general-bias-fourier}. \\
$\kappa$ & \eqref{eq:cumulative-intensity} and the same-sign product
  theorem \Cref{thm:intensity}. \\
$\tau(m)$ & The fixed-degree laws in
  \eqref{eq:fixed-degree-geometry-inputs-mas19}. \\
$K=\sum_iK_i$ & The independent Bernoulli model of
  \Cref{thm:real-intensity-mas13}, summing $-\log(1-q_{i,p})$
  over copies and primes. \\
\bottomrule
\end{tabularx}
\end{center}

For Bernoulli coordinates whose probabilities and jumps are both
$O(p^{-1})$, the expected absolute tail beyond $z$ is $O(z^{-1})$ and
expected coordinate information through $y$ is $O(\log y)$.
Activation probabilities alone do not give these bounds for general
countably atomic coordinates.

The Fourier results use additional inputs: distinguished first-jump
states in \Cref{thm:raw-product-fourier}, fixed periodic weights in
\Cref{thm:periodic-weight-dissipation,thm:periodic-raw-product-fourier},
and conditional cancellation in \Cref{lem:conditional-block-smoothing}.
Cumulative hazard alone does not supply these hypotheses, and summable
coupling does not transfer quantitative Fourier decay
(\Cref{rem:summable-coupling-scope}). The jump-scale activation mass,
summability, cancellation and sign conditions are specified in
\Cref{thm:activation-scale,prop:activation-scale-endpoint}.

\section{Prime-counting estimates}\label{sec:prime-counting-inputs}

Throughout, $\pi(x;m,a)=\#\{p\le x:p\textup{ prime},\ p\equiv a\pmod m\}$
and $\operatorname{Li}(x)=\int_2^xdt/\log t$.

We use the Siegel--Walfisz theorem~\cite[Corollary~11.21]{MV06},
\cite[Theorem~II.8.17]{Ten15}.

\begin{theorem}[Siegel--Walfisz]
\leavevmode\phantomsection\label{thm:sw}\ignorespaces
For every fixed $A>0$ there exists $c_A>0$ such that uniformly for integers $m\le (\log x)^A$ and reduced residue classes $a \pmod m$,
\[
\pi(x;m,a)
=
\frac{\operatorname{Li}(x)}{\varphi(m)}
+
O_A\!\bigl(xe^{-c_A\sqrt{\log x}}\bigr).
\]
The implied constant is ineffective.
\end{theorem}

We also use Mertens' third theorem~\cite{Mer74}, \cite[Theorem~2.7(e)]{MV06}.

\begin{theorem}[Mertens' third theorem]
\label{thm:mertens}
As $x\to\infty$,
\[
\prod_{\substack{\ell \le x \\ \ell\textup{ prime}}} \Bigl(1 - \frac{1}{\ell}\Bigr)
\;\sim\;
\frac{e^{-\gamma}}{\log x},
\]
where $\gamma \approx 0.5772$ is Euler's constant.
\end{theorem}

Let $c_{\mathrm M}$ be the Meissel--Mertens constant and, for
$U\ge\log2$, put
$\mathsf m(U)=\sum_{p\le e^U}1/p$ and
$\rho_{\mathrm M}(U)=\mathsf m(U)-\log U-c_{\mathrm M}$.

\begin{lemma}[Quantitative Mertens estimates]
\label{lem:prime-log-stieltjes}
There are effectively computable absolute constants $c,C>0$ such that
\begin{equation}\label{eq:mertens-dvp}
 |\rho_{\mathrm M}(U)|\le Ce^{-c\sqrt U}\qquad(U\ge\log2).
\end{equation}
Moreover, as $x\to\infty$,
\begin{equation}\label{eq:mertens-product-dvp}
 \prod_{p\le x}(1-1/p)
 =\frac{e^{-\gamma}}{\log x}
   \left\{1+O\!\left(e^{-c\sqrt{\log x}}\right)\right\},
\end{equation}
with an effective absolute implied constant.
For every $g\in C^1([U_0,U_1])$ with $[U_0,U_1]\subset[2,\infty)$,
\begin{equation}\label{eq:stieltjes}
 \sum_{e^{U_0}<p\le e^{U_1}}\frac{g(\log p)}p
 =\int_{U_0}^{U_1}g(U)\,\frac{dU}U
   +[g\rho_{\mathrm M}]_{U_0}^{U_1}
   -\int_{U_0}^{U_1}\rho_{\mathrm M}(U)g'(U)\,dU.
\end{equation}
\end{lemma}

\begin{proof}
By the prime number theorem with the de la Vall\'ee Poussin zero-free region
\cite[Theorem~6.9]{MV06}, \cite[Chapter~II.4]{Ten15},
$\Delta=\pi-\operatorname{Li}$ satisfies
$|\Delta(x)|\le C_0xe^{-c_0\sqrt{\log x}}$ $(x\ge2)$ with effectively
computable absolute constants $C_0,c_0>0$. Partial summation, with
$\operatorname{Li}(x)/x+\int_2^x\operatorname{Li}(u)u^{-2}\,du=\log\log x-\log\log2$
and
\[
 \frac{|\Delta(x)|}x+\int_x^\infty\frac{|\Delta(u)|}{u^2}\,du
 \ll\int_{\log x}^\infty e^{-c_0\sqrt s}\,ds
 \ll\sqrt{\log x}\,e^{-c_0\sqrt{\log x}},
\]
gives $|\rho_{\mathrm M}(\log x)|\ll\sqrt{\log x}\,e^{-c_0\sqrt{\log x}}$ with an
effective implied constant, since
$c_{\mathrm M}=-\log\log2+\int_2^\infty\Delta(u)u^{-2}\,du$. Taking
$0<c<c_0$ and $x=e^U$ proves \eqref{eq:mertens-dvp}. Also,
\[
 -\log\prod_{p\le x}(1-1/p)
 =\mathsf m(\log x)+\sum_{p\le x}\sum_{j\ge2}\frac1{jp^j}.
\]
The double sum converges with tail $O(1/x)$ and an effective implied
constant. Hence \eqref{eq:mertens-dvp}, with \Cref{thm:mertens}
identifying the constant, gives \eqref{eq:mertens-product-dvp}. Finally,
$d\mathsf m=dU/U+d\rho_{\mathrm M}$ as Stieltjes measures, and integration by parts
gives \eqref{eq:stieltjes}.
\end{proof}

\begin{lemma}[Dense subsums; even totient ratios]
\label{lem:phi-even-dense}
\textup{(i)} If $a_j>0$, $a_j\to0$ and $\sum_ja_j=\infty$, then the finite
subsums $\sum_{j\in J}a_j$ are dense in $[0,\infty)$.
\textup{(ii)} The set $\{\varphi(m)/m:m\ge1\textup{ even}\}$ is dense in
$[0,\tfrac12]$.
\end{lemma}

\begin{proof}
For~(i), let $t>0$ and $\varepsilon>0$, and choose $j_0$ such that
$a_j<\varepsilon$ for $j\ge j_0$; the first partial sum starting at $j_0$
that reaches $t$ lies in $[t,t+\varepsilon)$~\cite[no.~17, p.~194]{Sch28}.
For~(ii), enumerate the odd primes as $r_1<r_2<\cdots$ and put
$a_j=\log(r_j/(r_j-1))$. These satisfy~(i) by \Cref{thm:mertens}, and
\[
 \{\varphi(m)/m:m\ge1\textup{ even}\}
 =\{\tfrac12\exp(-\sum_{j\in J}a_j):J\subset\mathbb N\textup{ finite}\}.
\]
By~(i) and continuity of $u\mapsto\tfrac12e^{-u}$, this set is dense in
$(0,\tfrac12]$, hence in $[0,\tfrac12]$.
\end{proof}

\section{Random-series representation and pure-type classification}\label{sec:convolution-structure}

\begin{lemma}[Support of a convergent convolution]\label{lem:supp_conv}
Let $X_n$ be independent with laws $\mu_n$, and suppose $\sum_nX_n=X$
almost surely. If $a_n\in\supp\mu_n$ and $s=\sum_na_n$ converges, then
$s\in\supp\operatorname{Law}(X)$; this holds verbatim for $\R^k$-valued $X_n$. If $X_n\ge0$ almost surely,
$0\in\supp\mu_n$ for every $n$, and there are $b_n\in\supp\mu_n$ with
$b_n>0$, $b_n\to0$ and $\sum_nb_n=\infty$, then
$\supp\operatorname{Law}(X)=[0,\infty)$.
\end{lemma}

\begin{proof}
Fix $\varepsilon>0$. Since $T_N=\sum_{n>N}X_n\to0$ almost surely, for
large $N$ we have $\mathbb P(|T_N|<\varepsilon/3)>0$ and
$|\sum_{n>N}a_n|<\varepsilon/3$. As $a_n\in\supp\mu_n$, the independent
events $\{|X_n-a_n|<\varepsilon/(3N)\}$ $(n\le N)$ and
$\{|T_N|<\varepsilon/3\}$ have positive probability; on their
intersection $|X-s|<\varepsilon$, proving the
first assertion. Under the additional hypotheses, take $a_n=b_n$ at
finitely many indices and $a_n=0$ elsewhere. Every finite subsum then
belongs to the support; \Cref{lem:phi-even-dense}\textup{(i)} and
nonnegativity give the second assertion.
\end{proof}

\begin{lemma}[Atomless independent sums]\label{lem:levy-atomless}
Let $(Z_\ell)$ be independent, countably atomic random variables whose sum converges almost surely, and suppose $\mathbb P(Z_\ell=0)>\tfrac12$ for all but finitely many $\ell$. Then $\operatorname{Law}(\sum_\ell Z_\ell)$ is atomless if and only if $\sum_\ell\mathbb P(Z_\ell\ne0)=\infty$.
\end{lemma}
\begin{proof}
Let $m_\ell>0$ be the largest atom of $Z_\ell$; for all but finitely many $\ell$, $1-m_\ell=\mathbb P(Z_\ell\ne0)$. By L\'evy's criterion~\cite[Th.~XIII]{Lev31}, in the form \cite[Theorem~III.2.7(a), (2.17), p.~435]{Ten15}, the sum is atomless if and only if $\sum_\ell(1-m_\ell)=\infty$, and finitely many indices do not affect this.
\end{proof}

\begin{theorem}[Shifted-prime Erd\H{o}s--Wintner theorem and pure-type classification]
\label{thm:pure-type}
\index{Pure-type theorem}
Let $h$ be a real-valued additive arithmetical function, and put
$h^*(n)=h(n)\mathbf1_{|h(n)|\le1}$. The following are equivalent:
\begin{enumerate}
\item[\textup{(i)}] There is a distribution function $H$ such that, at
every continuity point $z$ of $H$,
\[
 \lim_{x\to\infty}\frac1{\pi(x)}
 \#\{p\le x:p\textup{ prime},\ h(p-1)\le z\}=H(z).
\]
\KeepHeadingSpace{6\baselineskip}
\item[\textup{(ii)}] The three series
\begin{equation}\label{eq:ew-three}
 \sum_{\substack{\ell\textup{ prime}\\|h(\ell)|>1}}\frac1\ell,
 \qquad \sum_{\ell\textup{ prime}}\frac{h^*(\ell)^2}{\ell},
 \qquad \sum_{\ell\textup{ prime}}\frac{h^*(\ell)}{\ell}
\end{equation}
converge.
\end{enumerate}
Under these conditions, let $\mu$ be the probability measure with
distribution function $H$. Then:
\begin{enumerate}
\item[\textup{(a)}] The law is purely discrete, purely singular continuous,
or purely absolutely continuous with respect to Lebesgue measure.
\item[\textup{(b)}] It is continuous if and only if
$\sum_{h(\ell)\ne0}1/\ell=\infty$.
\item[\textup{(c)}] Its characteristic function is
\begin{equation}\label{eq:sp-charfun}
 \widehat H(\tau)=\prod_{\ell\textup{ prime}}
 \left(1-\frac1{\ell-1}
 +\sum_{\nu\ge1}\frac{e^{i\tau h(\ell^\nu)}}{\ell^\nu}\right).
\end{equation}
\item[\textup{(d)}] Its support is
\[
 \supp\mu=\overline{\{h(2n):n\ge1\}}
 =\overline{\{h(p-1):p\ge3\textup{ prime}\}}.
\]
\item[\textup{(e)}] If the $V_\ell$ \textup($\ell$ prime\textup) are independent with
$\mathbb P(V_\ell=0)=\frac{\ell-2}{\ell-1}$ and
$\mathbb P(V_\ell=\nu)=\ell^{-\nu}$ $(\nu\ge1)$, and
$Z_\ell:=h(\ell^{V_\ell})$, then $\sum_\ell Z_\ell$ converges almost
surely and its law is $\mu$.
\end{enumerate}
\end{theorem}

The equivalence and Euler product in \Cref{thm:pure-type} are recorded
for every nonzero shift in \cite[\S~1 and formula~(3)]{Ten12}, and
Timofeev~\cite[Theorem~7, p.~118]{Tim91} gives an independent proof of necessity;
sufficiency is due to K\'atai~\cite[Theorem~3]{Kat69} and necessity to
Hildebrand~\cite[Theorem~5 and \S~1, p.~211]{Hil89}. For $p-1$, no prime
divides the shift, so all local factors occur in~\eqref{eq:sp-charfun}.
We prove~(a), (b), (d) and~(e).

\begin{proof}
For~(e), $Z_\ell=0$ when $V_\ell=0$, as $h(1)=0$. By~\eqref{eq:ew-three},
the variables $h(\ell)\mathbf 1_{\{V_\ell=1\}}$ satisfy Kolmogorov's
three-series criterion~\cite[Theorem~34, pp.~85--86]{JW35}. Since
$\sum_\ell\mathbb P(V_\ell\ge2)=\sum_\ell1/(\ell(\ell-1))<\infty$,
Borel--Cantelli gives $Z_\ell=h(\ell)\mathbf 1_{\{V_\ell=1\}}$ for all but
finitely many $\ell$ almost surely. Hence $\sum_\ell Z_\ell$ converges
almost surely, and by~\eqref{eq:sp-charfun} and uniqueness of
characteristic functions its law is $\mu$.

For~(d), let $n$ be even and let $v_\ell$ denote the $\ell$-adic
valuation. Since $\mathbb P(V_\ell=v_\ell(n))>0$ for every $\ell$
(at $\ell=2$ because $n$ is even), each $h(\ell^{v_\ell(n)})$ lies in
$\supp\operatorname{Law}(Z_\ell)$. These values vanish for $\ell\nmid n$
and sum to $h(n)$, so \Cref{lem:supp_conv} gives $h(n)\in\supp\mu$.
In particular, the closed set
$A:=\overline{\{h(p-1):p\ge3\textup{ prime}\}}$ is contained in
$\supp\mu$. Conversely, the empirical measure over $p\le x$ assigns
mass at least $1-1/\pi(x)$ to $A$, so weak convergence and the
Portmanteau theorem give $\mu(A)=1$; hence $\supp\mu=A$.
Since every $p-1$ with $p\ge3$ is even, the closure of the even-input
values lies between $A$ and $\supp\mu$, proving both identities.

For~(a), the $Z_\ell$ of~(e) are discrete, so the Jessen--Wintner theorem
\cite[Theorem~35, p.~86]{JW35} (see also
\cite[Theorem~III.2.7(b)]{Ten15}) gives pure type.

For~(b), with $Z_\ell$ as in~(e), for $\ell\ge5$ the atom of $Z_\ell$ at $0$ has mass at least $(\ell-2)/(\ell-1)>\tfrac12$, and $\mathbb P(Z_\ell\ne0)=\mathbf 1_{\{h(\ell)\ne0\}}/\ell+O\bigl(1/(\ell(\ell-1))\bigr)$. Hence \Cref{lem:levy-atomless} gives~(b).
\end{proof}

The following extends to arbitrary positive multiplicative functions
the criterion proved by Deshouillers for $0<g\le1$~\cite{Des76},
originally conjectured in~\cite[\S~4.6]{Des70}. The strongly multiplicative
$p+1$ analogue for $0<g\le1$, under a continuous limiting law,
is~\cite[Th\'eor\`eme~(0)]{Des75}.

\begin{corollary}[Deshouillers' criterion]\label{cor:deshouillers-criterion}
Let $g$ be positive and multiplicative, and suppose $g(p-1)$ over primes
has a continuous limiting probability law $\nu_g$ with distribution
function $G_g$. Then
\[
 \supp\nu_g=\overline{\{g(2n):n\ge1\}}
 =\overline{\{g(p-1):p\ge3\textup{ prime}\}}.
\]
For $x<y$, one has $G_g(x)<G_g(y)$ if and only if some even integer
$m\ge2$ satisfies $x<g(m)<y$, equivalently if some odd prime $p$
satisfies $x<g(p-1)<y$.
\end{corollary}

\begin{proof}
Put $h=\log g$. Positivity and continuity give $G_g(0)=0$, so
$H(t)=G_g(e^t)$ is a proper continuous limiting distribution for $h(p-1)$.
By \Cref{thm:pure-type}\textup{(d)}, $\supp dH$ is the closure of the
even-input values of $h$, and also of its odd-prime shifted values.
Since $\nu_g$ is the image of $dH$ under $u\mapsto e^u$, its support is
the closure in $\R$ of $\exp(\supp dH)$, giving both asserted closures.
Since $\nu_g$ is atomless, $G_g(y)-G_g(x)=\nu_g((x,y))$ is positive
exactly when $(x,y)$ meets the support, equivalently when it contains
an even-input value or an odd-prime shifted value.
\end{proof}

In particular, $G_g$ is strictly increasing on a nondegenerate interval
$I$ if and only if $I\subseteq\supp\nu_g$.

\begin{corollary}[A fixed nonzero shift]\label{cor:nonzero-shift-pure-type-mas17}
Fix $a\in\Z\setminus\{0\}$ and put
\[
 \mathcal A_a:=\{n\ge1:(n,a)=1,\quad 2\mid n\ \text{if }2\nmid a\},
 \qquad
 \mathcal P_a:=\{p>2:p\text{ prime},\ p\nmid a,\ p+a\ge1\}.
\]
For a real additive function $h$, a proper limiting law $\mu_a$ of
$h(p+a)$ over primes with positive argument exists if and only if
\eqref{eq:ew-three} holds. It is the law of the almost surely
convergent series
\[
 \sum_{\ell\nmid a}h(\ell^{V_\ell}),
 \qquad
 \widehat\mu_a(t)=\prod_{\ell\nmid a}
 \left(1-\frac1{\ell-1}
              +\sum_{k\ge1}\frac{e^{it h(\ell^k)}}{\ell^k}\right),
\]
with the independent valuations of \Cref{thm:pure-type}; equivalently,
set $V_\ell=0$ at primes dividing $a$.
The law is of pure type, and is atomless if and only if
$\sum_{\ell\nmid a,\ h(\ell)\ne0}\ell^{-1}=\infty$. Its support is
\begin{equation}\label{eq:nonzero-shift-support-mas17}
 \supp\mu_a
 =\overline{\{h(n):n\in\mathcal A_a\}}
 =\overline{\{h(p+a):p\in\mathcal P_a\}}.
\end{equation}
If $g$ is positive and multiplicative and $g(p+a)$ has a proper
continuous limiting probability law $\nu_{g,a}$, then
\begin{equation}\label{eq:shifted-deshouillers-mas17}
 \supp\nu_{g,a}
 =\overline{\{g(n):n\in\mathcal A_a\}}
 =\overline{\{g(p+a):p\in\mathcal P_a\}}.
\end{equation}
For $x<y$, its distribution function strictly increases from $x$ to
$y$ exactly when $(x,y)$ contains one of these admissible integer
values, equivalently one of these admissible shifted-prime values.
\end{corollary}

\begin{proof}
The fixed-shift Erd\H{o}s--Wintner theorem and its product are
\cite[\S~1, (3)]{Ten12}. Removing the finitely many primes dividing
$a$ does not change convergence of the three series. The
three-series argument in the proof of \Cref{thm:pure-type} applies
to the remaining coordinates: the higher-valuation probabilities
sum to $\sum_{\ell\nmid a}1/(\ell(\ell-1))<\infty$.
This proves almost-sure convergence and identifies the product law.
Jessen--Wintner gives pure type. At every sufficiently large free
prime the atom at zero has mass greater than $1/2$, and
\[
 \mathbb P\{h(\ell^{V_\ell})\ne0\}
 =\frac{\mathbf1_{\{h(\ell)\ne0\}}}{\ell}+O(\ell^{-2});
\]
\Cref{lem:levy-atomless} proves the continuity criterion.

For $n\in\mathcal A_a$, all its valuations belong to the respective
local supports: they vanish at primes dividing $a$, and the valuation
at $2$ is positive whenever $2\nmid a$. The resulting local values
have finite sum $h(n)$, so \Cref{lem:supp_conv} puts every such value
in $\supp\mu_a$. Every $p\in\mathcal P_a$ has $p+a\in\mathcal A_a$,
since $(p+a,a)=(p,a)=1$ and the required parity holds. Let $C$ be the
closed set generated by these shifted-prime values. The empirical
law gives $C$ mass $1-O_a(1/\pi(x))$: the excluded primes are $2$,
the prime divisors of $a$, and the finitely many primes with
$p+a<1$. Portmanteau therefore gives $\mu_a(C)=1$.
Consequently
\[
 C\subseteq\overline{h(\mathcal A_a)}\subseteq\supp\mu_a\subseteq C,
\]
which proves \eqref{eq:nonzero-shift-support-mas17}.

Finally put $h=\log g$. Positivity implies that $\nu_{g,a}$ is
supported on $[0,\infty)$; continuity excludes an atom at zero.
Thus its logarithmic image is a proper continuous probability law,
and the continuous-mapping theorem identifies it with the limiting
law of $h(p+a)$. Exponentiating the support identity and taking
closure in $\R$ proves \eqref{eq:shifted-deshouillers-mas17}.
Atomlessness identifies the increase from $x$ to $y$ with the mass
of $(x,y)$, which is positive exactly when that interval meets the
support. Density of either displayed value set gives the final claim.
\end{proof}

The omission of exceptional primes is essential to the support formula,
although it does not affect weak convergence. For example, if $a=1$
and $h(n)=\mathbf1_{\{2\mid n\}}$, the limiting law is the point mass
at $1$, whereas the exceptional prime $p=2$ gives $h(p+1)=0$.

\section{The distribution of \texorpdfstring{$\varphi(p-1)/(p-1)$}{phi(p-1)/(p-1)}}
\label{sec:dist-phi}

\begin{lemma}[Verification of the Erd\H{o}s--Wintner hypotheses]
\label{lem:ew-check}
The function $f$ of~\eqref{eq:f-def} is strongly additive. For
$h\in\{f,\gsig\}$, the three series~\eqref{eq:ew-three} converge and
$h(\ell)>0$ for every prime $\ell$; in particular
$\sum_{h(\ell)\ne0}1/\ell=\sum_\ell1/\ell=\infty$.
\end{lemma}

\begin{proof}
For every prime $\ell$ and $\nu\ge1$, $f(\ell^\nu)=f(\ell)=\log(\ell/(\ell-1))$,
so $f$ is strongly additive; also $\gsig(\ell)=\log(1+1/\ell)$. Hence, for
$h\in\{f,\gsig\}$, $0<h(\ell)\le\log2<1$ and $h(\ell)=1/\ell+O(\ell^{-2})$.
Thus $h^*(\ell)=h(\ell)$, the first series in~\eqref{eq:ew-three} is
empty, and the other two are $\ll\sum_\ell\ell^{-3}$ and
$\ll\sum_\ell\ell^{-2}$.
\end{proof}

Recall that, for odd primes $\ell$,
\begin{equation}\label{eq:def-omega}
 \omega_\ell=f(\ell)=\log\frac\ell{\ell-1}
 =\frac1\ell+O(\ell^{-2}),
\end{equation}
and put
\begin{equation}\label{eq:def-dell}
 \begin{aligned}
 d_\ell&=\frac{2(\ell-2)}{(\ell-1)^2}
        =\frac2\ell-\frac2{\ell(\ell-1)^2}=\frac2\ell+O(\ell^{-3}),\\
 S(\tau)&=\sum_{\ell\ge3}d_\ell(1-\cos(\tau\omega_\ell))\qquad(\tau\in\R).
 \end{aligned}
\end{equation}
The series defining $S(\tau)$ converges for every $\tau\in\R$, since
its summands are $O_\tau(\ell^{-3})$.

\begin{lemma}[Convolution representation and Fourier product of $\mu_f$]
\label{lem:factor}
\index{Euler products!Bernoulli representation}
\index{Bernoulli series!totient law}
For every odd prime $\ell$, put
$\nu_\ell=\frac{\ell-2}{\ell-1}\delta_0+\frac1{\ell-1}\delta_{\omega_\ell}$.
Then
\begin{equation}\label{eq:BC}
 \mu_f=\delta_{\log2}*\Conv_{\ell\ge3}\nu_\ell;
\end{equation}
equivalently, $\mu_f$ is the law of
$\log2+\sum_{\ell\ge3}B_\ell\omega_\ell=-\log X$, with $(B_\ell)$ and $X$ as in the introduction.
For every $\tau\in\R$,
\begin{equation}\label{eq:euler}
 |\widehat{\mu_f}(\tau)|^2
 =\prod_{\substack{\ell\ge3\\\ell\textup{ prime}}}
 \left(1-d_\ell(1-\cos(\tau\omega_\ell))\right).
\end{equation}
\end{lemma}

\begin{proof}
By \Cref{thm:pure-type,lem:ew-check}, the limiting law $\mu_f$ exists and
is the law of $\sum_\ell Z_\ell$ in \Cref{thm:pure-type}\textup{(e)} with
$h=f$. There $V_2\ge1$ almost surely and $f(\ell^\nu)=f(\ell)$ for
$\nu\ge1$, so $Z_2=\log2$ and, for odd $\ell$,
$Z_\ell=\omega_\ell\mathbf 1_{\{V_\ell\ge1\}}$ has law $\nu_\ell$, since
$\mathbb P(V_\ell\ge1)=1/(\ell-1)$. This proves~\eqref{eq:BC} and the
Bernoulli representation.
For nonnegative weights $p_m$ (finitely or countably many)
with $\sum_mp_m=1$ and real $\theta_m$, expanding and using $(\sum_mp_m)^2=1$ gives
\begin{equation}\label{eq:atomic-modulus}
 \Bigl|\sum_mp_me^{i\theta_m}\Bigr|^2
 =1-2\sum_{m<n}p_mp_n\bigl(1-\cos(\theta_n-\theta_m)\bigr)\in[0,1].
\end{equation}
For $\widehat\nu_\ell(\tau)$ the right side is $1-d_\ell(1-\cos(\tau\omega_\ell))$,
which proves~\eqref{eq:euler}.
\end{proof}

\begin{theorem}[Limiting distribution: support and strict monotonicity]
\label{thm:phi-dist}
\index{Totient laws!shifted primes}
\index{Limiting distributions!totient ratios}
The distribution function $G$ of $X$ is continuous, and for every
$\alpha\in\R$,
\[
 \lim_{x\to\infty}\frac1{\pi(x)}
 \#\{p\le x:p\textup{ prime},\ \varphi(p-1)/(p-1)\le\alpha\}=G(\alpha).
\]
Its law $\mu_G$ has support $[0,\tfrac12]$, on which $G$ is strictly
increasing. The same limit holds with $\varphi(p-1)/(p-1)$ replaced
by $c(p)$.
\end{theorem}

\begin{proof}
By \Cref{thm:pure-type,lem:ew-check}, $f(p-1)$ has a continuous limiting
law $\mu_f$; by \Cref{lem:factor}, $\mu_f=\operatorname{Law}(-\log X)$, so
its image under $u\mapsto e^{-u}$ has distribution function $G$.
\Cref{cor:deshouillers-criterion,lem:phi-even-dense} give
$\supp\mu_G=[0,\tfrac12]$, and consequently
$\supp\mu_f=[\log2,\infty)$. Every nondegenerate subinterval has
positive mass, so $G$ is strictly increasing there. Finally, for
$U(p)=\varphi(p-1)/(p-1)$,
$0\le U(p)-c(p)\le1-P(p)\le\sum_{j\ge1}p^{-j}=(p-1)^{-1}$.
The same limit for $c(p)$ follows by continuity of $G$.
\end{proof}

For the classical integer law of $\varphi(n)/n$, the support is
$[0,1]$~\cite[Theorem~1(2), p.~319, and \S~8, p.~323]{Sch36}.

\KeepHeadingSpace{5\baselineskip}
\section{From \texorpdfstring{$\varphi(p-1)/(p-1)$}{phi(p-1)/(p-1)} to \texorpdfstring{$c(p)$}{c(p)}}
\label{sec:transfer}

\begin{lemma}[Perturbation lemma]
\label{lem:transport}
Let $\mathcal S$ be an infinite set of primes, and let $X(p)\ge0$ and
$a(p)>0$ for $p\in\mathcal S$, with $a(p)\to1$ as $p\to\infty$ through
$\mathcal S$. If $X(p)$ has a continuous limiting distribution function
$H$ over $\mathcal S$, i.e.
\[
\lim_{x\to\infty}\frac{\#\{p\in\mathcal S:p\le x,\ X(p)\le\alpha\}}{\#\{p\in\mathcal S:p\le x\}}=H(\alpha)\qquad(\alpha\in\R),
\]
then $a(p)X(p)$ has the same limiting distribution $H$ over $\mathcal S$.
\end{lemma}

\begin{proof}
Both sides vanish for $\alpha<0$; let $\alpha\ge0$ and $0<\varepsilon<1$.
For all but finitely many $p\in\mathcal S$, $1-\varepsilon<a(p)<1+\varepsilon$, whence
\[
\{X(p)\le\alpha/(1+\varepsilon)\}
\;\subseteq\;
\{a(p)X(p)\le\alpha\}
\;\subseteq\;
\{X(p)\le\alpha/(1-\varepsilon)\}.
\]
Taking densities over $\mathcal S$ and letting $\varepsilon\downarrow0$, continuity of $H$ proves the claim.
\end{proof}

\begin{lemma}[Elementary matrix-product bounds]\label{lem:matrix-product-bounds}
For real $t>1$ put $P(t):=\prod_{j\ge1}(1-t^{-j})$, extending~\eqref{eq:def-P}.
Then $P$ is strictly increasing on $(1,\infty)$, and for every integer $n\ge1$,
\begin{equation}\label{eq:matrix-product-bounds}
 \begin{gathered}
 1-\frac{1-t^{-n}}{t-1}
 \le\prod_{j=1}^n(1-t^{-j})\le1-t^{-1},\\
 0<P(t)<\prod_{j=1}^n(1-t^{-j}),\qquad
 P(t)\ge1-\frac1{t-1}.
 \end{gathered}
\end{equation}
\end{lemma}

\begin{proof}
For $0\le a_j\le1$, induction gives
$\prod_{j=1}^n(1-a_j)\ge1-\sum_{j=1}^na_j$. Apply this with
$a_j=t^{-j}$ and then let $n\to\infty$. The product is positive
because $\sum_jt^{-j}<\infty$, and every omitted factor is less than
one, giving the strict upper bound by a finite product. The first
factor is strictly increasing, and the product of the remaining
factors is positive and nondecreasing; hence $P$ is strictly increasing.
\end{proof}

\section{The limiting law along an arithmetic progression}\label{sec:ap-law}

For the ranges in which arithmetic convergence is uniform, see
\Cref{ssec:progression-regimes}.

\begin{proposition}[Divisibility patterns of $p-1$ at polylogarithmic moduli]\label{prop:ap-bivariate}
\index{Divisibility patterns!polylogarithmic moduli}
Fix $A>0$ and let $c_A$ be as in \Cref{thm:sw}.
Let $k,m\ge1$ be integers, let $a\in\Z$ satisfy $\gcd(a,m)=1$,
let $\ell_1<\dots<\ell_k$ be odd primes not dividing $m$, and let
$S\subseteq\{1,\dots,k\}$.
Put
\[
\begin{aligned}
\pi_S(x;m,a)&:=\#\bigl\{p\le x:p\textup{ prime},\ p\equiv a\pmod m,\\
&\qquad\qquad \ell_j\mid p-1\ \forall j\in S,\ \ell_j\nmid p-1\ \forall j\notin S\bigr\},\\
c_S&:=\prod_{j\in S}\frac{1}{\ell_j-1}\,
\prod_{j\notin S}\frac{\ell_j-2}{\ell_j-1}>0.
\end{aligned}
\]
Uniformly for $m\ell_1\cdots\ell_k\le(\log x)^A$,
\begin{equation}\label{eq:ap-bivariate}
\pi_S(x;m,a)=c_S\,\frac{\operatorname{Li}(x)}{\varphi(m)}
+O_A\!\bigl(2^{k-|S|}x e^{-c_A\sqrt{\log x}}\bigr).
\end{equation}
\end{proposition}

\begin{proof}
Write $S^c:=\{1,\dots,k\}\setminus S$, and for $U\subseteq\{1,\dots,k\}$ let
$N_U(x)$ count the primes $p\le x$ with $p\equiv a\pmod m$ and
$\ell_j\mid p-1$ for all $j\in U$. By the Chinese Remainder Theorem these
conditions define one reduced class modulo
$m\prod_{j\in U}\ell_j\le(\log x)^A$, so \Cref{thm:sw} and multiplicativity
of $\varphi$ give
\[
 N_U(x)=\frac{\operatorname{Li}(x)}{\varphi(m)\prod_{j\in U}(\ell_j-1)}
 +O_A\!\bigl(xe^{-c_A\sqrt{\log x}}\bigr).
\]
Insert this into the inclusion--exclusion identity
$\pi_S(x;m,a)=\sum_{T\subseteq S^c}(-1)^{|T|}N_{S\cup T}(x)$, which has
$2^{k-|S|}$ terms; the main terms sum to
\[
 \frac{\operatorname{Li}(x)}{\varphi(m)}\prod_{j\in S}\frac1{\ell_j-1}
 \prod_{j\in S^c}\Bigl(1-\frac1{\ell_j-1}\Bigr)=c_S\frac{\operatorname{Li}(x)}{\varphi(m)}.
\]
\end{proof}

Retain $\gcd(a,m)=1$, but allow primes $\ell_j$ to divide $m$. The
reduced class $a\bmod m$ fixes
$\mathbf1_{\ell_j\mid p-1}=\mathbf1_{a\equiv1\pmod{\ell_j}}$ for each
$\ell_j\mid m$. Put
\[
 J_0:=\{j:\ell_j\mid m\},\qquad
 L_{\mathrm{free}}:=\prod_{j\notin J_0}\ell_j,\qquad
 r:=\#\{j:j\notin S\cup J_0\}.
\]
If $\mathbf1_{a\equiv1\pmod{\ell_j}}\ne\mathbf1_{j\in S}$ for some
$j\in J_0$, then $\pi_S(x;m,a)=0$ for every $x\ge2$.
Otherwise, deleting the fixed conditions and applying
\Cref{prop:ap-bivariate} gives
\[
 \pi_S(x;m,a)=\frac{\operatorname{Li}(x)}{\varphi(m)}
 \prod_{j\in S\setminus J_0}\frac1{\ell_j-1}
 \prod_{j\notin S\cup J_0}\frac{\ell_j-2}{\ell_j-1}
 +O_A\!\bigl(2^r x e^{-c_A\sqrt{\log x}}\bigr),
\]
uniformly for $mL_{\mathrm{free}}=[m,\ell_1\cdots\ell_k]\le(\log x)^A$.
When every $\ell_j$ divides $m$, the empty products equal $1$ and the
formula follows directly from \Cref{thm:sw}.

For example, if $m=\ell_1=3$ and $S=\{1\}$, then
$\pi_{\{1\}}(x;3,1)=\pi(x;3,1)\sim\operatorname{Li}(x)/2$,
whereas applying~\eqref{eq:ap-bivariate} without coprimality would give
$\operatorname{Li}(x)/4$. Also $\pi_{\{1\}}(x;3,2)=0$ identically.

\begin{theorem}[Limiting law along an arithmetic progression]\label{thm:ap-law}
\index{Arithmetic progressions!limiting laws}
Fix $Q\ge1$ and $\gcd(b,Q)=1$, and put
\[
 \mathcal P_Q:=\{2\}\cup\{\ell:\ell\mid Q\},\qquad
 e_\ell(b):=\mathbf1_{b\equiv1\pmod{\ell}},\qquad
 A_{Q,b}:=\frac12\prod_{\substack{\ell\mid Q\\\ell\ge3}}
 (1-1/\ell)^{e_\ell(b)}.
\]
Let $G_{Q,b}$ be the distribution function of
\begin{equation}\label{eq:XQ}
 X_{Q,b}:=A_{Q,b}\prod_{\ell\notin\mathcal P_Q}(1-1/\ell)^{B_\ell}.
\end{equation}
Then $G_{Q,b}$ is continuous, its law has support $[0,A_{Q,b}]$,
and $G_{Q,b}$ is strictly increasing on this interval. As $x\to\infty$,
\[
 \sup_{\alpha\in\R}\left|
 \frac{\#\{p\le x:p\textup{ prime},\ p\equiv b\pmod Q,
 c(p)\le\alpha\}}{\pi(x;Q,b)}-G_{Q,b}(\alpha)\right|\longrightarrow0.
\]
The same convergence holds with $c(p)$ replaced by $\varphi(p-1)/(p-1)$.
\end{theorem}

\begin{proof}
Discard $p=2$. Then $2\mid p-1$, and for each odd $\ell\mid Q$ the indicator
$\mathbf1_{\ell\mid p-1}$ equals $e_\ell(b)$. For every fixed nonempty finite set
$T$ of odd primes not dividing $Q$, \Cref{prop:ap-bivariate} (with $m=Q$, $a=b$)
and $\pi(x;Q,b)\sim\operatorname{Li}(x)/\varphi(Q)$ show that each exact
divisibility pattern of $p-1$ by the primes of $T$ has relative density equal
to the probability of the same pattern for $(B_\ell)_{\ell\in T}$ in~\eqref{eq:XQ}.

For the tail, take $y>\max(2,P^+(Q))$. For $y<\ell\le\sqrt x/Q$, the Brun--Titchmarsh inequality~\cite[Theorem~3.9]{MV06} in the reduced class modulo $Q\ell\le\sqrt x$, together with \Cref{thm:sw} for $\pi(x;Q,b)$, gives relative density $\ll1/(\ell-1)$. Moreover, once $x>Q^6$, the product of three primes exceeding $\sqrt x/Q$ is greater than $x$. Thus $p-1\le x$ has at most two such prime divisors, of total $f$-weight $O(Qx^{-1/2})$. Hence, for each $A>0$, uniformly for $Q\le(\log x)^A$ and $x\ge x_0(A)$,
\begin{equation}\label{eq:ap-tail}
 \frac1{\pi(x;Q,b)}\sum_{\substack{p\le x\\p\equiv b\pmod{Q}}}\sum_{\substack{\ell>y\\\ell\mid p-1}}f(\ell)
 \ll_A\sum_{\ell>y}\frac{f(\ell)}{\ell-1}+\frac Q{\sqrt x}.
\end{equation}
The sum $\sum_{\ell>y}f(\ell)/(\ell-1)$ in~\eqref{eq:ap-tail} also bounds the expected contribution of primes $\ell>y$ to the
nonnegative random series
$\log(A_{Q,b}/X_{Q,b})=\sum_{\ell\notin\mathcal P_Q}B_\ell f(\ell)$, which therefore
converges almost surely. The converging-together principle, applied to these tail bounds and
the finite-dimensional convergence, gives the limiting law of $f(p-1)$,
hence that of $\varphi(p-1)/(p-1)$.
Removing finitely many primes leaves $\sum_{\ell\notin\mathcal P_Q}(\ell-1)^{-1}=\infty$, so \Cref{lem:levy-atomless} makes the logarithmic law, and hence $G_{Q,b}$, continuous; the convergence therefore holds at every $\alpha$. Since $P(p)\to1$, \Cref{lem:transport} transfers it to $c(p)$.

The second assertion of
\Cref{lem:supp_conv}, applied to the summands indexed by odd primes
$\ell\nmid Q$, gives support $[0,\infty)$ for their logarithmic sum;
exponentiation, the factor $A_{Q,b}$ and closedness give $[0,A_{Q,b}]$.
Full support gives strict increase there, and convergence to a continuous
distribution function is uniform (P\'olya's theorem).
\end{proof}

For $Q$ a power of two, $X_{Q,b}\overset d=X$, so \Cref{thm:ap-law} contains
\Cref{thm:phi-dist}, with uniform convergence in $\alpha$.

\section{Truncation and its endpoint atom}\label{sec:truncation}
\index{Truncation!endpoint atom}

Here the approximation retains finitely many coordinates of the
limiting random product. Its error is distinct from the error in
replacing a limiting law by a finite arithmetic sample. To measure the
first error, the Wasserstein distance $\mathcal W_1$ takes the least possible
mean absolute displacement over all couplings of the finite and full
products. Using the same prime coordinates for both products will
compute this distance exactly in the lemma below.

A small mean displacement need not remove a visible jump in the
distribution function. Truncating the convolution of \Cref{lem:factor}
at $y$ leaves an atom at $\log 2$, which the map $u\mapsto e^{-u}$ sends
to the endpoint $1/2$ of $\supp(\mu_G)$. Its mass is
\begin{equation}\label{eq:def-Q}
Q(y):=\prod_{\substack{3\le\ell\le y\\\ell\textup{ prime}}}
\frac{\ell-2}{\ell-1}\asymp\frac1{\log y}\qquad(y\to\infty).
\end{equation}
The asymptotic follows from \Cref{thm:mertens} and the convergence of
$\prod_{\ell\ge3}(1-(\ell-1)^{-2})$.

\begin{lemma}[Truncation in Wasserstein distance]\label{lem:truncation-wasserstein}
For $N\ge1$ and the $N$th odd prime $y=\ell_N$, put, with the $B_\ell$ of $X$,
\[
 X^{(N)}:=\frac12\prod_{3\le\ell\le y}
 \left(1-\frac1\ell\right)^{B_\ell}.
\]
Then
\begin{equation}\label{eq:truncation-wasserstein-exact}
 \mathcal W_1\!\left(\operatorname{Law}(X^{(N)}),\operatorname{Law}(X)\right)
 =\mathbb E X^{(N)}-\mathbb E X<\frac1{2y}.
\end{equation}
As $N\to\infty$,
\begin{equation}\label{eq:truncation-wasserstein-asymp}
 \mathcal W_1\!\left(\operatorname{Law}(X^{(N)}),\operatorname{Law}(X)\right)
 =\frac{\mathbb E X}{y\log y}
       \left(1+O\!\left(\frac1{\log y}\right)\right).
\end{equation}
\end{lemma}

\begin{proof}
Through the common $B_\ell$,
$X=X^{(N)}\prod_{\ell>y}(1-1/\ell)^{B_\ell}\le X^{(N)}$, so this coupling
has cost $\mathbb E X^{(N)}-\mathbb E X$; it is optimal, since every
coupling has cost at least the absolute difference of the means.
Put $a_\ell=1/[\ell(\ell-1)]$, $S_y=\sum_{\ell>y}a_\ell$ and
$P_y=\prod_{\ell>y}(1-a_\ell)$. Since $\mathbb E(1-1/\ell)^{B_\ell}=1-a_\ell$,
independence and bounded convergence give
$\mathbb E X^{(N)}=\frac12\prod_{3\le\ell\le y}(1-a_\ell)\le\frac12$ and
$\mathbb E X=P_y\,\mathbb E X^{(N)}$. Since
$S_y-\frac12S_y^2\le1-e^{-S_y}\le1-P_y\le S_y$ and $P_y^{-1}=1+O(S_y)$,
\[
 \mathbb E X^{(N)}-\mathbb E X
 =(1-P_y)\,\mathbb E X^{(N)}
 =(P_y^{-1}-1)\,\mathbb E X
\]
is at most $\frac12S_y$ and equals $\mathbb E X\{S_y+O(S_y^2)\}$. As the
prime tail omits positive terms of the integer tail,
$S_y<\sum_{n>y}1/[n(n-1)]=1/y$; this proves
\eqref{eq:truncation-wasserstein-exact}. Finally,
$S_y=\sum_{\ell>y}\ell^{-2}+O(y^{-2})$, and partial summation with the
prime number theorem $\pi(t)=t/\log t+O(t/\log^2t)$ of \Cref{thm:sw} gives
$\sum_{\ell>y}\ell^{-2}=(y\log y)^{-1}(1+O(1/\log y))$; this proves
\eqref{eq:truncation-wasserstein-asymp}.
\end{proof}

\begin{corollary}[Truncation of the four laws]
\label{cor:truncation-four-laws}
\index{Wasserstein distance!product truncation}
Let $Y$ be any of the four raw ratios defined by
\eqref{eq:valuation-laws}, and let $C$ be a fixed positive-probability
event determined by finitely many valuations. For $y$ beyond these
primes, let $Y_y$ retain the full local factors at primes at most $y$.
Put $Z=|\log Y|$ and $Z_y=|\log Y_y|$. Then, as $y\to\infty$,
\begin{align}
 \mathcal W_1\!\left(\operatorname{Law}(Y_y\mid C),
                         \operatorname{Law}(Y\mid C)\right)
 &=|\mathbb E(Y_y\mid C)-\mathbb E(Y\mid C)|\notag\\
 &=\frac{\mathbb E(Y\mid C)}{y\log y}
       \left(1+O\!\left(\frac1{\log y}\right)\right),
 \label{eq:truncation-four-laws}\\
 \mathcal W_1\!\left(\operatorname{Law}(Z_y\mid C),
                         \operatorname{Law}(Z\mid C)\right)
 &=\mathbb E(Z-Z_y\mid C)
  =\frac1{y\log y}\left(1+O\!\left(\frac1{\log y}\right)\right).
 \notag
\end{align}
\end{corollary}

\begin{proof}
The common-coordinate coupling orders each pair, so its cost equals
the difference of means, the classical one-dimensional
$\mathcal W_1$ identity~\cite{Val74}. The tail is independent of $C$.
At a tail prime $\ell$, the expected raw factors, in the order integer
and shifted-prime totient, then integer and shifted-prime divisor sum, are
\[
 1-\ell^{-2},\qquad 1-\frac1{\ell(\ell-1)},\qquad
 (1-\ell^{-2})^{-1},\qquad
 1+\frac{\ell}{(\ell-1)^2(\ell+1)}.
\]
Each is $1\pm\ell^{-2}+O(\ell^{-3})$, and each positive logarithmic
local mean is $\ell^{-2}+O(\ell^{-3})$. Independence and the prime-tail
estimate in \Cref{lem:truncation-wasserstein} give the formulas.
\end{proof}

The cross-model coupling in the proof of \Cref{cor:arithmetic-entropy-invariance} also
orders the valuations of $\mathsf I,\mathsf A,\mathsf B$ coordinatewise.
Writing $\Phi_{\mathsf M},\Sigma_{\mathsf M}$ for their totient and
divisor-sum ratios, it gives
\[
 \Phi_{\mathsf I}\ge\Phi_{\mathsf A}\ge\Phi_{\mathsf B},\qquad
 \Sigma_{\mathsf I}\le\Sigma_{\mathsf A}\le\Sigma_{\mathsf B}
 \quad\hbox{almost surely}.
\]
Indeed, the odd-prime coordinates of $\mathsf I,\mathsf A$ agree,
and at $2$ one may take $J_2\le J_2^{\mathsf A}=V_2$.
Thus each pair has $\mathcal W_1$ distance equal to its mean difference.
For the totient pairs $(\mathsf I,\mathsf B)$,
$(\mathsf I,\mathsf A)$ and $(\mathsf A,\mathsf B)$ these distances are,
respectively,
\[
 \frac1{\zeta(2)}-\mathbb EX,\qquad
 \frac2{\pi^2},\qquad \frac4{\pi^2}-\mathbb EX.
\]

\begin{remark}[Two prime coordinates]\label{rem:prime35-example}
For \Cref{lem:truncation-wasserstein} with $N=2$ and $y=5$, the finite
head is
\[
 X^{(2)}=\frac12\left(\frac23\right)^{B_3}
                    \left(\frac45\right)^{B_5},\qquad
 -\log X^{(2)}=\log2+B_3\log\frac32+B_5\log\frac54.
\]
Under the shifted-prime law $\mathsf B$, the indicators $B_3,B_5$ are
independent with probabilities $1/2,1/4$. For comparison, under
$\mathsf A=\mathsf I(\,\cdot\mid J_2\ge1)$ the indicators
$\mathbf1_{\{J_3\ge1\}},\mathbf1_{\{J_5\ge1\}}$ are independent with
probabilities $1/3,1/5$. The same finite-head expression then gives
\[
\begin{array}{cc|c|cc}
 \text{indicator at }3&\text{indicator at }5&\text{head value}
       &\mathsf B\text{-probability}&\mathsf A\text{-probability}\\ \hline
 0&0&1/2 &3/8&8/15\\
 0&1&2/5 &1/8&2/15\\
 1&0&1/3 &3/8&4/15\\
 1&1&4/15&1/8&1/15
\end{array}
\]

In the shifted-prime model, conditioning on any one row leaves
\[
 X=x_{b_3,b_5}T_{>5},\qquad
 T_{>5}:=\prod_{\ell>5}(1-1/\ell)^{B_\ell},
\]
where $x_{b_3,b_5}$ is the displayed head value and the independent tail
keeps its original law. The table gives the atoms of the finite
truncation, while $X$ has the continuous law of \Cref{thm:phi-dist}.
\end{remark}

\begin{figure}[t]
\centering
\includegraphics[width=0.7\textwidth]{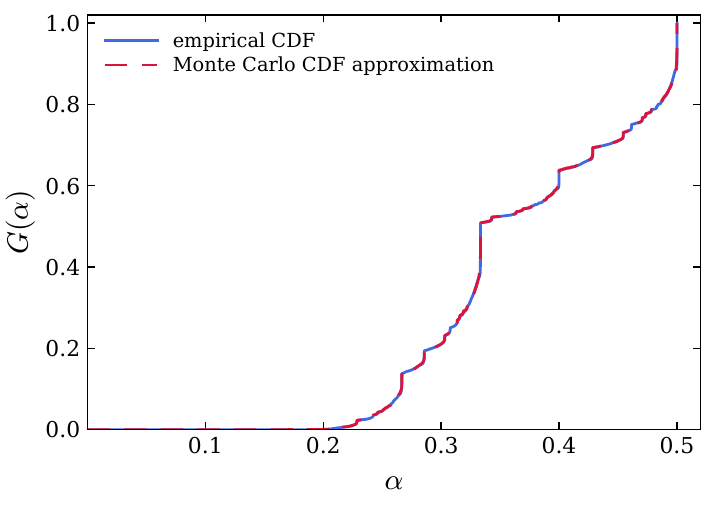}
\caption{Empirical distribution function of~$c(p)$ over the $78{,}497$
primes $3\le p\le10^6$ (solid), and empirical distribution function of
$10^7$ samples from $\operatorname{Law}(X^{(N)})$ with $N=10{,}000$ (dashed).
The empirical values use $\prod_{j=1}^{20}(1-p^{-j})$ in place of
$P(p)$. The simulation uses NumPy's PCG64 generator with seed~$42$;
the curves are thinned for display.}
\label{fig:cdf-overlay}
\end{figure}

\begin{remark}[Numerical truncation]\label{rem:truncation-numerical}
For $N=10{,}000$, one has $\ell_N=104{,}743$, and
\Cref{lem:truncation-wasserstein} gives
\[
 \mathcal W_1\!\left(\operatorname{Law}(X^{(N)}),\operatorname{Law}(X)\right)
 <\frac1{2\ell_N}<5\times10^{-6}.
\]
The leading term $\mathbb E X/(\ell_N\log\ell_N)$ in
\eqref{eq:truncation-wasserstein-asymp}, with $\mathbb E X=M(1)$ by
\Cref{thm:moments}, is approximately $3.09\times10^{-7}$.
The truncated law has an atom of mass $Q(\ell_N)\approx0.064$ at $1/2$,
whereas $G$ is continuous, so its Kolmogorov distance from $G$ is at least
$Q(\ell_N)$; the sharp asymptotic is \Cref{cor:sharp-truncation-inverse}.
The overlay in \Cref{fig:cdf-overlay} illustrates this truncation.
\end{remark}

\section{Limiting law for \texorpdfstring{$\csig$}{c-sigma}}\label{sec:dist-sigma}

For every prime $p$ the integer $p-1$ has prime-power factorization
$p-1=\prod_{\ell^{k}\,\|\,p-1}\ell^{k}$, and multiplicativity of $\sigma/\mathrm{id}$
gives the Euler-product expansion
\begin{equation}\label{eq:sigma-csigma}
\csig(p)
\;:=\;\frac{\sigma(p-1)}{p-1}
\;=\;\prod_{\ell^{k}\,\|\,p-1}\frac{\sigma(\ell^{k})}{\ell^{k}}
\;=\;\prod_{\ell^{k}\,\|\,p-1}\frac{\ell^{k+1}-1}{\ell^{k}(\ell-1)},
\end{equation}
where the empty product gives $\csig(2)=1$. Each nonempty local factor satisfies
\[
\frac{\sigma(\ell^{k})}{\ell^{k}}
\;=\;1+\frac{1}{\ell}+\cdots+\frac{1}{\ell^{k}}
\;\in\;\Bigl[\tfrac{\ell+1}{\ell},\tfrac{\ell}{\ell-1}\Bigr).
\]
For odd $p$, the $2$-part gives $\csig(p)\ge3/2$, with equality only at
$p=3$: if $v_2(p-1)\ge2$ it is at least $7/4$, while if $v_2(p-1)=1$
and $p>3$, an odd prime divisor contributes a factor greater than $1$.
The logarithmic $2$-part is
$\gsig(2^{v_2(p-1)})\in\{\log(3/2),\log(7/4),\log(15/8),\dots\}$.
Put $\wsig_\ell=\gsig(\ell)=\log((\ell+1)/\ell)$ for odd primes $\ell$.

Existence and continuity of the limiting law of $\csig(p)$ go back to
K\'atai~\cite{Kat69}; strict increase on its support also follows from
Deshouillers' theorem~\cite[Th\'eor\`eme~2, \S~4.1]{Des70} for
$n/\sigma(n)$, by reciprocation.
Recall the factors $\nusig_\ell$ from \eqref{eq:nusig}, and the
prime-$2$ head $\beta_2$ and odd-prime convolution $\mugodd$ from
\Cref{ssec:four-scalar-laws}.

\begin{lemma}[Atom locations and first moment]\label{lem:atoms-sigma}
For every odd prime $\ell$ and integer $k\ge1$,
\[
 \gsig(\ell^k)\in
 \Bigl[\log\tfrac{\ell+1}{\ell},\log\tfrac{\ell}{\ell-1}\Bigr),
 \qquad
 \gsig(\ell^k)=\ell^{-1}+O(\ell^{-2})
 \quad\text{uniformly in }k\ge1.
\]
Moreover,
\[
 \int_\R u\,d\nusig_\ell(u)=\ell^{-2}+O(\ell^{-3}).
\]
\end{lemma}
\begin{proof}
Taking logarithms in the local-factor bound above gives the interval
bound; both of its endpoints are
$\ell^{-1}+O(\ell^{-2})$, which gives the uniform expansion. Since the
nonzero atoms in \eqref{eq:nusig} have total mass
$\sum_{k\ge1}\ell^{-k}=1/(\ell-1)$, the first moment is
$(\ell^{-1}+O(\ell^{-2}))/(\ell-1)=\ell^{-2}+O(\ell^{-3})$.
\end{proof}

\begin{theorem}[Limiting law for $\csig$]\leavevmode\phantomsection\label{thm:dist-sigma}
\index{Divisor-sum laws!shifted primes}
\index{Limiting distributions!divisor-sum ratios}\ignorespaces
There is a continuous distribution function $H_\sigma$ such that, for
every $\alpha\in\R$,
\[
 \lim_{x\to\infty}\frac1{\pi(x)}
 \#\{p\le x:p\textup{ prime},\ \csig(p)\le\alpha\}=H_\sigma(\alpha).
\]
Its probability measure is $\muH=\exp_*(\beta_2*\mugodd)$, and
$\supp(\muH)=[3/2,\infty)$, so $H_\sigma$ is strictly increasing there.
\end{theorem}

\begin{proof}
By \Cref{lem:ew-check}, \Cref{thm:pure-type} applies to the additive
function $\gsig$ and gives weak convergence of the empirical laws of
$\gsig(p-1)$, with characteristic function~\eqref{eq:sp-charfun}.

In \Cref{thm:pure-type}\textup{(e)} with $h=\gsig$, $V_2\ge1$ almost
surely, so $\operatorname{Law}(Z_2)=\beta_2$, and
$\operatorname{Law}(Z_\ell)=\nusig_\ell$ for odd $\ell$
by~\eqref{eq:nusig}. Hence, for independent $Z_\ell^\sigma\sim\nusig_\ell$,
$\sum_{\ell\ge3}Z_\ell^\sigma$ converges almost surely, $\mugodd$ is its
law, and the limit law of $\gsig(p-1)$ is $\beta_2*\mugodd$. Since
$\csig(p)=e^{\gsig(p-1)}$, its exponential image is $\muH$.

For $\ell\ge5$, $\nusig_\ell(\{0\})=(\ell-2)/(\ell-1)>\tfrac12$.
The other atoms $\gsig(\ell^k)$ are positive, so
$\mathbb P(Z_\ell^\sigma\ne0)=1/(\ell-1)$ has divergent sum.
Thus \Cref{lem:levy-atomless} shows that $\mugodd$ is atomless.
Convolution with $\beta_2$ and the exponential map preserve atomlessness,
so $\muH$ is atomless and weak convergence gives the asserted limit
for every $\alpha\in\R$.

Each $\nusig_\ell$ is carried by $[0,\infty)$ and has atoms at $0$ and
$\wsig_\ell>0$, where $\wsig_\ell\to0$ and $\sum_{\ell\ge3}\wsig_\ell=\infty$.
The second assertion of \Cref{lem:supp_conv}, applied to the almost surely
convergent series above, gives $\supp(\mugodd)=[0,\infty)$. As
$\min\supp(\beta_2)=\log(3/2)$, we get
$\supp(\beta_2*\mugodd)=[\log(3/2),\infty)$ and
$\supp(\muH)=[3/2,\infty)$.
\end{proof}

\chapter{Carrier geometry and local dimensions}
\label{sec:dimension-liouville}

The four arithmetic laws have interval or ray support, yet concentrate
all their mass on sets of Hausdorff dimension zero. Support records
which neighborhoods have positive mass; a full carrier is a Borel set
on which the measure is concentrated, and need not be closed. Thus the
size of the support and the size of a full carrier answer different
questions about the same law.

This chapter constructs such carriers and then asks how mass scales
near a typical point. Local dimensions, packing dimension and regularity
dimensions distinguish features that the support alone cannot reveal.
The final comparison of valuation models explains which of these
features transfer between the arithmetic sampling laws. Rational
approximation is developed next in \Cref{sec:liouville-carriers},
followed by finite-resolution information in
\Cref{sec:information-dimensions}.

With the jumps $\omega_\ell$ and the Bernoulli variables $B_\ell$ of
\Cref{lem:factor}, put
\begin{equation}\label{eq:local-log-jump}
 Z_\ell:=B_\ell\omega_\ell.
\end{equation}

\section{A covering criterion and Hausdorff dimension zero}\label{sec:convolution}

For a Borel probability measure $\mu$ on $\mathbb{R}$, a Borel set
$E$ with $\mu(E)=1$ is called a \emph{full carrier}. We use
\begin{equation}\label{eq:measure-Hausdorff-dimension}
\dim_H(\mu):=\inf\{\dim_H(E):E\subset\mathbb{R}
 \text{ is a Borel full carrier of }\mu\}.
\end{equation}
Its Fourier transform and Fourier dimension are defined by
\begin{equation}\label{eq:measure-Fourier-dimension}
\begin{aligned}
\widehat\mu(\tau)&:=\int_{\mathbb R}e^{i\tau x}\,d\mu(x),\\
\dim_F(\mu)&:=
\sup\Bigl\{s\in[0,1]:
 |\widehat\mu(\tau)|\ll_{\mu,s}|\tau|^{-s/2}
 \text{ as }|\tau|\to\infty\Bigr\}.
\end{aligned}
\end{equation}
The quantity in \eqref{eq:measure-Hausdorff-dimension} is Falconer's upper
Hausdorff dimension $\dim_H^\ast(\mu)$
\cite[\S~10.1, Proposition~10.3]{Fal97}; the normalization by
$|\tau|^{-s/2}$ in \eqref{eq:measure-Fourier-dimension} follows K\"orner's
definition for closed sets~\cite[Definition~1.3]{Kor11}. The lower variant
$\inf\{\dim_H E:\mu(E)>0\}$~\cite[Chapter~17, equation~(17.35), p.~288]{Fal03}
agrees with \eqref{eq:measure-Hausdorff-dimension} for exact-dimensional
measures, and also when $\dim_H(\mu)=0$, as for $\mu_f$ in
\Cref{thm:hausdorff-zero}.

\begin{lemma}[Locally Lipschitz images]\label{lem:lipschitz-image}
If a Borel probability measure $\mu$ on $\R^d$ has a Borel full carrier
$E$ of Hausdorff dimension zero, $U\subset\R^d$ is open with $\mu(U)=1$,
and $F\colon U\to\R^{d'}$ is locally Lipschitz, then $F_\ast\mu$
has a Borel full carrier of Hausdorff dimension zero.
\end{lemma}

\begin{proof}
Inner regularity
gives compact $K_m\subset E\cap U$ with $\mu(K_m)\to1$.
The set $\bigcup_mF(K_m)$ is Borel and has full image measure; a countable
cover of $U$ by sets on which $F$ is Lipschitz shows that its dimension is
zero.
\end{proof}

\begin{lemma}[Fourier and Hausdorff dimensions]
\label{lem:fourier-Hausdorff}
\index{Fourier dimension}
\index{Hausdorff dimension!Fourier obstruction}
For every Borel probability measure $\mu$ on $\R$,
\[
 \dim_F(\mu)\le \dim_H(\mu).
\]
\end{lemma}

\begin{proof}
There is nothing to prove if $\dim_F(\mu)=0$. Otherwise let
$0<t<s<\dim_F(\mu)$ and let $E$ be a Borel full carrier of $\mu$. Choose $k$
with $\mu([-k,k])>0$ and $\psi\in C_c^\infty(\R)$ with $0\le\psi\le1$ and
$\psi=1$ on $[-k,k]$, and put $\nu:=\psi\mu\ne0$. As
$\widehat{\nu}=(2\pi)^{-1}\widehat{\psi}*\widehat{\mu}$ with $\widehat\psi$
rapidly decreasing, $|\widehat{\nu}(\tau)|\ll_{\mu,\psi,s}(1+|\tau|)^{-s/2}$;
since $t<s$, $\int_\R|\widehat{\nu}(\tau)|^2|\tau|^{t-1}\,d\tau<\infty$, so
$I_t(\nu)<\infty$ by the Fourier energy identity \cite[equation~(4.17), p.~73]{Fal03}.
Since $\nu(E)=\nu(\R)>0$, inner regularity gives a compact $K\subset E$ with
$\nu(K)>0$; the restriction of $\nu$ to $K$ has finite $t$-energy, so
$\dim_H(E)\ge t$ by \cite[Theorem~4.13(a), p.~70]{Fal03}. As $E$ and
$t<\dim_F(\mu)$ are arbitrary, the assertion follows.
\end{proof}

The covering construction separates the random sum into a finite head,
an intervening prime block, and a tail. A bound on the head information
limits the number of likely head values. An empty middle block makes
the head an accurate approximation when the remaining tail is small.
The hypotheses below balance these three requirements so that suitable
covers recur infinitely often almost surely; the successful scales may
depend on the realization. At those scales the covering cost is summable
for each positive Hausdorff exponent, which is enough to construct a
full carrier of dimension zero.

\begin{theorem}[Dimension-zero criterion]
\label{thm:dimension-zero-criterion}
\index{Hausdorff dimension!covering criterion}
Fix $d\ge1$ and let $(Z_\ell)$ be independent countably atomic
random vectors in $\R^d$, indexed by the primes. Write $\|\cdot\|$
for the Euclidean norm and put
\[
 \nu_\ell:=\operatorname{Law}(Z_\ell),\qquad
 I(y):=\sum_{\ell\le y}-\log\nu_\ell(\{Z_\ell\}).
\]
Suppose that, after deletion of finitely many factors (all sums and bounds
below then concern the remaining primes), there are $c_\ast,\beta>0$ and
$\alpha\ge0$ with $\alpha c_\ast\le1$ such that:
\begin{enumerate}
\item[\textup{(i)}] either
\begin{equation}\label{eq:covering-zero-hazard}
 \Theta(y):=\sum_{\ell\le y}-\log\mathbb P(Z_\ell=0)
 =c_\ast\log\log y+O(1),
\end{equation}
or, for some $K_0\ge0$,
\begin{equation}\label{eq:covering-activation-envelope}
 p_\ell:=\mathbb P(Z_\ell\ne0)
 \le\frac{c_\ast}{\ell}+\frac{K_0}{\ell^2};
\end{equation}
\item[\textup{(ii)}] for all sufficiently large $z$,
\begin{equation}\label{eq:covering-absolute-tail}
 \sum_{\ell>z}\mathbb E\|Z_\ell\|\ll z^{-\beta};
\end{equation}
\item[\textup{(iii)}]
\begin{equation}\label{eq:interpolating-information}
 \mathbb E I(y)\ll(\log y)(\log\log(3y))^\alpha.
\end{equation}
\end{enumerate}
Then $\sum_\ell Z_\ell$ converges absolutely almost surely, and its law
$\mu$ has a Borel full carrier of Hausdorff dimension zero.
\end{theorem}

\begin{proof}
The finitely many deleted variables have a countably atomic sum independent
of the others, and countably many translates of a set of Hausdorff
dimension zero form a set of dimension zero; so we may assume that no
factor is deleted. Then Tonelli's theorem and
\eqref{eq:covering-absolute-tail} give $\sum_\ell\|Z_\ell\|<\infty$ almost surely.
Under \eqref{eq:covering-activation-envelope}, omit a further finite
set of small primes so that $g_\ell:=c_\ast/\ell+K_0/\ell^2<1/2$.

Fix $s>0$, start at $n=3$ with $y_3\ge e$, and put
\[
 t_n:=(n\log n\log\log n)^{1/c_\ast},\quad
 a_n:=\left\lceil\frac{4t_n}{\beta s}\right\rceil,
 \quad z_n:=y_n^{a_n},\quad y_{n+1}:=z_n+1,
 \quad r_n:=z_n^{-\beta/2}.
\]
Since $\log y_{n+1}\le(a_n+1)\log y_n$ and $\log a_n=O(\log n)$, we have
$\log\log(3y_n)=O(n\log n)$; since eventually $a_n\ge2$, $\sum_n r_n<\infty$.

Under \eqref{eq:covering-zero-hazard}, let $F_n$ be the sure event.
Under \eqref{eq:covering-activation-envelope}, enlarge the probability
space by independent events $F_n$, independent of $(Z_\ell)$, with
$\mathbb P(F_n)=\prod_{y_n<\ell\le z_n}(1-g_\ell)/(1-p_\ell)\le1$.
Let $A_n:=\{I(y_n)\le t_n\log y_n\}$ and
$E_n:=\{Z_\ell=0\ (y_n<\ell\le z_n)\}\cap F_n$. Under
\eqref{eq:covering-zero-hazard}, subtracting the two endpoint estimates gives
\[
 \mathbb P(E_n)=\exp\{-\Theta(z_n)+\Theta(y_n)\}\asymp a_n^{-c_\ast}.
\]
Under \eqref{eq:covering-activation-envelope}, since
$-\log(1-g_\ell)=c_\ast/\ell+O(\ell^{-2})$, Mertens' theorem gives
\[
 \mathbb P(E_n)=\prod_{y_n<\ell\le z_n}(1-g_\ell)\asymp a_n^{-c_\ast}.
\]
In either case the $E_n$ are independent, and $A_n$, which depends only
on $(Z_\ell)_{\ell\le y_n}$, is independent of $E_n$.
As $a_n\asymp t_n$, $\mathbb P(E_n)\asymp t_n^{-c_\ast}=(n\log n\log\log n)^{-1}$,
so $\sum_n\mathbb P(E_n)=\infty$. Independence, Markov's inequality,
\eqref{eq:interpolating-information} and $\alpha c_\ast\le1$ give
\[
\begin{aligned}
 \mathbb P(A_n^c\cap E_n)
 &=\mathbb P(A_n^c)\mathbb P(E_n)\\
 &\ll\frac{(\log\log(3y_n))^\alpha}{t_n^{1+c_\ast}}\\
 &\ll\frac{(n\log n)^{\alpha-1/c_\ast}}{n\log n(\log\log n)^{1+1/c_\ast}}\\
 &\le\frac1{n\log n(\log\log n)^{1+1/c_\ast}},
\end{aligned}
\]
so $\sum_n\mathbb P(A_n^c\cap E_n)<\infty$. For $R_n:=\sum_{\ell>z_n}\|Z_\ell\|$, Markov's
inequality also gives $\mathbb P(R_n>r_n)\ll r_n$. The
Borel--Cantelli lemmas imply that
$A_n\cap E_n\cap\{R_n\le r_n\}$ occurs infinitely often almost surely.

By independence, the realized tuple $(Z_\ell)_{\ell\le y_n}$ has
probability $e^{-I(y_n)}$, so every value of it that occurs with positive
probability in $A_n$ has probability at least $y_n^{-t_n}$, and there are
at most $y_n^{t_n}$ such values. Let $U_n$ be the union of the closed
Euclidean balls of radius $r_n$ centered at the
corresponding values of $\sum_{\ell\le y_n}Z_\ell$. On
$A_n\cap E_n\cap\{R_n\le r_n\}$ the retained sum lies in $U_n$, and
\[
 \sum_n y_n^{t_n}(2r_n)^s
 \le2^s\sum_n y_n^{-t_n}<\infty.
\]
Since $r_n\to0$, the Borel set $\limsup_n U_n$ has full measure and zero
$s$-dimensional Hausdorff measure. Intersecting these full carriers for
$s=1/m$, $m\ge1$, gives a Borel full carrier $E$ of dimension zero.
\end{proof}

\begin{theorem}[Explicit zero-measure gauges]\label{thm:zero-gauges-mas13}
Retain the independent countably atomic vectors and the zero-hazard
alternative \eqref{eq:covering-zero-hazard} or activation envelope
\eqref{eq:covering-activation-envelope} of
\Cref{thm:dimension-zero-criterion}, as well as
\eqref{eq:interpolating-information}. Replace its tail hypothesis by
\begin{equation}\label{eq:gauge-tail-mas13}
 \sum_{\ell>z}\mathbb E\|Z_\ell\|
 \ll\exp\!\left\{-\frac{\beta\log z}
                         {(\log\log z)^\vartheta}\right\},
 \qquad \beta>0,
\end{equation}
where $\alpha,\vartheta,\theta\ge0$ and
\begin{equation}\label{eq:gauge-threshold-mas13}
 \alpha+\vartheta+\theta\le\frac1{c_\ast}.
\end{equation}
For sufficiently small $r>0$ define the Hausdorff gauge
\[
 h_{\eta,\theta}(r)
 :=r^{\eta/(\log\log(1/r))^\theta},\qquad \eta>0,
\]
and extend it monotonically to larger radii, with $h_{\eta,\theta}(0)=0$.
The sum converges absolutely almost surely. Its law has a Borel full
carrier $E$ such that $\mathcal H^{h_{\eta,\theta}}(E)=0$
simultaneously for every $\eta>0$.
\end{theorem}

\begin{proof}
As in \Cref{thm:dimension-zero-criterion}, the finitely many deleted
variables may be restored by countably many translates at the end.
The retained tail has finite expected absolute sum by
\eqref{eq:gauge-tail-mas13}; the remaining finite sum is finite almost
surely. Fix $\eta>0$. For $n\ge3$ put
\[
 b_n=n\log n,\qquad
 a_n=\left\lceil(n\log n\log\log n)^{1/c_\ast}\right\rceil,
 \qquad z_n=y_n^{a_n},\qquad y_{n+1}=z_n+1,
\]
starting with sufficiently large $y_3$. Eventually $a_n\ge2$, and
\[
 \log\log z_n=O(b_n),\qquad
 \log y_n\gg2^n.
\]
The first bound follows by summing $\log(a_j+1)=O(\log j)$ in
the recursion for $\log y_j$. Set
\[
 r_n=\exp\!\left\{-\frac{\beta\log z_n}
                              {2(\log\log z_n)^\vartheta}\right\}.
\]
Then $r_n\to0$ and $\sum_n r_n<\infty$. Moreover, for all sufficiently
large $n$, the logarithm of $1/(2r_n)$ is at least a positive constant
times $a_n\log y_n/b_n^\vartheta$, and its own logarithm is $O(b_n)$.
Consequently there is $c>0$, depending on the fixed parameters, with
\begin{equation}\label{eq:gauge-cost-mas13}
 -\log h_{\eta,\theta}(2r_n)
 \ge c\eta\frac{a_n\log y_n}{b_n^{\vartheta+\theta}}.
\end{equation}
Choose a fixed $\epsilon>0$ with $2\epsilon<c\eta$, and put
\[
 t_n=\epsilon\frac{a_n}{b_n^{\vartheta+\theta}},\qquad
 A_n=\{I(y_n)\le t_n\log y_n\}.
\]
Let $E_n$ be the empty middle-block event, thinned independently under
the activation-envelope alternative exactly as in the proof of
\Cref{thm:dimension-zero-criterion}. Thus the $E_n$ are independent,
$A_n$ is independent of $E_n$, and
\[
 \mathbb P(E_n)\asymp a_n^{-c_\ast}
 \asymp\frac1{n\log n\log\log n}.
\]
In particular their probabilities have divergent sum. The information
bound and independence give the crucial summable estimate
\begin{align*}
 \mathbb P(A_n^c\cap E_n)
 &\ll \frac{b_n^{\alpha+\vartheta+\theta}}{a_n^{1+c_\ast}}\\
 &\ll\frac{b_n^{\alpha+\vartheta+\theta-1/c_\ast}}
               {n\log n(\log\log n)^{1+1/c_\ast}}\\
 &\le\frac{C}{n\log n(\log\log n)^{1+1/c_\ast}}.
\end{align*}
This remains summable at equality in \eqref{eq:gauge-threshold-mas13}.
For $R_n=\sum_{\ell>z_n}\|Z_\ell\|$,
\eqref{eq:gauge-tail-mas13} and Markov's inequality give
$\mathbb P(R_n>r_n)\ll r_n$. The two Borel--Cantelli lemmas therefore
show that $A_n\cap E_n\cap\{R_n\le r_n\}$ occurs infinitely often
almost surely, on the enlarged probability space if thinning was used.

There are at most $\exp(t_n\log y_n)$ head tuples in $A_n$, since
each has probability at least $\exp(-t_n\log y_n)$. Cover their head
sums by closed balls of radius $r_n$, and call their union $U_n$.
Every successful sum lies in $U_n$. By \eqref{eq:gauge-cost-mas13},
the total gauge cost satisfies
\[
 \sum_n\exp(t_n\log y_n)h_{\eta,\theta}(2r_n)
 \le\sum_n\exp(-t_n\log y_n)<\infty.
\]
Indeed \eqref{eq:gauge-threshold-mas13} implies
$t_n\gg(\log\log n)^{1/c_\ast}$, and $\log y_n\gg2^n$.
Thus $\limsup_nU_n$ is a Borel full carrier with zero
$h_{\eta,\theta}$-Hausdorff measure: its covers from indices $n\ge N$
have diameters tending to zero and costs tending to zero as $N\to\infty$.
The full-measure conclusion descends from any auxiliary thinning space
because these $U_n$ are deterministic subsets of $\mathbb R^d$.
Restoring the deleted head gives a countable union of translates,
still a Borel full carrier of zero gauge measure. Finally intersect
these carriers for $\eta=1/j$, $j\ge1$. For any $\eta>0$, choose
$1/j\le\eta$; then $h_{\eta,\theta}(r)\le h_{1/j,\theta}(r)$ at
small radii, proving the simultaneous assertion.
\end{proof}

\begin{corollary}[Gauge carriers for the eight basic laws]
\label{cor:basic-gauges-mas13}
Each of the four basic logarithmic laws and its raw ratio image has
a Borel full carrier of zero $\mathcal H^{h_{\eta,1}}$ measure
simultaneously for all $\eta>0$.
More generally the conclusion of \Cref{thm:zero-gauges-mas13} passes
to locally Lipschitz images on open sets of full measure.
\end{corollary}

\begin{proof}
The \hyperref[par:four-law-valuation-bounds]{four-law valuation calculations}
(p.~\pageref{par:four-law-valuation-bounds})
give $c_\ast=1$, $\alpha=0$, and $\mathbb E|Z_\ell|\ll\ell^{-2}$
for each basic logarithmic law.
Thus \eqref{eq:gauge-tail-mas13} holds with $\beta=1$, $\vartheta=0$;
take $\theta=1$, at equality in the threshold.
For every fixed $C>0$, direct differentiation of
$t/(\log t)^\theta$, with $t=\log(1/r)$, shows that
$h_{\eta,\theta}(Cr)/h_{\eta,\theta}(r)$ stays bounded as $r\downarrow0$.
Hence Lipschitz maps preserve zero measure for this gauge.
For a locally Lipschitz map, apply this fact on countably many
Lipschitz neighborhoods. Choose compact subsets of the common full
carrier in its domain with probabilities tending to one; their images
form a Borel full carrier, as in \Cref{lem:lipschitz-image}.
The maps $x\mapsto e^{-x}$ and $x\mapsto e^x$ are locally Lipschitz,
so the same conclusion holds for the four raw laws.
\end{proof}

\begin{corollary}[A dimension-zero criterion for prime-indexed sums]
\label{cor:prime-series-zero-dimension}
\index{Prime series!dimension-zero criterion}
Let $(Z_\ell)$ be independent countably atomic real random variables,
indexed by the primes. Put $p_\ell:=\mathbb P(Z_\ell\ne0)$,
$\nu_\ell:=\operatorname{Law}(Z_\ell)$ and
$\iota_\ell:=-\log\nu_\ell(\{Z_\ell\})$. Suppose that, for some
$c_\ast>0$ and all sufficiently large primes $\ell$,
\[
 p_\ell=\frac{c_\ast}{\ell}+O(\ell^{-2}),\qquad
 \mathbb E|Z_\ell|\ll\ell^{-2},
 \qquad \mathbb E\iota_\ell\ll\frac{\log\ell}{\ell}.
\]
Then $\sum_\ell Z_\ell$ converges absolutely almost surely, and its law
has a Borel full carrier of Hausdorff dimension zero.
\end{corollary}

\begin{proof}
After deletion of the finitely many primes at which the hypotheses fail,
the bound on $p_\ell$ gives~\eqref{eq:covering-activation-envelope},
Mertens' theorem gives~\eqref{eq:interpolating-information} with
$\alpha=0$, and $\sum_{\ell>z}\ell^{-2}\ll z^{-1}$
gives~\eqref{eq:covering-absolute-tail} with $\beta=1$. Hence
\Cref{thm:dimension-zero-criterion} applies with $d=1$.
\end{proof}

\begin{remark}[Nonzero-coordinate probabilities and the tail hypothesis]\label{rem:dim-zero-robust}
Hypothesis~\eqref{eq:covering-zero-hazard} holds, after omission of
finitely many small primes, whenever $p_\ell=c_\ast/\ell+\rho_\ell$ with
$c_\ast>0$ and $\sum_\ell|\rho_\ell|<\infty$: eventually
$p_\ell,c_\ast/\ell<1/2$, and the Lipschitz bound for $-\log(1-u)$ on
$[0,1/2]$ and Mertens' theorem give~\eqref{eq:covering-zero-hazard}.

The power decay in~\eqref{eq:covering-absolute-tail} cannot be relaxed to
decay like a negative power of $\log z$, even when the bounds on $p_\ell$
and $\mathbb E I(y)$ hold. For the law $\mu_A$ of \Cref{thm:slow-jump-ac},
with $f_A(\ell)=(\log\ell)^{-A}$ and $A>0$, the odd-prime coordinates have
$p_\ell=1/(\ell-1)=1/\ell+O(\ell^{-2})$ and $\mathbb E I(y)\ll\log y$, so
these bounds hold with $c_\ast=1$ and $\alpha=0$, while partial summation
gives
\[
 \sum_{\substack{\ell>z\\\ell\textup{ prime}}}p_\ell f_A(\ell)
 \sim\frac{(\log z)^{-A}}{A}.
\]
Nevertheless, $\mu_A$ is absolutely continuous and $\dim_H(\mu_A)=1$
by \Cref{thm:slow-jump-ac}; the deterministic prime-$2$ shift does not
affect this.
\end{remark}

The failure persists for tails smaller than every negative power of
$\log z$. Fix $c_\ast>0$ and $0<\kappa<1/2$, use the canonical probabilities
$p_\ell=c_\ast/\ell$ beyond a sufficiently large cutoff, and put
$F(v)=e^{-v^\kappa}$. Then \eqref{eq:stieltjes} gives, with $U=\log z$,
\begin{align*}
 \sum_{\ell>z}\frac{c_\ast}{\ell}e^{-(\log\ell)^\kappa}
 &=c_\ast\int_U^\infty e^{-v^\kappa}\,\frac{dv}{v}
       +O(e^{-U^\kappa-c\sqrt U})\\
 &=\frac{c_\ast}{\kappa}
       \int_{U^\kappa}^\infty\frac{e^{-w}}w\,dw
       +O(e^{-U^\kappa-c\sqrt U})\\
 &\sim\frac{c_\ast}{\kappa}U^{-\kappa}e^{-U^\kappa}.
\end{align*}
The last equivalence follows by integration by parts, whose remainder
is $O(e^{-U^\kappa}U^{-2\kappa})$. These Bernoulli coordinates still
have $\mathbb E I(y)\ll\log y$ and the required activation hazard,
but their sum is absolutely continuous by
\Cref{thm:general-slow-jump-ac-mas03}. Thus the power tail hypothesis
cannot be replaced by this stretched-exponential decay in $\log z$.

\begin{theorem}[Hausdorff dimension zero of the limiting law]
\label{thm:hausdorff-zero}
\index{Hausdorff dimension!arithmetic laws}
$\dim_H(\mu_f)=\dim_H(\mu_G)=0$.
\end{theorem}

\begin{proof}
By \Cref{lem:factor}, the variables in \eqref{eq:local-log-jump} have
$p_\ell=1/(\ell-1)=1/\ell+O(\ell^{-2})$ for $\ell\ge3$.
For each realized value of $Z_\ell$, the negative logarithm of its
probability is
\[
 \iota_\ell=B_\ell\log(\ell-1)
 +(1-B_\ell)\log\frac{\ell-1}{\ell-2}.
\]
Consequently,
\[
 \mathbb E\iota_\ell\ll\frac{\log\ell}{\ell},
 \qquad
 \mathbb E|Z_\ell|=\frac{\omega_\ell}{\ell-1}\ll\ell^{-2}.
\]
\Cref{cor:prime-series-zero-dimension} and \Cref{lem:lipschitz-image}, first
for translation by $\log2$ and then for $T(u)=e^{-u}$, give the two
assertions.
\end{proof}

By \Cref{thm:phi-dist,thm:hausdorff-zero}, the limiting
law of $c(p)$ is singular continuous with support exactly
$[0,\tfrac12]$, and its distribution function $G$ is strictly increasing
on this interval.

\Needspace{8\baselineskip}
\section{Concentration and vanishing dyadic dimensions for \texorpdfstring{$q>1$}{q>1}}\label{ssec:integer-dim-zero}

For fixed $Q\ge1$ and $\gcd(b,Q)=1$, conditioning the model of
\Cref{lem:factor} on the positive-probability cylinder
\[
 D_{Q,b}:=\{B_\ell=e_\ell(b)\text{ for every odd prime }\ell\mid Q\},
 \qquad d_{Q,b}:=\mathbb P(D_{Q,b})>0,
\]
gives $\operatorname{Law}(X\mid D_{Q,b})=\operatorname{Law}(X_{Q,b})$
by \eqref{eq:XQ}. Thus, for every Borel set $B$,
\begin{equation}\label{eq:ap-cylinder-domination}
 \mathbb P(X_{Q,b}\in B)\le d_{Q,b}^{-1}\mu_G(B).
\end{equation}
The dimension and continuity conclusions are included in
\Cref{cor:ap-liouville} below.

For a Borel probability measure $\nu$ on $\R$, let
$Q_\nu(h):=\sup_x\nu([x,x+h])$ $(h>0)$ be its concentration function, put
$\mathcal D_n:=\{[k2^{-n},(k+1)2^{-n}):k\in\Z\}$, and for $q>0$, $q\ne1$, let
$\underline D_q(\nu)$ and $\overline D_q(\nu)$ be the $\liminf$ and
$\limsup$, as $n\to\infty$, of
\begin{equation}\label{eq:dyadic-Lq}
 \frac{-\log\sum_{I\in\mathcal D_n}\nu(I)^q}{(q-1)n\log2}.
\end{equation}
For $0<q<1$, the sum may be $+\infty$, in which case the quotient is
interpreted as $+\infty$. For $q>1$ these are the lower and upper $L^q$
dimensions of~\cite[p.~192]{FLR02}, and
$\underline D_q(\nu)\le\dim_H(\nu)$~\cite[Theorems~1.2 and~1.4]{FLR02},
with $\dim_H$ as in \eqref{eq:measure-Hausdorff-dimension}.

\begin{proposition}[Vanishing of the dyadic $L^q$ dimensions]
\label{prop:renyi-collapse}
\begin{enumerate}[label=\textup{(\alph*)}]
\item If $\log Q_\nu(2^{-n})=o(n)$ as $n\to\infty$, then
$\underline D_q(\nu)=\overline D_q(\nu)=0$ for every $q>1$.

\item Fix $L\ge2$ and let $H$ and $(Z_\ell)_{\ell>L}$ be
independent real random variables, with prime indices $\ell$. Suppose
that $\eta:=\mathbb P(H=a)>0$ for some $a\in\R$,
$\mathbb P(Z_\ell=0)>0$ for every $\ell>L$, and, for some $C,\beta>0$
and every $z\ge L$,
\[
 \sum_{\ell>z}\mathbb E|Z_\ell|\le Cz^{-\beta},\qquad
 \Theta(z):=\sum_{L<\ell\le z}-\log\mathbb P(Z_\ell=0)=o(\log z).
\]
Then $Y:=H+\sum_{\ell>L}Z_\ell$ converges absolutely almost surely.
Let $\mu:=\operatorname{Law}(Y)$ and let $T\colon\R\to\R$ be a Borel map that is
Lipschitz on a neighborhood of $a$. For every sufficiently small $h>0$ there
is $z=z(h)\ge L$ satisfying the \emph{empty-block estimate}
\begin{equation}\label{eq:lq-empty-block}
 \begin{gathered}
 Q_{T_\ast\mu}(2h)\ge T_\ast\mu([T(a)-h,T(a)+h])
 \ge\tfrac12\eta\,e^{-\Theta(z)},\\
 \log z=\beta^{-1}\log(1/h)+O(1).
 \end{gathered}
\end{equation}
Consequently $\log Q_{T_\ast\mu}(2^{-n})=o(n)$, and hence
$\underline D_q(T_\ast\mu)=\overline D_q(T_\ast\mu)=0$ for every $q>1$. If
moreover $\Theta(z)=c_\ast\log\log z+O(1)$ with $c_\ast>0$, then
\begin{equation}\label{eq:renyi-polylog-concentration}
 Q_{T_\ast\mu}(2h)\gg(\log(1/h))^{-c_\ast}
 \qquad(h\downarrow0),
\end{equation}
with implied constants, also in \eqref{eq:lq-empty-block}, allowed to depend
on $T$ and the fixed data.
\end{enumerate}
\end{proposition}

\begin{proof}
\textup{(a)} An interval of length $2^{-n}$ meets at most two cells of
$\mathcal D_n$. Convexity therefore gives
\[
 1\ge\sum_{I\in\mathcal D_n}\nu(I)^q
 \ge2^{1-q}Q_\nu(2^{-n})^q.
\]
Hence the negative logarithm of the middle term is $o(n)$, so both
limits in~\eqref{eq:dyadic-Lq} vanish.

\textup{(b)} Tonelli's theorem gives absolute convergence. Choose
$r_0>0$ such that $T$ is Lipschitz on $[a-r_0,a+r_0]$, with constant $K$,
and put $K_1:=\max(1,K)$. For $0<h\le r_0$ so small that
$z:=(2CK_1/h)^{1/\beta}\ge L$, Markov's inequality gives
$\mathbb P(\sum_{\ell>z}|Z_\ell|\le h/K_1)\ge1/2$. By independence, the
event that $H=a$, that $Z_\ell=0$ for $L<\ell\le z$, and that
$\sum_{\ell>z}|Z_\ell|\le h/K_1$ has probability at least
$\tfrac12\eta e^{-\Theta(z)}$; on it, $|Y-a|\le h/K_1\le r_0$, so
$|T(Y)-T(a)|\le h$. Hence \eqref{eq:lq-empty-block} holds with this $z$, for
which $\log z=\beta^{-1}\log(1/h)+O(1)$.
With $2h=2^{-n}$, the hypothesis $\Theta(z)=o(\log z)$ gives
$\log Q_{T_\ast\mu}(2^{-n})=o(n)$. If
$\Theta(z)=c_\ast\log\log z+O(1)$, then
$e^{-\Theta(z)}\asymp(\log z)^{-c_\ast}$, and
$\log z=\beta^{-1}\log(1/h)+O(1)$ in \eqref{eq:lq-empty-block}
gives \eqref{eq:renyi-polylog-concentration}. A sufficient pointwise
example is $p_\ell:=\mathbb P(Z_\ell\ne0)=c_\ast/\ell+O(\ell^{-2})$:
expansion of $-\log(1-p_\ell)$ and Mertens' theorem give this hazard,
with the finitely many $p_\ell>1/2$ contributing $O(1)$ because each
zero-state probability is positive.
\end{proof}

\begin{remark}[Dyadic and Hausdorff dimension]
Part~\textup{(b)} does not give Hausdorff dimension zero: if every
$Z_\ell=0$ and $H$ has an atom and an absolutely continuous part, then
\textup{(b)} applies, but every Borel full carrier of $\mu$ has Hausdorff
dimension one.
\end{remark}

\phantomsection\label{par:four-law-valuation-bounds}
The four scalar models are fixed in \Cref{ssec:four-scalar-laws}.
All four logarithmic laws have Hausdorff-zero full
carriers by \Cref{cor:prime-series-zero-dimension}, with $c_\ast=1$.
Indeed, for $Z_\ell=g(\ell^{K_\ell})$, $g\in\{f,\gsig\}$ and either
valuation law, the activation probability at odd primes is
$1/\ell+O(\ell^{-2})$, and $0\le Z_\ell\le\omega_\ell\ll1/\ell$ gives
$\mathbb E|Z_\ell|\ll\ell^{-2}$. The expected negative logarithm of
the realized atom probability is at most the entropy of $K_\ell$:
mapping atoms to their images can only increase their probabilities.
At odd primes these entropies are, respectively,
\[
 -\log(1-1/\ell)+\frac{\log\ell}{\ell-1},\qquad
 -z_\ell\log z_\ell+\frac{\ell\log\ell}{(\ell-1)^2},
 \quad z_\ell:=\frac{\ell-2}{\ell-1},
\]
both $O((\log\ell)/\ell)$. The ratio images have Hausdorff-zero full
carriers by \Cref{lem:lipschitz-image}; \Cref{thm:liouville-carriers}
below strengthens this to Liouville full carriers. Singularity of the integer
divisor-sum law is classical~\cite[Chapter~III.2, Exercise~257, p.~441]{Ten15}.

\section{Diffuse submeasures and dyadic dimensions}
\label{sec:diffuse-geometry}
\phantomsection\label{par:positive-order-quantization}
For a Borel probability $\lambda$ on $\R$ and $r>0$, use the
positive-order quantization convention of~\cite[(1.1)]{KNZ22}:
\[
 e_{N,r}(\lambda):=\inf_{1\le |A|\le N}
       \left(\int\operatorname{dist}(x,A)^r\,d\lambda(x)\right)^{1/r}.
\]

For the four arithmetic laws and their logarithmic images, including
arbitrary positive-probability conditioning in the valuation sigma-field,
\Cref{thm:quantization} collects the conclusions for each fixed $r>0$
and the rate--distortion comparison. Under finite-coordinate conditioning,
their order-zero errors vanish in the geometric-mean convention of the
\hyperref[par:order-zero-quantization]{order-zero discussion} in
\Cref{sec:information-quantization}.
For $B(x,r):=(x-r,x+r)$ and $x\in\supp\lambda$, put
\begin{equation}\label{eq:local-dimensions}
 \underline d_\lambda(x):=\liminf_{r\downarrow0}
 \frac{\log\lambda(B(x,r))}{\log r},\qquad
 \overline d_\lambda(x):=\limsup_{r\downarrow0}
 \frac{\log\lambda(B(x,r))}{\log r}.
\end{equation}

\begin{lemma}[Universal upper bound for local dimensions]
\label{lem:universal-upper-local-dimension}
For every Borel probability measure $\lambda$ on $\R$,
$\overline d_\lambda(x)\le1$ for $\lambda$-almost every $x$.
\end{lemma}

\begin{proof}
Fix a bounded interval and $\delta>0$, and put $h=2^{-n}$. Only
$O(h^{-1})$ cells of $\mathcal D_n$ meet the interval; those with
mass below $h^{1+\delta}$ have total mass $O(h^\delta)$.
Borel--Cantelli along $n$ shows that $\lambda$-almost every point $x$
of the interval eventually lies in a cell of mass at least
$h^{1+\delta}$. That cell is contained in $B(x,h)$. Monotonicity
between dyadic radii therefore gives
$\overline d_\lambda(x)\le1+\delta$ almost everywhere on the interval.
Exhausting $\R$ and letting $\delta\downarrow0$ proves the claim.
\end{proof}

\begin{proposition}[Dominated diffuse submeasures]
\label{prop:dominated-diffuse-submeasures}
Let $\lambda$ be a Borel probability on $\R$. Suppose that for every
$0<s<1$ there are $c_s\in(0,1]$, $C_s>0$ and $h_s>0$ such that,
for every $0<h<h_s$, a finite nonnegative Borel measure
$\sigma_{s,h}\le\lambda$ satisfies
\begin{equation}\label{eq:dominated-diffuse-submeasures}
 \sigma_{s,h}(\R)\ge c_s,\qquad
 \sup_x\sigma_{s,h}([x,x+3h])\le C_s h^s.
\end{equation}
Then the following conclusions hold.
\begin{enumerate}[label=\textup{(\alph*)}]
\item For $0<q<1$, $\underline D_q(\lambda)\ge1$. If
$\sum_{j\in\Z}\lambda([j,j+1))^q<\infty$, then
$\underline D_q(\lambda)=\overline D_q(\lambda)=1$.
\item For every $r>0$ and $0<s<1$, there is $k_{s,r}>0$ such that
$e_{N,r}(\lambda)\ge k_{s,r}N^{-1/s}$ for all sufficiently large $N$.
If $\lambda$ has every positive moment, then
$e_{N,r}(\lambda)=N^{-1+o(1)}$ for every $r>0$, and the summability
hypothesis in \textup{(a)} holds for every $0<q<1$.
\item The essential supremum of $\overline d_\lambda$, with respect
to $\lambda$, is one.
\item For each $0<s<1$, the mesh entropies satisfy
\[
 \liminf_{h\downarrow0}
 \frac{-\sum_{j\in\Z}\lambda([jh,(j+1)h))
                  \log\lambda([jh,(j+1)h))}{\log(1/h)}\ge c_s s,
\]
with $0\log0=0$ and infinite entropy allowed.
\end{enumerate}
\end{proposition}

\begin{proof}
\textup{(a)} Fix $s$ and put $h=2^{-n}$, $\sigma=\sigma_{s,h}$,
$m=\sigma(\R)$ and $M=C_s h^s$. Each dyadic $h$-cell has
$\sigma$-mass at most $M$. Since $q-1<0$,
\[
 \sum_{I\in\mathcal D_n}\lambda(I)^q
 \ge\sum_{I\in\mathcal D_n}\sigma(I)^q
 \ge m M^{q-1}\ge c_s C_s^{q-1}h^{-s(1-q)}.
\]
Taking logarithms in \eqref{eq:dyadic-Lq} and letting $s\uparrow1$
gives the lower bound, also when the sum is infinite. Write
$m_j=\lambda([j,j+1))$. Concavity within each unit interval gives
\[
 \sum_{I\in\mathcal D_n}\lambda(I)^q
 \le h^{q-1}\sum_{j\in\Z}m_j^q.
\]
Thus summability gives the reverse upper bound one. A finite moment
of order $a$ with $aq>1$ suffices: Markov's inequality gives
$m_j=O((1+|j|)^{-a})$ outside finitely many unit intervals.

\textup{(b)} The closed radius-$h$ neighborhoods of at most $N$
codepoints have total $\sigma_{s,h}$-mass at most $NC_s h^s$.
For $h=(c_s/(2C_sN))^{1/s}<h_s$, every such codebook $A$ satisfies
\[
 \int\operatorname{dist}(x,A)^r\,d\lambda(x)
 \ge h^r\sigma_{s,h}\{\operatorname{dist}(x,A)>h\}
 \ge\tfrac12c_s h^r.
\]
This proves the assertion with
$k_{s,r}=(c_s/2)^{1/r}(c_s/(2C_s))^{1/s}$, for all $r>0$.
For the upper bound, fix $0<\varepsilon<1$ and use $N\ge2$ equally
spaced points in $[-N^\varepsilon,N^\varepsilon]$, including the endpoints.
For $X\sim\lambda$, the central $r$th error is
$O(N^{-r(1-\varepsilon)})$, while for every $a>0$ the tail error is at most
\[
 \mathbb E\bigl(|X|^r\mathbf1_{\{|X|>N^\varepsilon\}}\bigr)
 \le N^{-\varepsilon a}\mathbb E|X|^{r+a}.
\]
Choose $a\ge r(1-\varepsilon)/\varepsilon$. Hence
$e_{N,r}\le C_{r,\varepsilon}N^{-1+\varepsilon}$; the lower bounds
for every $s<1$ and upper bounds for every $\varepsilon>0$ give
$\log e_{N,r}/\log N\to-1$. The moment implication in \textup{(a)}
proves the remaining summability assertion.

\textup{(c)} Fix $0<\varepsilon<s<1$. Put
$I_j=[jh,(j+1)h)$ and $J_j=[(j-1)h,(j+2)h)$.
Since $\sum_j\lambda(J_j)=3$, at most $3h^{-s+\varepsilon}$ indices
satisfy $\lambda(J_j)>h^{s-\varepsilon}$. Their cells $I_j$ have total
$\sigma_{s,h}$-mass at most $3C_s h^\varepsilon$.
For every $x$ in another cell, $B(x,h)\subset J_j$, so, with
$F_h=\{x:\lambda(B(x,h))\le h^{s-\varepsilon}\}$,
\[
 \lambda(F_h)\ge\sigma_{s,h}(F_h)\ge c_s-3C_s h^\varepsilon.
\]
Consequently $\lambda(\limsup_nF_{2^{-n}})\ge c_s$.
On this limsup, $\overline d_\lambda\ge s-\varepsilon$; letting
$s\uparrow1$ and $\varepsilon\downarrow0$ proves the essential-supremum
lower bound. The upper bound $\overline d_\lambda\le1$ almost
everywhere is \Cref{lem:universal-upper-local-dimension}.

\textup{(d)} For the fixed $h$-mesh, write $H(\alpha)$ for the entropy
of a probability $\alpha$, and use $m,M,\sigma$ as in \textup{(a)}.
If $m<1$, decompose
$\lambda=m(\sigma/m)+(1-m)((\lambda-\sigma)/(1-m))$.
Entropy concavity gives $H(\lambda)\ge mH(\sigma/m)$; for $m=1$
this is an equality. Every cell of $\sigma/m$ has mass at most $M/m$,
so $H(\sigma/m)\ge\log(m/M)$. Therefore
\[
 H(\lambda)\ge m\log(m/M)
 =ms\log(1/h)+m\log(m/C_s)
 \ge c_s s\log(1/h)-O_s(1).
\]
The entropy inequalities hold for countable partitions and extended
entropy by finite coarsening. Division by $\log(1/h)$ proves the claim.
\end{proof}

\begin{lemma}[Separated prime blocks]\label{lem:separated-prime-blocks}
Fix $k\ge1$ and numbers $0<a_j<b_j$ such that
\[
 2b_j<a_{j+1}\quad(j<k),\qquad 2b_k<1.
\]
Let $v$ be a $C^1$ function on $[t_0,\infty)$, for some $t_0\ge2$,
and suppose that constants $c_v,C_v>0$ satisfy
\begin{equation}\label{eq:diffuse-first-jump-regularity}
 0<v(t)\le C_v/t,\qquad -v'(t)\ge c_v/t^2\quad(t\ge t_0).
\end{equation}
For every subset $S\subset\{1,\ldots,k\}$, the sums
$\sum_{j\in S}v(p_j)$, with primes $h^{-a_j}<p_j\le h^{-b_j}$,
are pairwise more than $3h$ apart for all sufficiently small $h>0$.
The threshold is uniform in $S$ and in the prime choices.
\end{lemma}

\begin{proof}
For small $h$, all the block arguments exceed $t_0$.
If two tuples first differ at $j$, their $j$th contributions differ
by at least $c_vh^{2b_j}$, since the distinct prime arguments
differ by at least one. The later contributions differ by at most
$2C_vk h^{a_{j+1}}=o(h^{2b_j})$ when $j<k$, and by zero when $j=k$.
Their sums therefore differ by at least
$\tfrac12c_vh^{2b_j}\ge\tfrac12c_vh^{2b_k}>3h$. There are only finitely
many possible first differing indices.
\end{proof}

The arithmetic first-jump functions $\log(t/(t-1))$ and
$\log(1+1/t)$ are positive and $C^1$ on $[2,\infty)$. For $t\ge2$,
\[
 \log\frac{t}{t-1}\le\frac1{t-1}\le\frac2t,
 \qquad -\frac{d}{dt}\log\frac{t}{t-1}
       =\frac1{t(t-1)}\ge\frac1{t^2},
\]
\[
 \log\Bigl(1+\frac1t\Bigr)\le\frac1t,
 \qquad -\frac{d}{dt}\log\Bigl(1+\frac1t\Bigr)
       =\frac1{t(t+1)}\ge\frac{2}{3t^2}.
\]
Thus they satisfy \eqref{eq:diffuse-first-jump-regularity} with
$(t_0,c_v,C_v)=(2,1,2)$ and $(2,2/3,1)$, respectively.

For every $0<s<1$ there are $k\ge1$ and exponents $0<a_j<b_j$
satisfying
\begin{equation}\label{eq:selected-prime-exponents}
 2b_j<a_{j+1}\quad(j<k),\qquad 2b_k<1,\qquad
 A:=\sum_{j=1}^k a_j>s.
\end{equation}
Indeed, choose $R>2$ sufficiently close to $2$, then $k$ sufficiently
large, and put $a_j=R^{j-k-1}$ and $b_j=(R+2)a_j/4$. The separation
inequalities hold and $\sum_j a_j=(1-R^{-k})/(R-1)>s$.
Fix such exponents and put
$\mathcal I_{j,h}=\{p:h^{-a_j}<p\le h^{-b_j}\}$.
These blocks are disjoint and escape every fixed finite set of primes
as $h\downarrow0$.

\begin{lemma}[Selected prime blocks]\label{lem:selected-prime-blocks}
Let $(\xi_p)$ be independent coordinates indexed by all sufficiently
large primes, each with distinguished states $0$ and $1$; additional
states are allowed. Suppose
\[
 z_p:=\mathbb P(\xi_p=0)>0,\qquad
 r_p:=\frac{\mathbb P(\xi_p=1)}{z_p}=O(p^{-1}),
\]
and, for each block above, suppose that as $h\downarrow0$,
\[
 W_{j,h}:=\sum_{p\in\mathcal I_{j,h}}r_p\longrightarrow w_j>0,
 \qquad
 \prod_{p\in\mathcal I_{j,h}}z_p\longrightarrow z_j^*>0.
\]
Let $E_h$ require exactly one coordinate in state $1$ in each block
and every other coordinate in these blocks in state $0$. On $E_h$,
write $P_j$ for the selected prime in block $j$.
\begin{enumerate}[label=\textup{(\alph*)}]
\item \emph{Selection law.} One has
\[
 \mathbb P(E_h)\longrightarrow c:=\prod_{j=1}^k(w_jz_j^*)>0.
\]
Conditional on $E_h$ and all coordinates outside the blocks, the $P_j$
are independent, with probabilities $r_p/W_{j,h}=O(h^{a_j})$.
\item \emph{Separation.} Let $v$ satisfy
\eqref{eq:diffuse-first-jump-regularity}, and put
$S_h=\sum_jv(P_j)$. Distinct prime tuples give values of $S_h$ more
than $3h$ apart. Under the conditioning in \textup{(a)}, every atom
and every interval of length $3h$ have mass $O(h^A)$.
\end{enumerate}
The bounds hold for all sufficiently small $h$, uniformly in the
interval and, almost surely, the outside coordinates.
\end{lemma}

\begin{proof}
\textup{(a)} Independence gives
\[
 \mathbb P(E_h)=\prod_{j=1}^k
    \left(W_{j,h}\prod_{p\in\mathcal I_{j,h}}z_p\right)
 \longrightarrow\prod_{j=1}^k(w_jz_j^*).
\]
In particular, $E_h$ has positive probability for small $h$.
For a tuple $p_j\in\mathcal I_{j,h}$, the event selecting exactly
these primes has probability
$\prod_j r_{p_j}\prod_j\prod_{p\in\mathcal I_{j,h}}z_p$.
Dividing by $\mathbb P(E_h)$ gives $\prod_j(r_{p_j}/W_{j,h})$.
The outside coordinates are independent of the block coordinates,
so this conditional distribution is unchanged by conditioning on them.
This remains true after adjoining a head independent of the whole
coordinate family. Since $W_{j,h}\to w_j>0$ and $p_j>h^{-a_j}$,
each factor is $O(h^{a_j})$.

\textup{(b)} Each tuple has conditional probability $O(h^A)$.
By \Cref{lem:separated-prime-blocks}, distinct tuples have sums
more than $3h$ apart. Thus each interval of length $3h$ contains at
most one tuple value, proving the conditional atom and interval bounds.
\end{proof}

\begin{corollary}[Diffuse submeasures from selected prime blocks]
\label{cor:prime-block-diffuse}
Let the independent coordinates $(\xi_p)$ satisfy the hypotheses of
\Cref{lem:selected-prime-blocks} for every $s\in(0,1)$, with a choice of
blocks and exponent $A>s$ allowed to depend on $s$. Let $Z_p$ be real
functions of $\xi_p$ whose values at states $0,1$ are respectively
$0,v(p)$, where $v$ satisfies \eqref{eq:diffuse-first-jump-regularity}.
Suppose $\sum_p Z_p$ converges almost surely. Let $\Xi$ be a random
element independent of the entire coordinate family, and let $H,V$
be almost surely finite, real, $\sigma(\Xi)$-measurable random variables.
Set $U=H+\sum_p Z_p$. Let $F\colon\R^2\to\R$ be Borel and continuously
differentiable in its first variable, with continuous, nowhere-zero
partial derivative $\partial_1F$. If $D$ is any positive-probability
event in $\sigma(\Xi,(\xi_p)_p)$, then
$\lambda=\operatorname{Law}(F(U,V)\mid D)$ satisfies
\eqref{eq:dominated-diffuse-submeasures} for every $0<s<1$.
The constants may depend on $s,D,F$ and the full law. 
\end{corollary}

\begin{proof}
Fix $s$, and let $E_h$ be the event in \Cref{lem:selected-prime-blocks},
with $\mathbb P(E_h)\to c>0$. Its blocks eventually avoid every fixed
finite set. Put $\mathcal G_N=\sigma(\Xi,\xi_p:p\le N)$ and
$D_N=\{\mathbb E(\mathbf1_D\mid\mathcal G_N)>1/2\}$.
The increasing sigma-fields $\mathcal G_N$ generate
$\sigma(\Xi,(\xi_p)_p)$, so martingale convergence gives
$\mathbb P(D\mathbin\triangle D_N)\to0$.
Once the blocks avoid the primes at most $N$, $E_h$ is independent
of $\mathcal G_N$. Hence
\[
 \bigl|\mathbb P(E_h\cap D)-\mathbb P(E_h)\mathbb P(D)\bigr|
 \le2\mathbb P(D\mathbin\triangle D_N).
\]
First letting $h\downarrow0$ for fixed $N$, then $N\to\infty$,
gives $\mathbb P(E_h\mid D)\to c$.
Choose $K$ so that
$\mathbb P(\{|U|>K\}\cup\{|V|>K\}\mid D)<c/4$ and put
\[
 d_K=\min_{|u|,|v|\le K}|\partial_1F(u,v)|>0.
\]
At target scale $\rho$, use $h=\rho/d_K$ and define
\[
 \sigma_{s,\rho}(B)=
 \mathbb P(F(U,V)\in B,E_h,|U|\le K,|V|\le K\mid D).
\]
This is dominated by $\lambda$ and has mass at least $c/4$ for small
$\rho$. Let $\mathcal G_h$ be generated by $\Xi$ and the coordinates
outside the blocks, and let $C_h=H+\sum_{p\text{ outside the blocks}}Z_p$.
Both $C_h$ and $V$ are $\mathcal G_h$-measurable. For each of the
finitely many selected-prime tuples $\mathbf p=(p_1,\ldots,p_k)$, put
$s_{\mathbf p}=\sum_jv(p_j)$ and
$\pi_{\mathbf p}=\prod_j(r_{p_j}/W_{j,h})=O(h^A)$.
Its selection event is independent of $\mathcal G_h$ and has probability
$\mathbb P(E_h)\pi_{\mathbf p}$. Thus, for an interval $J$ of length
$3\rho$, finite summation and independence give
\[
\begin{split}
 &\mathbb P(F(U,V)\in J,E_h,|U|,|V|\le K)\\
 &\quad=\mathbb P(E_h)\,
 \mathbb E\sum_{\mathbf p}\pi_{\mathbf p}
 \mathbf1_{\{F(C_h+s_{\mathbf p},V)\in J,
                   |C_h+s_{\mathbf p}|\le K,\ |V|\le K\}}.
\end{split}
\]
For fixed $|V|\le K$, the derivative of $u\mapsto F(u,V)$ has
constant sign on $[-K,K]$. The inverse image there of $J$ is an
interval of length at most $3\rho/d_K=3h$.
\Cref{lem:separated-prime-blocks} therefore allows at most one tuple
in the displayed sum, giving an $O(h^A)$ bound.
Restricting to $D$ gives
\[
 \sigma_{s,\rho}([x,x+3\rho])
 \le\mathbb P(D)^{-1}\mathbb P(F(U,V)\in[x,x+3\rho],E_h,
                                  |U|,|V|\le K)
 \ll_{s,D,F}\rho^A\ll\rho^s.
\]
This proves the asserted submeasure condition.
\end{proof}

If $H$ and all $Z_p$ are countably
atomic, the identity-map case of \Cref{cor:prime-block-diffuse},
\Cref{prop:dominated-diffuse-submeasures}\textup{(c)}, and
\Cref{lem:atomic-series-dimensions}
give upper local dimension one almost everywhere for the law of $U$,
and packing dimension one for every Borel set of positive measure under
that law. If the $Z_p$ also satisfy the hypotheses of
\Cref{thm:dimension-zero-criterion}, the law of $U$ has a Hausdorff-zero
full carrier and lower local dimension zero almost everywhere.
These local and packing conclusions pass to dominated conditional
laws by differentiation of measures. In contrast, a fixed-anchor
concentration lower bound uses a head atom and an independent empty
block as in \Cref{prop:renyi-collapse}; arbitrary conditioning need not
preserve that argument.

\begin{remark}[Rare large states do not supply moment bounds]
\label{rem:diffuse-large-outliers}
Fix $\kappa>0$. At all sufficiently large primes let independent
$Z_p$ take the values $p^{-1},e^p,0$ with probabilities
$\kappa/p,p^{-2},1-\kappa/p-p^{-2}$, respectively. Their first jump is
$v(t)=1/t$, their cumulative zero hazard is
$\kappa\log\log y+O(1)$, and their coordinate entropies are
$O_\kappa((\log p)/p)$. The series $S=\sum_pZ_p$ converges almost
surely and satisfies the selected-block hypotheses. Nevertheless,
$\mathbb ES^r=\infty$ and $e_{N,r}(\operatorname{Law}(S))=\infty$
for every $r>0$ and finite $N$. Thus intensity, first-jump regularity
and coordinate entropy cannot replace full-law tail or moment
hypotheses in a geometry and quantization theorem.
\end{remark}

\begin{proof}
The expected sum of the $p^{-1}$ states is
$\kappa\sum_pp^{-2}<\infty$, and the $e^p$ states occur only finitely
many times almost surely by Borel--Cantelli. This proves convergence.
Expansion of $-\log(1-\kappa/p-p^{-2})$ gives
$\kappa/p+O_\kappa(p^{-2})$; Mertens' theorem gives the hazard.
The two nonzero contributions to entropy are
$(\kappa/p)\log(p/\kappa)$ and $2p^{-2}\log p$, while the zero-state
contribution is $O_\kappa(p^{-1})$. The selected-state odds are
$\kappa/p+O_\kappa(p^{-2})$, so on each block their sum tends to
$\kappa\log(b_j/a_j)$ and the zero-state product to
$(a_j/b_j)^\kappa>0$. For every large prime,
$\mathbb ES^r\ge e^{rp}/p^2$, which tends to infinity.
For any finite nonempty codebook $A$, put $M=\max_{a\in A}|a|$.
On $\{Z_p=e^p\}$ one has $S\ge e^p$, and hence
$\mathbb E\operatorname{dist}(S,A)^r\ge p^{-2}(e^p-M)^r$ whenever
$e^p>M$. Letting $p\to\infty$ proves that every such codebook has
infinite error. 
\end{proof}

Let the independent $K_p$ have either of the valuation laws in
\eqref{eq:valuation-laws}. \Cref{lem:selected-prime-blocks} applies to
$\xi_p=K_p$ at odd primes, with selected state $1$ and any exponents satisfying
\eqref{eq:selected-prime-exponents}; for small $h$ the blocks avoid $2$.
For odd primes write $z_p=\mathbb P(K_p=0)$ and
$q_p=\mathbb P(K_p=1)/z_p$.
In the integer model $q_p=1/p$; in the shifted-prime model
$q_p=(p-1)/(p(p-2))$. Thus $z_p>0$ and, in both models,
$q_p=p^{-1}+O(p^{-2})$ and $\log z_p=-p^{-1}+O(p^{-2})$.
Mertens' theorem gives the block limits required by the lemma,
which then gives the event probability:
\begin{equation}\label{eq:local-one-prime-event}
 \begin{gathered}
 W_{j,h}:=\sum_{p\in\mathcal I_{j,h}}q_p\longrightarrow\log(b_j/a_j),\\
 \prod_{p\in\mathcal I_{j,h}}z_p\longrightarrow a_j/b_j,\\
 \mathbb P(E_h)=\prod_{j=1}^k\left(W_{j,h}\prod_{p\in\mathcal I_{j,h}}z_p\right)
 \longrightarrow c:=\prod_{j=1}^k\frac{a_j}{b_j}\log\frac{b_j}{a_j}>0.
 \end{gathered}
\end{equation}

% Keep the lemma opening with its hypotheses after the preceding display.
\KeepHeadingSpace{3\baselineskip}
\begin{lemma}[Diffuse submeasures for the arithmetic laws]
\label{lem:arithmetic-diffuse-submeasures}
Let $g\in\{f,\gsig\}$, let the independent $K_p$ have either of the
valuation laws in \eqref{eq:valuation-laws}, and put
$Z=\sum_p g(p^{K_p})$. Let $D$ be an event of positive probability
in the sigma-field generated by all $K_p$.
Let $T$ be the identity or, respectively, $T(u)=e^{-u}$ for $g=f$
and $T(u)=e^u$ for $g=\gsig$, and put
$\lambda=\operatorname{Law}(T(Z)\mid D)$.
Then $\lambda$ has every positive moment. For every $0<s<1$ there
are $c_s,C_s,h_s>0$ such that, for every $0<h<h_s$, there is a finite
nonnegative Borel measure $\sigma_{s,h}\le\lambda$ satisfying
\[
 \sigma_{s,h}(\R)\ge c_s,\qquad
 \sup_x\sigma_{s,h}([x,x+3h])\le C_s h^s.
\]
The constants may depend on $s,D,g$ and the valuation model.
\end{lemma}
\begin{proof}
The local bounds $0\le g(p^{K_p})\le\omega_p=O(p^{-1})$ and
$\mathbb P(K_p>0)=O(p^{-1})$ give, for every fixed $t>0$,
$\mathbb E e^{t g(p^{K_p})}=1+O_t(p^{-2})$ at odd primes.
The prime-$2$ increment is bounded by $\log2$ in both models.
Independence and monotone convergence imply
$\mathbb E e^{tZ}<\infty$. Consequently $Z,e^Z,e^{-Z}$ have every
positive moment, and conditioning preserves these bounds because
$\mathbb E(U\mid D)\le\mathbb EU/\mathbb P(D)$ for nonnegative $U$.

Apply \Cref{cor:prime-block-diffuse} to the odd-prime coordinates
$\xi_p=K_p$, with $Z_p=g(p^{K_p})$, $\Xi=K_2$,
$H=g(2^{K_2})$, $V=0$, and $F(u,v)=T(u)$.
The head is bounded and independent of the odd-prime coordinates,
and the preceding moment bound gives almost-sure convergence.
The state-$0$ increments vanish; the state-$1$ increments are
$v(p)=\log(p/(p-1))$ for $g=f$ and $v(p)=\log(1+1/p)$ for $g=\gsig$,
which satisfy \eqref{eq:diffuse-first-jump-regularity}.
The preceding odds calculation and \eqref{eq:local-one-prime-event}
verify the selected-block hypotheses for every $s\in(0,1)$.
Moreover, $D\in\sigma(\Xi,(\xi_p)_{p>2})$, and
$\partial_1F=T'$ is continuous and nowhere zero.
The corollary supplies the required submeasures.
\end{proof}

The retained mass $c_s$ depends on $s$, so
\Cref{prop:dominated-diffuse-submeasures}\textup{(d)} yields the Shannon
lower bound $c_s s$. Sharper information bounds require the separate
argument in \Cref{sec:information-dimensions}.
The diffuse construction and moment bounds determine the spectrum for
$0<q<1$. For $q>1$ a concentration interval may be chosen anew at
each scale, even after joint valuation conditioning.

\KeepHeadingSpace{5\baselineskip}
\paragraph*{Dyadic $L^q$ dimensions}
\begin{proposition}[Concentration under joint valuation conditioning]
\label{prop:conditioned-renyi-images}
Let $Z_1,\ldots,Z_n$ be independent basic logarithmic samples, choose
a locally Lipschitz function $T:\R^n\to\R$, and let $D$ be any
event of positive probability in their joint valuation sigma-field.
If $\lambda$ is the conditional law of $T(Z_1,\ldots,Z_n)$ given $D$,
then
\[
 \log Q_\lambda(h)=o(\log(1/h)),\qquad
 \underline D_q(\lambda)=\overline D_q(\lambda)=0\quad(q>1).
\]
\end{proposition}

\begin{proof}
Work under the original product probability $\mathbb P$. Write
$\boldsymbol Z=(Z_1,\ldots,Z_n)$. Fix a compact cube $C$ such that
$E=D\cap\{\boldsymbol Z\in C\}$ has probability $e>0$.
On $C$ the function $T$ has a finite Lipschitz constant.
For $0<\eta<1$, $y=h^{-\eta}$ and $z=h^{-2}$, record the complete
valuation head $V_y$ through $y$. Its entropy is at most $A\log y$,
by the \hyperref[par:four-law-valuation-bounds]{local valuation
entropy bounds}, independence and $\sum_{p\le y}(\log p)/p\ll\log y$.
Choose $K>4A/e$ independently of $\eta$. Markov's inequality for
$-\log\mathbb P(V_y=v)$ at the realized head $v$ shows that head
atoms of mass at least $y^{-K}$ carry probability at least $1-e/4$.
There are at most $y^K$ such atoms.

Write $b=(\eta/2)^n$. The event that every copy has zero valuations
on $(y,z]$ has probability tending to $b$, by Mertens' theorem and
$\mathbb P(K_{i,p}\ne0)=1/p+O(p^{-2})$. The sum over all copies of
the absolute logarithmic contributions above $z$ has mean $O_n(1/z)$,
and is at most $h$ with probability $1-O_n(h)$. These two events
are independent of each other and of the head under $\mathbb P$.
Choose a finite-coordinate event $F$ with
$\mathbb P(F\mathbin{\triangle}E)< eb/16$; such events approximate
the joint valuation sigma-field in probability. For small $h$, $F$ is
head-measurable. Its intersection with the retained head atoms has
mass at least $e-eb/16-e/4\ge 11e/16$. Intersect further with the
empty middle and small tail, whose combined probability is at least
$b/2$. Replacing $F$ by $E$ loses at most $eb/16$, leaving mass at
least $eb/4$.

For each retained head the resulting vectors lie in $C$ and differ
by at most $2h$ in the $\ell^1$ norm. For a fixed $c_C>0$, their
images lie in an interval of length $c_Ch$. Some such interval has
conditional mass at least $eb\,y^{-K}/(4\mathbb P(D))$.
After a constant rescaling of $h$, this gives
$\limsup_{h\downarrow0}[-\log Q_\lambda(h)]/\log(1/h)\le K\eta$.
Let $\eta\downarrow0$ and apply \Cref{prop:renyi-collapse}\textup{(a)}.
\end{proof}

\begin{corollary}[Dyadic $L^q$ spectrum]\label{cor:renyi-spectrum}
\index{Renyi spectrum@R\'enyi spectrum}
\index{Lq spectrum@$L^q$ spectrum}
Let $\nu$ be any of the four integer or shifted-prime totient or
divisor-sum ratio laws, or its logarithmic law. Then
\[
 \underline D_q(\nu)=\overline D_q(\nu)=
 \begin{cases}1,&0<q<1,\\0,&q>1.\end{cases}
\]
Both assertions persist after conditioning on any positive-probability
event in the valuation sigma-field. Under positive-probability
conditioning on finitely many valuations the stronger bound
$Q_\nu(2h)\gg(\log(1/h))^{-1}$ holds, with the interval anchored
at a fixed image of a head atom.
\end{corollary}

\begin{proof}
For $0<q<1$, \Cref{lem:arithmetic-diffuse-submeasures} supplies
\eqref{eq:dominated-diffuse-submeasures} and all positive moments.
\Cref{prop:dominated-diffuse-submeasures}\textup{(a),(b)} gives both
dimensions equal to one, including the required unit-cell summability.
For $q>1$, use \Cref{prop:conditioned-renyi-images} with the identity
or the appropriate exponential map. For the fixed-anchor estimate,
write the logarithmic sum as $Z=\sum_p g(p^{K_p})$ and let $D$ be
determined by finitely many valuations, with $\mathbb P(D)>0$.
Choose $L\ge2$ above all conditioned primes. Under conditioning on $D$,
the head $H=\sum_{p\le L}g(p^{K_p})$ is countably atomic and has an
atom of positive mass; it is independent of the unchanged tail.
For $p>L$, put $Z_p=g(p^{K_p})$. In either valuation model,
\[
\begin{gathered}
 \mathbb P(Z_p\ne0)=p^{-1}+O(p^{-2}),\\
 \Theta(z)=\sum_{L<p\le z}-\log\mathbb P(Z_p=0)
          =\log\log z+O(1),\\
 \sum_{p>z}\mathbb E|Z_p|\ll z^{-1}.
\end{gathered}
\]
Here $0\le Z_p\le\omega_p\ll p^{-1}$ and
$\mathbb P(Z_p=0)>0$ for every $p>L$. Apply
\Cref{prop:renyi-collapse}\textup{(b)} with $T=\mathrm{id}$ or the
relevant exponential map, which is Lipschitz near every head atom.
Its empty-block estimate proves the displayed concentration bound
with $c_\ast=1$ and both zero dimensions for $q>1$.
\end{proof}

Independent sums and their concentration lower bounds are treated in
\Cref{cor:independent-sum-geometry,cor:independent-sum-concentration}.

\section{Local, packing and regularity dimensions}
\label{sec:local-packing-dimensions}
For a Borel probability measure $\lambda$ on $\R$, write
\begin{equation}\label{eq:packing-carrier-dimensions}
 P_{\mathrm{full}}(\lambda):=\inf_{\lambda(E)=1}\dim_P E,
 \qquad P_{\mathrm{pos}}(\lambda):=\inf_{\lambda(E)>0}\dim_P E,
\end{equation}
where both infima range over Borel sets. We use the characterization
\[
 \dim_P E=\inf_{E\subset\bigcup_j E_j}
             \sup_j\overline{\dim}_B E_j,
\]
with bounded closed cover sets and upper box dimension
$\overline{\dim}_B$; see~\cite[\S~4, (38)]{Fal21}.

Upper box dimension counts the intervals of a common small size needed
to cover a bounded set, taking an upper limit as the size tends to zero;
packing dimension allows the countable decomposition in the display.
Hausdorff dimension instead permits covering intervals of different
sizes. The local dimensions in \eqref{eq:local-dimensions} describe the
corresponding question for mass near a fixed point, by comparing ball
mass and radius on a logarithmic scale. For the arithmetic laws, the
lower local exponent will be zero and the upper exponent one almost
everywhere: different
sequences of radii at the same typical point give different limiting
exponents. The results below connect this oscillation to packing
dimension one, despite the Hausdorff-zero full carrier.
\Cref{fig:dimension-mechanisms} in \Cref{sec:information-dimensions}
compares the arguments behind these conclusions with those for the
information and quantization dimensions.

\begin{lemma}[Local dimensions of atomic series]
\label{lem:atomic-series-dimensions}
Let $S=\sum_{j\ge1}Z_j$ converge almost surely, where the real variables
$Z_j$ are independent and countably atomic, and let
$\lambda=\operatorname{Law}(S)$. There are constants $0\le a\le b\le1$
such that $\underline d_\lambda=a$ and $\overline d_\lambda=b$
$\lambda$-almost everywhere, and
$P_{\mathrm{pos}}(\lambda)=P_{\mathrm{full}}(\lambda)=b$.
Replacing finitely many summands by any independent, almost surely finite,
countably atomic real head preserves $a,b$. If $\lambda$ has a
Hausdorff-zero full carrier, then $a=0$.
\end{lemma}

\begin{proof}
For fixed $r$, $x\mapsto\lambda(B(x,r))$ is lower semicontinuous by
Fatou's lemma. Monotonicity in $r$ reduces both exponents to
$r=2^{-m}$, so they are Borel. Write $S=A_n+T_n$, with finite head
$A_n$, tail law $\tau_n$ and $\pi_n(u):=\mathbb P(A_n=u)$. For $u$ with $\pi_n(u)>0$,
\[
 \alpha_{n,u}(E):=\mathbb P(A_n=u,S\in E)
 =\pi_n(u)\tau_n(E-u)\le\lambda(E).
\]
Differentiation of measures~\cite[Theorem~2.10]{DeL08} gives
\[
 \frac{\alpha_{n,u}(B(x,r))}{\lambda(B(x,r))}
 \longrightarrow\frac{d\alpha_{n,u}}{d\lambda}(x)\in(0,1]
 \quad\text{for }\alpha_{n,u}\text{-almost every }x.
\]
If $N_{n,u}$ is the spatial exceptional set, the joint event
$\{A_n=u,S\in N_{n,u}\}$ has probability zero. Discarding their
countable union over $n,u$, using
$\alpha_{n,u}(B(x,r))=\pi_n(u)\tau_n(B(x-u,r))$, and dividing
logarithms by $\log r$ gives
\[
 \underline d_\lambda(S)=\underline d_{\tau_n}(T_n),\qquad
 \overline d_\lambda(S)=\overline d_{\tau_n}(T_n)
 \quad\text{for every }n.
\]
Each rational-threshold event is therefore independent of every finite
coordinate head, hence of the whole coordinate sigma-field and itself.
The zero--one law gives constants $a,b$. The same argument compares any
independent atomic head with the common tail, proving the finite-change
assertion. Nonnegativity is immediate, and $b\le1$ by
\Cref{lem:universal-upper-local-dimension}.

For $s>b$, almost every $x$ eventually satisfies
$\lambda(B(x,r))\ge r^s$. Partition these points into bounded Borel
sets with a common radius threshold. Disjoint-ball counting gives
covering numbers $O(h^{-s})$ on each set, so the countable-cover
characterization gives $P_{\mathrm{full}}(\lambda)\le b$.
Conversely, if a positive-mass Borel set had packing dimension below
$s<b$, some positive-mass piece $A$ would be contained in a bounded
closed set of upper box dimension below $s$. For
$\rho:=\lambda|_A/\lambda(A)$, at most $O(h^{-s})$ dyadic cells meet
its support. Those with $\rho$-mass below $h^{s+\epsilon}$ have total
mass $O(h^\epsilon)$. Borel--Cantelli and dyadic interpolation give
$\overline d_\rho\le s$ almost everywhere, after
$\epsilon\downarrow0$. Differentiation of $\rho\ll\lambda$ preserves
both local exponents at $\rho$-typical points, contradicting
$\overline d_\lambda=b$. Thus $P_{\mathrm{pos}}(\lambda)\ge b$.
Finally, if $a>s>0$, a positive-mass bounded set would satisfy
$\lambda(B(x,r))\le r^s$ uniformly for all sufficiently small $r$.
Intersecting it with a Hausdorff-zero full carrier and summing over
small covers gives positive $s$-dimensional Hausdorff measure, a
contradiction.
\end{proof}

\begin{corollary}[Summands at summable total variation distance]
\label{cor:summable-local-coupling}
\index{Coupling!summable local discrepancies}
Let $(U_n)_{n\ge1}$ and $(V_n)_{n\ge1}$ be sequences of countably atomic
real random variables, independent within each sequence, with
\[
 \sum_{n\ge1}d_{\mathrm{TV}}(\operatorname{Law}(U_n),\operatorname{Law}(V_n))<\infty,
 \qquad d_{\mathrm{TV}}(\alpha,\beta):=\sup_{A\textup{ Borel}}
 |\alpha(A)-\beta(A)|.
\]
Then $\sum_nU_n$ converges almost surely if and only if $\sum_nV_n$ does.
When they converge, their laws have the same upper Hausdorff dimension
\eqref{eq:measure-Hausdorff-dimension}, the same values of
$P_{\mathrm{full}}$ and $P_{\mathrm{pos}}$ in
\eqref{eq:packing-carrier-dimensions}, the same almost-sure lower and
upper local-dimension constants, and the same masses of the atomic,
singular-continuous and absolutely continuous components; one law is
Rajchman if and only if the other is.
\end{corollary}

The equivalences for almost sure convergence, absolute continuity and
discreteness are~\cite[Lemma~1, p.~194]{Bab92}.

\begin{proof}
Couple each pair maximally, independently across $n$. The
Borel--Cantelli lemma gives only finitely many mismatches almost surely,
proving the convergence equivalence. For the convergent sums $U,V$, let
$G_0$ be the countable additive group generated by all coordinate atoms.
Then $V-U\in G_0$ almost surely, so
$\mathbb P(U\in N)\le\mathbb P(V\in N+G_0)$ for every Borel set $N$. The
Borel set $N+G_0$ has the Hausdorff and packing dimensions of $N$,
and is Lebesgue-null or countable whenever $N$ is. Applying this to
full carriers and positive-mass sets, and reversing the roles, gives
equality of the upper Hausdorff dimensions and of both quantities in
\eqref{eq:packing-carrier-dimensions}. By
\Cref{lem:atomic-series-dimensions} and the local-dimension
characterization of upper Hausdorff dimension
\cite[Proposition~10.3]{Fal97}, these are respectively the almost-sure
lower and upper local-dimension constants. Applying it to Lebesgue-null
and countable sets gives equal singular and atomic masses; subtraction
gives the remaining component masses.

A finite positive measure absolutely continuous with respect to a finite
positive Rajchman measure $\mu$ is Rajchman (compare the circle case
\cite[\S~2, (2.1)]{Lyo95}): write it as $h\mu$ with $h\in L^1(\mu)$ and
approximate $h$ in $L^1(\mu)$ by $\psi\in C_c^\infty(\R)$. Fourier inversion
and dominated convergence give
$\widehat{\psi\mu}=(2\pi)^{-1}\widehat\psi\ast\widehat\mu\to0$, and the
Fourier error is at most $\|h-\psi\|_{L^1(\mu)}$.
Now $\alpha_g(A):=\mathbb P(U\in A,V-U=g)\le\operatorname{Law}(U)(A)$ and
\[
 \operatorname{Law}(V)=\sum_{g\in G_0}(x\mapsto x+g)_*\alpha_g.
\]
If $\operatorname{Law}(U)$ is Rajchman, so is every summand. Finite partial sums
converge in total variation, giving the assertion for $V$; reverse the
roles for the converse.
\end{proof}

\begin{remark}[Integer and shifted-prime local models]
\label{rem:summable-coupling-scope}
Since $\mathbb P(V_\ell=k)>\mathbb P(J_\ell=k)$ for every $k\ge1$ in
\eqref{eq:valuation-laws},
$d_{\mathrm{TV}}(\operatorname{Law}(J_\ell),\operatorname{Law}(V_\ell))
=\mathbb P(J_\ell=0)-\mathbb P(V_\ell=0)=1/(\ell(\ell-1))$.
A common real function of the valuation can only decrease this distance.
A fixed progression changes only the local laws at primes dividing its
modulus. Thus, once the arithmetic limits are identified with convergent
independent series, \Cref{cor:summable-local-coupling} transfers its
qualitative conclusions between the integer and shifted-prime versions
of each logarithmic $\varphi$ and $\sigma$ model and each model with
prime increments $(\log\ell)^{-A}$ from \Cref{ssec:slow-jump-ac}, and between each model and its
fixed-progression restrictions.
Moments and supports need not be preserved, even after one coordinate
changes, nor are quantitative Fourier decay or Fourier dimension.
For example, let $B_n\sim\operatorname{Bernoulli}(1/2)$ and
$C_n\sim\operatorname{Bernoulli}(1/2+(n+2)^{-2})$, independent within
each sequence, and put $U=\sum_{n\ge1}2^{-n}B_n$ and
$V=\sum_{n\ge1}2^{-n}C_n$. The local distances $(n+2)^{-2}$ are summable
and $U$ is uniform on $[0,1]$. At $t_n=2\pi2^{n-1}$ the
factors of $\widehat{\operatorname{Law}(V)}$ before $n$ equal $1$, the $n$th has
modulus $2/(n+2)^2$, and the remaining moduli have product at least
$\prod_{k\ge1}\cos(\pi/2^{k+1})=2/\pi$. Hence
\[
 |\widehat{\operatorname{Law}(V)}(t_n)|\ge\frac4{\pi(n+2)^2},\qquad
 \dim_F\operatorname{Law}(V)=0<1=\dim_F\operatorname{Law}(U).
\]
Both laws are absolutely continuous and Rajchman by
\Cref{cor:summable-local-coupling}.
\end{remark}

\begin{theorem}[Local-dimension oscillation and packing dimension]
\leavevmode\phantomsection\label{thm:local-dimension-oscillation}
\index{Local dimensions}
\index{Packing dimension}\ignorespaces
Let $g\in\{f,\gsig\}$ and let the independent $K_\ell$ have either
valuation law in \eqref{eq:valuation-laws}. Fix a finite set of primes
$F$ and an almost surely finite, countably atomic real variable $H$,
independent of $(K_\ell)_{\ell\notin F}$. Let $\lambda$ denote either
the law of
\[
 H+\sum_{\ell\notin F}g(\ell^{K_\ell})
\]
or its image under $u\mapsto e^{-u}$ when $g=f$ and $u\mapsto e^u$
when $g=\gsig$. Then
\begin{equation}\label{eq:local-dimension-oscillation}
 \underline d_\lambda(x)=0,\qquad \overline d_\lambda(x)=1
 \quad(\lambda\text{-a.e. }x),\qquad
 P_{\mathrm{pos}}(\lambda)=P_{\mathrm{full}}(\lambda)=1.
\end{equation}
Thus these conclusions hold for all four integer and shifted-prime
ratio laws and their logarithmic laws, and for every fixed-progression
totient law. Every Borel set of positive measure has packing dimension one.
Every probability measure $\nu\ll\lambda$ has Hausdorff dimension zero
and satisfies \eqref{eq:local-dimension-oscillation} with $\nu$ in place
of $\lambda$. In particular, these conclusions survive arbitrary
positive-probability conditioning.
\end{theorem}

\begin{proof}
First take $F=\varnothing$, $H=0$, and put
$Z=\sum_\ell g(\ell^{K_\ell})$, $\lambda=\operatorname{Law}(Z)$.
\Cref{cor:prime-series-zero-dimension,lem:atomic-series-dimensions} give
the lower local dimension, the packing identities, and an almost surely
constant upper local dimension $d\le1$.
\Cref{lem:arithmetic-diffuse-submeasures} and
\Cref{prop:dominated-diffuse-submeasures}\textup{(c)} give
$\operatorname*{ess\,sup}_{\lambda}\overline d_\lambda=1$.
Since this exponent is almost surely the constant $d$, we have $d=1$.

\Cref{lem:atomic-series-dimensions} preserves the local constants under
the stated atomic-head changes.
The exponential maps are locally bi-Lipschitz, and decomposition into
bounded intervals preserves the packing-carrier dimensions.
Finally, the conditioning identity preceding
\eqref{eq:ap-cylinder-domination} identifies a fixed-progression totient
law with finite-coordinate conditioning, leaving an independent atomic
head and the same tail.

Countably many translates of a Hausdorff-zero carrier give such a
carrier after any stated atomic-head change. If $\nu\ll\lambda$,
differentiation of measures gives
\[
 \frac{\nu(B(x,r))}{\lambda(B(x,r))}
 \longrightarrow\frac{d\nu}{d\lambda}(x)\in(0,\infty)
 \quad(\nu\text{-a.e. }x).
\]
Thus both local exponents agree $\nu$-almost everywhere. The full
Hausdorff-zero carrier is inherited, and every $\nu$-positive Borel set
is $\lambda$-positive, proving both packing assertions.
\end{proof}

For a Borel probability measure $\lambda$ on $\R$, its upper
regularity dimension is the infimum of $s\ge0$ for which
\[
 \frac{\lambda(B(x,R))}{\lambda(B(x,r))}\le C(R/r)^s
 \quad(x\in\operatorname{supp}\lambda,\ 0<r<R),
\]
with a uniform $C$. Its lower regularity dimension is the supremum of
$s\ge0$ for which the reverse inequality holds with a positive uniform
constant, with $R<\operatorname{diam}(\operatorname{supp}\lambda)$;
see~\cite{FH19,How19}.

\begin{corollary}[Regularity dimensions of the measures]
\label{cor:regularity}
\index{Regularity dimensions}
For every law $\lambda$ in \Cref{thm:local-dimension-oscillation},
and every probability measure equivalent to it, the lower regularity
dimension is zero and the upper regularity dimension is infinite.
In particular, none of these measures is doubling.
The supports of all these measures are closed intervals with nonempty
interior, possibly unbounded; as sets, these supports have lower and
Assouad dimension one.
\end{corollary}

\begin{proof}
Restricting further to
$R$ below a fixed positive scale gives the same conclusions here. A positive lower exponent, with
$x,R$ fixed, would imply $\lambda(B(x,r))\le C_xr^s$, contrary to
\Cref{thm:local-dimension-oscillation}. The lower dimension is therefore
zero.

Absorb the factor at $2$ into the atomic head. The remaining positive
atoms $g(p)$ tend to zero with divergent sum, and each local factor has
an atom at zero, so \Cref{lem:supp_conv} gives tail support $[0,\infty)$.
The logarithmic support is the closure of the union of its translates
by the head atoms, hence a ray or $\R$. Exponentiation gives the raw
interval supports. Equivalent measures have the same support and the
same almost-everywhere local dimensions by
\Cref{thm:local-dimension-oscillation}.
Suppose a uniform small-scale doubling constant $D$ existed. At an
interior support point $x$, for all sufficiently small $r$, put
$y=x+3r/4$. Then $B(y,r/8)$ is contained in
$B(x,r)\setminus B(x,r/2)$, while $B(x,r)\subset B(y,2r)$.
Four doublings show that
\[
 \lambda(B(x,r/2))\le(1-D^{-4})\lambda(B(x,r)).
\]
Iteration gives a strictly positive lower local dimension at every
interior $x$. Atomlessness and
\Cref{thm:local-dimension-oscillation} contradict this conclusion.
Finite upper regularity dimension would imply doubling by taking
$R=2r$, so that dimension is infinite. Finally, an interval or ray
requires and admits $\asymp R/r$ balls of radius $r$ inside a ball of
radius $R$ (with $R$ below the diameter in the bounded case), proving
the assertion about the support.
\end{proof}

\section{Comparing the valuation models}
\label{sec:valuation-comparison}
\index{Valuation models!comparison}

Recall from \eqref{eq:valuation-laws} that $\mathsf I$ is the integer
valuation law, $\mathsf B$ is the shifted-prime law, and
$\mathsf A=\mathsf I(\,\cdot\mid J_2\ge1)$. The comparison below applies
to the same measurable map of $\mathsf A$ and $\mathsf B$.
The support and pointwise local-dimension conclusions require this
conditioning at $2$.
\Cref{fig:four-laws} displays the three measures; its integer laws are images of
$\mathsf I$, not of $\mathsf A$.

\begin{lemma}[Comparison of valuation measures]
\label{lem:valuation-likelihood}
For the product measures $\mathsf A,\mathsf B$ defined after
\eqref{eq:valuation-laws}, put
\[
\begin{gathered}
 C_2:=\prod_{\ell>2}\left(1-\frac1{(\ell-1)^2}\right)
 =\frac{\Smer{}}{2},\\
 C_+:=\prod_{\ell>2}\left(1+\frac1{(\ell-1)^3}\right),\qquad
 C_-:=\prod_{\ell>2}\left(1+\frac1{\ell^2(\ell-2)}\right),
\end{gathered}
\]
where the products run over primes.
\begin{enumerate}[label=\textup{(\alph*)}]
\item \emph{Likelihood.} The Radon--Nikodym derivative is
\begin{equation}\label{eq:valuation-likelihood-exact}
 L_{\mathrm{val}}(v):=\frac{d\mathsf B}{d\mathsf A}(v)
 =C_2\prod_{\substack{\ell>2\\v_\ell\ge1}}
       \frac{\ell-1}{\ell-2}.
\end{equation}
The product converges to a positive finite value almost surely under
$\mathsf A$.
\item \emph{Event comparison.} Every measurable valuation event $E$ satisfies
\begin{equation}\label{eq:valuation-tail-sandwich}
 C_-^{-1}\mathsf A(E)^2\le\mathsf B(E)
 \le C_+^{1/2}\mathsf A(E)^{1/2}.
\end{equation}
For every fixed $q>1$, $L_{\mathrm{val}}\in L^q(\mathsf A)$ and
\begin{equation}\label{eq:valuation-allq}
 C_2\mathsf A(E)\le\mathsf B(E)
 \le\|L_{\mathrm{val}}\|_{L^q(\mathsf A)}\mathsf A(E)^{1-1/q}.
\end{equation}
\item \emph{Small events.} For every family of measurable valuation events
with $0<\mathsf A(E_t)\to0$,
\begin{equation}\label{eq:valuation-double-log}
 \frac{\log\mathsf B(E_t)}{\log\mathsf A(E_t)}\longrightarrow1,
 \qquad
 \log\log\frac1{\mathsf B(E_t)}
 =\log\log\frac1{\mathsf A(E_t)}+o(1).
\end{equation}
\item \emph{Reciprocal-totient comparison.} For every measurable valuation
event $E$ and the functional
$R_\varphi(v)=\prod_{v_\ell\ge1}(1-1/\ell)^{-1}$,
\begin{equation}\label{eq:valuation-reciprocal-tilt}
 \frac{C_2}{2}\,\mathbb E_{\mathsf A}[R_\varphi\mathbf1_E]
 \le\mathsf B(E)
 \le\frac12\,\mathbb E_{\mathsf A}[R_\varphi\mathbf1_E].
\end{equation}
\end{enumerate}
\end{lemma}

\begin{proof}
\textup{(a)} The laws at $2$ agree, with probabilities $2^{-k}$ for $k\ge1$.
For odd $\ell$, let $a_\ell(k),b_\ell(k)$ be the integer and shifted
probabilities in \eqref{eq:valuation-laws}. Their ratio is
\[
 r_\ell(k):=\frac{b_\ell(k)}{a_\ell(k)}
 =\begin{cases}
 1-(\ell-1)^{-2},&k=0,\\
 \ell/(\ell-1),&k\ge1.
 \end{cases}
\]
Summing the geometric tails gives
\[
 \sum_{k\ge0}a_\ell(k)r_\ell(k)^2
 =1+\frac1{(\ell-1)^3},\qquad
 \sum_{k\ge0}b_\ell(k)r_\ell(k)^{-2}
 =1+\frac1{\ell^2(\ell-2)}.
\]
The finite-coordinate likelihoods
$L_y=\prod_{2<\ell\le y}r_\ell(v_\ell)$ form a mean-one martingale
under $\mathsf A$, with $\mathbb E_{\mathsf A}L_y^2\le C_+$.
Consequently they converge in $L^2(\mathsf A)$ to a mean-one limit.
The identity
\[
 L_y=\prod_{2<\ell\le y}\left(1-\frac1{(\ell-1)^2}\right)
       \prod_{\substack{2<\ell\le y\\v_\ell\ge1}}
       \frac{\ell-1}{\ell-2}
\]
identifies that limit with \eqref{eq:valuation-likelihood-exact}:
the logarithm of the second product has finite expectation, since
$\mathsf A(v_\ell\ge1)=1/\ell$ for odd $\ell$ and
$\sum_{\ell>2}\ell^{-1}\log((\ell-1)/(\ell-2))<\infty$.
For every cylinder event $E$, $\int_E L_y\,d\mathsf A=\mathsf B(E)$
for all sufficiently large $y$. Passing to the limit and then using
uniqueness of measures proves the Radon--Nikodym identity for all
measurable events.

\textup{(b)} For a cylinder event, Cauchy--Schwarz applied to $L_y$ under $\mathsf A$
and to $L_y^{-1}$ under $\mathsf B$ gives
\eqref{eq:valuation-tail-sandwich}. Cylinder events approximate every
measurable event in $(\mathsf A+\mathsf B)$-measure, so the bounds hold
in general. For fixed $q>1$, the local ratios satisfy
$3/4\le r_\ell\le3/2$, $\mathbb E_{\mathsf A}r_\ell=1$ and
$\mathbb E_{\mathsf A}(r_\ell-1)^2=(\ell-1)^{-3}$.
Taylor's formula therefore gives
$\mathbb E_{\mathsf A}r_\ell^q=1+O_q(\ell^{-3})$.
Independence bounds $\mathbb E_{\mathsf A}L_y^q$ uniformly in $y$;
Fatou's lemma gives $L_{\mathrm{val}}\in L^q(\mathsf A)$.
The exact product gives $L_{\mathrm{val}}\ge C_2$, so H\"older's
inequality proves \eqref{eq:valuation-allq}.

\textup{(c)} Taking the rare-event limit in \eqref{eq:valuation-allq}
for each fixed $q$, and then $q\to\infty$, proves
\eqref{eq:valuation-double-log}.

\textup{(d)} Since $v_2\ge1$ under $\mathsf A$,
\[
 \frac{L_{\mathrm{val}}}{R_\varphi}
 =\frac{C_2}{2}\prod_{\substack{\ell>2\\v_\ell\ge1}}
       \frac{(\ell-1)^2}{\ell(\ell-2)}
 \in[C_2/2,1/2],
\]
because the product over all odd primes equals $C_2^{-1}$.
Integration over $E$ proves \eqref{eq:valuation-reciprocal-tilt}.
\end{proof}

\begin{corollary}[Local dimensions under the valuation likelihood]
\label{cor:valuation-local-dimensions}
Let $F$ be a measurable map from the valuation space to a metric space,
and put $\mu=F_*\mathsf A$, $\nu=F_*\mathsf B$. The measures are equivalent
and have the same support. At every point $x$ of this support,
\[
 \underline d_\mu(x)=\underline d_\nu(x),\qquad
 \overline d_\mu(x)=\overline d_\nu(x),
\]
where the local dimensions use metric balls.
\end{corollary}

\begin{proof}
The measures are equivalent by \eqref{eq:valuation-likelihood-exact}.
At a common atom both local dimensions are zero. Otherwise put
$a_r=-\log\mu(B(x,r))$ and $b_r=-\log\nu(B(x,r))$; then $a_r\to\infty$.
Applying \eqref{eq:valuation-allq} to $F^{-1}(B(x,r))$ gives
\[
 (1-1/q)a_r-O_q(1)\le b_r\le a_r+O(1).
\]
Thus $b_r/a_r\to1$, first letting $r\downarrow0$ and then
$q\to\infty$. Division by $-\log r$ proves both assertions, including
infinite local dimensions.
\end{proof}

\paragraph*{Transfer mechanisms and their limits}
The likelihood comparison gives logarithmic rare-event equivalence and
pointwise local dimensions by \eqref{eq:valuation-allq} and
\Cref{cor:valuation-local-dimensions}. The exact extreme-tail comparison
uses the likelihood-saturation argument of \Cref{thm:exact-extreme-tail-transfer},
whereas comparison with the unconditioned integer endpoint law uses
the head-product ratio and tail estimate in
\eqref{eq:endpoint-integer-transfer}.

Changes to the prime coordinates use three mechanisms.
\begin{itemize}
\item Positive-probability conditioning preserves full-measure events
by domination (\Cref{cor:ap-liouville}) and local dimensions by
differentiation (\Cref{thm:local-dimension-oscillation}). For arbitrary
valuation-measurable conditioning of the basic laws, preservation of
both information dimensions uses finite-entropy coupling and
cylinder approximation (\Cref{cor:arithmetic-entropy-invariance}).
\item Summable total-variation coupling of independent atomic summands
preserves the carrier dimensions, component masses and Rajchman property
listed in \Cref{cor:summable-local-coupling}, but does not in general
transfer quantitative Fourier decay or Fourier dimension
(\Cref{rem:summable-coupling-scope}). Shannon comparison requires the
finite-entropy record of \Cref{cor:arithmetic-entropy-invariance};
Liouville conclusions require rational approximation beyond countable
displacement.
\item Deterministic deletion sets selected prime coordinates to zero.
For totient and divisor-sum products, including mixed local choices,
convergence of the retained reciprocal-prime sum gives an almost surely
rational product; divergence gives an atomless Liouville law.
Both conclusions allow a dependent positive rational-valued head
(\Cref{thm:deterministic-deletion}). For arbitrary adaptive choices,
the simultaneous assertion is that every limit is rational or Liouville
(\Cref{cor:adaptive-deletion}).
\end{itemize}

\paragraph*{Rational approximation and entropy}
\Cref{sec:liouville-carriers} next constructs simultaneous rational
approximations to the raw ratios. Its continued-fraction conclusions
also use the packing dimension established above.
\Cref{sec:information-dimensions} subsequently studies finite-resolution
entropy and quantization. \Cref{sec:information-convolution} combines
the entropy bounds with the rational statistics.

\chapter{Diophantine approximation and rational statistics}
\label{sec:liouville-carriers}

Let $\mathcal L$ be the set of Liouville numbers: $x\in\R\setminus\Q$
belongs to $\mathcal L$ if, for every integer $m\ge1$, there are
$a\in\Z$ and an integer $q\ge2$ with $0<|x-a/q|<q^{-m}$.
Finite products of rational local factors provide rational
approximations. The question is whether the omitted tail is small
relative to the denominator of the finite product. An empty prime
block leaves the prefix and its denominator unchanged while moving the
start of the remaining tail to a much larger cutoff. A denominator bound
for the prefix and a separate bound for that tail then produce the
required approximation scales. Under the hypotheses of
\Cref{lem:rational-blocks}, this gives Liouville numbers whenever the
limits are irrational, even for countably valued local factors,
refining the finite-product coverings in~\cite[\S~3]{Erd39}.
The dimension-zero criterion of \Cref{thm:dimension-zero-criterion}
requires no rationality of the atoms. Arithmetic antecedents include
Erd\H{o}s's assertion, stated without proof, that divergence of the
reciprocal sum of primitive $\alpha$-abundant numbers forces $\alpha$
to be Liouville~\cite[pp.~25--26]{Erd59}, and his lacunary Euler product,
proved Liouville by Kobayashi~\cite[Proposition~5.18]{Kob10}.

The polynomial nonconcentration criterion
\Cref{prop:prime-product-polynomial-nonconcentration} allows dependent
coordinates but does not supply the rational-height estimates used
here. Simultaneous approximation in
\Cref{lem:rational-blocks,lem:joint-rational-carrier} yields
\Cref{thm:liouville-carriers,cor:rational-statistics}.
The prescribed-rate dichotomy is developed in
\Cref{sec:diophantine-fixed-functions}. Continued-fraction recurrence uses
\Cref{thm:local-dimension-oscillation}; \Cref{cor:convolution-information}
gives information bounds for independent logarithmic sums in
\Cref{sec:information-dimensions}.

\section{Simultaneous approximation and joint carriers}
\label{sec:diophantine-simultaneous}

\begin{lemma}[Simultaneous rational approximation]
\label{lem:rational-blocks}
\index{Rational approximation!prime blocks}
Let $d\ge1$, and let $\boldsymbol R_\ell\in\Q_{>0}^d$ be independent
random vectors indexed by primes $\ell>\ell_0$. Let $q_{\ell,j}$ be the
reduced denominator of $R_{\ell,j}$, and let
$\boldsymbol h\in\Q_{>0}^d$ be a random vector on the same probability
space. Suppose that
$\mathbb P(\boldsymbol R_\ell=\boldsymbol1)>0$ for every $\ell>\ell_0$ and,
for some $c_\ast,\beta>0$,
\begin{equation}\label{eq:rational-zero-hazard}
 \Theta(y):=\sum_{\ell_0<\ell\le y}
       -\log\mathbb P(\boldsymbol R_\ell=\boldsymbol1)
 =c_\ast\log\log y+O(1),
\end{equation}
\[
 \sum_{\ell>z}\mathbb E\sum_{j=1}^d|\log R_{\ell,j}|\ll z^{-\beta},
 \qquad
 \mathbb E\sum_{\ell_0<\ell\le y}\sum_{j=1}^d\log q_{\ell,j}\ll\log y.
\]
Put $X_j(y):=h_j\prod_{\ell_0<\ell\le y}R_{\ell,j}$.
The limits $X_j:=\lim_{y\to\infty}X_j(y)$ exist and are finite and
positive almost surely. Almost surely, for every integer $m\ge1$ there
are cutoffs $y_n\to\infty$ and integers $Q_n\to\infty$ with
$Q_nX_j(y_n)\in\Z$ for every $j$, such that
\[
 \max_j|X_j-X_j(y_n)|<Q_n^{-m}.
\]
\end{lemma}

\begin{proof}
It suffices to treat each fixed rational $\boldsymbol h$ and intersect
the resulting probability-one events over $\Q_{>0}^d$; fix $\boldsymbol h$.
Tonelli's theorem gives almost sure absolute convergence
of the logarithmic products. Choose $Q_0\in\N$ divisible by the
denominators of the $h_j$, and put
\[
 Q(y):=Q_0\prod_{\ell_0<\ell\le y}\prod_j q_{\ell,j},\qquad
 T(y):=\sum_{\ell>y}\sum_j|\log R_{\ell,j}|.
\]
Then $Q(y)X_j(y)\in\Z$ for every $j$, and $\mathbb E\log Q(y)\ll\log y$.
Moreover, $Q(y)\to\infty$ almost surely:
\eqref{eq:rational-zero-hazard} implies
$\sum_\ell\mathbb P(\boldsymbol R_\ell\ne\boldsymbol1)=\infty$
(otherwise these probabilities tend to zero and their negative
logarithmic complements have finite sum). The second Borel--Cantelli
lemma therefore makes $\boldsymbol R_\ell\ne\boldsymbol1$ for infinitely many $\ell$;
absolute convergence allows only finitely many of these to lie in
$\N^d$, since such vectors have $\sum_j|\log R_{\ell,j}|\ge\log2$, and
each of the others has some $q_{\ell,j}\ge2$.

Fix an integer $m\ge1$, put $t_n:=n^{1/c_\ast}$ and
$a_n:=\lceil4mt_n/\beta\rceil$, choose an integer
$u_1>\max(\ell_0,2)$, and set $y_n:=u_n^{a_n}$,
$u_{n+1}:=y_n+1$,
\[
 A_n:=\{\log Q(u_n)\le t_n\log u_n\},\qquad
 E_n:=\{\boldsymbol R_\ell=\boldsymbol1\ (u_n<\ell\le y_n)\}.
\]
The $E_n$ are independent, $A_n$ is independent of $E_n$, and
\eqref{eq:rational-zero-hazard} gives
\[
 \mathbb P(E_n)=\exp\{-\Theta(y_n)+\Theta(u_n)\}
 \asymp a_n^{-c_\ast}\asymp_m n^{-1}.
\]
Markov's inequality gives $\mathbb P(A_n^c)\ll1/t_n$, hence
$\mathbb P(A_n^c\cap E_n)\ll_m n^{-1-1/c_\ast}$, and
$\mathbb P(T(y_n)>y_n^{-\beta/2})\ll y_n^{-\beta/2}$, which is summable.
By the two Borel--Cantelli lemmas, almost surely
$A_n\cap E_n\cap\{T(y_n)\le y_n^{-\beta/2}\}$ occurs for infinitely many
$n$, say $n_1<n_2<\cdots$. For such $n$, $E_n$ gives $Q(y_n)=Q(u_n)$ and
$A_n$ gives $Q(y_n)\le u_n^{t_n}\le y_n^{\beta/(4m)}$. Since
$|\log(X_j/X_j(y_n))|\le T(y_n)$, once $n$ is large,
\[
 \max_j|X_j-X_j(y_n)|\le\max_jX_j\bigl(e^{y_n^{-\beta/2}}-1\bigr)
 <y_n^{-\beta/4}\le Q(y_n)^{-m}.
\]
After discarding finitely many $n_k$, the cutoffs $y_{n_k}$ and
$Q_k:=Q(y_{n_k})\to\infty$ satisfy the assertion for this $m$;
intersecting over $m$ completes the proof.
\end{proof}

After finitely many factors have been absorbed into the head
$\boldsymbol h$, the zero-hazard hypothesis
\eqref{eq:rational-zero-hazard} holds whenever
$\mathbb P(\boldsymbol R_\ell\ne\boldsymbol1)
=c_\ast/\ell+O(\ell^{-2})$.
This follows from $-\log(1-u)=u+O(u^2)$ and Mertens' theorem.

\begin{lemma}[Consequences of simultaneous rational approximation]\label{lem:joint-rational-carrier}
Let $\boldsymbol X\in\R^d$ satisfy the following condition almost surely:
for every integer $m\ge1$ there are $\boldsymbol a_n\in\Z^d$ and
$Q_n\in\N$ with $Q_n\to\infty$ and $\max_j|X_j-a_{n,j}/Q_n|<Q_n^{-m}$.
(For $\boldsymbol X\in\Q^d$ this holds with $Q_n$ the multiples of a
common denominator.)
\begin{enumerate}[label=\textup{(\alph*)}]
\item The law of $\boldsymbol X$ has a Borel full carrier of Hausdorff
dimension zero. If all coordinates are positive almost surely, its
logarithmic image also has such a carrier.
\item Almost surely, $R(\boldsymbol X)\in\Q\cup\mathcal L$
simultaneously for every $R\in\Q(T_1,\ldots,T_d)$ whose denominator does
not vanish at $\boldsymbol X$.
\end{enumerate}
\end{lemma}

\begin{proof}
In a fixed bounded box there are $O(q^d)$ rational centers with common
denominator $q$. Balls of radius $\sqrt d\,q^{-m}$ have summable
$s$-dimensional content when $ms>d+1$. Their limsup therefore has zero
$\mathcal H^s$-measure and, since $Q_n\to\infty$, almost surely contains
$\boldsymbol X$ whenever $\boldsymbol X$ is interior to the box. Intersecting
over $m$ and exhausting $\R^d$ by boxes gives the required Borel full carrier.
\Cref{lem:lipschitz-image} gives the logarithmic conclusion on $(0,\infty)^d$.

The constant case is immediate. Otherwise write $R=P/S$ with integer
polynomials and
$D=\max(\deg P,\deg S)\ge1$. At a non-pole, $R$ is locally Lipschitz.
Approximations of order $M>Dm+1$ give error $O(Q^{-M})$, while the
reduced denominator of $R(\boldsymbol a/Q)$ is at most $O(Q^D)$,
because $Q^D S(\boldsymbol a/Q)$ is a nonzero integer eventually.
If $R(\boldsymbol X)$ is irrational, these errors are nonzero and the
reduced denominators tend to infinity, so $R(\boldsymbol X)\in\mathcal L$.
Countability of the rational functions and orders gives the
simultaneous conclusion.
\end{proof}

\begin{corollary}[Rational images under polynomial nonconcentration]
\label{cor:nonconcentrated-rational-images}
Let $\boldsymbol X\in\R^d$ satisfy the almost-sure approximation
hypothesis of \Cref{lem:joint-rational-carrier}.
\begin{enumerate}[label=\textup{(\alph*)}]
\item \emph{Real coefficients.} Let $\boldsymbol Z$ be $\boldsymbol X$,
or $F(\boldsymbol X)$ for a locally Lipschitz map $F\colon U\to\R^{d'}$ on an open set $U\subset\R^d$
with $\mathbb P(\boldsymbol X\in U)=1$; for example
$\boldsymbol Z=\pm(\log X_1,\ldots,\log X_d)$ if all coordinates are
positive almost surely. Suppose that $B(\boldsymbol Z)\ne0$ almost
surely for each fixed nonzero real polynomial $B$ in as many variables
as $\boldsymbol Z$ has coordinates. Then, for each fixed nonconstant
real rational function $R$ in these variables, $R(\boldsymbol Z)$ is
defined almost surely; its law is atomless and has a Borel full carrier
of Hausdorff dimension zero, and hence is singular continuous with
Fourier dimension zero.
\item \emph{Rational coefficients.} Suppose that $B(\boldsymbol X)\ne0$
almost surely for each fixed nonzero $B\in\Q[T_1,\ldots,T_d]$. Then, almost surely,
$X_1,\ldots,X_d$ are algebraically independent over $\Q$, and
$R(\boldsymbol X)\in\mathcal L$ simultaneously for every nonconstant
$R\in\Q(T_1,\ldots,T_d)$.
\end{enumerate}
\end{corollary}

\begin{proof}
\textup{(a)} The law of $\boldsymbol Z$ has a Borel full carrier of
Hausdorff dimension zero, by \Cref{lem:joint-rational-carrier}\textup{(a)}
and, when $\boldsymbol Z=F(\boldsymbol X)$, by \Cref{lem:lipschitz-image}. Write
$R=P/S$ with real polynomials $P$ and $S\ne0$. By hypothesis
$S(\boldsymbol Z)\ne0$ almost surely, so the open set $\{S\ne0\}$ has
full measure under the law of $\boldsymbol Z$ and $R(\boldsymbol Z)$ is
defined almost surely. For each $t\in\R$ the polynomial $P-tS$ is
nonzero, because $R$ is nonconstant; hence
$\mathbb P(R(\boldsymbol Z)=t)=0$. Since $R$ is locally Lipschitz on
$\{S\ne0\}$, \Cref{lem:lipschitz-image} gives the carrier, which is
Lebesgue-null, and \Cref{lem:fourier-Hausdorff} gives Fourier dimension
zero.

\textup{(b)} Countability gives a probability-one event on which
$B(\boldsymbol X)\ne0$ for every nonzero $B\in\Q[T_1,\ldots,T_d]$; this
is algebraic independence. On this event, for
every nonconstant $R=P/S$ with $P,S\in\Q[T_1,\ldots,T_d]$, the
denominator does not vanish at $\boldsymbol X$, and
$R(\boldsymbol X)\ne t$ for every $t\in\Q$, since $P-tS$ is a nonzero
rational polynomial. Thus $R(\boldsymbol X)\notin\Q$, and
\Cref{lem:joint-rational-carrier}\textup{(b)} gives $R(\boldsymbol X)\in\mathcal L$.
\end{proof}

\KeepHeadingSpace{6\baselineskip}
\begin{proposition}[Polynomial nonconcentration for prime products]
\label{prop:prime-product-polynomial-nonconcentration}
\index{Prime products!polynomial nonconcentration}
Let $d\ge1$, and let $\boldsymbol H$ and
$(\boldsymbol R_p)_{p>p_0}$ be mutually independent random vectors in
$(0,\infty)^d$, each with a countably atomic law; the indices $p$ are
primes. Suppose that the coordinatewise product
\[
 \boldsymbol X=\boldsymbol H\prod_{p>p_0}\boldsymbol R_p
\]
converges almost surely to a vector in $(0,\infty)^d$.
Suppose there are constants $0<\beta\le\kappa$ and deterministic
vectors $\boldsymbol r_p\in(0,\infty)^d$ such that, as $p\to\infty$,
\begin{equation}\label{eq:polynomial-nonconcentration-states}
 a_p:=\mathbb P(\boldsymbol R_p=\boldsymbol1)
       =1-\frac\kappa p+O(p^{-2}),\qquad
 b_p:=\mathbb P(\boldsymbol R_p=\boldsymbol r_p)
       =\frac\beta p+O(p^{-2}).
\end{equation}
Assume also that, for every $\boldsymbol m\in\Z^d\setminus\{0\}$,
there is a threshold $p(\boldsymbol m)$ such that
\begin{equation}\label{eq:polynomial-nonconcentration-separation}
 \boldsymbol r_p^{\boldsymbol m}\ne1,\qquad
 \boldsymbol r_p^{\boldsymbol m}\ne\boldsymbol r_q^{\boldsymbol m}
 \quad\text{for distinct primes }p,q>p(\boldsymbol m),
 \qquad
 \boldsymbol r_p^{\boldsymbol m}:=\prod_{j=1}^d r_{p,j}^{m_j}.
\end{equation}
Then, for every fixed nonzero $P\in\R[T_1,\ldots,T_d]$,
$\mathbb P(P(\boldsymbol X)=0)=0$.
Consequently the coordinates of $\boldsymbol X$ are algebraically
independent over $\Q$ almost surely. Every fixed nonconstant
$R\in\R(T_1,\ldots,T_d)$ is defined at $\boldsymbol X$ almost surely
and has an atomless image law.
\end{proposition}
\begin{proof}
Products, quotients and powers of vectors below are coordinatewise.
Suppose that the first conclusion fails. Among the nonzero real
polynomials whose zero sets have positive probability under
$\boldsymbol X$, choose one, $P$, with the smallest number $s$ of
nonzero monomials. Since all coordinates of $\boldsymbol X$ are
strictly positive, a nonzero monomial never vanishes, so $s\ge2$.

For each cutoff $M>p_0$, put
\[
 \boldsymbol H_M=\boldsymbol H\prod_{p_0<p\le M}\boldsymbol R_p,
 \qquad \boldsymbol W_M=\prod_{p>M}\boldsymbol R_p.
\]
The tail converges to a positive finite vector almost surely, and
$\boldsymbol H_M$ is countably atomic and independent of
$\boldsymbol W_M$. The tail charges no zero set of a nonzero real
polynomial with fewer than $s$ monomials. Indeed, if such a polynomial
$Q$ vanished at $\boldsymbol W_M$ with positive probability, choose
any atom $\boldsymbol h$ of $\boldsymbol H_M$ with positive mass.
Independence would give positive probability to
$Q(\boldsymbol h^{-1}\boldsymbol X)=0$. Positive diagonal scaling
preserves the number of nonzero monomials, contradicting the choice
of $s$.

Write $P(\boldsymbol x)=\sum_{\alpha\in A}c_\alpha\boldsymbol x^\alpha$,
where $A\subset\mathbb N_0^d$, $|A|=s$ and every $c_\alpha\ne0$.
Choose distinct $\alpha_0,\alpha_1\in A$, and put
$\boldsymbol m=\alpha_1-\alpha_0$. Take $N>p(\boldsymbol m)$
sufficiently large that $a_p>0$ and $0<b_p/a_p\le1$ for $p>N$.
Let $M=M(N)>N$ be the smallest prime cutoff for which
$\sum_{N<p\le M}p^{-1}\ge1$. Divergence of the reciprocal-prime sum
ensures its existence; the last summand is at most $1/N$, so
\[
 \sum_{N<p\le M}\frac1p=1+O(N^{-1}),\qquad
 \sum_{N<p\le M}\frac1{p^2}=O(N^{-1}).
\]
In this block, let $w_0$ be the probability that every factor is
$\boldsymbol1$, and let $w_p$ be the probability that the factor at
$p$ is $\boldsymbol r_p$ and all the other factors are
$\boldsymbol1$. These configurations are disjoint, since
\eqref{eq:polynomial-nonconcentration-separation} implies
$\boldsymbol r_p\ne\boldsymbol1$. Independence gives
\begin{equation}\label{eq:polynomial-nonconcentration-block-mass}
 \begin{split}
 w_0&=\prod_{N<p\le M}a_p=e^{-\kappa}+O(N^{-1}),\\
 S_N:=\sum_{N<p\le M}w_p
     &=w_0\sum_{N<p\le M}\frac{b_p}{a_p}
       =\beta e^{-\kappa}+O(N^{-1}),\qquad w_p\le w_0.
 \end{split}
\end{equation}
Here $\log a_p=-\kappa/p+O(p^{-2})$ and
$b_p/a_p=\beta/p+O(p^{-2})$; the constants may depend on the fixed
local-law bounds, $\kappa$ and $\beta$.

Fix a positive-probability atom $\boldsymbol h$ of
$\boldsymbol H_N$. Under the independent law of $\boldsymbol W_M$,
consider the events
\[
 E_{\boldsymbol v}(\boldsymbol h)
 =\{P(\boldsymbol h\boldsymbol v\boldsymbol W_M)=0\},
 \qquad
 \boldsymbol v\in\{\boldsymbol1\}\cup
          \{\boldsymbol r_p:N<p\le M\}.
\]
They are pairwise disjoint modulo null sets. For distinct branches
$\boldsymbol v,\boldsymbol w$, form
\[
 Q_{\boldsymbol v,\boldsymbol w}(\boldsymbol x)
 =P(\boldsymbol h\boldsymbol v\boldsymbol x)
  -\frac{\boldsymbol v^{\alpha_0}}{\boldsymbol w^{\alpha_0}}
        P(\boldsymbol h\boldsymbol w\boldsymbol x).
\]
Its $\alpha_0$ coefficient is zero, whereas its $\alpha_1$ coefficient is
\[
 c_{\alpha_1}\boldsymbol h^{\alpha_1}
 \boldsymbol v^{\alpha_0}
 (\boldsymbol v^{\boldsymbol m}-\boldsymbol w^{\boldsymbol m})\ne0
\]
by \eqref{eq:polynomial-nonconcentration-separation}, including when
one branch is $\boldsymbol1$. Thus $Q_{\boldsymbol v,\boldsymbol w}$
is nonzero with fewer than $s$ monomials. The intersection of the two
events is contained in its tail zero set, which has probability zero
by the preceding minimality argument.

Set $A_P=\{P(\boldsymbol X)=0\}$ and
$\mathcal F_N=\sigma(\boldsymbol H,(\boldsymbol R_p)_{p_0<p\le N})$.
Conditioning on these finitely many independent factors leaves the
block and its tail independent with their original laws. The head
values form a countable set. For every positive-probability head value,
pairwise disjointness gives $\sum_{\boldsymbol v}
\mathbb P(E_{\boldsymbol v}(\boldsymbol h))\le1$.
The block configurations other than the displayed branches have total
mass $1-w_0-S_N$. Since all displayed branch weights are at most $w_0$,
\[
 \mathbb P(A_P\mid\mathcal F_N)
 \le 1-w_0-S_N+w_0
       \sum_{\boldsymbol v}\mathbb P(E_{\boldsymbol v}(\boldsymbol H_N))
 \le1-S_N
 \le1-\tfrac12\beta e^{-\kappa}<1
\]
almost surely for every sufficiently large integer $N$.
The event $A_P$ is measurable with respect to
$\sigma(\bigcup_N\mathcal F_N)$, because $\boldsymbol X$ is the
almost-sure limit of the finite products. Martingale convergence
therefore gives
$\mathbb P(A_P\mid\mathcal F_N)\to\mathbf1_{A_P}$ almost surely.
The uniform strict upper bound contradicts $\mathbb P(A_P)>0$.

Intersecting the resulting probability-one events over the countably
many nonzero rational polynomials proves algebraic independence over
$\Q$. For a fixed nonconstant real rational function $R=P/S$ with
$S\ne0$, the first conclusion gives
$\mathbb P(S(\boldsymbol X)=0)=0$. For each fixed $t\in\R$, $P-tS$
is a nonzero polynomial, so
$\mathbb P(R(\boldsymbol X)=t)=0$. This proves atomlessness and
almost-sure definition of the rational image.
\end{proof}
Every realized vector $\boldsymbol x$ is a zero of the nonzero real
polynomial $T_1-x_1$, so the real-polynomial assertion cannot hold
simultaneously for all such polynomials.
The proposition imposes no rationality or denominator-height
condition. If its factors and head also satisfy
\Cref{lem:rational-blocks}, then
\Cref{cor:nonconcentrated-rational-images} supplies the
Hausdorff-zero rational images and simultaneous rational-coefficient
Liouville conclusions.

A real-valued arithmetic function $F$ is \emph{nonclustering} if, for
each fixed $u\in\R$, the upper density of
$\{n:|F(n)-u|<\delta\}$ tends to zero as $\delta\downarrow0$.
Lebowitz-Lockard and Pollack~\cite[Theorem~1]{LLP17} give an arithmetic
nonclustering criterion for linear combinations with nonzero coefficients
of nowhere-zero real-valued multiplicative functions: one summand and every pairwise
quotient must be nonclustering, and no summand may cluster at zero.
Their argument reduces the number of summands.
The proposition above instead checks selected local states of an
independent prime product and reduces the number of monomials.
It supplies a probabilistic sufficient criterion for polynomial zero
sets without assuming nonclustering of arithmetic quotient functions.

\begin{corollary}[Joint law of the totient and divisor-sum ratios]
\label{cor:joint-phi-sigma}
\index{Joint laws!totient and divisor-sum ratios}
Let the $K_\ell$ be independent, with either the integer laws
$\operatorname{Law}(J_\ell)$ or the shifted-prime laws
$\operatorname{Law}(V_\ell)$ of \eqref{eq:valuation-laws}, and put
\[
 \Phi:=\prod_\ell(1-1/\ell)^{\mathbf1_{\{K_\ell>0\}}},\qquad
 \Sigma:=\prod_\ell\frac{\sigma(\ell^{K_\ell})}{\ell^{K_\ell}}.
\]
\begin{enumerate}[label=\textup{(\alph*)}]
\item The empirical laws of $(\varphi(n)/n,\sigma(n)/n)$ over integers,
respectively of $(\varphi(p-1)/(p-1),\sigma(p-1)/(p-1))$ over primes,
converge to the corresponding law of $(\Phi,\Sigma)$.
\item The joint law and its logarithmic image have Borel full carriers
of Hausdorff dimension zero. Almost surely, $(\Phi,\Sigma)$ has the
simultaneous rational prefix approximations of \Cref{lem:rational-blocks}.
\item For each fixed nonconstant $R\in\R(T_1,T_2)$, the variable
$R(\Phi,\Sigma)$ is defined almost surely and has a singular continuous
law of Hausdorff and Fourier dimension zero.
\item Almost surely, $\Phi,\Sigma$ are algebraically independent over
$\Q$, and $R(\Phi,\Sigma)\in\mathcal L$ simultaneously for every
nonconstant $R\in\Q(T_1,T_2)$.
\end{enumerate}
\end{corollary}

\begin{proof}
For fixed real $a,b$, the additive function
$a\log(\varphi(n)/n)+b\log(\sigma(n)/n)$ has prime value
$(b-a)/\ell+O_{a,b}(\ell^{-2})$, so its three series
\eqref{eq:ew-three} converge. In either model the local terms of
$\log \Phi$ and $\log \Sigma$ have expected absolute value $O(\ell^{-2})$, so
these products converge almost surely. The Erd\H{o}s--Wintner
theorem~\cite{EW39} over the integers and \Cref{thm:pure-type} over the
shifted primes give limiting characteristic functions equal to the Euler
products of the local characteristic functions induced by
\eqref{eq:valuation-laws}; thus each scalar
projection converges in law to $a\log \Phi+b\log \Sigma$ in the corresponding
model. The Cram\'er--Wold device and exponentiation give
part~\textup{(a)}.

For either model, \Cref{lem:rational-blocks} applies with $d=2$,
$\ell_0=2$, $\boldsymbol h$ the factor vector at $2$, and
$c_\ast=\beta=1$: at an odd prime $\ell$ the factor vector differs from
$(1,1)$ with probability $\ell^{-1}+O(\ell^{-2})$, the absolute
logarithms of its coordinates have expectation $O(\ell^{-2})$, and the
logarithm of the product of its reduced denominators is at most
$(\mathbf1_{\{K_\ell>0\}}+K_\ell)\log\ell$, of expectation
$O((\log\ell)/\ell)$. With \Cref{lem:joint-rational-carrier}, this
gives part~\textup{(b)}.

To prove polynomial nonconcentration, put
\[
 \boldsymbol R_\ell=
 \left((1-1/\ell)^{\mathbf1_{\{K_\ell>0\}}},
       \frac{\sigma(\ell^{K_\ell})}{\ell^{K_\ell}}\right),\qquad
 \boldsymbol r_\ell=(1-1/\ell,1+1/\ell).
\]
The local vectors and the factor at $2$ are independent and countably
atomic. For odd $\ell$,
the neutral state is $K_\ell=0$ and the selected state is $K_\ell=1$;
\eqref{eq:valuation-laws} gives
$a_\ell=1-\ell^{-1}+O(\ell^{-2})$ and
$b_\ell=\ell^{-1}+O(\ell^{-2})$ in either model.
For every fixed $(a,b)\in\Z^2\setminus\{(0,0)\}$,
\[
 \frac{d}{dt}\log\bigl((1-1/t)^a(1+1/t)^b\bigr)
 =\frac{(a-b)t+(a+b)}{t(t^2-1)}\qquad(t>1).
\]
The numerator is not identically zero, so this derivative has a fixed
nonzero sign for all sufficiently large $t$. The expression inside
the logarithm tends to $1$; eventual strict monotonicity therefore
makes its values distinct and different from $1$.
\Cref{prop:prime-product-polynomial-nonconcentration}, with
$d=2$ and $\kappa=\beta=1$, now shows that every fixed nonzero real
polynomial has a null zero set under $\boldsymbol X=(\Phi,\Sigma)$.

By part~\textup{(b)}, $\boldsymbol X$ satisfies the approximation
hypothesis of \Cref{lem:joint-rational-carrier}. The polynomial
zero-set assertion just proved is the hypothesis of
\Cref{cor:nonconcentrated-rational-images}\textup{(a)} with
$\boldsymbol Z=\boldsymbol X$ and, restricted to rational coefficients,
of its part~\textup{(b)}. These give parts~\textup{(c)} and~\textup{(d)}.
\end{proof}

Quantitative joint Mellin estimates for prime sampling, including
progressions, are given in \Cref{cor:joint-prime-mellin}.

\section{Liouville laws and independent rational statistics}
\label{sec:diophantine-qualitative}

\begin{theorem}[The ratio variables are almost surely Liouville numbers]
\label{thm:liouville-carriers}
\index{Liouville numbers!arithmetic laws}
\index{Carriers!Liouville}
The shifted-prime and integer totient and divisor-sum laws satisfy
\[
 \mu_G(\mathcal L)=\muH(\mathcal L)
 =\mu_F^{\mathrm{int}}(\mathcal L)
 =\mu_\sigma^{\mathrm{int}}(\mathcal L)=1.
\]
All four laws and their logarithmic images are singular continuous
and have Hausdorff dimension zero.
\end{theorem}

\begin{proof}
Apply \Cref{cor:joint-phi-sigma}\textup{(d)} to the two coordinate maps
in both models; the marginals are the four stated laws.
The one-dimensional rational-center covering in the proof of
\Cref{lem:joint-rational-carrier} gives $\dim_H\mathcal L=0$.
The logarithm is locally Lipschitz on $(0,\infty)$, so
\Cref{lem:lipschitz-image} gives dimension zero also for the logarithmic
images. Atomlessness follows from \Cref{cor:joint-phi-sigma}\textup{(c)}
and injectivity of the logarithm. This recovers \Cref{thm:hausdorff-zero}.
\end{proof}

The full-measure sets of Hausdorff dimension zero for the joint law and
its logarithmic image in \Cref{cor:joint-phi-sigma} follow from simultaneous
approximation; dimension zero of the marginal laws alone is insufficient.

Rational functions on common Liouville sets are treated in~\cite{KTW15};
for the independent models below, \Cref{lem:rational-blocks} gives
simultaneous rational approximations by finite products.

\begin{corollary}[Rational functions of independent limit variables]
\label{cor:rational-statistics}
\index{Rational statistics}
\index{Carriers!joint laws}
Let $d\ge1$ be fixed, and let $X_1,\ldots,X_d$ be independent, each with
one of the four laws in \Cref{thm:liouville-carriers}. Put
$\boldsymbol X:=(X_1,\ldots,X_d)$ and
$\log\boldsymbol X:=(\log X_1,\ldots,\log X_d)$.
\begin{enumerate}[label=\textup{(\alph*)}]
\item For $\boldsymbol Z\in\{\boldsymbol X,\log\boldsymbol X\}$, the law
of $\boldsymbol Z$ has a Borel full carrier of Hausdorff dimension zero,
and so has the law of $F(\boldsymbol Z)$ whenever $U\subset\R^d$ is open,
$\mathbb P(\boldsymbol Z\in U)=1$ and
$F\colon U\to\R^{d'}$ is locally Lipschitz.
\item For each fixed nonconstant $R\in\R(T_1,\ldots,T_d)$,
$R(\boldsymbol X)$ and $R(\log\boldsymbol X)$ are defined almost surely.
Each image law $\lambda$ is singular continuous, has Fourier dimension
zero, and satisfies
\[
 \underline d_\lambda=0,\qquad\overline d_\lambda=1
 \quad(\lambda\text{-a.e.}),\qquad
 P_{\mathrm{pos}}(\lambda)=P_{\mathrm{full}}(\lambda)=1.
\]
\item Almost surely, $X_1,\ldots,X_d$ are algebraically independent over
$\Q$, and $R(\boldsymbol X)\in\mathcal L$ simultaneously for every
nonconstant $R\in\Q(T_1,\ldots,T_d)$.
\end{enumerate}
\end{corollary}

\begin{proof}
Combine independent copies of the local models into one vector at each
odd prime. The probability that this vector of factors differs from
$\boldsymbol1$ is $d/\ell+O_d(\ell^{-2})$. Summing the bounds for the
individual coordinates gives the required bounds on logarithms of
denominators and on expected absolute logarithmic tails. Including the
independent rational factors at $2$,
\Cref{lem:rational-blocks,lem:joint-rational-carrier,lem:lipschitz-image}
give the full-measure sets of Hausdorff dimension zero asserted in
part~\textup{(a)} and show that each rational function with rational
coefficients takes a rational or Liouville value almost surely wherever
its denominator is nonzero.

Independence and atomlessness imply that every fixed nonzero real
polynomial vanishes at $\boldsymbol X$ with probability zero. Induct on
$d$, viewing it as a polynomial in the last variable. A nonzero
coefficient polynomial is nonzero almost surely by induction; conditional
on the other coordinates, finitely many roots each have zero mass.
The same argument applies to the independent atomless logarithms.
Apply \Cref{cor:nonconcentrated-rational-images}\textup{(a)} with
$\boldsymbol Z=\boldsymbol X$ and with $\boldsymbol Z=\log\boldsymbol X$,
using the locally Lipschitz map $\log$ on $(0,\infty)^d$.
Each image in part~\textup{(b)} is therefore defined almost surely,
and its law is atomless, has a Borel full carrier of Hausdorff dimension
zero, and is singular with Fourier dimension zero.

For either raw or logarithmic coordinates, choose a variable on which
$R=P/S$ depends, say the last. The nonzero polynomial
$S\partial_dP-P\partial_dS$ and the preceding null-set argument show
that almost every specialization of the other coordinates leaves a
nonconstant one-variable rational function. If a Borel set $A$ has
positive image measure, Fubini gives such a specialization whose
preimage of $A$ has positive last-coordinate measure. Outside the
finitely many poles and critical points, cover the line by
countably many compact intervals on which the specialized map is
bi-Lipschitz. One preimage piece has positive measure and hence packing
dimension one by \Cref{thm:local-dimension-oscillation}; its image lies
in $A$. Thus $P_{\mathrm{pos}}(\lambda)=P_{\mathrm{full}}(\lambda)=1$.
The covering and disjoint-ball arguments in the proof of
\Cref{lem:atomic-series-dimensions}, applied to countably many bounded
sets with a common radius threshold, give lower local dimension zero
from the Hausdorff-zero carrier and upper local dimension one from this
positive-mass packing assertion.

The null-set argument for $\boldsymbol X$ covers in particular every
nonzero $B\in\Q[T_1,\ldots,T_d]$; this is the hypothesis of
\Cref{cor:nonconcentrated-rational-images}\textup{(b)}, which gives
part~\textup{(c)}.
\end{proof}

\begin{corollary}[Independent pairs of totient and divisor-sum ratios]
\label{cor:independent-phi-sigma-pairs}
\index{Joint laws!independent pairs}
Let $(\Phi_i,\Sigma_i)$, $1\le i\le n$, be independent pairs with the
joint laws of \Cref{cor:joint-phi-sigma}, choosing either valuation model
for each pair, and put
$\boldsymbol X=(\Phi_1,\Sigma_1,\ldots,\Phi_n,\Sigma_n)$.
The vector has the simultaneous rational prefix approximations of
\Cref{lem:rational-blocks}. Its law and its logarithmic image have
Borel full carriers of Hausdorff dimension zero, as do their locally
Lipschitz images on open sets of full measure.
Every fixed nonconstant $R\in\R(T_1,\ldots,T_{2n})$ is defined at
$\boldsymbol X$ almost surely, and $R(\boldsymbol X)$ has a singular
continuous law of Hausdorff and Fourier dimension zero.
Almost surely, all $2n$ coordinates are algebraically independent over
$\Q$, and $R(\boldsymbol X)\in\mathcal L$ simultaneously for every
nonconstant $R\in\Q(T_1,\ldots,T_{2n})$.
\end{corollary}

\begin{proof}
Combine the independent pairs into one local vector at each odd prime.
Its activation probability is $n/\ell+O_n(\ell^{-2})$; the denominator
and tail estimates in \Cref{cor:joint-phi-sigma} add. Thus
\Cref{lem:rational-blocks,lem:joint-rational-carrier,lem:lipschitz-image}
give the approximation and carrier assertions.
For a nonzero real polynomial, induct on the number of pairs, viewing
it as a polynomial in the last two coordinates. A nonzero coefficient
polynomial survives almost surely by induction. Conditional on the
earlier pairs, the polynomial zero-set assertion proved in
\Cref{cor:joint-phi-sigma} applies to the independent last pair.
Thus every fixed nonzero real polynomial is nonzero at $\boldsymbol X$
almost surely. Together with the approximations obtained above, this
verifies the hypotheses of \Cref{cor:nonconcentrated-rational-images}\textup{(a)}
with $\boldsymbol Z=\boldsymbol X$ and, restricted to rational coefficients,
of its part~\textup{(b)}; these parts give the remaining assertions.
\end{proof}

Independent samples permit
conditioning on all but one coordinate in the polynomial zero-set
argument. The totient and divisor-sum ratios of one input instead share
their valuation coordinates: \Cref{cor:joint-phi-sigma} verifies the
selected-state criterion of
\Cref{prop:prime-product-polynomial-nonconcentration} for this joint
model. Conditioning on independent pairs then gives
\Cref{cor:independent-phi-sigma-pairs}. For totient
tuples at the fixed affine forms of \Cref{thm:fixed-shift-tuple}, the coordinates at a prime
share a reduced residue. The domination argument in
\Cref{cor:fixed-shift-rational-statistics} transfers the required
full-measure events from an independent reference model. Normality and Rajchman conclusions require the
additional analytic input used in
\Cref{cor:normal-liouville-field,lem:rational-orbit-domination}.
\Cref{fig:reference-models} displays the reference model and the direction of the
transfer, together with the separate inputs of the entropy conclusions and of the
quantitative Fourier rates for the same tuples.

\Cref{fig:rational-nonconcentration} records how the arguments of
\Cref{sec:diophantine-simultaneous,sec:diophantine-qualitative} fit together:
simultaneous rational approximation and polynomial nonconcentration are separate
inputs, needed jointly in \Cref{cor:nonconcentrated-rational-images}, and
nonconcentration is established by different arguments for the same-input pair and
for independent coordinates.

% Diagram (design brief F3): Rational approximation and nonconcentration: two different inputs.
\begin{figure}[t]
\centering
\begin{tikzpicture}[figdiagram,
  % local: centre box lines with flush glue, so that short lines keep
  % their natural math spacing (align=center would stretch relations)
  fbox/.append style={align=flush center}]
% Grid (pt). Outer edges x=0 and x=383 are shared by the two lanes, the
% dotted group frame and the bottom row.  Lanes: [0,184.5] and [198.5,383]
% (gutter 14pt, centre line x=191.5).  Fork: PA [198.5,287.25],
% PB [294.25,383].  Inside the frame the two result boxes are inset by 5pt.
% ---- lane tags (text aligned with the box frames below)
\node[ftag, fill=none, inner xsep=0pt, anchor=south west] (tl) at (0pt,1.5pt)
  {Simultaneous rational approximation};
\node[ftag, fill=none, inner xsep=0pt, anchor=south west] (tr) at (198.5pt,1.5pt)
  {Polynomial nonconcentration};
% ---- left lane: Lemma 4.1 -> Lemma 4.2
\node[finp, anchor=north west, text width=177.5pt] (L0) at (0pt,0pt)
  {Independent rational factors $\boldsymbol R_\ell\in\Q_{>0}^d$;\\
   zero hazard \eqref{eq:rational-zero-hazard}; in mean, absolute log-tails\\
   $\ll z^{-\beta}$ and summed log-denominators $\ll\log y$};
\node[fres, anchor=north west, text width=177.5pt] (L1) at ($(L0.south west)+(0,-14pt)$)
  {a.s., for every $m\ge1$: $y_n,Q_n\to\infty$ with\\
   $Q_nX_j(y_n)\in\Z$, $\max_j|X_j-X_j(y_n)|<Q_n^{-m}$\\
   \figref{\Cref{lem:rational-blocks}}};
\node[fres, anchor=north west, text width=177.5pt] (L2) at ($(L1.south west)+(0,-14pt)$)
  {(a) Hausdorff-zero full carrier of $\operatorname{Law}(\boldsymbol X)$;\\
   (b) a.s. $R(\boldsymbol X)\in\Q\cup\mathcal L$ simultaneously for all\\
   $R\in\Q(T_1,\ldots,T_d)$ defined at $\boldsymbol X$\\
   \figref{\Cref{lem:joint-rational-carrier}}};
\draw[fuse] (L0) -- (L1);
\draw[fuse] (L1) -- (L2);
% ---- right lane: two alternative routes (i), (ii) to nonconcentration
\node[fres, anchor=north west, text width=81.75pt, minimum height=70.5pt] (Pa) at (198.5pt,0pt)
  {Independent prime\\ factors; neutral and\\
   selected states \eqref{eq:polynomial-nonconcentration-states},\\ eventual separation\\
   \eqref{eq:polynomial-nonconcentration-separation}; coordinates\\ may be dependent\\
   \figref{\Cref{prop:prime-product-polynomial-nonconcentration}}};
\node[fmech, anchor=north east, text width=81.75pt, minimum height=70.5pt] (Pb) at (383pt,0pt)
  {Independent atomless\\ coordinates:\\ conditioning on the\\ other coordinates,\\ induction on $d$\\
   \figref{proof of \Cref{cor:rational-statistics}}};
\node[finp, anchor=south west, text width=177.5pt] (N) at (198.5pt,0 |- L2.south)
  {Polynomial nonconcentration: $B(\boldsymbol Z)\ne0$ a.s.\\
   for each fixed nonzero real polynomial $B$\\
   \figref{hypothesis of \Cref{cor:nonconcentrated-rational-images}}};
\draw[fuse] (219pt,0 |- Pa.south) -- coordinate[midway] (ma) (219pt,0 |- N.north);
\node[fhyp, anchor=west, align=left] (hi) at ($(ma)+(1.5pt,0)$)
  {(i) same-input pair $(\Phi,\Sigma)$\\ of \Cref{cor:joint-phi-sigma}; $\boldsymbol Z=\boldsymbol X$};
\draw[fuse] (329pt,0 |- Pb.south) -- coordinate[midway] (mb) (329pt,0 |- N.north);
\node[fhyp, anchor=west, align=left] (hii) at ($(mb)+(1.5pt,0)$)
  {(ii) $\boldsymbol Z=\boldsymbol X$\\ or $\log\boldsymbol X$};
% ---- joint junction J: Lemma 4.2 and nonconcentration -> Corollary 4.3
\coordinate (jy) at ($(L2.south)+(0,-12.7pt)$);
\node[fjoin] (J) at (191.5pt,0 |- jy) {};
\draw[ffeed] (31.5pt,0 |- L2.south) |- (J);
\draw[ffeed] (351.5pt,0 |- N.south) |- (J);
\node[fhyp, anchor=south, inner sep=0.6pt] (jl) at ($(J)+(0,2.8pt)$) {jointly};
\node[flab, anchor=north] (la) at (111.5pt,0 |- jy) {gives the carrier and $\Q\cup\mathcal L$};
\node[flab, anchor=north] (lb) at (271.5pt,0 |- jy) {excludes atoms and rational values};
% ---- Corollary 4.3: the two quantifiers in two boxes
\node[fres, anchor=north west, text width=172.5pt] (O1) at ($(5pt,0 |- J)+(0,-26.95pt)$)
  {Each fixed nonconstant real rational\\ function $R$: $R(\boldsymbol Z)$ is defined a.s., with\\
   atomless law and Hausdorff-zero full carrier,\\ hence singular continuous with $\dim_F=0$\\
   \figref{\Cref{cor:nonconcentrated-rational-images}\textup{(a)}; uses \Cref{lem:lipschitz-image,lem:fourier-Hausdorff}}};
\node[fres, anchor=north east, text width=172.5pt, minimum height=51.7pt] (O2) at (378pt,0 |- O1.north)
  {Almost surely, simultaneously for every\\ nonconstant $R\in\Q(T_1,\ldots,T_d)$: $R(\boldsymbol X)\in\mathcal L$;\\
   $X_1,\ldots,X_d$ algebraically independent over $\Q$\\
   \figref{\Cref{cor:nonconcentrated-rational-images}\textup{(b)}; rational $B$ and $\boldsymbol Z=\boldsymbol X$ suffice}};
% frame: 6.5pt above the boxes (room for the tag), 4.5pt below
\coordinate (gtop) at ($(O1.north)+(0,2pt)$);
\begin{scope}[on background layer]
  \node[fgroup, fit=(O1)(O2)(gtop), inner xsep=5pt, inner ysep=4.5pt] (G) {};
\end{scope}
\node[ftag] (gt) at (191.5pt,0 |- G.north) {Rational images under polynomial nonconcentration};
\draw[fuse] (J) -- (gt.north);
% ---- bottom row: Corollary 4.3(a) and Theorem 3.22 jointly -> Corollary 4.7(b)
\node[fres, anchor=north west, text width=119pt, minimum height=51.7pt] (LP) at ($(0pt,0 |- G.south)+(0,-7pt)$)
  {Independent coordinates with\\ ratio laws, in addition:\\
   $\underline d_\lambda=0$, $\overline d_\lambda=1$ $\lambda$-a.e.,\\
   $P_{\mathrm{pos}}(\lambda)=P_{\mathrm{full}}(\lambda)=1$\\ \figref{\Cref{cor:rational-statistics}\textup{(b)}}};
\node[fres, anchor=north east, text width=119pt, minimum height=51.7pt] (T) at (383pt,0 |- LP.north)
  {Each ratio law and its\\ logarithmic law: Borel sets\\ of positive measure have\\ packing dimension one\\
   \figref{\Cref{thm:local-dimension-oscillation}}};
\node[fjoin] (K) at (146pt,0 |- LP.east) {};
\draw[ffeed] (146pt,0 |- O1.south) -- (K);
\draw[ffeed] (T.west) -- node[flab, above] (lk) {specialization of the other\\ coordinates (Fubini);}
  node[flab, below] (lk2) {bi-Lipschitz pieces} (K);
\draw[fuse] (K) -- (LP.east);
\end{tikzpicture}
\caption{Simultaneous rational prefix approximations give Hausdorff-zero carriers and
rational-or-Liouville values. Polynomial nonconcentration excludes atoms and rational values; it is
established by different arguments for coordinates sharing their valuations~\textup{(i)} and for
independent coordinates~\textup{(ii)}. Here $\boldsymbol X\in(0,\infty)^d$ is the limit of the prime products
$\boldsymbol X(y)$ over primes up to~$y$, $\boldsymbol Z$ is $\boldsymbol X$ or a locally Lipschitz image such as
$\log\boldsymbol X$, and $\lambda$ is the law of $R(\boldsymbol X)$ or $R(\log\boldsymbol X)$.}
\label{fig:rational-nonconcentration}
\end{figure}

\section{Approximation with prescribed rate functions}
\label{sec:diophantine-fixed-functions}

Convergence of the prime-series test excludes sufficiently accurate
approximations at every sufficiently large denominator; divergence
produces them along the actual reduced prime prefixes. The function is
fixed before the almost-sure assertion; the simultaneous form allows
a prescribed countable family.

\subsection{Prime-series tests and denominator control}

For a fixed finite positive function $\psi$ on the positive integers, put
\[
 h_\psi(q):=q^{-\psi(q)},\qquad
 I(\psi):=\sum_{p\ \mathrm{prime}}\frac1{p\psi(p)}.
\]
When $\psi$ is eventually nondecreasing, use its nondecreasing step
extension at sufficiently large real arguments and put
$\Psi(u):=\psi(\exp(\exp u))$. Changing finitely many values is immaterial.

\begin{lemma}[The prime-series test]\label{lem:prime-test}
For every fixed finite positive eventually nondecreasing $\psi$,
\begin{equation}\label{eq:prime-integral}
 I(\psi)<\infty
 \quad\Longleftrightarrow\quad
 \int^\infty\frac{dx}{x\log x\,\psi(x)}<\infty
 \quad\Longleftrightarrow\quad
 \int^\infty\frac{du}{\Psi(u)}<\infty.
\end{equation}
In the convergent case,
$\psi(x)/\log\log x\longrightarrow\infty$.
\end{lemma}

\begin{proof}
By \Cref{lem:prime-log-stieltjes}, the sum of $1/p$ over
$\exp(\exp m)<p\le\exp(\exp(m+1))$ is $1+o(1)$.
Monotonicity bounds the contribution of this block to $I(\psi)$
between $(1+o(1))/\Psi(m+1)$ and $(1+o(1))/\Psi(m)$.
The corresponding bounds for the integral of $1/\Psi$ over $[m,m+1]$
prove the equivalence, including functions with jumps; the substitution
$u=\log\log x$ gives the other integral. In the convergent case,
\[
 \frac{u}{2\Psi(u)}
 \le\int_{u/2}^{u}\frac{dv}{\Psi(v)}\longrightarrow0,
\]
which proves the last assertion.
\end{proof}

\Needspace{8\baselineskip}
\begin{lemma}[Denominator envelopes at general intensity]
\label{lem:denominator-coupling}
Let $\kappa>0$, and let independent Bernoulli variables $B_p$ at odd
primes $p\ge p_0\ge3$ have probabilities $\alpha_p$. Suppose that,
for some $p_1\ge p_0$,
\begin{equation}\label{eq:cumulative-intensity}
 \begin{gathered}
 0\le\alpha_p<1,\qquad \alpha_p\ll p^{-1}\quad(p\ge p_1),\\
 \sum_{p_1\le p\le y}-\log(1-\alpha_p)
 =\kappa\log\log y+O(1).
 \end{gathered}
\end{equation}
Enumerate their active primes as $P_1<P_2<\cdots$.
Let $C_0$ be an almost surely finite nonnegative random variable and
let $q_n$ be positive-integer random variables satisfying, almost surely
eventually,
\begin{equation}\label{eq:denominator-envelope}
\begin{split}
 \log P_n&\le\log q_n\le C_0+\sum_{i\le n}\log P_i,\\
 L_n&:=\frac{C_0+\sum_{i\le n}\log P_i}{\log P_n}.
\end{split}
\end{equation}
For every fixed finite positive eventually nondecreasing $\psi\ge1$ with
$\sum_p[p\psi(p)^\kappa]^{-1}=\infty$, almost surely
\begin{equation}\label{eq:weighted-divergence}
 \frac{\log\log q_n}{n}\longrightarrow\kappa^{-1},
 \qquad \sum_n[L_n\psi(q_n)]^{-\kappa}=\infty.
\end{equation}
\end{lemma}

\begin{proof}
Independence and $\sum_p\alpha_p=\infty$ give infinitely many active
primes almost surely. Choose a deterministic cutoff $b\ge p_1$ with
$0\le\alpha_p<1/2$ for $p\ge b$, and put
$K=\sum_{p_0\le p<b}B_p$. For $p\ge b$ set
$\lambda_p=-\log(1-\alpha_p)$ and
$H(x)=\sum_{b\le p\le x}\lambda_p$.
By \eqref{eq:cumulative-intensity},
$H(x)=\kappa\log\log x+O(1)$, while
$\lambda_p=O(p^{-1})$ and $\lambda_p-\alpha_p=O(p^{-2})$.
On an extension of the original space, put $N_p=0$ when $\alpha_p=0$.
At the remaining primes, conditionally on the original variables,
put $N_p=0$ if $B_p=0$ and otherwise sample a Poisson variable of mean
$\lambda_p$ conditioned to be positive, using independent auxiliary
randomness at different primes. The $N_p$ are independent Poisson
variables and $B_p=\mathbf1_{\{N_p\ge1\}}$.
Uniform independent placement of their points in the consecutive cells
$(H(p^-),H(p)]$, omitting empty cells, gives a rate-one Poisson process,
with arrival times $\Gamma_m$ and independent mean-one exponential
increments $E_m$.
Because $\sum_p(\lambda_p-\alpha_p)<\infty$, its total excess
$D=\sum_{p\ge b}(N_p-B_p)$ is finite almost surely.
Beyond the last excess cell the cell of $P_n$ contains exactly arrival
$n+J$, where $J=D-K$ is a finite, possibly negative, integer. Hence
\begin{align}
 |H(P_n)-\Gamma_{n+J}|&\le\lambda_{P_n},
 \label{eq:intensity-coupling}\\
 \log\log P_n&=\frac{\Gamma_{n+J}}{\kappa}+\delta_n,
 \quad |\delta_n|\le C_1\quad\hbox{eventually},
 \label{eq:pathwise-coupling}
\end{align}
Here $C_1<\infty$ is deterministic and depends on the activation law.

Extend the exponential increments independently to all integer indices
and define
\begin{equation}\label{eq:perpetuity}
 R_m^*:=1+\sum_{j\ge1}
 \exp\{-(E_m+\cdots+E_{m-j+1})/\kappa\}.
\end{equation}
Tonelli gives
\begin{equation}\label{eq:intensity-perpetuity}
 \mathbb E R_m^*=\sum_{j\ge0}
       \left(\frac\kappa{\kappa+1}\right)^j=\kappa+1,
\end{equation}
so these series are finite almost surely simultaneously for all $m$.
The two-sided independent shift is mixing: approximate measurable
events by finite-coordinate cylinders, which become independent under
sufficiently separated shifts. Thus its stationary factor $(R_m^*)$
is ergodic. The ergodic theorem \cite[Theorem~6.2.1]{Dur19} and Markov's
inequality give asymptotic density at least $1/2$ to the indices with
$R_m^*\le2(\kappa+1)$.

Discard finitely many indices so that
$|\delta_i|\le C_1$ and $i+J\ge1$ on the remainder.
The discarded contribution and $C_0$, divided by $\log P_n$, tend to
zero. The coupling therefore gives
\[
 L_n\le o(1)+e^{2C_1}
 \sum_{\substack{i\le n\\i\text{ retained}}}
 e^{(\Gamma_{i+J}-\Gamma_{n+J})/\kappa}
 \le o(1)+e^{2C_1}R_{n+J}^*.
\]
We obtain a set $A$ of positive lower density
on which $L_n\le M:=2e^{2C_1}(\kappa+1)+1$ eventually: a finite integer
shift changes each realization's counting function by $O(|J|)$.
The exponential strong law gives $\log\log P_n/n\to1/\kappa$.
The envelope and $\sum_{i\le n}\log P_i\le n\log P_n$ yield
\[
 \log\log P_n\le\log\log q_n
 \le\log\log P_n+\log(n+C_0/\log P_n),
\]
proving the first conclusion of \eqref{eq:weighted-divergence}.
For fixed $a_0>1/\kappa$, eventually $\psi(q_n)\le\Psi(a_0n)$.
By \Cref{lem:prime-test} applied to $\psi^\kappa$, the
eventually nonincreasing weights $w_n=\Psi(a_0n)^{-\kappa}$ have
divergent sum. Discarding finitely many terms, make them nonincreasing
throughout. With $A(N)=\#(A\cap[1,N])\ge\delta N$ eventually,
summation by parts gives
\[
 \sum_{\substack{n\le N\\n\in A}}w_n
 =A(N)w_N+\sum_{n<N}A(n)(w_n-w_{n+1})
 \ge\delta\sum_{n\le N}w_n-O(1)\longrightarrow\infty.
\]
On $A$ the desired summand is eventually at least
$M^{-\kappa}w_n$. This proves the second conclusion, which
projects to the original marginal law because both conclusions involve
only original random variables.
\end{proof}

\subsection{The same-sign dichotomy and its specializations}

For convergence, close rational centers are forced to be finite-product
anchors whose probabilities are summable. For divergence, the denominator
estimate gives infinitely many long gaps, and a separate tail estimate
converts those gaps into accurate prefix approximations.

\begin{theorem}[Same-sign products of general intensity]
\label{thm:intensity}
\index{Activation intensity!prime-index, Diophantine dichotomy}
Fix $\kappa>0$, $s\in\{-1,1\}$ and $c\in\mathbb Q_{>0}$.
Let $(B_p)_{p\ge p_0}$ be independent Bernoulli variables at odd primes,
where $p_0\ge3$ and $\alpha_p:=\mathbb P(B_p=1)$ satisfies
\eqref{eq:cumulative-intensity}. In particular, this holds if
$\alpha_p=\kappa/p+O(p^{-2})$. Put
\begin{equation}\label{eq:bernoulli-model}
 X=cZ_s,\qquad Z_s:=\prod_{p\ge p_0}(1+s/p)^{B_p}.
\end{equation}
For each fixed finite positive eventually nondecreasing $\psi$, or
fixed positive function comparable to such a function, put
\begin{equation}\label{eq:intensity-test}
 I_\kappa(\psi):=\sum_{p\ \mathrm{prime}}\frac1{p\psi(p)^\kappa}.
\end{equation}
Let $r_n=j_n/q_n$ be the actual prefix through the $n$th active prime,
including $c$, reduced to lowest terms. Almost surely $X\in(0,\infty)$,
$q_n\to\infty$, $|X-r_n|>0$ for every $n$, and
\begin{align}
 I_\kappa(\psi)<\infty&\ \Longrightarrow\
 \lim_{q\to\infty}q^{\psi(q)}\min_{j\in\mathbb Z}|X-j/q|=\infty,
 \label{eq:scaled-exclusion}\\
 I_\kappa(\psi)=\infty&\ \Longrightarrow\
 \liminf_{n\to\infty}q_n^{\psi(q_n)}|X-r_n|=0.
 \label{eq:prefix-liminf}
\end{align}
The assertions hold simultaneously for all positive rational $c$ and
each prescribed countable family of functions. Consequently a positive
rational random multiplier, with arbitrary dependence on the $B_p$,
is allowed. They also hold after an almost surely finite alteration
of positive rational local factors, using the altered actual prefixes.
\end{theorem}

\begin{proof}
Choose a deterministic cutoff beyond which $0\le\alpha_p<1/2$.
We first prove the assertions for that tail and every fixed rational
multiplier; the countable intersection at the end will reinsert the
finite head. Thus temporarily assume $0\le\alpha_p<1/2$ throughout.
Write $Z=X/c$, $k_p=\alpha_p/(1-\alpha_p)$, and, for finite prime sets,
$K(S)=\prod_{p\in S}k_p$, $D_S=\prod_{p\in S}p$ and
$x_S=c\prod_{p\in S}(1+s/p)$.
The hypothesis and the prime estimates in
\Cref{lem:prime-log-stieltjes} give
\begin{align}
 &\sum_{p_0\le p\le y}\alpha_p\log p\ll\log y,\qquad
 \prod_{p_0\le p<y}(1+k_p)\asymp(\log y)^\kappa,
 \label{eq:mertens-inputs}\\
 &\mathcal H(y):=\sum_{p_0\le p\le y}-\log(1-\alpha_p)
       =\kappa\log\log y+O(1).
 \label{eq:intensity-inputs}
\end{align}
Indeed $\alpha_p\ll p^{-1}$ and partial summation give the first
estimate, while $1+k_p=(1-\alpha_p)^{-1}$ gives the product estimate.
Constants may depend on the fixed activation law. The product $Z$ is almost surely positive and
finite because $\sum_p\alpha_p|\log(1+s/p)|<\infty$.

Finite subsets give distinct anchors: after canceling $c$ and common
factors, the largest prime $P$ in a symmetric difference divides one
denominator and no numerator, since $p+s<P$ for every smaller odd
prime $p$ and $P\nmid P+s$. The largest atom of the finite head through
$y$ is therefore $\prod_{p_0\le p\le y}(1-\alpha_p)\to0$.
Conditioning on the independent tail bounds every full-product atom
by this number. Thus $X$ is atomless, and hence irrational almost surely.

First suppose $\psi$ is eventually nondecreasing. Applying
\Cref{lem:prime-test} to $\psi^\kappa$ gives
\begin{equation}\label{eq:approximation-intensity-integral}
 I_\kappa(\psi)<\infty
 \quad\Longleftrightarrow\quad
 \int^\infty\frac{du}{\Psi(u)^\kappa}<\infty,
 \qquad \frac{\Psi(u)^\kappa}{u}\longrightarrow\infty
 \quad\hbox{in the convergent case}.
\end{equation}

\emph{Convergence.} Fix $0<\eta<M<\infty$ and let $E_m$ be the union
of windows $|X-j/q|<h_\psi(q)$ over reduced centers $j/q\in[\eta,M]$
in the block
\begin{equation}\label{eq:coarse-blocks}
 Q=\exp(\exp m),\quad R=Q^e,\quad Q\le q<R,
 \quad\Xi=CR,\quad H=Q^{-\psi(Q)},\quad T=H^{-d},
\end{equation}
where $C>2\max(1,M)$ and $0<a<b<d<1/2$ are fixed.
These adjacent blocks cover all sufficiently large denominators.
Set $z=\log(1/H)$ and $L=\log\Xi$; each window retains its own
radius $h_\psi(q)\le H$. In particular $T>\Xi$ eventually.
For $D=\prod_{p_0\le p\le\Xi}p^{B_p}$, convexity on $[0,1]$ gives
\[
 \log\mathbb E e^{\theta\log D/L}
 \le\sum_{p_0\le p\le\Xi}\alpha_p(p^{\theta/L}-1)
 \le K_0(e^\theta-1),\qquad\theta\ge0.
\]
With $w=az/L\sim a\Psi(m)/e$, optimization at
$\theta=\log(w/K_0)$ gives eventually
\begin{align}
 \mathbb P(D\ge H^{-a})
 &\le\exp\{-w\log(w/K_0)+w-K_0\}
 \label{eq:head}\\
 &\le\exp(-c_1m^{1/\kappa}\log m).
 \label{eq:intensity-head}
\end{align}
Here \eqref{eq:approximation-intensity-integral} gives
$\Psi(m)\ge m^{1/\kappa}$ eventually. This bound is summable for
every fixed $\kappa>0$.

The original windows are disjoint for large $m$: distinct reduced
centers are separated by at least $R^{-2}$ and $R^2H\to0$.
Partition their union with an active $\Xi<p\le T$ by its center $r=j/q$
and least such $p$. Deleting $p$ sends that window to center
$jp/[q(p+s)]$ and radius $h_\psi(q)p/(p+s)\le2H$.
These centers are distinct. For two distinct primes choose the larger
as $p$ in $jpq'(p'+s)=j'p'q(p+s)$; reduction modulo $p$ contradicts
$0<j'<MR<p$, $q<R<p$, $p'<p$, and $p\nmid p+s$.
For equal primes equality forces the same original center.
The displayed denominators are at most $R(T+1)$, so distinct deleted
centers are separated by at least $[R(T+1)]^{-2}$.
Since $R^2H(T+1)^2\to0$, all deleted windows are also disjoint.

Write a partition event as $G_{r,p}=\{B_p=1\}\cap A_{r,p}$, where
$A_{r,p}$ specifies the earlier zero middle coordinates and the
image-window condition for the product with the $p$-factor omitted.
It depends only on coordinates other than $p$. Its deleted image
$G'_{r,p}=\{B_p=0\}\cap A_{r,p}$ has
$\mathbb P(G_{r,p})=k_p\mathbb P(G'_{r,p})$ by independence.
On that image the full product $X$ is in its deleted window; thus the
images are disjoint over both $r$ and $p$. Consequently
\begin{equation}\label{eq:middle}
 \mathbb P(E_m,\ \exists\,\Xi<p\le T:B_p=1)
 \le\sup_{p>\Xi}k_p\sum_{r,p}\mathbb P(G'_{r,p})\ll R^{-1}.
\end{equation}
For $W_T=\sum_{p>T}B_p|\log(1+s/p)|$ we have
$\mathbb E W_T\ll\sum_{n>T}n^{-2}\ll T^{-1}$, and therefore
\begin{equation}\label{eq:tail}
 \mathbb P(W_T>H^b)\ll H^{d-b}.
\end{equation}
Both error series converge. Outside the latter event the tail product
$\Pi_T$ satisfies $|\Pi_T-1|\le2H^b$ and $\Pi_T\ge e^{-H^b}$.

Outside these three exceptions, a sample in a window has
$X=x_S\Pi_T$, where $S=\{p_0\le p\le\Xi:B_p=1\}$ and
$D_S<H^{-a}$. Since $X\le M+H$,
$x_S\le(M+H)e^{H^b}\le2(M+1)$, so
$|x_S-j/q|\le H+4(M+1)H^b$.
Writing $c=u/v$ in lowest terms, the reduced denominator of $x_S$
divides $vD_S$. If $x_S\ne j/q$, rational separation gives
$|x_S-j/q|\ge1/(qvD_S)>H^a/(vR)$, a contradiction because
$RH^{b-a}\to0$. Thus the center equals its unique anchor $x_S$.

For the full product $Z$, any active $p\le1/(2t)$ makes
$|Z-1|\ge1/p\ge2t$, since all factors move in the same direction.
It follows from \eqref{eq:intensity-inputs} that
\begin{equation}\label{eq:endpoint}
 \mathbb P(|Z-1|<t)
 \le\prod_{p_0\le p\le1/(2t)}(1-\alpha_p)
 \ll[\log(1/t)]^{-\kappa}\qquad(t\downarrow0).
\end{equation}
For a fixed surviving $S$ of positive probability, $K(S)>0$, and all
remaining conditions depend only on coordinates outside $S$, together
with the deterministic condition $D_S<H^{-a}$. Deleting $S$ thus changes
its probability by the factor
$K(S)^{-1}$. At the image point the full product $Z$ is within
$h_\psi(q(S))/x_S$ of one, where $q(S)$ is the reduced denominator of
$x_S$. Discarding the image's additional coordinate restrictions only
after this deletion gives, uniformly for $x_S\in[\eta,M]$,
\begin{equation}\label{eq:anchor-probability}
 \mathbb P(\hbox{surviving event for }S)
 \ll_{\eta,M}\frac{K(S)}{[\psi(q(S))\log q(S)]^\kappa}.
\end{equation}
If $P=\max S$ exceeds all prime divisors of $u$, no numerator can
cancel $P$; thus $P\mid q(S)$. Restrict the sum to positive-probability
anchors with sufficiently large $q(S)$. Apart from finitely many
subsets with bounded largest prime, monotonicity and the product estimate give
\begin{align}
 \sum_S\frac{K(S)}{[\psi(q(S))\log q(S)]^\kappa}
 &\ll 1+\sum_P\frac{k_P\prod_{p_0\le p<P}(1+k_p)}
                       {[\psi(P)\log P]^\kappa}
 \label{eq:anchor-sum}\\
 &\ll1+I_\kappa(\psi)<\infty.
 \label{eq:intensity-anchors}
\end{align}
Each anchor belongs to exactly one block. The three error sums and
the anchor sum imply $\sum_m\mathbb P(E_m)<\infty$; Borel--Cantelli
therefore gives eventual exclusion on $[\eta,M]$.
Intersect over $[\eta,M]=[1/k,k]$, $k\ge2$, to cover all centers
near $X\in(0,\infty)$. For an unreduced $j/q$, reduce to denominator
$q'\le q$. Large $q'$ use $h_\psi(q)\le h_\psi(q')$; bounded $q'$
give only finitely many centers in a fixed compact, all at positive
distance from the irrational $X$. Since $h_\psi(q)\to0$, these too
are excluded eventually.
Apply it to $\psi/2$ to get
\[
 q^{\psi(q)}\min_j|X-j/q|
 \ge q^{\psi(q)/2}\longrightarrow\infty.
\]

\emph{Divergence.} Replacing $\psi$ by $\max(1,\psi)$ preserves
divergence of $I_\kappa$ and decreases target radii, since an eventually
nondecreasing positive function is bounded below eventually.
Assume therefore $\psi\ge1$. Enumerate the active primes as $P_n$
and put $F_p=(1+s/p)^{B_p}$. Partial summation gives
\begin{equation}\label{eq:local-tail}
 |\log F_p|\le2B_p/p,\qquad
 \sum_{p>t}\mathbb E|\log F_p|\ll(t\log t)^{-1}\quad(t\ge3).
\end{equation}
The actual prefixes satisfy $r_n\to X$. For $c=u/v$, once $P_n$
exceeds the prime divisors of $u$, its denominator prime survives
exactly as above. Thus \eqref{eq:denominator-envelope} holds
with $C_0=\log v$, and in particular $q_n\to\infty$.
By \Cref{lem:denominator-coupling}, with its corresponding
$L_n$, we have $\sum_n[L_n\psi(q_n)]^{-\kappa}=\infty$ almost surely.

Enumerate the deterministic primes in the tail as $p_j$, set
$\mathcal D_j=\sigma(B_{p_i}:i\le j)$, and let $\tau_n$ be the index
of the $n$th active prime. This is a stopping time; put
$\mathcal F_n=\mathcal D_{\tau_n}$ and take $\mathcal F_0$ trivial.
Partitioning by $\{\tau_n=j\}$ shows that the unexposed original
coordinates retain their independent laws. Both $r_n$ and its actual
reduced denominator $q_n$ are $\mathcal F_n$-measurable.
For fixed $\eta>0$ put
\begin{equation}\label{eq:adapted-threshold}
\begin{split}
 K_\eta&:=\max(4,4/\eta),\\
 T_n&:=\max\{3,P_n,K_\eta\max(1,r_n)q_n^{\psi(q_n)}\},\qquad
 G_n:=\{P_{n+1}>T_n\}.
\end{split}
\end{equation}
Then $T_n$ is $\mathcal F_n$-measurable and
$G_n\in\mathcal F_{n+1}$. Conditional independence gives
\begin{align}
 g_n:=\mathbb P(G_n\mid\mathcal F_n)
 &=\prod_{P_n<p\le T_n}(1-\alpha_p)
   =e^{\mathcal H(P_n)-\mathcal H(T_n)}
 \label{eq:gap-probability}\\
 &\asymp\left(\frac{\log P_n}{\log T_n}\right)^\kappa
   \gg_{\omega,\eta}[L_n\psi(q_n)]^{-\kappa}.
 \label{eq:intensity-gap}
\end{align}
The comparison is uniform for $T_n\ge P_n\to\infty$, since the
remainder in \eqref{eq:intensity-inputs} is bounded.
The last inequality uses the denominator envelope,
$\psi\ge1$, and $r_n\to X\in(0,\infty)$.
In particular $\sum_ng_n=\infty$ almost surely.
Apply conditional Borel--Cantelli
\cite[Theorem~4.3.4]{Dur19} with the filtration
$(\mathcal F_n)_{n\ge0}$, $A_1=\varnothing$, and $A_{n+1}=G_n$
for $n\ge1$. The conditional probabilities are $g_n$, so infinitely
many $G_n$ occur.

The entire omitted tail needs a separate estimate. For
$W_n=\sum_{p>T_n}|\log F_p|$, conditional Tonelli and
\eqref{eq:local-tail} give
$\mathbb E(W_n\mid\mathcal F_n)\ll(T_n\log T_n)^{-1}$.
Here each indicator $\mathbf1_{\{p>T_n\}}$ is $\mathcal F_n$-measurable
and forces $p>P_n$. Conditional Markov and Tonelli yield
\begin{align}
 \sum_n\mathbb P(W_n>T_n^{-1})
 &\ll\sum_n\mathbb E\frac1{\log P_n}
 \label{eq:tail-summability}\\
 \sum_n\mathbb E\frac1{\log P_n}
 &=\sum_{p\ge p_0}\frac{\alpha_p}{\log p}<\infty.
 \label{eq:intensity-tail}
\end{align}
The equality enumerates each active prime once; the last convergence
follows by partial summation from $\pi(x)\ll x/\log x$.
First Borel--Cantelli makes all sufficiently late tails good.
On $G_n$ all omitted factors lie above $T_n$; on a good tail also,
\[
 0<|X-r_n|\le r_n(e^{W_n}-1)
 \le2r_n/T_n<\eta q_n^{-\psi(q_n)}.
\]
Infinitely many nonidentity factors remain, all on the same side of $1$,
so $|X-r_n|>0$.
Intersecting the probability-one events for $\eta=1/j$, $j\ge1$,
proves the zero liminf.

For a fixed comparable function $a_0g\le\psi\le b_0g$ eventually,
where $a_0,b_0>0$ and $g$ is finite, positive and eventually
nondecreasing, $I_\kappa(g)$ has the same convergence status as
$I_\kappa(\psi)$. In the convergent case, exclusion for $a_0g/2$ gives
$q^{\psi(q)}\min_j|X-j/q|\ge q^{a_0g(q)/2}\to\infty$.
In the divergent case apply the result to $\max(1,b_0g)$ and use
$q^{\psi(q)}\le q^{\max(1,b_0g(q))}$ eventually.

For any prescribed countable family, intersect the resulting events
over the family and all $c\in\mathbb Q_{>0}$ before selecting a
possibly dependent rational multiplier. To check altered actual
prefixes, write $F_p=(1+s/p)^{B_p}$ and suppose positive rational
$\widetilde F_p$ differ from $F_p$ at only finitely many primes.
Put $C=c\prod_p(\widetilde F_p/F_p)\in\mathbb Q_{>0}$ and choose $Y$
beyond all alterations. At every cutoff $y\ge Y$ the altered prefix
equals $C\prod_{p\le y}F_p$. Define the finite integer
\[
 \Delta=\#\{p_0\le p\le Y:\widetilde F_p\ne1\}
          -\#\{p_0\le p\le Y:B_p=1\}.
\]
The active sets agree above $Y$, so for every sufficiently large $n$
their actual reduced prefixes and denominators satisfy exactly
\[
 \widetilde r_{n+\Delta}=r_{n,C},\qquad
 \widetilde q_{n+\Delta}=q_{n,C}.
\]
This eventual fixed integer shift preserves the full liminf and
denominator growth. The same argument reinserts the initially omitted
primes, with their finite active count as the index shift.
The altered full product is irrational, being a positive rational
multiple of the irrational tail product, whereas every altered prefix
is rational; hence the error is positive for every prefix, including
those before $Y$.
\end{proof}

\begin{lemma}[Rational rigidity for canonical prime factors]
\label{lem:canonical-rational-rigidity}
Fix $s\in\{-1,1\}$, $c\in\Q_{>0}$ and a prime cutoff $p_0\ge3$;
also allow $p_0=2$ when $s=-1$. Let $(B_p)_{p\ge p_0}$ be independent
Bernoulli variables whose probabilities satisfy
\eqref{eq:cumulative-intensity} on a tail, with a finite intensity
$\kappa>0$, and are arbitrary, including $0$ and $1$, at the finitely
many earlier primes. Put
\[
 X=c\prod_{p\ge p_0}(1+s/p)^{B_p},
\]
and fix a finite positive eventually nondecreasing function $\psi_0$ with
\[
 \psi_0(q)=(\log\log\log q)^2
 \quad\hbox{for all sufficiently large }q.
\]
Almost surely, every reduced positive rational $j/q$ with sufficiently
large denominator and
\[
 |X-j/q|<q^{-\psi_0(q)}
\]
equals an actual finite prime prefix
$c\prod_{p_0\le p\le y}(1+s/p)^{B_p}$. The cutoff $y$ can be chosen
to tend to infinity with $q$; the denominator threshold may depend on
$c$ and the sample.
\end{lemma}

\begin{proof}
Fix a compact interval $[\eta,M]\subset(0,\infty)$ and use the
blocks and parameters
$Q,R,\Xi,H,T,a,b,d$ in \eqref{eq:coarse-blocks}, with $\psi=\psi_0$.
For $D=\prod_{p_0\le p\le\Xi}p^{B_p}$, the head estimate before its
convergence-only simplification is \eqref{eq:head}. Since
\[
 w=\frac{a\log(1/H)}{\log\Xi}
       \sim\frac ae(\log m)^2,
\]
it gives, for some constant $c_1>0$ and all sufficiently large $m$,
\[
 \mathbb P(D\ge H^{-a})
 \le \exp\{-c_1(\log m)^2\log\log m\}.
\]
This is summable in $m$. Finitely many forced head coordinates do not
change the estimate: their logarithmic contribution is bounded, and
the same exponential-moment argument applies after subtracting that
constant. The middle-deletion and tail estimates
\eqref{eq:middle} and \eqref{eq:tail} give respectively
$O(R^{-1})$ and $O(H^{d-b})$; both are summable. Their proofs require
only $\psi_0(Q)\to\infty$, which ensures
$T>\Xi$, $R^2H(T+1)^2\to0$ and $RH^{b-a}\to0$.

Outside these three exceptional events, the rational-separation
argument in the proof of \Cref{thm:intensity} applies verbatim at the
individual window radius: an approximant $j/q\in[\eta,M]$ in that
block equals
\[
 x_S=c\prod_{p\in S}(1+s/p),
 \qquad S=\{p_0\le p\le\Xi:B_p=1\}.
\]
The omitted middle coordinates are zero and the
tail changes this head by $O_M(H^b)$, while
$D_S:=\prod_{p\in S}p=D<H^{-a}$.
If $c=u/v$ in lowest terms and $j/q\ne x_S$, their distance is at
least $(qvD_S)^{-1}>H^a/(vR)$, contradicting
$|x_S-j/q|\le H+O_M(H^b)$. Distinct finite sets give distinct anchors:
the largest prime in their symmetric difference survives in one
denominator after cancellation of $c$ and common factors. For $s=-1$
this argument also applies when $2$ is present; its only other effect
is in the finite head. Thus the identified set is unique.
Borel--Cantelli, followed by intersection over $[\eta,M]=[1/k,k]$,
proves the assertion for every sufficiently large denominator near $X$.
The identified cutoff $\Xi=CR$ tends to infinity with $q$.
\end{proof}

\begin{proposition}[A common Borel decoder for canonical prime factors]
\label{prop:canonical-decoder-mas08}
Fix $s\in\{-1,1\}$, $c\in\Q_{>0}$ and a prime cutoff $p_0\ge3$;
also allow $p_0=2$ when $s=-1$. There is a Borel map
\[
 \mathcal D_{c,s}: (0,\infty)\longrightarrow
                  \{0,1\}^{\{p\ge p_0:p\text{ prime}\}}
\]
with the following property. For every independent Bernoulli array
$(B_p)$ whose probabilities satisfy \eqref{eq:cumulative-intensity}
on a tail, with any finite intensity $\kappa>0$, and with arbitrary
Bernoulli probabilities (including $0$ and $1$) at the finitely many
earlier primes,
\[
 X=c\prod_{p\ge p_0}(1+s/p)^{B_p}
 \quad\Longrightarrow\quad
 \mathcal D_{c,s}(X)=(B_p)_p\quad\hbox{almost surely}.
\]
\end{proposition}

The map $\mathcal D_{c,s}$ depends on the fixed canonical factors and
$c$, but not on the activation probabilities or their intensity. In
particular it is a common observation decoder when comparing any two
such laws with the same $c,s,p_0$. For $s=-1,c=1,p_0=2$, this includes
all prime divisibility indicators in each of the named totient laws
$\mu_F^{\mathrm{int}}$ and $\mu_G$; the latter has its coordinate at
$2$ forced to one.

\begin{proof}
Choose $\psi_0$ as in \Cref{lem:canonical-rational-rigidity}.
For every fixed $\kappa>0$, \Cref{lem:prime-test} gives
\[
 I_\kappa(\psi_0)=\infty,
 \qquad
 \int^\infty\frac{du}{(\log u)^{2\kappa}}=\infty.
\]
The divergent assertion of \Cref{thm:intensity}, after reinserting the
finite canonical head, supplies infinitely many distinct actual prefix
anchors $r=j/q$ with $|X-r|<q^{-\psi_0(q)}$ and $q\to\infty$.

By \Cref{lem:canonical-rational-rigidity}, every sufficiently late
approximant in these windows is the actual prefix through a cutoff
tending to infinity with its denominator.

Here is an explicit Borel construction. Enumerate the positive reduced
rationals $r_i=j_i/q_i$ in increasing $j_i+q_i$, breaking ties by $j_i$.
For each one that is a finite canonical anchor, let $S_i$ be its unique
finite prime set, as in \Cref{lem:canonical-rational-rigidity}, and set
\[
 U_i=\{x>0:|x-r_i|<q_i^{-\psi_0(q_i)}\}.
\]
Discard the other rationals and all finitely many denominator values
below a fixed threshold where $\psi_0\ge1$. Equivalently their windows
are empty. Finite-anchor recognition is a countable lookup among the
exact rationals $c\prod_{p\in S}(1+s/p)$; it makes no reference to the
probability law. For each prime $p$ define the $p$th decoded coordinate
as the indicator of the Borel set
\[
 \left(\bigcap_{N\ge1}\bigcup_{i\ge N}U_i\right)
 \cap
 \left(\bigcup_{N\ge1}
        \bigcap_{\substack{i\ge N\\p\notin S_i}}U_i^c\right).
\]
Empty windows may be assigned $S_i=\varnothing$. The first condition
requires infinitely many surviving windows; the second says that their
$p$th indicators are eventually one. For a fixed $x$, only finitely
many selected windows with bounded denominator can contain $x$, since
their centers then lie in a bounded interval. Almost surely infinitely
many windows survive by the prefix approximation already proved, and
all sufficiently late ones identify the true head through cutoffs
tending to infinity.
Consequently each fixed coordinate stabilizes to $B_p$. This proves
both Borel measurability and simultaneous recovery of all coordinates.
\end{proof}

\begin{corollary}[Equivalence and singularity of canonical product laws]
\label{cor:canonical-profile-equivalence}
Fix the same $c,s,p_0$ as in \Cref{prop:canonical-decoder-mas08}.
Let $(\alpha_p)$ and $(\beta_p)$ be two profiles satisfying its
hypotheses, possibly with different intensities. Suppose the two
Bernoulli coordinate laws are equivalent at every prime.
The laws of the two observed products are equivalent precisely when
\[
 \sum_{p\ge p_0}\left[(\sqrt{\alpha_p}-\sqrt{\beta_p})^2
 +(\sqrt{1-\alpha_p}-\sqrt{1-\beta_p})^2\right]<\infty .
\]
If the series diverges, the observed laws are mutually singular.
\end{corollary}

\begin{proof}
Common deterministic coordinates contribute zero. On every other
coordinate the Hellinger affinity is
$\rho_p=\sqrt{\alpha_p\beta_p}+\sqrt{(1-\alpha_p)(1-\beta_p)}>0$.
The displayed summand is $2(1-\rho_p)$, so convergence is equivalent
to $\prod_p\rho_p>0$. Kakutani's product criterion~\cite{Kak48}
gives equivalence of the indicator laws in this case and mutual
singularity otherwise. Equivalence passes through the common product
map. In the singular case, pull a separating Borel set of indicator
arrays back through the common decoder in
\Cref{prop:canonical-decoder-mas08}. Its preimage has probability one
for the first observation and zero for the second.
\end{proof}

The fixed known multiplier and the canonical local factors are essential
for this recovery statement. A dependent rational multiplier
$(1+s/p)^{-B_p}$ erases the $p$th coordinate. Thus the decoder does not
inherit the arbitrary dependent rational multipliers or finite factor
alterations allowed by the Diophantine theorem \Cref{thm:intensity}.

\begin{corollary}[The fixed-function approximation dichotomy]
\label{thm:approximation}
\index{Rational approximation!fixed-function refinements}
Realize $X$ by the independent valuation model \eqref{eq:valuation-laws}
for any of the four raw laws
$\mu_G,\muH,\mu_F^{\mathrm{int}},\mu_\sigma^{\mathrm{int}}$.
Let $r_n=j_n/q_n$ be its actual prefix through the $n$th active odd
prime, including the full prime-$2$ coordinate, reduced to lowest terms.
Fix a finite positive eventually nondecreasing function $\psi$, or a
fixed positive function comparable to such a function. Almost surely,
$q_n\to\infty$, $|X-r_n|>0$ for every $n$, and
\begin{equation}\label{eq:dichotomy}
\begin{split}
 I(\psi)<\infty&\ \Longrightarrow\
 \lim_{q\to\infty}q^{\psi(q)}\min_{j\in\mathbb Z}|X-j/q|=\infty,\\
 I(\psi)=\infty&\ \Longrightarrow\
 \liminf_{n\to\infty}q_n^{\psi(q_n)}|X-r_n|=0.
\end{split}
\end{equation}
The same assertions hold for $X=cZ_s$ in
\eqref{eq:bernoulli-model} with $p_0=3$ and either
$\alpha_p=1/p$ for all odd $p$ or $\alpha_p=1/(p-1)$ for all odd $p$.
In these models, $r_n$ is the actual prefix through the $n$th active
odd prime, including $c$, reduced to lowest terms.
They hold simultaneously for all positive rational $c$ and every
prescribed countable family of functions, allowing dependent positive
rational multipliers.
\end{corollary}

\begin{proof}
For either valuation law in \eqref{eq:valuation-laws}, write $K_p$ for
the valuation and $B_p=\mathbf1_{\{K_p\ge1\}}$. The independent odd
prime indicators have probabilities $1/p$ or $1/(p-1)$, respectively.
Totient factors are $(1-1/p)^{B_p}$; their prime-$2$ factor is positive
rational. For divisor sums put $F_p=\sum_{i=0}^{K_p}p^{-i}$.
The prime-$2$ valuation is finite almost surely in both models, and
\[
 \mathbb P(K_p\ge2)=p^{-2}\quad\hbox{or}\quad[p(p-1)]^{-1}.
\]
Their sum over odd primes converges, so only finitely many higher
valuations occur almost surely. Consequently the exact correction is
\begin{equation}\label{eq:sigma-correction}
 C=F_2\prod_{\substack{p\ge3\\K_p\ge2}}
       \frac{F_p}{1+1/p}\in\mathbb Q_{>0},\qquad
 X=C\prod_{p\ge3}(1+1/p)^{B_p}.
\end{equation}
The actual prefixes coincide at every cutoff beyond the exceptional
primes. Their active sets are identical, so the eventual integer index
shift is zero and their reduced denominators agree exactly.
Apply \Cref{thm:intensity} with $\kappa=1$ through its countable
rational-multiplier assertion. The Bernoulli specializations follow
directly. Positive errors and $q_n\to\infty$ in the divergent case
give infinitely many distinct reduced approximants; the all-denominator
conclusion in the convergent case excludes every such sequence.
\end{proof}

\begin{lemma}[Critical-scale rational rigidity]\label{lem:critical-rigidity-mas10}
Let $B_p$, at odd primes, be independent Bernoulli variables with
probabilities either $1/p$ throughout or $1/(p-1)$ throughout. Almost
surely, simultaneously for every $d\in\Q_{>0}$, the following holds for
$Y=d\prod_{p\ge3}(1-1/p)^{B_p}$: every reduced rational $b/a$ with
sufficiently large denominator and
\[
 |Y-b/a|<a^{-\log\log a}
\]
equals an actual finite prime prefix
$d\prod_{t\in T}(1-1/t)$. Here $T$ consists of the active odd primes
through some cutoff, and the lower threshold for $a$ may depend on
$d$ and the sample.
\end{lemma}

\begin{proof}
Fix $d\in\Q_{>0}$. Both activation laws satisfy
\eqref{eq:cumulative-intensity} with $\kappa=1$, so
\Cref{lem:canonical-rational-rigidity} applies with $s=-1$, $c=d$
and $p_0=3$. For all sufficiently large $a$,
\[
 (\log\log\log a)^2<\log\log a,
 \qquad a^{-\log\log a}\le a^{-(\log\log\log a)^2}.
\]
Its conclusion therefore applies to every approximant in the stated
window. Intersect over the countably many positive rational $d$.
\end{proof}

\begin{corollary}[Reciprocal-totient versus divisor-sum laws]
\label{cor:reciprocal-sigma-singularity-mas10}
Let $\iota(u)=1/u$ for $u>0$. Each of
$\iota_*\mu_G$ and $\iota_*\mu_F^{\mathrm{int}}$ is mutually singular
with each of $\muH$ and $\mu_\sigma^{\mathrm{int}}$.
\end{corollary}

\begin{proof}
For $c\in\Q_{>0}$ let $\mathcal A_c$ be the countable collection of
reduced positive fractions
\[
 \frac jq=c\prod_{p\in S}(1+1/p),
 \qquad S\text{ a finite set of odd primes}.
\]
Define the Borel sets
\begin{equation}\label{eq:separation-limsup-mas10}
 E_c=\bigcap_{Q\ge3}\ \bigcup_{\substack{j/q\in\mathcal A_c\\q\ge Q}}
 \left(\frac jq-q^{-8\log\log q},\frac jq+q^{-8\log\log q}\right),
 \qquad E=\bigcup_{c\in\Q_{>0}}E_c.
\end{equation}
The prime test for $\psi(q)=8\log\log q$ diverges by
\Cref{lem:prime-test}. Therefore \Cref{thm:approximation} gives
infinitely many corresponding prefix windows for either divisor-sum
law, with unbounded reduced denominators. Beyond the finitely many
higher valuations, \eqref{eq:sigma-correction} identifies these
prefixes with members of $\mathcal A_C$, where $C$ is a positive
rational random variable. The countable union over $c$ therefore gives
$\muH(E)=\mu_\sigma^{\mathrm{int}}(E)=1$.

For either totient model, write
$Y=d\prod_{p\ge3}(1-1/p)^{B_p}$, where its prime-$2$ factor
$d$ is positive rational, and work on the probability-one event of
\Cref{lem:critical-rigidity-mas10}. Put $x=1/Y$ and suppose $x\in E_c$
for some fixed rational $c>0$. There are fractions $j/q\in\mathcal A_c$
with $q\to\infty$ and the stated error. Since $j/q\to x>0$,
$j\asymp_x q$ and inversion gives
\[
 \left|Y-\frac qj\right|
 \ll_x q^{-8\log\log q}<j^{-\log\log j}
 \quad\text{for all sufficiently large }q.
\]
The reciprocal $q/j$ is reduced. Rigidity makes it
$d\prod_{t\in T}(1-1/t)$ for a finite active prefix $T$.
Writing the fixed rational $cd=u/v$ in lowest terms gives
\begin{equation}\label{eq:largest-prime-separation-mas10}
 u\prod_{p\in S}(p+1)\prod_{t\in T}(t-1)
 =v\prod_{p\in S}p\prod_{t\in T}t.
\end{equation}
Because $c$ is fixed and $q\to\infty$, the largest prime in $S$
tends to infinity along a subsequence: bounded prime sets yield
only finitely many anchors. Thus the largest prime $P$ in the
\emph{union} $S\cup T$ eventually exceeds every prime divisor of $u$.
It divides the right side of \eqref{eq:largest-prime-separation-mas10}.
It divides none of its left side: if $p<P$ is odd then $p+1<P$,
while $P\nmid P+1$, and every $t-1<P$.
The contradiction excludes $x\in E_c$. It holds for every rational
$c$, and rigidity was simultaneous over the rational head $d$.
Hence both reciprocal-totient laws give $E$ mass zero, proving all
four mutual singularities.
\end{proof}

\begin{corollary}[Products of independent variables]
\label{cor:scalar-products}
Let $m\ge1$ be fixed. The product of $m$ independent totient
raw variables, or of $m$ independent divisor-sum raw variables, obeys
\Cref{thm:intensity} with $\kappa=m$. Each copy may use either
the integer or the shifted-prime model. Prefixes include all prime-$2$
coordinates and are reduced after multiplying the full local factors.
\end{corollary}

\begin{proof}
Let $B_p$ indicate that at least one copy is active at the odd prime
$p$. Independence across copies and primes gives independent $B_p$
with
\[
 \mathbb P(B_p=1)=1-\prod_{i=1}^m(1-\alpha_{p,i})
                  =\frac m p+O_m(p^{-2}).
\]
The probability that two copies are active at the same prime is
$O_m(p^{-2})$. For divisor sums the probability of any valuation
exceeding one is also $O_m(p^{-2})$, by \eqref{eq:valuation-laws}.
Only finitely many such primes occur almost surely. The combined
prime-$2$ factor is rational and finite. Beyond the finite exceptions
the local factor is exactly $(1+s/p)^{B_p}$, with $s=-1$ for totients
and $s=1$ for divisor sums. The actual prefixes coincide at every
prime cutoff beyond the exceptional set with those of this Bernoulli
product times the finite rational correction. The active sets agree
at every odd prime, so their eventual integer index shift is zero;
the reduced prefixes and denominators therefore coincide for all
sufficiently large active indices. Apply \Cref{thm:intensity}
through its countable rational-multiplier assertion.
\end{proof}

\begin{corollary}[Coarse scales and the iterated-logarithm boundary]
\label{cor:boundaries}
For any variable covered by \Cref{thm:intensity}, almost surely
simultaneously for every $A>0$,
\begin{equation}\label{eq:all-positive-a}
 \lim_{q\to\infty}
 \exp\left(\frac{A(\log q)^2}{\log\log q}\right)
 \min_{j\in\mathbb Z}|X-j/q|=\infty.
\end{equation}
For fixed $C>0$ and $\beta\in\mathbb R$, put, at sufficiently large
arguments,
\begin{equation}\label{eq:log-boundary}
 \psi(q)=C(\log\log q)^{1/\kappa}
                  (\log\log\log q)^\beta.
\end{equation}
The scaled eventual exclusion holds when $\kappa\beta>1$, and the
scaled prefix liminf is zero when $\kappa\beta\le1$.
In particular, for each of the four raw laws $\kappa=1$ and the
boundary is $\beta=1$, with divergence for every fixed $C>0$ there.
\end{corollary}

\begin{proof}
For \eqref{eq:log-boundary}, the test in
\eqref{eq:approximation-intensity-integral} is
\[
 C^{-\kappa}\int^\infty\frac{du}{u(\log u)^{\kappa\beta}},
\]
which converges precisely when $\kappa\beta>1$. These functions are
eventually nondecreasing, so \Cref{thm:intensity} applies.
Fix one such boundary, for instance $C=1$, $\beta=2/\kappa$.
Its exponent is $o(\log q/\log\log q)$; comparison with this
single probability-one scaled-exclusion event proves
\eqref{eq:all-positive-a} simultaneously for every $A>0$.
At $A=0$ the nearest rational with denominator $q$ has error at most
$1/(2q)$, so zero cannot be included.
\end{proof}

\begin{corollary}[Singularity at different intensities]
\label{cor:intensity-singularity}
\index{Activation intensity!prime-index, mutual singularity}
Two laws covered by \Cref{thm:intensity} with distinct intensities are
mutually singular, as are their images under either map $u\mapsto\log u$ or
$u\mapsto-\log u$. In particular,
the additive convolution powers of each basic logarithmic law are
pairwise mutually singular.
\end{corollary}

\begin{proof}
For $\kappa_1<\kappa_2$, choose $1/\kappa_2<a<1/\kappa_1$.
The Borel set of $x$ for which
\[
 |x-j/q|<q^{-(\log\log q)^a}
 \quad\hbox{for infinitely many }(j,q)\in\Z\times\N_{\ge3}
\]
has probability one at intensity $\kappa_1$ and zero at intensity
$\kappa_2$, by \Cref{thm:intensity}. The logarithm is injective;
\Cref{cor:scalar-products} gives the convolution specialization.
\end{proof}

The fixed-degree field laws are mutually singular even when their
intensities agree; see \Cref{cor:fixed-degree-field-singularity}.

By \Cref{lem:valuation-likelihood},
$\mathsf A^{\otimes m}$ and $\mathsf B^{\otimes m}$ are equivalent for each
fixed $m$, and hence also their images under any common measurable map.
This equivalence uses the even conditioning in $\mathsf A$; equal
intensity alone does not imply equivalence.

\section{Continued fractions and packing dimension}
\label{sec:diophantine-continued-fractions}

Liouville approximation forces exceptionally large continued-fraction
digits along subsequences. For the four raw laws and their four
logarithmic laws, packing dimension gives the complementary recurrence:
every finite word of positive digits occurs infinitely often almost
surely. Each eventual-avoidance exceptional set has packing dimension
below one. In particular, successive denominator ratios are below two
infinitely often, while Liouville approximation for the raw laws
produces unbounded ratios of their logarithms. Sufficiently accurate
reduced prime prefixes also become convergents, linking the
fixed-function hierarchy to the next continued-fraction digit.

\begin{lemma}[Large continued-fraction digits and packing]
\label{lem:digit-packing}
The Borel set of irrational real numbers whose continued-fraction
digits tend to infinity has packing dimension at most $1/2$.
\end{lemma}

\begin{proof}
Fix $N\ge2$ and $s>1/2$ with
$S_N:=\sum_{a\ge N}a^{-2s}<1$. Let $K_N$ be the closure in $[0,1/N]$
of the numbers all of whose continued-fraction digits are at least $N$.
For $f_a(u)=1/(a+u)$, compactness gives
\[
 K_N=\{0\}\cup\bigcup_{a\ge N}f_a(K_N).
\]
Indeed, unbounded first digits tend to zero; bounded first digits admit
a constant subsequence with limiting tail in $K_N$. Conversely, zero
is a limit of branches, and continuity of $f_a$ gives
$f_a(K_N)\subset K_N$.
Let $C_N(\delta)$ be the least number of intervals of length $\delta$
covering $K_N$. For $0<\delta<N^{-2}$ put
$M=\lfloor\delta^{-1/2}\rfloor$. The branches with $a\ge M$ lie in
$[0,1/M]$, covered by at most $3\delta^{-1/2}$ such intervals.
Since $f_a$ is $a^{-2}$-Lipschitz,
\[
 C_N(\delta)\le3\delta^{-1/2}
                 +\sum_{a=N}^{M-1}C_N(\delta a^2).
\]
Choose $C\ge3/(1-S_N)$ also large enough for
$C_N(\delta)\le C\delta^{-s}$ on $[N^{-2},1]$.
Induction over the bands $[N^{-2(k+1)},N^{-2k}]$ is valid because each
$\delta a^2\ge N^2\delta$ is at a previously bounded larger scale.
The displayed recurrence gives
$C_N(\delta)\le(3+CS_N)\delta^{-s}\le C\delta^{-s}$.
Thus $\overline{\dim}_BK_N\le s$.
Every irrational whose digits are eventually at least $N$ belongs to
a countable union of finite-prefix M\"obius images of $K_N$ and their
integer translates. These maps are Lipschitz on $K_N$. The digit
conditions are Borel, since finite continued-fraction cylinders are
Borel. Countable stability of packing dimension gives the bound $s$.
For each $s>1/2$ one can choose $N$ as above, proving the claim.
\end{proof}

\begin{lemma}[Forbidden continued-fraction words and packing dimension]
\label{lem:forbidden-word-packing-mas10}
Fix a nonempty finite word $w$ of positive integers. The Borel set of
irrational real numbers whose continued-fraction digits eventually
avoid $w$ has packing dimension strictly less than one.
\end{lemma}

\begin{proof}
Write $f_a(x)=1/(a+x)$ on $[0,1]$, and for a word
$v=(a_1,\ldots,a_n)$ put $f_v=f_{a_1}\circ\cdots\circ f_{a_n}$ and
$I_v=f_v([0,1])$. Let $q_j$ be the reduced denominator of
$[0;a_1,\ldots,a_j]$ for $1\le j\le n$, with $q_0=1$.
The convergent-denominator formula gives
\[
 |f_v'(x)|=(q_n+q_{n-1}x)^{-2},\qquad
 \frac{\sup_{[0,1]}|f_v'|}{\inf_{[0,1]}|f_v'|}\le4.
\]
In particular, prescribing the next $k=|w|$ digits to equal $w$
removes at least the fraction $\varepsilon=|I_w|/4>0$ of every
parent cylinder's length. Cylinders of a fixed length have disjoint
interiors and cover $[0,1]$ up to endpoints. Therefore, if
$\mathcal V_m$ consists of words of length $N=km$ whose $m$
aligned $k$-blocks all differ from $w$, then
\[
 \sum_{v\in\mathcal V_m}|I_v|\le(1-\varepsilon)^m,
 \qquad
 \sum_{v\in\mathcal V_m}L_v\le4(1-\varepsilon)^m,
 \quad L_v:=\operatorname{Lip}(f_v).
\]
Choose $m$ so large that $N\ge2$ and the latter sum is less than one.
There is then a uniform contraction bound $L_v\le\rho<1$, since
$q_N$ is bounded below by the corresponding Fibonacci denominator.
Fix $1/2<s_0<1$ before varying the exponent. The chain rule gives
\[
 \sum_{|v|=N}L_v^{s_0}
 \le\left(\sum_{a\ge1}a^{-2s_0}\right)^N<\infty.
\]
Dominated convergence as $s\uparrow1$, with $s\ge s_0$, now supplies
$s<1$ such that
\begin{equation}\label{eq:word-pressure-mas10}
 \theta:=\sum_{v\in\mathcal V_m}L_v^s<1.
\end{equation}
Choose $t$ with $1/2<t<s$.

We record the endpoint estimate needed for this countable family.
Let $D_j$ consist of $0$ and all finite continued fractions
$[0;a_1,\ldots,a_i]$ with $i\le j$. These sets are compact: a sequence
with unbounded first digit tends to zero, while a bounded first digit
admits a constant subsequence, and induction applies to its tail.
For the least number $N_\delta$ of intervals of length $\delta$
covering a set, we claim
\begin{equation}\label{eq:finite-cf-endpoints-mas10}
 N_\delta(D_j)\ll_{j,t}\delta^{-t}\qquad(0<\delta\le1).
\end{equation}
The assertion is immediate for $j=0$. For the induction step split
$f_a(D_{j-1})$ at $A=\lceil\delta^{-1/2}\rceil$. All branches with
$a\ge A$, together with zero, lie in $[0,1/A]$, requiring
$O(\delta^{-1/2})$ intervals. For $a<A$, the map is $a^{-2}$-Lipschitz,
so the inductive bound
gives
\[
 N_\delta(D_j)\ll\delta^{-1/2}
  +C_{j-1,t}\delta^{-t}\sum_{a<A}a^{-2t}
 \ll_{j,t}\delta^{-t}.
\]
This proves \eqref{eq:finite-cf-endpoints-mas10}.

Let $K$ be the compact closure of the numbers obtained by infinitely
concatenating words from $\mathcal V_m$, and put
$P=\{f_v(0),f_v(1):v\in\mathcal V_m\}$. Both endpoints belong to
$D_N$, since replacing the final digit $a_N$ by $a_N+1$ represents
$f_v(1)$. Crucially,
\begin{equation}\label{eq:word-closure-mas10}
 K\subset\overline P\ \cup\ \bigcup_{v\in\mathcal V_m}f_v(K).
\end{equation}
Indeed, take a convergent sequence of infinite concatenations.
If its first branches have a constant subsequence, compactness of
the tails gives a limit in $f_v(K)$. Otherwise choose distinct first
branches. By \eqref{eq:word-pressure-mas10} their Lipschitz constants
tend to zero, and the sequence approaches their endpoints in $P$.

For $L_v\le\delta$, the whole interval $f_v([0,1])$ lies within
$\delta$ of $P$. By \eqref{eq:finite-cf-endpoints-mas10}, these small
branches and $\overline P$ can jointly be covered by
$C\delta^{-t}$ intervals of length $\delta$: enlarge each interval
in a $\delta$-cover of $D_N$ by $\delta$ at both ends and divide it
into three intervals. The remaining branches in
\eqref{eq:word-closure-mas10} are finite in number and give
\[
 N_\delta(K)\le C\delta^{-t}
   +\sum_{\substack{v\in\mathcal V_m\\L_v>\delta}}
          N_{\delta/L_v}(K).
\]
Choose $A\ge C/(1-\theta)$, enlarged to cover scales in $[\rho,1]$.
Scale induction on $[\rho^{j+1},\rho^j]$ is valid because
$\delta/L_v\ge\delta/\rho$. Using $t<s$ and
\eqref{eq:word-pressure-mas10}, it yields
\[
 N_\delta(K)\le(C+A\theta)\delta^{-s}\le A\delta^{-s}.
\]
Thus $K$ has upper box dimension at most $s<1$.

Every continued fraction avoiding $w$ everywhere belongs to $K$,
since it avoids $w$ at the aligned positions in particular.
Eventual avoidance is contained in the countable union of finite-prefix
M\"obius images of $K$ and their integer translates. Each map is
Lipschitz on $[0,1]$, so countable stability of packing dimension gives
the same bound $s<1$. Finally these digit conditions are Borel,
being countable combinations of continued-fraction cylinders.
\end{proof}

\begin{corollary}[Packing dimension and continued-fraction recurrence]
\label{cor:packing-word-recurrence}
Let $\lambda$ be a Borel probability measure on $\R$ such that every
positive-$\lambda$-mass Borel set has packing dimension one. Then
$\lambda$ gives full mass to irrational numbers, and almost surely
every finite word of positive continued-fraction digits occurs
infinitely often. In particular this holds for the four basic raw
laws and their four logarithmic laws.
\end{corollary}

\begin{proof}
Every singleton has packing dimension zero, so the hypothesis excludes
rational atoms, and countability gives $\lambda(\Q)=0$.
For each fixed finite word, the Borel exceptional set of eventual
avoidance has packing dimension below one by
\Cref{lem:forbidden-word-packing-mas10}, hence has zero $\lambda$-mass.
There are countably many finite words; intersecting their probability-one
recurrence events proves the simultaneous assertion. The eight-law
application follows from \Cref{thm:local-dimension-oscillation}.
\end{proof}

\begin{corollary}[Continued-fraction approximation and selected prefixes]
\label{cor:continued-fractions}
\index{Continued fractions}
For each of the four raw laws in \Cref{thm:liouville-carriers}, write $X=[a_0;a_1,a_2,\ldots]$ and let $p_n/q_n$ be
its reduced convergents. Almost surely,
\begin{equation}\label{eq:cf-oscillation}
 \liminf_n a_n=1,\qquad
 \liminf_n\frac{\log q_{n+1}}{\log q_n}=1,\qquad
 \limsup_n\frac{\log q_{n+1}}{\log q_n}=\infty.
\end{equation}
There are infinitely many indices with $q_{n+1}<2q_n$.
Every reduced prefix $b/q$ with $|X-b/q|<1/(2q^2)$ is a convergent;
if it is $p_n/q_n$, its next digit satisfies
\begin{equation}\label{eq:next-digit}
 a_{n+1}>\frac1{q_n^2|X-p_n/q_n|}-2.
\end{equation}
For each fixed function in the divergent scope of
\Cref{thm:approximation}, and simultaneously for any prescribed
countable family of such functions, its approximation hierarchy is
attained along selected prefix convergents, with next digits as in
\eqref{eq:next-digit}.
The convergent part of that theorem applies to every sufficiently
large convergent denominator.
\end{corollary}

In particular $\log q_{n+1}/\log q_n$ does not tend to infinity;
in this explicitly stated sense these variables are not strong Liouville
numbers.

\begin{proof}
By \Cref{cor:packing-word-recurrence}, digit one occurs infinitely
often, giving the first assertion of
\eqref{eq:cf-oscillation} for a raw law.
The recurrence $q_{n+1}=a_{n+1}q_n+q_{n-1}$ gives
$q_{n+1}<2q_n$ whenever $a_{n+1}=1$ and $n$ is sufficiently large.
Since $q_n\to\infty$ and $q_{n+1}>q_n$ eventually, the lower limit is one.

The determinant and recurrence identities for
convergents give
\[
 \frac1{a_{n+1}+2}
 <q_n^2|X-p_n/q_n|<\frac1{a_{n+1}};
\]
indeed the reciprocal of the middle expression is
$[a_{n+1};a_{n+2},\ldots]+q_{n-1}/q_n$
(cf.~\cite[\S~1.12(ii)]{DLMF}). For $q=1$, an integer $b$ with $|X-b|<1/2$ is either $a_0$ or
$a_0+1$; in the latter case $a_1=1$, so it is again a convergent.
For $q\ge2$, the two finite continued-fraction expansions of a reduced
$b/q$ give adjacent cylinders of lengths
$1/[q(q+s)]$ and $1/[q(2q-s)]$, where $0<s<q$ is a preceding
convergent denominator. Both lengths exceed $1/(2q^2)$, and in either
cylinder $b/q$ is a convergent. This proves the criterion and
\eqref{eq:next-digit}.
By \Cref{thm:liouville-carriers}, approximations of arbitrarily large
fixed power exist with unbounded reduced denominators. They are
convergents by this criterion, and the lower error bound forces
$a_{n+1}>q_n^{m-2}-2$ along approximations of order $m$.
The recurrence therefore gives the infinite upper limit in
\eqref{eq:cf-oscillation}.

Finally fix a function $\psi$ in the divergent scope of
\Cref{thm:approximation}. Replacing it by
$\widehat\psi=\max(\psi,3)$ preserves divergence of its prime series:
a function comparable to an eventually nondecreasing positive function
is either eventually bounded above or eventually bounded below by $3$.
The theorem supplies actual reduced prefixes $b_j/q_j$, $q_j\to\infty$,
with $q_j^{\widehat\psi(q_j)}|X-b_j/q_j|\to0$.
Thus $q_j^2|X-b_j/q_j|\to0$, so eventually these are convergents and
\eqref{eq:next-digit} applies. This proves the stated hierarchy
for each fixed $\psi$, and simultaneously for any prescribed countable
family.
\end{proof}

For each rational image $Y=R(\boldsymbol X)$ or
$R(\log\boldsymbol X)$ in \Cref{cor:rational-statistics}\textup{(b)},
the same packing argument gives $\liminf a_n=1$,
$\liminf\log q_{n+1}/\log q_n=1$ and $q_{n+1}<2q_n$ infinitely often.
For $R\in\Q(T_1,\ldots,T_d)$ applied to the raw vector, part~\textup{(c)}
also gives the infinite upper limit in \eqref{eq:cf-oscillation}.
These almost-sure assertions survive positive-probability conditioning.

\section{Deletion of ratio factors}
\label{sec:diophantine-deletion}

\begin{theorem}[Deterministic deletion of ratio factors]
\label{thm:deterministic-deletion}
\index{Deletion of prime factors!deterministic}
Let $\mathcal P$ be any deterministic set of odd primes, let the
independent $K_p$ have either valuation law in \eqref{eq:valuation-laws},
and let $h$ be any positive rational-valued random variable, possibly
dependent on the $K_p$. Both products
\[
 X_{\mathcal P}:=h\prod_{p\in\mathcal P}
 (1-1/p)^{\mathbf1_{\{K_p>0\}}},\qquad
 Y_{\mathcal P}:=h\prod_{p\in\mathcal P}
 \frac{\sigma(p^{K_p})}{p^{K_p}}
\]
exist and are finite and positive almost surely. If
$\sum_{p\in\mathcal P}p^{-1}<\infty$, both are rational almost surely.
If that sum diverges, both laws are atomless and carried by $\mathcal L$.
The same dichotomy holds for
$Z_{\mathcal P}=h\prod_{p\in\mathcal P}R_p(K_p)$, where each local
function $R_p(k)$ is fixed as either $(1-1/p)^{\mathbf1_{\{k>0\}}}$ or
$\sigma(p^k)/p^k$; the choice may vary with $p$.
\end{theorem}

\begin{proof}
First take $h=1$. Put $B_p=\mathbf1_{\{K_p>0\}}$ and
$\alpha_p=\mathbb P(B_p=1)$, equal to $1/p$ or $1/(p-1)$.
The logarithmic increments satisfy
\[
 0\le\gsig(p^{K_p})\le B_p\omega_p,\qquad
 \sum_p\alpha_p\omega_p<\infty.
\]
Thus every stated logarithmic product converges absolutely almost surely.
If the retained reciprocal sum converges, Borel--Cantelli leaves only
finitely many active retained primes, giving rational products.
If it diverges, the independent increments of each logarithmic
product are nonzero with probabilities $\alpha_p$ of divergent sum;
their zero atoms have mass $1-\alpha_p>1/2$ for $p\ge5$.
\Cref{lem:levy-atomless} gives atomlessness, hence almost sure
irrationality of every stated raw product.

Extend the valuations independently to all odd primes and put
\[
 U:=\prod_{p\ge3}\sigma(p^{K_p})/p^{K_p}.
\]
Each local reduced denominator is $p^{K_p}$. The zero hazard is
$\log\log y+O(1)$, the expected logarithmic tail is $O(y^{-1})$,
and $\mathbb E\sum_{3\le p\le y}K_p\log p\ll\log y$.
Thus \Cref{lem:rational-blocks}, for each fixed $m\ge1$, supplies
cutoffs $y_j\to\infty$ with
\[
 Q_j=\prod_{3\le p\le y_j}p^{K_p}\to\infty,
 \qquad 0\le U-U(y_j)<Q_j^{-2m}.
\]
Since $U(y_j)\ge1$, its positive logarithmic tail satisfies
\[
 T_\sigma(y_j):=\log\frac U{U(y_j)}
 \le\frac{U-U(y_j)}{U(y_j)}<Q_j^{-2m}.
\]
Deletion decreases this tail. Moreover,
$\omega_p\le2\log(1+1/p)\le2\gsig(p^{K_p})$ when $K_p>0$ and $p\ge3$,
so every retained absolute logarithmic tail is at most
$2T_\sigma(y_j)$. Consequently, for each stated unit-head product $Z$,
\[
 |Z-Z(y_j)|\le Z\bigl(e^{2T_\sigma(y_j)}-1\bigr)
 \ll_Z Q_j^{-2m}.
\]
Each retained reduced denominator $q_j$ divides $Q_j$. The prefixes
converge to irrational limits, so their errors are nonzero and their
reduced denominators tend to infinity. The displayed errors are
therefore eventually less than $Q_j^{-m}\le q_j^{-m}$. Intersecting
over $m$ proves the Liouville assertions for unit heads.

Multiplication by a nonzero rational preserves rationality and the
Liouville property, by multiplying denominators and increasing the
approximation order. This applies pointwise to any stated head $h$.
For each unit-head variable $Z$ and every $t\in\R$,
\[
 \mathbb P(hZ=t)\le\sum_{r\in\Q_{>0}}\mathbb P(Z=t/r)=0
\]
in the divergent case.
\end{proof}

If the deterministic retained set satisfies
\[
 \sum_{\substack{p\le y\\p\in\mathcal P}}\frac1p
 =\delta\log\log y+O(1),\qquad\delta>0,
\]
then $X_{\mathcal P}$ and $Y_{\mathcal P}$ obey \Cref{thm:intensity}
with $\kappa=\delta$, using their actual reduced prefixes through the
active retained primes, including $h$. Indeed, the activation
probabilities $\mathbf1_{\mathcal P}(p)/p$ or
$\mathbf1_{\mathcal P}(p)/(p-1)$ satisfy
\eqref{eq:cumulative-intensity}. For $Y_{\mathcal P}$, only finitely
many higher valuations occur almost surely; their rational correction,
as in \eqref{eq:sigma-correction}, is covered by the theorem's
finite-alteration clause.

\begin{corollary}[Simultaneous and adaptive deletion]
\label{cor:adaptive-deletion}
\index{Deletion of prime factors!adaptive}
Let the independent $K_p$ at all odd primes have either valuation law
in \eqref{eq:valuation-laws}. Almost surely, simultaneously for every
choice
\[
 R_p\in\left\{1,\ (1-1/p)^{\mathbf1_{\{K_p>0\}}},
                   \sigma(p^{K_p})/p^{K_p}\right\},
 \qquad h\in\Q_{>0},
\]
the product $h\prod_{p\ge3}R_p$ exists in $(0,\infty)$ and belongs
to $\Q\cup\mathcal L$. Every finite vector of such products has the
simultaneous approximations of \Cref{lem:joint-rational-carrier};
every rational-coefficient rational function defined at that vector
also belongs to $\Q\cup\mathcal L$.
Thus the choices and rational heads may be measurable functions of all
the valuations and of additional randomness. The selected vector and
its logarithmic image have laws of Hausdorff dimension zero.
\end{corollary}

\begin{proof}
On the common probability-one event in the preceding proof, the product
$U$ is finite and, for every $m\ge1$, there are common cutoffs and
denominators $Q_j\to\infty$ with $T_\sigma(y_j)<Q_j^{-2m}$.
For every unit-head selection $Z$, its absolute logarithmic tail is
at most $2T_\sigma(y_j)$, its prefix denominator divides $Q_j$, and
$Z\le U$. Hence, uniformly over all selections,
\[
 |Z-Z(y_j)|\le U\bigl(e^{2T_\sigma(y_j)}-1\bigr)
 \ll_U Q_j^{-2m}.
\]
Absolute logarithmic convergence follows from the same domination.
For any finite vector, multiply $Q_j$ by a common denominator of its
rational heads; increasing $m$ absorbs the fixed factors.
\Cref{lem:joint-rational-carrier} gives all remaining assertions,
including the coordinate conclusion by its rational maps.
\end{proof}

\section{Rational images and conditioning}
\label{sec:diophantine-convolution}

\Cref{cor:rational-statistics}\textup{(a)--(b)} applies in particular to
each fixed nonzero real linear combination and each finite independent
convolution power of the four raw laws or their logarithmic images.
For $d=2$ and integer laws, singularity of the law of
$a_1\log X_1+a_2\log X_2$ $(a_1,a_2\in\R)$ already follows
from~\cite[Theorem~3, p.~198]{Bab92}.
General locally Lipschitz images may have atoms. The Liouville conclusion
fails for almost every real coefficient: since $X_1>0$ and $\mathcal L$
is Lebesgue-null, Fubini's theorem gives
$\mathbb P(aX_1\in\mathcal L)=0$ for almost every $a\in\R$.

\KeepHeadingSpace{4\baselineskip}
Positive-probability conditioning uses domination of the original joint
law; the conditional samples need not remain independent.

\begin{corollary}[Liouville values after positive-probability conditioning]
\label{cor:ap-liouville}
\index{Conditioning!Liouville carriers}
Let $\boldsymbol X$ have either joint model of
\Cref{cor:rational-statistics,cor:independent-phi-sigma-pairs}, and let
$E$ be any measurable event with $\mathbb P(E)>0$.
All conclusions of the corresponding corollary hold under
$\mathbb P(\,\cdot\mid E)$, including when $E$ involves infinitely many
valuation coordinates or couples the samples. In particular,
$\mathbb P(X_{Q,b}\in\mathcal L)=1$ for fixed $Q\ge1$ and
$\gcd(b,Q)=1$; its law is singular continuous with Hausdorff dimension zero.
\end{corollary}

\enlargethispage{6pt}
\begin{proof}
For the original and conditional joint laws $\lambda$ and $\lambda_E$,
\[
 \lambda_E(A)=\frac{\mathbb P(\boldsymbol X\in A,E)}{\mathbb P(E)}
 \le\mathbb P(E)^{-1}\lambda(A)
\]
for every Borel set $A$. The same domination holds for measurable
images. It preserves the Hausdorff-zero full carriers, null pole and
level sets for each fixed real rational map, and the simultaneous
full-measure assertions for the countable rational-coefficient family.
Conditioning also preserves the probability-one prefix approximation
event in \Cref{cor:independent-phi-sigma-pairs}.
Locally Lipschitz images retain the inherited carrier dimension;
\Cref{lem:fourier-Hausdorff} gives Fourier dimension zero.
For the independent model, positive-mass packing dimensions persist
by domination, and local dimensions persist by differentiation of
measures as in \Cref{thm:local-dimension-oscillation}.
For $X_{Q,b}$, combine \eqref{eq:ap-cylinder-domination} with
\Cref{thm:liouville-carriers}. Information dimensions of logarithmic
sums under valuation-measurable conditioning are treated in
\Cref{cor:arithmetic-entropy-invariance}.
\end{proof}

\FloatBarrier
\chapter{Information dimensions and quantization}
\label{sec:information-dimensions}
\index{Information dimension!definition}
\index{Shannon entropy!quantized observations}

At resolution $h$, an observation records the mesh-cell index
$\lfloor Y/h\rfloor$. How much information does this reveal about a
sample of an arithmetic law? Shannon entropy averages the information
in the cell index according to its probability. A Hausdorff-zero full
carrier does not determine this average at each resolution.

The lower estimates recover information about hidden prime choices
from the observed sum. The upper estimate records a prime head and a quantized tail
with an adaptive cutoff. The quantization theorem combines geometric
codebook bounds with the rate--distortion identity of Kawabata and
Dembo \cite[Proposition~3.3]{KD94}. The final section treats weighted
independent partial sums and convolution powers
(\Cref{cor:convolution-information}), then infinite sums
(\Cref{thm:infinite-weighted-sums}). The Fourier estimate
\Cref{prop:infinite-sum-fourier-count} and moment bounds yield the
Schwartz conclusion in \Cref{cor:infinite-sum-schwartz}.

Write $\mathsf H$ for discrete entropy with natural logarithms and set
\[
 H_h(Y):=\mathsf H(\lfloor Y/h\rfloor),\qquad
 \underline d_1(Y):=\liminf_{h\downarrow0}\frac{H_h(Y)}{\log(1/h)},
 \qquad
 \overline d_1(Y):=\limsup_{h\downarrow0}\frac{H_h(Y)}{\log(1/h)}.
\]
These are the lower and upper information dimensions; see~\cite{WV10}.
The basic logarithmic laws are $Z_g=\sum_p g(p^{K_p})$, with
$g\in\{f,\gsig\}$ and either valuation model in \eqref{eq:valuation-laws}.
Their raw images are $e^{-Z_f}$ and $e^{Z_{\gsig}}$.
\Cref{fig:four-laws} displays these logarithmic laws, their raw images and the two
valuation models.
Throughout, $x_+:=\max(x,0)$.

For digit prefixes, this average information differs from min-entropy,
which depends only on the most probable prefix. Almost-sure normality
asks about digit frequencies along one infinite expansion. The finite
prefix comparison in \Cref{cor:quantized-randomness} keeps these three
questions separate.

\paragraph*{Scalar information bounds}
The scalar constants $c_L,d_*$ are defined in \eqref{eq:scalar-functional};
\eqref{eq:scalar-strengthened-lower} gives the strengthened lower bound $c_L^+$.

\begin{theorem}[Two-sided bounds for information dimension]
\label{thm:shannon-dimension}
\index{Information dimension!scalar bounds}
Let $g\in\{f,\gsig\}$ and let $Z_g=\sum_\ell g(\ell^{K_\ell})$
use either valuation model in \eqref{eq:valuation-laws}. Put
$Y_g=e^{-Z_g}$ for $g=f$ and $Y_g=e^{Z_g}$ for $g=\gsig$.
All their quantized entropies are finite, and
\begin{equation}\label{eq:positive-information-dimension}
 \begin{gathered}
 c_L^+\le\underline d_1(Z_g)=\underline d_1(Y_g)
 \le\overline d_1(Z_g)=\overline d_1(Y_g)\le d_*,\\
 0.55043653<c_L^+,\qquad d_*<0.570489495.
 \end{gathered}
\end{equation}
For each $g$, both dimensions agree between the integer and shifted-prime
models. They are unchanged by conditioning on any positive-probability
event in the valuation sigma-field. In particular this
holds for the fixed-progression totient laws.
\end{theorem}

For the lower bound, reveal most prime coordinates and leave a finite
collection of prime choices hidden. The observation can distinguish
these choices only to the extent that different contributions avoid
falling within the same tolerance. Bounding such near-collisions
therefore bounds the information still hidden after observation.
Single-prime recovery gives a baseline; joint recovery of pairs,
clusters and selected triples improves it. These gains must belong
to one compatible partition, so no hidden choice is counted twice.

The ordered limits in \Cref{lem:ordered-limits} turn these finite
recoveries into a bound for the full lower limit. Cutoffs, finite
partitions and strict arithmetic margins are fixed first as the mesh
tends to zero, then removed in the stated order. Weak convergence
alone does not give entropy convergence.

\KeepHeadingSpace{4\baselineskip}
\section{Entropy transfers and ordered limits}
\label{sec:information-transfers}

\begin{lemma}[Finite entropy comparisons]\label{lem:transfers}
\begin{enumerate}[label=\textup{(\roman*)},leftmargin=*]
\item Let $Z$ be real with $\mathbb E|Z|<\infty$. Its mesh entropies and
those of $e^Z$ are finite. With
$J=\lfloor|Z|\rfloor$ and $E=\mathbf1_{\{Z<0\}}$,
\begin{equation}\label{eq:entropy-exponential-comparison}
 |H_h(e^Z)-H_h(Z)|\le\mathsf H(J,E)+\mathbb EJ+O(1)
 \qquad(0<h\le1).
\end{equation}
Moreover, $\underline d_1(e^Z)=\underline d_1(Z)$ and
$\overline d_1(e^Z)=\overline d_1(Z)$.
\item Let $U,D$ be real random variables on the same probability space,
with $D$ countably valued and $\mathsf H(D)<\infty$.
Whenever one of the two entropies is finite,
\begin{equation}\label{eq:entropy-displacement}
 |H_h(U+D)-H_h(U)|\le\mathsf H(D)+\log2.
\end{equation}
The information dimensions satisfy
$\underline d_1(U+D)=\underline d_1(U)$ and
$\overline d_1(U+D)=\overline d_1(U)$.
\end{enumerate}
\end{lemma}

\begin{proof}
Comparison with the probabilities $2^{-j-1}$ gives
$\mathsf H(J)\le(\mathbb EJ+1)\log2$. After recording $(J,E)$,
a unit interval meets at most $h^{-1}+2$ mesh cells. Hence
$H_h(Z)\le\mathsf H(J,E)+\log(h^{-1}+2)<\infty$.
On either signed unit interval specified by $(J,E)=(j,e)$,
the derivatives of the exponential map and its inverse on the image
are at most $e^{j+1}$. Each input mesh cell therefore meets at most
$e^{j+1}+2$ output cells, in either direction. The entropy chain rule
proves \eqref{eq:entropy-exponential-comparison}, including finiteness
of the output entropy. Replacing $Z$ by $-Z$ gives the same comparison
for $e^{-Z}$; reflection changes mesh entropy by at most $\log2$.
Given $D$ and either quantizer in \eqref{eq:entropy-displacement},
the other has at most two possible values. Applying the chain rule
in both directions proves that inequality and simultaneous entropy
finiteness. The information-dimension identities follow from these
bounded entropy differences.
\end{proof}

\begin{corollary}[Entropy invariance for arithmetic laws]
\label{cor:arithmetic-entropy-invariance}
For fixed $n\ge1$, let $Z=b+\sum_{i=1}^n a_iZ_i$, where
$Z_i=\sum_p g_i(p^{K_{i,p}})$ are independent basic logarithmic
variables and $b,a_1,\ldots,a_n$ are fixed real numbers.
Both information dimensions of $Z$ and $e^Z$ are unchanged under
any of the following operations:
\begin{enumerate}[label=\textup{(\roman*)}]
\item replacing any valuation model with the other model for the same
arithmetic function;
\item replacing any $Z_i$ with $g_i=\gsig$ by
$\sum_p\mathbf1_{\{K_{i,p}\ge1\}}\log(1+p^{-1})$;
\item conditioning on any positive-probability event
$D\in\sigma(K_{i,p}:i,p)$.
\end{enumerate}
All resulting mesh entropies are finite. For the conditioning in
\textup{(iii)},
\[
 H_h(Z\mid D)-H_h(Z)=o(\log(1/h))\qquad(h\downarrow0).
\]
\end{corollary}

\begin{proof}
The mean absolute prime increments in \eqref{eq:valuation-laws}
are $O(p^{-2})$. All the stated finite combinations therefore have
finite absolute mean. To compare the two valuation models, couple
them independently over primes by
\[
\begin{aligned}
 \mathbb P(J_p=V_p=0)&=(p-2)/(p-1),\\
 \mathbb P(J_p=V_p=r)&=(p-1)p^{-r-1} &&(r\ge1),\\
 \mathbb P(J_p=0,V_p=r)&=p^{-r-1} &&(r\ge1).
\end{aligned}
\]
These have the required marginals, also at $p=2$. Record zero on a
match and $r$ on a mismatch. The mismatch probability is
$\delta_p=1/[p(p-1)]$, and its record entropy is at most
\[
 -(1-\delta_p)\log(1-\delta_p)
 +\sum_{r\ge1}(r+1)p^{-r-1}\log p
 \ll\frac{\log p}{p^2}.
\]
The whole record has almost surely finite support, is countably
valued, and has entropy at most the sum of these entropies. For the
last assertion, finite-coordinate entropies increase to the entropy
of the full countable record. Its resulting logarithmic displacement
thus has finite entropy, so \eqref{eq:entropy-displacement} applies.

For either valuation model, set $T_p=K_p\mathbf1_{\{K_p\ge2\}}$.
Writing $\epsilon_p=\mathbb P(K_p\ge2)$, one has
$\epsilon_p=p^{-2}$ or $1/[p(p-1)]$. Given $T_p\ne0$,
$T_p=2+G_p$, where
$\mathbb P(G_p=j)=(1-1/p)p^{-j}$ for $j\ge0$. Hence
\[
 \mathsf H(T_p)
 =-\epsilon_p\log\epsilon_p-(1-\epsilon_p)\log(1-\epsilon_p)
   +\epsilon_p\mathsf H(G_p)
 \ll\frac{\log p}{p^2}.
\]
Thus the full exceptional record has finite entropy. With
$B_p=\mathbf1_{\{K_p\ge1\}}$, the difference
\[
 \sum_p\log(1+p^{-1}+\cdots+p^{-K_p})
       -\sum_pB_p\log(1+p^{-1})
\]
is a function of this record and has finite entropy. The totient
first-jump representation is already exact.

For a positive finite-coordinate event $C$, couple the unchanged tails
identically. The original and conditioned heads have finite entropy,
so their displacement $\Delta_C$ has finite entropy. By
\eqref{eq:entropy-displacement},
\[
 |H_h(Z\mid C)-H_h(Z)|\le\mathsf H(\Delta_C)+\log2=O_C(1).
\]
Finite concatenations of these records allow all fixed coefficients
and mixed choices of the basic laws.

If real laws $P,Q$ have finite first moments and total-variation
distance $0<\delta<1$, write
$P=(1-\delta)R+\delta P'$ and $Q=(1-\delta)R+\delta Q'$ using their
common part. The entropy bounds for mixtures and the unit-interval
estimate above give
\[
 |H_h(P)-H_h(Q)|\le\delta\log(1/h)+O_{P,Q,\delta}(1).
\]
Here $H_h$ denotes mesh entropy of a law as well as of a variable.
Approximate $D$ in probability by positive finite-coordinate events
$C_m$. Their conditional laws of $Z$ converge in total variation to
$\operatorname{Law}(Z\mid D)$, with distances $\delta_m\to0$.
Combining the two displays, first letting $h\downarrow0$ for fixed $m$
and then $m\to\infty$, proves the asserted $o(\log(1/h))$ difference.
Conditioning preserves finite absolute mean, so
\eqref{eq:entropy-exponential-comparison} gives the same conclusion
for exponential images. The fixed-progression totient laws are the
finite-coordinate conditionings by $D_{Q,b}$ identified before
\eqref{eq:ap-cylinder-domination}.
\end{proof}

\begin{lemma}[Entropy from a collision count]\label{lem:energy-entropy}
Let $W$ have finite support $\Omega$, let
$p_*:=\max_{x\in\Omega}\mathbb P(W=x)$, and let $F\colon\Omega\to\mathbb R$.
Suppose every conditional support of $F(W)$ given an observation $Y$
has diameter at most $\delta$. If
$R_\delta=\#\{(x,x')\in\Omega^2:|F(x)-F(x')|\le\delta\}$, then
\[
 \mathsf H(W)-\mathsf H(W\mid Y)
 \ge[\mathsf H(W)-\log(p_*R_\delta)]_+.
\]
\end{lemma}

\begin{proof}
Put $A(x)=\#\{x':|F(x')-F(x)|\le\delta\}$. For each realized
$x$ in a conditional support, that support has size at most $A(x)$.
The support entropy bound, the tower property and Jensen's inequality give
\[
 \mathsf H(W\mid Y)\le\mathbb E\log A(W)
 \le\log\mathbb E A(W)\le\log(p_*R_\delta).
\]
The entropy loss is nonnegative, which permits taking the positive part.
\end{proof}

For the next two lemmas, let $Z=b+\sum_{i=1}^n a_iZ_i$, where the
$Z_i$ are independent basic logarithmic variables and
$b,a_1,\ldots,a_n$ are fixed real numbers with $a_i\ne0$.
Put $h=e^{-L}$ and assign the coordinate $(i,p)$ the exponent
$\log p/L$. Fix $a>0$ and partition $(a,1)$ into $m$ exponent
intervals. Retain an interval $I$ only when exactly one of its
coordinates $(i,p)$ has valuation one and all its other coordinates
have valuation zero. Reveal the copy label $i$ and the unique interval
$(P,2P]$ containing $p$, where $P=2^k$, $k\in\Z_{\ge0}$, but not
the valuations of copy $i$ in that range.
If $(\log P/L,\log(2P)/L]$ is not contained in $I$, reveal
all valuations in $I$ instead and do not retain its prime choice.
Reveal all other valuations. Let $\mathcal D$ be the resulting
sigma-field and $W$ the finite vector of retained prime choices,
with a dummy entry for every interval not retained.

\begin{lemma}[Conditional prime-block entropy]
\label{lem:hidden-vector}
For any $\mathcal D$-measurable partition of the retained choices
into consecutive blocks, write $N_B=\prod_{j\in B}P_j$.
Let $P_+$ be the first retained dyadic scale after $B$, and set
$\delta_B=C_m(h+P_+^{-1})$, with $P_+^{-1}=0$ for the final block.
Here $C_m$ may depend on the fixed coefficients. Let
$F_B(\boldsymbol p)$ be the actual logarithmic contribution of $B$,
and let $R_B$ be an upper bound for the ordered number of pairs in
its dyadic prime product satisfying
$|F_B(\boldsymbol p)-F_B(\boldsymbol p')|\le\delta_B$.
Then, uniformly over the retained ranges,
\begin{equation}\label{eq:block-reward}
 H_h(Z)\ge
 \mathbb E\sum_B[\,2\log N_B-\log R_B\,]_+
 -O_m(\log L).
\end{equation}
\end{lemma}

\begin{proof}
For sufficiently large $L$, the retained primes are odd. In each coarse
interval the record specifies retention status and,
for a retained choice, its copy label and dyadic range, while suppressing
all valuations of that copy in the selected range. These recording
rules factor over the coarse intervals. Consequently, given
$\mathcal D$, the hidden primes are independent, with
\begin{equation}\label{eq:prior}
 \mathbb P(W_j=p\mid\mathcal D)
 =\frac{q_{i,p}}{\sum_{P_j<r\le2P_j}q_{i,r}},\qquad
 q_{i,p}:=\frac{\mathbb P(K_{i,p}=1)}{\mathbb P(K_{i,p}=0)}
 =\frac1p+O(p^{-2}).
\end{equation}
The revealed infinite series converges absolutely and is
$\mathcal D$-measurable.
Thus $Q=\lfloor Z/h\rfloor$ is determined by $(\mathcal D,W)$.
Since $W$ is finite-valued for every fixed $L$,
\begin{equation}\label{eq:finite-identity}
 \mathsf H(Q)\ge\mathsf H(Q\mid\mathcal D)
 =\mathsf H(W\mid\mathcal D)-\mathsf H(W\mid Q,\mathcal D).
\end{equation}

Uniformly for $P\ge e^{aL}/2$, the prime number theorem, or
\Cref{lem:prime-log-stieltjes}, gives
$\sum_{P<p\le2P}q_{i,p}\asymp1/\log P$. It follows that
\[
 \max_p\mathbb P(W_j=p\mid\mathcal D)\ll L/P_j,
 \qquad
 \mathsf H(W_j\mid\mathcal D)=\log P_j+O(\log L).
\]
At the ends of each coarse interval, at most two partial dyadic
intervals are discarded. Enlarging each to a full dyadic interval gives
total odds $O(1/L)$, so the probability of discarding a singleton
there is $O(1/L)$. There are at most $m$ choices, each contributing
at most $L+O(1)$ entropy; these boundary events therefore cost $o(L)$
for fixed $m$.

Fix a record and apply the chain rule in increasing block order.
After $Q$ and earlier blocks are known, all later hidden increments
have absolute sum at most $C_mP_+^{-1}$. Thus any two feasible values
of the current block differ by at most $\delta_B$, enlarging $C_m$
if necessary. The conditional prior satisfies
\[
 \max_x\mathbb P(W_B=x\mid\mathcal D)
 \le (CL)^{|B|}/N_B,\qquad
 \mathsf H(W_B\mid\mathcal D)=\log N_B+O_m(\log L).
\]
Given $\mathcal D$, this prior is independent of the earlier blocks.
The prime collision count is bounded by the corresponding count in
any containing integer product.
Apply \Cref{lem:energy-entropy} to $W_B$ with observation
$(Q,W_{<B})$. It bounds the block's entropy loss below by
$[2\log N_B-\log R_B]_+-O_m(\log L)$.
Summing in \eqref{eq:finite-identity} proves
\eqref{eq:block-reward}.
\end{proof}

\begin{lemma}[Passage to exponent configurations]
\label{lem:ordered-limits}
In the preceding prime-block construction, fix $0<a<1$ and a
deterministic nested sequence of finite partitions of $(a,1)$ whose
mesh tends to zero. Let $R_a$ be a Borel function of the retained
marked exponents. For each partition, let $\mathcal D$ be the
corresponding sigma-field and choose a $\mathcal D$-measurable
consecutive block rule using each arithmetic estimate only on its
validity domain. For every fixed partition and every $\eta>0$,
assume that its arithmetic bounds in \eqref{eq:block-reward}
give a block-reward sum at least $L(R_a-\eta)$ for all sufficiently
large $L$, uniformly on compact subsets of their strict validity
domains. The implied constants and threshold for $L$ may depend on
the partition; $L\to\infty$ is taken before partition refinement.
On each compact family of marked configurations of bounded
size, assume that $R_a$ is continuous outside finitely many equality
surfaces of Poisson measure zero and satisfies
$0\le R_a\le\sum_j u_j$.
Let $\Pi_n$ be a Poisson process on $(0,1)$ with intensity $n\,du/u$
and independent uniform marks in $\{1,\ldots,n\}$, and write
$\Pi_n^{(a)}$ for its restriction to $(a,1)$. Then
\[
 \underline d_1(Z)\ge\mathbb E R_a(\Pi_n^{(a)}).
\]
If these rewards are nondecreasing under extension to smaller
cutoffs, then $R(\Pi_n):=\lim_{a\downarrow0}R_a(\Pi_n^{(a)})$
exists almost surely and
$\underline d_1(Z)\ge\mathbb E R(\Pi_n)$.
\end{lemma}

\begin{proof}
First fix the cutoff, coarse partition, maximal configuration size,
and positive arithmetic margins. In every coarse interval $(u,v)$,
\eqref{eq:valuation-laws} and \Cref{lem:prime-log-stieltjes} give
\[
 \prod_{i,p}(\mathbb P(K_{i,p}=0))\longrightarrow(u/v)^n,
 \qquad \sum_pq_{i,p}\longrightarrow\log(v/u).
\]
More generally, the probability generating functions of the
independent rare coordinates converge to those of independent
Poisson counts, with equal intensity $du/u$ for each copy. Indeed,
the sum of the squared activation probabilities on $(a,1)$ tends
to zero, and their sums on each fixed subinterval converge as above.
Higher valuations in this range have total probability tending to
zero. The limiting process therefore has intensity $n\,du/u$ with
independent uniform copy marks. Dyadic exponents differ from the
actual exponents by at most $(\log2)/L$.

Divide \eqref{eq:block-reward} by $L$ and take
$\liminf_{L\to\infty}$ with all these choices fixed to obtain a lower
bound for $\underline d_1(Z)$. Before division by $L$, a subpower
arithmetic factor contributes $O_m(\varepsilon L)+O_{m,\varepsilon}(1)$; let
$\varepsilon\downarrow0$ only after this limit. Refine the coarse
partition along the fixed sequence at fixed $a$, using the assumed
lower estimates and then letting $\eta\downarrow0$. The limiting process has finitely many
distinct points almost surely, so every occupied bin is eventually
a singleton. The boundary loss just proved vanishes before this
refinement. The limiting rewards are bounded by the total exponent mass
and hence, at fixed $a$, by the point count. The latter has finite first
moment, which permits removal of the configuration-size truncation. Equality
surfaces have probability zero, since the points have a joint
density conditional on their number. Dominated convergence thus
removes the strict margins.

Under the cutoff-extension hypothesis,
the rewards increase as the cutoff decreases and remain bounded by the total
exponent mass. This mass has expectation
$\int_0^1u\,n\,du/u=n$, and the omitted mass below $a$ has mean
$na$. Monotone convergence proves the final assertion.
\end{proof}

\section{Pair energy estimates}
\label{sec:information-energy}
\index{Entropy bounds!pair energy}

A pair of hidden primes may be more informative jointly than two
separate recoveries. The obstacle is cancellation between their two
logarithmic differences. The estimates below count these cancellations;
the resulting pair rewards are compared with the singleton baseline.
The divisor argument uses equal absolute coefficient magnitudes,
whereas the later energy estimates allow arbitrary fixed nonzero
coefficients.

Put $f_s(x)=\log(1+s/x)$ for $s\in\{-1,1\}$ and
$x>\max(0,-s)$.

\begin{lemma}[Divisor bound for logarithmic pairs]
\label{lem:divisor-energy}
Let $3\le P\le Q$, let $s,t\in\{-1,1\}$, and let
$a,b\in\mathbb R$ satisfy $|a|=|b|=c>0$.
Let $R_\delta$ count ordered integer quadruples in fixed comparable
ranges $p,p'\asymp P$, $q,q'\asymp Q$, all exceeding $1$, for which
\[
 \bigl|a(f_s(p)-f_s(p'))+b(f_t(q)-f_t(q'))\bigr|\le\delta.
\]
For a sufficiently small $c_0>0$, depending only on $c$ and the range
constants, $\delta Q\le c_0$ implies
\begin{equation}\label{eq:divisor-energy}
 R_\delta\ll_{c,\varepsilon}(PQ)^\varepsilon
       \{PQ+\delta PQ^3+\delta P^5Q\}
 \qquad(\varepsilon>0).
\end{equation}
The implied constant may also depend on the fixed range constants.
The same estimate holds on any subsets of these integer ranges.
\end{lemma}

\begin{proof}
First take $a=b=1$. Since $|f_s'(x)|\asymp x^{-2}$,
near equality gives $|d|\ll P^2/Q$, where $d=p'-p$.
Put
\[
 A=p'(p+s),\qquad B=p(p'+s),\qquad
 K=dqq'+stAq'-stBq.
\]
The ratio of the two Euler products is
$A(q+t)q'/\{Bq(q'+t)\}$. Its logarithm has absolute value at most
$\delta$, and the difference of its numerator and denominator is $sK$.
Thus $K$ is an integer and $|K|\ll\delta P^2Q^2$. Exactly,
\begin{equation}\label{eq:factorization}
 (dq+stA)(dq'-stB)=dK-AB.
\end{equation}
Here $AB\asymp P^4$ and $|dK|\ll\delta P^4Q$. On decreasing $c_0$,
the right side is therefore a nonzero integer of absolute value
$O(P^4)$. For fixed $p,p',K$ with $d\ne0$, each signed factorization
determines $q,q'$ uniquely. The divisor bound
$\tau(m)\ll_\eta m^\eta$ gives $O_\varepsilon((PQ)^\varepsilon)$
possibilities. Indeed, $r+1\le\ell^{\eta r}$ for all sufficiently large
primes $\ell$, and the finitely many remaining primes contribute a
bounded factor to $\prod_{\ell^r\Vert m}(r+1)m^{-\eta}$.
There are $O(P^3/Q)$ pairs $p,p'$ with $d\ne0$: if the bound for
$|d|$ is below $1$ there are none, and otherwise the number of nonzero
integer offsets is $O(P^2/Q)$. There are
$O(1+\delta P^2Q^2)$ possible $K$. This gives
\[
 O_\varepsilon\bigl((PQ)^\varepsilon
          (P^3/Q+\delta P^5Q)\bigr).
\]
For $d=0$, derivative separation gives
$|q-q'|\ll\delta Q^2$, contributing $O(PQ(1+\delta Q^2))$.
Since $P\le Q$, these estimates prove
\eqref{eq:divisor-energy}.

For general $a,b$, divide the tolerance by $c$ and use the exact identity
\[
 -f_s(x)=f_{-s}(x+s).
\]
Each negative coefficient is thereby removed by a fixed integer shift
and a change of branch. The shifted coordinates remain in fixed
comparable ranges. The only new endpoint is $x=1$ in the plus branch,
where the same derivative and algebraic estimates hold. The preceding
argument applies with unchanged powers of $P,Q,\delta$.
\end{proof}

\begin{corollary}[Pair entropy reward]
\label{cor:divisor-reward}
Fix $a,b\in\mathbb R$ with $|a|=|b|>0$ and
$s,t\in\{-1,1\}$. Let $L\to\infty$, $0<u<v<w$, and suppose
$P=e^{uL+o(L)}$, $Q=e^{vL+o(L)}$, $\delta=e^{-wL+o(L)}$.
Let $X=(p,q)$ have a prior supported on the ranges of
\Cref{lem:divisor-energy}, with
\[
 \max_x\mathbb P(X=x)\le e^{-(u+v)L+o(L)}.
\]
If, for each outcome of an observation $Y$, the conditional support
of $af_s(p)+bf_t(q)$ has diameter at most $\delta$, then
\begin{equation}\label{eq:divisor-reward}
 \mathsf H(X)-\mathsf H(X\mid Y)\ge F_D(u,v;w)L-o(L),
 \quad F_D(u,v;w)=\min\{u+v,w+u-v,w-3u+v\}.
\end{equation}
Against $b(u,v;w)=\min(u,v-u)+\min(v,w-v)$, its positive gain is
\begin{equation}\label{eq:divisor-gain}
 [F_D-b]_+=[\min(2u,w-2u)-v]_+,
 \qquad x_+:=\max(x,0).
\end{equation}
\end{corollary}

\begin{proof}
The maximal-atom bound gives
\[
 \mathsf H(X)\ge-\log\max_x\mathbb P(X=x)
 \ge(u+v)L-o(L).
\]
Apply \Cref{lem:energy-entropy} with the stated prior and
observation. The entropy loss is at least
$2(u+v)L-\log R_\delta-o(L)$.
The three exponents on the right of
\eqref{eq:divisor-energy} are $u+v$, $u+3v-w$ and
$5u+v-w$. First take $L\to\infty$ and then $\varepsilon\downarrow0$.
Subtracting their maximum from $2(u+v)$ proves
\eqref{eq:divisor-reward}.
If $v\ge2u$ or $w\le2v$, both sides of
\eqref{eq:divisor-gain} vanish. Otherwise $b=2v-u$, and the
middle entry in $F_D-b$ is no smaller than its last entry, giving
the displayed identity. In particular, positive gain requires
$v<2u$ and $w>2u+v>2v$.
\end{proof}

\begin{lemma}[Logarithmic pair energies]\label{lem:rs-energy}
Fix
$a,b\in\mathbb R\setminus\{0\}$ and $s,t\in\{-1,1\}$.
For integers $P,Q\ge2$, let $R(P,Q;\delta)$ count the integer
quadruples $P<p,p'\le2P$, $Q<q,q'\le2Q$ satisfying
\[
 |a f_s(p)-a f_s(p')+b f_t(q)-b f_t(q')|\le\delta.
\]
For every $\epsilon>0$ and $\delta>0$,
\begin{equation}\label{eq:rs-energy}
 R(P,Q;\delta)\ll_{a,b,\epsilon}
 (PQ)^\epsilon PQ
 \bigl((1+\delta P^3)(1+\delta Q^3)\bigr)^{1/2}.
\end{equation}
The same estimate holds on arbitrary subsets, in particular on primes.
\end{lemma}

\begin{proof}
Robert--Sargos~\cite[Theorem~2]{RS06} give, for fixed
$\alpha\notin\{0,1\}$, integer $M\ge2$ and every $\eta>0$,
\[
 \#\{M<m_i\le2M:
 |m_1^\alpha+m_2^\alpha-m_3^\alpha-m_4^\alpha|
 \le\eta M^\alpha\}
 \ll_{\alpha,\epsilon}M^\epsilon(M^2+\eta M^4).
\]
The exact identity
\begin{equation}\label{eq:midpoint}
 f_s(n)=2s\operatorname{arctanh}\frac1{2n+s}
       =\frac{2s}{2n+s}+O(N^{-3}),\qquad N<n\le2N,
\end{equation}
puts the reciprocal variables in $(M,2M]$, with $M=2N+1$ for
$s=1$ and $M=2N$ for $s=-1$.
Applying the cited theorem with $\alpha=-1$ and
$\eta\ll N(\delta+N^{-3})$ proves
\[
 \begin{split}
 E_{a,s}(N,\delta)&:=
 \#\bigl\{(n_1,n_2,n_3,n_4)\in(N,2N]^4\cap\mathbb Z^4:\\
 &\hspace{12mm}|a f_s(n_1)+a f_s(n_2)-a f_s(n_3)-a f_s(n_4)|
       \le\delta\bigr\}\\
 &\ll_{a,\epsilon}N^\epsilon(N^2+\delta N^5).
 \end{split}
\]
Here division by $|a|$ only changes the constant, and the cubic error
contributes $O(N^2)$.
Bin the differences $a f_s(p)-a f_s(p')$ and
$b f_t(q)-b f_t(q')$ in intervals $[k\delta,(k+1)\delta)$,
with occupancies $A_k,B_k$. A collision joins bins with $k+l$ in
a fixed set of at most four integers. Cauchy--Schwarz gives
\[
 R(P,Q;\delta)\le4\Bigl(\sum_k A_k^2\Bigr)^{1/2}
                         \Bigl(\sum_k B_k^2\Bigr)^{1/2}.
\]
The two sums are at most $E_{a,s}(P,\delta)$ and
$E_{b,t}(Q,\delta)$ after relabelling the variables.
This proves \eqref{eq:rs-energy}.
\end{proof}

\begin{lemma}[Restricted differences]\label{lem:liu-energy}
Fix $a,b\in\mathbb R\setminus\{0\}$, $s,t\in\{-1,1\}$,
integers $P,Q\ge2$ and $\delta>0$, and use the count
$R(P,Q;\delta)$ from \Cref{lem:rs-energy}.
Let $E_{a,s}(P,H;\delta)$ count
\[
 \begin{gathered}
 P<p,p',r,r'\le2P,\qquad
 0<|p-p'|,|r-r'|\le H,\\
 |a f_s(p)-a f_s(p')-a f_s(r)+a f_s(r')|\le\delta.
 \end{gathered}
\]
There is an absolute $c>0$ such that, for $1\le H<cP^{3/4}$,
\begin{equation}\label{eq:liu-restricted}
 E_{a,s}(P,H;\delta)\ll_a
 (\log 2P)^5(PH+\delta P^4H+H^{8/3}).
\end{equation}
For some $C_{a,b}>0$ depending only on $a,b$, and every $\epsilon>0$,
if $P\le Q$, $0<\delta\le Q^{-1}$ and
$H=\max(1,C_{a,b}P^2/Q)<cP^{3/4}$, then
\begin{equation}\label{eq:liu-rectangular}
 \begin{split}
 R(P,Q;\delta)\ll_{a,b,\epsilon}(PQ)^\epsilon
 \bigl\{&PQ(1+\delta Q^2)\\
 &+[(PH+\delta P^4H+H^{8/3})(Q^2+\delta Q^5)]^{1/2}\bigr\}.
 \end{split}
\end{equation}
\end{lemma}

\begin{proof}
For fixed $\alpha\notin\{0,1\}$, put
$t(m,q)=(m+q)^\alpha-(m-q)^\alpha$ and $T=M^{\alpha-1}h$.
Liu~\cite[Theorem~2]{Liu93} proves that the number of
$M<m,m_1\le2M$, $h<q,q_1\le2h$ with
$|t(m,q)-t(m_1,q_1)|<\Delta T$ is
\[
 \ll(Mh+\Delta M^2h^2+h^{8/3})(\log 2M)^4,
 \qquad 1\le h<\eta M^{3/4},\quad\Delta>0,
\]
for sufficiently small fixed $\eta>0$; the constant depends on
$\alpha,\eta$.
Writing $\rho(z)=\operatorname{arctanh}(1/z)-1/z$, we have
\[
 \rho'(z)=-\frac1{z^2(z^2-1)}.
\]
Thus the remainder in \eqref{eq:midpoint}, taken across a
difference of length at most $2h$, is $O(hP^{-4})$.
For $p>p'$, set $m=p+p'+s$, $q=p-p'$; then
$m+q=2p+s$, $m-q=2p'+s$. These are integer variables, and
$M<m\le2M$ with $M=2P+1$ or $2P$ as above. Parity and endpoint
restrictions only decrease the count. On each shell $h<q\le2h$,
apply Liu with $\alpha=-1$ and
$\Delta\ll_a(\delta+hP^{-4})M^2/h$, enlarging its constant to
replace a weak inequality by a strict one. The resulting bound is
\[
 \ll_a(\log 2P)^4
       (Ph+\delta P^4h+h^2+h^{8/3}).
\]
Absorb $h^2$ into $Ph$. The shell $q=1$ has bound
$O_a(P+\delta P^4)$: the derivative of
$x\mapsto a(f_s(x+1)-f_s(x))$ has constant sign and absolute
value comparable to $P^{-3}$. Negative differences follow by reversal.
For the $J=O(\log 2P)$ signed shells, let $A_{j,k}$ be their
difference-bin occupancies. The inequality
\[
 \sum_k\Bigl(\sum_j A_{j,k}\Bigr)^2
 \le J\sum_j\sum_k A_{j,k}^2
\]
and the same inequality for adjacent bins include all comparisons
between different shells. Summing their geometric bounds proves
\eqref{eq:liu-restricted}.

Since $|a f_s'|\asymp_a P^{-2}$, the rectangular inequality forces
$|p-p'|\le C_{a,b}P^2/Q$. The terms $p=p'$ contribute at most
$O_{a,b}(PQ(1+\delta Q^2))$, since $|b f_t'|\asymp_b Q^{-2}$.
For $p\ne p'$, use the restricted estimate for the first
difference-bin square sum and the preceding unrestricted estimate
for the second. Cauchy--Schwarz proves
\eqref{eq:liu-rectangular}, absorbing logarithms into
$(PQ)^\epsilon$.
\end{proof}

\begin{corollary}[Universal pair rewards]\label{cor:pair-rewards}
Fix $a,b\in\mathbb R\setminus\{0\}$ and $s,t\in\{-1,1\}$.
Let $P=e^{uL+o(L)}$, $Q=e^{vL+o(L)}$ and
$\delta=e^{-wL+o(L)}$, where $0<u<v<w$.
Suppose independent prime choices $p,q$ in the preceding ranges have
point masses at most $e^{o(L)}/P$ and $e^{o(L)}/Q$.
Put $W=(p,q)$. If every conditional support of
$a f_s(p)+b f_t(q)$ given an observation $Y$ has diameter at most
$\delta$, then the lower limit of
$\{\mathsf H(W)-\mathsf H(W\mid Y)\}/L$ is at least
\begin{equation}\label{eq:rs-reward}
 F_R(u,v;w)=u+v-\tfrac12(3u-w)_+-\tfrac12(3v-w)_+.
\end{equation}
On the strict region $5u/4<v<2u$ it is also at least
\begin{equation}\label{eq:liu-reward}
 \begin{split}
 F_L(u,v;w)=\min\bigl\{&u+v,\ w+u-v,\\
 &\tfrac12u+\tfrac32v
 -\tfrac12\max(0,3u-w,(7u-5v)/3)
 -\tfrac12(3v-w)_+\bigr\}.
 \end{split}
\end{equation}
\end{corollary}

\begin{proof}
The prior hypotheses give
$\mathsf H(W)=(u+v)L+o(L)$ and
$p_*\le e^{-(u+v)L+o(L)}$. Thus
\Cref{lem:energy-entropy} bounds the entropy loss below by
$2(u+v)L-\log R(P,Q;\delta)-o(L)$.
Apply \eqref{eq:rs-energy}, let
$L\to\infty$, and then let $\epsilon\downarrow0$.
For \eqref{eq:liu-rectangular}, $H=e^{(2u-v)L+o(L)}$.
Its diagonal term has exponent
$u+v+(2v-w)_+$; its other term has exponent
\[
 \tfrac12\{3u+v+\max(0,3u-w,(7u-5v)/3)+(3v-w)_+\}.
\]
Subtracting these exponents from $2u+2v$ proves
\eqref{eq:liu-reward}. The condition $H<cP^{3/4}$ holds
uniformly on compact subsets of $5u/4<v<2u$.

\end{proof}

\section{Compatible pairs, clusters and triples}
\label{sec:information-compatible}
\index{Entropy bounds!compatible prime clusters}

A useful local gain must fit into the recovery of the whole hidden
vector. This section arranges the gains into consecutive blocks:
profitable pairs are disjoint, suitable common-magnitude clusters
can be recovered in full, and a further family of triples can replace
existing blocks without overlapping them. Prime-product separation
supplies the full-recovery step. The worked three-point example below
shows explicitly how the gains add within one block.

\begin{lemma}[Prime-product separation]\label{lem:prime-injectivity}
Fix pairwise disjoint ranges of odd primes $p_j\asymp P_j$, and
$(a_j,b_j)\in\{(1,0),(-1,0),(0,1),(0,-1)\}$, $1\le j\le k$.
Put $N=\prod_jP_j$ and
\[
 T(\boldsymbol p)=\sum_{j=1}^k\log\frac{p_j+a_j}{p_j+b_j}.
\]
The map $\boldsymbol p\mapsto T(\boldsymbol p)$ is injective, and
distinct values are separated by $\gg_kN^{-2}$. The constants may
also depend on the fixed range constants. In particular, a block
with exponents $u_1,\ldots,u_k$ has full leading entropy recovery
at resolution $e^{-wL+o(L)}$ whenever $w>2\sum_j u_j$.
This applies to arbitrary fixed signs and logarithmic branches
whose coefficient magnitudes have a common positive value.
\end{lemma}

\begin{proof}
Cancel every prime occurring in the same labeled coordinate of two
tuples, together with its entire rational factor. If any prime
remains, let $\ell$ be the largest. It occurs in exactly one tuple
and contributes valuation $1$ or $-1$ to the quotient of the
products. Every smaller remaining odd prime $p$ satisfies
$p+1<\ell$, and $\ell\pm1$ is coprime to $\ell$. Thus its valuation
cannot cancel, excluding equality. Disjoint ranges ensure that a
prime shared by the tuples has the same label and factor.

In the natural representation $e^T=A/B$, both positive integers
$A,B$ are comparable to $N$. Distinct rational values differ by at
least $1/(BB')\gg_kN^{-2}$. All such values lie in a fixed compact
subset of $(0,\infty)$, giving the same separation for their
logarithms. The energy at the asserted resolution is therefore
diagonal; \Cref{lem:hidden-vector} gives the full leading
entropy. Finally divide by the common coefficient magnitude.
Each signed branch is the logarithm of one of the four displayed
factors, so the argument includes negative coefficients.
\end{proof}

For the pair rewards above, put
\begin{equation}\label{eq:pair-gains}
 \begin{split}
 G_U(u,v;w)&=[\max\{F_R,F_L\}-b(u,v;w)]_+,\\
 G_C(u,v;w)&=[\max\{F_D,F_R,F_L\}-b(u,v;w)]_+.
 \end{split}
\end{equation}
\Needspace{3\baselineskip}
Here $F_L$ is omitted unless $5u/4<v<2u$. The subscript $C$
permits the divisor estimate for equal absolute coefficient magnitudes;
the subscript $U$ permits arbitrary fixed nonzero coefficients.

\begin{lemma}[Compatible pair and cluster rewards]
\label{lem:compatible-selection}
Let $0<u_1<\cdots<u_k<1$, put $u_{k+1}=1$, and attach a fixed
nonzero coefficient magnitude to each point. Write
$s(u,w)=\min(u,w-u)$. A pair of consecutive points receives
$G_C$ if its magnitudes agree and $G_U$ otherwise.
Call a consecutive block $(u_i,\ldots,u_j)$ eligible if its
length is at least three, its magnitudes agree, and
\begin{equation}\label{eq:cluster-condition}
 u_{r+1}<2u_r\quad(i\le r<j),\qquad
 2\sum_{r=i}^j u_r<u_{j+1}.
\end{equation}
The following is the reward of a single compatible partition into
singletons, pairs and fully recovered blocks:
\begin{equation}\label{eq:compatible-functional}
 \sum_{i=1}^k s(u_i,u_{i+1})
 +\sum_{i=1}^{k-1}G_i(u_i,u_{i+1};u_{i+2})
 +\sum_{\substack{(i,j)\text{ eligible}}}(2u_i-u_{i+1}),
\end{equation}
where $G_i$ is the applicable envelope. This nonnegative reward is
at most $\sum_i u_i$. With the prime-block construction above it
is an admissible lower information reward.
\end{lemma}

\begin{proof}
The singleton energy is $O(P(1+\delta P^2))$, by derivative
separation $|f_s'(x)|\asymp P^{-2}$, also after multiplication by
a fixed nonzero coefficient. Thus
\Cref{lem:hidden-vector} gives $s(u,w)$.
For every pair method a positive gain requires
\begin{equation}\label{eq:pair-separation}
 v<2u,\qquad w>2v.
\end{equation}
Indeed, if $v\ge2u$, then
$b=\min(u+v,w+u-v)$. The divisor and Liu rewards are bounded by
both entries. So is $F_R$: for $w<2v$, its penalty
$(3v-w)_+/2$ is at least $2v-w$, and for $w\ge2v$ the claim is
immediate. If $v<2u$ and $w\le2v$, then $b=w-u$.
The divisor entry $w-3u+v$ is smaller than $b$;
$F_R\le w-(u+v)/2<b$; and the third Liu entry is at most
$w-u$, using $\max(0,3u-w,(7u-5v)/3)\ge3u-w$ and
$(3v-w)_+=3v-w$. This proves
\eqref{eq:pair-separation}. A profitable predecessor
would require $v>2u$, and a profitable successor would require
$w<2v$. All profitable pairs are consequently disjoint, even
when different pairs use different envelopes.

Every suffix of length at least three of an eligible block is
eligible. Two overlapping eligible blocks have the same last
point: otherwise the following gap of the earlier-ending block,
which is larger than twice its last point, would be an internal
gap of the other block. Hence maximal eligible blocks are
disjoint, and their eligible subblocks are precisely their
suffixes of length at least three. Within a maximal block all
internal pairs except its last pair have zero gain, by
\eqref{eq:pair-separation}. Its last pair is fully
recovered: $w>2\sum u_r>3u_j$, so $F_R=u_{j-1}+u_j$.
No profitable pair enters the block on the left, since its first
internal gap is below twice the first point; none leaves on the
right, since the following gap exceeds twice the last point.

For a common-magnitude block, divide by that magnitude and use
\Cref{lem:prime-injectivity}. The rational products have
numerator and denominator $e^{(S+o(1))L}$, where $S$ is the sum
of its exponents. Distinct products, and hence their logarithms
in a fixed compact interval, are separated by
$e^{-(2S+o(1))L}$. The strict condition $2S<w$ therefore gives
full reward $S$ by \Cref{lem:hidden-vector}.
Starting with the recovered last pair, extend its recovered
suffix one point to the left. The previous singleton reward was
$u_{i+1}-u_i$, and full recovery of the extension increases the
reward by exactly $2u_i-u_{i+1}$. Summing these increments over
all suffixes telescopes to the reward of the maximal block.
Retain every exterior profitable pair and all remaining
singletons. This proves \eqref{eq:compatible-functional}
and its asserted bound. Apply
\Cref{lem:ordered-limits}, keeping strict compact margins
for all recovery conditions and for Liu's domain before the
quantization limit, then removing them by dominated convergence.
\end{proof}

\phantomsection
\paragraph*{A compatible three-point reward}\label{par:compatible-reward}
Take three retained exponents $u_1=2/25$, $u_2=3/25$, $u_3=1/5$
with equal nonzero coefficient magnitudes, and set the terminal
resolution boundary $u_4=1$. The cluster
conditions in \eqref{eq:cluster-condition} hold strictly:
\[
 \frac3{25}<\frac4{25},\qquad \frac5{25}<\frac6{25},\qquad
 2\left(\frac2{25}+\frac3{25}+\frac5{25}\right)=\frac45<1.
\]
The singleton rewards are $1/25$, $2/25$ and $5/25$.
The last pair has $F_R(3/25,1/5;1)=8/25$, hence gains $1/25$
over its singleton reward $7/25$. The first pair has no gain by
\eqref{eq:pair-separation}. The cluster extension adds
$2u_1-u_2=1/25$, giving
\[
 \frac8{25}+\frac1{25}+\frac1{25}=\frac25=u_1+u_2+u_3.
\]
This is the full reward of one compatible block in
\Cref{lem:compatible-selection}, counting its last pair once.
In the independent-sum application, the marked expectations in
\Cref{prop:campbell} supply the factors $\rho_m$ in
\Cref{thm:independent-sums}.

\begin{proposition}[Expected compatible reward]
\label{prop:campbell}
Let the points in $(0,1)$ form a Poisson process of fixed intensity
$\kappa\,du/u$, where $\kappa>0$. Give them independent magnitude
classes with probabilities $\theta_1,\ldots,\theta_\ell$, and put
$\rho_m=\sum_g\theta_g^m$. Set
\begin{equation}\label{eq:pair-integral}
 J_A(\kappa)=\kappa^2\int_{0<u<v<1}
 u^{\kappa-1}G_A(u,v;1)\,\frac{du\,dv}{v},\qquad A\in\{U,C\}.
\end{equation}
For $m\ge3$, put $r_1=1$,
$D_m=\{r_{j-1}<r_j<2r_{j-1}:2\le j\le m\}$ and
\begin{equation}\label{eq:cluster-integral}
 I_m(\kappa)=\kappa^m\int_{D_m}
 \frac{2-r_2}{(2\sum_{j=1}^m r_j)^{\kappa+1}}
 \prod_{j=2}^m\frac{dr_j}{r_j}.
\end{equation}
The expectation of \eqref{eq:compatible-functional}, with
the cutoff decreased to zero, is
\begin{equation}\label{eq:pair-cluster-mean}
 1-2^{-\kappa}
 +(\kappa+1)\{J_U(\kappa)+\rho_2[J_C(\kappa)-J_U(\kappa)]\}
 +\sum_{m\ge3}\rho_m I_m(\kappa).
\end{equation}
The series converges. For $n$ independent basic laws with nonzero
coefficients, the arithmetic application has $\kappa=n$ and
$\theta_g=n_g/n$, where $n_g$ is the size of a magnitude class.
\end{proposition}

\begin{proof}
At a point $u$, the next point $w$, capped at $1$, satisfies
$\mathbb P(w>u+t)=(u/(u+t))^\kappa$ for $0<t<1-u$.
Integration of the survival function and then $t=ux$ gives
\[
 \mathbb E\sum_u s(u,w)
 =\kappa\int_0^1\frac{du}{u}
       \int_0^{\min(u,1-u)}\left(\frac{u}{u+t}\right)^\kappa dt
 =\kappa\int_0^1(1+x)^{-\kappa-1}\,dx
 =1-2^{-\kappa}.
\]
The density of a terminal consecutive $m$-tuple is
\[
 \kappa^m u_1^{\kappa-1}\,du_1
       \prod_{j=2}^m\frac{du_j}{u_j},
\]
since the product intensity is multiplied by the void probability
$u_1^\kappa$. For a tuple followed by a point $w<1$, the density is
\[
 \kappa^{m+1}(u_1/w)^\kappa
       \prod_{j=1}^m\frac{du_j}{u_j}\,\frac{dw}{w}.
\]
For any nonnegative degree-one homogeneous local reward, the
substitution $u_j=wv_j$ cancels $dw/w$ against its factor $w$.
Its total interior expectation is therefore $\kappa$ times its
terminal expectation. These density formulas follow directly
by requiring one point in each of disjoint small intervals and
no other point between the first and last specified endpoints;
nonnegative approximation extends them to the displayed rewards.

The pair terminal expectation is exactly $J_A(\kappa)$.
Two marks agree with probability $\rho_2$, giving the pair term
in \eqref{eq:pair-cluster-mean}. For a cluster put
$u_j=ur_j$. Its gain is $u(2-r_2)$ and its terminal recovery
condition is $0<u<(2\sum r_j)^{-1}$. Integrating $u^\kappa du$
gives a factor $1/(\kappa+1)$, which cancels the multiplier
$\kappa+1$ from terminal and interior contributions. This proves
\eqref{eq:cluster-integral}. Its marks agree with
probability $\rho_m$. Linearity of expectation gives the cluster term.
At a positive cutoff the construction is finite. Decreasing it
retains all existing terms and prepends nonnegative terms;
\Cref{lem:compatible-selection} bounds the result by the
total exponent mass, whose mean is $\kappa$. Monotone convergence
proves the assertion. In the arithmetic application, the
independent uniform copy marks of
\Cref{lem:ordered-limits} give the stated $\theta_g$.
\end{proof}

\begin{lemma}[Cluster series and its tail]\label{lem:resolvent}
Put $p=\kappa+1$ and, for $0\le z\le1$, define
\[
 A_\kappa(z)=\frac{\kappa^2}{2^p}
       \int_1^2\frac{2-y}{y(z+y)^p}\,dy,\qquad
 (K_\kappa f)(z)=\kappa\int_1^2
       \frac{f(z/(z+y))}{y(z+y)^p}\,dy.
\]
Write
\[
 q_\kappa=\frac{\kappa(1-2^{-p})}{p}<1,\qquad
 a_\kappa=A_\kappa(0)
       =\frac{\kappa^2}{2^p}\int_1^2(2-y)y^{-p-1}\,dy.
\]
The unique bounded solution of $V_\kappa=A_\kappa+K_\kappa V_\kappa$
satisfies
\begin{equation}\label{eq:resolvent}
 \sum_{m\ge3}I_m(\kappa)=V_\kappa(1)-A_\kappa(1),\qquad
 V_\kappa(0)=\frac{a_\kappa}{1-q_\kappa}.
\end{equation}
For every integer $M\ge2$,
\begin{equation}\label{eq:cluster-tail}
 \sum_{m>M}I_m(\kappa)
 \le\frac{a_\kappa q_\kappa^{M-1}}{1-q_\kappa}.
\end{equation}
For a mark of probability $\theta$, let $V_{\kappa,\theta}$ be the unique
bounded solution of
$V_{\kappa,\theta}=\theta^2A_\kappa+\theta K_\kappa V_{\kappa,\theta}$.
Then
\[
 \sum_{m\ge3}\theta^m I_m(\kappa)
 =V_{\kappa,\theta}(1)-\theta^2A_\kappa(1),\qquad
 \sum_{m>M}\theta^m I_m(\kappa)
 \le\frac{\theta^2a_\kappa(\theta q_\kappa)^{M-1}}
              {1-\theta q_\kappa}.
\]
\end{lemma}

\begin{proof}
The positive operator has sup norm exactly $q_\kappa$, since
its row mass is largest at $z=0$. Also
$0\le A_\kappa(z)\le a_\kappa$. The Neumann series therefore
converges uniformly to the unique bounded solution and gives
the displayed endpoint by evaluating the equation at zero.
In \eqref{eq:cluster-integral}, put
$y_j=r_j/r_{j-1}\in(1,2)$. The Jacobian in logarithmic
coordinates is one and
\[
 \sum_{j=1}^m r_j=1+y_2(1+y_3(1+\cdots+y_m)).
\]
Repeated substitution in $K_\kappa$, with the independent ratio
variables named in reverse order, consequently gives
$I_m(\kappa)=(K_\kappa^{m-2}A_\kappa)(1)$.
The geometric remainder of the Neumann series proves
\eqref{eq:cluster-tail}. Replacing $A_\kappa,K_\kappa$
by $\theta^2A_\kappa,\theta K_\kappa$ proves the marked claims;
summing them over classes gives the marked cluster series.

At $\kappa=1$, $a_1=1/16$, $q_1=3/8$, and $V_1(0)=1/10$.
Equivalently, for $t\ge1$, put $U(t)=t^{-2}V_1(1/t)$. Then
\begin{equation}\label{eq:scalar-resolvent}
 \begin{split}
 H(t)&=\tfrac12\log\frac{2(t+1)}{2t+1}-\frac1{4(t+1)},\\
 U(t)&=H(t)+\int_1^2U(1+yt)\,\frac{dy}{y},\qquad
 \sum_{m\ge3}I_m(1)=U(1)-H(1).
 \end{split}
\end{equation}
This follows by substitution and elementary integration of
$t^{-2}A_1(1/t)$. Thus the scalar tail in
\eqref{eq:cluster-tail} is at most
$\frac1{10}(3/8)^{M-1}$.
\end{proof}

\begin{lemma}[Prime-specific recursive energy]
\label{lem:recursive-energy}
Use the fixed factors and disjoint prime ranges of
\Cref{lem:prime-injectivity}. Let $k\ge2$,
$M=\prod_{j<k}P_j$, $Q=P_k$ and $N=MQ$.
Write $R_k(\delta)$ for the number of ordered tuple pairs with
$|T(\boldsymbol p)-T(\boldsymbol p')|\le\delta$, and
$R_{k-1}(\eta)$ for the corresponding head energy.
There are constants $c,C>0$ such that $\delta Q\le c$ implies
\begin{equation}\label{eq:recursive-energy}
 R_k(\delta)\ll_{k,\epsilon}N^\epsilon
 \{N(1+\delta Q^2)+\delta N^2R_{k-1}(C/Q)\}
 \qquad(\epsilon>0).
\end{equation}
The constants may depend on the fixed factors and range constants.
\end{lemma}

\begin{proof}
Write the natural head products as $A/B,A'/B'$, with all four
integers comparable to $M$, and put
$\alpha=AB'$, $\beta=A'B$, $d=\alpha-\beta$.
For the last coordinates $q=p_k$ and $q'=p'_k$, put
$x=q+b_k$, $x'=q'+b_k$, and $c_k=a_k-b_k\in\{-1,1\}$.
The last factor is $(x+c_k)/x$. Clearing the full difference gives the integer
\[
 K=dxx'+c_k\alpha x'-c_k\beta x,
 \qquad |K|\ll\delta M^2Q^2.
\]
The two head logarithms differ by $O(Q^{-1})$, since the last
logarithms have that size and $\delta\ll Q^{-1}$. Hence
$|d|\ll M^2/Q$. Exact multiplication gives
\begin{equation}\label{eq:factor-completion}
 (dx+c_k\alpha)(dx'-c_k\beta)=dK-\alpha\beta.
\end{equation}
Here $\alpha\beta\asymp M^4$ and
$|dK|\ll\delta QM^4$. Decreasing $c$ makes the right side a
nonzero integer of size $O(M^4)$.

If $d\ne0$, then $K\ne0$: otherwise the full products are equal,
and \Cref{lem:prime-injectivity} forces identical tuples,
contradicting $d\ne0$. There are consequently at most
$2\lfloor C\delta N^2\rfloor\le2C\delta N^2$ possible residuals. For each head pair and residual,
the divisor bound used in \Cref{lem:divisor-energy} gives
$O_{k,\epsilon}(N^\epsilon)$ signed factorizations in
\eqref{eq:factor-completion}; each determines $x,x'$ uniquely.
Only the $R_{k-1}(C/Q)$ compatible head pairs occur. This proves
the second term of \eqref{eq:recursive-energy}.

If $d=0$, prime-product injectivity makes the heads identical.
There are $O_k(M)$ possible heads. The last logarithm has derivative
of constant sign and absolute value comparable to $Q^{-2}$, so
each of the $O(Q)$ last primes has $O(1+\delta Q^2)$ partners.
This proves the first term.
\end{proof}

\begin{proposition}[Additional recoverable triples]
\label{prop:triples}
Consider consecutive retained exponents $u<v<r$, with next exponent
$w$ (or $w=1$ for a final triple). Suppose their coefficient
magnitudes agree. Their available entropy reward includes
\begin{equation}\label{eq:triple-reward}
 F_3(u,v,r;w)=\min\left\{
 u+v+r,\ w+u+v-r,\
 w-u-v-\frac{(3u-r)_++(3v-r)_+}{2}\right\}.
\end{equation}
Every triple satisfying
\begin{equation}\label{eq:triple-region}
 1<\frac vu<\frac54,\qquad
 \frac{5v-u}{2}<r<3u,\qquad
 \frac72(u+v)<w
\end{equation}
has full leading entropy reward $u+v+r$. Such triples are
mutually disjoint, no profitable pair crosses their boundaries,
and no eligible cluster from \Cref{lem:compatible-selection}
intersects them. Replacing their head pair and final singleton
therefore adds exactly
\begin{equation}\label{eq:triple-gain}
 h_3(u,v,r)=\frac32(u+v)-r>0
\end{equation}
to the compatible pair and cluster reward.
\end{proposition}

\begin{proof}
Put $S=u+v$. At head tolerance $e^{-rL+o(L)}$,
\eqref{eq:rs-energy} gives energy exponent
$S+\{(3u-r)_++(3v-r)_+\}/2$.
The three terms in \eqref{eq:recursive-energy} have exponents
\[
 S+r,\qquad S+3r-w,\qquad
 3S+2r-w+\frac{(3u-r)_++(3v-r)_+}{2}.
\]
Subtract their maximum from $2(S+r)$, using
\Cref{lem:hidden-vector}, and then let the subpower
parameter decrease to zero. This proves
\eqref{eq:triple-reward}.

On \eqref{eq:triple-region} one has
$v<2u$, $r>2v$, $w>2r$ and $r<2S$.
Both positive parts in the preceding display are active, so the
energy exponents reduce to $S+r$, $S+3r-w$ and $9S/2+r-w$.
The last two are strictly below $S+r$, proving full recovery.
The head pair has singleton reward $2v-u$ and Robert--Sargos reward
$F_R(u,v;r)=r-S/2$. This reward exceeds the singleton reward by
$r-(5v-u)/2>0$. Also
\[
 F_D(u,v;r)\le r-3u+v<r-S/2,
\]
since $v<5u/4<5u/3$. Liu's estimate does not apply because $v<5u/4$.
Thus the head pair earns exactly $r-S/2$ in the chosen pair
envelope. The final singleton earns $r$, and the other internal
pair has zero gain because $r>2v$. Replacing these rewards by
$S+r$ adds $3S/2-r>0$.

Every profitable pair requires its second exponent to be less
than twice its first, and its next exponent to exceed twice its
second, by \Cref{lem:compatible-selection}.
A pair crossing the left boundary would require $v>2u$;
one crossing the right boundary would require $w<2r$.
A second selected triple starting at $v$ would require $r<2v$;
one starting at $r$ would require $w<2r$. Applying the same
argument to the earlier of two triples excludes all overlaps.

An eligible cluster cannot cross either gap $(v,r)$ or $(r,w)$,
since all its consecutive ratios are below two. If it intersects
the triple before $r$, its endpoint is consequently $u$ or $v$.
An endpoint $u$ would require $v>2u$ for full recovery.
An endpoint $v$ forces the cluster to contain $u$ and would require
$r>2(u+v)$, contrary to $r<3u<2(u+v)$.
A cluster of length at least three containing $r$ must cross one
of the two forbidden gaps. Thus no eligible cluster
intersects the triple. The replacements preserve the first subsequent
exponent for every exterior block, so all other rewards remain
valid. Compact strict margins and the ordered limits in
\Cref{lem:ordered-limits} justify the stated asymptotic rewards.
\end{proof}

\begin{proposition}[Mean triple increment]
\label{prop:triple-mean}
For Poisson points of intensity $\kappa\,du/u$ on $(0,1)$,
$\kappa>0$, the expected total increment
\eqref{eq:triple-gain} over
\eqref{eq:triple-region} equals
\begin{equation}\label{eq:delta-three}
 \Delta_3(\kappa)=\kappa^3\left(\frac27\right)^{\kappa+1}
 \int_1^{5/4}\int_{(5t-1)/2}^{3}
 \frac{3(1+t)/2-z}{tz(1+t)^{\kappa+1}}\,dz\,dt.
\end{equation}
With independent magnitude-class marks of probabilities $\theta_g$,
retaining only triples of one class multiplies this mean by
$\rho_3=\sum_g\theta_g^3$.
\end{proposition}

\begin{proof}
The density of a final consecutive triple $u<v<r$ is
$\kappa^3u^{\kappa-1}\,du\,dv\,dr/(vr)$: the three intensity
factors are multiplied by the probability $u^\kappa$ of no
other points in $(u,1)$. Substituting $v=ut$, $r=uz$, its expected
increment is
\[
 \kappa^3\int_1^{5/4}\int_{(5t-1)/2}^{3}
 \int_0^{2/[7(1+t)]}
 u^\kappa\frac{3(1+t)/2-z}{tz}\,du\,dz\,dt.
\]
The bound on $u$ implies $r<1$ throughout this domain.
For an interior triple followed by $w$, the density is
$\kappa^4u^{\kappa-1}\,du\,dv\,dr\,dw/(vrw^{\kappa+1})$.
Scaling its first three points by $w$ and using degree-one
homogeneity of the increment shows that the interior expectation
is $\kappa$ times the final expectation. The $u$ integral contributes
$1/(\kappa+1)$, canceled by the factor $\kappa+1$ from the two
types of triple. This gives \eqref{eq:delta-three}.
Decreasing the cutoff retains every existing triple and adds
nonnegative increments. The total reward is bounded by the total
exponent mass, whose mean is $\kappa$, so monotone convergence and
domination justify the cutoff and margin limits. Independent marks
make any specified triple
monochromatic with probability $\sum_g\theta_g^3$. Linearity of
expectation supplies this factor in the mean reward.
After removing zero coefficients, let $n\ge1$ be the number of
remaining independent logarithmic copies. Then $\kappa=n$ and
$\theta_g=n_g/n$, where $n_g$ counts equal absolute coefficient
magnitudes. Thus the addition is $\rho_3\Delta_3(n)$, also when
the signs or branches within a class differ.
\end{proof}

\section{Adaptive coding}
\label{sec:information-coding}
\index{Adaptive coding}
\index{Entropy bounds!upper bound}

For an upper entropy bound, it suffices to describe a record that
determines the observed mesh cell. Recording more of the prime head
costs information but leaves a smaller tail to specify at resolution
$h$. The code below chooses its cutoff from a fixed finite grid after
seeing the prime configuration. Its leading cost is the information
in the selected head plus the remaining tail resolution. Passing to
the limiting exponent process and then refining the grid yields the
coding functional below. 

For $\kappa>0$, let $\Pi_\kappa$ be a Poisson point process on $(0,1)$
with intensity $\kappa\,du/u$, and put
\begin{equation}\label{eq:coding-functional}
 S_\kappa(t)=\sum_{u\in\Pi_\kappa,\ u\le t}u,\qquad
 M_\kappa=\sup_{0\le t\le1}\{t-S_\kappa(t)\},\qquad
 d_\kappa=1-\mathbb E M_\kappa.
\end{equation}
Here $S_\kappa(0)=0$ and $\mathbb ES_\kappa(t)=\kappa t$;
in particular the sum is finite almost surely. For each jump time $t$,
taking $s\uparrow t$ in the defining supremum gives
$M_\kappa\ge t-S_\kappa(t-)$.

\begin{lemma}[Adaptive entropy bound]\label{lem:adaptive-code}
Let $Z=b+\sum_{i=1}^n a_iZ_i$, where $n\ge1$, the $Z_i$ are
independent choices among the four basic logarithmic laws, and
$b,a_1,\ldots,a_n$ are fixed real numbers with $a_i\ne0$. Then
\[
 \overline d_1(Z)=\overline d_1(e^Z)\le d_n.
\]
The bound and the equality remain valid after conditioning on any
positive-probability event in the joint valuation sigma-field.
\end{lemma}

\begin{proof}
By \Cref{cor:arithmetic-entropy-invariance,lem:transfers}, it suffices
to use the first-jump variables and to omit prime $2$. Thus write
\[
 Z=b+\sum_{i=1}^n\sum_{p\ge3}a_iB_{i,p}v_i(p),
 \qquad
 v_i(p)\in\{\log(p/(p-1)),\log(1+1/p)\},
\]
where the independent Bernoulli probabilities $q_{i,p}$ are $1/p$ or
$1/(p-1)$ according to the valuation model. Let $h=e^{-L}$ and fix a
finite grid $T\subset[0,1]$ containing $0,1$. Define
\[
 S_L(t)=\frac1L\sum_{i=1}^n\sum_{3\le p\le e^{tL}}B_{i,p}\log p.
\]
Choose $J\in T$ minimizing $1-t+S_L(t)$, with a fixed tie rule.
For each fixed $t$, let $V_t$ be the full labelled Bernoulli vector
through $e^{tL}$, let $A_t$ be its contribution together with $b$,
and let $R_t=Z-A_t$. The head law in the reference distribution $Q$
below is the unconditional law of $V_t$.

Set $s_t=e^{(1-t)L}$, $r_t=e^{-1/s_t}$, and
\[
 G_t(k)=\frac{1-r_t}{1+r_t}r_t^{|k|}\quad(k\in\mathbb Z),\qquad
 Q(t,v,k,c)=\frac{\mathbb P(V_t=v)G_t(k)}{2|T|}
 \quad(c\in\{0,1\}).
\]
Since $\sum_{k\in\mathbb Z}r_t^{|k|}=(1+r_t)/(1-r_t)$,
successive summation over $c,k,v,t$ gives $\sum Q=1$ on the
disjoint record space. The selected record
$(J,V_J,\lfloor R_J/h\rfloor,C_J)$ determines $\lfloor Z/h\rfloor$,
where
\[
 C_t=\lfloor Z/h\rfloor-\lfloor A_t/h\rfloor-\lfloor R_t/h\rfloor
 \in\{0,1\}.
\]
Cross entropy against $Q$ therefore bounds the output entropy.

Uniformly over all head configurations and $t\in[0,1]$,
\begin{align*}
 -\log\mathbb P(V_t)
 &=\sum_{i,\,3\le p\le e^{tL}}-\log(1-q_{i,p})
   +\sum_{i,\,3\le p\le e^{tL}}B_{i,p}\log\frac{1-q_{i,p}}{q_{i,p}}\\
 &=LS_L(t)+O_n(\log(L+2)).
\end{align*}
Indeed, the two log odds are $\log(p-1)$ and $\log(p-2)$;
their errors from $\log p$, as well as the absent-coordinate costs,
are $O(1/p)$, and \Cref{lem:prime-log-stieltjes} bounds their sum.
Since $s_t\ge1$, the two-sided geometric law satisfies
\[
 -\log G_t(\lfloor R_t/h\rfloor)
 \le(1-t)L+O(1)+e^{tL}|R_t|.
\]
The jump bound $|a_iv_i(p)|\ll_{a_i}p^{-1}$ gives, even by summing
$p^{-2}$ over all integers,
\[
 \mathbb E|R_t|\le C_{a,n}e^{-tL},\qquad
 \mathbb E[e^{JL}|R_J|]
 \le\sum_{t\in T}e^{tL}\mathbb E|R_t|=O_{a,n,T}(1).
\]
Thus the cutoff, head, signed tail and carry costs give
\begin{equation}\label{eq:grid-code}
 H_{e^{-L}}(Z)
 \le L\,\mathbb E\min_{t\in T}\{1-t+S_L(t)\}
      +O_{a,n,T}(\log(L+2)).
\end{equation}

Fix $\varepsilon>0$. For every nonnegative step
function $f$ supported on $(\varepsilon,1)$, independence gives
\begin{align*}
 \log\mathbb E\exp\!\left(-\sum_{i,p}
 B_{i,p}f(\log p/L)\right)
 &=\sum_{i,p}q_{i,p}(e^{-f(\log p/L)}-1)+o(1)\\
 &\longrightarrow n\int_\varepsilon^1(e^{-f(u)}-1)\,\frac{du}{u}.
\end{align*}
The error is bounded by a constant times
$\sum_{i,p>e^{\varepsilon L}}q_{i,p}^2=o(1)$; the limit follows
from \Cref{lem:prime-log-stieltjes}. Approximation on this compact
interval gives convergence of its weighted point sums. The discarded
mass satisfies
\[
 \mathbb ES_L(\varepsilon)\le C_n\varepsilon+O_n(L^{-1}),
 \qquad \mathbb ES_n(\varepsilon)=n\varepsilon.
\]
It changes any fixed-grid minimum by at most that mass. First passing
$L\to\infty$ at fixed $\varepsilon$, then letting
$\varepsilon\downarrow0$, proves convergence of the grid expectation
in \eqref{eq:grid-code}: its integrand is in $[0,1]$ because
$0\in T$. Finally take nested dyadic grids. Right continuity and
approximation from either side of each jump give
\[
 \min_{t\in T}\{1-t+S_n(t)\}\downarrow
 \inf_{0\le t\le1}\{1-t+S_n(t)\}=1-M_n.
\]
Bounded convergence proves $\overline d_1(Z)\le d_n$.
All finite-entropy corrections, positive-probability conditioning in
the joint valuation sigma-field and exponential images follow from
\Cref{lem:transfers,cor:arithmetic-entropy-invariance}.
\end{proof}

\begin{lemma}[Evaluation of the coding constant]
\label{lem:reflected-integral}
For every $\kappa>0$,
\begin{equation}\label{eq:coding-constant-integral}
 d_\kappa=1-\left[\int_0^\infty
       \exp\{\kappa\operatorname{Ein}(t)-t\}\,dt\right]^{-1},
 \qquad
 \operatorname{Ein}(t)=\int_0^t\frac{1-e^{-s}}s\,ds.
\end{equation}
Moreover,
\begin{equation}\label{eq:coding-deficit}
 \begin{gathered}
 \frac{e^{-\gamma\kappa-1}}{\Gamma(\kappa+1)}
 <1-d_\kappa<\frac{e^{-\gamma\kappa}}{\Gamma(\kappa+1)},
 \\
 1-d_\kappa\sim\frac{e^{-\gamma\kappa}}{\Gamma(\kappa+1)}
 \quad(\kappa\to\infty).
 \end{gathered}
\end{equation}
\end{lemma}

\begin{proof}
Put $D=S_\kappa(1)$, $M=M_\kappa$ and $W=D+M$. The largest point
$U$ has distribution $\mathbb P(U\le u)=u^\kappa$. Under
$u=e^{-x}$ the process is a homogeneous Poisson process of rate
$\kappa$; independence after its first arrival shows that the points
below $U$, rescaled by $U$, form an independent copy of $\Pi_\kappa$.
Write $(D',M')$ for the corresponding copy of the pair $(D,M)$.
Separating the intervals before and after the largest jump yields
\begin{equation}\label{eq:reflection}
 D=U(1+D'),\qquad M=\max\{1-D,UM'\},\qquad
 W=\max\{1,U(1+W')\},
\end{equation}
where $U$ and $W'=D'+M'$ are independent. The supremum over
$0\le t<U$ is $UM'$, with the left limit at $U$ obtained by
$t\uparrow U$; on $[U,1]$ the maximum is at $1$.
Since $0\le M\le1$ and $\mathbb ED=\kappa$, one has
$1\le W\le1+D$ and $\mathbb EW<\infty$.

Let $\alpha=\mathbb P(W=1)$. Conditional on $W'=w$, the reflected
law has an atom $(1+w)^{-\kappa}$ at $1$ and density
$\kappa x^{\kappa-1}/(1+w)^\kappa$ on $1<x<1+w$. Consequently
\[
 \alpha=\mathbb E(1+W)^{-\kappa},\qquad
 \mathbb E(W\mid W'=w)
 =\frac{\kappa(1+w)+(1+w)^{-\kappa}}{\kappa+1}.
\]
Taking expectations gives $\mathbb EW=\kappa+\alpha$, whence
$\mathbb EM=\alpha$. Conditional integration by parts also gives,
for bounded differentiable $\phi$ with bounded $x\phi'(x)$,
\[
 \mathbb E\{W\phi'(W)\mathbf1_{\{W>1\}}\}
 =\kappa\mathbb E\{\phi(W+1)-\phi(W)\}.
\]
Apply this to $\phi(x)=e^{-tx}$ and set $F(t)=\mathbb Ee^{-tW}$.
Retaining the atom at $1$ gives
\[
 tF'(t)+\kappa(1-e^{-t})F(t)=-\alpha t e^{-t}\qquad(t>0).
\]
The integrating factor is $e^{\kappa\operatorname{Ein}(t)}$.
Its product with $F$ tends to zero at infinity, since $F(t)\le e^{-t}$
and $\operatorname{Ein}(t)\le1+\log t$ for $t\ge1$.
Integration from $0$ to infinity, using $F(0)=1$, therefore gives
\[
 1=\alpha\int_0^\infty e^{\kappa\operatorname{Ein}(t)-t}\,dt.
\]
Together with $d_\kappa=1-\mathbb EM$ this proves
\eqref{eq:coding-constant-integral}.

Write its integral as $I_\kappa$. The identity
\[
 \operatorname{Ein}(t)=\gamma+\log t+E_1(t),\qquad
 E_1(t):=\int_t^\infty\frac{e^{-u}}u\,du>0
\]
gives $I_\kappa>e^{\gamma\kappa}\Gamma(\kappa+1)$.
The function $\Delta(t)=\gamma+\log(1+t)-\operatorname{Ein}(t)$ tends
to zero at infinity and has derivative
$((1+t)e^{-t}-1)/(t(1+t))<0$. Hence
\[
 I_\kappa<e^{\gamma\kappa}\int_0^\infty(1+t)^\kappa e^{-t}\,dt
 <e^{\gamma\kappa+1}\Gamma(\kappa+1),
\]
proving the two strict bounds. For fixed $\varepsilon>0$, the estimate
$E_1(t)\le e^{-t}/t$ gives
$\operatorname{Ein}(t)\le\gamma+\log(t+\varepsilon)$ for $t\ge T_\varepsilon$.
Thus the tail of $I_\kappa$ is at most
$e^{\gamma\kappa+\varepsilon}\Gamma(\kappa+1)$; its initial segment is
$O_\varepsilon(e^{\kappa\operatorname{Ein}(T_\varepsilon)})
=o(e^{\gamma\kappa}\Gamma(\kappa+1))$.
Letting $\varepsilon\downarrow0$ proves the asymptotic in
\eqref{eq:coding-deficit}.
\end{proof}

\subsection{Real activation intensities}\label{sec:real-intensity-mas13}

The Poisson functionals above are defined for every positive real
intensity. The following prime model supplies the arithmetic bridge to
those functionals.

\begin{theorem}[Information bounds at real intensity]
\label{thm:real-intensity-mas13}
Fix $m\ge1$, $K_1,\ldots,K_m>0$, and nonzero real coefficients
$a_1,\ldots,a_m$. For all primes $p>p_0$ and $1\le i\le m$ let $B_{i,p}$
be independent Bernoulli variables, and write
\[
 q_{i,p}:=\mathbb P(B_{i,p}=1),\qquad
 v_i(p)\in\{\log(p/(p-1)),\log(1+1/p)\},
\]
with one fixed choice of branch for each $i$. Suppose that there are
constants $C_i>0$ and $\mathfrak b_i\in\mathbb R$ such that
\begin{align}
 0\le q_{i,p}&\le\frac{C_i}{p}\qquad(p>p_0),
 \label{eq:real-intensity-envelope-mas13}\\
 \sum_{p_0<p\le x}q_{i,p}&=K_i\log\log x+\mathfrak b_i
   +o\!\left(\frac1{\log x}\right)\qquad(x\to\infty).
 \label{eq:real-intensity-mertens-mas13}
\end{align}
Let $H$ be an independent countably valued random variable with
$\mathsf H(H)<\infty$ and $\mathbb E|H|<\infty$. Put
\[
 Z=H+\sum_{i=1}^m a_i\sum_{p>p_0}B_{i,p}v_i(p),\qquad
 K=\sum_iK_i.
\]
For each class $g$ of equal absolute coefficient magnitudes, set
$\theta_g=K^{-1}\sum_{i\in g}K_i$ and $\rho_j=\sum_g\theta_g^j$.
Then all mesh entropies of $Z,e^Z,e^{-Z}$ are finite, and
\begin{equation}\label{eq:real-intensity-bounds-mas13}
 \begin{gathered}
 c(K,\boldsymbol\theta)
 \le\underline d_1(Z)=\underline d_1(e^Z)=\underline d_1(e^{-Z})\\
 \le\overline d_1(Z)=\overline d_1(e^Z)=\overline d_1(e^{-Z})
 \le d_K,\\
 c(K,\boldsymbol\theta)
 =1-2^{-K}+(K+1)\{J_U(K)+\rho_2[J_C(K)-J_U(K)]\}
       +\sum_{j\ge3}\rho_j I_j(K)+\rho_3\Delta_3(K).
 \end{gathered}
\end{equation}
In particular, for a single intensity-$K$ copy the lower bound is
\[
 1-2^{-K}+(K+1)J_C(K)+\sum_{j\ge3}I_j(K)+\Delta_3(K).
\]
The functionals are those in \eqref{eq:pair-integral},
\eqref{eq:cluster-integral}, \eqref{eq:delta-three}, and
\eqref{eq:coding-functional}.
\end{theorem}

\paragraph*{Cutoff convention}
The sufficient special case $q_{i,p}=K_i/p+O_i(p^{-2})$ follows from
\eqref{eq:mertens-dvp}: the summable correction has an $O_i(1/x)$
tail. Any finite Bernoulli head, with probabilities in $[0,1]$, may be
included in $H$. Enlarging the fixed cutoff changes only the constants
$\mathfrak b_i$ and moves finitely many independent coordinates into
$H$; it does not weaken the remainder $o(1/\log x)$.

\begin{proof}
First enlarge $p_0$, if necessary, so that $q_{i,p}\le1/2$ for all
remaining coordinates. The enlarged head is still independent of the
tail, has finite absolute mean, and has finite entropy because it is
the sum of $H$ and a finite-valued independent variable. Relabel this
head and the constants $\mathfrak b_i$ as before.
Since $v_i(p)=O(p^{-1})$ and $q_{i,p}=O(p^{-1})$ by
\eqref{eq:real-intensity-envelope-mas13}, the series of expected absolute
increments converges. The sum is absolutely convergent almost surely and has finite
absolute mean. \Cref{lem:transfers}\textup{(i),(ii)} therefore gives
entropy finiteness, permits removal of $H$ at an $O(1)$ entropy cost,
and transfers both dimensions to the exponential images. Reflection
changes mesh entropy by at most $\log2$. We prove the remaining bounds
for $H=0$.

For the lower bound put $L=\log(1/h)$ and use the prime-block
construction preceding \Cref{lem:hidden-vector}, with a coordinate
active when $B_{i,p}=1$ and no higher valuations. Retain exponents
$u=\log p/L$ in a fixed interval $(a,1)$ with $a>0$, first fixing the
coarse partition and a bound on the number of retained choices.
Conditional on a singleton of copy $i$ in $(P,2P]$, its hidden prime
has probability $r_{i,p}/\sum_{P<\ell\le2P}r_{i,\ell}$, where
\[
 r_{i,p}:=\frac{q_{i,p}}{1-q_{i,p}},\qquad
 q_{i,p}\le r_{i,p}\le2q_{i,p}\le\frac{2C_i}{p}.
\]
Subtracting \eqref{eq:real-intensity-mertens-mas13} at $P$ from its
value at $2P$ gives
\begin{equation}\label{eq:real-intensity-dyadic-mas13}
 \sum_{P<p\le2P}q_{i,p}
 =K_i\log\!\left(1+\frac{\log2}{\log P}\right)
   +o\!\left(\frac1{\log P}\right)
 =\frac{K_i\log2+o(1)}{\log P}.
\end{equation}
Moreover, $0\le r_{i,p}-q_{i,p}\le2q_{i,p}^2$, so the sum of the
odds differs from this sum by $O(P^{-1})$. For sufficiently large $P$
it is at least $K_i\log2/(2\log P)$. The largest conditional atom is
therefore at most $(4C_i/(K_i\log2))(\log P)/P$, and the conditional entropy is
$\log P+O(\log L)$, uniformly for $e^{aL}/2\le P\le e^L$.
The constants and threshold may depend on $C_i,K_i$ and the remainder
in \eqref{eq:real-intensity-mertens-mas13}.
The priors may vanish at some primes. The collision counts of
\Cref{lem:divisor-energy,lem:rs-energy,lem:liu-energy,lem:recursive-energy}
hold on subsets of their full ranges, and the atom bound gives the
point-mass hypotheses of \Cref{cor:divisor-reward,cor:pair-rewards}.
Each partial dyadic interval at a coarse-bin endpoint has total odds
at most $2C_i\sum_{P<p\le2P}1/p\ll1/L$, by
\eqref{eq:real-intensity-envelope-mas13} and \eqref{eq:mertens-dvp}.
There are finitely many such intervals at a fixed partition, and each
lost choice has entropy at most $L+O(1)$; their expected entropy cost
is $O(1)$ and vanishes after division by $L$. Conditioning on the revealed record
still factors over disjoint prime ranges. Consequently the entropy
identity and collision argument of \Cref{lem:hidden-vector} give
the same block rewards, with error $O(\log L)$ for a fixed number
of choices: the integer collision counts concern the unchanged jumps
and fixed coefficients, while only these conditional priors enter
their entropy application.

For every nonnegative step function $f$ supported
on $(a,1)$ the point process of copy $i$ satisfies
\begin{align*}
 \log\mathbb E\exp\!\left\{-\sum_p B_{i,p}f(\log p/L)\right\}
 &=\sum_pq_{i,p}(e^{-f(\log p/L)}-1)+o(1)\\
 &\longrightarrow K_i\int_a^1(e^{-f(u)}-1)\,\frac{du}{u}.
\end{align*}
Indeed the error is $O(\sum_{p>e^{aL}}q_{i,p}^2)=O(e^{-aL})$ by
\eqref{eq:real-intensity-envelope-mas13}. On each interval
$(u,v]\subset(a,1]$ where $f=c$, \eqref{eq:real-intensity-mertens-mas13}
gives the linear contribution $(e^{-c}-1)\{K_i\log(v/u)+o(1)\}$.
Independence over $i$ gives a superposed process of intensity $K\,du/u$,
whose independent copy marks have probabilities $K_i/K$; hence its
magnitude-class probabilities are precisely $\theta_g$.
At fixed cutoff the point counts have bounded second moments, which
justify removal of the configuration-size truncation. Conditional on
their number, the limiting exponents have a joint density, so the
arithmetic equality boundaries have probability zero. Taking the
quantization limit before partition refinement and margin removal
therefore gives the ordered-limit argument of
\Cref{lem:ordered-limits} for this marked process. The compatible
pair, cluster and triple rewards of
\Cref{lem:compatible-selection,prop:triples} are nonnegative and bounded
by total exponent mass. This mass has mean $K$, and the omitted mass
below $a$ has mean $Ka$. Monotone convergence removes the cutoff.
\Cref{prop:campbell,prop:triple-mean}, already stated for real positive
intensity, give exactly $c(K,\boldsymbol\theta)$, with convergence of
the cluster series also supplied by \Cref{lem:resolvent}.

For the upper bound use the finite-grid record and reference law in
\Cref{lem:adaptive-code}, now with
\[
 S_L(t)=\frac1L\sum_i\sum_{p_0<p\le e^{tL}}B_{i,p}\log p.
\]
For $q_{i,p}>0$ put
\[
 e_{i,p}:=\log\frac{1-q_{i,p}}{q_{i,p}}-\log p
 =\log(1-q_{i,p})-\log(pq_{i,p}),
\]
and set $e_{i,p}=0$ when $q_{i,p}=0$, since that coordinate is almost
surely inactive. For the full head $V_t$ through $e^{tL}$, its
information $\mathcal I_t=-\log\mathbb P(V_t)$ satisfies
\[
 \mathcal I_t-LS_L(t)
 =\sum_i\sum_{p_0<p\le e^{tL}}-\log(1-q_{i,p})
  +\sum_i\sum_{p_0<p\le e^{tL}}B_{i,p}e_{i,p}.
\]
For $0<y\le C_i$ one has
$y|\log y|\le D_i:=\max\{e^{-1},C_i\log C_i\}$.
Taking $y=pq_{i,p}$, and using $q_{i,p}\le1/2$, gives
\[
 -\log(1-q_{i,p})\le\frac{2C_i}{p},\qquad
 q_{i,p}|e_{i,p}|\le\frac{C_i\log2+D_i}{p}.
\]
Both bounds also hold when $q_{i,p}=0$. Therefore
\begin{equation}\label{eq:expected-code-error-mas13}
 \mathbb E\sup_{0\le t\le1}|\mathcal I_t-LS_L(t)|
 \le\sum_i\sum_{p_0<p\le e^L}
 \{-\log(1-q_{i,p})+q_{i,p}|e_{i,p}|\}
 =O(\log(L+2)),
\end{equation}
by \eqref{eq:mertens-dvp}. This expected estimate also holds at the
random selected cutoff, with no lower bound on positive $pq_{i,p}$.

The remaining tail $R_t$ has
$\mathbb E|R_t|\ll e^{-tL}$, by summing $O(p^{-2})$ over integers.
For a fixed finite grid $T$ containing $0,1$, its adaptively selected
cutoff $J$ thus obeys
$\mathbb E(e^{JL}|R_J|)\le\sum_{t\in T}e^{tL}\mathbb E|R_t|=O_T(1)$.
The signed geometric tail code and carry bit used in
\Cref{lem:adaptive-code}, combined with
\eqref{eq:expected-code-error-mas13}, give
\[
 H_{e^{-L}}(Z)
 \le L\,\mathbb E\min_{t\in T}\{1-t+S_L(t)\}+O_T(\log(L+2)).
\]
The displayed Laplace-functional convergence gives weighted-process
convergence away from zero. Below a cutoff $\epsilon$ the expected
discarded mass is at most $C\epsilon+O(L^{-1})$, whereas for the
limiting process it is $K\epsilon$. These bounds give convergence
of each grid expectation, first taking $L\to\infty$ and then
$\epsilon\downarrow0$. The minima lie in $[0,1]$ because $0\in T$.
Refining the grids as in the proof of \Cref{lem:adaptive-code}
therefore gives $\overline d_1(Z)\le d_K$.
\end{proof}

\begin{remark}[Scope of the intensity extension]
\label{rem:intensity-scope-mas13}
 The products $pq_{i,p}$ need
not converge, and the probabilities may vanish, for example outside
a fixed reduced residue class. The lower bound uses the remainder
$o(1/\log x)$ for \eqref{eq:real-intensity-dyadic-mas13}; the upper
bound $d_K$ needs only the remainder $o(1)$.
\end{remark}

\section{Certified bounds for scalar information dimensions}
\label{sec:information-certification}

For the scalar case set
\begin{equation}\label{eq:scalar-functional}
 \begin{gathered}
 B_C:=\tfrac12+2J_C(1),\qquad I_m:=I_m(1),\qquad
 \Delta_3:=\Delta_3(1),\\
 c_L:=B_C+\sum_{m\ge3}I_m+\Delta_3,\qquad d_*:=d_1.
 \end{gathered}
\end{equation}

The singleton baseline and common-magnitude pair gains form $B_C$;
$\sum_{m\ge3}I_m$ adds compatible cluster increments, and $\Delta_3$
adds the disjoint triple improvement. Their additivity follows from
\Cref{lem:compatible-selection,prop:triples}. The upper constant $d_*$
comes from the adaptive code and its reflected-process integral in
\Cref{lem:adaptive-code,lem:reflected-integral}.
\Cref{prop:certificates} gives the intensity-one enclosures.

\begin{proposition}[Two further scalar triple increments]
\label{prop:scalar-extra-triples}
For one basic logarithmic law at intensity one, the lower information
dimension is at least
\begin{equation}\label{eq:scalar-strengthened-lower}
 c_L^+:=c_L+\frac8{138915}+\frac1{57600}>0.55043653.
\end{equation}
\end{proposition}

\begin{proof}
Use the consecutive exponents $u<v<r$ and their successor $w$ (with
$w=1$ at the endpoint) as in \Cref{prop:triples}. Put $t=v/u$,
$z=r/u$, $S=u+v$. In addition to \eqref{eq:triple-region}, select
the following two disjoint ranges for $1<t<5/4$:
\[
\begin{array}{c|c|c|c}
 & z & w & \text{extra reward}\\ \hline
 A &2t<z<(5t-1)/2 & w>7S/2 & 2u-v\\
 B &3<z<3t & w>2(S+r) & (3v-r)/2 .
\end{array}
\]
In $A$, $r<3u$, $w>2r$, and the third entry of
\eqref{eq:triple-reward} is $w-5S/2+r>S+r$. The second entry
exceeds $S+r$ as well. In $B$ distinct prime products are separated
at scale $e^{-(2(S+r)+o(1))L}$ by \Cref{lem:prime-injectivity},
so the strict cutoff again gives full recovery. The final singleton
earns $r$ in either range, and the two head singletons earn $2v-u$.
Liu's bound is outside its range since $v<5u/4$. In $A$ the
Robert--Sargos reward $F_R=r-S/2$ falls below $2v-u$, and
$F_D\le r-3u+v<F_R$ since $3v<5u$. In $B$ one has
$F_R=S-(3v-r)/2>2v-u$ since $z>5t-4$, and again
$F_D\le r-3u+v<F_R$ since $z<3t<8-3t$.
Subtracting the head reward and $r$ gives the last column.

All three selected ranges satisfy $v<2u$, $r>2v$, $w>2r$ and
$r<2S$. The first three inequalities exclude an overlapping selected
triple or a profitable pair crossing a boundary, as in
\Cref{prop:triples}. An eligible cluster from
\Cref{lem:compatible-selection} cannot cross $(v,r)$ or $(r,w)$.
Ending at $u$ would require $v>2u$, and ending at $v$ would require
$r>2S$. This preserves all exterior rewards.

At intensity one the density for a terminal triple is
$du\,dv\,dr/(vr)$, as in \Cref{prop:triple-mean}. After substituting
$v=ut$, $r=uz$, the integration in $u$ of a gain $uf(t,z)$ gives
$U^2f(t,z)/2$, with $U=2/[7(1+t)]$ in $A$ and
$U=1/[2(1+t+z)]$ in $B$. For an interior triple, scaling its first
three points by $w$ gives the same contribution by homogeneity and
the successor density in \Cref{prop:triple-mean}. Thus the additional
expected gains are
\[
 \Delta_A=\frac4{49}\int_1^{5/4}\!\int_{2t}^{(5t-1)/2}
 \frac{2-t}{tz(1+t)^2}\,dz\,dt,\qquad
 \Delta_B=\int_1^{5/4}\!\int_3^{3t}
 \frac{3t-z}{8tz(1+t+z)^2}\,dz\,dt.
\]
The first domain has area $1/64$, with $2-t\ge3/4$ and
$tz(1+t)^2\le(5/4)(21/8)(9/4)^2$. In the second,
$8tz(1+t+z)^2\le1350$ and integrating $3t-z$ over the triangle
gives $3/128$. These give $\Delta_A\ge8/138915$ and
$\Delta_B\ge1/57600$. Apply the compact strict margins and ordered
limits of \Cref{lem:ordered-limits} to each range, and exhaust the
open triangles by compact subsets. Their boundaries have zero measure
and the total gain is bounded by the integrable total exponent mass.
Adding to \eqref{eq:scalar-functional} and using the rational lower
endpoint of \eqref{eq:lower-certificate} proves the result.
\end{proof}

\begin{lemma}[Evaluation of the pair integrals]
\label{lem:pair-evaluation}
For the integrals in \eqref{eq:pair-integral}, set
$B_U=1/2+2J_U(1)$. Then
\begin{align}
 B_C={}&\frac8{15}-\frac{62}{21}\log2+\frac{887}{420}\log3
 -\frac{10}{7}\log5+\frac{53}{30}\log7\notag\\
 &\hspace{12mm}-\frac{13}{84}\log11-\frac7{20}\log19,
 \label{eq:common-pair}\\
 B_U={}&\frac8{15}-\frac{13}{3}\log2+\frac{73}{30}\log3
 -\frac16\log5+\frac{53}{30}\log7\notag\\
 &\hspace{12mm}-\frac34\log11-\frac7{20}\log19.
 \label{eq:universal-pair}
\end{align}
\end{lemma}

\begin{proof}
Write $G_A(u,v;w)=u g_A(x,t)$, where $x=v/u$, $t=w/u$.
The gains vanish unless $1<x<2$ and $t>2x$.
On this region the singleton reward divided by $u$ is $2x-1$,
and the divisor and Robert--Sargos gains are
\[
 g_D=[\min(2-x,t-2-x)]_+,
 \qquad
 g_R=[2-x-\tfrac12(3-t)_+-\tfrac12(3x-t)_+]_+.
\]
For $5/4<x<2$ the Liu gain is
\[
 g_L=\begin{cases}
 [\min(2-x,t-2x,t/2+(2-7x)/6)]_+,&x<7/5,\\
 [\min(2-x,t-2x,t/2+3/2-2x)]_+,&7/5\le x<3/2,\\
 [\min(2-x,t/2+3/2-2x)]_+,&3/2\le x<2.
 \end{cases}
\]
These formulas follow by subtracting $2x-1$ from the pair
rewards. The Liu entry $t+2-3x$ is inactive: for $t\le3x$
it exceeds its third entry by at least $(t+1-2x)/2>0$, and
for $t>3x$ it exceeds the cap $2-x$.
Comparison of the affine pieces shows that $g_R\ge g_D$
for $x\le5/4$, whereas $g_L\ge g_R$ on its domain and
$g_L\ge g_D$ up to $x=3/2$. For $3/2<x<5/3$, the Liu
and divisor ramps cross at $t=7-2x$, after which the divisor
ramp reaches its cap at $t=4$. For $5/3\le x<2$, $g_D$
dominates. These comparisons determine both envelopes completely.

Substitution in the terminal pair integral and integration by parts give
\[
 J_A(1)=\int_1^2 P_A(x)\,\frac{dx}{x},\qquad
 P_A(x)=\int_{2x}^{\infty}\frac{g_A(x,t)}{t^3}\,dt
       =\frac12\int_{2x}^{\infty}\frac{\partial_tg_A(x,t)}{t^2}\,dt.
\]
The boundary terms vanish. Integrating the constant ramp slopes yields
\[
 P_C(x)=\begin{cases}
 \dfrac1{5x-1}-\dfrac1{12}-\dfrac1{12x},&1<x<5/4,\\[2pt]
 \dfrac1{4x}-\dfrac3{4(2+5x)}-\dfrac3{4(10+x)},&5/4<x<7/5,\\[2pt]
 \dfrac1{4x}-\dfrac1{12}-\dfrac1{4(2x+1)},&7/5<x<3/2,\\[2pt]
 \dfrac1{4(4x-3)}+\dfrac1{4(7-2x)}-\dfrac18,&3/2<x<5/3,\\[2pt]
 \dfrac1{2(2+x)}-\dfrac18,&5/3<x<2.
 \end{cases}
\]
Here $P_U=P_C$ below $3/2$, while
\[
 P_U(x)=\frac1{4(4x-3)}-\frac1{4(2x+1)}
 \quad(3/2<x<2).
\]
For example, the second row has slopes $1$ on
$(2x,(2+5x)/3)$ and $1/2$ on $((2+5x)/3,(10+x)/3)$.
The other rows follow from the same ramp endpoints just listed.
Endpoint values do not affect the integrals. Partial fractions, using
\[
 \int\frac{dx}{x(ax+b)}=\frac1b\log\left|\frac{x}{ax+b}\right|
 \quad(b\ne0),\qquad
 \int\frac{dx}{ax^2}=-\frac1{ax},
\]
give \eqref{eq:common-pair} and
\eqref{eq:universal-pair}.
\end{proof}

\begin{proposition}[Certified scalar constants]
\label{prop:certificates}
\index{Information dimension!numerical certification}
At intensity one, the constructed lower functional $c_L$ and coding
constant $d_*=d_1$ satisfy
\begin{gather}
 0.55036158455538428674\le c_L
 \le0.55036178355883781299,
 \label{eq:lower-certificate}\\
 0.5704894927181160<d_*<0.5704894946732035
 <0.570489495.
 \label{eq:upper-certificate}
\end{gather}
\end{proposition}

\begin{proof}
All logarithmic enclosures use the following elementary inequality.
For rational $r\ge1$, put $q=(r-1)/(r+1)$. For every $M\ge1$,
\begin{equation}\label{eq:log-remainder}
 0\le\log r-2\sum_{j=0}^{M-1}\frac{q^{2j+1}}{2j+1}
 \le\frac{2q^{2M+1}}{(2M+1)(1-q^2)}.
\end{equation}
It follows by bounding the positive tail of the power series for
$2\operatorname{arctanh}q$. Each rational operation and each
displayed decimal endpoint below is rounded outwards. Negative
coefficients interchange the lower and upper logarithmic endpoints.
Reducing the prime arguments in \eqref{eq:common-pair}
by powers of two into $[1,2]$, and using $M=180$, gives
\begin{equation}\label{eq:pair-certificate}
 0.543984420532148874363279
 < B_C <
 0.543984420532148874363280.
\end{equation}

For the entire cluster sum use $V=V_1$, $A=A_1$ and $K=K_1$
from \Cref{lem:resolvent}. Thus $V=A+KV$,
$0\le V\le1/10$, and $V(0)=1/10$. The function $A$ decreases.
The operator $K$ preserves nonnegative decreasing functions:
$z/(z+y)$ increases with $z$, while $[y(z+y)^2]^{-1}$ decreases.
Iteration from zero consequently proves that $V$ decreases.
For $z>0$, substitution $t=z/(z+y)$ gives
\begin{equation}\label{eq:kernel-primitive}
 (KV)(z)=z^{-2}\int_{z/(z+2)}^{z/(z+1)}V(t)\,dW(t),
 \qquad W(t)=-t-\log(1-t).
\end{equation}
In particular, all primitive arguments belong to $[0,1/2]$.

The following finite procedure encloses this fixed point.
Set $N=65536$, $Q=10^{20}$, $R=10^{40}$, $z_j=j/N$.
If integers $\ell_j,u_j$ enclose $QV(z_j)$, monotonicity gives
the cell bounds $\ell_{j+1}/Q\le V(t)\le u_j/Q$ on
$[z_j,z_{j+1}]$. Enclose each nonnegative weight
$R[W(b)-W(a)]$ on an intersected cell by integers $w^-,w^+$,
using \eqref{eq:log-remainder} with $M=48$.
For row $i>0$, sum $S_i^- =\sum\ell_{j+1}w^-$ and
$S_i^+=\sum u_jw^+$ over the cells intersecting
$[z_i/(z_i+2),z_i/(z_i+1)]$. Partial cells use their exact
rational endpoints. Then certified new bounds for $QKV(z_i)$ are
\[
 \left\lfloor\frac{N^2S_i^-}{Ri^2}\right\rfloor,
 \qquad
 \left\lceil\frac{N^2S_i^+}{Ri^2}\right\rceil.
\]
For $i>0$ the inhomogeneous term is evaluated as
\[
 A(z_i)=\frac{N^2}{i^2}
 \left\{\frac12\log\frac{2(i+N)}{i+2N}
              -\frac{i}{4(i+N)}\right\},
\]
with $M=32$ and working scale $R$, retaining the exact factor
$N^2/i^2$. Its logarithmic series has $q\le1/7$; the
primitive series has $q\le1/3$.

Start with lower bounds zero and upper bounds $Q/10$, fixing
both endpoint bounds at $j=0$ to $Q/10$. Add the bounds for
$QA(z_i)$ to each new row. Taking the maximum with a previous
lower bound and the minimum with a previous upper bound preserves
the enclosure, as does propagating lower bounds to smaller indices
and upper bounds to larger indices. Induction proves that every
iterate encloses the fixed point. Whole-cell sums can be stored
as integer prefix sums; their differences are sums of the same
bounded nonnegative terms. Forty iterations, followed by
subtraction of the interval for $A(1)$, give
\begin{equation}\label{eq:cluster-certificate}
 0.00577226072909446735
 \le\sum_{m\ge3}I_m\le
 0.00577244660956974973.
\end{equation}

Integrating the inner variable in \eqref{eq:delta-three}
at intensity one gives
\[
 \Delta_3=\frac4{49}\int_1^{5/4} f(t)\,dt,\qquad
 f(t)=\frac{N_3(t)}{t(1+t)^2},\qquad
 N_3(t)=\frac32(1+t)\log\frac6{5t-1}-\frac{7-5t}{2}.
\]
The defining inner integral shows $N_3(t)\ge0$. Moreover,
using $\log x\le x-1$ gives
\[
 N_3'(t)=\frac32\log\frac6{5t-1}+\frac52
                  -\frac{15(1+t)}{2(5t-1)}\le-\frac12.
\]
Thus $f$ decreases. On $N=65536$ equal subintervals of
$[1,5/4]$, its right and left endpoint sums bracket the
integral. At each endpoint evaluate the logarithm by
\eqref{eq:log-remainder}, with $M=24$ and scale
$10^{35}$; here $q\le1/5$. Exact integer division for every
term and the remainder yields
\begin{equation}\label{eq:triple-certificate}
 0.000604903294140945029963
 \le\Delta_3\le
 0.000604916417119188894005.
\end{equation}
Adding the rational endpoints for the pair, cluster and triple
terms, and rounding outwards to twenty decimal places, proves
\eqref{eq:lower-certificate}.

Finally, let $F(t)=\exp(\operatorname{Ein}(t)-t)$, so that
\eqref{eq:coding-constant-integral} gives
$d_*=1-(\int_0^\infty F)^{-1}$.
Writing $a(t)=\operatorname{Ein}(t)-t$, one has
\[
 a'(t)=-1+\int_0^1e^{-tu}\,du,
 \quad |a'|\le1,
 \quad |a^{(j)}|\le1/j\ (2\le j\le4),
 \quad F\le1.
\]
The fourth derivative of $e^a$ therefore satisfies
\[
 |F^{(4)}|
 \le1+6\cdot\tfrac12+3\cdot\tfrac14
                  +4\cdot\tfrac13+\tfrac14=\frac{19}{3}.
\]
Composite Simpson quadrature on $[0,40]$ with mesh $1/128$
has error at most $40\cdot19/(540\cdot128^4)$.
For $t\ge1$, $\operatorname{Ein}(t)\le1+\log t$, whence
\[
 \int_{40}^\infty F(t)\,dt
 \le 41e^{1-40}\le\frac{123}{2^{40}}.
\]
At each mesh point, use the first $180$ terms of
\[
 \operatorname{Ein}(t)=\sum_{k\ge1}\frac{(-1)^{k+1}t^k}{k\,k!}.
\]
The omitted alternating tail is between zero and its first
positive term, since $t\le40$ makes the omitted term magnitudes
decrease. Directed $70$-digit decimal arithmetic encloses all
finite operations; each correctly rounded exponential is enlarged
to its two adjacent representable values. Add the Simpson error
in both directions and the nonnegative tail in the upper direction,
then apply the increasing map $x\mapsto1-1/x$. This gives
\eqref{eq:upper-certificate}.
\end{proof}

\section{Independent sums and scalar specialization}
\label{sec:information-sums}

Each independent nonzero summand supplies prime choices whose copy
label becomes a mark in the limiting exponent process. Universal pair
estimates allow different coefficient magnitudes; the additional
divisor, cluster and triple gains require equal magnitudes. The
following specialization of the real-intensity theorem averages these
gains with their mark probabilities. The number of summands and all
coefficients remain fixed as the resolution tends to zero;
\Cref{cor:convolution-information} then treats the limit over weighted
partial sums.

\begin{corollary}[Independent logarithmic sums]
\label{thm:independent-sums}
\index{Information dimension!independent sums}
Let $Z_1,\ldots,Z_N$ be independent choices among the four basic
logarithmic laws, and put $Z=b+\sum_{i=1}^N a_iZ_i$, where $b,a_1,\ldots,a_N$
are fixed real numbers. If $Y_i$ is the corresponding raw variable, put
$\epsilon_i=-1$ for a totient law and $\epsilon_i=1$ for a divisor-sum law.
Then $Z_i=\epsilon_i\log Y_i$, and the exponential image is
\[
 e^Z=e^b\prod_{i=1}^N Y_i^{\epsilon_i a_i}.
\]
Let $I=\{i\in\{1,\ldots,N\}:a_i\ne0\}$ and $n=|I|$.
If $n=0$, both information dimensions of $Z$ and $e^Z$ are zero.
For $n\ge1$, let $n_g$ be the sizes of the classes in $I$ determined by
equal absolute coefficient magnitudes, and put
$\rho_m=\sum_g(n_g/n)^m$. Define
\begin{equation}\label{eq:marked-lower}
 \begin{split}
 L(\boldsymbol a)={}&1-2^{-n}
 +(n+1)\{J_U(n)+\rho_2[J_C(n)-J_U(n)]\}\\
 &+\sum_{m\ge3}\rho_m I_m(n)+\rho_3\Delta_3(n).
 \end{split}
\end{equation}
Then all quantized entropies are finite, and
\begin{equation}\label{eq:independent-bounds}
 L(\boldsymbol a)
 \le\underline d_1(Z)=\underline d_1(e^Z)
 \le\overline d_1(Z)=\overline d_1(e^Z)\le d_n.
\end{equation}
Both dimensions are preserved by replacing any integer model with its
shifted-prime model for the same arithmetic function, or by conditioning
on any positive-probability event in the joint valuation sigma-field.
\end{corollary}

\begin{proof}
If $n=0$, then $Z=b$ and $e^Z=e^b$ are constant, with zero mesh
entropy. Suppose $n\ge1$ and retain only $i\in I$. Realize the copies
by independent valuations $K_{i,p}$, and put
$B_{i,p}=\mathbf1_{\{K_{i,p}\ge1\}}$. The probabilities are
$q_{i,p}=1/p$ or $1/(p-1)$ at odd primes, according to the sampling
model. Replace each divisor-sum increment by
$B_{i,p}\log(1+1/p)$; totient logarithmic increments already equal
$B_{i,p}\log(p/(p-1))$. Denote the resulting sum, including $b$, by
$\widetilde Z$.

For every copy, $q_{i,p}=p^{-1}+O(p^{-2})$ and $q_{i,p}\le2/p$.
The prime Mertens estimate \eqref{eq:mertens-dvp}, together with the
convergent correction sum and its $O(1/x)$ tail, gives
\[
 \sum_{p_0<p\le x}q_{i,p}
 =\log\log x+\mathfrak b_i+o(1/\log x).
\]
Choose a fixed $p_0\ge3$ so that all remaining probabilities are at
most $1/2$. In \Cref{thm:real-intensity-mas13} take $m=n$, $K_i=1$,
and let $H$ be $b$ plus all contributions at $p\le p_0$ after the
first-jump replacement. This head is finite-valued, independent of the
remaining Bernoulli array, and has finite entropy and finite absolute
mean. The two canonical jump branches and the fixed signed
coefficients are exactly those of that theorem. Thus $K=n$;
for each class of equal \emph{absolute} coefficient magnitude,
$\theta_g=n_g/n$ and $\rho_j=\sum_g(n_g/n)^j$.
Substitution into \eqref{eq:real-intensity-bounds-mas13} gives
$c(n,\boldsymbol\theta)=L(\boldsymbol a)$ and $d_K=d_n$,
including convergence of the displayed reward series.

All original and replacement logarithmic sums have finite absolute
mean, since their expected absolute prime increments are summable.
The finite-entropy displacement $Z-\widetilde Z$, which may depend on
the first-jump array, is covered by \Cref{lem:transfers}\textup{(ii)}.
Thus \Cref{cor:arithmetic-entropy-invariance}\textup{(ii)} transfers
\eqref{eq:independent-bounds} and entropy finiteness from $\widetilde Z$
to $Z$ and their exponential images. Parts \textup{(i),(iii)} of the
corollary give model replacement and arbitrary positive-probability
valuation conditioning, including entropy finiteness and the
exponential-image identities.
\end{proof}

For $n\ge1$, when all nonzero coefficient magnitudes agree, $\rho_m=1$.
When these magnitudes are all distinct,
$\rho_m=n^{1-m}>0$, so repeated marks from one copy still contribute
the same-magnitude gains. 

\begin{proof}[Proof of \Cref{thm:shannon-dimension}]
For $n=1$, one has $\rho_m=1$, so
\eqref{eq:marked-lower} and \eqref{eq:scalar-functional} give
$L(a_1)=\tfrac12+2J_C(1)+\sum_{m\ge3}I_m(1)+\Delta_3(1)=c_L$
and $d_n=d_*$. The two compatible scalar replacements in
\Cref{prop:scalar-extra-triples} raise the lower bound to $c_L^+$
for the same first-jump array. The finite-entropy displacement
in \Cref{cor:arithmetic-entropy-invariance}\textup{(ii)}
transfers this improvement to the original sum. Apply \Cref{thm:independent-sums} with
$b=0$, taking $a_1=1$ for $g=\gsig$ and $a_1=-1$ for $g=f$.
Reflection changes mesh entropy by at most $\log2$, so the latter
choice also gives the bounds for $Z_f$.
\Cref{cor:arithmetic-entropy-invariance} supplies entropy finiteness and the model
and conditioning equalities, including the fixed-progression assertion.
The strict decimal inequalities follow from \Cref{prop:scalar-extra-triples,prop:certificates}.
\end{proof}

 Existence of the information-dimension limit,
equality with the coding
constant, optimality of that constant, and equality between the totient
and sigma information dimensions remain open. 

\section{Positive-order quantization and rate--distortion}
\label{sec:information-quantization}

Use the \hyperref[par:positive-order-quantization]{codebook convention}
from the geometry chapter. A codebook offers at most $N$ reconstruction
values and minimizes the mean $r$th-power error. Rate--distortion
instead minimizes the mutual information between a sample and a
reconstruction under a mean-error constraint. For the four basic laws
and their logarithmic images, these constraints give different dimensions:
the positive-order codebook
dimension is one, whereas the lower and upper rate--distortion
dimensions equal the corresponding Shannon dimensions, bounded
strictly below one.

The codebook lower bound uses diffuse submeasures: for each fixed
$s\in(0,1)$, a positive amount of mass persists at all sufficiently
small scales and cannot be covered by too few small neighborhoods of
codepoints. Together with the moment
bounds, \Cref{lem:arithmetic-diffuse-submeasures,prop:dominated-diffuse-submeasures}
give the positive-order codebook estimate. For the compactly supported
laws in \Cref{thm:quantization}, the positive-order quantization assertion
also follows from \Cref{cor:renyi-spectrum} and
\cite[Theorems~1.1 and~1.11]{KNZ22}; the direct proof below also treats
the unbounded laws.

The rate--distortion
identities are applications of Kawabata--Dembo
\cite[Proposition~3.3]{KD94}, restated in~\cite[Appendix~III, Theorem~23]{WV10};
they transfer the arithmetic information-dimension bounds in
\Cref{thm:shannon-dimension} to rate--distortion dimensions.
With $D=h^r$, the normalization $rR(D)/\log(1/D)$ in those references
is exactly $R(h^r)/\log(1/h)$, as used below.
The direct proof compares the rate--distortion function with mesh
entropy up to a bounded additive error. The order $r>0$ is fixed.
For the laws and finite-coordinate conditionings in
\Cref{thm:quantization}, the
\hyperref[par:order-zero-quantization]{order-zero errors} vanish under
the geometric-mean convention.

\Needspace{3\baselineskip}
\begin{lemma}[Mesh entropy and rate--distortion]
\label{lem:mesh-entropy-rate-distortion}
Let $X$ have a Borel probability law $\lambda$ on $\R$, and suppose
$H_h(X):=\mathsf H(\lfloor X/h\rfloor)<\infty$ for $0<h\le1$.
For fixed $r>0$, define
\[
 R_{\lambda,r}(D)=\inf_{\operatorname{Law}(X)=\lambda,
                         \,\mathbb E|X-Y|^r\le D}I(X;Y).
\]
There is a finite $C_r$, depending only on $r$, such that
\[
 H_h(X)-C_r\le R_{\lambda,r}(h^r)\le H_h(X)\qquad(0<h\le1).
\]
Consequently the lower and upper rate--distortion dimensions equal,
separately, the lower and upper information dimensions. 
\end{lemma}

\begin{proof}
The deterministic reconstruction $Y=h\lfloor X/h\rfloor$ has
$|X-Y|<h$ and mutual information $H_h(X)$, proving the upper bound.
For any feasible coupling, set
$K_h=\lfloor X/h\rfloor-\lfloor Y/h\rfloor$. Then
$|K_h|\le|X-Y|/h+1$, so $\mathbb E|K_h|^r\le c_r$.
The probability law on $\Z$ proportional to $(1+|k|)^{-1-r}$ has
finite normalizing constant. Nonnegativity of relative entropy,
first on finite coarsenings if necessary, bounds the entropy of $K_h$
by a constant times $1+\mathbb E\log(1+|K_h|)$, hence by a constant
$C_r$ because $\log(1+t)\le c'_r(1+t^r)$ for $t\ge0$.
Knowing $Y$ and $K_h$ determines $\lfloor X/h\rfloor$, so data
processing and the entropy chain rule give
\[
 H_h(X)\le I(X;Y)+\mathsf H(\lfloor X/h\rfloor\mid Y)
          \le I(X;Y)+\mathsf H(K_h)\le I(X;Y)+C_r.
\]
Take the infimum over feasible couplings, divide by $\log(1/h)$,
and take lower and upper limits separately.
\end{proof}

\begin{theorem}[Quantization and rate--distortion dimensions]
\label{thm:quantization}
\index{Quantization!positive-order dimensions}
\index{Rate--distortion}
Let $\lambda$ be any of the four integer or shifted-prime totient or
divisor-sum laws, or its logarithmic law, possibly conditioned on any
positive-probability event in the valuation sigma-field.
Then, for every fixed $r>0$,
\begin{equation}\label{eq:quantization}
 e_{N,r}(\lambda)=N^{-1+o(1)},\qquad
 \lim_{N\to\infty}\frac{\log N}{-\log e_{N,r}(\lambda)}=1.
\end{equation}
For the same order, put
\[
 R_{\lambda,r}(D):=\inf\{I(X;Y):\operatorname{Law}(X)=\lambda,
                                      \ \mathbb E|X-Y|^r\le D\},
\]
where mutual information uses natural logarithms. The lower and upper
rate--distortion dimensions, defined by the respective lower and upper
limits of $R_{\lambda,r}(h^r)/\log(1/h)$ as $h\downarrow0$, satisfy separately
\begin{equation}\label{eq:rate-distortion}
 \underline d_{\mathrm{RD}}^{(r)}(\lambda)=\underline d_1(\lambda),\qquad
 \overline d_{\mathrm{RD}}^{(r)}(\lambda)=\overline d_1(\lambda).
\end{equation}
In particular, with the constants of \Cref{thm:shannon-dimension},
$c_L^+\le\underline d_{\mathrm{RD}}^{(r)}(\lambda)
\le\overline d_{\mathrm{RD}}^{(r)}(\lambda)\le d_*<1$.
\end{theorem}

\begin{proof}
By \Cref{lem:arithmetic-diffuse-submeasures}, $\lambda$ satisfies
\eqref{eq:dominated-diffuse-submeasures} and has every positive moment.
\Cref{prop:dominated-diffuse-submeasures}\textup{(b)} proves
\eqref{eq:quantization}, including the unbounded laws.

Put $H_h(X)=\mathsf H(\lfloor X/h\rfloor)$; it is finite by the moment
bound, comparing the integer distribution with a normalized geometric
law. \Cref{lem:mesh-entropy-rate-distortion} gives the comparison
$H_h(X)-C_r\le R_{\lambda,r}(h^r)\le H_h(X)$.
Divide by $\log(1/h)$ and take the lower and upper limits separately.
\Cref{thm:shannon-dimension} supplies the stated bounds.
\end{proof}

\Cref{fig:dimension-mechanisms} collects the dimensional quantities
established in \Cref{sec:dimension-liouville} and in this chapter for an
unconditioned basic law, and records the argument that controls each of
them.

\begin{figure}[t]
\centering
\begin{tikzpicture}[figdiagram,
  % local helpers: the two columns of panel (a) have common widths
  mcol/.style={fmech, anchor=west, minimum width=\wm cm},
  rcol/.style={fres, anchor=east, minimum width=\wr cm}]
% ---- grid (cm) ------------------------------------------------------
\def\xR{13.45}% common right edge of the result column (and of the rules)
\def\wm{6.35}% width of the mechanism boxes M1, Z, M2, M3
\def\wr{5.65}% width of the result boxes R1, R3, R2, R4 (left edge at 7.80)
\def\xd{6.90}% junction axis
\def\yA{-1.16}% lane 1: M1, R1
\def\yZ{-2.47}% lane 2: Z, j1, R3
\def\yM{-3.215}% lane 2: M2
\def\yN{-4.11}% lane 2: Mom, j2, R2
\def\yC{-5.26}% lane 3: M3, R4
\def\yD{-6.40}% lane 4: M4, R5, R6
\def\yB{-7.43}% panel (b) heading
% ---- panel (a) -------------------------------------------------------
\node[fhead, anchor=west] (hA) at (0,0) {(a) Mechanisms};
\draw[frule] (hA.south west) ++(0,-1.5pt) -- ++(\xR,0);
% ---- lane 1: covering -> Hausdorff-zero carrier ----------------------
\node[mcol, minimum height=14.6mm] (M1) at (0,\yA)
  {Covering: head information,\\ empty middle block, small tail\\
   \figref{\Cref{thm:dimension-zero-criterion,cor:prime-series-zero-dimension}}};
\node[rcol] (R1) at (\xR,\yA)
  {Hausdorff-zero full carrier,\\ so $\dim_H(\lambda)=\dim_F(\lambda)=0$\\
   \figref{\Cref{thm:hausdorff-zero} and p.~\pageref{par:four-law-valuation-bounds};}\\
   \figref{raw laws: \Cref{lem:lipschitz-image}; $\dim_F$: \Cref{lem:fourier-Hausdorff}}};
\draw[fuse] (M1) -- (R1);
% ---- lane 2: two junctions -------------------------------------------
\node[mcol] (Z) at (0,\yZ)
  {Zero--one law for atomic series \figref{\Cref{lem:atomic-series-dimensions}}};
\node[mcol] (M2) at (0,\yM)
  {Selected prime blocks: diffuse submeasures\\
   \figref{\Cref{prop:dominated-diffuse-submeasures}; \Cref{cor:prime-block-diffuse,lem:arithmetic-diffuse-submeasures}}};
\node[fmech, anchor=west] (Mom) at (0,\yN)
  {Every positive moment\\ \figref{\Cref{lem:arithmetic-diffuse-submeasures}}};
\node[fjoin] (j1) at (\xd,\yZ) {};
\node[fjoin] (j2) at (\xd,\yN) {};
\node[rcol] (R3) at (\xR,\yZ)
  {$\underline d_\lambda=0$,\ $\overline d_\lambda=1$\ $\lambda$-a.e.;\ \ $P_{\mathrm{pos}}=P_{\mathrm{full}}=1$\\
   \figref{\Cref{thm:local-dimension-oscillation}}};
\node[rcol] (R2) at (\xR,\yN)
  {Quantization dimension $1$, order $r>0$\\ Dyadic $L^q$ dimensions $1$, $0<q<1$\\
   \figref{\eqref{eq:quantization} in \Cref{thm:quantization}; \Cref{cor:renyi-spectrum}}};
% feeds into j1: carrier (from above), zero-one law (left), submeasures (below)
\draw[ffeed] ([yshift=-4.6mm]R1.west) -| (j1);
\draw[ffeed] (Z.east) -- (j1);
\draw[ffeed] ([yshift=1.2mm]M2.east) -| (j1);
% feeds into j2: submeasures (above), moments (left)
\draw[ffeed] ([yshift=-1.2mm]M2.east) -| (j2);
\draw[ffeed] (Mom.east) -- (j2);
\draw[fuse] (j1) -- (R3);
\draw[fuse] (j2) -- (R2);
% each label names the bound its own feed supplies; vertical feeds: right of the line
\node[flab, anchor=west] (lA) at ($(j1)+(0.8mm,4.4mm)$) {$\underline d_\lambda$};
\node[flab, anchor=west] (lB) at ($(j1)+(0.8mm,-3.1mm)$) {$\overline d_\lambda$};
\node[flab, anchor=west] (lC) at ($(j2)+(0.8mm,3.1mm)$) {$\ge1$};
\node[flab, anchor=south] (lD) at ($(Mom.east)!0.5!(j2)+(0,0.3pt)$) {$\le1$};
\node[fnote, anchor=west] (n1) at (\xd+0.13,\yM-0.045)
  {in \eqref{eq:dominated-diffuse-submeasures} $c_s$ depends on $s$: this gives only $\underline d_1\ge c_ss$};
% ---- lane 3: concentration -> high-order L^q collapse -----------------
\node[mcol, minimum height=8mm] (M3) at (0,\yC)
  {Head atom and empty block: concentration\\
   \figref{empty-block estimate \eqref{eq:lq-empty-block}, \Cref{prop:renyi-collapse}}};
\node[rcol, minimum height=8mm] (R4) at (\xR,\yC)
  {Dyadic $L^q$ dimensions $0$, $q>1$\\ \figref{\Cref{cor:renyi-spectrum}}};
\draw[fuse] (M3) -- (R4);
% ---- lane 4: recovery and coding -> information -> rate distortion ----
\node[fmech, anchor=west] (M4) at (0,\yD)
  {Recovery of hidden primes ($\ge c_L^+$)\\ Adaptive head--tail code ($\le d_*$)\\
   \figref{\Cref{lem:hidden-vector,lem:ordered-limits,lem:adaptive-code}}};
\node[fres, right=5.5mm of M4, minimum width=3.05cm, minimum height=11.4mm] (R5)
  {Information bounds\\ $c_L^+\le\underline d_1\le\overline d_1\le d_*$\\ \figref{\Cref{thm:shannon-dimension}}};
\node[fres, anchor=east, minimum width=3.05cm, minimum height=11.4mm] (R6) at (\xR,\yD)
  {Rate--distortion\\ dimensions in $[c_L^+,d_*]$\\ \figref{\Cref{thm:quantization}}};
\draw[fuse] (M4) -- (R5);
\draw[fuse] (R5) -- node[flab, above] (lE) {\Cref{lem:mesh-entropy-rate-distortion}}
  node[fhyp, below] (hE) {fixed $r>0$} (R6);
% ---- panel (b) -------------------------------------------------------
\node[fhead, anchor=west] (hB) at (0,\yB) {(b) Dimensions of an unconditioned basic law $\lambda$, not of its interval or ray support};
\draw[frule] (hB.south west) ++(0,-1.5pt) -- ++(\xR,0);
% three columns over the full panel width; references flush right with the result boxes
\node[anchor=north west, inner sep=0pt, font=\figfont] (tab) at (0,\yB-0.34) {%
  \begin{tabular*}{\xR cm}{@{\extracolsep{\fill}}llr@{}}
  Hausdorff and Fourier dimensions & $\dim_H(\lambda)=\dim_F(\lambda)=0$ & \llap{\figref{\Cref{thm:hausdorff-zero} and p.~\pageref{par:four-law-valuation-bounds}; \Cref{lem:fourier-Hausdorff}}}\\
  lower / upper local dimension & $\underline d_\lambda(x)=0$,\ \ $\overline d_\lambda(x)=1$\ \ for $\lambda$-a.e.\ $x$ & \figref{\Cref{thm:local-dimension-oscillation}}\\
  packing-carrier dimensions & $P_{\mathrm{pos}}(\lambda)=P_{\mathrm{full}}(\lambda)=1$ & \figref{\Cref{thm:local-dimension-oscillation}}\\
  dyadic $L^q$ dimensions, $0<q<1$ & $1$ & \figref{\Cref{cor:renyi-spectrum}}\\
  dyadic $L^q$ dimensions, $q>1$ & $0$ & \figref{\Cref{cor:renyi-spectrum}}\\
  information dimensions & $0.55043653<\underline d_1\le\overline d_1<0.570489495$ & \figref{\Cref{thm:shannon-dimension}; $\underline d_1=\overline d_1$ open}\\
  quantization dimension & $1$, for every order $r>0$ & \figref{\Cref{thm:quantization}}\\
  rate--distortion dimensions & $\underline d_1$ and $\overline d_1$, for every order $r>0$ & \figref{\Cref{thm:quantization}}
  \end{tabular*}};
\draw[frule] (tab.south west) ++(0,-1.1mm) -- ++(\xR,0);
\end{tikzpicture}
\caption{Different mechanisms control different dimensional quantities of an unconditioned basic
law $\lambda$ (ratio or logarithmic). Empty prime blocks give Hausdorff-zero carriers and
concentration, while selected prime blocks provide diffuse submeasures. Information bounds require
recovery and coding; the rate--distortion dimensions equal $\underline d_1,\overline d_1\le d_*<1$,
whereas the positive-order quantization dimension is $1$. The law $\lambda$ is Rajchman although
$\dim_F(\lambda)=0$ (\Cref{cor:raw-ratio-analytic-images}).}
\label{fig:dimension-mechanisms}
\end{figure}

\paragraph*{Order-zero quantization}\phantomsection\label{par:order-zero-quantization}
For a Borel probability law $\lambda$ with
$\int\log(1+|x|)\,d\lambda(x)<\infty$ and $N\ge1$, define
\[
 e_{N,0}(\lambda):=\inf_{\substack{A\subset\R\\1\le |A|\le N}}
 \exp\!\left(\int\log\operatorname{dist}(x,A)\,d\lambda(x)\right),
\]
with $\log0=-\infty$ and $\exp(-\infty)=0$.
For the laws in \Cref{thm:quantization}, with conditioning restricted
to positive-probability events determined by finitely many valuations,
$e_{N,0}(\lambda)=0$ for all $N\ge1$.
Indeed, take $L\ge2$ beyond the conditioned primes and write the
conditional logarithmic variable as $Z=H+\sum_{\ell>L}Z_\ell$, where
$H$ is countably supported and independent of the remaining prime
increments. Choose $a$ with $\mathbb P(H=a)>0$, and let $T$ be the
identity or the relevant ratio map, so that $\lambda=\operatorname{Law}(T(Z))$.
The remaining increments satisfy
$\mathbb P(Z_\ell\ne0)=\ell^{-1}+O(\ell^{-2})$ and
$\sum_{\ell>y}\mathbb E Z_\ell\ll y^{-1}$.
The anchored form of \eqref{eq:lq-empty-block} gives
\[
 \lambda(B(T(a),h))\gg\frac1{\log(1/h)}\qquad(h\downarrow0).
\]
By \Cref{lem:arithmetic-diffuse-submeasures},
$\int\log^+|x-T(a)|\,d\lambda(x)<\infty$.
Layer cake therefore gives
$\int\log|x-T(a)|\,d\lambda(x)=-\infty$; take $A=\{T(a)\}$.

For the logarithmic sums in \Cref{cor:independent-sum-concentration}\textup{(b)}, under
separate finite-coordinate conditioning and with $n\ge2$, one instead
has $e_{N,0}(\lambda)\ge\exp(-C_\lambda N^{1/n})>0$
for every $N\ge1$. Indeed, the concentration bound and layer cake give,
uniformly over $1\le|A|\le N$,
\[
 \int\log^-\operatorname{dist}(x,A)\,d\lambda(x)
 \le\int_0^\infty\min\{1,CN(1+u)^{-n}\}\,du
 \ll_\lambda N^{1/n},
\]
and taking the infimum proves the claim.

\paragraph*{Independent sums: common setup}
Let $W_1,\ldots,W_n$ be independent choices among the four basic ratio
laws and their logarithmic laws, each possibly conditioned on a fixed
positive-probability event determined by finitely many of its valuations.
Fix $n\ge1$, $b\in\R$ and $a_i\in\R\setminus\{0\}$, and put
$Z=b+\sum_{i=1}^n a_iW_i$ and $\lambda=\operatorname{Law}(Z)$.
This setup is used by the next lemma and three corollaries. Any further
conditioning is stated separately in each conclusion.

\begin{lemma}[Diffuse and moment inputs for independent sums]
\label{lem:independent-sum-diffuse-inputs}
In the common setup, $\lambda$ has every positive moment and satisfies
\eqref{eq:dominated-diffuse-submeasures}. Both assertions persist under
arbitrary positive-probability conditioning in the joint valuation
sigma-field. If every $W_i$ is logarithmic, then
$\mathbb E e^{t|Z|}<\infty$ for every $t>0$, also after such conditioning,
and the diffuse and positive-moment assertions hold for $e^Z$ as well.
\end{lemma}

\begin{proof}
Work under the original product valuation law and combine all separate
conditionings and any additional joint conditioning into a single
positive-probability event $D$. Denote the valuation coordinates of
sample $i$ by $K_p^{(i)}$. Write $W_1=T(U_1)$ with
$U_1=\sum_p g(p^{K_p^{(1)}})\ge0$, $g\in\{f,\gsig\}$, and
$T(u)=u,e^{-u}$ or $e^u$. Put
\[
 H=g(2^{K_2^{(1)}}),\qquad
 \Xi=\bigl(K_2^{(1)},(K_p^{(i)})_{p,\,2\le i\le n}\bigr),\qquad
 V=b+\sum_{i=2}^na_iW_i.
\]
Then $H,V$ are $\sigma(\Xi)$-measurable, $\Xi$ is independent of
$\xi_p:=K_p^{(1)}$ for odd $p$, and $U_1=H+\sum_{p>2}g(p^{\xi_p})$.
Moreover, $\sigma(\Xi,(\xi_p)_{p>2})$ is the full joint valuation
sigma-field, so it contains $D$. The odds and block limits are exactly
\eqref{eq:local-one-prime-event}.
The maps $F(u,v)=a_1T(u)+v$ satisfy the derivative condition of
\Cref{cor:prime-block-diffuse}, which therefore gives the diffuse
condition for $\operatorname{Law}(Z\mid D)$.
If all summands are logarithmic, use instead
$F(u,v)=\exp(a_1u+v)$ to obtain the same condition for $e^Z$.

For every basic logarithmic sum $U_i$ and every fixed $t>0$, its
local bounds give $\mathbb E e^{t g(p^{K_p})}=1+O_t(p^{-2})$
at odd primes; the prime-$2$ contribution is bounded. Independence
and monotone convergence give $\mathbb E e^{tU_i}<\infty$.
Thus every raw or logarithmic $W_i$ has all positive moments, and a
finite weighted sum has the same property. If all are logarithmic,
$|Z|\le|b|+\sum_i|a_i|U_i$ and independence give every absolute
exponential moment of $Z$, and hence every positive moment of $e^Z$.
For a nonnegative random variable $B$,
$\mathbb E(B\mid D)\le\mathbb EB/\mathbb P(D)$, which proves all
conditional moment assertions.
\end{proof}

\begin{corollary}[Geometry of independent sums]
\label{cor:independent-sum-geometry}
\index{Independent sums!geometry}
In the common setup,
\[
 \underline d_\lambda=0,\quad\overline d_\lambda=1
 \quad(\lambda\text{-a.e.}),\qquad
 \dim_H\lambda=0,\qquad
 P_{\mathrm{pos}}(\lambda)=P_{\mathrm{full}}(\lambda)=1.
\]
These geometric assertions persist under arbitrary positive-probability
conditioning of the joint sample. Moreover,
$\underline D_q(\lambda)=\overline D_q(\lambda)=1$ for $0<q<1$,
and both dimensions vanish for $q>1$. These statements hold also
under arbitrary positive-probability conditioning in the joint
valuation sigma-field. If every summand is logarithmic, all the
corresponding assertions hold for $\operatorname{Law}(e^Z)$.
\end{corollary}

\begin{proof}
The joint Hausdorff-zero carrier in
\Cref{cor:rational-statistics}\textup{(a)} survives the separate
conditionings by domination. The mixed raw and logarithmic sum is a
locally Lipschitz function of the positive raw coordinates, so
\Cref{lem:lipschitz-image} gives a Hausdorff-zero full carrier for
$\lambda$. If the lower local dimension were positive on a set of
positive mass, a positive-mass piece would satisfy
$\lambda(B(x,h))\le h^s$ uniformly for all sufficiently small $h$,
for some $s>0$. Intersecting it with that carrier and summing over
small covers would give positive $s$-dimensional Hausdorff measure,
a contradiction.

Write $\lambda=\mu*\tau$, where $\mu=\operatorname{Law}(a_1W_1)$ and
$\tau=\operatorname{Law}(b+\sum_{i=2}^n a_iW_i)$.
Every $\mu$-positive Borel set has packing dimension one by
\Cref{thm:local-dimension-oscillation}. If $\lambda(A)>0$, then
$\int\mu(A-y)\,d\tau(y)>0$, so a translate of $A$ has positive
$\mu$-mass. Hence both packing-carrier dimensions of $\lambda$ are one.
For rational $0<t<1$ and integers $m,k\ge1$, put
\[
 E_{t,m,k}=\{x\in[-k,k]:\lambda(B(x,2^{-j}))\ge2^{-jt}
                                  \text{ for every }j\ge m\}.
\]
At $h=2^{-j}$, $j\ge m$, a maximal $2h$-separated subset has at most
$h^{-t}$ points, since its radius-$h$ balls are disjoint and each has
mass at least $h^t$. Radius-$2h$ balls cover $E_{t,m,k}$, so this set
has upper box dimension at most $t$. The packing conclusion forces
$\lambda(E_{t,m,k})=0$. Their countable union contains every point
with upper local dimension below one. The universal almost-everywhere
upper bound one (\Cref{lem:universal-upper-local-dimension}) gives
$\overline d_\lambda=1$ almost everywhere.

The diffuse condition and moments from
\Cref{lem:independent-sum-diffuse-inputs}, together with
\Cref{prop:dominated-diffuse-submeasures}\textup{(a),(b)}, give the
$q<1$ spectrum, including unit-cell summability, under the stated
valuation conditioning. For $q>1$, the mixed sum is a locally
Lipschitz scalar function of the vector of logarithmic samples.
Apply \Cref{prop:conditioned-renyi-images} under the original product
law, using the intersection of the separate conditioning events and
the further joint event. The same argument applies after composition
with the exponential map. For all-logarithmic sums the exponential map
is locally bi-Lipschitz, preserving the carrier and local dimensions;
restriction to countably many bounded intervals preserves the
positive-mass packing assertion. The lemma gives the diffuse and
moment inputs for its $q<1$ spectrum as well.
Finally any positive-probability conditioning of the joint sample
produces a law dominated by a constant multiple of the original law.
Differentiation of measures preserves its two local exponents at
conditional-almost-every point, and it inherits the full carrier.
Every positive-mass set for the conditional law has positive original
mass, proving the conditional packing assertions.
\end{proof}

\begin{corollary}[Concentration of independent sums]
\label{cor:independent-sum-concentration}
\index{Concentration functions!mixed logarithmic laws}
Use the common setup, including its separate finite-coordinate
conditioning, with no further arbitrary joint conditioning.
\begin{enumerate}[label=\textup{(\alph*)},leftmargin=*]
\item There is a fixed $z_0\in\R$ such that
\begin{equation}\label{eq:independent-sum-anchor}
 \lambda([z_0-h/2,z_0+h/2])\gg(\log(1/h))^{-n}
 \qquad(h\downarrow0).
\end{equation}
In particular $Q_\lambda(h)\gg(\log(1/h))^{-n}$ and
$\underline D_q(\lambda)=\overline D_q(\lambda)=0$ for every $q>1$.
If all summands are logarithmic, the fixed-anchor lower bound and
$q>1$ conclusion also hold for $\operatorname{Law}(e^Z)$.
\item If all summands are logarithmic, then
$Q_\lambda(h)\asymp(\log(1/h))^{-n}$.
\end{enumerate}
\end{corollary}

\begin{proof}
\textup{(a)} For each summand write $W_i=T_i(U_i)$, where $U_i$ is
its basic logarithmic sum and $T_i$ is the identity or the appropriate
exponential map. Under the separate finite-coordinate conditioning,
$U_i$ has an independent countably atomic finite-coordinate head with
some positive-mass atom $u_i$. The unchanged tail has expected absolute
sum beyond $y$ of size $O(y^{-1})$ and zero hazard
$\log\log y+O(1)$. Since $T_i$ is Lipschitz near $u_i$,
\eqref{eq:lq-empty-block} gives, with $x_i=T_i(u_i)$,
\[
 \mathbb P(|W_i-x_i|\le\delta)\gg(\log(1/\delta))^{-1}
\]
for all small $\delta$. Set $A=\sum_i|a_i|>0$,
$z_0=b+\sum_i a_ix_i$, and $\delta=h/(2A)$. On the intersection
of these $n$ events, $|Z-z_0|\le h/2$. The summands are still
independent, so multiplication proves \eqref{eq:independent-sum-anchor}.
\Cref{prop:renyi-collapse}\textup{(a)} gives the $q>1$ assertion.
For all-logarithmic sums, $u\mapsto e^u$ is Lipschitz near $z_0$;
a fixed rescaling of $h$ in the same anchored event gives the lower
bound near $e^{z_0}$ and the same $q>1$ conclusion.

\textup{(b)} The upper estimate follows from the Fourier-moment argument
in \Cref{rem:mixed-logarithmic-concentration}.
\end{proof}

\begin{corollary}[Quantization and information transfer for independent sums]
\label{cor:independent-sum-quantization}
\index{Independent sums!quantization}
In the common setup, for every fixed $r>0$, \eqref{eq:quantization}
and \eqref{eq:rate-distortion} hold and all mesh entropies are finite.
These assertions persist under arbitrary positive-probability
conditioning in the joint valuation sigma-field. If every summand is
logarithmic, they also hold for $\operatorname{Law}(e^Z)$; the
information dimensions obey \Cref{thm:independent-sums}. In this all-logarithmic case, the lower
information dimensions of $Z$ and $e^Z$ agree, and separately their
upper information dimensions agree. Each is unchanged by the stated
joint valuation conditioning. 
\end{corollary}

\begin{proof}
\Cref{lem:independent-sum-diffuse-inputs} supplies diffuse submeasures
and every positive moment for each law asserted here. Hence
\Cref{prop:dominated-diffuse-submeasures}\textup{(b)} proves
\eqref{eq:quantization}. Finite first absolute moment bounds each
mesh entropy by comparison with a normalized geometric law on $\Z$.
\Cref{lem:mesh-entropy-rate-distortion} gives
\eqref{eq:rate-distortion}, taking the lower and upper limits separately.
For all-logarithmic sums the information bounds are
\Cref{thm:independent-sums}; the exponential-image and conditioning
transfers are exactly \Cref{lem:transfers,cor:arithmetic-entropy-invariance},
with the absolute exponential moments verified in
\Cref{lem:independent-sum-diffuse-inputs}. 
\end{proof}

\section{Finite and infinite weighted sums}
\label{sec:information-convolution}

For logarithmic sums and the corresponding products of raw ratios,
the information bounds below keep the number $n$ of independent summands
and their coefficients fixed as the quantization scale tends to zero,
before considering $n\to\infty$. Every finite partial sum has a
Hausdorff-zero full carrier.

For integers $n\ge1$ put
\begin{equation}\label{eq:convolution-lower-functional}
 c_n^{\mathrm{low}}
 :=1-2^{-n}+(n+1)J_C(n)+\sum_{m\ge3}I_m(n)+\Delta_3(n).
\end{equation}
This is the value of $L(\boldsymbol a)$ in \eqref{eq:marked-lower} when
$\boldsymbol a$ has $n$ nonzero coefficients, all of the same absolute
value, so that $\rho_m=1$ for every $m$, and it is the single-copy lower
bound of \Cref{thm:real-intensity-mas13} at intensity $K=n$.
By \eqref{eq:scalar-functional},
$c_1^{\mathrm{low}}=c_L$. For a single summand,
\Cref{thm:shannon-dimension} and affine invariance give the
stronger lower bound $c_L^+$.
\Cref{fig:information-reuse} records the route from the recovery and
coding arguments through \Cref{thm:real-intensity-mas13} to independent
sums and, through them, the basic laws in this chapter, and to the
fixed-degree field laws in
\Cref{cor:fixed-degree-information-quantization}.
\begin{figure}[t]
\centering
\begin{tikzpicture}[figdiagram]
% Grid: width 13.46 cm (with the frame lines 384pt). Row 1 (the two arguments) has
% its top at y=0, the bridge theorem T spans the full width from y=-1.8, and the
% three applications of row 3 share the top line y=-4.75.
% ---- row 1: the two arguments (grey), joined at a dot -------------------
\node[fmech, anchor=north west, minimum width=5.9cm, minimum height=1.16cm] (mL) at (0,0)
  {Recovery of hidden primes (lower bound):\\ prime blocks $\to$ exponent configurations\\
   \figref{\Cref{lem:hidden-vector,lem:ordered-limits}; \Cref{sec:information-energy,sec:information-compatible}}};
\node[fmech, anchor=north east, minimum width=5.9cm, minimum height=1.16cm] (mU) at (13.46,0)
  {Adaptive coding (upper bound):\\ prime head, adaptive cutoff, quantized tail\\
   \figref{\Cref{lem:adaptive-code}; $d_\kappa$: \eqref{eq:coding-functional}}};
\node[fjoin] (j1) at (6.73,0 |- mL.east) {};
\draw[ffeed] (mL.east) -- (j1);
\draw[ffeed] (mU.west |- j1) -- (j1);
% ---- row 2: the bridge theorem (full width); the labels of the joint arrow sit
% beside its shaft, clear of the tip --------------------------------------------
\node[fres, anchor=north, minimum width=13.46cm] (T) at (6.73,-1.8)
  {Information bounds at real intensity\\
   if $q_{i,p}\le C_i/p$ \eqref{eq:real-intensity-envelope-mas13} and
   $\sum_{p_0<p\le x}q_{i,p}=K_i\log\log x+\mathfrak b_i+o(1/\log x)$ \eqref{eq:real-intensity-mertens-mas13},\\
   with independent head $H$, then $Z=H+\sum_ia_i\sum_{p>p_0}B_{i,p}v_i(p)$ has
   $c(K,\boldsymbol\theta)\le\underline d_1\le\overline d_1\le d_K$\\[0.8pt]
   \figref{\Cref{thm:real-intensity-mas13}; $c(K,\boldsymbol\theta)$: \eqref{eq:real-intensity-bounds-mas13}}};
\draw[fuse] (j1) -- (T.north);
\node[flab, anchor=east] (lP) at (6.55,-1.43)
  {limiting exponent process: $K\,du/u$, marks $K_i/K$};
\node[flab, anchor=west] (lR) at (6.91,-1.43)
  {\Cref{lem:hidden-vector,lem:ordered-limits,lem:adaptive-code} rerun at real intensity};
% ---- row 3: the applications, on one top line --------------------------------
\node[fres, anchor=north west] (c525) at (0,-4.75)
  {Independent\\ logarithmic sums\\ $L(\boldsymbol a)\le\underline d_1\le\overline d_1\le d_n$\\
   \figref{\Cref{thm:independent-sums}}};
\node[fres, anchor=north west] (t51) at (4.83,-4.75)
  {The four basic laws\\ $0.55043653<c_L^+\le\underline d_1$\\ $\le\overline d_1\le d_*<0.570489495$\\
   \figref{\Cref{thm:shannon-dimension}}};
\node[fres, anchor=north east] (c168) at (13.28,-4.75)
  {Fixed degree $m$: $X_m$, $-\log X_m$\\
   $c^{\mathrm{low}}_{\tau(m)}\le\underline d_1\le\overline d_1\le d_{\tau(m)}$\\
   \figref{\Cref{cor:fixed-degree-information-quantization}}};
\node[fnote, anchor=north, align=center] (qn) at ($(c168.south)+(0,-0.06)$)
  {quantization \eqref{eq:quantization}: a separate\\ argument, diffuse submeasures,\\
   moments, \figref{\Cref{prop:dominated-diffuse-submeasures}\textup{(b)}}};
% theorem -> independent sums, jointly with first-jump invariance
\coordinate (a1) at (2.4,0);
\node[fjoin] (j3) at ($(a1 |- T.south)!0.5!(a1 |- c525.north)$) {};
\draw[ffeed] (a1 |- T.south) -- (j3);
\draw[fuse] (j3) -- (a1 |- c525.north);
\node[fhyp, anchor=east, align=right] (lA) at ($(j3)+(-0.15,0)$)
  {$n=\#\{a_i\ne0\}$,\\ $K_i=1$: $K=n$,\\ $\theta_g=n_g/n$};
\node[fres, anchor=west, minimum width=3.095cm] (c53) at ($(j3)+(0.5,0)$)
  {First-jump invariance\\ \figref{\Cref{cor:arithmetic-entropy-invariance}\textup{(ii)}}};
\draw[ffeed] (c53.west) -- (j3);
% theorem -> fixed degree: the long branch into the Chapter 16 frame
\draw[fuse] (T.south -| c168.north) -- (c168.north);
\node[fhyp, anchor=north east, align=right] (lB) at ($(T.south -| c168.north)+(-0.15,-0.07)$)
  {one copy, $K=\tau(m)$, head $H_m$;\\
   \eqref{eq:real-intensity-mertens-mas13} from \eqref{eq:fixed-degree-progression-mertens} over $d\mid m$;\\
   \eqref{eq:real-intensity-envelope-mas13} with constant $2m$};
% independent sums -> basic laws, jointly with the certificates
\node[fjoin] (j2) at (4.22,0 |- c525.east) {};
\draw[ffeed] (c525.east) -- node[fhyp, above] (lN) {$n=1$} (j2);
\node[fres, anchor=north] (p524) at (4.22,-6.52)
  {Certified scalar constants\\ \figref{\Cref{prop:certificates}}};
\draw[ffeed] (p524.north) -- (j2);
\draw[fuse] (j2) -- (t51.west |- j2);
% the Chapter 16 frame around c168 and qn; its bottom edge is aligned with the
% bottom of p524 (7pt = the frame's inner ysep)
\coordinate (gb) at ($(qn.south |- p524.south)+(0,7pt)$);
\begin{scope}[on background layer]
  \node[fgroup, fit=(c168)(qn)(gb)] (g) {};
\end{scope}
\node[ftag, anchor=east] (gt) at ($(g.north east)+(-2mm,0)$) {\Cref{sec:composite-modulus}};
% pointer to the figure on sums
\coordinate (pp) at (1.0,0);
\node[fnote, anchor=north, align=center] (pn) at ($(pp |- c525.south)+(0,-0.48)$)
  {weighted sums,\\ $n\to\infty$: \Cref{fig:sum-regularity}};
\draw[fptr] (pp |- c525.south) -- (pn.north);
\end{tikzpicture}
\caption{Reuse of the information estimates. The recovery and coding arguments are rerun for
independent Bernoulli $B_{i,p}$ of mean $q_{i,p}$, $a_i\ne0$ and $v_i(p)$ on one fixed branch,
$\log(p/(p-1))$ or $\log(1+1/p)$, once the head $H$ (finite entropy and absolute mean) is removed;
$\theta_g$ is the share of $K=\sum_iK_i$ in the class $g$ of equal $|a_i|$. Independent sums and the
field laws $X_m$ \eqref{eq:fixed-degree-field-law} supply their intensities and finite heads;
$L(\boldsymbol a)$, $c_L$, $c^{\mathrm{low}}_{\tau(m)}$ are the resulting values of
$c(K,\boldsymbol\theta)$. For one basic law the two further scalar
recoveries in \Cref{prop:scalar-extra-triples} give $c_L^+$.}
\label{fig:information-reuse}
\end{figure}

\KeepHeadingSpace{10\baselineskip}
\begin{corollary}[Information dimensions of weighted partial sums]
\label{cor:convolution-information}
\index{Convolution powers!information dimension}
\index{Independent sums!information dimension}
Let $Z_1,Z_2,\ldots$ be independent, each with one of the four basic
logarithmic laws. Fix $a_i\in\R\setminus\{0\}$, $i\ge1$, and put
$S_n=\sum_{i=1}^n a_iZ_i$ and $\boldsymbol a_n=(a_1,\ldots,a_n)$.
Let $L$ be as in \eqref{eq:marked-lower}.
\begin{enumerate}[label=\textup{(\alph*)},leftmargin=*]
\item \emph{Bounds and limits.}
\begin{equation}\label{eq:convolution-information-dimension}
 \begin{gathered}
 L(\boldsymbol a_n)\le\underline d_1(S_n)
 \le\overline d_1(S_n)\le d_n<1,\\
 1-d_n\le1-\overline d_1(S_n)\le1-\underline d_1(S_n)\le2^{-n}
 \quad(n\ge1),\\
 \underline d_1(S_n)\uparrow1,\qquad
 \overline d_1(S_n)\uparrow1\quad(n\to\infty).
 \end{gathered}
\end{equation}
\item \emph{Coding deficit.}
One has $1-d_n\sim e^{-\gamma n}/n!$ as $n\to\infty$, and
\[
 1-\overline d_1(S_n)\ge1-d_n>\frac{e^{-\gamma n-1}}{n!}
 \qquad(n\ge1).
\]
\end{enumerate}
\end{corollary}

\begin{proof}
Apply \Cref{thm:independent-sums} to each finite partial sum.
The pair, cluster and triple increments are nonnegative, so
$L(\boldsymbol a_n)\ge1-2^{-n}$. Equation~\eqref{eq:coding-deficit} gives $d_n<1$, the explicit lower
bound and the asymptotic for $1-d_n$.

If $X,Y$ are independent with finite relevant entropies, knowing $Y$
and $\lfloor(X+Y)/h\rfloor$ leaves at most two values of
$\lfloor X/h\rfloor$. Independence and the chain rule give
\[
 H_h(X)=\mathsf H(\lfloor X/h\rfloor\mid Y)
 \le H_h(X+Y)+\log2.
\]
Divide by $\log(1/h)$ and take lower or upper limits at fixed
$X,Y$. This proves monotonicity under independent addition. Apply this with $X=S_n$ and $Y=a_{n+1}Z_{n+1}$; the lower
bound $1-2^{-n}$ gives both limits in
\eqref{eq:convolution-information-dimension}.
\end{proof}

\begin{remark}[Equal weights and finite-sum invariants]
\label{rem:weighted-partial-sum-consequences}
For the partial sums in \Cref{cor:convolution-information}, the
following additional consequences hold at each fixed finite $n$.
\begin{enumerate}[label=\textup{(\alph*)},leftmargin=*]
\item \emph{Equal magnitudes.}
If $|a_1|=\cdots=|a_n|$, then
$L(\boldsymbol a_n)=c_n^{\mathrm{low}}$.
In particular, this lower bound applies to every convolution power of
any one of the four laws.
\item \emph{Invariance.}
Both information dimensions of $S_n$ are unchanged by a nondegenerate
affine transformation or by replacing $S_n$ with
$e^{S_n}=\prod_{i=1}^nX_i^{\epsilon_i a_i}$, where
$Z_i=\epsilon_i\log X_i$ as in \Cref{thm:independent-sums}.
\item \emph{Geometry.}
For every finite $n$, the laws of $S_n$ and of $e^{S_n}$ are singular
continuous with a Hausdorff-zero full carrier. Both satisfy
\Cref{cor:independent-sum-geometry,cor:independent-sum-quantization}.
The concentration conclusions of \Cref{cor:independent-sum-concentration}
apply to $S_n$; its fixed-anchor lower bound and $q>1$ conclusion
also apply to $e^{S_n}$.
\end{enumerate}
\end{remark}

\begin{proof}
Equal coefficient magnitudes give $\rho_m=1$, so
\eqref{eq:marked-lower} reduces to
$L(\boldsymbol a_n)=c_n^{\mathrm{low}}$.
Translation changes mesh entropy by at most $\log2$. Scaling by a
fixed nonzero constant replaces $h$ by a constant multiple, with at
most a bounded quantizer discrepancy; reflection also costs at most
$\log2$. Both information dimensions are therefore affine invariant.
\Cref{thm:independent-sums} gives the exponential-image equalities.
\Cref{cor:independent-sum-geometry,cor:independent-sum-quantization}
apply to both laws; \Cref{cor:independent-sum-concentration} gives the
stated concentration conclusions for each.
\end{proof}

\Needspace{20\baselineskip}
Let $Z_i$ be independent, each with one of the four basic logarithmic laws,
and let $a_i\in\R$ have infinitely many nonzero terms. Consider either
\[
 \begin{aligned}
 S&=\sum_i a_iZ_i,&\qquad \sum_i|a_i|&<\infty,\\
 S&=\sum_i a_i(Z_i-\mathbb EZ_i),&\qquad \sum_i a_i^2&<\infty,
 \end{aligned}
\]
the uncentered and centered cases, respectively. Put
$\nu=\operatorname{Law}(S)$ and
$N(\varepsilon)=\#\{i:|a_i|\ge\varepsilon\}$.

\begin{theorem}[Convergence and information of infinite weighted sums]
\label{thm:infinite-weighted-sums}
\index{Independent sums!infinite series}
\index{Information dimension!infinite sums}
In either case the series converges almost surely and in every $L^p$,
$1\le p<\infty$; in the uncentered case it also converges absolutely
almost surely. Moreover,
\[
 \underline d_1(S)=\overline d_1(S)
 =\underline d_1(e^S)=\overline d_1(e^S)=1.
\]
Both laws satisfy \eqref{eq:quantization} for every $r>0$, and their
lower and upper rate--distortion dimensions are one.
Moreover, $\mathbb E e^{r|S|}<\infty$ for every $r>0$.
\end{theorem}

\begin{proof}
In the uncentered case put $A=\sum_i|a_i|$ and let $M(r)$ be the
largest of $\mathbb E e^{rZ}$ over the four logarithmic laws; it is
finite by \Cref{lem:four-ratio-moments}. Tonelli's theorem proves
absolute convergence. Jensen's inequality with weights $|a_i|/A$
gives $\sup_n\mathbb E e^{r|S_n|}\le M(rA)$ for the partial sums $S_n$.

In the centered case put $Y_i=Z_i-\mathbb EZ_i$. The same moment
bounds and Taylor's formula give constants $C,\delta>0$ such that
\[
 \log\mathbb E e^{uY_i}\le Cu^2\qquad(|u|\le\delta)
\]
uniformly in $i$. Since $\sum_i\Var(a_iY_i)<\infty$, Kolmogorov's
variance criterion gives almost-sure convergence, and orthogonality gives
$L^2$ convergence. For fixed $r>0$, only finitely many indices satisfy
$r|a_i|>\delta$; their exponential moments are finite. Multiplication
of the remaining moment generating functions therefore gives
$\sup_n\mathbb E e^{r|S_n|}<\infty$. In both cases Fatou's lemma and
uniform integrability give all absolute exponential moments of $S$
and convergence in every finite $L^p$.

After discarding zero coefficients, write $S=S_n+T_n$ with $T_n$
independent of $S_n$. Centering only translates each finite head.
The entropy inequality in the proof of \Cref{cor:convolution-information} gives
$H_h(S)\ge H_h(S_n)-\log2$. After division by $\log(1/h)$, taking
the lower limit as $h\downarrow0$ with $n$ fixed yields
$\underline d_1(S)\ge1-2^{-n}$; letting $n\to\infty$ gives
$\underline d_1(S)\ge1$.
Finite mean gives $\overline d_1(S)\le1$ by the unit-interval entropy
bound in \Cref{lem:transfers}; that lemma also transfers the dimensions
to $e^S$.
For each $0<s<1$, \Cref{lem:independent-sum-diffuse-inputs}
supplies the diffuse submeasures of one nonzero summand. If
$\sigma_h$ is one of these submeasures and $\tau$ is the remainder
law, then $\sigma_h*\tau\le\nu$, its mass is unchanged, and
$(\sigma_h*\tau)([x,x+3h])\le C_sh^s$ by integration.
Choose a compact interval $I$ whose complement has $\nu$-mass less
than half the positive uniform submeasure mass. Restriction to $I$
leaves positive uniform mass. On $e^I$, the inverse map $\log$ has
bounded derivative. Thus the preimage in $I$ of an interval of length
$3h$ is covered by a fixed number of intervals of length $3h$, giving
the same $O_s(h^s)$ bound for the exponential pushforward.
 The moment bounds and
\Cref{prop:dominated-diffuse-submeasures}\textup{(b)} give
\eqref{eq:quantization}; \Cref{lem:mesh-entropy-rate-distortion}
gives the rate--distortion assertion from the already finite mesh
entropies.
\end{proof}

The next proposition gives Fourier bounds from coefficient counts.
The following corollary combines them with the exponential moments
above to obtain Schwartz densities.

\begin{proposition}[Fourier bounds from coefficient counts]
\label{prop:infinite-sum-fourier-count}
In either summability regime above, let $S$ and $N$ be as in
\Cref{thm:infinite-weighted-sums}.
With $\rho$ as in \eqref{eq:rho-def}, there are absolute constants
$C,T_0>0$ such that $C\rho(T)<1$ for $T\ge T_0$ and
\begin{equation}\label{eq:infinite-sum-counting}
 |\widehat\nu(t)|\le\{C\rho(T)\}^{N(T/|t|)}
 \qquad(T\ge T_0,\ t\ne0).
\end{equation}
In particular,
\begin{equation}\label{eq:infinite-sum-fourier}
 |\widehat\nu(t)|\le
 \exp\!\left\{-\frac{N(|t|^{-1/2})}{3}
 \bigl(\log\log|t|-\log\log\log|t|-O(1)\bigr)\right\}
 \qquad(|t|\to\infty).
\end{equation}
In particular, $|\widehat\nu(t)|=O_M((\log|t|)^{-M})$ for every
$M>0$.
\end{proposition}

\begin{proof}
The four logarithmic laws satisfy the same bound $O(\rho(t))$ by
\Cref{thm:vk-fourier,cor:rajchman-integer,cor:rajchman-sigma}; centering
changes their transforms only by phases.
For sufficiently large $T$, eventual monotonicity of $\rho$ gives
$|\mathbb E e^{ita_iZ_i}|\le C\rho(T)<1$ whenever $|a_it|\ge T$.
Multiplication over these finitely many indices, bounding the
independent remainder's transform by one, proves
\eqref{eq:infinite-sum-counting}. Taking $T=|t|^{1/2}$ gives
\eqref{eq:infinite-sum-fourier}; $N(\varepsilon)\to\infty$ gives
every inverse-logarithmic saving.
This factor-counting argument is classical; see Wintner
\cite{Win35a,Win35b}.
\end{proof}

\begin{corollary}[Schwartz densities for infinite weighted sums]
\label{cor:infinite-sum-schwartz}
Under the summability and independence hypotheses of
\Cref{thm:infinite-weighted-sums}, retain the counting function $N$
from \Cref{prop:infinite-sum-fourier-count}.
If
\begin{equation}\label{eq:infinite-sum-smoothing-criterion}
 \frac{N(\varepsilon)\log\log(1/\varepsilon)}
      {\log(1/\varepsilon)}\longrightarrow\infty
 \qquad(\varepsilon\downarrow0),
\end{equation}
then $|\widehat\nu(t)|=O_M(|t|^{-M})$ for every $M>0$;
both $S$ and $e^S$ have Schwartz densities on $\R$ and Hausdorff
dimension one.
\end{corollary}

\begin{proof}
Taking $T=|t|^{1/2}$ in
\Cref{prop:infinite-sum-fourier-count}, condition
\eqref{eq:infinite-sum-smoothing-criterion} makes the negative
exponent in \eqref{eq:infinite-sum-fourier} divided by $\log|t|$
tend to infinity. Thus every power saving holds. Fourier inversion
then gives a density $f\in C^\infty(\R)$ of $S$ with every derivative
bounded: $t^j\widehat\nu(t)$ is integrable for every $j\ge0$.
The absolute exponential moments from
\Cref{thm:infinite-weighted-sums} give, for every $A>0$,
$\mathbb P(|S|\ge u)\le C_Ae^{-Au}$.

We claim that, for every $j\ge0$ and $B>0$,
\[
 |f^{(j)}(x)|\le C_{j,B}e^{-B|x|}.
\]
For $j=0$, nonnegativity of the density and its bounded first
derivative imply, for $0<h\le1$,
\[
 f(x)\le\frac{\mathbb P(S\in[x-h,x+h])}{2h}
             +h\|f'\|_\infty
 \le\frac{C_Ae^{-A(|x|-1)}}{2h}+h\|f'\|_\infty.
\]
Choose $h=e^{-B|x|}$ and $A=3B$, absorbing bounded $x$ into the
constant. This proves the claim for $j=0$. If the claim holds for
$j$, Taylor's formula gives
\[
 |f^{(j+1)}(x)|\le
 h^{-1}\bigl(|f^{(j)}(x+h)|+|f^{(j)}(x)|\bigr)
 +\tfrac12h\|f^{(j+2)}\|_\infty.
\]
Use the induction hypothesis with $3B$ and again
$h=e^{-B|x|}$, noting $|x+h|\ge|x|-1$. This proves the claim for
$j+1$. In particular $f$ is Schwartz.

For $y>0$ the density of $e^S$ is $g(y)=y^{-1}f(\log y)$.
Repeated differentiation expresses $g^{(j)}(y)$ as a finite linear
combination of $y^{-j-1}f^{(i)}(\log y)$ with $0\le i\le j$.
For $y\ge1$, the claim with arbitrarily large $B$ bounds each of
these terms by every prescribed inverse power of $y$. For
$0<y\le1$, it bounds them by $O(y^{B-j-1})$ for every $B$.
Hence every derivative tends to zero faster than every positive
power at zero. Defining $g(y)=0$ for $y\le0$ therefore gives a
$C^\infty$ function, flat at zero, and all its derivatives are
rapidly decreasing at both infinities. Thus this zero extension is
Schwartz. Absolute continuity gives Hausdorff dimension one for
both probability laws.
\end{proof}

\begin{corollary}[Absolute continuity of infinite weighted sums]
\label{cor:infinite-weighted-ac}
In either the uncentered or the centered model above, if
\begin{equation}\label{eq:infinite-sum-ac-criterion-mas01}
 \liminf_{\varepsilon\downarrow0}
 \frac{N(\varepsilon)}{\log\log\log(1/\varepsilon)}>0,
\end{equation}
then the laws of $S$ and $e^S$ are absolutely continuous and have
Hausdorff dimension one. In particular, for each fixed $c>0$ this
holds for $a_i=\exp(-\exp(\exp(ci)))$, since for small $\varepsilon$
their count is $\lfloor c^{-1}\log\log\log(1/\varepsilon)\rfloor$.
\end{corollary}

\begin{proof}
We verify the general-shell criterion in
\Cref{lem:conditional-block-smoothing}\textup{(a)} by grouping prime
coordinates into dyadic blocks. Realize the independent
copies by independent valuations $K_{i,p}$ from
\eqref{eq:valuation-laws}, and write
\[
 Z_i=\sum_p g_i(p^{K_{i,p}}),\qquad g_i\in\{f,\gsig\}.
\]
Let $I_0=\{2\}$ and $I_k=\{p:2^k<p\le2^{k+1}\}$ for $k\ge1$,
and put $X_{i,k}=\sum_{p\in I_k}g_i(p^{K_{i,p}})$.
The independent block variables used below are $Y_{i,k}=a_iX_{i,k}$
in the uncentered case and
$Y_{i,k}=a_i(X_{i,k}-\mathbb EX_{i,k})$ in the centered case.
In the first case,
\[
 \sum_{i,k}\mathbb E|Y_{i,k}|
 \le\bigl(\max_Z\mathbb EZ\bigr)\sum_i|a_i|<\infty,
\]
where the maximum is over the four laws. Thus every enumeration of
the double array is absolutely convergent with sum $S$.
In the centered case, independence within each copy and its $L^2$
prime-series convergence give
\[
 \sum_{i,k}\Var(Y_{i,k})
 =\sum_i a_i^2\Var(Z_i)
 \le\bigl(\max_Z\Var Z\bigr)\sum_i a_i^2<\infty.
\]
Kolmogorov's variance criterion gives almost-sure convergence in any
fixed enumeration of the double array. Orthogonality gives the same
$L^2$ limit for all increasing finite sets exhausting the array; by
first completing finitely many copies and then letting their number
increase, this limit is $S$.

For $k\ge1$, declare a block successful when its total valuation is
exactly one:
\[
 C_{i,k}=\mathbf1_{\{\sum_{p\in I_k}K_{i,p}=1\}}.
\]
This requires one valuation equal to one and all others zero, also
for the divisor-sum law. Set
$z_{i,p}=\mathbb P(K_{i,p}=0)$ and
$w_{i,p}=\mathbb P(K_{i,p}=1)/z_{i,p}$. For odd primes,
\[
 w_{i,p}=\begin{cases}
 1/p,&\text{integer sampling},\\
 (p-1)/(p(p-2)),&\text{shifted-prime sampling}.
 \end{cases}
\]
Writing $W_{i,k}=\sum_{p\in I_k}w_{i,p}$, the prime number theorem
and $w_{i,p}=p^{-1}+O(p^{-2})$ imply, uniformly over the four types,
\begin{equation}\label{eq:weighted-block-success-mas01}
 q_{i,k}:=\mathbb P(C_{i,k}=1)
 =W_{i,k}\prod_{p\in I_k}z_{i,p}\sim\frac1k.
\end{equation}
Conditional on success, the prime has probabilities $w_{i,p}/W_{i,k}$
and contributes either $\omega_p$ or $\log(1+1/p)$.
Both weights above are decreasing and $O(2^{-k})$ on $I_k$, whereas
$W_{i,k}\asymp k^{-1}$. Fix $d>0$ strictly below the liminf in
\eqref{eq:infinite-sum-ac-criterion-mas01}. Choose an integer $r\ge3$
sufficiently large, then $0<\delta<1/[4r(r+1)]$, so that with
$a=1/(r+1)+\delta$ and $b=1-\delta$ one has
$d\log(3b/(8a))>8$. This is possible since $b/a$ can be arbitrarily
large. Choose $j\ge2$ with $a(j+1-j/r)>1$, and fix
$0<\gamma<\gamma_0$ as in \Cref{thm:annulus}. With these parameters
fixed before the frequency limit, Abel summation gives
\begin{equation}\label{eq:weighted-block-fourier-mas01}
 \left|\mathbb E(e^{itY_{i,k}}\mid C_{i,k}=1)\right|
 \le C|a_it|^{-\gamma}
 \quad\text{if }|a_it|^a\le2^k\le|a_it|^b
 \text{ and }|a_it|\ge T_0.
\end{equation}
Indeed the weighted numerator is bounded by
$O(k^{-1}|a_it|^{-\gamma})$; division by $W_{i,k}$ proves the claim.
Centering contributes only a phase, and negative $a_it$ follow by
complex conjugation, so the constants are common to all copies.

Take $T_n=2^{2^n}$ and retain the finite deterministic set of blocks
\[
 J_n=\left\{(i,k):T_n^{-1/4}\le|a_i|\le1,\quad
 \left\lceil2a\,2^n\right\rceil\le k\le
 \left\lfloor(3b/4)2^n\right\rfloor\right\}.
\]
For every $T_n\le|t|<T_{n+1}=T_n^2$ and $(i,k)\in J_n$,
\[
 T_n^{3/4}\le|a_it|\le T_n^2,\qquad
 |a_it|^a\le T_n^{2a}\le2^k
 \le T_n^{3b/4}\le|a_it|^b.
\]
Also $|a_it|\ge|t|^{3/4}$, hence
\eqref{eq:weighted-block-fourier-mas01} bounds the conditional
transform by $C|t|^{-3\gamma/4}$ throughout the whole shell.
By \eqref{eq:weighted-block-success-mas01}, each retained copy
contributes $\log(3b/(8a))+o(1)$, uniformly in the copy, by the
harmonic sum over the displayed interval. Only finitely many
coefficients exceed one in absolute value. Since
$\log\log\log(T_n^{1/4})=\log n+O(1)$, there are at least
$(d+o(1))\log n$ retained copies. Consequently
\[
 \liminf_{n\to\infty}\frac{M_n}{\log n}
 \ge d\log(3b/(8a))>8,\qquad M_n=\sum_{(i,k)\in J_n}q_{i,k},
\]
and hence
$\sum_n e^{-M_n/8}<\infty$.
Apply \Cref{lem:conditional-block-smoothing}\textup{(a)} to a fixed enumeration
of the independent block pairs $(Y_{i,k},C_{i,k})$, assigning an
arbitrary count to the unused $k=0$ blocks. It proves absolute
continuity of $S$. The map $x\mapsto e^x$ is a smooth bijection
from $\R$ onto $(0,\infty)$ with locally Lipschitz inverse, so its
image law is also absolutely continuous. A probability measure
absolutely continuous with respect to one-dimensional Lebesgue
measure has Hausdorff dimension one, giving both dimension claims.
\end{proof}

\begin{corollary}[Decay of the coefficients]
\label{cor:infinite-sum-coefficients}
Under the hypotheses of \Cref{thm:infinite-weighted-sums}, both
$S$ and $e^S$ have Schwartz
densities and Hausdorff dimension one if $|a_i|\asymp\lambda^i$ with
$0<\lambda<1$, or if $|a_i|\asymp i^{-\alpha}$ with $\alpha>1$ in
the uncentered case and $\alpha>1/2$ in the centered case.
In the geometric case,
\[
 |\widehat\nu(t)|\le
 \exp\{-c_\lambda\log|t|\log\log|t|\}
\]
for all sufficiently large $|t|$, with some $c_\lambda>0$.
If $\liminf_{i\to\infty}i^\alpha|a_i|>0$ for some $\alpha>1/2$, then
there are $c_0,C_0,A_0>0$ such that the density $f$ of $S$ satisfies
\begin{equation}\label{eq:infinite-sum-gevrey}
 |\widehat\nu(t)|\le C_0e^{-c_0|t|^{1/\alpha}},\qquad
 \|f^{(j)}\|_\infty\le C_0A_0^j(j!)^\alpha
 \quad(t\in\R,\ j\ge0).
\end{equation}
\index{Gevrey regularity}
For $\alpha\ge1$, $f$ is Gevrey of order $\alpha$, as is the density
of $e^S$ on every compact subinterval of $(0,\infty)$.
For $\alpha=1$, $f$ extends holomorphically to a strip about $\R$;
for $1/2<\alpha<1$, it extends to an entire function. In both cases
the density of $e^S$ is real analytic on $(0,\infty)$.
\end{corollary}

\begin{proof}
Both coefficient families satisfy
\eqref{eq:infinite-sum-smoothing-criterion}, so
\Cref{cor:infinite-sum-schwartz} gives their Schwartz densities and
Hausdorff dimension one. In the geometric case,
counting the coefficients in \eqref{eq:infinite-sum-fourier} gives
the displayed decay bound.

For the polynomial lower bound, apply
\Cref{prop:infinite-sum-fourier-count}: fixing $T\ge T_0$ gives
$N(T/|t|)\ge c|t|^{1/\alpha}-O(1)$, hence the first estimate in
\eqref{eq:infinite-sum-gevrey}. Fourier inversion bounds
$\|f^{(j)}\|_\infty$ by a constant times
\[
 \int_0^\infty t^je^{-c_0t^{1/\alpha}}\,dt
 =\alpha c_0^{-\alpha(j+1)}\Gamma(\alpha(j+1)).
\]
Stirling's formula gives the second estimate. For $\alpha\ge1$,
composition with the analytic map $y\mapsto\log y$ gives the stated
local Gevrey regularity of $y^{-1}f(\log y)$. For $\alpha=1$ the
Fourier inversion integral converges locally uniformly in a strip;
for $1/2<\alpha<1$ it does so throughout $\C$. This proves the
holomorphic extensions and the local analyticity of the exponential image.
\end{proof}

\Cref{fig:sum-regularity} records how the results of this section on infinite sums depend on one
another and on \Cref{ssec:slow-jump-ac,sec:raj-annulus,ssec:annulus}. It separates the conclusions
available under the summability hypotheses alone from the two routes, each with its own
coefficient-count condition, to Schwartz densities and to absolute continuity.

\begin{figure}[p]
\centering
\begin{tikzpicture}[figdiagram,
  % three equal columns under the full-width assumptions box
  colbox/.style={minimum width=42.2mm},
  % the two inputs of route B, between the two feed lanes of that column
  lanebox/.style={minimum width=33.2mm}]
% Grid (cm): width 13.46; columns A [0,4.22], core [4.62,8.84], B [9.24,13.46];
% feed lanes of column B at x=9.465 (summability) and x=13.235 (Theorem 13.1).
\def\W{13.46}\def\xA{2.11}\def\xC{6.73}\def\xB{11.35}
\def\laneL{9.465}\def\laneR{13.235}
% ---- (a) finite partial sums: a self-contained strip ------------------
\node[fhead, anchor=west] (hA) at (0,0) {(a) Finite sums $S_n=\sum_{i\le n}a_iZ_i$: independent basic logarithmic $Z_i$, $a_i\neq0$};
\draw[frule] (hA.south west) ++(0,-1.5pt) -- ++(\W,0);
\node[fres, anchor=north west, minimum width=6.58cm] (geo) at (0,-0.46)
  {Every $S_n$ and $e^{S_n}$ has a singular continuous law\\ with a Hausdorff-zero full carrier ($\dim_H=0$)\\
   \figref{\Cref{cor:independent-sum-geometry,rem:weighted-partial-sum-consequences}}};
\node[fres, anchor=north east, minimum width=6.58cm] (inf) at (\W,-0.46)
  {$1-2^{-n}\le\underline d_1(S_n)\le\overline d_1(S_n)<1$ for every $n$;\\ both increase to $1$ as $n\to\infty$\\
   \figref{\Cref{cor:convolution-information}}};
% ---- (b) the infinite sum: assumptions ---------------------------------
\node[fhead, anchor=west] (hB) at (0,-2.0) {(b) The infinite sum $S$};
\draw[frule] (hB.south west) ++(0,-1.5pt) -- ++(\W,0);
\node[finp, anchor=north west, minimum width=\W cm] (setup) at (0,-2.38)
  {$Z_1,Z_2,\ldots$ independent, each with one of the four basic logarithmic laws;\\
   $S=\sum_i a_iZ_i$ with $\sum_i|a_i|<\infty$ (uncentered)\quad or\quad $S=\sum_i a_i(Z_i-\mathbb EZ_i)$ with $\sum_i a_i^2<\infty$ (centered);\\
   infinitely many $a_i\neq0$;\quad $\nu=\operatorname{Law}(S)$,\quad $N(\varepsilon)=\#\{i:|a_i|\ge\varepsilon\}$};
% ---- row 1: Chapter 11 bounds | Theorem 5.34 | Theorem 13.1 ------------
\path (setup.south) ++(0,-0.64) coordinate (r1);
\path let \p1=(r1) in node[fres, colbox, anchor=north] (ch11) at (\xA,\y1)
  {Four logarithmic laws:\\ Fourier transforms $O(\rho(\tau))$,\\ {\figsmall $\rho(\tau)=(\log\log|\tau|/\log|\tau|)^{1/3}$}\\
   \figref{\Cref{thm:vk-fourier},}\\ \figref{\Cref{cor:rajchman-integer,cor:rajchman-sigma}}};
\node[ftag, anchor=south west] (tag11) at ($(ch11.north west)+(0,2pt)$) {\Cref{sec:raj-annulus}};
\path let \p1=(r1) in node[fres, colbox, anchor=north] (t534) at (\xC,\y1)
  {The series converges a.s.\\ and in $L^p$, $p<\infty$;\\ $\mathbb Ee^{r|S|}<\infty$\ \ $(r>0)$;\\
   $\underline d_1=\overline d_1=1$ for $S$ and for $e^S$\\ \figref{\Cref{thm:infinite-weighted-sums}}};
\draw[fuse] (setup.south -| t534.north) -- node[flab, left=1pt] {no further hypothesis} (t534.north);
\path let \p1=(r1) in node[fres, lanebox, anchor=north] (t131) at (\xB,\y1)
  {Cancellation in dyadic\\ prime intervals\\ \figref{\Cref{thm:annulus}}};
\node[ftag, anchor=south west] (tag13) at ($(t131.north west)+(0,2pt)$) {\Cref{ssec:annulus}};
\node[fnote, align=center, anchor=north] (nB1) at ($(t131.south)+(0,-0.07)$)
  {blocks: dyadic prime\\ intervals of each copy; a\\ block is successful when\\ its total valuation is one};
% ---- row 2: Proposition 5.35 | -- | Lemma 10.3(a) ------------------------
\path let \p1=(nB1.south), \p2=(ch11.south) in coordinate (r2) at (0,{min(\y1-11.8pt,\y2-16pt)});
\path let \p1=(r2) in node[fres, colbox, anchor=north] (p535) at (\xA,\y1)
  {Coefficient-count bound:\\ $|\widehat\nu(t)|\le\{C\rho(T)\}^{N(T/|t|)}$\\
   {\figsmall absolute $C,T_0$, $C\rho(T)<1$;}\\ {\figsmall all $T\ge T_0$, $t\neq0$}\\
   \figref{\Cref{prop:infinite-sum-fourier-count}}};
\draw[fuse] (ch11) -- node[flab, right=1pt] {factor counting} (p535);
\path let \p1=(r2) in node[fres, lanebox, anchor=north] (l103) at (\xB,\y1)
  {Conditional block\\ smoothing,\\ general shells\\ \figref{\Cref{lem:conditional-block-smoothing}\textup{(a)}}};
\node[ftag, anchor=south east] (tag10) at ($(l103.north east)+(0,2pt)$) {\Cref{ssec:slow-jump-ac}};
% ---- route B junction: Theorem 13.1, Lemma 10.3(a) and the summability hypotheses
\node[fjoin] (jB) at ($(l103.south)+(0,-0.25)$) {};
\draw[ffeed] (\laneL,0 |- setup.south) |- (jB);
\draw[ffeed] (t131.east) -| (\laneR,0 |- jB) -- (jB);
\draw[ffeed] (l103.south) -- (jB);
% ---- the two coefficient-count conditions, at one height --------------
\path (p535.south) ++(0,-0.13) coordinate (rh);
\path let \p1=(rh) in node[fhyp, anchor=north] (hypA) at (\xA,\y1) {if, as $\varepsilon\downarrow0$,\\[1pt]
   $\dfrac{N(\varepsilon)\log\log(1/\varepsilon)}{\log(1/\varepsilon)}\to\infty$:\\[1pt]
   $|\widehat\nu(t)|=O_M(|t|^{-M})$\ \,$(M>0)$};
\path let \p1=(rh) in node[fhyp, anchor=north] (hypB) at (\xB,\y1) {if, as $\varepsilon\downarrow0$,\\[1pt]
   $\displaystyle\liminf_{\varepsilon\downarrow0}\dfrac{N(\varepsilon)}{\log\log\log(1/\varepsilon)}>0$};
\node[fnote, align=right, anchor=north east] (nB2) at ($(hypB.south)+(-0.15,-0.08)$)
  {almost every conditional\\ law is smooth; absolute\\ continuity passes to\\ the mixture};
% ---- row 4: the two conclusions -------------------------------------------
\path let \p1=(hypA.south), \p2=(nB2.south) in coordinate (r4) at (0,{min(\y1-20pt,\y2-3pt)});
\path let \p1=(r4) in node[fres, colbox, minimum height=11.8mm, anchor=north] (c536) at (\xA,\y1)
  {Schwartz densities\\ of $S$ and $e^S$; $\dim_H=1$\\ \figref{\Cref{cor:infinite-sum-schwartz}}};
\path let \p1=(r4) in node[fres, colbox, minimum height=11.8mm, anchor=north] (c537) at (\xB,\y1)
  {$S$ and $e^S$ absolutely\\ continuous; $\dim_H=1$\\ \figref{\Cref{cor:infinite-weighted-ac}}};
% route A junction: the count bound under (5.62) and the moments of Theorem 5.34
\node[fjoin] (jA) at ($(c536.north)+(0,0.55)$) {};
\draw[ffeed] (p535.south) -- (hypA.north);
\draw[ffeed] (hypA.south) -- (jA);
\draw[ffeed] ($(t534.south west)+(0.78,0)$) |- node[flab, pos=0.25] {absolute\\ exponential\\ moments of $S$} (jA);
\draw[fuse] (jA) -- (c536);
% route B: the junction arrow carries the weaker condition
\draw[fuse, -] (jB) -- (hypB.north);
\draw[fuse] (hypB.south) -- (c537.north);
% ---- example strip ------------------------------------------------------------
\path (c536.south) ++(0,-0.70) coordinate (r5);
\path let \p1=(r5) in node[fres, colbox, anchor=north] (e538) at (\xA,\y1)
  {Schwartz densities if\\ $|a_i|\asymp\lambda^i$, $0<\lambda<1$, or\\ $|a_i|\asymp i^{-\alpha}$, $\alpha>1$\\ ($\alpha>\frac12$ if centered)\\
   \figref{\Cref{cor:infinite-sum-coefficients}}};
\path let \p1=(r5) in node[fres, colbox, anchor=north] (e541) at (\xC,\y1)
  {suitable sparse $a_i>0$\\ with $\sum_ia_i<\infty$:\\ Hausdorff-zero full carrier\\
   \figref{\Cref{prop:sparse-infinite-carrier-mas02}}};
\path let \p1=(r5) in node[fres, colbox, inner xsep=2pt, anchor=north] (e540) at (\xB,\y1)
  {$a_i=2^{-2^i}$, $N(\varepsilon)\asymp\log\log(1/\varepsilon)$:\\ absolutely continuous, but\\ $\dim_F=0$ and no local $L^p$\\
   density, $1<p\le\infty$\\ \figref{\Cref{thm:double-exponential-companions-mas02}}};
\draw[fmap] (c536.south) -- node[flab, right=1pt] {specialization} (e538.north);
\draw[fmap] (c537.south) -- node[flab, left=1pt] {absolute continuity\\ by specialization} (e540.north);
\draw[fmap] (t534.south) -- node[flab, right=1pt, pos=0.12] {comparison:\\ same hypotheses} (e541.north);
\begin{scope}[on background layer]
  \node[fband, inner xsep=0pt, inner ysep=2pt, fit=(e538)(e541)(e540)] (band) {};
\end{scope}
\end{tikzpicture}
\caption{Under the standing assumptions, $S$ and $e^S$ have information dimension one,
and $S$ has all absolute exponential moments; finite partial sums remain singular.
Schwartz densities follow from these moments and the Fourier bounds of
\Cref{sec:raj-annulus} under condition~\eqref{eq:infinite-sum-smoothing-criterion};
absolute continuity follows from conditional block smoothing and prime cancellation
under \eqref{eq:infinite-sum-ac-criterion-mas01}, for logarithmic summands.
For the contrasting generic behavior in coefficient norm see
\Cref{cor:generic-coefficient-carriers}.}
\label{fig:sum-regularity}
\end{figure}

\begin{remark}[Extension to raw laws]
\label{rem:infinite-sum-raw-laws}
Let the independent $Z_i$ be chosen among the four basic ratio laws
and their logarithmic laws. Retain infinitely many nonzero coefficients
and either the uncentered regime $\sum_i|a_i|<\infty$ or the centered
regime $\sum_i a_i^2<\infty$ above.
The convergence and absolute exponential-moment conclusions of
\Cref{thm:infinite-weighted-sums} and the Fourier bounds of
\Cref{prop:infinite-sum-fourier-count} remain valid.
Under their respective additional coefficient hypotheses, the density,
regularity and Hausdorff-dimension conclusions of
\Cref{cor:infinite-sum-schwartz,cor:infinite-sum-coefficients}
also hold for $S$ and $e^S$.
\end{remark}

\begin{proof}
The raw totient laws are bounded, and
\Cref{prop:sigma-mellin-growth} gives every positive exponential
moment of the shifted-prime divisor-sum law. The integer divisor-sum
law is dominated by it under a coordinatewise coupling of the valuations
in \eqref{eq:valuation-laws}. Together with
\Cref{lem:four-ratio-moments}, this gives every absolute exponential
moment uniformly over the eight laws. The preceding Jensen and
quadratic log-mgf arguments therefore apply, as does the density
argument. Finally, \Cref{thm:raw-ratio-normality} gives the Fourier bound
for the four raw laws, and
\Cref{thm:vk-fourier,cor:rajchman-integer,cor:rajchman-sigma} give it for
their logarithmic laws. Thus all eight satisfy the common $O(\rho(t))$
bound; centering changes the transforms only by phases.
\end{proof}

\begin{theorem}[Double-exponential coefficients]
\label{thm:double-exponential-companions-mas02}
Let $Z_i$, $i\ge1$, be independent choices among the four unconditioned,
nonnegative basic logarithmic laws, and put
\[
 a_i=2^{-2^i},\qquad S=\sum_{i\ge1}a_iZ_i,\qquad
 \nu=\operatorname{Law}(S),\quad\eta=\operatorname{Law}(e^S).
\]
Both laws are absolutely continuous and have Hausdorff dimension one,
but
\begin{equation}\label{eq:double-exponential-concentration-mas02}
 \log\frac1{Q_\nu(h)}\sim\log\frac1{Q_\eta(h)}
 \sim\frac{(\log\log(1/h))^2}{\log2}\qquad(h\downarrow0).
\end{equation}
They have Fourier dimension zero. For either $\lambda=\nu$ or
$\lambda=\eta$ and every open set $U$ with $\lambda(U)>0$, the
restriction $\lambda|_U$ has no density in $L^p(U)$ for any
$1<p\le\infty$.
\end{theorem}

\begin{proof}
Here $N(\varepsilon)\asymp\log\log(1/\varepsilon)$, so
\eqref{eq:infinite-sum-ac-criterion-mas01} in
\Cref{cor:infinite-weighted-ac} gives the absolute continuity and
Hausdorff dimensions. 

There is a single constant $A\ge1$ such that, for each of the four
logarithmic laws with characteristic function $\phi$,
\begin{equation}\label{eq:four-law-all-moments-mas02}
 \frac1{2T}\int_{-T}^T|\phi(t)|^r\,dt
 \le\left(\frac{Ar}{\log T}\right)^r
 \qquad(r\ge1,\ T\ge3).
\end{equation}
Indeed, for each of the four laws \Cref{thm:valuation-fourier-moments}\textup{(a)}
gives this, with the first-jump function $v_\varphi(t)=\log(t/(t-1))$
for the two totient laws and $v_\sigma(t)=\log(1+1/t)$ for the two
divisor-sum laws, the finite prime set $\{2\}$, the prime-$2$
coordinate as head and $C_2=2$. Taking the
maximum of these four constants proves the uniform assertion for all $r$.

Set $L=\log(1/h)$, $T=1/h$, and
$n=\lfloor\log_2(L/(2\log2))\rfloor$. For $h$ sufficiently small,
$n\ge1$ and $\log(a_iT)\ge L/2$ for $1\le i\le n$.
The nonnegative concentration kernel in the proof of
\Cref{thm:valuation-fourier-moments}, independence and H\"older's inequality give
\begin{align*}
 Q_\nu(h)
 &\le C\frac1{2T}\int_{-T}^T
                    \prod_{i=1}^n|\phi_i(a_it)|\,dt\\
 &\le C\prod_{i=1}^n
 \left(\frac1{2a_iT}\int_{-a_iT}^{a_iT}|\phi_i(v)|^n\,dv\right)^{1/n}
 \le C\left(\frac{2An}{L}\right)^n.
\end{align*}
Since $n=(\log L)/\log2+O(1)$, this proves
\begin{equation}\label{eq:double-exponential-upper-mas02}
 \log\frac1{Q_\nu(h)}\ge
 \frac{(\log L)^2}{\log2}-O(\log L\log\log L).
\end{equation}

For the reverse bound, let $b_i=\min\supp(Z_i)$ and $X_i=Z_i-b_i$.
The possible values of $b_i$ are $0$, $\log2$, and $\log(3/2)$;
thus $X_i\ge0$ and $B=\sum_i a_ib_i$ is finite. There are constants
$d>0$ and $M\ge1$, uniform over the four types, such that
\begin{equation}\label{eq:four-law-endpoint-lower-mas02}
 \mathbb P(X_i\le u)\ge\frac d{1+\log^+(1/u)}\quad(u>0),
 \qquad \mathbb EX_i\le M,
\end{equation}
where $\log^+v=\max(0,\log v)$. For the lower bound, use the
argument for the empty-block estimate \eqref{eq:lq-empty-block}:
fix the prime-$2$ coordinate at its minimal contribution and suppress
all odd-prime contributions up to $y=1/u$. This event has probability
$\gg1/\log y$, since the zero probabilities are
$1-1/p+O(p^{-2})$. The independent tail has mean
$O(1/(y\log y))$, so Markov's inequality places it in $[0,u]$ with
probability $1-O(1/\log y)$. This proves the bound for small $u$;
monotonicity and a smaller $d$ cover the remaining $u$. The moment
bound follows from \Cref{lem:four-ratio-moments}.

Put $k=\lceil\log_2(L/\log2)\rceil$, so $a_k\le h$ and
$\sum_{i>k}a_i\le2h^2$. Consequently the nonnegative tail
$R_k=\sum_{i>k}a_iX_i$ satisfies
$\mathbb P(R_k\le h/2)\ge1-4Mh$. Requiring, independently,
$a_iX_i\le h/(2k)$ for $1\le i\le k$ gives
\begin{equation}\label{eq:double-exponential-endpoint-mas02}
 \mathbb P(B\le S\le B+h)
 \ge(1-4Mh)\left(\frac d{1+L+\log(2k)}\right)^k.
\end{equation}
Here $k=(\log L)/\log2+O(1)$, so the negative logarithm of the
right-hand side is $(\log L)^2/\log2+O(\log L\log\log L)$.
Together with \eqref{eq:double-exponential-upper-mas02}, this proves
the assertion for $\nu$.

The same lower estimate, with a constant depending on the anchor,
holds in every open set of positive mass. To see this explicitly,
fix the prime-$2$ contribution of $X_1$ at zero, choose any finite
pattern of odd-prime valuations, and let $c$ be its logarithmic sum.
Suppressing the other primes up to $1/u$ as above gives
\[
 \mathbb P(c\le X_1\le c+u)\ge
 \frac{d_c}{1+\log^+(1/u)}\qquad(0<u<u_c)
\]
for some $d_c,u_c>0$. Such $c$ are dense in $[0,\infty)$: it suffices
to use valuations zero and one, whose positive jumps tend to zero
and have divergent sum. Finite subsums of any tail then approximate
every nonnegative target from below with error at most its largest
jump. Hence the anchors $B+a_1c$ are dense in $[B,\infty)$.
Replacing only the first factor in the proof of
\eqref{eq:double-exponential-endpoint-mas02} by this cylinder estimate
shows, for each such anchor $v$,
\begin{equation}\label{eq:double-exponential-anchor-mas02}
 \nu([v,v+h])\ge
 \exp\{-O_v((\log\log(1/h))^2)\}=h^{o(1)}.
\end{equation}
The last notation records a subpower lower bound. Every open $U$ of
positive $\nu$-mass contains an anchor and its sufficiently short
right-hand intervals. If $\nu|_U=f\,dx$ with $f\in L^p(U)$,
H\"older would bound their masses by
$\|f\|_{L^p(U)}h^{1-1/p}$, contradicting
\eqref{eq:double-exponential-anchor-mas02}; this also covers
$p=\infty$.

For exponentiation, $S\ge B$ implies that the logarithmic preimage
of any interval of length $h$ in $[e^B,\infty)$ has length at most
$e^{-B}h$. Thus $Q_\eta(h)\le Q_\nu(e^{-B}h)$. Conversely,
\[
 Q_\eta(h)\ge
 \mathbb P\bigl(B\le S\le B+\log(1+e^{-B}h)\bigr).
\]
Use \eqref{eq:double-exponential-endpoint-mas02}, observing that
$\log(1+e^{-B}h)\asymp h$, to obtain the second equivalent in
\eqref{eq:double-exponential-concentration-mas02}. On a small fixed
neighborhood of an anchor the exponential map changes interval lengths
by bounded factors, so \eqref{eq:double-exponential-anchor-mas02} also
gives the asserted local $L^p$ obstruction for $\eta$.

Finally, if either transform had a bound
$|\widehat\lambda(t)|\ll|t|^{-\delta}$ for some $\delta>0$, replacing
$\delta$ by a smaller number in $(0,1)$ and applying the same
concentration kernel would give $Q_\lambda(h)\ll h^\delta$.
This contradicts \eqref{eq:double-exponential-concentration-mas02}.
Thus both Fourier dimensions are zero.
\end{proof}

\begin{proposition}[Sparse coefficients with a Hausdorff-zero carrier]
\label{prop:sparse-infinite-carrier-mas02}
For any sequence of independent choices $Z_i$ among the four basic
logarithmic laws, there are strictly positive coefficients $a_i$ with
$\sum_i a_i<\infty$ such that $S=\sum_i a_iZ_i$ has a Borel full
carrier of Hausdorff dimension zero. Its exponential image has the
same property.
\end{proposition}

\begin{proof}
Let $M\ge1$ bound all $\mathbb EZ_i$, and start with $a_1=1$.
Suppose $a_1,\ldots,a_n$ have been selected. By
\Cref{cor:independent-sum-geometry}, the finite sum
$S_n=\sum_{i\le n}a_iZ_i$ has a Borel Hausdorff-zero full carrier.
Inner regularity gives a compact subset $K_n$ of this carrier with
$\mathbb P(S_n\in K_n)\ge1-2^{-n-1}$. Cover $K_n$ by finitely many
open intervals $I_{n,r}$ of positive lengths $\ell_{n,r}$ such that
\[
 \max_r\ell_{n,r}<2^{-n},\qquad
 \sum_r\ell_{n,r}^{1/n}<2^{-n}.
\]
Such a finite cover follows from Hausdorff dimension zero and
compactness. Write $\ell_n=\min_r\ell_{n,r}>0$ and set
$b_n=2^{-n-2}\ell_n/M$. At the next step choose any positive
$a_{n+1}$ satisfying
\[
 a_{n+1}\le
 \min\left\{2^{-n-1},\frac{a_n}{2},
       \min_{1\le j\le n}2^{-((n+1)-j)}b_j\right\}.
\]
Every stage imposes only finitely many strictly positive bounds,
so the recursion continues. Crucially, for every earlier $n$ it
controls all subsequent coefficients:
\[
 \sum_{j>n}a_j\le
 b_n\sum_{j>n}2^{-(j-n)}=b_n.
\]
The series converges absolutely almost surely by Tonelli's theorem.
For its nonnegative remainder $R_n=S-S_n$, Markov's inequality gives
\[
 \mathbb P(R_n>\ell_n/2)
 \le\frac{2M b_n}{\ell_n}=2^{-n-1}.
\]
Enlarge each $I_{n,r}$ by $\ell_n/2$ at both ends, and let $U_n$ be
the union of the enlarged intervals. Each new length is at most
$2\ell_{n,r}$, and $S_n\in K_n$, $R_n\le\ell_n/2$ imply $S\in U_n$.
Thus $\mathbb P(S\notin U_n)\le2^{-n}$. Borel--Cantelli shows that
\[
 E=\bigcup_{N\ge1}\bigcap_{n\ge N}U_n
\]
is a Borel full carrier. For every fixed $s>0$ and all large $n$,
$1/n\le s$, the covering lengths are below one and their $s$-cost is
at most $2^{1/n}2^{-n}$. Their maximum length tends to zero. For each
$N$, these covers give
$\mathcal H^s(\bigcap_{n\ge N}U_n)=0$. Countable subadditivity gives
$\mathcal H^s(E)=0$ for every $s>0$, hence $\dim_H E=0$.
The exponential map is Lipschitz on each bounded interval, so its
image of $E$ also has Hausdorff dimension zero.
\end{proof}

\begin{corollary}[Generic singularity in coefficient space]
\label{cor:generic-coefficient-carriers}
Fix independent basic logarithmic samples $Z_i$. In real $\ell^1$
use $S_a=\sum_i a_i Z_i$; in real $\ell^2$ use
$S_a=\sum_i a_i(Z_i-\mathbb EZ_i)$. In each space, equipped with its norm topology, there is a dense
$G_\delta$ set of coefficients with every $a_i\ne0$ for which both
$S_a$ and $e^{S_a}$ have Hausdorff-dimension-zero full Borel carriers
and information dimension one. Coefficients for which both laws have
Schwartz densities are also dense.
\end{corollary}

\begin{proof}
The map $a\mapsto S_a$ is continuous in $L^1$ in the first case,
by the uniform first moment, and in $L^2$ in the second, by
independence and the uniform variance. For $n\ge1$, take the union
over finite families $\mathcal I$ of open rational intervals with
$\max_{I\in\mathcal I}|I|<2^{-n}$ and
$\sum_{I\in\mathcal I}|I|^{1/n}<2^{-n}$ of the coefficient sets
\[
 \left\{a:\mathbb P\left(S_a\in\bigcup_{I\in\mathcal I}I\right)
                          >1-2^{-n}\right\}.
\]
This union is open by weak convergence and the open-set inequality.
It contains every finite-support vector: the finite sum has a
Hausdorff-zero full carrier by \Cref{cor:independent-sum-geometry}
(or is constant). Cover that carrier with countably many open
rational intervals with the stated strict mesh and cost bounds,
then keep finitely many to exceed the required probability.
Finite-support vectors are dense. Intersect these open dense unions
and all open dense sets $\{a:a_i\ne0\}$; the Baire theorem gives a
dense $G_\delta$.

For each resulting $a$, choose one qualifying family at each $n$
and let $U_n$ be its union. Since
$\sum_n\mathbb P(S_a\notin U_n)<\infty$, the Borel set
$E=\limsup_n U_n$ is a full carrier. For $s>0$, the families with
$n\ge N>1/s$ cover $E$ with mesh at most $2^{-N}$ and total
$s$-cost at most $\sum_{n\ge N}2^{-n}$. Thus $\dim_H E=0$.
The exponential map is locally Lipschitz and is a homeomorphism
onto $(0,\infty)$, so $e^E$ is a Hausdorff-zero full Borel carrier.
\Cref{thm:infinite-weighted-sums} gives both information dimensions.
Finally any finite-support vector may be extended past its support
by a geometric tail with arbitrarily small norm. This satisfies the
coefficient-count condition of \Cref{cor:infinite-sum-schwartz},
which supplies both Schwartz densities.
\end{proof}

\FloatBarrier
\chapter{Mellin transforms and endpoint asymptotics}
\label{sec:moments}

How much mass does the law of $X$ place near zero or near its largest
value $1/2$? The Mellin transform $M(s)=\mathbb E[X^s]$ records powers
of the same random product. In the representation
$X=\tfrac12e^{-W}$, positive real powers give
$(2X)^r=e^{-rW}$ and suppress values away from $1/2$, whereas reciprocal
powers $X^{-r}$ magnify small values of $X$. Independence turns their
expectations into the Euler product of \Cref{thm:moments}.

The high reciprocal moments in \Cref{prop:hadamard} and the
limiting-model comparisons in
\Cref{cor:small-x-tail,thm:exact-extreme-tail-transfer,prop:moment-ratio}
control small values of $X$. The finite endpoint and arithmetic anchors
in \Cref{prop:endpoint-half,sec:endpoint} instead describe short-window
mass, providing the coefficients used in sharp concentration and its
arithmetic applications.

For a fixed compact set $K$ and derivative order $j$,
\Cref{thm:prime-mellin} controls arithmetic totient moments and derivatives
with arbitrary logarithmic savings in its stated modulus ranges;
its constants may be ineffective. The large-moment and extreme-tail
limits concern the limiting laws. Applications to residue number
systems, with their sampling and population-size restrictions, are
given in \Cref{sec:rns-prime-bases}.

We use the summands $Z_\ell$ of \eqref{eq:local-log-jump}.

\section{Moments and Mellin transform}\label{sec:mellin}

For $s\in\C$ put
\[
\begin{aligned}
 F_s(\ell)&:=\frac{\ell-2}{\ell-1}
 +\frac{1}{\ell-1}\Bigl(\frac{\ell-1}{\ell}\Bigr)^{\!s},\qquad
 \ell\ge3\textup{ prime},\\
 M(s)&:=2^{-s}\!\prod_{\substack{\ell\ge3\\\ell\textup{ prime}}}F_s(\ell).
\end{aligned}
\]

\begin{theorem}[Mellin transform]\label{thm:moments}
\index{Mellin transforms!totient law}
The Euler product $M(s)$ converges absolutely and uniformly on compact
subsets of $\C$, $M$ is entire, and
\begin{equation}\label{eq:moment-product}
 \mathbb E[X^s]=M(s)\qquad(s\in\C).
\end{equation}
\end{theorem}

\begin{proof}
In the independent representation of \Cref{lem:factor},
$\log(1/X)=\log2+W$, where
$W_L:=\sum_{3\le\ell\le L}Z_\ell\nearrow W:=\sum_{\ell\ge3}Z_\ell$
almost surely and $\mathbb E[e^{-sZ_\ell}]=F_s(\ell)$.
Since $F_s(\ell)-1=(e^{-s\omega_\ell}-1)/(\ell-1)$, the bound
$|e^w-1|\le|w|e^{|w|}$ gives, for $|s|\le R$,
\[
 |F_s(\ell)-1|\le\frac{R\omega_\ell e^{R\omega_\ell}}{\ell-1}\ll_R\ell^{-2}.
\]
Hence the product converges absolutely and uniformly on compact
subsets of $\C$, and $M$ is entire, by the Weierstrass theorem for
products of analytic functions~\cite[Theorem~VII.5.9]{Conway78}.
For $t\ge0$, independence and monotone convergence give
$\mathbb E[e^{tW}]=\prod_{\ell\ge3}F_{-t}(\ell)=2^{-t}M(-t)<\infty$.
For $s\in\C$, $|e^{-sW_L}|\le e^{|\operatorname{Re}s|W}$, so dominated
convergence gives
$\mathbb E[X^s]=2^{-s}\mathbb E[e^{-sW}]
=2^{-s}\lim_{L\to\infty}\prod_{3\le\ell\le L}F_s(\ell)=M(s)$.
\end{proof}

\subsection{Arithmetic moments and fixed derivatives}
\label{ssec:mellin-arithmetic-moments}

All complex powers use real logarithms. For $Q\ge1$ and
$\gcd(b,Q)=1$, let $A_{Q,b}$ be as in \Cref{thm:ap-law} and put
\begin{equation}\label{eq:ap-mellin-product}
 M_{Q,b}(z):=A_{Q,b}^{z}
 \prod_{\substack{\ell\ge3\textup{ prime}\\\ell\nmid Q}}
 \frac{\ell-2+(1-1/\ell)^z}{\ell-1}.
\end{equation}
For the $\sigma$ law, put
\[
 S_k(\ell):=\frac{1-\ell^{-k-1}}{1-\ell^{-1}}\quad(k\ge0),\qquad
 a_\ell:=v_\ell(Q),\qquad c_\ell:=v_\ell(b-1),
\]
allowing $c_\ell=\infty$, and define
\begin{equation}\label{eq:ap-sigma-local}
 L_{\ell,Q,b}(z):=
 \begin{cases}
 \displaystyle\frac{\ell-2}{\ell-1}
       +\sum_{k\ge1}\ell^{-k}S_k(\ell)^z,&a_\ell=0,\\[5pt]
 S_{c_\ell}(\ell)^z,&c_\ell<a_\ell,\\[3pt]
 \displaystyle(\ell-1)\ell^{a_\ell-1}
       \sum_{k\ge a_\ell}\ell^{-k}S_k(\ell)^z,
       &c_\ell\ge a_\ell\ge1,
 \end{cases}
\end{equation}
\begin{equation}\label{eq:ap-sigma-mellin}
 N_{\sigma,Q,b}(z):=\prod_{\ell\textup{ prime}}L_{\ell,Q,b}(z).
\end{equation}
If $Q$ is a power of $2$, then $M_{Q,b}=M$, whereas $L_{2,Q,b}$ depends on $b$ when $4\mid Q$.

\begin{lemma}[Uniform Mellin perturbations]\label{lem:mellin-perturbation}
Put $U(p)=\varphi(p-1)/(p-1)$. Fix $C_0>0$ and let $r_x(p)>0$
satisfy $|\log r_x(p)|\le C_0/p$ for every $x\ge2$ and prime
$p\le x$. Put $V_x(p)=U(p)r_x(p)$. For every compact $K\subset\C$
and integer $j\ge0$, uniformly over these arrays and all reduced
classes $b\pmod Q$ with $n:=\pi(x;Q,b)>0$,
\[
 \sup_{z\in K}\left|
 \frac{d^j}{dz^j}\frac1n
 \sum_{\substack{p\le x\\p\equiv b\pmod Q}}
 \bigl(V_x(p)^z-U(p)^z\bigr)\right|
 \ll_{K,j,C_0}\frac{\sqrt x}{n}.
\]
The hypothesis holds with $C_0=4$ for
$r_x(p)=\prod_{k\le m_x(p)}(1-p^{-k})$, uniformly over
$m_x(p)\in\Z_{\ge1}\cup\{\infty\}$.
\end{lemma}

\begin{proof}
On a fixed compact neighborhood of $K$, the hypothesis gives
$r_x(p)^z-1=O_{K,C_0}(p^{-1})$. Choose $R$ larger than the absolute
values of the real parts there. The classical bound
$m/\varphi(m)\ll\log\log3m$ gives
\[
 |U(p)^z|\le U(p)^{-R}
 \ll_R(\log\log3p)^R\ll_Rp^{1/2}.
\]
Thus $|U(p)^z(r_x(p)^z-1)|\ll_{K,C_0}p^{-1/2}$.
Summing this bound over the progression, then over all integers at most $x$,
bounds the difference of transforms by $O_{K,C_0}(\sqrt x/n)$.
Cauchy's formula on that neighborhood proves the derivative estimate.
For every finite or infinite matrix product,
\[
 0\le-\log\prod_{k\le m}(1-p^{-k})
 \le\sum_{k\ge1}\frac{p^{-k}}{1-p^{-k}}
 \le\frac{p}{(p-1)^2}\le\frac4p.
\]
\end{proof}

\begin{theorem}[Mellin moments over the primes]\label{thm:prime-mellin}
\index{Mellin transforms!prime averages}
The products \eqref{eq:ap-mellin-product} and
\eqref{eq:ap-sigma-mellin} converge uniformly on compact subsets of
$\C$ and are entire. Fix a compact $K\subset\C$, $B>0$, and an integer
$j\ge0$. Let $U(p)$ be either $c(p)$ or $\varphi(p-1)/(p-1)$.
All estimates are uniform over reduced classes $b\pmod Q$.
\begin{enumerate}[label=\textup{(\alph*)},leftmargin=*]
\item For fixed $A>0$, uniformly for $Q\le(\log x)^A$,
\begin{equation}\label{eq:prime-mellin-ap}
 \sup_{z\in K}\left|
 \frac1{\pi(x;Q,b)}\sum_{\substack{p\le x\\p\equiv b\, (Q)}}
 U(p)^z(\log U(p))^j-M_{Q,b}^{(j)}(z)\right|
 \ll_{A,B,K,j}(\log x)^{-B},
\end{equation}
\begin{equation}\label{eq:prime-mellin-sigma}
 \sup_{z\in K}\left|
 \frac1{\pi(x;Q,b)}\sum_{\substack{p\le x\\p\equiv b\, (Q)}}
 \csig(p)^z(\log\csig(p))^j-N_{\sigma,Q,b}^{(j)}(z)\right|
 \ll_{A,B,K,j}(\log x)^{-B}.
\end{equation}
\item For every fixed integer $D_0\ge2$, \eqref{eq:prime-mellin-ap} holds
uniformly for
\[
 Q=D_0^a\le x^{1/3}\exp\{-(\log\log x)^3\},\qquad a\in\Z_{\ge0},
\]
with the implied constant depending on $D_0,B,K,j$.
\item For fixed $0<\delta<1/2$ and $E>0$,
\eqref{eq:prime-mellin-ap} holds for all but
$O_{\delta,E,B,K,j}(H(\log x)^{-E})$ integers
$Q\le H\le x^{1/2-\delta}$. This exceptional set may depend on
the displayed fixed parameters but is independent of $b$.
\end{enumerate}
In all three totient estimates one may also take
$U(p)=c_{n_x(p)}(p)$, uniformly over
$n_x(p)\in\Z_{\ge1}\cup\{\infty\}$, with the same exceptional set
in \textup{(c)}.
The implied constants depend on the corresponding fixed parameters;
they and the threshold in $x$ may be ineffective.
\end{theorem}

\begin{proof}
For part~\textup{(a)}, first take $U(p)=\varphi(p-1)/(p-1)$ and $j=0$. Let $a_z$ be the
multiplicative function supported on squarefree integers with
$a_z(\ell)=(1-1/\ell)^z-1$.
Then $(\varphi(n)/n)^z=\sum_{d\mid n}a_z(d)$ and
$|a_z(\ell)|\le C_K/\ell$ uniformly on $K$. Let $\beta$ be
multiplicative, supported on squarefree integers, with
$\beta(\ell)=C_K/\ell$. Thus $|a_z(d)|\le\beta(d)$ and
\begin{equation}\label{eq:mellin-rankin}
 \sum_{d\ge1}\frac{\beta(d)d^{1/2}}{\varphi(d)}
 =\prod_\ell\left(1+\frac{C_K}{\sqrt\ell(\ell-1)}\right)<\infty,
 \qquad
 \sum_{d>D}\frac{|a_z(d)|}{\varphi(d)}\ll_K D^{-1/2}.
\end{equation}
Since $\beta(d)\le C_K^{\omega(d)}/d$ and $C_K^{\omega(d)}\ll_Kd^{1/2}$
(only the finitely many primes $\ell<C_K^2$ have $C_K>\ell^{1/2}$), we
have $|a_z(d)|\le\beta(d)\ll_Kd^{-1/2}$.

Put $D=(\log x)^C$, where $C>2(A+B+1)$ is fixed. By the Chinese
remainder theorem, $p\equiv b\pmod Q$ and $p\equiv1\pmod d$ are
compatible precisely when $(d,Q)\mid b-1$, and then define a class $r_d$
modulo $[Q,d]$, reduced since $(b,Q)=1$. Set
\begin{equation}\label{eq:mellin-crt-density}
 w_{Q,b}(d):=\mathbf1_{(d,Q)\mid b-1}
       \frac{\varphi(Q)}{\varphi([Q,d])}
 \le\frac{\varphi(Q)}{\varphi(d)}.
\end{equation}
For squarefree $d$ and power-of-two $Q$ every $d$ is compatible, and
\begin{equation}\label{eq:mellin-pow2-density}
 w_{Q,b}(d)=\frac1{\varphi(d)}.
\end{equation}
The exact divisor expansion is
\[
 \frac1{\pi(x;Q,b)}\sum_{\substack{p\le x\\p\equiv b\, (Q)}}U(p)^z
 =\sum_{\substack{d\le x\\(d,Q)\mid b-1}}
 a_z(d)\frac{\pi(x;[Q,d],r_d)}{\pi(x;Q,b)}.
\]
For $d\le D$, \Cref{thm:sw} applies to the moduli $Q$ and
$[Q,d]\le QD\le(\log x)^{A+C}$. It gives
$\pi(x;Q,b)\gg_A x/(\varphi(Q)\log x)$ and, since $w_{Q,b}(d)\le1$,
\[
 \frac{\pi(x;[Q,d],r_d)}{\pi(x;Q,b)}-w_{Q,b}(d)
 \ll_{A,B}\varphi(Q)(\log x)e^{-c_{A+C}\sqrt{\log x}};
\]
as $\sum_{d\le D}|a_z(d)|\ll_KD$, these errors sum to less than any
fixed negative power of $\log x$. For $D<d\le x$, the trivial bound
$\pi(x;[Q,d],r_d)\le x/d+1$ bounds the contribution of this range by
\[
 \ll_A\varphi(Q)\log x\Bigl(\sum_{d>D}\frac{|a_z(d)|}{\varphi(d)}
 +\frac1x\sum_{d\le x}|a_z(d)|\Bigr)
 \ll_{A,K}\varphi(Q)\log x\,(D^{-1/2}+x^{-1/2}),
\]
by \eqref{eq:mellin-rankin} and $|a_z(d)|\ll_Kd^{-1/2}$. Completing the
main sum to all $d$ costs $O_K(\varphi(Q)D^{-1/2})$, by
\eqref{eq:mellin-rankin} and \eqref{eq:mellin-crt-density}. Both bounds
are $O_{A,B,K}((\log x)^{-B})$, since $\varphi(Q)\le(\log x)^A$ and
$C>2(A+B+1)$. The Euler factors of the completed main sum
$\sum_da_z(d)w_{Q,b}(d)$ at $\ell\mid Q$
are $1+a_z(\ell)$ if $b\equiv1\pmod\ell$, and $1$ otherwise; at
$\ell\nmid Q$ they are $1+a_z(\ell)/(\ell-1)$. The factor at $2$
is always $2^{-z}$. Their product is \eqref{eq:ap-mellin-product}.
This proves \eqref{eq:prime-mellin-ap} for the totient ratio.

For $c(p)=U(p)P(p)$ and the dimension arrays in the statement,
\Cref{lem:mellin-perturbation} and the same population lower bound give an error
$O_{A,K}(\varphi(Q)\log x/\sqrt x)$, smaller than every fixed negative
power of $\log x$.

For \eqref{eq:prime-mellin-sigma}, define multiplicatively
\[
 g_z(\ell^\nu)=S_\nu(\ell)^z-S_{\nu-1}(\ell)^z\quad(\nu\ge1),
 \qquad g_z(1)=1.
\]
Then $(\sigma(n)/n)^z=\sum_{d\mid n}g_z(d)$ and, since
$1\le S_\nu(\ell)\le2$, one has
$|g_z(\ell^\nu)|\le C_K\ell^{-\nu}$. The multiplicative majorant
$\beta_\sigma(\ell^\nu):=C_K\ell^{-\nu}$, i.e.\ $\beta_\sigma(d)=C_K^{\omega(d)}/d$,
satisfies $\beta_\sigma(d)\ll_Kd^{-1/2}$, as for $\beta$, and the analogue
of \eqref{eq:mellin-rankin}, since the local factors of
$\sum_d\beta_\sigma(d)d^{1/2}/\varphi(d)$ are $1+O_K(\ell^{-3/2})$.
The same divisor split at $D=(\log x)^C$ and
\eqref{eq:mellin-crt-density} give the required error with limit
$\sum_dg_z(d)w_{Q,b}(d)$. Write $a=a_\ell$. If $a=0$, the local sum is
\[
 1+\sum_{\nu\ge1}\frac{g_z(\ell^\nu)}{\varphi(\ell^\nu)}
 =\frac{\ell-2}{\ell-1}+\sum_{k\ge1}\ell^{-k}S_k(\ell)^z.
\]
If $c_\ell<a$, compatibility cuts the sum off at $\nu=c_\ell$ and
all its weights equal $1$, giving $S_{c_\ell}(\ell)^z$. If
$c_\ell\ge a\ge1$, its weights are $1$ for $\nu\le a$ and
$\ell^{a-\nu}$ for $\nu>a$, so summation by parts gives
\[
 S_a(\ell)^z+\sum_{\nu>a}\ell^{a-\nu}
       (S_\nu(\ell)^z-S_{\nu-1}(\ell)^z)
 =(\ell-1)\ell^{a-1}\sum_{k\ge a}\ell^{-k}S_k(\ell)^z.
\]
These are exactly \eqref{eq:ap-sigma-local}. Each local series
converges normally on compact sets, and for $\ell\nmid2Q$ both it and
the factor $1+a_z(\ell)/(\ell-1)$ of \eqref{eq:ap-mellin-product} are
$1+O_K(\ell^{-2})$. Hence both products converge uniformly on compact
sets and are entire.
All empirical transforms are entire. Apply Cauchy's formula on a fixed
positive-radius neighborhood of $K$ to obtain every fixed derivative
with the same rate.

For parts~\textup{(b),(c)}, put $L=\log x$ and use
\Cref{lem:ap-finite-head} with a fixed saving $M$ chosen below, in the
ranges of \Cref{cor:ap-finite-head-ranges}: part~\textup{(iii)}, with
$D_0$ as the base, for \textup{(b)}, and part~\textup{(i)}, with $E$ as
the exponent of the exceptional count, for \textup{(c)}. These results
are proved in \Cref{ssec:sharp-prime-convergence} from
\Cref{lem:sparse-bonferroni,lem:bv-weighted} and Elliott's theorem.
The logarithmic heads
$W_y(p),W_y$ retain every forced coordinate in
$a_{Q,b}=-\log A_{Q,b}$, their laws are within total variation
$O(L^{-M})$ by \eqref{eq:ap-truncated-tv}, and
$\log y\asymp L/\log L$.
Choose $R>1+\sup_{z\in K}|z|$. Uniformly in the permitted moduli,
\[
 e^{R a_{Q,b}}\ll_R(\log\log(3Q))^R,\qquad
 \max\{e^{R W_y(p)},e^{R W_y}\}\ll_R L^{2R}.
\]
The head-transform error is therefore $O_K(L^{2R-M})$.
For odd $p$, the empirical omitted logarithm is at most
$2L/(y\log y)$, giving a transform error
$O_K(L^{2R}L/(y\log y))$. In the independent model, the remainder
$T_y$ is independent of the head and satisfies
$\mathbb E(e^{RT_y}-1)\ll_R y^{-1}$, giving a transform error
$O_K(L^{2R}y^{-1})$.
The possible prime $2$ contributes $O_K(L^{2R}/n)$, where
$n=\pi(x;Q,b)\gg x/(\varphi(Q)L)$ by
\eqref{eq:ap-polylog-population}. Taking $M>2R+B+2$ proves the
claimed saving. The matrix-array error in
\Cref{lem:mellin-perturbation} is $O_K(\varphi(Q)L/\sqrt x)$,
again negligible in both ranges. Applying this argument on a fixed
neighborhood of $K$ and then Cauchy's formula gives every fixed
derivative. The exceptional set consequently depends on $B,K,j$ as
well as $\delta,E$.
\end{proof}

\AddToHookNext{build/column/before}{\raggedbottom}
Taking $j=0$ and $z=-1,-2$ in \eqref{eq:prime-mellin-ap} shows that, in
the totient ranges of \Cref{thm:prime-mellin}, the mean and variance of
$U(p)^{-1}$ over $p\le x$, $p\equiv b\pmod Q$, are $M_{Q,b}(-1)$ and
$M_{Q,b}(-2)-M_{Q,b}(-1)^2$ up to an arbitrary fixed inverse power
of $\log x$. For the variance, increase the saving in the first moment,
using $M_{Q,b}(-1)\ll Q/\varphi(Q)\ll\log\log(3Q)$ from
\eqref{eq:ap-mellin-product}.

For fixed even $m$, $Q=m$, $b=1$, $j=0$ and $z=k\in\mathbb N$ (so that
$A_{m,1}=\varphi(m)/m$), the limiting moment in the totient case of
\eqref{eq:prime-mellin-ap} agrees with the moment computation of
Deshouillers and Hassani~\cite[\S~3]{DH12}, obtained there for fixed $k\ge2$
from Bombieri--Vinogradov.
For $m=2$, where $M_{2,1}=M$, these moments were obtained earlier, with an
error term explicit in $k$, by Vaughan~\cite[Lemma~4.4, p.~72]{Vau73}:
$\sum_{2<p\le x}(\varphi(p-1)/(p-1))^k=M(k)\,x/\log x
+O(x(\log x)^{-2}e^{C(k\log k+k)})$ with an absolute constant $C$.
For each fixed $k$, \eqref{eq:prime-mellin-ap} gives the main term
$M(k)\pi(x)$ with error $O_{k,B}(\pi(x)(\log x)^{-B})$ for every fixed
$B>0$.
At $Q=1$, $z=1$ and $j=0$, \eqref{eq:prime-mellin-ap} gives, for every
$B>0$,
\[
\begin{aligned}
 \frac1{\pi(x)}\sum_{p\le x}c(p)
 &=A_{\mathrm{Art}}+O_B((\log x)^{-B}),\\
 A_{\mathrm{Art}}:=\prod_{\ell\textup{ prime}}
 \left(1-\frac1{\ell(\ell-1)}\right)
 &=M(1)=\mathbb E[X]\approx0.3739558136.
\end{aligned}
\]
The corresponding totient prime mean is classical: it is equivalent
to~\cite[Lemma~1]{Ste69}, which concerns $\sum_{p\le x}\varphi(p-1)/p$,
and is the case $k=1$ of~\cite[Lemma~4.4]{Vau73}; the numerical evaluation is due to
Wrench~\cite{Wre61}.

\begin{corollary}[Joint Mellin moments]\label{cor:joint-prime-mellin}
\index{Mellin transforms!joint laws}
For $Q\ge1$ and $(b,Q)=1$, put
\[
 T_k(\ell;z,w):=(1-1/\ell)^{z\mathbf1_{\{k>0\}}}S_k(\ell)^w
 \qquad(k\ge0),
\]
and define $\mathcal M_{Q,b}(z,w)$ by
\eqref{eq:ap-sigma-local}--\eqref{eq:ap-sigma-mellin}, replacing each
$S_k(\ell)^z$ by $T_k(\ell;z,w)$. This product converges uniformly on
compact subsets of $\C^2$, is entire, and equals
$\mathbb E[\Phi_{Q,b}^z\Sigma_{Q,b}^w]$, where $(\Phi_{Q,b},\Sigma_{Q,b})$
is the shifted-prime pair of \Cref{cor:joint-phi-sigma} conditioned on
$V_\ell=c_\ell$ if $c_\ell<a_\ell$, and on $V_\ell\ge a_\ell$
otherwise, for every $\ell\mid Q$.

Fix a compact $K\subset\C^2$, $A,B>0$ and integers $j,k\ge0$.
For $U(p)=\varphi(p-1)/(p-1)$ or $c(p)$, uniformly for
$Q\le(\log x)^A$ and $(b,Q)=1$,
\begin{equation}\label{eq:joint-prime-mellin}
\begin{split}
 \sup_{(z,w)\in K}\Bigg|
 &\frac1{\pi(x;Q,b)}\sum_{\substack{p\le x\\p\equiv b\pmod Q}}
 U(p)^z\csig(p)^w(\log U(p))^j(\log\csig(p))^k\\
 &\hspace{12mm}-\partial_z^j\partial_w^k\mathcal M_{Q,b}(z,w)
 \Bigg|\ll_{A,B,K,j,k}(\log x)^{-B}.
\end{split}
\end{equation}
The constants and threshold may be ineffective. In particular, for
the unrestricted shifted-prime pair,
\begin{equation}\label{eq:joint-phi-sigma-mean}
 \mathbb E[\Phi\Sigma]
 =\frac56\prod_{\ell\ge3}
       \left(1-\frac1{\ell(\ell^2-1)}\right).
\end{equation}
\end{corollary}

\begin{proof}
Define $h_{z,w}$ multiplicatively by $h_{z,w}(1)=1$ and
\[
 h_{z,w}(\ell^\nu)
 =T_\nu(\ell;z,w)-T_{\nu-1}(\ell;z,w)\qquad(\nu\ge1).
\]
Then $(\varphi(n)/n)^z(\sigma(n)/n)^w=\sum_{d\mid n}h_{z,w}(d)$.
Uniformly on compact subsets of $\C^2$, the first local difference is
$(w-z)/\ell+O_K(\ell^{-2})$, and all subsequent differences are
$O_K(\ell^{-\nu})$, since the totient factor is constant for
$\nu\ge1$. Thus the majorant and both ranges in the divisor-sum
proof of \Cref{thm:prime-mellin} apply unchanged. The completed sum
$\sum_dh_{z,w}(d)w_{Q,b}(d)$ has exactly the stated local factors.
They are entire and, outside $2Q$, equal $1+O_K(\ell^{-2})$.
The conditional valuation model identifies their product with the
joint expectation: $1\le\Sigma_{Q,b}\le\Phi_{Q,b}^{-1}$ and the
finite reciprocal moments of \Cref{cor:ap-moments} justify dominated
convergence on every compact set.

For $U(p)=c(p)$, use
$\csig(p)\le(p-1)/\varphi(p-1)\ll\log\log(3p)$ and
$P(p)^z-1=O_K(p^{-1})$. As in \Cref{lem:mellin-perturbation}, the
normalized perturbation is $O_K(\sqrt x/\pi(x;Q,b))$, which is smaller
than every fixed negative power of $\log x$ in the stated range.
Cauchy's formula in both variables gives the mixed derivatives.
Finally $T_k(\ell;1,1)=1-\ell^{-k-1}$ for $k\ge1$, so the
unconditioned local mean is $1-1/(\ell(\ell^2-1))$; at $\ell=2$
it equals $5/6$. This proves \eqref{eq:joint-phi-sigma-mean}.
\end{proof}

\begin{corollary}[Normalized divisor-sum distributions in fixed progressions]\label{cor:ap-sigma-law}
\index{Arithmetic progressions!divisor-sum laws}
Fix $Q\ge1$ and $\gcd(b,Q)=1$, and put $m=[2,\gcd(b-1,Q)]$.
The distribution functions of $\csig(p)$ over primes $p\equiv b\pmod Q$
converge uniformly to that of a law with Mellin transform
$N_{\sigma,Q,b}$ and support $[\sigma(m)/m,\infty)$.
This law and its image under $\log$ are singular continuous with
Hausdorff dimension zero.
\end{corollary}

\begin{proof}
Let $(V_\ell)$ be the independent shifted-prime valuations of
\eqref{eq:valuation-laws}. Condition on the
positive-probability cylinder
\[
 D:=\bigcap_{\ell\mid Q}C_\ell,\qquad
 C_\ell:=
 \begin{cases}
 \{V_\ell=c_\ell\},&c_\ell<a_\ell,\\
 \{V_\ell\ge a_\ell\},&c_\ell\ge a_\ell.
 \end{cases}
\]
The conditional coordinates remain independent, and their factors
$S_{V_\ell}(\ell)$ have Mellin transforms given by
\eqref{eq:ap-sigma-local}, using $\mathbb P(V_\ell=k)=\ell^{-k}$ for $k\ge1$ before conditioning.
The product $Y_{Q,b}:=\prod_\ell S_{V_\ell}(\ell)$ under this
conditioning has Mellin transform $N_{\sigma,Q,b}$ on $\C$: the proof of
\Cref{thm:moments} applies to $\log Y_{Q,b}$, since
$L_{\ell,Q,b}(z)=1+O_K(\ell^{-2})$ uniformly on compact $K\subset\C$ for
$\ell\nmid2Q$ (proof of \Cref{thm:prime-mellin}). Equation~\eqref{eq:prime-mellin-sigma} with
$j=0$ and $z=it$, followed by L\'evy's continuity theorem and
exponentiation, identifies the arithmetic limit.

Domination by $\mathbb P(D)^{-1}$ times the unrestricted laws
transfers atomlessness from \Cref{thm:dist-sigma}. The dimension-zero
conclusion for the conditional product and its logarithmic image
follows from \Cref{cor:ap-liouville}. The product of the factors at primes dividing $2Q$
has a countably atomic law whose least value $\sigma(m)/m$ is an atom.
By \Cref{lem:supp_conv}, the sum of the logarithms of the remaining
factors has support $[0,\infty)$: its summands are nonnegative with atoms
at $0$, and the increments $\log(1+1/\ell)$ tend to zero with divergent
sum. Hence the support is $[\sigma(m)/m,\infty)$, and weak convergence to the
continuous limit distribution function is uniform.
\end{proof}

\begin{corollary}[Reciprocal moments and tail bounds]\label{cor:avg-loss}
\index{Reciprocal moments}
\[
\begin{aligned}
 \mathbb E[X^{-1}]&=M(-1)
 =2\prod_{\substack{\ell\ge3\\\ell\textup{ prime}}}
 \left(1+\frac1{(\ell-1)^2}\right),\\
 \mathbb E[\log(1/X)]&=-M'(0)
 =\log2+\sum_{\substack{\ell\ge3\\\ell\textup{ prime}}}
 \frac1{\ell-1}\log\frac\ell{\ell-1}.
\end{aligned}
\]
For all $\varepsilon,\sigma>0$,
\[
 \mathbb P(X<\varepsilon)\le\varepsilon^\sigma M(-\sigma).
\]
For fixed $\sigma,B>0$, uniformly for $T>0$ as $x\to\infty$,
\[
 \frac{\#\{p\le x:1/c(p)>T\}}{\pi(x)}
 \le T^{-\sigma}\left(M(-\sigma)
       +O_{\sigma,B}((\log x)^{-B})\right).
\]
\end{corollary}

\begin{proof}
\Cref{thm:moments} gives every negative moment and the displayed product.
Differentiation at $s=0$ gives the logarithmic mean. Markov's inequality
applied to $X^{-\sigma}$ gives the limiting-law bound; its empirical
version follows from \Cref{thm:prime-mellin} at $Q=1$, $z=-\sigma$ and
$j=0$.
\end{proof}

Truncating the products and series at primes $\ell\le10^7$ gives
\[
 M(-1)\approx2.82642,\qquad
 M(-2)-M(-1)^2\approx0.45385,\qquad
 -M'(0)\approx1.01106.
\]

For comparison, let $X_\sigma$ have the full shifted-prime divisor-sum
law $\muH$. Its mean is
\begin{equation}\label{eq:sigma-mean-mas17}
 \mathbb E X_\sigma
 =\frac53\prod_{\ell\ge3}
       \left(1+\frac{\ell}{(\ell-1)^2(\ell+1)}\right).
\end{equation}
Here the prime-$2$ local mean is
$\sum_{k\ge1}2^{-k}(2-2^{-k})=5/3$. At an odd prime,
\[
 \mathbb E\left[\frac{\sigma(\ell^{V_\ell})}{\ell^{V_\ell}}\right]-1
 =\sum_{k\ge1}\ell^{-k}\sum_{j=1}^k\ell^{-j}
 =\frac{\ell}{(\ell-1)^2(\ell+1)}.
\]
Independence and monotone convergence give \eqref{eq:sigma-mean-mas17};
the product converges because its deviations from $1$ are summable.
The arithmetic mean converges to the same value by
\Cref{thm:prime-mellin} at $Q=1$ and exponent $1$.

\begin{corollary}[Mellin transform in a progression]\label{cor:ap-moments}
Let $Q\ge1$ and $\gcd(b,Q)=1$, with the notation of \Cref{thm:ap-law}.
Then
\begin{equation}\label{eq:ap-mellin-identity}
 \mathbb E[X_{Q,b}^{z}]=M_{Q,b}(z)\qquad(z\in\C).
\end{equation}
In particular,
\begin{equation}\label{eq:ap-mean}
 \mathbb E[X_{Q,b}^{-1}]
 =M(-1)\prod_{\substack{\ell\mid Q\\\ell\textup{ odd}}}
 \frac{(\ell/(\ell-1))^{e_\ell(b)}}{1+(\ell-1)^{-2}}.
\end{equation}
If $Q$ is a power of two, then $X_{Q,b}\overset d=X$ for every reduced
class $b$.
\end{corollary}

\begin{proof}
In \eqref{eq:XQ}, write
\[
 X_{Q,b}=A_{Q,b}\exp\left(-\sum_{\substack{\ell\ge3\\\ell\nmid Q}}
 Z_\ell\right).
\]
In the common Bernoulli realization, the nonnegative sum is at most
$\log(1/X)-\log2$, which has every positive exponential moment by
\Cref{thm:moments}. Finite independence and dominated convergence
therefore give \eqref{eq:ap-mellin-identity} for every $z\in\C$.
At $z=-1$, each odd prime $\ell\nmid Q$ contributes
\[
 \mathbb E[(\ell/(\ell-1))^{B_\ell}]
 =\frac{\ell-2}{\ell-1}+\frac\ell{(\ell-1)^2}
 =1+\frac1{(\ell-1)^2}.
\]
Inserting the deterministic factors at primes dividing $Q$ gives
\eqref{eq:ap-mean}. A power-of-two modulus imposes no condition on the
odd-prime Bernoulli variables, which proves the distribution identity.
\end{proof}

For $b=1$ and odd $\ell\mid Q$, the
multiplier in \eqref{eq:ap-mean} is
$\ell(\ell-1)/((\ell-1)^2+1)$.

\subsection{Growth and reciprocal tails}
\label{ssec:mellin-growth-tails}

\begin{proposition}[Order, maximum modulus and zeros of $M$]
\label{prop:hadamard}
$M$ is entire of order exactly~$1$ and infinite type. For every $r\ge0$,
\begin{equation}\label{eq:mellin-max-modulus}
 \max_{|s|=r}|M(s)|=M(-r),
\end{equation}
and, as $r\to\infty$,
\begin{equation}\label{eq:high-reciprocal-moments}
 \log M(-r)=r(\log\log r+\gamma)
       +O\!\left(\frac{r\log\log r}{\log r}\right).
\end{equation}
All its zeros are simple and are precisely the points
\[
   s_{\ell,k}
   =\frac{\log(\ell-2)+(2k+1)\pi i}{\log((\ell-1)/\ell)},
   \qquad \ell\ge3\textup{ prime},\quad k\in\mathbb Z.
\]
\end{proposition}

\begin{proof}
Since $0<X\le1/2$ almost surely, $|M(s)|\le M(\operatorname{Re}s)
\le M(-r)$ for $|s|=r$; equality holds at $s=-r$.
Put $m(v):=-\sum_{\ell\le v}\log(1-1/\ell)$, including the prime $2$.
By Mertens' theorem in the form \cite[Theorem~2.7(e)]{MV06},
$m(v)=\log\log v+\gamma+O(1/\log v)$.
For $r\ge3$, $F_{-r}(\ell)\le e^{r\omega_\ell}$ when $\ell\le r$,
whereas $\log F_{-r}(\ell)\ll r/\ell^2$ when $\ell>r$. Consequently
\[
 \log M(-r)\le r m(r)+O(1)
 =r(\log\log r+\gamma)+O(r/\log r).
\]
For the lower bound take $y=r/(\log r)^2$. Retain the term
$e^{r\omega_\ell}/(\ell-1)$ in each factor with $3\le\ell\le y$ and use
$F_{-r}(\ell)\ge1$ for the rest:
\[
 \log M(-r)\ge r m(y)-\sum_{3\le\ell\le y}\log(\ell-1)
 \ge r m(y)-y\log y.
\]
Since $y\log y\le r/\log r$ and
$\log\log y=\log\log r+O(\log\log r/\log r)$, this gives
\eqref{eq:high-reciprocal-moments}. Together with
\eqref{eq:mellin-max-modulus}, this proves order $1$ and infinite type.

The zeros of $F_s(\ell)$ are the displayed $s_{\ell,k}$; they are simple,
since there
$\frac{d}{ds}F_s(\ell)=-\frac{\ell-2}{\ell-1}\log\frac{\ell-1}{\ell}\ne0$.
As $\log(\ell-2)\ge0$ strictly increases and $\log(\ell/(\ell-1))>0$
strictly decreases with $\ell$,
$\operatorname{Re}s_{\ell,k}=-\log(\ell-2)/\log(\ell/(\ell-1))$ strictly
decreases with $\ell$; hence distinct primes have disjoint zero sets.

Since $\sum_\ell|F_s(\ell)-1|$ converges uniformly on compact subsets of
$\C$ (proof of \Cref{thm:moments}) and $2^{-s}$ never vanishes,
\cite[Theorem~VII.5.9]{Conway78} shows that the zeros of $M$ are those of
the factors $F_s(\ell)$, with multiplicities added. As the local zero
sets are disjoint and consist of simple zeros, the zeros of $M$ are
exactly the $s_{\ell,k}$, and they are simple.
\end{proof}

The zero description also gives the canonical factorization
\begin{equation}\label{eq:hadamard-product-mas17}
 M(s)=\exp\!\left\{-s\left(\log2+
                \sum_{\ell\ge3}\frac{\omega_\ell}{\ell-1}\right)\right\}
 \prod_{\substack{\ell\ge3\textup{ prime}\\k\in\Z}}
 \left(1-\frac{s}{s_{\ell,k}}\right)e^{s/s_{\ell,k}}.
\end{equation}
This is a genus-one product, convergent locally uniformly in $s$. Indeed,
$\sum_{k\in\Z}|s_{\ell,k}|^{-2}\ll\omega_\ell^2$, so the sum over
all zeros converges. Hadamard factorization for an entire function
of order one supplies an exponential of degree at most one.
The normalization $M(0)=1$ and
$M'(0)=-\log2-\sum_{\ell\ge3}\omega_\ell/(\ell-1)$ determine it,
since every regularized zero factor has derivative zero at the origin.

\begin{proposition}[Zero count and directional growth]
\label{prop:zero-growth}
Let $n_M(r)$ count the zeros of $M$ in $|s|\le r$, with multiplicity.
Then
\begin{equation}\label{eq:zero-count}
 n_M(r)=\frac r\pi(\log\log r+\gamma-\log2)
       +O\left(\frac{r\log\log r}{\log r}\right).
\end{equation}
For each fixed real $\theta$,
\begin{equation}\label{eq:directional-growth}
 \limsup_{r\to\infty}
 \frac{\log|M(re^{i\theta})|}{r\log\log r}
 =\max\{0,-\cos\theta\}.
\end{equation}
More precisely, the indicator
$h_M(\theta)=\limsup_{r\to\infty}r^{-1}\log|M(re^{i\theta})|$
equals $-\log2\cos\theta$ if $\cos\theta>0$, equals $0$ if
$\cos\theta=0$, and equals $+\infty$ if $\cos\theta<0$.
On the imaginary axis,
\begin{equation}\label{eq:imaginary-lower}
 |M(it)|\ge
 \frac{c\,|\cos(t\omega_3/2)|}
 {\bigl(\log(t+3)\bigr)^{5/3}
  \bigl(\log\log(t+e^e)\bigr)^{1/3}}
 \qquad(t\ge0)
\end{equation}
for an absolute $c>0$.
\end{proposition}

\begin{proof}
The zero set in \Cref{prop:hadamard} shows that the contribution of
prime $\ell$ is
\[
 \frac1\pi\sqrt{(r\omega_\ell)^2-\log^2(\ell-2)}_++O(1)
\]
when $\log(\ell-2)\le r\omega_\ell$, and is zero otherwise.
For large $r$ the contributing primes are at most $r$.
Thus the upper bound is
$\frac r\pi\sum_{3\le\ell\le r}\omega_\ell+O(\pi(r))$.
For the lower bound retain $\ell\le y=r/(\log r)^2$.
All these primes contribute for large $r$. Using
$\sqrt{u^2-v^2}\ge u-v^2/u$ for $0\le v\le u$, the total loss is
\[
 O\left(\frac1r\sum_{\ell\le y}\ell(\log\ell)^2+\pi(y)\right)
 =O(r/(\log r)^3).
\]
Mertens' product estimate gives
$\sum_{3\le\ell\le v}\omega_\ell
=\log\log v+\gamma-\log2+O(1/\log v)$.
Since $\log\log y=\log\log r+O(\log\log r/\log r)$,
the two bounds prove~\eqref{eq:zero-count}.

Write
$F_s(\ell)=(\ell-2+e^{-s\omega_\ell})/(\ell-1)$, so that
$M(s)=2^{-s}\prod_{\ell\ge3}F_s(\ell)$.
If $\cos\theta>0$, the factors with $5\le\ell\le r$ have modulus
between $(\ell-3)/(\ell-1)$ and $1$. Their logarithmic sum is
$O(\log\log r)$, and the factor at $3$ is bounded away from zero for
large $r$. For $\ell>r$,
$F_{re^{i\theta}}(\ell)=1+O(r/\ell^2)$, whose total logarithmic
contribution is $O(1/\log r)$. Hence
\[
 \log|M(re^{i\theta})|
 =-r\log2\cos\theta+O_\theta(\log\log r).
\]
On the imaginary axis the factor at $3$ has modulus
$|\cos(t\omega_3/2)|$. The same head lower bound and tail estimate give
$|M(it)|\gg|\cos(t\omega_3/2)|(\log(t+3))^{-2}$.
The stronger estimate~\eqref{eq:imaginary-lower} follows from
\Cref{prop:dissipation-upper-mas11} below, since
$|M(it)|=|\FT(t)|$. Enlarging the denominator as displayed and
decreasing $c$ extends that estimate to the compact remaining range:
the product with the factor at $3$ removed is continuous and never
zero there. The upper bound is one.
Taking $t=2\pi j/\omega_3$ proves both limsup assertions there.

It remains to prove a lower bound in the left half-plane. Put
$a=-\cos\theta>0$. The upper bound
$|M(re^{i\theta})|\le M(-ar)$ and \Cref{prop:hadamard} give
\[
 \log|M(re^{i\theta})|\le ar\log\log r+O_\theta(r).
\]
For every sufficiently large $R$, choose $r\in[R,2R]$ at distance at
least one from every number
\[
 r_\ell=\frac{\log(\ell-2)}{a\omega_\ell}\qquad(\ell\ge5).
\]
This is possible: centers whose unit neighborhoods meet $[R,2R]$
have $\ell\ll_\theta R/\log R$, and their number is
$O_\theta(R/(\log R)^2)=o(R)$.
For $\ell\le y=r/(\log r)^2$, the exponential term dominates
$\ell-2$, uniformly as $r\to\infty$. Including the deterministic
factor at $2$, these factors contribute at least
\[
 ar\sum_{\ell\le y}\omega_\ell
   -\sum_{3\le\ell\le y}\log(2(\ell-1))
 =ar\log\log r+O_\theta(r).
\]
Fix a sufficiently large $C=C_\theta$. For $y<\ell\le Cr$, reverse
triangle inequality and $|r-r_\ell|\ge1$ imply
\[
 |F_{re^{i\theta}}(\ell)|
 \ge\frac{\ell-2}{\ell-1}
          |e^{a\omega_\ell(r-r_\ell)}-1|
 \ge c_\theta/\ell.
\]
Their total logarithmic loss is $O_\theta(r)$ by the prime number
theorem. Finally the factors with $\ell>Cr$ have logarithmic loss
$O_\theta(r\sum_{\ell>Cr}\ell^{-2})=O_\theta(1/\log r)$.
Along the chosen radii this gives
$\log|M(re^{i\theta})|\ge ar\log\log r-O_\theta(r)$, proving
both remaining limsup statements.
\end{proof}

Since $\operatorname{Re}s_{3,k}=0$ and $\operatorname{Re}s_{\ell,k}<0$ for
$\ell\ge5$, $M$ is zero-free on $\{\operatorname{Re}s>0\}$, and its zeros
on $i\R$ are exactly the $s_{3,k}$.

\begin{corollary}[The small-$X$ tail]\label{cor:small-x-tail}
\index{Tail estimates!small totient ratios}
The limiting distribution satisfies
\begin{equation}\label{eq:small-x-tail}
 \lim_{\varepsilon\downarrow0}
 \varepsilon\log\log\frac1{G(\varepsilon)}=e^{-\gamma}.
\end{equation}
\end{corollary}

\begin{proof}
Put $Z=1/X$ and fix $0<\delta<1$. For
$r=\exp((1-\delta)e^{-\gamma}t)$, Markov's inequality and
\eqref{eq:high-reciprocal-moments} give
\[
 \mathbb P(Z>t)\le t^{-r}M(-r)
 =\exp\{r(\log(1-\delta)+o(1))\},
\]
so $\log\log(1/\mathbb P(Z>t))\ge(1-\delta)e^{-\gamma}t+O_\delta(1)$.
Conversely, let $y=\exp((1+\delta)e^{-\gamma}t)$. The event that
$B_\ell=1$ for every odd prime $\ell\le y$ has probability
$\prod_{3\le\ell\le y}(\ell-1)^{-1}\ge e^{-Cy}$ for an absolute $C>0$,
and on it Mertens' theorem gives
$Z\ge\prod_{\ell\le y}(1-1/\ell)^{-1}=(1+\delta)t(1+o(1))>t$ for large
$t$. Hence $\log\log(1/\mathbb P(Z>t))\le(1+\delta)e^{-\gamma}t+O(1)$.
Letting $\delta\downarrow0$ gives
$t^{-1}\log\log(1/\mathbb P(Z>t))\to e^{-\gamma}$. Since $G$ is
continuous, $G(\varepsilon)=\mathbb P(Z>1/\varepsilon)$, and
\eqref{eq:small-x-tail} follows.
\end{proof}

A fixed-order refinement of \eqref{eq:small-x-tail} is given in
\eqref{eq:small-x-full-expansion-mas20}, after
\Cref{thm:exact-extreme-tail-transfer}.

\subsection{Comparison on the valuation space}
\label{ssec:mellin-valuation-comparison}

On the valuation space of \eqref{eq:valuation-laws}, let
\[
 R_\varphi(v)=\prod_{v_\ell\ge1}(1-1/\ell)^{-1},\qquad
 R_\sigma(v)=\prod_\ell S_{v_\ell}(\ell),\qquad
 U_R(t):=\mathsf I(R>t).
\]

\begin{theorem}[Quantitative comparison of integer and shifted-prime extreme tails]
\label{thm:exact-extreme-tail-transfer}
\index{Tail estimates!transfer between valuation models}
For either $R=R_\varphi$ or $R=R_\sigma$ and every fixed $N>0$,
\begin{equation}\label{eq:exact-extreme-tail-transfer}
 \mathsf B(R>t)=tU_R(t)\{1+O_N(t^{-N})\}\qquad(t\to\infty),
\end{equation}
\begin{equation}\label{eq:extreme-conditional-tv}
 d_{\mathrm{TV}}\bigl(\mathsf B(\,\cdot\mid R>t),
                       \mathsf I(\,\cdot\mid R>t)\bigr)=O_N(t^{-N}),
\end{equation}
where total variation is taken on the entire valuation space.
In particular,
\[
 G(1/t)=tU_{R_\varphi}(t)\{1+O_N(t^{-N})\},\qquad
 \muH((t,\infty))=tU_{R_\sigma}(t)\{1+O_N(t^{-N})\},
\]
and $\log\mathsf B(R>t)=\log U_R(t)+\log t+O_N(t^{-N})$.
For the corresponding moments,
\begin{equation}\label{eq:integer-shifted-high-moment}
 M(-r)\sim\mathbb E_{\mathsf I}R_\varphi^{r+1},\qquad
 \mathbb E_{\mathsf B}R_\sigma^r
 \sim\mathbb E_{\mathsf I}R_\sigma^{r+1}
 \qquad(r\to\infty).
\end{equation}
\end{theorem}

\begin{proof}
We use the exact likelihood relative to
$\mathsf A=\mathsf I(\,\cdot\mid J_2\ge1)$ from
\Cref{lem:valuation-likelihood}. We compare the conditional laws through
$I_t\to Q_t\to P_t\to B_t$: the integer law conditioned on
$R>t$, its even-conditioned version, the $R$-size-biased law on
$\{R>t\}$ built from $\mathsf A$, and the shifted-prime law
conditioned on $R>t$. The factor $2$ from even
conditioning combines with $\mathbb E_{P_t}D=1/2+O_N(t^{-N})$,
where $D=L_{\mathrm{val}}/R$, in the tail comparison. Rapid variation and likelihood
saturation control the three steps on the entire valuation space.

We use the full fixed-order expansion of Weingartner~\cite[Theorem~1]{Wei10}:
for $c=e^{-\gamma}$ and each fixed integer $m\ge2$,
\begin{equation}\label{eq:integer-extreme-leading}
 -\log U_R(t)=e^{ct}\left(1+\sum_{j=2}^m \alpha_j^{\mathrm{tail}}t^{-j}
                              +O_m(t^{-m-1})\right).
\end{equation}
The integer limiting distributions are continuous, so the strict tail
agrees with the convention in that theorem. Fix $A>0$ and take $m>A+1$.
The derivative of the displayed main term is $e^{ct}(c+O_m(t^{-2}))$;
subtracting its values at $t+h$ and $t$, and bounding the two remainders,
gives, for some $c_A>0$,
\begin{equation}\label{eq:extreme-tail-increment}
 \frac{U_R(t+h)}{U_R(t)}\le\exp(-c_Ae^{ct}h)
 \qquad(t^{-A}\le h\le1).
\end{equation}
For every fixed $a>1$ the same expansion gives
$U_R(at)/U_R(t)\le\exp(-e^{cat}/4)$ for sufficiently large $t$.
Moreover,
\begin{equation}\label{eq:extreme-overshoot}
 \frac1{U_R(t)}\int_t^\infty U_R(u)\,du=O_A(t^{-A}).
\end{equation}
Indeed, the intervals $[t,t+t^{-A}]$ and $[t+t^{-A},t+1]$
contribute at most $t^{-A}$ and $\exp(-c_Ae^{ct}t^{-A})$.
For the remaining interval, \eqref{eq:integer-extreme-leading} gives
$U_R(u)\le e^{-g(u)}$, where $g(u)=e^{cu}(1-Cu^{-2})$ and
$g'(u)\ge(c/2)e^{cu}$ for large $u$. Hence its contribution is at most
$2e^{-g(t+1)}/(ce^{c(t+1)}U_R(t))=O(\exp(-c'e^{ct}))$ for some $c'>0$.
This also proves finiteness of every positive moment.

Put $F(t)=\mathsf A(R>t)$ and
$\delta_t=\mathsf I(R>t,v_2=0)/U_R(t)$. Since
$\mathsf A=\mathsf I(\,\cdot\mid v_2\ge1)$,
\[
 F(t)=2U_R(t)(1-\delta_t),\qquad
 \delta_t\le
 \begin{cases}
 U_R(2t)/U_R(t),&R=R_\varphi,\\
 2U_R(3t/2)/U_R(t),&R=R_\sigma.
 \end{cases}
\]
For the first bound, expose the odd-prime product and compare the
multipliers $1$ and $2$ at $2$. For the second, compare $v_2=0,1$,
whose integer probabilities are $1/2,1/4$ and multipliers $1,3/2$.
Thus $\delta_t=O_N(t^{-N})$ for every fixed $N$. By
\eqref{eq:extreme-overshoot} and $F(u)\le2U_R(u)$,
\begin{equation}\label{eq:extreme-rapid-variation}
 H_R(t):=\mathbb E_{\mathsf A}[R\mathbf1_{R>t}]
       =tF(t)\{1+O_N(t^{-N})\}.
\end{equation}
In particular $H_R(s)/H_R(t)\ll U_R(s)/U_R(t)$ for $t\le s\le2t$.

Let $P_t$ have density $R\mathbf1_{R>t}/H_R(t)$ relative to $\mathsf A$,
and put $D=L_{\mathrm{val}}/R$. The exact likelihood
\eqref{eq:valuation-likelihood-exact}, together with
$1\le R_\varphi/R_\sigma\le\prod_\ell(1-\ell^{-2})^{-1}=\zeta(2)$,
gives fixed positive upper and lower bounds for $D$ in both cases. We prove
\begin{equation}\label{eq:extreme-likelihood-saturation}
 \mathbb E_{P_t}|D-1/2|=O_N(t^{-N})\qquad(N>0).
\end{equation}
For $R=R_\varphi$, independence at an odd prime $\ell$ gives, with
$a_\ell=\ell/(\ell-1)$,
\[
 P_t(v_\ell=0)\le\frac{(\ell-1)^2}{\ell}
                       \frac{H_R(a_\ell t)}{H_R(t)}.
\]
For $\ell\le t^K$ the threshold increment is at least $t^{1-K}$.
Monotonicity and \eqref{eq:extreme-tail-increment}, with a sufficiently
large fixed $A$, show that the probability of any inactive odd prime
up to $t^K$ is $O_{K,N}(t^{-N})$ for every fixed $K,N>0$.
On the complementary event, the exact likelihood yields
\[
 2D=\prod_{\substack{\ell>2\\v_\ell=0}}
          (1-(\ell-1)^{-2})=1+O(t^{-K}).
\]

For $R=R_\sigma$, fix $K,L>0$ and put
$h_\ell=\lceil L\log t/\log\ell\rceil$ for $\ell\le t^K$.
Writing $R=S_{v_\ell}(\ell)W$ with $W$ independent of $v_\ell$, comparison
with the state $v_\ell=h_\ell$ gives
\[
 P_t(v_\ell<h_\ell)\le
 \frac{H_R(a t)}{a\mathsf A(v_\ell=h_\ell)H_R(t)},\qquad
 a=\frac{S_{h_\ell}(\ell)}{S_{h_\ell-1}(\ell)}.
\]
Here $a-1\ge\tfrac12t^{-K-L}$ and
$\mathsf A(v_\ell=h_\ell)\gg t^{-K-L}$, also at $2$ for large $t$.
The union of these events therefore has probability
$O_{K,L,N}(t^{-N})$ by \eqref{eq:extreme-tail-increment}.
On its complement,
\[
 2D=\prod_{\substack{\ell>2\\v_\ell=0}}(1-(\ell-1)^{-2})
       \prod_{v_\ell\ge1}(1-\ell^{-(v_\ell+1)})^{-1},
 \qquad |D-1/2|\ll t^{-K}+t^{-L}\log\log t.
\]
Indeed, above $t^K$ the logarithmic errors sum to $O(t^{-K})$, and
below it $\ell^{-(v_\ell+1)}\le t^{-L}/\ell$. Taking $K,L>N+2$ and
using boundedness on the exceptional events proves
\eqref{eq:extreme-likelihood-saturation}.

Set $m_t=\mathbb E_{P_t}D=1/2+O_N(t^{-N})$.
The likelihood identity gives
\[
 \frac{\mathsf B(R>t)}{tU_R(t)}
 =(1-\delta_t)\frac{H_R(t)}{tF(t)}\,2m_t=1+O_N(t^{-N}).
\]
Continuity of $G$ identifies $\mathsf B(R_\varphi>t)=G(1/t)$.
To prove \eqref{eq:extreme-conditional-tv}, put
$Q_t=\mathsf A(\,\cdot\mid R>t)$,
$B_t=\mathsf B(\,\cdot\mid R>t)$ and
$I_t=\mathsf I(\,\cdot\mid R>t)$. Then
\[
 d_{\mathrm{TV}}(P_t,Q_t)\le1-\frac{tF(t)}{H_R(t)},\qquad
 d_{\mathrm{TV}}(B_t,P_t)\le m_t^{-1}\mathbb E_{P_t}|D-1/2|,
 \qquad d_{\mathrm{TV}}(I_t,Q_t)=\delta_t.
\]
The first inequality follows by subtracting the common submeasure
$(tF(t)/H_R(t))Q_t$; the second follows from the density $D/m_t$;
the last follows from the disjoint even and odd restrictions of $I_t$.
The triangle inequality completes the proof of the two tail estimates.

For the moment comparison, at each odd prime write $a=\ell/(\ell-1)$. The Euler factor
of $\mathbb E_{\mathsf B}R_\varphi^r=M(-r)$ is
$(\ell-2)/(\ell-1)+a^r/(\ell-1)$, whereas that of
$\mathbb E_{\mathsf I}R_\varphi^{r+1}$ is
$(\ell-1)/\ell+a^r/(\ell-1)$. Their difference is
$-1/(\ell(\ell-1))$. For $r\ge0$ their ratio has logarithm
$O(\ell^{-2})$ uniformly, and tends to one as $r\to\infty$
for each fixed prime. Dominated convergence of the logarithmic
Euler products, together with the ratio
$2^{r+1}/(1+2^{r+1})\to1$ at $2$, proves the first equivalence in
\eqref{eq:integer-shifted-high-moment}. For $R=R_\sigma$, layer-cake
integration and \eqref{eq:exact-extreme-tail-transfer} give
\[
 \mathbb E_{\mathsf B}R^r
 =r\int_0^\infty t^{r-1}\mathsf B(R>t)\,dt
 \sim r\int_0^\infty t^rU_R(t)\,dt
 =\frac r{r+1}\mathbb E_{\mathsf I}R^{r+1}.
\]
Indeed, given $\eta>0$, choose $T$ so that
$|\mathsf B(R>t)/(tU_R(t))-1|\le\eta$ for every $t\ge T$.
As $r\to\infty$, the integrals over $[0,T]$ are negligible because
$R$ has unbounded support. Letting $\eta\downarrow0$ proves the
comparison. All moments are finite by the Euler products.
\end{proof}

\Cref{thm:exact-extreme-tail-transfer} and
\eqref{eq:integer-extreme-leading}, for $R=R_\varphi$ and
$t=1/\varepsilon$, give the full fixed-order refinement of
\eqref{eq:small-x-tail}: for every fixed integer $m\ge2$, as
$\varepsilon\downarrow0$,
\begin{equation}\label{eq:small-x-full-expansion-mas20}
 -\log G(\varepsilon)
 =\exp(e^{-\gamma}/\varepsilon)
   \left\{1+\sum_{j=2}^m\alpha_j^{\mathrm{tail}}\varepsilon^j
                    +O_m(\varepsilon^{m+1})\right\}.
\end{equation}
Indeed the transfer adds only $-\log(1/\varepsilon)+O_N(\varepsilon^N)$
to the integer negative log-tail $-\log U_{R_\varphi}(1/\varepsilon)$,
which is absorbed by the displayed remainder. This is an expansion of
$-\log G$, not a multiplicative expansion of $G$.

Let $K_\ell^+$ be independent with
$\mathbb P(K_\ell^+=j)=(\ell-1)\ell^{-j}$ for $j\ge1$, and put
\[
 \mathcal R_\infty=\prod_\ell(1-\ell^{-(K_\ell^++1)}).
\]

\begin{corollary}[Joint laws under extreme conditioning]
\label{cor:joint-extreme-conditioning}
Under either valuation law $\mathsf M\in\{\mathsf I,\mathsf B\}$,
\[
\begin{aligned}
 \operatorname{Law}_{\mathsf M}\!\left((R_\varphi/t,R_\sigma/t)
                    \mid R_\varphi>t\right)
     &\Rightarrow\operatorname{Law}(1,\mathcal R_\infty),\\
 \operatorname{Law}_{\mathsf M}\!\left((R_\varphi/t,R_\sigma/t)
                    \mid R_\sigma>t\right)
     &\Rightarrow\delta_{(1,1)}
 \qquad(t\to\infty).
\end{aligned}
\]
\end{corollary}

\begin{proof}
The saturation estimates in the proof of
\Cref{thm:exact-extreme-tail-transfer} transfer to the corresponding
integer conditional laws by its total-variation comparisons.
Under $\mathsf I(\,\cdot\mid R_\varphi>t)$, conditioning concerns only
activation indicators, so positive valuation marks retain the independent
laws of $K_\ell^+$, and every fixed prime is active with probability
tending to one. In the product
\[
 \frac{R_\sigma}{R_\varphi}
 =\prod_{v_\ell\ge1}(1-\ell^{-(v_\ell+1)}),
\]
the logarithmic tail above $y$ is uniformly $O(\sum_{\ell>y}\ell^{-2})$.
Finite-coordinate convergence therefore gives
$R_\sigma/R_\varphi\Rightarrow\mathcal R_\infty$, whereas
\eqref{eq:extreme-overshoot} gives $R_\varphi/t\to1$ in probability.
Under $\mathsf I(\,\cdot\mid R_\sigma>t)$, every fixed valuation tends
to infinity in probability. The same tail bound gives
$R_\sigma/R_\varphi\to1$, and \eqref{eq:extreme-overshoot} gives
$R_\sigma/t\to1$, both in probability. These prove the joint limits
under $\mathsf I$; \eqref{eq:extreme-conditional-tv} transfers them to
$\mathsf B$.
\end{proof}

\begin{proposition}[A refined reciprocal-moment comparison]
\label{prop:moment-ratio}
For the shifted-prime totient Mellin transform and the integer
valuation law of \Cref{thm:exact-extreme-tail-transfer}, as real
$r\to\infty$,
\begin{equation}\label{eq:moment-ratio}
 \frac{M(-r)}{\mathbb E_{\mathsf I}R_\varphi^{r+1}}
 =1-\frac1r+o(1/r),\qquad
 \log\frac{M(-r)}{\mathbb E_{\mathsf I}R_\varphi^{r+1}}
 =-\frac1r+o(1/r).
\end{equation}
For $Y>1$ satisfying $Y\log Y=r$ and $L=\log Y$,
\begin{equation}\label{eq:mellin-saddle}
 \log M(-r)=r(\log\log Y+\gamma)-Y
   +\frac{\pi^2Y}{6L(L+1)}
   +\frac{7\pi^4Y}{60L^4}+O(Y/L^5).
\end{equation}
\end{proposition}

\begin{proof}
Put $a_\ell=\ell/(\ell-1)$. Subtraction of the local moment factors,
as in the proof of \eqref{eq:integer-shifted-high-moment}, gives the
exact formula
\[
 Q(r):=\frac{M(-r)}{\mathbb E_{\mathsf I}R_\varphi^{r+1}}
 =\frac{2^{r+1}}{1+2^{r+1}}\prod_{\ell\ge3}(1-d_\ell(r)),
 \qquad d_\ell(r)=\frac1{(\ell-1)^2+\ell a_\ell^r}.
\]
Fix $0<\eta<1/2$ and put $A=(1-\eta)Y$, $B=(1+\eta)Y$.
For $\ell\le A$,
$d_\ell(r)\le e^{-r/\ell}/\ell$. Since $e^{-r/x}/x$ increases for
$x<r$, the prime-counting bound gives
\[
 \sum_{3\le\ell\le A}d_\ell(r)
 \le\pi(A)e^{-r/A}/A
 \ll_\eta Y^{-1/(1-\eta)}/L=o(1/r).
\]
For $\ell\ge B$, monotonicity of
$r\log(x/(x-1))-\log x$ gives
\[
 \sup_{\ell\ge B}a_\ell^r/\ell
 \le B^{-1}\exp(r\log(B/(B-1)))
 \ll_\eta Y^{-\eta/(1+\eta)}=o(1).
\]
Hence $d_\ell(r)=(1+o(1))\ell^{-2}$ uniformly in that tail.
Partial summation of the prime number theorem yields
\[
 \sum_{\ell>z}\ell^{-2}\sim\sum_{\ell>z}(\ell-1)^{-2}
 \sim(z\log z)^{-1}.
\]
Writing $S(r)=\sum_{\ell\ge3}d_\ell(r)$, the upper estimate
$d_\ell(r)\le(\ell-1)^{-2}$ and the preceding lower tail therefore give
\[
 \frac1{1+\eta}\le\liminf rS(r)
 \le\limsup rS(r)\le\frac1{1-\eta}.
\]
Letting $\eta\downarrow0$ after taking the limits proves
$S(r)=r^{-1}+o(r^{-1})$. Since
$1-S\le\prod(1-d_\ell)\le e^{-S}=1-S+O(S^2)$ and the factor at
$2$ is $1+O(2^{-r})$, this proves~\eqref{eq:moment-ratio}.

Write
$M_{\mathsf I,\varphi}(s):=\mathbb E_{\mathsf I}R_\varphi^s
=\prod_\ell(1+(a_\ell^s-1)/\ell)$.
For $s=z\log z$, Weingartner~\cite[equation~(5), Lemma~5]{Wei10} gives
\[
 \log M_{\mathsf I,\varphi}(s)=s(\log\log z+\gamma)-z
 +z\left(\frac{b_2}{(\log z)^2}
        +\frac{b_3}{(\log z)^3}
        +\frac{b_4}{(\log z)^4}\right)
 +O\left(\frac z{(\log z)^5}\right),
\]
where $b_2=\pi^2/6$, $b_3=-\pi^2/6$ and
$b_4=\pi^2/6+7\pi^4/60$.
Let $z_s\log z_s=s$. The derivative of
$s(\log\log z_s+\gamma)-z_s$ is $\log\log z_s+\gamma$.
Thus replacing $s=r+1$ by $r$ changes this main expression by
$O(\log L)$; moreover $z_{r+1}-Y=O(1/L)$ and the three correction
terms change by $O(L^{-3})$. All these errors, and
$\log Q(r)=O(1/r)$, are absorbed by $O(Y/L^5)$.
Finally
$1/(L(L+1))=L^{-2}-L^{-3}+L^{-4}+O(L^{-5})$,
which gives~\eqref{eq:mellin-saddle}.
\end{proof}

\section{Endpoint and anchor asymptotics; nondifferentiability}\label{sec:endpoint}

The two ends use different events in the early prime coordinates.
Small $X$ is produced by many active primes, as in the lower bound of
\Cref{cor:small-x-tail}. Near $X=1/2$, the sum $W$ must be small, so a
prime whose jump exceeds the permitted deficit cannot be active.
The empty-cylinder argument, \Cref{lem:empty-cylinder} below, separates
the endpoint event into the absence of early activations and a condition
on the remaining tail.
Positivity and independence give the exact factorization
\eqref{eq:endpoint-cylinder}; the resulting expansion supplies the
coefficients for sharp concentration. The empty-block estimate
\eqref{eq:lq-empty-block} gives a sufficient-event lower bound under
the hypotheses of \Cref{prop:renyi-collapse}\textup{(b)}, also for
signed increments.
The vector endpoint proofs in
\Cref{prop:deleted-corners,cor:joint-degree-endpoint} use
coordinate-dependent cutoffs, through part~\textup{(b)} of the lemma.

Partial summation and Chebyshev's bound give the prime tail bound
\begin{equation}\label{eq:endpoint-tail-square}
 \sum_{\substack{\ell>y\\\ell\textup{ prime}}}\frac1{(\ell-1)^2}
 \ll\frac1{y\log y}\qquad(y\ge3).
\end{equation}

\begin{lemma}[Empty-cylinder comparison]
\label{lem:empty-cylinder}
\index{Endpoint asymptotics!empty-cylinder comparison}
Let $d\ge1$, and write $\omega_\ell=\log(\ell/(\ell-1))$ for every prime
$\ell$, including $\ell=2$. For each prime $\ell$ let
$\boldsymbol\xi_\ell=(\xi_{\ell,1},\ldots,\xi_{\ell,d})$ be a random
vector in $\{0,1\}^d$. Suppose that the vectors $\boldsymbol\xi_\ell$
are independent; the coordinates at one prime may be dependent, and
any of them may vanish identically. Put
\[
 W_j:=\sum_\ell\xi_{\ell,j}\omega_\ell\in[0,\infty]\qquad(1\le j\le d).
\]
For $\boldsymbol t=(t_1,\ldots,t_d)\in(0,1)^d$, let
$\mathcal A_{\boldsymbol t}$ be the event that $\xi_{\ell,j}=0$ for
every $j$ and every prime $\ell\le1/t_j$, and put
$T_j:=\sum_{\ell>1/t_j}\xi_{\ell,j}\omega_\ell$. Let $\diamond$ be
either $<$ or $\le$.
\begin{enumerate}[label=\textup{(\alph*)}]
\item One has
\[
\begin{aligned}
 \{W_j\diamond t_j\text{ for every }j\}
 &=\mathcal A_{\boldsymbol t}\cap\{T_j\diamond t_j\text{ for every }j\},\\
 \mathbb P(\mathcal A_{\boldsymbol t})
 &=\prod_\ell\mathbb P\bigl(\xi_{\ell,j}=0\text{ for every }j
                    \text{ with }\ell\le1/t_j\bigr).
\end{aligned}
\]
If $t_1=\cdots=t_d$, the two events on the right of the first identity
are independent. In particular, for $d=1$ and $0<t<1$,
\begin{equation}\label{eq:endpoint-cylinder}
 \mathbb P(W_1\diamond t)
 =\prod_{\ell\le1/t}\mathbb P(\xi_{\ell,1}=0)\;
  \mathbb P\Bigl(\sum_{\ell>1/t}\xi_{\ell,1}\omega_\ell\diamond t\Bigr).
\end{equation}
\item Suppose that, for some constant $C>0$,
$\mathbb P(\xi_{\ell,j}=1)\le C/\ell$ for all $\ell$ and $j$. Then,
uniformly in $\boldsymbol t\in(0,1)^d$,
\begin{equation}\label{eq:empty-cylinder-comparison}
 \mathbb P\bigl(W_j\diamond t_j\text{ for every }j\bigr)
 =\mathbb P(\mathcal A_{\boldsymbol t})
  \left\{1+O_{C,d}\!\left(\sum_{j=1}^d\frac1{\log(1/t_j)}\right)\right\}.
\end{equation}
\end{enumerate}
\end{lemma}

\begin{proof}
All summands are nonnegative, and $\omega_\ell>1/\ell$. If
$\xi_{\ell,j}=1$ for some $j$ and some $\ell\le1/t_j$, then
$W_j\ge\omega_\ell>t_j$. Hence $W_j\diamond t_j$ for every $j$ implies
$\mathcal A_{\boldsymbol t}$, and on $\mathcal A_{\boldsymbol t}$ one
has $W_j=T_j$ for every $j$; this is the identity of events. Write
$\mathcal A_{\boldsymbol t}=\bigcap_\ell A_\ell$, where $A_\ell$ is the
event that $\xi_{\ell,j}=0$ for every $j$ with $\ell\le1/t_j$. It
depends only on $\boldsymbol\xi_\ell$ and is the sure event for
$\ell>\max_j1/t_j$, so independence gives the product formula. If all
$t_j$ equal $t$, then $\mathcal A_{\boldsymbol t}$ is determined by the
vectors with $\ell\le1/t$ and the $T_j$ by those with $\ell>1/t$; this
gives the independence and \eqref{eq:endpoint-cylinder}.

For \textup{(b)}, put $y_0:=\max(3,2dC)$, and let
$\Delta_{\boldsymbol t}\ge0$ be the difference between
$\mathbb P(\mathcal A_{\boldsymbol t})$ and the left side of
\eqref{eq:empty-cylinder-comparison}. By \textup{(a)} and Markov's
inequality,
\[
 \Delta_{\boldsymbol t}\le\mathbb P(\mathcal A_{\boldsymbol t}),\qquad
 \Delta_{\boldsymbol t}\le\sum_{j=1}^d
   \mathbb P\bigl(\mathcal A_{\boldsymbol t}\cap\{T_j\ge t_j\}\bigr)
 \le\sum_{j=1}^d
   \frac{\mathbb E(T_j\mathbf1_{\mathcal A_{\boldsymbol t}})}{t_j}.
\]
If $t_j>1/y_0$ for some $j$, then $\sum_i1/\log(1/t_i)>1/\log y_0$ and
the first bound suffices. Otherwise let $1\le j\le d$ and
$\ell>1/t_j$. Then $\ell>2dC$, so
$\mathbb P(A_\ell)\ge1-\sum_i\mathbb P(\xi_{\ell,i}=1)\ge1/2$, and
independence across primes gives
\[
 \mathbb P\bigl(\{\xi_{\ell,j}=1\}\cap\mathcal A_{\boldsymbol t}\bigr)
 \le\mathbb P(\xi_{\ell,j}=1)\prod_{\ell'\ne\ell}\mathbb P(A_{\ell'})
 \le\frac{2C}\ell\,\mathbb P(\mathcal A_{\boldsymbol t}).
\]
Since $\omega_\ell<1/(\ell-1)$, Tonelli's theorem and summation over
$\ell>1/t_j\ge3$, together with
\eqref{eq:endpoint-tail-square} yield
\[
 \mathbb E\bigl(T_j\mathbf1_{\mathcal A_{\boldsymbol t}}\bigr)
 \le2C\,\mathbb P(\mathcal A_{\boldsymbol t})
     \sum_{\ell>1/t_j}\frac1{(\ell-1)^2}
 \ll\frac{C\,t_j}{\log(1/t_j)}\,\mathbb P(\mathcal A_{\boldsymbol t}).
\]
Hence $\Delta_{\boldsymbol t}\ll C\,\mathbb P(\mathcal A_{\boldsymbol t})
\sum_j1/\log(1/t_j)$.
\end{proof}

The constants in \eqref{eq:intro-endpoint-constant} are
$\Smer{}\approx1.3203236$ and $\kappa_{\mathrm{end}}\approx0.7413$; in the
convention of~\cite{Wre61}, $\Smer{}=2C_2$ with $C_2\approx0.6601618$.

\begin{proposition}[Quantitative endpoint asymptotic at $1/2$]
\label{prop:endpoint-half}
\index{Endpoint asymptotics!totient law}
Let $W:=\sum_{\ell\ge3}B_\ell\omega_\ell$, so that
$X\overset d=\tfrac12e^{-W}$, and let $Q$ be as in \eqref{eq:def-Q}.
As $y\to\infty$,
\begin{equation}\label{eq:endpoint-Q-quantitative}
 Q(y)=\frac{\kappa_{\mathrm{end}}}{\log y}
       \left\{1+O\!\left(e^{-c\sqrt{\log y}}\right)\right\}
       \qquad(c>0\text{ absolute}).
\end{equation}
There are real constants $a_j$, with $a_1=1$, $a_2=0$ and
$a_3=-\pi^2/12$, such that, for every fixed integer $N\ge1$ and either
$\diamond\in\{<,\le\}$,
\begin{equation}\label{eq:endpoint-W-quantitative}
 \mathbb P(W\diamond t)
 =\kappa_{\mathrm{end}}\sum_{j=1}^N
       \frac{a_j}{\log^j(1/t)}
   +O_N\!\left(\log^{-N-1}(1/t)\right)\qquad(t\downarrow0).
\end{equation}
The same coefficients give, as $s\downarrow0$,
\begin{equation}\label{eq:hend-asymp}
 H_{\mathrm{end}}(s)=1-G\bigl(\tfrac12(1-s)\bigr)
 =\kappa_{\mathrm{end}}\left\{\frac1L-\frac{\pi^2}{12L^3}
                         +O(L^{-4})\right\},\qquad L=\log(1/s),
\end{equation}
and the expansion to every fixed order is obtained by replacing the
braces by $\sum_{j=1}^Na_jL^{-j}+O_N(L^{-N-1})$.
More precisely, let $H_I(s)=\mathbb P(Y_I>1-s)$ for the integer model
$Y_I=\prod_\ell(1-1/\ell)^{C_\ell}$ with independent
$C_\ell\sim\operatorname{Bernoulli}(1/\ell)$. Then
\begin{equation}\label{eq:endpoint-integer-transfer}
 H_{\mathrm{end}}(s)=\Smer{}H_I(s)
       +O\!\left(\frac{s}{\log^2(1/s)}\right).
\end{equation}
\end{proposition}

\begin{proof}
Put $T_y:=\sum_{\ell>y}B_\ell\omega_\ell$. For $0<t<1/3$ and either
$\diamond\in\{<,\le\}$, \eqref{eq:endpoint-cylinder}, applied with
the independent indicators
$\xi_{\ell,1}=B_\ell$ for $\ell\ge3$ and $\xi_{2,1}=0$, gives
\[
 \mathbb P(W\diamond t)=Q(1/t)\,\mathbb P(T_{1/t}\diamond t).
\]
Let $Q_I(y)=\prod_{\ell\le y}(1-1/\ell)$. The local factor identity and
\eqref{eq:endpoint-tail-square} give
\[
 \frac{Q(y)}{Q_I(y)}
 =2\prod_{3\le\ell\le y}\left(1-\frac1{(\ell-1)^2}\right)
 =\Smer{}\left\{1+O\!\left(\frac1{y\log y}\right)\right\}.
\]
By \eqref{eq:mertens-product-dvp},
$Q_I(y)=e^{-\gamma}(\log y)^{-1}
\{1+O(e^{-c\sqrt{\log y}})\}$ for some absolute $c>0$.
The head-ratio error $O(1/(y\log y))$ is smaller, proving
\eqref{eq:endpoint-Q-quantitative}.

Set $t=-\log(1-s)$ and $y=1/t$. The integer endpoint event also forces
every coordinate at $\ell\le y$ to vanish. Couple the two sequences
above $y$ using independent common uniform random variables. The
probability of any mismatch is at most
\[
 \sum_{\ell>y}\left|\frac1{\ell-1}-\frac1\ell\right|
 \ll\frac1{y\log y}.
\]
Their tail sums use identical jumps $\omega_\ell$, so the corresponding
tail-event probabilities differ by at most this quantity. Multiplying
by $Q_I(y)\ll1/\log y$ and using the head ratio proves
\eqref{eq:endpoint-integer-transfer}, since $t\asymp s$. 

Tenenbaum and Toulmonde \cite[Theorem~1.2, (1.7)--(1.9)]{TT06}
give $H_I(s)=\sum_{j=1}^Ng_jL^{-j}+O_N(L^{-N-1})$ for every fixed
$N$, where $g_1=e^{-\gamma}$, $g_2=0$ and
$g_3=-\pi^2e^{-\gamma}/12$. Thus $a_j=e^\gamma g_j$ and
\eqref{eq:endpoint-integer-transfer} prove \eqref{eq:hend-asymp}
to every fixed order. Finally,
$\mathbb P(W<t)=H_{\mathrm{end}}(1-e^{-t})$ and
$\log(1/(1-e^{-t}))=\log(1/t)+O(t)$ give
\eqref{eq:endpoint-W-quantitative}; atomlessness gives the same formula
with $\le$.
\end{proof}

The leading term is the shifted-prime analogue of
Erd\H{o}s's integer endpoint asymptotic
\cite[Theorem~3, p.~529]{Erd46}. The higher coefficients are classical;
\eqref{eq:endpoint-integer-transfer} transfers them to the present law.

Fixing a finite set of prime coordinates at their active values moves
the conditional right endpoint to an \emph{arithmetic anchor}. Mass
just below this endpoint again asks for a small remaining sum. For the
same event in the remaining coordinates, changing the prescribed
coordinates from inactive to active multiplies the probability by
their activation odds; this is the role
of $\mathcal K(S)$ below. \Cref{lem:anchor-endpoint} identifies the
contribution from this prescribed configuration. Other configurations
can also contribute to the same window, a distinction needed when
comparing anchors in \Cref{sec:global-modulus}.

\begin{definition}[Arithmetic anchors]\label{def:anchor-weights}
\index{Arithmetic anchors!definition}
For a finite set $S$ of odd primes put
\[
\begin{aligned}
 N_S&:=\prod_{\ell\in S}(\ell-1),& D_S&:=\prod_{\ell\in S}\ell,&
 x_S&:=\frac{N_S}{2D_S},\\
 \mathcal K(S)&:=\prod_{\ell\in S}\frac1{\ell-2},&
 \kappa_S&:=\kappa_{\mathrm{end}}\mathcal K(S),&
 q_S&:=\prod_{\ell\in S}\frac{\ell-2}{\ell-1},
\end{aligned}
\]
and
\begin{equation}\label{eq:deleted-bernoulli-sum}
 \mathcal C_S:=\{B_\ell=1\text{ for every }\ell\in S\},\qquad
 W_S:=\sum_{\substack{\ell\ge3\\\ell\notin S}}B_\ell\omega_\ell,
\end{equation}
so that $W_\varnothing$ is the variable $W$ of \Cref{prop:endpoint-half}.
For $S'\subseteq S$, the \emph{stratum} $\mathcal C_{S'}^{\mathrm{ex}}$ is
the event that $B_\ell=1$ on $S'$ and $B_\ell=0$ on $S\setminus S'$. The
strata partition the sample space, and
$\mathcal C_S^{\mathrm{ex}}=\mathcal C_S$ is the \emph{main stratum}.
\end{definition}

\begin{lemma}[Main stratum at an arithmetic anchor]
\label{lem:anchor-endpoint}
Let $S$ be a finite set of odd primes and $0<\varepsilon<1$, with
$\varepsilon\le1/\max S$ if $S\ne\varnothing$. Then
\begin{equation}\label{eq:anchor-main-exact}
 \mathbb P\bigl(\mathcal C_S,\ x_S(1-\varepsilon)<X\le x_S\bigr)
 =\mathcal K(S)H_{\mathrm{end}}(\varepsilon).
\end{equation}
For each fixed $S$, as $\varepsilon\downarrow0$,
$\mathcal K(S)H_{\mathrm{end}}(\varepsilon)=\kappa_S/\log(1/\varepsilon)
+O_S(\log^{-2}(1/\varepsilon))$.
\end{lemma}

\begin{proof}
For every event $\mathcal A$ measurable in the coordinates outside $S$,
independence and $\mathbb P(B_\ell=1)/\mathbb P(B_\ell=0)=1/(\ell-2)$
$(\ell\in S)$ give the odds identity
\begin{equation}\label{eq:anchor-odds}
 \mathbb P(\mathcal C_S\cap\mathcal A)
 =\mathcal K(S)\,
   \mathbb P(\mathcal C_{\varnothing}^{\mathrm{ex}}\cap\mathcal A).
\end{equation}
Put $t:=-\log(1-\varepsilon)$. On $\mathcal C_S$, $X=x_Se^{-W_S}$, so the
event in \eqref{eq:anchor-main-exact} is $\mathcal C_S\cap\{W_S<t\}$.
If $S\ne\varnothing$, then $\varepsilon\le1/\max S$ gives
$t\le\min_{\ell\in S}\omega_\ell$; since
$W_\varnothing=W_S+\sum_{\ell\in S}B_\ell\omega_\ell$ and $W_S\ge0$, the
event $W_\varnothing<t$ forces every coordinate in $S$ to vanish (this is
vacuous if $S=\varnothing$). Hence
$\mathcal C_\varnothing^{\mathrm{ex}}\cap\{W_S<t\}=\{W_\varnothing<t\}
=\{X>\tfrac12(1-\varepsilon)\}$, and \eqref{eq:anchor-odds} with
$\mathcal A=\{W_S<t\}$ gives \eqref{eq:anchor-main-exact}.
The asymptotic follows from \Cref{prop:endpoint-half}.
\end{proof}

\begin{lemma}[Anchor separation]\label{lem:anchor-sep}
Let $S\ne T$ be finite sets of odd primes, and put
$T\triangle S:=(T\setminus S)\cup(S\setminus T)$. Then
\begin{equation}\label{eq:anchor-sep}
 |x_T-x_S|\ge x_S\Bigl(\prod_{\ell\in T\triangle S}\ell\Bigr)^{-1},
\end{equation}
with equality when $T=S\cup\{p\}$ for a prime $p\notin S$.
\end{lemma}

\begin{proof}
Put $A:=T\setminus S$ and $B:=S\setminus T$. After removing common
factors,
\[
 |x_T-x_S|
 =\prod_{\ell\in T\cap S}(1-1/\ell)
   \frac{|N_AD_B-N_BD_A|}{2D_AD_B}.
\]
The integer numerator is nonzero: if $p=\max(A\cup B)$ and, without loss
of generality, $p\in A$, then $p\nmid N_AD_B$ and $p\mid N_BD_A$.
Thus its absolute value is at least $1$. Since
$\prod_{\ell\in T\cap S}(1-1/\ell)\ge2x_S$ and
$D_AD_B=\prod_{\ell\in T\triangle S}\ell$, this proves
\eqref{eq:anchor-sep}. For $T=S\cup\{p\}$,
$x_T=x_S(1-1/p)$, giving equality.
\end{proof}

\begin{lemma}[Concentration after removing finitely many prime coordinates]\label{lem:mass-split}
Let $S$ be a finite set of odd primes. For every interval $I$,
$\mathbb P(W_\varnothing\in I)\ge q_S\,\mathbb P(W_S\in I)$. Moreover,
for every $h>0$,
\[
 Q_{\muF}(h)\le\sup_{a\in\R}\mathbb P\bigl(W_S\in[a,a+h)\bigr)
 \le q_S^{-1}Q_{\muF}(h).
\]
\end{lemma}

\begin{proof}
The event $\mathcal C_\varnothing^{\mathrm{ex}}$ has probability $q_S$, is
independent of $W_S$, and on it $W_\varnothing=W_S$; this gives the first
inequality. Since $W_\varnothing-W_S=\sum_{\ell\in S}B_\ell\omega_\ell$ is
independent of $W_S$, conditioning on it gives
$\sup_a\mathbb P(W_\varnothing\in[a,a+h))\le\sup_a\mathbb P(W_S\in[a,a+h))$.
As $\muF$ is the continuous law of $\log2+W_\varnothing$, the left side is
$Q_{\muF}(h)$, and the first inequality with $I=[a,a+h)$ bounds the middle
term by $q_S^{-1}Q_{\muF}(h)$.
\end{proof}

\begin{remark}[Differentiability of the totient distribution functions]\label{rem:erdos-partial}
Erd\H{o}s asked whether the integer totient distribution function has
derivative zero wherever it is differentiable~\cite[p.~167]{Dia13};
see~\cite[pp.~1--3]{Tou06b} and the finite-positive-derivative question for
$\sigma(n)/n$ in~\cite[p.~59]{Erd74}.
At integer totient values, its infinite left derivative follows
from~\cite[Th\'eor\`emes~A,~1]{Tou06}. Erd\H{o}s~\cite[p.~529]{Erd46}
asserted infinite derivatives of the reciprocal-law distribution function
at every $n/\varphi(n)$.

For the shifted-prime law, $G'=0$ Lebesgue-almost everywhere on
$[0,\tfrac12]$ by singularity (\Cref{thm:hausdorff-zero}) and Lebesgue's
differentiation theorem. Since the main stratum is part of the window,
\Cref{lem:anchor-endpoint}, with parameter $\varepsilon/x_S$, gives
$G(x_S)-G(x_S-\varepsilon)\gg_S1/\log(1/\varepsilon)$ $(\varepsilon\downarrow0)$
at every arithmetic anchor $x_S$. As $\varepsilon^\alpha\log(1/\varepsilon)\to0$
for every $\alpha>0$, $G$ is not H\"older of any positive order at
any arithmetic anchor. Precisely, its left difference quotient satisfies
\[
 \lim_{h\downarrow0}\frac{G(x_S)-G(x_S-h)}h=+\infty.
\]
The anchors are dense in
$[0,\tfrac12]$ by \Cref{lem:phi-even-dense}.
The non-anchor derivative questions are recorded in \Cref{ssec:open_modulus}.
\end{remark}

\section{The \texorpdfstring{$\sigma$}{sigma}-case Mellin transform and left-endpoint asymptotic}\label{sec:mellin-sigma}

For the divisor-sum law, positive powers probe large values and
reciprocal powers probe the finite left endpoint $3/2$. Its odd-prime
activation probabilities agree with those of the totient model, but an
active jump depends on the valuation, and the prime-$2$ head is random.
A common realization of the activation indicators will compare the two
laws while keeping these differences explicit.

On a product probability space, let $K$ and the variables
$(Z_\ell^\sigma)_{\ell\ge3,\ \ell\textup{ prime}}$ be independent, with
\[
 \mathbb P(K=k)=2^{-k}\quad(k\ge1),\qquad
 Z_\ell^\sigma\sim\nusig_\ell,
\]
and put
\[
 W_{\mathrm{odd}}:=\sum_{\ell\ge3}Z_\ell^\sigma,\qquad
 X_\sigma:=\exp\!\bigl(\gsig(2^K)+W_{\mathrm{odd}}\bigr).
\]
By \Cref{thm:dist-sigma}, the series converges almost surely,
$W_{\mathrm{odd}}\sim\mugodd$, and $X_\sigma$ has law~$\muH$.
Put
\[
 B_\ell:=\mathbf1_{\{Z_\ell^\sigma>0\}},\qquad
 V:=\sum_{\ell\ge3}B_\ell\omega_\ell.
\]
By \eqref{eq:nusig}, the $B_\ell$ are independent with
$\mathbb P(B_\ell=1)=1/(\ell-1)$. Thus $V$ has the law of $W$ in
\Cref{prop:endpoint-half}, and $X:=\tfrac12e^{-V}$ has distribution
function $G$. The atom bounds in \Cref{lem:atoms-sigma} give
$\log(1+1/\ell)\le Z_\ell^\sigma\le\omega_\ell$ on
$\{B_\ell=1\}$. Since
$\log(1+1/\ell)\ge1/(\ell+1)\ge\omega_\ell/2$ for $\ell\ge3$,
we obtain, almost surely,
\begin{equation}\label{eq:sigma-coupling}
 V/2\le W_{\mathrm{odd}}\le V,\qquad X_\sigma\le2e^{V}=1/X.
\end{equation}
For odd primes $\ell$ and $s\in\C$ put
\[
 F_s^\sigma(\ell):=L_{\ell,1,1}(s)
 =\frac{\ell-2}{\ell-1}+\sum_{k\ge1}\ell^{-k}S_k(\ell)^s.
\]
Define, for $s\in\C$,
\begin{equation}\label{eq:full-sigma-mellin}
\begin{aligned}
 M_\sigma(s)&:=\prod_{\substack{\ell\ge3\\\ell\textup{ prime}}}F_s^\sigma(\ell),\\
 H_2(s)&:=\sum_{k\ge1}2^{-k}(2-2^{-k})^s,\\
 N_\sigma(s)&:=N_{\sigma,1,1}(s).
\end{aligned}
\end{equation}

\begin{theorem}[Mellin transforms of the odd-prime and full $\sigma$ laws]\label{thm:mellin-sigma}
\index{Mellin transforms!divisor-sum laws}
For every $s\in\C$,
\[
 M_\sigma(s)=\int_\R e^{su}\,d\mugodd(u),\qquad
 \mathbb E[X_\sigma^s\mid K=k]=S_k(2)^sM_\sigma(s)\quad(k\ge1).
\]
The Euler product converges uniformly on compact subsets of $\C$, and
$M_\sigma$ is entire of order exactly~$1$.

For every $s\in\C$,
\[
 N_\sigma(s)=H_2(s)M_\sigma(s)=\mathbb E[X_\sigma^s].
\]
The functions $H_2$ and $N_\sigma$ are entire of order exactly~$1$,
and $H_2$ has finite exponential type.
\end{theorem}

\begin{proof}
Normal convergence and entireness follow from
\Cref{thm:prime-mellin}. For the growth estimates, write
$F_s^\sigma(\ell)=1+a_\ell^\sigma(s)$ for $|s|\le R$.
The atom bounds of \Cref{lem:atoms-sigma} give
\[
 |a_\ell^\sigma(s)|
 \le\sum_{k\ge1}\ell^{-k}|e^{s\gsig(\ell^k)}-1|
 \le\frac{R\omega_\ell e^{R\omega_\ell}}{\ell-1}
 \ll_R\ell^{-2}.
\]
Independence gives
\[
 \mathbb E e^{sW_L^\sigma}=
 \prod_{3\le\ell\le L}F_s^\sigma(\ell),\qquad
 W_L^\sigma:=\sum_{3\le\ell\le L}Z_\ell^\sigma.
\]
For $a=|\operatorname{Re}s|$, monotone convergence gives
$\mathbb E e^{aW_{\mathrm{odd}}}=M_\sigma(a)<\infty$.
Dominated convergence then proves the integral identity; independence
of $K$ proves the conditional identity.

Positive coefficients give
$\max_{|s|\le r}|M_\sigma(s)|\le M_\sigma(r)$.
For $\ell\le r$, $\log F_r^\sigma(\ell)\le r\omega_\ell$; for
$\ell>r$, the preceding estimate gives
$\log F_r^\sigma(\ell)\ll r/\ell^2$.
Mertens' estimate therefore yields
$\log\max_{|s|=r}|M_\sigma(s)|\ll r\log\log r$.
Since $F_r^\sigma(3)\ge(4/3)^r/3$ and all other real factors are at
least~$1$, the order is exactly~$1$.

By \Cref{cor:ap-sigma-law}, $N_\sigma(s)=\mathbb E[X_\sigma^s]$.
At $Q=1$, the factor $L_{2,1,1}(s)$ in \eqref{eq:ap-sigma-local} is
$H_2(s)$, since its constant term vanishes and $S_k(2)=2-2^{-k}$.
Hence $N_\sigma=H_2M_\sigma$.
The series for $H_2$ converges normally and satisfies
$|H_2(s)|\le2^{|s|}$, so it has finite exponential type.
The lower bounds $H_2(r)\ge(3/2)^r/2$ and $M_\sigma(r)\ge1$,
together with the upper bounds already proved, give order exactly~$1$
for $H_2$ and $N_\sigma$.
\end{proof}

\begin{proposition}[Zeros of the divisor-sum Mellin transforms]\label{prop:sigma-mellin-zeros}
\begin{enumerate}[label=\textup{(\alph*)},leftmargin=*]
\item $M_\sigma$ is zero-free on $\{\operatorname{Re}s\le0\}$.
\item Every odd local factor $F_s^\sigma(\ell)$ has infinitely
many zeros, all in $\{\operatorname{Re}s>0\}$, and so does $M_\sigma$.
For $\ell\ge5$, put $\lambda_1=\log((\ell+1)/\ell)$ and
$A=\ell(\ell-2)/(\ell-1)$. Each disk
\[
 \left|s-\frac{\log A+(2m+1)\pi i}{\lambda_1}\right|
 <\frac{\log2}{\lambda_1}\qquad(m\in\Z)
\]
contains exactly one zero of $F_s^\sigma(\ell)$, which is simple.
\item The functions $H_2$ and $N_\sigma$ are zero-free on
$\{\operatorname{Re}s\le0\}$. The function $H_2$ has infinitely many
zeros, all in $\{\operatorname{Re}s>0\}$. The zeros of $N_\sigma$
are those of $H_2$ and $M_\sigma$, with multiplicities added at common zeros.
\item Each of $M_\sigma$ and $N_\sigma$ has exactly one zero in
$|s-10i|<1/2$. This is the same simple zero of $F_s^\sigma(3)$.
\end{enumerate}
\end{proposition}

\begin{proof}
For $u=\operatorname{Re}s<0$,
\[
 \sum_{k\ge1}\ell^{-k}e^{u\gsig(\ell^k)}
 <\frac1{\ell-1}\le\frac{\ell-2}{\ell-1}.
\]
The constant term strictly dominates the remaining terms. For $u=0$
and $\ell\ge5$, $|F_{it}^\sigma(\ell)|\ge(\ell-3)/(\ell-1)>0$.
At $\ell=3$, vanishing would imply
\[
 \sum_{k\ge1}3^{-k}(1+\cos(t\gsig(3^k)))=0,
\]
and hence $t\gsig(3^k)\in\pi+2\pi\Z$ for every $k\ge1$.
For $t\ne0$ this is a distinct convergent sequence in a discrete set,
which is impossible; for $t=0$ the factor equals~$1$.
The normal-convergence estimate in the proof of \Cref{thm:mellin-sigma} makes $\sum_\ell\log(1+a_\ell^\sigma(s))$
normally convergent after finitely many factors, with
$|a_\ell^\sigma(s)|<1/2$ on each fixed compact set. The tail product
is nonvanishing, proving the assertion for $M_\sigma$.

Each local factor is entire of exponential type at most
$\omega_\ell$ and, as $r\to\infty$,
\[
 F_{-r}^\sigma(\ell)=c_0+O_\ell(e^{-\lambda_1r}),\qquad
 c_0=\frac{\ell-2}{\ell-1},\quad
 \lambda_1=\log\frac{\ell+1}{\ell}.
\]
If it had finitely many zeros, Hadamard factorization would give
$F_s^\sigma(\ell)=e^{as+b}P(s)$ with $P$ a polynomial. Its finite
nonzero limit on the negative real axis forces $P$ constant and $a=0$,
first by moduli and then by convergence of the phase. This contradicts
$F_0^\sigma(\ell)=1$, since $c_0<1$. Part~\textup{(a)} locates these zeros.

For $\ell\ge5$, put $\theta=\omega_\ell/\lambda_1$ and write
\[
\begin{aligned}
 F_s^\sigma(\ell)&=P_0(s)+R(s),\\
 P_0(s)&=c_0+\ell^{-1}e^{\lambda_1s},\qquad
 R(s)=\sum_{k\ge2}\ell^{-k}e^{s\gsig(\ell^k)}.
\end{aligned}
\]
On the circle centered at
$s_m=(\log A+(2m+1)\pi i)/\lambda_1$ with radius
$\log2/\lambda_1$, set $w=\lambda_1(s-s_m)$. For $|w|=\log2$,
$|1-e^{-w}|\ge1/2$; otherwise the principal-log series would give
$|w|<\log2$. Hence
$|P_0(s)|=c_0|1-e^w|\ge c_0e^{\operatorname{Re}w}/2$.
Since $A>2$, the circle lies in $\operatorname{Re}s>0$, and
\[
 \frac{|R(s)|}{|P_0(s)|}
 \le\frac{2A^{\theta-1}e^{(\theta-1)\operatorname{Re}w}}{\ell-1}
 \le\frac{A^{\theta-1}2^\theta}{\ell-1}.
\]
The bounds
\[
 \omega_\ell-\lambda_1=-\log(1-\ell^{-2})\le\frac1{\ell^2-1},
 \qquad \lambda_1\ge\frac1{\ell+1}
\]
give $0<\theta-1\le1/(\ell-1)$. Since $A<\ell$ and
$\log(2u)/(u-1)$ decreases for $u\ge5$,
\[
 \frac{|R(s)|}{|P_0(s)|}
 \le\frac{2(2\ell)^{1/(\ell-1)}}{\ell-1}
 \le\frac{10^{1/4}}2<1.
\]
Rouch\'e's theorem~\cite[Theorem~V.3.8]{Conway78} gives one zero counting
multiplicity in each disk, hence one simple zero of the specified local
factor. The disks are disjoint because $\log2<\pi$.
Normal convergence makes every local zero a zero of $M_\sigma$.

After division by $(3/2)^s$, the first term of $H_2$ has mass $1/2$.
For $u=\operatorname{Re}s<0$, the remaining moduli sum to
\[
 \sum_{k\ge2}2^{-k}
 \left(\frac{2-2^{-k}}{3/2}\right)^u<\frac12.
\]
For $s=it$, cancellation would require
$t\log((2-2^{-k})/(3/2))\in\pi+2\pi\Z$ for every $k\ge2$.
If $t\ne0$, this is again a distinct convergent sequence in a discrete
set; $H_2(0)=1$ handles $t=0$. Thus $H_2$, and by part~\textup{(a)} also
$N_\sigma$, are zero-free on the closed left half-plane.

If $H_2$ had finitely many zeros, Hadamard factorization would give
$H_2(s)=e^{as+b}P(s)$. But
\[
 \frac{H_2(-r)}{(3/2)^{-r}}
 =\frac12+O((6/7)^r).
\]
Taking moduli forces $P$ constant and
$\operatorname{Re}a=\log(3/2)$; convergence of the phase then forces
$a=\log(3/2)$ and $e^bP=1/2$, contradicting $H_2(0)=1$.
This proves that $H_2$ has infinitely many zeros; the product identity
gives the zeros of $N_\sigma$ with their multiplicities.

For part~\textup{(d)}, write $s_0=10i$ and
\[
 F(s)=F_s^\sigma(3)=\tfrac12+\sum_{j\ge1}3^{-j}e^{s\lambda_j},
 \qquad \lambda_j=\log((3-3^{-j})/2).
\]
For the truncation $F_5$ through $j=5$, the following strict enclosures
hold at $s_0$:
\[
 \begin{aligned}
 0.0461314&<\operatorname{Re}F_5(s_0)<0.0461315,\\
 -0.0084334&<\operatorname{Im}F_5(s_0)<-0.0084333,\\
 0.1421684&<-\operatorname{Re}F'_5(s_0)<0.1421686 .
 \end{aligned}
\]
For an exact calculation, put $x=(3-3^{-j})/2$ and $u=(x-1)/(x+1)$.
Evaluate $2\sum_{n=0}^8u^{2n+1}/(2n+1)$ for $\log x$ with error
at most $2u^{19}/(19(1-u^2))<10^{-14}$. At ten times this rational
approximation use the sine and cosine Taylor polynomials through
degree $26$. The argument error is at most $10^{-13}$ and the
Taylor remainder is bounded by $4.1^{27}/27!<4\cdot10^{-12}$.
The omitted tails of $F(s_0)$ and $F'(s_0)$ have moduli at most
$1/486$ and $\log(3/2)/486$, respectively. Consequently
$|F(s_0)|<0.05$ and $|F'(s_0)|>0.14$. On the closed disk,
\[
 |F''(s)|\le \tfrac12\sqrt{3/2}\,\log^2(3/2)<0.102 .
\]
On its boundary the difference between $F(s)$ and $F'(s_0)(s-s_0)$
has modulus at most $0.05+0.102/8<0.07$. Rouch\'e's theorem gives
one zero counting multiplicity. For each odd $\ell\ge5$, the sum of
the nonconstant terms in modulus is at most
$(\ell/(\ell-1))^{1/2}/(\ell-1)<(\ell-2)/(\ell-1)$ throughout the disk.
The normally convergent product of these factors is nonzero there by the
tail-log estimate in part~\textup{(a)}. The arguments of the summands of
$H_2$ lie in an arc of width at most $10.5\log(4/3)<\pi$, so their sum
is nonzero as well. This proves the claimed common simple zero.
\end{proof}

\begin{proposition}[Positive moment growth of the divisor-sum laws]\label{prop:sigma-mellin-growth}
As real $r\to\infty$,
\begin{equation}\label{eq:sigma-signed-mellin-comparison-mas16}
 0\le\log M(-r)-\log N_\sigma(r)\ll\sqrt r.
\end{equation}
For $Y>1$ defined by $Y\log Y=r$ and $L=\log Y$, this gives
\begin{equation}\label{eq:sigma-mellin-saddle-mas16}
\begin{aligned}
 \log N_\sigma(r)
 &=r(\log\log Y+\gamma)-Y
   +\frac{\pi^2Y}{6L(L+1)}
   +\frac{7\pi^4Y}{60L^4}+O(Y/L^5),\\
 \log M_\sigma(r)
 &=r(\log\log Y+\gamma-\log2)-Y
   +\frac{\pi^2Y}{6L(L+1)}
   +\frac{7\pi^4Y}{60L^4}+O(Y/L^5).
\end{aligned}
\end{equation}
Thus $M_\sigma$ and $N_\sigma$ have infinite type, and
$\mathbb E e^{aX_\sigma}<\infty$ for every $a>0$.
\end{proposition}

\begin{proof}
Realize the coupling by shifted-prime valuations
$K_2=K$ and $K_\ell$ for odd primes, with
$Z_\ell^\sigma=\gsig(\ell^{K_\ell})$ and
$B_\ell=\mathbf1_{\{K_\ell\ge1\}}$. Put
$\mathcal A=\{\ell:K_\ell\ge1\}$, which contains $2$, and write
\[
 R:=1/X=\prod_{\ell\in\mathcal A}(1-1/\ell)^{-1},
 \qquad
 X_\sigma=R e^{-D},\qquad
 D:=\sum_{\ell\in\mathcal A}
       -\log(1-\ell^{-K_\ell-1})\ge0.
\]
The series defining $D$ converges, since its summands are
$O(\ell^{-2})$. For $r\ge4$, define a probability measure by
\[
 \frac{d\mathbb P_r}{d\mathbb P}=\frac{R^r}{M(-r)}.
\]
The density factors coordinatewise: its finite-dimensional marginals
show that the coordinates remain independent under $\mathbb P_r$.
Only the activation probabilities change, since $R$ depends on
$K_\ell$ only through $\mathbf1_{\{K_\ell\ge1\}}$. In particular,
including $\ell=2$,
\[
 \mathbb P_r(K_\ell\ge k\mid K_\ell\ge1)=\ell^{1-k}
 \qquad(k\ge1).
\]
For $\ell\le\sqrt r$, set
$k_\ell=\lceil\log r/\log\ell\rceil$, and let
\[
 E_r=\bigcap_{\ell\le\sqrt r}
       \bigl(\{K_\ell=0\}\cup\{K_\ell\ge k_\ell\}\bigr).
\]
Each local event has tilted probability at least $\ell^{1-k_\ell}$.
Independence and the prime-counting bound therefore give
\[
 \mathbb P_r(E_r)\ge\prod_{\ell\le\sqrt r}\ell^{1-k_\ell}
 \ge\exp\{-\pi(\sqrt r)\log r\}
 \ge\exp\{-C\sqrt r\}.
\]
On $E_r$, an active small prime has
$\ell^{-K_\ell-1}\le1/(r\ell)$, whereas every active large prime has
$\ell^{-K_\ell-1}\le\ell^{-2}$. Consequently
\[
 rD\ll\sum_{\ell\le\sqrt r}\frac1\ell
        +r\sum_{\ell>\sqrt r}\frac1{\ell^2}
 \ll\sqrt r.
\]
Here even the sum over all integers suffices for the second estimate.
Thus
\[
 1\ge\frac{N_\sigma(r)}{M(-r)}
   =\mathbb E_r e^{-rD}
   \ge\mathbb P_r(E_r)e^{-C\sqrt r}
   \ge e^{-C'\sqrt r},
\]
which proves \eqref{eq:sigma-signed-mellin-comparison-mas16}.

Equation~\eqref{eq:mellin-saddle} now gives the first expansion in
\eqref{eq:sigma-mellin-saddle-mas16}, because
$\sqrt r=o(Y/L^5)$ when $r=YL$. For the second expansion,
$H_2(r)\le2^r$, while the terms with
$k\ge\lceil\log_2r\rceil+1$ give $H_2(r)\gg2^r/r$.
Hence $\log H_2(r)=r\log2+O(\log r)$, and \Cref{thm:mellin-sigma} gives precisely
the stated shift of $\gamma$. The expansions imply infinite type.
Finally, $X_\sigma\le R$ and \Cref{cor:small-x-tail} give
$\mathbb E e^{aX_\sigma}<\infty$ for every $a>0$.
\end{proof}

At the finite endpoint, only the smallest prime-$2$ head contributes,
with probability $\mathbb P(K=1)=1/2$. This gives the factor $1/2$ in
\Cref{prop:endpoint-sigma}; \Cref{cor:opposite-end-moments} records the
corresponding normalizations for $M(r)$ and $N_\sigma(-r)$.

\begin{proposition}[Quantitative endpoint asymptotic at $3/2$]\label{prop:endpoint-sigma}
\index{Endpoint asymptotics!divisor-sum law}
With the coefficients $a_j$ of \Cref{prop:endpoint-half}, for every
fixed integer $N\ge1$, as $t\downarrow0$,
\begin{equation}\label{eq:sigma-odd-endpoint}
 \mathbb P(W_{\mathrm{odd}}<t)=\mathbb P(W_{\mathrm{odd}}\le t)
 =\kappa_{\mathrm{end}}\sum_{j=1}^N
     \frac{a_j}{\log^j(1/t)}+O_N\!\left(\log^{-N-1}(1/t)\right).
\end{equation}
In particular, with $L=\log(1/\varepsilon)$, as $\varepsilon\downarrow0$,
\[
 H_\sigma(3/2+\varepsilon)
 =\frac{\kappa_{\mathrm{end}}}{2}
 \left\{\frac1{L+\log(3/2)}
 -\frac{\pi^2}{12(L+\log(3/2))^3}+O(L^{-4})\right\}.
\]
The same shifted logarithm gives the expansion to every fixed order.
\end{proposition}

\begin{proof}
In the coupling \eqref{eq:sigma-coupling}, an active prime has jump
$a_{\ell,k}=\log(1+\ell^{-1}+\cdots+\ell^{-k})$, $k\ge1$, and
\[
 \log(1+1/\ell)\le a_{\ell,k}\le\omega_\ell,\qquad
 0\le\omega_\ell-a_{\ell,k}
 \le-\log(1-\ell^{-2})\ll\ell^{-2}.
\]
On $W_{\mathrm{odd}}\le t$, every active prime is at least
$(e^t-1)^{-1}\gg1/t$. Consequently
$\omega_\ell\le(1+Ct)a_{\ell,k}$ there, for an absolute $C$ and all
small $t$. Summing yields
\[
 \mathbb P(V\le t)\le\mathbb P(W_{\mathrm{odd}}\le t)
                  \le\mathbb P(V\le t+Ct^2).
\]
Apply \eqref{eq:endpoint-W-quantitative} at both ends.
The logarithms differ by $O(t)$, so every fixed coefficient agrees.
Atomlessness from \Cref{thm:dist-sigma} permits strict inequalities.
For $0<\varepsilon<1/4$,
\[
 \{X_\sigma<3/2+\varepsilon\}
 =\{K=1,\ W_{\mathrm{odd}}<\log(1+2\varepsilon/3)\}.
\]
Independence, $\mathbb P(K=1)=1/2$, and
$\log(1/\log(1+2\varepsilon/3))
=L+\log(3/2)+O(\varepsilon)$ prove the remaining assertions.
\end{proof}

\begin{corollary}[Endpoint expansions in fixed progressions]
\label{cor:ap-endpoint-expansions}
\index{Endpoint asymptotics!arithmetic progressions}
Fix $Q\ge1$ and $(b,Q)=1$, and put
\[
\begin{aligned}
 S&:=\{\ell\ge3:\ell\textup{ prime},\ \ell\mid Q\},&
 q_S&:=\prod_{\ell\in S}\frac{\ell-2}{\ell-1},\\
 m&:=[2,\gcd(b-1,Q)],&
 a_\sigma&:=\sigma(m)/m,\\
 \rho_{Q,b}&:=2^{-\mathbf1_{\{2\nmid Q\}}}
       \prod_{\substack{\ell\mid Q\\v_\ell(b-1)\ge v_\ell(Q)}}
                    (1-1/\ell).
\end{aligned}
\]
Let $H_{\sigma,Q,b}$ be the distribution function in
\Cref{cor:ap-sigma-law}. With the coefficients $a_j$ of
\Cref{prop:endpoint-half}, for each fixed integer $N\ge1$, as
$\varepsilon\downarrow0$ and $L=\log(1/\varepsilon)$,
\begin{equation}\label{eq:ap-endpoint-series}
\begin{aligned}
 1-G_{Q,b}(A_{Q,b}-\varepsilon)
 &=\frac{\kappa_{\mathrm{end}}}{q_S}
   \sum_{j=1}^N\frac{a_j}{(L+\log A_{Q,b})^j}
   +O_{N,Q,b}(L^{-N-1}),\\
 H_{\sigma,Q,b}(a_\sigma+\varepsilon)
 &=\frac{\kappa_{\mathrm{end}}\rho_{Q,b}}{q_S}
   \sum_{j=1}^N\frac{a_j}{(L+\log a_\sigma)^j}
   +O_{N,Q,b}(L^{-N-1}).
\end{aligned}
\end{equation}
\end{corollary}

\begin{proof}
For the totient law write $X_{Q,b}=A_{Q,b}e^{-W_S}$ as in
\eqref{eq:XQ}. For all sufficiently small $t>0$, the event
$W<t$ forces every coordinate in $S$ to vanish, so
\[
 \mathbb P(W_S<t)=q_S^{-1}\mathbb P(W<t).
\]
Take $t=-\log(1-\varepsilon/A_{Q,b})$ and apply
\eqref{eq:endpoint-W-quantitative}.

For the divisor-sum law, the product of the factors at primes dividing
$2Q$ is an independent countably atomic head with isolated minimum
$a_\sigma$. At $\ell\mid Q$ its valuation is fixed when
$v_\ell(b-1)<v_\ell(Q)$; otherwise its least possible valuation is
$v_\ell(Q)$, with conditional probability $1-1/\ell$. If $2\nmid Q$, the least
valuation at $2$ is one, with probability $1/2$. Hence the head equals
$a_\sigma$ with probability $\rho_{Q,b}$. Choose $\varepsilon>0$
small enough that only this head value contributes and
$t:=\log(1+\varepsilon/a_\sigma)$ lies below every deleted positive jump.
Writing $W_{\mathrm{odd},S}$ for $W_{\mathrm{odd}}$ with the coordinates
in $S$ removed gives
\[
 H_{\sigma,Q,b}(a_\sigma+\varepsilon)
 =\rho_{Q,b}\mathbb P(W_{\mathrm{odd},S}\le t)
 =\frac{\rho_{Q,b}}{q_S}\mathbb P(W_{\mathrm{odd}}\le t).
\]
Apply \eqref{eq:sigma-odd-endpoint}. In the two cases respectively,
$\log(1/t)=L+\log A_{Q,b}+O_{Q,b}(\varepsilon)$ and
$\log(1/t)=L+\log a_\sigma+O_{Q,b}(\varepsilon)$, proving
\eqref{eq:ap-endpoint-series}.
\end{proof}

\begin{corollary}[Large powers and reciprocal endpoint moments]\label{cor:opposite-end-moments}
As real $r\to\infty$, put $L=\log r$ and
\[
 A(r)=\kappa_{\mathrm{end}}\left\{\frac1L-\frac\gamma{L^2}
                   +\frac{\gamma^2+\pi^2/12}{L^3}\right\}.
\]
Then
\[
 \begin{aligned}
 2^rM(r)&=A(r)+O(L^{-4}),\\
 M_\sigma(-r)&=A(r)+O(L^{-4}),\\
 (3/2)^rN_\sigma(-r)&=\tfrac12A(r)+O(L^{-4}).
 \end{aligned}
\]
Each quantity admits an expansion to every fixed inverse-logarithmic order.

For fixed $Q,b$, use the notation of \Cref{cor:ap-endpoint-expansions}.
Then
\[
 A_{Q,b}^{-r}M_{Q,b}(r)=q_S^{-1}A(r)+O_{Q,b}(L^{-4}),\qquad
 a_\sigma^rN_{\sigma,Q,b}(-r)
 =\rho_{Q,b}q_S^{-1}A(r)+O_{Q,b}(L^{-4}),
\]
with expansions to every fixed order.
\end{corollary}

\begin{proof}
For $Y=W$ or $Y=W_{\mathrm{odd}}$, integration by parts gives
\[
 \mathbb E e^{-rY}=\int_0^\infty e^{-u}\mathbb P(Y\le u/r)\,du.
\]
The intervals $u<r^{-1/2}$ and $u>r^{1/2}$ contribute
$O(r^{-1/2})$ and $O(e^{-r^{1/2}})$, respectively. On the intervening
interval, \eqref{eq:endpoint-W-quantitative} and
\eqref{eq:sigma-odd-endpoint}, with $L=\log r$, give
\[
 \mathbb P(Y\le u/r)=\kappa_{\mathrm{end}}\left\{
 \frac1L+\frac{\log u}{L^2}
 +\frac{(\log u)^2-\pi^2/12}{L^3}\right\}
 +O\!\left(\frac{1+|\log u|^3}{L^4}\right).
\]
This follows by expanding $(L-\log u)^{-1}$ and
$(L-\log u)^{-3}$, whose denominators are at least $L/2$.
Use $\int_0^\infty e^{-u}\log u\,du=-\gamma$ and
$\int_0^\infty e^{-u}(\log u)^2\,du=\gamma^2+\pi^2/6$.
The same expansion to arbitrary fixed order is justified by
$\int_0^\infty e^{-u}|\log u|^j\,du<\infty$.
Finally $H_2(-r)=(3/2)^{-r}(1/2+O((6/7)^r))$ proves the three displayed asymptotics.

For the progression assertions, put
$Y_\varphi=\log(A_{Q,b}/X_{Q,b})$ and
$Y_\sigma=\log(Y_{Q,b}/a_\sigma)$, where $Y_{Q,b}$ is the
divisor-sum variable of \Cref{cor:ap-sigma-law}.
The proof of \Cref{cor:ap-endpoint-expansions} gives the same
small-$t$ inverse-logarithmic expansion for their distribution functions,
multiplied respectively by $q_S^{-1}$ and $\rho_{Q,b}q_S^{-1}$.
Applying the displayed Laplace integral to these nonnegative variables
therefore multiplies every coefficient by the same constants. Its left
side is respectively $A_{Q,b}^{-r}M_{Q,b}(r)$ and
$a_\sigma^rN_{\sigma,Q,b}(-r)$, which proves the stated normalizations.
\end{proof}

The first asymptotic sharpens the lower bound
$M(k)\gg2^{-k}/\log k$ of \cite[(3.1)]{DH12} (the case $m=2$ there).

\begin{corollary}[Endpoint order statistics]\label{cor:endpoint-order-statistics}
Let $X_1,\ldots,X_n$ be independent with distribution
function $G$, and let $Y_1,\ldots,Y_n$ be independent with distribution
function $H_\sigma$. Write their order statistics as
$X_{n,1}\ge\cdots\ge X_{n,n}$ and $Y_{n,1}\le\cdots\le Y_{n,n}$.
For each fixed $j\ge1$ and $t>0$, as $n\to\infty$,
\begin{equation}\label{eq:endpoint-order-statistics}
 \begin{aligned}
 \mathbb P\left(\frac1n\log\frac1{1/2-X_{n,j}}\le t\right)
 &=e^{-\kappa_{\mathrm{end}}/t}
   \sum_{r=0}^{j-1}\frac{(\kappa_{\mathrm{end}}/t)^r}{r!}
   +O_{j,t}(n^{-1}),\\
 \mathbb P\left(\frac1n\log\frac1{Y_{n,j}-3/2}\le t\right)
 &=e^{-\kappa_{\mathrm{end}}/(2t)}
   \sum_{r=0}^{j-1}\frac{(\kappa_{\mathrm{end}}/(2t))^r}{r!}
   +O_{j,t}(n^{-1}).
 \end{aligned}
\end{equation}

For samples from $G_{Q,b}$ and $H_{\sigma,Q,b}$ in any fixed
progression, \eqref{eq:endpoint-order-statistics} holds with endpoints
$A_{Q,b},a_\sigma$ and coefficients
$\kappa_{\mathrm{end}}/q_S,\kappa_{\mathrm{end}}\rho_{Q,b}/q_S$, respectively;
the errors are $O_{Q,b,j,t}(n^{-1})$. Here the endpoint and coefficient
notation is that of \Cref{cor:ap-endpoint-expansions}.
\end{corollary}

\begin{proof}
\eqref{eq:hend-asymp} with $s=2e^{-nt}$ and
\Cref{prop:endpoint-sigma} give
\[
 \begin{aligned}
 1-G(1/2-e^{-nt})&=\frac{\kappa_{\mathrm{end}}}{nt}+O_t(n^{-2}),\\
 H_\sigma(3/2+e^{-nt})&=\frac{\kappa_{\mathrm{end}}}{2nt}+O_t(n^{-2}).
 \end{aligned}
\]
The events in \eqref{eq:endpoint-order-statistics} respectively require
at most $j-1$ samples above or below these thresholds. Summing the
first $j$ binomial probabilities, with success probability
$c/n+O_t(n^{-2})$, gives $e^{-c}\sum_{r<j}c^r/r!+O_{j,t}(n^{-1})$.

For the fixed-progression assertions, \eqref{eq:ap-endpoint-series}
at the additive displacement $e^{-nt}$ gives success probabilities
$\kappa_{\mathrm{end}}/(q_Snt)+O_{Q,b,t}(n^{-2})$ and
$\kappa_{\mathrm{end}}\rho_{Q,b}/(q_Snt)+O_{Q,b,t}(n^{-2})$.
The same finite binomial sum gives the asserted coefficients and errors.
\end{proof}

\chapter{Sharp concentration and localization}
\label{sec:global-modulus}

Where should a short interval be placed to capture the most probability?
For small relative widths, the endpoint $1/2$ supplies a natural
benchmark, but the window ending at $1/3$ carries strictly more mass.
The two windows share the same leading asymptotic constant, so the
leading endpoint estimate alone cannot locate an exact maximizer.

The first part develops Fourier moment bounds that remain available
after finitely many prime coordinates are removed. For the shifted-prime
totient law, the later argument identifies which small-prime
configurations can contribute near the maximum, compares their windows
with endpoint windows, and uses the coordinate at $3$ to distinguish
the two leading candidates. This yields both sharp concentration and
localization. The estimates also supply inputs to prime convergence
and shrinking-window counts in \Cref{ssec:sharp-prime-convergence}.
\Cref{thm:sparse-count-concentration} isolates a leading-order
concentration theorem for fixed bounded sparse prime counts with a
deterministic shift. The binomial specialization
\Cref{thm:convolution-modes} retains its sharper remainder, exact tie
counts and endpoint-uniqueness assertions.

Uniform concentration bounds for general additive functions are due
to Ruzsa~\cite{Ruz80}. On shifted primes, Elliott's bound
\cite[Theorem~1, pp.~2,~25]{Ell94} is uniform in the window location,
the function and $x\ge2$, with a constant depending only on the fixed
nonzero shift; it removes the squared logarithm in
\cite[Theorem~2, p.~116]{Tim91}.

The Fourier moment bounds below rest on a bounded mean oscillation (BMO)
estimate for prime cosine sums
(\Cref{lem:euler-majorant-bmo}), proved using the Montgomery--Vaughan mean-value inequality for small
primes and the Brun--Titchmarsh inequality for the remaining blocks. For
comparison, we also record the bound given by the method of~\cite{DR84}
in the form of~\cite[Lemme~2.2, \S~2.3]{dlBT12}.

Use $f$, $\omega_\ell$ and $\nu_\ell$ as in \eqref{eq:f-def} and
\Cref{lem:factor}, $Q_\nu$ as in \Cref{ssec:integer-dim-zero},
and the anchor data of \Cref{def:anchor-weights}. For $0<\varepsilon<1$ put
\[
 q_\varepsilon(t):=G(t)-G\bigl(t(1-\varepsilon)\bigr)
 =\mathbb P\bigl(t(1-\varepsilon)<X\le t\bigr),\qquad z:=\log(1/\varepsilon),
\]
so that $Q_G(\varepsilon)=\sup_{t\in\R}q_\varepsilon(t)$.
For a finite set $S$ of odd primes let $\lambda_S$ be the law of $W_S$ and put
\[
  \mathcal N_S(\tau):=\widehat{\lambda_S}(\tau)
   =\mathbb E\,e^{i\tau W_S}
   =\prod_{\ell\ge3,\ \ell\notin S}\widehat{\nu_\ell}(\tau).
\]
Thus $|\mathcal N_\varnothing|=|\FT|$.

\KeepHeadingSpace{6\baselineskip}
\section{Fourier moments and concentration}\label{ssec:global-modulus-order}

\paragraph*{First-jump functions}
Throughout this section $v\colon[2,\infty)\to\R$ is a $C^1$ function for which there
are constants $c_v,C_v>0$ with
\begin{equation}\label{eq:first-jump-hypotheses}
 0<v(t)\le\frac{C_v}{t},\qquad -v'(t)\ge\frac{c_v}{t^2}\qquad(t\ge2).
\end{equation}
Since $v(t)\to0$ as $t\to\infty$, integration gives
\begin{equation}\label{eq:first-jump-spacing}
 v(s)-v(t)\ge c_v\Bigl(\frac1s-\frac1t\Bigr)\quad(2\le s<t),\qquad
 t\,v(t)\ge c_v\quad(t\ge2).
\end{equation}
In particular $v$ is strictly decreasing. Only the values $v(p)$ at primes enter the
statements below. The first-jump functions of the basic laws are
\[
 v_\varphi(t):=\log\frac t{t-1},\qquad v_\sigma(t):=\log\Bigl(1+\frac1t\Bigr),
\]
so that $v_\varphi(\ell)=\omega_\ell=f(\ell)$ and $v_\sigma(\ell)=\gsig(\ell)$; they
satisfy \eqref{eq:first-jump-hypotheses} with $(c_v,C_v)=(1,2)$ and $(2/3,1)$,
respectively. For real $3\le y_0<y$ put
\[
 \mathcal T^{v}_{y_0,y}(\tau):=\sum_{y_0<p\le y}\frac{\cos(\tau v(p))}p,\qquad
 \mathcal T_{y_0,y}:=\mathcal T^{v_\varphi}_{y_0,y},\qquad
 \mathcal T^{\sigma}_{y_0,y}:=\mathcal T^{v_\sigma}_{y_0,y},
\]
where the sums run over primes.

\begin{lemma}[Euler-product majorant]
\label{lem:euler-majorant}
Let $3\le y_0<y$ and $C_2\ge0$. Let $Y$ and $\xi_p$ $(y_0<p\le y)$ be
independent real random variables. Suppose that each $\xi_p$ is countably atomic and
that
\begin{equation}\label{eq:two-atom-mass}
 \mathbb P(\xi_p=0)\,\mathbb P\bigl(\xi_p=v(p)\bigr)\ge\frac1p-\frac{C_2}{p^2}
 \qquad(y_0<p\le y).
\end{equation}
Then the characteristic function $\phi$ of $Y+\sum_{y_0<p\le y}\xi_p$ satisfies, for
every $\tau\in\R$,
\begin{equation}\label{eq:euler-majorant}
 |\phi(\tau)|\le\mathfrak C_2\,e^{\mathcal T^{v}_{y_0,y}(\tau)}
 \exp\Bigl\{-\sum_{y_0<p\le y}\frac1p\Bigr\},\qquad
 \mathfrak C_2:=\exp\Bigl\{2C_2\sum_{p\ge5}\frac1{p^2}\Bigr\}.
\end{equation}
\end{lemma}

\begin{proof}
By independence and $|\mathbb Ee^{i\tau Y}|\le1$,
$|\phi(\tau)|\le\prod_{y_0<p\le y}|\mathbb Ee^{i\tau\xi_p}|$. Fix such a $p$, and let
$\alpha_p,\beta_p$ be the masses of the atoms $0$ and $v(p)$ of $\xi_p$; these atoms
are distinct because $v(p)>0$. All terms of the sum in \eqref{eq:atomic-modulus},
applied to the atoms of $\xi_p$, are nonnegative. Retaining only the pair
$\{0,v(p)\}$ and using $1-x\le e^{-x}$ gives
\[
 |\mathbb Ee^{i\tau\xi_p}|^2\le1-2\alpha_p\beta_p\bigl(1-\cos(\tau v(p))\bigr)
 \le\exp\bigl\{-2\alpha_p\beta_p\bigl(1-\cos(\tau v(p))\bigr)\bigr\}.
\]
By \eqref{eq:two-atom-mass} and $0\le1-\cos\le2$,
$-\alpha_p\beta_p(1-\cos(\tau v(p)))\le-(1-\cos(\tau v(p)))/p+2C_2/p^2$.
Multiplying over $y_0<p\le y$, all of which are at least $5$, gives
\eqref{eq:euler-majorant}.
\end{proof}

\KeepHeadingSpace{3\baselineskip}
\begin{lemma}[BMO bound for the prime cosine sums $\mathcal T^v_{y_0,y}$]
\label{lem:euler-majorant-bmo}
For a locally integrable real function $A$ on $\R$ and every bounded
interval $I$ of positive length, put
\[
 E_IA:=\frac1{|I|}\int_I A,\qquad
 \|A\|_*:=\sup_I\frac1{|I|}\int_I|A-E_IA|.
\]
There are absolute constants $\mathfrak b,C_{\mathrm{JN}}>0$ and a constant $B_v>0$
depending only on $c_v$ and $C_v$ such that
\begin{equation}\label{eq:euler-majorant-bmo}
 \|\mathcal T^v_{y_0,y}\|_*\le\frac{B_v}{\log y_0}
 \qquad(y>y_0\ge3),
\end{equation}
and, whenever $\|A\|_*\le\mathfrak b$,
\begin{equation}\label{eq:bmo-exponential-integrability}
 \frac1{|I|}\int_I e^{A-E_IA}\le C_{\mathrm{JN}}.
\end{equation}
Consequently, if $y_0\ge3$, $b\in\R$, $|b|B_v\le\mathfrak b\log y_0$
and $y_0<y\le T$, then
\begin{equation}\label{eq:general-exp-moment}
 \frac1{2T}\int_{-T}^{T}e^{b\mathcal T^v_{y_0,y}(\tau)}\,d\tau
 \le C_{\mathrm{JN}}e^{|b|/m_v},\qquad m_v:=\inf_{p\ge5}p\,v(p)\ge c_v.
\end{equation}
Write $B:=B_{v_\varphi}$ and $B_\sigma:=B_{v_\sigma}$; these are absolute constants.
If $y_0\ge3$, $b\in\R$, $|b|B\le\mathfrak b\log y_0$ and $y_0<y\le T$, then
\begin{equation}\label{eq:cosine-exp-moment}
 \frac1{2T}\int_{-T}^{T}e^{b\mathcal T_{y_0,y}(\tau)}\,d\tau
 \le C_{\mathrm{JN}}e^{|b|}.
\end{equation}
If $|b|B_\sigma\le\mathfrak b\log y_0$ and $3\le y_0<y\le T$, then
\begin{equation}\label{eq:sigma-exponential}
 \frac1{2T}\int_{-T}^T e^{b\mathcal T^\sigma_{y_0,y}(t)}\,dt
 \le C_{\mathrm{JN}}e^{4|b|/3}.
\end{equation}
\end{lemma}

\begin{proof}
Let $I$ be a bounded interval of length $\lambda>0$, fix $\tau_0\in I$, and write
$\mathcal T^v_{y_0,y}=D_1+D_2$, where $D_1$ is the sum over the primes
$y_0<p\le\min(y,\lambda)$ and $D_2$ the sum over the primes $\max(y_0,\lambda)<p\le y$.

\emph{(i) Large primes.} For $\tau\in I$ and every prime $p$,
$|\cos(\tau v(p))-\cos(\tau_0v(p))|\le|\tau-\tau_0|\,v(p)\le\lambda C_v/p$. With
$M:=\max(y_0,\lambda)\ge3$, \eqref{eq:endpoint-tail-square} gives
\[
 \sup_{\tau\in I}|D_2(\tau)-D_2(\tau_0)|
 \le C_v\lambda\sum_{p>M}\frac1{p^2}
 \ll\frac{C_v\lambda}{M\log M}\le\frac{C_v}{\log y_0}.
\]

\emph{(ii) Small primes.} When $\lambda\le y_0$, $D_1=0$.
Otherwise first sum $e^{i\tau v(p)}/p$ over
$P=\{p:y_0<p\le\min(y,\lambda^{3/4})\}$. By
\eqref{eq:first-jump-spacing}, the distance of $v(p)$ from the
other frequencies in $P$ is at least $c_v/(p(p+1))$. Translating $I$
only multiplies the coefficients by unit complex numbers. Thus the
weighted mean-value inequality~\cite[Corollary~2, (1.9)]{MV74} gives
\[
 E_I\Big|\sum_{p\in P}e^{i\tau v(p)}/p\Big|^2
 \ll_v \sum_{p>y_0}p^{-2}+\pi(\lambda^{3/4})/\lambda
 \ll_v (\log y_0)^{-2}.
\]
The same bound is immediate if $|P|\le1$.

Partition the remaining primes up to $\min(y,\lambda)$ into dyadic
blocks $N<p\le2N$, truncating at either cutoff. In a nonempty block
$\lambda^{3/4}/2\le N\le\lambda$. If $t_I$ is the midpoint of $I$,
a constant times $w((\tau-t_I)/\lambda)$, $w(x)=(\sin x/x)^2$,
majorizes its indicator. The Fourier transform of $w$ is bounded
and supported in $[-2,2]$. Expanding the square and taking absolute
values bounds the normalized second moment of the block by $O(N^{-2})$
times the number of pairs with $|v(p)-v(q)|\le2/\lambda$.
By \eqref{eq:first-jump-spacing} this forces
$|p-q|\ll_v N^2/\lambda$. For large $\lambda$, the interval
length here is $\asymp_v N^2/\lambda\gg_v\sqrt\lambda$.
Brun--Titchmarsh in a short interval, with modulus one
\cite[Theorem~3.9]{MV06}, bounds the number of choices of $q$ per $p$
by $O_v(N^2/(\lambda\log\lambda))$. There are
$O(N/\log\lambda)$ choices of $p$. The normalized $L^2(I)$ norm
is therefore $O_v(\sqrt{N/\lambda}/\log\lambda)$.
The triangle inequality and $\sum_N\sqrt{N/\lambda}\ll1$ now give
$E_I|D_1|\ll_v1/\log y_0$. Bounded $\lambda$ is absorbed by enlarging
the constant. All estimates use only $c_v,C_v$ and are uniform for
truncated blocks.

\emph{(iii) Conclusion.} Taking $c=D_2(\tau_0)$ gives
$E_I|\mathcal T^v_{y_0,y}-c|\ll_v1/\log y_0$. For every real $c$,
$\frac1{|I|}\int_I|A-E_IA|\le\frac2{|I|}\int_I|A-c|$. Taking the supremum over $I$
proves \eqref{eq:euler-majorant-bmo}, with $B_v$ depending only on $c_v$ and $C_v$.

The John--Nirenberg inequality~\cite{JN61} gives absolute constants
$c_0,C_0>0$ such that, when $\|A\|_*>0$,
\[
 \frac{|\{t\in I:|A(t)-E_IA|>s\}|}{|I|}
 \le C_0e^{-c_0s/\|A\|_*}\qquad(s>0).
\]
Take $\mathfrak b=c_0/2$. If $0<\|A\|_*\le\mathfrak b$, integration of this
bound gives $|I|^{-1}\int_I e^{|A-E_IA|}\le1+C_0$; if $\|A\|_*=0$,
$A=E_IA$ almost everywhere on $I$. Thus
\eqref{eq:bmo-exponential-integrability} holds with $C_{\mathrm{JN}}=1+C_0$.
For \eqref{eq:general-exp-moment}, take $I=[-T,T]$ and
$A=b\mathcal T^v_{y_0,y}$. Then $\|A\|_*\le|b|B_v/(\log y_0)\le\mathfrak b$
by \eqref{eq:euler-majorant-bmo}. By \eqref{eq:first-jump-spacing},
$p\,v(p)\ge c_v$, so $m_v\ge c_v$; as every prime $p>y_0$ is at least $5$,
$|E_IA|=|b|\bigl|\sum_{y_0<p\le y}\sin(Tv(p))/(Tp\,v(p))\bigr|
\le|b|\pi(y)/(m_vT)\le|b|/m_v$; multiply \eqref{eq:bmo-exponential-integrability}
by $e^{E_IA}\le e^{|b|/m_v}$. Since $p\,\omega_p>1$, $m_{v_\varphi}\ge1$, and
\eqref{eq:general-exp-moment} gives \eqref{eq:cosine-exp-moment}; since
$p\,v_\sigma(p)\ge p/(p+1)\ge3/4$ at odd primes, $m_{v_\sigma}\ge3/4$, and it gives
\eqref{eq:sigma-exponential}.

\end{proof}

\paragraph*{Comparison with the general admissible-function estimate}
Following \cite[\S~1]{dlBT12}, put $u_j:=2^{2^j}$ $(j\ge0)$.
For a prime $p$ and $u\ge2$, let
$\rho_1(p,u^2)\le\rho_2(p,u^2)\le\cdots$ be the increasing rearrangement of
$\{|v(p)-v(q)|:q\le u^2\textup{ prime}\}$, and put
\[
  \mathfrak e(u):=\sum_{p>u}\frac{v(p)}p,\qquad
  \mathfrak d_p(u):=\frac{\rho_{\lfloor u\rfloor}(p,u^2)}u,
\]
\[
\begin{aligned}
 \mathfrak s^*(j,u)&:=
 \sum_{u_{j-1}<p\le\min(u_j,u)}\frac1{p^2\mathfrak d_p(u_{j-1})}
 \quad(j\ge1),\\
 \mathfrak s(u)&:=\sum_{\substack{j\ge1\\u_j\le u^2}}
 \mathfrak s^*(j,u)^{1/2},\\
 \mathfrak m(u)&:=\mathfrak e(u)\mathfrak s(u)^2.
\end{aligned}
\]
\begin{lemma}[Spacing and rates for first-jump functions]
\label{lem:f-admissible}
Let $f^{[v]}$ be the strongly additive function with $f^{[v]}(p)=v(p)$ for every
prime $p$, and define $\mathfrak e,\mathfrak d_p,\mathfrak s^*,\mathfrak s,\mathfrak m$
as above. Then $f^{[v]}$ satisfies \cite[(1.6)]{dlBT12}, its prime values $v(p)$ are
pairwise distinct, and
\begin{equation}\label{eq:admissibility-rates}
  \mathfrak e(u)\ll_v\frac1{u\log u},\qquad
  \mathfrak s(u)^2\ll_v\frac u{\log u},\qquad
  \mathfrak m(u)\ll_v\frac1{(\log u)^2}
  \qquad(u\ge2).
\end{equation}
Moreover
\begin{equation}\label{eq:delta-p-lower}
 \mathfrak d_p(w)\ge\frac{c_v}{6p^2}\qquad(w\ge6,\ w<p\le w^2).
\end{equation}
For $v=v_\varphi$ the function $f^{[v]}$ is $f$; for $v=v_\sigma$ it is the auxiliary
strongly additive function with prime values $\gsig(p)$.
\end{lemma}

\begin{proof}
\emph{(i) Condition \cite[(1.6)]{dlBT12} and distinctness.} Since
$v(p)\le C_v/p$, \eqref{eq:endpoint-tail-square} gives (trivially for $u<3$)
\[
  \sum_p\frac{\min(1,v(p))}{p}\le C_v\sum_p\frac1{p^2}<\infty,
  \qquad
  \mathfrak e(u)\le C_v\sum_{p>u}\frac1{p^2}\ll_v\frac1{u\log u},
\]
the first of which is \cite[(1.6)]{dlBT12}. Since $v$ is strictly decreasing, the
values $v(p)$ are pairwise distinct.

\emph{(ii) Spacing.} Let $w\ge6$, let $p$ be a prime with $w<p\le w^2$, and put
$s:=c_vw/(6p^2)$, so that $s<c_v/(6p)$. Let $q\ne p$ be a prime with
$|v(p)-v(q)|<s$. If $q>p$, \eqref{eq:first-jump-spacing} gives
$c_v(1/p-1/q)<s<c_v/(6p)$, hence $q<6p/5$, and then $q-p<spq/c_v=wq/(6p)<w/5$.
If $q<p$, it gives $c_v(p-q)/p^2\le c_v(1/q-1/p)<s$, hence $p-q<w/6$. Thus every
prime $q\le w^2$ with $|v(p)-v(q)|<s$, including $q=p$, lies in the open interval
$(p-w/6,\,p+w/5)$, which contains at most $11w/30+1$ integers. For $w\ge6$,
$11w/30+1\le w-1<\lfloor w\rfloor$, so fewer than $\lfloor w\rfloor$ of the numbers
$|v(p)-v(q)|$, $q\le w^2$, are smaller than $s$. Hence
$\rho_{\lfloor w\rfloor}(p,w^2)\ge s$, that is, $\mathfrak d_p(w)\ge s/w$, which is
\eqref{eq:delta-p-lower}.

\emph{(iii) Summation.} By \eqref{eq:delta-p-lower},
$p^{2}\mathfrak d_p(w)\ge c_v/6$ for $w<p\le w^{2}$, $w\ge6$.
Hence, for $u_{j-1}\ge6$ and $m_j:=\min(u_j,u)$, Chebyshev's bound gives
$\mathfrak s^{*}(j,u)\le(6/c_v)\pi(m_j)\ll_v m_j/\log m_j$; the finitely many blocks
with $u_{j-1}<6$ have finitely many positive spacing denominators and
contribute $O_v(1)$ to $\mathfrak s(u)$. Indeed, the only zero distance is the
one with $q=p$, whereas the indices are $2$ and $4$. If $J$ is maximal with $u_J\le u^{2}$,
then $u_{J-1}\le u<u_J$, so $m_J=u$ and $m_j=u_j$ for $j<J$; since
$a_j:=\sqrt{u_j/\log u_j}$ satisfies $a_{j-1}/a_j=\sqrt{2/u_{j-1}}\le1/\sqrt2$
$(j\ge2)$,
\[
  \mathfrak s(u)\ll_v1+\sum_{j<J}a_j+\sqrt{\frac u{\log u}}
  \ll_v\sqrt{\frac u{\log u}}.
\]
Together with $\mathfrak e(u)\ll_v1/(u\log u)$, this proves
\eqref{eq:admissibility-rates}; in particular $\mathfrak m$ is bounded.
\end{proof}

For $v\in\{v_\varphi,v_\sigma\}$, where moreover $0<v(p)\le\log2<1$ for every prime
$p$, the function $f^{[v]}$ is admissible in the sense of \cite[\S~1]{dlBT12} by
\Cref{lem:f-admissible}, and \cite[Lemme~2.2]{dlBT12} gives the following comparison estimate, with an absolute constant $\mathfrak A$:
\[
 \|\mathcal T^v_{y_0,y}\|_*
 \le\mathfrak A\left\{\frac1{\sqrt{\log y_0}}
       +\sup_{u\ge y_0}\mathfrak m(u)^{1/3}\right\}
 \qquad(y>y_0\ge3),
\]
and \eqref{eq:admissibility-rates} bounds the supremum by $O((\log y_0)^{-2/3})$.

\begin{remark}[Relation to {\cite[Corollaire~1.3]{dlBT12}}]
\leavevmode\phantomsection\label{rem:dlbt-cor13}\ignorespaces
Here $v=v_\varphi$, so that $f^{[v]}=f$, $c_v=1$, and
$\mathfrak d_p,\mathfrak s,\mathfrak m$ are formed with the values $\omega_p$.
Since $\omega_p\asymp p^{-1}$ and
$|\omega_p-\omega_q|\ge\min\{|p-q|,p\}/(4p^2)$ $(p\ne q)$, $f$ satisfies
the hypotheses of \cite[Corollaire~1.3]{dlBT12} with $c=1$.
Both conclusions of that corollary remain valid for the integer limiting
law of this $f$, but two intermediate estimates in its proof
(\cite[after (1.11)]{dlBT12}) require adjustment for~$f$.
Let $u$ be sufficiently large and
$p\le u^2$ prime. Every prime $q\in(p/2,u^2]$ satisfies
$|\omega_p-\omega_q|<\omega_p$ (by monotonicity if $q>p$; if $q<p$, since
$\omega_q\le\log(1+2/(p-1))<2\omega_p$), and there are at least
$\pi(u^2)-\pi(u^2/2)\gg u^2/\log u>\lfloor u\rfloor$ of them. Hence
$\mathfrak d_p(u)\le\omega_p/u<1/((p-1)u)$, which rules out the estimate
$\mathfrak d_p(u)\gg(\log u)/p$ used there for $u<p\le u^2$.
The lower bound $\mathfrak d_p(u)\gg(\log u)/p^2$ would also fail uniformly:
for $u<p\le2u$, the cap is $O(u^{-2})$, whereas that lower bound
has order $(\log u)u^{-2}$.

At
$u=e\,u_{j-1}$, this cap and \eqref{eq:delta-p-lower} give
$\mathfrak d_p(u_{j-1})\asymp1/(pu_{j-1})$ for $u_{j-1}<p\le u$, so
$\mathfrak s(u)^2\ge\mathfrak s^*(j,u)\gg u/\log u$, which rules out the
subsequent estimate $\mathfrak s(u)^2\ll u/(\log u)^2$.
The direct
verification \eqref{eq:admissibility-rates} gives
$\lambda_f:=\limsup_{u\to\infty}\mathfrak m(u)^{1/3}=0$ in the notation of
\cite{dlBT12}. Together with the tail bound in
\eqref{eq:admissibility-rates}, this recovers the upper and lower
concentration conclusions for this integer law from
\cite[Th\'eor\`emes~1.1 and~1.2]{dlBT12}, respectively.

\end{remark}

\begin{theorem}[Fourier moments and concentration for independent valuation models]
\label{thm:valuation-fourier-moments}
\index{Fourier transforms!moment estimates}
Let $v$ satisfy \eqref{eq:first-jump-hypotheses}, let $F$ be a finite set of primes,
and let $C_2\ge0$. For the primes $p\notin F$ let $K_p$ be independent random
variables with values in $\{0,1,2,\ldots\}$ such that
\begin{equation}\label{eq:valuation-mass}
 \mathbb P(K_p=0)\,\mathbb P(K_p=1)\ge\frac1p-\frac{C_2}{p^2}
 \qquad(p\notin F,\ p\ge5).
\end{equation}
Let $g(p^k)$ $(p\notin F,\ k\ge1)$ be real numbers with $g(p)=v(p)$, put
$g(p^0):=0$, and assume that $\sum_{p\notin F}g(p^{K_p})$ converges almost surely.
Let $H$ be a real random variable independent of $(K_p)_{p\notin F}$, and set
\[
 Z=H+\sum_{p\notin F}g(p^{K_p}),\qquad\nu=\operatorname{Law}(Z).
\]
\begin{enumerate}[label=\textup{(\alph*)},leftmargin=*]
\item There is $A_{v,F}\ge e$, depending only on $v$, $F$ and $C_2$, such that, with
$a_{v,F}:=(eA_{v,F})^{-1}$,
\begin{equation}\label{eq:valuation-moments}
 \frac1{2T}\int_{-T}^{T}|\widehat\nu(t)|^r\,dt
 \le\left(\frac{A_{v,F}r}{\log T}\right)^r
 \qquad(r\ge1,\ T\ge3),
\end{equation}
\begin{equation}\label{eq:valuation-fourier-tail}
 \frac1T\Bigl|\Bigl\{t\in[0,T]:|\widehat\nu(t)|>\frac K{\log T}\Bigr\}\Bigr|
 \le A_{v,F}e^{-a_{v,F} K}
 \qquad(T\ge3,\ K\ge1).
\end{equation}
If $U_T$ is uniform on $[0,T]$ and
$V_T=(\log T)|\widehat\nu(U_T)|$, then for every fixed law in
the theorem
\begin{equation}\label{eq:valuation-moment-optimality}
 \sup_{T\ge3}\|V_T\|_r\asymp_{\nu,v,F,C_2}r\quad(r\ge1),
 \qquad
 \sup_{T\ge3}\mathbb E e^{cV_T}<\infty\quad(0<c<a_{v,F}).
\end{equation}
\item With the same constant,
\begin{equation}\label{eq:valuation-concentration}
 Q_\nu(h)\le\frac{3A_{v,F}}{\log(1/h)}\qquad(0<h\le1/3).
\end{equation}
\item Suppose in addition that $H$ has an atom and that, for some $C_3>0$,
\[
 \mathbb P(K_p\ge1)\le\frac1p+\frac{C_2}{p^2},\qquad|g(p^k)|\le\frac{C_3}p
 \qquad(p\notin F,\ k\ge1).
\]
Then the series converges absolutely almost surely, and
$Q_\nu(h)\gg1/\log(1/h)$ as $h\downarrow0$, with an implied constant that may depend
on all the data, including the law of $H$.
\end{enumerate}
\end{theorem}

\begin{proof}
\textup{(a)} Put $\xi_p:=g(p^{K_p})$ for $p\notin F$. Since $g(p^0)=0$ and
$g(p)=v(p)$, $\mathbb P(\xi_p=0)\ge\mathbb P(K_p=0)$ and
$\mathbb P(\xi_p=v(p))\ge\mathbb P(K_p=1)$, so \eqref{eq:valuation-mass} gives
\eqref{eq:two-atom-mass} for $p\notin F$, $p\ge5$. With $B_v,\mathfrak b$ as in
\Cref{lem:euler-majorant-bmo}, choose $H_F\ge1$ so that
$e^{H_F}>\max(F\cup\{3\})$ and $B_v/H_F\le\mathfrak b$. For $r\ge1$ put
$L:=H_Fr$ and $y_0:=e^L$, so that $y_0>\max(F\cup\{3\})$ and
$rB_v\le\mathfrak b\log y_0$. Let $T>y_0$. The primes $y_0<p\le T$ lie outside
$F$ and are at least $5$, and $Y:=Z-\sum_{y_0<p\le T}\xi_p$ is a function of $H$ and of
the $K_p$ with $p\notin(y_0,T]$, hence independent of these $\xi_p$. So
\Cref{lem:euler-majorant}, with $y=T$, gives
$|\widehat\nu(t)|\le\mathfrak C_2\exp\{\mathcal T^v_{y_0,T}(t)-\sum_{y_0<p\le T}1/p\}$.
By \eqref{eq:mertens-dvp},
$\sum_{y_0<p\le T}1/p=\log(\log T/L)+O(L^{-1})$. Hence \eqref{eq:general-exp-moment},
with $y=T$ and $b=r$, gives
\[
 \frac1{2T}\int_{-T}^{T}|\widehat\nu(t)|^r\,dt
 \le C_{\mathrm{JN}}\mathfrak C_2^re^{r/m_v}
     \left(\frac L{\log T}\right)^r e^{O(r/L)}
 \le\left(\frac{A_{v,F}r}{\log T}\right)^r
\]
for $A_{v,F}\ge e$ sufficiently large in terms of $v$, $C_2$ and $H_F$; here
$r,L\ge1$, $C_{\mathrm{JN}}\ge1$ and $m_v\ge c_v$. Taking also $A_{v,F}\ge H_F$ covers
$3\le T\le y_0$, since then $\log T\le H_Fr$ and $|\widehat\nu|\le1$.

By evenness of $|\widehat\nu|$ and Markov's inequality, the left-hand side of
\eqref{eq:valuation-fourier-tail} is at most $(A_{v,F}r/K)^r$ for every $r\ge1$.
For $K\ge eA_{v,F}$, the choice $r=K/(eA_{v,F})$ gives $e^{-a_{v,F} K}$;
for $1\le K<eA_{v,F}$, the trivial bound $1$ is at most $A_{v,F}e^{-a_{v,F} K}$,
since $a_{v,F} K<1$ and $A_{v,F}\ge e$.

Integrating the tail proves the exponential-moment assertion. For the
lower bound, continuity at zero gives $0<\delta\le1$ with
$|\widehat\nu|\ge1/2$ on $[0,\delta]$. Taking $T=e^r$ for
$r\ge\log3$ gives $\|V_T\|_r\ge (r/2)(\delta e^{-r})^{1/r}
\gg_\nu r$. Taking $T=3$ covers the remaining bounded values of $r$.
This establishes \eqref{eq:valuation-moment-optimality}.

\textup{(b)} The kernel
$k(x)=(\sin(x/2)/(x/2))^2=\int_{-1}^1(1-|u|)e^{iux}\,du$
is at least $k(1)>0.9$ on $[-1,1]$. For $a\in\R$ and $0<h\le1/3$, applying it to
$(Z-a)/h$ gives
\[
 \mathbb P\bigl(Z\in[a,a+h]\bigr)\le\frac{\mathbb Ek((Z-a)/h)}{k(1)}
 \le\frac{h}{k(1)}\int_{-1/h}^{1/h}|\widehat\nu(t)|\,dt
 \le\frac{2A_{v,F}}{k(1)\log(1/h)},
\]
by \eqref{eq:valuation-moments} with $r=1$ and $T=1/h$. As $2/k(1)<3$, this proves
\eqref{eq:valuation-concentration}.

\textup{(c)} Under these hypotheses
$\sum_{p\notin F}\mathbb E|\xi_p|\le C_3\sum_p(p^{-2}+C_2p^{-3})<\infty$, so the
series converges absolutely almost surely. Put
$L:=\max(F\cup\{2\})+\lceil C_2\rceil+1$,
$H':=H+\sum_{p\le L,\,p\notin F}\xi_p$ and $Z_\ell:=\xi_\ell$ for primes $\ell>L$.
Then $H'$ is independent of $(Z_\ell)_{\ell>L}$ and has an atom, since $H$ has one
and the finitely many $\xi_p$ are discrete; and $Z=H'+\sum_{\ell>L}Z_\ell$. For
$\ell>L$ one has $\ell\notin F$, $\ell\ge5$ and $\ell>C_2$, so
\eqref{eq:valuation-mass} gives $\mathbb P(K_\ell=1)\ge1/\ell-C_2/\ell^2>0$ and
$\mathbb P(Z_\ell=0)\ge\mathbb P(K_\ell=0)>0$. Next,
$\sum_{\ell>z}\mathbb E|Z_\ell|\le C_3\sum_{\ell>z}(\ell^{-2}+C_2\ell^{-3})\ll z^{-1}$
for $z\ge L$. Finally, $g(\ell^0)=0$ and $g(\ell)=v(\ell)>0$ give
$\{K_\ell=1\}\subseteq\{Z_\ell\ne0\}\subseteq\{K_\ell\ge1\}$, hence
$\mathbb P(Z_\ell\ne0)=1/\ell+O(\ell^{-2})$. In particular,
with $\Theta(z):=\sum_{L<\ell\le z}-\log\mathbb P(Z_\ell=0)$,
Mertens' theorem gives $\Theta(z)=\log\log z+O(1)=o(\log z)$. Thus \Cref{prop:renyi-collapse}\textup{(b)},
with $T$ the identity, $\beta=1$ and $c_\ast=1$, gives $Q_\nu(2h)\gg1/\log(1/h)$,
which implies the assertion.
\end{proof}

\begin{corollary}[Fourier bounds with finitely many prime coordinates removed]
\label{thm:uniform-fourier-moments}
\index{Fourier transforms!moment estimates}
For every finite set $S$ of odd primes there is $A_S\ge e$ such that,
with $a_S:=(eA_S)^{-1}$,
\begin{equation}\label{eq:uniform-fourier-moments}
  \frac1{2T}\int_{-T}^{T}|\mathcal N_S(\tau)|^r\,d\tau
  \le\left(\frac{A_Sr}{\log T}\right)^r
  \qquad(r\ge1,\ T\ge3),
\end{equation}
\begin{equation}\label{eq:uniform-fourier-tail}
  \frac1T\Bigl|\Bigl\{\tau\in[0,T]:
    |\mathcal N_S(\tau)|>\frac K{\log T}\Bigr\}\Bigr|
  \le A_Se^{-a_S K}
  \qquad(T\ge3,\ K\ge1).
\end{equation}
\end{corollary}

\begin{proof}
Apply \Cref{thm:valuation-fourier-moments}\textup{(a)} with $v=v_\varphi$,
$F=S\cup\{2\}$, $K_\ell=V_\ell$ (the shifted-prime law in \eqref{eq:valuation-laws}),
$g=f$, $H=0$ and $C_2=2$. By the proof of \Cref{lem:factor}, for odd $\ell$ the
variable $f(\ell^{V_\ell})=\omega_\ell\mathbf 1_{\{V_\ell\ge1\}}$ has law $\nu_\ell$;
the series converges almost surely because its nonnegative summands have expectations
$\omega_\ell/(\ell-1)=O(\ell^{-2})$. Hence $\nu=\lambda_S$ and
$\widehat\nu=\mathcal N_S$. Further
$\mathbb P(V_\ell=0)\,\mathbb P(V_\ell=1)=(\ell-2)/(\ell(\ell-1))\ge1/\ell-2/\ell^2$.
With $A_S:=A_{v_\varphi,S\cup\{2\}}$ and $a_S=a_{v_\varphi,S\cup\{2\}}$,
\eqref{eq:valuation-moments} and \eqref{eq:valuation-fourier-tail} are
\eqref{eq:uniform-fourier-moments} and \eqref{eq:uniform-fourier-tail}.
\end{proof}

\begin{corollary}[Divisor-sum Fourier moments and concentration]
\label{thm:sigma-moments}
\index{Concentration functions!divisor-sum laws}
Let $F$ be a fixed finite set of primes containing $2$. Outside $F$,
let the independent valuations $K_p$ have either the integer law or
the shifted-prime law in \eqref{eq:valuation-laws}. Let $H$ be an
independent, almost surely finite, countably atomic real variable, and set
\[
 Z=H+\sum_{p\notin F}\gsig(p^{K_p}),\qquad \nu=\operatorname{Law}(Z).
\]
There is $A_F\ge e$, independent of the distribution of $H$, such that
\begin{equation}\label{eq:sigma-moments}
 \frac1{2T}\int_{-T}^T|\widehat\nu(t)|^r\,dt
 \le\left(\frac{A_Fr}{\log T}\right)^r
 \qquad(r\ge1,\ T\ge3).
\end{equation}
With $a_F=(eA_F)^{-1}$, the relative measure in $[0,T]$ where
$|\widehat\nu|>K/\log T$ is at most $A_Fe^{-a_F K}$ for $K\ge1$.
Moreover,
\begin{equation}\label{eq:sigma-concentration}
 Q_\nu(h)\asymp_{F,H}\frac1{\log(1/h)}\qquad(h\downarrow0).
\end{equation}
The upper constant in this last estimate is independent of $H$.
\end{corollary}

\begin{proof}
The series converges almost surely since its expected nonnegative
summands are $O(p^{-2})$. Apply \Cref{thm:valuation-fourier-moments} with
$v=v_\sigma$, $g=\gsig$ and $C_2=C_3=2$; here $\gsig(p)=v_\sigma(p)$. The products
$\mathbb P(K_p=0)\,\mathbb P(K_p=1)$ are $(p-1)^2/p^3$ for the integer law and
$(p-2)/(p(p-1))$ for the shifted-prime law, and both are at least $1/p-2/p^2$. Put
$A_F:=A_{v_\sigma,F}$; it does not depend on the law of $H$. Part~\textup{(a)} gives
\eqref{eq:sigma-moments} and the tail estimate, and part~\textup{(b)} gives the upper
bound in \eqref{eq:sigma-concentration}, with the constant $3A_F$. For the lower
bound, $\mathbb P(K_p\ge1)$ is $1/p$ or $1/(p-1)$, both at most $1/p+2/p^2$;
$0\le\gsig(p^k)\le\omega_p\le2/p$ for $k\ge1$; and the countably atomic variable $H$
has an atom. Part~\textup{(c)} gives the lower bound.
\end{proof}

For the shifted-totient logarithmic law and its finite deletions,
sharp fixed-order moment asymptotics are given in
\Cref{thm:fourier-amplitude-law,cor:deleted-amplitude}; the sharp
concentration asymptotic for $\muF$ is in \Cref{thm:sharp-global-modulus}.
For $3\notin S$, \eqref{eq:uniform-fourier-moments} follows from the case
$S=\varnothing$, as $|\widehat{\nu_\ell}|\ge(\ell-3)/(\ell-1)$; the case $3\in S$,
used in \Cref{cor:deleted-amplitude}, does not, since $\widehat{\nu_3}$ has zeros.

\paragraph*{Interval widths and their uses}
The three concentration functions below maximize different interval masses.
\begin{center}
\begin{tabularx}{\linewidth}{@{}p{.20\linewidth}p{.32\linewidth}X@{}}
\toprule
Quantity & Interval & Local use \\
\midrule
$Q_G(\varepsilon)$ & $(t(1-\varepsilon),t]$ in the raw variable
 & Relative windows and \Cref{thm:sharp-global-modulus}. \\
$Q_{\mu_f}(h)$ & $[a,a+h]$ in the logarithmic variable
 & \raggedright\arraybackslash Progression perturbations in \Cref{lem:ap-uniform-concentration}
 and the transfer to $c(p)$ in \Cref{thm:sharp-prime-convergence}. \\
$C(h)$ & $(t-h,t]$ in the raw variable
 & Additive perturbations and shortest intervals in
 \Cref{cor:sharp-additive-modulus,cor:shortest-mass-intervals}. \\
\bottomrule
\end{tabularx}
\end{center}
For $U_T$ uniform on $[0,T]$, \Cref{thm:uniform-fourier-moments}
implies uniform integrability of every fixed positive power of
$(\log T)|\FT(U_T)|$ as $T\to\infty$.
This supplies the positive-moment input to \Cref{thm:fourier-amplitude-law};
inverse moments are treated in \Cref{lem:totient-inverse-moments}.
\Cref{fig:fourier-questions} shows two uses of \Cref{thm:uniform-fourier-moments}
in \Cref{ssec:typical-decay}: exceptional sets and moment convergence.
The arithmetic CDF upper bound uses logarithmic concentration, while
its matching lower bound uses endpoint mass
(\Cref{thm:sharp-prime-convergence}).

Since $\muF$ is the atomless law of $-\log X$
(\Cref{lem:factor,thm:phi-dist,thm:pure-type}), the change of variables
$a=-\log t$ gives
\begin{equation}\label{eq:modulus-transport}
  Q_G(\varepsilon)=Q_{\muF}(h)\quad\bigl(h>0,\ \varepsilon:=1-e^{-h}\bigr),
  \qquad\tfrac12h\le\varepsilon\le h\quad(0<h\le1).
\end{equation}
For $t>0$ and $h>0$, the exact interval correspondence is
\[
 \{te^{-h}<X\le t\}
 =\{-\log t\le-\log X<-\log t+h\},
 \qquad \varepsilon=1-e^{-h}.
\]
Thus the raw interval $(te^{-h},t]$ becomes
$[-\log t,-\log t+h)$; atomlessness permits either endpoint convention
in the concentration function. The decreasing map sends the raw upper
endpoint $t$ to the logarithmic lower endpoint $a=-\log t$, explaining
the opposite localization signs in \Cref{thm:sharp-global-modulus}.
For $t\le0$ the raw event has probability zero. The bounds in
\eqref{eq:modulus-transport} follow from
$1-e^{-h}\le h$ and $1-e^{-h}\ge h-h^2/2$.

\begin{remark}[Fourier proof of the upper bound for independent sums]
\label{rem:mixed-logarithmic-concentration}
Use the all-logarithmic setup of
\Cref{cor:independent-sum-concentration}, including its separate
finite-coordinate conditioning. Write $Y_1,\ldots,Y_k$ for the summands,
$\nu_i=\operatorname{Law}(Y_i)$ and
$\nu=\operatorname{Law}(\sum_i a_iY_i)$; the fixed translation does not
affect concentration. Each conditioning leaves an
independent countably atomic finite-coordinate head and an unchanged
prime tail. Hence \Cref{thm:valuation-fourier-moments}\textup{(a)}, with
$v=v_\varphi$ for the two totient laws and $v=v_\sigma$ for the two divisor-sum
laws, $F$ consisting of $2$ and the conditioned primes, $C_2=2$ and that head as $H$,
bounds every fixed positive Fourier moment of $\nu_i$. Thus Esseen's
inequality~\cite{Ess68} and H\"older's inequality give
\[
 Q_\nu(h)\ll h\prod_{i=1}^k
 \left(\int_{-1/h}^{1/h}|\widehat\nu_i(a_it)|^k\,dt\right)^{1/k}
 \ll\log^{-k}(1/h).
\]
For the unconditioned law $\muF$, write $Y_i=\log2+W_i$ with
$W_i\overset d=W\ge0$ as in
\Cref{prop:endpoint-half}, and put $A:=\sum_i|a_i|$.
On $E_h:=\{W_i\le h/A\ (1\le i\le k)\}$ the sum lies in a fixed
interval of length $h$, and independence and
\eqref{eq:endpoint-W-quantitative} give
$\mathbb P(E_h)\asymp_{a_1,\ldots,a_k}\log^{-k}(1/h)$.
If every $a_i>0$ and $b:=(\sum_i a_i)\log2$, then
\[
 \nu([b,b+h])=\frac{\kappa_{\mathrm{end}}^k}{\log^k(1/h)}
 +O_{a_1,\ldots,a_k}\bigl(\log^{-k-1}(1/h)\bigr),
\]
since
$E_h\subseteq\{b\le\sum_i a_iY_i\le b+h\}
\subseteq\{W_i\le h/a_i\ (1\le i\le k)\}$
and \eqref{eq:endpoint-W-quantitative} expands both outer probabilities.
\end{remark}

\section{Uniform lower bounds for short windows}
\label{ssec:uniform-window-lower}

The following result uses the explicit inverse-jump estimates
\eqref{eq:window-inverse-bounds} and the short-prime-interval input
\eqref{eq:window-prime-count}. The first-jump assumptions
\eqref{eq:first-jump-hypotheses} alone give an upper bound on the
absolute inverse derivative, not the lower bound used here.

\begin{lemma}[A lower bound for primes in short intervals]
\label{lem:short-prime-window-lower}
Fix $21/40<\theta<1$. For all sufficiently large $a$ and all $b>a$
with $b-a\ge2b^\theta$,
\begin{equation}\label{eq:window-prime-count}
 \pi(b)-\pi(a)\gg_\theta\frac{b-a}{\log b}.
\end{equation}
\end{lemma}

\begin{proof}
Baker--Harman--Pintz~\cite[p.~562]{BHP01} prove, for all sufficiently large $x$,
\[
 \pi(x+x^{21/40})-\pi(x)\ge\frac9{100}\frac{x^{21/40}}{\log x}.
\]
Tile $(a,b]$ forwards by intervals $(y,y+y^{21/40}]$ until the next
tile would cross $b$. The omitted length is less than $b^{21/40}$,
so the complete tiles account for at least $(b-a)/2$. Sum the
displayed lower bound over these disjoint tiles, using $\log y\le\log b$.
\end{proof}

\begin{theorem}[Uniform windows after finitely many prime deletions]
\label{thm:cofinite-window-lower}
\index{Short windows!uniform lower bounds}
Let $\mathcal P$ be a set of odd primes containing all sufficiently large
primes, let $g\in\{f,\gsig\}$, and let the independent $K_\ell$
have the valuation laws of $\mathsf M\in\{\mathsf I,\mathsf B\}$ in
\eqref{eq:valuation-laws}. Put
$W_{\mathcal P}:=\sum_{\ell\in\mathcal P}g(\ell^{K_\ell})$.
For every $k>0$ and $C>1/(2\log(59/40))$, there is
$\eta_0=\eta_0(\mathcal P,k,C,g,\mathsf M)>0$ such that
\begin{equation}\label{eq:cofinite-window-lower}
 \mathbb P\{w\le W_{\mathcal P}<w+\eta\}
 \ge \eta\exp\{-C(\log\log(1/\eta))^2\}
 \qquad(0\le w\le k,\ 0<\eta<\eta_0).
\end{equation}
In the shifted-prime totient case, for every fixed $0<\delta<1/2$
and all sufficiently small $\varepsilon>0$,
\begin{equation}\label{eq:uniform-relative-window-lower}
 G(t)-G(t(1-\varepsilon))
 \ge\varepsilon\exp\{-C(\log\log(1/\varepsilon))^2\}
 \qquad(\delta\le t\le1/2).
\end{equation}
The threshold in the last assertion depends on $\delta$ and $C$.
\end{theorem}

\begin{proof}
Choose $21/40<\theta<1$ with $1/(2\log(2-\theta))<C$, and put
$\rho=(2-\theta)^{-1}$ and $L=\log(2-\theta)$. We use
\eqref{eq:window-prime-count} with this value of $\theta$.

Write $a_S=\sum_{\ell\in S}g(\ell)$ and put
\[
 \mathfrak w_{01}(S)=\prod_{\ell\in S}
       \frac{\mathbb P(K_\ell=1)}{\mathbb P(K_\ell=0)}.
\]
The individual odds are $1/\ell$ for $\mathsf I$ and
$(\ell-1)/(\ell(\ell-2))$ for $\mathsf B$, both at least $1/\ell$.
Define the finite sum
\[
 \mathfrak N(w,\eta):=
 \sum_{S\subset\mathcal P\cap[1,1/(6\eta))}
 \mathfrak w_{01}(S)\,\mathbf1_{\{w\le a_S<w+\eta\}}.
\]
We first estimate this deterministic sum; the zero-valuation
probability is inserted at the end. The inverse $\psi=\psi_g$ of the
first-jump function satisfies, for $0<v\le1/2$,
\begin{equation}\label{eq:window-inverse-bounds}
\begin{gathered}
 \psi_f(v)=(1-e^{-v})^{-1},\qquad
 \psi_{\gsig}(v)=(e^v-1)^{-1}=\psi_f(v)-1,\\
 \frac1v\le\psi_f(v)\le\frac1v+1,\qquad
 -\psi_f'(v)=-\psi_{\gsig}'(v)\ge\frac1{2v^2}.
\end{gathered}
\end{equation}

First we need a finite seed. For each $k$ there is $N_1$ such that,
for $N\ge N_1$ and $3/N\le w\le k$, some
$S\subset\mathcal P\cap[1,N)$ satisfies
\begin{equation}\label{eq:window-seed}
 w\le a_S<w+1/(10N).
\end{equation}
To prove this, choose $p_0$ beyond the deleted primes so that consecutive
primes $q<q'<q''$ above $p_0$ have $q''<1.21q$; the prime number theorem
permits this choice. Thus $g(q)<g(q')+g(q'')$.
For a decreasing finite sequence $a_1>\cdots>a_n>0$ satisfying
$a_i<a_{i+1}+a_{i+2}$ for $i\le n-2$, greedy approximation from below to a target smaller than
$\sum a_i$ leaves a defect less than $a_{n-1}$: if the last rejected
index were at most $n-2$, its next two terms could not both have been
accepted. Apply this to the primes in $[p_0,N/4)$ with target $w-3/N$.
Their sum exceeds $k$ for large $N$, and their second largest prime exceeds
$N/5$. The defect from $w$ therefore lies in $[3/N,8.1/N)$.
Adjoin distinct primes in $(3N/4,0.99N)$ to the selected set, subtracting
their jumps from the defect until it is at most $3.5/N$.
Each such jump lies in $(1/N,1.4/N)$; at most five suffice,
and the remaining defect $d$ lies in $(2.1/N,3.5/N]$.
By \eqref{eq:window-inverse-bounds}, the interval
$(\psi(d+1/(10N)),\psi(d)]$ lies in $(N/4,N/2)$ and has length at least $N/260$.
The prime number theorem supplies a prime there, uniformly for these $d$.
Adjoining this prime proves \eqref{eq:window-seed}; the three prime ranges are
disjoint.

For small $\eta$, put $P=P(\eta)=(1000/\eta)^\rho$ and
$h(\eta)=1/(2P)$. We claim that
\begin{equation}\label{eq:window-recursion}
 \mathfrak N(w,\eta)\ge
 c\,\frac{\eta P}{\log P}\,
 \mathfrak N(w-2/P,1/(2P)),
\end{equation}
where $c>0$ depends only on $\theta$. A set $S'$ counted on the right
uses only primes below $P/3$ and has defect
$d=w-a_{S'}\in(3/(2P),2/P]$. The primes with
$d\le g(\ell)<d+\eta$ lie in
$(\psi(d+\eta),\psi(d)]\subset(P/3,P)$.
Its length is at least $\eta P^2/10=100P^\theta$, so
\eqref{eq:window-prime-count} supplies $\gg_\theta\eta P^2/\log P$
primes. Each has odds at least $1/\ell\ge1/P$. Moreover
$P<1/(6\eta)$ for small $\eta$, so $S'\cup\{\ell\}$ is counted on the
left. This map is injective: $\ell$ is its unique prime at least $P/3$.
Since $\mathfrak w_{01}(S'\cup\{\ell\})\ge\mathfrak w_{01}(S')/P$,
summing proves \eqref{eq:window-recursion}.

There is also a terminal estimate
\begin{equation}\label{eq:window-singleton}
 \mathfrak N(w,\eta)\gg_\theta\frac{\eta P}{\log P}
 \qquad(12\eta\le w<8/P).
\end{equation}
In fact, count singletons with primes in $(\psi(w+\eta),\psi(w)]$.
This interval lies in $(P/10,1/(6\eta))$, has length at least
$0.4\eta/w^2$, and satisfies the hypothesis of
\eqref{eq:window-prime-count}: if $b=\psi(w)$, then $b\le2/w$ and
\[
 \frac{0.4\eta}{w^{2-\theta}}
 >\frac{400}{8^{2-\theta}}>4\ge2^{1+\theta}.
\]
The resulting sum of odds is
$\gg_\theta\eta/(w\log(2/w))\gg_\theta\eta P/\log P$.
The last inequality follows because $1/(w\log(2/w))$ decreases with $w$
in this range. All primes used in these two estimates belong to
$\mathcal P$ once $\eta$ is sufficiently small.

Choose $\eta_*>0$ small enough for both estimates to hold up to
$h(\eta_*)$, for $1/(6h(\eta_*))\ge N_1$, and for
$\eta P(\eta)<1/100$ throughout that range.
Starting from $\eta_0<\eta_*$ and $12\eta_0\le w_0\le k$, iterate
\[
 \eta_{j+1}=h(\eta_j),\qquad w_{j+1}=w_j-2/P(\eta_j),
\]
until either $w_J<8/P(\eta_J)$ or $\eta_J\ge\eta_*$.
The invariant $12\eta_j\le w_j\le k$ follows from
$w_j\ge8/P(\eta_j)$ at each nonterminal step.
In the first terminal case use \eqref{eq:window-singleton}.
In the second case $\eta_*\le\eta_J\le h(\eta_*)$ and
$w_J\ge8/P(\eta_J)\ge18\eta_J$, so the seed with
$N=1/(6\eta_J)$ applies. Its primes lie below $1/(6\eta_*)$; hence its
weight has a positive lower bound depending only on
$\mathcal P,k,\theta,g,\mathsf M$.
In either case,
\[
 \mathfrak N(w_J,\eta_J)\ge
 \kappa\frac{\eta_J}{\log P(\eta_J)}
\]
for some fixed $\kappa>0$.

Put $z_j=\log(1/\eta_j)$ and
$z_*=(\rho\log1000+\log2)/(1-\rho)$. Then
$z_j-z_*=\rho^j(z_0-z_*)$. Decrease $\eta_*$ further so that
$\log(1/h(\eta_*))-z_*\ge1$.
It follows that $J\le(\log z_0)/L$ and
$\log P(\eta_j)\le A\rho^jz_0$ for $0\le j\le J$, with $A$ fixed.
The factors in \eqref{eq:window-recursion} telescope, since
$\eta_jP(\eta_j)=\eta_j/(2\eta_{j+1})$. Therefore
\[
 \mathfrak N(w_0,\eta_0)
 \ge\kappa\eta_0(c/2)^J
       \prod_{j=0}^J(\log P(\eta_j))^{-1}.
\]
Writing $n=J+1$ gives
\[
 \sum_{j=0}^J\log\log P(\eta_j)
 \le n\log z_0-\frac L2n(n-1)+n\log A
 \le\frac{(\log z_0)^2}{2L}+O_\theta(\log z_0).
\]
Consequently, uniformly for $12\eta\le w\le k$,
\begin{equation}\label{eq:window-weighted-lower}
 \mathfrak N(w,\eta)\ge
 \kappa'\eta\exp\left\{-\frac{(\log\log(1/\eta))^2}{2L}
                  -O_\theta(\log\log(1/\eta))\right\}.
\end{equation}

We now pass to probabilities. For finite sets $S\subseteq R\subseteq\mathcal P$,
\[
 \mathbb P\{K_\ell=\mathbf1_{\{\ell\in S\}}\ (\ell\in R)\}
 =\prod_{\ell\in R}\mathbb P(K_\ell=0)\,\mathfrak w_{01}(S).
\]
Mertens' theorem gives
$\prod_{\ell\in\mathcal P,\,\ell\le M}\mathbb P(K_\ell=0)
\gg_{\mathcal P,\mathsf M}1/\log M$.
Also the independent tail $T_M=\sum_{\ell\in\mathcal P,\,\ell>M}
g(\ell^{K_\ell})$ satisfies
$\mathbb ET_M\le\sum_{n>M}(n-1)^{-2}\le2/M$ for large $M$, since
$g(\ell^{K_\ell})\le\omega_\ell\mathbf1_{\{K_\ell>0\}}$.
Take $M=16/\eta$. Every set counted by $\mathfrak N(w,\eta/2)$ lies
below $M$; requiring valuation zero at all other primes in $\mathcal P$
up to $M$ and
$T_M<\eta/2$ gives disjoint events in $[w,w+\eta)$.
Markov's inequality and independence therefore give
\begin{equation}\label{eq:window-probability-transfer}
 \mathbb P\{w\le W_{\mathcal P}<w+\eta\}
 \gg_{\mathcal P,\mathsf M}\frac{\mathfrak N(w,\eta/2)}{\log(1/\eta)}.
\end{equation}
For $w\ge6\eta$, \eqref{eq:window-weighted-lower} applies.

If $0\le w<6\eta$, let $u=\max(w,\eta/4)$ and require exactly one
prime in $\mathcal P$ up to $M=32/\eta$ to have valuation one with
jump in $[u,u+\eta/4)$, all others to have valuation zero, and
$T_M<\eta/4$.
These events lie in $[w,w+\eta)$. The admissible primes occupy
$(\psi(u+\eta/4),\psi(u)]$, an interval of length $\gg1/\eta$ with endpoints
$\asymp1/\eta$. The prime number theorem gives total odds
$\gg1/\log(1/\eta)$, uniformly in $w$. The same inactivity and tail
estimates give probability $\gg_{\mathcal P,\mathsf M}\log^{-2}(1/\eta)$.
Combining both cases and using $1/(2L)<C$ proves
\eqref{eq:cofinite-window-lower}.
Finally, in the shifted-prime totient case $X=\frac12e^{-W}$, and a
relative left window corresponds to
$[w,w-\log(1-\varepsilon))$, where $w=\log(1/(2t))$.
It contains $[w,w+\varepsilon)$; applying the proved estimate with width
$\varepsilon$ proves \eqref{eq:uniform-relative-window-lower}.
\end{proof}

\section{Reduction to endpoint windows}
\label{ssec:global-modulus-deletion}

The comparison changes a finite prime configuration while leaving the
remaining coordinates fixed. Outside exceptional events of controlled
total probability, all surviving points in a short window have the
same small-prime configuration. Setting its active coordinates to zero
turns the surviving event into one in a window near $1/2$; the
probability ratio is $\mathcal K(S)$, the product of the activation
odds. The deletion bound below makes this comparison uniform in the
window location. Its error must be controlled more finely than the
leading inverse-logarithmic term, because the two largest anchor
weights coincide.

The arguments adapt Toulmonde's integer-law methods as follows.
\begin{itemize}[leftmargin=*]
\item Uniform deletion, \Cref{lem:shifted-uniform-deletion}:
  \cite[Lemmes~6,~8,~9]{Tou06}.
\item Anchor asymptotics, \Cref{thm:anchor-sharp}:
  \cite[Th\'eor\`eme~1 and Remarque~2]{Tou06}.
\item The prime-$3$ comparison, \Cref{lem:prime-three-split}\textup{(c)}:
  \cite[Lemme~5]{Tou06}.
\item Fixed non-anchor decay, \Cref{cor:non-anchor-decay}\textup{(a)}:
  the remark after \cite[Lemme~9]{Tou06}.
\item Two-window localization, \Cref{prop:two-window}\textup{(d)}:
  \cite[\S~7.1.1]{Tou06}.
\item Endpoint comparisons, \cite[\S\S~7.1.2--7.1.3]{Tou06}:
  \Cref{prop:endpoint-window}\textup{(b)} corresponds to \S~7.1.2,
  and \Cref{lem:endpoint-small-comparison} to \S~7.1.3.
\end{itemize}
The anchor asymptotic below is uniform over an explicit growing family
of shifted-prime anchors; Toulmonde's Th\'eor\`eme~1 gives growing-family
uniformity for integer-law anchors.

For $0<\varepsilon<e^{-1}$ and $c>0$, put
\[
\begin{aligned}
 \delta&:=\varepsilon^{2/5},\qquad
 \Xi:=\exp\sqrt{z\log z/20},\\
 \mathcal R_c(\varepsilon)&:=e^{-c\sqrt{z\log z}},\qquad
 \mathcal R(\varepsilon):=\mathcal R_{1/5}(\varepsilon),
\end{aligned}
\]
and define
\[
 \begin{aligned}
 Q_{\mathrm{end}}(\varepsilon)&:=
 \sup_{\frac12(1-\delta)\le t\le\frac1{2(1-\varepsilon)}}q_\varepsilon(t).
 \end{aligned}
\]
Extend the endpoint mass $H_{\mathrm{end}}$ of \Cref{sec:intro} to all $s\in\R$
by $H_{\mathrm{end}}(s):=\mathbb P\bigl(X>\tfrac12(1-s)\bigr)$.
Since $X\le\tfrac12$ almost surely, $H_{\mathrm{end}}=0$ on $(-\infty,0]$
and, for $\theta<1$ and $0<\varepsilon<1$,
\begin{equation}\label{eq:q-H-identity}
 q_\varepsilon\bigl(\tfrac12(1-\theta)\bigr)
 =H_{\mathrm{end}}\bigl(\theta+\varepsilon(1-\theta)\bigr)-H_{\mathrm{end}}(\theta);
\end{equation}
in particular $q_\varepsilon(\tfrac12)=H_{\mathrm{end}}(\varepsilon)$.
By \eqref{eq:hend-asymp}, for fixed $b\ge1$, uniformly for $0<s\le s'\le bs$,
\begin{equation}\label{eq:hend-increment}
 0\le H_{\mathrm{end}}(s')-H_{\mathrm{end}}(s)
 \ll_b\frac1{\log^2(1/s)}\qquad(s\downarrow0).
\end{equation}
Since the factor $1/(\ell-2)$ of $\mathcal K(S)$ equals $1$ at $\ell=3$ and
is at most $\tfrac13$ for $\ell\ge5$,
\begin{equation}\label{eq:weight-trichotomy}
 \mathcal K(S)=1\iff S=\varnothing\ \textup{or}\ S=\{3\},\qquad
 \mathcal K(S)\le\tfrac13\ \textup{for every other }S.
\end{equation}
For every fixed $A>0$ and every fixed $0<d_0\le d_1<\infty$,
\begin{equation}\label{eq:residual-properties}
 \mathcal R(\varepsilon)=o(z^{-A}),\qquad
 \sup_{d_0\le d\le d_1}
 \left|\frac{\mathcal R(d\varepsilon)}{\mathcal R(\varepsilon)}-1\right|
 \longrightarrow0\qquad(\varepsilon\downarrow0).
\end{equation}
These follow from $\sqrt{z\log z}/\log z\to\infty$ and
$(\sqrt{z\log z})'=(\log z+1)/(2\sqrt{z\log z})\to0$, because
$\varepsilon\mapsto d\varepsilon$ is the bounded shift $z\mapsto z-\log d$.

The deletion mechanism also applies to bounded prime counts with controlled
successive odds. The following formulation uses the same canonical jumps.

\begin{lemma}[Uniform deletion for bounded sparse prime counts]
\label{lem:sparse-count-uniform-deletion}
Fix an integer $m\ge1$, $C>0$ and $A>0$. For every prime $p$, let
$(\pi_p(0),\ldots,\pi_p(m))$ be a probability vector with
\[
 \pi_p(0)>0,\qquad \pi_p(j)\le\frac Cp\pi_p(j-1)
       \quad(1\le j\le m).
\]
Let the independent $J_p$ have these probabilities, put
$\omega_p=\log(p/(p-1))$, and define
\[
 R=A\exp\left(-\sum_pJ_p\omega_p\right),\qquad
 P_\varepsilon(t)=\mathbb P\{t(1-\varepsilon)<R\le t\}.
\]
Fix $0<2\alpha<\beta<1/2$ and $0<c<\sqrt{\alpha/2}$, and put
\[
 z=\log(1/\varepsilon),\quad L=\sqrt{\alpha z\log z/2},\quad
 \Xi_*=e^L,\quad\delta_*=\varepsilon^\beta,\quad
 K(\mathbf j)=\prod_p\frac{\pi_p(j_p)}{\pi_p(0)}.
\]
For sufficiently small $\varepsilon$, uniformly for
$0<t<A/(1-\varepsilon)$, either
$P_\varepsilon(t)\ll\exp\{-c\sqrt{z\log z}\}$, or there is a unique
vector $\mathbf j\in\{0,\ldots,m\}^{\{p\le\Xi_*\}}$, extended by zero,
with positive coordinate probabilities, such that
\[
 D_{\mathbf j}:=\prod_pp^{j_p}<\varepsilon^{-\alpha},\qquad
 t(1-\varepsilon)<a_{\mathbf j}:=
 A\prod_p(1-1/p)^{j_p}<\frac{t}{1-\delta_*}.
\]
In the latter case,
\begin{equation}\label{eq:sparse-count-uniform-deletion}
 \begin{aligned}
 P_\varepsilon(t)&\le K(\mathbf j)P_\varepsilon(At/a_{\mathbf j})
                  +O(e^{-c\sqrt{z\log z}}),\\
 A(1-\delta_*)&<At/a_{\mathbf j}<A/(1-\varepsilon).
 \end{aligned}
\end{equation}
Outside an event of the displayed error probability, every point of the
original window has this head vector and has zero coordinates for
$\Xi_*<p<\varepsilon^{-\rho}$, where $\rho=(\beta+1/2)/2$.
The constants and threshold depend only on the fixed parameters.
\end{lemma}

\begin{proof}
Iteration of the odds bound gives $\pi_p(j)\le(C/p)^j$.
For $p\ge2C$, therefore,
$\mathbb P(J_p\ge r)\le2(C/p)^r$ for $1\le r\le m$.
Choose $A_*=2^{m+2}\max(C,1)$. For $p\ge2A_*$, if
$B_p\sim\operatorname{Bin}(m,A_*/p)$, then
\[
 \mathbb P(B_p\ge r)\ge(A_*/p)^r(1-A_*/p)^{m-r}
 \ge2^{-m}(A_*/p)^r\ge2(C/p)^r.
\]
Thus $J_p$ is stochastically dominated by $B_p$ for all such $p$;
in particular, for $v\ge1$,
\begin{equation}\label{eq:sparse-count-pgf-bound}
 \mathbb E v^{J_p}\le\left(1+\frac{A_*}{p}(v-1)\right)^m.
\end{equation}
The finitely many smaller primes have $J_p\le m$.
Also $\mathbb E J_p\ll_{m,C}p^{-1}$ for large $p$, so the
nonnegative defining series converges almost surely.

Put $T=\varepsilon^{-\rho}$. For the head product
$D=\prod_{p\le\Xi_*}p^{J_p}$ choose $s=\log z/(2L)$.
Then $\Xi_*^s=\sqrt z$, $\alpha sz=L$, and the finitely many
primes below $2A_*$ contribute $O_{m,C}(s)$ to $\log\mathbb ED^s$.
For the others, \eqref{eq:sparse-count-pgf-bound}, convexity on
$[0,L]$ and Mertens' first theorem~\cite{Mer74} give
\[
 \log\mathbb ED^s\le O_{m,C}(s)
  +mA_*\sum_{p\le\Xi_*}\frac{p^s-1}{p}
 \le O_{m,C}(s)+\frac{mA_*(\sqrt z-1)}L
          \sum_{p\le\Xi_*}\frac{\log p}{p}
 \ll_{m,C}\sqrt z.
\]
Consequently $\mathbb P(D\ge\varepsilon^{-\alpha})
\le\exp\{-L+O_{m,C}(\sqrt z)\}
\ll e^{-c\sqrt{z\log z}}$.

Partition the window event having a positive middle coordinate
$\Xi_*<p<T$ by its least such prime, and decrement that coordinate.
For each fixed $p$ the map is injective, and its probability ratio is
at most $C/p$; zero-probability coordinates are omitted. Images for
different primes $p<p'$ are disjoint: two preimages of a common image
would have products with ratio
\[
 \frac{1-1/p'}{1-1/p}
 =1+\frac{p'-p}{p'(p-1)}>1+T^{-2}>(1-\varepsilon)^{-1}
\]
for small $\varepsilon$, since $2\rho<1$.
Summing the image probabilities bounds this entire part by $C/\Xi_*$.
For $\Pi=\prod_{p\ge T}(1-1/p)^{J_p}$,
\[
 \mathbb E[-\log\Pi]\ll_{m,C}T^{-1},\qquad
 \mathbb P(\Pi\le1-\delta_*)\ll_{m,C}\varepsilon^{\rho-\beta}.
\]
Both errors are $O(e^{-c\sqrt{z\log z}})$.

Each remaining point has $R=a_{\mathbf j}\Pi$, with
$D_{\mathbf j}<\varepsilon^{-\alpha}$ and $1-\delta_*<\Pi\le1$.
Its anchor lies in the stated window, of length
$O_A(\varepsilon^\beta)$. Distinct vectors give distinct rational
products: at the largest prime where their exponents differ, the
prime valuation of their ratio is nonzero. With denominators
$D_{\mathbf j},D_{\mathbf j'}$ this gives
\[
 |a_{\mathbf j}-a_{\mathbf j'}|
 \ge A/(D_{\mathbf j}D_{\mathbf j'})>A\varepsilon^{2\alpha}.
\]
Since $\beta>2\alpha$, at most one vector is admissible.
If no points remain, the first alternative holds. Otherwise remove
the common head vector. Independence and $\pi_p(0)>0$ give the exact
probability ratio $K(\mathbf j)$ between the surviving event and its
image. That image lies in the relative window ending at
$At/a_{\mathbf j}$. Adding the discarded probability proves the
inequality and the anchor bounds give its second line.
\end{proof}

For binomial prime counts, the successive odds give the following
specialization.

\begin{lemma}[Uniform deletion for binomial prime coordinates]
\label{lem:binomial-uniform-deletion}
Fix an integer $k\ge1$, a finite set $E$ of odd primes, $A>0$, and
\[
 0<2\alpha<\beta<\tfrac12,
 \qquad 0<c<\sqrt{\alpha/2}.
\]
For odd primes $p\notin E$, let $J_p$ be independent with distribution
$\operatorname{Bin}(k,1/(p-1))$, and put
\[
 R=A\exp\left(-\sum_{p\notin E}J_p\omega_p\right),\qquad
 P_\varepsilon(t)=\mathbb P\{t(1-\varepsilon)<R\le t\}.
\]
All prime sums and products in this lemma are over odd primes outside $E$.
For $z=\log(1/\varepsilon)>1$, set
\[
 L=\sqrt{\alpha z\log z/2},\qquad
 \Xi_* =e^L,\qquad \delta_* =\varepsilon^\beta.
\]
For all sufficiently small $\varepsilon>0$, uniformly for
$0<t<A/(1-\varepsilon)$, either
$P_\varepsilon(t)\ll\mathcal R_c(\varepsilon)$, or there is a unique
vector $\mathbf j=(j_p)$, with $j_p\in\{0,\ldots,k\}$, supported on $p\le\Xi_*$,
such that
\[
 D_{\mathbf j}:=\prod_p p^{j_p}<\varepsilon^{-\alpha},\qquad
 t(1-\varepsilon)<a_{\mathbf j}:=
 A\prod_p(1-1/p)^{j_p}<\frac{t}{1-\delta_*}.
\]
In the latter case, with
$K_k(\mathbf j):=\prod_p\binom{k}{j_p}(p-2)^{-j_p}$,
\begin{equation}\label{eq:binomial-uniform-deletion}
\begin{aligned}
 P_\varepsilon(t)
 &\le K_k(\mathbf j)P_\varepsilon(At/a_{\mathbf j})
       +O\bigl(\mathcal R_c(\varepsilon)\bigr),\\
 A(1-\delta_*)&<\frac{At}{a_{\mathbf j}}<\frac{A}{1-\varepsilon}.
\end{aligned}
\end{equation}
The thresholds and implied constants may depend on the fixed parameters.
\end{lemma}

\begin{proof}
Apply \Cref{lem:sparse-count-uniform-deletion} with $m=k$, $C=3k$,
zero counts at $2$ and at the primes in $E$, and the stated binomial
counts elsewhere. At an odd prime $p\notin E$,
\[
\begin{aligned}
 \frac{\pi_p(j)}{\pi_p(j-1)}
 &=\frac{k-j+1}{j(p-2)}\le\frac{3k}{p}
 \quad(1\le j\le k),\\
 \frac{\pi_p(j)}{\pi_p(0)}&=\binom{k}{j}(p-2)^{-j}.
\end{aligned}
\]
Thus $K(\mathbf j)=K_k(\mathbf j)$, and the general lemma gives
\eqref{eq:binomial-uniform-deletion} uniformly in $t$, with the
stated parameter ranges. Its proof also gives the surviving event
used below: outside a set of probability $O(\mathcal R_c(\varepsilon))$,
every point of the window has the same head vector $\mathbf j$,
zero coordinates for $\Xi_*<p<\varepsilon^{-\rho}$, and
\[
 R=a_{\mathbf j}\Pi,\qquad
 \Pi=\prod_{p\ge\varepsilon^{-\rho}}(1-1/p)^{J_p}
       \in(1-\delta_*,1],
\]
where $\rho=(\beta+1/2)/2$.
\end{proof}

\begin{lemma}[Removing small-prime factors uniformly over interval locations]
\label{lem:shifted-uniform-deletion}
Fix $0<c<1/\sqrt{20}$. For all sufficiently small $\varepsilon>0$
and every real $t$ with $0<t<1/\bigl(2(1-\varepsilon)\bigr)$,
one of the following holds, with the threshold and implied constants
depending only on $c$:
\begin{enumerate}[label=\textup{(\roman*)},leftmargin=*]
\item $q_\varepsilon(t)\ll_c\mathcal R_c(\varepsilon)$;
\item there is a unique finite set $S=S(t,\varepsilon)$ of odd primes
$\ell\le\Xi$ with $D_S<\varepsilon^{-1/10}$ such that
\begin{equation}\label{eq:deletion-anchor-window}
 t(1-\varepsilon)<x_S<\frac{t}{1-\delta},
\end{equation}
an interval of length at most $3\delta$, and, with $t_S:=t/(2x_S)$,
\begin{equation}\label{eq:shifted-uniform-deletion}
 q_\varepsilon(t)\le\mathcal K(S)q_\varepsilon(t_S)
 +O_c\bigl(\mathcal R_c(\varepsilon)\bigr)
 \le\mathcal K(S)Q_{\mathrm{end}}(\varepsilon)
 +O_c\bigl(\mathcal R_c(\varepsilon)\bigr).
\end{equation}
\end{enumerate}
\end{lemma}

\begin{proof}
Apply \Cref{lem:binomial-uniform-deletion} with
$k=1$, $E=\varnothing$, $A=1/2$, and
$(\alpha,\beta)=(1/10,2/5)$.
By \Cref{lem:factor}, $R=X$; the vectors are the indicators of finite
sets $S$, and $a_{\mathbf j}=x_S$, $K_1(\mathbf j)=\mathcal K(S)$,
$\Xi_* =\Xi$ and $\delta_* =\delta$.
On the surviving event in that proof,
\begin{equation}\label{eq:deletion-factorisation}
 X=x_S\Pi,\qquad \Pi\in(1-\delta,1].
\end{equation}
The anchor window has length at most
$\delta/((1-\varepsilon)(1-\delta))\le3\delta$ for
$\varepsilon<6^{-5}$. Its normalized endpoint satisfies
$(1-\delta)/2<t_S<1/(2(1-\varepsilon))$, so
$q_\varepsilon(t_S)\le Q_{\mathrm{end}}(\varepsilon)$.
\end{proof}

\section{The prime \texorpdfstring{$3$}{3}, the endpoint \texorpdfstring{$1/2$}{1/2}, and arithmetic anchors}
\label{ssec:global-modulus-anchors}

Put
\[
 \widetilde X:=\prod_{\ell\ge5}(1-1/\ell)^{B_\ell},\qquad
 \widetilde q_\varepsilon(u):=
 \mathbb P\bigl(u(1-\varepsilon)<\widetilde X\le u\bigr)
 \quad(u>0,\ 0<\varepsilon<1).
\]

The coordinate at $3$ splits the law into two branches of equal
probability. For $0<\varepsilon\le1/3$, the window ending at $1/2$
receives only the endpoint contribution from the branch $B_3=0$.
The window ending at $1/3$ receives the identical endpoint contribution
from the branch $B_3=1$, together with a strictly positive interior
contribution from $B_3=0$. Thus equality of the leading asymptotic
constants is compatible with a strict difference between the window
masses. The lemma gives the exact comparison.

\begin{lemma}[The prime $3$: splitting, support, and rescaling by $2/3$]
\label{lem:prime-three-split}\leavevmode\par
\begin{enumerate}[label=\textup{(\alph*)},leftmargin=*]
\item The coordinate $B_3$ is a fair coin independent of $\widetilde X$,
and $X=\tfrac12\widetilde X$ on $\{B_3=0\}$ and
$X=\tfrac13\widetilde X$ on $\{B_3=1\}$. For every $t>0$ and
$0<\varepsilon<1$,
\begin{equation}\label{eq:b3-split}
 q_\varepsilon(t)=\tfrac12\widetilde q_\varepsilon(2t)
 +\tfrac12\widetilde q_\varepsilon(3t).
\end{equation}
\item $\supp\bigl(\operatorname{Law}(\widetilde X)\bigr)=[0,1]$, and
$\widetilde q_\varepsilon(u)>0$ for $0<u<1/(1-\varepsilon)$.
\item If $\tfrac13\le\alpha<\beta$ and $\alpha<\tfrac34$, then
\begin{equation}\label{eq:prime-three-translate}
 \mathbb P\bigl(\tfrac23\alpha<X\le\tfrac23\beta\bigr)
 =\mathbb P(\alpha<X\le\beta)+\tfrac12\mathbb P\bigl(\tfrac43\alpha<\widetilde X\le\tfrac43\beta\bigr)
 >\mathbb P(\alpha<X\le\beta).
\end{equation}
In particular, if $0<\varepsilon<1$ and $1/(3(1-\varepsilon))\le t<3/(4(1-\varepsilon))$, then $q_\varepsilon(\tfrac23t)=q_\varepsilon(t)+\tfrac12\widetilde q_\varepsilon(\tfrac43t)>q_\varepsilon(t)$.
\end{enumerate}
\end{lemma}

\begin{proof}
\emph{(a)} The first assertion is \Cref{lem:factor} at $\ell=3$, where
$\mathbb P(B_3=1)=\tfrac12$; conditioning on $B_3$ gives \eqref{eq:b3-split}.

\emph{(b)} By \Cref{lem:factor,thm:mertens}, the summands of the almost surely
convergent series $-\log\widetilde X=\sum_{\ell\ge5}B_\ell\omega_\ell$ have
supports $\{0,\omega_\ell\}$ with $\omega_\ell>0$, $\omega_\ell\to0$ and
$\sum_\ell\omega_\ell=\infty$. The second assertion of \Cref{lem:supp_conv}
gives $\supp\operatorname{Law}(-\log\widetilde X)=[0,\infty)$, hence
$\supp\operatorname{Law}(\widetilde X)=[0,1]$. For $0<u<1/(1-\varepsilon)$
the nonempty open interval $\bigl(u(1-\varepsilon),\min(u,1)\bigr)$ lies in
$[0,1]$, so $\widetilde q_\varepsilon(u)>0$.

\emph{(c)} Since $3\alpha\ge1\ge\widetilde X$, the branch $X=\tfrac13\widetilde X$ contributes nothing to $\{\alpha<X\le\beta\}$. Conditioning on $B_3$ at $(\tfrac23\alpha,\tfrac23\beta]$ therefore gives the identity. Its last term is positive by~(b), because $(\tfrac43\alpha,\min(\tfrac43\beta,1))$ is a nonempty open subinterval of $[0,1]$.
\end{proof}

\begin{proposition}[Concentration near the endpoint $1/2$]
\label{prop:endpoint-window}
As $\varepsilon\downarrow0$,
\begin{enumerate}[label=\textup{(\alph*)},leftmargin=*]
\item $Q_{\mathrm{end}}(\varepsilon)=\kappa_{\mathrm{end}}/z+O(z^{-2})$;
\item uniformly for $\tfrac12(1-\delta)\le t\le\tfrac12(1-\varepsilon)$,
\begin{equation}\label{eq:endpoint-window-interior}
 q_\varepsilon(t)=O(z^{-2}).
\end{equation}
\end{enumerate}
\end{proposition}

\begin{proof}
For $\tfrac12(1-\delta)\le t\le1/(2(1-\varepsilon))$ write $t=\tfrac12(1-\theta)$, so $\theta\le\delta$, and $\theta':=\theta+\varepsilon(1-\theta)$; by \eqref{eq:q-H-identity}, $q_\varepsilon(t)=H_{\mathrm{end}}(\theta')-H_{\mathrm{end}}(\theta)$. If $\theta<\varepsilon$, then $\theta'<2\varepsilon$ and, since $H_{\mathrm{end}}$ is nonnegative and nondecreasing, $q_\varepsilon(t)\le H_{\mathrm{end}}(2\varepsilon)=\kappa_{\mathrm{end}}/z+O(z^{-2})$ by \eqref{eq:hend-asymp}. If $\varepsilon\le\theta\le\delta$, then $\theta<\theta'\le2\theta$ and $\log(1/\theta)\ge\tfrac25z$, so \eqref{eq:hend-increment} gives $q_\varepsilon(t)=O(z^{-2})$. This proves~(b) and the upper bound in~(a); the lower bound is $Q_{\mathrm{end}}(\varepsilon)\ge q_\varepsilon(\tfrac12)=H_{\mathrm{end}}(\varepsilon)=\kappa_{\mathrm{end}}/z+O(z^{-2})$ by \eqref{eq:hend-asymp}.
\end{proof}

\begin{lemma}[Comparing endpoint probability with an increment of $H_{\mathrm{end}}$]
\label{lem:endpoint-small-comparison}
Fix $0<a\le b<\infty$. There is $c_0=c_0(a,b)>0$ such that, for every
fixed $0<c\le c_0$ and all sufficiently small $s>0$ in terms of $a,b,c$,
\begin{equation}\label{eq:endpoint-small-comparison}
 H_{\mathrm{end}}\bigl(e^{-c\log^2(1/s)}\bigr)
 -\bigl[H_{\mathrm{end}}(vs)-H_{\mathrm{end}}(us)\bigr]
 \ge\frac{\kappa_{\mathrm{end}}}{2c\log^2(1/s)}
\end{equation}
uniformly for $a\le u\le v\le b$.
\end{lemma}

\begin{proof}
Write $z_s:=\log(1/s)$. By \eqref{eq:hend-asymp}, uniformly for
$a\le u\le v\le b$,
\[
 0\le H_{\mathrm{end}}(vs)-H_{\mathrm{end}}(us)\le C_{a,b}z_s^{-2},
 \qquad
 H_{\mathrm{end}}(e^{-cz_s^2})
 =\frac{\kappa_{\mathrm{end}}}{cz_s^2}+O_c(z_s^{-4}).
\]
Choose $c_0>0$ so that $\kappa_{\mathrm{end}}/c_0>2C_{a,b}$.
For each fixed $0<c\le c_0$, subtraction proves
\eqref{eq:endpoint-small-comparison} when $s$ is sufficiently small.
\end{proof}

\begin{theorem}[Quantitative asymptotic at every arithmetic anchor]
\label{thm:anchor-sharp}
\index{Arithmetic anchors!sharp constants}
For every fixed finite set $S$ of odd primes, as $\varepsilon\downarrow0$,
\begin{equation}\label{eq:anchor-residual-upper}
 0\le q_\varepsilon(x_S)-\mathcal K(S)H_{\mathrm{end}}(\varepsilon)
 \ll_S\mathcal R(\varepsilon).
\end{equation}
Fix $0<b<1/\sqrt{20}$ and an integer $N\ge1$, and put
$\varepsilon=h/x_S$ and $z=\log(x_S/h)$.
For some $\varepsilon_0(b)\in(0,e^{-1})$, uniformly over finite sets $S$
of odd primes with
$D_S\le\exp\{b\sqrt{z\log z}\}$ and
$0<\varepsilon\le\varepsilon_0(b)$,
\begin{equation}\label{eq:anchor-sharp}
 G(x_S)-G(x_S-h)
 =\kappa_S\left\{\sum_{j=1}^N\frac{a_j}{z^j}
 +O_{b,N}(z^{-N-1})\right\},
\end{equation}
where $a_j$ are the coefficients of \Cref{prop:endpoint-half}.
\end{theorem}

\begin{proof}
Fix $0<c<1/\sqrt{20}$ and take $\varepsilon$ sufficiently small in terms
of $c$. For any $S$ with every $\ell\in S$ at most $\Xi$ and
$D_S<\varepsilon^{-1/10}$, one has $\varepsilon\le1/\max S$ if
$S\ne\varnothing$. Thus \eqref{eq:anchor-main-exact} gives
$q_\varepsilon(x_S)\ge\mathcal K(S)H_{\mathrm{end}}(\varepsilon)$.
At $t=x_S$, the set in alternative~(ii) of \Cref{lem:shifted-uniform-deletion}
is $S$ by uniqueness, since $x_S$ lies in \eqref{eq:deletion-anchor-window};
thus $t_S=\tfrac12$ and $q_\varepsilon(t_S)=H_{\mathrm{end}}(\varepsilon)$.
In either alternative, uniformly over these $S$,
\[
 0\le q_\varepsilon(x_S)-\mathcal K(S)H_{\mathrm{end}}(\varepsilon)
 \ll_c\mathcal R_c(\varepsilon).
\]
Taking $c=1/5$ proves \eqref{eq:anchor-residual-upper} for fixed $S$.
For \eqref{eq:anchor-sharp}, take $c=(b+1/\sqrt{20})/2$.
The restriction $D_S\le\exp\{b\sqrt{z\log z}\}$ ensures the above
conditions for all sufficiently large $z$ in terms of $b$.
Since $\mathcal K(S)\ge D_S^{-1}$,
\[
 \frac{\mathcal R_c(\varepsilon)}{\kappa_S}
 \ll e^{-(c-b)\sqrt{z\log z}}\ll_{b,N}z^{-N-1}.
\]
Now \eqref{eq:hend-asymp} and
$G(x_S)-G(x_S-h)=q_{h/x_S}(x_S)$ give \eqref{eq:anchor-sharp}.
\end{proof}

\begin{remark}[Conditional logarithmic deficits]\label{rem:anchor-pareto}
Fix $S$, put $\Delta=1-X/x_S$ and $L=\log(1/\varepsilon)$, and let
$\nu_\varepsilon$ be the law of $L^{-1}\log(1/\Delta)$ conditional on
$0<\Delta\le\varepsilon$. The peaks-over-threshold form of regular
variation~\cite{BdH74,Pic75} gives the Pareto limit $P_1$, with survival
function $P_1([t,\infty))=t^{-1}$ for $t\ge1$. Here the missing second
anchor coefficient gives the sharper estimate
\[
 d_K(\nu_\varepsilon,P_1)
 =\frac{\pi^2}{18\sqrt3\,L^2}+O_S(L^{-3}).
\]
Indeed, atomlessness (\Cref{thm:pure-type}) identifies the conditional
tail with $q_{\varepsilon^t}(x_S)/q_\varepsilon(x_S)$, and
\eqref{eq:anchor-sharp}, with $a_2=0$ and $a_3=-\pi^2/12$, gives uniformly
for $t\ge1$
\[
 \frac{q_{\varepsilon^t}(x_S)}{q_\varepsilon(x_S)}
 =\frac1t+\frac{\pi^2}{12L^2}\left(\frac1t-\frac1{t^3}\right)
   +O_S(L^{-3}).
\]
The maximum of $t^{-1}-t^{-3}$ is $2/(3\sqrt3)$.
The same ratio shows $\Delta/\varepsilon\to0$ in conditional probability.
For the endpoint $x_\varnothing=1/2$, \Cref{cor:endpoint-order-statistics}
gives the related order-statistic limit. For fixed $t\ge1$, the Pareto limit also holds for primes in the
shrinking anchor window when $\varepsilon=\varepsilon(x)\downarrow0$ and
$\log(1/\varepsilon)=o(\log x)$, by
\Cref{cor:prime-shrinking-anchor}\textup{(i)}. More generally the same
fixed-$t$ ratio holds for moving $S$ under the size condition of
\eqref{eq:prime-moving-anchor} when $L/(\kappa_S\log x)\to0$.
Apply that estimate at $L$ and $tL$. Each empirical endpoint atom
has mass $O(1/\log x)$ by the continuous model and the uniform CDF
bound, so the choice of weak or strict inequalities has vanishing
relative effect.
\end{remark}

\begin{corollary}[Two-sided bounds for fixed anchor residuals]
\leavevmode\phantomsection\label{cor:anchor-residual}\ignorespaces
Fix a nonempty finite set $S$ of odd primes and
$C>1/(2\log(59/40))$. For all sufficiently small $\varepsilon>0$,
\begin{equation}\label{eq:anchor-residual-two-sided}
 \varepsilon\exp\{-C(\log\log(1/\varepsilon))^2\}
 \le q_\varepsilon(x_S)-\mathcal K(S)H_{\mathrm{end}}(\varepsilon)
 \ll_S\mathcal R(\varepsilon).
\end{equation}
\end{corollary}

\begin{proof}
Put $r=\max S$, $v=(r-1)/r$, and
$Y_r=\prod_{\ell>r}(1-1/\ell)^{B_\ell}$.
On the positive-probability cylinder
\[
 \mathcal Z=\{B_\ell=\mathbf1_{\{\ell\in S\setminus\{r\}\}}
                    \text{ for all odd primes }\ell\le r\},
\]
which is disjoint from $\mathcal C_S$, one has $X=(x_S/v)Y_r$.
The tail is independent of $\mathcal Z$. By
\eqref{eq:anchor-main-exact} and \Cref{thm:cofinite-window-lower},
\[
 q_\varepsilon(x_S)-\mathcal K(S)H_{\mathrm{end}}(\varepsilon)
 \ge\mathbb P(\mathcal Z)\,
       \mathbb P\{v(1-\varepsilon)<Y_r\le v\}
 \ge\mathbb P(\mathcal Z)\,\varepsilon
       e^{-C'(\log\log(1/\varepsilon))^2},
\]
where $1/(2\log(59/40))<C'<C$ and $\varepsilon$ is sufficiently small.
Absorbing the fixed cylinder probability proves the lower bound;
the upper bound is \eqref{eq:anchor-residual-upper}.
\end{proof}
\Needspace{6\baselineskip}
\section{Sharp constant and localization}
\label{ssec:global-modulus-localization}

\subsection{Sharp concentration and localization}
\label{ssec:sharp-concentration-localization}

\begin{proposition}[Uniform upper bound and localization of superlevel sets]
\label{prop:two-window}
Write
\[
  I_2(\varepsilon):=\Bigl(\tfrac12(1-\delta),\ \frac1{2(1-\varepsilon)}\Bigr),
  \qquad
  I_3(\varepsilon):=\Bigl(\tfrac13(1-\delta),\ \frac1{3(1-\varepsilon)}\Bigr).
\]
\begin{enumerate}[label=\textup{(\alph*)},leftmargin=*]
\item \emph{Support.}  For every $0<\varepsilon<1$, $q_\varepsilon(t)=0$
whenever $t\le0$ or $t\ge1/\bigl(2(1-\varepsilon)\bigr)$.
\item \emph{Uniform upper bound.}  As $\varepsilon\downarrow0$, uniformly for
$t\in\R$,
\begin{equation}\label{eq:sharp-upper-uniform}
  q_\varepsilon(t)\;\le\;\frac{\kappa_{\mathrm{end}}}{z}+O(z^{-2}).
\end{equation}
\item \emph{Separation.} For $0<\varepsilon<6^{-5}$,
\[
 \sup I_3(\varepsilon)=\frac1{3(1-\varepsilon)}
 <\tfrac12(1-\delta)=\inf I_2(\varepsilon).
\]
\item \emph{Confinement.} For every fixed $a>0$, the set
$\mathcal A_a:=\{x_S:\mathcal K(S)\ge a\}$ is finite, and, for all
sufficiently small $\varepsilon>0$ in terms of $a$,
\begin{equation}\label{eq:finite-anchor-superlevels}
 \left\{t:q_\varepsilon(t)\ge\frac{a\kappa_{\mathrm{end}}}{z}\right\}
 \subseteq\bigcup_{x_S\in\mathcal A_a}
 \left(x_S(1-\delta),\frac{x_S}{1-\varepsilon}\right).
\end{equation}
In particular $\mathcal A_{1/3}=\{1/2,1/3,2/5,4/15\}$.

For every fixed $a>\tfrac13$, one has
$\mathcal A_a\subseteq\{1/2,1/3\}$, with equality for
$\tfrac13<a\le1$ and $\mathcal A_a=\varnothing$ for $a>1$.
For every fixed $a>\tfrac13$, there is $\varepsilon_1(a)\in(0,6^{-5})$ such that,
for $0<\varepsilon<\varepsilon_1(a)$, every $t$ with
\begin{equation}\label{eq:near-max}
 q_\varepsilon(t)\ge\frac{a\kappa_{\mathrm{end}}}{z}
\end{equation}
belongs to $I_2(\varepsilon)\cup I_3(\varepsilon)$.
\end{enumerate}
\end{proposition}

\begin{proof}
Fix an admissible threshold $\varepsilon_{\mathrm{del}}>0$ for
\Cref{lem:shifted-uniform-deletion} with $c=1/5$, and let
$C$ bound the constants in both alternatives uniformly in $t$ for
$0<\varepsilon<\varepsilon_{\mathrm{del}}$.  Use \eqref{eq:residual-properties} and the bound
$\mathcal K(S)\le1$ from \eqref{eq:weight-trichotomy}.

\emph{(a)}  If $t\le0$ then $\{t(1-\varepsilon)<X\le t\}$ is contained in
$\{X\le0\}$, which is null.  If $t\ge1/\bigl(2(1-\varepsilon)\bigr)$ then
$t(1-\varepsilon)\ge\tfrac12$, so
$\{t(1-\varepsilon)<X\le t\}\subseteq\{X>\tfrac12\}$, again a null event
because $X\le\tfrac12$ almost surely (\Cref{thm:phi-dist}).

\emph{(b)}  By~(a) we may assume $0<t<1/\bigl(2(1-\varepsilon)\bigr)$ and
apply \Cref{lem:shifted-uniform-deletion}.  In alternative~(i),
$q_\varepsilon(t)\le C\,\mathcal R(\varepsilon)=o(z^{-2})$.  In
alternative~(ii), \eqref{eq:shifted-uniform-deletion}, the bound
$\mathcal K(S)\le1$ and \Cref{prop:endpoint-window}(a) give
\[
  q_\varepsilon(t)
  \;\le\;\mathcal K(S)\,Q_{\mathrm{end}}(\varepsilon)
   +C\,\mathcal R(\varepsilon)
  \;\le\;Q_{\mathrm{end}}(\varepsilon)+C\,\mathcal R(\varepsilon)
   \;=\;\frac{\kappa_{\mathrm{end}}}{z}+O(z^{-2}).
\]

\emph{(c)} On $0<\varepsilon<6^{-5}$, $\delta<1/36$ and
$3(1-\varepsilon)(1-\delta)>3(1-6^{-5})(1-1/36)>2$.
Thus $1/(3(1-\varepsilon))<\tfrac12(1-\delta)$, which gives the stated
separation.

\emph{(d)} Fix $a>0$. By \Cref{lem:shifted-uniform-deletion,prop:endpoint-window}
and \eqref{eq:residual-properties}, at a point of the displayed
superlevel set alternative~\textup{(i)} is impossible for sufficiently
small $\varepsilon$, and alternative~\textup{(ii)} gives
$\mathcal K(S)\ge a-o(1)$.
There are only finitely many sets with $\mathcal K(S)\ge a/2$:
every prime $\ell\ge5$ in such a set satisfies $\ell-2\le2/a$,
and $3$ may be either present or absent. Among this finite family,
the weights strictly below $a$ have a fixed positive gap from $a$.
Consequently $\mathcal K(S)\ge a$ eventually. Solving
\eqref{eq:deletion-anchor-window} for $t$ gives precisely the interval
in \eqref{eq:finite-anchor-superlevels}.
For $a=1/3$ the sets are $\varnothing,\{3\},\{5\},\{3,5\}$.
For $1/3<a\le1$, \eqref{eq:weight-trichotomy} leaves exactly
$S=\varnothing$ and $S=\{3\}$, which give the two stated windows;
for $a>1$ there are no such sets.
Choose $\varepsilon_1(a)<6^{-5}$ within the sufficiently small range
for the fixed level $a$ and within the range where \Cref{lem:shifted-uniform-deletion} applies.
This proves the stated specialization.
\end{proof}

\Needspace{5\baselineskip}
\begin{corollary}[Decay away from arithmetic anchors]
\label{cor:non-anchor-decay}\leavevmode\par
\begin{enumerate}[label=\textup{(\alph*)},leftmargin=*]
\item For every fixed non-anchor $t\in\R$,
$q_\varepsilon(t)=o_t(1/z)$ as $\varepsilon\downarrow0$.
\item For every fixed $0<c<1/\sqrt{20}$ and all sufficiently small
$\varepsilon>0$, uniformly over rational non-anchors $t=m/b$ in lowest
terms with $1\le b\le\varepsilon^{-2/5}$,
\begin{equation}\label{eq:rational-nonanchor-decay}
 q_\varepsilon(t)\ll_c e^{-c\sqrt{z\log z}}.
\end{equation}
\item Let $t$ be a fixed irrational satisfying
\begin{equation}\label{eq:nonanchor-diophantine}
 |t-m/n|\ge c_tn^{-\eta}\quad(m\in\Z,\ n\ge1),
 \qquad c_t>0,\quad\eta\ge2.
\end{equation}
For every $0<c<1/(2\sqrt\eta)$,
\begin{equation}\label{eq:diophantine-nonanchor-decay}
 q_\varepsilon(t)\ll_{t,c,\eta}e^{-c\sqrt{z\log z}}
 \qquad(\varepsilon\downarrow0).
\end{equation}
\end{enumerate}
\end{corollary}

\begin{proof}
For (a), fix $a>0$. A fixed non-anchor $t$ eventually lies outside
the shrinking intervals around the finite set $\mathcal A_a$.
By \Cref{prop:two-window}(d), $q_\varepsilon(t)<a\kappa_{\mathrm{end}}/z$
for all sufficiently small $\varepsilon$. Letting $a\downarrow0$ proves (a).

For (b), choose $2c^2<\alpha<1/10$ and
$\alpha+2/5<\beta<1/2$. By \Cref{prop:two-window}(a), we may assume
$0<t<1/(2(1-\varepsilon))$. Apply
\Cref{lem:binomial-uniform-deletion} with $k=1$, $E=\varnothing$ and $A=1/2$.
In its second alternative, $x_S=N_S/(2D_S)\ne t$ and
$D_S<\varepsilon^{-\alpha}$, so
\[
 |t-x_S|\ge\frac1{2bD_S}>
 \tfrac12\varepsilon^{\alpha+2/5}>3\varepsilon^\beta
\]
for small $\varepsilon$, whereas the anchor window has length at most
$3\varepsilon^\beta$ and contains both points. Thus only the first
alternative occurs, giving \eqref{eq:rational-nonanchor-decay}.

For \eqref{eq:diophantine-nonanchor-decay}, choose constants
\[
 2c^2<\alpha<\frac1{2\eta},\qquad
 \alpha\eta<\beta<\frac12.
\]
By \Cref{prop:two-window}(a) we may restrict to
$0<t<1/(2(1-\varepsilon))$.
Since $\eta\ge2$, these choices also give $2\alpha<\beta$.
Apply \Cref{lem:binomial-uniform-deletion} with $k=1$, $E=\varnothing$
and $A=1/2$. In its second alternative the anchor $x_S$ has
$D_S<\varepsilon^{-\alpha}$ and belongs to an interval containing $t$
of length at most $3\varepsilon^\beta$ for small $\varepsilon$.
However, \eqref{eq:nonanchor-diophantine} at the reduced denominator
of $x_S$, which is at most $2D_S$, gives
\[
 |t-x_S|\ge c_t(2D_S)^{-\eta}
 >c_t2^{-\eta}\varepsilon^{\alpha\eta}
 >3\varepsilon^\beta.
\]
Thus the second alternative is impossible, and the first gives
\eqref{eq:diophantine-nonanchor-decay}, since $c<\sqrt{\alpha/2}$.
\end{proof}

\begin{lemma}[Comparison of $q_\varepsilon$ on $I_3(\varepsilon)$ and $I_2(\varepsilon)$]
\label{lem:two-window-comparison}
For all sufficiently small $\varepsilon>0$, uniformly for
$t\in I_3(\varepsilon)$,
\begin{equation}\label{eq:two-window-comparison}
 0<q_\varepsilon(t)-q_\varepsilon(\tfrac32t)
 =\tfrac12\widetilde q_\varepsilon(2t)
 \ll\mathcal R(\varepsilon).
\end{equation}
\end{lemma}

\begin{proof}
Since $\tfrac32t\in I_2(\varepsilon)$,
\Cref{prop:two-window}(c) and \Cref{lem:prime-three-split}(c) give the
identity and strict positivity. Apply \Cref{lem:shifted-uniform-deletion}
at $t$. In alternative~(i), the difference is at most
$q_\varepsilon(t)\ll\mathcal R(\varepsilon)$. In alternative~(ii),
the window \eqref{eq:deletion-anchor-window} contains $1/3$;
since $3<\Xi$ and $D_{\{3\}}=3<\varepsilon^{-1/10}$, uniqueness gives
$S=\{3\}$ and $t_S=\tfrac32t$. As $\mathcal K(\{3\})=1$,
\eqref{eq:shifted-uniform-deletion} gives the upper bound.
\end{proof}

\begin{lemma}[Negligible increments of the endpoint probability]
\leavevmode\phantomsection\label{lem:microscopic-endpoint-increment}\ignorespaces
Fix $c>0$ and put $\theta_0:=e^{-cz^2}$. Then
\begin{equation}\label{eq:microscopic-endpoint-increment}
 H_{\mathrm{end}}(\varepsilon+\theta_0)-H_{\mathrm{end}}(\varepsilon)
 =o\bigl(\mathcal R(\varepsilon)\bigr)
 \qquad(\varepsilon\downarrow0).
\end{equation}
\end{lemma}

\begin{proof}
Put $u:=\theta_0/(1-\varepsilon)$ and $a:=(1-\varepsilon)/2$.
Since $a(1-u)=(1-\varepsilon-\theta_0)/2$, the endpoint increment
equals $q_u(a)$.
Apply \Cref{lem:shifted-uniform-deletion}, with parameter $1/5$, at width $u$ and
point $a$; there $\delta_u:=u^{2/5}=o(\varepsilon)$.
Alternative~(i) gives $q_u(a)\ll\mathcal R(u)$. In alternative~(ii),
\eqref{eq:deletion-anchor-window} gives
\[
 1-\varepsilon-\theta_0<2x_S
 <\frac{1-\varepsilon}{1-\delta_u}<1.
\]
Thus $S\ne\varnothing$, and every $\ell\in S$ satisfies
$1-1/\ell\ge2x_S>1-\varepsilon-\theta_0$, so $\ell>1/(\varepsilon+\theta_0)$;
as every factor of $\mathcal K(S)$ is at most one,
$\mathcal K(S)\le(\ell-2)^{-1}=O(\varepsilon)$. With $q_u(t_S)\le1$,
\eqref{eq:shifted-uniform-deletion} gives $q_u(a)=O(\varepsilon+\mathcal R(u))$
in either alternative.
Since $\log(1/u)\ge cz^2/2$, one has $\mathcal R(u)\le e^{-z}=\varepsilon$ for
small $\varepsilon$; as $\varepsilon=o(\mathcal R(\varepsilon))$, this proves
\eqref{eq:microscopic-endpoint-increment}.
\end{proof}

\begin{lemma}[A lower bound for endpoint increments]
\label{lem:endpoint-increment-lower}
There are absolute constants $c>0$ and $\varepsilon_0>0$ such that
\begin{equation}\label{eq:endpoint-increment-lower}
 H_{\mathrm{end}}(\varepsilon)
 -H_{\mathrm{end}}(\varepsilon(1-d))
 \ge \frac{cd}{\log^2(1/\varepsilon)}
 \qquad
 (0<\varepsilon<\varepsilon_0,\quad
  \varepsilon^{1/4}\le d\le\tfrac12).
\end{equation}
\end{lemma}

\begin{proof}
Write $z=\log(1/\varepsilon)$ and
\[
 a=-\log(1-\varepsilon),\qquad
 b=-\log(1-\varepsilon(1-d)),\qquad
 \Delta=a-b
 =\log\left(1+\frac{\varepsilon d}{1-\varepsilon}\right).
\]
Uniformly in the stated range, $a,b\asymp\varepsilon$ and
$\Delta\asymp\varepsilon d$. For
$W=\sum_{p\ge3}B_p\omega_p$ as in \Cref{prop:endpoint-half},
the left side of \eqref{eq:endpoint-increment-lower} equals
$\mathbb P(b\le W<a)$.

Let $g(v)=(1-e^{-v})^{-1}$, the inverse of
$x\mapsto\log(x/(x-1))$, and put
\[
 Y=4/\Delta,\qquad
 J=\bigl(g(b+\Delta/2),g(b+\Delta/4)\bigr].
\]
The endpoints of $J$ are comparable to $\varepsilon^{-1}$, and
$-g'(v)=e^{-v}(1-e^{-v})^{-2}\asymp\varepsilon^{-2}$ on the
intervening range. Hence $|J|\asymp d/\varepsilon$. For
$\varepsilon\le1/4$ one has $b\ge\varepsilon/2$,
$\Delta\le2\varepsilon/3$ and $g(v)\le1+1/v$, so $J\subset(3,Y]$
when $\varepsilon$ is sufficiently small.
Moreover
\[
 \frac{|J|}{(\sup J)^{2/3}}
 \gg d\varepsilon^{-1/3}\ge\varepsilon^{-1/12}\longrightarrow\infty.
\]
Thus \Cref{lem:short-prime-window-lower}, with $\theta=2/3$, gives
uniformly
\[
 \#\{p\in J:p\textup{ prime}\}\gg\frac{d}{\varepsilon z},\qquad
 \sum_{\substack{p\in J\\p\textup{ prime}}}\frac1{p-2}
 \gg\frac d z.
\]

Require exactly one odd prime at most $Y$ to be active, with that prime
in $J$. Since the activation probability at $p$ is $1/(p-1)$, this
event has probability
\[
 Q(Y)\sum_{\substack{p\in J\\p\textup{ prime}}}\frac1{p-2}.
\]
Here $Q$ is defined in \eqref{eq:def-Q}, and
$\log Y=z+\log(1/d)+O(1)\asymp z$ uniformly, because
$d\ge\varepsilon^{1/4}$. Therefore
\eqref{eq:endpoint-Q-quantitative} gives $Q(Y)\gg z^{-1}$.
The independent tail $T_Y=\sum_{p>Y}B_p\omega_p$ satisfies
\[
 \mathbb ET_Y
 \le\sum_{p>Y}\frac1{(p-1)^2}
 \ll\frac1{Y\log Y}
 \ll\frac{\Delta}{\log Y}
\]
by \eqref{eq:endpoint-tail-square}. Markov's inequality gives
$\mathbb P(T_Y\le\Delta/4)\ge1/2$ for all sufficiently small
$\varepsilon$, uniformly in $d$. On the intersection of these
independent events,
\[
 b+\Delta/4\le W\le b+3\Delta/4<a.
\]
Multiplying their probabilities proves
\eqref{eq:endpoint-increment-lower}.
\end{proof}

\begin{theorem}[Sharp global modulus and localization of maximizers]
\label{thm:sharp-global-modulus}
\index{Concentration functions!sharp logarithmic expansion}
\index{Maximizing intervals!localization}\leavevmode\par
Put $L=\log(1/h)$.
\begin{enumerate}[label=\textup{(\alph*)},leftmargin=*]
\item The relative concentration function of $G$ and the concentration
function of $\muF$ satisfy
\begin{equation}\label{eq:sharp-global-modulus}
\begin{aligned}
 Q_G(\varepsilon)
 &=H_{\mathrm{end}}(\varepsilon)+O\bigl(\mathcal R(\varepsilon)\bigr)\\
 &=\frac{\kappa_{\mathrm{end}}}{z}
   -\frac{\kappa_{\mathrm{end}}\pi^2}{12z^3}+O(z^{-4})
 &&(\varepsilon\downarrow0),\\
 Q_{\muF}(h)
 &=\frac{\kappa_{\mathrm{end}}}{L}
   -\frac{\kappa_{\mathrm{end}}\pi^2}{12L^3}
   +O\!\left(L^{-4}\right)
 &&(h\downarrow0).
\end{aligned}
\end{equation}
More generally, the expansions to every fixed order $N\ge1$ have
coefficients $\kappa_{\mathrm{end}}a_j$ from \Cref{prop:endpoint-half},
with respective remainders $O_N(z^{-N-1})$ and
$O_N(L^{-N-1})$.
Moreover, for every fixed $C>1/(2\log(59/40))$ and all sufficiently
small $\varepsilon>0$,
\begin{equation}\label{eq:global-endpoint-residual}
 Q_G(\varepsilon)-H_{\mathrm{end}}(\varepsilon)
 \ge\varepsilon e^{-C(\log z)^2}.
\end{equation}
\item For every $0<\varepsilon<1$, the supremum defining
$Q_G(\varepsilon)$ is attained, and every maximizing upper endpoint
$t_\varepsilon$ satisfies
\begin{equation}\label{eq:sharp-localisation-right}
 t_\varepsilon<\frac1{3(1-\varepsilon)}.
\end{equation}
For all sufficiently small $\varepsilon>0$, every maximizer satisfies
\begin{equation}\label{eq:sharp-localisation}
 0\le t_\varepsilon-\tfrac13
 \ll\varepsilon z^2\mathcal R(\varepsilon).
\end{equation}
\item For every $h>0$, the supremum defining $Q_{\muF}(h)$ is
attained. For all sufficiently small $h>0$, every $a_h$ satisfying
$\muF([a_h,a_h+h])=Q_{\muF}(h)$ satisfies
\begin{equation}\label{eq:sharp-log-localisation}
 0\le\log3-a_h
 \ll hL^2\mathcal R(h).
\end{equation}
\end{enumerate}
\end{theorem}

\begin{proof}
\emph{Step 1: the leading terms in \eqref{eq:sharp-global-modulus}.}
Since $q_\varepsilon(\tfrac12)=H_{\mathrm{end}}(\varepsilon)$, \eqref{eq:hend-asymp}
and \eqref{eq:sharp-upper-uniform} give
$\kappa_{\mathrm{end}}/z+O(z^{-2})=H_{\mathrm{end}}(\varepsilon)\le Q_G(\varepsilon)
\le\kappa_{\mathrm{end}}/z+O(z^{-2})$. The leading term of $Q_{\muF}$ follows, since
$Q_{\muF}(h)=Q_G(1-e^{-h})$ and $\tfrac12h\le1-e^{-h}\le h$ by \eqref{eq:modulus-transport}.

\emph{Step 2: attainment, and \eqref{eq:sharp-localisation-right}.}
By \Cref{thm:phi-dist}, $G$ is continuous and $\supp(\mu_G)=[0,\tfrac12]$.
Hence $q_\varepsilon$ is continuous, vanishes off
$\bigl[0,1/(2(1-\varepsilon))\bigr]$ by \Cref{prop:two-window}(a), and
$Q_G(\varepsilon)\ge q_\varepsilon(\tfrac12)
=\mathbb P\bigl(X>\tfrac12(1-\varepsilon)\bigr)>0$. Thus, for every
$0<\varepsilon<1$, $Q_G(\varepsilon)$ is attained and every maximizer satisfies
$t_\varepsilon<1/(2(1-\varepsilon))<3/(4(1-\varepsilon))$. If
$t_\varepsilon\ge1/(3(1-\varepsilon))$, then \eqref{eq:prime-three-translate}
gives $q_\varepsilon(\tfrac23t_\varepsilon)>q_\varepsilon(t_\varepsilon)$,
contradicting maximality. This proves \eqref{eq:sharp-localisation-right}.

\emph{Step 3: localization in $I_3(\varepsilon)$.} By Step~1,
$Q_G(\varepsilon)\ge\tfrac23\kappa_{\mathrm{end}}/z$ for all sufficiently small
$\varepsilon$, so \Cref{prop:two-window}(d) with $a=\tfrac23$ places every
maximizer $t_\varepsilon$ in $I_2(\varepsilon)\cup I_3(\varepsilon)$. Every
point of $I_2(\varepsilon)$ exceeds $1/(3(1-\varepsilon))$ by
\Cref{prop:two-window}(c), so \eqref{eq:sharp-localisation-right} gives
$t_\varepsilon\in I_3(\varepsilon)$.

\emph{Step 4: shrinking $I_3(\varepsilon)$, and the first line of
\eqref{eq:sharp-global-modulus}.} Fix $c>0$ as in
\Cref{lem:endpoint-small-comparison} for $[a,b]=[1,2]$, put
$\theta_0:=e^{-cz^2}$. Recall that $H_{\mathrm{end}}(s)=0$ for $s\le0$,
since $X\le\tfrac12$. Write $t\in I_3(\varepsilon)$ as
$t=\tfrac13(1-\theta)$, so that $-\varepsilon/(1-\varepsilon)<\theta<\delta$,
and put $\theta':=\theta+\varepsilon(1-\theta)>0$. By
\eqref{eq:two-window-comparison},
\begin{equation}\label{eq:sharp-localisation-transfer}
 q_\varepsilon(t)
 =q_\varepsilon\bigl(\tfrac12(1-\theta)\bigr)+O\bigl(\mathcal R(\varepsilon)\bigr)
 =H_{\mathrm{end}}(\theta')-H_{\mathrm{end}}(\theta)
  +O\bigl(\mathcal R(\varepsilon)\bigr).
\end{equation}
If $\varepsilon\le\theta<\delta$, the main term is $O(z^{-2})$ by
\eqref{eq:endpoint-window-interior}. If $\theta_0\le\theta<\varepsilon$, then
$\theta'<2\varepsilon$, and \Cref{lem:endpoint-small-comparison} with
$s=\varepsilon$, $u=1$, $v=2$ bounds the main term by
$H_{\mathrm{end}}(2\varepsilon)-H_{\mathrm{end}}(\theta_0)
\le H_{\mathrm{end}}(\varepsilon)-\kappa_{\mathrm{end}}/(2cz^2)$. By
\eqref{eq:hend-asymp} and \eqref{eq:residual-properties}, in both ranges
$q_\varepsilon(t)<H_{\mathrm{end}}(\varepsilon)\le Q_G(\varepsilon)$ for small
$\varepsilon$; with Step~3, every maximizer therefore satisfies the
intermediate Gaussian bound $t_\varepsilon>\tfrac13(1-e^{-cz^2})$.
For such $\theta$,
$0<\theta'<\varepsilon+\theta_0$, so the main term in
\eqref{eq:sharp-localisation-transfer} is at most
$H_{\mathrm{end}}(\varepsilon+\theta_0)$, and
\Cref{lem:microscopic-endpoint-increment} gives
$Q_G(\varepsilon)\le H_{\mathrm{end}}(\varepsilon)+O\bigl(\mathcal R(\varepsilon)\bigr)$.
With Step~1 this is the first line of \eqref{eq:sharp-global-modulus}.
Since $\mathcal R(\varepsilon)=o(z^{-N})$ for every fixed $N$,
\eqref{eq:hend-asymp} gives its displayed expansion, and indeed every
fixed logarithmic order. The identity
$Q_{\muF}(h)=Q_G(1-e^{-h})$ gives the same coefficients for $Q_{\muF}$,
because $\log(1/(1-e^{-h}))=L+O(h)$.

\emph{Step 5: anchor containment.}
Put $a=1/3$ and $\kappa=\kappa_{\mathrm{end}}$.
By \Cref{lem:anchor-endpoint} with $S=\{3\}$, \eqref{eq:hend-asymp}
and \Cref{prop:two-window}(d), for all sufficiently small $\eta>0$,
\begin{equation}\label{eq:anchor-exchange-bounds}
 q_\eta(a)\ge\frac{3\kappa}{4\log(1/\eta)},\qquad
 q_\eta(t)<\frac{\kappa}{2\log(1/\eta)}
 \quad\bigl(t\notin I_2(\eta)\cup I_3(\eta)\bigr).
\end{equation}
Choose $\eta_0>0$ so that these bounds hold and $\inf I_2(\eta)>a$
for $0<\eta<\eta_0$. Direct subtraction gives
\[
 q_\varepsilon(a)-q_\varepsilon(a(1-\eta))
 =q_\eta(a)-q_\eta(a(1-\varepsilon)).
\]
If $0<\eta<\min(\eta_0,\varepsilon^{5/2})$, then
$a(1-\varepsilon)<a(1-\eta^{2/5})=\inf I_3(\eta)$, so the right side
is greater than $\kappa/(4\log(1/\eta))>0$.
A maximizer below $a$ has the form $a(1-\eta)$, where Step~4 gives
$0<\eta<e^{-c\log^2(1/\varepsilon)}<\min(\eta_0,\varepsilon^{5/2})$
for small $\varepsilon$, a contradiction. Together with Step~2, this gives
$1/3\le t_\varepsilon<1/(3(1-\varepsilon))$ for every maximizer.

\emph{Step 6: the displacement bound.}
Put $H=H_{\mathrm{end}}$, write $t_\varepsilon=(1+u)/3$, and set
\[
 d=\frac{(1-\varepsilon)u}{\varepsilon}\in[0,1).
\]
Since $t_\varepsilon\in I_3(\varepsilon)$, the support endpoint
$X\le1/2$ and \eqref{eq:two-window-comparison} give
\[
 q_\varepsilon(t_\varepsilon)
 =H\bigl(\varepsilon(1-d)\bigr)+r_\varepsilon,
 \qquad 0<r_\varepsilon\ll\mathcal R(\varepsilon).
\]
The same comparison at $t=1/3$ gives $q_\varepsilon(1/3)\ge H(\varepsilon)$.
Maximality therefore implies, uniformly over all maximizers,
\begin{equation}\label{eq:maximizer-endpoint-decrement}
 0\le H(\varepsilon)-H\bigl(\varepsilon(1-d)\bigr)
 \ll\mathcal R(\varepsilon).
\end{equation}
If $d\ge1/2$, monotonicity and \Cref{lem:endpoint-increment-lower}
at $d=1/2$ contradict \eqref{eq:maximizer-endpoint-decrement}, since
$\mathcal R(\varepsilon)=o(z^{-2})$. Hence $d<1/2$.
For $d\ge\varepsilon^{1/4}$ the same lemma yields
$d\ll z^2\mathcal R(\varepsilon)$. This also holds for
$d<\varepsilon^{1/4}$, because
$\varepsilon^{1/4}=o(z^2\mathcal R(\varepsilon))$.
Finally,
$t_\varepsilon-1/3=\varepsilon d/(3(1-\varepsilon))$,
which proves \eqref{eq:sharp-localisation}.

\emph{Step 7: $\muF$, and \eqref{eq:global-endpoint-residual}.}
By \eqref{eq:modulus-transport}, maximizers correspond under
$\varepsilon=1-e^{-h}$ and $t_\varepsilon=e^{-a_h}$. Hence $Q_{\muF}(h)$ is
attained for every $h>0$. For all sufficiently small $h$, the lower
bound in \eqref{eq:sharp-localisation} gives
\[
 0\le\log3-a_h=\log(3t_\varepsilon)
 \le3(t_\varepsilon-1/3).
\]
Since $\varepsilon\sim h$, one has
$\log(1/\varepsilon)\sim L$ and
$\mathcal R(\varepsilon)\sim\mathcal R(h)$ by
\eqref{eq:residual-properties}. Thus \eqref{eq:sharp-localisation}
proves \eqref{eq:sharp-log-localisation}.
Finally, \Cref{cor:anchor-residual} at the fixed nonempty set $S=\{3\}$,
for which $x_S=\tfrac13$ and $\mathcal K(S)=1$, gives
\[
 Q_G(\varepsilon)-H_{\mathrm{end}}(\varepsilon)
 \ge q_\varepsilon(\tfrac13)-H_{\mathrm{end}}(\varepsilon)
 \ge\varepsilon e^{-C(\log z)^2}
\]
for every fixed $C>1/(2\log(59/40))$ and sufficiently small
$\varepsilon$. This proves \eqref{eq:global-endpoint-residual}.
\end{proof}

\begin{remark}[Exact and near maximizers]\label{rem:sharp-margin}
By \eqref{eq:weight-trichotomy}, \Cref{thm:anchor-sharp} and part~(c) of
\Cref{lem:prime-three-split}, the anchors $1/2$ and $1/3$ have the
same leading constant $\kappa_{\mathrm{end}}$, every other anchor has
constant at most $\kappa_{\mathrm{end}}/3$, and
$q_\varepsilon(1/3)>q_\varepsilon(1/2)=H_{\mathrm{end}}(\varepsilon)$
for $0<\varepsilon\le1/3$.
The exclusion of $I_2(\varepsilon)$ in \eqref{eq:sharp-localisation}
concerns exact maximizers: $\tfrac12\in I_2(\varepsilon)$ and
$q_\varepsilon(\tfrac12)=Q_G(\varepsilon)-O(\mathcal R(\varepsilon))$.
The exclusion has a positive margin uniform in the window endpoint:
for every $C>1/(2\log(59/40))$ and all sufficiently small $\varepsilon$,
\begin{equation}\label{eq:uniform-exclusion-margin}
 \sup_{t\ge1/(3(1-\varepsilon))}q_\varepsilon(t)
 \le Q_G(\varepsilon)
      -\tfrac12\varepsilon e^{-C(\log z)^2}.
\end{equation}
Indeed, \Cref{thm:cofinite-window-lower} for $\mathcal P=\{\ell\ge5\}$
gives
$\widetilde q_\varepsilon(4t/3)\ge
\varepsilon e^{-C(\log z)^2}$ uniformly for
$1/(3(1-\varepsilon))\le t\le3/4$, since its logarithmic target
$\log(3/(4t))$ lies in $[0,\log(9/4)]$.
Now apply \Cref{lem:prime-three-split}(c) where $q_\varepsilon$ is
nonzero. Outside that range use \Cref{prop:two-window}(a) and
$Q_G(\varepsilon)\ge H_{\mathrm{end}}(\varepsilon)
\sim\kappa_{\mathrm{end}}/z$.
For $t\in I_2(\varepsilon)$ the same residual is
$\ll\mathcal R(\varepsilon)$ by \eqref{eq:two-window-comparison}.
\end{remark}

\subsection{Convolution constants and mode anchors}
\label{ssec:convolution-mode-anchors}

\begin{theorem}[Leading concentration for bounded sparse prime counts]
\label{thm:sparse-count-concentration}
Let the fixed independent counts $J_p$ satisfy the hypotheses of
\Cref{lem:sparse-count-uniform-deletion}, and suppose in addition that,
for a fixed $\kappa>0$,
\[
 \pi_p(0)=1-\kappa/p+O(p^{-2}).
\]
For a deterministic $a\in\R$, define
\[
 Y=a+\sum_pJ_p\omega_p,\quad\nu=\operatorname{Law}(Y),\quad
 C_0=\lim_{y\to\infty}(\log y)^\kappa\prod_{p\le y}\pi_p(0),
 \quad M=\prod_p\max_{0\le j\le m}\frac{\pi_p(j)}{\pi_p(0)}.
\]
The constant $C_0$ exists in $(0,\infty)$, and only finitely many factors
in $M$ differ from one. As $h\downarrow0$,
\begin{equation}\label{eq:sparse-count-leading-constants}
 \mathbb P(Y\le a+h)\sim C_0\log^{-\kappa}(1/h),\qquad
 Q_\nu(h)\sim MC_0\log^{-\kappa}(1/h).
\end{equation}
The law is atomless, and the concentration supremum is attained for
every $h>0$. Define the finite set of modal anchors
\[
 \mathcal B(a)=\left\{a+\sum_pj_p\omega_p:
       j_p\in\mathop{\rm argmax}_{0\le j\le m}\pi_p(j)\right\}.
\]
There is $\eta>0$ such that every finite vector not formed from local
maxima satisfies $K(\mathbf j)\le(1-\eta)M$.
For all sufficiently small $h$, the head selected by uniform deletion
for every maximizing interval is formed from local maxima; if that
interval has lower endpoint $b_h$, then, for some $b\in\mathcal B(a)$,
\begin{equation}\label{eq:sparse-count-modal-localisation}
 b-h\le b_h\le b+o(h).
\end{equation}
The $o(h)$ is uniform over maximizing intervals for the fixed law.
\end{theorem}

\begin{proof}
Write $S=Y-a$ and $F(u)=\mathbb P(S\le u)$, with $F(u)=0$ for $u<0$.
Since $\log\pi_p(0)-\kappa\log(1-1/p)=O(p^{-2})$, Mertens'
product theorem (\Cref{thm:mertens}) gives the positive, finite constant
\[
 C_0=e^{-\kappa\gamma}
       \prod_p\frac{\pi_p(0)}{(1-1/p)^\kappa}.
\]
The decrement bound implies $\pi_p(j)/\pi_p(0)\le(C/p)^j$;
thus zero is the unique local maximum for $p>C$.
In particular $M$ and $\mathcal B(a)$ are finite constructions.

For $u\downarrow0$, put $L_u=\log(1/u)$. Since $\omega_p>1/p$,
$S\le u$ forces $J_p=0$ for $p\le1/u$. Hence
$F(u)\le\prod_{p\le1/u}\pi_p(0)$.
Conversely take $y=L_u/u$. The estimate $\mathbb EJ_p\ll p^{-1}$
and \eqref{eq:endpoint-tail-square} give
\[
 \mathbb E\sum_{p>y}J_p\omega_p\ll\frac1{y\log y}=o(u).
\]
Independence and Markov's inequality give
$F(u)\ge\prod_{p\le y}\pi_p(0)(1-o(1))$.
Since $\log y/L_u\to1$, this proves the endpoint equivalent.
It also proves, for every fixed $D>1$,
\begin{equation}\label{eq:sparse-count-endpoint-increment}
 \sup_{u\le v\le Du}\{F(v)-F(u)\}=o(L_u^{-\kappa}).
\end{equation}
Indeed the equivalent is uniform for all arguments sufficiently close
to zero, and $\log(1/v)=L_u+O_D(1)$.

Finite head sums are distinct by the largest-prime valuation argument
in the deletion proof. Conditioning on the independent remaining tail
bounds every atom of $S$ by the largest head atom. For large cutoffs
that atom is $M\prod_{p\le y}\pi_p(0)\to0$. Thus $S$ is atomless.
The mass of $[x,x+h]$ is continuous in $x$ and tends to zero as
$|x|\to\infty$, so its positive supremum is attained.

Apply uniform deletion to $R=e^{-Y}$ with $A=e^{-a}$,
$(\alpha,\beta)=(1/10,2/5)$ and $\varepsilon=1-e^{-h}$.
Put $z=\log(1/\varepsilon)$ and $\delta=\varepsilon^{2/5}$.
The normalized relative endpoint $t'$ lies between
$A(1-\delta)$ and $A/(1-\varepsilon)$. If $t'\ge A$, its
window mass is at most $F(h)$. If $t'=A(1-\vartheta)<A$,
$0\le\vartheta<\delta$, set $u=-\log(1-\vartheta)$.
The normalized mass is $F(u+h)-F(u)$.
For $0\le\vartheta<\varepsilon$ it is at most $F(3h)$.
For $\varepsilon\le\vartheta<\delta$, one has $h\le u$,
$u+h\le2u$ and $\log(1/u)\ge(2/5+o(1))z$.
Equation~\eqref{eq:sparse-count-endpoint-increment} therefore gives
$o(z^{-\kappa})$ uniformly in this range. Deletion and $K(\mathbf j)\le M$
give the global upper bound $(MC_0+o(1))z^{-\kappa}$.

Fix a modal vector $\mathbf j$. For $h$ below every jump in its finite
support, $S\le h$ forces zero at those coordinates. Replacing these
zeros by $j_p$ multiplies probability by $M$ and produces points in
$[b,b+h]$, where $b=a+\sum j_p\omega_p$. Thus this interval has mass
at least $MF(h)$. Since $z=\log(1/h)+o(1)$, the matching equivalent
follows.

To prove the gap choose $P>2C$. At primes $p>P$ every nonzero
choice has ratio at most $1/2$ relative to the local maximum.
For primes $p\le P$ there are finitely many ratios
$\pi_p(j)/\max_r\pi_p(r)$ strictly below one. Let $r_*<1$ be the maximum of $1/2$ and these ratios.
The product normalized by $M$ has factors at most one, and for
any nonmodal vector at least one is at most $r_*$. Set $\eta=1-r_*$.
The upper comparison and modal lower bound then exclude every
nonmodal vector for maximizing windows once $h$ is small.

A maximizing window has positive surviving mass because the deletion
error is $o(\log^{-\kappa}(1/h))$. Its surviving points lie above its
modal anchor $b$, so $b_h+h\ge b$. Put $s=b_h-b$.
The anchor bounds imply $s=O(h^{2/5})$. If $s>0$, head removal
bounds its mass by $M\{F(s+h)-F(s)\}$ plus the deletion error.
For any fixed $d>0$, if $s\ge dh$ then $s+h\le(1+1/d)s$ and
$\log(1/s)\ge(2/5+o(1))\log(1/h)$.
The increment estimate makes this $o(\log^{-\kappa}(1/h))$,
contradicting the modal lower bound. This holds uniformly for each
$d$, proving \eqref{eq:sparse-count-modal-localisation}.
\end{proof}

\begin{corollary}[Leading truncation for bounded sparse prime counts]
\label{cor:sparse-count-truncation}
With the fixed laws and deterministic shift in
\Cref{thm:sparse-count-concentration}, put
$Y_y=a+\sum_{p\le y}J_p\omega_p$. Then
\begin{equation}\label{eq:sparse-count-truncation}
 d_K(\operatorname{Law}(Y_y),\operatorname{Law}(Y))
       \sim MC_0(\log y)^{-\kappa}\qquad(y\to\infty).
\end{equation}
The same equivalent holds after a common strictly monotone continuous
transformation of the two variables.
\end{corollary}

\begin{proof}
The tail $T_y=Y-Y_y$ is independent of the head, nonnegative, and
positive almost surely: the probability that all its counts vanish
is $\prod_{p>y}\pi_p(0)=0$. Distinct head vectors give distinct sums,
and for large $y$ the largest head atom is
$M\prod_{p\le y}\pi_p(0)\sim MC_0(\log y)^{-\kappa}$.
At an atom $v$ of this size,
\[
 \mathbb P(Y_y\le v)-\mathbb P(Y\le v)
 =\mathbb P(Y_y\le v<Y)\ge\mathbb P(Y_y=v),
\]
which proves the lower bound.
For $y$ above the domination threshold in the deletion proof,
\eqref{eq:sparse-count-pgf-bound} and $y\omega_p\le1+2/y=:b_y$ for
$p>y$ imply, by convexity,
\[
 \begin{aligned}
 \log\mathbb E e^{s yT_y}
 &\le mA_*\sum_{p>y}\frac{e^{s y\omega_p}-1}{p}\\
 &\le\frac{mA_*(e^{s b_y}-1)}{b_y}
             \sum_{p>y}\frac{y\omega_p}{p}
 \ll_{m,C}\frac{e^{s b_y}-1}{\log y}.
 \end{aligned}
\]
Take $s=\log\log y$ and a fixed $d>\kappa+1$; the logarithmic
moment bound is $O_{m,C}(1)$, and Chernoff gives
$\mathbb P(T_y>d/y)=O_{m,C,d}((\log y)^{-d})$.
On the coupling event $Y_y\le t<Y$ with $T_y\le d/y$ one has
$t<Y\le t+d/y$. Consequently
\[
 0\le\mathbb P(Y_y\le t)-\mathbb P(Y\le t)
 \le Q_\nu(d/y)+O((\log y)^{-d}).
\]
The leading concentration equivalent proves the upper bound.
For a strictly increasing continuous transformation, thresholds in its
range correspond to original thresholds; thresholds in any omitted
range contribute the common extreme values. For a decreasing map,
use complementary probabilities and left limits. Suprema of the
left-limit and right-continuous distribution-function differences
agree.
\end{proof}

By \Cref{thm:sharp-global-modulus}(b), every maximizing relative
interval $(t_\varepsilon(1-\varepsilon),t_\varepsilon]$ contains $1/3$
for all sufficiently small $\varepsilon$.
Uniqueness and the exact equality of a maximizing upper endpoint to
$1/3$ remain open. For fixed convolution powers,
\Cref{thm:convolution-modes} gives a finite set of mode anchors and,
after deleting all sufficiently small prime coordinates, exact
endpoint maximization.

Fix an integer $k\ge1$, a finite set $E$ of odd primes and $a\in\R$.
Let $J_\ell$, for odd primes $\ell\notin E$, be independent with law
$\operatorname{Bin}(k,1/(\ell-1))$, and put
\begin{equation}\label{eq:deleted-convolution-law}
 Y=a+\sum_{\substack{\ell\ge3\\\ell\notin E}}J_\ell\omega_\ell,
 \qquad \nu_{k,E,a}=\operatorname{Law}(Y),\qquad
 q_E=\prod_{\ell\in E}\frac{\ell-2}{\ell-1}.
\end{equation}
Define
\[
\begin{aligned}
 \mathfrak c_{k,E}&=\prod_{\substack{\ell\ge3\\\ell\notin E}}
   \max_{0\le j\le k}\frac{\binom{k}{j}}{(\ell-2)^j},\\
 \mathcal B_{k,E}(a)&=\left\{a+\sum_{\substack{\ell\ge3\\\ell\notin E}}
 j_\ell\omega_\ell:
 j_\ell\in\mathop{\rm argmax}_{0\le j\le k}
 \frac{\binom{k}{j}}{(\ell-2)^j}\right\}.
\end{aligned}
\]
Both definitions involve finitely many nontrivial choices.

For this binomial model, set $\pi_2(0)=1$, also set
$\pi_p(0)=1$ at each deleted prime, and use
$\operatorname{Bin}(k,1/(p-1))$ at the other odd primes.
The decrement bound holds with $m=k$ and $C=3k$, while
$\kappa=k$, $C_0=(\kappa_{\mathrm{end}}/q_E)^k$, and
$M=\mathfrak c_{k,E}$. The binomial specialization gives a sharper
error and displacement bound, an exact tie count, and endpoint
uniqueness after suitable prime deletions.

\Needspace{16\baselineskip}
\begin{theorem}[Fixed convolution constants and mode anchors]
\label{thm:convolution-modes}
\index{Convolution powers!concentration}
With this notation, as $h\downarrow0$,
\begin{equation}\label{eq:convolution-constant}
 Q_{\nu_{k,E,a}}(h)=
 \frac{\mathfrak c_{k,E}(\kappa_{\mathrm{end}}/q_E)^k}{\log^k(1/h)}
 +O_{k,E}(\log^{-k-2}(1/h)).
\end{equation}
The supremum is attained for every $h>0$. For all sufficiently small
$h$, every maximizing interval with lower endpoint $b_h$ has, for
some $b\in\mathcal B_{k,E}(a)$,
\begin{equation}\label{eq:convolution-displacement-mas20}
 b-h\le b_h\le b+\exp\{-c_{k,E}\log^{1+1/k}(1/h)\},
 \qquad c_{k,E}>0.
\end{equation}
Moreover,
\[
 |\mathcal B_{k,E}(a)|=2^{r_{k,E}},\qquad
 r_{k,E}:=\#\{\ell\ge3:\ell\textup{ prime},\ \ell\notin E,
                       \ \ell-1\mid k+1\}.
\]
Thus the mode anchor is unique precisely when every odd prime
$\ell$ with $\ell-1\mid k+1$ belongs to $E$.
If $E$ contains every odd prime $\ell\le k+2$, then, for all sufficiently
small $h>0$, the unique maximizing interval is $[a,a+h]$, and
\[
 Q_{\nu_{k,E,a}}(h)=\nu_{k,E,a}([a,a+h])
                  =q_E^{-k}\nu_{k,\varnothing,0}([0,h]).
\]
The threshold depends only on $k,E$.
\end{theorem}

\begin{proof}
Let $W_E=\sum_{\ell\notin E}B_\ell\omega_\ell$, with independent
$B_\ell\sim\operatorname{Bernoulli}(1/(\ell-1))$, and let $V_{k,E}$
be the sum of $k$ independent copies of $W_E$. All prime indices in
this proof are odd. For sufficiently small $u>0$, the event
$W_\varnothing\le u$ forces $B_\ell=0$ for every $\ell\in E$; hence
\[
 \mathbb P(W_E\le u)=q_E^{-1}\mathbb P(W_\varnothing\le u).
\]
Write $F_{k,E}(u)=\mathbb P(V_{k,E}\le u)$. The same argument gives
$F_{k,E}(u)=q_E^{-k}F_{k,\varnothing}(u)$ below every deleted jump.
Write $F(u)=\mathbb P(W_E\le u)$ and $K=\kappa_{\mathrm{end}}/q_E$.
We claim
\begin{equation}\label{eq:convolution-endpoint-refined-mas20}
 F_{k,E}(u)=K^k\log^{-k}(1/u)
               +O_{k,E}(\log^{-k-2}(1/u)).
\end{equation}
For $k=1$ this is \Cref{prop:endpoint-half}, since $a_2=0$.
For $k\ge2$, compare $F(u)^k$ with $F_{k,E}(u)$.
On their difference event every coordinate is at most $u$, their sum
exceeds $u$, and the largest coordinate is at least $u/k$.
Put $L_u=\log(1/u)$ and $J=\lceil3\log_2L_u\rceil$.
If the largest coordinate is at most $u/2$, two coordinates lie in
$[u/(2k),u]$: the maximum is at least $u/k$, and another coordinate
is greater than $u/(2(k-1))$. Their shell probabilities are
$O_{k,E}(L_u^{-2})$.
Otherwise partition its distance from $u$ into
$2^{-j-1}u\le u-W_i<2^{-j}u$, $1\le j\le J$.
Then another coordinate exceeds $2^{-j-1}u/(k-1)$.
The fixed-order endpoint expansion, used through order four, gives
uniformly for these $j$
\[
 \begin{aligned}
 \mathbb P\{(1-2^{-j})u<W_E\le(1-2^{-j-1})u\}
       &\ll_E2^{-j}L_u^{-2}+L_u^{-5},\\
 \mathbb P\{2^{-j-1}u/(k-1)<W_E\le u\}
       &\ll_{k,E}(j+1)L_u^{-2}.
 \end{aligned}
\]
The other $k-2$ coordinates contribute $O_{k,E}(L_u^{-k+2})$.
Summation uses $\sum_j(j+1)2^{-j}<\infty$; the remainder
$L_u^{-5}\sum_{j\le J}(j+1)L_u^{-2}$ is smaller.
In the last thin strip $u-W_i<2^{-J-1}u$, the same expansion bounds
its probability by $O_E(L_u^{-5})$, and all other coordinates are at
most $u$. A union bound over ordered pairs of coordinates thus gives
$0\le F(u)^k-F_{k,E}(u)\ll_{k,E}L_u^{-k-2}$.
Finally $F(u)=K/L_u+O_E(L_u^{-3})$ proves
\eqref{eq:convolution-endpoint-refined-mas20}.
In particular, for fixed $C$, the increment
$F_{k,E}(v)-F_{k,E}(u)$ is $O_{k,E,C}(\log^{-k-1}(1/u))$
uniformly for $u\le v\le Cu$. The law of $W_E$ is atomless by
\Cref{lem:levy-atomless}; thus the width-$h$ interval mass of $Y$
is continuous in its endpoint and tends to zero at infinity.
Its supremum is therefore attained.

Put $A=e^{-a}$ and $Z=e^{-Y}=A\prod_{\ell\notin E}(1-1/\ell)^{J_\ell}$.
Apply \Cref{lem:binomial-uniform-deletion} with this $E,A$ and
$(\alpha,\beta)=(1/10,2/5)$.
Windows with $t\le0$ or $t\ge A/(1-\varepsilon)$ have zero mass.
For any remaining relative window $(t(1-\varepsilon),t]$, put
$z=\log(1/\varepsilon)$ and $\delta=\varepsilon^{2/5}$.
Either its mass is $O_{k,E,c}(e^{-c\sqrt{z\log z}})$, or it is at most
$K_k(\mathbf j)$ times the mass of the relative window ending at
$t'=At/A_j$, plus that error, where
\[
\begin{aligned}
 A_j&=A\prod_{\ell\notin E}(1-1/\ell)^{j_\ell},\\
 K_k(\mathbf j)&=\prod_{\ell\notin E}
 \binom{k}{j_\ell}(\ell-2)^{-j_\ell}\le\mathfrak c_{k,E},\\
 A(1-\delta)&<t'<\frac A{1-\varepsilon}.
\end{aligned}
\]
Here $0<c<1/\sqrt{20}$ is fixed. Translation by $a$ does not change
concentration or localization errors, so all constants can be taken
independent of $a$. The surviving event in the lemma's proof has
this single head vector and satisfies $Z=A_j\Pi\le A_j$.

If $t'\ge A$, its mass is at most
$F_{k,E}(-\log(1-\varepsilon))$.
Otherwise put $t'=A(1-\vartheta)$, $0\le\vartheta<\delta$,
and $H_{k,E}(s)=F_{k,E}(-\log(1-s))$.
The window mass is
$H_{k,E}(\vartheta+\varepsilon(1-\vartheta))-H_{k,E}(\vartheta)$.
For $\vartheta<\varepsilon$, bound it by $H_{k,E}(2\varepsilon)$.
For $\varepsilon\le\vartheta<\delta$, the two positive logarithmic
arguments have ratio at most two, so the increment estimate bounds
it by $O_{k,E}(z^{-k-1})$. This proves the preliminary global
upper bound with error $O_{k,E}(z^{-k-1})$, which is sufficient for
the localization argument below.
For the lower bound take a fixed maximizing vector $\mathbf j$.
For sufficiently small $h$, the event $V_{k,E}<h$ forces $J_\ell=0$
at every prime with $j_\ell>0$. Replacing these zero counts by
$j_\ell$ multiplies the event's probability by $\mathfrak c_{k,E}$ and
produces a width-$h$ interval beginning at $-\log A_j$.
With $\varepsilon=1-e^{-h}$, this supplies the matching lower bound
$\mathfrak c_{k,E}F_{k,E}(h)$ and the preliminary leading asymptotic.

Successive local weights have ratio $(k-j+1)/(j(\ell-2))$.
The unique mode is $\lfloor(k+1)/(\ell-1)\rfloor$ unless
$(k+1)/(\ell-1)$ is an integer $n$, in which case the two modes
are $n-1$ and $n$. Thus the maximal factor is one for $\ell\ge k+2$,
but $\ell=k+2$, when prime and not deleted, allows both $j=0,1$.
By \Cref{thm:linind}, the $r_{k,E}$ independent tie choices give
$2^{r_{k,E}}$ distinct anchors. This proves the uniqueness criterion.

For fixed $k,E$, nonmaximal odds products are uniformly separated
below $\mathfrak c_{k,E}$: there are finitely many choices at primes at
most $k+2$, and a nonzero choice at a larger prime has weight at most
$k/(\ell-2)$, uniformly below one. Hence a maximizing window must
select a maximal-mode anchor for small $h$.
The surviving event intersects $[b,\infty)$, so $b_h\ge b-h$.
If $b_h\ge b+\eta h$ for some fixed $\eta>0$, the endpoint
increment estimate gives only $O_{k,E,\eta}(\log^{-k-1}(1/h))$
mass plus the residual, a contradiction. Thus $b_h\le b+o_{k,E}(h)$.
For the stronger bound put $s=b_h-b>0$, $L=\log(1/h)$ and
$R_h=O_{k,E}(e^{-c\sqrt{L\log L}})$.
The same head-removal comparison, now in logarithmic coordinates,
bounds the maximizing interval by
\[
 \mathfrak c_{k,E}\{F_{k,E}(h+s)-F_{k,E}(s)\}+R_h,
\]
whereas the modal cylinder gives at least
$\mathfrak c_{k,E}F_{k,E}(h)$. Since $s<h$ for small $h$,
\[
 F_{k,E}(s)\le F_{k,E}(2h)-F_{k,E}(h)+O_{k,E}(R_h)
                    \ll_{k,E}L^{-k-1}.
\]
The endpoint lower bound $F_{k,E}(s)\gg_{k,E}\log^{-k}(1/s)$
therefore implies $\log(1/s)\ge c_{k,E}L^{1+1/k}$, proving
\eqref{eq:convolution-displacement-mas20} after reducing the constant.
Apply \eqref{eq:convolution-endpoint-refined-mas20} at $h+s$ and $h$.
Their main terms differ by $O_{k,E}((s/h)L^{-k-1})$, smaller than every
fixed inverse power of $L$, and their remainders are $O_{k,E}(L^{-k-2})$.
If $b_h\le b$, head removal instead bounds the interval by
$\mathfrak c_{k,E}F_{k,E}(h)+R_h$ directly. Together with the modal
lower bound these comparisons prove \eqref{eq:convolution-constant}.

Suppose now that $E$ contains every odd prime through $k+2$.
Translate to $a=0$ and put $F=F_{k,E}$, $L_u=\log(1/u)$.
Every surviving local mode is zero, so $\mathcal B_{k,E}(0)=\{0\}$.
Full support on $[0,\infty)$, by \Cref{lem:supp_conv}, excludes negative
maximizing lower endpoints. Localization reduces the remaining comparison to
\[
 F(h+s)-F(h)<F(s)\qquad(0<s\le h).
\]
For $h^3\le s\le h$, the left side is at most
$F(2h)-F(h)=O_{k,E}(L_h^{-k-1})$, whereas $F(s)\gg_{k,E}L_h^{-k}$.
For $0<s<h^3$, apply \Cref{lem:binomial-uniform-deletion} with
$A=1$, $t=e^{-h}$ and $\varepsilon=1-e^{-s}$, retaining
$(\alpha,\beta)=(1/10,2/5)$.
Since $\delta=\varepsilon^{2/5}=o(h)$, any surviving head is nonempty;
its anchor exceeds $e^{-(h+s)}\ge e^{-2h}$, so every selected prime
exceeds $1/(2h)$. The odds satisfy
$\binom{k}{j}(p-2)^{-j}\le(k/(p-2))^j$, hence
$K_k(\mathbf j)=O_k(h)$. Equations~\eqref{eq:binomial-uniform-deletion}
and~\eqref{eq:convolution-constant} give
\[
 F(h+s)-F(h)\ll_{k,E}hL_s^{-k}
       +e^{-c\sqrt{L_s\log L_s}}=o(F(s))
\]
uniformly for $0<s<h^3$. This proves uniqueness and, with the exact
endpoint identity above, the final assertion.
\end{proof}

For $E=\varnothing$ and $a=k\log2$, write
$\mathfrak c_k=\mathfrak c_{k,\varnothing}$ and
$\mathcal B_k=\mathcal B_{k,\varnothing}(k\log2)$; then
$\nu_{k,\varnothing,k\log2}=\mu_f^{*k}$, and the mode anchor is unique
precisely when $k$ is even. Here
$\mathfrak c_2=2$, $\mathfrak c_3=3$, $\mathfrak c_4=8$,
$\mathcal B_2=\{\log6\}$, $\mathcal B_4=\{\log45\}$, and
$\mathcal B_3=\{\log12,\log15,\log18,\log(45/2)\}$.
The uniqueness assertion follows since $\ell-1$ is even and
$\ell=3$ is available; the displayed values follow from the local modes.

\KeepHeadingSpace{4\baselineskip}
\begin{corollary}[Growth of the convolution constants]
\label{cor:convolution-constant-growth}
As $k\to\infty$, the constants satisfy
\[
 \log\mathfrak c_k
 =k\log\log k-k\log\kappa_{\mathrm{end}}+O(k/\log k),
 \qquad
 \frac{\mathfrak c_k^{1/k}}{\log k}\longrightarrow
 \kappa_{\mathrm{end}}^{-1}.
\]
\end{corollary}

\begin{proof}
For odd primes $\ell$, let $m_{k,\ell}$ be the maximal atom of
$\operatorname{Bin}(k,1/(\ell-1))$. Then
\[
 \mathfrak c_k=Q(k+2)^{-k}\prod_{3\le\ell\le k+2}m_{k,\ell}.
\]
Chebyshev's inequality gives
$m_{k,\ell}\gg(1+k/\ell)^{-1/2}$.
Chebyshev's prime-counting bound and partial summation yield
\[
 0\le-\sum_{3\le\ell\le k+2}\log m_{k,\ell}
 \ll\sum_{3\le\ell\le k+2}\{1+\log(1+k/\ell)\}
 \ll\frac{k}{\log k}.
\]
Now \eqref{eq:endpoint-Q-quantitative} proves the stated asymptotic
and limit.
\end{proof}

\subsection{Additive concentration and progression laws}
\label{ssec:additive-concentration-progressions}

For the raw additive width $h$, the window ending at $1/3$ has relative
width $3h$. The corollary below proves
$C(h)=H_{\mathrm{end}}(3h)+O(\mathcal R(h))$; the shift
$\log(1/(3h))=\log(1/h)-\log3$ supplies its second-order coefficient.

\begin{corollary}[Sharp additive concentration and localization]
\leavevmode\phantomsection\label{cor:sharp-additive-modulus}
\index{Concentration functions!raw totient law}\ignorespaces
As $h\downarrow0$, the concentration function
$C(h)=\sup_{t\in\R}\{G(t)-G(t-h)\}$ of $\mu_G$ satisfies
\begin{equation}\label{eq:sharp-additive-modulus}
\begin{aligned}
 C(h)&=H_{\mathrm{end}}(3h)+O\bigl(\mathcal R(h)\bigr)\\
     &=\frac{\kappa_{\mathrm{end}}}{L}
       +\frac{\kappa_{\mathrm{end}}\log3}{L^2}
       +\frac{\kappa_{\mathrm{end}}((\log3)^2-\pi^2/12)}{L^3}
       +O(L^{-4}),\qquad L=\log(1/h).
\end{aligned}
\end{equation}
More generally, for every fixed integer $N\ge1$,
$C(h)=\kappa_{\mathrm{end}}\sum_{j=1}^N a_j(L-\log3)^{-j}
+O_N(L^{-N-1})$, with $a_j$ as in \Cref{prop:endpoint-half}.
For every $h>0$ the supremum is attained, and every maximizing upper
endpoint $t_h$ satisfies $t_h<\tfrac13+h$. For all sufficiently small
$h>0$, every such endpoint satisfies
\begin{equation}\label{eq:additive-localisation}
 0\le t_h-\tfrac13\ll hL^2\mathcal R(h).
\end{equation}
\end{corollary}

\begin{proof}
Write $A_h(t):=G(t)-G(t-h)$, $H:=H_{\mathrm{end}}$ and $\kappa:=\kappa_{\mathrm{end}}$. As in Step~2 of the proof of \Cref{thm:sharp-global-modulus}, $C(h)$ is attained, and every maximizer lies in $(0,\tfrac12+h)$.

Fix $h>0$ and let $\tfrac13+h\le t<\tfrac12+h$. Then $\eta:=h/t\in(0,1)$
satisfies $A_h(t)=q_\eta(t)$ and $\tfrac13\le t(1-\eta)=t-h<\tfrac12$, so
\eqref{eq:prime-three-translate} applies at width $\eta$. As $G$ is
nondecreasing and $\tfrac23t-h<\tfrac23(t-h)$,
\[
 A_h(\tfrac23t)\ge q_\eta(\tfrac23t)>q_\eta(t)=A_h(t).
\]
This excludes $t_h\ge\tfrac13+h$ for every $h>0$.

Now let $h\downarrow0$ and put $L:=\log(1/h)$. The branch $\{B_3=1\}$ in
\Cref{lem:prime-three-split}(a) and \eqref{eq:hend-asymp} give
\[
 C(h)\ge A_h(\tfrac13)\ge H(3h)=\kappa/L+O(L^{-2}).
\]
If $t\le\sqrt h$, then
$A_h(t)\le G(\sqrt h)\le M(-1)\sqrt h$ by \Cref{cor:avg-loss}, so such
$t$ cannot maximize for small $h$. At a maximizer put
$\eta:=h/t_h$ and $z_\eta:=\log(1/\eta)$. Then
$\eta\le\sqrt h$ and $z_\eta\ge L/2$, whence, for small $h$,
\[
 q_\eta(t_h)=C(h)\ge\frac{(2/5)\kappa}{z_\eta}.
\]
By \Cref{prop:two-window}(d) with $a=\tfrac25$,
$t_h\in I_2(\eta)\cup I_3(\eta)$, in particular $t_h\ge\tfrac14$.
If $t_h\in I_2(\eta)$, \Cref{prop:two-window}(c) gives
$t_h(1-\eta)>\tfrac13$, contrary to $t_h<\tfrac13+h$.
Thus $t_h\in I_3(\eta)$.

First we prove an intermediate Gaussian bound, for which we may assume
$t_h\le\tfrac13$. Write $t_h=\tfrac13(1-\theta)$, so $0\le\theta<\eta^{2/5}$ and
$3h\le\eta\le4h$. By \Cref{lem:two-window-comparison}, uniformly in this range,
\begin{equation}\label{eq:additive-endpoint-transfer}
 A_h(t_h)=H(\theta+3h)-H(\theta)+O(\mathcal R(\eta)),
\end{equation}
where $H(0)=0$; the error is $o(L^{-2})$ by \eqref{eq:residual-properties}
with $[d_0,d_1]=[3,6]$. If $3h\le\theta<\eta^{2/5}$, then $\theta+3h\le2\theta$
and $\log(1/\theta)\ge\tfrac25L-O(1)$, so \eqref{eq:hend-asymp} makes the right
side of \eqref{eq:additive-endpoint-transfer} $O(L^{-2})$, contradicting
$C(h)\ge H(3h)$. Fix $0<c\le c_0(3,6)$ as in \Cref{lem:endpoint-small-comparison}
and put $\theta_0:=e^{-cL^2}$. If $\theta_0\le\theta<3h$, monotonicity of $H$ in
\eqref{eq:additive-endpoint-transfer} and \Cref{lem:endpoint-small-comparison} give
\[
 C(h)-A_h(t_h)\ge H(\theta_0)-[H(6h)-H(3h)]-O(\mathcal R(\eta))>0,
\]
again impossible. Hence $\theta<\theta_0$, so
$t_h>\tfrac13(1-e^{-c\log^2(1/h)})$.

By this Gaussian bound, every maximizer satisfies
$\eta=h/t_h<3h/(1-\theta_0)\le3h+\theta_0$ for small $h$. Since
$q_\varepsilon(t)$ is nondecreasing in $\varepsilon$, so is $Q_G$; hence
\Cref{thm:sharp-global-modulus}(a) and \eqref{eq:residual-properties} give
\[
 C(h)=q_\eta(t_h)\le Q_G(3h+\theta_0)
 =H(3h+\theta_0)+O\bigl(\mathcal R(h)\bigr).
\]
As $\theta_0\le e^{-c\log^2(1/(3h))}$, \eqref{eq:microscopic-endpoint-increment}
at $\varepsilon=3h$ and \eqref{eq:residual-properties} give
$H(3h+\theta_0)-H(3h)=o(\mathcal R(h))$. With $C(h)\ge H(3h)$, this proves the
first line of \eqref{eq:sharp-additive-modulus}. The second follows
from \eqref{eq:hend-asymp}, $\log(1/(3h))=\log(1/h)-\log3$, and
$\mathcal R(h)=o(\log^{-N}(1/h))$ for every fixed $N$.
The same substitution gives the expansion to every fixed order.

For the lower bound in \eqref{eq:additive-localisation}, put $a=1/3$ and suppose
$t_h=a-s<a$. The Gaussian bound gives $0<s<ae^{-c\log^2(1/h)}$.
Set $b=a-h>0$, $\rho=s/a$ and $\eta=s/b$. Then
\[
 A_h(a)-A_h(a-s)=q_\rho(a)-q_\eta(b).
\]
For small $h$, one has $\eta^{2/5}<3h$, so $b=a(1-3h)$ lies below
both $I_2(\eta)$ and $I_3(\eta)$. Moreover
\[
 \log(1/\eta)=\log(1/\rho)+\log(1-3h),\qquad
 \log(1/\rho)>c\log^2(1/h).
\]
Thus $\log(1/\eta)/\log(1/\rho)\ge3/4$ uniformly for such $s$ when $h$
is small. Applying \eqref{eq:anchor-exchange-bounds} at both widths gives
\[
 q_\eta(b)<\frac{\kappa}{2\log(1/\eta)}
 \le\frac{2\kappa}{3\log(1/\rho)}
 <\frac{3\kappa}{4\log(1/\rho)}\le q_\rho(a),
\]
contradicting maximality. Hence $t_h\ge a$.

Write now $t_h=1/3+s$, where $0\le s<h$, and put $\eta=h/t_h$.
We have $t_h\in I_3(\eta)$, as proved above, and $\eta\asymp h$.
Since $3t_h/2\ge1/2$ and $t_h-h<1/3$,
\eqref{eq:two-window-comparison} gives
\[
 A_h(t_h)=q_\eta(t_h)
 =H\bigl(3h-3s\bigr)+r_h,
 \qquad 0<r_h\ll\mathcal R(\eta)\ll\mathcal R(h),
\]
where the last bound follows from \eqref{eq:residual-properties}.
Comparing with $A_h(1/3)\ge H(3h)$ yields
\begin{equation}\label{eq:additive-maximizer-decrement}
 0\le H(3h)-H\bigl(3h(1-s/h)\bigr)\ll\mathcal R(h).
\end{equation}
Put $d=s/h\in[0,1)$. If $d\ge1/2$, monotonicity and
\Cref{lem:endpoint-increment-lower}, applied at width $3h$ and
decrement $1/2$, contradict \eqref{eq:additive-maximizer-decrement},
since $\mathcal R(h)=o(\log^{-2}(1/h))$. Thus $d<1/2$.
For $d\ge(3h)^{1/4}$, the same lemma gives
$d\ll\log^2(1/h)\mathcal R(h)$; for smaller $d$ this bound follows
from $(3h)^{1/4}=o(\log^2(1/h)\mathcal R(h))$.
Multiplication by $h$ proves \eqref{eq:additive-localisation}.
\end{proof}

Fix $Q\ge1$ and $\gcd(b,Q)=1$. Put
\[
 S:=\{\ell\ge3:\ell\textup{ prime},\ \ell\mid Q\},\qquad
 F:=\prod_{\substack{\ell\in S\\b\not\equiv1\pmod{\ell}}}(1-1/\ell).
\]
Use $q_S$ from \Cref{def:anchor-weights}. Let $\mu_{f,Q,b}$ be the law
of $-\log X_{Q,b}$ and put
$C_{Q,b}(h):=\sup_t\{G_{Q,b}(t)-G_{Q,b}(t-h)\}$.

\begin{corollary}[Concentration in a fixed progression]
\label{cor:ap-analytic-structure}
Put $A=A_{Q,b}$, $K=\kappa_{\mathrm{end}}/q_S$ and $L=\log(1/h)$.
For these fixed progression parameters, as $h\downarrow0$,
\begin{equation}\label{eq:ap-sharp-concentration}
 \begin{aligned}
 Q_{\mu_{f,Q,b}}(h)&=K/L+O_Q(L^{-3}),\\
 1-G_{Q,b}(A-h)&=K/L-K\log A/L^2+O_Q(L^{-3}),\\
 C_{Q,b}(h)&=
 \begin{cases}
 K/L+K\log(3/(2A))/L^2+O_Q(L^{-3}),&3\nmid Q,\\
 K/L-K\log A/L^2+O_Q(L^{-3}),&3\mid Q.
 \end{cases}
 \end{aligned}
\end{equation}
More generally, to every fixed order $N$, these quantities equal
$K\sum_{j=1}^Na_j(L+d)^{-j}+O_{N,Q}(L^{-N-1})$, respectively with
$d=0$, $d=\log A$, and $d=\log(2A/3)$ or $\log A$ according as
$3\nmid Q$ or $3\mid Q$. The $a_j$ are those of
\Cref{prop:endpoint-half}. The raw endpoint expansion also appears in
\Cref{cor:ap-endpoint-expansions}.
If $3\mid Q$, the unique maximizing interval for $\mu_{f,Q,b}$,
for all sufficiently small $h>0$, is
$[-\log A_{Q,b},-\log A_{Q,b}+h]$, and
\[
 Q_{\mu_{f,Q,b}}(h)=q_S^{-1}H_{\mathrm{end}}(1-e^{-h}).
\]
\end{corollary}

\begin{proof}
Write $A=A_{Q,b}$, so that $X_{Q,b}=Ae^{-W_S}$ by \eqref{eq:XQ}, with $W_S$
as in \eqref{eq:deleted-bernoulli-sum}, while $X=\tfrac12e^{-W_\varnothing}$.
For all sufficiently small $u>0$, deleting the forced-zero
coordinates gives the exact identity
\[
 \mathbb P(W_S\le u)=q_S^{-1}\mathbb P(W_\varnothing\le u).
\]
Also \Cref{lem:mass-split} gives
\[
 C_{Q,b}(h)\le q_S^{-1}C(h/(2A)),\qquad
 Q_{\mu_{f,Q,b}}(h)\le q_S^{-1}Q_{\muF}(h).
\]
The logarithmic endpoint lower bound and
\Cref{thm:sharp-global-modulus} sandwich logarithmic concentration
between quantities with the same expansion to every fixed order.
The raw endpoint follows by putting $u=-\log(1-h/A)$, whose logarithmic
argument is $L+\log A+O_Q(h)$.
If $3\nmid Q$, the cylinder obtained by activating $3$ in the
endpoint event has equal probability (its odds are one), so the
interval ending at $2A/3$ has mass at least
$q_S^{-1}H_{\mathrm{end}}(3h/(2A))$.
The additive upper bound and \Cref{cor:sharp-additive-modulus}
have precisely this expansion, with denominator $L+\log(2A/3)$.

For $3\mid Q$, \Cref{thm:convolution-modes} with $k=1$, $E=S$ and
$a=-\log A$ proves the stated exact logarithmic maximization.
For raw additive concentration, the supremum is attained by continuity
and compact support. An upper endpoint $t\le\sqrt h$ has mass at most
$\mathbb P(X_{Q,b}\le\sqrt h)\ll_Q\sqrt h$, using the finite first
reciprocal moment. It cannot maximize, because the endpoint mass is
$K/L+O_Q(L^{-2})$. At a maximizing $t>\sqrt h$, put
$\varepsilon=h/t$ and $z=\log(1/\varepsilon)$; then
$L/2\le z\le L+O_Q(1)$.
Apply \Cref{lem:binomial-uniform-deletion} with $k=1$, $E=S$ and
$(\alpha,\beta)=(1/10,2/5)$.
A nonzero head has odds product at most $1/3$, since every available
prime is at least $5$. Its interval mass is consequently at most
\[
 \tfrac13 Q_{\mu_{f,Q,b}}(-\log(1-\varepsilon))
       +O_Q(e^{-c\sqrt{z\log z}})
 \le(2/3+o_Q(1))K/L,
\]
by the logarithmic expansion already proved. The small-residual
alternative is excluded as well. Thus the surviving head is zero;
the lemma gives $A(1-\varepsilon^{2/5})<t<A/(1-\varepsilon)$.
First $t\to A$, then $\varepsilon\asymp_Q h$, so
$t=A(1+O_Q(h^{2/5}))$. The additive interval is a logarithmic interval
of width $-\log(1-h/t)$, with
\[
 \log\frac1{-\log(1-h/t)}=L+\log A+O_Q(h^{2/5}).
\]
Its logarithmic concentration bound, to any fixed order, and the raw
endpoint lower bound prove the claimed additive expansion.
Finally $a_1=1$ and $a_2=0$ give all second coefficients in
\eqref{eq:ap-sharp-concentration}.
\end{proof}

\begin{corollary}[Mellin transforms and lower tails in a fixed progression]
\label{cor:ap-mellin-tail-structure}
For the fixed progression above, as $r\to\infty$,
\begin{equation}\label{eq:ap-high-moments}
 \log M_{Q,b}(-r)
 =r(\log\log r+\gamma+\log F)
   +O_Q\!\left(\frac{r\log\log r}{\log r}\right),
\end{equation}
and
\begin{equation}\label{eq:ap-small-tail}
 \lim_{\varepsilon\downarrow0}
 \varepsilon\log\log\frac1{G_{Q,b}(\varepsilon)}
 =\frac{e^{-\gamma}}F.
\end{equation}
The zeros of $M_{Q,b}$ are simple and are exactly the zeros $s_{\ell,k}$
of $M$ in \Cref{prop:hadamard} with $\ell\nmid Q$. Zeros on the imaginary
axis occur if and only if $3\nmid Q$.
\end{corollary}

\begin{proof}
Put $A=A_{Q,b}$. For the reciprocal moments, \eqref{eq:ap-mellin-product} and the definition
of $M$ in \Cref{sec:mellin} give
\[
 M_{Q,b}(z)\prod_{\ell\in S}F_z(\ell)=(2A)^zM(z)\qquad(z\in\C).
\]
For fixed $\ell$, $\log F_{-r}(\ell)=r\omega_\ell+O_\ell(1)$. Since
$-\log(2A)=\sum_{\ell\in S}e_\ell(b)\omega_\ell$, the linear terms total
$-r\sum_{\ell\in S}(1-e_\ell(b))\omega_\ell=r\log F$, so
\eqref{eq:high-reciprocal-moments} proves \eqref{eq:ap-high-moments}.

Put $Z=1/X_{Q,b}$, so that $\mathbb E[Z^r]=M_{Q,b}(-r)$ by
\eqref{eq:ap-mellin-identity}, and fix $0<\eta<1$. Repeat the
argument of \Cref{cor:small-x-tail} with scale $e^{-\gamma}/F$ in
both $r$ and $y$, using \eqref{eq:ap-high-moments} for the Markov
bound and continuity of $G_{Q,b}$ from \Cref{thm:ap-law} to identify
the strict reciprocal tail.
In the lower bound, require $B_\ell=1$ only for odd primes $\ell\le y$ with
$\ell\notin S$; this cylinder has probability at least
$\prod_{3\le\ell\le y}(\ell-1)^{-1}\ge e^{-Cy}$. Once $y$ exceeds every prime
in $S$, $A^{-1}=2\prod_{\ell\in S}(1-1/\ell)^{-e_\ell(b)}$ gives, on it,
$Z\ge F\prod_{\ell\le y}(1-1/\ell)^{-1}=(1+\eta)t(1+o_Q(1))>t$ for large $t$.
The two bounds and $\eta\downarrow0$ prove \eqref{eq:ap-small-tail}.

Finally, by \eqref{eq:ap-mellin-product}, $M_{Q,b}$ is the nonvanishing
factor $A^z$ times the product of the $F_z(\ell)$ over odd primes $\ell\nmid Q$,
so the proof of \Cref{prop:hadamard} applies with $A^z$ in place of $2^{-s}$;
the zeros on $i\R$ are the $s_{3,k}$, which occur if and only if $3\nmid Q$.
\end{proof}

\begin{remark}[Convolution roots]\label{rem:iid-convolution-roots}
Let $\nu$ be $\mu_{f,Q,b}$, $\mugodd$ or $\mug$.
For positive integers $k,m$, the law $\nu^{*k}$ has an $m$th iid
convolution root if and only if $m\mid k$.
This follows from the analytic-factor argument of L\'evy--Raikov and
the zero-multiplicity argument of Lukacs--Sz\'asz~\cite[\S\S~3--4 and Theorem~4]{LS53}.
Indeed, if $\eta^{*m}=\nu^{*k}$, Tonelli's theorem shows that $\eta$ has
all real exponential moments. The entire Laplace transforms then satisfy
$H_\eta(z)^m=H_\nu(z)^k$ by the identity theorem. Here $H_\nu(z)$ is
$M_{Q,b}(-z)$, $M_\sigma(z)$ or $N_\sigma(z)$, respectively. Each has
a simple zero by \Cref{cor:ap-mellin-tail-structure,prop:sigma-mellin-zeros},
so comparison of multiplicities forces $m\mid k$. Conversely,
$\eta=\nu^{*(k/m)}$ is a root when $m\mid k$. In particular none of
these laws is infinitely divisible.
\end{remark}

\subsection{Truncation error, shortest intervals and quantiles}
\label{ssec:truncation-shortest-quantiles}

\Needspace{6\baselineskip}
\begin{corollary}[Sharp truncation]
\label{cor:sharp-truncation-inverse}
\index{Truncation!sharp error}
With $k,E,a,Y$ as in \Cref{thm:convolution-modes}, put
$Y_y=a+\sum_{3\le\ell\le y,\,\ell\notin E}J_\ell\omega_\ell$.
As $y\to\infty$, with $y\ge\max(E\cup\{k+2\})$,
\begin{equation}\label{eq:sharp-convolution-truncation}
 d_K(\operatorname{Law}(Y_y),\operatorname{Law}(Y))
 =\frac{\mathfrak c_{k,E}(\kappa_{\mathrm{end}}/q_E)^k}{\log^k y}
       +O_{k,E}(\log^{-k-1}y).
\end{equation}
The same formula holds after applying $u\mapsto e^{-u}$ to both laws.
In particular, $E=\varnothing$, $k=1$, $a=\log2$ and $y=\ell_N$
give the sharper truncation error, with $L=\log y$,
\begin{equation}\label{eq:undeleted-truncation-refined-mas20}
 d_K(\operatorname{Law}(X^{(N)}),\mu_G)
 =\frac{\kappa_{\mathrm{end}}}{L}
       +O\!\left(\frac{(\log L)^2}{L^3}\right).
\end{equation}
Here $y=\ell_N$ is the last retained prime and $X^{(N)}$ is as in
\Cref{lem:truncation-wasserstein}; the same absolute error holds for
the logarithmic truncations.
Taking $E=\{\ell\ge3:\ell\mid Q\}$ and $a=-k\log A_{Q,b}$ gives
the product of $k$ independent copies of $X_{Q,b}$, all truncated
at the same $y$.
\end{corollary}

\begin{proof}
Put $T_y=Y-Y_y=\sum_{\ell>y}J_\ell\omega_\ell\ge0$.
It is independent of $Y_y$ and positive almost surely, since the
activation probabilities have divergent sum. By \Cref{thm:linind},
distinct head vectors give distinct values of $Y_y$. Its largest atom is
\[
 \mathfrak c_{k,E}\left(\frac{Q(y)}{q_E}\right)^k
 =\frac{\mathfrak c_{k,E}(\kappa_{\mathrm{end}}/q_E)^k}{\log^k y}
     +O_{k,E}(\log^{-k-1}y),
\]
by \eqref{eq:endpoint-Q-quantitative}; the cutoff includes every
nontrivial local mode. At an atom $v$ attaining this mass,
\[
 \mathbb P(Y_y\le v)-\mathbb P(Y\le v)
 =\mathbb P(Y_y\le v<Y)\ge\mathbb P(Y_y=v),
\]
which proves the lower bound.

By \eqref{eq:endpoint-tail-square}, $y\mathbb ET_y\ll_k1/\log y$.
For $\ell>y\ge2$, $y\omega_\ell\le1+2/y$. Decomposing the binomial
counts into $k$ Bernoulli copies, convexity and independence give
\[
 \log\mathbb E e^{syT_y}
 \ll_k\frac{e^{s(1+2/y)}-1}{\log y}\qquad(s>0).
\]
Taking $s=\log\log y$ and $d=k+2$ yields
$\mathbb P(T_y>d/y)\ll_k\log^{-k-2}y$.
Consequently, uniformly in $t$,
\[
 0\le\mathbb P(Y_y\le t)-\mathbb P(Y\le t)
 \le Q_{\nu_{k,E,a}}(d/y)+O_k(\log^{-k-2}y).
\]
Equation \eqref{eq:convolution-constant} proves the upper bound in
\eqref{eq:sharp-convolution-truncation}. Kolmogorov distance is invariant
under a common strictly monotone continuous map (using left limits
when the order is reversed), which gives the product assertion.

For the refinement \eqref{eq:undeleted-truncation-refined-mas20},
we first record a smoothing inequality. Let $H$ and $T\ge0$ be
independent, $V=H+T$, $p=\mathbb P(T\le h)>0$, and write $F_H,F_V$
for their distribution functions. Since $F_H\ge F_V$,
\[
 F_V(t+h)\ge pF_H(t)+\mathbb E[F_V(t+h-T)\mathbf1_{\{T>h\}}].
\]
Rearranging gives
\[
 p\{F_H(t)-F_V(t)\}
 \le p\{F_V(t+h)-F_V(t)\}
 +\mathbb E[\{F_V(t+h)-F_V(t+h-T)\}\mathbf1_{\{T>h\}}].
\]
Bound the two increments by their concentration functions and take
suprema in $t$ to obtain
\begin{equation}\label{eq:tail-smoothing-mas20}
 p\,d_K(\operatorname{Law}(H),\operatorname{Law}(V))
 \le pQ_{\operatorname{Law}(V)}(h)
       +\mathbb E[Q_{\operatorname{Law}(V)}(T)\mathbf1_{\{T>h\}}].
\end{equation}
Apply this with the undeleted Bernoulli head, $T=T_y$, $L=\log y$
and $h=(1+2/L)/y$. Split $yT_y$ into jumps exceeding $1/L$ and the
remaining sum $R$. If $N$ is the number of the former, prime partial
summation gives
\[
 \lambda:=\mathbb EN\ll\frac{\log L}{L},\qquad
 \mathbb ER\ll L^{-2},
\]
since the two prime ranges are $y<p\ll yL$ and $p\gg yL$,
respectively. They are independent, and each large jump is at most
$1+2/y$. If $N=0$, Markov bounds the exceptional probability by
$O(L^{-2})$; if $N=1$, exceeding $yh$ requires $R>1/L$ for large $y$,
so independence bounds it by $O(\lambda/L)$.
Finally $\mathbb P(N\ge2)\le\lambda^2/2$, by the second factorial moment.
Thus
\[
 1-p=\mathbb P(T_y>h)\ll (\log L)^2/L^2.
\]
The preceding exponential-moment bound with $s=\log L$ also gives
$\mathbb P(T_y>y^{-1/2})\le\exp\{-\sqrt y\log L+O(1)\}$.
On $h<T_y\le y^{-1/2}$, \Cref{thm:sharp-global-modulus} bounds the
concentration by $O(1/L)$; beyond this range use the bound one.
Consequently the expectation in \eqref{eq:tail-smoothing-mas20} is
$O((\log L)^2/L^3)$. Since
$\log(1/h)=L-\log(1+2/L)=L+O(L^{-1})$, the same theorem gives
$Q_{\operatorname{Law}(Y)}(h)=\kappa_{\mathrm{end}}/L+O(L^{-3})$.
The lower bound is the largest head atom $Q(y)$, whose relative
error in \eqref{eq:endpoint-Q-quantitative} is exponentially small
in $\sqrt L$. This proves the refinement and its logarithmic version.

\end{proof}

\Needspace{6\baselineskip}
\begin{corollary}[Shortest intervals of prescribed probability]
\label{cor:shortest-mass-intervals}
For $0<u<1$, the minimum width $w(u)$ of an interval of
$\mu_G$-mass at least $u$ is attained and satisfies $C(w(u))=u$.
Every minimizing interval has mass exactly $u$, and its upper endpoint
$b_u$ satisfies $b_u<\tfrac13+w(u)$. As $u\downarrow0$,
\[
 w(u)=\frac13\exp\!\left\{-\frac{\kappa_{\mathrm{end}}}{u}
             +\frac{\pi^2}{12\kappa_{\mathrm{end}}}u+O(u^2)\right\}.
\]
For all sufficiently small $u>0$, every minimizing upper endpoint
also satisfies $b_u\ge\tfrac13$.
\end{corollary}

\begin{proof}
Uniform continuity of $G$ implies continuity of $C$, with
$C(0)=0$ and $C(\tfrac12)=1$. If $0<h<h'\le\tfrac12$, a maximizing
interval of width $h$ leaves an open part of $[0,\tfrac12]$ uncovered.
Extending it by $h'-h$ on a side with uncovered support adds an open
interval of positive mass, since $\supp(\mu_G)=[0,\tfrac12]$.
Hence $C(h')>C(h)$; the case $h=0$ follows from the same support property.
Thus $C$ has a continuous inverse from $(0,1)$ to $(0,\tfrac12)$.
Put $w(u)=C^{-1}(u)$. Attainment of $C(w(u))$ supplies an interval of
width $w(u)$ and mass $u$, whereas strict increase excludes any shorter
interval of mass at least $u$. Every minimizing interval has mass $u$
because its mass is bounded above by $C(w(u))$.

Put $v_u=\log(1/(3w(u)))$.
Equations \eqref{eq:sharp-additive-modulus} and \eqref{eq:hend-asymp} give
$u=\kappa_{\mathrm{end}}/v_u
-\kappa_{\mathrm{end}}\pi^2/(12v_u^3)+O(v_u^{-4})$.
First $v_u\sim\kappa_{\mathrm{end}}/u$; substitution back gives
$v_u=\kappa_{\mathrm{end}}/u
-\pi^2u/(12\kappa_{\mathrm{end}})+O(u^2)$, proving the width formula.
Every minimizing interval maximizes concentration at width $w(u)$,
so the all-width bound in \Cref{cor:sharp-additive-modulus} gives
$b_u<\tfrac13+w(u)$. Since $w(u)\to0$ as $u\downarrow0$,
\eqref{eq:additive-localisation} also gives $b_u\ge\tfrac13$ for small $u$.
\end{proof}

The shortest-interval result inverts the maximum mass $C(h)$ and
retains the prefactor $1/3$ in $w(u)$. The raw-quantile modulus below
inverts the minimum mass $m(h)$ over intervals inside the support;
its scale is determined by the small-$X$ tail.

\KeepHeadingSpace{3\baselineskip}
\begin{theorem}[Global modulus of the raw quantile function]
\label{thm:quantile-modulus}
\index{Quantiles!modulus of continuity}
Let $q=G^{-1}:[0,1]\to[0,1/2]$, with $q(0)=0$ and $q(1)=1/2$.
Then
\begin{equation}\label{eq:quantile-modulus}
 \sup_{|u-v|\le\eta}|q(u)-q(v)|
 \sim\frac{e^{-\gamma}}{\log\log(1/\eta)}
 \qquad(\eta\downarrow0).
\end{equation}
More precisely, put
$m(h)=\min_{0\le a\le1/2-h}\{G(a+h)-G(a)\}$ for $0\le h\le1/2$.
The left side of \eqref{eq:quantile-modulus} is $m^{-1}(\eta)$ for
$0\le\eta\le1$, and $h\log\log(1/m(h))\to e^{-\gamma}$ as $h\downarrow0$.
\end{theorem}

\begin{proof}
We first construct a uniform finite-head mesh. For all sufficiently
large consecutive odd primes $p<p'<p''$, the prime number theorem
gives $\omega_p<\omega_{p'}+\omega_{p''}$.
Consider the decreasing list of these jumps up to $y$ and approximate
any target between zero and their total greedily from below.
If the last rejected term is $a_i$ and is not among the last two,
all later terms were accepted. The remaining target $r$ at that
rejection would satisfy
$a_i>r\ge\sum_{j>i}a_j\ge a_{i+1}+a_{i+2}>a_i$, a contradiction.
Thus the final defect is $O(1/y)$; if no term was rejected, the
defect is zero. The total of these late jumps diverges and eventually
exceeds each of the finitely many omitted early jumps. Adjoining one
early jump gives the union of a mesh and its translate on overlapping
intervals, preserving the same error. Iteration gives a one-sided
$O(1/y)$ mesh of the whole interval of logarithmic subsums.

Consequently, every $x_0\in[m_y,1/2]$, where
$m_y=\prod_{\ell\le y}(1-1/\ell)\sim e^{-\gamma}/\log y$,
has a head product $x_H\ge x_0$ with $x_H-x_0=O(1/y)$.
Each prescribed head has probability at least
$\prod_{3\le\ell\le y}(\ell-1)^{-1}\ge e^{-Cy}$.
The independent tail logarithm $T_y$ has mean $O((y\log y)^{-1})$,
so $\mathbb P(T_y\le1/y)\ge1/2$ for large $y$.

Fix $c'>e^{-\gamma}$ and set $y=\exp(c'/h)$.
For any $[a,a+h]\subset[0,1/2]$, choose
$x_0=\max(m_y,a+h/2)$ and its approximating head $x_H$.
The lower margin is at least $h/2$, and the upper margin is at least
$\min(h/2,h-m_y)$, a positive fixed multiple of $h$ uniformly in $a$.
The mesh error and the displacement
$x_H(1-e^{-T_y})\le1/(2y)$ on the good-tail event fit these margins.
Thus
\[
 m(h)\ge\tfrac12\exp\{-C\exp(c'/h)\}.
\]
On the other hand $m(h)\le G(h)$, and
\Cref{cor:small-x-tail} gives
$h\log\log(1/G(h))\to e^{-\gamma}$.
Letting $c'\downarrow e^{-\gamma}$ proves the assertion for $m$.

Continuity and full support of $G$ imply that $m(h)>0$ for $h>0$,
that the minimum is attained, and that $m$ is continuous.
Splitting an interval into lengths $h$ and $h'-h$ gives
$m(h')\ge m(h)+m(h'-h)>m(h)$ when $h'>h$.
Hence $m$ maps $[0,1/2]$ continuously and strictly increasingly onto
$[0,1]$. A minimizing interval of width $h=m^{-1}(\eta)$ has mass
$\eta$ and supplies a quantile increment of length $h$; every longer
interval has mass greater than $\eta$. This proves the exact inverse
identity and \eqref{eq:quantile-modulus}.
By \Cref{cor:small-x-tail}, the endpoint increment $q(\eta)-q(0)$ has the same leading size.
\end{proof}

\begin{corollary}[Quantile modulus in a fixed progression]
\label{cor:ap-quantile-modulus}
Fix $Q\ge1$ and $(b,Q)=1$, and put
\[
 F_{Q,b}:=\prod_{\substack{\ell\mid Q,\ \ell>2\\
                           b\not\equiv1\pmod\ell}}(1-1/\ell),
 \qquad q_{Q,b}:=G_{Q,b}^{-1}:[0,1]\to[0,A_{Q,b}].
\]
Then
\begin{equation}\label{eq:ap-quantile-modulus}
 \sup_{|u-v|\le\eta}|q_{Q,b}(u)-q_{Q,b}(v)|
 \sim\frac{e^{-\gamma}}{F_{Q,b}\log\log(1/\eta)}
 \qquad(\eta\downarrow0).
\end{equation}
More precisely, if
\[
 m_{Q,b}(h):=\min_{0\le a\le A_{Q,b}-h}
                  \{G_{Q,b}(a+h)-G_{Q,b}(a)\},
\]
then the left side of \eqref{eq:ap-quantile-modulus} is
$m_{Q,b}^{-1}(\eta)$, and
$h\log\log(1/m_{Q,b}(h))\to e^{-\gamma}/F_{Q,b}$.
\end{corollary}

\begin{proof}
Once $y$ exceeds all prime divisors of $2Q$, the smallest head product
in \eqref{eq:XQ} is
\[
 A_{Q,b}\prod_{\substack{3\le\ell\le y\\\ell\nmid Q}}(1-1/\ell)
 =\frac{m_y}{F_{Q,b}},
 \qquad m_y=\prod_{\ell\le y}(1-1/\ell).
\]
Deleting finitely many jumps preserves the one-sided $O_Q(1/y)$ mesh
in the proof of \Cref{thm:quantile-modulus}. Each prescribed head has
probability at least $e^{-C_Qy}$, and the remaining logarithmic tail
has mean $O((y\log y)^{-1})$. For
$c'>e^{-\gamma}/F_{Q,b}$, use $y=\exp(c'/h)$ in that proof to obtain
\[
 m_{Q,b}(h)\ge\tfrac12\exp\{-C_Q\exp(c'/h)\}.
\]
The reverse comparison $m_{Q,b}(h)\le G_{Q,b}(h)$ and
\eqref{eq:ap-small-tail} give the asserted limit after
$c'\downarrow e^{-\gamma}/F_{Q,b}$. Continuity and full support on
$[0,A_{Q,b}]$ from \Cref{thm:ap-law} give continuity, strict
monotonicity and the exact inverse identity, as in
\Cref{thm:quantile-modulus}.
\end{proof}

\chapter{Quantitative convergence over the primes}
\label{ssec:sharp-prime-convergence}

How closely do the values of $X(p)$ or $c(p)$ over a finite prime
population follow their limiting law? The answer depends on what is
measured. Kolmogorov distance is the largest discrepancy in the
proportion of values below a threshold. The $1$-Wasserstein distance
is the least average displacement among couplings of the two laws.
The finite-product example in \Cref{rem:truncation-numerical} already
shows how different these errors can be. That example approximates an
infinite random product by finitely many prime coordinates; the present
chapter studies the further question of approximation by arithmetic
values sampled over primes.

In each fixed reduced progression, the largest threshold discrepancy
has sharp order $1/\log x$. For the raw and negative-logarithmic laws,
the Wasserstein error of every fixed finite order has arbitrary
inverse-logarithmic savings in the polylogarithmic, fixed-base power
and almost-all modulus ranges of \Cref{prop:prime-wasserstein}.
These estimates may have
ineffective constants and thresholds. Their proofs compare finite
prime-divisibility patterns and control the omitted coordinates, but
threshold counts also require the concentration estimates of the
preceding chapter. Allowing the progression modulus to grow requires
separate arithmetic input and retains the ranges stated below.

\section{Prime statistics and comparison estimates}

For prime $p$, we shall also use the finite products
\begin{equation}\label{eq:prime-finite-product}
 c_n(p)=X(p)\prod_{j=1}^{n}(1-p^{-j}),\qquad
 X(p)=\frac{\varphi(p-1)}{p-1},\qquad c_\infty(p)=c(p).
\end{equation}
A dimension array means an arbitrary choice
$n_x(p)\in\Z_{\ge1}\cup\{\infty\}$ for each sampled prime; it may
vary with both $x$ and $p$. The product bounds in
\Cref{lem:matrix-product-bounds} and the logarithmic estimate in the
proof of \Cref{lem:mellin-perturbation} give, uniformly in $n$,
\[
 c(p)\le c_n(p)\le X(p),\qquad
 0\le-\log c_n(p)-f(p-1)\le\frac4p.
\]
Here $n$ truncates the matrix product at a fixed prime $p$; it is
separate from a cutoff in the prime coordinates of the random model.
The determinant-counting interpretation is derived in
\Cref{sec:det-density}. The progression ranges and the different
error metrics are compared in \Cref{ssec:progression-regimes}.

For a real prime statistic $U$, write
\[
 F^U_{x,Q,b}(t):=
 \frac{\#\{p\le x:p\equiv b\pmod Q,\ U(p)\le t\}}{\pi(x;Q,b)},
 \qquad D_x^U:=\|F^U_{x,1,1}-G\|_\infty.
\]
Write $\mathbb P_{x,Q,b}$, $\mathbb E_{x,Q,b}$ and
$\operatorname{Law}_{x,Q,b}$ for uniform sampling, expectation and law
over these primes. Use $A_{Q,b},X_{Q,b},G_{Q,b}$ from
\Cref{thm:ap-law}, and put
\[
 D(Q):=\prod_{\substack{\ell\mid Q\\\ell>2}}
          \frac{\ell-1}{\ell-2}=q_S^{-1},
\]
where $S=\{\ell>2:\ell\textup{ prime},\ \ell\mid Q\}$ and $q_S$ is defined
in \Cref{def:anchor-weights}.
The following elementary form of truncated inclusion--exclusion uses
only joint inclusions of a controlled number of primes. For more general
shifted-prime and sieve-sequence models, see \cite{For25,Gor26Kub}.

\begin{lemma}[Truncated inclusion--exclusion in total variation]
\label{lem:sparse-bonferroni}
Let $\mu,\nu$ be laws of subsets of a finite set $\mathcal P$, and put
$a_\mu(T)=\mu\{S:T\subseteq S\}$, with analogous notation for $\nu$.
For every integer $r\ge0$,
\begin{equation}\label{eq:sparse-bonferroni}
\begin{split}
 2d_{\mathrm{TV}}(\mu,\nu)\le{}&
 \sum_{j=0}^r2^j\sum_{\substack{T\subseteq\mathcal P\\|T|=j}}
                  |a_\mu(T)-a_\nu(T)|\\
 &+2^{r+1}\sum_{\substack{T\subseteq\mathcal P\\|T|=r+1}}
                  \bigl(a_\mu(T)+a_\nu(T)\bigr).
\end{split}
\end{equation}
If $|\mathcal P|\le y$, $y\ge3$, $\nu$ has independent indicators with
probabilities $p_\ell$, $\lambda=\sum_{\ell\in\mathcal P}p_\ell$, and
$|a_\mu(T)-a_\nu(T)|\le\epsilon$ whenever $|T|\le r+1$, then
\begin{equation}\label{eq:sparse-bonferroni-bound}
 d_{\mathrm{TV}}(\mu,\nu)
 \ll(r+2)(2y)^{r+1}\epsilon+\frac{(2\lambda)^{r+1}}{(r+1)!}.
\end{equation}
\end{lemma}

\begin{proof}
For a fixed $S\subseteq\mathcal P$ with $|S|\le r$, truncate the
inclusion--exclusion formula for $\mu(\{S\})$ after $r-|S|$ exclusions.
The Bonferroni inequalities bound its remainder in absolute value by
\[
 \sum_{\substack{T\supseteq S\\|T|=r+1}}a_\mu(T).
\]
The same holds for $\nu$. Summing these remainders over $|S|\le r$
counts each $(r+1)$-set $2^{r+1}-1$ times. The total probability of
$|S|>r$ is at most $\sum_{|T|=r+1}a_\mu(T)$, and similarly for $\nu$.
In the difference of the truncated formulas, the coefficient of the
error attached to a $j$-set is bounded in absolute sum by $2^j$.
Since twice total variation is the sum of the absolute point-mass
differences, these observations prove \eqref{eq:sparse-bonferroni}.
For independent indicators,
$\sum_{|T|=r+1}a_\nu(T)\le\lambda^{r+1}/(r+1)!$.
Replace each $a_\mu(T)$ in the last sum by $a_\nu(T)+\epsilon$,
and use $\binom{|\mathcal P|}{j}\le y^j$ to obtain
\eqref{eq:sparse-bonferroni-bound}.
\end{proof}

\begin{lemma}[Conditioning in total variation]\leavevmode\phantomsection\label{lem:conditional-tv}\ignorespaces
Let $\mu,\nu$ be probability measures on the same measurable space,
with $d_{\mathrm{TV}}(\mu,\nu)\le\varepsilon$. If
$\mu(C),\nu(C)>0$, then
\[
 d_{\mathrm{TV}}(\mu(\,\cdot\mid C),\nu(\,\cdot\mid C))
 \le\frac{2\varepsilon}{\mu(C)}.
\]
\end{lemma}

\begin{proof}
For every measurable $A$, subtract the two conditional probabilities:
\[
 \left|\frac{\mu(A\cap C)}{\mu(C)}-
             \frac{\nu(A\cap C)}{\nu(C)}\right|
 \le\frac{|\mu(A\cap C)-\nu(A\cap C)|}{\mu(C)}
 +\frac{\nu(A\cap C)|\mu(C)-\nu(C)|}{\mu(C)\nu(C)}
 \le\frac{2\varepsilon}{\mu(C)}.
\]
Take the supremum over $A$.
\end{proof}

\begin{lemma}[Concentration after fixing progression coordinates]
\label{lem:ap-uniform-concentration}
The factor $D(Q)$ satisfies
\begin{equation}\label{eq:ap-concentration-factor}
 D(Q)\ll\log\log(3Q)\qquad(Q\ge1).
\end{equation}
Uniformly in reduced classes $b\pmod Q$ and $0<h<1/2$,
\begin{equation}\label{eq:ap-uniform-concentration}
 Q_{\operatorname{Law}(-\log X_{Q,b})}(h)
 \le D(Q)Q_{\mu_f}(h)
 \ll\frac{D(Q)}{\log(1/h)}.
\end{equation}
\end{lemma}

\begin{proof}
The unrestricted logarithmic law is the convolution of the sum over
odd primes not dividing $Q$ with the remaining finite prime sum and
the constant $\log2$. That finite sum has an atom at zero of mass
$D(Q)^{-1}$. Retaining this component and translating intervals by
the deterministic shift $-\log A_{Q,b}$ gives the first inequality
in \eqref{eq:ap-uniform-concentration};
\Cref{thm:sharp-global-modulus} gives the second.
For \eqref{eq:ap-concentration-factor}, put
$z=\max\{3,\log(3Q)\}$. Mertens' theorem and convergence of
$\prod_{\ell>2}(1-(\ell-1)^{-2})$ give
$\prod_{3\le\ell\le z}(\ell-1)/(\ell-2)\ll\log z$.
The logarithm of the remaining factors is
\[
 \ll\sum_{\substack{\ell\mid Q\\\ell>z}}\frac1\ell
 \le\frac{\log Q}{z\log z}\ll1.
\]
This proves the asserted uniform bound, with the bounded small values
of $Q$ absorbed into the constant.
\end{proof}

\begin{lemma}[A weighted Bombieri--Vinogradov bound]\label{lem:bv-weighted}
Put
\[
 E_x(m)=\max_{(a,m)=1}
 \left|\pi(x;m,a)-\frac{\operatorname{Li}(x)}{\varphi(m)}\right|.
\]
For every $K>0$ there is $B_K>0$ such that
\begin{equation}\label{eq:bv-weighted}
 \sum_{m\le x^{1/2}(\log x)^{-B_K}}3^{\omega(m)}E_x(m)
 \ll_K\frac{x}{(\log x)^K}.
\end{equation}
\end{lemma}

\begin{proof}
For $m\le\sqrt x$, counting integers in one class and using
$m/\varphi(m)\ll\log\log(3m)$ gives
$E_x(m)\ll x\log\log(3x)/m$. Positivity and Mertens' theorem give
\[
 \sum_{m\le M}\frac{9^{\omega(m)}}m
 \le\prod_{p\le M}\left(1+\frac9{p-1}\right)
 \ll(\log(2M))^9.
\]
Hence $\sum_{m\le\sqrt x}9^{\omega(m)}E_x(m)\ll x(\log x)^{10}$.
For $M\le\sqrt x$, Cauchy--Schwarz gives
\[
 \sum_{m\le M}3^{\omega(m)}E_x(m)
 \le\left(\sum_{m\le M}E_x(m)\right)^{1/2}
     \left(\sum_{m\le M}9^{\omega(m)}E_x(m)\right)^{1/2}.
\]
Apply the ordinary Bombieri--Vinogradov theorem~\cite{Bom65,Vin65}
in the $\pi$-form of~\cite[Lemma~4]{Kat68}, with saving $2K+10$.
\end{proof}

\begin{lemma}[Finite-prime sample heads]\label{lem:finite-prime-sample-head}
Let $Q\ge1$, $(b,Q)=1$, let $\mathcal N$ be a finite set of primes,
let $J>0$, and let $\mathcal P$ be a finite set of odd primes not
dividing $Q$. Fix an integer $k\ge0$. Let $\mathcal D$ consist of
the squarefree products of at most $k+1$ elements of $\mathcal P$,
including $1$. Suppose nonnegative errors $E(Qd)$ satisfy
\[
 \left|\#\{p\in\mathcal N:p\equiv b\pmod Q,\ d\mid p-1\}
             -\frac{J}{\varphi(Qd)}\right|\le E(Qd)
 \quad(d\in\mathcal D),
\]
and $E(Q)\le J/(2\varphi(Q))$.
The progression population $n=\#\{p\in\mathcal N:p\equiv b\pmod Q\}$
then satisfies $|n-J/\varphi(Q)|\le E(Q)$ and
$n\ge J/(2\varphi(Q))>0$.
Let $\mu$ be the law of $\{\ell\in\mathcal P:\ell\mid p-1\}$ for a
uniform prime from this population, and let $\nu$ have independent
indicators of probabilities $(\ell-1)^{-1}$. Put
$\lambda=\sum_{\ell\in\mathcal P}(\ell-1)^{-1}$. Then
\begin{equation}\label{eq:finite-prime-sample-head}
 d_{\mathrm{TV}}(\mu,\nu)\le\frac{\varphi(Q)}{J}
 \sum_{d\in\mathcal D}2^{\omega(d)}
 \left(E(Qd)+\frac{E(Q)}{\varphi(d)}\right)
 +\frac{(2\lambda)^{k+1}}{(k+1)!}.
\end{equation}
The same bound holds after applying any common measurable map to the
heads. Here $d_{\mathrm{TV}}=\sup_A|\mu(A)-\nu(A)|$, so twice total
variation is the sum of the absolute point-mass differences.
\end{lemma}

\begin{proof}
Let $n_d$ be the inclusion count in the hypothesis, and let $T$ be
the subset with product $d$. The case $d=1$ gives the asserted
population estimate. Since $(d,Q)=1$,
$\varphi(Qd)=\varphi(Q)\varphi(d)$. Therefore
$a_\mu(T)=n_d/n$, $a_\nu(T)=1/\varphi(d)$ and
\[
 |a_\mu(T)-a_\nu(T)|\le\frac1n
 \left(E(Qd)+\frac{E(Q)}{\varphi(d)}\right)
 \le\frac{2\varphi(Q)}J
 \left(E(Qd)+\frac{E(Q)}{\varphi(d)}\right).
\]
Apply \eqref{eq:sparse-bonferroni} with $r=k$. In its last sum use
$a_\mu(T)\le a_\nu(T)+|a_\mu(T)-a_\nu(T)|$.
Independence gives
$\sum_{|T|=k+1}a_\nu(T)\le\lambda^{k+1}/(k+1)!$:
the expansion of $\lambda^{k+1}$ contains each product over distinct
indices exactly $(k+1)!$ times, and its other terms are nonnegative.
Thus the error terms for $0\le|T|\le k+1$, after division by $2$,
are bounded by the first term in \eqref{eq:finite-prime-sample-head}.
The two model contributions at level $k+1$, multiplied by $2^{k+1}$
and divided by $2$, give exactly $(2\lambda)^{k+1}/(k+1)!$.
If $k+1>|\mathcal P|$ the relevant sum is empty and the same bound
remains valid. Taking preimages of measurable sets proves contraction
under a common map.
\end{proof}

\begin{proposition}[Head and tail comparison for distribution functions]
\label{prop:head-tail-cdf}
Let $(U,H)$ and $(V,K)$ be pairs of real random variables; the pairs
may be defined on different probability spaces. Write
$\alpha=\operatorname{Law}(H)$, $\beta=\operatorname{Law}(K)$,
$\nu=\operatorname{Law}(V)$ and $\delta=d_{\mathrm{TV}}(\alpha,\beta)$.
Suppose $\eta_{\rm emp}=\mathbb E|U-H|<\infty$ and
$\eta_{\rm mod}=\mathbb E|V-K|<\infty$. With
$Q_\nu(a)=\sup_t\nu([t,t+a])$ as above, for every $h>0$,
\begin{equation}\label{eq:head-tail-cdf}
 d_K(\operatorname{Law}(U),\nu)
 \le\delta+\frac{\eta_{\rm emp}+\eta_{\rm mod}}h+Q_\nu(2h).
\end{equation}
 If, outside an exceptional event of
probability $\epsilon$, $|U-H|\le a$ deterministically, one may instead
use a coupling error probability
$\delta+\epsilon+\mathbb P(|V-K|>h-a)$ for $h>a$.
\end{proposition}

\begin{proof}
Couple the heads so that $\mathbb P(H\ne K)=\delta$. Explicitly,
with respect to the dominating measure $\alpha+\beta$, their common
measure has density the minimum of the two densities and mass
$1-\delta$. Use this measure on the diagonal, and couple the two
residual measures, each of mass $\delta$, by their normalized product.
This also defines the coupling when $\delta=0$ or $1$ by omitting
zero-mass components. Extend it by the conditional laws of $U$ given
$H$ and of $V$ given $K$, independently conditional on the heads.
Real variables admit these conditional laws; the two original pair
marginals, and hence their remainder bounds, are preserved.
On $H=K$, $|U-V|\le|U-H|+|V-K|$. Consequently Markov's inequality
and the union bound give
\[
 \mathbb P(|U-V|>h)\le\delta+
             (\eta_{\rm emp}+\eta_{\rm mod})/h=:e_h.
\]
For every $t$,
$F_V(t-h)-e_h\le F_U(t)\le F_V(t+h)+e_h$.
The masses of $(t-h,t]$ and $(t,t+h]$ are at most
$Q_\nu(2h)$, so taking the supremum proves the assertion.
Under the deterministic alternative, failure
of $|U-V|\le h$ requires head mismatch, the empirical exceptional
event, or $|V-K|>h-a$, proving the stated variant. 
\end{proof}

\begin{proposition}[Head and tail comparison for transport]
\label{prop:head-tail-transport}
Use the pairs and head laws of \Cref{prop:head-tail-cdf}. Fix
$1\le r<\infty$, and suppose the full variables and heads have finite
$r$th absolute moments. Put
$\tau_{\rm emp,r}=\|U-H\|_r$ and
$\tau_{\rm mod,r}=\|V-K\|_r$. Then
\begin{equation}\label{eq:head-tail-transport}
 \mathcal W_r(\operatorname{Law}(U),\operatorname{Law}(V))
 \le\tau_{\rm emp,r}+\tau_{\rm mod,r}+\mathcal C_r,
\end{equation}
where any one of the following valid head-cost bounds can be used:
\begin{enumerate}[label=\textup{(\alph*)},leftmargin=*]
\item If both heads belong almost surely to a common set of diameter
$D<\infty$, take $\mathcal C_r=D\delta^{1/r}$.
\item For any common deterministic center $a\in\mathbb R$, take
\[
 \mathcal C_r=\left(2^{r-1}\int|t-a|^r\,d|\alpha-\beta|(t)\right)^{1/r}.
\]
Here $|\alpha-\beta|$ is the variation measure, of total mass $2\delta$.
\item If $p>r$ and $m_p(\alpha,a)=\int|t-a|^p\,d\alpha(t)$ and
$m_p(\beta,a)$ are finite, take
\[
 \mathcal C_r=\bigl(m_p(\alpha,a)^{1/p}+m_p(\beta,a)^{1/p}\bigr)
                  \delta^{1/r-1/p}.
\]
\end{enumerate}
For a family, uniform integrability of the centered $r$th powers is
another sufficient condition for the head cost to tend to zero as
$\delta\to0$. More precisely, for every $T>0$, (b) is at most
\begin{equation}\label{eq:head-tail-uniform-integrability}
 \left[2^{r-1}\left\{2T^r\delta+
  \int_{|t-a|>T}|t-a|^r\,d(\alpha+\beta)(t)\right\}\right]^{1/r}.
\end{equation}
A bounded $r$th moment alone supplies neither this uniform
integrability nor a $\delta^{1/r}$ head rate on unbounded supports.
Common deterministic translations cancel from these estimates.
\end{proposition}

\begin{proof}
Use the same extended maximal coupling. Minkowski's inequality gives
$\|U-V\|_r\le\tau_{\rm emp,r}+\tau_{\rm mod,r}+\|H-K\|_r$.
Taking an infimum over full couplings gives
\eqref{eq:head-tail-transport} for every upper bound below on the
last term. On the mismatch event the distance is at most $D$ in (a),
and elsewhere it is zero. For (b), put
$\alpha_0=\alpha-(\alpha\wedge\beta)$ and
$\beta_0=\beta-(\alpha\wedge\beta)$; then
$\alpha_0+\beta_0=|\alpha-\beta|$. The residual coupling and
$|s-t|^r\le2^{r-1}(|s-a|^r+|t-a|^r)$ give
\[
 \mathbb E|H-K|^r\le
 2^{r-1}\left(\int|s-a|^r\,d\alpha_0(s)+
                   \int|t-a|^r\,d\beta_0(t)\right).
\]
This is (b). For (c), H\"older's inequality on the mismatch event
and then Minkowski in $L^p$ give
\[
 \|H-K\|_r\le
 \bigl\||H-a|+|K-a|\bigr\|_p\delta^{1/r-1/p}
 \le\bigl(m_p(\alpha,a)^{1/p}+m_p(\beta,a)^{1/p}\bigr)
                  \delta^{1/r-1/p}.
\]
Split the integral in (b) at $|t-a|=T$. Its bounded part is at most
$2T^r\delta$ since $|\alpha-\beta|(\mathbb R)=2\delta$;
its other part is bounded by the corresponding integral against
$\alpha+\beta$. This proves \eqref{eq:head-tail-uniform-integrability}.
First choose $T$ using uniform integrability, then let $\delta$ tend
to zero. For a family with bounded $r$th moments, take
$\alpha_n=(1-1/n)\delta_0+n^{-1}\delta_{n^{1/r}}$ and
$\beta_n=\delta_0$. Their head TV tends to zero and their $r$th
moments are bounded by $1$, but $\mathcal W_r(\alpha_n,\beta_n)=1$.
Even uniform integrability does not imply the bounded-diameter rate:
replacing $n^{1/r}$ by $n^{1/(2r)}$ makes the $r$th moments tend to
zero, while $\mathcal W_r=n^{-1/(2r)}$, not $O(n^{-1/r})$.
Subtracting a common center proves the translation assertion.
\end{proof}

\begin{proposition}[Weighted comparison for Mellin tests]
\label{prop:head-tail-weighted-mellin}
Let $(U,H)$ and $(V,K)$ now denote real logarithmic variables and
their heads, with head laws $\alpha,\beta$. Fix an integer $j\ge0$
and $R\ge1$, and put
$w(t)=e^{R|t|}(1+|t|^{j+1})$. Suppose
$\mathbb Ew(U)+\mathbb Ew(H)+\mathbb Ew(V)+\mathbb Ew(K)<\infty$,
and let
\[
 \begin{split}
 \epsilon_{\rm head}&=\int w(t)\,d|\alpha-\beta|(t),\\
 \epsilon_{\rm emp}&=\mathbb E[|U-H|\{w(U)+w(H)\}],\\
 \epsilon_{\rm mod}&=\mathbb E[|V-K|\{w(V)+w(K)\}]
 \end{split}
\]
be finite. Uniformly for $|z|\le R$,
\begin{equation}\label{eq:head-tail-weighted-mellin}
 \left|\mathbb E[(-U)^je^{-zU}]-\mathbb E[(-V)^je^{-zV}]\right|
 \le\epsilon_{\rm head}+(R+j)(\epsilon_{\rm emp}+\epsilon_{\rm mod}).
\end{equation}
For $|z|<R$ these tests are the $j$th derivatives of the Mellin
transforms of $e^{-U}$ and $e^{-V}$. 
\end{proposition}

\begin{proof}
Put $g_{z,j}(t)=(-t)^je^{-zt}$. For $|z|\le R$,
$|g_{z,j}(t)|\le w(t)$ and
$|g'_{z,j}(t)|\le(R+j)w(t)$; for $j=0$ the polynomial derivative
term is zero. On the interval between two real numbers $s,t$, the
absolute value of every point is at most $\max(|s|,|t|)$, and $w$
is increasing with absolute value. The fundamental theorem of
calculus therefore gives
\[
 |g_{z,j}(s)-g_{z,j}(t)|
 \le(R+j)|s-t|\{w(s)+w(t)\}.
\]
Insert the two head expectations between the two full expectations.
The first and last differences are bounded by the weighted remainder
terms above. The head difference is the integral of $g_{z,j}$ against
$\alpha-\beta$, with absolute value at most $\epsilon_{\rm head}$.
This proves the inequality. On a compact subset of $|z|<R$, every
fixed derivative of $e^{-zU}$ is dominated by a constant times
$e^{R|U|}$, using the positive margin to $R$; the assumed integrability
justifies differentiation under expectation. The same holds for $V$.
Polynomial moments cannot replace this domination: if
$U_n=n$ with probability $e^{-n/2}$ and $0$ otherwise, and $V=H=K=0$,
then all fixed polynomial remainder moments tend to zero, but
$\mathbb Ee^{U_n}=1-e^{-n/2}+e^{n/2}\to\infty$.
\end{proof}

The following comparison and modulus ranges combine
\Cref{lem:sparse-bonferroni,lem:bv-weighted} and, for powers of a fixed
base, Elliott's theorem~\cite{Elliott07}. They compare the small free
prime divisors of $p-1$ in a progression with the independent model
and supply the arithmetic input for
\Cref{thm:prime-mellin}\textup{(b),(c)} and
\Cref{prop:prime-wasserstein,prop:ap-kolmogorov-pow2}.

\paragraph*{Finite-head notation}
Fix $M>0$ and $0<\delta<1/2$, and choose $c_M>2e$ with
$c_M\log(c_M/(2e))>M+2$. For $Q\ge1$, let $\ell$ range over primes and set
\[
 r=\lceil c_M\log\log x\rceil,\qquad
 y=\exp\{\delta\log x/(4(r+1))\},\qquad
 \mathcal P_y(Q)=\{\ell:3\le\ell\le y,\ \ell\nmid Q\},
\]
so that $y^{r+1}=x^{\delta/4}$ and
$\log y\asymp_{\delta,M}\log x/\log\log x$.
Let $\mathcal D_Q$ consist of squarefree products of elements of
$\mathcal P_y(Q)$ with at most $r+1$ factors,
including $1$, and, with $E_x$ as in \Cref{lem:bv-weighted}, put
\[
\begin{aligned}
 P_y&=\prod_{3\le\ell\le y}\left(1+\frac2{\ell-1}\right)
 \ll(\log(2y))^2,
 \\
 R_Q&=\frac{\varphi(Q)}{\operatorname{Li}(x)}
 \left\{\sum_{d\in\mathcal D_Q}2^{\omega(d)}E_x(Qd)
                         +P_yE_x(Q)\right\}.
\end{aligned}
\]
For a reduced class $b\pmod Q$, let $\mu$ be the law of the subset of
$\mathcal P_y(Q)$ dividing $p-1$ for a uniform prime $p\le x$,
$p\equiv b\pmod Q$, and let $\nu$ be the law of
$\{\ell\in\mathcal P_y(Q):B_\ell=1\}$; its indicators are independent
with probabilities $(\ell-1)^{-1}$. Put
$\lambda=\sum_{\ell\in\mathcal P_y(Q)}(\ell-1)^{-1}
\le\log\log x+O(1)$, $a_{Q,b}=-\log A_{Q,b}$ and
\[
 W_y(p)=a_{Q,b}+\sum_{\substack{\ell\in\mathcal P_y(Q)\\\ell\mid p-1}}f(\ell),
 \qquad
 W_y=a_{Q,b}+\sum_{\ell\in\mathcal P_y(Q)}B_\ell f(\ell).
\]

\begin{lemma}[Finite-head comparison in a progression]
\label{lem:ap-finite-head}
With the notation above, for $x\ge x_0(M)$ and every modulus $Q$ with
$R_Q\le(\log x)^{-M}$, uniformly in $Q$ and reduced $b$,
\begin{equation}\label{eq:ap-polylog-population}
 n:=\pi(x;Q,b)=\frac{\operatorname{Li}(x)}{\varphi(Q)}
       \{1+O((\log x)^{-M})\}
 \gg\frac{x}{\varphi(Q)\log x},
\end{equation}
\begin{equation}\label{eq:ap-truncated-tv}
 d_{\mathrm{TV}}(\mu,\nu)
 \ll R_Q+\frac{(2\lambda)^{r+1}}{(r+1)!}
 \ll(\log x)^{-M}.
\end{equation}
The implied constants and $x_0(M)$ depend at most on $M$ and on the
choice of $c_M$.
\end{lemma}

\begin{proof}
For the original prime population, let $R_Q\le(\log x)^{-M}$.
Since $1\in\mathcal D_Q$, we have
$\varphi(Q)E_x(Q)\le R_Q\operatorname{Li}(x)$, which gives
\eqref{eq:ap-polylog-population} with implied constant $1$; in
particular $n\ge\operatorname{Li}(x)/(2\varphi(Q))$ as soon as
$(\log x)^{-M}\le1/2$. CRT identifies the conditions
$p\equiv b\pmod Q$ and $d\mid p-1$, for $d\in\mathcal D_Q$, with one
reduced class modulo $Qd$. \Cref{lem:finite-prime-sample-head} therefore applies
with $\mathcal N=\{p\le x:p\textup{ prime}\}$,
$J=\operatorname{Li}(x)$, $E(m)=E_x(m)$,
$\mathcal P=\mathcal P_y(Q)$ and $k=r$. Since
$\sum_{d\in\mathcal D_Q}2^{\omega(d)}/\varphi(d)\le P_y$, it yields
the first bound in \eqref{eq:ap-truncated-tv}, with implied constant $1$. As
$r+1\ge c_M\log\log x$ and $\lambda\le\log\log x+O(1)$, the ratio
$2e\lambda/(r+1)$ is at most $(2e/c_M)(1+O(1/\log\log x))<1$ for large
$x$. Stirling's inequality and the choice of $c_M$ therefore give
\begin{equation}\label{eq:ap-head-factorial}
 \frac{(2\lambda)^{r+1}}{(r+1)!}
 \le\left(\frac{2e\lambda}{r+1}\right)^{r+1}
 \le\left(\frac{2e\lambda}{r+1}\right)^{c_M\log\log x}
 \ll_M(\log x)^{-c_M\log(c_M/(2e))}\le(\log x)^{-M-2},
\end{equation}
which is the second bound.
\end{proof}

The laws of $W_y(p)$ under $\mathbb P_{x,Q,b}$ and of
$W_y$ are the images of $\mu$ and $\nu$ under one map, so they satisfy
the same total-variation bound.

\begin{corollary}[Modulus ranges for finite-head comparison]
\label{cor:ap-finite-head-ranges}
With the parameters above, the condition $R_Q\le(\log x)^{-M}$ holds
in the following ranges.
\begin{enumerate}[\upshape(i)]
\item \emph{Almost all moduli.} For every fixed $A>0$, uniformly for
$1\le H\le x^{1/2-\delta}$, it holds for all but
$O_{\delta,A,M}(H(\log x)^{-A})$ integers $Q\le H$. The exceptional
moduli are those with $R_Q>(\log x)^{-M}$; they depend on
$x,\delta,M$ and the choice of $c_M$, but not on $b$, and no
dimension array enters.
\item \emph{Polylogarithmic moduli.} For fixed $C>0$ and
$\delta=1/4$, it holds for every $Q\le(\log x)^C$ once
$x\ge x_0(C,M)$.
\item \emph{Powers of a fixed base.} For a fixed integer $B\ge2$ and
$\delta=1/3$, that is, $y=\exp\{\log x/(12(r+1))\}$, it holds for
every $Q=B^a\le x^{1/3}\exp\{-(\log\log x)^3\}$, $a\in\Z_{\ge0}$,
once $x\ge x_0(B,M)$.
\end{enumerate}
The constants and thresholds in \textup{(i)--(iii)} may be ineffective.
\end{corollary}

\begin{proof}
For \textup{(i)}, every modulus $Qd$ with $Q\le H$ and
$d\in\mathcal D_Q$ is at most $Hy^{r+1}\le x^{1/2-3\delta/4}$.
Repetitions are bounded by
\[
 \sum_{\substack{Qd=m\\d\text{ squarefree}}}2^{\omega(d)}
 \le\sum_{\substack{d\mid m\\d\text{ squarefree}}}2^{\omega(d)}
 =3^{\omega(m)}.
\]
Thus \Cref{lem:bv-weighted}, with arbitrarily large $K$, together with
$\varphi(Q)\le H$ and $P_y\ll(\log x)^2$, gives
$\sum_{Q\le H}R_Q\ll_{\delta,K}H(\log x)^{3-K}$.
With $K>A+M+3$, all but $O_{\delta,A,M}(H(\log x)^{-A})$
moduli $Q\le H$ have $R_Q\le(\log x)^{-M}$. For \textup{(ii)}, take
$H=(\log x)^C$ and $A>C$ in \textup{(i)}; the exceptional count is
eventually less than one.

For \textup{(iii)}, put $L=\log x$. Elliott's theorem~\cite{Elliott07},
in the form restated in \cite[Theorem~B, pp.~25--26]{Guo13}, gives,
for every $K>0$, a constant $C_0=C_0(B,K)>0$ such that
\[
 \sum_{\substack{d\le x^{1/2}Q^{-1}L^{-C_0}\\(d,Q)=1}}
 E_x(Qd)\ll_{B,K}\frac{x}{\varphi(Q)L^K}
\]
uniformly for $Q=B^a\le x^{1/3}\exp\{-(\log\log x)^3\}$, $a\ge1$.
The cited estimate has maxima
over endpoints up to $x$ and reduced residues modulo $Qd$;
evaluation at $x$ gives the displayed form. Every
$d\in\mathcal D_Q$ is coprime to $Q$ and satisfies
$d\le y^{r+1}=x^{1/12}\le x^{1/2}Q^{-1}L^{-C_0}$ for sufficiently
large $x$, depending on $B,K$. Moreover,
$2^{\omega(d)}\le2^{r+1}\ll_M L^{c_M\log2}$ and $P_y\ll L^2$.
The definition of $R_Q$ therefore gives
\[
 R_Q\ll_{B,K,M}(2^{r+1}+P_y)L^{1-K}
 \ll_{B,K,M}L^{1+c_M\log2-K}.
\]
Choose $K>M+c_M\log2+2$; then $R_Q\le L^{-M}$ for all sufficiently
large $x$. For $Q=1$ use \textup{(i)} with $H=1$ and the same
$\delta=1/3$: its exceptional count is eventually less than one.
\end{proof}

\section{Smooth parts and sharp threshold bounds}

To obtain the sharp CDF bound, the proof compares successive smooth
cutoffs. A small added block can change a threshold count only when
the initial sum lies sufficiently close to that threshold. This
localizes the comparison error, so the concentration estimate supplies
a gain; unusually large smooth parts are controlled separately. The
resulting recursion reaches the order $1/\log x$ for fixed progressions.

For the proofs of \Cref{thm:addfn-sharp-order,thm:sharp-prime-convergence},
we compare the smooth parts of $p-1$ with Ford's model~\cite{For25}. Its independent
exponents $W_q$ satisfy
$\mathbb P(W_2=v)=2^{-v}$ for $v\ge1$, and, for odd $q$,
\[
 \mathbb P(W_q=0)=\frac{q-2}{q-1},\qquad
 \mathbb P(W_q=v)=q^{-v}\quad(v\ge1).
\]
Write $s_z(n)=\prod_{q\le z}q^{v_q(n)}$,
$S_z=\prod_{q\le z}q^{W_q}$, $Y_z=f(S_z)$, and
$Y=\lim_zY_z$. Thus $Y$ has law $\muF$. Put
$g_z(m)=\mathbb P(S_z=m)$ for every $z$-smooth $m$; it is zero for
odd $m$. For even $m$, by~\cite[Lemma~3.2 and (3.4)]{For25},
\begin{equation}\label{eq:prime-model-weights}
 g_z(m)=\frac1m\prod_{\substack{3\le q\le z\\q\nmid m}}
       \left(1-\frac1{q-1}\right)
 \asymp\frac1{\varphi(m)\log z}.
\end{equation}

Fix $Q\ge1$ and a reduced residue $b\pmod Q$. For $\ell\mid Q$, put
$a_\ell=v_\ell(Q)$ and
$c_\ell=\min\{v_\ell(b-1),a_\ell\}$, with $v_\ell(0)=\infty$.
Condition the model on the event $\mathcal C_{Q,b}$ requiring
$W_\ell=c_\ell$ if $c_\ell<a_\ell$, and $W_\ell\ge a_\ell$
otherwise. Its probability $\beta_{Q,b}$ is positive; in particular,
$2\mid Q$ implies $c_2\ge1$. Denote the conditioned variables by
$S_z^*,Y_z^*$ and $Y^*$. For $z>P^+(2Q)$ call a smooth integer
$m$ compatible if it is even and satisfies these valuation conditions.
Then
\begin{equation}\label{eq:ap-model-weights}
 g_z^*(m):=\mathbb P(S_z^*=m)
 =\frac{g_z(m)}{\beta_{Q,b}}\asymp_Q\frac1{\varphi(m)\log z}
 \quad(m\text{ compatible}).
\end{equation}
The law of $Y^*$ is that of $-\log X_{Q,b}$, and coordinates outside
$Q$ remain independent. Only the model is conditioned: the empirical
condition $p\equiv b\pmod Q$ need not be determined by these valuations.

\begin{lemma}[Smooth parts in a fixed progression]\label{lem:ap-smooth-pattern}
There are absolute constants $c>0$ and $\delta>0$, with $\delta$
sufficiently small, such that, for each fixed $Q$, all sufficiently
large $x$ and $P^+(2Q)<z\le x^\delta$, there is a set $E_z$ of
$z$-smooth integers at most $x^{2/5}$,
independent of $b$, satisfying
\begin{equation}\label{eq:prime-model-exceptional}
 \sum_{m\in E_z}\frac1{\varphi(m)}\ll_Q(\log x)^{-12}.
\end{equation}
For compatible $m\le x^{11/30}$ outside $E_z$, with
$v_m=\log(x^{2/5}/m)/\log z$,
\begin{equation}\label{eq:prime-model-pattern}
\begin{split}
 \Phi_{Q,b;m}(x,z)&:=\#\{2<p\le x:p\equiv b\pmod Q,\ s_z(p-1)=m\}\\
 &=\frac{\operatorname{Li}(x)}{\varphi(Q)}g_z^*(m)
 \{1+O_Q(e^{-cv_m\log(v_m+1)}+(\log x)^{-10})\}.
\end{split}
\end{equation}
The first factor can be replaced by the odd-prime population in the
progression, with the same error.
\end{lemma}

\begin{proof}
Put $E_x(r)=\max_{(a,r)=1}|\pi(x;r,a)-\operatorname{Li}(x)/\varphi(r)|$
and
\[
\begin{gathered}
 R_z(m)=\sum_{\substack{d\le x^{2/5}/m\\P^+(d)\le z}}E_x([Q,md]),\\
 E_z=\left\{m\le x^{2/5}:P^+(m)\le z,\
 R_z(m)>\frac{\operatorname{Li}(x)}{\varphi(m)(\log x)^{11}}\right\}.
\end{gathered}
\]
Here $[a,b]$ denotes the least common multiple. With $\tau$ the divisor
function, grouping by $n=md$ gives
\[
 \sum_{\substack{m\le x^{2/5}\\P^+(m)\le z}}R_z(m)
 \le\sum_{n\le x^{2/5}}\tau(n)E_x([Q,n])
 \le\tau(Q)\sum_{r\le Qx^{2/5}}\tau(r)E_x(r).
\]
Indeed $n\mid[Q,n]=r$ and this map has at most $\tau(Q)$ preimages.
For $r\le\sqrt x$, elementary counting gives
$E_x(r)\ll x\log\log(3x)/r$. Since $\tau(r)^2\le d_4(r)$,
where $d_4$ counts ordered factorizations into four factors,
$\sum_{r\le\sqrt x}\tau(r)^2/r\ll(\log x)^4$. Thus
\begin{equation}\label{eq:ap-divisor-weighted-bv}
\begin{split}
 \sum_{r\le R}\tau(r)E_x(r)
 &\ll\left(\sum_{r\le R}E_x(r)\right)^{1/2}
       \{x(\log x)^4\log\log(3x)\}^{1/2}
 \\&\ll_{Q,K}\frac{x}{(\log x)^K},\qquad R=Qx^{2/5}.
\end{split}
\end{equation}
The last step is ordinary max-residue Bombieri--Vinogradov
\cite{Bom65,Vin65}, in the $\pi$-form of~\cite[Lemma~4]{Kat68},
with saving greater than $2K+5$: for fixed $Q$, the range $R$ is
eventually below $\sqrt x/(\log x)^B$ for every fixed $B$.
Taking $K=24$ proves \eqref{eq:prime-model-exceptional}.

For a compatible $m$, the conditions $p\equiv b\pmod Q$ and
$p\equiv1\pmod m$ define one reduced class modulo $M=[Q,m]$.
Sieve the quotient $(p-1)/m$. For squarefree $d$, its divisibility
by $d$ is either incompatible, or defines one reduced class modulo
$[Q,md]$. Relative to $\operatorname{Li}(x)/\varphi(M)$ the
multiplicative density is
\[
 k_m(\ell)=
 \begin{cases}
 0,&\ell\mid Q,\ c_\ell<a_\ell,\\
 1/\ell,&\ell\mid Q,\ v_\ell(m)\ge a_\ell=c_\ell,\\
 1/\ell,&\ell\nmid Q,\ \ell\mid m,\\
 1/(\ell-1),&\text{otherwise }(\ell>2).
 \end{cases}
\]
Incompatible $d$ have zero count; the
other remainders are bounded by $E_x([Q,md])+O_Q(1)$, allowing
the omission of small primes. Put $h(2)=1/2$ and
$h(\ell)=1/(\ell-1)$ for odd primes. Uniformly in $m,Q,b$,
$0\le k_m(\ell)\le h(\ell)\le1/2$, and Mertens' theorem gives
\begin{equation}\label{eq:ap-sieve-density}
 \prod_{t\le\ell\le v}(1-k_m(\ell))^{-1}
 \le\prod_{t\le\ell\le v}(1-h(\ell))^{-1}
 \le\frac{\log v}{\log t}
          \exp\left(\frac{B_0}{\log t}\right)
 \qquad(2\le t\le v)
\end{equation}
with an absolute $B_0$. For $t\ge3$, replacing $1/\ell$ by
$1/(\ell-1)$ changes the logarithm by $O(\sum_{\ell\ge t}\ell^{-2})
=O(1/t)$; bounded $t$ is absorbed by increasing $B_0$.

Apply the dimension-one fundamental lemma
\cite[Theorem~3.6(a),(b), pp.~38--39]{For23Sieve} at
$D_m=\lfloor x^{2/5}/m\rfloor$. Its squarefree weights satisfy
$|\lambda_d^\pm|\le1$, are supported on $d\le D_m$, $P^+(d)\le z$,
and have relative error
$\exp\{-s\log s+s\log\log\log s+O_{B_0}(s)\}$,
where $s=\log D_m/\log z$, provided $z\le\sqrt{D_m}$.
Since $D_m\ge x^{1/30}-1$, choosing $\delta$ sufficiently small
ensures this condition and a sufficiently large absolute depth.
The error is therefore $O(e^{-cv_m\log(v_m+1)})$.
The exact main term is
\begin{equation}\label{eq:ap-pattern-normalization}
 \frac{\operatorname{Li}(x)}{\varphi([Q,m])}
       \prod_{\ell\le z}(1-k_m(\ell))
 =\frac{\operatorname{Li}(x)}{\varphi(Q)}g_z^*(m).
\end{equation}
Indeed the local factor of $\varphi(Q)/\varphi([Q,m])$ times the
sieve product is $1$ in a forced valuation $c_\ell<a_\ell$,
and $\varphi(\ell^{a_\ell})\ell^{-v}$ when
$v=v_\ell(m)\ge a_\ell=c_\ell$; outside $Q$ it is the original
mass of $W_\ell=v$.
The bounded weights leave additive error at most $R_z(m)+O_Q(D_m)$.
By \eqref{eq:ap-model-weights}, outside $E_z$ its relative size is
\[
 O_Q\left(\frac{\log z}{(\log x)^{11}}
              +x^{-3/5}\log x\log z\right)
 =O_Q((\log x)^{-10}).
\]
Finally fixed-modulus Siegel--Walfisz replaces
$\operatorname{Li}(x)/\varphi(Q)$ by the odd-prime population.
\end{proof}

The unconditional upper-sieve bound
\cite[Lemma~3.4]{For25}, with shift $a=-1$, gives, for $2\le z\le x$ and
every integer $m\le(x-1)/z$, without a smoothness restriction,
\begin{equation}\label{eq:prime-model-sieve}
 \#\{2<p\le x:m\mid p-1,\ P^-((p-1)/m)>z\}
 \ll\frac{x}{\varphi(m)\log^2z}.
\end{equation}
Its absolute constant is effective. For later comparisons, the
distribution hypothesis $Z(1/2)$ is supplied by Bombieri--Vinogradov.
Taking $a=-1$ and $\alpha=1/4$ in~\cite[Theorem~1, p.~3]{For25} gives,
for every fixed $B>0$, uniformly for $2\le z\le x$,
\begin{equation}\label{eq:prime-model-base}
 d_{\mathrm{TV}}\bigl((v_q(p-1))_{q\le z},(W_q)_{q\le z}\bigr)
 \ll_B e^{-u\log u/4}+(\log x)^{-B},
 \qquad u=\frac{\log x}{\log z},
\end{equation}
where $p$ is uniform on the odd primes at most $x$.
The estimate is used in
\Cref{cor:ap-local-windows}; the fixed-progression initialization
follows from \Cref{lem:ap-smooth-pattern}.
The proof of \Cref{prop:prime-wasserstein} uses
\Cref{lem:ap-finite-head,cor:ap-finite-head-ranges}.

\paragraph*{Additive functions with regularly spaced prime values}
Let $\psi$ be a real-valued additive function, and put
$Y^\psi_z=\psi(S_z)$ and $Y^{\psi,*}_z=\psi(S_z^*)$; thus $Y_z=Y^f_z$.
Suppose that there are $q_0\ge2$, $C_\psi\ge1$, $c_\psi>0$ and a
function $v_\psi\in C^1([q_0,\infty))$ such that
\begin{align}
 0\le\psi(q^j)&\le\frac{C_\psi}{q}
 \qquad(q>q_0\text{ prime},\ j\ge1),\label{eq:addfn-size}\\
 \psi(q)&=v_\psi(q)\quad(q>q_0\text{ prime}),\qquad
 -v_\psi'(t)\ge\frac{c_\psi}{t^2}\quad(t\ge q_0).
 \label{eq:addfn-first-jump}
\end{align}
The finitely many summands at $q\le q_0$ are finite almost surely.
By \eqref{eq:addfn-size}, for $q>q_0$,
\[
 0\le\mathbb E\psi(q^{W_q})
 \le\frac{C_\psi}{q(q-1)}.
\]
These means are summable, so $Y^\psi=\lim_zY^\psi_z$ is finite almost
surely. Since $\mathbb P(\mathcal C_{Q,b})=\beta_{Q,b}>0$, the same
holds for $Y^{\psi,*}=\lim_zY^{\psi,*}_z$.
The function $v_\psi$ is decreasing and is nonnegative at arbitrarily
large primes, hence throughout $[q_0,\infty)$. Integrating
$-v_\psi'$ and then letting the upper endpoint tend to infinity gives
\begin{equation}\label{eq:addfn-spacing}
 v_\psi(t)\ge\frac{c_\psi}{t}\quad(t\ge q_0),\qquad
 v_\psi(n)-v_\psi(n')\ge\frac{c_\psi(n'-n)}{4H^2}
\end{equation}
for $H\ge1$ and all integers $q_0<n<n'$ in $(H,2H]$. Thus, for every
bounded interval $J$, at most $1+4H^2|J|/c_\psi$ integers
$n\in(H,2H]$ with $n>q_0$ have $v_\psi(n)\in J$. The second
inequality and this counting statement use only
\eqref{eq:addfn-first-jump}.
Both $f$ and $\gsig$ satisfy \eqref{eq:addfn-size} and
\eqref{eq:addfn-first-jump} with $q_0=2$; see the proofs of
\Cref{thm:sharp-prime-convergence,thm:sharp-sigma-convergence-mas06}.

\begin{lemma}[Concentration from the spacing of the prime values]
\label{lem:addfn-concentration}
Let $\psi$ be a real-valued additive function satisfying
\eqref{eq:addfn-first-jump}. For $2\le z<\infty$ and $0<h\le1/2$,
\begin{equation}\label{eq:addfn-concentration}
 \sup_{a\in\R}\mathbb P\bigl(Y^\psi_z\in[a,a+h]\bigr)
 \ll\frac1{\log\min\{z,1/h\}},
\end{equation}
and the same bound holds for $Y^\psi$ with $z=\infty$ whenever
$Y^\psi$ exists and is finite almost surely. In particular,
\begin{equation}\label{eq:prime-model-concentration}
 \sup_a\mathbb P(Y^\psi_z\in[a,a+h])\ll\frac1{\log z}
 \quad(z\ge2,\ 0<h\le z^{-1/2}).
\end{equation}
The implied constants depend only on $q_0$ and $c_\psi$. For
$z>P^+(2Q)$ the same bounds hold for $Y^{\psi,*}_z$, and for
$Y^{\psi,*}$, with the constants multiplied by $\beta_{Q,b}^{-1}$.
\end{lemma}

\begin{proof}
Conditioning gives, for every interval $J$,
\[
 \mathbb P(Y^{\psi,*}_z\in J)
 =\frac{\mathbb P(Y^\psi_z\in J,\ \mathcal C_{Q,b})}{\beta_{Q,b}}
 \le\beta_{Q,b}^{-1}\mathbb P(Y^\psi_z\in J),
\]
and likewise for the almost-sure limits. It suffices to treat
$Y^\psi_z$ for $2\le z\le\infty$, where $Y^\psi_\infty=Y^\psi$.
Fix an interval $J$ of length $h\in(0,1/2]$. Let $H_0$ be the least
power of $2$ with $H_0\ge q_0$, and put $y=\min\{z,h^{-1/2}\}$. If
$y<2H_0$, then $\min\{z,1/h\}<4H_0^2$, and the bound is trivial.

Otherwise let $H_1$ be the largest power of $2$ with $2H_1\le y$, so
that $H_0\le H_1$ and $2H_1>y/2$, and put
$\mathcal H=\{H_0,2H_0,\dots,H_1\}$. For $H\in\mathcal H$ set
$T_H=\sum_{H<q\le2H}\psi(q^{W_q})$, and let $R$ be the sum of the
remaining terms, at $q\le H_0$ or $2H_1<q\le z$. When $z=\infty$,
this complementary sum converges almost surely by the assumed
convergence after deletion of finitely many coordinates. Thus $R$ is
a function of the complementary coordinates and is independent of
the $T_H$, which are also mutually independent. Call the block $H$
active if $W_q\ge1$ for some prime $q\in(H,2H]$; an inactive block
has $T_H=0$. Every prime in these blocks is odd and exceeds $q_0$.

\emph{Inactive blocks.} For $H\in\mathcal H\cup\{H_0/2\}$,
\Cref{thm:mertens} and $(q-2)/(q-1)\le1-1/q$ give
\[
 \mathbb P(\text{no block }H'>H\text{ is active})
 =\prod_{2H<q\le2H_1}\frac{q-2}{q-1}
 \le C_0\,\frac{\log(4H)}{\log(2H_1)}
\]
with an absolute constant $C_0$; this includes the empty product.

\emph{One block.} For $H\in\mathcal H$ and any interval $J'$ of
length $h$,
\[
 \mathbb P(\text{block }H\text{ is active},\ T_H\in J')
 \le\frac3H+\frac{8hH}{c_\psi}.
\]
Indeed, the event that some $W_q\ge2$ in this block has probability
at most
\[
 \sum_{H<q\le2H}\frac1{q(q-1)}
 \le\sum_{n>H}\frac1{n(n-1)}=\frac1H.
\]
On the remaining active event, every block coordinate is $0$ or $1$.
Let $q^*$ be the largest active prime and let $\omega'$ be the
configuration $\omega=(W_q)_{H<q\le2H}$ with its $q^*$ coordinate
replaced by $0$. Then
$T_H(\omega)=T_H(\omega')+v_\psi(q^*)$, even when several primes are
active. The map $\omega\mapsto(\omega',q^*)$ is injective and
\eqref{eq:valuation-laws} gives
\[
 \mathbb P(\omega)
 =\frac{q^*-1}{q^*(q^*-2)}\,\mathbb P(\omega')
 \le\frac2H\,\mathbb P(\omega').
\]
For fixed $\omega'$, the requirement
$v_\psi(q^*)\in J'-T_H(\omega')$ allows at most
$1+4H^2h/c_\psi$ primes by the spacing bound. Summing over all
$0$--$1$ configurations $\omega'$ therefore bounds this part by
$(2/H)(1+4H^2h/c_\psi)$.

\emph{Summation.} Split according to the largest active block. If no
block is active, the inactive-block estimate with $H=H_0/2$ bounds
its probability by $C_0\log(2H_0)/\log(2H_1)$. If $H$ is the largest
active block, the event that all higher blocks are inactive is
independent of $R$, of the blocks below $H$ and of block $H$.
Condition on $R$ and the lower blocks, and apply the one-block
estimate to $J'=J-R-\sum_{H'<H}T_{H'}$. This gives
\[
 \mathbb P(Y^\psi_z\in J)
 \le C_0\frac{\log(2H_0)}{\log(2H_1)}
 +\sum_{H\in\mathcal H}C_0\frac{\log(4H)}{\log(2H_1)}
   \left(\frac3H+\frac{8hH}{c_\psi}\right).
\]
The dyadic sums satisfy
\[
 \sum_{H\in\mathcal H}\frac{\log(4H)}H\ll1,\qquad
 \sum_{H\in\mathcal H}H\log(4H)
 \le2H_1\log(4H_1)\le4H_1\log(2H_1).
\]
Since $2H_1\le y$, the resulting bound is
\[
 \ll\frac{\log(2H_0)}{\log(2H_1)}+\frac{hy}{c_\psi}.
\]
Now $y\ge2H_0\ge4$, so
$\log(2H_1)\ge\tfrac12\log y$ and
$\log y\ge\tfrac12\log\min\{z,1/h\}$. Moreover,
$hy\le\sqrt h\ll1/\log(1/h)\le1/\log\min\{z,1/h\}$.
This proves \eqref{eq:addfn-concentration}.

For \eqref{eq:prime-model-concentration}, if $h\le1/2$ then
$1/h\ge z^{1/2}$, so $\log\min\{z,1/h\}\ge\tfrac12\log z$.
If $h>1/2$, then $z<4$ and the trivial probability bound suffices.
\end{proof}

\begin{lemma}[Localized model tail]\label{lem:prime-model-local-tail}
Let $\psi$ be a real-valued additive function satisfying
\eqref{eq:addfn-first-jump}. For sufficiently large $M$, $z\ge2$,
$v=\log M/\log z\ge16$, and an interval $I$ of length at most
$z^{-1/2}$,
\begin{equation}\label{eq:prime-model-local-tail}
 \mathbb P(S_z>M,\ Y^\psi_z\in I)
 \ll\frac{2^{-v/4}}{\log z}
       +\exp\{-c(\log M)^{0.58}\},
\end{equation}
for some absolute $c>0$. The threshold for $M$ is absolute, and the
implied constant depends only on $q_0$ and $c_\psi$.
\end{lemma}

\begin{proof}

Put $z_k=z^{2^{-k}}$, $M_k=M^{(3/4)^k}$ and
$v_k=(3/2)^kv$. Stop at the first $k=k_*$ with
$\log z_k\le\log\log M$; if this holds initially, stop at zero.
For $k<k_*$, every prime in the active block satisfies
\[
 q>z_{k+1}=\sqrt{z_k}>\sqrt{\log M}>2.
\]
For $q>2$, the valuation law \eqref{eq:valuation-laws} gives exactly
\[
 \mathbb E2^{W_q}
 =\frac{q-2}{q-1}+\sum_{j\ge1}(2/q)^j
 =1+\frac{q}{(q-1)(q-2)}.
\]
Mertens' theorem bounds $\sum_{z_{k+1}<q\le z_k}q^{-1}$ uniformly,
and independence therefore gives
\[
 \mathbb E2^{\sum_{z_{k+1}<q\le z_k}W_q}
 =\prod_{z_{k+1}<q\le z_k}
       \left(1+\frac{q}{(q-1)(q-2)}\right)\ll1.
\]
The cutoff is essential here, since $\mathbb E2^{W_2}=\infty$.
The block product therefore exceeds $M_k^{1/4}$ with
probability $O(2^{-v_k/4})$. Independence and
\eqref{eq:prime-model-concentration} give, if $F_k$ is the supremum
of $\mathbb P(S_{z_k}>M_k,\ Y^\psi_{z_k}\in I')$ over translates
$I'$ of $I$,
\[
 F_k\le C\frac{2^k2^{-v_k/4}}{\log z}+F_{k+1}.
\]
Indeed $S_{z_k}>M_k$ forces either this large block product or
$S_{z_{k+1}}>M_{k+1}$; conditioning on the block translates $I$.
The original length bound $|I|\le z^{-1/2}$ implies
$|I|\le z_{k+1}^{-1/2}$, so concentration applies at every step.
The sum of the displayed terms is $O(2^{-v/4}/\log z)$.
Finally $z_{k_*}\le\log M$, while
$\log M_{k_*}\ge(\log M)^{0.584}$ for large $M$
($0.584<\log(3/2)/\log2$).
The Euler product gives $\mathbb E S_t^{1/2}\le\exp(C\sqrt t)$.
Markov's inequality at the stopping point bounds $F_{k_*}$ by
$\exp\{-c(\log M)^{0.58}\}$, proving the assertion.
\end{proof}

\begin{lemma}[Large smooth parts in short value windows]
\label{lem:prime-large-smooth-window}
Let $\psi$ be a real-valued additive function satisfying
\eqref{eq:addfn-size} and \eqref{eq:addfn-first-jump}.
Let $u=\log x/\log w$, where
$U\le u\le2(\log\log x)^2$ and $w\le x^\delta$; here $\delta>0$ is a
sufficiently small absolute constant and $U$ is a sufficiently large
absolute constant, with $U\ge48$. Uniformly for intervals $I$ of
length at most $w^{-1}(\log x)^{1/2}$,
\begin{equation}\label{eq:prime-large-smooth-window}
 \mathbb P_x\{s_w(p-1)>x^{1/3},\ \psi(s_w(p-1))\in I\}
 \ll\frac{u e^{-cu}}{\log w}+(\log x)^{-3},
\end{equation}
for some absolute $c>0$, where $\mathbb P_x$ denotes uniform measure
on odd primes at most $x$. The implied constant depends only on
$q_0$, $C_\psi$ and $c_\psi$.
\end{lemma}

\begin{proof}
It suffices to consider sufficiently large $x$, depending on $q_0$,
$C_\psi$ and $c_\psi$: for bounded $x$, the probability is at most one
and is absorbed by the $(\log x)^{-3}$ term after enlarging the
implied constant.
If $p-1\ne s=s_w(p-1)$, then $s\le(x-1)/w$. By
\eqref{eq:prime-model-sieve} and \eqref{eq:prime-model-weights},
the number of such primes in the stated window is at most
\[
 \frac{Cx}{\log w}\,
 \mathbb P\{S_w>x^{1/3},\ Y^\psi_w\in I\}.
\]
Apply \Cref{lem:prime-model-local-tail} with $M=x^{1/3}$.
The interval hypothesis there holds for all sufficiently large $x$,
uniformly in the present range. After division by $\pi(x)-1$, this
proves \eqref{eq:prime-large-smooth-window} for these primes.

It remains to count $n=p-1>x^{1/3}$ with $P^+(n)\le w$ and
$\psi(n)\in I$. Set $h=|I|$, $Q_0=(\log x)^4$, and let
$\mathcal E(n)=\{q:q_0<q\le w^{3/4},\ q\Vert n\}$.
If some $q\in\mathcal E(n)$ is at least $Q_0$, write $n=qB$, so that
$\psi(n)=\psi(B)+v_\psi(q)$. For fixed $B$ and a dyadic interval
$(H,2H]$, \eqref{eq:addfn-spacing} allows at most
$1+4H^2h/c_\psi$ integers $q\in(H,2H]$ with $q>q_0$ that put
$\psi(B)+v_\psi(q)$ in $I$. Ignoring primality, the number of these
$n$ is therefore
\[
 \ll\sum_{\substack{H\text{ dyadic}\\H\ge Q_0/2}}^{H\le w^{3/4}}
       \frac{x}{H}\Bigl(1+\frac{4H^2h}{c_\psi}\Bigr)
 \ll_\psi\frac{x}{Q_0}+xw^{3/4}h
 \ll_\psi\frac{\pi(x)}{(\log x)^3}.
\]

Suppose next that $q=\max\mathcal E(n)<Q_0$, and choose dyadic
$H$ with $H<q\le2H$. There is a unique factorization
\[
 n=a q b d,
 \quad P^+(a)<q,\quad
 q<r\le w^{3/4}\ (r\mid b),\quad
 w^{3/4}<r\le w\ (r\mid d),
\]
where $b$ is squarefull and $q\nmid abd$. Thus the factors are
disjoint; a factor may equal one, with $P^+(1)=1$. We may discard $a>x^{1/20}$ or
$b>x^{1/20}$ at a cost $O(x^{0.99})$. To see this, the reciprocal
tail of squarefull integers is $O(T^{-1/2})$, and Rankin's inequality
gives
\[
 \sum_{\substack{a>x^{1/20}\\P^+(a)\le2Q_0}}\frac1a
 \le x^{-1/80}\prod_{r\le2Q_0}(1-r^{-3/4})^{-1}
 =x^{-1/80+o(1)}.
\]
Multiplying these tails by $x$ bounds all integers having such a
divisor. If $d=1$, then $n$ is a product of a $(2Q_0)$-smooth integer
and a squarefull integer. The number of these is at most
\[
 x^{3/4}
 \prod_{r\le2Q_0}(1-r^{-3/4})^{-1}
 \sum_{b\text{ squarefull}}b^{-3/4}
 =x^{3/4+o(1)}.
\]

Otherwise let $r=P^+(n)>w^{3/4}$ and $B=abd/r$. Since
$a,b\le x^{1/20}$ and $q<Q_0$, we have
\[
 \Omega(d/r)\ge\frac{\log(d/r)}{\log w}
 \ge\frac{7u}{30}-\frac{\log Q_0}{\log w}-1
 \ge\frac u5-1
\]
for large $x$, uniformly in the stated range. Prime factors are
counted with multiplicity. Also, if $r^j\Vert n$, then
\[
 \psi(n)=\psi(B)+v_\psi(q)+e_r,\qquad
 e_r=\psi(r^j)-\psi(r^{j-1}),\qquad |e_r|\le C_\psi w^{-3/4},
\]
by \eqref{eq:addfn-size}, with $\psi(r^0)=0$. The increment $e_r$
may have either sign. Since
$4Q_0^2(h+2C_\psi w^{-3/4})<c_\psi$ for large $x$,
\eqref{eq:addfn-spacing} shows that the window fixes at most one
integer $q=q^*(B,H)\in(H,2H]$ with $q>q_0$; every
$q\in\mathcal E(n)$ is of this kind. For this $q$, any candidate has
$p=Bqr+1$ with prime $r>w^{3/4}$. If such a candidate exists, then
$m=Bq=(p-1)/r<(x-1)/w^{3/4}$, so applying
\eqref{eq:prime-model-sieve} to this $m$ at level $w^{3/4}$
bounds their number by $Cx/(H\varphi(B)\log^2w)$. The empty case
has count zero.

The reciprocal sums needed to sum this bound are
\[
 \sum_{P^+(a)\le2H}\frac1{\varphi(a)}\ll\log(4H),\qquad
 \sum_{b\text{ squarefull}}\frac1{\varphi(b)}\ll1,
\]
and
\[
 \sum_{\substack{w^{3/4}<r\le w\ (r\mid d')\\
                  \Omega(d')\ge u/5-1}}
       \frac1{\varphi(d')}
 \le3^{1-u/5}
   \prod_{w^{3/4}<r\le w}
       \left(1+\sum_{j\ge1}\frac{3^j}{\varphi(r^j)}\right)
 \ll3^{-u/5}.
\]
The last product is bounded by Mertens' theorem. These bounds remain
upper bounds when the disjointness restrictions are dropped, since
$\varphi(abc)\ge\varphi(a)\varphi(b)\varphi(c)$.
As $\sum_{H\text{ dyadic}}\log(4H)/H<\infty$, this case contributes
$O(x3^{-u/5}/\log^2w)$.

If $\mathcal E(n)$ is empty, write $n=a_0bd$ with $P^+(a_0)\le q_0$,
$b$ squarefull over primes in $(q_0,w^{3/4}]$ and $d$ supported above
$w^{3/4}$. Discard $a_0>x^{1/20}$, $b>x^{1/20}$ and $d=1$ by the
preceding estimates, which apply since $q_0\le2Q_0$ for large $x$.
Removing $r=P^+(n)$ again leaves $\Omega(d/r)\ge u/5-1$. Apply
\eqref{eq:prime-model-sieve} to $m=n/r$ at level $w^{3/4}$; its range
condition again holds whenever a candidate exists. Sum the same Euler
products, using $\sum_{P^+(a_0)\le q_0}1/\varphi(a_0)\ll_{q_0}1$ in
place of the first sum. This gives $O_\psi(x3^{-u/5}/\log^2w)$ again.
Division by $\pi(x)-1\asymp x/\log x$ completes the proof.
\end{proof}

\paragraph*{The head at the primes dividing $2Q$}
Let $\psi\ge0$ be additive, with $\psi(\ell^0)=\psi(1)=0$. For
$\ell\mid2Q$ let $\mathcal J_\ell$ be the set of valuations permitted
by $\mathcal C_{Q,b}$: it is $\{c_\ell\}$ if $\ell\mid Q$ and
$c_\ell<a_\ell$, and $\{j:j\ge a_\ell\}$ if $\ell\mid Q$ and
$c_\ell=a_\ell$; if $2\nmid Q$, put $\mathcal J_2=\{j:j\ge1\}$.
Define
\[
 a_{Q,b}(\psi):=\sum_{\ell\mid2Q}
                 \inf_{j\in\mathcal J_\ell}\psi(\ell^j),
 \qquad
 \rho_{Q,b}(\psi):=
 \mathbb P\left(\sum_{\ell\mid2Q}\psi(\ell^{W_\ell})
       =a_{Q,b}(\psi)\,\middle|\,\mathcal C_{Q,b}\right).
\]
The infima are finite and nonnegative. Conditioning preserves the
independence of these coordinates, and every permitted valuation has
positive conditional probability. Thus $\rho_{Q,b}(\psi)>0$ exactly
when every local infimum is attained; otherwise $\rho_{Q,b}(\psi)=0$.
For $f$ the head is constant on $\mathcal C_{Q,b}$, so
$\rho_{Q,b}(f)=1$ and $a_{Q,b}(f)=-\log A_{Q,b}$.

\begin{theorem}[Kolmogorov order for additive functions in fixed progressions]
\label{thm:addfn-sharp-order}
\index{Kolmogorov distance!additive functions of p-1@additive functions of $p-1$}
Let $\psi\ge0$ be an additive function satisfying \eqref{eq:addfn-size}
and \eqref{eq:addfn-first-jump}. For each fixed $Q\ge1$, uniformly over
reduced residues $b\pmod Q$,
\begin{equation}\label{eq:addfn-sharp-order}
 \frac{\kappa_{\mathrm{end}}\rho_{Q,b}(\psi)D(Q)+o_{\psi,Q}(1)}{\log x}
 \le d_K\bigl(\operatorname{Law}_{x,Q,b}(\psi(p-1)),
               \operatorname{Law}(Y^{\psi,*})\bigr)
 \ll_{\psi,Q}\frac1{\log x}.
\end{equation}
The upper constants and thresholds may be ineffective. The upper bound also
holds for every real-valued additive $\psi$ satisfying
\eqref{eq:addfn-size} and \eqref{eq:addfn-first-jump}, whatever its
values at the prime powers $q^j$ with $q\le q_0$.
\end{theorem}

When $\rho_{Q,b}(\psi)>0$, the two bounds in \eqref{eq:addfn-sharp-order}
have the same order $1/\log x$.

\begin{proof}[Proof of the upper bound]
Fix $Q$ and a reduced residue $b$, and choose $\delta>0$ admissible
for both \Cref{lem:ap-smooth-pattern,lem:prime-large-smooth-window}.
Let $\mathbb P^*_{x,Q,b}$ denote uniform sampling from its odd primes
at most $x$, with corresponding
law $\operatorname{Law}^*_{x,Q,b}$, and put
\[
 \Delta(z)=\sup_t\left|\mathbb P^*_{x,Q,b}(\psi(s_z(p-1))\le t)
                           -\mathbb P(Y^{\psi,*}_z\le t)\right|.
\]
All cutoffs below exceed $P^+(2Q)$ and $q_0$. First take
$w_0=x^{1/u_0}$ with $u_0\asymp(\log\log x)^2$.
For $t=1/\log z$ and $z\to\infty$, the Euler product and Mertens'
estimate give
\[
 \log\mathbb E S_z^t
 \ll1+\sum_{3\le q\le z}\frac{q^t-1}{q}
 \ll1+\frac1{\log z}\sum_{q\le z}\frac{\log q}{q}\ll1.
\]
Since $\mathbb E(S_z^*)^t\le\beta_{Q,b}^{-1}\mathbb E S_z^t$,
Markov's inequality gives
$\mathbb P(S_{w_0}^*>x^{1/3})\ll_Q e^{-u_0/3}$.
Also \eqref{eq:prime-model-exceptional} and
\eqref{eq:ap-model-weights} give model mass
$O_Q((\log x)^{-12}/\log w_0)$ to $E_{w_0}$.
On every remaining head at most $x^{1/3}$,
\eqref{eq:prime-model-pattern} gives relative error
$O_Q(e^{-cu_0}+(\log x)^{-10})$. Summing the minima of corresponding
point masses, and using equality of total masses, proves
\begin{equation}\label{eq:ap-pattern-initialization}
 d_{\mathrm{TV}}\bigl(\operatorname{Law}^*_{x,Q,b}(s_{w_0}(p-1)),
                     \operatorname{Law}(S_{w_0}^*)\bigr)
 \ll_Q(\log x)^{-3}.
\end{equation}
Pushing the two laws in \eqref{eq:ap-pattern-initialization} forward
by $\psi$ gives $\Delta(w_0)\ll_Q(\log x)^{-3}$.

Compare consecutive levels $w',w=w'^2$, with
$u'=\log x/\log w'$, $U\le u'\le2(\log\log x)^2$ and
$w\le x^\delta$. Call a compatible $w'$-smooth integer $s$ good if
$s\le x^{1/3}$ and $s\notin E_{w'}$.
Conditional on $s_{w'}(p-1)=s$ in the progression, let $K$ be the
$(w',w]$-part of $p-1$, and put $K^W=\prod_{w'<q\le w}q^{W_q}$.
This model block is independent of $S_{w'}^*$, since its primes do
not divide $Q$. For good $s$,
\begin{equation}\label{eq:prime-conditional-swap}
 d_{\mathrm{TV}}(K\mid s,K^W)
 \ll_Q e^{-cu'}+(\log x)^{-10}+\epsilon(s),
 \qquad
 \epsilon(s)=\sum_{k:sk\in E_w}\mathbb P(K^W=k).
\end{equation}
Indeed, for $k\le K_0=x^{1/30}$ and $sk\notin E_w$, divide
\eqref{eq:prime-model-pattern} at $(sk,w)$ by its value at $(s,w')$.
Here $sk\le x^{11/30}$, the two sieve depths are at least $u'/60$
and $u'/15$, and $g_w^*(sk)=g_{w'}^*(s)\mathbb P(K^W=k)$.
The relative main-term factor at $(s,w')$ is at least $1/2$ for
$U$ sufficiently large depending on $Q$. The same Euler-product bound gives
$\mathbb E(K^W)^{1/\log w}\ll1$, and consequently
$\mathbb P(K^W>K_0)\ll e^{-u'/60}$.
Summing the minima on the remaining indices proves
\eqref{eq:prime-conditional-swap}.

The exceptional terms satisfy
\begin{equation}\label{eq:prime-swap-exceptional-mass}
 \sum_{s\in E_{w'}}\mathbb P^*_{x,Q,b}(s_{w'}(p-1)=s)
 +\sum_{s\text{ good}}\mathbb P^*_{x,Q,b}(s_{w'}(p-1)=s)\epsilon(s)
 \ll_Q(\log x)^{-10}.
\end{equation}
For the first sum, counting integers gives
$\mathbb P^*_{x,Q,b}(s_{w'}(p-1)=s)\ll_Q\log x/\varphi(s)$;
use \eqref{eq:prime-model-exceptional}.
For the second, \eqref{eq:prime-model-pattern} bounds the empirical
mass by $2g_{w'}^*(s)$. Unique factorization across the two prime
ranges bounds its sum by $2\sum_{m\in E_w}g_w^*(m)$, to which
\eqref{eq:ap-model-weights} and \eqref{eq:prime-model-exceptional} apply.

Set $L=\lceil3\log\log x\rceil$ and $\eta=(u'+L)C_\psi/w'$.
The arithmetic block $K$ has at most $u'$ distinct prime factors, all
exceeding $w'>q_0$, so $0\le\psi(K)\le u'C_\psi/w'\le\eta$ by
\eqref{eq:addfn-size}. Likewise $\psi(K^W)\ge0$, and
$\psi(K^W)>\eta$ forces more than $L$ active primes in the model
block. Their number $N$ is a Bernoulli sum of mean at most one for
large $x$; thus
$\mathbb P(\psi(K^W)>\eta)\le\mathbb E\binom NL\le1/L!
\ll(\log x)^{-3}$.
Add and subtract the law of $\psi(s_{w'}(p-1))+\psi(K^W)$, with $K^W$
independent of the empirical head: adding an independent model block
contracts Kolmogorov distance. For a head $s$, compare
$\mathbb P(\psi(s)+\psi(K)\le t\mid s)$ with
$\mathbb P(\psi(s)+\psi(K^W)\le t)$: both vanish if $\psi(s)>t$;
they differ by at most $1/L!$ if $\psi(s)\le t-\eta$; and they differ
by at most $d_{\mathrm{TV}}(K\mid s,K^W)$ if
$\psi(s)\in(t-\eta,t]$.
Conditioning \eqref{eq:prime-model-concentration} costs at most
$\beta_{Q,b}^{-1}$, and $\eta\le(w')^{-1/2}$ for large $x$, so this
interval has empirical mass at most $2\Delta(w')+C_{\psi,Q}/\log w'$.
For the large heads, discard the progression restriction in the
numerator of \Cref{lem:prime-large-smooth-window}; normalization
costs $O_Q(1)$. Its interval hypothesis holds because
$(u'+L)C_\psi\le(\log x)^{1/2}$ for large $x$. The other bad heads and the averaged
$\epsilon(s)$ are covered by \eqref{eq:prime-swap-exceptional-mass}.
Putting $\tau(u')=e^{-cu'}+(\log x)^{-10}$, we obtain
\begin{equation}\label{eq:prime-swap-recursion}
 \Delta(w)\le(1+C_{\psi,Q}\tau(u'))\Delta(w')
       +\frac{C_{\psi,Q}\tau(u')+C_{\psi,Q}u'e^{-cu'}}{\log w'}
       +\frac{C_{\psi,Q}}{(\log x)^3}.
\end{equation}

Choose a fixed sufficiently large $U_0=U_0(Q)\ge\max(U,1/\delta)$,
and put
$J=\lceil\log((\log\log x)^2/U_0)/\log2\rceil$,
$u_j=U_02^{J-j}$, $w_j=x^{1/u_j}$ for $0\le j\le J$.
For large $x$, $(\log\log x)^2\le u_0<2(\log\log x)^2$,
$u_J=U_0$, and $w_j\le x^\delta$ throughout. Thus the initialization,
the pattern estimates and the window estimate apply at every step.
Then $\Delta(w_0)\ll_Q(\log x)^{-3}$ by
\eqref{eq:ap-pattern-initialization}. Since $1/\log w_j=u_j/\log x$,
\eqref{eq:prime-swap-recursion} becomes
\begin{equation}\label{eq:ap-conditioned-recursion}
 \Delta(w_{j+1})\le(1+C_{\psi,Q}\tau(u_j))\Delta(w_j)
 +\frac{C_{\psi,Q}\{u_j\tau(u_j)+u_j^2e^{-cu_j}\}}{\log x}
 +\frac{C_{\psi,Q}}{(\log x)^3}.
\end{equation}
The geometric sequence satisfies
\[
 \sum_{j<J}\tau(u_j)\ll1,\qquad
 \sum_{j<J}(u_j+u_j^2)e^{-cu_j}\ll1,\qquad
 J\ll\log\log\log x.
\]
The additional sum $(\log x)^{-10}\sum_{j<J}u_j$ is
$O((\log\log x)^2(\log x)^{-10})$. Hence iteration of
\eqref{eq:ap-conditioned-recursion} gives
$\Delta(w_J)\ll_{\psi,Q}1/\log x$.
At $w=w_J$, the omitted arithmetic sum lies in $[0,C_\psi U_0/w]$
by \eqref{eq:addfn-size}, whereas
\[
 \mathbb E(Y^{\psi,*}-Y^{\psi,*}_w)
 \le C_\psi\sum_{q>w}\frac1{q(q-1)}
 \ll\frac{C_\psi}{w\log w},
\]
because all omitted coordinates are outside $Q$. For large $x$ the
arithmetic remainder is at most $w^{-1/2}$, and Markov's inequality
bounds the probability that the model remainder exceeds $w^{-1/2}$
by $O(C_\psi/(\sqrt w\log w))$. Sandwich the two distribution
functions at distance $w^{-1/2}$, and use the conditioned version of
\eqref{eq:prime-model-concentration} for the intervening intervals.
Together with the bound for $\Delta(w_J)$, this gives
\[
 d_K\bigl(\operatorname{Law}^*_{x,Q,b}(\psi(p-1)),
               \operatorname{Law}(Y^{\psi,*})\bigr)
 \ll_{\psi,Q}\frac1{\log x}.
\]
Restoring $p=2$ costs $O_Q(\log x/x)$, proving the upper bound.
Ineffectivity enters through Siegel--Walfisz in the prime-distribution
inputs, including Bombieri--Vinogradov and fixed-$Q$ normalization.
All constants may be maximized over the finitely many reduced classes
for fixed $Q$, giving the asserted uniformity.
\end{proof}

\begin{proof}[Proof of the lower bound]
If $\rho_{Q,b}(\psi)=0$, the lower bound follows from nonnegativity
of the distance. Otherwise put $\rho=\rho_{Q,b}(\psi)$,
$\alpha=a_{Q,b}(\psi)$, $q_1=\max\{q_0,P^+(2Q)\}$ and
$t_x=\alpha+c_\psi/x$.
For an odd prime $p\le x$ with $p\equiv b\pmod Q$, the valuations
at $\ell\mid2Q$ lie in $\mathcal J_\ell$, so their contribution to
$\psi(p-1)$ is at least $\alpha$. Every other contribution is
nonnegative. If a prime $q>q_1$ divides $p-1$ exactly once, then
\[
 \psi(p-1)\ge\alpha+v_\psi(q)
             \ge\alpha+\frac{c_\psi}{q}>t_x,
\]
by \eqref{eq:addfn-spacing}, since $q\le p-1<x$.
Consequently $\psi(p-1)\le t_x$ forces
$p-1=a_0d$, where $P^+(a_0)\le q_1$ and $d$ is squarefull.
Rankin's inequality bounds the number of such integers up to $x$ by
\[
 x^{3/4}\prod_{r\le q_1}(1-r^{-3/4})^{-1}
             \sum_{d\text{ squarefull}}d^{-3/4}
 \ll_{\psi,Q}x^{3/4}.
\]
The sum converges: every squarefull integer has the form $a^2b^3$
with $b$ squarefree, so it is bounded by
$\zeta(3/2)\zeta(9/4)$. Since
$\pi(x;Q,b)\asymp_Q x/\log x$, inclusion of the possible prime $2$
gives
\[
 \mathbb P_{x,Q,b}(\psi(p-1)\le t_x)
 \ll_{\psi,Q}x^{-1/4}\log x=o_{\psi,Q}(1/\log x).
\]

For $x\ge q_1$, consider the model event on which the head at
$\ell\mid2Q$ equals $\alpha$ and $W_q=0$ for every prime
$q\le x$ outside $2Q$.
The head, these coordinates and the remaining tail are independent
under the conditioned model. Thus
\[
 \mathbb P(Y^{\psi,*}\le t_x)\ge\rho
 \prod_{\substack{3\le q\le x\\q\nmid Q}}\frac{q-2}{q-1}
 \;\mathbb P\left(\sum_{q>x}\psi(q^{W_q})\le\frac{c_\psi}{x}\right).
\]
By \eqref{eq:endpoint-Q-quantitative}, the product is
\[
 D(Q)\prod_{3\le q\le x}\frac{q-2}{q-1}
 =\frac{\kappa_{\mathrm{end}}D(Q)+o_Q(1)}{\log x}.
\]
All primes in the remaining tail exceed $q_0$ and do not divide
$Q$. By \eqref{eq:addfn-size} and \eqref{eq:endpoint-tail-square},
\[
 0\le\mathbb E\sum_{q>x}\psi(q^{W_q})
 \le C_\psi\sum_{q>x}\frac1{q(q-1)}
 \ll\frac{C_\psi}{x\log x}.
\]
Markov's inequality makes the tail probability
$1-O_\psi(1/\log x)$. Subtracting the empirical distribution
function at $t_x$ proves the lower bound with coefficient
$\kappa_{\mathrm{end}}\rho_{Q,b}(\psi)D(Q)$. There are finitely many
reduced classes for fixed $Q$, so all remainders are uniform over
these classes.
\end{proof}

\begin{theorem}[Sharp order in fixed progressions]\label{thm:sharp-prime-convergence}
\index{Kolmogorov distance!sharp order}
\index{Arithmetic progressions!fixed-modulus convergence}
For $U(p)=X(p)=\varphi(p-1)/(p-1)$ or $U(p)=c(p)$ and each fixed
$Q\ge1$, uniformly over reduced residues $b\pmod Q$,
\[
 \frac{\kappa_{\mathrm{end}}D(Q)+o_Q(1)}{\log x}
 \le\|F^U_{x,Q,b}-G_{Q,b}\|_\infty\ll_Q\frac1{\log x}.
\]
In particular, $(\kappa_{\mathrm{end}}+o(1))/\log x
\le D_x^U\ll1/\log x$. The upper constants and thresholds may be
ineffective. If $\varepsilon\downarrow0$ and
$\log(1/\varepsilon)=o(\log x)$, then
\begin{equation}\label{eq:ap-endpoint-transfer}
\begin{split}
 &\mathbb P_{x,Q,b}\{A_{Q,b}(1-\varepsilon)<U(p)\le A_{Q,b}\}\\
 &\quad=
 \{1-G_{Q,b}(A_{Q,b}(1-\varepsilon))\}
 \left\{1+O_Q\left(\frac{\log(1/\varepsilon)}{\log x}\right)\right\}.
\end{split}
\end{equation}
The model mass here is asymptotic to
$\kappa_{\mathrm{end}}D(Q)/\log(1/\varepsilon)$.
\end{theorem}

\begin{proof}
The function $f$ is nonnegative and strongly additive, with
$f(q^j)=\omega_q=v_\varphi(q)$ for $j\ge1$, where
$v_\varphi(t)=\log(t/(t-1))$. For primes $q>2$,
$\omega_q\le1/(q-1)\le3/(2q)$, and
$-v_\varphi'(t)=1/(t(t-1))\ge t^{-2}$ for $t\ge2$. Thus
\eqref{eq:addfn-size} and \eqref{eq:addfn-first-jump} hold with
$q_0=2$, $C_f=3/2$ and $c_f=1$. The law of $Y^{f,*}=Y^*$ is that of
$-\log X_{Q,b}$, so \Cref{thm:addfn-sharp-order} gives
$d_K(\operatorname{Law}_{x,Q,b}(f(p-1)),\operatorname{Law}(Y^*))
\ll_Q1/\log x$.
The decreasing map $v\mapsto e^{-v}$ preserves Kolmogorov distance
from an atomless law, proving the upper bound for $X(p)$.

For $p>\sqrt x$, \Cref{lem:matrix-product-bounds} gives
$0\le-\log P(p)\le3x^{-1/2}$, where $c(p)=X(p)P(p)$.
The smaller primes have progression mass $O_Q(x^{-1/2})$.
The preceding CDF bound and the concentration of $Y^*$, obtained
from \Cref{thm:sharp-global-modulus} by conditioning at cost
$\beta_{Q,b}^{-1}$, give the same upper bound for $c(p)$.

For the lower bound put $A=A_{Q,b}$. If an odd progression prime
has a prime factor of $p-1$ outside $2Q$, then $X(p)\le A(1-1/x)$.
The remaining $p-1$ are supported on the fixed prime divisors of
$2Q$, so there are $O_Q((\log x)^{\omega(2Q)})$ exceptions.
Also every odd progression prime satisfies
$c(p)\le A(1-1/p)\le A(1-1/x)$.
Their exceptional masses, including $p=2$, are $o_Q(1/\log x)$.
Evaluate the CDF at $A(1-1/x)$ and use
\eqref{eq:ap-sharp-concentration}, with $q_S^{-1}=D(Q)$.
Finally, the interval-probability error is at most twice the CDF
error. Dividing by the endpoint mass in that same formula proves
\eqref{eq:ap-endpoint-transfer}.
\end{proof}

\Needspace{5\baselineskip}
\begin{corollary}[Sharp convergence in every fixed reduced progression]
\label{cor:ap-sharp-convergence}
For fixed $Q\ge1$, uniformly over reduced residues $b\pmod Q$ and
all choices $U(p)=X(p)$ or $c_{n_x(p)}(p)$, with arbitrary arrays
$n_x(p)\in\Z_{\ge1}\cup\{\infty\}$,
\begin{equation}\label{eq:ap-sharp-prime-convergence}
 \frac{\kappa_{\mathrm{end}}D(Q)+o_Q(1)}{\log x}
 \le\|F^U_{x,Q,b}-G_{Q,b}\|_\infty\ll_Q\frac1{\log x}.
\end{equation}
The constants and threshold may be ineffective. If
$\varepsilon\downarrow0$ and $\log(1/\varepsilon)=o(\log x)$, the
endpoint transfer \eqref{eq:ap-endpoint-transfer} also holds uniformly
over these arrays.
\end{corollary}

\begin{proof}
For $X(p)$ this is \Cref{thm:sharp-prime-convergence}.
For every array, the product formula gives
$c(p)\le c_{n_x(p)}(p)\le X(p)$, hence
$F^X_{x,Q,b}\le F^{c_{n_x(p)}}_{x,Q,b}\le F^c_{x,Q,b}$.
The same theorem bounds both outer CDFs, proving the uniform upper
bound. Every odd progression prime also satisfies
$c_{n_x(p)}(p)\le A_{Q,b}(1-1/p)\le A_{Q,b}(1-1/x)$.
The endpoint argument in that theorem therefore gives the stated
lower bound uniformly in the arrays. Dividing the interval error by
the model endpoint mass proves the final assertion.
\end{proof}

For $Q=1$, the dimension-array assertion is
\Cref{cor:finite-dimension}\textup{(i)}; the present corollary gives
its counterpart in every fixed reduced progression.

\begin{corollary}[Sharp order for divisor sums in fixed progressions]
\label{thm:sharp-sigma-convergence-mas06}
\index{Kolmogorov distance!divisor-sum laws}
For fixed $Q\ge1$ and reduced $b\pmod Q$, define
\[
 F^\sigma_{x,Q,b}(t)=\frac1{\pi(x;Q,b)}
       \#\{p\le x:p\equiv b\pmod Q,\ \sigma(p-1)/(p-1)\le t\}.
\]
Let $H_{\sigma,Q,b}$ be the model CDF of
\Cref{cor:ap-sigma-law}, and put
\[
 D(Q)=\prod_{\substack{q\mid Q\\q>2}}\frac{q-1}{q-2},\qquad
 \rho_{Q,b}=2^{-\mathbf1_{\{2\nmid Q\}}}
   \prod_{\substack{q\mid Q\\v_q(b-1)\ge v_q(Q)}}(1-1/q).
\]
Uniformly over the reduced classes for this fixed modulus,
\begin{equation}\label{eq:sharp-sigma-convergence-mas06}
 \frac{\kappa_{\mathrm{end}}\rho_{Q,b}D(Q)+o_Q(1)}{\log x}
 \le\|F^\sigma_{x,Q,b}-H_{\sigma,Q,b}\|_\infty
 \ll_Q\frac1{\log x}.
\end{equation}
For $Q=1$ the displayed lower coefficient is
$\kappa_{\mathrm{end}}/2$.
The upper constants and thresholds may be ineffective.
\end{corollary}

\begin{proof}
For every prime $q$ and integer $j\ge1$, write
$S_j(q)=\sum_{i=0}^jq^{-i}$, so
\begin{equation}\label{eq:sigma-local-increments-mas06}
 0\le\gsig(q^j)=\log S_j(q)\le\omega_q,
 \qquad
 0<\log\frac{S_j(q)}{S_{j-1}(q)}\le\frac1q.
\end{equation}
The second inequality follows from
$S_j/S_{j-1}=1+q^{-j}/S_{j-1}$ and $\log(1+t)\le t$.
Since $\omega_q\le3/(2q)$ for $q>2$, the function $\gsig\ge0$ satisfies
\eqref{eq:addfn-size} with $q_0=2$ and $C_{\gsig}=3/2$. Also
$\gsig(q)=v_\sigma(q)$ with $v_\sigma(t)=\log(1+1/t)$ and
$-v_\sigma'(t)=1/(t(t+1))\ge2/(3t^2)$ for $t\ge2$, so
\eqref{eq:addfn-first-jump} holds with $c_{\gsig}=2/3$.

Condition the full valuation model on $\mathcal C_{Q,b}$ as in
\eqref{eq:ap-model-weights}, and set
$T_z^*=\gsig(S_z^*)$, $T^*=\lim_zT_z^*$.
The law of $e^{T^*}$ is $H_{\sigma,Q,b}$.
At $q\mid Q$ with $c_q<a_q$ the valuation is fixed at $c_q$;
if $c_q=a_q$, its conditional probabilities are
\[
 \mathbb P(W_q=a_q+j\mid W_q\ge a_q)
       =(1-1/q)q^{-j}\qquad(j\ge0).
\]
When $2\nmid Q$, $W_2$ retains the law $\mathbb P(W_2=j)=2^{-j}$,
$j\ge1$. Consequently the finite head at primes dividing $2Q$
remains random wherever only a valuation lower bound is imposed.

By the second inequality in \eqref{eq:sigma-local-increments-mas06},
$j\mapsto\gsig(\ell^j)$ is strictly increasing. Hence the head
$\sum_{\ell\mid2Q}\gsig(\ell^{W_\ell})$ attains its infimum
$\gsig(m)=\log a_\sigma$, where $m=[2,\gcd(b-1,Q)]$ and
$a_\sigma=\sigma(m)/m$, exactly when every $W_\ell$ with $\ell\mid2Q$
takes the least value of $\mathcal J_\ell$. By the conditional laws
above this has probability $\rho_{Q,b}$, so
$\rho_{Q,b}(\gsig)=\rho_{Q,b}$. Since $T^*=Y^{\gsig,*}$,
\Cref{thm:addfn-sharp-order} gives, uniformly over reduced residues
$b\pmod Q$,
\[
 \frac{\kappa_{\mathrm{end}}\rho_{Q,b}D(Q)+o_Q(1)}{\log x}
 \le d_K\bigl(\operatorname{Law}_{x,Q,b}(\gsig(p-1)),
              \operatorname{Law}(T^*)\bigr)
 \ll_Q\frac1{\log x}.
\]
The increasing logarithmic change of variable preserves
Kolmogorov distance.
\end{proof}

\begin{proposition}[Threshold estimates and displacements from rational anchors]
\label{prop:fixed-threshold}
\index{Prime sampling!rational thresholds}
Fix $Q\ge1$ and $(b,Q)=1$. Let $U_x(p)=X(p)$ or $c_{n_x(p)}(p)$
with arbitrary $n_x(p)\in\mathbb Z_{\ge1}\cup\{\infty\}$.
For each fixed rational $t$ and every $B>0$,
\begin{equation}\label{eq:fixed-threshold}
 \bigl|F^{U_x}_{x,Q,b}(t)-G_{Q,b}(t)\bigr|
 \ll_{Q,t,B}(\log x)^{-B}.
\end{equation}
The same holds for each fixed irrational $t$ satisfying
$|t-a/d|\ge c_t d^{-\eta}$ for all integers $a$, $d\ge1$, with
fixed $c_t>0$ and finite $\eta\ge2$.
For each compact interval $I\subset(0,A_{Q,b})$ and $B>0$, there is
$c=c(Q,b,I,B)>0$ such that \eqref{eq:fixed-threshold} holds uniformly
for reduced rational $t=a/d\in I$ with
\[
 d\le\exp\{c\log x/\log\log x\},
\]
with implied constant depending only on $Q,b,I,B$.
For unrestricted primes ($Q=1$), fix a rational $t=a/d\in(0,1/2]$
in lowest terms and set $h=1/(dx)$. If $t=x_S$ is an arithmetic
anchor, then, for every fixed integer $N\ge1$,
\begin{equation}\label{eq:moving-rational-anchor}
 F^{U_x}_{x,1,1}(t-h)-G(t-h)
 =\kappa_S\sum_{j=1}^N\frac{a_j}{(\log(ax))^j}
       +O_{S,N}\bigl((\log x)^{-N-1}\bigr).
\end{equation}
At a fixed rational non-anchor the left-hand side is
$O_{t,B}((\log x)^{-B})$ for every fixed $B>0$.
In particular, for $t_x=(1/3)(1-1/x)$ and $L=\log x$ one has
\begin{equation}\label{eq:moving-third}
 F^{U_x}_{x,1,1}(t_x)-G(t_x)
 =\frac{\kappa_{\mathrm{end}}}{L}
  -\frac{\kappa_{\mathrm{end}}\pi^2}{12L^3}+O(L^{-4}).
\end{equation}
All estimates are uniform over the dimension arrays.
\end{proposition}

\begin{proof}
Put $A_0=A_{Q,b}$ and $E=\{\ell>2:\ell\mid Q\}$. We require a
one-sided estimate for the finite product
$Z_y=A_0\prod_{\ell\le y,\,\ell\nmid2Q}(1-1/\ell)^{B_\ell}$.
For rational $t$, restrict to a fixed compact interval
$I\subset(0,A_0)$; in the irrational case fix $t$ as in the statement.
For small $h$, put $\varepsilon=h/(t+h)$ and $z=\log(1/\varepsilon)$.
In the rational case take $(\alpha,\beta,\rho)=(1/20,1/5,3/10)$.
In the irrational case choose fixed positive exponents with
$2\alpha<\beta<\rho<1/2$ and $\alpha\eta<\beta$.
Use the proof of
\Cref{lem:binomial-uniform-deletion} with $k=1$, this $E$, $A=A_0$
and the truncated product, taking
$D<\varepsilon^{-\alpha}$, $\delta=\varepsilon^\beta$,
$T=\varepsilon^{-\rho}$ and
$\log\Xi=\lambda\sqrt{z\log z}$ with fixed $\lambda>0$.
For $y\ge T$, the head moment estimate, disjoint images obtained by
deleting the least active middle prime, and the tail Markov bound
are unchanged for $Z_y$: removing tail coordinates only decreases
its expected logarithm. With $s=\log z/(2\log\Xi)$, their bounds are
\[
 \exp\{-(\alpha/(2\lambda)+o(1))\sqrt{z\log z}\},
 \quad O(e^{-\lambda\sqrt{z\log z}}),
 \quad O(\varepsilon^{\rho-\beta}),
\]
respectively. The disjointness follows because restoring different
middle primes gives products with ratio exceeding
$1+2\varepsilon^{2\rho}>(1-\varepsilon)^{-1}$.
Since $A_0$ is rational, any surviving anchor
$a_S=A_0\prod_{\ell\in S}(1-1/\ell)$ has denominator at most
$C_Q\varepsilon^{-\alpha}$ and satisfies
$t<a_S\le(t+h)/(1-\delta)$.
An anchor equal to $t$ is excluded by the strict left inequality.
If $t=a/d\in I$ is reduced and $d\le h^{-1/20}$, rational separation
and $\varepsilon\asymp_I h$ give
\[
 |a_S-t|\gg_Q d^{-1}\varepsilon^{1/20}
          \gg_{Q,I}\varepsilon^{1/10}.
\]
For the specified irrational $t$, separation is
$\gg_{Q,t}\varepsilon^{\alpha\eta}$. In both cases it exceeds the
bracket width $O(\varepsilon^\beta)$, with constants uniform in the
stated rational range. Thus no survivor is possible, and
\begin{equation}\label{eq:finite-right-window}
 \mathbb P(t<Z_y\le t+h)
 \ll_{Q,t}\exp\{-c_{Q,t}\sqrt{\log(1/h)\log\log(1/h)}\}.
\end{equation}
Its constants are uniform for rational $t\in I$ with $d\le h^{-1/20}$.
Put $L=\log x$ and use the head comparison \eqref{eq:ap-truncated-tv}
in the range of \Cref{cor:ap-finite-head-ranges}\textup{(ii)} for the
fixed modulus $Q$,
with arbitrary saving $O_{Q,B}(L^{-B})$
and $\log y\asymp_B L/\log L$. Retain all forced coordinates in
$A_0$ and take $h=y^{-1/2}$. In the uniform rational case choose $c>0$ so that
$\exp(cL/\log L)\le y^{1/80}$ for all sufficiently large $x$.
Then $d\le(Ch)^{-1/20}$ for every fixed $C>0$, so the same uniform
bound applies to all fixed multiples of $h$ below.
The arithmetic downward displacement is $O(L/(y\log y))=o(h)$;
the model displacement exceeds $h$ with probability
$O(y^{-1/2}/\log y)$.
The CDF changes when these tails are restored are bounded by
\eqref{eq:finite-right-window}, the head total variation and that
exceptional probability. The exponent in
\eqref{eq:finite-right-window} has order $\sqrt L$, exceeding every
fixed multiple of $\log L$.
Discarding $p\le\sqrt x$ costs $O_Q(L/\sqrt x)$ and makes the uniform
determinant displacement at most $1/(2(p-1))<2h$; apply the same
bound with a fixed multiple of $h$. Restoring $p=2$ costs $O_Q(L/x)$.
The endpoints and thresholds outside $(0,A_0)$ are immediate.
This proves \eqref{eq:fixed-threshold}.

For the moving threshold put $Q=1$ and $t=a/d$ as in the statement.
For odd $p$ with $X(p)\ne t$, rational separation gives
$|X(p)-t|\ge1/(d(p-1))>h$. Equality $X(p)=t$ can occur only at an
anchor, where the odd prime support of $p-1$ is exactly $S$ by
uniqueness of the finite product. There are at most
$O_S((\log x)^{1+|S|})$ integers $p-1$ supported on $\{2\}\cup S$.

For determinant densities use $c_1(p)=\varphi(p-1)/p$. For $p\nmid d$
one has $|c_1(p)-t|\ge1/(dp)$ because $0<\varphi(p-1)<p$.
For every allowed $n$, the tail of the matrix product gives
$0\le c_1(p)-c_n(p)\le\sum_{j\ge2}p^{-j}\ll p^{-2}$.
Thus for $p$ sufficiently large in terms of $d$, a $c_1(p)>t$
leaves $c_n(p)>t$; if $c_1(p)<t$, then $c_n(p)\le t-1/(dp)\le t-h$.
Consequently the empirical mass of $(t-h,t]$ is smaller than every
fixed inverse power of $\log x$ in both models. Subtract that mass
from the empirical CDF at $t$ and apply \eqref{eq:fixed-threshold}.
The model increment is given by \Cref{thm:anchor-sharp} with
$\log(t/h)=\log(ax)$, or by \eqref{eq:rational-nonanchor-decay} at a
non-anchor. For $t=1/3$, $\kappa_{\{3\}}=\kappa_{\mathrm{end}}$ and
$a_1=1$, $a_2=0$, $a_3=-\pi^2/12$ give \eqref{eq:moving-third}.
\end{proof}

\KeepHeadingSpace{5\baselineskip}
\begin{proposition}[A uniform endpoint obstruction]
\label{prop:ap-uniform-endpoint-mas07}
Fix $0<\delta<1/2$ and $c_{\mathrm{pop}}>0$. For sufficiently large
$x$, depending only on these constants, suppose
\[
 Q\le x^{1/2-\delta},\qquad (b,Q)=1,\qquad
 \pi(x;Q,b)\ge\frac{c_{\mathrm{pop}}x}{\varphi(Q)\log x}.
\]
For $U(p)=X(p)$ or $c_{n_x(p)}(p)$ with an arbitrary allowed
dimension array, uniformly over these data,
\begin{equation}\label{eq:ap-uniform-endpoint-mas07}
 \|F^U_{x,Q,b}-G_{Q,b}\|_\infty
 \ge D(Q)H_{\mathrm{end}}(1/x)-x^{-1/2}.
\end{equation}
\end{proposition}

\begin{proof}
We first give an integer bound also used below for interval populations
and Wasserstein lower bounds. For $z\ge3$, $Q\le z$, and fixed
$0<\epsilon<1$, let
$N_Q(z)$ count the positive integers at most $z$ all of whose prime
divisors divide $2Q$; this includes $1$. Rankin's inequality gives
\begin{equation}\label{eq:supported-rankin-mas07}
 N_Q(z)\le z^\epsilon\prod_{\ell\mid2Q}(1-\ell^{-\epsilon})^{-1}
 \le z^\epsilon
       \exp\{O_\epsilon((\log z)^{1-\epsilon})\}.
\end{equation}
Indeed, the logarithm of the product is at most
$C_\epsilon\sum_{\ell\mid2Q}\ell^{-\epsilon}$. The primes
$\ell\le\log z$ contribute $O_\epsilon((\log z)^{1-\epsilon})$
by comparison with the sum over all positive integers. There are
$O(\log z/\log\log z)$ prime divisors exceeding $\log z$, whose
contribution is smaller. Enlarging the constant covers bounded $z$.
Thus $N_Q(z)\le z^{\epsilon+o(1)}$, uniformly for $Q\le z$.

Put $A=A_{Q,b}$ and $t=A(1-1/x)$. In the unrestricted product
$2X=\prod_{\ell>2}(1-1/\ell)^{B_\ell}$, the event
$2X>1-1/x$ forces $B_\ell=0$ for every odd $\ell\mid Q$:
all these primes are smaller than $x$. Their joint inactivity
probability is $D(Q)^{-1}$, independently of the remaining
coordinates. Consequently the exact identity is
\begin{equation}\label{eq:open-endpoint-identity-mas07}
 1-G_{Q,b}(t)
 =\mathbb P\{X_{Q,b}>t\}=D(Q)H_{\mathrm{end}}(1/x).
\end{equation}
For an odd progression prime, the factors at primes dividing $2Q$
are precisely those in $A$. If a prime $\ell\nmid2Q$ divides
$p-1$, then $\ell<x$ and
$U(p)\le X(p)\le A(1-1/\ell)<t$.
Hence every observation in the full open upper tail $(t,\infty)$
has $p-1$ supported on $2Q$. This includes the possible prime $2$,
whose predecessor is $1$, even when its value lies above the model
support. For $X(p)$ these are exactly the observations in that tail;
for a determinant statistic they form a containing exceptional set.
It follows from \eqref{eq:supported-rankin-mas07}, with
$\epsilon=\delta/2$, and the population hypothesis that
\[
 1-F^U_{x,Q,b}(t)
 \le\frac{N_Q(x)}{\pi(x;Q,b)}
 \ll_{\delta,c_{\mathrm{pop}}}
 x^{-1/2-\delta/2}
       (\log x)\exp\{O_\delta((\log x)^{1-\delta/2})\}
 \le x^{-1/2}
\]
for sufficiently large $x$, uniformly in $Q,b$ and the array.
Subtracting the two full upper-tail probabilities proves
\eqref{eq:ap-uniform-endpoint-mas07}.
\end{proof}

\begin{theorem}[Short-interval prime laws]
\label{thm:short-interval}
\index{Short intervals!prime sampling}
Fix $Q\ge1$, $(b,Q)=1$, and $7/12<\theta\le1$.
Put $H=x^\theta$, $L=\log x$, and sample primes in $(x,x+H]$
congruent to $b\pmod Q$ uniformly.
For $U_x(p)=X(p)$ or $c_{n_x(p)}(p)$ with arbitrary
$n_x(p)\in\mathbb Z_{\ge1}\cup\{\infty\}$,
write $F^U_{x,H,Q,b}$ for the empirical CDF and
$\nu=\operatorname{Law}(X_{Q,b})$,
$\nu_{\log}=\operatorname{Law}(-\log X_{Q,b})$.
Then
\begin{equation}\label{eq:short-cdf}
 \frac{\kappa_{\mathrm{end}}D(Q)+o_{Q,\theta}(1)}{L}
 \le\|F^U_{x,H,Q,b}-G_{Q,b}\|_\infty
 \ll_{Q,\theta}\frac{D(Q)\log L}{L}.
\end{equation}
For every fixed $1\le r<\infty$ and $B>0$, the raw and logarithmic
empirical laws satisfy
\begin{equation}\label{eq:short-wasserstein}
 \mathcal W_r(\operatorname{Law}(U_x),\nu)+
 \mathcal W_r(\operatorname{Law}(-\log U_x),\nu_{\log})
 \ll_{Q,\theta,r,B}L^{-B}.
\end{equation}
The bound \eqref{eq:fixed-threshold}, including its uniform rational
range in \Cref{prop:fixed-threshold}, also holds with this short-interval
CDF, with constants, including $c$, depending additionally on $\theta$.
\end{theorem}

\begin{proof}
We record the distribution input and the weights needed for the
transfer. Fix $v=(\theta-11/20)/2$ and put $M=x^v$.
Timofeev's maximal-prefix theorem~\cite{Tim88} gives, for every $K>0$,
\[
 \sum_{m\le M}E_m^\psi\ll_{K,\theta,v}HL^{-K},\qquad
 E_m^\psi=\max_{(a,m)=1}\max_{0\le s\le H}
 \left|\sum_{\substack{x<n\le x+s\\n\equiv a\pmod m}}\Lambda(n)
                 -\frac{s}{\varphi(m)}\right|.
\]
Indeed apply its displayed estimate with outer parameter $2x$ and
initial point $x$, leaving a fixed margin in its level
$H(2x)^{-11/20-\delta}$. Its interval condition follows from
$\theta>7/12$.
Removing prime powers costs $O(\sqrt x L^2)$ per modulus, uniformly
in every prefix. If $\Delta_m(s)$ is the resulting weighted-prime
error, partial summation gives the prime-counting error against
$J(s)/\varphi(m)$, $J(s)=\operatorname{Li}(x+s)-\operatorname{Li}(x)$,
as
\[
 \frac{\Delta_m(s)}{\log(x+s)}+
 \int_0^s\frac{\Delta_m(t)}{(x+t)\log^2(x+t)}\,dt.
\]
Consequently its maximum $E_m$ satisfies
$\sum_{m\le M}E_m\ll_K HL^{-K}$ for every $K$;
the extra cost $M\sqrt x L$ is negligible since $v+1/2<\theta$.
Integer counting gives $E_m\ll H/m$ for $m\le M\ll H$.
Thus
\[
 \sum_{m\le M}9^{\omega(m)}E_m\ll HL^{10},\qquad
 \sum_{m\le M}3^{\omega(m)}E_m\ll_K HL^{-K},
\]
where the second estimate follows by Cauchy--Schwarz from the first
and the arbitrary unweighted savings. For the first, bound
$\sum_{m\le M}9^{\omega(m)}/m$ by
$\prod_{\ell\le M}(1+9/(\ell-1))\ll(\log M)^9$.

Given $A>0$, choose $k=\lceil c_A\log L\rceil$ with
$c_A\log(c_A/(2e))>A+3$, and set
$y=\exp(vL/(2(k+1)))$.
Retain all coordinates forced by $Q$, and apply
\Cref{lem:finite-prime-sample-head} to the primes in $(x,x+H]$, with
$J=J(H)$, $E(m)=E_m$, the free odd primes at most $y$ and this $k$.
Every inclusion modulus is $Qd\le Qy^{k+1}<M$.
The progression population is asymptotic to
$J(H)/\varphi(Q)\asymp_Q H/L$. Since $E_Q\ll_KHL^{-K}$ for every
$K$, also $E_Q\le J(H)/(2\varphi(Q))$ for large $x$.
By CRT each inclusion error is at most
\[
 O_Q\!\left(\frac L H
              \left(E_{Qd}+\frac{E_Q}{\varphi(d)}\right)\right).
\]
The first errors, summed with coefficients $2^{\omega(d)}$, are
controlled by the weighted estimate. The second cost at most
$O((\log y)^2)E_Q$, since
$\prod_{3\le\ell\le y}(1+2/(\ell-1))\ll(\log y)^2$.
The independent head has
$\lambda=\log L-\log\log L+O_{Q,\theta,A}(1)$, so Stirling gives
$(2\lambda)^{k+1}/(k+1)!\ll L^{-A}$.
The head total variation is therefore $O_{Q,\theta,A}(L^{-A})$.

The empirical omitted logarithm is at most
$2\log(2x)/(y\log y)$, and its model mean is $O(1/y)$.
The head diameter is $O_Q(\log\log y)$; determinant logarithmic
corrections are $O(1/x)$ uniformly in dimension.
A maximal coupling, with $A>B+2$, proves the logarithmic
bound for $r=1$. For $r>1$, the empirical logarithms are $O(L)$
uniformly in the arrays, and \Cref{cor:ap-moments} gives finite
moments of every order for $\nu_{\log}$. Cauchy--Schwarz on an
optimal $\mathcal W_1$ coupling yields
$\mathcal W_r^{2r}\ll_{Q,r}L^{2r-1}\mathcal W_1$.
The arbitrary logarithmic savings prove \eqref{eq:short-wasserstein};
the map $u\mapsto e^{-u}$ is $1$-Lipschitz on the logarithmic supports.
For CDFs take $A=10$, use width $y^{-1/2}$ in
\Cref{prop:head-tail-cdf}, and apply
\Cref{lem:ap-uniform-concentration}. The total error is
\[
 O_{Q,\theta}(L^{-10}+y^{-1/2}+D(Q)/\log y)
 \ll_{Q,\theta}D(Q)\log L/L.
\]
For the lower bound put $A_0=A_{Q,b}$ and evaluate at
$A_0(1-1/(x+H))$. Every determinant value lies below this point.
For $X(p)$ the only exceptions have $p-1$ supported on primes
of $2Q$, giving $O_Q(L^{\omega(2Q)})$ exceptions.
The model endpoint mass is
$(\kappa_{\mathrm{end}}D(Q)+o(1))/\log(x+H)$ by
\eqref{eq:ap-sharp-concentration}; the exceptions are negligible
compared with the population $\asymp_Q H/L$.
This proves \eqref{eq:short-cdf}.
Finally use the preceding head approximation and $h=y^{-1/2}$
in \eqref{eq:finite-right-window}, with the fixed rational factor
$A_{Q,b}$ and deletion set $\{\ell>2:\ell\mid Q\}$.
Its exponent has order $\sqrt L$; the tails and determinant
corrections are smaller than every $L^{-B}$. For the uniform rational
range choose $c$ so that $\exp(cL/\log L)\le y^{1/80}$.
The same threshold comparison proves the assertion.
\end{proof}

\begin{corollary}[Uniform short-interval CDFs for polylogarithmic moduli]
\label{cor:short-uniform-cdf-mas07}
Fix $C>0$ and $7/12<\theta\le1$, and put $H=x^\theta$ and
$L=\log x$. Uniformly for $Q\le L^C$, $(b,Q)=1$, and every
statistic and dimension array in \Cref{thm:short-interval},
\begin{equation}\label{eq:short-uniform-cdf-mas07}
 \begin{split}
 D(Q)H_{\mathrm{end}}(1/(x+H))-x^{-1/2}
 &\le\|F^U_{x,H,Q,b}-G_{Q,b}\|_\infty\\
 &\ll_{C,\theta}D(Q)\frac{\log L}{L}.
 \end{split}
\end{equation}
\end{corollary}

\begin{proof}
Put $v=(\theta-11/20)/2$ and $M=x^v$. For
$J(s)=\operatorname{Li}(x+s)-\operatorname{Li}(x)$, define
\[
 E_m=\max_{(a,m)=1}\max_{0\le s\le H}
 \left|\pi(x+s;m,a)-\pi(x;m,a)-\frac{J(s)}{\varphi(m)}\right|.
\]
The Timofeev input and its weighted consequence proved in
\Cref{thm:short-interval} give, for every $K>0$,
\begin{equation}\label{eq:short-weighted-input-mas07}
 \sum_{m\le M}3^{\omega(m)}E_m\ll_{K,\theta}HL^{-K}.
\end{equation}
In particular, for $Q\le L^C<M$ the population $N=\pi(x+H;Q,b)-\pi(x;Q,b)$ satisfies
\begin{equation}\label{eq:short-uniform-population-mas07}
 N=\frac{J(H)}{\varphi(Q)}
       \{1+O_{K,\theta}(L^{C+1-K})\}
 \asymp\frac{H}{\varphi(Q)L},
\end{equation}
uniformly in $b$, provided $K>C+2$.

For an arbitrary fixed head accuracy $R>0$, choose $c_R>2e$ with
$c_R\log(c_R/(2e))>R+2$, and put
\[
 k=\lceil c_R\log L\rceil,\qquad
 y=\exp\{vL/(2(k+1))\}.
\]
All forced coordinates are retained. For the free odd primes
$\ell\le y$, $\ell\nmid Q$, compare the subset dividing $p-1$
with independent indicators of probabilities $(\ell-1)^{-1}$.
If $d$ is a squarefree product of at most $k+1$ such primes,
the inclusion modulus obeys the necessary level restriction
\[
 Qd\le Qy^{k+1}\le L^C x^{v/2}<M
\]
for sufficiently large $x$. CRT and
\eqref{eq:short-uniform-population-mas07} bound the difference
of the arithmetic and independent inclusion probabilities by
\[
 \frac{2\varphi(Q)}{J(H)}
       \left(E_{Qd}+\frac{E_Q}{\varphi(d)}\right).
\]
With $\mathcal D$ the above products,
\[
 \begin{split}
 &\sum_{d\in\mathcal D}2^{\omega(d)}
       |a_{\mathrm{arith}}(d)-a_{\mathrm{model}}(d)|\\
 &\quad\ll\frac{\varphi(Q)L}{H}
 \left\{\sum_{m\le M}3^{\omega(m)}E_m+
 E_Q\prod_{3\le\ell\le y}\left(1+\frac2{\ell-1}\right)\right\}
 \ll_{K,\theta}L^{C+3-K}.
 \end{split}
\]
Here the first sum is bounded because $d\mapsto Qd$ is injective
and $2^{\omega(d)}\le3^{\omega(Qd)}$; the product is
$O((\log y)^2)$. Choose the saving exponent
$K>C+R+4$. The head mean has the uniform upper bound
\[
 \lambda=\sum_{\substack{3\le\ell\le y\\\ell\nmid Q}}
                      \frac1{\ell-1}
 \le\log\log y+O(1)\le\log L+O(1).
\]
Stirling's inequality and our choice of $c_R$ give
$(2\lambda)^{k+1}/(k+1)!\ll_R L^{-R}$.
By \eqref{eq:short-weighted-input-mas07} with $K>C+2$,
$E_Q\le J(H)/(2\varphi(Q))$ for large $x$, uniformly in $Q\le L^C$.
\Cref{lem:finite-prime-sample-head}, applied to the
primes in $(x,x+H]$ and the free odd primes $\ell\le y$,
$\ell\nmid Q$, with $J=J(H)$, $E(m)=E_m$ and truncation level $k$,
therefore proves head total variation $O_{C,\theta,R}(L^{-R})$.

Take $R=10$. The empirical logarithmic tail is deterministically
at most $2\log(2x)/(y\log y)$, and the model tail has mean
$O(1/y)$, uniformly in $Q$. The determinant correction to the
logarithm is $O(1/x)$ uniformly over the arrays. Couple the heads
maximally and use Markov's inequality on the model tail at
$h=y^{-1/2}$. The two full logarithmic variables then differ by at
most a fixed multiple of $h$, outside probability
$O_{C,\theta}(L^{-10}+y^{-1/2})$. Bracketing their CDFs and using
\Cref{lem:ap-uniform-concentration} gives
\[
 \|F^U_{x,H,Q,b}-G_{Q,b}\|_\infty
 \ll_{C,\theta}L^{-10}+y^{-1/2}+\frac{D(Q)}{\log y}
 \ll_{C,\theta}D(Q)\frac{\log L}{L}.
\]
The logarithmic comparison transfers to the raw variables because
the model law is continuous and exponentiation is strictly monotone.

For the lower bound, put $z=x+H$ and $t=A_{Q,b}(1-1/z)$.
Every deleted odd prime is smaller than $z$, so the independence
argument in \eqref{eq:open-endpoint-identity-mas07} gives
$1-G_{Q,b}(t)=D(Q)H_{\mathrm{end}}(1/z)$ exactly.
Any interval observation exceeding $t$ has $p-1$ supported on
$2Q$, since a free prime divisor $\ell<z$ would force
$U(p)\le A_{Q,b}(1-1/\ell)<t$.
By \eqref{eq:supported-rankin-mas07}, with
$\epsilon=(\theta-1/2)/2>0$, and
\eqref{eq:short-uniform-population-mas07}, the proportion of such
observations is at most
\[
 \frac{N_Q(2x)}{N}
 \ll_{C,\theta} x^{\epsilon-\theta+o(1)}L^{C+1}
 \le x^{-1/2}.
\]
Taking the difference of the full open upper tails proves the
lower bound, with a threshold depending only on $C,\theta$.
\end{proof}

\begin{corollary}[Shrinking-window counts]
\label{cor:prime-shrinking-anchor}
\index{Shrinking windows!relative counts}
For unrestricted primes let $U(p)=X(p)$ or $U(p)=c_{n_x(p)}(p)$,
where $n_x(p)\in\mathbb Z_{\ge1}\cup\{\infty\}$ may vary with $x,p$.
The estimates below are uniform over these arrays.
\begin{enumerate}[\upshape(i)]
\item Let $z=\log(1/\varepsilon)\to\infty$ and fix
$0<b<1/\sqrt{20}$. Uniformly over finite sets $S=S(x)$ of odd primes
with $D_S\le\exp\{b\sqrt{z\log z}\}$, including the empty set,
\begin{equation}\label{eq:prime-moving-anchor}
 \#\{p\le x:x_S(1-\varepsilon)<U(p)\le x_S\}
 =\frac{\kappa_S\pi(x)}{z}
   \left\{1+O_b\!\left(z^{-2}+\frac{z}{\kappa_S\log x}\right)\right\}.
\end{equation}
In particular this is a relative asymptotic if $z/(\kappa_S\log x)\to0$.
For fixed $S$, if $\varepsilon=\varepsilon(x)\downarrow0$ and
$\log(1/\varepsilon)=o(\log x)$, then
\[
 \#\{p\le x:x_S(1-\varepsilon)<U(p)\le x_S\}
 =\frac{\kappa_S\pi(x)}{\log(1/\varepsilon)}
  \left\{1+O_S\left(\frac1{\log^2(1/\varepsilon)}
                  +\frac{\log(1/\varepsilon)}{\log x}\right)\right\}.
\]
\item Fix $0<\delta<1/2$ and $0<a<1$, and put
$\varepsilon=(\log x)^{-a}$. Uniformly for $\delta\le t\le1/2$ as $x\to\infty$,
\[
\begin{split}
 \#\{p\le x:t(1-\varepsilon)<U(p)\le t\}
 &=\pi(x)\{G(t)-G(t(1-\varepsilon))\}\\
 &\quad\times\left\{1+O\bigl((\log x)^{-(1-a)+o(1)}\bigr)\right\}.
\end{split}
\]
\end{enumerate}
\end{corollary}

\begin{proof}
At $Q=1$, \Cref{cor:ap-sharp-convergence} gives an $O(1/\log x)$ CDF
error uniformly over the specified arrays. The absolute error for an
interval is at most twice this bound.
For \textup{(i)}, apply the uniform expansion in \Cref{thm:anchor-sharp}
with $h=x_S\varepsilon$ and $N=2$. Since $a_1=1$ and $a_2=0$, the
model increment is $(\kappa_S/z)\{1+O_b(z^{-2})\}$.
Dividing the absolute error $O(1/\log x)$ by $\kappa_S/z$ gives \eqref{eq:prime-moving-anchor}.
For fixed $S$, both $D_S$ and $\kappa_S>0$ are constant.
For \textup{(ii)}, fix $C>1/(2\log(59/40))$ and divide the same
uniform CDF bound (with the factor two for an interval) by
\eqref{eq:uniform-relative-window-lower}. The relative error is at most
\[
 O\left((\log x)^{a-1}
       \exp\{C(\log(a\log\log x))^2\}\right)
 =(\log x)^{-(1-a)+o(1)}.
\]
The constants and thresholds in \textup{(ii)} may depend on $\delta,a,C$.
\end{proof}

More generally, for fixed $0<\delta<1/2$ and
$C>1/(2\log(59/40))$, the same division proves the uniform relative
asymptotic in \textup{(ii)} for any $\varepsilon(x)\downarrow0$ satisfying
\begin{equation}\label{eq:general-shrinking-condition-mas17}
 \frac{\exp\{C(\log\log(1/\varepsilon))^2\}}
      {\varepsilon\log x}\longrightarrow0.
\end{equation}
The displayed fraction bounds its relative error up to a constant
depending on $\delta,C$. This uses the uniform lower bound
\eqref{eq:uniform-relative-window-lower} and the $O(1/\log x)$ CDF
error. 

\section{Transport in arithmetic progressions}

For order-$r$ transport, a head coupling with mismatch probability
$\delta$ and common diameter $D$ contributes at most $D\delta^{1/r}$.
The omitted logarithmic parts contribute their $L^r$ norms. This gives
arbitrary logarithmic savings in the polylogarithmic, fixed-base power
and almost-all modulus ranges of \Cref{thm:prime-mellin}.
The subsequent lower bound holds uniformly under the corresponding
progression-population lower bound; under GRH,
\Cref{cor:ap-grh-wasserstein} gives stretched-exponential upper bounds.

\begin{proposition}[Empirical Wasserstein convergence]\label{prop:prime-wasserstein}
\index{Wasserstein distance!empirical convergence}
Fix $A,B>0$ and $1\le r<\infty$. Let $U(p)$ be
$\varphi(p-1)/(p-1)$ or $c(p)$, and let $\operatorname{Law}_{x,Q,b}$
sample primes $p\le x$, $p\equiv b\pmod Q$, uniformly.
Put $\nu_{Q,b}=\operatorname{Law}(X_{Q,b})$ and
$\nu_{Q,b}^{\log}=\operatorname{Law}(-\log X_{Q,b})$.
Uniformly for $Q\le(\log x)^A$ and $(b,Q)=1$,
\begin{equation}\label{eq:prime-wasserstein}
 \mathcal W_r\!\left(\operatorname{Law}_{x,Q,b}(U),\nu_{Q,b}\right)
 +\mathcal W_r\!\left(\operatorname{Law}_{x,Q,b}(-\log U),
                         \nu_{Q,b}^{\log}\right)
 \ll_{A,B,r}(\log x)^{-B}.
\end{equation}
Here $\mathcal W_r(\alpha,\beta)=(\inf\mathbb E|V-W|^r)^{1/r}$,
over couplings of the two laws. The same estimate holds uniformly for
$U(p)=c_{n_x(p)}(p)$ with arbitrary
$n_x(p)\in\mathbb Z_{\ge1}\cup\{\infty\}$.
It also holds in both larger unconditional modulus ranges of
\eqref{eq:prime-mellin-ap} in \Cref{thm:prime-mellin}. In the
almost-all range the exceptional set has size
$O_{\delta,E,B,r}(H(\log x)^{-E})$ for every fixed $E>0$;
it may depend on $\delta,E,B,r$, but not on $b$ or the dimension array.
The constants depend on the fixed base, or on $\delta,E,B,r$, as
applicable; they and the thresholds may be ineffective.
\end{proposition}

\begin{proof}
Put $L=\log x$. Use \Cref{lem:ap-finite-head} with arbitrary fixed
$M\ge10$, in each range of \Cref{cor:ap-finite-head-ranges}, and retain
all forced coordinates in
$a_{Q,b}=-\log A_{Q,b}$. With its free-prime set $\mathcal P_y(Q)$, write
\[
 Y_y(p)=a_{Q,b}+\sum_{\substack{\ell\in\mathcal P_y(Q)\\\ell\mid p-1}}
 \omega_\ell,\qquad
 Y_y=a_{Q,b}+\sum_{\ell\in\mathcal P_y(Q)}B_\ell\omega_\ell,
 \qquad Y=-\log X_{Q,b}.
\]
The common forced shift cancels in a coupling; the free-head diameter
is $O(1+\log\log y)$. For odd $p$,
\begin{equation}\label{eq:ap-empirical-tail}
 0\le f(p-1)-Y_y(p)\le\frac{2L}{y\log y}.
\end{equation}
For the model tail,
\[
 \log\mathbb E e^{y(Y-Y_y)}
 \le\sum_{\substack{\ell>y\\\ell\nmid Q}}
       \frac{e^{y\omega_\ell}-1}{\ell-1}\ll1.
\]
Hence $\mathbb E(Y-Y_y)^r\ll_r y^{-r}$. \Cref{prop:head-tail-transport}\textup{(a)}, applied to the
heads and these $L^r$ remainders, gives $r$th cost at most
\[
 \ll_r(1+\log\log y)^rL^{-M}
       +\left(\frac{L}{y\log y}\right)^r+y^{-r}
       +\frac{L^{r+1}}{n},\qquad n=\pi(x;Q,b).
\]
The final term allows the possible prime $2$ and the matrix correction:
uniformly over all dimension arrays,
\[
 0\le-\log c_{n_x(p)}(p)-f(p-1)\le-\log P(p)\le4/p.
\]
Since $\log y\asymp L/\log L$ and
$n\gg x/(\varphi(Q)L)$, choosing $M>rB+2$ proves the logarithmic
bound. The map $u\mapsto e^{-u}$ is $1$-Lipschitz on $[0,\infty)$,
giving the raw bound. The polylogarithmic case follows by taking
$H=L^A$, $\delta=1/4$ and $E>A$ in the almost-all estimate, so that
eventually no exceptional modulus remains.
\end{proof}

\begin{corollary}[Uniform empirical quantiles on compact level intervals]
\label{cor:compact-empirical-quantiles-mas12}
Fix a reduced progression $(Q,b)$, $B>0$ and $0<\eta<1/2$.
For $U(p)=\varphi(p-1)/(p-1)$ or
$U(p)=c_{n_x(p)}(p)$, with arbitrary
$n_x(p)\in\Z_{\ge1}\cup\{\infty\}$, let
$\widehat q_{x,Q,b}(u)$ and $q_{Q,b}(u)$ be the generalized inverse
CDFs of $\operatorname{Law}_{x,Q,b}(U)$ and $\nu_{Q,b}$, respectively:
$q(u)=\inf\{t:F(t)\ge u\}$. Then, uniformly over these arrays,
\begin{equation}\label{eq:compact-empirical-quantiles-mas12}
 \sup_{\eta\le u\le1-\eta}
 |\widehat q_{x,Q,b}(u)-q_{Q,b}(u)|
 \ll_{Q,b,\eta,B}(\log x)^{-B}.
\end{equation}
The same statement holds for the empirical laws of $-\log U(p)$ and
the corresponding limiting law $\nu_{Q,b}^{\log}$.
\end{corollary}

\begin{proof}
Write $\nu$ for either limiting law, $F$ for its CDF, and $q$ for its
quantile function; let $\widehat\nu_x$ and $\widehat F_x$ denote the
corresponding empirical law and CDF. Both limiting
laws are continuous and strictly increasing on the interiors of their
supports. Choose a fixed compact interval $J$ inside that interior
whose interior contains $[q(\eta),q(1-\eta)]$.
The representation $X_{Q,b}=A_{Q,b}e^{-W_S}$ in
\eqref{eq:XQ}, with only finitely many odd primes deleted, and
\Cref{thm:cofinite-window-lower} give constants $c,C,h_0>0$ such that
\begin{equation}\label{eq:compact-interval-mass-mas12}
 F(v+h)-F(v)
 \ge c h\exp\{-C(\log\log(1/h))^2\}
\end{equation}
whenever $[v,v+h]\subset J$ and $0<h<h_0$.
For the logarithmic law this is the cofinite-window bound after its
fixed shift. For the raw law, $t\mapsto-\log t$ sends such intervals
to intervals of comparable length inside a fixed bounded interval;
changing $c,C,h_0$ therefore gives the same bound.

Let $u\in[\eta,1-\eta]$, put $q=q(u)$, and take $h>0$ small enough
that $[q-h,q+h]\subset J$ uniformly in $u$. Continuity gives $F(q)=u$.
If $\widehat q_x(u)\ge q+h$, then $\widehat F_x(t)<u$ for
$t<q+h$, while for $q+h/2\le t<q+h$,
\[
 F(t)-\widehat F_x(t)\ge F(q+h/2)-F(q)
 \ge c' h\exp\{-C'(\log\log(1/h))^2\}.
\]
If $\widehat q_x(u)\le q-h$, then $\widehat F_x(t)\ge u$ for
$q-h\le t\le q-h/2$, and the same lower bound holds for
$\widehat F_x(t)-F(t)$ by using the interval $[q-h/2,q]$.
Thus either discrepancy of magnitude at least $h$ forces, by the
one-dimensional CDF formula for Wasserstein distance,
\begin{equation}\label{eq:quantile-transport-obstruction-mas12}
 \mathcal W_1(\widehat\nu_x,\nu)
 =\int_{\mathbb R}|\widehat F_x(t)-F(t)|\,dt
 \ge c''h^2\exp\{-C'(\log\log(1/h))^2\}.
\end{equation}
Take $h=(\log x)^{-B}$. The last lower bound is
$(\log x)^{-2B-o(1)}$, whereas \Cref{prop:prime-wasserstein}, with
saving $2B+1$, bounds this distance by $O_{Q,B}((\log x)^{-2B-1})$
uniformly over all arrays. For sufficiently large $x$ the two are
incompatible. No $u$ in the indicated compact level interval can
therefore have such a discrepancy, proving the corollary.
\end{proof}

Extending the quantile conclusion uniformly to growing $Q$ requires
uniform constants and thresholds in \eqref{eq:compact-interval-mass-mas12};
the cofinite-window theorem presently permits dependence on the
deleted primes, so the wider Wasserstein ranges alone do not give
that extension.

Let $U(p)=X(p)$ or $c_{n_x(p)}(p)$, where
$n_x(p)\in\Z_{\ge1}\cup\{\infty\}$. For $1\le r<\infty$,
put $L=\log x$ and set
\[
 C_r=\frac1{\sqrt2}\left(4\kappa_{\mathrm{end}}
              \int_0^1\frac{\min(u,1-u)^r}{u}\,du\right)^{1/r}.
\]

\Needspace{12\baselineskip}
\begin{proposition}[Empirical Wasserstein lower bounds]
\label{prop:wasserstein-bounds}
\index{Wasserstein distance!upper and lower bounds}
Fix $0<\delta<1/2$ and $c_{\mathrm{pop}}>0$.
Suppose $Q\le x^{1/2-\delta}$, $(b,Q)=1$, and
$\pi(x;Q,b)\ge c_{\mathrm{pop}}x/(\varphi(Q)L)$.
With $A=A_{Q,b}$, $D=D(Q)$ and the model laws of
\Cref{prop:prime-wasserstein}, uniformly over these data and the dimension arrays,
\begin{equation}\label{eq:wasserstein-lower}
 \begin{aligned}
 \mathcal W_r(\operatorname{Law}_{x,Q,b}(U),\nu_{Q,b})
 &\ge(4C_r+o(1))A D^{1/r}x^{-1/2}L^{-2/r},\\
 \mathcal W_r(\operatorname{Law}_{x,Q,b}(-\log U),\nu_{Q,b}^{\log})
 &\ge(4C_r+o(1))D^{1/r}x^{-1/2}L^{-2/r}.
 \end{aligned}
\end{equation}
The errors depend only on $\delta,r,c_{\mathrm{pop}}$.
\end{proposition}

\begin{proof}
Fix $0<a_0<b_0<2\sqrt2$, and write
\[
 I_x=(1-b_0/\sqrt x,1-a_0/\sqrt x),\qquad
 \widetilde\mu_x=\operatorname{Law}_{x,Q,b}(U/A),\quad
 \widetilde\nu=\operatorname{Law}(X_{Q,b}/A).
\]
For every odd progression prime, $X(p)\le A$ and
$p-1=dm$, where $d\ge2$ is supported on primes dividing $2Q$
and $(m,2Q)=1$. If $m$ has two distinct prime divisors $q,r$, then $qr\le m\le(x-1)/2$
and
\[
 1-U(p)/A\ge1-X(p)/A
 \ge\frac1q+\frac1r-\frac1{qr}
 \ge\frac2{\sqrt{qr}}-\frac1{qr}
 \ge2\sqrt{2/x}-2/x.
\]
The last comparison uses that $2v^{-1/2}-v^{-1}$ decreases for $v>1$.
Such values cannot lie in $I_x$ for sufficiently large $x$.
The remaining free cofactors are $m=1$ and prime powers.
For $m=q$ prime, \eqref{eq:matrix-product-bounds} gives
$1-U(p)/A\le1/q+1/(p-1)\le3/(2q)$, so a candidate has
$q\le3\sqrt x/(2a_0)$.
For each allowed $d$, the number of higher powers $q^j\le x/d$, $j\ge2$,
is at most $O(\sqrt x\log x)$: sum the bound $x^{1/j}\le\sqrt x$
over $2\le j\le\log_2x$.
There is in addition the possible prime $2$.
Let $\mathcal D$ be the positive integers at most $x$ supported on
primes dividing $2Q$. Apply \eqref{eq:supported-rankin-mas07} with
$z=x$ and $\epsilon=\delta/2\in(0,1)$. Since $Q\le\sqrt x\le x$,
this gives $\#\mathcal D\le x^{\delta/2+o(1)}$ uniformly.
Multiplying by the $O(\sqrt x\log x)$ choices per $d$ gives at most
$x^{1/2+\delta/2+o(1)}$ candidates. The population hypothesis
therefore yields
$\widetilde\mu_x(I_x)\le x^{-\delta/2+o(1)}=o(L^{-2})$ uniformly.
Every coupling of $V\sim\widetilde\nu$ and $W\sim\widetilde\mu_x$ satisfies
\[
 \mathbb E|V-W|^r\ge
 \int_{I_x}\operatorname{dist}(t,I_x^c)^r\,d\widetilde\nu(t)
       -|I_x|^r\widetilde\mu_x(I_x).
\]
All deleted primes are at most $Q=o(\sqrt x)$, so for
$a_0\le s\le b_0$ the undeleted endpoint event forces them inactive.
Consequently \Cref{prop:endpoint-half} gives, uniformly in the progression data,
\[
 \begin{aligned}
 D^{-1}\widetilde\nu((1-s/\sqrt x,1])
 &=H_{\mathrm{end}}(s/\sqrt x)\\
 &=\frac{2\kappa_{\mathrm{end}}}{L}
   +\frac{4\kappa_{\mathrm{end}}\log s}{L^2}+O_{a_0}(L^{-3}).
 \end{aligned}
\]
Taking differences and integrating the continuous function
$\min(s-a_0,b_0-s)^r$ gives the uniform lower limit
\[
 \liminf_{x\to\infty}\frac{x^{r/2}L^2}{D}
       \mathcal W_r(\widetilde\mu_x,\widetilde\nu)^r
 \ge4\kappa_{\mathrm{end}}
       \int_{a_0}^{b_0}\frac{\min(s-a_0,b_0-s)^r}{s}\,ds;
\]
$D\ge1$ makes the subtracted empirical term negligible.
Let $a_0\downarrow0$ and then $b_0\uparrow2\sqrt2$ after this
uniform limit, and substitute $s=b_0u$. The resulting constant is
$4C_r$, with $C_r$ as defined above. Scaling by $A$ proves the raw bound.
For the logarithmic laws, use $-\log I_x$ and
$-\log(1-s/\sqrt x)=s/\sqrt x+O(x^{-1})$ uniformly on
$[a_0,b_0]$. The common shift $-\log A$ cancels and boundary distances
are multiplied by $1+o(1)$, giving the second bound.
\end{proof}

For $Q=1$, the raw and logarithmic constants are $2C_r$ and $4C_r$;
in particular $C_1=2\sqrt2\,\kappa_{\mathrm{end}}\log2$.
The population estimates \eqref{eq:ap-polylog-population} and
\eqref{eq:ap-grh-population} give these lower bounds in the corresponding
unconditional and GRH modulus ranges, with their exceptional sets.

\Needspace{10\baselineskip}
\begin{corollary}[Global Wasserstein bound under GRH]
\label{cor:wasserstein-grh-global}
Let $U(p)$ be either of the statistics above and put $L=\log x$.
Under Dirichlet GRH, for every fixed $1\le r<\infty$ and
$0<\epsilon<1/(2\sqrt r)$, uniformly over the dimension arrays,
\begin{equation}\label{eq:wasserstein-grh}
\begin{split}
 &\mathcal W_r(\operatorname{Law}_{x,1,1}(U),\mu_G)
 +\mathcal W_r(\operatorname{Law}_{x,1,1}(-\log U),\mu_f)\\
 &\hspace{12mm}\ll_{r,\epsilon}
 \exp\{-(1/(2\sqrt r)-\epsilon)\sqrt{L\log L}\}.
\end{split}
\end{equation}
\end{corollary}

\begin{proof}
Set $T=\sqrt{L\log L}$,
$a=1/(2\sqrt r)-\epsilon/2$, $b_1=2ra$,
$\log y=aT$, and $k=\lfloor b_1\sqrt{L/\log L}\rfloor$.
By \Cref{thm:bv-grh}, each inclusion probability of at most $k+1$
odd-prime coordinates has error $O(x^{-1/2}L^2)$, since its modulus
is at most $y^{k+1}=x^{ab_1+o(1)}$ with $ab_1<1/2$.
Apply \Cref{lem:sparse-bonferroni}, retaining these errors before
summing. For $\lambda=\sum_{3\le\ell\le y}(\ell-1)^{-1}
=\log\log y+O(1)$, the head total variation $d_y$ satisfies
\[
 d_y\ll(k+2)(2y)^{k+1}x^{-1/2}L^2+
                 \frac{(2\lambda)^{k+1}}{(k+1)!}.
\]
The logarithms of these two terms are at most
$-(1/2-ab_1)L+O(T)$ and $-(b_1/2+o(1))T$, respectively.
The logarithmic head diameter is $O(1+\log\log y)$, and the
empirical omitted logarithm is at most $2L/(y\log y)$.
For the model omitted sum $R_y=\sum_{\ell>y}B_\ell\omega_\ell$,
$y\omega_\ell\le2$ gives
\[
 \log\mathbb E e^{yR_y}
 \le\sum_{\ell>y}\frac{e^{y\omega_\ell}-1}{\ell-1}
 \ll y\sum_{\ell>y}\frac{\omega_\ell}{\ell-1}\ll1.
\]
Hence $\mathbb E R_y^r\ll_r y^{-r}$. Couple the heads maximally, then sample the empirical tail conditionally
on its head and the model tail independently. The $r$th transport cost
is therefore at most a constant depending on $r$ times
\[
 (1+\log\log y)^r d_y+
 \left(\frac{L}{y\log y}\right)^r+y^{-r}.
\]
Taking the $r$th root gives exponent
$\min(a,b_1/(2r))=1/(2\sqrt r)-\epsilon/2$; the margin absorbs the
subleading factors. Restoring $p=2$ and the determinant product adds
$r$th cost $O_r(L^{r+2}/x)$, which is negligible.
The map $u\mapsto e^{-u}$ is $1$-Lipschitz on the nonnegative
logarithmic support, proving the raw bound.
\end{proof}

\section{Uniformity in the progression modulus}

As $Q$ grows, two features of the comparison change. Fixing its prime
coordinates changes the concentration of the model, with loss measured
by $D(Q)$. Counting joint inclusions of the remaining small primes
requires arithmetic estimates at moduli containing both $Q$ and those
primes. The available distribution input therefore determines the
uniform range in $Q$, in addition to the concentration loss.

\begin{proposition}[Kolmogorov bounds for growing progressions]
\label{prop:ap-kolmogorov-pow2}
\index{Arithmetic progressions!varying moduli}
Let $U(p)$ be $X(p)=\varphi(p-1)/(p-1)$ or
$c_{n_x(p)}(p)$, where $n_x(p)\in\Z_{\ge1}\cup\{\infty\}$.
\begin{enumerate}[label=\textup{(\alph*)},leftmargin=*]
\item Fix $0<\delta<1/2$ and $A>0$. Uniformly for
$1\le H\le x^{1/2-\delta}$, all but
$O_{\delta,A}(H(\log x)^{-A})$ integers $Q\le H$ satisfy
\begin{equation}\label{eq:ap-average-cdf}
 \sup_{(b,Q)=1}\|F^U_{x,Q,b}-G_{Q,b}\|_\infty
 \ll_\delta D(Q)\frac{\log\log x}{\log x}.
\end{equation}
The exceptional set is independent of $b$ and of the dimension array.
\item For every fixed $C>0$, uniformly for $Q\le(\log x)^C$ and $(b,Q)=1$,
\begin{equation}\label{eq:ap-kolmogorov-general}
 \|F^U_{x,Q,b}-G_{Q,b}\|_\infty
 \ll_C D(Q)\frac{\log\log x}{\log x}.
\end{equation}
Consequently the supremum over these $Q,b$ is
$O_C((\log\log x)(\log\log\log x)/\log x)$.
\item Elliott's fixed-base power-modulus estimate~\cite{Elliott07}
gives, for every fixed integer $B\ge2$,
\begin{equation}\label{eq:ap-kolmogorov-power-range}
 \sup_{\substack{Q=B^a\le x^{1/3}\exp\{-(\log\log x)^3\},\ a\in\Z_{\ge0}\\
 b\bmod Q,\ (b,Q)=1}}
 \|F^U_{x,Q,b}-G_{Q,b}\|_\infty
 \ll_B\frac{\log\log x}{\log x}.
\end{equation}
\end{enumerate}
All bounds are uniform over the dimension arrays. The constants and
thresholds may be ineffective.
\end{proposition}

\begin{proof}
We first prove \textup{(a)}; $\delta$ and $A$ are its parameters.
Apply \Cref{lem:ap-finite-head} with $M=10$ and the given $\delta$.
By \Cref{cor:ap-finite-head-ranges}\textup{(i)}, all but
$O_{\delta,A}(H(\log x)^{-A})$ integers $Q\le H$ satisfy
$R_Q\le(\log x)^{-10}$. Fix such a
modulus; then \eqref{eq:ap-polylog-population} and
\eqref{eq:ap-truncated-tv} hold with $M=10$, uniformly in reduced $b$.

With $a_{Q,b}$, $W_y(p)$ and $W_y$ as in that lemma, put
$W_{Q,b}=-\log X_{Q,b}$.
All factors forced by $Q$, including those above $y$, are retained
in $a_{Q,b}$. For odd progression primes the omitted sum is
\[
 L_y(p):=f(p-1)-W_y(p)
       =\sum_{\substack{\ell>y,\ \ell\nmid Q\\\ell\mid p-1}}f(\ell)\ge0.
\]
Set $L_y(2)=0$. The deterministic estimate
\eqref{eq:ap-empirical-tail} applies to this head, while
\begin{equation}\label{eq:ap-model-tail}
 \mathbb E(W_{Q,b}-W_y)
 =\sum_{\substack{\ell>y\\\ell\nmid Q}}\frac{f(\ell)}{\ell-1}
 \ll y^{-1}.
\end{equation}
By \eqref{eq:ap-polylog-population}, primes at most $\sqrt x$ have
relative mass $O_\delta(x^{-\delta}\log x)$. On the complement,
$0\le-\log U(p)-f(p-1)\ll x^{-1/2}$ for every allowed $U$.
At $h=y^{-1/2}$, the empirical remainder is at most
$2\log x/(y\log y)+O(x^{-1/2})=o(h)$, uniformly over the arrays.
The deterministic alternative of \Cref{prop:head-tail-cdf} and
Markov's inequality for \eqref{eq:ap-model-tail} therefore give
\[
\begin{split}
 d_K(\operatorname{Law}_{x,Q,b}(-\log U(p)),\operatorname{Law}(W_{Q,b}))
 \ll_\delta{}& (\log x)^{-10}+y^{-1/2}+x^{-\delta}\log x\\
 &+Q_{\operatorname{Law}(W_{Q,b})}(3h)
 \ll_\delta D(Q)\frac{\log\log x}{\log x},
\end{split}
\]
by \Cref{lem:ap-uniform-concentration}. Exponentiation preserves
Kolmogorov distance from the continuous comparison law, proving
\eqref{eq:ap-average-cdf}. For $H=(\log x)^C$, take $\delta=1/4$
and $A>C$; the exceptional count is eventually less than one, which
proves \eqref{eq:ap-kolmogorov-general}.
Finally \eqref{eq:ap-concentration-factor} gives
the asserted supremum over polylogarithmic moduli.

For \eqref{eq:ap-kolmogorov-power-range}, fix $B\ge2$, put $L=\log x$,
and apply \Cref{cor:ap-finite-head-ranges}\textup{(iii)} with $M=10$.
Thus $y=\exp\{\log x/(12(r+1))\}$, and \eqref{eq:ap-polylog-population} and
\eqref{eq:ap-truncated-tv} hold, for all sufficiently large $x$, for
every $Q=B^a\le x^{1/3}\exp\{-(\log\log x)^3\}$, $a\in\Z_{\ge0}$, and
every reduced $b$.
The same deterministic and model-tail estimates apply. Since $Q\le x^{1/3}$,
the small-prime exceptional mass is $O_B(x^{-1/6}L)$ by
\eqref{eq:ap-polylog-population}. Bracketing at $h=y^{-1/2}$ and using
$D(B^a)=D(B)$ for $a\ge1$ gives
\[
 \|F^U_{x,Q,b}-G_{Q,b}\|_\infty
 \ll_B L^{-10}+y^{-1/2}+x^{-1/6}L
       +\frac{D(B)}{\log y}
 \ll_B\frac{\log\log x}{\log x}.
\]
The dimension-array correction is the same as above, completing the
proof of \eqref{eq:ap-kolmogorov-power-range}.
\end{proof}

For $Q=2^a$ one has $D(Q)=1$, and $G_{Q,b}=G$ by \Cref{cor:ap-moments}.
Hence the case $B=2$ of \eqref{eq:ap-kolmogorov-power-range} compares
$F^U_{x,2^a,b}$ with $G$ itself, with error $O(\log\log x/\log x)$,
uniformly for $2^a\le x^{1/3}\exp\{-(\log\log x)^3\}$ and reduced
$b\pmod{2^a}$. 

\begin{corollary}[Global and local sampling in residue-one progressions]
\label{cor:ap-local-windows}
Fix $A,C>0$ and put
\[
 F^U_{(x,x+h],Q,1}(t):=
 \frac{\#\{x<p\le x+h:p\equiv1\pmod Q,\ U(p)\le t\}}
      {\pi(x+h;Q,1)-\pi(x;Q,1)}.
\]
Uniformly for $x/(\log x)^A\le h\le x$, $Q\le(\log x)^C$, and
every $U$ in \Cref{prop:ap-kolmogorov-pow2},
\begin{equation}\label{eq:ap-local-cdf}
 \|F^U_{(x,x+h],Q,1}-G_{Q,1}\|_\infty
 \ll_{A,C}D(Q)\frac{\log\log x}{\log x\log\log\log x}.
\end{equation}
The same interval populations satisfy the uniform lower bound
\begin{equation}\label{eq:ap-local-endpoint-mas07}
 \|F^U_{(x,x+h],Q,1}-G_{Q,1}\|_\infty
 \ge D(Q)H_{\mathrm{end}}(1/(x+h))-x^{-1/2}.
\end{equation}
For the same moduli and $U$,
\begin{equation}\label{eq:ap-residue-one-polylog}
 \|F^U_{x,Q,1}-G_{Q,1}\|_\infty
 \ll_C D(Q)\frac{\log\log x}{\log x\log\log\log x}.
\end{equation}
In particular, if $Q$ is a power of two with
$Q\le(\log x)^C$, the comparison law is $G$ and $D(Q)=1$.
The constants and threshold may be ineffective.
\end{corollary}

\begin{proof}
Write $L_1=\log x$, $L_2=\log L_1$, $L_3=\log L_2$, and choose
$u=c_{A,C}L_2/L_3$, $y=x^{1/u}$, with $c_{A,C}$ sufficiently large.
Eventually $y>Q$. Apply \eqref{eq:prime-model-base} with
$B=A+C+4$ at both $x$ and $x+h$, with this common cutoff $y$.
If $N_t=\pi(t)-1$ and $\mu_{t,y}$ is the truncated valuation law
on the odd primes at most $t$, the window law is exactly
\[
 \frac{N_{x+h}\mu_{x+h,y}-N_x\mu_{x,y}}{N_{x+h}-N_x}.
\]
Siegel--Walfisz (\Cref{thm:sw}) at modulus $1$ gives
$N_{x+h}-N_x\sim h/\log x$ uniformly in the stated range.
Subtracting the model measure therefore multiplies its total
variation error by at most $O(x/h)$. The same theorem at modulus
$Q$, applied at both endpoints, gives
\[
 \pi(x+h;Q,1)-\pi(x;Q,1)
 \sim\frac{h}{\varphi(Q)\log x}.
\]
The event
$\mathcal C_Q=\{v_\ell(p-1)\ge v_\ell(Q):\ell\mid Q\}$ is
measurable at cutoff $y$ and has model probability $1/\varphi(Q)$.
Its window probability is asymptotic to the same value, so
\Cref{lem:conditional-tv} costs only $O(\varphi(Q))$. The resulting head error is
\[
 \ll_{A,C}L_1^{A+C}
       \{e^{-u\log u/4}+L_1^{-B}\}\ll_{A,C}L_1^{-4}.
\]
The conditioned logarithmic model is that of $-\log X_{Q,1}$, truncated
at $y$. Every $p\le2x$ has at most $\log(2x)/\log y=O(u)$ prime
factors above $y$, so its omitted logarithmic weight is $O(u/y)$.
The model tail has mean $O(1/y)$, since all primes dividing $Q$
are below $y$. For $p>x$, the matrix correction is $O(1/x)$,
uniformly over the arrays. Bracketing at width $y^{-1/2}$ and
using Markov's inequality and \Cref{lem:ap-uniform-concentration}
now costs
\[
 O_{A,C}\left(L_1^{-4}+u y^{-1/2}+\frac{D(Q)}{\log y}\right)
 \ll_{A,C}D(Q)\frac{L_2}{L_1L_3}.
\]
Exponentiation proves \eqref{eq:ap-local-cdf}.

For \eqref{eq:ap-residue-one-polylog}, apply
\eqref{eq:prime-model-base} only at $x$, with
$u=c_C L_2/L_3$ and $B=C+4$. Siegel--Walfisz gives
$\mathbb P_x(\mathcal C_Q)\sim1/\varphi(Q)$, so
\Cref{lem:conditional-tv} bounds the head error by
\[
 O_C\left(L_1^C\{e^{-u\log u/4}+L_1^{-B}\}\right)
 =O_C(L_1^{-4}).
\]
The arithmetic and model tail bounds remain $O(u/y)$ and
$O(1/y)$. Primes $p\le\sqrt x$ have relative mass
$O_C(Q\log x/\sqrt x)$ in the progression. Outside them the
matrix correction is $O(x^{-1/2})=o(y^{-1/2})$. The preceding
bracketing argument, with the additional error
$O_C(Q\log x/\sqrt x)$, proves the assertion uniformly in $Q$
and the dimension arrays.

For \eqref{eq:ap-local-endpoint-mas07}, put
$t=A_{Q,1}(1-1/(x+h))$. Every odd prime dividing $Q$ is smaller
than $x+h$, so independence as in
\eqref{eq:open-endpoint-identity-mas07} gives the exact model
upper tail $D(Q)H_{\mathrm{end}}(1/(x+h))$. An observation above
$t$ must have $p-1$ supported on $2Q$. The interval population
estimate already proved and \eqref{eq:supported-rankin-mas07}
with $\epsilon=1/4$ bound its proportion by
\[
 \frac{N_Q(2x)}{\pi(x+h;Q,1)-\pi(x;Q,1)}
 \ll_{A,C}x^{-3/4+o(1)}(\log x)^{A+C+1}
 \le x^{-1/2}.
\]
Subtract the full open upper-tail probabilities. 
\end{proof}

Equation~\eqref{eq:ap-residue-one-polylog} improves
\eqref{eq:ap-kolmogorov-general} in the residue-one class by a
factor $\log\log\log x$; for fixed $Q$,
\Cref{cor:ap-sharp-convergence} is sharper still.

\begin{theorem}[Distribution and moment estimates in growing progressions under GRH]
\label{thm:ap-grh-law}
\index{Generalized Riemann hypothesis!progression laws}
Assume GRH for Dirichlet $L$-functions and fix $0<\delta<1/2$.
For every reduced class $b\pmod Q$
with $Q\le x^{1/2-\delta}$, put $n=\pi(x;Q,b)$. Uniformly in $Q,b$,
\begin{equation}\label{eq:ap-grh-population}
 n=\frac{\operatorname{Li}(x)}{\varphi(Q)}
   \bigl(1+O(x^{-\delta}(\log x)^2)\bigr)
 \gg_\delta\frac{x}{\varphi(Q)\log x}.
\end{equation}
Let $U(p)=\varphi(p-1)/(p-1)$ or $c_{n_x(p)}(p)$, where
$n_x(p)\in\Z_{\ge1}\cup\{\infty\}$ is an arbitrary dimension array
(with $c_\infty=c$).
The following bounds are uniform over these arrays.
\begin{equation}\label{eq:ap-grh-cdf}
 \|F^U_{x,Q,b}-G_{Q,b}\|_\infty
 \ll_\delta D(Q)\frac{\log\log x}{\log x}
 \ll_\delta\frac{(\log\log x)^2}{\log x}.
\end{equation}
For every compact $K\subset\C$ and fixed integer $j\ge0$ there is
$E_K>0$ such that
\begin{equation}\label{eq:ap-grh-mellin}
 \sup_{\substack{Q\le x^{1/2-\delta}\\(b,Q)=1}}\sup_{z\in K}\left|
 \frac1n\sum_{\substack{p\le x\\p\equiv b\pmod{Q}}}
 U(p)^z(\log U(p))^j-M_{Q,b}^{(j)}(z)\right|
 \ll_{\delta,K,j}x^{-\delta}(\log x)^{E_K}.
\end{equation}
\end{theorem}

\begin{proof}
The case $k=0$ of \Cref{thm:bv-grh}, with $m=Q$, gives
\eqref{eq:ap-grh-population}. For the distribution estimate choose
\[
 r=\lceil32\log\log x\rceil,\qquad
 y=\exp\{\delta\log x/(4(r+1))\},
\]
and use the free-prime set $\mathcal P_y(Q)$ and the subset laws
$\mu,\nu$ of \Cref{lem:ap-finite-head}, formed with these $r,y$
(its case $M=10$, $c_M=32$); the omitted sum $L_y(p)$ and the
bracketing used below are those of the proof of
\Cref{prop:ap-kolmogorov-pow2}.
Every inclusion event with at most $r+1$ coordinates has modulus at
most $Qy^{r+1}\le x^{1/2-3\delta/4}$. The same GRH estimate and
\eqref{eq:ap-grh-population} bound each normalized inclusion error
by $O_\delta(x^{-\delta}(\log x)^2)$. Here
$\lambda\le\log\log y+O(1)\le\log\log x+O(1)$, so
\eqref{eq:sparse-bonferroni-bound} gives
\[
 d_{\mathrm{TV}}(\mu,\nu)
 \ll_\delta x^{-\delta/2}+(\log x)^{-10}.
\]
Indeed, $(2y)^{r+1}=x^{\delta/4}(\log x)^{O(1)}$, and the factorial
remainder is estimated by \eqref{eq:ap-head-factorial}.
The forced coordinates, including those above $y$, remain in the
head. By \eqref{eq:ap-empirical-tail}, the empirical tail is at most
$2\log x/(y\log y)$. The population estimate
\eqref{eq:ap-grh-population} bounds the relative mass of primes up to
$\sqrt x$ by $O_\delta(x^{-\delta}\log x)$; outside this set the
dimension-array correction is $O(x^{-1/2})$.
At $h=y^{-1/2}$ the empirical remainder outside this set is $o(h)$.
Equation \eqref{eq:ap-model-tail} and Markov's inequality bound the model
remainder probability by $O(y^{-1/2})$.
Since $\log y\asymp_\delta\log x/\log\log x$, the same bracketing and
\Cref{lem:ap-uniform-concentration} prove \eqref{eq:ap-grh-cdf}.

For \eqref{eq:ap-grh-mellin}, first take the totient ratio and $j=0$.
Freeze the primes dividing $Q$. For $p>2$, the divisor expansion in
\Cref{thm:prime-mellin} becomes
\[
 U(p)^z=A_{Q,b}^z
 \sum_{\substack{d\mid p-1\\(d,2Q)=1}}a_z(d).
\]
Every contributing $d$ is squarefree and at most $x$. Its inclusion
class is reduced modulo $Qd\le x^{3/2}$, has density $1/\varphi(d)$
relative to the progression, and satisfies, by \Cref{thm:bv-grh},
\[
 \pi(x;Qd,r_d)=\frac{\operatorname{Li}(x)}{\varphi(Qd)}
                  +O(\sqrt x\log x).
\]
On a fixed compact neighborhood of $K$, for some $R_K>0$,
\[
 |A_{Q,b}^z|\le\left(\frac{2Q}{\varphi(2Q)}\right)^{R_K}
 \ll_K(\log\log(3Q))^{R_K}\ll_K\log x.
\]
With $\beta,C_K$ as in the same proof,
$\sum_{d\le x}\beta(d)\ll_K(\log x)^{C_K}$. Hence the normalized
summed errors, including the frozen factor, are
$O_{\delta,K}(x^{-\delta}(\log x)^{C_K+3})$.
Completing the main sum costs $O_K(x^{-3/4}\log x)$, since
$\sum_d\beta(d)d^{3/4}/\varphi(d)<\infty$, and gives
\[
 A_{Q,b}^z\sum_{(d,2Q)=1}\frac{a_z(d)}{\varphi(d)}=M_{Q,b}(z).
\]
The denominator correction is $O_{\delta,K}(x^{-\delta}(\log x)^3)$,
since $\sum_d|a_z(d)|/\varphi(d)\ll_K1$.
The possible prime $2$ contributes $O_K((1+|A_{Q,b}^z|)/n)$,
which is smaller. By \Cref{lem:mellin-perturbation} and
\eqref{eq:ap-grh-population}, replacing the totient ratio by any
$c_{n_x(p)}(p)$ costs
$O_{\delta,K}(\varphi(Q)\log x/\sqrt x)
=O_{\delta,K}(x^{-\delta}\log x)$, uniformly over the arrays.
Cauchy's formula on that neighborhood gives every fixed derivative,
after enlarging $E_K$ if necessary.
\end{proof}

\begin{corollary}[Transport in growing progressions under GRH]
\label{cor:ap-grh-wasserstein}
\index{Wasserstein distance!growing progressions}
Assume GRH for Dirichlet $L$-functions. Fix $1\le r<\infty$,
$0<\delta<1/2$ and $0<\varepsilon<\sqrt{\delta/(2r)}$, and put
$\gamma=\sqrt{\delta/(2r)}-\varepsilon$ and
$T=\sqrt{\log x\log\log x}$.
With the laws and dimension arrays of \Cref{prop:prime-wasserstein},
uniformly for $Q\le x^{1/2-\delta}$ and $(b,Q)=1$,
\begin{equation}\label{eq:ap-grh-wasserstein}
 \begin{split}
 &\mathcal W_r(\operatorname{Law}_{x,Q,b}(U),\nu_{Q,b})
 +\mathcal W_r(\operatorname{Law}_{x,Q,b}(-\log U),\nu_{Q,b}^{\log})\\
 &\hspace{15mm}\ll_{\delta,r,\varepsilon}e^{-\gamma T}.
 \end{split}
\end{equation}
\end{corollary}

\begin{proof}
Put $L=\log x$, $a=\sqrt{\delta/(2r)}-\varepsilon/2$,
$b_0=2ra$, $\log y=aT$ and
$k=\lfloor b_0\sqrt{L/\log L}\rfloor$; then $ab_0<\delta$.
Retain all forced coordinates in $-\log A_{Q,b}$ and apply
\Cref{lem:sparse-bonferroni} to the free odd primes at most $y$.
Every inclusion modulus is at most
$Qy^{k+1}\le x^{1/2-\delta+ab_0+o(1)}$.
By \Cref{thm:bv-grh} and \eqref{eq:ap-grh-population}, each normalized
inclusion error is $O_\delta(x^{-\delta}L^2)$. Thus, with
$\lambda\le\log\log y+O(1)$, the head total variation is at most
\[
 \ll_\delta (k+2)(2y)^{k+1}x^{-\delta}L^2
       +\frac{(2\lambda)^{k+1}}{(k+1)!}.
\]
The logarithms of these terms are at most
$-(\delta-ab_0)L+O(T)$ and $-(b_0/2+o(1))T$.
The common forced shift cancels, so the head diameter is
$O(1+\log\log y)$. The empirical omitted logarithm is at most
$2L/(y\log y)$. Deleting the primes dividing $Q$ only decreases the
positive model tail, so the proof of \Cref{cor:wasserstein-grh-global}
gives its $r$th moment $O_r(y^{-r})$, uniformly in $Q$.
The same maximal coupling therefore gives logarithmic $\mathcal W_r$
exponent $\min(a,b_0/(2r))=\sqrt{\delta/(2r)}-\varepsilon/2$.
By \eqref{eq:ap-grh-population}, the dimension-array correction and
restoration of the possible prime $2$ have $r$th cost
$O_{\delta,r}(\varphi(Q)L^{r+2}/x)$, negligible on the $T$ scale.
The map $u\mapsto e^{-u}$ is $1$-Lipschitz on $[0,\infty)$,
giving the raw estimate. The margin $\varepsilon$ absorbs all
subleading factors.
\end{proof}

\section{Progression regimes and error metrics}
\label{ssec:progression-regimes}

The following guide records bounds for the CDF error $\eta_x$ between
the indicated prime population and $G_{Q,b}$. Every row permits
$U_x(p)=X(p)$ or any dimension array $c_{n_x(p)}(p)$ from
\eqref{eq:prime-finite-product}. Write $L_1=\log x$,
$L_2=\log\log x$ and $L_3=\log\log\log x$; the parameters
$A,C>0$, $0<\delta<1/2$, $7/12<\theta\le1$ and the integer $B\ge2$
are fixed. For power-of-two $Q$, $G_{Q,b}=G$ and $D(Q)=1$.

\begingroup\small
\let\vss\vfil
\begin{longtable}{@{}>{\raggedright\arraybackslash}p{0.69\textwidth}>{\raggedright\arraybackslash}p{0.26\textwidth}@{}}
\toprule
Prime population and modulus range & CDF error $\eta_x$ \\
\midrule
\endhead
$p\le x$, fixed $Q$, any reduced $b$:
\Cref{cor:ap-sharp-convergence} & $O_Q(L_1^{-1})$, of sharp order \\
\addlinespace
$p\le x$, every $Q\le L_1^C$, any reduced $b$:
\Cref{prop:ap-kolmogorov-pow2}\textup{(b)} & $O_C\!\left(\frac{D(Q)L_2}{L_1}\right)$ \\
\addlinespace
$p\le x$, almost all $Q\le H\le x^{1/2-\delta}$, any reduced $b$:
\Cref{prop:ap-kolmogorov-pow2}\textup{(a)} & $O_\delta\!\left(\frac{D(Q)L_2}{L_1}\right)$ \\
\addlinespace
$p\le x$, every $Q=B^a\le x^{1/3}e^{-L_2^3}$, any reduced $b$:
\eqref{eq:ap-kolmogorov-power-range} & $O_B(L_2/L_1)$ \\
\addlinespace
$p\le x$, or $x<p\le x+h$ with $x/L_1^A\le h\le x$;
$Q\le L_1^C$, $b=1$:
\Cref{cor:ap-local-windows} & $O_{A,C}\!\left(\frac{D(Q)L_2}{L_1L_3}\right)$ \\
\addlinespace
Under GRH: $p\le x$, every $Q\le x^{1/2-\delta}$, any reduced $b$:
\Cref{thm:ap-grh-law} & $O_\delta\!\left(\frac{D(Q)L_2}{L_1}\right)$ \\
\addlinespace
$x<p\le x+x^\theta$, every $Q\le L_1^C$, any reduced $b$:
\Cref{cor:short-uniform-cdf-mas07} & $O_{C,\theta}\!\left(\frac{D(Q)L_2}{L_1}\right)$ \\
\bottomrule
\end{longtable}
\endgroup
In the almost-all row, for every fixed $A>0$ there are at most
$O_{\delta,A}(HL_1^{-A})$ exceptional integers $Q\le H$; this set
is independent of the reduced residue and the dimension array.
The sharp fixed-$Q$, polylogarithmic, fixed-base power and almost-all estimates
may have ineffective constants and thresholds. For the global rows, the population
estimates \eqref{eq:ap-polylog-population}, \eqref{eq:ap-grh-population}
and the exceptional sets in the cited results allow
\Cref{prop:ap-uniform-endpoint-mas07} to supply the lower bound
$D(Q)H_{\mathrm{end}}(1/x)-x^{-1/2}$. The two interval rows use their
own endpoint bounds \eqref{eq:ap-local-endpoint-mas07} and
\eqref{eq:short-uniform-cdf-mas07}. In every fixed reduced
progression, \Cref{prop:fixed-threshold} gives stronger estimates for
fixed Diophantine irrational thresholds and uniformly for rational
thresholds in its stated denominator range, also for the short-interval population.
The latter threshold assertion and the short-interval transport bounds
retain fixed $Q$.

\paragraph*{Inputs for each observable}
The finite-population normalization in
\Cref{lem:finite-prime-sample-head} supplies a head comparison only.
\Cref{prop:head-tail-cdf,prop:head-tail-transport,prop:head-tail-weighted-mellin}
separate its uses. 
\begingroup\small
\begin{longtable}{@{}>{\raggedright\arraybackslash}p{0.22\textwidth}>{\raggedright\arraybackslash}p{0.72\textwidth}@{}}
\toprule
Observable & Additional input and applications \\
\midrule\endhead
CDF upper bound & Both remainder means and model concentration;
\Cref{prop:ap-kolmogorov-pow2,thm:short-interval,thm:ap-grh-law}.
The sharp fixed-$Q$ order $1/\log x$ uses the localized
smooth-pattern recursion and large-smooth-window estimate in
\Cref{thm:addfn-sharp-order}. \\
\addlinespace
Order-$r$ transport & $L^r$ remainders and a bounded head
diameter, weighted variation, higher moment, or uniform-integrability
bound as specified in \Cref{prop:head-tail-transport}.
\Cref{prop:prime-wasserstein} uses the bounded free-head diameter and
an exponential estimate for the model remainder.
\Cref{cor:ap-grh-wasserstein} chooses a separate optimized cutoff. \\
\addlinespace
Mellin derivatives & Exponential or weighted control on a compact
complex neighborhood, supplied by the divisor expansions and denominator
estimates in \Cref{thm:prime-mellin,thm:ap-grh-law}. \\
\addlinespace
Fixed thresholds & The directional window $(t,t+h]$, separation from
surviving anchors, and downward perturbations in
\Cref{prop:fixed-threshold}. Rational anchors are included among the
thresholds: an anchor equal to $t$ is excluded from this window. \\
\addlinespace
Lower bounds & Arithmetic support exclusions, counts of integers
supported on $2Q$, and the actual population lower bound in
\Cref{prop:ap-uniform-endpoint-mas07,prop:wasserstein-bounds}.
The interval populations use their own denominators and endpoint
arguments. \\
\addlinespace
Certified search & Distribution transfer, deterministic certificate
enclosures, one-sided model windows and positive limiting acceptance;
sampling and correctness are established in
\Cref{cor:gencert-polylog,thm:gencert}. \\
\bottomrule
\end{longtable}
\endgroup
Uniformity over dimension arrays uses the deterministic estimate
$0\le-\log c_{n_x(p)}(p)-f(p-1)\le4/p$ and the relevant population
normalization. In almost-all modulus assertions the exceptional set
may depend on the required saving, compact set and derivative or
transport order. 

\paragraph*{Choosing the metric}
For real probability laws, let $d_{\mathrm K}$ be the supremum CDF
error and $d_{\mathrm L}$ the L\'evy distance. Directly from its defining
horizontal and vertical shifts, $d_{\mathrm L}\le d_{\mathrm K}$.
A Wasserstein bound alone instead gives, for $r\ge1$,
\begin{equation}\label{eq:levy-wasserstein-mas17}
 d_{\mathrm L}(\mu,\nu)\le
       \mathcal W_r(\mu,\nu)^{r/(r+1)}.
\end{equation}
Indeed, a coupling $(Y,Z)$ satisfies
$\mathbb P(|Y-Z|>\eta)\le\mathbb E|Y-Z|^r/\eta^r$; on its complement,
$\{Y\le t\}\subseteq\{Z\le t+\eta\}$ and the reverse inclusion with
$t-\eta$ holds. Take $\eta=\mathcal W_r^{r/(r+1)}$ using optimal, or
arbitrarily near optimal, couplings. Thus a fixed transport rate incurs
this exponent, whereas arbitrary logarithmic savings absorb it by
first requesting a larger saving.
CDF error controls events with one threshold. The modulus ranges,
exceptional sets and uniformity for Mellin and fixed-order transport
estimates are specified in \Cref{thm:prime-mellin,prop:prime-wasserstein};
the dimension-array extensions are
\Cref{cor:finite-dimension}\textup{(ii),(iii)}.
Divisor-sum and joint Mellin estimates have the polylogarithmic range.
Under GRH, \Cref{thm:ap-grh-law,cor:ap-grh-wasserstein} give Mellin
and stretched-exponential transport bounds for every reduced class
with $Q\le x^{1/2-\delta}$, for fixed $0<\delta<1/2$.

\Cref{prop:wasserstein-bounds} gives lower bounds in these same admissible
population regimes, with raw factor $A_{Q,b}D(Q)^{1/r}$ and logarithmic
factor $D(Q)^{1/r}$. Its lower-bound scale
$x^{-1/2}(\log x)^{-2/r}$ is smaller than the available unconditional
and GRH upper-bound scales, already at $Q=1$.

\paragraph*{From absolute error to a window count}
For a value window $(a_x,b_x]$ in any one of these prime populations,
put $m_x=G_{Q,b}(b_x)-G_{Q,b}(a_x)$. If its uniform CDF error is at
most $\eta_x$ and $m_x>0$, then
\[
 \left|\frac{\mathbb P\{a_x<U_x\le b_x\}}{m_x}-1\right|
 \le\frac{2\eta_x}{m_x}.
\]
This bound gives a relative asymptotic when $\eta_x=o(m_x)$ with the
actual model mass. This value-window width is distinct from the
prime-sampling interval length $h$ in the table. For fixed $Q$, the
endpoint mass in \eqref{eq:ap-endpoint-transfer} gives the range
$\log(1/\varepsilon)=o(\log x)$; the compact-interior assertion of
\Cref{cor:prime-shrinking-anchor}\textup{(ii)} instead fixes
$0<\delta<1/2$, restricts $\delta\le t\le1/2$ at $Q=1$, and uses
$\varepsilon=(\log x)^{-a}$ with $0<a<1$.
For moving moduli the endpoint window has an explicit model mass.
For real $y\ge3$ put
\[
 D_{\le y}(Q)=\prod_{\substack{2<\ell\le y\\ \ell\mid Q}}
                    \frac{\ell-1}{\ell-2},
\]
where the product runs over primes. Uniformly in $Q$ and $(b,Q)=1$,
as $y\to\infty$,
\begin{equation}\label{eq:moving-modulus-endpoint}
 \mathbb P\{A_{Q,b}(1-y^{-1})<X_{Q,b}\le A_{Q,b}\}
 =\frac{\kappa_{\mathrm{end}}D_{\le y}(Q)}{\log y}
       \{1+O(1/\log y)\}.
\end{equation}
Indeed, a free prime $\ell\le y$ must be inactive; its activation would
give $X_{Q,b}\le A_{Q,b}(1-1/y)$. Independence and Mertens' estimate
for $\prod_{2<\ell\le y}(1-1/(\ell-1))$ give their joint inactivity
probability as the right side of \eqref{eq:moving-modulus-endpoint}.
The independent remaining logarithmic sum has expectation
$\ll\sum_{\ell>y}\ell^{-2}\ll1/(y\log y)$, so Markov's inequality
at $-\log(1-1/y)$ costs a relative $O(1/\log y)$.

In every row of the preceding table with
$\eta_x\ll D(Q)L_2/L_1$ or a smaller error, the empirical proportion
in the same window is asymptotic to
$\kappa_{\mathrm{end}}D_{\le y}(Q)/\log y$ if
$Q\le x$, $y\to\infty$ and $\log y=o(L_1/L_2)$. Here the sample is normalized
by the number of primes in the indicated progression and sampling
interval, with the same hypotheses and exceptional moduli as in its
row. To see the uniformity, for $Q\le x$ split the primes dividing
$Q$ above $y$ at $L_1$. The product through $L_1$ is
$\ll\max(1,L_2/\log y)$ by Mertens' theorem. There are
$\le L_1/\log L_1$ prime divisors above $L_1$, so their reciprocal
sum is $O(1/L_2)$ and their product is bounded. Hence the relative
CDF error is $O((L_2/L_1)\max(\log y,L_2))=o(1)$.
The preceding fixed-$Q$ result also allows $\log y=o(L_1)$.

Endpoint consequences and the transition at power-sized widths are
considered in \Cref{ssec:prime-cofactor-detectors}. Certified-search
applications, including acceptance bounds, costs and effectivity, are
given in \Cref{sec:certified-prime-search}.
\Cref{fig:certified-search} shows how the distribution function estimates of
this chapter and the concentration bounds of \Cref{sec:global-modulus} enter
these applications.

\chapter{Joint laws of \texorpdfstring{$f$}{f} at affine prime arguments}\label{sec:difference-law}

Sampling $f$ at fixed affine forms of the same prime retains arithmetic
dependence between the coordinates. The residue of the sampled prime
modulo $\ell$ determines their whole vector of contributions. Independence
holds across primes in the limiting model. The construction below
separates a finite exceptional head from the remaining local factors.

The chapter first constructs the laws for fixed polynomial prime
arguments in \Cref{thm:polynomial-root-law}, with corner estimates for
admissible joint finite heads in \Cref{cor:polynomial-minimal-head-corners}.
For the affine models treated subsequently, the joint laws have
Hausdorff dimension zero and supports that are finite unions of
translated orthants. A fixed linear projection with $r\ge1$ nonzero
coefficients has concentration of order $(\log(1/h))^{-r}$ as
$h\downarrow0$. Differences at two distinct nonzero shifts have
support $\R$.

\section{Polynomial root models}
\label{sec:polynomial-root-models}
Aivazidis--Sofos
\cite[Theorem~6.3 and the proof of Theorem~1.3(i)]{AS15} state and apply
the polynomial-prime distribution theorem attributed there to
Toleuov--Fa\u{\i}nle\u{\i}b. We construct the tuple law directly,
including all exceptional local factors.

\begin{definition}[Fixed polynomial root model]
\label{def:polynomial-root-model}
\index{Polynomial root models}
Fix $d\ge1$ and nonzero polynomials $F_1,\ldots,F_d\in\Z[T]$.
Write, by Gauss's lemma,
\[
 F_j(T)=c_jT^{a_j}\prod_{h\in\mathcal H}h(T)^{e_{j,h}},
 \qquad c_j\in\Z\setminus\{0\},\quad a_j,e_{j,h}\in\Z_{\ge0},
\]
where $\mathcal H$ is the set of distinct primitive irreducible factors
other than $T$, each chosen with positive leading coefficient and each
occurring in at least one coordinate. Put
$\mathcal H_j=\{h:e_{j,h}>0\}$, $r_j=|\mathcal H_j|$,
$D=\sum_{h\in\mathcal H}\deg h$, and
$v_h=(\mathbf1_{\{h\in\mathcal H_j\}})_{j=1}^d$.
Let $E$ be the finite set consisting of the primes at most $D+1$ and the prime divisors of the
following nonzero integer:
\begin{equation}\label{eq:polynomial-root-exceptions}
 \left(\prod_j c_j\right)
 \left(\prod_{h\in\mathcal H}\operatorname{lc}(h)h(0)
                    \operatorname{disc}(h)\right)
 \left(\prod_{h<h'}\operatorname{Res}(h,h')\right).
\end{equation}
Here the last product uses any ordering of $\mathcal H$, the
discriminant of a linear polynomial is $1$, and empty products are $1$.
Independently for each prime $\ell$, let $U_\ell$ be uniform on
$\mathbb F_\ell^\times$, and set
\begin{equation}\label{eq:polynomial-root-law-model}
 \begin{gathered}
 B_{\ell,j}=\mathbf1_{\{F_j(U_\ell)=0\}},\qquad
 Z_j=\sum_\ell\omega_\ell B_{\ell,j},\\
 X_j=\prod_\ell(1-1/\ell)^{B_{\ell,j}}=e^{-Z_j},
 \qquad \omega_\ell=\log\frac\ell{\ell-1}.
 \end{gathered}
\end{equation}
The congruence is evaluated using the original $F_j$, including its
content. For a fixed cutoff $b\ge\max(E\cup\{2\})$, write
$\boldsymbol H_b=\sum_{\ell\le b}\omega_\ell\boldsymbol B_\ell$.
This is a joint finite head, with the same $U_\ell$ used in every
coordinate. Finally put
$\rho_h(\ell)=\#\{u\in\mathbb F_\ell^\times:h(u)=0\}$.
\end{definition}

\begin{lemma}[Fixed-polynomial root intensity]
\label{lem:polynomial-root-intensity}
\index{Polynomial root intensity}
For each $h\in\mathcal H$ and fixed $b$ as in
\Cref{def:polynomial-root-model}, there are real constants
$C_{h,b},C'_{h,b}$ and $c_h>0$ such that
\begin{equation}\label{eq:polynomial-root-intensity}
 \begin{split}
 \sum_{b<\ell\le y}\frac{\rho_h(\ell)}\ell
   &=\log\log y+C_{h,b}
                  +O_{h,b}(e^{-c_h\sqrt{\log y}}),\\
 \sum_{b<\ell\le y}\frac{\rho_h(\ell)}{\ell-1}
   &=\log\log y+C'_{h,b}
                  +O_{h,b}(e^{-c_h\sqrt{\log y}}).
 \end{split}
\end{equation}
In particular both remainders are $o(1/\log y)$. For every
$\mathcal A\subseteq\mathcal H$, put
$q_{\ell,\mathcal A}=\sum_{h\in\mathcal A}\rho_h(\ell)/(\ell-1)$
for $\ell>b$. Then $0\le q_{\ell,\mathcal A}<1$ and
\begin{equation}\label{eq:polynomial-root-hazard}
 \sum_{b<\ell\le y}-\log(1-q_{\ell,\mathcal A})
   =|\mathcal A|\log\log y+C_{\mathcal A,b}
      +o(1/\log y).
\end{equation}
The same conclusion holds with the summand replaced by
$q_{\ell,\mathcal A}$, with its own constant. When $\mathcal A$ is
empty these sums and constants are zero. 
\end{lemma}

\begin{proof}[Proof of \Cref{lem:polynomial-root-intensity}]
We first specify the number-field input, including its error term.
For a fixed number field $K$ of degree $n>1$ and absolute discriminant
$\Delta_K$, let $\pi_K(x)$ count its prime ideals of norm at most $x$.
Apply Lagarias--Odlyzko \cite[Theorem~1.3, (1.6)--(1.7)]{LO77}
to the trivial extension $K/K$. For
$x\ge\exp(10n(\log\Delta_K)^2)$ it gives
\[
 |\pi_K(x)-\operatorname{Li}(x)|
 \le \operatorname{Li}(x^{\beta_K})
          +c_3x\exp(-c_4\sqrt{\log x/n}),
\]
where $c_3,c_4>0$ are absolute, and the first term occurs only if the
fixed Dedekind zeta function has the exceptional real zero
$1-(4\log\Delta_K)^{-1}\le\beta_K<1$.
For fixed $K$ this term is absorbed, for sufficiently large $x$, into
$x\exp(-c_K\sqrt{\log x})$: indeed
$(1-\beta_K)\log x-c_K\sqrt{\log x}\to\infty$.
Increasing the constant covers the bounded initial interval. For
$K=\Q$ use the prime number theorem in the proof of
\Cref{lem:prime-log-stieltjes}. Thus in every case
\begin{equation}\label{eq:polynomial-fixed-field-pit}
 \pi_K(x)=\operatorname{Li}(x)
              +O_K(xe^{-c_K\sqrt{\log x}})\qquad(x\ge2).
\end{equation}
The implied constant in this unconditional estimate may depend
ineffectively on $K$.

Write $h(T)=aT^n+\cdots$ and choose a root $\alpha$, with
$K=\Q(\alpha)$. The algebraic integer $\beta=a\alpha$ has monic
minimal polynomial $g(V)=a^{n-1}h(V/a)\in\Z[V]$.
Its discriminant is
$a^{(n-1)(n-2)}\operatorname{disc}(h)$. For primes outside the
leading-coefficient and discriminant exclusions,
$\ell\nmid[\mathcal O_K:\Z[\beta]]$, since the square of that index
divides $\operatorname{disc}(g)$. Reduction therefore identifies
$\mathcal O_K/\ell\mathcal O_K$ with $\mathbb F_\ell[V]/(g)$.
The substitution $V=aT$ is invertible modulo $\ell$, so linear
factors of $h$ correspond exactly to prime ideals of $K$ of norm
$\ell$. The exclusion of $h(0)$ makes every such root nonzero.
Each rational prime has at most $n$ prime ideals above it; any prime
ideal with residue degree at least two and norm at most $x$ lies over
a rational prime at most $\sqrt x$. Consequently
\[
 A_h(x):=\sum_{\ell\le x}\rho_h(\ell)
    =\pi_K(x)+O_h(\sqrt x+1)
    =\operatorname{Li}(x)+O_h(xe^{-c_h\sqrt{\log x}}).
\]
The finitely many excluded primes change the first equality by a
bounded constant. This also treats degree one.

Set $\Delta_h=A_h-\operatorname{Li}$. Stieltjes partial summation gives
\[
 \sum_{\ell\le y}\frac{\rho_h(\ell)}\ell
 =\log\log y-\log\log2
       +\frac{\Delta_h(y)}y+\int_2^y\frac{\Delta_h(t)}{t^2}\,dt.
\]
The integral converges absolutely at infinity. After absorbing its
limit into a constant, the error is
\[
 \frac{\Delta_h(y)}y-\int_y^\infty\frac{\Delta_h(t)}{t^2}\,dt
 \ll_h (1+\sqrt{\log y})e^{-c_h\sqrt{\log y}}
 \ll_h e^{-c'_h\sqrt{\log y}},
\]
where the integral bound follows by $u=\sqrt{\log t}$.
Deleting the primes through $b$ changes only the constant.
The series of differences
$\rho_h(\ell)((\ell-1)^{-1}-\ell^{-1})$ converges absolutely with
remaining tail $O_h(y^{-1})$, proving the second estimate.
For $\ell>b$ the reductions of distinct $h$ have disjoint roots by
the resultant exclusion, each has at most $\deg h$ roots, and
$\ell-1>D$. Thus $q_{\ell,\mathcal A}<1$ and
$-\log(1-q_{\ell,\mathcal A})-q_{\ell,\mathcal A}=O_{\boldsymbol F}(\ell^{-2})$.
Its remaining tail is $O_{\boldsymbol F}(y^{-1})$. Summing the
individual estimates proves the hazard assertion, including its
strong remainder and the empty case.
\end{proof}

\begin{theorem}[Totient laws at fixed polynomial prime arguments]
\label{thm:polynomial-root-law}
Use \Cref{def:polynomial-root-model}, and let
$\mathcal P_{\boldsymbol F}(x)$ be the primes $p\le x$ for which
$\prod_jF_j(p)\ne0$. Put $N_{\boldsymbol F}(x)=|\mathcal P_{\boldsymbol F}(x)|$.
The series in \eqref{eq:polynomial-root-law-model} converge absolutely
almost surely; in particular $0<X_j\le1$ almost surely. As $x\to\infty$,
\begin{equation}\label{eq:polynomial-root-empirical-law}
 \frac1{N_{\boldsymbol F}(x)}\sum_{p\in\mathcal P_{\boldsymbol F}(x)}
 \delta_{(\varphi(|F_j(p)|)/|F_j(p)|)_{j=1}^d}
 \ \Longrightarrow\ \operatorname{Law}(X_1,\ldots,X_d).
\end{equation}
For the vector of negative logarithms, convergence holds in every
fixed Wasserstein order $1\le r<\infty$. For every fixed
$\boldsymbol s\in\R^d$ the corresponding mixed Mellin moment
converges to $\mathbb E\prod_j X_j^{s_j}$.
Beyond $b$, the local indicator vector has neutral state $0$ and
root-type states $v_h$, of respective probabilities
\begin{equation}\label{eq:polynomial-root-local-types}
 1-\sum_{h\in\mathcal H}\frac{\rho_h(\ell)}{\ell-1},
 \qquad \frac{\rho_h(\ell)}{\ell-1}\quad(h\in\mathcal H).
\end{equation}
If two $v_h$ coincide, their probabilities are added to obtain the
probability of that vector atom. Repeated factors have no effect on
these indicators. Removing $T^{a_j}$ has no effect on the model and
changes the sampled negative logarithm by a number in $[0,\omega_p]$.
If $r_j=0$, then $F_j=c_jT^{a_j}$ and
$X_j=\varphi(|c_j|)/|c_j|$ deterministically. If $r_j>0$, its tail
activation intensity is exactly $r_j$, in the strong sense of
\eqref{eq:polynomial-root-hazard} with $\mathcal A=\mathcal H_j$.
\end{theorem}

\begin{proof}[Proof of \Cref{thm:polynomial-root-law}]
All integers in \eqref{eq:polynomial-root-exceptions} are nonzero:
the $h$ are distinct and irreducible in characteristic zero, none is
$T$, and resultants of distinct primitive irreducibles are nonzero.
Gauss's lemma gives the displayed factorization with
$|c_j|$ the content. Outside $E$, no content or leading coefficient
vanishes, no root is zero, and distinct factors have no common root.
The vanishing of $F_j$ is therefore exactly membership in the union
of the root sets of $\mathcal H_j$. Multiplicities do not change this
union. This proves \eqref{eq:polynomial-root-local-types} and, with
\Cref{lem:polynomial-root-intensity}, the claimed marginal intensities.
At primes in $E$ the original congruences retain every forced or
coincident state.

For $\ell>b$,
$\mathbb P(B_{\ell,j}=1)\le D/(\ell-1)$ and
$\omega_\ell\le2/\ell$. Hence
\[
 \sum_{\ell>y}\mathbb E\sum_j\omega_\ell B_{\ell,j}
 \ll_{\boldsymbol F}\sum_{\ell>y}\ell^{-2}
 \ll_{\boldsymbol F}y^{-1}\qquad(y\ge b).
\]
The head is finite. Tonelli's theorem proves almost sure absolute
convergence and strict positivity of every product.

Let $G_j(T)=c_j\prod_hh(T)^{e_{j,h}}$, so $F_j=T^{a_j}G_j$.
For a prime $p$ with $F_j(p)\ne0$, the only possible extra prime
divisor introduced by $p^{a_j}$ is $p$ itself. Consequently, writing
$f(n)=-\log(\varphi(n)/n)$ for $n\ge1$,
\[
 0\le f(|F_j(p)|)-f(|G_j(p)|)\le\omega_p.
\]
In particular the correction vanishes as $p\to\infty$ and its prime
average vanishes: split at any fixed $M$, then use
$\sup_{p>M}\omega_p\to0$. On reduced residues $U_\ell\ne0$,
$F_j(U_\ell)=0$ and $G_j(U_\ell)=0$ are equivalent even at exceptional
primes. We may thus prove empirical convergence using $G_j$.
Their finitely many integer zeros remove only finitely many primes,
so $N_{\boldsymbol F}(x)=\pi(x)+O_{\boldsymbol F}(1)$.

For each fixed finite prime set, the Chinese remainder theorem and
\Cref{thm:sw} make the reduced residues of the sampled prime
asymptotically uniform and independent. The sampled primes equal to
a modulus prime form a finite set. The truncated logarithmic vectors
therefore converge to their model counterparts, including all
exceptional factors.

For $y>b$ and $y<\ell\le\sqrt x$, all roots of $G_j$ are reduced,
since $G_j(0)\ne0$ and its prime divisors lie in $E$; there are at most
$\deg G_j$ such root classes. Brun--Titchmarsh
\cite[Theorem~3.9]{MV06}, together with $\pi(x)\asymp x/\log x$,
gives
\[
 \frac1{\pi(x)}\sum_{p\in\mathcal P_{\boldsymbol F}(x)}
 \sum_{\substack{y<\ell\le\sqrt x\\\ell\mid G_j(p)}}\omega_\ell
 \ll_{\boldsymbol F}\sum_{\ell>y}\frac{\omega_\ell}{\ell-1}
 \ll_{\boldsymbol F}y^{-1}.
\]
For all $p\le x$ with $G_j(p)\ne0$,
$1\le|G_j(p)|\le C_{\boldsymbol F}(1+x)^{\deg G_j}$.
The number of its distinct prime divisors exceeding $\sqrt x$ is
at most $\log|G_j(p)|/\log\sqrt x=O_{\boldsymbol F}(1)$.
Each has weight $O(x^{-1/2})$. Thus
\begin{equation}\label{eq:polynomial-root-empirical-tail}
 \limsup_{x\to\infty}\frac1{N_{\boldsymbol F}(x)}
 \sum_{p\in\mathcal P_{\boldsymbol F}(x)}
 \sum_{\substack{\ell>y\\\ell\mid G_j(p)}}\omega_\ell
 \ll_{\boldsymbol F}y^{-1}.
\end{equation}
A constant $G_j$ has no such tail once $y>b$, so is included.
For any bounded Lipschitz test function of the logarithmic vector,
the difference between its full and truncated averages is bounded
by its Lipschitz constant times the sum of these coordinate tails;
the model satisfies the same bound. First let $x\to\infty$ at fixed
$y$, then let $y\to\infty$. The bounded Lipschitz characterization
proves joint weak convergence for the logarithms. Applying the
continuous coordinate map $z\mapsto e^{-z}$ proves
\eqref{eq:polynomial-root-empirical-law}. The vanishing $T$ corrections
do not change either limit.

For moment convergence fix $t>0$ and one polynomial $F_j$. Write
$F_j(T)=T^{a_j}G_j(T)$ as above, and put $y=(\log x)^2$.
The extra summand $\omega_p$ and the contributions of the finitely
many exceptional primes are bounded. For the remaining primes the
part of $f(|G_j(p)|)$ above $y$ is
$O_{\boldsymbol F}(\log x/(y\log y))$ uniformly, since
$|G_j(p)|\ll_{\boldsymbol F}(1+x)^{\deg G_j}$ on the sampling set.
For good primes $\ell\le y$ put $b_\ell=e^{t\omega_\ell}-1$.
Expanding the exponential of the truncated log statistic gives the
positive coefficients $b(m)=\prod_{\ell\mid m}b_\ell$ over squarefree
products $m$ of these primes. Their indicator is $m\mid G_j(p)$.
For $m\le\sqrt x$, the Chinese remainder theorem gives at most
$D_0^{\omega(m)}$ reduced root classes, with
$D_0=\max(1,\deg G_j)$. The constant term of $G_j$ is a unit at every
good prime, so all roots are reduced. Brun--Titchmarsh, and
$N_{\boldsymbol F}(x)\sim\pi(x)$, bound the divisibility probability by
$O(D_0^{\omega(m)}/\varphi(m))$, also for $m=1$. Summing costs at most
a constant times $\prod_{\ell\le y}(1+D_0b_\ell/(\ell-1))=O_{t,\boldsymbol F}(1)$.
For larger products the bound by one suffices, since
\[
 \sum_{m>\sqrt x}b(m)
 \le x^{-1/(2\log y)}\prod_{\ell\le y}
       (1+\ell^{1/\log y}b_\ell)
 \ll_t x^{-1/(2\log y)}(\log y)^{O_t(1)}=o(1).
\]
Products here use only good primes. This proves uniform exponential
integrability at every fixed order for each sampled log coordinate.
H\"older's inequality gives the corresponding joint bound; the limit
law inherits it by truncation. Weak convergence now gives convergence
of all mixed real Mellin moments and of every positive moment of the
norm of the log vector. The moment characterization of Wasserstein
convergence proves the assertion for each $r$.
Finally, when $\mathcal H_j=\varnothing$, precisely the prime divisors
of $c_j$ have $B_{\ell,j}=1$; every other coordinate indicator is zero.
Euler's product formula gives the stated deterministic value.
\end{proof}

\begin{corollary}[Support and carriers of the polynomial model]
\label{cor:polynomial-root-geometry}
In \Cref{thm:polynomial-root-law}, let
$\mathcal C_{\mathcal H}=\{\sum_{h\in\mathcal H}t_hv_h:t_h\ge0\}$,
with $\mathcal C_\varnothing=\{0\}$. Then
\begin{equation}\label{eq:polynomial-root-support}
 \supp\operatorname{Law}(\boldsymbol Z)
   =\supp\operatorname{Law}(\boldsymbol H_b)+\mathcal C_{\mathcal H}.
\end{equation}
Both $\boldsymbol X$ and $\boldsymbol Z$ have Borel full carriers of
Hausdorff dimension zero. Almost surely,
$R(\boldsymbol X)\in\Q\cup\mathcal L$ simultaneously for all
$R\in\Q(T_1,\ldots,T_d)$ whose denominator is nonzero at
$\boldsymbol X$. Every ratio coordinate $X_j$ with $r_j>0$ is singular continuous,
is Liouville almost surely, and has support $[0,A_j]$, where
$A_j=\exp(-\min\supp\operatorname{Law}(H_{b,j}))$.
The $X_j$ with $r_j=0$ are the deterministic rational variables
specified in \Cref{thm:polynomial-root-law}.
For a fixed $a\in\R^d$, the law of $a\cdot\boldsymbol Z$ is atomless
if some $a\cdot v_h\ne0$; otherwise it is the finite law of
$a\cdot\boldsymbol H_b$. In the atomless case it is singular continuous
and has Hausdorff dimension zero.
\end{corollary}

\begin{proof}[Proof of \Cref{cor:polynomial-root-geometry}]
The cone $\mathcal C_{\mathcal H}$ is closed. Indeed all its generating
vectors are nonnegative and nonzero, so the sum of their coefficients
is bounded by the sum of the coordinates of their combination; a
convergent sequence of combinations therefore has a convergent
subsequence of coefficient vectors. Every tail vector belongs to
this cone, proving one inclusion in \eqref{eq:polynomial-root-support}.
For each $h$, \Cref{lem:polynomial-root-intensity} and
$\rho_h(\ell)\le\deg h$ show that
\[
 \sum_{\substack{\ell>b\\\rho_h(\ell)>0}}\omega_\ell=\infty,
 \qquad \omega_\ell\to0.
\]
Given $t_h\ge0$ and any error tolerance, finite sums of the available
$\omega_\ell$ approximate $t_h$ from below: starting beyond any
prescribed cutoff, add terms until the next would first exceed $t_h$;
the gap is at most the largest permitted term. Apply this successively
to the finitely many $h$, always beyond the preceding finite set.
The resulting distinct primes realize a finite configuration whose
sum approximates $\sum_ht_hv_h$.
Each selected root-type state and every neutral state has positive
probability. Any such finite configuration lies in the support of the
whole tail law: force its states and zero at the other primes through
a larger cutoff, an event of positive probability; choose that cutoff
so the independent remaining tail has expected norm less than half
the required neighborhood radius, and apply Markov's inequality.
Adding any head atom and using closedness proves the reverse inclusion.

If $\mathcal H$ is empty, both vectors are deterministic and rational
in ratio coordinates, so the carrier and rational-function claims
follow directly. Otherwise use the rational local vector
$\boldsymbol R_\ell=((1-1/\ell)^{B_{\ell,j}})_j$ for $\ell>b$ and
the rational head $e^{-\boldsymbol H_b}$. Its neutral hazard is
$|\mathcal H|\log\log y+O_{\boldsymbol F}(1)$ by
\eqref{eq:polynomial-root-hazard}. The expected sum of local denominator
logarithms through $y$ is at most
\[
 \sum_{b<\ell\le y}\sum_j\mathbb P(B_{\ell,j}=1)\log\ell
 \ll_{\boldsymbol F}\sum_{\ell\le y}\frac{\log\ell}\ell
 \ll_{\boldsymbol F}\log y,
\]
and the expected absolute logarithmic tail is $O_{\boldsymbol F}(z^{-1})$,
as proved above. All hypotheses of \Cref{lem:rational-blocks} hold
with $c_\ast=|\mathcal H|$ and $\beta=1$.
\Cref{lem:joint-rational-carrier}\textup{(a),(b)} gives both
zero-dimensional carriers and the simultaneous rational-function claim.

For a fixed $a$, the tail scalar local value is
$\omega_\ell(a\cdot v_h)$ on root type $h$.
For sufficiently large $\ell$ its zero atom is largest, and its
nonzero mass is
$\sum_{h:a\cdot v_h\ne0}\rho_h(\ell)/(\ell-1)$.
If that set is nonempty, its sum over primes diverges by
\Cref{lem:polynomial-root-intensity}; \Cref{lem:levy-atomless} proves
atomlessness. If it is empty, the tail projection is identically zero.
The locally Lipschitz image of the already obtained carrier has
Hausdorff dimension zero, proving the singularity assertion in the
atomless case. Taking $a$ to be a coordinate vector proves continuity
for every $r_j>0$. Its ratio image is also atomless, and the
rational-function claim then makes it Liouville almost surely.
Projection of the support formula onto this coordinate gives
$[\min\supp(H_{b,j}),\infty)$; exponentiation and closure give
$[0,A_j]$. The other coordinates were identified in the theorem.
\end{proof}

\paragraph*{Specializations}
For the affine forms in \Cref{thm:fixed-shift-tuple}, dividing each
form by its content gives one primitive linear factor. The
nonproportionality hypothesis makes the factors distinct, so
$v_h$ runs through the coordinate vectors and every good prime has
$\rho_h(\ell)=1$. The root and affine models agree at every prime,
including exceptional primes; their finite-head conventions differ only
by regrouping finitely many coordinates. The affine estimates
\eqref{eq:tuple-fourier-bound}, \eqref{eq:tuple-projection-concentration}
and \eqref{eq:joint-shifted-tail} are proved below.
For $F_j=T^{m_j}-1$, the factors are the cyclotomic polynomials
$\Phi_a$ with $a\mid m_j$, so $r_j=\tau(m_j)$; the cyclic-group identity
\eqref{eq:joint-degree-local} computes their intersections.
The results in \Cref{cor:fixed-degree-fields,cor:joint-fixed-degree-fields,prop:fixed-degree-endpoint,cor:joint-degree-endpoint}
also give uniformity over matrix dimension arrays, Fourier estimates
and explicit endpoint constants.
For example $T^2+1$ has one factor and intensity one, even though it
has no rational root: apart from $2$, its root count is $2$ when
$\ell\equiv1\pmod4$ and zero when $\ell\equiv3\pmod4$.
Shared factors can force relations such as $X_1=X_2$.

\section{Affine tuple models and supports}
\label{sec:affine-tuple-models}

Fix $k\ge1$ and integer forms $L_j(n)=\alpha_jn+b_j$, where
$\alpha_j\ge1$, $b_j\ne0$, and
$\alpha_i b_j-\alpha_j b_i\ne0$ for $i\ne j$.
Write $\boldsymbol L=(L_1,\ldots,L_k)$.
Let $(U_\ell)_\ell$ be independent, with $U_\ell$ uniform on
$(\Z/\ell\Z)^*$, and put
\[
 \boldsymbol Z_\ell
 =f(\ell)(\mathbf1_{\{L_1(U_\ell)=0\}},\ldots,
                    \mathbf1_{\{L_k(U_\ell)=0\}}),\qquad
 \boldsymbol Z=\sum_\ell\boldsymbol Z_\ell.
\]
The equalities in the indicators are modulo $\ell$.
Let $E$ consist of the primes at most $k+1$ and the prime divisors of
$\prod_j\alpha_jb_j\prod_{i<j}(\alpha_i b_j-\alpha_j b_i)$, and set
$\boldsymbol H=\sum_{\ell\in E}\boldsymbol Z_\ell$.
For $\ell\notin E$, the local atoms are
$0,\omega_\ell e_1,\ldots,\omega_\ell e_k$, with probabilities
$1-k/(\ell-1),1/(\ell-1),\ldots,1/(\ell-1)$.

\paragraph*{Exceptional primes and the finite head}
The vectors $\boldsymbol Z_\ell$ are independent across primes; their
coordinates share $U_\ell$ at each prime. The $j$th indicator is one
if $\ell\mid\gcd(\alpha_j,b_j)$, zero if $\ell$ divides exactly one
of $\alpha_j,b_j$, and otherwise has its root at
$-b_j\alpha_j^{-1}\pmod\ell$. If two roots coincide at a nonzero residue,
several coordinates can activate together. The set $E$ removes these
possibilities and the small primes: outside $E$ the $k$ roots are distinct
and nonzero, and $\ell-1>k$, so the zero atom has positive mass.
This convenient exceptional set need not be minimal. The finite head
must be restored when computing support: the formula in
\Cref{thm:fixed-shift-tuple} is a finite union of translated orthants,
although the law itself has Hausdorff dimension zero.

For example, take $L_1(n)=n-1$, $L_2(n)=n+1$, so $E=\{2,3\}$.
At $2$ both coordinates receive $\log2$; at $3$ the reduced residues
$1,2$ give $(\omega_3,0)$ and $(0,\omega_3)$, each with probability
$1/2$. After subtracting $(\log2,\log2)$, the support is therefore
$((\omega_3,0)+[0,\infty)^2)\cup((0,\omega_3)+[0,\infty)^2)$,
as proved in \Cref{thm:diff-dist}. The centered law gives the square
$[0,\varepsilon]^2$ mass zero for $\varepsilon<\omega_3$.
For corner asymptotics, \Cref{prop:deleted-corners} uses
coordinatewise minimal atoms of the joint head and their full probabilities;
in this example both minimal atoms have probability $1/2$.

\KeepHeadingSpace{3\baselineskip}
\begin{theorem}[Fixed affine tuples]\label{thm:fixed-shift-tuple}
\index{Affine prime arguments}
\index{Joint laws!fixed affine tuples}
Let $\boldsymbol L$, $\boldsymbol Z$, $E$ and $\boldsymbol H$ be as above.
Put $B=\max_j|b_j|$ and
$\mu_{\boldsymbol L}=\operatorname{Law}(\boldsymbol Z)$; when all
$\alpha_j=1$, write also $\mu_{\boldsymbol b}=\mu_{\boldsymbol L}$.
\begin{enumerate}[\upshape(a)]
\item The series $\boldsymbol Z$ converges absolutely almost surely,
and the empirical laws of $(f(L_1(p)),\ldots,f(L_k(p)))$, over primes
$p>B$, converge to $\mu_{\boldsymbol L}$.
\item
\[
 \supp\mu_{\boldsymbol L}=\supp(\boldsymbol H)+[0,\infty)^k,
 \qquad \dim_H\mu_{\boldsymbol L}=0.
\]
For every $\boldsymbol a\in\R^k\setminus\{0\}$, the image of
$\mu_{\boldsymbol L}$ under $\boldsymbol u\mapsto\boldsymbol a\cdot\boldsymbol u$
is continuous and singular of Hausdorff dimension zero.
\item For every $\boldsymbol t\in\R^k$,
\begin{equation}\label{eq:tuple-fourier-bound}
 |\widehat\mu_{\boldsymbol L}(\boldsymbol t)|
 \le C_{\boldsymbol L}\exp\left(-\frac12\sum_{j=1}^kS(t_j)\right),
\end{equation}
where $C_{\boldsymbol L}$ depends only on the fixed forms.
\item For fixed $\boldsymbol a\in\R^k\setminus\{0\}$, put
$r=\#\{j:a_j\ne0\}$ and
$\nu=\operatorname{Law}(\boldsymbol a\cdot\boldsymbol Z)$. Then
\begin{equation}\label{eq:tuple-projection-concentration}
 Q_\nu(h)\asymp_{\boldsymbol a,\boldsymbol L}
       (\log(1/h))^{-r}\qquad(h\downarrow0).
\end{equation}
\end{enumerate}
\end{theorem}

\begin{proof}
Each $L_j$ has one primitive irreducible factor other than $T$,
and nonproportionality makes these factors distinct. Thus $v_h=e_j$,
and \Cref{thm:polynomial-root-law,cor:polynomial-root-geometry}
give \textup{(a),(b)}. Here $L_j(p)>0$ for $p>B$, and excluding the
finitely many smaller primes does not change the empirical limit.
To identify the support, choose a cutoff $b$ containing both exceptional
sets. The polynomial head consists of $\boldsymbol H$ and finitely many
affine good-prime vectors. Their values are nonnegative and each has a
zero atom, so their support added to $[0,\infty)^k$ is $[0,\infty)^k$.
The cone generated by the $v_h$ is this orthant, and every nonzero
projection has $\boldsymbol a\cdot v_h\ne0$ for some $h$.

Put $C=\max_j\alpha_j$. For $y>\max E$, the classes
$-b_j\alpha_j^{-1}\pmod\ell$, $\ell>y$, are reduced.
Brun--Titchmarsh up to $\sqrt{Cx+B}$ gives
\begin{equation}\label{eq:joint-shifted-tail}
 \limsup_{x\to\infty}\frac1{\pi(x)}
 \sum_{B<p\le x}\sum_{\substack{\ell>y\\\ell\mid L_j(p)}}\omega_\ell
 \ll_{\boldsymbol L}\sum_{\ell>y}\frac{\omega_\ell}{\ell-1}
 \ll_{\boldsymbol L}y^{-1}\qquad(1\le j\le k).
\end{equation}
Above $\sqrt{Cx+B}$, each positive $L_j(p)$ has at most one prime
divisor, whose contribution is $O_{\boldsymbol L}(x^{-1/2})$.

For a prime $\ell\notin E$, put $q=(\ell-1)^{-1}$, $A=1-kq$, and
$u_j=t_j\omega_\ell$. Direct expansion gives
\[
 \left|A+q\sum_je^{iu_j}\right|^2
 =1-2Aq\sum_j(1-\cos u_j)
       -2q^2\sum_{i<j}(1-\cos(u_i-u_j)).
\]
Since $2Aq=d_\ell-2(k-1)q^2$ and $0\le1-\cos u_j\le2$, discarding the final
nonnegative sum and using $1+x\le e^x$ give
\[
 \left|A+q\sum_je^{iu_j}\right|^2
 \le\exp\left(-d_\ell\sum_j(1-\cos u_j)+4k(k-1)q^2\right).
\]
Local factors at primes in $E$ have modulus at most one, and these primes
contribute at most $k\sum_{\ell\in E,\,\ell\ge3}d_\ell$ to
$\tfrac12\sum_jS(t_j)$. Multiplying the estimates and taking square roots
proves~\eqref{eq:tuple-fourier-bound} with
$\log C_{\boldsymbol L}=k\sum_{\ell\in E,\,\ell\ge3}d_\ell
+2k(k-1)\sum_{\ell\notin E}(\ell-1)^{-2}$.

For \textup{(d)}, put $J=\{j:a_j\ne0\}$. For $\ell\ge5$,
$x_\ell=d_\ell(1-\cos(t\omega_\ell))\le3/4$ and
$\log(1-x_\ell)=-x_\ell+O(x_\ell^2)$. Since $\sum_{\ell\ge5}d_\ell^2<\infty$,
\eqref{eq:euler} gives
$e^{-S(t)/2}\ll|\mathcal N_{\{3\}}(t)|$, uniformly in $t$.
Thus \eqref{eq:tuple-fourier-bound}, Esseen's inequality and H\"older's
inequality give
\[
 Q_\nu(h)\ll_{\boldsymbol L}h\prod_{j\in J}
 \left(\int_{-1/h}^{1/h}|\mathcal N_{\{3\}}(a_jt)|^r\,dt\right)^{1/r}
 \ll_{\boldsymbol a,\boldsymbol L}(\log(1/h))^{-r}
\]
by \Cref{thm:uniform-fourier-moments}. The lower bound uses a sufficient
event and absolute-tail control, as in \eqref{eq:lq-empty-block}.
Fix an atom $x_0$ of $\boldsymbol a\cdot\boldsymbol H$ of mass $w>0$. For
$y=K/h>\max E$, require that no coordinate in $J$ activate at any
prime $\ell\le y$ outside $E$. Together with the fixed head, this event
has probability
\[
 w\prod_{\substack{\ell\le y\\\ell\notin E}}
       \left(1-\frac r{\ell-1}\right)
 \asymp_{\boldsymbol a,\boldsymbol L}(\log y)^{-r}.
\]
The independent tail has expected absolute value $O_{\boldsymbol a,\boldsymbol L}(1/y)$.
For sufficiently large fixed $K$, it lies in $[-h/2,h/2]$ with
probability at least $1/2$. The interval $[x_0-h/2,x_0+h/2]$ proves
the lower bound.
\end{proof}

\subsection*{Polynomial arrays and full observations}
Use the fixed polynomial model of \Cref{def:polynomial-root-model}.
Write $\mathcal H_j$ for the set of distinct primitive irreducible
factors other than $T$ of $F_j$, and $r_j=|\mathcal H_j|$.
On $\Omega=\{0,1\}^{\{\ell\text{ prime}\}\times\{1,\ldots,d\}}$,
let $P$ be the law of $(B_{\ell,j})$, and let $Q$ make all entries
independent with the same one-entry marginals as $P$. Thus
\[
 q_{\ell,j}:=P(B_{\ell,j}=1),\qquad
 Q=\bigotimes_{\ell,j}\operatorname{Bernoulli}(q_{\ell,j}).
\]
At every prime, including exceptional ones, denote the local vector
laws by $P_\ell,Q_\ell$. For $a\in\{0,1\}^d$, put
$|a|:=\sum_{j=1}^d a_j$ and
$L_\ell(a)=P_\ell(a)/Q_\ell(a)$ if $Q_\ell(a)>0$, and $L_\ell(a)=0$
otherwise. The observation maps are
\[
 T(A)_j=\sum_\ell\omega_\ell A_{\ell,j},\qquad
 F(A)_j=\exp(-T(A)_j).
\]
They are defined arbitrarily on the null set where a sum diverges.
In this paragraph and the next two results, $P,Q,T,F$ refer to the
polynomial model; the affine notation is introduced separately below.

\begin{proposition}[Likelihood and singularity of polynomial root arrays]
\label{prop:root-array-likelihood}
\index{Polynomial prime arguments!array likelihood}
\begin{enumerate}[label=\textup{(\alph*)}]
\item If $\mathcal H_i\cap\mathcal H_j=\varnothing$ for $i\ne j$,
then $P\ll Q$. More precisely,
\[
 L_y:=\prod_{\ell\le y}L_\ell(A_\ell)
 \longrightarrow L=\frac{dP}{dQ}.
\]
The convergence holds $Q$-almost surely and in every $L^s(Q)$,
$1\le s<\infty$.
For all sufficiently large $\ell$, put $S_\ell=\sum_jq_{\ell,j}<1$.
The local likelihood is
\begin{equation}\label{eq:polynomial-array-local-likelihood}
 \begin{split}
 L_\ell(0)&=\frac{1-S_\ell}{\prod_j(1-q_{\ell,j})},\\
 L_\ell(e_j)&=\prod_{i\ne j}(1-q_{\ell,i})^{-1}
                   \quad(q_{\ell,j}>0),\\
 L_\ell(a)&=0\quad\bigl(|a|\ge2\text{ or }a=e_j, q_{\ell,j}=0\bigr).
 \end{split}
\end{equation}
\item If $\mathcal H_i\cap\mathcal H_j\ne\varnothing$ for some
$i\ne j$, then $P\perp Q$. A separating Borel event is
\begin{equation}\label{eq:polynomial-array-singularity-event}
 \begin{gathered}
 \mathcal V_{ij}:=\{A:A_{\ell,i}=A_{\ell,j}=1
                         \text{ at infinitely many primes }\ell\},\\
 P(\mathcal V_{ij})=1,\qquad Q(\mathcal V_{ij})=0.
 \end{gathered}
\end{equation}
\end{enumerate}
\end{proposition}

\begin{proof}
Every local atom of positive $P_\ell$-mass has positive mass in each
of its coordinate marginals, hence positive $Q_\ell$-mass. Thus
$P_\ell\ll Q_\ell$ at every prime, including primes where a coordinate
is forced. There are finitely many local atoms, so each exceptional
prime has finite likelihood moments of all finite orders.

Suppose first that the factor sets are pairwise disjoint. Outside $E$,
distinct factors have disjoint nonzero root sets; hence only $0$ and
the singleton vectors $e_j$ occur, of probabilities $1-S_\ell$ and
$q_{\ell,j}$ respectively. Also $0\le q_{\ell,j}\ll_{\boldsymbol F}\ell^{-1}$.
The product of the actual marginal laws gives
$Q_\ell(0)=\prod_j(1-q_{\ell,j})$ and
$Q_\ell(e_j)=q_{\ell,j}\prod_{i\ne j}(1-q_{\ell,i})$.
This proves \eqref{eq:polynomial-array-local-likelihood}, with zero
likelihood assigned on zero-$Q_\ell$-mass atoms.
For every fixed $s>1$, direct summation gives
\[
 \mathbb E_Q L_\ell^s
 =(1-S_\ell)^s\prod_j(1-q_{\ell,j})^{1-s}
       +\sum_jq_{\ell,j}\prod_{i\ne j}(1-q_{\ell,i})^{1-s}
 =1+O_{s,\boldsymbol F}(\ell^{-2}).
\]
Indeed, uniformly on $S_\ell\le1/2$, the first term is
$1-sS_\ell+(s-1)S_\ell+O_{s,d}(S_\ell^2)=1-S_\ell+O_{s,d}(S_\ell^2)$,
and the sum is $S_\ell+O_{s,d}(S_\ell^2)$.

Since $\mathbb E_Q L_\ell=1$ and the local vectors are independent
under $Q$, the finite products $L_y$ form a nonnegative mean-one
martingale in the increasing finite-prime sigma fields. Independence,
the displayed estimate, and the finite exceptional factors give
$\sup_y\mathbb E_Q L_y^s<\infty$ for each fixed $s>1$.
The martingale $L^s$ convergence theorem
\cite[Theorem~4.4.6]{Dur19} gives a common almost-sure limit $L$ and
convergence in every finite $L^s$, also in $L^1$ by H\"older.
In particular $\mathbb E_Q L=1$.
For a cylinder $C$ depending on primes at most $y_0$, one has
$\mathbb E_Q[L_y\mathbf1_C]=P(C)$ whenever $y\ge y_0$.
Pass to the $L^1$ limit and apply uniqueness on the cylinder pi-system
to obtain $P=LQ$. The reverse absolute continuity can fail: for example,
$F_1=T-1,F_2=T+1$ have
$P(B_{5,1}=B_{5,2}=1)=0$ but $Q(B_{5,1}=B_{5,2}=1)=1/16$.

If a factor $h$ belongs to both $\mathcal H_i$ and $\mathcal H_j$,
then at every prime beyond $E$,
\[
 P(B_{\ell,i}=B_{\ell,j}=1)\ge
       \frac{\rho_h(\ell)}{\ell-1},\qquad
 Q(B_{\ell,i}=B_{\ell,j}=1)=q_{\ell,i}q_{\ell,j}
                                    \ll_{\boldsymbol F}\ell^{-2}.
\]
The first series diverges by \Cref{lem:polynomial-root-intensity},
and its events are independent across primes under $P$.
The second Borel--Cantelli lemma gives $P(\mathcal V_{ij})=1$;
the first Borel--Cantelli lemma gives $Q(\mathcal V_{ij})=0$.
The event is Borel, since it is a countable intersection of countable
unions of cylinder events. It therefore proves $P\perp Q$.
\end{proof}

\begin{corollary}[Observed likelihood and singularity for polynomial products]
\label{cor:polynomial-observed-likelihood}
\index{Polynomial prime arguments!common decoder}
There is a single Borel map $\mathcal D:(0,\infty)^d\to\Omega$
such that $\mathcal D(F(A))=A$ both $P$-almost surely and
$Q$-almost surely. Put $\lambda=F_*P$ and $\eta=F_*Q$; thus
$\eta$ is the product of the actual scalar product laws.
In the case of disjoint non-$T$ factors in
\Cref{prop:root-array-likelihood}\textup{(a)},
\begin{equation}\label{eq:polynomial-observed-density}
 \begin{gathered}
 \lambda\ll\eta,\qquad
 \frac{d\lambda}{d\eta}(F(A))=L(A)\quad Q\text{-almost surely},\\
 \int\left(\frac{d\lambda}{d\eta}\right)^s\,d\eta
       =\int L^s\,dQ<\infty\qquad(1\le s<\infty).
 \end{gathered}
\end{equation}
In the shared-factor case of \Cref{prop:root-array-likelihood}\textup{(b)},
$\lambda\perp\eta$. The same assertions hold for the logarithmic
observations $T_*P,T_*Q$, and the total-variation distances between
these full observations equal the corresponding array distance.
\end{corollary}

\begin{proof}
For each $j$, the coordinates $(B_{\ell,j})_\ell$ are independent
under $P$, and their law under $Q$ is identical. For large $\ell$,
\[
 q_{\ell,j}=\sum_{h\in\mathcal H_j}\frac{\rho_h(\ell)}{\ell-1}
       \ll_{\boldsymbol F}\ell^{-1}.
\]
Thus $\sum_\ell\omega_\ell q_{\ell,j}<\infty$, proving almost-sure
convergence of every observed sum under both laws by Tonelli.
If $r_j>0$, \Cref{lem:polynomial-root-intensity} and the convergent
quadratic corrections in $-\log(1-q)=q+O(q^2)$ give, beyond a fixed
cutoff $b$, with a fixed constant $C_j$,
\[
 \sum_{b<\ell\le y}-\log(1-q_{\ell,j})
 =r_j\log\log y+C_j+o(1/\log y).
\]
In particular the envelope and cumulative-intensity hypotheses
\eqref{eq:cumulative-intensity} hold with $\kappa=r_j>0$.
Apply \Cref{prop:canonical-decoder-mas08} with the unchanged canonical
factors, $s=-1,c=1,p_0=2$. It provides the same Borel scalar decoder
$\mathcal D_{1,-1}$ under the two marginal laws, allowing all forced
or suppressed exceptional coordinates. If $r_j=0$, the polynomial
has the form $F_j=c_jT^{a_j}$ and its indicator array is exactly the
deterministic vector $(\mathbf1_{\ell\mid c_j})_\ell$; use that
constant Borel decoder for this coordinate. Combining the scalar
maps and intersecting their finitely many full-measure events gives
one $\mathcal D$ recovering the entire array under both $P$ and $Q$.

In the case of disjoint non-$T$ factors, $P=LQ$, so integration
against bounded measurable
functions of $F$ gives
\[
 \frac{d\lambda}{d\eta}(F)=\mathbb E_Q[L\mid\sigma(F)].
\]
Choose a Borel representative of $L$ (for example its finite-product
limit where that limit exists, and zero elsewhere). Since
$\mathcal D\circ F$ is the identity $Q$-almost surely,
$L=(L\circ\mathcal D)\circ F$ there. The conditional expectation is
therefore $L$, and integrating its $s$th power proves
\eqref{eq:polynomial-observed-density}. In the shared-factor case,
$\mathcal D^{-1}(\mathcal V_{ij})$ is a Borel set with $\lambda$-mass
one and $\eta$-mass zero. This proves observed singularity using the
common decoder explicitly. Composing $\mathcal D$ with $\boldsymbol z\mapsto(e^{-z_j})_{j=1}^d$ gives the logarithmic assertions.
Finally, a measurable map cannot increase total variation, and the
decoder is a measurable inverse on full sets for both measures.
Applying that inequality to the observation and then to the decoder
gives equality of the array and observed distances.
\end{proof}

\section{Independent reference models and rational statistics}
\label{sec:affine-reference-statistics}

To compare affine prime arguments with independent samples, we account for
their actual marginals as well as their dependence. The vectors above
share residues at each prime; the same-input totient and divisor-sum pair
of \Cref{cor:joint-phi-sigma} instead shares valuations, while
\Cref{cor:rational-statistics} uses independent limit variables.
We compare the affine array first with the
product of its actual marginals, then with an independent reference
using finite conditioning and the prescribed parity at $2$.
The resulting density transfers full-measure events from the independent
reference to the affine model, giving simultaneous rational approximation
below. A finite-entropy coupling also compares joint and scalar mesh entropies.

Let $P$ be the law of the indicator array
$A_{\ell,j}=\mathbf1_{\{L_j(U_\ell)=0\}}$ in
\Cref{thm:fixed-shift-tuple}, and let $Q$ be the product of all its
one-coordinate marginals. For each prime $\ell$, let $P_\ell(a)$ and
$Q_\ell(a)$ denote the masses of the local vector
$(A_{\ell,1},\ldots,A_{\ell,k})$ under $P$ and $Q$. For
$a\in\{0,1\}^k$, put $|a|:=\sum_{j=1}^k a_j$ and define
\[
 L_\ell(a)=
 \begin{cases}
 P_\ell(a)/Q_\ell(a),&Q_\ell(a)>0,\\
 0,&Q_\ell(a)=0.
 \end{cases}
\]
This definition includes the exceptional primes in $E$;
$Q_\ell(a)=0$ implies $P_\ell(a)=0$.
Put $T(A)=\sum_\ell\omega_\ell(A_{\ell,1},\ldots,A_{\ell,k})$,
$\rho=T_*P$ and $\nu=T_*Q$.

\begin{proposition}[Likelihood for affine forms]
\label{prop:observed-likelihood}
\index{Joint laws!observed images}
There is $L\in\bigcap_{1\le s<\infty}L^s(Q)$ such that
\begin{equation}\label{eq:observed-density}
 dP=L\,dQ,\qquad
 \frac{d\rho}{d\nu}(T)=\mathbb E_Q[L\mid\sigma(T)]=L
 \quad Q\hbox{-almost surely}.
\end{equation}
Outside the exceptional set $E$, with $q=(\ell-1)^{-1}$, the local
factors are
\[
 L_\ell(0)=\frac{1-kq}{(1-q)^k},\qquad
 L_\ell(e_j)=(1-q)^{1-k},\qquad
 L_\ell(a)=0\quad(|a|\ge2).
\]
With each factor evaluated at its local vector, the finite products
$L_y=\prod_{\ell\le y}L_\ell$ converge to $L$ $Q$-almost surely and
in every fixed $L^s(Q)$, $1\le s<\infty$. The observed density
$d\rho/d\nu$ also belongs to every finite $L^s(\nu)$, with
\[
 \left\|\frac{d\rho}{d\nu}\right\|_{L^s(\nu)}^s
 =\|L\|_{L^s(Q)}^s\qquad(1\le s<\infty).
\]
\end{proposition}

\begin{proof}
The primitive factors of the $L_j$ are distinct and different from $T$.
Thus \Cref{prop:root-array-likelihood}\textup{(a)} gives $P=LQ$ and
all the asserted convergence statements, with the actual marginals
at every prime. For $\ell\notin E$, substituting
$q_{\ell,j}=(\ell-1)^{-1}$ in
\eqref{eq:polynomial-array-local-likelihood} gives the displayed factors.
\Cref{cor:polynomial-observed-likelihood} gives the density and moment
equalities for $T$. Its common decoder makes $L$ measurable from $T$
modulo a $Q$-null set, so the conditional expectation in
\eqref{eq:observed-density} also equals $L$.
The same decoder gives equality of the array and observed total-variation
distances and chi-square divergences.
\end{proof}

For comparison on the full infinite-prime array space of this fixed
finite affine tuple, let $R$ have mutually independent odd-prime entries
$A_{\ell,j}\sim\operatorname{Bernoulli}(1/(\ell-1))$ and deterministic
$A_{2,j}=\mathbf1_{\{\alpha_j+b_j\text{ even}\}}$. Put
\[
 \varepsilon_{\ell,j}=\mathbf1_{\{\ell\mid\gcd(\alpha_j,b_j)\}},\qquad
 C=\bigcap_{j=1}^k\bigcap_{\substack{\ell\mid\alpha_jb_j\\\ell>2}}
       \{A_{\ell,j}=\varepsilon_{\ell,j}\}.
\]
Write $r=R(C)>0$ and
\[
 c_j=2^{\mathbf1_{\{\alpha_j+b_j\text{ odd}\}}}.
\]
Put $F_y(A)_j=\prod_{\ell\le y}(1-1/\ell)^{A_{\ell,j}}$,
$F(A)=(e^{-T(A)_j})_{j=1}^k$ and
$T_{\boldsymbol c}(\boldsymbol x)=(c_jx_j)_{j=1}^k$.
Put $\lambda=F_*P$ and $\eta=F_*R$.

\begin{corollary}[Independent reference for affine forms]
\label{cor:affine-independent-reference}
One has $Q=R(\,\cdot\mid C)$ and, extending $L$ by zero off $C$,
\begin{equation}\label{eq:fixed-shift-reference-density}
 H:=\frac{dP}{dR}=\frac{\mathbf1_C L}{r},\qquad
 \|H\|_{L^s(R)}^s=r^{1-s}\|L\|_{L^s(Q)}^s
 \quad(1\le s<\infty).
\end{equation}
Moreover, $\eta=(T_{\boldsymbol c})_*\mu_G^{\otimes k}$ and
\begin{equation}\label{eq:fixed-shift-reference-observation}
 \frac{d\lambda}{d\eta}(F)=\mathbb E_R[H\mid\sigma(F)],\qquad
 \left\|\frac{d\lambda}{d\eta}\right\|_{L^s(\eta)}^s
 =r^{1-s}\|L\|_{L^s(Q)}^s\quad(1\le s<\infty).
\end{equation}
For every fixed $0<\epsilon<1$ and all measurable array events $B$,
\begin{equation}\label{eq:fixed-shift-event-transfer}
 P(B)\le C_{\epsilon,\boldsymbol L}R(B)^{1-\epsilon}.
\end{equation}
The same bound holds for the pairs $(P,Q)$, $(\rho,\nu)$ and
$(\lambda,\eta)$ on their respective measurable spaces.
Under $R$ the actual prefixes satisfy $F_y=(c_jY_j(y))_{j=1}^k$
for every $y\ge2$, where the $Y_j(y)$ are independent canonical
$\mu_G$ prefixes, with prime-$2$ factor $1/2$.
\end{corollary}

\begin{proof}
At an odd coefficient prime $\ell\mid\alpha_jb_j$, the $j$th marginal
is the constant $\varepsilon_{\ell,j}$; otherwise it is
Bernoulli$(1/(\ell-1))$. Its prime-$2$ value is
$\mathbf1_{\{\alpha_j+b_j\text{ even}\}}$. Since $\alpha_jb_j\ne0$,
$C$ contains only finitely many constraints. Independence under $R$ gives
\[
 r=\prod_{j=1}^k\prod_{\substack{\ell\mid\alpha_jb_j\\\ell>2}}
       \left(\frac1{\ell-1}\right)^{\varepsilon_{\ell,j}}
       \left(1-\frac1{\ell-1}\right)^{1-\varepsilon_{\ell,j}}>0,
 \qquad Q=R(\,\cdot\mid C).
\]
Combining this identity with $P=LQ$ proves
\eqref{eq:fixed-shift-reference-density}, including its exact $s$th moment.
Under $R$ the weighted indicator sum has finite expectation, so $F$
is defined and positive almost surely. Define the independent canonical
prefixes and their limits by
\[
 Y_j(y)=\tfrac12\prod_{3\le\ell\le y}(1-1/\ell)^{A_{\ell,j}},
 \qquad Y_j=\lim_{y\to\infty}Y_j(y)\sim\mu_G.
\]
The actual prefixes satisfy $F_y(A)_j=c_jY_j(y)$ for every $y\ge2$;
passage to the limit gives
$F(A)_j=c_jY_j$. Integration against bounded functions of $F$ identifies
the observed density as the displayed conditional expectation.
Under $R$, \Cref{prop:canonical-decoder-mas08} with the known multiplier
$c_j/2$ recovers every odd-prime coordinate from $F(A)_j$.
Thus $H$ is $\sigma(F)$-measurable modulo an $R$-null set, and
$\mathbb E_R[H\mid\sigma(F)]=H$. This proves the exact norm identity
in \eqref{eq:fixed-shift-reference-observation}.
H\"older's inequality with $s=1/\epsilon$ proves
\eqref{eq:fixed-shift-event-transfer}, with
$C_{\epsilon,\boldsymbol L}=\|H\|_{L^{1/\epsilon}(R)}$.
Since $r\le1$, this constant also bounds the corresponding moments
under $Q$, and conditional Jensen gives both observed versions.
\end{proof}

\KeepHeadingSpace{12\baselineskip}
Retain the affine vector $\boldsymbol Z$ of \Cref{thm:fixed-shift-tuple}.
For a real random vector $\boldsymbol V=(V_j)_{j=1}^k$, write
$H_h^{(k)}(\boldsymbol V)=\mathsf H((\lfloor V_j/h\rfloor)_{j=1}^k)$,
and define its lower and upper joint information dimensions by the
liminf and limsup of $H_h^{(k)}(\boldsymbol V)/\log(1/h)$ as $h\downarrow0$.
Let $\boldsymbol W=(W_j)_{j=1}^k$ have independent coordinates of law
$\mu_f$, and put $\boldsymbol X=(e^{-Z_j})_{j=1}^k$.

\begin{corollary}[Information dimensions of affine laws]
\label{cor:affine-information-dimensions}
With this notation,
\[
 H_h^{(k)}(\boldsymbol Z)
 =H_h^{(k)}(\boldsymbol X)+O_{\boldsymbol L}(1)
 =kH_h(W_1)+O_{\boldsymbol L}(1)\qquad(0<h\le1).
\]
The joint lower and upper dimensions of both vectors are respectively
$k\underline d_1(W_1)$ and $k\overline d_1(W_1)$.
For fixed $b\in\R$ and $\boldsymbol a\in\R^k$, let
$U=b+\boldsymbol a\cdot\boldsymbol Z$ and
$U_0=b+\sum_{j=1}^k a_jW_j$. Then
\[
 |H_h(U)-H_h(U_0)|+|H_h(e^U)-H_h(e^{U_0})|
 \ll_{\boldsymbol L,\boldsymbol a,b}1\qquad(0<h\le1).
\]
Thus the corresponding lower and upper information dimensions agree,
and \Cref{thm:independent-sums} gives the information bounds for $U$ and $e^U$.
\end{corollary}

\begin{proof}
Use $P,R,\boldsymbol c$ from \Cref{cor:affine-independent-reference}.
Couple $P_\ell$ and $R_\ell$ maximally,
independently across primes. Outside $E$, with $q=(\ell-1)^{-1}$,
their total variation distance is
\[
 \delta_\ell=kq\{1-(1-q)^{k-1}\}=O_k(\ell^{-2}).
\]
Record zero on a match and the ordered pair of local vectors on a
mismatch. There are at most $4^k$ nonzero records, so the coordinate
entropy is $O_k(\ell^{-2}\log\ell)$; the finitely many exceptional
coordinates have finite entropy. The full record $M$ is almost surely finitely supported,
countably valued, and has entropy bounded by the convergent sum of the
coordinate entropies. Under $R$, write
$\boldsymbol Z_0=(W_j-\log c_j)_{j=1}^k$.
The record determines $\boldsymbol Z-\boldsymbol Z_0$, hence also
$\boldsymbol Z-\boldsymbol W$. Given $M$ and either cubical quantizer,
the other has at most $2^k$ values. The chain rule in both directions gives
\[
 |H_h^{(k)}(\boldsymbol Z)-kH_h(W_1)|
 \le\mathsf H(M)+k\log2.
\]
For the raw image, put $\boldsymbol J=(\lfloor Z_j\rfloor)_{j=1}^k$.
Since $Z_j\ge0$ and $\sum_j\mathbb EZ_j<\infty$,
$\mathsf H(\boldsymbol J)+\sum_j\mathbb EJ_j<\infty$.
On each shell, a cubical cell meets at most
$\prod_j(e^{J_j+1}+2)$ cells in either direction under the exponential
map $\boldsymbol z\mapsto(e^{-z_j})_j$. The two entropy chain rules give
$|H_h^{(k)}(\boldsymbol X)-H_h^{(k)}(\boldsymbol Z)|\ll_{\boldsymbol L}1$,
as in \Cref{lem:transfers}(i).
Finally, $U-U_0$ is determined by $M$ and both sums have finite absolute
mean. Equations~\eqref{eq:entropy-displacement}
and~\eqref{eq:entropy-exponential-comparison} prove the scalar comparison;
\Cref{thm:independent-sums} gives its information-dimension consequences.
\end{proof}

\paragraph*{Opposite shifts}
For $L_1(n)=n-1$, $L_2(n)=n+1$, the observed joint law and the product of its
marginal laws satisfy the exact identities
\begin{equation}\label{eq:dependence-bounds}
\begin{split}
 d_{\mathrm{TV}}(\rho,\nu)
 &=1-\frac12\prod_{\ell\ge5}\left(1-\frac1{(\ell-1)^2}\right),\\
 \chi^2(\rho,\nu)
 &=\prod_{\ell\ge3}\left(1+\frac1{(\ell-2)^2}\right)-1.
\end{split}
\end{equation}
Their values rounded to seven decimal places are respectively
$0.5598921$ and $1.4050052$.
Here $\chi^2(\rho,\nu)=\int(d\rho/d\nu-1)^2\,d\nu$.

Indeed, the prime $2$ is deterministic in both coordinates, and
at every odd prime the only joint states are $00,10,01$, of masses
$1-2q,q,q$, where $q=(\ell-1)^{-1}$. Direct calculation gives
$\mathbb E_Q L_\ell^2=1+(\ell-2)^{-2}$.
Let $S$ be the array event that precisely one coordinate is active at
$3$, and that no larger prime activates both coordinates.
Off $S$, $L=0$; on $S$,
\[
 L\ge2\prod_{\ell\ge5}(1-(\ell-2)^{-2})>1.
\]
Indeed the sum of $(\ell-2)^{-2}$ is at most
$\sum_{n\ge3,\ n\ {\textup{odd}}}n^{-2}<1/2$, and a product of
$1-u_j$ is at least $1-\sum u_j$. Consequently
\[
 d_{\mathrm{TV}}(P,Q)=1-Q(S)
 =1-\frac12\prod_{\ell\ge5}(1-(\ell-1)^{-2}),
 \qquad
 \chi^2(P,Q)=\prod_{\ell\ge3}(1+(\ell-2)^{-2})-1.
\]
The second equality follows by $L^2$ convergence of the finite
likelihood products in \Cref{prop:observed-likelihood}.
By \eqref{eq:observed-density}, $L$ is measurable from the full
observed vector.
The formulas $d_{\mathrm{TV}}=\tfrac12\mathbb E_Q|L-1|$ and
$\chi^2=\mathbb E_Q(L-1)^2$ prove equality between the observed and
array distances, yielding \eqref{eq:dependence-bounds}.

The stated rounding has a finite rational certificate. With
$N=10^8$, put
\[
 V_N=\prod_{5\le\ell\le N}(1-(\ell-1)^{-2}),\qquad
 C_N=\prod_{3\le\ell\le N}(1+(\ell-2)^{-2}).
\]
Integer square-sum tails and $e^u\le(1-u)^{-1}$ for $0\le u<1$ give
\[
 1-\tfrac12V_N\le d_{\mathrm{TV}}(\rho,\nu)
 \le1-\tfrac12V_N\frac{N-2}{N-1},
 \qquad
 C_N-1\le\chi^2(\rho,\nu)
 \le C_N\frac{N-2}{N-3}-1.
\]
For example, the omitted negative-product terms have sum at most
$1/(N-1)$, while the omitted positive-product terms have sum at most
$1/(N-2)$. Exact rational multiplication with outward rounding gives
\[
\begin{split}
 0.5598921225&<d_{\mathrm{TV}}(\rho,\nu)<0.5598921270,\\
 1.4050051639&<\chi^2(\rho,\nu)<1.4050051881.
\end{split}
\]
One reproducible integer procedure starts both lower and upper product
numerators at $10^{30}$ and, for each rational factor $a/b$, replaces
them by $\lfloor a\,\mathrm{lower}/b\rfloor$ and
$\lceil a\,\mathrm{upper}/b\rceil$, respectively. The exact prime sieve
through $N$, these recurrences and the displayed rational tail bounds
certify the enclosures without floating-point assumptions.

\paragraph*{Arrays and their observed images}
The observations $T$ and $F$ give logarithmic sums and raw products,
while $F_y$ retains the actual finite products. Prefix events transfer
through the array density \eqref{eq:fixed-shift-reference-density}
and $F_y=(c_jY_j(y))_j$, $c_j\in\{1,2\}$, as in
\Cref{cor:fixed-shift-rational-statistics}; observed null sets transfer
through \eqref{eq:fixed-shift-reference-observation}.

The countable-field argument in \Cref{cor:fixed-shift-normal-field}
applies this comparison to each finite tuple, then intersects over
tuple lengths and rational functions over $\Q$; each real rational
function is treated as fixed. Qualitative Rajchman transfer also uses
\Cref{lem:rational-orbit-domination} and its band-property step.
The full affine observation recovers the canonical prime array by
\Cref{prop:canonical-decoder-mas08}, so its likelihood moments and
dependence distances equal those on array space.
The quantitative rate for linear logarithmic images comes from
\eqref{eq:tuple-fourier-bound}, as used in \Cref{cor:rajchman-diff}.
For a raw monomial $\prod_je^{-a_jZ_j}$ with fixed
$\boldsymbol a\in\R^k\setminus\{0\}$,
\Cref{cor:affine-raw-monomial-fourier} gives the same power
$\rho(t)^{\#\{j:a_j\ne0\}}$ as \Cref{cor:rajchman-diff}, where
$\rho(t)$ is the rate in \eqref{eq:rho-def}. This follows from the local
atoms outside $E$ and \Cref{thm:raw-product-fourier}. 
\Cref{fig:reference-models} displays the array comparisons and their observed images,
and separates transfer by domination from the finite-entropy coupling of
\Cref{cor:affine-information-dimensions} and from the direct Fourier estimates.
\begin{figure}[t]
\centering
\begin{tikzpicture}[figdiagram,
  % local: one grid of three columns, each 4.05 cm wide, with gaps of 0.65 cm;
  % column centres x = 2.025, 6.725, 11.425; right edge x = 13.45.
  % The extra inner xsep of the thin-framed and the unframed boxes makes the
  % outer widths of all boxes equal to that of the heavy-framed ones.
  fcol/.style={text width=3.78cm},
  fresc/.style={fres, fcol},
  finpc/.style={finp, fcol, inner xsep=3.675pt},
  fmechc/.style={fmech, fcol, inner xsep=3.875pt},
  % local: common heights of the boxes in one row or lane
  hrowA/.style={minimum height=14.83mm}, hrowB/.style={minimum height=9.20mm},
  hlane1/.style={minimum height=21.40mm}, hlane2/.style={minimum height=19.64mm},
  hlane3/.style={minimum height=18.25mm}]
% ================= panel (a) =================================================
\node[fhead, anchor=west] (hA) at (0,0) {(a) Array laws and observed laws};
\draw[frule] (hA.south west) ++(0,-1.5pt) -- ++(13.45,0);
\node[fnote, anchor=east] (nC) at (13.45,0)
  {$C$: finite cylinder fixing $A_{\ell,j}$ at the odd primes $\ell\mid\alpha_jb_j$};
% ---- row A: the three array laws (common centre line y = -1.862) -------------
\node[finpc, hrowA] (P) at (2.025,-1.862)
  {$P$: law of the array\\ $A_{\ell,j}=\mathbf 1_{\{L_j(U_\ell)=0\}}$;\\ one $U_\ell$ shared by all $j\le k$\\
   \figref{\Cref{thm:fixed-shift-tuple}}};
\node[finpc, hrowA] (Q) at (6.725,-1.862)
  {$Q=R(\,\cdot\mid C)$: product of\\ the actual one-coordinate\\ marginals of $P$;\ \ $dP=L\,dQ$\\
   \figref{\Cref{prop:observed-likelihood}, \Cref{cor:affine-independent-reference}}};
\node[finpc, hrowA] (R) at (11.425,-1.862)
  {$R$: independent entries,\\ $\operatorname{Bernoulli}\bigl(\tfrac1{\ell-1}\bigr)$ at odd $\ell$,\\
   $A_{2,j}=\mathbf 1_{\{\alpha_j+b_j\ \mathrm{even}\}}$\\
   \figref{\Cref{cor:affine-independent-reference}}};
% the chain P << Q = R(.|C) << R: undirected dashed comparisons
\draw[fmap,-] (P) -- node[fhyp, above=0.4pt, fill=none] (lPQ) {$\ll$} (Q);
\draw[fmap,-] (Q) -- node[fhyp, above=0.4pt, fill=none] (lQR) {$\ll$} (R);
% consequence arrow above the row, opposite to the domination
\draw[fuse] (R.north) -- ++(0,0.65) -| node[pos=0.25, fhyp] (lRP)
  {$R$-full event $\Rightarrow$ $P$-full event,\ \ by\ \ $H:=dP/dR=\mathbf 1_C\,L/R(C)$
   \figref{\eqref{eq:fixed-shift-reference-density}}} (P.north);
% ---- row B: the observed laws -------------------------------------------------
\node[finpc, hrowB] (oP) at (2.025,-3.662)
  {$T_*P=\mu_{\boldsymbol L}=\operatorname{Law}(\boldsymbol Z)$\\ $F_*P=\operatorname{Law}(\boldsymbol X)$};
\node[finpc, hrowB] (oQ) at (6.725,-3.662)
  {$T_*Q$;\quad $Q$-a.s.\\ $\bigl(d\,T_*P/d\,T_*Q\bigr)\circ T=L$};
\node[finpc, hrowB] (oR) at (11.425,-3.662)
  {$F_*R=(T_{\boldsymbol c})_*\mu_G^{\otimes k}$;\quad $R$-a.s.\\ $\bigl(d\,F_*P/d\,F_*R\bigr)\circ F=H$};
% decoder (up, left of the centre) and observation map (down, right of it)
\foreach \a/\b/\dn/\up in {P/oP/{$T_*,\,F_*$}/{decoder $\mathcal D$},
                           Q/oQ/{$T_*$}/{$\mathcal D$, \figref{\eqref{eq:observed-density}}},
                           R/oR/{$F_*$}/{decoder, \figref{\eqref{eq:fixed-shift-reference-observation}}}}{
  \draw[fmap] ($(\b.north)+(-0.35,0)$) -- node[flab, left=1pt] {\up} ($(\a.south)+(-0.35,0)$);
  \draw[fmap] ($(\a.south)+(0.35,0)$) -- node[flab, right=1pt] {\dn} ($(\b.north)+(0.35,0)$);
}
% ================= panel (b) =================================================
\node[fhead, anchor=west] (hB) at (0,-4.548) {(b) Three inputs and their conclusions for the tuple};
\draw[frule] (hB.south west) ++(0,-1.5pt) -- ++(13.45,0);
\node[fnote, anchor=east] (nR) at (13.45,-4.548) {$r=\#\{j:a_j\ne0\}$ in the Fourier rate};
% ---- lane 1: transfer by domination -----------------------------------------
\node[fmechc, hlane1] (M1) at (6.725,-6.097)
  {Transfer by domination\\ from independent\\ $c_jY_j$, $Y_j\sim\mu_G$; Rajchman\\ via the band property\\
   \figref{\Cref{lem:rational-blocks,lem:rational-orbit-domination},}\\
   \figref{\Cref{cor:summable-local-coupling,cor:rational-statistics,cor:multivariable-analytic-images}}};
\node[fresc, hlane1] (R1) at (2.025,-6.097)
  {a.s.: simultaneous rational\\ prefix approximations, and\\ $X_1,\ldots,X_k$ algebraically\\ independent over $\Q$\\
   \figref{\Cref{cor:fixed-shift-rational-statistics}}};
\node[fresc, hlane1] (R2) at (11.425,-6.097)
  {fixed nonconstant rational\\ or analytic images of $\boldsymbol X$:\\ a.s.\ in $\mathfrak N$; Rajchman law\\ (transfer gives no rate)\\
   \figref{\Cref{cor:fixed-shift-normal-field}; \Cref{sec:affine-reference-statistics}}};
\draw[fuse] (M1) -- (R1);
\draw[fuse] (M1) -- (R2);
\draw[fptr] ($(R1.south west)+(0.35,0)$) |- ++(0.4,-0.3)
  node[fnote, right] (nF3) {rational images: \Cref{fig:rational-nonconcentration}};
% ---- lane 2: finite-entropy coupling ----------------------------------------
\node[fmechc, hlane2] (M2) at (6.725,-8.369)
  {Finite-entropy coupling:\\ $P_\ell$, $R_\ell$ coupled maximally\\ at each prime; record $M$\\ with $\mathsf H(M)<\infty$\\
   \figref{\Cref{lem:transfers}}};
\node[fresc, hlane2] (R3) at (11.425,-8.369)
  {mesh entropy $H_h^{(k)}(\boldsymbol Z)$\\ $=k\,H_h(W_1)+O_{\boldsymbol L}(1)$, $h\le1$;\\ joint dimensions, $W_1\sim\muF$:\\
   $k\,\underline d_1(W_1)$ and $k\,\overline d_1(W_1)$\\
   \figref{\Cref{cor:affine-information-dimensions}}};
\draw[fuse] (M2) -- (R3);
% ---- lane 3: direct Fourier estimate ------------------------------------------
\node[fmechc, hlane3] (M3) at (6.725,-10.483)
  {Direct estimates from the\\ local atoms of $P$, $\ell\notin E$:\\
   $|\widehat\mu_{\boldsymbol L}(\boldsymbol t)|\le C_{\boldsymbol L}\,e^{-\frac12\sum_jS(t_j)}$\\
   \figref{\Cref{thm:fixed-shift-tuple}\textup{(c)}, \eqref{eq:tuple-fourier-bound}}};
\node[fresc, hlane3] (R4) at (11.425,-10.483)
  {fixed $\boldsymbol a\ne\boldsymbol 0$: rate $\rho(\tau)^r$ for\\ $\boldsymbol a\cdot\boldsymbol Z$ from this bound;\\
   for $\prod_jX_j^{a_j}$ from the atoms\\[1.5pt] and \Cref{thm:raw-product-fourier}\\
   \figref{\Cref{cor:rajchman-diff,cor:affine-raw-monomial-fourier}}};
\draw[fuse] (M3) -- (R4);
% ================= inset (c) =================================================
% the inset fills the left column below lane 1: its heading sits on the centre line of
% lane 2, its box ends on the bottom line of lane 3
\node[fhead, anchor=west] (hC) at (0,-8.369) {(c) Polynomial arrays};
\draw[frule] (hC.south west) ++(0,-1.5pt) -- ++(4.05,0);
\node[fresc, minimum height=25.48mm] (PI) at (2.025,-10.122)
  {$\mathcal H_i\cap\mathcal H_j=\varnothing$ for all $i\ne j$:\\ $P\ll Q$ and $F_*P\ll F_*Q$\\
   $\mathcal H_i\cap\mathcal H_j\ne\varnothing$ for some $i\ne j$:\\ $P\perp Q$; by $\mathcal D$, $F_*P\perp F_*Q$\\
   \figref{\Cref{def:polynomial-root-model},}\\
   \figref{\Cref{prop:root-array-likelihood}, \Cref{cor:polynomial-observed-likelihood}}};
\end{tikzpicture}
\caption{Arithmetic tuples retain local dependence even when full-measure properties
transfer from an independent reference. A common Borel decoder $\mathcal D$
(\Cref{prop:canonical-decoder-mas08,cor:polynomial-observed-likelihood}) preserves the
array comparisons under observation. Entropy comparison and quantitative Fourier estimates
require separate arguments. Notation as in \Cref{sec:difference-law}; $\rho(\tau)$ is the
rate \eqref{eq:rho-def}. In (c), $P,Q,F$ refer to the polynomial model of
\Cref{def:polynomial-root-model}, as in
\Cref{prop:root-array-likelihood,cor:polynomial-observed-likelihood}.}
\label{fig:reference-models}
\end{figure}

\begin{corollary}[Rational statistics at affine prime arguments]
\label{cor:fixed-shift-rational-statistics}
For the affine forms of \Cref{thm:fixed-shift-tuple}, put
$\boldsymbol X=(e^{-Z_1},\ldots,e^{-Z_k})$. The empirical laws of
$(\varphi(L_j(p))/L_j(p))_{j=1}^k$, over primes
$p>\max_j|b_j|$, converge to the law of
$\boldsymbol X$. Moreover:
\begin{enumerate}[label=\textup{(\alph*)}]
\item Almost surely, $\boldsymbol X$ has the simultaneous rational
prefix approximations of \Cref{lem:rational-blocks}. The laws of
$\boldsymbol X$ and $\boldsymbol Z$ have Borel full carriers of
Hausdorff dimension zero, as do their locally Lipschitz images on
full-measure open sets.
\item For each fixed nonconstant $R\in\R(T_1,\ldots,T_k)$,
$R(\boldsymbol X)$ and $R(\boldsymbol Z)$ are defined almost surely
and have singular continuous laws of Hausdorff and Fourier dimension zero.
\item Almost surely, $X_1,\ldots,X_k$ are algebraically independent
over $\Q$, and $R(\boldsymbol X)\in\mathcal L$ simultaneously for every
nonconstant $R\in\Q(T_1,\ldots,T_k)$.
\end{enumerate}
\end{corollary}

\begin{proof}
The empirical convergence follows by exponentiation in
\Cref{thm:fixed-shift-tuple}. Use $P,R,\boldsymbol c$ from
\Cref{cor:affine-independent-reference}, with actual prefixes
$F_y(A)_j=\prod_{\ell\le y}(1-1/\ell)^{A_{\ell,j}}$.
Under $R$ these equal $c_jY_j(y)$, where the $Y_j(y)$ are independent
canonical $\mu_G$ prefixes. Their joint factor vector is nontrivial
with probability $k/\ell+O_k(\ell^{-2})$. Its expected sum of
denominator logarithms is $O_k((\log\ell)/\ell)$, and its expected
sum of absolute logarithms is $O_k(\ell^{-2})$. Hence \Cref{lem:rational-blocks} gives, for every
integer $m\ge1$, $y_n,Q_n\to\infty$ with $Q_nY_j(y_n)\in\Z$ and
$\max_j|Y_j-Y_j(y_n)|<Q_n^{-(m+1)}$.
Since $c_j\in\{1,2\}$, the same denominators make $Q_nF_{y_n}(A)_j$
integral, and their errors are at most $2Q_n^{-(m+1)}<Q_n^{-m}$
eventually. This is a Borel event on array space: restrict cutoffs to
integers and express it through the countably many approximation orders,
thresholds, cutoffs and integer denominators. It is $R$-full, hence
$P$-full by \eqref{eq:fixed-shift-reference-density}.
\Cref{lem:joint-rational-carrier,lem:lipschitz-image}, including the
map $\boldsymbol x\mapsto-\log\boldsymbol x$ on the positive orthant,
now give part~\textup{(a)}.

For each fixed nonzero real polynomial $B$, independence and atomlessness
imply $B(c_1Y_1,\ldots,c_kY_k)\ne0$ almost surely. Indeed, induction on
$k$ makes one nonzero coefficient in the last variable nonzero almost
surely; its finitely many roots then have conditional probability zero.
The same argument applies to the independent atomless coordinates
$-\log Y_j-\log c_j$. Absolute continuity therefore gives a null zero
set for each fixed nonzero real polynomial under both $\boldsymbol X$
and $\boldsymbol Z$.
The first paragraph supplies the approximation hypothesis of
\Cref{lem:joint-rational-carrier} for $\boldsymbol X$ almost surely
under $P$. Its coordinates are positive, and
$\boldsymbol Z=-\log\boldsymbol X$ is its locally Lipschitz image on
the positive orthant. Applying
\Cref{cor:nonconcentrated-rational-images}\textup{(a)} to these two
vectors, with the null zero sets just proved, gives part~\textup{(b)}.
Applying its part~\textup{(b)} to $\boldsymbol X$ and the rational-coefficient
null zero sets gives part~\textup{(c)}.
\end{proof}

\begingroup
\widowpenalty=10000
For each fixed quotient $F=P/Q$, with $P,Q$ real analytic on
$(0,\infty)^k$, $Q\not\equiv0$ and $F$ nonconstant,
$F(\boldsymbol X)$ is defined and absolutely normal almost surely.
Its law $\lambda$ is Rajchman and singular continuous, with Hausdorff
and Fourier dimensions zero. Its lower and upper local dimensions are
$0$ and $1$, respectively, $\lambda$-almost everywhere, and its
packing-carrier dimensions satisfy
$P_{\mathrm{pos}}(\lambda)=P_{\mathrm{full}}(\lambda)=1$.
The same assertions hold for $F(\boldsymbol Z)$ for each fixed quotient
$F=P/Q$ with $P,Q$ real analytic on $\R^k$, $Q\not\equiv0$ and $F$
nonconstant. All these conclusions persist under positive-probability
conditioning of the tuple.
Indeed, \eqref{eq:fixed-shift-reference-observation} gives
$\operatorname{Law}(\boldsymbol X)\ll
\operatorname{Law}(c_1Y_1,\ldots,c_kY_k)$ for independent $Y_j\sim\mu_G$.
Compose $F$ with this scaling, or with
$\boldsymbol y\mapsto(-\log(c_jy_j))_j$, and apply
\Cref{cor:multivariable-analytic-images}. Null sets and positive-mass
packing dimension transfer by domination, local dimensions by
differentiation of measures, and the Rajchman property by the band
argument of \Cref{cor:summable-local-coupling}.
\par\endgroup

\section{Differences at fixed shifts}
\label{sec:fixed-shift-differences}

For odd primes $p$, put
\[
 \Delta f(p):=f(p+1)-f(p-1)
 =\sum_{\substack{\ell\mid p+1\\\ell\ge3}}\omega_\ell
  -\sum_{\substack{\ell\mid p-1\\\ell\ge3}}\omega_\ell.
\]
The common prime-$2$ contribution cancels.

\begin{theorem}[Joint law at $p\pm1$]\label{thm:diff-dist}
\index{Joint laws!opposite shifts}
The empirical laws of $(f(p-1)-\log2,f(p+1)-\log2)$ converge to
\[
 \mu_{\mathrm{joint}}
 =\Conv_{\ell\ge3}
 \left(\frac{\ell-3}{\ell-1}\delta_{(0,0)}
 +\frac{\delta_{(\omega_\ell,0)}+\delta_{(0,\omega_\ell)}}{\ell-1}\right),
\]
which has Hausdorff dimension zero and support
\begin{equation}\label{eq:joint-support}
 \bigl([\omega_3,\infty)\times[0,\infty)\bigr)
 \cup\bigl([0,\infty)\times[\omega_3,\infty)\bigr).
\end{equation}
Its image under $(u_1,u_2)\mapsto u_2-u_1$ is the symmetric continuous
law $\mu_{\Delta f}$, singular of dimension zero and of support $\R$,
with
\[
 \mu_{\Delta f}=\Conv_{\ell\ge3}
 \left(\frac{\ell-3}{\ell-1}\delta_0+
       \frac{\delta_{-\omega_\ell}+\delta_{\omega_\ell}}{\ell-1}\right).
\]
\end{theorem}

\begin{proof}
Apply \Cref{thm:fixed-shift-tuple} with $(b_1,b_2)=(-1,1)$, so $E=\{2,3\}$.
The prime-$2$ vector is the constant $(\log2,\log2)$, and the prime-$3$
vector takes $(\omega_3,0)$ and $(0,\omega_3)$ with probability $1/2$ each;
subtracting $(\log2,\log2)$ gives the displayed joint convolution and
support. The difference map gives the scalar formulas, with symmetric
local factors. Part~(b) of \Cref{thm:fixed-shift-tuple} with
$\boldsymbol a=(-1,1)$ gives continuity and dimension zero, and either
quadrant in~\eqref{eq:joint-support} has difference set $\R$.
\end{proof}

By \Cref{prop:renyi-collapse}\textup{(b)}, the lower and upper dyadic $L^q$ dimensions
($q>1$) of $\mu_{\Delta f}$ and of its images under $u\mapsto e^{\pm u}$
vanish: take the prime-$3$ factor as
head; for $\ell\ge5$ the local factor is nonzero with probability
$2/(\ell-1)$ and has absolute mean $2\omega_\ell/(\ell-1)\ll\ell^{-2}$,
so $\Theta(z)\ll\log\log z$ and $\beta=1$.

For distinct nonzero integers $b,c$, define
\begin{equation}\label{eq:fixed-shift-difference}
 D_{b,c}:=\sum_\ell f(\ell)
       (\mathbf1_{U_\ell=-b}-\mathbf1_{U_\ell=-c}),\qquad
 Q_{b,c}(h):=\sup_u\mathbb P(D_{b,c}\in[u,u+h]).
\end{equation}
For an integer $a\ge1$ and primes $p>a$, put $\Delta_a f(p):=f(p+a)-f(p-a)$.

\begin{corollary}[Distinct fixed shifts]\label{cor:fixed-symmetric-shifts}
\index{Radical symmetry}
Let $b,c$ be distinct nonzero integers. Empirical laws below are taken
over primes with $p+b,p+c>0$.
\begin{enumerate}[\upshape(a)]
\item The empirical laws of $f(p+b)-f(p+c)$ converge to
$\operatorname{Law}(D_{b,c})$, which is continuous, singular of dimension zero,
and of support $\R$.
\item The empirical laws of $\varphi(p+b)/\varphi(p+c)$ converge to
$\operatorname{Law}(e^{-D_{b,c}})$, which gives full measure to the
Liouville numbers.
\item As $h\downarrow0$,
\[
 Q_{b,c}(h)\asymp_{b,c}(\log(1/h))^{-2}.
\]
\item The law of $D_{b,c}$ is symmetric about zero if and only if
$\operatorname{rad}(|b|)=\operatorname{rad}(|c|)$, where
$\operatorname{rad}(n)=\prod_{\ell\mid n}\ell$. For $b=a,c=-a$, $a\ge1$,
the limiting law of $\Delta_a f$ is
\begin{equation}\label{eq:fixed-shift-law}
 \mu_{\Delta_a f}
 =\Conv_{\substack{\ell\ge3\\\ell\nmid a}}
 \left(\frac{\ell-3}{\ell-1}\delta_0+
       \frac{\delta_{-\omega_\ell}+\delta_{\omega_\ell}}{\ell-1}\right).
\end{equation}
\item Put $\delta_{b,c}=\mathbb P(D_{b,c}>0)$. Each of the comparisons
\[
 \varphi(p+c)>\varphi(p+b),\qquad
 \frac{\varphi(p+c)}{p+c}>\frac{\varphi(p+b)}{p+b}
\]
has relative prime density $\delta_{b,c}$; the reverse comparisons
have density $1-\delta_{b,c}$, and ties have density zero.
If $\operatorname{rad}(|b|)=\operatorname{rad}(|c|)$, then
$\delta_{b,c}=1/2$. If $\operatorname{rad}(|b|)$ is a proper divisor of
$\operatorname{rad}(|c|)$, then $1/2<\delta_{b,c}<1$.
\end{enumerate}
\end{corollary}

\begin{proof}
Let $E$ consist of the primes at most $3$ and the prime divisors of
$bc(b-c)$. Outside $E$, the two roots are distinct and nonzero, giving
the ternary factors of \Cref{thm:diff-dist}, with positive mass at zero.
The sum of the coordinates indexed by $E$ is independent of the
coordinates outside $E$.
\Cref{thm:fixed-shift-tuple} gives convergence, continuity and dimension
zero. The difference map sends
$\supp(\boldsymbol H)+[0,\infty)^2$ onto $\R$.

The concentration assertion is
\Cref{thm:fixed-shift-tuple}\textup{(d)} with $\boldsymbol a=(1,-1)$.

The identity
\[
 \log\frac{\varphi(p+b)}{\varphi(p+c)}
 =-f(p+b)+f(p+c)+\log\frac{p+b}{p+c}
 =-f(p+b)+f(p+c)+O_{b,c}(p^{-1})
\]
and continuous mapping prove ratio convergence. For the tuple $(b,c)$,
the limiting ratio is $X_1/X_2=e^{-D_{b,c}}$.
\Cref{cor:fixed-shift-rational-statistics}\textup{(c)} applied to
$R(T_1,T_2)=T_1/T_2$ proves the Liouville assertion.

If the radicals agree, each local difference is zero or has a symmetric
ternary law; this proves symmetry and \eqref{eq:fixed-shift-law}.
Conversely, absolute convergence in mean gives
\[
 \mathbb ED_{b,c}
 =\sum_{\ell\mid c}\frac{f(\ell)}{\ell-1}
  -\sum_{\ell\mid b}\frac{f(\ell)}{\ell-1}.
\]
Symmetry makes this zero. The rational independence in \Cref{thm:linind},
including $f(2)=\log2$, forces the radicals to agree.

Under proper radical containment, let
$J=\{\ell:\ell\mid c,\ \ell\nmid b\}$ and write $D_{b,c}=Z+H$, where
\[
 H=\sum_{\ell\in J}f(\ell)\mathbf1_{\{U_\ell=-b\}}.
\]
Then $H\ge0$ is finite, $\mathbb P(H>0)>0$, and the independent remainder
$Z$ is symmetric. The support and atomlessness arguments above, unchanged
by finite deletions, give $\supp\operatorname{Law}(Z)=\R$ and a continuous
strictly increasing distribution function $F_Z$, with $F_Z(0)=1/2$.
Consequently $\delta_{b,c}=\mathbb EF_Z(H)\in(1/2,1)$.
Finally, atomlessness at zero and the logarithmic-ratio identity give
all sign frequencies and the zero density of ties.
\end{proof}

\KeepHeadingSpace{6\baselineskip}
\section{Corner asymptotics}
\label{sec:affine-corner-asymptotics}

\index{Polynomial prime arguments!joint corners}
Use \Cref{def:polynomial-root-model}, with the finite set
$\mathcal H=\bigcup_j\mathcal H_j$ of distinct factors other than $T$.
Choose a fixed cutoff $b\ge\max(E\cup\{2\})$, increasing it if necessary
so that $D/(\ell-1)\le1/2$ for every prime $\ell>b$, where
$D=\sum_{h\in\mathcal H}\deg h$. Put
\[
 H_{b,j}=\sum_{\ell\le b}\omega_\ell B_{\ell,j},\qquad
 W_{b,j}=Z_j-H_{b,j},\qquad
 q_{\ell,h}=\frac{\rho_h(\ell)}{\ell-1}\quad(\ell>b),\qquad
 r=|\mathcal H|.
\]

\paragraph*{Admissible finite heads}
The admissible head atoms and their masses are explicit finite data.
For $a\in\{0,1\}^d$, put
\[
 n_\ell(a)=\#\{u\in\mathbb F_\ell^\times:
        (\mathbf1_{F_j(u)=0})_{j=1}^d=a\}.
\]
The head support consists of the vectors
$\sum_{\ell\le b}\omega_\ell a_\ell$ with every $n_\ell(a_\ell)>0$,
and, for any such vector $h$,
\begin{equation}\label{eq:polynomial-head-mass}
 P\{\boldsymbol H_b=h\}
 =\sum_{\substack{(a_\ell)_{\ell\le b}:\,
                  \sum_{\ell\le b}\omega_\ell a_\ell=h}}
       \prod_{\ell\le b}\frac{n_\ell(a_\ell)}{\ell-1}.
\end{equation}
Minimal atoms are selected from this finite set by the coordinatewise
order; distinct residue configurations giving the same atom are
included in its full mass.
Independence and uniformity of the finitely many reduced residues give
\eqref{eq:polynomial-head-mass} by grouping equal weighted indicator
vectors.

\begin{proposition}[Polynomial root corners]
\label{prop:polynomial-root-corners}
With the notation above, the constant
\begin{equation}\label{eq:polynomial-corner-constant}
 K_b:=\lim_{y\to\infty}(\log y)^r
          \prod_{b<\ell\le y}\left(1-\sum_{h\in\mathcal H}q_{\ell,h}\right)
\end{equation}
exists in $(0,\infty)$. For $t_j\in(0,1)$, put
$L_j=\log(1/t_j)$ and
$L_h^*=\max_{j:h\in\mathcal H_j}L_j$.
Uniformly as $\min_jL_j\to\infty$, with no restriction on their ratios,
for either choice $\diamond\in\{<,\le\}$,
\begin{equation}\label{eq:polynomial-root-corners}
 P\{W_{b,j}\diamond t_j\text{ for every }j\}
 =\frac{K_b}{\prod_{h\in\mathcal H}L_h^*}
       \left(1+O_{\boldsymbol F,b}\left(\sum_jL_j^{-1}\right)\right).
\end{equation}
Empty products equal one; thus the assertion includes constant or
$T$-only coordinates and the completely deterministic tuple.
\end{proposition}

\begin{proof}
Write $q_\ell=\sum_{h\in\mathcal H}q_{\ell,h}$, so
$0\le q_\ell\le D/(\ell-1)\le1/2$ beyond $b$.
For each $h$, the strong root-intensity expansion gives
\[
 \sum_{b<\ell\le y}q_{\ell,h}
 =\log\log y+C_{h,b}+o(1/\log y).
\]
Since $\log(1-u)+u=O(u^2)$ on $[0,1/2]$ and the omitted prime
square tail is $O(1/(y\log y))$, there is a positive constant $k_{h,b}$
such that
\begin{equation}\label{eq:polynomial-factor-neutral-product}
 \prod_{b<\ell\le y}(1-q_{\ell,h})
 =\frac{k_{h,b}}{\log y}\{1+o(1/\log y)\}.
\end{equation}
For a subset $J\subseteq\mathcal H$, set
\[
 R_\ell(J)=\frac{1-\sum_{h\in J}q_{\ell,h}}
                         {\prod_{h\in J}(1-q_{\ell,h})}.
\]
All these factors are positive and
$|\log R_\ell(J)|\ll_{\boldsymbol F}\ell^{-2}$ uniformly in $J$.
Consequently $R_b:=\prod_{\ell>b}R_\ell(\mathcal H)$ exists in
$(0,\infty)$, and
$K_b=R_b\prod_{h\in\mathcal H}k_{h,b}$ proves
\eqref{eq:polynomial-corner-constant}.

Put $y_j=1/t_j$, $y_*:=\min_jy_j$, and
$Y_h:=\max_{j:h\in\mathcal H_j}y_j$; hence
$\log Y_h=L_h^*$. Let $\mathcal A$ prohibit $B_{\ell,j}=1$ whenever
$b<\ell\le y_j$. At a prime outside $E$ a residue lies in at most
one distinct factor root set, though that factor can activate several
coordinates. Thus, with $J_\ell=\{h:\ell\le Y_h\}$, the local
probability of $\mathcal A$ is $1-\sum_{h\in J_\ell}q_{\ell,h}$.
It follows exactly that
\[
 P(\mathcal A)
 =\left(\prod_{\ell>b}R_\ell(J_\ell)\right)
        \prod_{h\in\mathcal H}\prod_{b<\ell\le Y_h}(1-q_{\ell,h}).
\]
If $\ell\le y_*$, then $J_\ell=\mathcal H$. For larger primes,
\[
 \left|\sum_{\ell>b}\log R_\ell(J_\ell)-\log R_b\right|
 \ll_{\boldsymbol F}\sum_{\ell>y_*}\ell^{-2}
 \ll_{\boldsymbol F}(y_*\log y_*)^{-1}.
\]
Apply
\eqref{eq:polynomial-factor-neutral-product} at each $Y_h$.
Because $\mathcal H$ is fixed and $\log Y_h\ge\min_j\log y_j$,
the resulting relative error is $O_{\boldsymbol F,b}(\sum_jL_j^{-1})$,
uniformly even if the $y_j$ have unbounded ratios. We obtain
\begin{equation}\label{eq:polynomial-empty-cylinder}
 P(\mathcal A)=\frac{K_b}{\prod_h\log Y_h}
       \left\{1+O_{\boldsymbol F,b}\left(\sum_jL_j^{-1}\right)\right\}.
\end{equation}

Apply \Cref{lem:empty-cylinder}\textup{(b)} with
$\xi_{\ell,j}=0$ for $\ell\le b$ and
$\xi_{\ell,j}=B_{\ell,j}$ for $\ell>b$.
The vectors are independent across primes, and
\[
 P(\xi_{\ell,j}=1)\le\frac{D}{\ell-1}\le\frac{2D}{\ell}
 \qquad(\ell>b).
\]
Thus the lemma applies with $C=\max(1,2D)$ and
$\mathcal A_{\boldsymbol t}=\mathcal A$. Combining its uniform comparison
for unequal widths and either choice of $\diamond$ with
\eqref{eq:polynomial-empty-cylinder} proves
\eqref{eq:polynomial-root-corners}. If $\mathcal H$ is empty all tails
vanish, and the same statements hold with $K_b=1$.
\end{proof}

\begin{corollary}[Corners at minimal finite-head atoms]
\label{cor:polynomial-minimal-head-corners}
In the notation of \Cref{prop:polynomial-root-corners}, let $h$ be a
coordinatewise minimal atom of the \emph{joint} finite head
$\boldsymbol H_b$: no distinct head atom $h'$ satisfies
$h'\le h$ in all coordinates. For $t_j>0$ with sufficiently small
$\max_jt_j$,
\begin{equation}\label{eq:polynomial-minimal-head}
 P\!\left\{\boldsymbol Z\in h+\prod_j[0,t_j]\right\}
 =P\{\boldsymbol H_b=h\}\,
       P\{W_{b,j}\le t_j\text{ for every }j\}.
\end{equation}
For either $\diamond\in\{<,\le\}$, the same equality holds for the
event $Z_j\diamond h_j+t_j$ for every $j$, with the corresponding tail
inequalities. In particular, for
$\varepsilon_j\downarrow0$ and $M_j=\log(1/\varepsilon_j)$,
\begin{equation}\label{eq:polynomial-raw-head-corner}
 P\{X_j>e^{-h_j}(1-\varepsilon_j)\text{ for every }j\}
 =\frac{P\{\boldsymbol H_b=h\}K_b}
        {\prod_{g\in\mathcal H}\max_{j:g\in\mathcal H_j}M_j}
 \left(1+O_{\boldsymbol F,b}\left(\sum_jM_j^{-1}\right)\right).
\end{equation}
Non-strict raw inequalities give the same formula. The head probability
is its full joint mass. The vector of the separate coordinate minima
need not be a head atom. 
\end{corollary}

\begin{proof}
The head has finite support. For each atom $h'\ne h$, minimality of
$h$ provides a coordinate $j$ with $h'_j>h_j$. The minimum, over the
finitely many other atoms, of $\max_j(h'_j-h_j)$ is positive
(unless there are no other atoms, in which case no restriction is
needed). Once every $t_j$ is smaller than this minimum, the
nonnegative tail prevents any such $h'$ from contributing to either
specified event. On $\{\boldsymbol H_b=h\}$, each event is exactly
the corresponding tail rectangle, and the head and tail are
independent. This proves \eqref{eq:polynomial-minimal-head} and its
one-sided variant. Finally apply \eqref{eq:polynomial-root-corners}
with $t_j=-\log(1-\varepsilon_j)$. Then $\log(1/t_j)=\log(1/\varepsilon_j)+O(\varepsilon_j)$, uniformly
as all widths shrink. The maxima change by at most
$O(\max_j\varepsilon_j)$, so replacement in the finite denominator
preserves the displayed relative error and proves
\eqref{eq:polynomial-raw-head-corner} with both boundary conventions.
\end{proof}

\paragraph*{Finite compatibility and repeated factors}
For $F_1=T-1,F_2=T+1$, take $b=3$. The prime $2$ contributes
$(\log2,\log2)$, and the prime $3$ contributes
$(\log(3/2),0)$ or $(0,\log(3/2))$, each with probability $1/2$.
These are the two minimal joint head atoms after the common translation.
The vector $(\log2,\log2)$ of separate coordinate minima is not an atom;
indeed $P\{Z_j\le\log2+t_j\ (j=1,2)\}=0$ if
$\max(t_1,t_2)<\log(3/2)$. Here
\[
 K_3=\lim_{y\to\infty}(\log y)^2
              \prod_{3<\ell\le y}(1-2/(\ell-1))
     =2e^{-2\gamma}\mathfrak S_3,
\]
with $\mathfrak S_3$ as in \Cref{prop:deleted-corners}; each admissible
corner consequently has coefficient $e^{-2\gamma}\mathfrak S_3$.
More generally the affine case of \Cref{prop:polynomial-root-corners}
has $K_b=e^{-k\gamma}\kappa_{E_b}$ for
$E_b=\{\ell:\ell\le b\}$, by Mertens' product formula. Removing only
the original affine exceptional set $E$ gives exactly
\Cref{prop:deleted-corners}, including its sharper explicit constants
and separate concentration lower bound.

Repeated powers change neither $\mathcal H_j$ nor any root indicator.
For example $(T-1)^2$ and $(T-1)^3$ have equal products in the common
model, one factor in the corner denominator, and joint law singular
relative to the product of their actual marginals. Shared factors $T$
and constant content alone only change the sampled-prime correction
and finite head described in \Cref{def:polynomial-root-model}.
For $F=T^2+1$, the nonzero root count at an odd prime is $2$ for
$\ell\equiv1\pmod4$ and zero for $\ell\equiv3\pmod4$.
Its one-factor corner exponent is still one by
\Cref{lem:polynomial-root-intensity}; its marginal probabilities are
not the affine probabilities $1/(\ell-1)$. 

\begin{proposition}[Corners after deletion of the exceptional primes]
\label{prop:deleted-corners}
In the affine tuple model of \Cref{thm:fixed-shift-tuple}, put
$\boldsymbol W=\boldsymbol Z-\boldsymbol H$ and
\[
 \kappa_E=
 \prod_{\ell\in E}(1-1/\ell)^{-k}
 \prod_{\ell\notin E}
 \frac{1-k/(\ell-1)}{(1-1/\ell)^k}.
\]
Then $0<\kappa_E<\infty$. Uniformly for positive widths with
$\max_j\varepsilon_j\to0$,
\begin{equation}\label{eq:deleted-corner}
\begin{aligned}
 \mathbb P\!\left(\boldsymbol W\in\prod_{j=1}^k[0,\varepsilon_j]\right)
 &=\frac{e^{-k\gamma}\kappa_E}{\prod_{j=1}^k\log(1/\varepsilon_j)}\\
 &\quad\times\left(1+O_{\boldsymbol L}\!\left(
       \sum_{j=1}^k\frac1{\log(1/\varepsilon_j)}\right)\right).
\end{aligned}
\end{equation}
If $h$ is an atom of $\boldsymbol H$ and no distinct atom $h'$ satisfies
$h'\le h$ coordinatewise, the same formula for
$\mathbb P(\boldsymbol Z\in h+\prod_j[0,\varepsilon_j])$ has the additional
factor $\mathbb P(\boldsymbol H=h)$.

For the centered $(-1,1)$ tuple, set $a=\log(3/2)$ and
\[
 \mathfrak S_3=\frac92\prod_{\ell\ge5}
                 \frac{\ell^2(\ell-3)}{(\ell-1)^3}.
\]
Its two minimal corners are $(a,0)$ and $(0,a)$, each with coefficient
$e^{-2\gamma}\mathfrak S_3$ in~\eqref{eq:deleted-corner}.
For equal widths $\varepsilon_1=\varepsilon_2=\varepsilon$, its origin
square has mass zero when $\varepsilon<a$. For every pair
of distinct nonzero shifts,
\begin{equation}\label{eq:difference-lower}
 \liminf_{h\downarrow0}(\log(1/h))^2Q_{b,c}(h)
 \ge e^{-2\gamma}\mathfrak S_3.
\end{equation}
\end{proposition}

\begin{proof}
The infinite product defining $\kappa_E$ has positive factors
$1+O_k(\ell^{-2})$. Put $y_j=1/\varepsilon_j$ and
$y_*:=\min_jy_j>\max E$. For $\ell\notin E$, set
$r_\ell=\#\{j:\ell\le y_j\}$, and let $A$ exclude each local state
$\omega_\ell e_j$ with $\ell\le y_j$.
This is the event $\mathcal A_{\boldsymbol t}$ of \Cref{lem:empty-cylinder}
for $\boldsymbol t=(\varepsilon_1,\ldots,\varepsilon_k)$ and the indicators
$\xi_{\ell,j}=\mathbf1_{\{L_j(U_\ell)=0\}}$ for $\ell\notin E$,
$\xi_{\ell,j}=0$ for $\ell\in E$, whose weighted sums are the coordinates
of $\boldsymbol W$. The indicator vectors are independent across primes
because the $U_\ell$ are independent. The local states outside $E$
are mutually exclusive, so
\[
 \mathbb P(A)=\prod_{\ell\notin E}(1-r_\ell/(\ell-1)).
\]
The ratio of each factor to $(1-1/\ell)^{r_\ell}$ is
$1+O_k(\ell^{-2})$. Below $y_*$ it agrees with the ratio defining
$\kappa_E$; the changed tail contributes $1+O_k(y_*^{-1})$.
Mertens' product estimate in \Cref{lem:prime-log-stieltjes}, applied
at each $y_j$, therefore gives
\[
 \mathbb P(A)=\frac{e^{-k\gamma}\kappa_E}{\prod_j\log y_j}
 \left(1+O_{\boldsymbol L}\!\left(\sum_j\frac1{\log y_j}\right)\right).
\]
Since $\mathbb P(\xi_{\ell,j}=1)\le1/(\ell-1)\le2/\ell$,
\Cref{lem:empty-cylinder}\textup{(b)} shows that the rectangle
probability is $\mathbb P(A)\{1+O_k(\sum_j1/\log y_j)\}$,
proving~\eqref{eq:deleted-corner}.
For a minimal head atom every other head atom has a coordinate
strictly larger than the corresponding coordinate of $h$. Finiteness
of the head and nonnegativity of the tail exclude all those atoms
from $h+\prod_j[0,\varepsilon_j]$ once $\max_j\varepsilon_j$ is small.

For $(-1,1)$, deletion of $2,3$ leaves the stated product with
$\kappa_{\{2,3\}}=2\mathfrak S_3$. After subtracting $(\log2,\log2)$,
the head takes $(a,0),(0,a)$ with probability $1/2$ each. This proves
the corner and origin assertions.

For the last statement, use a sufficient event and absolute-tail
control as in \eqref{eq:lq-empty-block}. The prime-$2$ difference is
deterministic. At $3$ choose a difference atom of mass at least $1/2$.
At any $\ell\ge5$ the probability of zero difference is at least
$1-2/(\ell-1)$: there are at most two unit residue classes on which
the two indicators differ. Prescribe these zero states through
$y=1/h$. The probability of the resulting cylinder is at least
\[
 \frac12\prod_{5\le\ell\le y}(1-2/(\ell-1))
 \sim\frac{e^{-2\gamma}\mathfrak S_3}{(\log y)^2}.
\]
The independent absolute difference tail has mean
$O((y\log y)^{-1})$. With probability $1-O(1/\log y)$ it is at most
$h/2$, so an interval of length $h$ centered at the prescribed head
has the asserted mass, proving~\eqref{eq:difference-lower}.
\end{proof}

\paragraph*{Relation to earlier work}
For the present $f$, existence and continuity of the law of
$f(p+b)-f(p+c)$ for distinct nonzero shifts go back to
K\'atai~\cite[Theorem~2]{Kat69}, who also proved existence of the joint law
\cite[Theorems~1 and~3]{Kat69}. K\'atai's joint mean-value theorem for coprime linear forms at prime
arguments is~\cite[Theorem~1, (1.3)]{Kat69}. De Koninck and
K\'atai~\cite[Theorem~B]{DK19} state a version for distinct affine forms
without proof, as a consequence of the methods of~\cite{Kat68}; its criterion
covers $g_j=e^{i\tau_jf}$, since
$\sum_\ell|1-g_j(\ell)|/\ell\le|\tau_j|\sum_\ell f(\ell)/\ell<\infty$.
In the cited author manuscript, enlarge the exceptional set of
\cite[Theorem~B]{DK19} by the prime divisors of the $\alpha_jb_j$ and
replace $1-1/(q-1)$ in the remaining local factors by $1-k/(q-1)$.
The local factors for $f$ are those of the model preceding
\Cref{thm:fixed-shift-tuple}. They also give the symmetric scalar
formulas~\cite[Theorem~5]{DK19}.
The sign asymptotics are due to Garcia and Luca~\cite[arXiv v2, Theorem~1.1]{GL18}; De Koninck
and K\'atai~\cite[Theorem~5 and the remark following it; \S~3 for $a=1$]{DK19} also deduce them at every fixed
positive symmetric shift $a$, using the limiting logarithmic law and
$(p+a)/(p-a)\to1$.

The order $(\log(1/h))^{-2}$ for the logarithmic ratio at shift $1$
was obtained by De Koninck and K\'atai~\cite[Theorem~3]{DK19},
using the comparison of Indlekofer and K\'atai~\cite{IK92}.
Equation~\eqref{eq:tuple-projection-concentration} gives this order for
differences at admissible affine forms, and the order
$(\log(1/h))^{-r}$ for fixed projections with $r$ nonzero coefficients.

\chapter{Additive laws with logarithmically decreasing prime values}
\label{ssec:slow-jump-ac}

Replace the totient increments $\omega_\ell\asymp1/\ell$ by
$f_A(\ell)=(\log\ell)^{-A}$, for fixed $A>0$, while keeping the
shifted-prime activation probabilities $1/(\ell-1)$. The resulting
law has a density for every $A>0$, with distinct strict thresholds
for stronger regularity. The convolution example at $A=3$
illustrates these thresholds. The families in
\Cref{cor:heavier-activation,cor:positive-activation} instead keep
$\omega_\ell$ and vary the activation probabilities.

The integer model provides the historical comparison. Babu proved
an $L^2$ density for $0<A<2$~\cite[pp.~401--403]{Bab75}.
The all-$A$ absolute-continuity result below contradicts the singularity
conjecture stated for $A>2$ in
\cite[Remark~2 after Theorem~3, p.~297]{EK79}.
This is distinct from the concentration conjecture in Remark~1 there,
whose order $h^{1/A}$ for fixed $A>1$ was established
in~\cite[Theorem~1.3]{Kou14}. The integer laws are discussed in
\hyperref[par:slow-jump-integer]{Sampling over the integers}.
For common prime activation of finitely many distinct exponents,
\Cref{thm:shared-prime-anisotropic,prop:shared-prime-global-regularity,prop:shared-prime-sobolev}
give absolute continuity, Fourier bounds, and differentiability
and Sobolev thresholds under their prime-series hypotheses;
\Cref{thm:planar-regularity} treats the planar case.

\section{Scalar laws and conditional smoothing}

Throughout this chapter $A>0$ is fixed and $\ell$ ranges over odd primes
unless stated otherwise. Define the prime increments
\begin{equation}\label{eq:slow-aell}
  a_\ell:=(\log\ell)^{-A}\quad(\ell\ge3),\qquad a_2:=(\log2)^{-A},
\end{equation}
the Bernoulli probabilities $p_\ell:=1/(\ell-1)$, the two-point local laws
\begin{equation}\label{eq:slow-nu}
  \nu_{A,\ell}:=(1-p_\ell)\,\delta_0+p_\ell\,\delta_{a_\ell},
\end{equation}
and the shifted-prime Erd\H{o}s--Wintner law of the strongly additive function
$f_A(n)=\sum_{\substack{q\mid n\\q\textup{ prime}}}(\log q)^{-A}$,
\begin{equation}\label{eq:slow-mu}
  \mu_A\;:=\;\delta_{a_2}\;*\;\Conv_{\substack{\ell\ge3\\\ell\textup{ prime}}}\nu_{A,\ell},
\end{equation}
the limiting distribution of $f_A(p-1)$ over primes
(\Cref{thm:pure-type}); the atom $\delta_{a_2}$
records $2\mid p-1$ for odd primes $p$.
Write $\Phi_A(\tau)=\widehat\mu_A(\tau)$ and use the weights
$d_\ell=2(\ell-2)/(\ell-1)^2$ from \eqref{eq:def-dell}. Define the
nonnegative sum
\begin{equation}\label{eq:slow-dissip}
 S_A(\tau)=\sum_{\substack{\ell\ge3\\\ell\textup{ prime}}}
 d_\ell(1-\cos(\tau a_\ell)).
\end{equation}
We also use the reference sum
\begin{equation}\label{eq:slow-reference-dissip}
 \Sigma_A(\tau):=\sum_{\substack{\ell\ge3\\\ell\textup{ prime}}}
 \frac{1-\cos(\tau a_\ell)}{\ell}.
\end{equation}

\paragraph*{Density and regularity}
For the basic law $\mu_A$, absolute continuity holds for every $A>0$
(\Cref{lem:slow-jump-all-ac,thm:slow-jump-ac}). Write $h_A$ for its density.
For $s\in\R$, let $H^s(\R)$ be the space of tempered distributions $u$
whose Fourier transform is a function satisfying
\[
 \int_{\R}(1+t^2)^s|\widehat u(t)|^2\,dt<\infty.
\]
 The exact
Fourier and Sobolev thresholds are
\[
 \Phi_A\in L^q(\R)\ \Longleftrightarrow\ q>A\quad(q>0),\qquad
 h_A\in H^s(\R)\ \Longleftrightarrow\ s<1/A-1/2\quad(s\in\R).
\]
They follow from the cancellation and shell-average estimates in
\Cref{prop:slow-jump-regularity}. By
\Cref{prop:slow-jump-positive-endpoint}, a globally $C^k$ density exists
exactly when $A<1/(k+1)$, for integers $k\ge0$. When $A>1$,
\Cref{prop:scalar-density}\textup{(a)} gives essential unboundedness
on every open interval in the interior of the support.

\paragraph*{Factorization and support}
Partial summation from Mertens' theorem~\cite{Mer74} gives
\[
 \sum_{\ell\ge3}p_\ell a_\ell
 \le2\sum_{\ell\ge3}\frac{(\log\ell)^{-A}}\ell<\infty.
\]
Hence the independent nonnegative prime series defining $\mu_A$
converges almost surely. Since $a_\ell\to0$, $\sum_\ell a_\ell=\infty$
and every local law has atoms at $0$ and $a_\ell$, \Cref{lem:supp_conv}
gives
\begin{equation}\label{eq:slow-support}
 \supp\mu_A=[a_2,\infty).
\end{equation}
The law is atomless by \Cref{lem:levy-atomless}, since
$\sum_\ell p_\ell=\infty$. Independence and bounded convergence give
$\Phi_A(\tau)=e^{i\tau a_2}\prod_{\ell\ge3}\widehat\nu_{A,\ell}(\tau)$.
The local squared modulus is
$1-2p_\ell(1-p_\ell)(1-\cos(\tau a_\ell))$, so $1-x\le e^{-x}$ gives
\begin{equation}\label{eq:slow-euler}
 |\Phi_A(\tau)|^2
 =\prod_{\substack{\ell\ge3\\\ell\textup{ prime}}}
 \bigl(1-d_\ell(1-\cos(\tau a_\ell))\bigr)
 \le e^{-S_A(\tau)}\qquad(\tau\in\R).
\end{equation}
The series for $S_A$ converges locally uniformly, since its terms are
$O_\tau(\ell^{-1}(\log\ell)^{-2A})$.

\begin{lemma}[Bounds for $S_A$ and $\Sigma_A$]
\label{lem:slow-jump-dissipation}
Uniformly in $\tau\in\R$,
\begin{equation}\label{eq:slow-diss-comparison}
 S_A(\tau)=2\Sigma_A(\tau)+O(1).
\end{equation}
For all sufficiently large $|\tau|$, with constants depending only on $A$,
\begin{equation}\label{eq:slow-diss-sharp}
 \frac1A\log|\tau|-\log\log|\tau|-O_A(1)
 \le\Sigma_A(\tau)
 \le\frac1A\log|\tau|+\log\log|\tau|+O_A(1).
\end{equation}
\end{lemma}

\begin{proof}
Since $d_\ell=2/\ell+O(\ell^{-3})$ and $0\le1-\cos v\le2$,
the difference in \eqref{eq:slow-diss-comparison} is uniformly bounded.
By evenness take $\tau>0$. For fixed $A>0$,
\cite[Lemma~3, p.~300]{EK79} gives
\[
 \sum_{\tau^{10}\le\ell\le\exp(\tau^{1/A})}
 \frac{\cos(\tau(\log\ell)^{-A})}{\ell}=O_A(1).
\]
Mertens' theorem therefore evaluates the contribution of these primes
to $\Sigma_A$ as
\[
 \sum_{\tau^{10}\le\ell\le\exp(\tau^{1/A})}
 \frac{1-\cos(\tau a_\ell)}{\ell}
 =\frac1A\log\tau-\log\log\tau+O_A(1).
\]
The smaller primes contribute between zero and
$2\log\log\tau+O(1)$. On the remaining range, $1-\cos v\le v^2/2$
and partial summation from Mertens' theorem give
\[
 \sum_{\log\ell>\tau^{1/A}}
 \frac{1-\cos(\tau a_\ell)}{\ell}
 \ll\tau^2\sum_{\log\ell>\tau^{1/A}}
 \frac1{\ell(\log\ell)^{2A}}=O_A(1).
\]
These two tails prove \eqref{eq:slow-diss-sharp}.
\end{proof}

\begin{lemma}[$L^2$ inversion]\label{lem:l2-inversion}
Let $\mu$ be a finite positive Borel measure on $\R$ with
$\widehat\mu\in L^2(\R)$, where $\widehat\mu(\tau)=\int_\R e^{i\tau x}\,d\mu(x)$.
Then $\mu$ is absolutely continuous with a density $h\in L^1(\R)\cap L^2(\R)$,
equal in $L^2$ to $\lim_{R\to\infty}\frac1{2\pi}\int_{-R}^{R}e^{-i\tau\,\cdot}\,\widehat\mu(\tau)\,d\tau$.
If in addition $\widehat\mu\in L^1(\R)$, then $h$ has a bounded continuous
version $h(x)=\frac1{2\pi}\int_\R e^{-i\tau x}\widehat\mu(\tau)\,d\tau$.
\end{lemma}

\begin{proof}
Let $g_\varepsilon$ be the centered Gaussian density of variance
$\varepsilon^2$. Fourier inversion gives the continuous density
$h_\varepsilon=\mathcal F^{-1}(\widehat\mu\,\widehat g_\varepsilon)$
of $\mu*g_\varepsilon$. Dominated convergence and Plancherel give
$h_\varepsilon\to h:=\mathcal F^{-1}\widehat\mu$ in $L^2$, with
$\|h\|_2^2=(2\pi)^{-1}\|\widehat\mu\|_2^2$.
Since $\mu*g_\varepsilon\Rightarrow\mu$, testing against $C_c(\R)$ gives
$\mu=h\,dx$, hence $h\ge0$ and $h\in L^1\cap L^2$.
The truncated inverses converge in $L^2$ by Plancherel.
If $\widehat\mu\in L^1$, the ordinary inversion formula gives the
bounded continuous version.
\end{proof}

The obstacle to proving absolute continuity for every $A$ is that
the unconditioned Fourier transform need not be square integrable.
The following argument instead conditions on the number of activated
primes in each block. A block containing exactly one activation still
has randomness in the location of that prime. At a given large
frequency, many such blocks provide small conditional Fourier factors;
their product yields a smooth conditional density.

This adapts the block-conditioning method of Peled--Peres--Pitman--Tanaka
\cite[Theorem~4.1]{PPPT15}, for increments $n^{-s}$ indexed by all positive
integers, to logarithmic increments indexed by primes. The prime number
theorem supplies the conditional block estimate. The frequency threshold
may depend on the conditioned count sequence, so the argument does not
give a uniform smoothness bound for the mixture. What passes to the
unconditioned law is absolute continuity: every Lebesgue-null set has
zero mass in almost every conditional law, hence also in their mixture.

The following lemma is also applied in \Cref{cor:infinite-weighted-ac}, to infinite weighted sums of
independent basic logarithmic variables; \Cref{fig:sum-regularity} shows its place in that argument.

\begin{lemma}[Conditional block smoothing]
\label{lem:conditional-block-smoothing}
Let $d\ge1$ be an integer and let $(Y_k,N_k)_{k\ge1}$ be independent
pairs, with $Y_k\in\R^d$ and $N_k$ countable-valued. Suppose
$\sum_kY_k$ converges almost surely.
Put $q_k=\mathbb P(N_k=1)$ and, when $q_k>0$, put
$\psi_k(t)=\mathbb E(e^{i\langle t,Y_k\rangle}\mid N_k=1)$.
\begin{enumerate}[label=\textup{(\alph*)},leftmargin=*]
\item \emph{General shells.}
Let $1<T_1<T_2<\cdots\to\infty$. Suppose finite deterministic sets
$J_m$ and fixed constants $C,\delta>0$ satisfy the following conditions,
with the Fourier bound required for all sufficiently large $m$:
\begin{equation}\label{eq:conditional-general-shells-mas01}
 \begin{gathered}
 M_m:=\sum_{k\in J_m}q_k,\qquad
 \sum_m e^{-M_m/8}<\infty,\\
 |\psi_k(t)|\le C\|t\|_\infty^{-\delta}
 \quad\bigl(T_m\le\|t\|_\infty<T_{m+1},\ k\in J_m,\ q_k>0\bigr).
 \end{gathered}
\end{equation}
For almost every count sequence $\mathbf N=(N_k)$, the conditional sum
law $\nu_{\mathbf N}$ satisfies, on every sufficiently large shell,
\begin{equation}\label{eq:conditional-general-decay-mas01}
 |\widehat\nu_{\mathbf N}(t)|
 \le\exp\{-\tfrac14\delta M_m\log\|t\|_\infty\}.
\end{equation}
Each such law has a smooth density with every derivative bounded.
The unconditioned sum law, and its convolution with any independent
law on $\R^d$, are absolutely continuous.
\item \emph{Dyadic specialization.}
In the setup of part~\textup{(a)}, if $T_m=2^m$,
$M_m\ge cm^\rho$ for all sufficiently large $m$, with fixed $c,\rho>0$,
and the singleton bound is written as
\[
 \sup_{2^m\le\|t\|_\infty<2^{m+1}}|\psi_k(t)|
 \le C2^{-\delta m}\quad(k\in J_m,\ q_k>0),
\]
then, almost surely for all sufficiently large $\|t\|_\infty$,
\[
 |\widehat\nu_{\mathbf N}(t)|
 \le\exp\{-c'(\log\|t\|_\infty)^{1+\rho}\},
\]
where $c'>0$ is deterministic.
\end{enumerate}
The threshold in each part may depend on $\mathbf N$.
\end{lemma}

\begin{proof}
For part~\textup{(a)}, the indicators in $Z_m=\sum_{k\in J_m}\mathbf1_{\{N_k=1\}}$ are
independent and have mean sum $M_m$. With $s=\log2$,
\[
 \mathbb P(Z_m<M_m/2)
 \le e^{sM_m/2}\mathbb Ee^{-sZ_m}
 \le\exp\{(s/2-1/2)M_m\}\le e^{-M_m/8}.
\]
These bounds are summable. Borel--Cantelli gives one event on which
$Z_m\ge M_m/2$ for every sufficiently large $m$.

For each $k$ let $K_k(n,\cdot)$ be the law of $Y_k$ given $N_k=n$,
chosen arbitrarily when $\mathbb P(N_k=n)=0$. The product kernels
$\bigotimes_kK_k(n_k,\cdot)$ are measurable in $(n_k)$ on cylinder
sets, hence on the product $\sigma$-field by the monotone-class theorem.
Independence gives the conditional-law identities on cylinder sets;
the same theorem extends these identities to the product $\sigma$-field.
Thus this product kernel is a conditional law of $(Y_k)$ given
$\sigma(N_k:k\ge1)$. The Borel set on which $\sum_kY_k$ converges
has conditional probability one for almost every sequence.
For each such sequence, bounded convergence gives, for every $t$,
\[
 \widehat\nu_{\mathbf N}(t)
 =\lim_{r\to\infty}\prod_{k=1}^r
       \int_{\R^d}e^{i\langle t,y\rangle}K_k(N_k,dy).
\]
All factors have modulus at most one. For sufficiently large $m$,
$C\|t\|_\infty^{-\delta}\le\|t\|_\infty^{-\delta/2}$ throughout
the shell. Retaining the successful factors from the finite set $J_m$
therefore bounds the product by
\[
 \|t\|_\infty^{-\delta Z_m/2}
 \le\exp\{-\delta M_m\log\|t\|_\infty/4\}.
\]
This holds on all sufficiently large shells on one event of
probability one. Summability forces $M_m\to\infty$, so every
polynomially weighted conditional transform is integrable.
Differentiated Fourier inversion gives the asserted conditional
densities and derivatives. For any Lebesgue-null Borel set $E$,
integration of $\nu_{\mathbf N}(E)=0$ proves absolute continuity of
the mixture. Integrating the same assertion for the translates $E-y$
proves the independent-convolution assertion.

For part~\textup{(b)},
$C2^{-\delta m}\le C2^\delta\|t\|_\infty^{-\delta}$ on the shell,
and $\sum_m e^{-cm^\rho/8}<\infty$. Applying the preceding argument
and $m\asymp\log\|t\|_\infty$ proves its claimed estimate.
For arbitrary shells a bound stated only as $CT_m^{-\delta}$ does
not give \eqref{eq:conditional-general-shells-mas01} without additional
control; for example, a bounded ratio
$\log T_{m+1}/\log T_m$ would suffice after reducing $\delta$.
\end{proof}

\begin{lemma}[Absolute continuity for logarithmic prime increments]
\leavevmode\phantomsection\label{lem:slow-jump-all-ac}\ignorespaces
Let $A,c_\ast>0$, $\beta\ge0$, and let $(B_\ell)_{\ell\ge3}$ be independent Bernoulli
variables, indexed by odd primes, with probabilities $p_\ell\in[0,1]$
satisfying
\[
 \sum_{\ell\ge3}\left|p_\ell-
       \frac{c_\ast}{\ell(\log\log\ell)^\beta}\right|<\infty.
\]
Then $\sum_{\ell\ge3}B_\ell(\log\ell)^{-A}$ converges almost surely and
has an absolutely continuous law for $0\le\beta\le1$, and a countably
atomic law for $\beta>1$.
\end{lemma}

\begin{proof}
For $\beta>1$, Mertens' theorem and partial summation give
$\sum_\ell p_\ell<\infty$. Borel--Cantelli then gives only finitely
many active coordinates, proving convergence and countable atomicity.
Suppose $0\le\beta\le1$. First take
$p_\ell=c_\ast/[\ell(\log\log\ell)^\beta]$ for $\ell>e^{2^{k_0}}$, with $k_0$
large enough that these probabilities are less than $1/2$. The series
has finite mean by partial summation from Mertens' theorem. For
$k\ge k_0$, put $U_k=2^k$ and
\[
 I_k=\{\ell:e^{U_k}<\ell\le e^{2U_k}\},\qquad
 N_k=\sum_{\ell\in I_k}B_\ell,\qquad
 Y_k=\sum_{\ell\in I_k}B_\ell(\log\ell)^{-A}.
\]
Set $w_\ell=p_\ell/(1-p_\ell)$ and
$W_k=\sum_{\ell\in I_k}w_\ell$. Weighted Mertens summation and
$w_\ell=p_\ell+O(p_\ell^2)$ give
\[
 W_k\sim\frac{c_\ast\log2}{(\log U_k)^\beta},\qquad
 q_k:=\mathbb P(N_k=1)
 =W_k\prod_{\ell\in I_k}(1-p_\ell)\asymp k^{-\beta}.
\]
In particular, $q_k\sim c_\ast/k$ when $\beta=1$.
Conditional on $N_k=1$, the selected prime has probabilities
$w_\ell/W_k$. Its characteristic function is therefore
\[
 \psi_k(\tau)=W_k^{-1}\sum_{\ell\in I_k}
                  w_\ell e^{i\tau(\log\ell)^{-A}}.
\]
For $U=U_k$ and $\tau\ge1$, \eqref{eq:stieltjes} gives
\[
 \begin{aligned}
 \sum_{\ell\in I_k}\frac{e^{i\tau(\log\ell)^{-A}}}
                           {\ell(\log\log\ell)^\beta}
 &=\int_U^{2U}e^{i\tau v^{-A}}\,\frac{dv}{v(\log v)^\beta}\\
 &\quad+O_{A,\beta}\!\left(
      \frac{e^{-c\sqrt U}(1+\tau U^{-A})}{(\log U)^\beta}\right).
 \end{aligned}
\]
\AddToHookNext{build/column/before}{\raggedbottom}
After the substitution $w=v^{-A}$, integration by parts bounds the
integral by $O_{A,\beta}(U^A/[\tau(\log U)^\beta])$. The odds-weight
error is $O_{c_\ast}(e^{-U}(\log U)^{-2\beta})$.
Division by $W_k$ therefore gives
\begin{equation}\label{eq:slow-singleton-transform}
 |\psi_k(\tau)|\ll_{A,c_\ast,\beta}
       \frac{U_k^A}{\tau}+(1+\tau)e^{-c\sqrt{U_k}}.
\end{equation}
In particular, for all sufficiently large $|\tau|$, uniformly over
\[
 \frac{\log|\tau|}{4A\log2}\le k\le
 \frac{\log|\tau|}{2A\log2},
\]
one has $|\psi_k(\tau)|\le|\tau|^{-1/3}$; negative frequencies follow
by conjugation. Put
\[
 J_m=\{k\ge k_0:\lceil(m+1)/(4A)\rceil\le k\le\lfloor m/(2A)\rfloor\}.
\]
For every $2^m\le|\tau|<2^{m+1}$ these blocks lie in the displayed
window, and $\sum_{k\in J_m}q_k\asymp m^{1-\beta}$.
For $\beta<1$, \Cref{lem:conditional-block-smoothing}\textup{(b)}, with $d=1$,
$\rho=1-\beta$ and $\delta=1/3$, proves absolute continuity.

For $\beta=1$, put $T_n=2^{2^n}$ and, for a fixed sufficiently large
integer $K$, retain the blocks
\[
 J_n'=\{k:2(n+1)+K\le k\le\lfloor2^n/(2A)\rfloor\}.
\]
On $T_n\le|\tau|<T_{n+1}$, their first term in
\eqref{eq:slow-singleton-transform} is at most $|\tau|^{-1/2}$;
since $\sqrt{U_k}\ge2^{n+1+K/2}$, choosing $K$ large makes the
second at most $O(|\tau|^{-3})$. Thus $|\psi_k(\tau)|\le|\tau|^{-1/3}$
uniformly on the shell for all large $n$. Here
$M_n:=\sum_{k\in J_n'}q_k\sim c_\ast n\log2$. The Chernoff estimate
in \Cref{lem:conditional-block-smoothing} gives
\[
 \mathbb P\!\left(\sum_{k\in J_n'}\mathbf1_{\{N_k=1\}}<M_n/2\right)
 \le e^{-M_n/8},
\]
which is summable in $n$. Its conditional product argument therefore
gives, almost surely for all sufficiently large $|\tau|$,
$|\widehat\nu_{\mathbf N}(\tau)|
\le\exp\{-c'\log|\tau|\log\log|\tau|\}$.
Differentiated inversion gives smooth conditional densities; their
mixture is absolutely continuous.

For general $p_\ell$, couple the large-prime coordinates with the
canonical ones by \Cref{cor:summable-local-coupling}; their total
variation distances are
$|p_\ell-c_\ast/[\ell(\log\log\ell)^\beta]|$. The coupled sums differ
almost surely by an element of the countable additive group generated
by $\{(\log\ell)^{-A}:\ell\ge3\}$. A countable union of translates of
a Lebesgue-null set is null, so absolute continuity transfers.
The finitely many omitted coordinates enter by convolution. The same
coupling proves convergence, completing the proof.
\end{proof}

\begin{theorem}[Absolute continuity for slowly decreasing prime jumps]
\label{thm:general-slow-jump-ac-mas03}
Let $c_\ast>0$ and $0<\kappa<1/2$. Let $F$ be a positive $C^1$
function on $[v_0,\infty)$ which is eventually decreasing and satisfies
\[
 \int_{v_0}^\infty F(v)\,\frac{dv}{v}<\infty.
\]
Suppose that $g(v):=v|F'(v)|$ is eventually nonincreasing and
$g(v)\ge e^{-v^\kappa}$ for all sufficiently large $v$.
For a sufficiently large prime cutoff $L$, let $(B_\ell)_{\ell>L}$
be independent Bernoulli variables with probabilities $p_\ell\in[0,1]$
such that
\[
 \sum_{\ell>L}\left|p_\ell-\frac{c_\ast}{\ell}\right|<\infty.
\]
Then $S_F:=\sum_{\ell>L}B_\ell F(\log\ell)$ converges almost surely.
Its law, and its convolution with every independent probability law
on $\R$, are absolutely continuous and have Hausdorff dimension one.
\end{theorem}

\begin{proof}
First take the canonical probabilities $p_\ell=c_\ast/\ell$, enlarging
$L$ so that $p_\ell<1/2$ and all the assumptions on $F,g$ hold.
Monotonicity and the integral assumption give $F(v)\to0$.
Applying \eqref{eq:stieltjes} on finite intervals and then letting
the upper endpoint tend to infinity shows that
$\sum_{\ell>L}F(\log\ell)/\ell<\infty$: the error is finite since
$F$ is bounded and $\int_{\log L}^\infty|F'(v)|\,dv=F(\log L)$.
Tonelli's theorem proves almost-sure convergence.

For all sufficiently large integers $k$, set $U_k=2^k$ and
\[
 I_k=\{\ell:e^{U_k}<\ell\le e^{2U_k}\},\qquad
 N_k=\sum_{\ell\in I_k}B_\ell,\qquad
 Y_k=\sum_{\ell\in I_k}B_\ell F(\log\ell).
\]
With $w_\ell=p_\ell/(1-p_\ell)$ and $W_k=\sum_{I_k}w_\ell$,
Mertens' theorem and $w_\ell=c_\ast/\ell+O_{c_\ast}(\ell^{-2})$
give
\[
 W_k\longrightarrow c_\ast\log2,\qquad
 q_k:=\mathbb P(N_k=1)
 =W_k\prod_{\ell\in I_k}(1-p_\ell)
 \longrightarrow q_\ast:=c_\ast(\log2)2^{-c_\ast}>0.
\]
The conditional transform is
$\psi_k(t)=W_k^{-1}\sum_{I_k}w_\ell e^{itF(\log\ell)}$.
For $U=U_k$, \eqref{eq:stieltjes} bounds the error in replacing
$\sum_{I_k}\ell^{-1}e^{itF(\log\ell)}$ by
$\int_U^{2U}e^{itF(v)}\,dv/v$ by
\[
 O\!\left(e^{-c\sqrt U}
       \{1+|t|[F(U)-F(2U)]\}\right)
 =O_F((1+|t|)e^{-c\sqrt U}).
\]
To estimate the integral, put $s=\log v$. The phase derivative
$\frac{d}{ds}F(e^s)=-g(e^s)$ is nonzero and monotone, with modulus
at least $g(2U)$. More explicitly, the substitution $u=F(e^s)$
gives a monotone amplitude $1/g(e^s)$ of supremum $1/g(2U)$;
Stieltjes integration by parts against $e^{itu}/(it)$ bounds the
integral by $3/(|t|g(2U))$. The odds-weight error is $O_{c_\ast}(e^{-U})$.
Consequently, uniformly for $|t|\ge1$ and large $k$,
\begin{equation}\label{eq:general-slow-singleton-mas03}
 |\psi_k(t)|\ll_{F,c_\ast}
 \frac1{|t|g(2U_k)}+(1+|t|)e^{-c\sqrt{U_k}}.
\end{equation}

Put $T_n=2^{2^n}$. Fix positive integers $K,D$ so large that
$c2^{K/2}\ge4\log2$ and
$2^{\kappa(1-D)}\le(\log2)/4$, and retain
\[
 J_n=\{k:2(n+1)+K\le k\le\lfloor n/\kappa-D\rfloor\}.
\]
For every $T_n\le|t|<T_{n+1}$ and $k\in J_n$,
\[
 (2U_k)^\kappa\le\tfrac14\log T_n\le\tfrac14\log|t|,
 \qquad c\sqrt{U_k}\ge4\log T_{n+1}>4\log|t|.
\]
Thus \eqref{eq:general-slow-singleton-mas03} and the lower bound on
$g$ give $|\psi_k(t)|\ll |t|^{-3/4}$ throughout the entire shell.
Moreover,
\[
 M_n:=\sum_{k\in J_n}q_k
 \sim q_\ast(1/\kappa-2)n,
 \qquad \sum_n e^{-M_n/8}<\infty.
\]
\Cref{lem:conditional-block-smoothing}\textup{(a)} now proves absolute continuity
of the canonical sum. Finitely many omitted blocks enter by convolution.

For the stated probabilities, \Cref{cor:summable-local-coupling}
couples the Bernoulli coordinates to the canonical ones with only
finitely many disagreements almost surely. The two sums therefore
both converge, and their difference belongs almost surely to the
countable additive group generated by $\{F(\log\ell):\ell>L\}$.
For a Lebesgue-null set $E$, membership of the perturbed sum in $E$
forces the canonical sum into a countable union of translates of $E$.
This proves absolute continuity without requiring the difference to
be independent. Finally, independent convolution preserves absolute
continuity, and every absolutely continuous probability law on $\R$
has Hausdorff dimension one.
\end{proof}

The hypotheses of \Cref{thm:general-slow-jump-ac-mas03} include
\[
 F(v)=v^{-A}(\log v)^B\quad(A>0,\ B\in\R),
 \qquad F(v)=e^{-v^\kappa}\quad(0<\kappa<1/2).
\]
For the first family,
$g(v)=v^{-A}(\log v)^B(A-B/\log v)$ is positive and decreasing
eventually and exceeds $e^{-v^\kappa}$ for any fixed $\kappa>0$;
the integral condition follows from $A>0$.
For the second, $g(v)=\kappa v^\kappa e^{-v^\kappa}$ is eventually
decreasing and at least $e^{-v^\kappa}$, and the integral is finite.

\begin{corollary}[Stronger prime-counting inputs for slow jumps]
\label{cor:slow-jump-pnt-ranges-mas24}
Under the other hypotheses of
\Cref{thm:general-slow-jump-ac-mas03}, its conclusions hold throughout
$0<\kappa<3/5$. Under the Riemann hypothesis for $\zeta(s)$ they hold
throughout $0<\kappa<1$. In particular these ranges apply to
$F(v)=e^{-v^\kappa}$.
\end{corollary}

\begin{proof}
We first state the extra prime-counting inputs. The
Vinogradov--Korobov estimate of Johnston and Yang
\cite[Theorem~1.4, (1.8)]{JY22} gives, for an absolute $c>0$,
\[
 \vartheta(x)-x
 \ll x(\log x)^{1.801}
       \exp\{-c(\log x)^{3/5}(\log\log x)^{-1/5}\}.
\]
Partial summation therefore gives, for every fixed $0<\xi<3/5$,
\begin{equation}\label{eq:slow-jump-pnt-input-mas24}
 \pi(x)-\operatorname{Li}(x)
 \ll_\xi x\exp\{-c_\xi(\log x)^\xi\},\qquad c_\xi>0.
\end{equation}
Under RH, Schoenfeld's bound
$|\pi(x)-\operatorname{li}(x)|<\sqrt{x}\log x/(8\pi)$ for
$x\ge2657$, as recorded in \cite[(1.1)]{Joh22}, gives
\eqref{eq:slow-jump-pnt-input-mas24} for every fixed $0<\xi<1$.
The two conventions for the logarithmic integral differ by a constant,
which is absorbed in the error.

Fix $\xi>\kappa$ in the applicable range. Applying partial summation
once more to \eqref{eq:slow-jump-pnt-input-mas24}, and reducing
$c_\xi$ if necessary, gives a constant $b_1$ such that
\[
 \sum_{p\le e^v}\frac1p
 =\log v+b_1+O_\xi(e^{-c_\xi v^\xi}).
\]
Indeed the remaining error tail is bounded by a constant times
$\int_v^\infty e^{-c_\xi u^\xi}\,du$, and its polynomial factor
can be absorbed by reducing the positive exponential constant.
For the canonical probabilities and the blocks $U_k=2^k$ in the
preceding proof, Stieltjes integration by parts against
$e^{itF(v)}$ now bounds the reciprocal-prime sum's error on
$[U_k,2U_k]$ by
\[
 O_{F,\xi}\!\left((1+|t|)e^{-c_\xi U_k^\xi}\right).
\]
The odds-weight error remains $O_{c_\ast}(e^{-U_k})$ and is absorbed
in this bound because $\xi<1$. The same first-derivative argument
therefore proves
\begin{equation}\label{eq:slow-jump-pnt-singleton-mas24}
 |\psi_k(t)|\ll_{F,c_\ast,\xi}
 \frac1{|t|g(2U_k)}+(1+|t|)e^{-c_\xi U_k^\xi}.
\end{equation}

Use $T_n=2^{2^n}$ and choose fixed positive integers $K,D$ with
$c_\xi2^{\xi K}\ge4\log2$ and
$2^{\kappa(1-D)}\le(\log2)/4$. Retain precisely
\[
 J_n=\{k:\lceil(n+1)/\xi+K\rceil
                 \le k\le\lfloor n/\kappa-D\rfloor\}.
\]
For every $k\in J_n$ and $T_n\le|t|<T_{n+1}$,
\[
 c_\xi U_k^\xi\ge4\log T_{n+1}>4\log|t|,
 \qquad (2U_k)^\kappa\le\tfrac14\log T_n
                              \le\tfrac14\log|t|.
\]
Thus \eqref{eq:slow-jump-pnt-singleton-mas24} is
$O(|t|^{-3/4})$ on the entire shell. The singleton probabilities
still tend to $q_\ast=c_\ast(\log2)2^{-c_\ast}>0$, so
\[
 \sum_{k\in J_n}q_k
 \sim q_\ast(1/\kappa-1/\xi)n.
\]
Since $\xi>\kappa$, the coefficient is positive and the Chernoff
failures are summable. Thus
\Cref{lem:conditional-block-smoothing}\textup{(a)} gives absolute continuity.
The convergence, summable-perturbation coupling and independent-head
arguments in the preceding proof apply unchanged.
\end{proof}

\section{Independent convolutions and exact regularity}

For independent scalar sums, Fourier transforms multiply. The next
theorem measures their combined activation above the jump scale
$|t|^{-1}$. Cancellation gives an upper bound for the transform, while
averages over frequency shells detect failure at the boundary of
integrability. For positive sums, mass near the least support point
provides the additional obstruction needed for an exact global $C^k$
criterion.
\Cref{fig:activation-scale} records how the increments $(\log\ell)^{-A}$ of this
chapter and the activation probabilities $K\log\ell/\ell$ of \Cref{ssec:annulus}
enter \Cref{thm:activation-scale} through a single coefficient.

\begin{theorem}[Regularity from activation at the jump scale]
\label{thm:activation-scale}
\index{Activation intensity!jump-scale, regularity thresholds}
Let $a_n>0$, $\epsilon_n\in\{-1,1\}$, and let $B_n$ be independent
Bernoulli variables with probabilities $p_n\in[0,1]$ such that
$\sum_n p_na_n<\infty$ and $\sum_n p_n^2<\infty$. Let $H$ be an
independent random variable equal to a constant plus a finite sum of
independent variables, each supported on a finite arithmetic progression,
and put
\[
 \begin{gathered}
 X=H+\sum_n\epsilon_na_nB_n,\qquad \mu=\operatorname{Law}(X),\\
 \Lambda(T)=\sum_{a_n\ge T^{-1}}p_n,\qquad
 \mathcal D(t)=\sum_np_n(1-\cos(ta_n)).
 \end{gathered}
\]
Suppose that, for some $b>0$, $C_0\ge0$ and $E(t)=o(\log|t|)$,
\begin{equation}\label{eq:activation-scale-hypotheses}
 \Lambda(T)=b\log T+O(1),\qquad
 \mathcal D(t)\ge b\log|t|-E(t)-C_0
 \quad(|t|\text{ sufficiently large}).
\end{equation}
Then the following hold.
\begin{enumerate}[label=\textup{(\alph*)},leftmargin=*]
\item \emph{Pointwise decay and logarithmic shell mean.}
\begin{equation}\label{eq:activation-scale-bounds}
 |\widehat\mu(t)|\ll |t|^{-b}e^{E(t)},\qquad
 \frac1T\int_T^{2T}-\log|\widehat\mu(t)|^2\,dt
       =2b\log T+O(1).
\end{equation}
\item \emph{Shell lower bound.} For each fixed $q>0$,
\begin{equation}\label{eq:activation-scale-shell}
 \int_T^{2T}|\widehat\mu(t)|^q\,dt\gg_q T^{1-qb}.
\end{equation}
\item \emph{Weighted integrability.} For $q>0$ and $s\in\R$,
\begin{equation}\label{eq:activation-scale-integrability}
 \int_{\R}(1+t^2)^s|\widehat\mu(t)|^q\,dt<\infty
 \quad\Longleftrightarrow\quad qb>2s+1.
\end{equation}
\item \emph{Fourier dimension.} $\dim_F\mu=\min(1,2b)$.
\item \emph{Sobolev regularity.} For every $s\in\R$, as a tempered
distribution, $\mu\in H^s(\R)$ if and only if $s<b-\tfrac12$.
For $s\ge0$, this is equivalent to having a density in $H^s(\R)$.
\item \emph{Derivatives.} If $k\ge0$ is an integer and $k+1<b$, the
density of $\mu$ has bounded continuous derivatives through order $k$.
Its $k$th derivative is globally $\alpha$-H\"older continuous
if $0<\alpha\le1$ and $k+\alpha+1<b$.
\item \emph{One sign.} If all $\epsilon_n$ have the same sign, then, for
every integer $k\ge0$, $\mu$ has a globally $C^k$ density if and only if
$b>k+1$.
\end{enumerate}
\end{theorem}

\begin{proof}
The series converges absolutely almost surely by its finite expected
absolute sum. Remove the finitely many coordinates with $p_n>1/4$
into the finite head and write $\mu_0$ for the remaining sum law.
Since $\sum_np_n\min(1,a_n^2)<\infty$, its logarithm series converges
locally uniformly and its Fourier transform is nonzero. Taylor's formula and $\sum_np_n^2<\infty$ give
\begin{equation}\label{eq:activation-scale-log}
 -\log|\widehat\mu_0(t)|^2=2\mathcal D(t)+O(1)
\end{equation}
uniformly in real $t$; omitting the finite coordinates changes
$\mathcal D$ by $O(1)$. This proves the pointwise bound.
The first hypothesis in \eqref{eq:activation-scale-hypotheses}
bounds the activation mass in each dyadic interval of sufficiently
small jumps. Geometric summation, and $\sum_np_na_n<\infty$ for
jumps bounded away from zero, give
\[
 \frac1T\sum_{a_n\ge T^{-1}}\frac{p_n}{a_n}=O(1),\qquad
 T^2\sum_{a_n<T^{-1}}p_na_n^2=O(1).
\]
Direct integration of the cosines therefore yields
$T^{-1}\int_T^{2T}\mathcal D(t)\,dt=\Lambda(T)+O(1)$.
Each nonconstant finite head factor has transform
$e^{izt}P(e^{iat})$, where $a\ne0$, $P$ is a polynomial and $P(1)=1$.
Factoring $P$ shows that $-\log|P(e^{iu})|$ is periodic and locally
integrable: unit-circle roots give integrable logarithmic singularities,
and the other roots give bounded terms. It is nonnegative and has
bounded averages on $[T,2T]$ after the fixed scaling $u=at$.
Adding the finitely many factors preserves this bound.
Thus \eqref{eq:activation-scale-log} proves the logarithmic mean
in \eqref{eq:activation-scale-bounds}; Jensen's inequality gives
\eqref{eq:activation-scale-shell}.
Since $(1+t^2)^s\asymp_s T^{2s}$ on $[T,2T]$ for every real $s$,
summing dyadic shells proves divergence in
\eqref{eq:activation-scale-integrability} when $qb\le2s+1$,
including equality. The pointwise bound and $E(t)=o(\log|t|)$ prove
convergence in the strict complementary range. The upper bound gives
$\dim_F\mu\ge\min(1,2b)$, while \eqref{eq:activation-scale-shell}
with $q=2$ precludes every Fourier exponent exceeding $2b$.
The Fourier-weight definition gives the Sobolev assertion;
for $s\ge0$, Plancherel and \Cref{lem:l2-inversion} give the density.
Fourier inversion gives the bounded derivatives; the inequality
$|e^{iv}-1|\le2|v|^\alpha$ gives their stated H\"older regularity.
For $\epsilon_n=1$, let $z$ be the least value of the enlarged finite
head, which has positive mass there. Put $N(u)=\sum_{a_n\ge u}p_n$
for the remaining coordinates. Since $N(u)=b\log(1/u)+O(1)$,
\[
 \begin{gathered}
 \sum_{a_n<u}p_na_n=\int_0^u\bigl(N(v)-N(u)\bigr)\,dv\ll u,\\
 \prod_{a_n\ge u}(1-p_n)\asymp u^b.
 \end{gathered}
\]
Take $u=\theta h$ with fixed $\theta>0$ sufficiently small.
Independence and Markov's inequality give
$\mathbb P(X\le z+h)\gg h^b$. All derivatives through order $k$ of a
globally $C^k$ density vanish at $z$, giving
$\mathbb P(X\le z+h)=o(h^{k+1})$.
Thus $b>k+1$; reflect for $\epsilon_n=-1$.
\end{proof}

The cancellation hypothesis cannot be omitted. For each $k\ge1$,
take $M_k=\lceil 4b(\log2)2^k\rceil$ independent copies of the jump
$2^{-k}$, each with probability $b\log2/M_k$. Then both summability
conditions hold and $\Lambda(T)=b\log T+O(1)$, but
\[
 \mathcal D(2\pi2^n)
 =b\log2\sum_{j\ge1}(1-\cos(2\pi2^{-j}))=O(1).
\]
Equation~\eqref{eq:activation-scale-log} shows that the corresponding
Fourier transform stays bounded away from zero along this sequence.

\begin{proposition}[Positive endpoint at the activation scale]
\label{prop:activation-scale-endpoint}
Let $(B_n)$ be independent Bernoulli variables with probabilities
$p_n\in[0,1)$, and let $a_n>0$ satisfy
$\sum_n p_na_n<\infty$ and $\sum_n p_n^2<\infty$. Put
$X=\sum_n a_nB_n$ and $\Lambda(T)=\sum_{a_n\ge T^{-1}}p_n$. Suppose
\[
 \Lambda(T)=b\log T+d+o(1),\qquad b>0,
 \qquad R=\sum_n\bigl(\log(1-p_n)+p_n\bigr).
\]
Then $R$ is finite and
\begin{equation}\label{eq:activation-scale-endpoint}
 \mathbb P(X\le h)\sim
 \frac{e^{-d-\gamma b+R}}{\Gamma(1+b)}h^b
 \qquad(h\downarrow0).
\end{equation}
If the cancellation hypothesis in \eqref{eq:activation-scale-hypotheses}
also holds, then, for every integer $k\ge0$, the law has a globally
$C^k$ density if and only if $b>k+1$; its derivatives through order
$k$ are then bounded. This equivalence persists after addition of an
independent random variable $H$ with $H\ge z$ almost surely and
$\mathbb P(H=z)=w>0$, and restoration of probability-one coordinates
as a translation. For such $H$,
$\mathbb P(X+H\le z+h)\sim w\mathbb P(X\le h)$.
\end{proposition}

\begin{proof}
Put $N(x)=\sum_{a_n\ge x}p_n$. Tonelli's theorem and
$\int_0^\infty N(x)\,dx=\sum_np_na_n<\infty$ give
\[
 \sum_np_n(1-e^{-sa_n})
 =s\int_0^\infty e^{-sx}N(x)\,dx
 =b\log s+d+\gamma b+o(1).
\]
For the last equality substitute $u=sx$, use
$N(x)=b\log(1/x)+d+o(1)$ at zero, and
$-\int_0^\infty e^{-u}\log u\,du=\gamma$; the part $x\ge\delta>0$
is at most $se^{-s\delta}\int_\delta^\infty N(x)\,dx$.
The correction
$\log(1-p_n(1-e^{-sa_n}))+p_n(1-e^{-sa_n})$
is dominated outside a finite set by $O(p_n^2)$ and tends to
$\log(1-p_n)+p_n$. Dominated convergence therefore gives
$\mathbb Ee^{-sX}\sim e^{-d-\gamma b+R}s^{-b}$.
Karamata's Tauberian theorem at zero
\cite[Chapter~XIII, \S~5, Theorems~2 and~3, p.~445]{Fel71} proves
\eqref{eq:activation-scale-endpoint}.
A globally $C^k$ density vanishes with its first $k$ derivatives at
zero, and hence gives $\mathbb P(X\le h)=o(h^{k+1})$.
This excludes $b\le k+1$; the reverse implication follows from
\Cref{thm:activation-scale}. For the stated independent head,
\[
 w\mathbb P(X\le h)\le\mathbb P(X+H\le z+h)
 \le\mathbb P(H\le z+h)\mathbb P(X\le h).
\]
Since $\mathbb P(H\le z+h)\to w$, the same necessity argument applies.
Convolution with any probability law preserves bounded continuous
derivatives, proving sufficiency.
\end{proof}

Fix integers $m\ge1$ and $n_j\ge0$ $(1\le j\le m)$, real translations
$z_j$, positive parameters $A_j,c_j$, nonzero coefficients
$\lambda_j\in\R$, real jumps $\eta_{j,r}$ $(1\le r\le n_j)$ and
probabilities $p_{j,\ell}\in[0,1]$.
Let all variables $B_{j,\ell}$ and $H_{j,r}$ below be independent.
The $B_{j,\ell}$ are Bernoulli with probabilities $p_{j,\ell}$, and each
$H_{j,r}$ is supported on a finite arithmetic progression. Put
\[
 \begin{aligned}
 Y_j&=z_j+\sum_{r=1}^{n_j}\eta_{j,r}H_{j,r}
       +\sum_{\substack{\ell\ge3\\\ell\textup{ prime}}}
          (\log\ell)^{-A_j}B_{j,\ell},\\
 \nu&=\operatorname{Law}\left(\sum_{j=1}^m\lambda_jY_j\right),\qquad
 b=\sum_{j=1}^m\frac{c_j}{A_j},\qquad C=\sum_{j=1}^mc_j.
 \end{aligned}
\]
\AddToHookNext{build/column/before}{\raggedbottom}
\begin{proposition}[Exact conditions for Fourier integrability and Sobolev regularity]
\label{prop:slow-jump-regularity}
\index{Sobolev regularity!scalar laws}
Assume
\[
 \sum_{\substack{\ell\ge3\\\ell\textup{ prime}}}
 \left|p_{j,\ell}-\frac{c_j}{\ell}\right|<\infty
 \qquad(1\le j\le m),
\]
and let $Y_j,\nu,b,C$ be as above. The series defining the $Y_j$ converge
almost surely. Each $Y_j$ has an absolutely continuous law, and $\nu$ is
absolutely continuous. Moreover:
\begin{enumerate}[label=\textup{(\alph*)},leftmargin=*]
\item For $|\tau|\ge3$,
\begin{equation}\label{eq:slow-convolution-decay}
 |\widehat\nu(\tau)|\ll |\tau|^{-b}(\log|\tau|)^C,
 \qquad \dim_F\nu=\min(1,2b),\qquad
 \dim_H\nu=1.
\end{equation}
\item For every fixed $q>0$ and all sufficiently large $T$,
\begin{equation}\label{eq:slow-shell-lower}
 \int_T^{2T}|\widehat\nu(\tau)|^q\,d\tau\gg_q T^{1-qb}.
\end{equation}
Moreover, $\widehat\nu\in L^q(\R)$ if and only if $qb>1$.
\item For every $s\in\R$, the density of $\nu$ belongs to $H^s(\R)$ if and only if
$s<b-\tfrac12$.
\item If $k\ge0$ is an integer with $k+1<b$, the density has bounded
continuous derivatives through order $k$. If also $0<\alpha\le1$ and
$k+\alpha+1<b$, its $k$th derivative is globally
$\alpha$-H\"older continuous.
\end{enumerate}
The constants may depend on the fixed data. For $m=1$, $A_1=A$,
$c_1=c_\ast$, $\lambda_1=1$, $z_1=a_2$ and $n_1=0$, suppose that
$|p_{1,\ell}-c_\ast/\ell|\le B/\ell^2$ with $B\ge0$, and denote the
law by $\mu_A^{(c_\ast)}$. For this family the Fourier bound in
\textup{(a)} holds with constant and threshold depending only on
$A,c_\ast,B$, uniformly over the probability sequence.
\end{proposition}

\begin{proof}
Put $a_{j,\ell}=(\log\ell)^{-A_j}$ and
$d_{j,\ell}=2p_{j,\ell}(1-p_{j,\ell})$.
The summability assumptions give $p_{j,\ell}\to0$,
$\sum_\ell p_{j,\ell}^2<\infty$ and
$\sum_\ell|d_{j,\ell}-2c_j/\ell|<\infty$.
Partial summation from Mertens' theorem gives
$\sum_\ell p_{j,\ell}a_{j,\ell}<\infty$, proving almost sure convergence.
By \Cref{lem:slow-jump-all-ac} with $\beta=0$, each prime series is absolutely
continuous. Finite convolution and nonzero scaling preserve absolute
continuity, proving this assertion for $Y_j$ and $\nu$, and
$\dim_H\nu=1$.
Choose $L\ge3$ so that $p_{j,\ell}\le1/4$ for $\ell>L$, and let
$Y_j^{\mathrm{tail}}$ denote the corresponding prime tail. The logarithm
expansion in \eqref{eq:activation-scale-log} gives, uniformly in $u$,
\begin{equation}\label{eq:slow-summable-log}
 \begin{aligned}
 -\log|\widehat{\operatorname{Law}(Y_j^{\mathrm{tail}})}(u)|^2
 &=\sum_{\ell>L}d_{j,\ell}(1-\cos(ua_{j,\ell}))+O(1)\\
 &=2c_j\sum_{\ell>L}\frac{1-\cos(ua_{j,\ell})}{\ell}+O(1).
 \end{aligned}
\end{equation}
For the jumps $|\lambda_j|a_{j,\ell}$, Mertens' theorem gives
$\Lambda_j(T)=(c_j/A_j)\log T+O(1)$.
By \eqref{eq:slow-diss-sharp}, their dissipation is at least
$(c_j/A_j)\log|t|-c_j\log\log|t|-O(1)$.
Thus \Cref{thm:activation-scale} applies with
$b=\sum_jc_j/A_j$ and $E(t)=C\log\log|t|$, proving all the
Fourier, Sobolev and derivative assertions. In particular,
\begin{equation}\label{eq:slow-log-average}
 \frac1T\int_T^{2T}-\log|\widehat\nu(t)|^2\,dt
 =2b\log T+O(1).
\end{equation}
For the final uniformity assertion take
$L=\max\{3,8c_\ast,\sqrt{8B}\}$. Then $p_{1,\ell}\le1/4$ for $\ell>L$,
and $\sum_{\ell>L}p_{1,\ell}^2$,
$\sum_{\ell>L}|d_{1,\ell}-2c_\ast/\ell|$ and
$\sum_{\ell\le L}1/\ell$ are $O_{c_\ast,B}(1)$.
Equation~\eqref{eq:slow-diss-sharp}, followed by
\eqref{eq:slow-summable-log}, therefore gives the Fourier bound with
constant and threshold depending only on $A,c_\ast,B$.
\end{proof}

\begin{proposition}[Positive convolutions: endpoint and exact smoothness]
\label{prop:slow-jump-positive-endpoint}
Let $m\ge1$, $A_j,c_j,\lambda_j>0$, and let the Bernoulli variables
$B_{j,\ell}$ be mutually independent over $j$ and the odd primes $\ell$.
Suppose their probabilities $p_{j,\ell}<1$ satisfy
$\sum_\ell|p_{j,\ell}-c_j/\ell|<\infty$. Put
\[
 X=\sum_{j=1}^m\lambda_j\sum_{\ell\ge3}
                B_{j,\ell}(\log\ell)^{-A_j},\qquad
 b_j=c_j/A_j,\qquad b=\sum_jb_j,
\]
\[
 C_j=\lim_{y\to\infty}(\log y)^{c_j}
              \prod_{3\le\ell\le y}(1-p_{j,\ell}).
\]
Then $C_j\in(0,\infty)$ and
\begin{equation}\label{eq:slow-positive-endpoint}
 \mathbb P(X\le h)\sim
 \frac{\prod_j C_je^{-\gamma b_j}\lambda_j^{-b_j}}
      {\Gamma(1+b)}h^b\qquad(h\downarrow0).
\end{equation}
For every integer $k\ge0$, the law of $X$ has a globally $C^k$ density
if and only if $b>k+1$; in that case its derivatives through order $k$
are bounded. The same equivalence holds after addition of any independent
$H\ge z$ with $\mathbb P(H=z)>0$, and after restoration of coordinates
with probability one.
\end{proposition}

\begin{proof}
The summable discrepancies give $p_{j,\ell}\to0$ and
$\sum_\ell p_{j,\ell}^2<\infty$. Partial summation from Mertens' theorem
gives $\sum_\ell p_{j,\ell}(\log\ell)^{-A_j}<\infty$, so $X$ has finite
mean. Put
\[
 R_j=\sum_{\ell\ge3}
       \bigl(\log(1-p_{j,\ell})+p_{j,\ell}\bigr).
\]
These sums converge: the tail terms are $O(p_{j,\ell}^2)$, and the
finitely many large probabilities are less than one. Mertens' theorem
and the summable discrepancies show that
$\sum_{3\le\ell\le y}p_{j,\ell}-c_j\log\log y$ has a finite limit.
Thus the logarithm of the defining expression for $C_j$ has a finite
real limit, proving $C_j\in(0,\infty)$ and
\[
 \sum_{3\le\ell\le y}p_{j,\ell}
 =c_j\log\log y-\log C_j+R_j+o(1).
\]
For the scaled jumps, the cutoff at size $T^{-1}$ is exactly
$\ell\le\exp((\lambda_jT)^{1/A_j})$. Hence
\[
 \Lambda_j(T)=b_j\log T+b_j\log\lambda_j-\log C_j+R_j+o(1).
\]
Summing gives $\Lambda(T)=b\log T+d+o(1)$, where
\[
 \begin{gathered}
 d=\sum_j(b_j\log\lambda_j-\log C_j+R_j),\qquad
 R=\sum_jR_j,\\
 e^{-d+R}=\prod_jC_j\lambda_j^{-b_j}.
 \end{gathered}
\]
The cancellation estimate in the proof of
\Cref{prop:slow-jump-regularity} gives
\[
 \begin{aligned}
 \mathcal D(t)&:=\sum_{j,\ell}p_{j,\ell}
       \bigl(1-\cos(t\lambda_j(\log\ell)^{-A_j})\bigr)\\
 &\ge b\log|t|-\Bigl(\sum_jc_j\Bigr)\log\log|t|-O(1).
 \end{aligned}
\]
Since $\log\log|t|=o(\log|t|)$, all hypotheses of
\Cref{prop:activation-scale-endpoint} hold. It proves
\eqref{eq:slow-positive-endpoint}, using
$e^{-\gamma b}=\prod_je^{-\gamma b_j}$, and the exact $C^k$ criterion,
with bounded derivatives when $b>k+1$. Its minimum-atom assertion gives
\[
 \mathbb P(X+H\le z+h)
 \sim\mathbb P(H=z)\mathbb P(X\le h),
\]
and the same criterion for $X+H$, including failure at $b=k+1$.
Summability permits only finitely many probability-one coordinates;
restoring them translates the minimum and preserves the criterion.
The endpoint constant is computed for the remaining coordinates and
the endpoint is measured from the translated minimum.
\end{proof}

For $A>0$, let
\[
 M_{\mathrm{odd}}=\lim_{v\to\infty}
 \left(\sum_{3\le\ell\le e^v}\ell^{-1}-\log v\right)
 =c_{\mathrm M}-\tfrac12.
\]
The corresponding linear prime sum satisfies
\begin{equation}\label{eq:slow-laplace-prime-sum}
 \sum_{\ell\ge3}\frac{1-e^{-s(\log\ell)^{-A}}}{\ell}
 =\frac{\log s+\gamma}{A}+M_{\mathrm{odd}}+o(1)
 \qquad(s\to\infty).
\end{equation}
Indeed, $N(x)=\sum_{(\log\ell)^{-A}\ge x}\ell^{-1}$ satisfies
$N(x)=A^{-1}\log(1/x)+M_{\mathrm{odd}}+o(1)$ as $x\downarrow0$.
Tonelli's theorem gives the left side as
$s\int_0^\infty e^{-sx}N(x)\,dx$. Substituting $u=sx$ and using
$-\int_0^\infty e^{-u}\log u\,du=\gamma$ proves the identity;
the tail is negligible since
$\int_0^\infty N(x)\,dx=\sum_{\ell\ge3}\ell^{-1}(\log\ell)^{-A}<\infty$,
as in the proof of \Cref{prop:activation-scale-endpoint}.

The dense-anchor obstruction and $L^p$ bounds for increments $n^{-s}$
appear in \cite[Theorem~4.2, Remark~4.3 and Corollary~1.2]{PPPT15}.
The following proposition treats weighted sums of independent
logarithmic prime series and independent shifts having an atom;
\Cref{prop:slow-jump-positive-endpoint} supplies an exact endpoint
asymptotic in place of the Markov tail bound.

\begin{proposition}[Local density obstruction for convolutions]
\label{prop:scalar-density}
Let $X,b$ be as in \Cref{prop:slow-jump-positive-endpoint}, allowing
nonzero real coefficients $\lambda_j$. Let $H$ be independent and put
$\mu=\operatorname{Law}(X+H)$. Probability-one coordinates may be
absorbed into $H$.
\begin{enumerate}[label=\textup{(\alph*)},leftmargin=*]
\item Suppose all $\lambda_j>0$, $H\ge z_0$ almost surely and
$\mathbb P(H=z_0)>0$. Then $\supp\mu=[z_0,\infty)$.
If $0<b<1$, its density $g$ is essentially unbounded on every nonempty
open interval in $(z_0,\infty)$ and has no finite-valued representative
continuous at any point of that interval. Moreover, on every such
interval $I$,
\begin{equation}\label{eq:scalar-local-exclusion}
 g\notin L^p(I)\qquad\left(\frac1{1-b}\le p\le\infty\right).
\end{equation}
For $1/2<b<1$, the exact global range is
\[
 g\in L^p(\R)\quad\Longleftrightarrow\quad
 1\le p<\frac1{1-b}.
\]
\item Suppose the $\lambda_j$ include both signs and $H$ has an atom.
Then $\mu$ is absolutely continuous and $\supp\mu=\R$ for every $b>0$.
For $0<b<1$, the local density conclusions of \textup{(a)} hold on
every nonempty open interval $I\subset\R$; for $1/2<b<1$, its global
$L^p$ range also holds.
\end{enumerate}
\end{proposition}

Part~\textup{(a)} applies with $b=1/A$ to the shifted-prime law
$\mu_A$ and the integer law with activation probabilities $1/\ell$,
for $A>1$.

\begin{proof}
For part~\textup{(a)}, all coefficients are positive. For each $j$, the primes
with $p_{j,\ell}=0$ have finite reciprocal
sum, whereas $\sum_{\ell:p_{j,\ell}>0}1/\ell=\infty$.
The available jumps $\lambda_j(\log\ell)^{-A_j}$ therefore tend to zero
and have divergent unweighted sum. Their finite subsums are dense in
$[0,\infty)$: partial sums of sufficiently small available jumps
approximate any prescribed positive value with arbitrarily small
overshoot. Each finite pattern has positive probability, since
$p_{j,\ell}<1$ and every selected jump has positive activation
probability. Fix such a pattern through a prime cutoff, and condition
also on $H=z_0$. Its anchor $z$ is $z_0$ plus the selected finite sum.
The independent remainder satisfies
\Cref{prop:slow-jump-positive-endpoint} with the same $b$, since deleting
finitely many coordinates preserves its hypotheses. Consequently,
\begin{equation}\label{eq:scalar-anchor}
 \mu([z,z+h])\ge c_z h^b\qquad(0<h<h_z),
\end{equation}
where $c_z,h_z>0$. This places every anchor in the support; their
density and positivity identify the support as $[z_0,\infty)$.
Absolute continuity follows from \Cref{prop:slow-jump-regularity}
and convolution. The summability hypothesis permits only finitely
many probability-one coordinates, which give a translation.

For part~\textup{(b)}, first absorb probability-one
coordinates into $H$ and fix an atom $z_0$ of $H$. The positive and
negative finite subsums are dense in their respective half-lines,
so the signed anchors are dense in $\R$. Fix a pattern through a
common prime cutoff in every component. Its absolute weighted tail
$U$ satisfies \Cref{prop:slow-jump-positive-endpoint} with the same
$b$, and the signed remainder $R$ obeys $|R|\le U$. Hence
\[
 \mu([z-h,z+h])\ge c_z h^b\qquad(0<h<h_z).
\]
This identifies the support as $\R$; absolute continuity again follows
from \Cref{prop:slow-jump-regularity} and convolution.

Every nonempty open interval in the support interior contains an anchor
and its sufficiently small one-sided or symmetric windows $J_h$.
When $b<1$, their mass lower bounds exclude a
finite essential supremum on that interval. Continuity of a
finite-valued representative at any interior point would make it
bounded in some neighborhood, which is impossible.
If $g\in L^p(I)$ for finite $p\ge1/(1-b)$, choose an anchor $z\in I$.
Absolute continuity of the integral of $|g|^p$ and H\"older's inequality
would give
\[
 \mu(J_h)
 \le (2h)^{1-1/p}\left(\int_{J_h}|g(x)|^p\,dx\right)^{1/p}
 =o(h^{1-1/p}),
\]
contradicting the mass lower bound, including at $1-1/p=b$.

If $b>1/2$, \Cref{prop:slow-jump-regularity} gives
$\widehat{\operatorname{Law}(X)}\in L^q$ whenever $qb>1$.
Since $|\widehat\mu|\le|\widehat{\operatorname{Law}(X)}|$, this also
holds for $\widehat\mu$, and $g\in L^2$ by Plancherel.
For $2\le p<1/(1-b)$, its conjugate exponent satisfies
$1/b<q\le2$, so inverse Hausdorff--Young gives $g\in L^p$.
Interpolation between $L^1$ and $L^2$ supplies $1\le p\le2$.
\end{proof}

\begin{corollary}[Fourier decay and absolute continuity]
\label{thm:slow-jump-ac}
\index{Absolute continuity!logarithmic prime increments}
\index{Erdos--Katai conjecture@Erd\H{o}s--K\'atai conjecture}
Let $A>0$. The law $\mu_A$ is absolutely continuous, with a density
$h_A\in L^1(\R)$ carried by $[a_2,\infty)$. There are constants
$C_A>0$ and $\tau_0(A)\ge2$ such that
\begin{equation}\label{eq:slow-decay}
 |\Phi_A(\tau)|^2\le C_A|\tau|^{-2/A}(\log|\tau|)^2
 \qquad(|\tau|\ge\tau_0(A)).
\end{equation}
Moreover,
\[
 \dim_F(\mu_A)=\min(1,2/A),\qquad \dim_H(\mu_A)=1.
\]
The density $h_A$ belongs to $L^2(\R)$ if and only if $A<2$.
For every integer $k\ge0$, $h_A$ has a globally $C^k$ version if and
only if $A<1/(k+1)$; its derivatives through order $k$ are then bounded.
In particular, $\Phi_A\in L^1(\R)$ and $h_A$ has a bounded continuous
version exactly when $A<1$.
\end{corollary}

\begin{proof}
Apply \Cref{prop:slow-jump-regularity} with $m=1$, $c_1=1$, $A_1=A$,
$\lambda_1=1$, $z_1=a_2$, $n_1=0$ and $p_{1,\ell}=1/(\ell-1)$.
Since $|p_{1,\ell}-1/\ell|=1/(\ell(\ell-1))\le2/\ell^2$,
part \textup{(a)} and its uniformity clause give \eqref{eq:slow-decay}
with constants depending only on $A$. Parts \textup{(a)--(c)} give the
dimensions and the exact $L^2$ condition.
\Cref{prop:slow-jump-positive-endpoint} gives the exact $C^k$ conditions,
and \eqref{eq:slow-support} gives the support.
\end{proof}

For an integer $m\ge1$, the $m$-fold convolution power of
$\mu_A^{(c_\ast)}$ is the law $\nu$ of \Cref{prop:slow-jump-regularity}
with $m$ independent copies of the single-component data above, so
$b=mc_\ast/A$ and $C=mc_\ast$. Thus
$\dim_F((\mu_A^{(c_\ast)})^{*m})=\min(1,2mc_\ast/A)$; for each $s\in\R$ its
density belongs to $H^s(\R)$ exactly when $mc_\ast/A>s+1/2$; and, by
\Cref{prop:slow-jump-positive-endpoint}, it has a globally $C^k$ density
exactly when $mc_\ast/A>k+1$.
\paragraph*{Four positive convolution powers at $A=3$}
Let $Y_1,Y_2,\ldots$ be independent with the shifted-prime law $\mu_3$
(so $c_\ast=1$), and put $S_m=Y_1+\cdots+Y_m$.
All coefficients are positive and the scalar activation coefficient is
$b=m/3$. The preceding criteria give the following exact conclusions;
the Sobolev column covers every real $s$.
\begin{center}
\begin{tabular}{cclc}
$m$ & $b$ & density in $H^s(\R)$ & globally $C^0$ density \\
\hline
$1$ & $1/3$ & $s<-1/6$ & no \\
$2$ & $2/3$ & $s<1/6$ & no \\
$3$ & $1$ & $s<1/2$ & no \\
$4$ & $4/3$ & $s<5/6$ & yes \\
\end{tabular}
\end{center}
Every row has an absolutely continuous law; none has a globally $C^1$
density. In the first two rows, \Cref{prop:scalar-density}\textup{(a)} also gives
essential unboundedness on every support-interior interval. For $m=2$,
the density belongs to $L^p(\R)$ exactly when $1\le p<3$.
For $m=3$, the positive endpoint mass excludes global continuity
at $b=1$; failure of Fourier $L^1$ integrability alone would not suffice.

\KeepHeadingSpace{6\baselineskip}
\section{Joint laws and dependence}

\paragraph*{Dependence structures}
\begin{itemize}[leftmargin=*]
\item \emph{Independent scalar sums.}
In \Cref{prop:slow-jump-regularity}, the entries $B_{j,\ell}$ are
independent over both indices, and $\sum_j\lambda_jY_j$ takes values in
$\R$. Its scalar activation coefficient is $b=\sum_jc_j/A_j$.
The exact Sobolev statement holds for every real $s$. For coefficients
of one sign, \Cref{prop:slow-jump-positive-endpoint}, with reflection
if necessary, gives the exact global $C^k$ criterion. For mixed signs,
\Cref{prop:scalar-density}\textup{(b)} excludes bounded or continuous
densities when $0<b<1$; for $b\ge1$, the derivative condition in
\Cref{prop:slow-jump-regularity}\textup{(d)} is sufficient.
The finite arithmetic-progression head hypothesis is needed for
\eqref{eq:slow-shell-lower}: an arbitrary independent summand may
have a Fourier transform vanishing on intervals.
\item \emph{Common activation.}
A single $B_\ell$ activates every coordinate of
$((\log\ell)^{-A_j})_{j=1}^d$. For fixed $0<A_1<\cdots<A_d$,
let the $B_\ell$ be independent with probabilities $q_\ell\in[0,1]$
satisfying $\sum_\ell|q_\ell-c/\ell|<\infty$ for some $c>0$.
The joint law is absolutely continuous by
\Cref{thm:shared-prime-anisotropic}. The integer and shifted-prime
models of \Cref{thm:shared-prime-vector-mas14} have $c=1$.
Equal exponents give identical coordinates, so the law is supported
on a proper hyperplane.
\item \emph{Affine forms.}
For the forms of \Cref{thm:fixed-shift-tuple} and exponents $A_j>0$,
\Cref{cor:joint-arithmetic} gives an absolutely continuous joint law,
with a globally $C^r$ density exactly when $\max_jA_j<1/(r+1)$,
and the sharp Fourier and Sobolev thresholds in
\eqref{eq:affine-fourier-sobolev-thresholds}.
For $(f_A(p-1),f_B(p+1))$ and $\ell>3$, the local states are
$(0,0),((\log\ell)^{-A},0),(0,(\log\ell)^{-B})$; the two activations are mutually
exclusive. With the finite head at $2,3$ restored,
\Cref{cor:opposite-shift-regularity} gives its exact Fourier and Sobolev
thresholds by specializing the general affine formula.
\end{itemize}

\begin{lemma}[Uniform directional oscillation for distinct exponents]
\label{lem:directional-exponentials-mas14}
Fix pairwise distinct positive numbers $A_1,\ldots,A_d$ and $L_0>0$.
For real $u_1,\ldots,u_d$, put
\[
 F(w)=\sum_{j=1}^du_je^{-A_jw},\qquad L=\max_j|u_j|.
\]
There is a constant $C$, depending only on this fixed family and $L_0$,
such that, for $L\ge1$ and every subinterval $I\subset[0,L_0]$,
\begin{equation}\label{eq:directional-exponentials-mas14}
 \left|\int_I e^{iF(w)}\,dw\right|\le CL^{-1/d}.
\end{equation}
\end{lemma}

\begin{proof}
The matrix taking $(u_je^{-A_jw})_{j=1}^d$ to
$(F^{(k)}(w))_{k=1}^d$ has entries $(-A_j)^k$. Its determinant has
absolute value
\[
 \left(\prod_{j=1}^dA_j\right)
 \prod_{1\le i<j\le d}|A_j-A_i|>0.
\]
Its inverse is fixed, while $e^{-A_jw}$ is bounded below on
$[0,L_0]$. Hence, for some $c>0$,
\begin{equation}\label{eq:directional-derivative-mas14}
 \max_{1\le k\le d}|F^{(k)}(w)|\ge cL
 \qquad(0\le w\le L_0).
\end{equation}
A nonzero linear combination of $n$ exponentials with distinct real
exponents has at most $n-1$ distinct zeros on an interval. To see
this, discard zero coefficients, divide by one exponential, and
differentiate. The derivative has $n-1$ terms, so induction and
Rolle's theorem give the assertion, starting with $n=1$.
Consequently each equation $F^{(k)}=\pm cL/2$ has at most $d$
solutions, and $F''$ has at most $d-1$ zeros. None of these functions
is identically zero, since $L>0$ and all $A_j>0$.
Partition $I$ at these points for $1\le k\le d$ and at the zeros
of $F''$. There are $O_d(1)$ pieces. On each piece, $F'$ is monotone
and some fixed derivative has $|F^{(k)}|\ge\lambda:=cL/2$ throughout,
by \eqref{eq:directional-derivative-mas14} and the threshold partition.

The required derivative estimate follows from an
elementary sublevel argument. If $g$ has $|g^{(n)}|\ge\lambda$ on an
interval, then
\begin{equation}\label{eq:derivative-sublevel-mas14}
 \bigl|\{w:|g(w)|\le s\}\bigr|
 \le C_n(s/\lambda)^{1/n}\qquad(n\ge1).
\end{equation}
Indeed, if this set has measure $m>0$, one can choose $n+1$ of its
points with pairwise distances at least $m/[2(n+1)]$: at each step,
the forbidden intervals have total length less than $m$.
The divided-difference formula at these points, followed by the
mean-value formula for divided differences, gives
$\lambda/n!\le C_ns/m^n$, proving
\eqref{eq:derivative-sublevel-mas14}.
On a piece with $k\ge2$, apply this bound to $g=F'$ and $n=k-1$.
The part where $|F'|\le s$ contributes at most
$C_k(s/\lambda)^{1/(k-1)}$ to the integral. Its complement has at
most two intervals, since $F'$ is monotone. Integration by parts
there gives $O(s^{-1})$: the boundary terms are at most $2/s$ and
the total variation of $1/F'$ is at most $2/s$ on each interval.
Taking $s=\lambda^{1/k}$ gives $O_k(\lambda^{-1/k})$.
For $k=1$, the same integration by parts gives $O(\lambda^{-1})$
directly. Summing the bounded number of pieces, and using
$L^{-1/k}\le L^{-1/d}$ for $L\ge1$ and $k\le d$, proves the lemma.
\end{proof}

\begin{theorem}[Common prime activation and anisotropic Fourier thresholds]
\label{thm:shared-prime-anisotropic}
\index{Absolute continuity!common prime activations}
\index{Fourier transforms!anisotropic integrability}
Fix an integer $d\ge1$, a constant $c>0$, and
$0<A_1<\cdots<A_d$. Let $(B_p)_p$ be independent Bernoulli variables
with legal probabilities $q_p\in[0,1]$ satisfying
\begin{equation}\label{eq:shared-prime-summable-activation}
 \sum_{p\ \mathrm{prime}}|q_p-c/p|<\infty.
\end{equation}
Put $Q=\sum_{j=1}^dA_j$ and
\[
 Z=\sum_pB_p v_p,\qquad
 v_p=((\log p)^{-A_1},\ldots,(\log p)^{-A_d}),\qquad
 \mu=\operatorname{Law}(Z),\quad \Phi=\widehat\mu.
\]
The series converges almost surely, and $\mu$ is absolutely continuous
on $\R^d$. With $R(\xi)=\max_j|\xi_j|^{1/A_j}$,
\begin{equation}\label{eq:shared-prime-anisotropic-decay}
 |\Phi(\xi)|\le C R(\xi)^{-c}(\log R(\xi))^{2c}
 \qquad(R(\xi)\ge3).
\end{equation}
For every $q>0$,
\begin{equation}\label{eq:shared-prime-fourier-threshold}
 \Phi\in L^q(\R^d)\quad\Longleftrightarrow\quad cq>Q.
\end{equation}
For $1\le k\le d$, set
\begin{equation}\label{eq:shared-prime-mixed-boxes}
 \begin{split}
 \mathcal B_{k,R}
 &=\prod_{j<k}[R^{A_j},2R^{A_j}]
       \times\prod_{j\ge k}[R^{A_k},2R^{A_k}],\qquad R\ge3,\\
 V_k&=\sum_{j<k}A_j+(d-k+1)A_k.
 \end{split}
\end{equation}
Writing $\langle\cdot\rangle_{\mathcal B}$ for normalized Lebesgue
average, one has
\begin{equation}\label{eq:shared-prime-log-boxes}
 \langle-\log|\Phi|^2\rangle_{\mathcal B_{k,R}}
       =2c\log R+O(1),\qquad
 \int_{\mathcal B_{k,R}}|\Phi(\xi)|^q\,d\xi
       \gg_q R^{V_k-cq}.
\end{equation}
All constants may depend on the fixed exponents, $c$ and the entire
probability sequence; those in the lower integral may also depend on
$q$. Equal exponents force identical coordinates and
destroy absolute continuity in the ambient space.
\end{theorem}

\begin{proof}
Write $u_p=\log p$ and $\delta_p=q_p-c/p$. Partial summation from
\Cref{lem:prime-log-stieltjes} gives, for every fixed $a>0$,
\begin{equation}\label{eq:shared-prime-weighted-tail}
 \sum_{u_p>U}\frac{u_p^{-a}}p
       =\frac{U^{-a}}a+o(U^{-a})\qquad(U\to\infty).
\end{equation}
Indeed, integrate $u^{-a}$ against
$d(\log u+c_{\mathrm M}+\rho_{\mathrm M}(u))$ on $(U,\infty)$;
the error has size $O_a(U^{-a}e^{-b\sqrt U})$ after decreasing
$b>0$ if necessary. The perturbation contributes at most
$U^{-a}\sum_{u_p>U}|\delta_p|=o(U^{-a})$.
Thus $\sum_p q_p\|v_p\|_1<\infty$, which proves almost sure
convergence by nonnegativity. Also $q_p\to0$ and
$\sum_pq_p^2<\infty$, since
$q_p^2\le2c^2p^{-2}+2\delta_p^2$ and $\sum_p|\delta_p|<\infty$.

\emph{Absolute continuity.}
First use the reference probabilities $b_p=c/p$ beyond a fixed
$p_0>\max(2,4c)$, and $b_p=0$ for $p\le p_0$.
For $U_k=2^k$ sufficiently large put
\[
 N_k=\sum_{e^{U_k}<p\le e^{2U_k}}B_p,\qquad
 Y_k=\sum_{e^{U_k}<p\le e^{2U_k}}B_pv_p,\qquad
 W_k=\sum_{e^{U_k}<p\le e^{2U_k}}\frac{c}{p-c}.
\]
Here and in this paragraph the Bernoulli variables have probabilities
$b_p$. The pairs $(Y_k,N_k)$ are independent. Mertens' theorem and
$c/(p-c)=c/p+O_c(p^{-2})$ give
\[
 W_k\longrightarrow c\log2,\qquad
 \prod_{e^{U_k}<p\le e^{2U_k}}(1-c/p)\longrightarrow2^{-c},
 \qquad \mathbb P(N_k=1)\longrightarrow c(\log2)2^{-c}>0.
\]
Given $N_k=1$, the selected prime has probability
$c/((p-c)W_k)$. For $T=\|t\|_\infty$ and
$L_k(t)=\max_j|t_j|U_k^{-A_j}\ge1$, Stieltjes summation gives
\begin{equation}\label{eq:shared-prime-intensity-block}
 \left|\mathbb E(e^{i\langle t,Y_k\rangle}\mid N_k=1)\right|
 \le C L_k(t)^{-1/d}+C(1+T)e^{-b\sqrt{U_k}}.
\end{equation}
Here is the error calculation. Replacing $c/(p-c)$ by $c/p$
in the numerator costs $O_c(e^{-U_k})$. In
\eqref{eq:stieltjes} apply
$g(u)=\exp(i\sum_jt_ju^{-A_j})$ on $[U_k,2U_k]$.
The endpoint terms and the integral of $|\rho_{\mathrm M}g'|$
are bounded by
$C e^{-b\sqrt{U_k}}(1+\sum_j|t_j|U_k^{-A_j})$.
The main integral is
$c\int_{U_k}^{2U_k}g(u)\,du/u$; the substitution
$u=U_ke^w$ and \Cref{lem:directional-exponentials-mas14}, with
$L_0=\log2$, bound its modulus by $C L_k(t)^{-1/d}$.
Division by $W_k\asymp1$ proves
\eqref{eq:shared-prime-intensity-block}.

For $T^{1/(4A_d)}\le U_k\le T^{1/(2A_d)}$, the coordinate with
$|t_j|=T$ gives $L_k(t)\ge T^{1/2}$. Thus the last bound is at most
$T^{-1/(3d)}$ for all sufficiently large $T$, uniformly in direction.
On $2^m\le T<2^{m+1}$ use the deterministic set
\[
 J_m=\left\{k:
  \left\lceil\frac{m+1}{4A_d}\right\rceil\le k
        \le\left\lfloor\frac{m}{2A_d}\right\rfloor\right\}.
\]
All its blocks satisfy that estimate throughout the shell, and
\[
 \sum_{k\in J_m}\mathbb P(N_k=1)
 \sim\frac{c(\log2)2^{-c}}{4A_d}\,m.
\]
The hypotheses of \Cref{lem:conditional-block-smoothing}\textup{(b)}
hold with $\rho=1$ and $\delta=1/(3d)$. Its conclusion proves
absolute continuity for the reference tail and, after restoring its
finite independent head, for the reference series. 

Couple the reference and original variables with the same independent
uniform random variable at each prime. The probabilities of disagreement
sum to $\sum_p|q_p-b_p|<\infty$. By Borel--Cantelli, the two vector sums
differ almost surely by an element of the countable group
\[
 \Gamma=\left\{\sum_{p\in F}n_pv_p:
                   F\text{ finite},\ n_p\in\Z\right\}.
\]
For every Lebesgue-null Borel set $E$, membership of the original sum
in $E$ implies membership of the reference sum in
$\bigcup_{\gamma\in\Gamma}(E-\gamma)$, outside one null event.
The latter set is null, proving absolute continuity. The random
translation is not required to be independent. 

\emph{Directional decay.}
For $\xi\in\R^d$ put
\[
 D(\xi)=\sum_p\frac{1-\cos(\sum_j\xi_j u_p^{-A_j})}{p}.
\]
This series is finite for every $\xi$, by the quadratic bound and
\eqref{eq:shared-prime-weighted-tail}. Let $R=R(\xi)$ be large.
For $2\le U_0\le R$, divide $[U_0,R]$ into its intersections with
$[e^{-n-1}R,e^{-n}R]$, $n\ge0$. The substitution
$u=e^{-n-1}Re^w$ puts each integral of
$\exp(i\sum_j\xi_ju^{-A_j})\,du/u$ on a subinterval of $[0,1]$.
The new coefficients are $\xi_jR^{-A_j}e^{A_j(n+1)}$; their maximum
modulus is at least $e^{A_1(n+1)}$, because
$\max_j|\xi_j|R^{-A_j}=1$. Thus
\Cref{lem:directional-exponentials-mas14} gives
\[
 \left|\int_{U_0}^R\exp(i\sum_j\xi_ju^{-A_j})\,\frac{du}{u}\right|
 \le C\sum_{n\ge0}e^{-A_1(n+1)/d}\le C'.
\]
Take $U_0=K(\log R)^2$, where $K$ is fixed and sufficiently large.
For large $R$ this is at most $R$.
The Stieltjes error, as calculated above, is
\[
 O\left(e^{-b\sqrt{U_0}}
                  (1+\sum_j|\xi_j|U_0^{-A_j})\right)=O(1),
\]
since $|\xi_j|\le R^{A_j}\le R^{A_d}$ and we can take
$b\sqrt K>A_d+1$. Mertens' theorem and nonnegativity of omitted
terms consequently give
\begin{equation}\label{eq:shared-prime-directional-dissipation}
 D(\xi)\ge\log R-2\log\log R-O(1).
\end{equation}

Choose a finite initial prime set $F$ so that $q_p\le1/4$ off $F$.
Write $\Phi=\Phi_F\Phi_{F^c}$. Since
\[
 |1-q_p+q_pe^{it}|^2=1-2q_p(1-q_p)(1-\cos t),
\]
the expansion $-\log(1-x)=x+O(x^2)$, uniformly for $0\le x\le3/4$,
and the sums of $q_p^2$ and $|\delta_p|$ prove the uniform identity
\begin{equation}\label{eq:shared-prime-log-comparison}
 -\log|\Phi_{F^c}(\xi)|^2=2cD(\xi)+O(1).
\end{equation}
The finitely many terms of $D$ at $F$ are absorbed in $O(1)$.
Since $|\Phi_F|\le1$, combining this identity with
\eqref{eq:shared-prime-directional-dissipation} proves
\eqref{eq:shared-prime-anisotropic-decay} for large $R$; enlarging
$C$ handles the remaining compact range $3\le R\le R_0$.

\emph{Lower boxes, including the finite head.}
Fix $k$ and average on $\mathcal B_{k,R}$.
For $u\le R$, integration in coordinate $k$ gives
\[
 \left|\left\langle e^{i\sum_j\xi_ju^{-A_j}}
                   \right\rangle_{\mathcal B_{k,R}}\right|
 \le\min(1,2u^{A_k}/R^{A_k}).
\]
All other one-dimensional average factors have modulus at most one.
Partial summation gives
$\sum_{u_p\le R}p^{-1}(u_p/R)^{A_k}=O(1)$.
Also every coordinate on this box satisfies $|\xi_j|\le2R^{A_j}$
for $R\ge1$. Hence the tail $u_p>R$ has
\[
 \sum_{u_p>R}\frac{1-\cos(\sum_j\xi_ju_p^{-A_j})}{p}
 \le\frac d2\sum_j\xi_j^2
                       \sum_{u_p>R}\frac{u_p^{-2A_j}}p=O(1),
\]
uniformly on the box. Mertens' theorem yields
\begin{equation}\label{eq:shared-prime-dissipation-box-average}
 \langle D\rangle_{\mathcal B_{k,R}}=\log R+O(1).
\end{equation}
To restore the finite head in \eqref{eq:shared-prime-log-comparison},
note that a factor with $q_p\ne1/2$ has modulus at least
$|1-2q_p|>0$, and therefore bounded negative logarithmic modulus.
This includes $q_p=0,1$, whose moduli equal one. For $q_p=1/2$,
the squared modulus is
$\cos^2((\sum_j\xi_j u_p^{-A_j})/2)$.
The nonnegative periodic function $-\log\cos^2(t/2)$ is locally
integrable and has bounded averages on every interval of length at
least one. Indeed, split such an interval into full periods and at
most two remainders, each bounded by one period's integral.
With other coordinates fixed, the phase interval swept out by
coordinate $k$ has length $R^{A_k}u_p^{-A_k}\to\infty$.
Its average negative logarithm is therefore uniformly bounded for
large $R$. The zero hyperplanes of the finitely many such factors
have measure zero. Fubini and
\eqref{eq:shared-prime-log-comparison} and
\eqref{eq:shared-prime-dissipation-box-average}
prove the logarithmic identity in
\eqref{eq:shared-prime-log-boxes}, with an integrable logarithm.
Jensen's inequality then gives
\[
 \frac1{|\mathcal B_{k,R}|}
       \int_{\mathcal B_{k,R}}|\Phi|^q
 \ge\exp\left\{-\frac q2
        \langle-\log|\Phi|^2\rangle_{\mathcal B_{k,R}}\right\}
 \gg_q R^{-cq}.
\]
Since $|\mathcal B_{k,R}|=R^{V_k}$, this proves the lower bound.

Finally, $|\{\xi:R(\xi)\le r\}|=2^d r^Q$.
The bound \eqref{eq:shared-prime-anisotropic-decay}, integrated over
dyadic anisotropic shells, proves $\Phi\in L^q$ if $cq>Q$.
For necessity take $k=d$, so $V_d=Q$, and $R_n=\lambda^n$ with
$\lambda^{A_d}>2$. The boxes $\mathcal B_{d,R_n}$ are disjoint
in their last coordinate. If $cq\le Q$, each integral is bounded
below by a positive constant for all large $n$, including at equality.
Their sum diverges. Equal exponents force equality of the corresponding
coordinates term by term, completing the proof.
\end{proof}

\begin{proposition}[Exact global differentiability for common activation]
\label{prop:shared-prime-global-regularity}
\index{Density!global differentiability}
Under the hypotheses of \Cref{thm:shared-prime-anisotropic}, for every
integer $r\ge0$,
\begin{equation}\label{eq:shared-prime-global-cr}
 \mu\text{ has a globally }C^r(\R^d)\text{ density}
 \quad\Longleftrightarrow\quad Q+rA_d<c.
\end{equation}
In this range all derivatives through order $r$ are bounded.
Let $z_0=\sum_{p:q_p=1}v_p$, a finite sum. Then
\begin{equation}\label{eq:shared-prime-endpoint-rectangle}
 \mu\left(z_0+\prod_{j=1}^d[0,R^{-A_j}]\right)\gg R^{-c}
 \qquad(R\ge R_0).
\end{equation}
In particular, $Q>c$ excludes a bounded density, whereas $Q=c$
excludes a globally continuous density without deciding boundedness
alone. 
\end{proposition}

\begin{proof}
The almost sure convergence and $q_p\to0$ proved above imply that
$D_1=\{p:q_p=1\}$ is finite. Subtracting $z_0$ leaves a measure
supported on $[0,\infty)^d$.
For a fixed $K\ge1$, the probability of no nondeterministic activation
at primes with $u_p\le KR$ satisfies
\begin{equation}\label{eq:shared-prime-empty-prefix}
 \prod_{\substack{u_p\le KR\\q_p<1}}(1-q_p)\asymp R^{-c}.
\end{equation}
Remove a fixed initial
prime set containing $D_1$ and all $q_p>1/4$. The remaining logarithm
is $-\sum q_p+O(\sum q_p^2)=-c\log(KR)+O(1)$ by Mertens and
$\sum_p|q_p-c/p|<\infty$. Each removed nondeterministic factor
$1-q_p$ is a positive fixed number.

For the remaining tail $T_{R,j}=\sum_{u_p>KR}B_pu_p^{-A_j}$,
\eqref{eq:shared-prime-weighted-tail} gives
\[
 \mathbb ET_{R,j}
 =\frac{c}{A_j}(KR)^{-A_j}+o((KR)^{-A_j}).
\]
The perturbation error is at most
$(KR)^{-A_j}\sum_{u_p>KR}|q_p-c/p|=o((KR)^{-A_j})$
for fixed $K$. Choose $K$ large enough that
$\sum_j(c/A_j)K^{-A_j}<1/4$.
For all large $R$, Markov's inequality and a union bound give
\[
 \mathbb P(T_{R,j}\le R^{-A_j}\text{ for every }j)
 \ge1-\sum_jR^{A_j}\mathbb ET_{R,j}\ge1/2.
\]
Independence of this tail and the prefix event in
\eqref{eq:shared-prime-empty-prefix} proves
\eqref{eq:shared-prime-endpoint-rectangle}.

For sufficiency in \eqref{eq:shared-prime-global-cr}, let
$\alpha\in\Z_{\ge0}^d$ with $|\alpha|\le r$.
On $2^n\le R(\xi)<2^{n+1}$ one has
$|\xi^\alpha|\le2^{(n+1)\sum_j\alpha_j A_j}
\le2^{(n+1)rA_d}$. The shell has volume at most
$2^d2^{(n+1)Q}$. Thus \eqref{eq:shared-prime-anisotropic-decay}
gives
\[
 \int_{R(\xi)\ge3}|\xi^\alpha\Phi(\xi)|\,d\xi
 \ll\sum_{n\ge1}2^{n(Q+rA_d-c)}(1+n)^{2c}<\infty
\]
when $Q+rA_d<c$. The bounded-frequency integral is finite as well.
Differentiated Fourier inversion gives a globally $C^r$ density with
bounded derivatives of all orders at most $r$.

Conversely suppose there is a globally $C^r$ density $h$, and write
$H(x)=h(z_0+x)$. Since its measure vanishes in $x_d<0$, continuity
forces $H=0$ throughout that open half-space. All its derivatives
of order at most $r$ vanish there and, by continuity, on $x_d=0$.
Write $x'=(x_1,\ldots,x_{d-1})$; for $d=1$ this tuple is empty.
For $r=0$, uniform continuity on a compact rectangle gives
$H(x',y)=o(1)$ uniformly for $x'\in[0,1]^{d-1}$ as $y\downarrow0$.
For $r\ge1$, Taylor's formula and the vanished boundary derivatives give
\[
 H(x',y)=\frac{y^r}{(r-1)!}
       \int_0^1(1-t)^{r-1}\partial_d^rH(x',ty)\,dt=o(y^r),
\]
uniformly for $x'\in[0,1]^{d-1}$, since $\partial_d^rH(x',0)=0$
and this derivative is uniformly continuous on a compact rectangle.
Integration over $\prod_j[0,R^{-A_j}]$ yields
\[
 \mu\left(z_0+\prod_j[0,R^{-A_j}]\right)
       =o(R^{-Q-rA_d}).
\]
If $Q+rA_d\ge c$, multiplying by $R^c$ contradicts
\eqref{eq:shared-prime-endpoint-rectangle}; this also excludes equality.
Finally, a bounded density would assign at most $O(R^{-Q})$ mass
to the rectangle, which contradicts that lower bound when $Q>c$.
For $Q=c$ the boundedness comparison alone is inconclusive, while
the $r=0$ argument still excludes global continuity.
\end{proof}

\begin{proposition}[Exact isotropic Sobolev threshold for common activation]
\label{prop:shared-prime-sobolev}
\index{Sobolev regularity!common prime activations}
Under the hypotheses of \Cref{thm:shared-prime-anisotropic}, let
$h\in L^1(\R^d)$ be the density of $\mu$. For every real $s$,
\begin{equation}\label{eq:shared-prime-sobolev-threshold}
 \mu\in H^s(\R^d)
 \quad\Longleftrightarrow\quad h\in H^s(\R^d)
 \quad\Longleftrightarrow\quad
 s<\min_{1\le k\le d}
 \left\{\frac{2c-\sum_{j<k}A_j}{2A_k}
                       -\frac{d-k+1}{2}\right\}.
\end{equation}
Here a finite measure, or its density, is regarded as a tempered
distribution, and the inhomogeneous Sobolev convention is
\[
 \mu\in H^s(\R^d)
 \quad\Longleftrightarrow\quad
 \int_{\R^d}(1+|\xi|^2)^s|\widehat\mu(\xi)|^2\,d\xi<\infty.
\]
 In particular,
$h\in L^2(\R^d)$ exactly when $Q<2c$.
\end{proposition}

\begin{proof}
Absolute continuity is supplied by
\Cref{thm:shared-prime-anisotropic}. Since $\mu=h\,dx$ is finite,
both define the same tempered distribution and the same Fourier
transform $\Phi$. It remains to decide finiteness of the displayed
weighted integral.

\emph{Integration of the weight in every sector.}
For $r\ge2$ put
\[
 K_r=\prod_{j=1}^d[-r^{A_j},r^{A_j}],\qquad
 M_s(r)=\int_{K_r}(1+|\xi|^2)^s\,d\xi,
 \qquad A_0=0.
\]
Set $S_0=0$, $S_k=\sum_{j=1}^kA_j$ for $1\le k\le d$, and
\begin{equation}\label{eq:shared-prime-sobolev-sector-powers}
 E_0=0,\qquad
 E_k=S_{k-1}+(d-k+1+2s)A_k=V_k+2sA_k
 \quad(1\le k\le d).
\end{equation}
We will prove the bound, valid for every real $s$,
\begin{equation}\label{eq:shared-prime-sobolev-box-weight}
 M_s(r)\le C_s(1+\log r)\sum_{k=0}^d r^{E_k}.
\end{equation}
For $s<0$, replacing the weight by its value at the largest corner
of $K_r$ would underestimate the contribution from smaller $|\xi|$.
The sector decomposition below keeps these contributions.

To prove the bound, write $t=\|\xi\|_\infty$.
On $t\ge1$ the inequalities
$t^2\le1+|\xi|^2\le(d+1)t^2$ imply
$(1+|\xi|^2)^s\asymp_{d,s}t^{2s}$ for either sign of $s$.
The part $t\le1$ contributes at most a constant depending on $d,s$.
The exact volume distribution of $t$ on $K_r$ is
\[
 F_r(t)=|K_r\cap[-t,t]^d|
       =2^d\prod_{j=1}^d\min(t,r^{A_j}).
\]
For $r^{A_{k-1}}<t<r^{A_k}$ its derivative is
\[
 F_r'(t)=2^d(d-k+1)r^{S_{k-1}}t^{d-k}.
\]
The function $F_r$ is continuous and piecewise continuously
differentiable, so there is no atom at a sector boundary. Integration
against this volume distribution gives
\begin{equation}\label{eq:shared-prime-sobolev-sector-integrals}
 M_s(r)\ll_{d,s}
 1+\sum_{k=1}^d r^{S_{k-1}}
       \int_{r^{A_{k-1}}}^{r^{A_k}}t^{2s+d-k}\,dt.
\end{equation}

Put $\sigma_k=2s+d-k+1$. Since
$S_{k-1}+\sigma_kA_{k-1}=E_{k-1}$, direct integration gives
\begin{equation}\label{eq:shared-prime-sobolev-sector-evaluation}
 r^{S_{k-1}}
       \int_{r^{A_{k-1}}}^{r^{A_k}}t^{2s+d-k}\,dt
 =\begin{cases}
 \displaystyle\frac{r^{E_k}-r^{E_{k-1}}}{\sigma_k},
                                     &\sigma_k\ne0,\\[6pt]
 (A_k-A_{k-1})(\log r)r^{E_k},&\sigma_k=0.
 \end{cases}
\end{equation}
In the second case $E_k=E_{k-1}=S_{k-1}$.
For $\sigma_k\ne0$ the displayed nonnegative quantity is at most
$|\sigma_k|^{-1}(r^{E_k}+r^{E_{k-1}})$.
This proves \eqref{eq:shared-prime-sobolev-box-weight}, including
all logarithmic boundaries $s=-(d-k+1)/2$.

\emph{Sufficiency.}
Suppose that all inequalities in
\eqref{eq:shared-prime-sobolev-threshold} hold. They are equivalent
to $E_k<2c$ for $1\le k\le d$; also $E_0=0<2c$ because $c>0$.
Decompose the unbounded frequencies into the disjoint anisotropic
shells
\[
 \mathcal S_n=\{\xi:2^n\le R(\xi)<2^{n+1}\},\qquad n\ge2.
\]
On each shell \eqref{eq:shared-prime-anisotropic-decay} gives
$|\Phi(\xi)|^2\ll 2^{-2cn}(1+n)^{4c}$, while
$\mathcal S_n\subset K_{2^{n+1}}$.
The integrand is nonnegative, so, for every real $s$,
\begin{align*}
 \int_{\mathcal S_n}(1+|\xi|^2)^s|\Phi(\xi)|^2\,d\xi
 &\ll 2^{-2cn}(1+n)^{4c}M_s(2^{n+1})\\
 &\ll_s(1+n)^{4c+1}
                    \sum_{k=0}^d2^{n(E_k-2c)}.
\end{align*}
Each geometric series remains summable after multiplication by
$(1+n)^{4c+1}$, since every $E_k-2c$ is strictly negative.
The bounded region $R(\xi)<4$ contributes finitely because
$|\Phi|\le1$ and the inhomogeneous weight is bounded on compact
sets. This proves sufficiency.

\emph{Necessity and equality.}
Fix $1\le k\le d$. On the mixed box $\mathcal B_{k,R}$ of
\eqref{eq:shared-prime-mixed-boxes}, one has
$R^{A_k}\le|\xi|\le2\sqrt d\,R^{A_k}$, so
$(1+|\xi|^2)^s\asymp_{d,s}R^{2sA_k}$ for every real $s$.
Taking $q=2$ in \eqref{eq:shared-prime-log-boxes} therefore yields
\begin{equation}\label{eq:shared-prime-sobolev-lower-box}
 \int_{\mathcal B_{k,R}}(1+|\xi|^2)^s|\Phi(\xi)|^2\,d\xi
       \gg_s R^{V_k+2sA_k-2c}=R^{E_k-2c}.
\end{equation}
Choose a fixed $\lambda>1$ with $\lambda^{A_k}>2$ and let
$R_n=\lambda^n$. For all sufficiently large $n$, these boxes are
pairwise disjoint in coordinate $k$. If $E_k\ge2c$, each integral
in \eqref{eq:shared-prime-sobolev-lower-box} is bounded below by a
fixed positive number. Their sum diverges, including when
$E_k=2c$. Thus weighted integrability forces $E_k<2c$ for every
$k$, which is exactly the asserted criterion. 

Finally, at $s=0$ the powers $V_k$ increase with $k$, because
$V_{k+1}-V_k=(d-k)(A_{k+1}-A_k)>0$ for $k<d$, and $V_d=Q$.
The criterion is consequently $Q<2c$; $H^0=L^2$ by Plancherel.
\end{proof}

\paragraph*{Scalar and planar Sobolev specializations}
For $d=1$, \eqref{eq:shared-prime-sobolev-threshold} reads
$s<c/A_1-1/2$, exactly the scalar threshold with activation
coefficient $b=c/A_1$ in \Cref{prop:slow-jump-regularity}.
The prime-$2$ factor is a finite Bernoulli head allowed there.
For $d=2$, writing $a_-=A_1<a_+=A_2$, it reads
\[
 s<\min\left\{\frac{c}{a_-}-1,
                  \frac{2c-a_--a_+}{2a_+}\right\}.
\]
With $c=1$, these are exactly the two strict inequalities of
\eqref{eq:planar-thresholds}, including every negative order and
both equality exclusions. For arbitrary distinct planar exponents,
interchanging the coordinates is orthogonal and preserves the
Sobolev integral. \Cref{thm:planar-regularity} below also records the
Fourier-integrability threshold and global differentiability assertions.

\paragraph*{Canonical activation models}
The integer probabilities $q_p=1/p$ satisfy
\eqref{eq:shared-prime-summable-activation} with $c=1$ and zero
deviation. For the shifted-prime probabilities $q_2=1$ and
$q_p=1/(p-1)$ at odd primes,
\[
 \sum_p|q_p-1/p|
       =\frac12+\sum_{p\ge3}\frac1{p(p-1)}<\infty.
\]
Their deterministic minima are respectively $0$ and $v_2$.
Thus \Cref{thm:shared-prime-anisotropic,prop:shared-prime-global-regularity,prop:shared-prime-sobolev}
apply to both models with $c=1$, including their finite
probability-$1/2$ factor (at $2$ and $3$, respectively).
Their Sobolev thresholds are given by
\eqref{eq:shared-prime-sobolev-threshold} with $c=1$.

\begin{theorem}[Shared prime activations in finite dimension]
\label{thm:shared-prime-vector-mas14}
Let $d\ge1$ and let $\boldsymbol A=(A_1,\ldots,A_d)$ be a fixed family
of pairwise distinct positive exponents. Put
\[
 v_\ell=\bigl((\log\ell)^{-A_1},\ldots,(\log\ell)^{-A_d}\bigr).
\]
In the integer model take independent Bernoulli variables with
$\mathbb P(B_\ell=1)=1/\ell$ for all primes. In the shifted-prime
model take $B_2=1$ and independent odd-prime variables with
$\mathbb P(B_\ell=1)=1/(\ell-1)$. In either case the same $B_\ell$
drives every coordinate of the almost surely convergent vector series
\[
 Z_{\boldsymbol A}=\sum_{\ell\ \mathrm{prime}}B_\ell v_\ell.
\]
Its two laws, denoted by $\mu_{\boldsymbol A}^{\mathrm{int}}$ and
$\mu_{\boldsymbol A}^{\mathrm{sh}}$, are absolutely continuous on
$\R^d$. If two exponents are equal, the corresponding coordinates agree,
so the law is supported on a proper hyperplane and is not absolutely
continuous on $\R^d$.
\end{theorem}

\begin{proof}
Order the exponents by a coordinate permutation. As shown above, both
probability sequences satisfy \eqref{eq:shared-prime-summable-activation}
with $c=1$, so \Cref{thm:shared-prime-anisotropic} gives convergence
and absolute continuity for both models.
Equal exponents give identical coordinates term by term.
\end{proof}

\begin{theorem}[Joint logarithmic increments]
\label{thm:planar-regularity}
Let $A,B>0$ be distinct and let
\[
 Z=\sum_{\ell\ \mathrm{prime}}B_\ell
       \bigl((\log\ell)^{-A},(\log\ell)^{-B}\bigr),\qquad
 \mu=\operatorname{Law}(Z).
\]
The variables $B_\ell$ are independent, with either
$\mathbb P(B_\ell=1)=1/\ell$ for all primes, or $B_2=1$ and
$\mathbb P(B_\ell=1)=1/(\ell-1)$ for odd primes. Put
$a_-=\min(A,B)$, $a_+=\max(A,B)$ and $Q=a_-+a_+$. Then $\mu$ has a density
$h\in L^1(\R^2)$. For every $q>0$ and every real $s$,
\begin{equation}\label{eq:planar-thresholds}
 \begin{split}
 \widehat\mu\in L^q(\R^2)&\quad\Longleftrightarrow\quad q>Q,\\
 h\in H^s(\R^2)&\quad\Longleftrightarrow\quad
 s<\min\left\{\frac{2-a_- - a_+}{2a_+},\frac1{a_-}-1\right\}.
 \end{split}
\end{equation}
Here $H^s$ is defined by
$\int_{\R^2}(1+|\zeta|^2)^s|\widehat h(\zeta)|^2\,d\zeta<\infty$.
In particular, $h\in L^2$ exactly when $Q<2$. For every integer
$r\ge0$,
\begin{equation}\label{eq:planar-global-cr-mas14}
 \mu\text{ has a globally }C^r(\R^2)\text{ density}
 \quad\Longleftrightarrow\quad a_-+(r+1)a_+<1.
\end{equation}
Under this inequality all derivatives through order $r$ are bounded.
Thus a bounded continuous density exists exactly when $Q<1$.
\end{theorem}

\begin{proof}
Interchange the coordinates so that $A=a_-<B=a_+$. The canonical
activation checks above allow
\Cref{thm:shared-prime-anisotropic,prop:shared-prime-global-regularity,prop:shared-prime-sobolev}
with $d=2$ and $c=1$. The Fourier criterion
\eqref{eq:shared-prime-fourier-threshold} reads $q>Q$. The Sobolev
conditions for $k=1,2$ are $s<1/a_--1$ and $s<(2-Q)/(2a_+)$,
respectively; both are strict for every real $s$. The global
differentiability condition is $Q+ra_+<1$. This proves the stated
equivalences and derivative bounds.

As in \Cref{prop:shared-prime-global-regularity}, $Q>1$ excludes
bounded densities, whereas $Q=1$ excludes global continuity without
deciding boundedness alone.
\end{proof}

\paragraph*{A shared vector and an independent sum}
Take $(A,B)=(3/4,5/4)$ with the shifted-prime probabilities.
For the shared vector $Z$ of \Cref{thm:planar-regularity}, both
scalar marginals have $L^2$ densities, but $Q=2$ and
\[
 \min\left\{\frac{2-A-B}{2B},\frac1A-1\right\}=0.
\]
Thus its planar density belongs to $H^s(\R^2)$ exactly for real $s<0$,
and does not belong to $L^2(\R^2)$.
For comparison, take independent $Y_A\sim\mu_A$ and $Y_B\sim\mu_B$
and form the scalar sum $S=Y_A+Y_B$. Its activation coefficient is
$1/A+1/B=32/15$. By
\Cref{prop:slow-jump-regularity,prop:slow-jump-positive-endpoint},
$S$ has a globally $C^1$ density, and its density belongs to
$H^s(\R)$ exactly for real $s<49/30$.

\begin{corollary}[Arithmetic limits with a common input]
\label{cor:common-input-arithmetic}
\index{Absolute continuity!joint laws}
For every fixed finite family of pairwise distinct positive
exponents $\boldsymbol A=(A_1,\ldots,A_d)$, the empirical laws of
$(f_{A_j}(n))_{j=1}^d$ over the integers and
$(f_{A_j}(p-1))_{j=1}^d$ over primes converge, respectively, to
$\mu_{\boldsymbol A}^{\mathrm{int}}$ and
$\mu_{\boldsymbol A}^{\mathrm{sh}}$ from
\Cref{thm:shared-prime-vector-mas14}. Both limits are absolutely
continuous on $\R^d$. Put $Q=\sum_jA_j$ and $A_{\max}=\max_jA_j$.
For either limit $\mu$ and every $q>0$,
$\widehat\mu\in L^q(\R^d)$ exactly when $q>Q$.
For every integer $r\ge0$, a globally $C^r$ density exists exactly
when $Q+rA_{\max}<1$, and its derivatives through order $r$ are
then bounded. Write $A_{(1)}<\cdots<A_{(d)}$ for the exponents in
increasing order. For every real $s$, the density $h$ of either limit
satisfies
\begin{equation}\label{eq:common-input-sobolev}
 h\in H^s(\R^d)\quad\Longleftrightarrow\quad
 s<\min_{1\le k\le d}
 \left\{\frac{2-\sum_{j<k}A_{(j)}}{2A_{(k)}}
                         -\frac{d-k+1}{2}\right\}.
\end{equation}
In dimension two these are exactly the Sobolev conclusions of
\Cref{thm:planar-regularity}.
\end{corollary}

\begin{corollary}[Affine arithmetic laws and smoothness]
\label{cor:joint-arithmetic}
Let $d\ge1$ be an integer, let $L_j(n)=\alpha_jn+b_j$ $(1\le j\le d)$
satisfy the hypotheses of \Cref{thm:fixed-shift-tuple}, in particular
$\alpha_i b_j-\alpha_j b_i\ne0$ for $i\ne j$, and let
$A_1,\ldots,A_d>0$. The exponents may coincide.
The empirical laws of
$(f_{A_j}(L_j(p)))_{j=1}^d$, over primes $p>\max_j|b_j|$, converge to
the absolutely continuous law $\mu_{\boldsymbol A,\boldsymbol L}$ of
\[
 V_j=\sum_{\ell\ \mathrm{prime}}
       (\log\ell)^{-A_j}\mathbf1_{\{L_j(U_\ell)\equiv0\pmod\ell\}}
 \quad(1\le j\le d),
\]
where the $U_\ell$ are independent and uniform on $(\Z/\ell\Z)^*$.
For every integer $r\ge0$,
\begin{equation}\label{eq:log-shift-smoothness}
 \mu_{\boldsymbol A,\boldsymbol L}\text{ has a globally }C^r(\R^d)
 \text{ density}
 \quad\Longleftrightarrow\quad
 \max_jA_j<\frac1{r+1}.
\end{equation}
Its derivatives through order $r$ are then bounded. Write
$g$ for the density and $\mu=\mu_{\boldsymbol A,\boldsymbol L}$.
For every $q>0$ and real $s$,
\begin{equation}\label{eq:affine-fourier-sobolev-thresholds}
 \begin{aligned}
 \widehat\mu\in L^q(\R^d)&\quad\Longleftrightarrow\quad q>\max_j A_j,\\
 g\in H^s(\R^d)&\quad\Longleftrightarrow\quad
 s<\min_{\varnothing\ne J\subseteq\{1,\ldots,d\}}
        \sum_{j\in J}(1/A_j-1/2).
 \end{aligned}
\end{equation}
\end{corollary}

\begin{proof}[Proof of \Cref{cor:common-input-arithmetic,cor:joint-arithmetic}]
\emph{Empirical convergence for both models.}
For any fixed finite set of primes, the Chinese remainder theorem
gives independent uniform residue classes in integer sampling. Over
primes, \Cref{thm:sw}, applied to their fixed product modulus, gives
independent uniform reduced residue classes in the limit. Thus each
common-input vector has one shared activation per prime, with
probability $1/\ell$ in integer sampling and $1/(\ell-1)$ in
shifted-prime sampling; at $2$ the latter activation is deterministic.
The same finite-pattern argument gives exactly the displayed
indicator-array law for the affine forms.

Here is the required tail control. In integer sampling, for every
$C>0$, counting multiples gives
\[
 \frac1{\lfloor x\rfloor}\sum_{n\le x}
 \sum_{\substack{\ell\mid n\\\ell>y}}(\log\ell)^{-C}
 \le\sum_{\ell>y}\frac{(\log\ell)^{-C}}\ell.
\]
For the prime models, use the given affine forms or the single form
$L(n)=n-1$ for the common-input vector. For the chosen family set
$C_0=\max_j\alpha_j$, $B_0=\max(1,\max_j|b_j|)$, and let $E$ be
the exceptional set of \Cref{thm:fixed-shift-tuple}. For $y>\max E$, the divisors
$y<\ell\le\sqrt{C_0x+B_0}$ contribute to a coordinate's mean at most
\[
 O\left(\sum_{\ell>y}\frac{(\log\ell)^{-C}}{\ell-1}\right)+o(1),
\]
by Brun--Titchmarsh as in \eqref{eq:joint-shifted-tail}.
Each positive integer at most $C_0x+B_0$ has at most one prime
divisor exceeding $\sqrt{C_0x+B_0}$, whose jump is
$O((\log x)^{-C})$. All these tail bounds tend to zero as
$y\to\infty$, after taking $\limsup_{x\to\infty}$, by partial
summation from Mertens' theorem. Each limiting random series has
the same vanishing finite-mean tail bound.
For a bounded Lipschitz test function on the finite-dimensional
space, replacing the full vector by its truncation at $y$ changes
its expectation by at most the Lipschitz constant times the sum
of these coordinate tail means. Fixed-pattern convergence applies
to the truncated vector. First let $x\to\infty$ and then
$y\to\infty$; this proves weak convergence of all the asserted
empirical laws. Absolute continuity of the common-input limits
follows from \Cref{thm:shared-prime-vector-mas14}.
The integer and shifted-prime activation sequences satisfy
\eqref{eq:shared-prime-summable-activation} with $c=1$, as checked
above the canonical vector theorem. After ordering the exponents,
\Cref{thm:shared-prime-anisotropic,prop:shared-prime-global-regularity,prop:shared-prime-sobolev}
give all the asserted common-input Fourier, differentiability and
Sobolev thresholds. Coordinate permutations preserve the Euclidean
Sobolev weight, which justifies the ordering in
\eqref{eq:common-input-sobolev}. 

\emph{Absolute continuity and smoothness for the affine forms.}
For the affine forms, \Cref{prop:observed-likelihood}
gives $P\ll Q$ on the indicator array, where $Q$
is the product of its coordinate marginals. Under $Q$, the $V_j$ are
independent scalar laws with finitely modified activation probabilities.
They are absolutely continuous by \Cref{lem:slow-jump-all-ac}; pushing
forward $P\ll Q$ proves absolute continuity of
$\mu_{\boldsymbol A,\boldsymbol L}$.

Put $q_\ell=1/(\ell-1)$ and $a_{j,\ell}=(\log\ell)^{-A_j}$.
Outside $E$ the roots $-b_j\alpha_j^{-1}\pmod\ell$ are distinct
and nonzero. For all sufficiently large such primes, uniformly in
$\boldsymbol t$,
\[
 \log\left|1+q_\ell\sum_j(e^{it_ja_{j,\ell}}-1)\right|
 =-q_\ell\sum_j(1-\cos(t_ja_{j,\ell}))+O_d(q_\ell^2).
\]
Since $q_\ell=1/\ell+O(\ell^{-2})$, summation and
\eqref{eq:slow-diss-sharp} give
\begin{equation}\label{eq:log-shift-fourier-bound}
 |\widehat\mu_{\boldsymbol A,\boldsymbol L}(\boldsymbol t)|
 \ll_{\boldsymbol A,\boldsymbol L}
 \prod_{j=1}^d\frac{\log(e+|t_j|)}{(1+|t_j|)^{1/A_j}}.
\end{equation}
If $1/A_j>r+1$ for every $j$, every monomial of total degree at most
$r$ is integrable against this majorant. Differentiated Fourier
inversion proves sufficiency in \eqref{eq:log-shift-smoothness}.

For necessity, fix $j$. Let $D_j$ be the set of primes dividing
$\gcd(\alpha_j,b_j)$, with $2$ adjoined if $\alpha_j+b_j$ is even, and put
\[
 m_j=\sum_{\ell\in D_j}(\log\ell)^{-A_j},\qquad
 Y=\exp(Kh^{-1/A_j}),
\]
with $K$ a sufficiently large fixed constant. The primes in $D_j$ are
exactly the deterministic activations of coordinate $j$.
Let $E_{j,Y}$ exclude its activations at odd primes
$\ell\le Y$ with $\ell\nmid\alpha_jb_j$. Mertens' theorem gives
\[
 \mathbb P(E_{j,Y})
 =\prod_{\substack{3\le\ell\le Y\\\ell\nmid\alpha_jb_j}}
       \left(1-\frac1{\ell-1}\right)
 \asymp_{\boldsymbol A,\boldsymbol L,K}h^{1/A_j}.
\]
Conditional on $E_{j,Y}$, each other coordinate has activation
probability at most $1/(\ell-2)$ outside the fixed finite set $E$.
Consequently,
$\mathbb E(\sum_{i\ne j}V_i\mid E_{j,Y})$ is uniformly bounded.
For $Y>\max E$, the independent tail in coordinate $j$ has mean
$O((\log Y)^{-A_j})=O(K^{-A_j}h)$. Markov's inequality gives a fixed
$M>0$ such that
\[
 \mu_{\boldsymbol A,\boldsymbol L}
 \{\boldsymbol x:m_j\le x_j\le m_j+h,\ 0\le x_i\le M\ (i\ne j)\}
 \gg h^{1/A_j}.
\]
The law is supported in $x_j\ge m_j$. A globally $C^r$ density and
its normal derivatives through order $r$ vanish on $x_j=m_j$.
Uniform Taylor expansion on the displayed compact box therefore
makes its mass $o(h^{r+1})$. This excludes $1/A_j\le r+1$, proving
\eqref{eq:log-shift-smoothness}.

For $q>\max_j A_j$, the product in \eqref{eq:log-shift-fourier-bound}
is in $L^q$. On rectangular dyadic shells $1+|t_j|\asymp2^{n_j}$
with $m=\max_j n_j$, the weighted square integral is bounded by a
fixed polynomial in $m+1$ times
$2^{2sm+\sum_j(1-2/A_j)n_j}$. For $m>0$ the maximum of this
exponent on $0\le n_j\le m$, $\max n_j=m$, occurs at a vertex
whose coordinates equal $m$ on a nonempty set $J$ and zero elsewhere.
The stated strict bound on $s$ makes this exponent at most
$-c m$ for a fixed $c>0$. Summing the polynomially many shells per
$m$ gives Sobolev sufficiency for all real $s$.

For necessity, fix a nonempty $J$ and let $D_J(T)$ have intervals
$[T,2T]$ in its coordinates and $[0,1]$ elsewhere. Write
$\langle\cdot\rangle_D$ for normalized Lebesgue average on $D$.
The local expansion used for \eqref{eq:log-shift-fourier-bound} gives
$-\log|\widehat\mu_{\rm tail}|^2=2\sum_j\Sigma_{A_j}(t_j)+O(1)$
after removing a fixed finite head. The error is uniform since
$\sum_\ell q_\ell^2<\infty$. On $[0,1]$ each $\Sigma_{A_j}$ is
bounded, and on $[T,2T]$ its mean is $A_j^{-1}\log T+O(1)$ by
\eqref{eq:activation-scale-bounds} applied to the scalar tail
(together with its same uniform local expansion).

For each removed prime the characteristic factor is
$P(e^{ia_{1,\ell}t_1},\ldots,e^{ia_{d,\ell}t_d})$, where $P$ is
the multiaffine generating polynomial of the indicator vector, with
nonnegative coefficients summing to one. Choose an atom of mass $c>0$.
Reflect the circle coordinates occupied by that atom and multiply by
their monomial. On the torus this preserves the modulus after the
measure-preserving reflections and gives a polynomial with constant
coefficient $c$. Iterating the subharmonic mean inequality in each
coordinate shows that the torus mean of $\log|P|$ is at least $\log c$.
For instance, apply it first to
$\tfrac12\log(|P|^2+\varepsilon)$ for the transformed polynomial and
then decrease $\varepsilon$ to zero. Thus $-\log|P|$ is nonnegative
and integrable over the torus, including its zero locus in the
integral. Cover $D_J(T)$ by rectangles of the individual periods
$2\pi/a_{j,\ell}$. The ratio of the covered volume to $|D_J(T)|$
is bounded, since every side of the box has length at least one.
Consequently each finite-head factor contributes $O(1)$ to the
normalized mean of its negative log modulus, uniformly in $T\ge1$.

Combining head and tail gives
\[
 \langle-\log|\widehat\mu|^2\rangle_{D_J(T)}
    =2\sum_{j\in J}A_j^{-1}\log T+O(1),\qquad
 \int_{D_J(T)}|\widehat\mu|^q
    \gg_q T^{|J|-q\sum_{j\in J}1/A_j},
\]
the latter by Jensen's inequality. Dyadic boxes for fixed $J$ are
disjoint up to their boundaries. Singletons give the necessity of
$q>\max_j A_j$. With $q=2$ the Sobolev weight on $D_J(T)$ is
comparable to $T^{2s}$, so summing proves failure whenever
$s\ge\sum_{j\in J}(1/A_j-1/2)$, including equality.
\end{proof}

\begin{corollary}[Opposite-shift Fourier and Sobolev thresholds]
\label{cor:opposite-shift-regularity}
\index{Sobolev regularity!opposite shifts}
Let $A,B>0$ and specialize \Cref{cor:joint-arithmetic} to
$d=2$, $L_1(n)=n-1$, $L_2(n)=n+1$, $A_1=A$ and $A_2=B$.
Write $\mu_{A,B}^{\mathrm{opp}}$ for this law and
$g_{A,B}^{\mathrm{opp}}$ for its density.
For every $q>0$ and every real $s$,
\begin{equation}\label{eq:opposite-shift-thresholds}
 \begin{split}
 \widehat{\mu_{A,B}^{\mathrm{opp}}}\in L^q(\R^2)
   &\quad\Longleftrightarrow\quad q>\max(A,B),\\
 g_{A,B}^{\mathrm{opp}}\in H^s(\R^2)
   &\quad\Longleftrightarrow\quad
     s<\min\left\{\frac1A-\frac12,\frac1B-\frac12,
                         \frac1A+\frac1B-1\right\}.
 \end{split}
\end{equation}
In particular, $g_{A,B}^{\mathrm{opp}}\in L^2$ exactly when
$\max(A,B)<2$, and it has a bounded continuous version exactly when
$\max(A,B)<1$. Setting $B=A$ gives the equal-exponent thresholds
$q>A$ and $s<\min\{2/A-1,1/A-1/2\}$.
\end{corollary}

\begin{proof}
Put $q_\ell=1/(\ell-1)$, $a_\ell=(\log\ell)^{-A}$ and
$b_\ell=(\log\ell)^{-B}$, and write $\Psi=\widehat{\mu_{A,B}^{\mathrm{opp}}}$ and
$F_\ell(\xi,\eta)=1-2q_\ell+q_\ell e^{ia_\ell\xi}+q_\ell e^{ib_\ell\eta}$.
For $\ell\ge7$, Taylor expansion gives, uniformly in $(\xi,\eta)$,
\[
 \log|F_\ell(\xi,\eta)|
 =-q_\ell\bigl(1-\cos(a_\ell\xi)+1-\cos(b_\ell\eta)\bigr)
       +O(q_\ell^2).
\]
Thus the product $\Psi_{\ge7}$ of these factors satisfies
\begin{equation}\label{eq:opposite-shift-log}
 -\log|\Psi_{\ge7}(\xi,\eta)|^2
 =2\Sigma_A(\xi)+2\Sigma_B(\eta)+O_{A,B}(1).
\end{equation}
By \eqref{eq:slow-diss-sharp},
\begin{equation}\label{eq:opposite-shift-decay}
 |\Psi(\xi,\eta)|\ll_{A,B}
 \frac{\log(e+|\xi|)\log(e+|\eta|)}
      {(1+|\xi|)^{1/A}(1+|\eta|)^{1/B}}.
\end{equation}
The forms are nonproportional, and the three nonempty subsets of the
two coordinates give exactly the three Sobolev bounds in
\eqref{eq:opposite-shift-thresholds} by
\eqref{eq:affine-fourier-sobolev-thresholds}. The bounded continuous
criterion is \eqref{eq:log-shift-smoothness} with $r=0$.
\end{proof}

\section{Sampling over the integers}\label{par:slow-jump-integer}
The integer Erd\H{o}s--Wintner law of $f_A$ is
\[
 \mu_A^{\mathrm{int}}
 =\left(\tfrac12\delta_0+\tfrac12\delta_{a_2}\right)
 *\Conv_{\substack{\ell\ge3\\\ell\textup{ prime}}}
 \left(\left(1-\frac1\ell\right)\delta_0+\frac1\ell\delta_{a_\ell}\right).
\]
Write $\mu_{A,\mathrm{odd}}^{(1)}$ for the odd-prime convolution in this
display.
The full integer law is the case $m=1$, $c_1=1$ of
\Cref{prop:slow-jump-regularity}, with
one additional independent Bernoulli term of probability $1/2$ and
increment $a_2$.
Hence $\mu_A^{\mathrm{int}}$ is absolutely continuous for every $A>0$,
with $\dim_F(\mu_A^{\mathrm{int}})=\min(1,2/A)$ and
$\dim_H(\mu_A^{\mathrm{int}})=1$; it has an $L^2$ density exactly when
$A<2$. By \Cref{prop:slow-jump-positive-endpoint}, it has a globally
$C^k$ density exactly when $A<1/(k+1)$.
More generally, \Cref{prop:slow-jump-regularity} applies to
$(\mu_A^{\mathrm{int}})^{*m}$ $(m\ge1)$ with $b=m/A$ and $C=m$;
\Cref{prop:slow-jump-positive-endpoint} gives a globally $C^k$ density
exactly when $m/A>k+1$.
For $s=0$ and integer data with $m=2$, $c_1=c_2=1$ and
$0<A_1,A_2<2$, the sufficiency in
\Cref{prop:slow-jump-regularity}\textup{(c)} is contained
in~\cite[Theorem~2 and its proof, pp.~193--194]{Bab92}.
More generally, under the printed Condition~K, Babu obtains a
square-integrable Fourier transform for the convolution of two integer
limiting laws, when they exist, with
$|g_j(p)|=\psi_j(p)$ outside an exceptional prime set with convergent
reciprocal sum, where $\psi_j$ eventually decreases to $0$ and
$\psi_j(x)(\log x)^{A_j}$ is eventually nondecreasing for
$0<A_j<2$. The example following Theorem~2 takes exponents $3$ and $4$,
although these lie outside the exponent restriction printed in Condition~K.

For $A>1$, the concentration function satisfies
$Q_{\mu_A^{\mathrm{int}}}(h)\asymp_A h^{1/A}$ as $h\downarrow0$ by
\cite[Theorem~1.3]{Kou14}, settling the concentration conjecture in
\cite[Remark~1 after Theorem~3, p.~297]{EK79};
\cite[Corollaire~1.4]{dlBT12} proved this order for sufficiently large $A$.
For the shifted-prime law one likewise has
$Q_{\mu_A}(h)\asymp_A h^{1/A}$ as $h\downarrow0$ for $A>1$;
the lower bound is given by \cite[Theorem~1]{Ten12}.
Indeed, put $\nu=\mu_{A,\mathrm{odd}}^{(1)}$ and take mutually independent
$I_\ell\sim\operatorname{Bernoulli}(1/\ell)$ and
$E_\ell\sim\operatorname{Bernoulli}((\ell-1)^{-2})$ over odd primes.
Then $B_\ell=\max(I_\ell,E_\ell)$ has probability $1/(\ell-1)$ of being one,
and $S=\{\ell:E_\ell=1\}$ is almost surely finite.
Conditionally on $S$, the shifted-prime sum is a translate of the law
$\nu_S$ obtained from $\nu$ by deleting the coordinates in $S$.
As in \eqref{eq:ap-cylinder-domination},
$\nu\ge w_S\nu_S$, where $w_S=\prod_{\ell\in S}(1-1/\ell)$.
Consequently
\[
 Q_{\mu_A}(h)\le Q_\nu(h)\mathbb E w_S^{-1}
 =Q_\nu(h)\prod_{\ell\ge3}\bigl(1+(\ell-1)^{-3}\bigr).
\]
The component $S=\varnothing$ gives
$Q_{\mu_A}(h)\ge Q_\nu(h)\prod_{\ell\ge3}(1-(\ell-1)^{-2})$.
Both products are finite and positive. Since the full integer law is
the equal mixture of $\nu$ and its translate by $a_2$,
$Q_{\mu_A^{\mathrm{int}}}(h)\le Q_\nu(h)\le2Q_{\mu_A^{\mathrm{int}}}(h)$.
Thus $Q_{\mu_A}(h)\asymp Q_{\mu_A^{\mathrm{int}}}(h)$ with absolute
constants for every $A>0$ and $h>0$, proving the asserted order.
For $A>2$, the concentration lower bound and Esseen's inequality
also give $\dim_F\mu\le2/A$ for
$\mu\in\{\mu_A,\mu_A^{\mathrm{int}}\}$: a bound
$|\widehat\mu(t)|\ll|t|^{-\sigma/2}$ with $2/A<\sigma<1$ would imply
$Q_\mu(h)\ll h^{\sigma/2}=o(h^{1/A})$.

For integer sampling, absolute continuity in the range $0<A<2$ is
classical. Erd\H{o}s \cite[\S~2, p.~724]{Erd39} stated it for $A=1$,
and Babu proved the existence of an $L^2$ density for $0<A<2$
\cite[pp.~401--403]{Bab75}. His equation~(5) gives a squared Fourier
modulus bounded by a constant times $|\tau|^{-\delta}$ for every
$\delta<2/A$. Babu later treated additive functions $g$ with an existing
integer limiting law and $|g(p)|=\psi(p)$ outside a set of primes with
convergent reciprocal sum. He proved absolute continuity when $\psi$
eventually decreases to $0$ and $\psi(x)(\log x)^\alpha$ is eventually
nondecreasing for some $0<\alpha<2$~\cite[Theorem~1, p.~30]{Bab95}.
For $\psi(x)=(\log x)^{-A}$ this again requires $A<2$.
Integrability of the Fourier transform for $0<A<1$ is recorded in
\cite[p.~296 and (4.2), p.~302]{EK79}.

For the odd-prime integer law $\mu_{A,\mathrm{odd}}^{(1)}$, Saffari's
two-sided estimate
$|\widehat{\mu_{A,\mathrm{odd}}^{(1)}}(\tau)|=|\tau|^{-1/A+o(1)}$
is recorded in \cite[Chapter~III.2, Exercise~259(c), pp.~441--442]{Ten15}.
The full integer and shifted-prime transforms carry the additional modulus
factors $|\cos(\tau a_2/2)|$ and $|\cos(\tau a_3/2)|$, which vanish
infinitely often, so this two-sided estimate fails for the full laws.
\Cref{prop:slow-jump-regularity}\textup{(a)} nevertheless gives their exact
Fourier dimension. By \eqref{eq:slow-summable-log}, after discarding the
primes $\ell\le L$ for a fixed $L\ge3$, the negative logarithms of the two
remaining Fourier moduli differ by $O(1)$, uniformly in the frequency.

The all-$A$ conclusion of \Cref{lem:slow-jump-all-ac} also applies to
the integer law. In particular, the choice
$g(\ell)=(\log\ell)^{-A}$, $A>2$, has an absolutely continuous
integer limiting law, contrary to the singularity conjecture in
\cite[Remark~2 after Theorem~3, p.~297]{EK79}.

\chapter{Fourier decay and normality}
\label{sec:raj-annulus}

The four basic totient and divisor-sum laws, and their logarithmic
images, have Hausdorff dimension zero. This excludes Fourier bounds $O(|t|^{-\delta})$ with $\delta>0$
(\Cref{lem:fourier-Hausdorff}). This chapter proves quantitative
logarithmic decay. For logarithmic laws, the proof combines Euler
products with cancellation in prime sums. For raw ratios, conditioning
on a prime block produces a reciprocal phase to which the same
cancellation estimates apply. The resulting bounds imply almost-sure
normality. The basic-law estimates in
\Cref{thm:vk-fourier,thm:raw-ratio-normality} are effective.
For fixed periodic weights,
\Cref{thm:periodic-weight-dissipation,thm:periodic-raw-product-fourier}
allow ineffective constants through their progression-mass input.
The next chapter studies Fourier amplitudes over long frequency intervals
and selected resonant frequencies.

Throughout, $e(z):=e^{2\pi iz}$. 
Here $\mathbf d_k$ denotes the $k$-fold divisor function.

\KeepHeadingSpace{4\baselineskip}

\section{Fourier products and independent prime phases}
\label{sec:raj-prelim}

For the logarithmic totient law $\mu_f$, the Fourier product of
\Cref{lem:factor} turns a lower bound on the nonnegative sum $S(\tau)$
into an upper bound on $|\FT(\tau)|$. After removing the potentially vanishing factor at $3$, the squared
modulus is comparable to the exponential of the negative dissipation
sum with the same prime omitted.

A second input concerns phase averages. By the $\Q$-linear independence
of the $\omega_\ell$ (\Cref{thm:linind}) and
Kronecker--Weyl~\cite[Satz~5, pp.~319--320]{Wey16}, the phases
$\tau\omega_\ell$ for each fixed finite set of primes are equidistributed
modulo $2\pi$ as the frequency interval grows; in particular, they can
approach prescribed values simultaneously. This supplies the independent
uniform angles in the later totient amplitude laws. Passing to the
infinite product still requires tail estimates (\Cref{lem:amplitude-tails}). Uniform decay at every
large frequency instead requires cancellation over a prime range that
grows with the frequency.

\paragraph*{Notation}\phantomsection\label{sec:raj-notation}
Throughout \Cref{sec:raj-annulus,ssec:typical-decay,ssec:annulus},
$\ell$ ranges over odd primes unless otherwise stated. Recall
$\omega_\ell$, $d_\ell$ and $S$ from~\eqref{eq:def-omega}
and~\eqref{eq:def-dell}.

\subsection{Linear independence of the jumps \texorpdfstring{$\omega_\ell$}{omega-ell}}
We recall Schoenberg's independence lemma
\cite[no.~19, p.~196]{Sch28}.

\begin{theorem}[Linear independence]\leavevmode\phantomsection\label{thm:linind}
\index{Logarithmic jumps!linear independence}\ignorespaces
For primes $3\le\ell_1<\cdots<\ell_k$, the numbers
$\log2,\omega_{\ell_1},\ldots,\omega_{\ell_k}$ are linearly independent over~$\Q$.
\end{theorem}

\begin{proof}
Clear denominators in a rational relation, obtaining
$a_0\log2+\sum_i a_i\omega_{\ell_i}=0$ with $a_i\in\Z$.
If some $a_i\ne0$ with $i\ge1$, choose the largest such $\ell_j$.
Exponentiation gives $2^{a_0}\prod_{i\le j}(\ell_i/(\ell_i-1))^{a_i}=1$.
The $\ell_j$-adic valuation of the left side is $a_j$, since
$\ell_i-1<\ell_j$ for $i\le j$ and $\ell_i<\ell_j$ for $i<j$.
The right side has valuation zero, a contradiction.
Thus $a_i=0$ for $i\ge1$, and then $a_0=0$.
\end{proof}

\Needspace{5\baselineskip}
\subsection{The cosine sum and the Fourier product}\label{sec:raj-series}

\begin{lemma}[Comparing the cosine sum with the Fourier product]\label{lem:dissipation-comparisons}
Put
\[
 S_5(\tau):=\sum_{\ell\ge5}d_\ell(1-\cos(\tau\omega_\ell)),\qquad
 C_{\mathrm{sh}}:=\sum_{\ell\ge5}[-\log(1-2d_\ell)-2d_\ell].
\]
For the product $\mathcal N_{\{3\}}$ of
\Cref{thm:uniform-fourier-moments}, obtained by removing the factor at
$\ell=3$, one has $C_{\mathrm{sh}}<\infty$
and, for every $\tau\in\R$,
\begin{equation}\label{eq:deleted-comparison}
 e^{-C_{\mathrm{sh}}-S_5(\tau)}
 \le|\mathcal N_{\{3\}}(\tau)|^2\le e^{-S_5(\tau)},\qquad
 |\widehat{\mu_f}(\tau)|^2
 =\cos^2(\tau\omega_3/2)|\mathcal N_{\{3\}}(\tau)|^2.
\end{equation}
Both multiplicative constants are optimal.

For every $\tau\in\R$,
\begin{equation}\label{eq:forward}
 |\widehat{\mu_f}(\tau)|^2\le e^{-S(\tau)}.
\end{equation}
\end{lemma}

\begin{proof}
Put $r(v):=-\log(1-v)-v=\int_0^v u(1-u)^{-1}\,du$ and
$x_\ell:=d_\ell(1-\cos(\tau\omega_\ell))$.
For $\ell\ge5$, $0\le x_\ell\le2d_\ell\le2d_5=3/4$, so
\[
 0\le r(x_\ell)\le r(2d_\ell)\le8d_\ell^2.
\]
Thus $R(\tau):=\sum_{\ell\ge5}r(x_\ell)$ converges uniformly and
$0\le R(\tau)\le C_{\mathrm{sh}}<\infty$.
The product in~\eqref{eq:euler} gives
$|\mathcal N_{\{3\}}(\tau)|^2=e^{-S_5(\tau)-R(\tau)}$, and the
prime-$3$ factor of the full product is $\cos^2(\tau\omega_3/2)$, proving
\eqref{eq:deleted-comparison}. The upper constant is attained at
$\tau=0$, where $R(0)=0$.
By \Cref{thm:linind} and Kronecker--Weyl, the phases approach $\pi$
simultaneously for each fixed finite set of primes. The remaining terms
of the uniformly convergent series for $R$ can be made uniformly small.
It follows that $\sup_\tau R(\tau)=C_{\mathrm{sh}}$, proving
optimality of the lower constant.

Each factor $1-d_\ell(1-\cos(\tau\omega_\ell))$ in
\eqref{eq:euler} lies in $[0,1]$. Apply $1-x\le e^{-x}$ for $x\ge0$
to every factor of a finite product. By the convergence noted after
\eqref{eq:def-dell}, the corresponding partial sum converges to $S(\tau)$;
passing to the limit
gives~\eqref{eq:forward}.
\end{proof}

\section{A Vinogradov bound for logarithmic and reciprocal phases}
\label{sec:vinogradov-log-phase}

The arithmetic input must control every partial prime sum in a dyadic
interval, so that weights can be inserted by Abel summation. Vaughan's
identity reduces these prime sums to linear and bilinear sums. The
derivative estimate must then apply both to the original phases and
to differences of phases created by Cauchy--Schwarz. The logarithmic
phases feed the Fourier products; the reciprocal phase will be used
when a raw product is conditioned on a single active prime.

\begin{lemma}[Vinogradov's derivative estimate {\cite[Lemma~4.1]{MRSTT22}}]\label{lem:vk-input}
Let $X\ge2$, $F\ge X^4$, $\alpha\ge1$, let $I\subset[X,2X]$ be
an interval, and let a smooth real phase $g$ satisfy
\[
 \alpha^{-j^3}F\le\frac{u^j}{j!}|g^{(j)}(u)|
 \le\alpha^{j^3}F\quad(u\in I),\qquad
 1\le j\le10\lceil\log F/\log X\rceil+1.
\]
If
\begin{equation}\label{eq:vk-extra-condition}
 (\log\alpha)\frac{(\log F)^2}{(\log X)^3}<10^{-3},
\end{equation}
one has, with absolute effective constants and some absolute $c_0>0$,
\begin{equation}\label{eq:vk-derivative-input}
 \left|\sum_{n\in I\cap\Z}e(g(n))\right|
 \ll\alpha X\exp\left(-c_0\frac{(\log X)^3}{(\log F)^2}\right).
\end{equation}
\end{lemma}

\begin{lemma}[Vaughan reduction with uniform interval bounds]
\label{lem:vk-uniform-reduction}
Let $N\ge64$, $0<E\le1$, and let $\phi:[N/4,8N]\to\R$.
Let $\chi$ be a Dirichlet character modulo an integer $D\ge1$, with
the constant character $1$ when $D=1$. Suppose, with a common constant
$C_0$, that the following bounds hold on every indicated interval:
\begin{enumerate}[label=\textup{(\alph*)}]
\item For every integer $1\le m\le N^{2/3}$, with $W=N/m$,
\[
 \left|\sum_{n\in J\cap\Z}\chi(n)e(\phi(mn))\right|
 \le C_0WE\qquad(J\subseteq(W,2W]).
\]
\item Whenever
\[
 N^{1/3}/2\le V\le U\le2N^{2/3},\qquad N/4<UV\le2N,
\]
and $V<v_1<v_2\le2V$ are integers,
\[
 \left|\sum_{u\in J\cap\Z}
 e(\phi(uv_1)-\phi(uv_2))\right|
 \le C_0UE\qquad(J\subseteq(U,2U]).
\]
\end{enumerate}
Then, uniformly for $N\le x\le2N$,
\[
 \left|\sum_{N<p\le x}\chi(p)e(\phi(p))\right|
 \ll_{C_0}N(\log N)^8\{E^{1/2}+N^{-1/6}\}.
\]
The implied constant is effective in $C_0$, with no further dependence
on $D$.
\end{lemma}

\begin{proof}
Put $z=N^{1/3}$ and use subscripts for truncation at $z$. On $n>z$,
\begin{equation}\label{eq:vk-vaughan}
 \Lambda=\mu_{\le z}*\log
 -\mu_{\le z}*1*\Lambda_{\le z}
 +\mu_{>z}*1*\Lambda_{>z}.
\end{equation}
Indeed $\log=1*\Lambda$ and $\mu*1$ is the convolution identity;
the omitted term $\Lambda_{\le z}$ vanishes on $n>z$.
Complete multiplicativity, including zeros on nonunits, factors $\chi$
exactly in each convolution.

The first two terms have outer variable $m\le z^2$, inner interval
$(N/m,x/m]$, and coefficient magnitudes at most $1$ and
\[
 \left|\sum_{\substack{ab=m\\a,b\le z}}\mu(a)\Lambda(b)\right|
 \le\sum_{b\mid m}\Lambda(b)=\log m.
\]
Only the first inner sum has weight $\log n$. Abel summation using
\textup{(a)} costs $O(\log N)$ for this weight. The estimate
$\sum_{m\le z^2}(1+\log m)/m\ll(\log N)^2$ therefore bounds the
combined Type~I contribution by $O_{C_0}(N(\log N)^2E)$.

The third term has variables $u,v>z$, coefficients
$a(u)=\sum_{d\mid u,\,d>z}\mu(d)$ and $b(v)=\Lambda(v)$, and the
exact cutoff $N<uv\le x$. Dyadic subdivision independent of $x$,
followed if necessary by interchange, gives $O((\log N)^2)$ blocks
with the scales in \textup{(b)}. Absorb the characters into the
coefficients before Cauchy--Schwarz, writing
$A_u=a_u\chi(u)$, $B_v=b_v\chi(v)$ also after interchange, and put
$A_2=\sum|A_u|^2$, $B_2=\sum|B_v|^2$. A block satisfies
\[
 |T|^2\le A_2\sum_{U<u\le2U}
 \left|\sum_{\substack{V<v\le2V\\N<uv\le x}}
 B_ve(\phi(uv))\right|^2.
\]
The outer sum here contains all integers in its interval; no character
factor or coprimality restriction survives. The diagonal contributes
$O(A_2B_2U)$. For $v_1<v_2$ the exact remaining interval is
\[
 \bigl(\max(U,N/v_1,N/v_2),
        \min(2U,x/v_1,x/v_2)\bigr],
\]
possibly empty. Hypothesis \textup{(b)} and
$(\sum_v|B_v|)^2\ll VB_2$ give
$|T|^2\ll_{C_0}A_2B_2U(1+VE)$.
Since $|\chi|\le1$, $|a(u)|\le\mathbf d_2(u)$ and
$|b(v)|\le\log(2N)$, one has
\[
 A_2B_2\ll UV(\log N)^5,\qquad
 |T|\ll_{C_0}N(\log N)^{5/2}\{N^{-1/6}+E^{1/2}\}.
\]
Here $\mathbf d_2(n)^2\le\mathbf d_4(n)$ follows prime by prime from
$(a+1)^2\le\binom{a+3}{3}$, and ordered factorization gives
$\sum_{n\le y}\mathbf d_4(n)\le y(1+\log y)^3$; these estimates
also cover interchanged blocks.

Summing the blocks gives, uniformly in $x$,
\[
 \left|\sum_{N<n\le x}\Lambda(n)\chi(n)e(\phi(n))\right|
 \ll_{C_0}N(\log N)^5\{N^{-1/6}+E^{1/2}\}.
\]
Prime powers of exponent at least two number $O(N^{1/2}\log N)$
up to $2N$. Their weights are at most $\log(2N)$, so their contribution
is $O(N^{1/2}(\log N)^2)$. Put
\[
 \mathcal B(x)=\sum_{N<p\le x}(\log p)\chi(p)e(\phi(p)).
\]
Abel summation gives
\[
 \sum_{N<p\le x}\chi(p)e(\phi(p))
 =\frac{\mathcal B(x)}{\log x}
   +\int_N^x\frac{\mathcal B(u)}{u(\log u)^2}\,du.
\]
This proves the asserted bound, uniformly in the real endpoint.
\end{proof}

\begin{proposition}[Bounds for partial prime sums]
\label{prop:vk-prime-prefix}
\index{Prime cancellation!logarithmic and reciprocal phases}
There are absolute effective constants $b,c,C>0$ such that, uniformly for
real $\tau,N,x$ with $\tau>1$, $64\le N\le b\tau^{1/5}$ and
$N\le x\le2N$, and for any of the phases
$w(u)=\log(u/(u-1))$, $w(u)=\log(1+1/u)$ or $w(u)=1/u$,
\begin{equation}\label{eq:vk-prime-prefix}
 \left|\sum_{N<p\le x}e^{i\tau w(p)}\right|
 \le CN(\log N)^8\left\{
 \exp\left(-c\frac{(\log N)^3}{(\log\tau)^2}\right)+N^{-1/6}\right\}.
\end{equation}
\end{proposition}

\begin{proof}
Put $H=\log\tau$ and $\phi(u)=\tau w(u)/(2\pi)$.
For the first phase, the exact identities
\[
 \phi(mu)=\frac{\tau}{2\pi}\int_0^{1/m}\frac{ds}{u-s},\qquad
 \phi(uv_1)-\phi(uv_2)
 =\frac{\tau}{2\pi}\int_{1/v_2}^{1/v_1}\frac{ds}{u-s}
 \quad(v_1<v_2)
\]
give the two-sided derivative bounds required in \Cref{lem:vk-input}
at every order, as follows. On $[W,2W]$ take
$F_m=\tau/(2\pi mW)$; on $[U,2U]$ take
$F_\Delta=\tau\Delta/(2\pi U)$, where
$\Delta=1/v_1-1/v_2$. The ratio of $u^j|g^{(j)}(u)|/j!$ to
its respective $F$ is $(W/u)$ or $(U/u)$ times the average of
$(1-s/u)^{-j-1}$ over the indicated interval. If $mW\ge2$ or
$UV\ge2$, respectively, and $v_i\in(V,2V]$, these ratios lie in
$[1/2,2^{j+1}]$. Thus $\alpha=8$ works for every $j\ge1$.
For $w(u)=\log(1+1/u)$, replace $u-s$ by $u+s$ in both
integrals. The ratios then lie in $[2^{-j-2},1]$ when $mW,UV\ge1$.
For $w(u)=1/u$, the phases are $\tau/(2\pi mu)$ and
$\tau\Delta/(2\pi u)$, respectively; their normalized derivative ratios
are exactly $W/u$ and $U/u$, in $[1/2,1]$ at every order.
The derivative signs are immaterial; $\alpha=32$ is a common choice
for all three phases, at every order and on every subinterval.

Use the scales of \Cref{lem:vk-uniform-reduction}. In Type~I,
$W=N/m\ge N^{1/3}$ and $F_m=\tau/(2\pi N)$, so
$F_m\ge W^4$ if $\tau\ge2\pi N^5$. In Type~II,
\[
 N^{1/3}/2\le V\le U\le2N^{2/3},\qquad N/4<UV\le2N.
\]
For every off-diagonal pair $v_1<v_2$, including $v_2-v_1=1$,
\[
 F_\Delta=\frac{\tau(v_2-v_1)}{2\pi Uv_1v_2}
 \ge\frac{\tau}{8\pi UV^2},\qquad
 U^5V^2=U^3(UV)^2\le32N^4.
\]
Thus $F_\Delta\ge U^4$ when $\tau\ge256\pi N^4$.
Both size conditions hold with $b=(1024\pi)^{-1/5}$.
Both derivative-size parameters $F_m$ and $F_\Delta$ are less than $\tau$, and the scale parameter $X$
in \Cref{lem:vk-input} is at least $N^{1/3}$: in Type~II,
$U\ge\sqrt N/2\ge N^{1/3}$ for $N\ge64$.
Writing $q=(\log N)^3/H^2$, condition~\eqref{eq:vk-extra-condition}
therefore follows from $27\log32/q<10^{-3}$. When this fails, $q$
is bounded absolutely, so the trivial bound $O(X)$ supplies the same
estimate after increasing its absolute constant. Applying
\eqref{eq:vk-derivative-input} with $\alpha=32$, every inner interval
sum above is
\begin{equation}\label{eq:vk-inner-saving}
 O(XE),\qquad E=\exp(-c_1q),
\end{equation}
for some absolute effective $c_1>0$.

These are the two interval hypotheses of
\Cref{lem:vk-uniform-reduction} with $\chi=1$. That lemma proves
\eqref{eq:vk-prime-prefix}, after decreasing the exponential constant.
\end{proof}

\begin{proposition}[Partial prime sums with a fixed character]
\label{prop:vk-prime-prefix-character}
Fix an integer $D\ge1$. There are effective constants $c_D,C_D>0$
such that, for every Dirichlet character $\chi$ modulo $D$, every
$\tau>1$, $64\le N\le b\tau^{1/5}$ and $N\le x\le2N$,
\begin{equation}\label{eq:vk-prime-prefix-character}
 \left|\sum_{N<p\le x}\chi(p)e^{i\tau w(p)}\right|
 \le C_DN(\log N)^8
 \left\{\exp\left(-c_D\frac{(\log N)^3}{(\log\tau)^2}\right)
              +N^{-1/6}\right\},
\end{equation}
where $b=(1024\pi)^{-1/5}$ and $w(u)$ is any one of
$\log(u/(u-1))$, $\log(1+1/u)$ and $1/u$.
\end{proposition}

\begin{proof}
Write $H=\log\tau$, $Q=(\log N)^3/H^2$ and
$\phi(u)=\tau w(u)/(2\pi)$. For Type~I sums, split into residue
classes on which the character is constant. For Type~II sums, absorb
the characters into the coefficients before Cauchy--Schwarz.

First assume $N\ge N_D:=\max\{64,(2D)^{12}\}$. Put $z=N^{1/3}$.
We verify \Cref{lem:vk-uniform-reduction}\textup{(a)} for
$m\le z^2$ and every subinterval of $(W,2W]$, where
$W=N/m\ge N^{1/3}$.

For each residue $0\le r<D$, set $n=Dv+r$. Since $W\ge2D$, the
permitted $v$ lie between $W/(2D)$ and $2W/D$. Split this interval
at $W/D$; each resulting interval is contained in $[X,2X]$, with
$X=W/(2D)$ or $X=W/D$. The character $\chi(n)=\chi(r)$ is constant.
For $g(v)=\phi(m(Dv+r))$, use
\[
 F_X=\frac{\tau}{2\pi mDX},\qquad
 \frac{\tau}{2\pi N}\le F_X\le\frac{\tau}{\pi N}<\tau.
\]
Here $X\le W$, and $\tau\ge1024\pi N^5$, so $F_X\ge X^4$.
Let $n=Dv+r$. For the first logarithmic phase, differentiation of
\[
 g(v)=\frac{\tau}{2\pi}\int_0^{1/m}\frac{ds}{Dv+r-s}
\]
shows that the ratio of $v^j|g^{(j)}(v)|/j!$ to $F_X$ equals
\[
 \frac{DX}{n}\left(\frac{Dv}{n}\right)^j
       m\int_0^{1/m}(1-s/n)^{-j-1}\,ds.
\]
For the second logarithmic phase replace $1-s/n$ by $1+s/n$;
for the reciprocal phase replace the integral average by $1$.
On the indicated intervals $1/3\le DX/n\le1$,
$1/2\le Dv/n\le1$, and $0\le s/n\le1/2$.
Thus, for every integer $j\ge1$, all three ratios lie between
$2^{-2j-1}/3$ and $2^{j+1}$. These bounds are contained in
$[32^{-j^3},32^{j^3}]$.
Moreover
\[
 \log X\ge\tfrac13\log N-\log(2D)\ge\tfrac14\log N.
\]
Therefore the logarithmic condition in \Cref{lem:vk-input} holds
whenever $64\log32/Q<10^{-3}$. If it fails, $Q$ is bounded
absolutely and the trivial bound $O(X)$ gives the same conclusion
after increasing the constant. The lemma in the other case gives,
for some effective absolute $c_1>0$, the uniform bound
$O(XE)$ with $E=\exp(-c_1Q)$ on every such subinterval.
Summing the two intervals and the $D$ residues gives $O(WE)$.
This proves the required Type~I interval bound.

For Type~II use the scales in
\Cref{lem:vk-uniform-reduction}\textup{(b)} and distinct integers
$V<v_1<v_2\le2V$. Put $\Delta=v_1^{-1}-v_2^{-1}$ and
$F_\Delta=\tau\Delta/(2\pi U)$.
For the first logarithmic phase the difference phase is
$\tau(2\pi)^{-1}\int_{1/v_2}^{1/v_1}(u-s)^{-1}\,ds$;
for the second it is the same integral with $u+s$; for the reciprocal
phase it is $\tau\Delta/(2\pi u)$.
The normalized derivative ratios at scale $U$ lie in
$[2^{-j-2},2^{j+1}]$ for every $j\ge1$, so again $\alpha=32$ works.
Even when $v_2-v_1=1$,
\[
 F_\Delta\ge\frac{\tau}{8\pi UV^2},\qquad
 U^5V^2=U^3(UV)^2\le32N^4.
\]
Consequently $F_\Delta\ge U^4$; also $F_\Delta<\tau$ and
$U\ge\sqrt N/2\ge N^{1/3}$. The derivative lemma and the same
bounded-$Q$ alternative give $O(UE)$ on every off-diagonal interval.
Applying \Cref{lem:vk-uniform-reduction} gives
\eqref{eq:vk-prime-prefix-character} for $N\ge N_D$; the modulus
enters only through the lower threshold $N_D$.

Finally, if $64\le N<N_D$, the assumed upper range implies
$H\ge5\log N-5\log b\ge5\log N$, and hence
$Q\le(\log N_D)/25$. The trivial estimate $2N$ is therefore
absorbed by the asserted right side on increasing $C_D$ effectively.
\end{proof}

\begin{lemma}[Prime mass with fixed periodic weights]
\label{lem:periodic-prime-mass}
For each fixed integer $d\ge1$, there are constants $B_d\in\mathbb R$
and $c_d>0$ such that
\begin{equation}\label{eq:fixed-degree-progression-mertens}
 \sum_{\substack{\ell\le y\\\ell\equiv1\pmod d}}\frac1\ell
 =\frac{\log\log y}{\varphi(d)}+B_d
    +O_d(\sqrt{\log y}\,e^{-c_d\sqrt{\log y}}).
\end{equation}
More generally, fix $D\ge1$ and a function
$a:(\mathbb Z/D\mathbb Z)^\times\to\mathbb R$, extended by zero on
nonunits, and put
$\kappa=\varphi(D)^{-1}\sum_{b\in(\mathbb Z/D\mathbb Z)^\times}a(b)$.
There are constants $B_{D,a}\in\mathbb R$ and $c_D>0$ such that
\begin{equation}\label{eq:periodic-prime-mass}
 \sum_{p\le y}\frac{a(p)}p
 =\kappa\log\log y+B_{D,a}
   +O_{D,a}(\sqrt{\log y}\,e^{-c_D\sqrt{\log y}}).
\end{equation}
In particular, as $N\to\infty$,
\begin{equation}\label{eq:periodic-prime-dyadic-mass}
 \sum_{N<p\le2N}\frac{a(p)}p
 =\frac{\kappa\log2}{\log N}+O_{D,a}((\log N)^{-2}).
\end{equation}
Finitely many omitted primes change only the constant in the cumulative
formula. Effectivity of the constants and thresholds in this lemma is
not asserted.
\end{lemma}

\begin{proof}
For any reduced class $b$ modulo the fixed integer $D$, let
$E_b(u)=\pi(u;D,b)-\operatorname{Li}(u)/\varphi(D)$.
After enlarging a constant on a bounded interval, \Cref{thm:sw} gives
$|E_b(u)|\ll_D u\exp(-c_D\sqrt{\log u})$ for $u\ge2$,
simultaneously in all reduced classes. Partial summation gives
\[
 \sum_{\substack{p\le y\\p\equiv b\pmod D}}\frac1p
 =\frac{\pi(y;D,b)}y+\int_2^y\frac{\pi(u;D,b)}{u^2}\,du.
\]
The main term is
$(\log\log y-\log\log2)/\varphi(D)$, since
$(\operatorname{Li}(u)/u)'=(u\log u)^{-1}
-\operatorname{Li}(u)/u^2$ and $\operatorname{Li}(2)=0$.
The error integral converges absolutely, and its remainder is bounded by
\[
 \frac{|E_b(y)|}y+\int_y^\infty\frac{|E_b(u)|}{u^2}\,du
 \ll_D\sqrt{\log y}\,e^{-c_D\sqrt{\log y}}.
\]
Indeed, the substitution $v=\sqrt{\log u}$ turns the integral into
$2\int_{\sqrt{\log y}}^\infty ve^{-c_Dv}\,dv$.
This proves the formula for each class, including
\eqref{eq:fixed-degree-progression-mertens} by taking $D=d,b=1$.
Multiply the finitely many class formulas by $a(b)$ and add to obtain
\eqref{eq:periodic-prime-mass}. Subtracting it at $2N$ and $N$, and
using
$\log\log(2N)-\log\log N=(\log2)/\log N+O((\log N)^{-2})$,
gives \eqref{eq:periodic-prime-dyadic-mass}. Removing finitely many
primes changes only the limiting constant. 
\end{proof}

For $|\tau|>e$, put
\begin{equation}\label{eq:rho-def}
 \rho(\tau):=\frac{(\log\log|\tau|)^{1/3}}
                   {(\log|\tau|)^{1/3}}.
\end{equation}
For the logarithmic divisor-sum law $\mug$, put
\[
 S_\sigma(\tau):=\sum_{\substack{\ell\ge3\\\ell\textup{ prime}}}
 d_\ell^\sigma(1-\cos(\tau\wsig_\ell)),\qquad
 d_\ell^\sigma:=\frac{2(\ell-2)}{(\ell-1)\ell}.
\]

\begin{theorem}[Uniform Fourier decay at a logarithmic rate]\label{thm:vk-fourier}
\index{Fourier decay!logarithmic laws}
\index{Rajchman measures!logarithmic laws}
For all sufficiently large real $|\tau|$,
\begin{align}
 S(\tau)&\ge\frac23\log\log|\tau|
       -\frac23\log\log\log|\tau|-O(1),\label{eq:vk-dissipation}\\
 |\FT(\tau)|&\ll\rho(\tau).\label{eq:vk-fourier}
\end{align}
The same lower bound holds for $S_\sigma(\tau)$, and all implied constants
and the threshold are effective. In particular, $\muF$ is Rajchman
(cf.\ \Cref{cor:raw-ratio-analytic-images} with $F(x)=-\log x$
for this qualitative assertion).
\end{theorem}

\begin{proof}
By evenness take $\tau>0$ and put $H=\log\tau$. Let $(\kappa_p,\theta_p)$
be $(d_p,\omega_p)$ for $S$ and $(d_p^\sigma,\wsig_p)$ for $S_\sigma$, and use
the first, resp.\ second, phase of \Cref{prop:vk-prime-prefix}. Since
$d_p=2/p-2/[p(p-1)^2]$ and $d_p^\sigma=2/p-2/[p(p-1)]$, one has
$\sum_p|\kappa_p-2/p|<\infty$. Fix $K>0$ with $cK^3>10$, where $c$ is as in
\Cref{prop:vk-prime-prefix}. Let $N_-$ be the next dyadic number above
$\exp(KH^{2/3}(\log H)^{1/3})$, and put $N_+:=b\tau^{1/5}$; this window
is nonempty eventually. On any full or truncated dyadic block
$(N,x]\subset(N_-,N_+]$, Abel summation with the decreasing weight $2/t$
bounds $|\sum_{N<p\le x}(2/p)e^{i\tau\theta_p}|$ by
$(2/N)\sup_{y\in[N,2N]}|\sum_{N<p\le y}e^{i\tau\theta_p}|$.
Summing~\eqref{eq:vk-prime-prefix} over $O(H)$ blocks gives
\[
 \Bigl|\sum_{N_-<p\le N_+}\kappa_pe^{i\tau\theta_p}\Bigr|
 \ll H^9\exp\Bigl(-c\frac{(\log N_-)^3}{H^2}\Bigr)+H^9N_-^{-1/6}+1
 =O(1)
\]
by the choice of $K$, and Mertens' theorem gives
\[
 \sum_{N_-<p\le N_+}\kappa_p
 =2\log\frac{\log N_+}{\log N_-}+O(1)
 =\frac23\log H-\frac23\log\log H-O(1).
\]
Discarding the remaining nonnegative terms of $S$ and $S_\sigma$, as in
Wintner's method \cite[\S~1, (7)--(10)]{Erd39}, proves
\eqref{eq:vk-dissipation} and the same bound for $S_\sigma$.
By~\eqref{eq:forward}, $|\FT(\tau)|\le e^{-S(\tau)/2}$,
proving~\eqref{eq:vk-fourier}.
\end{proof}

The bound \eqref{eq:vk-fourier} holds at every sufficiently large frequency.
\Cref{fig:fourier-questions} sets it beside the statements of
\Cref{ssec:typical-decay} for a uniformly sampled frequency and for selected
resonant frequencies.

\begin{theorem}[Dissipation with fixed periodic weights]
\label{thm:periodic-weight-dissipation}
Fix $D\ge1$ and a nonnegative function
$a:(\mathbb Z/D\mathbb Z)^\times\to[0,\infty)$, extended by zero on
nonunits, whose mean
$\kappa=\varphi(D)^{-1}\sum_{b\in(\mathbb Z/D\mathbb Z)^\times}a(b)$
is positive. Suppose that $\lambda_p\ge0$ for every prime $p$ and
\[
 \lambda_p=\frac{2a(p)}p+r_p,\qquad \sum_p|r_p|<\infty.
\]
For any of the three phases in \Cref{prop:vk-prime-prefix-character},
\begin{equation}\label{eq:periodic-weight-dissipation}
 \sum_p\lambda_p\{1-\cos(tw(p))\}
 \ge\frac{2\kappa}{3}
       \{\log\log|t|-\log\log\log|t|\}-O(1)
 \qquad(|t|\to\infty).
\end{equation}
Consequently any function $\Phi$ satisfying, for some fixed $C_0>0$,
\[
 |\Phi(t)|^2\le C_0
 \exp\left(-\sum_p\lambda_p\{1-\cos(tw(p))\}\right)
\]
satisfies $|\Phi(t)|\ll\rho(t)^\kappa$.
Constants may depend on $D,a,\sum_p|r_p|$ and, in the last conclusion,
on $C_0$. The progression-mass input makes this assertion ineffective
as proved here. Finitely many factors or omitted primes are allowed
through $C_0$ and the summable remainder.
\end{theorem}

\begin{proof}
The sum defining dissipation converges for every $t$: its main terms
are $O_{D,a}(t^2p^{-3})$ for large $p$, and the remainder is bounded
absolutely by $2\sum|r_p|$. Since cosine is even, assume $t>0$ and
put $H=\log t$. Fourier expansion on the finite unit group gives
\[
 a(n)=\sum_{\chi\bmod D}\widehat a(\chi)\chi(n),\qquad
 \widehat a(\chi)=\frac1{\varphi(D)}
       \sum_{b\in(\mathbb Z/D\mathbb Z)^\times}
                    a(b)\overline{\chi(b)}.
\]
This identity also holds at nonunits, by the zero-extension convention.
Consequently \Cref{prop:vk-prime-prefix-character} bounds every
$a(p)$-weighted prime prefix by the right side of
\eqref{eq:vk-prime-prefix-character} times
$\sum_\chi|\widehat a(\chi)|$.

Choose a fixed $K$ such that $c_DK^3>10$, let $N_-$ be the least
dyadic number at least $\exp\{KH^{2/3}(\log H)^{1/3}\}$, and put
$N_+=bt^{1/5}$. These cutoffs form a nonempty window for large $t$.
On each full or terminal partial dyadic block, Abel summation against
$2/u$ bounds the weighted oscillatory sum by $2/N$ times the
supremum of the corresponding weighted prime prefix.
There are $O(H)$ blocks, with $\log N\le H$; thus
\[
 \left|\sum_{N_-<p\le N_+}\frac{2a(p)}p e^{itw(p)}\right|
 \ll_{D,a}H^9\{\exp(-c_D(\log N_-)^3/H^2)+N_-^{-1/6}\}
 =O_{D,a}(1).
\]
Adding the $r_p$ changes this by at most $\sum|r_p|$.
Independently, \Cref{lem:periodic-prime-mass} gives
\[
 \sum_{N_-<p\le N_+}\lambda_p
 =2\kappa\log\frac{\log N_+}{\log N_-}+O(1)
 =\frac{2\kappa}{3}(\log H-\log\log H)+O(1).
\]
Subtract the real part of the oscillatory sum. Since every omitted
term $\lambda_p(1-\cos(tw(p)))$ is nonnegative, this proves
\eqref{eq:periodic-weight-dissipation}. Taking square roots in the
assumed bound for $\Phi$, and using \eqref{eq:rho-def}, proves the
last assertion.
\end{proof}

\begin{proposition}[Upper dissipation and a cosine-preserving Fourier lower bound]
\label{prop:dissipation-upper-mas11}
For all sufficiently large $|\tau|$,
\begin{align}
 S(\tau)&\le\frac{10}{3}\log\log|\tau|
       +\frac23\log\log\log|\tau|+O(1),
       \label{eq:dissipation-upper-mas11}\\
 |\FT(\tau)|&\ge c\,|\cos(\tau\omega_3/2)|
       (\log|\tau|)^{-5/3}(\log\log|\tau|)^{-1/3},
       \label{eq:fourier-cosine-lower-mas11}
\end{align}
where $c>0$ is absolute. Consequently,
\begin{equation}\label{eq:dissipation-limit-ranges-mas11}
 \frac23\le\liminf_{|\tau|\to\infty}
       \frac{S(\tau)}{\log\log|\tau|}\le2
 \le\limsup_{|\tau|\to\infty}
       \frac{S(\tau)}{\log\log|\tau|}\le\frac{10}{3}.
\end{equation}
\end{proposition}

\begin{proof}
By evenness take $\tau>0$, write $H=\log\tau$, and use exactly the
cutoffs $N_-,N_+$ in the proof of \Cref{thm:vk-fourier}. Since
$d_p=2/p+O(p^{-3})$ and $1-\cos u\le2$, Mertens' theorem bounds the
small-prime contribution by
\[
 \sum_{3\le p\le N_-}d_p(1-\cos(\tau\omega_p))
 \le4\log\log N_-+O(1)
 =\frac83\log H+\frac43\log\log H+O(1).
\]
The weighted cancellation already established in that proof gives
\[
 \sum_{N_-<p\le N_+}d_p(1-\cos(\tau\omega_p))
 =\sum_{N_-<p\le N_+}d_p+O(1)
 =\frac23\log H-\frac23\log\log H+O(1).
\]
For $N_+<p\le\tau$, the trivial bound contributes at most
$4\log(\log\tau/\log N_+)+O(1)=O(1)$, since
$N_+=b\tau^{1/5}$. For $p>\tau$, use
$\omega_p\le1/(p-1)$ and $1-\cos u\le u^2/2$ to obtain
\[
 \sum_{p>\tau}d_p(1-\cos(\tau\omega_p))
 \ll\tau^2\sum_{n>\tau}n^{-3}=O(1).
\]
Adding these four estimates proves~\eqref{eq:dissipation-upper-mas11}.
By \Cref{lem:dissipation-comparisons}, and because $S_5\le S$,
\[
 |\FT(\tau)|\ge e^{-C_{\mathrm{sh}}/2}
 |\cos(\tau\omega_3/2)|e^{-S_5(\tau)/2}
 \ge e^{-C_{\mathrm{sh}}/2}|\cos(\tau\omega_3/2)|e^{-S(\tau)/2},
\]
which proves~\eqref{eq:fourier-cosine-lower-mas11}.

The outer inequalities in~\eqref{eq:dissipation-limit-ranges-mas11}
follow from~\eqref{eq:vk-dissipation} and
\eqref{eq:dissipation-upper-mas11}. For the inner inequalities use
$\overline S_T:=T^{-1}\int_0^T S(t)\,dt=2\log\log T+O(1)$ from
\Cref{thm:variance}.
If the upper limit were below $2$, an eventual bound
$S(t)\le(2-\eta)\log\log t$ would contradict this mean.
If the lower limit were above $2$, integration of
$S(t)\ge(2+\eta)\log\log t$ over $[\sqrt T,T]$, where
$\log\log t=\log\log T+O(1)$, would give the opposite contradiction.
\end{proof}

The window in the proof of \Cref{thm:vk-fourier} lies below $|\tau|^{1/4}$
eventually and $\sum_p|d_p-2/p|<\infty$, so the same argument gives
\begin{equation}\label{eq:vk-bbac-mass}
 \sum_{3\le p\le|\tau|^{1/4}}\frac{1-\cos(\tau\omega_p)}p
 \ge\frac13\log\log|\tau|-\frac13\log\log\log|\tau|-O(1).
\end{equation}
Since $\sum_{3\le p\le|\tau|^{1/4}}1/p=\log\log|\tau|+O(1)$, for each
fixed $0<\delta<1/3$ and all sufficiently large $|\tau|$,
$\operatorname{Re}\sum_{3\le p\le|\tau|^{1/4}}e^{i\tau\omega_p}/p
\le(1-\delta)\sum_{3\le p\le|\tau|^{1/4}}1/p$.

\section{Fourier decay, analytic images and normality}
\label{ssec:vk-consequences}

\subsection{Raw-law decay and analytic images}
\label{ssec:raw-decay-analytic-images}

Let $\mathfrak N$ denote the set of reals normal to every integer base $b\ge2$.

Normality concerns the frequencies of digit blocks along one infinite
expansion. The criterion below converts this question into a mean-square
estimate for Weyl sums. On expanding that square, the Fourier transform
is sampled at differences of powers of the base. Their rapid growth
makes even the slow decay obtained here sufficient for almost-sure
normality.

\begin{lemma}[Normality from bounds involving iterated logarithms]
\label{lem:iterlog-normality}
Let $\nu$ be a probability measure on $\R$. Suppose that for every pair of
integers $b\ge2$ and $h\ne0$ there is $q=q(b,h)>1$ such that
\[
 |\widehat\nu(\tau)|\ll_{b,h}
 \frac1{\log\log|\tau|\,(\log\log\log|\tau|)^q}
\]
at $\tau=2\pi h(b^n-b^m)$, uniformly in $1\le m<n$ with $|\tau|$
sufficiently large. Then $\nu(\mathfrak N)=1$. In particular, it suffices that, for all sufficiently large $|\tau|$, $|\widehat\nu(\tau)|\ll(\log\log|\tau|)^{-c}$ for some $c>1$ or $|\widehat\nu(\tau)|\ll\rho(\tau)^\kappa$ for some $\kappa>0$.
\end{lemma}

\begin{proof}
Fix $b,h$ and put $W_N(y):=N^{-1}\sum_{n=1}^Ne^{2\pi ihb^ny}$ and
$t_{m,n}=2\pi h(b^n-b^m)$, so that $|t_{m,n}|\ge\pi b^n$ for $m<n$.
Expanding the square and splitting the sum at $n=\sqrt N$ gives
\[
 \int|W_N|^2\,d\nu
 =\frac1N+\frac2{N^2}\sum_{m<n\le N}\operatorname{Re}\widehat\nu(t_{m,n})
 \ll\frac1{\log N\,(\log\log N)^q}.
\]
Choose $0<\beta<q-1$ and $N_j=\lfloor\exp(j/(\log j)^\beta)\rfloor$ for
large $j$. Then $\int|W_{N_j}|^2\,d\nu\ll1/[j(\log j)^{q-\beta}]$ is
summable, so Tonelli gives $W_{N_j}\to0$ $\nu$-almost everywhere.
Since $N_{j+1}/N_j\to1$ and $|W_N-W_{N_j}|\le2(N-N_j)/N$ for
$N_j\le N<N_{j+1}$, also $W_N\to0$ $\nu$-almost everywhere.
Intersecting over $b,h$ and applying Weyl's criterion and Wall's theorem
\cite[Chapter~1, Theorem~8.1]{KN74} proves the assertion, as in~\cite{DEL63}.
\end{proof}

\begin{lemma}[Moments of the ratio laws]\label{lem:four-ratio-moments}
Let $K_p=J_p$ or $K_p=V_p$ in \eqref{eq:valuation-laws}, and put
\[
 \Phi=\prod_{p:K_p\ge1}(1-1/p),\qquad
 \Sigma=\prod_p(1+p^{-1}+\cdots+p^{-K_p}).
\]
The products converge almost surely, $0<\Phi\le1\le\Sigma<\infty$, and
$\mathbb E(\Phi^r+\Phi^{-r}+\Sigma^r+\Sigma^{-r})<\infty$ for every
$r>0$.
\end{lemma}

\begin{proof}
Put $q_p=\mathbb P(K_p\ge1)$ and
$L=\sum_p\mathbf1_{\{K_p\ge1\}}\omega_p$. In the integer model $q_p=1/p$;
in the shifted-prime model $q_2=1$ and $q_p=1/(p-1)$ for $p\ge3$.
Thus $\mathbb EL<\infty$, since $\omega_p\le2/p$ and $q_p\le2/p$.
Moreover $\Phi=e^{-L}$ and the increasing partial products for $\Sigma$
lie between $1$ and $e^L$. Independence and monotone convergence give
\[
 \mathbb Ee^{rL}=\prod_p\{1+q_p[(1-1/p)^{-r}-1]\}<\infty,
\]
since $0\le(1-1/p)^{-r}-1\le r2^{r+1}/p$.
This proves all four moment assertions.
\end{proof}

\begin{lemma}[Localized analytic prime phases]\label{lem:analytic-prime-phase}
Let $K\subset(0,\infty)$ be a compact interval, and let $F$ be real
analytic on a neighborhood of $K$, with $F'$ nowhere zero on $K$.
There are constants $c,C>0$ such that, for all sufficiently large $t$,
uniformly for $A\in K$, $d\in\{-1,1\}$, $N\le x\le2N$ and
\[
 \exp\{(\log t)^{2/3}(\log\log t)^{1/3}\}\le N\le t^{1/11},
\]
one has
\begin{equation}\label{eq:analytic-prime-prefix}
 \left|\sum_{N<p\le x}e^{itF(A(1+d/p))}\right|
 \le CN(\log N)^8
 \left\{e^{-c(\log N)^3/(\log t)^2}+N^{-1/6}\right\}.
\end{equation}
The constants and threshold may depend on $F,K$.
\end{lemma}

\begin{proof}
Compactness and real analyticity give, uniformly for $A\in K$ and
$d=\pm1$, a convergent expansion for sufficiently large $|z|$,
\[
 \begin{gathered}
 F(A(1+d/z))=\sum_{k\ge0}c_k(A,d)z^{-k},\qquad
 |c_k(A,d)|\le M B^k,\\
 0<c_-\le|c_1(A,d)|=|AF'(A)|\le c_+.
 \end{gathered}
\]
Indeed, finitely many complex disks of analyticity covering $K$, with
smaller concentric disks, give a common Taylor radius and Cauchy bounds.
We verify the interval hypotheses of
\Cref{lem:vk-uniform-reduction}. For a Type~I phase
$g(u)=tF(A(1+d/(mu)))/(2\pi)$, termwise differentiation gives
\[
 \frac{u^j}{j!}g^{(j)}(u)
 =\frac{t(-1)^j}{2\pi}\sum_{k\ge1}
 c_k(A,d)\binom{j+k-1}{j}(mu)^{-k}.
\]
Relative to its $k=1$ term, the sum of the absolute values of the
remaining terms is at most
\begin{equation}\label{eq:analytic-typeI-remainder}
 C_1\sum_{h\ge1}\binom{j+h}{j}r^h
 =C_1\{(1-r)^{-j-1}-1\},\qquad r\le C_2/N,
\end{equation}
since $mu\asymp N$ throughout the inner interval.
For a Type~II difference at $V<v_1<v_2\le2V$, put
$\Delta_k=v_1^{-k}-v_2^{-k}$. Then
\[
 0<\frac{\Delta_k}{\Delta_1}
 =\sum_{a=0}^{k-1}v_1^{-(k-1-a)}v_2^{-a}
 \le kV^{-(k-1)}.
\]
The corresponding relative remainder, on $u\in[U,2U]$, is therefore
at most
\begin{equation}\label{eq:analytic-typeII-remainder}
 C_1\sum_{h\ge1}(h+1)\binom{j+h}{j}r^h
 =C_1\{(1+jr)(1-r)^{-j-2}-1\},\qquad r\le C_2/N,
\end{equation}
because $UV\asymp N$. Both remainders are $O((j+1)/N)$ when
$(j+1)/N$ is sufficiently small, uniformly even for $v_2-v_1=1$.

Take the derivative-size parameters
\[
 F_m=\frac{t|c_1(A,d)|}{2\pi mW},\qquad
 F_\Delta=\frac{t|c_1(A,d)|\Delta_1}{2\pi U}
\]
with the scales $W,U,V$ of \Cref{prop:vk-prime-prefix}.
The size inequalities there are $t|c_1|\ge2\pi N^5$ and
$t|c_1|\ge256\pi N^4$; both hold uniformly in the present range.
Thus each size parameter is at least the fourth power of its inner
scale and is less than $t$ for large $t$. Every derivative order
required in \Cref{lem:vk-input} is $O(\log t/\log N)$, so
\eqref{eq:analytic-typeI-remainder}--\eqref{eq:analytic-typeII-remainder}
are less than $1/2$ uniformly. The normalized leading derivative ratios
are $W/u$ and $U/u$, respectively. Hence the full ratios lie in
$[1/4,3/2]$, and $\alpha=8$ suffices in \Cref{lem:vk-input}.
Its logarithmic condition is treated exactly as in
\Cref{prop:vk-prime-prefix}: the inner scales are at least $N^{1/3}$;
if the condition fails, $(\log N)^3/(\log t)^2$ is bounded and the
trivial estimate suffices. We obtain \eqref{eq:vk-inner-saving} with
$\tau=t$ and constants depending only on $F,K$.
\Cref{lem:vk-uniform-reduction} with $\chi=1$ now gives
\eqref{eq:analytic-prime-prefix}, uniformly in $A,d,x$.
\end{proof}

To obtain a nontrivial saving after logarithmic losses, take a
sufficiently large fixed constant in the lower cutoff. Indeed, with $H=\log t$
and $y=K_0H^{2/3}(\log H)^{1/3}$,
\[
 y^8e^{-cy^3/H^2}
 =K_0^8H^{16/3-cK_0^3}(\log H)^{8/3}.
\]
The conditional averages below cost one further factor $\log N$ from
their normalizing denominator; their proof chooses $K_0$ accordingly.

For later periodic weights we also need character twists of the
localized analytic estimate. The following companion retains uniformity
in the prime-sum endpoint and keeps the modulus fixed.

\begin{lemma}[Character twists of localized analytic prime phases]
\label{lem:analytic-prime-phase-twisted}
Let $K\subset(0,\infty)$ be a compact interval, let $F$ be real analytic
on a neighbourhood of $K$, and suppose that $F'$ is nowhere zero on $K$.
Fix an integer $D\ge1$. There are constants $c,C>0$ and $t_0>e$,
depending only on $F,K,D$, such that for every $t\ge t_0$, every Dirichlet
character $\chi$ modulo $D$, every $A\in K$, every $d\in\{-1,1\}$, and all real $N,x$ with
\[
 \exp\{(\log t)^{2/3}(\log\log t)^{1/3}\}\le N\le t^{1/11},
 \qquad N\le x\le2N,
\]
one has
\begin{equation}\label{eq:analytic-prime-prefix-twisted}
 \left|\sum_{N<p\le x}\chi(p)e^{itF(A(1+d/p))}\right|
 \le CN(\log N)^8
 \left\{\exp\left(-c\frac{(\log N)^3}{(\log t)^2}\right)
                  +N^{-1/6}\right\}.
\end{equation}
For $D=1$ the character means the constant function $1$.
The constants are effective in $D$ and in supplied uniform Taylor
bounds $|c_k(A,d)|\le MB^k$ and $0<c_-\le|c_1(A,d)|\le c_+$ for
$F(A(1+d/z))=\sum_{k\ge0}c_k(A,d)z^{-k}$, valid for $|z|>B$.
\end{lemma}

\begin{proof}
Uniform Taylor estimates follow from compactness and analyticity:
there are $M,B\ge1$ such that, for all $A\in K$ and $d=\pm1$,
\begin{equation}\label{eq:twisted-analytic-taylor}
 f(z):=F(A(1+d/z))=\sum_{k\ge0}c_kz^{-k},\qquad
 |c_k|\le MB^k,\qquad 0<c_-\le|c_1|\le c_+.
\end{equation}
Indeed the Taylor series of $F$ about points of $K$ have a common
positive radius and uniform Cauchy bounds after taking a finite cover
by smaller analytic disks; multiply their $k$th coefficients by
$(dA)^k$. Here $c_1=dAF'(A)$, so its lower bound follows from the
hypothesis. Increase $B$ if necessary. All constants below are uniform
in $A,d,\chi$. Put $H=\log t$, $y=\log N$, $q=y^3/H^2$, and
$L=\max(1,MB/c_-)$.

We first check derivatives including every order needed by
\Cref{lem:vk-input}. For $m\ge1$, put $W=N/m$, and let
$J\subseteq(W,2W]$ be any interval. Write the inner integer variable as
$n=Dv+r$, $0\le r<D$. When $W\ge2D$, all corresponding $v$ lie in
$[W/(2D),2W/D]$, so their interval can be partitioned into at most two
intervals inside $[X,2X]$, with $X=W/(2D)$ or $X=W/D$; assign a shared
endpoint to just one interval. On either interval put
\[
 g_r(v)=\frac{t}{2\pi}f(m(Dv+r)),\qquad
 \mathcal F_{m,X}=\frac{t|c_1|}{2\pi mDX}.
\]
Termwise differentiation of the uniformly convergent series gives, for
every integer $j\ge1$,
\begin{equation}\label{eq:twisted-analytic-affine-derivative}
 \frac{v^j g_r^{(j)}(v)}{j!}
 =\frac{t(-1)^j}{2\pi}
 \left(\frac{Dv}{Dv+r}\right)^j
 \sum_{k\ge1}c_k\binom{j+k-1}{j}[m(Dv+r)]^{-k}.
\end{equation}
Relative to its nonzero $k=1$ term, the absolute remainder is at most
\begin{equation}\label{eq:twisted-analytic-typeI-remainder}
 L\sum_{h\ge1}\binom{j+h}{j}\varrho^h
 =L\{(1-\varrho)^{-j-1}-1\},\qquad
 \varrho=\frac{B}{m(Dv+r)}\le\frac BN.
\end{equation}
If this is at most $1/2$, the ratio of
$v^j|g_r^{(j)}(v)|/j!$ to $\mathcal F_{m,X}$ lies between
\begin{equation}\label{eq:twisted-analytic-affine-profile}
 \frac{1}{6\,2^j}\quad\hbox{and}\quad\frac32.
\end{equation}
In fact $X\le v\le2X$, $r<D$, and $X\ge1$ imply
$DX/(Dv+r)\in[1/3,1]$ and $Dv/(Dv+r)\in[1/2,1]$;
the leading ratio in \eqref{eq:twisted-analytic-affine-derivative} is
the product of the first factor and the $j$th power of the second.
Thus $\alpha=32$ supplies both derivative inequalities of
\Cref{lem:vk-input}, simultaneously for all such orders.

For Type~II, let $v_1,v_2$ be distinct integers with
$V<v_1<v_2\le2V$, and $u\in[U,2U]$. Write
$\Delta_k=v_1^{-k}-v_2^{-k}$. The phase
\[
 g_\Delta(u)=\frac{t}{2\pi}\{f(uv_1)-f(uv_2)\},\qquad
 \mathcal F_\Delta=\frac{t|c_1|\Delta_1}{2\pi U},
\]
has the exact differentiated series
\[
 \frac{u^j g_\Delta^{(j)}(u)}{j!}
 =\frac{t(-1)^j}{2\pi}\sum_{k\ge1}
       c_k\Delta_k\binom{j+k-1}{j}u^{-k}\qquad(j\ge1).
\]
The exact identity
\[
 \frac{\Delta_k}{\Delta_1}
 =\sum_{a=0}^{k-1}v_1^{-(k-1-a)}v_2^{-a}
 \le kV^{-(k-1)}
\]
shows that its relative remainder is at most
\begin{equation}\label{eq:twisted-analytic-typeII-remainder}
 L\sum_{h\ge1}(h+1)\binom{j+h}{j}\varrho^h
 =L\{(1+j\varrho)(1-\varrho)^{-j-2}-1\},\qquad
 \varrho=\frac{B}{uV}\le\frac{4B}{N},
\end{equation}
whenever $UV>N/4$. This bound is independent of $v_2-v_1$, including
the case $v_2-v_1=1$. If it is at most $1/2$, the normalized derivative
ratio lies in $[1/4,3/2]$, because the leading ratio is $U/u$.
Again $\alpha=32$ works.

In Type~I, $m\le N^{2/3}$, so
$W\ge N^{1/3}$ and, once $N\ge(2D)^{12}$,
\[
 X\ge N^{1/3}/(2D)\ge N^{1/4},\quad X\le W,\quad
 \frac{t c_-}{2\pi N}\le\mathcal F_{m,X}
 \le\frac{t c_+}{\pi N}.
\]
In Type~II the block scales will satisfy
\begin{equation}\label{eq:twisted-analytic-bilinear-scales}
 N^{1/3}/2\le V\le U\le2N^{2/3},\qquad N/4<UV\le2N.
\end{equation}
Consequently $U\ge\sqrt N/2\ge N^{1/3}$ for $N\ge64$ and
\[
 \mathcal F_\Delta\ge\frac{t c_-}{8\pi UV^2},\qquad
 U^5V^2=U^3(UV)^2\le32N^4,\qquad
 \mathcal F_\Delta\le\frac{2t c_+}{\pi N}.
\]
Since $t\ge N^{11}$, by enlarging $t_0$ the inequalities
$t c_-\ge2\pi N^5$ and $t c_-\ge256\pi N^4$ hold uniformly.
They give $\mathcal F_{m,X}\ge W^4\ge X^4$ and
$\mathcal F_\Delta\ge U^4$. Both size parameters are less than $t$.
For either inner scale $X_*$, therefore, every required order obeys
\[
 j\le10\left\lceil\frac{\log\mathcal F}{\log X_*}\right\rceil+1
 \le40H/y+11\le50H/y,
\]
where $H/y\ge11$. For $0\le\varrho\le1/2$ and
$(j+2)\varrho\le1/4$, both braces in
\eqref{eq:twisted-analytic-typeI-remainder} and
\eqref{eq:twisted-analytic-typeII-remainder} are at most
$6(j+2)\varrho$: use $-\log(1-\varrho)\le2\varrho$ and
$e^a-1\le2a$ for $0\le a\le1/2$.
Thus both remainders are at most $24LB(j+2)/N$. Since
$y\ge H^{2/3}(\log H)^{1/3}$, their maximum over all required orders
tends effectively to zero. Increasing $t_0$ makes it at most $1/2$.

Because $\log X_*\ge y/4$ and $\log\mathcal F\le H$,
the additional logarithmic condition in \Cref{lem:vk-input} follows
from $64\log32/q<10^{-3}$. When this sufficient condition holds,
that lemma gives, for an absolute effective $c_1'>0$,
$O(X_*\exp(-c_1'q))$ for each unweighted inner interval sum.
When it fails, $q\le64000\log32$; the trivial bound $O(X_*)$
gives the same conclusion after an absolute effective enlargement
of the implied constant. In what follows put
\begin{equation}\label{eq:twisted-analytic-inner-saving}
 E=\exp(-c_1'q).
\end{equation}
Summing the two affine pieces and the $D$ residue classes gives
\begin{equation}\label{eq:twisted-analytic-typeI-inner}
 \left|\sum_{n\in J\cap\mathbb Z}\chi(n)e^{itf(mn)}\right|
 \ll WE\qquad(J\subseteq(W,2W]),
\end{equation}
since $\chi(Dv+r)=\chi(r)$ and the sum of the scales for each residue
is at most $3W/(2D)$. The Type~II difference bound is $O(UE)$ on every
subinterval of $[U,2U]$.

Apply \Cref{lem:vk-uniform-reduction} with
$\phi(u)=tf(u)/(2\pi)$ and the given character.
Equation~\eqref{eq:twisted-analytic-typeI-inner} is its Type~I
hypothesis, and the preceding untwisted difference bound is its
Type~II hypothesis on the scales
\eqref{eq:twisted-analytic-bilinear-scales}. Decreasing $c$ from
$c_1'$ to $c_1'/2$ proves \eqref{eq:analytic-prime-prefix-twisted}.
The constants and thresholds depend effectively on
$D,M,B,c_-,c_+$ and the constants of \Cref{lem:vk-input}.
\end{proof}

\begin{lemma}[Singleton blocks with constant or fixed periodic weights]
\label{lem:periodic-singleton-blocks}
Fix integers $D\ge1$, $r\ge1$ and $p_0\ge2$. For $1\le j\le r$ let
$a_j:(\mathbb Z/D\mathbb Z)^\times\to[0,\infty)$, extended by zero
on nonunits, and suppose that
\[
 \kappa:=\frac1{\varphi(D)}\sum_{b\in(\mathbb Z/D\mathbb Z)^\times}
                  \sum_{j=1}^r a_j(b)>0.
\]
Let independent coordinates $\xi_p$, for primes $p>p_0$, have distinct
states $0,1,\ldots,r$ and possibly additional states, with
\begin{equation}\label{eq:periodic-raw-probabilities}
 \alpha_p:=\mathbb P(\xi_p=0)\ge1-C/p,\qquad
 b_{p,j}:=\mathbb P(\xi_p=j)=a_j(p)/p+O(p^{-2}).
\end{equation}
The error constants are uniform in $p$. For every sufficiently large
dyadic $N$, put $I=(N,2N]$, $s_I=\sum_{p\in I}p^{-1}$, and call $I$
successful when exactly one prime has a state in $\{1,\ldots,r\}$ and
every other prime has state $0$. With $r_{p,j}=b_{p,j}/\alpha_p$ and
$D_I=\sum_{p\in I}\sum_jr_{p,j}$, the following hold.
\begin{enumerate}[label=\textup{(\alph*)}]
\item \emph{Block probabilities.} One has
\begin{align}
 r_{p,j}&=a_j(p)/p+O(p^{-2}),\notag\\
 D_I&=\frac{\kappa\log2}{\log N}+O((\log N)^{-2}),
       \label{eq:periodic-singleton-denominator}\\
 q_I:=\mathbb P(I\text{ successful})
 &=\sum_{p\in I}\frac{\sum_ja_j(p)}p+O(s_I^2+N^{-1}).\notag
\end{align}
\item \emph{Absence of success.} For sufficiently large dyadic $Y<Z$,
the full dyadic blocks covering $(Y,Z]$ satisfy
\begin{equation}\label{eq:periodic-no-singleton}
 \mathbb P(\text{no successful block in }(Y,Z])
 \ll(\log Y/\log Z)^\kappa.
\end{equation}
\item \emph{Conditional selection.} Conditional on success in a block
$I$ and on all coordinates outside it, its unique selected prime $P$
and state $J$ have law
\begin{equation}\label{eq:periodic-singleton-conditioning}
 \mathbb P(P=p,J=j\mid I\text{ successful},\text{outside})
       =r_{p,j}/D_I.
\end{equation}
The same formula holds when $I$ is selected as the first successful
block in increasing order.
\end{enumerate}
The constants may depend on all fixed data. For $D=1$ and constant
weights $a_j=\kappa_j$, they and a sufficient threshold are effective
in $p_0,r,C$, the weights and the probability-error constants.
For general fixed $D$, effectivity is not asserted.
\end{lemma}

\begin{proof}
\textup{(a)} Increase the lower prime cutoff beyond $2C$, so $\alpha_p\ge1/2$.
Then $1\le\alpha_p^{-1}\le1+2C/p$, and boundedness of the finitely
many periodic weights proves the odds expansion.
Write $a=\sum_ja_j$. By \Cref{lem:periodic-prime-mass},
\[
 \sum_{p\in I}\frac{a(p)}p
   =\frac{\kappa\log2}{\log N}+O((\log N)^{-2}),\qquad
 s_I\ll1/\log N.
\]
Since $\sum_{N<p\le2N}p^{-2}\ll N^{-1}$, the first assertion follows.
Independence gives the exact identity
\[
 q_I=\left(\prod_{p\in I}\alpha_p\right)D_I.
\]
The union bound gives $1-Cs_I\le\prod_{p\in I}\alpha_p\le1$,
and $D_I\ll s_I+N^{-1}$, proving the asserted expansion for $q_I$.

\textup{(b)} For $N=2^kY$ the errors sum to
\[
 \sum_{k\ge0}\{(\log(2^kY))^{-2}+(2^kY)^{-1}\}
       \ll1/\log Y.
\]
The cumulative formula of \Cref{lem:periodic-prime-mass} therefore
gives, uniformly in dyadic $Z>Y$,
\[
 \sum_{I\subset(Y,Z]}q_I
       =\kappa\log(\log Z/\log Y)+O(1/\log Y).
\]
Disjoint blocks are independent. Thus $\prod_I(1-q_I)\le
\exp(-\sum_Iq_I)$ proves \eqref{eq:periodic-no-singleton}.

\textup{(c)} The event selecting $p,j$ within $I$ has probability
$b_{p,j}\prod_{q\in I,q\ne p}\alpha_q
=r_{p,j}\prod_{q\in I}\alpha_q$; division by $q_I$ proves
\eqref{eq:periodic-singleton-conditioning}. When $I$ is the first
successful block, the failures in all earlier blocks are measurable
outside $I$. Independence therefore leaves this conditional law
unchanged. The denominators
are positive for all sufficiently large blocks by $\kappa>0$.
For $D=1$, the mass estimates used in \textup{(a)} and \textup{(b)}
are the effective prime number and Mertens estimates. All remaining
bounds are elementary, proving the effective clause.
\end{proof}

\begin{lemma}[Conditional integration over singleton blocks]
\label{lem:raw-singleton-integration}
Assume the hypotheses of \Cref{lem:periodic-singleton-blocks}.
Let $H_0>0$ be independent of the coordinates, and let $R_p$ be a positive
function of $\xi_p$, equal to $1$ in state $0$ and $\vartheta_j(1/p)$ in
state $j$. Suppose that the product converges almost surely to a positive
limit, put $X=H_0\prod_{p>p_0}R_p$, and assume
$\mathbb E(X+X^{-1})<\infty$ and
$2/3\le\vartheta_j(1/p)\le4/3$ for all sufficiently large primes.
For $t>e$, put $H=\log t$. For any fixed $K_0\ge1$, let $N_-$ be
the least dyadic number at least
$\exp\{K_0H^{2/3}(\log H)^{1/3}\}$ and $N_+$ the largest dyadic
number at most $t^{1/24}$.

Suppose that for some $c>0$ and all sufficiently large $t$,
uniformly for every full dyadic block
$I=(N,2N]\subset(N_-,N_+]$ and $t^{1/2}\le\tau\le t^{3/2}$,
\begin{equation}\label{eq:raw-singleton-numerator-input}
 \left|\sum_j\sum_{p\in I}r_{p,j}e^{i\tau\vartheta_j(1/p)}\right|
 \ll(\log N)^8
 \left\{e^{-c(\log N)^3/(\log\tau)^2}+N^{-1/6}\right\}+N^{-1}.
\end{equation}
If $(4c/9)K_0^3\ge10+\kappa$, then
$|\mathbb Ee^{itX}|\ll\rho(t)^\kappa$.
The constants are effective whenever the singleton-block and numerator
bounds, the factor cutoff, and an upper bound for
$\mathbb E(X+X^{-1})$ are effective.
\end{lemma}

\begin{proof}
For large $t$, the window is nonempty and above all fixed cutoffs.
By \eqref{eq:periodic-no-singleton}, the probability of no successful
block is $O((\log N_-/\log N_+)^\kappa)=O(\rho(t)^\kappa)$.
Select the first successful block and reveal $H_0$ and its outside
coordinates. The outside product
$A=H_0\prod_{q>p_0,q\notin I}R_q$ is positive and finite almost surely.
The conditional law in \eqref{eq:periodic-singleton-conditioning}
gives $X=A\vartheta_J(1/P)$.
On $t^{-1/2}\le A\le t^{1/2}$, set $\tau=tA$. Dividing
\eqref{eq:raw-singleton-numerator-input} by the \emph{full-block}
denominator $D_I\asymp1/\log N$ gives
\begin{align*}
 |\mathbb E(e^{itX}\mid I\text{ selected},\text{outside})|
 &\ll(\log N)^9
 \left\{e^{-c(\log N)^3/(\log\tau)^2}+N^{-1/6}\right\}
       +\frac{\log N}{N}\\
 &\ll H^{-1-\kappa}.
\end{align*}
Indeed $\log\tau\le3H/2$, $\log N\le H$ and
$(\log N)^3\ge K_0^3H^2\log H$ bound the exponential term by
$H^{9-(4c/9)K_0^3}\le H^{-1-\kappa}$; the other terms are smaller
than every fixed inverse power of $H$.

On success $2/3\le X/A\le4/3$. The union of selected-block events
with $A\notin[t^{-1/2},t^{1/2}]$ therefore has probability at most
\[
 \mathbb P(X<(4/3)t^{-1/2})+\mathbb P(X>(2/3)t^{1/2})
 \le t^{-1/2}\{(4/3)\mathbb EX^{-1}+(3/2)\mathbb EX\}.
\]
Integrate over the disjoint selected-block events, whose total
probability is at most one, and add the no-success probability.
Since $H^{-1-\kappa}+t^{-1/2}\ll\rho(t)^\kappa$, the result follows.
\end{proof}

\begin{theorem}[Fourier decay of raw products with first-jump states]
\label{thm:raw-product-fourier}
\index{Fourier decay!raw products}
Let $p_0\ge2$ and $m\ge1$ be integers, let $\kappa_1,\ldots,\kappa_m>0$, and
let $\vartheta_1,\ldots,\vartheta_m$ be positive real-analytic functions on a
neighbourhood of $[0,1/p_0]$ with
\[
 \vartheta_j(0)=1,\qquad \vartheta_j'(0)\ne0\qquad(1\le j\le m).
\]
Let $H_0$ be a positive random variable and let $(\xi_p)_{p>p_0}$ be
coordinates indexed by the primes $p>p_0$, each with distinguished states
$0,1,\ldots,m$; additional states are allowed. Suppose that $H_0$ and the
coordinates $\xi_p$, $p>p_0$, are mutually independent and that, for some
constant $C>0$ and every prime $p>p_0$,
\begin{equation}\label{eq:raw-product-probabilities}
 \mathbb P(\xi_p=0)\ge1-\frac Cp,\qquad
 \mathbb P(\xi_p=j)=\frac{\kappa_j}{p}+O(p^{-2})\quad(1\le j\le m).
\end{equation}
For each $p>p_0$ let $R_p$ be a positive function of $\xi_p$ such that
\begin{equation}\label{eq:raw-product-states}
 R_p=1\ \text{in state }0,\qquad
 R_p=\vartheta_j(1/p)\ \text{in state }j\quad(1\le j\le m),
\end{equation}
and suppose that $\prod_{p_0<p\le y}R_p$ converges almost surely to a
positive limit as $y\to\infty$. Put
\[
 X:=H_0\prod_{p>p_0}R_p,\qquad \kappa:=\kappa_1+\cdots+\kappa_m .
\]
Then, for every fixed real $s\ne0$ with $\mathbb E(X^s+X^{-s})<\infty$,
\begin{equation}\label{eq:raw-product-fourier}
 \left|\mathbb E e^{itX^s}\right|\ll\rho(t)^{\kappa}
 \qquad(|t|\to\infty),
\end{equation}
and the law of $X^s$ is Rajchman and gives full mass to $\mathfrak N$.
The implied constant and the threshold may depend on $s$, on the
functions $\vartheta_j$, and on the laws of $H_0$ and of the pairs
$(\xi_p,R_p)$. If $s=1$ and $\vartheta_j(z)=1+d_jz$ with $d_j\in\{-1,1\}$
for every $j$, the implied constant and a sufficient threshold are
effective: they are bounded in terms of $p_0$, $m$,
$\kappa_1,\ldots,\kappa_m$, $C$, the constants implied in
\eqref{eq:raw-product-probabilities}, and an upper bound for
$\mathbb E(X+X^{-1})$.
\end{theorem}

\begin{proof}
Replacing $H_0$, $R_p$, $\vartheta_j$ and $X$ by $H_0^s$, $R_p^s$,
$\vartheta_j^s$ and $X^s$ preserves the hypotheses:
$\vartheta_j^s=\exp(s\log\vartheta_j)$ is positive and real analytic on a
neighbourhood of $[0,1/p_0]$, $\vartheta_j^s(0)=1$,
$(\vartheta_j^s)'(0)=s\vartheta_j'(0)\ne0$, the state probabilities are
unchanged, and $\prod_pR_p^s$ converges almost surely to a positive limit.
We may therefore assume $s=1$, so that $\mathbb E(X+X^{-1})<\infty$.

Apply \Cref{lem:periodic-singleton-blocks} with $D=1$, $r=m$ and
$a_j=\kappa_j$. Write $a_p=\mathbb P(\xi_p=0)$ and
$r_{p,j}=\mathbb P(\xi_p=j)/a_p=\kappa_j/p+O(p^{-2})$.
By conjugation take $t>0$ and put $H=\log t$. For any fixed $K_0\ge1$,
use the window $N_-,N_+$ of \Cref{lem:raw-singleton-integration}.
Its primes eventually exceed $\max(p_0,2C)$ and satisfy
$2/3\le\vartheta_j(1/p)\le4/3$. The singleton lemma gives
\begin{equation}\label{eq:raw-no-singleton}
 \mathbb P(\text{no successful block})
 \ll\Bigl(\frac{\log N_-}{\log N_+}\Bigr)^{\kappa}
 \ll\rho(t)^{\kappa}.
\end{equation}
Select the first successful block $I=(N,2N]$ and reveal $H_0$ and
all outside coordinates. With
$A=H_0\prod_{q>p_0,\,q\notin I}R_q$, its conditional law is
\begin{equation}\label{eq:raw-singleton-conditioning}
 \begin{gathered}
 \mathbb P(P=p,\ \xi_P=j\mid I,\text{outside})=\frac{r_{p,j}}{D_I},
 \qquad D_I=\sum_{p\in I}\sum_jr_{p,j}\asymp\frac1{\log N},\\
 \qquad X=A\,\vartheta_{\xi_P}(1/P).
 \end{gathered}
\end{equation}
The outside product is positive and finite almost surely.

Put $F_j(u):=\vartheta_j(u-1)$ and fix $\eta\in(0,1)$ so small that
every $F_j$ is real analytic on a neighbourhood of
$K:=[1-\eta,1+\eta]$ with $F_j'$ nowhere zero on $K$; this is possible
because $F_j'(1)=\vartheta_j'(0)\ne0$. Let $c>0$ be the least of the
constants $c$ that \Cref{lem:analytic-prime-phase} supplies for these
finitely many pairs $(F_j,K)$, and use the largest of its constants and
thresholds. Choose
\[
 K_0\ge2\qquad\text{with}\qquad \tfrac49\,cK_0^3\ge10+\kappa .
\]
On $t^{-1/2}\le A\le t^{1/2}$ put $\tau=tA$, so that
$t^{1/2}\le\tau\le t^{3/2}$. The phase of $e^{itX}$ is
$\tau\vartheta_j(1/p)=\tau F_j(1+1/p)$. Apply
\Cref{lem:analytic-prime-phase} to $F_j$ and $K$, with the frequency
$\tau$ in place of its $t$, with its parameter $A$ equal to $1$, and with
$d=1$. Its lower cutoff holds because $K_0\ge2$ and
$(3/2)^{2/3}(\log(3H/2)/\log H)^{1/3}\le2$ for $H\ge e$. Its upper cutoff
holds since
\[
 N\le t^{1/24}\le\tau^{1/11}.
\]
The choice $1/24<1/22$ makes this cutoff valid throughout the
effective-frequency range.

Abel summation against $\kappa_j/p$ on each branch, followed by
$\sum_{p\in I}|r_{p,j}-\kappa_j/p|\ll N^{-1}$, gives
\eqref{eq:raw-singleton-numerator-input}. The moment hypothesis and
the chosen $K_0$ therefore verify every assumption of
\Cref{lem:raw-singleton-integration}, proving
\eqref{eq:raw-product-fourier}. Decay gives the Rajchman property,
and \Cref{lem:iterlog-normality} gives absolute normality.

Now let $s=1$ and $\vartheta_j(z)=1+d_jz$ with $d_j\in\{-1,1\}$. Then
$e^{i\tau F_j(1+1/p)}=e^{i\tau}e^{id_j\tau/p}$. The reciprocal phase of
\Cref{prop:vk-prime-prefix} bounds the prefix sums of $e^{i\tau/p}$, and
of their conjugates, for $64\le N\le b\tau^{1/5}$, with absolute effective
constants; this range contains every block, since
$N\le t^{1/24}\le b\tau^{1/5}$ eventually. Use this bound in place of
\Cref{lem:analytic-prime-phase}, with $c$ the constant of
\Cref{prop:vk-prime-prefix} and any $K_0\ge1$ with
$(4c/9)K_0^3\ge10+\kappa$. All other
thresholds and constants above are effective, by the effective prime
number and Mertens estimates. They depend only on $p_0$, $m$,
$\kappa_1,\ldots,\kappa_m$, $C$, the constants implied in
\eqref{eq:raw-product-probabilities}, and an upper bound for
$\mathbb E(X+X^{-1})$.
\end{proof}

\begin{theorem}[Raw products with fixed periodic first-jump weights]
\label{thm:periodic-raw-product-fourier}
Assume the hypotheses of \Cref{lem:periodic-singleton-blocks}.
Let $H_0>0$ be independent of the coordinates, and let $R_p>0$ be a
function of $\xi_p$ with $R_p=1$ in state $0$ and
$R_p=\vartheta_j(1/p)$ in state $j$, where every $\vartheta_j$ is
positive and real analytic on a neighbourhood of $[0,1/p_0]$, with
$\vartheta_j(0)=1$ and $\vartheta_j'(0)\ne0$.
Suppose that the product converges almost surely to a positive limit,
and put $X=H_0\prod_{p>p_0}R_p$.
For every fixed real $s\ne0$ such that
$\mathbb E(X^s+X^{-s})<\infty$,
\begin{equation}\label{eq:periodic-raw-product-fourier}
 \left|\mathbb E e^{itX^s}\right|\ll\rho(t)^\kappa
 \qquad(|t|\to\infty).
\end{equation}
The law of $X^s$ is Rajchman and gives full mass to $\mathfrak N$.
The constant and threshold may depend on the fixed modulus, weights,
jump functions, exponent, coordinate laws and head law. Effectivity is not asserted.
\end{theorem}

\begin{proof}
The power substitution in the proof of
\Cref{thm:raw-product-fourier} preserves the hypotheses and reduces
the assertion to $s=1$ with $\mathbb E(X+X^{-1})<\infty$.
By conjugation take $t>0$ and put $H=\log t$.

For each $j$, put $F_j(u)=\vartheta_j(u-1)$ and choose $\eta>0$
so small that all these functions are analytic near
$K=[1-\eta,1+\eta]\subset(0,\infty)$, with no zero derivative on
$K$. Let $c>0$ be the least constant supplied by
\Cref{lem:analytic-prime-phase-twisted} for the finitely many pairs
$(F_j,K)$ and the fixed modulus $D$; take the largest constants and
thresholds in that lemma. Choose $K_0\ge2$ with
$(4c/9)K_0^3\ge10+\kappa$.
Use the window of \Cref{lem:raw-singleton-integration}; its primes
eventually exceed all finite cutoffs and satisfy
$2/3\le\vartheta_j(1/p)\le4/3$. By
\eqref{eq:periodic-no-singleton},
\begin{equation}\label{eq:periodic-raw-failure}
 \mathbb P(\text{no successful block})
 \ll(\log N_-/\log N_+)^\kappa\ll\rho(t)^\kappa.
\end{equation}
For every $t^{1/2}\le\tau\le t^{3/2}$ the twisted analytic estimate
applies to $\tau F_j(1+1/p)$ with its parameter $A=1$ and $d=1$:
the cutoff checks in the proof of \Cref{thm:raw-product-fourier}
use only $K_0\ge2$, $N\le t^{1/24}$ and this frequency range,
and the twisted estimate has the same cutoffs.

Expand each periodic weight on the finite unit group:
\[
 a_j(n)=\sum_{\chi\bmod D}\widehat a_j(\chi)\chi(n),\qquad
 \widehat a_j(\chi)=\frac1{\varphi(D)}
       \sum_{b\in(\mathbb Z/D\mathbb Z)^\times}
                        a_j(b)\overline{\chi(b)}.
\]
This also holds at nonunits by zero extension. The sum
$\sum_{j,\chi}|\widehat a_j(\chi)|$ is a fixed finite constant.
Applying \eqref{eq:analytic-prime-prefix-twisted} to each character
and then Abel summation against $1/p$, uniformly in the endpoint,
gives the following conditional numerator estimate:
\begin{equation}\label{eq:periodic-singleton-numerator}
 \left|\sum_{j=1}^r\sum_{p\in I}
                 r_{p,j}e^{i\tau F_j(1+1/p)}\right|
 \ll(\log N)^8
       \left\{e^{-c(\log N)^3/(\log\tau)^2}+N^{-1/6}\right\}
       +N^{-1}.
\end{equation}
Indeed Abel summation costs at most $1/N$ times a constant times
the largest prime prefix; replacing $r_{p,j}$ by $a_j(p)/p$ costs
at most $\sum_{j,p\in I}O(p^{-2})\ll N^{-1}$.
This is precisely \eqref{eq:raw-singleton-numerator-input}.
\Cref{lem:raw-singleton-integration}, using the full-block denominator
\eqref{eq:periodic-singleton-denominator} and the assumed moments,
proves \eqref{eq:periodic-raw-product-fourier}.
Decay gives the Rajchman property, and
\Cref{lem:iterlog-normality} gives absolute normality.
\end{proof}

\begin{theorem}[Fourier decay and normality of the ratio laws]
\label{thm:raw-ratio-normality}
\index{Fourier decay!raw ratio laws}
\index{Normal numbers!arithmetic laws}
Let $\mu$ be any of
$\mu_G,\muH,\mu_F^{\mathrm{int}},\mu_\sigma^{\mathrm{int}}$, and let
$X\sim\mu$. Then
\begin{equation}\label{eq:raw-ratio-fourier}
 |\widehat\mu(t)|\ll\rho(t),\qquad
 \mu(\mathfrak N\cap\mathcal L)=1.
\end{equation}
The implied constant and a sufficient threshold for the Fourier bound
are absolute and effective.
\end{theorem}

\begin{proof}
Use the product model of \Cref{cor:joint-phi-sigma}, and write $Z_p$ for
its local factor. Put $d=-1$ in the totient case and $d=1$ in the
divisor-sum case. For $p\ge3$, \eqref{eq:valuation-laws} gives
\[
 a_p:=\mathbb P(Z_p=1)=1-p^{-1}+O(p^{-2}),\qquad
 b_p:=\mathbb P(Z_p=1+d/p)=p^{-1}+O(p^{-2}).
\]
For divisor sums the second event is $K_p=1$; its probability is
$(p-1)/p^2$ in the integer model and $1/p$ in the shifted-prime model.
Put $r_p=b_p/a_p=p^{-1}+O(p^{-2})$.
Take as coordinate $\xi_p$, $p\ge3$, the state of $Z_p$: $\xi_p=0$ on
$\{Z_p=1\}$, $\xi_p=1$ on $\{Z_p=1+d/p\}$ and, for divisor sums,
$\xi_p=K_p$ on $\{K_p\ge2\}$. These additional states have total
probability $O(p^{-2})$. Thus $R_p=Z_p$ is a positive function of
$\xi_p$, $\mathbb P(\xi_p=0)=a_p\ge1-2/p$ and
$\mathbb P(\xi_p=1)=b_p$, so $r_{p,1}=r_p$.
The coordinates $\xi_p$, $p\ge3$, and the head $H_0=Z_2$ are mutually
independent. Apply \Cref{thm:raw-product-fourier} with $p_0=2$, $m=1$,
$\kappa_1=1$, $\vartheta_1(z)=1+dz$, $C=2$ and $s=1$.
The jump function is positive and real analytic near $[0,1/2]$,
with $\vartheta_1(0)=1$ and $\vartheta_1'(0)=d\ne0$;
\Cref{lem:four-ratio-moments} supplies positive almost-sure convergence
and the required moments. This proves the Fourier estimate in
\eqref{eq:raw-ratio-fourier}.
The effective special case of \Cref{thm:raw-product-fourier} applies.
The constants implied in \eqref{eq:raw-product-probabilities} are
absolute, and the moment bound is explicit: the proof of
\Cref{lem:four-ratio-moments} gives
$\mathbb Ee^L\le\exp(4\sum_{n\ge2}n^{-2})<e^4$, so
$\mathbb E(X+X^{-1})\le1+e^4$ for all four laws. Hence the implied
constant and a sufficient threshold for the Fourier bound are absolute
and effective.
\Cref{lem:iterlog-normality,thm:liouville-carriers} give the remaining
assertion in \eqref{eq:raw-ratio-fourier}.
\end{proof}

\begin{proposition}[Collisions and Fourier means of the ratio laws]
\label{prop:raw-ratio-fourier-means}
Let $\mu$ be any of the four laws in \Cref{thm:raw-ratio-normality},
and let $X\sim\mu$.
If $X'$ is an independent copy of $X$, then
\begin{equation}\label{eq:raw-collision}
 \mathbb P(|X-X'|\le h)\asymp\frac1{\log^2(1/h)}
 \qquad(h\downarrow0),
\end{equation}
and
\begin{equation}\label{eq:raw-means}
 \frac1T\int_0^T|\widehat\mu(t)|^2\,dt\asymp\frac1{\log^2T},
 \qquad
 \frac1T\int_0^T|\widehat\mu(t)|\,dt\asymp\frac1{\log T}
 \qquad(T\to\infty).
\end{equation}
In particular, the root mean square has order $1/\log T$, and
\begin{equation}\label{eq:raw-large-values}
 \frac1T\left|\left\{t\in[0,T]:
       |\widehat\mu(t)|>\frac K{\log T}\right\}\right|
 \ll K^{-2}\qquad(T\ge3,\ K\ge1).
\end{equation}
\end{proposition}

\begin{proof}
Put $Y=\log X$ and
$\nu=\operatorname{Law}(Y)$. First,
\begin{equation}\label{eq:log-second-mean}
 \frac1T\int_0^T|\widehat\nu(t)|^2\,dt\ll(\log T)^{-2}
 \qquad(T\ge3).
\end{equation}
For each of the four laws this is \Cref{thm:valuation-fourier-moments}\textup{(a)}
with $r=2$, the first-jump function $v_\varphi$ for the two totient laws and $v_\sigma$
for the two divisor-sum laws, the finite prime set $\{2\}$, $C_2=2$ and the prime-$2$
coordinate as head, together
with evenness; reflection of the logarithm does not affect the modulus.

\AddToHookNext{build/column/before}{\raggedbottom}
Write $k(v)=(\sin(v/2)/(v/2))^2$, with $k(0)=1$. For any real
random variable $D$ and $u>0$, Fubini's theorem gives
\begin{equation}\label{eq:triangular-kernel}
 \mathbb E k(D/u)
 =u\int_{-1/u}^{1/u}(1-u|t|)\mathbb E e^{itD}\,dt.
\end{equation}
Since $k$ is nonnegative and bounded below on $[-1,1]$, apply this
identity to $D=Y-Y'$, where $Y'$ is an independent copy.
Equation~\eqref{eq:log-second-mean} yields
$\mathbb P(|Y-Y'|\le u)\ll\log^{-2}(1/u)$.
For totients, on $X,X'\ge\sqrt h$, the inequality
$|X-X'|\le h$ implies $|Y-Y'|\le\sqrt h$. The omitted probability is
at most $2h\mathbb EX^{-2}=O(h)$ by \Cref{lem:four-ratio-moments}.
For divisor sums, $X,X'\ge1$ and
$|Y-Y'|\le|X-X'|$. This proves the upper bound in
\eqref{eq:raw-collision}.

For the lower bounds use the anchors
\[
 a=\begin{cases}
 1,&\mu=\mu_F^{\mathrm{int}}\text{ or }\mu_\sigma^{\mathrm{int}},\\
 1/2,&\mu=\mu_G,\\
 3/2,&\mu=\muH.
 \end{cases}
\]
For $y=1/h$, set all integer-model coordinates at $p\le y$ to zero;
in the shifted-prime models set those at $3\le p\le y$ to zero and,
for divisor sums, also set $V_2=1$. The totient factor at $2$ is
already $1/2$. These events have probability $\asymp1/\log y$ by
\eqref{eq:valuation-laws} and Mertens' theorem. Their raw head equals
$a$. Independently, the absolute logarithmic tail $U_y$ has mean
\[
 \mathbb EU_y\ll\sum_{p>y}p^{-2}\ll\frac1{y\log y}.
\]
Markov's inequality gives $U_y\le h/(2a)$ with probability
$1-O(1/\log y)$. The inequalities $1-e^{-u}\le u$ and
$e^u-1\le2u$ for small $u\ge0$ therefore put $X$ in the one-sided
interval of length $h$ adjacent to $a$, below $a$ for totients and
above $a$ for divisor sums. In particular,
\begin{equation}\label{eq:endpoint-mass}
 \mathbb P(|X-a|\le h)\gg\frac1{\log(1/h)},
\end{equation}
with this mass in a single interval of length $h$. Two independent
copies in that interval prove the lower collision bound.

Set $D=X-X'$ in \eqref{eq:triangular-kernel} with $u=1/T$.
The collision lower bound proves the lower second-mean estimate.
Conversely, Gaussian Fourier inversion and positivity give
\begin{align*}
 \frac1T\int_{-T}^T|\widehat\mu(t)|^2\,dt
 &\ll\frac1T\int_{\mathbb R}e^{-t^2/(2T^2)}
                   |\widehat\mu(t)|^2\,dt\\
 &\ll\mathbb E e^{-T^2(X-X')^2/2}\\
 &\le\mathbb P(|X-X'|\le T^{-1/2})+e^{-T/2}.
\end{align*}
The collision upper bound proves the other second-mean inequality,
and Cauchy--Schwarz gives the first-absolute-mean upper bound.
For its lower bound, apply \eqref{eq:triangular-kernel} to
$D=X-a$, again with $u=1/T$. By
\eqref{eq:endpoint-mass}, its nonnegative left side is
$\gg1/\log T$, whereas its absolute value is at most
$2T^{-1}\int_0^T|\widehat\mu(t)|\,dt$.
This proves \eqref{eq:raw-means}; Markov's inequality with the
second mean gives \eqref{eq:raw-large-values}.
\end{proof}

\begin{remark}
The collision and second-mean order bounds are sharpened, with their
exact common constants, in \Cref{thm:exact-collision-mas04}.
\end{remark}

\KeepHeadingSpace{5\baselineskip}
\begin{corollary}[Fourier means of the ratio laws]
\label{cor:raw-fourier-means}
Let $\mu$ be any of the four laws in \Cref{thm:raw-ratio-normality}.
For every fixed $0<r\le2$,
\begin{equation}\label{eq:raw-fractional-means}
 \frac1T\int_0^T|\widehat\mu(t)|^r\,dt
 \asymp_r(\log T)^{-r}\qquad(T\to\infty).
\end{equation}
The lower bound holds for every fixed $r>0$.
There are absolute constants $0<c<C$ and $\delta>0$ such that,
for all sufficiently large $T$,
\begin{equation}\label{eq:raw-amplitude-band}
 \frac1T\left|\left\{t\in[0,T]:
 \frac c{\log T}\le|\widehat\mu(t)|\le\frac C{\log T}
 \right\}\right|\ge\delta.
\end{equation}
\end{corollary}

\begin{proof}
Write $m_r=T^{-1}\int_0^T|\widehat\mu(t)|^r\,dt$.
For $0<r<1$, H\"older's inequality gives
$m_r\ge m_1^{2-r}m_2^{-(1-r)}$; for $r\ge1$, Jensen's inequality
gives $m_r\ge m_1^r$. For $0<r\le2$, Jensen also gives
$m_r\le m_2^{r/2}$. Equation~\eqref{eq:raw-means} proves the moment bounds.
Let $U_T$ be uniform on $[0,T]$ and put
$V_T=(\log T)|\widehat\mu(U_T)|$. The same equation gives
$\mathbb EV_T\ge c_0>0$ and $\mathbb EV_T^2\le C_0$ for large $T$,
with absolute constants. Paley--Zygmund yields
$\mathbb P(V_T\ge c_0/2)\ge c_0^2/(4C_0)$, whereas
$\mathbb P(V_T>C)\le C_0/C^2$. Taking $C$ sufficiently large proves
\eqref{eq:raw-amplitude-band}.
\end{proof}

\begin{corollary}[Analytic images of the ratio laws]
\label{cor:raw-ratio-analytic-images}
\index{Rajchman measures!analytic images}
Let $\mu$ be any of the four laws in \Cref{thm:raw-ratio-normality},
and let $X\sim\mu$.
Let $F$ be a fixed real-analytic function on an open set
$U\subset(0,\infty)$ with $\mu(U)=1$, nonconstant on every connected
component. Then $\operatorname{Law}(F(X))$ is Rajchman and
$\mathbb P(F(X)\in\mathfrak N)=1$. More precisely, for every compact
interval $J\subset U$ contained in an interval on which $F'$ is
nowhere zero,
\begin{equation}\label{eq:analytic-image-local-fourier}
 \left|\mathbb E\bigl[\mathbf1_{\{X\in J\}}e^{itF(X)}\bigr]\right|
 \ll_{F,J}\rho(t).
\end{equation}
\end{corollary}

\begin{proof}
Use the local factors and coordinates constructed in the proof of
\Cref{thm:raw-ratio-normality}: $d=-1$ for totients, $d=1$ for
divisor sums, and $r_p=b_p/a_p=p^{-1}+O(p^{-2})$.
For \eqref{eq:analytic-image-local-fourier}, choose a compact interval
$K$ whose interior contains $J$, contained in a noncritical analytic
interval for $F$. By conjugation take $t>0$ and put $H=\log t$.
Use the successful blocks of the proof of \Cref{thm:raw-product-fourier}
for the present coordinates: $m=1$, $\vartheta_1(z)=1+dz$,
$r_{p,1}=r_p$ and $D_I=\sum_{p\in I}r_p$.
The estimates \eqref{eq:raw-no-singleton} and
\eqref{eq:raw-singleton-conditioning} hold for every fixed $K_0\ge1$;
choose $K_0$ in terms of $F,K$ so that $cK_0^3\ge11$, with $c$ the
constant of \Cref{lem:analytic-prime-phase} for $F,K$.
On success $X=A(1+d/P)$ by \eqref{eq:raw-singleton-conditioning}.
If $A(1+d/p)\in J$ for some selected prime, then $A\in K$ for large
$t$. For each such $A$, the restriction $A(1+d/p)\in J$ selects an
interval of primes. Its conditional contribution is exactly
\[
 \frac1{D_I}\sum_{\substack{p\in I\\A(1+d/p)\in J}}
 r_pe^{itF(A(1+d/p))}.
\]
Differences of the prefix estimates \eqref{eq:analytic-prime-prefix}
control every such subinterval. Abel summation as in that proof bounds this
expression by $O_{F,J}(H^{-2})$; its denominator remains the full-block
$D_I$ of \eqref{eq:raw-singleton-conditioning}.
Integrating over selected blocks and adding
\eqref{eq:raw-no-singleton} proves
\eqref{eq:analytic-image-local-fourier}.

The law $\mu$ is atomless by \Cref{cor:rational-statistics}\textup{(b)}.
The critical set of $F$ is discrete in each of the countably many
components of $U$, hence is $\mu$-null.
Up to countably many endpoints, partition its complement into
countably many compact intervals $J_n$ with disjoint interiors,
each contained in a noncritical analytic interval.
For every $n$ of positive mass, the probability measure
$\mu(J_n)^{-1}F_*(\mu|_{J_n})$ satisfies
\Cref{lem:iterlog-normality} by
\eqref{eq:analytic-image-local-fourier} and is Rajchman.
Their countable probability mixture therefore gives full mass to
$\mathfrak N$. Finite partial mixtures converge to it in total
variation, which also proves the Rajchman assertion.
\end{proof}

\begin{corollary}[Fixed powers and progressions of the raw ratio laws]
\label{thm:raw-power-progressions-mas16}
Let $X$ have any of the four laws in \Cref{thm:raw-ratio-normality}.
For every fixed real $s\ne0$,
\begin{equation}\label{eq:raw-power-fourier-mas16}
 \left|\mathbb E e^{itX^s}\right|
 \ll_s\left(\frac{\log\log|t|}{\log|t|}\right)^{1/3}
 \qquad(|t|\to\infty).
\end{equation}
The same estimate holds after deleting a fixed finite set of prime
factors and multiplying the remaining product by an independent
positive random variable $H_0$ satisfying
$\mathbb E(H_0^s+H_0^{-s})<\infty$; the constants and threshold may
then depend on the deleted set and the head law.

In particular, for fixed $Q\ge1$ and $\gcd(b,Q)=1$, the estimate
holds for both limiting ratios over primes $p\equiv b\pmod Q$
in \Cref{thm:ap-law,cor:ap-sigma-law}, with constants and thresholds
depending on $Q,s$. Each power law just described is Rajchman and
gives full mass to $\mathfrak N$. The case $s=-1$ includes the
reciprocal ratios.
\end{corollary}

\begin{proof}
We prove the assertion with a finite deleted set and independent head;
the unmodified laws are included by taking the set empty and $H_0=1$.
By \Cref{lem:four-ratio-moments}, the remaining prime product has all
positive and negative moments. Independence and the assumption on
$H_0$ imply
\[
 \mathbb E(X^s+X^{-s})<\infty.
\]
Write $Z_p$ for the undeleted local factors and take $d=-1$ for
totients and $d=1$ for divisor sums. Outside the finite head,
\[
 \begin{aligned}
 a_p&=\mathbb P(Z_p=1)=1-p^{-1}+O(p^{-2}),\\
 b_p&=\mathbb P(Z_p=1+d/p)=p^{-1}+O(p^{-2}),
 \end{aligned}
\]
and $a_p\ge1-2/p$. Use the coordinates $\xi_p$ from the proof of
\Cref{thm:raw-ratio-normality}: states $0$ and $1$ correspond to
$Z_p=1$ and $Z_p=1+d/p$;
for divisor sums, valuations $K_p\ge2$ are additional states.
Choose an integer $p_0\ge2$ larger than every deleted prime and take
the product of $H_0$ and the undeleted factors at $p\le p_0$ as the
head. This head is positive and independent of the remaining
coordinates. Finite deletion preserves the positive almost-sure
convergence in \Cref{lem:four-ratio-moments}.
Thus \Cref{thm:raw-product-fourier} applies with $m=1$,
$\vartheta_1(z)=1+dz$, $\kappa_1=1$, $C=2$ and the present exponent
$s$. This proves
\eqref{eq:raw-power-fourier-mas16}.

For a fixed reduced progression, the conditional valuation models
in \Cref{thm:ap-law,cor:ap-sigma-law} leave all prime coordinates
outside $2Q$ independent with their original laws. Their factors
at primes dividing $2Q$ form an independent positive head bounded
above and away from zero: each totient factor is either $1$ or
$1-1/p$, and each divisor-sum factor lies in $[1,p/(p-1)]$.
Both required head moments are therefore finite, uniformly over
the finitely many reduced classes for fixed $Q$. This proves the
progression assertion, including the random valuation head for
divisor sums. Decay gives the Rajchman property, and
\Cref{lem:iterlog-normality} gives absolute normality.
\end{proof}

The exponent in \eqref{eq:raw-product-fourier} is the sum of the
coefficients of the distinguished first-jump states. Additional states
of probability $O(1/p)$ are permitted but do not contribute to this sum.

The proof of \Cref{thm:raw-product-fourier} uses $X^s=A^sR_P^s$ on a successful block to absorb
the outside product into the frequency. It gives a global rate for
monomials in independent variables (\Cref{cor:raw-monomial-fourier}),
but not for every analytic image. For other analytic images of the
four ratio laws, \Cref{cor:raw-ratio-analytic-images} gives a local
estimate and qualitative conclusions.

\begin{corollary}[Monomials in independent ratio variables]
\label{cor:raw-monomial-fourier}
Let $X_1,\ldots,X_k$ be independent, each with one of the four laws in
\Cref{thm:raw-ratio-normality}, let
$\boldsymbol\alpha\in\R^k\setminus\{0\}$, and put
$k'=\#\{j:\alpha_j\ne0\}$. Then
\begin{equation}\label{eq:raw-monomial-fourier}
 \left|\mathbb E\exp\Bigl(it\prod_{j=1}^kX_j^{\alpha_j}\Bigr)\right|
 \ll_{\boldsymbol\alpha}\rho(t)^{k'}\qquad(|t|\to\infty).
\end{equation}
The law of this monomial is Rajchman and gives full mass to $\mathfrak N$.
\end{corollary}

\begin{proof}
Discard the variables with $\alpha_j=0$ and renumber the others, so that
$\alpha_j\ne0$ for $1\le j\le k'$. Realize these variables by mutually
independent copies of the product model of \Cref{cor:joint-phi-sigma},
with local factors $Z_{p,j}$ and signs $d_j\in\{-1,1\}$ as in the proof
of \Cref{thm:raw-ratio-normality}. For $p\ge3$,
\[
 a_{p,j}:=\mathbb P(Z_{p,j}=1)\ge1-2/p,\qquad
 b_{p,j}:=\mathbb P(Z_{p,j}=1+d_j/p)=p^{-1}+O(p^{-2}).
\]
Put $H_0=\prod_{j\le k'}Z_{2,j}^{\alpha_j}$ and
$R_p=\prod_{j\le k'}Z_{p,j}^{\alpha_j}$ for $p\ge3$. Define $\xi_p=0$
when every local factor is $1$, and $\xi_p=j$ when
$Z_{p,j}=1+d_j/p$ and $Z_{p,i}=1$ for all $i\ne j$. In every other case
record the full local vector as an additional, separately tagged state.
Thus $R_p$ is a function of $\xi_p$, and $H_0$ and the $\xi_p$ are
mutually independent. In state $0$ the factor is $1$; in state $j$ it is
$\vartheta_j(1/p)$, where
\[
 \vartheta_j(z)=(1+d_jz)^{\alpha_j},\qquad
 \vartheta_j(0)=1,\qquad \vartheta_j'(0)=d_j\alpha_j\ne0.
\]
These functions are positive and real analytic on a neighbourhood of
$[0,1/2]$. Independence gives
\[
 \mathbb P(\xi_p=0)=\prod_i a_{p,i}\ge1-2k'/p,\qquad
 \mathbb P(\xi_p=j)=b_{p,j}\prod_{i\ne j}a_{p,i}
                    =p^{-1}+O_k(p^{-2}).
\]
Equal functions $\vartheta_i=\vartheta_j$ remain distinct states,
distinguished by their coordinate.

By \Cref{lem:four-ratio-moments}, each product converges almost surely
to a positive limit. Hence $H_0\prod_{p\ge3}R_p$ converges almost surely
to $\prod_{j\le k'}X_j^{\alpha_j}>0$, and independence gives
\[
 \mathbb E\prod_{j\le k'}X_j^{\pm\alpha_j}
 =\prod_{j\le k'}\mathbb E X_j^{\pm\alpha_j}<\infty.
\]
Apply \Cref{thm:raw-product-fourier} with $s=1$, $p_0=2$, $m=k'$ and
$\kappa_j=1$ for every distinguished active state.
\end{proof}

The Rajchman property is not invariant under analytic diffeomorphisms:
Kaufman's nonlinear-image result, recalled in \cite[\S~1.1]{BB25},
gives Fourier decay for the image of the middle-thirds Cantor law under
$x\mapsto x+x^2$, whereas the inverse map restores the non-Rajchman law.

\begin{corollary}[Analytic images of independent ratio variables]
\label{cor:multivariable-analytic-images}
\index{Rajchman measures!analytic images}
\index{Normal numbers!analytic images}
Let $X_1,\ldots,X_d$ be independent, each with one of the four laws
in \Cref{thm:raw-ratio-normality}. Let $P,Q$ be real analytic on
$(0,\infty)^d$, with $Q$ not identically zero and $F=P/Q$
nonconstant. Then $F(\boldsymbol X)$ is defined almost surely. Its law
$\lambda$ is Rajchman, gives full mass to $\mathfrak N$, and is
singular continuous, with Hausdorff and Fourier dimensions zero and
\begin{equation}\label{eq:multivariable-image-dimensions}
 \underline d_\lambda=0,\qquad \overline d_\lambda=1
 \quad(\lambda\text{-a.e.}),\qquad
 P_{\mathrm{pos}}(\lambda)=P_{\mathrm{full}}(\lambda)=1.
\end{equation}
In particular, these assertions hold for every nonconstant
real-analytic $F:(0,\infty)^d\to\R$.
\end{corollary}

\begin{proof}
A product of atomless laws on $(0,\infty)$ charges no zero set of a
nonzero real-analytic function there. Induct on the number of variables,
using a nonzero Taylor coefficient in the last variable to exclude
identically zero slices, and then the discreteness of one-variable zeros.
This proves that $Q(\boldsymbol X)\ne0$ almost surely and, by applying
the same assertion to $P-tQ$ for each fixed $t$, that $\lambda$ is
atomless.

If every $Q\partial_jP-P\partial_jQ$ vanished identically, then
$\nabla(P/Q)=0$ on the open set $\{Q\ne0\}$, so $P/Q$ would equal
a constant $c$ on a ball in that set. Then $P-cQ$ would vanish on
that ball, hence on the connected domain $(0,\infty)^d$ by the
identity theorem for real-analytic functions, contradicting the
nonconstancy of $F$. After relabeling, we may therefore assume that
$Q\partial_dP-P\partial_dQ$ is not identically zero.
The same slice argument shows that, for almost every
$(X_1,\ldots,X_{d-1})$, both this function and $Q$ have nonzero
one-variable specializations. The specialized quotient is therefore
analytic and nonconstant on every component off its discrete set of
poles. Independence and \Cref{cor:raw-ratio-analytic-images} give normality
conditionally; Fubini gives $\lambda(\mathfrak N)=1$. Dominated
convergence of the conditional Fourier transforms proves that $\lambda$
is Rajchman.

The joint Hausdorff-zero carrier in
\Cref{cor:rational-statistics}\textup{(a)} survives the locally
Lipschitz map $F$ on $\{Q\ne0\}$. Thus $\lambda$ has Hausdorff
dimension zero, is singular, and has Fourier dimension zero by
\Cref{lem:fourier-Hausdorff}. The packing and local-dimension
argument of \Cref{cor:rational-statistics}\textup{(b)} applies to the
nonconstant slices established above: their poles and critical points
are discrete, so their complement has a countable cover by compact
bi-Lipschitz intervals. This proves
\eqref{eq:multivariable-image-dimensions}.
\end{proof}

\begin{corollary}[Joint normality with nonconstant integer combinations]
\label{cor:joint-normality-scope-mas17}
Let $\boldsymbol X$ be the independent vector of
\Cref{cor:multivariable-analytic-images}, and let $F_1,\ldots,F_k$
be quotients of real-analytic functions on $(0,\infty)^d$ whose
denominators are not identically zero. Suppose that
$\sum_jm_jF_j$ is nonconstant for every
$\boldsymbol m\in\Z^k\setminus\{0\}$. Then almost surely, for every
integer $b\ge2$,
\[
 \bigl(b^nF_1(\boldsymbol X),\ldots,b^nF_k(\boldsymbol X)\bigr)_{n\ge1}
 \quad\text{is uniformly distributed modulo one in }[0,1)^k.
\]
\end{corollary}

\begin{proof}
Every integer combination is an analytic quotient with a denominator
not identically zero. Its poles are null by
\Cref{cor:multivariable-analytic-images}, and its value is absolutely
normal almost surely by that corollary. Intersect these probability-one
events over the countably many nonzero integer vectors and bases.
The scalar Weyl sums for each such vector tend to zero; the
multidimensional Weyl criterion proves the assertion.
\end{proof}

The hypothesis matters: $(X,2X)$ has the constant integer combination
$2X-2X=0$, and its base-$b$ orbit lies in the proper subtorus
$2u-v=0\pmod1$. \Cref{cor:multivariable-analytic-images} gives Fourier decay
along each fixed real direction whose combination is nonconstant.
It does not give decay uniformly over directions that vary with the
frequency; the vector conclusion in \Cref{cor:rajchman-diff} instead
uses the explicit uniform product bound \eqref{eq:tuple-fourier-bound}.

\subsection{Normal fields and linear images}
\label{ssec:normal-fields-linear-images}

The next field statements distinguish a fixed real-coefficient rational
function from simultaneous rational operations over $\Q$: only the
latter form a countable family. The generator models are independent
samples in \Cref{cor:normal-liouville-field} and the affine forms in
\Cref{cor:fixed-shift-normal-field}. The second route uses
\Cref{lem:rational-orbit-domination} and
\Cref{cor:affine-independent-reference}.
\Cref{fig:reference-models} sets this route by domination beside the direct estimate
\eqref{eq:tuple-fourier-bound}, from which \Cref{cor:rajchman-diff} derives its rates
for the affine tuples.

\begin{corollary}[Normal fields of ratios and logarithms]\label{cor:normal-liouville-field}
\index{Liouville numbers!normal fields}
\index{Normal numbers!logarithmic fields}
Let $X_1,X_2,\ldots$ be independent, each with one of the four laws in
\Cref{thm:raw-ratio-normality}. For each fixed nonconstant
$R\in\R(T_1,U_1,\ldots,T_d,U_d)$, the variable
$R(X_1,\log X_1,\ldots,X_d,\log X_d)$ is defined almost surely and
has a Rajchman law giving full mass to $\mathfrak N$.
Almost surely, $X_1,\log X_1,X_2,\log X_2,\ldots$ are algebraically
independent over $\Q$, and the fields
\[
 \begin{gathered}
 K=\Q(X_1,X_2,\ldots),\qquad
 K_{\log}=\Q(\log X_1,\log X_2,\ldots),\\
 K_{\mathrm{mix}}=\Q(X_1,\log X_1,X_2,\log X_2,\ldots)
 \end{gathered}
\]
have infinite transcendence degree and satisfy
\[
 K\setminus\Q\subset\mathfrak N\cap\mathcal L,
 \qquad K_{\log}\setminus\Q\subset
 K_{\mathrm{mix}}\setminus\Q\subset\mathfrak N.
\]
In particular, $\Q(X_1)\setminus\Q\subset\mathfrak N\cap\mathcal L$
almost surely.
\end{corollary}

\begin{proof}
For a nonzero real polynomial $P(T,U)$, the function $P(x,\log x)$
is not identically zero on $(0,\infty)$. Indeed, substitution $x=e^t$
gives $\sum_j e^{jt}P_j(t)$, whose largest nonzero exponential term
dominates as $t\to\infty$. Iteration proves the same assertion for
$P(x_1,\log x_1,\ldots,x_d,\log x_d)$.
The analytic zero-set argument in the proof of
\Cref{cor:multivariable-analytic-images} therefore shows that every
nonzero polynomial in the paired generators vanishes with probability zero.

Write $R=P/Q$ and put
$p(\boldsymbol x)=P(x_1,\log x_1,\ldots,x_d,\log x_d)$, with $q$
defined similarly. The preceding argument makes $q$ nonzero almost
surely and shows that $p/q$ is nonconstant.
\Cref{cor:multivariable-analytic-images} gives normality and the
Rajchman assertion.

Intersect the probability-one events for the countably many nonzero
rational polynomials and nonconstant rational functions over $\Q$ in
finitely many paired generators. This gives algebraic independence and
the normality inclusion for $K_{\mathrm{mix}}$, hence for $K_{\log}$.
The Liouville inclusion for $K$ is
\Cref{cor:rational-statistics}\textup{(c)}.
\end{proof}

\begin{lemma}[Rational-orbit domination]\label{lem:rational-orbit-domination}
Let $\mu$ be a product of $d$ laws from \Cref{thm:raw-ratio-normality},
let $\Gamma\subset\Q_{>0}^d$ be countable, and put
$T_{\boldsymbol r}(\boldsymbol x)=(r_1x_1,\ldots,r_dx_d)$.
If a probability measure $\nu$ satisfies
\[
 \nu\ll\sum_{\boldsymbol r\in\Gamma}
 w_{\boldsymbol r}(T_{\boldsymbol r})_*\mu,
 \qquad w_{\boldsymbol r}>0,\quad\sum_{\boldsymbol r}w_{\boldsymbol r}=1,
\]
then all conclusions of \Cref{cor:rational-statistics} hold for a vector
of law $\nu$. Moreover, for each fixed nonconstant
$R\in\R(T_1,\ldots,T_d)$, $R_*\nu$ is Rajchman and gives full mass to
$\mathfrak N$.
\end{lemma}

\begin{proof}
The simultaneous approximations in \Cref{lem:joint-rational-carrier}
are preserved by each fixed rational scaling: multiply their common
denominator by a fixed integer and use an approximation of larger order.
Each nonzero real polynomial has a null zero set under every scaled
product law, and under every translated logarithmic product law, by the
independent atomless-coordinate argument in
\Cref{cor:rational-statistics}. Countable unions and domination preserve
these null sets and the dimension-zero carriers. A positive-mass Borel
set under a rational image of $\nu$, raw or logarithmic, has positive
mass under one of the corresponding images of the scaled product laws.
Each such image has positive-mass packing dimension one by
\Cref{cor:rational-statistics}; hence so does the image of $\nu$.
Its Hausdorff-zero carrier and this packing assertion give local
dimensions $(0,1)$ as in that corollary. The remaining conclusions
follow from these null sets, the preserved approximations and
\Cref{cor:nonconcentrated-rational-images}.
For each $\boldsymbol r$, $R\circ T_{\boldsymbol r}$ is nonconstant;
\Cref{cor:normal-liouville-field} gives a Rajchman image of full mass on
$\mathfrak N$. Their countable probability mixture is Rajchman, since
finite partial sums converge in total variation. The image $R_*\nu$ is
absolutely continuous with respect to this mixture; poles are null.
The Rajchman band property proved in
\Cref{cor:summable-local-coupling} and preservation of null sets complete
the proof.
\end{proof}

\begin{corollary}[Normal Liouville fields from affine forms]
\label{cor:fixed-shift-normal-field}
Let $L_j(n)=\alpha_jn+b_j$, $j\ge1$, be integer forms with
$\alpha_j\ge1$, $b_j\ne0$, and
$\alpha_i b_j-\alpha_j b_i\ne0$ for $i\ne j$.
Let $U_\ell$ be independent uniform elements of $(\Z/\ell\Z)^*$. Put
\[
 X_j=\prod_{\ell\textup{ prime}}(1-1/\ell)^{\mathbf1_{\{L_j(U_\ell)=0\}}},
 \qquad K=\Q(X_1,X_2,\ldots).
\]
The products converge to positive limits almost surely. Their finite
tuples have the ratio laws of \Cref{cor:fixed-shift-rational-statistics}.
For each fixed nonconstant $R\in\R(T_1,\ldots,T_d)$, the law of
$R(X_1,\ldots,X_d)$ is Rajchman and gives full mass to $\mathfrak N$.
Almost surely $K$ has infinite transcendence degree and
\[
 K\setminus\Q\subset\mathfrak N\cap\mathcal L.
\]
\end{corollary}

\begin{proof}
For each $j$, the expected sum of absolute logarithms is finite;
a countable intersection gives simultaneous positive convergence.
The finite tuples are precisely the models of
\Cref{cor:fixed-shift-rational-statistics}. Fix $d$ and put
$c_j=2^{\mathbf1_{\{\alpha_j+b_j\text{ odd}\}}}$.
By \Cref{cor:affine-independent-reference},
$\operatorname{Law}(X_1,\ldots,X_d)
\ll(T_{\boldsymbol c})_*\mu_G^{\otimes d}$.
Apply \Cref{lem:rational-orbit-domination} with
$\Gamma=\{\boldsymbol c\}$ and weight one. Intersect the finite-tuple arithmetic events and these
normality events over $d$ and the nonconstant rational functions over $\Q$.
Every field element uses finitely many generators. Algebraic independence
therefore gives infinite transcendence degree, and the simultaneous
normality and Liouville conclusions give the stated inclusion.
\end{proof}

\begin{corollary}[Fourier decay and normality of linear images]
\label{cor:rajchman-diff}
\begin{enumerate}[label=\textup{(\roman*)}]
\item Fix $\boldsymbol a\in\R^k\setminus\{0\}$ and put
$r=\#\{j:a_j\ne0\}$. Let $\nu$ be either the image of
$\mu_{\boldsymbol L}$ from \Cref{thm:fixed-shift-tuple} under
$\boldsymbol u\mapsto\boldsymbol a\cdot\boldsymbol u$, or the law of
$\sum_ja_jY_j$ for independent $Y_j$ of law $\muF$. Then
\[
 |\widehat\nu(\tau)|=O_{\boldsymbol L,\boldsymbol a}(\rho(\tau)^r),
 \qquad \nu(\mathfrak N)=1,
\]
with dependence only on $\boldsymbol a$ in the independent model.
In particular, $\muF(\mathfrak N)=1$.
\item The full vector transform satisfies
$\widehat\mu_{\boldsymbol L}(\boldsymbol t)\to0$ as
$\|\boldsymbol t\|_2\to\infty$.
\item For distinct nonzero integers $b,c$, let $D_{b,c}$ be as in
\eqref{eq:fixed-shift-difference}. Then
\[
 |\widehat{\operatorname{Law}(D_{b,c})}(\tau)|
 =O_{b,c}(\rho(\tau)^2),\qquad
 \operatorname{Law}(D_{b,c})(\mathfrak N)=1.
\]
This includes $\mu_{\Delta f}=\operatorname{Law}(D_{1,-1})$
of \Cref{thm:diff-dist}.
\end{enumerate}
\end{corollary}

\begin{proof}
\textup{(i)} In the two models,
$|\widehat\nu(\tau)|\le C\prod_{a_j\ne0}e^{-S(a_j\tau)/2}$,
by \eqref{eq:tuple-fourier-bound}, respectively by
\eqref{eq:forward} and independence. Equation~\eqref{eq:vk-dissipation}
and $\rho(a\tau)\asymp_a\rho(\tau)$ for fixed $a\ne0$ bound each
factor by $O_a(\rho(\tau))$. Apply \Cref{lem:iterlog-normality}.
The one-variable assertion also follows from
\Cref{cor:raw-ratio-analytic-images} with $F(x)=-\log x$.

\textup{(ii)} Since $S\ge0$ and $S(t)\to\infty$ as $|t|\to\infty$,
\eqref{eq:tuple-fourier-bound} gives the vector assertion.

\textup{(iii)} In \textup{(i)} take $L_1(n)=n+b$, $L_2(n)=n+c$
and $\boldsymbol a=(1,-1)$; the case $(b,c)=(1,-1)$ is
\Cref{thm:diff-dist}.
\end{proof}

\begin{corollary}[Raw monomials at affine prime arguments]
\label{cor:affine-raw-monomial-fourier}
Let $\boldsymbol L$ and $\boldsymbol Z$ be as in
\Cref{thm:fixed-shift-tuple}, put $X_j=e^{-Z_j}$ as in
\Cref{cor:fixed-shift-rational-statistics}, fix
$\boldsymbol a\in\R^k\setminus\{0\}$ and put $r=\#\{j:a_j\ne0\}$.
The empirical laws of
$\prod_{j=1}^k(\varphi(L_j(p))/L_j(p))^{a_j}$, over primes
$p>\max_j|b_j|$, converge to the law of
\[
 Y_{\boldsymbol a}:=\prod_{j=1}^kX_j^{a_j}
                   =e^{-\boldsymbol a\cdot\boldsymbol Z},
\]
and
\begin{equation}\label{eq:affine-raw-monomial-fourier}
 \left|\mathbb E e^{itY_{\boldsymbol a}}\right|
 =O_{\boldsymbol L,\boldsymbol a}(\rho(t)^r)
 \quad(|t|\to\infty),\qquad
 \mathbb P(Y_{\boldsymbol a}\in\mathfrak N)=1.
\end{equation}
In particular, for distinct nonzero integers $b,c$ and $D_{b,c}$ as in
\eqref{eq:fixed-shift-difference}, the limiting law of
$\varphi(p+b)/\varphi(p+c)$ in
\Cref{cor:fixed-symmetric-shifts}\textup{(b)} satisfies
\[
 \left|\mathbb E\exp\bigl(ite^{-D_{b,c}}\bigr)\right|
 =O_{b,c}(\rho(t)^2),\qquad
 \mathbb P\bigl(e^{-D_{b,c}}\in\mathfrak N\cap\mathcal L\bigr)=1.
\]
\end{corollary}

\begin{proof}
The hypotheses on the forms are $L_j(n)=\alpha_jn+b_j$ with
$\alpha_j\ge1$, $b_j\ne0$, and
$\alpha_i b_j-\alpha_j b_i\ne0$ for $i\ne j$.
For $p>\max_j|b_j|$ all $L_j(p)$ are positive, so their totient ratios
lie in $(0,1]$. By \Cref{thm:fixed-shift-tuple}\textup{(a)},
$\boldsymbol Z$ is finite almost surely, whence
$\boldsymbol X\in(0,\infty)^k$ almost surely. Thus
\Cref{cor:fixed-shift-rational-statistics} and the continuous mapping
theorem on $(0,\infty)^k$, applied to
$\boldsymbol x\mapsto\prod_jx_j^{a_j}$, give the empirical convergence
for these real exponents.

Renumber the forms so that $a_j\ne0$ exactly for $1\le j\le r$.
Let $E$ be the exceptional set of \Cref{sec:affine-tuple-models} and
put $p_0=\max E\ge2$. Every prime $\ell>p_0$ lies outside $E$;
the roots $-b_j\alpha_j^{-1}$ modulo $\ell$ are therefore distinct
and nonzero. Put $R_\ell=e^{-\boldsymbol a\cdot\boldsymbol Z_\ell}$,
and define $\xi_\ell=j$ when $L_j(U_\ell)=0$ for $j\le r$ and
$\xi_\ell=0$ otherwise. These states are mutually exclusive.
A root of a form with $a_j=0$ belongs to the neutral state, where
$R_\ell=1$. In state $j$,
\[
 R_\ell=(1-1/\ell)^{a_j}=\vartheta_j(1/\ell),\qquad
 \vartheta_j(z)=(1-z)^{a_j},\qquad \vartheta_j'(0)=-a_j\ne0.
\]
Each $\vartheta_j$ is positive and real analytic near $[0,1/p_0]$,
with $\vartheta_j(0)=1$. Uniformity of $U_\ell$ gives
\[
 \mathbb P(\xi_\ell=j)=\frac1{\ell-1}
       =\ell^{-1}+O(\ell^{-2}),\qquad
 \mathbb P(\xi_\ell=0)=1-\frac r{\ell-1}\ge1-\frac{2r}\ell.
\]
The head
$H_0=\exp(-\boldsymbol a\cdot\sum_{\ell\le p_0}\boldsymbol Z_\ell)$
is independent of the remaining coordinates. It is bounded above and
away from zero, since
$|\boldsymbol a\cdot\boldsymbol Z_\ell|\le\omega_\ell\sum_j|a_j|$.
Absolute almost sure convergence of $\boldsymbol Z$ gives positive
convergence of the product defining $Y_{\boldsymbol a}$. Moreover,
\[
 \mathbb ER_\ell^{\pm1}
 =1+\frac1{\ell-1}\sum_{j\le r}
       \bigl((1-1/\ell)^{\pm a_j}-1\bigr)
 =1+O_{\boldsymbol a}(\ell^{-2}).
\]
Independence bounds both moments of every finite product uniformly;
Fatou's lemma gives
$\mathbb E(Y_{\boldsymbol a}+Y_{\boldsymbol a}^{-1})<\infty$.
Apply \Cref{thm:raw-product-fourier} with $s=1$, $m=r$,
$\kappa_j=1$ and $C=2r$.

For the ratio take $L_1(n)=n+b$, $L_2(n)=n+c$ and
$\boldsymbol a=(1,-1)$, so that $Y_{\boldsymbol a}=e^{-D_{b,c}}$.
The factor $(p+b)/(p+c)\to1$ converts the ratio of normalized
totients to $\varphi(p+b)/\varphi(p+c)$, as in
\Cref{cor:fixed-symmetric-shifts}\textup{(b)}, which also supplies
the Liouville assertion.
\end{proof}

\subsection{Finite prefixes and defining-base nonnormality}
\label{ssec:finite-prefixes-nonnormality}

Almost-sure normality concerns block frequencies along an individual
infinite expansion. The next corollary instead asks how finite prefixes
are distributed across the law. For $Y=2X$, normality comes from the
analytic-image conclusion of \Cref{cor:raw-ratio-analytic-images}.
Comparing the endpoint cell with the cell containing $1/3$ determines
the largest prefix atom and the min-entropy. The Shannon bounds come
from \Cref{thm:shannon-dimension}. Its extraction
bound concerns one quantized observation, processed deterministically
conditional on an independent finite or countable public seed.

\begin{corollary}[Binary prefixes and deterministic extraction]
\label{cor:quantized-randomness}
\index{Digit statistics!finite prefixes}
\index{Min-entropy}
\index{Shannon entropy!digit prefixes}
Let $X\sim\mu_G$, $Y:=2X$, and, for integers $B\ge1$, put
$Z_B:=\lfloor2^BY\rfloor$, with an arbitrary value in
$\{0,\ldots,2^B-1\}$ on the null event $Y=1$. Identify $Z_B$ with
its $B$-digit binary expansion, including leading zeroes.
\begin{enumerate}[\upshape(i)]
\item $Y$ is absolutely normal almost surely. Put
$h=2^{-B-1}$, $z=B\log 2$, $q_B=\mathbb P(Z_B=2^B-1)=H_{\rm end}(2h)$
and $p_B=\max_j\mathbb P(Z_B=j)$. For all sufficiently large $B$ the
unique most likely prefix has index
\begin{equation}\label{eq:binary-prefix-maximizer}
 j_*=
 \begin{cases}
 2^B-1,&B\text{ odd},\\
 \lfloor2^{B+1}/3\rfloor,&B\text{ even}.
 \end{cases}
\end{equation}
For all such $B$, one has $p_B=q_B$ in odd length and
$0<p_B-q_B\ll\mathcal R(h)$ in even length.
Thus both parities share the full endpoint expansion, to every
fixed inverse-logarithmic order. In particular,
\begin{equation}\label{eq:binary-prefix-atom}
 \begin{aligned}
 p_B&=\kappa_{\rm end}\{z^{-1}-\pi^2/(12z^3)+O(z^{-4})\},\\
 H_\infty(Z_B):=-\log_2p_B
 &=\log_2(z/\kappa_{\rm end})+\frac{\pi^2}{12(\log2)z^2}+O(z^{-3}).
 \end{aligned}
\end{equation}
The Shannon entropy of $Z_B$ in bits satisfies
\begin{equation}\label{eq:binary-prefix-shannon}
 \begin{gathered}
 \frac{\mathsf H(Z_B)}{\log2}
 =-\sum_j\mathbb P(Z_B=j)\log_2\mathbb P(Z_B=j),\\
 (c_L^+-o(1))B\le\frac{\mathsf H(Z_B)}{\log2}\le(d_*+o(1))B.
 \end{gathered}
\end{equation}
The test $T_B(z)=\mathbf1_{\{z=2^B-1\}}$ is evaluable in $O(B)$ digit
operations and has acceptance probabilities $q_B$ under $Z_B$ and
$2^{-B}$ under the uniform $B$-bit law.

\item Let $S$ be a random seed independent of $X$, with values in a finite
or countable set $\mathcal S$. For every integer $m\ge1$ and every family
of deterministic maps
$E_s\colon\{0,\ldots,2^B-1\}\to\{0,1\}^m$, one has
\begin{equation}\label{eq:binary-prefix-extraction}
 d_{\mathrm{TV}}\!\left(
   \operatorname{Law}(S,E_S(Z_B)),
   \operatorname{Law}(S,U_m)\right)
 \ge \max\{p_B-2^{-m},0\},
\end{equation}
where $U_m$ is uniform on $\{0,1\}^m$ and independent of $S$.
\end{enumerate}
\end{corollary}

\begin{proof}
Apply \Cref{cor:raw-ratio-analytic-images} with $F(x)=2x$ for normality.
Continuity makes grid endpoints null, so each cell in $X$ may be
written $(t-h,t]$. With $H=H_{\rm end}$ and $L=\log(1/h)$ the last
cell has mass $q_B=H(2h)\sim\kappa_{\rm end}/L$. If $t\le\sqrt h$,
its mass is at most $G(\sqrt h)\le M(-1)\sqrt h$ by \Cref{cor:avg-loss}.
Otherwise $\eta=h/t\le\sqrt h$. Since $\log(1/\eta)\ge L/2$, a cell
of mass at least $q_B$ satisfies \eqref{eq:near-max} with $a=2/5$.
By \Cref{prop:two-window}(d) its upper edge is in $I_k(\eta)$ for
$k=2$ or $3$. Thus $t-h<1/k$, $t\asymp1$ and $\eta\asymp h$.
If $t<1/k$, the grid spacing gives $1/k-t\ge h/3$. Put
$\theta=1-kt$; then $h\ll\theta\ll h^{2/5}$. By
\eqref{eq:q-H-identity} and \eqref{eq:two-window-comparison} the cell mass is
$H(\theta+kh)-H(\theta)+O(\mathcal R(\eta))\ll L^{-2}$,
using \eqref{eq:hend-increment} and \eqref{eq:residual-properties}. This excludes
all cells except the last and the one containing $1/3$.

In the cell containing $1/3$, the lower edge is $1/3-fh$ with
$f=\{2^{B+1}/3\}$. Scaling the cell by $3/2$ therefore gives endpoint
mass $H(3fh)$ by \eqref{eq:q-H-identity}. The original mass exceeds
this by a strictly positive $O(\mathcal R(h))$ by
\Cref{lem:two-window-comparison} and \eqref{eq:residual-properties}.
For even $B$, $f=2/3$. For odd $B$, $f=1/3$ and
$H(2h)-H(h)\sim\kappa_{\rm end}\log2/L^2$ by \eqref{eq:hend-asymp}.
This proves uniqueness in both cases. The endpoint expansion and
\eqref{eq:residual-properties} give \eqref{eq:binary-prefix-atom}
and its extension to all orders. The all-ones test inspects $B$ digits.
Apply \Cref{thm:shannon-dimension} with $h=2^{-B-1}$.
Then $Z_B=\lfloor X/h\rfloor$ and
$\log(1/h)=(B+1)\log2$. Dividing the entropy bounds by $\log2$
and using $B+1=B+O(1)$ proves \eqref{eq:binary-prefix-shannon}.

Choose $j_*$ with $\mathbb P(Z_B=j_*)=p_B$ and set
$A:=\{(s,E_s(j_*)):s\in\mathcal S\}$.
Independence gives
\[
 \mathbb P((S,E_S(Z_B))\in A)\ge p_B,\qquad
 \mathbb P((S,U_m)\in A)=2^{-m}.
\]
The definition of total variation proves
\eqref{eq:binary-prefix-extraction}.
The seedless case follows by taking $S$ constant.
\end{proof}

For a fixed integer base $\beta\ge2$ and reduced progression $b\bmod Q$,
let $X$ have distribution function $G_{Q,b}$, put $Y=X/A_{Q,b}$ and
$Z_B=\lfloor\beta^B Y\rfloor$, with the same null-endpoint convention. Set
$S_Q=\{\ell\ge3:\ell\mid Q\}$,
$q_{S_Q}=\prod_{\ell\in S_Q}(\ell-2)/(\ell-1)$
and $K_Q=\kappa_{\mathrm{end}}/q_{S_Q}$. In this progression $Y$ is
absolutely normal almost surely, and the last-prefix mass, largest
atom and min-entropy satisfy
\[
 \begin{aligned}
 q_B=\mathbb P(Z_B=\beta^B-1)
   &=\frac{K_Q}{B\log\beta}+O_{Q,\beta}(B^{-3}),\\
 p_B=\max_j\mathbb P(Z_B=j)
   &=\frac{K_Q}{B\log\beta}+O_{Q,\beta}(B^{-2}),\\
 H_\infty(Z_B)
   &=\log_2 B+\log_2\frac{\log\beta}{K_Q}+O_{Q,\beta}(B^{-1}).
 \end{aligned}
\]
In \eqref{eq:binary-prefix-shannon}, replace $B$ by $B\log_2\beta$;
the last-prefix test has uniform acceptance probability $\beta^{-B}$,
and \eqref{eq:binary-prefix-extraction} is unchanged, with the same
proof, once the maps $E_s$ are defined on $\{0,\ldots,\beta^B-1\}$.
Indeed, $Y$ has the law of $2X_0$, $X_0\sim\mu_G$, conditional on
$B_\ell=0$ for $\ell\in S_Q$. Normality therefore persists, and
\Cref{cor:ap-endpoint-expansions,cor:ap-analytic-structure,thm:shannon-dimension}
give the displayed substitutions at mesh $\beta^{-B}$.

\begin{remark}
If $B=B(\lambda)\to\infty$, $B\le\lambda^C$ for a fixed $C>0$,
and $m\ge\lambda$, then the total variation distance in
\eqref{eq:binary-prefix-extraction} is $\Omega(\lambda^{-C})$.
Thus one quantized observation with an
independent public seed cannot give negligible statistical error in
this regime, where negligible means $o(\lambda^{-A})$ as
$\lambda\to\infty$ for every fixed $A>0$. 
\end{remark}

A fixed determinant density $c(p)$ is a different object from a random
sample of the limiting law. The following proposition concerns its
expansion in the defining base.

\begin{proposition}[Nonnormality of the Euler product in its defining base]
\label{prop:natural-base-nonnormality}
\index{Normal numbers!natural-base obstruction}
For every integer $b\ge2$ and $r\in\Q$, the number
$r\prod_{j\ge1}(1-b^{-j})$ is not normal in base $b$.
In particular, for every prime $p$, the determinant density $c(p)$
of \eqref{eq:def-P} is not normal in base $p$. Put
$r_p=\varphi(p-1)/(p-1)$. Uniformly for $m\in\Z$,
\[
 \frac1N\sum_{n=0}^{N-1}e^{2\pi i m p^n c(p)}
 =\frac{1+2\cos(2\pi m r_p)}3
   +O_p\!\left(\frac{1+|m|}{\sqrt N}\right).
\]
Thus the empirical orbit measures converge weakly on $\R/\Z$ to
$(\delta_0+\delta_{r_p}+\delta_{-r_p})/3$, with coincident atoms
counted with multiplicity.
\end{proposition}

\begin{proof}
Euler's pentagonal identity~\cite[\S~27.14(ii), (27.14.4)--(27.14.5)]{DLMF}
gives $P(p)=\sum_{e\in E}\sigma_ep^{-e}$, where
\[
 u_k=\frac{k(3k-1)}2,\quad v_k=\frac{k(3k+1)}2,\quad
 E=\{0\}\cup\{u_k,v_k:k\ge1\},
\]
and $\sigma_0=1$, $\sigma_{u_k}=\sigma_{v_k}=(-1)^k$.
The partial sums $C_n=\sum_{e\le n}\sigma_e$ are zero for
$u_k\le n<v_k$ and $(-1)^k$ for $v_k\le n<u_{k+1}$.
These blocks have lengths $k$ and $2k+1$; summing over even and odd $k$
and allowing the final incomplete block shows that each of $-1,0,1$
occurs $N/3+O(\sqrt N)$ times among $0\le n<N$.
Since $(p-1)r_p\in\Z$,
$p^nc(p)\equiv r_pC_n+T_n\pmod1$, where
$T_n=r_p\sum_{e>n}\sigma_ep^{n-e}$. Geometric summation gives
\[
 \sum_{n<N}|T_n|\le\frac{r_p}{p-1}\#(E\cap[0,N])
                  +\frac{r_p}{(p-1)^2}=O_p(\sqrt N).
\]
The occupation counts and $|e^{2\pi iu}-1|\le2\pi|u|$ prove the
displayed average and hence the circle limit. Taking $m=p-1$ gives
an average tending to $1$, contradicting Weyl's criterion for
base-$p$ normality.
For general $b,r$, choose an integer $h\ge1$ with $a=hr\in\Z$.
The truncated head of $hb^nrP(b)$ is integral. The same tail estimate,
with $p$ replaced by $b$ and $r_p$ by $|a|$, gives an average tending
to $1$ for the nonzero Fourier index $h$, proving the first assertion.
\end{proof}

Thus, for every prime power $q$, the limiting determinant density over
$\mathbb F_q$ is not normal in base $q$, while finite-dimensional densities
are rational.
The equivalence of normality in multiplicatively dependent bases is
classical~\cite{Max53}. More explicitly, if $b=c^u$, $d=c^v$ with
integers $c\ge2$, $u,v\ge1$, and $r,s\in\Q$, then, with $h\ge1$ the
least integer clearing their denominators, the same pentagonal-tail estimate gives
\[
 \frac1N\sum_{n<N}e^{2\pi i h d^n(rP(b)+s)}
 =1+O_{b,d,r,s}(N^{-1/2});
\]
the terms with $ue\le vn$ and the rational shift contribute integers,
and the remaining geometric tails have total size $O(\sqrt N)$.
Thus $rP(b)+s$ is not normal in base $d$.

\subsection{Signed series, progressions and cancellation limits}
\label{ssec:signed-series-progressions-cancellation}

\begin{corollary}[Fourier decay for signed Bernoulli series]
\label{cor:general-bias-fourier}
Fix $c_\ast>0$, an integer $\ell_0\ge2$, and signs
$\epsilon_\ell\in\{-1,1\}$ for primes $\ell>\ell_0$. Let $Z$ be a
real random variable and let $(B_\ell)_{\ell>\ell_0}$ be independent
Bernoulli variables, independent of $Z$, with
\[
 p_\ell:=\mathbb P(B_\ell=1),\qquad
 \Delta:=\sum_{\ell>\ell_0}|p_\ell-c_\ast/\ell|<\infty.
\]
The sum $Y=Z+\sum_{\ell>\ell_0}\epsilon_\ell B_\ell\omega_\ell$
converges absolutely almost surely. Its law $\nu$ is Rajchman and satisfies
\[
 |\widehat\nu(\tau)|=O(\rho(\tau)^{c_\ast}),\qquad
 \nu(\mathfrak N)=1,
\]
with implied constant and a sufficient threshold effectively bounded in
terms of $c_\ast,\Delta,\ell_0$.
If $Z$ is countably atomic, then $\nu$ is singular continuous and
$\dim_H\nu=0$.
\end{corollary}

\begin{proof}
Since $\omega_\ell\le2/\ell$ and
$p_\ell\le c_\ast/\ell+|p_\ell-c_\ast/\ell|$, the expected absolute
sum is finite, so Tonelli gives absolute convergence. Put
\[
 S_p(\tau):=\sum_{\ell>\ell_0}2p_\ell(1-p_\ell)
                 (1-\cos(\tau\omega_\ell)).
\]
The local factor of $\epsilon_\ell B_\ell\omega_\ell$ has squared modulus
$1-2p_\ell(1-p_\ell)(1-\cos(\tau\omega_\ell))$ and $|\mathbb Ee^{i\tau Z}|\le1$, so
$1-u\le e^{-u}$ and the limit in finite products, as in \eqref{eq:forward}, give
$|\widehat\nu(\tau)|^2\le e^{-S_p(\tau)}$. Put
\[
 C_0:=c_\ast\sum_{3\le\ell\le\ell_0}d_\ell
       +\sum_{\ell>\ell_0}|2p_\ell(1-p_\ell)-c_\ast d_\ell|<\infty.
\]
Then $S_p(\tau)\ge c_\ast S(\tau)-2C_0$, so
$|\widehat\nu(\tau)|\le e^{C_0}e^{-c_\ast S(\tau)/2}$.
To verify finiteness and the effective bound for $C_0$, take
$L=\max(\ell_0,\lceil2c_\ast\rceil,3)$. For $\ell>L$, the Lipschitz
bound for $u\mapsto2u(1-u)$ on $[0,1]$ and
$d_\ell=2/\ell-2/[\ell(\ell-1)^2]$ give
\[
 |2p_\ell(1-p_\ell)-c_\ast d_\ell|
 \le2|p_\ell-c_\ast/\ell|+2c_\ast^2/\ell^2
                 +2c_\ast/[\ell(\ell-1)^2].
\]
The finitely many smaller primes are bounded in terms of $c_\ast,\ell_0$.
Equation~\eqref{eq:vk-dissipation} now proves the rate with the asserted
dependence, and \Cref{lem:iterlog-normality} proves normality.

Now let $Z$ be countably atomic. Compare the coordinates above $L$
with independent Bernoulli variables of probabilities $c_\ast/\ell$,
using the same signs and jumps. The canonical signed series has
Hausdorff dimension zero by \Cref{cor:prime-series-zero-dimension}:
its expected absolute prime terms are $O(\ell^{-2})$ and its coordinate
entropies are $O(\log\ell/\ell)$. The coordinate total variation
distances sum to at most $\Delta$, so
\Cref{cor:summable-local-coupling} transfers dimension zero to the
original tail. Restoring the finite coordinates and the countably
atomic $Z$ gives a countable union of translates of a dimension-zero
full carrier. Finally, $\nu$ is atomless,
since $\sum_x\nu(\{x\})^2=\lim_{T\to\infty}(2T)^{-1}\int_{-T}^T|\widehat\nu|^2=0$
by Wiener's lemma and the Rajchman property.
\end{proof}

\begin{corollary}[Integer logarithmic totient law]
\label{cor:rajchman-integer}
The limiting law $\mu_F^{\mathrm{int},\log}$ of
$\log(n/\varphi(n))$ over the integers is Rajchman and satisfies
\[
 |\widehat{\mu_F^{\mathrm{int},\log}}(\tau)|=O(\rho(\tau)),\qquad
 \mu_F^{\mathrm{int},\log}(\mathfrak N)=1.
\]
\end{corollary}

\begin{proof}
Apply \Cref{cor:general-bias-fourier} to
\eqref{eq:integer-totient-factorization}, with $c_\ast=1$, $\ell_0=2$, $\epsilon_\ell=1$,
$p_\ell=1/\ell$, $\Delta=0$ and $Z$ the factor at $2$, of law $\frac12(\delta_0+\delta_{\log2})$.
The qualitative assertions also follow from
\Cref{cor:raw-ratio-analytic-images} with $F(x)=-\log x$.
\end{proof}

\begin{corollary}[Fourier decay of the logarithmic law in a fixed progression]
\label{cor:ap-rajchman}
\index{Arithmetic progressions!Fourier decay}
For fixed $Q\ge1$ and $\gcd(b,Q)=1$, the law $\mu_{f,Q,b}$ of
$-\log X_{Q,b}$ (see \eqref{eq:XQ}) is Rajchman and satisfies
\[
 |\widehat{\mu_{f,Q,b}}(\tau)|=O_Q(\rho(\tau)),\qquad
 \mu_{f,Q,b}(\mathfrak N)=1.
\]
\end{corollary}

\begin{proof}
By \eqref{eq:XQ}, apply \Cref{cor:general-bias-fourier} to $-\log X_{Q,b}$ with
$c_\ast=1$, $\epsilon_\ell=1$, $\ell_0$ the largest prime factor of $2Q$,
$p_\ell=1/(\ell-1)=1/\ell+O(\ell^{-2})$ and
$Z:=-\log A_{Q,b}+\sum_{\ell\le\ell_0,\ \ell\nmid2Q}B_\ell\omega_\ell$.
\end{proof}

\begin{proposition}[Failure of uniform cancellation when $N$ is of order $\log\tau$]
\label{prop:log-scale-obstruction}
For every $A>1$ and $0<\delta<1$, no estimate
\[
 \left|\sum_{N<p\le2N}e^{i\tau\omega_p}\right|
 \ll_{A,\delta}\frac{N}{(\log N)^A}
\]
holds uniformly for sufficiently large $\tau$ and dyadic
$N\in[\log\tau,\tau^{1-\delta}]$.
\end{proposition}

\begin{proof}
Dirichlet's simultaneous approximation theorem gives, for every $y\ge2$,
an integer
\[
 1\le\tau_y\le100^{\pi(y)},\qquad
 \left\|\frac{\tau_y\omega_p}{2\pi}\right\|\le\frac1{100}
 \quad(p\le y),
\]
where $\|x\|:=\min_{m\in\Z}|x-m|$.
If $\tau_y\le y/4$, Bertrand's theorem supplies a prime
$2\tau_y<p<4\tau_y\le y$, for which $1/p<\omega_p<1/(p-1)$ gives
\[
 \frac1{8\pi}<\frac{\tau_y\omega_p}{2\pi}\le\frac1{4\pi},
\]
contradicting the approximation. Thus $\tau_y>y/4$, while
$\log\tau_y\le\pi(y)\log100=O(y/\log y)$.
Choose a dyadic $N_y$ with $\log\tau_y\le N_y<2\log\tau_y$.
Then $2N_y\le y$ eventually, and
\[
 \operatorname{Re}\sum_{N_y<p\le2N_y}e^{i\tau_y\omega_p}
 \ge\cos(2\pi/100)(\pi(2N_y)-\pi(N_y))
 \asymp\frac{N_y}{\log N_y}.
\]
Since $\tau_y\to\infty$ and $N_y\le\tau_y^{1-\delta}$ eventually,
this contradicts the proposed estimate for $A>1$.
\end{proof}

\section{Fourier decay of the divisor-sum laws}
\label{sec:rajchman-sigma}

\begin{corollary}[Logarithmic divisor-sum laws]
\leavevmode\phantomsection\label{cor:rajchman-sigma}\ignorespaces
The limiting law $\mug$ of $\log(\sigma(p-1)/(p-1))$ over primes and the limiting
law $\mu_\sigma^{\mathrm{int},\log}$ of $\log(\sigma(n)/n)$ over the integers are
Rajchman, have Fourier transforms $O(\rho(\tau))$, and give full mass to $\mathfrak N$.
\end{corollary}
\begin{proof}
For an atomic law,
$|\sum_mp_me^{i\tau\phi_m}|^2=1-2\sum_{m<n}p_mp_n(1-\cos(\tau(\phi_n-\phi_m)))$.
In each odd-prime factor retain the atoms at $0$ and $\wsig_\ell$. In
\eqref{eq:nusig} their masses are $(\ell-2)/(\ell-1)$ and $1/\ell$, and twice
their product is $d_\ell^\sigma$. In
\eqref{eq:integer-sigma-factorization} they are $(\ell-1)/\ell$ and
$(\ell-1)/\ell^2$, and twice their product is $d_\ell^\sigma+O(\ell^{-2})$.
With $1-u\le e^{-u}$, and with the factors at $2$ bounded by one, both
transforms have squared modulus at most $e^{-S_\sigma(\tau)+O(1)}$.
\Cref{thm:vk-fourier,lem:iterlog-normality} give the rate and normality.
The qualitative assertions also follow from
\Cref{cor:raw-ratio-analytic-images} with $F(x)=\log x$.
\end{proof}

\begin{corollary}[Finite-coordinate and progression variants]
\label{cor:sigma-variants}
The bounds of \Cref{thm:sigma-moments} hold for
$\mugodd$, $\mug$ and $\mu_\sigma^{\mathrm{int},\log}$, after deletion
of any fixed finite collection of prime coordinates, and after
conditioning on any positive-probability event determined by finitely
many valuation coordinates. They also hold for the logarithmic laws
in \Cref{cor:ap-sigma-law}, for each fixed reduced progression.
For these laws, their exponential images also have additive
concentration of order $1/\log(1/h)$.
\end{corollary}
\begin{proof}
Place $2$ and all modified coordinates in $F$. Their sum is an
independent countably atomic head $H\ge0$; the remaining coordinates
have their original laws. For a progression, write
$a=v_p(Q)$ and $c=\min(v_p(b-1),a)$ at $p\mid Q$.
The conditional valuation is $c$ when $c<a$, and otherwise has law
\[
 \mathbb P(K_p=k)=(p-1)p^{a-k-1}\qquad(k\ge a).
\]
At $p\nmid Q$ it has the shifted-prime law, including the usual
positive valuation at $2$ when $2\nmid Q$; this is precisely the
model of \Cref{cor:ap-sigma-law}. Apply
\Cref{thm:sigma-moments} with $F\supseteq\{p:p\mid2Q\}$.
Since the raw variable $X=e^Z$ satisfies $X\ge1$, the intersection
of any interval of length $h$ with $[1,\infty)$ maps under $\log$
into an interval of length at most $h$, proving the upper bound.
For the lower bound, fix an atom $h_0$ of the head and repeat the
argument for the empty-block estimate \eqref{eq:lq-empty-block},
with tail width $\log(1+h e^{-h_0})\asymp_{h_0}h$. It places $X$ in
$[e^{h_0},e^{h_0}+h]$ with the required probability.
\end{proof}

\begin{remark}[Abundancy classes]\label{rem:friendly-lower}
For the full shifted-prime law and $r>0$, put
\[
 \Omega(r):=\sum_{\substack{m\ge2\ \mathrm{even}\\\sigma(m)/m=r}}
 2^{1-v_2(m)}
 \prod_{\substack{p^k\Vert m\\p\ge3}}
       \frac{p-1}{p^k(p-2)}.
\]
These weights satisfy
\[
 \liminf_{h\downarrow0}\log(1/h)\,
 \mug([\log r,\log r+h])\ge\frac{\kappa_{\mathrm{end}}}2\Omega(r),
 \qquad \sup_r\Omega(r)<\infty.
\]
Indeed, for any finite set of such integers choose a cutoff containing
all their prime factors. Their distinct valuation cylinders are
disjoint. The zero-tail endpoint asymptotic in
\Cref{prop:endpoint-sigma}, with the omitted zero-coordinate
probabilities divided out, gives the displayed coefficient for each
cylinder. Summing and then increasing the finite set proves the first
bound; \eqref{eq:sigma-concentration} proves the second.
The matching fixed-anchor limits and the uniform saturation argument
are proved in \Cref{thm:sigma-saturation-mas04}; the arithmetic
identification of the maximal class is given in
\Cref{thm:sigma-concentration-constants-mas05}. 
\end{remark}

\begin{lemma}[Uniform gaps between abundancy-class weights]
\label{lem:abundancy-class-gap-mas05}
Put $R(m):=\sigma(m)/m$, and, for odd primes $p$, put
\[
 a_p(0):=1,\qquad a_p(j):=\frac{p-1}{p^j(p-2)}\quad(j\ge1),
 \qquad a(u):=\prod_{p^j\Vert u}a_p(j)\quad(u\text{ odd}).
\]
Then the nonnegative class sums
\[
 F(r):=\sum_{\substack{u\ge1\ {\rm odd}\\R(u)=r}}a(u),\qquad
 \Omega(r)=\sum_{\substack{k\ge1,\ u\ge1\ {\rm odd}\\R(2^ku)=r}}
                   2^{1-k}a(u)
\]
satisfy
\begin{equation}\label{eq:abundancy-class-gap-mas05}
 \begin{aligned}
 F(1)&=1,\quad F(4/3)=\tfrac23, &
 F(r)&\le\tfrac4{15}\quad(r\notin\{1,4/3\}),\\
 \Omega(3/2)&=1, &
 \Omega(r)&\le g_1:=\max\{11/15,\Omega(2)\}<4/5\quad(r\ne3/2).
 \end{aligned}
\end{equation}
In particular, each endpoint class is uniquely maximal.
\end{lemma}
\begin{proof}
We first prove all estimates with the odd prime factors restricted to
a finite set $S$. Write the corresponding sums as $F_S$ and $\Omega_S$.
The bounds below are independent of $S$; increasing $S$ then proves
the assertions by monotone convergence. A sum over an empty class is
zero. Only positive rational $r$ need be considered.

We record three elementary facts used in the finite-support inductions.
If $r=A/B$ in lowest terms and $R(m)=r$, then $B\mid m$.
Consequently, a denominator prime $q$ with $v_q(B)=e$ must have
valuation at least $e$ in every member, and its total possible weight is
\begin{equation}\label{eq:denominator-cost-mas05}
 \sum_{j\ge e}a_q(j)=\frac{q^{1-e}}{q-2}.
\end{equation}
For example, if $q\in S$ is odd, multiplicativity gives
\[
 F_S(r)=\sum_{j\ge e}a_q(j)
             F_{S\setminus\{q\}}\bigl(r/R(q^j)\bigr),
\]
and the same identity holds with $F$ replaced by $\Omega$.
If $q\notin S$, the class is empty.
Also, every prime power $p^j$ is solitary: the reduced denominator of
$R(p^j)$ is $p^j$, so every member of its class is divisible by $p^j$.
The function $R$ strictly increases under proper divisibility, whence
that member must be $p^j$ itself.

Finally, fix the set $H$ of odd primes having valuation at least two
and fix those valuations. The remaining squarefree odd completion in
a prescribed class is unique if it exists. Indeed, the numbers
$(p+1)/p$, for odd primes $p$, are multiplicatively independent: in a
finite relation the largest prime occurs only in its own denominator,
and descent finishes the argument. For $\Omega_S$, fix $k$ as well.
A squarefree completion has weight at most $1$, at most $2/3$ if
nonempty, at most $4/15$ if nonempty and excluding $3$, and at most
$8/45$ if it contains at least two primes. Here $a_p(1)$ decreases with
$p$, and $a_3(1)=2/3$, $a_5(1)=4/15$.
The total weights of higher-valuation patterns are controlled by
\[
 D_5:=\prod_{p\ge5}\left(1+\frac1{p(p-2)}\right),\qquad d:=D_5-1,
 \qquad D_3:=\prod_{p\ge3}\left(1+\frac1{p(p-2)}\right),
\]
because $\sum_{j\ge2}a_p(j)=1/[p(p-2)]$.
Wallis's product over all odd integers gives
\begin{equation}\label{eq:pattern-constant-mas05}
 D_3\le\frac\pi2,\qquad
 D_5\le\frac{3\pi}{8}\frac{63}{64}<\frac{297}{256},
 \qquad d<\frac{41}{256}.
\end{equation}
For the second inequality we delete the factors at $3$ and at the
composite integer $9$ from the full product; its remaining factors
include all prime factors defining $D_5$. The strict numerical bound
uses $\pi<22/7$.

\emph{Odd classes without $3$.}
We prove $F_S(1)=1$ and $F_S(r)\le1/3$ for $r\ne1$, when $3\notin S$,
by induction on $|S|$. The assertion is immediate for $S=\varnothing$.
For a noninteger class choose a denominator prime $q\ge5$.
The induction hypothesis bounds every residual class by $1$, so
\eqref{eq:denominator-cost-mas05} bounds this class by
$1/(q-2)\le1/3$. For an integer class $r>1$, the higher-valuation
pattern is nonempty: a nonempty squarefree odd integer has its largest
prime in its reduced abundancy denominator. Uniqueness of the
squarefree completion bounds the class by $d<1/3$.
The class $1$ contains only $u=1$, and classes below $1$ are empty.
This proves the induction, including the bound by $1$ needed at its
next step.

\emph{All odd classes.}
Induct on $|S|$ again. Integer classes above $1$ have a nonempty
higher-valuation pattern and weight at most
$D_3-1\le\pi/2-1<2/3$.
A denominator prime $q\ge5$ gives at most $1/3$, using the bound by
$1$ for smaller supports. If the denominator contains $3$, remove that
coordinate. Unless a residual class is $1$, the preceding no-$3$
estimate bounds all residual classes by $1/3$, giving at most $1/3$
since $\sum_{j\ge1}a_3(j)=1$.
A residual class $1$ means $r=R(3^j)$; solitude makes the entire class
the singleton $3^j$, of weight $a_3(j)\le2/3$.
These cases and the endpoint establish the assertions about $F_S$.

\emph{The sharpened odd bounds.}
In the subsystem without $3$, a denominator prime $q\ge5$ gives
at most $4/15$: if a residual class is $1$, the member is the solitary
power $q^j$ of weight at most $4/15$; otherwise the no-$3$ bound
$1/3$ and \eqref{eq:denominator-cost-mas05} give at most $1/9$.
Integer classes above $1$ still cost at most $d<4/15$.
With $3$ allowed, a denominator prime $q\ge5$ gives at most $4/15$
by the same argument, using $2/3$ for nonendpoint residuals.
If $3$ is the only denominator prime, removal of its coordinate and
the improved no-$3$ bound give at most $4/15$, unless the class is a
solitary $3^j$. Its weight is $2/3$ for $j=1$ and at most $2/9$
for $j\ge2$. An integer class $r>1$ must have a nonempty higher-valuation
pattern. Patterns containing a prime at least $5$ cost at most $(4/3)d$,
allowing also the higher powers of $3$. If $3$ alone has valuation at least
two, the squarefree completion contains at least two primes:
$R(3^j)<3/2$ and each other squarefree prime contributes at most $6/5$.
Its weight is at most $a_5(1)a_7(1)=(4/15)(6/35)$. Thus the integer
class costs at most
\[
 (4/3)(41/256)+(1/3)(4/15)(6/35)<4/15 .
\]
Letting the finite prime support increase proves the bounds for $F$.

\emph{The dyadic factor.}
Write $\Omega(r)=\sum_{k\ge1}2^{1-k}F(r/R(2^k))$.
If any residual equals $1$, solitude gives a single term, of weight
at most $1/2$ off the endpoint. Otherwise at most one residual equals
$4/3$, since $R(2^k)$ increases strictly. In the subsystem excluding
$3$ every nonendpoint odd residual costs at most $4/15$, whence
\begin{equation}\label{eq:no-three-shifted-gap-mas05}
 \Omega_S(3/2)=1,\qquad \Omega_S(r)\le8/15\quad(r\ne3/2,\ 3\notin S).
\end{equation}
With all odd primes allowed, the same $8/15$ bound holds if no residual
equals $4/3$. If this exceptional residual occurs at $k\ge2$, its
coefficient is at most $1/2$, giving at most
$8/15+(2/3-4/15)/2=11/15$. At $k=1$ the exceptional case is $r=2$.

For $r=2$, write a class member as $2^ku$
with $u$ odd and put $M=2^{k+1}-1$. The equation
$M\sigma(u)=2^{k+1}u$ implies $M\mid u$ and
$R(u)=1+1/M$. Since
\[
 R(u)\ge R(M)\ge1+1/M,
\]
equality forces $u=M$ and $M$ prime: the first inequality is strict
under proper divisibility, and a composite $M$ has an extra divisor.
The member $6$ has weight $2/3$. Writing $a=k+1$, the remaining
prime candidates have total weight less than
\[
 \sum_{a\ge3}\frac{2^{2-a}}{2^a-3}
 \le\frac{32}{5}\sum_{a\ge3}4^{-a}=\frac2{15},
\]
where $2^a-3\ge(5/8)2^a$ for $a\ge3$ and
$a_{2^a-1}(1)<1/(2^a-3)$. The same upper bound holds with all odd primes allowed, so $\Omega(2)<4/5$.
\end{proof}

\begin{lemma}[Endpoint maximization from saturation]\label{lem:saturated-endpoint-maximization}
Let $\nu$ be atomless with support $[a,\infty)$, and write
$F(h)=\nu([a,a+h])$. Suppose $F(h)\sim Q_\nu(h)\sim w/\log(1/h)$,
$w>0$, and every family of relative near-maximizers of width $h$
eventually meets $[a,a+h]$. For all sufficiently small $h>0$, the
unique closed maximizing intervals for $\nu$ and $\exp_*\nu$ are
$[a,a+h]$ and $[e^a,e^a+h]$, respectively.
\end{lemma}
\begin{proof}
At fixed width the interval mass is continuous in its left endpoint
and tends to zero at infinity, so the maximum is attained. Full support
excludes any maximizer starting below $a$. If distinct maximizers exist
at widths $h\downarrow0$, take one of them as $[a+s,a+s+h]$ with $s>0$;
the same argument applies if the endpoint interval itself fails to maximize.
The intersection hypothesis forces $s\le h$ eventually. At $s=h$ the mass
is $F(2h)-F(h)=o(F(h))$. Thus $0<s<h$ eventually, and cancellation
of the overlap with $[a,a+h]$ gives
\[
  \nu([a+h,a+h+s])\ge F(s).
\]
Since $s\to0$ and $F(s)\sim Q_\nu(s)$, the interval on the left is a
relative near-maximizer at width $s$ but misses $[a,a+s]$. This
contradicts the hypothesis at width $s$ (pass to a decreasing subsequence
and use actual maximizers at unused widths). For the raw law put $A=e^a$.
An interval $[t,t+h]$ with $t>A$ has logarithmic width
$u=\log(1+h/t)<\log(1+h/A)$. Its mass is at most $F(u)$, strictly below
$F(\log(1+h/A))$ by full support. Intervals starting below $A$ likewise
give strictly less mass than the endpoint interval.
\end{proof}

\begin{theorem}[Exact divisor-sum concentration at the endpoint]
\label{thm:sigma-concentration-constants-mas05}
As $h\downarrow0$, the logarithmic laws and their exponential images
have the following concentration asymptotics:
\begin{equation}\label{eq:sigma-concentration-constants-mas05}
 \begin{aligned}
 Q_{\mugodd}(h)&\sim Q_{\exp_*\mugodd}(h)
                  \sim\frac{\kappa_{\mathrm{end}}}{\log(1/h)},\\
 Q_{\mug}(h)&\sim Q_{\muH}(h)
                  \sim\frac{\kappa_{\mathrm{end}}}{2\log(1/h)},\\
 Q_{\mu_\sigma^{\mathrm{int},\log}}(h)
   &\sim Q_{\mu_\sigma^{\mathrm{int}}}(h)
                  \sim\frac{e^{-\gamma}}{\log(1/h)}.
 \end{aligned}
\end{equation}
The uniquely maximal logarithmic anchor classes are respectively
$0$, $\log(3/2)$ and $0$. For any of these logarithmic laws $\nu$,
if intervals $I_h$ of length $h$ satisfy
$\nu(I_h)/Q_\nu(h)\to1$, then $I_h$ intersects
$[a_*,a_*+h]$ for all sufficiently small $h$, where $a_*$ is the
corresponding anchor. For all sufficiently small $h>0$ the unique closed
maximizing intervals for $\nu$ and $\exp_*\nu$ are $[a_*,a_*+h]$ and
$[e^{a_*},e^{a_*}+h]$. In particular, for the shifted odd and full laws
and every fixed $N\ge1$, writing $L=\log(1/h)$ and
$w=\kappa_{\mathrm{end}}$ or $\kappa_{\mathrm{end}}/2$, respectively,
\[
 Q_\nu(h)=w\sum_{j=1}^N a_j L^{-j}+O_N(L^{-N-1}),\qquad
 Q_{\exp_*\nu}(h)=w\sum_{j=1}^N\frac{a_j}{(L+a_*)^j}+O_N(L^{-N-1}).
\]
\end{theorem}

\Cref{rem:moving-anchor-windows-mas04} exhibits relative
near-maximizers that omit the anchor. 

\begin{proof}
For the integer law, Toulmonde's class function is
\[
 K_\sigma(1/r)=\sum_{R(m)=r}\frac1m.
\]
His Lemma~15 gives $K_\sigma(1)=1$ and
$K_\sigma(x)<0.9883$ for $x\ne1$; \S\S~10.1--10.2 and
Th\'eor\`eme~4 already identify the integer leading constant and
localize the reciprocal-law maximizing windows near its endpoint
\cite{Tou06}. The change $\varepsilon=1-e^{-h}$ converts reciprocal
multiplicative windows to logarithmic windows, giving the integer
logarithmic constant in \eqref{eq:sigma-concentration-constants-mas05}.

For the shifted odd and full laws, the anchor weights in
\Cref{lem:sigma-abundancy-coefficients-mas04} are respectively
$\kappa_{\mathrm{end}}F(r)$ and
$(\kappa_{\mathrm{end}}/2)\Omega(r)$.
\Cref{lem:abundancy-class-gap-mas05} identifies their unique maxima,
and \Cref{thm:sigma-saturation-mas04} gives the first two logarithmic
asymptotics. Its near-maximizer conclusion applies to all three laws,
using also the integer class gap just cited.

The fixed-anchor limit in \Cref{thm:sigma-saturation-mas04} gives
$\nu([a_*,a_*+h])\sim Q_\nu(h)$. All three laws are atomless with
support $[a_*,\infty)$: the odd free tail has atoms at $0$ and at the
first jump in each coordinate, and the support argument of
\Cref{thm:dist-sigma} applies also to the integer valuations; the
least dyadic term is an atom. Now \Cref{lem:saturated-endpoint-maximization}
gives the two exact maximizers. For the shifted laws,
\Cref{prop:endpoint-sigma} gives the expansion at the odd logarithmic
endpoint. For the full law the least dyadic head has probability $1/2$
and is isolated. For either raw law replace the logarithmic width by
$\log(1+he^{-a_*})$, whose inverse logarithm parameter is $L+a_*+O(h)$.
The leading raw equivalents follow in all three cases.
\end{proof}

\chapter{Fourier amplitude laws}
\label{ssec:typical-decay}

Let $U_T$ be uniform on $[0,T]$. For the shifted-prime logarithmic
totient law $\mu_f$, the normalized modulus $(\log T)|\FT(U_T)|$
has a nondegenerate limiting law. Its moments and tails describe the
proportions of frequencies with different amplitudes. The integer
logarithmic totient law has the same normalization
(\Cref{cor:integer-totient-amplitude}).
Under \Cref{def:rational-amplitude-data},
\Cref{thm:rational-haar-amplitude} constructs complex amplitude limits
from one Haar character on the positive rationals.

These sampling limits complement the pointwise bound of
\Cref{thm:vk-fourier} and the large values of
\Cref{prop:fourier-resonance}. They concern logarithmic laws;
pointwise and resonance results for the raw laws are stated separately.

% Diagram (design brief F8): Three Fourier questions: uniform, typical and exceptional.
\begin{figure}[!t]
\centering
\begin{tikzpicture}[figdiagram, x=1pt, y=1pt,
  % local helpers
  lanetag/.style={ftag, anchor=west},
  striprule/.style={frule, dash pattern=on 1.2pt off 1.4pt},
  rowA/.style={minimum height=36pt},
  rowB/.style={minimum height=33pt},
  rowBb/.style={minimum height=34pt},
  rowC/.style={minimum height=37pt},
  rowD/.style={minimum height=35pt},
  rowDb/.style={minimum height=33pt},
  leftcol/.style={minimum width=131pt},
  rightcol/.style={minimum width=172pt}]
% Grid: box frames run from x=1 to x=382.5; lane rules at y=0, -49.6, -143;
% the strip below the lanes sits under a dotted rule at y=-195.
% ================= lane 1: every sufficiently large |tau| =================
\draw[frule] (0.6,0) -- (382.9,0);
\node[lanetag] at (-0.5,0) {At every sufficiently large $|\tau|$:\ \ dissipation and pointwise decay};
\node[fres, rowA, leftcol, anchor=north west] (A1) at (1,-6.8)
  {Bounds for partial prime sums\\ $\sum_{N<p\le x}e^{i\tau\omega_p}$,\ \ $64\le N\le b|\tau|^{1/5}$\\ \figref{\Cref{prop:vk-prime-prefix}}};
\node[fres, rowA, rightcol, anchor=north east] (A2) at (382.5,-6.8)
  {$S(\tau)\ge\tfrac23\log\log|\tau|-\tfrac23\log\log\log|\tau|-O(1)$,\\ hence $|\FT(\tau)|\le e^{-S(\tau)/2}\ll\rho(\tau)$\\ \figref{\Cref{thm:vk-fourier}; \eqref{eq:forward}}};
\draw[fuse] (A1.east) -- node[flab, above] {Abel summation,} node[flab, below] {Mertens' theorem} (A2.west);
% ================= lane 2: U_T uniform on [0,T] =================
\draw[frule] (0.6,-49.6) -- (382.9,-49.6);
\node[lanetag] at (-0.5,-49.6) {At a frequency $U_T$ uniform on $[0,T]$:\ \ amplitude law, moments and exceptional sets};
% row a: equidistribution -> amplitude law
\node[fmech, rowB, leftcol, anchor=north west] (B3) at (1,-56.4)
  {Kronecker--Weyl equidistribution\\ of $(U_T\omega_\ell)_{\ell\le y}$, fixed $y$; tails $\ell>y$\\ \figref{\Cref{thm:linind}; \Cref{lem:amplitude-tails}}};
\node[fres, rowB, rightcol, anchor=north east] (B4) at (382.5,-56.4)
  {$(\log T)\,|\FT(U_T)|\Rightarrow W_F$;\\ $\mathbb E\,|\FT(U_T)|^{s}\sim M_F(s)/(\log T)^s$, fixed $s>-1$\\ \figref{\Cref{thm:fourier-amplitude-law}; $s<0$: \Cref{lem:totient-inverse-moments}}};
% row b: uniform moments -> tail bound
\node[fres, rowBb, anchor=north west] (B1) at (1,-102.2)
  {Finite-window moments, all $r\ge1$, $T\ge3$:\\ $\mathbb E\,|\FT(U_T)|^r\le\bigl(Ar/\log T\bigr)^r$\\ \figref{\Cref{thm:uniform-fourier-moments}; BMO: \Cref{lem:euler-majorant-bmo}}};
\node[fres, rowBb, anchor=north east] (B2) at (382.5,-102.2)
  {Tail bound, all $K\ge1$, $T\ge3$:\\ $\mathbb P\bigl(|\FT(U_T)|>K/\log T\bigr)\le Ae^{-aK}$\\ \figref{\Cref{thm:uniform-fourier-moments}; $K=(\log\log T)^2$: \eqref{eq:typical-decay}}};
\draw[fuse] (B1.east) -- node[flab, above] {Markov optimization} node[flab, below] {bounded $K$:\\ enlarge $A$} (B2.west);
% joint input to the moment part of the amplitude law
\node[fjoin] (j2) at (152,-72.9) {};
\draw[ffeed] (B3.east) -- (j2);
\draw[ffeed] (j2 |- B1.north) -- (j2);
\draw[fuse] (j2) -- (B4.west);
\node[fhyp, anchor=west] at (155,-95.8) {uniform integrability, $s>0$};
% ================= lane 3: selected frequencies t_T =================
\draw[frule] (0.6,-143) -- (382.9,-143);
\node[lanetag] at (-0.5,-143) {At selected frequencies $t_T\in(0,T]$, $\log t_T/\log T\to1$:\ \ resonant values, no uniform $O(1/\log|\tau|)$};
\node[fmech, rowC, anchor=north west] (C1) at (1,-149.8)
  {Positive-kernel\\ (Fej\'er) resonator\\ \figref{\Cref{thm:linind}}};
\node[fres, rowC, anchor=north east] (C2) at (242.3,-149.8)
  {$|\FT(t_T)|\ge(1-o(1))\,\log\log T/\log T$,\\ so $\limsup_{|\tau|\to\infty}\tfrac{\log|\tau|}{\log\log|\tau|}\,|\FT(\tau)|\ge1$\\ \figref{\Cref{prop:fourier-resonance}; \eqref{eq:fourier-endpoint-obstruction}}};
\draw[fuse] (C1.east) -- (C2.west);
\node[fopen, rowC, anchor=north east] (O) at (382.5,-149.8)
  {Open: is $|\FT(\tau)|\ll{}$\\ $\log\log|\tau|/\log|\tau|$\,?\\ \figref{\Cref{ssec:open_effective}}};
\draw[fmap, -] (C2.east) -- node[flab, above] {would be} node[flab, below] {sharp} (O.west);
% ================= strip: Section 12.5, second means and collisions =================
\draw[striprule] (0.6,-195) -- (382.9,-195);
\node[lanetag] at (-0.5,-195) {\Cref{sec:collision-saturation-mas04}:\ \ $\nu=\operatorname{Law}(\log R)$ as in \Cref{def:rational-amplitude-data}, e.g.\ $\muF$;\ \ $Y,Y'\sim\nu$ independent};
% row 1: second mean <-> collisions, then jointly with the anchors -> saturation
\node[fres, rowD, anchor=north west] (D1) at (1,-202.2)
  {Second Fourier mean\\ $\frac1T\int_0^T|\widehat\nu(t)|^2\,dt\sim\frac{C}{\log^2T}$\\[1.5pt] \figref{\Cref{thm:rational-haar-amplitude}, order $2$}};
\node[fres, rowD, anchor=north west] (D2) at (142.1,-202.2)
  {Collision probability\\ $\mathbb P(|Y-Y'|\le h)\sim\frac{C}{\log^2(1/h)}$\\[1.5pt] \figref{\Cref{lem:collision-tauberian-mas04}; \Cref{cor:rational-log-saturation}}};
\draw[fuse, <->] (D1.east) -- node[flab, above] {same $C$} (D2.west);
\node[fres, rowD, anchor=north east] (D4) at (382.5,-202.2)
  {Saturation\\ $Q_\nu(h)\sim\frac{\max_\rho w(\rho)}{\log(1/h)}$\\[1.5pt] \figref{\Cref{cor:rational-log-saturation}}};
\node[fjoin] (jb) at ($(D2.east)+(20,0)$) {};
\draw[ffeed] (D2.east) -- (jb);
\draw[fuse] (jb) -- (D4.west);
% row 2: raw law (from collisions) and anchor coefficients (into the joint input)
\node[fres, rowDb, anchor=north west] (D5) at (1,-251.3)
  {Raw law $\operatorname{Law}(R)$: same $C$,\\ if $\mathbb E(R^r+R^{-r})<\infty$ for some $r>0$\\ \figref{\Cref{lem:raw-log-collision-mas04}; \eqref{eq:rational-raw-transfer}}};
\node[fres, rowDb, anchor=north east] (D3) at (382.5,-251.3)
  {Anchor coefficients $w(\rho)$, Parseval:\\ $Z=\sum_\rho w(\rho)\chi(\rho)$ in $L^2$,\ \ $C=\mathbb E|Z|^2=\sum_\rho w(\rho)^2$\\ \figref{\Cref{prop:rational-anchor-parseval}}};
\draw[ffeed] (jb |- D3.north) -- (jb);
\draw[fmap] (D2.south west) ++(24,0) |- (D5.east);
\node[flab, anchor=east] at (163,-244.25) {raw/logarithmic transfer};
\end{tikzpicture}
\caption{Three questions about the Fourier transform $\FT$ of the shifted-prime logarithmic
totient law (raw laws: \Cref{thm:raw-ratio-normality,prop:raw-resonance}); in the top lane
$b$ is absolute and $x\le2N$. Bounds uniform in the order $r$, with absolute $A$, $a$,
control exceptional sets; moment limits hold for fixed orders. Below, $\rho$ ranges
over $\Q_{>0}$, $h\downarrow0$, and $Z$ is the Haar amplitude of $\nu$ for one Haar
character $\chi$ of $\Q_{>0}$.}
\label{fig:fourier-questions}
\end{figure}

\section{Sampling and moment domains}
For the shifted-prime logarithmic totient modulus,
\Cref{thm:fourier-amplitude-law} gives moment asymptotics for each
fixed real $s>-1$. The factor at $3$ causes divergence for $s\le-1$
once the sampling interval contains its first zero. Deleting a fixed
finite set $S$ permits every fixed real order precisely when $3\in S$
(\Cref{cor:deleted-amplitude}); otherwise $s>-1$ remains necessary.
The progression specialization concerns the logarithmic law with fixed $Q$.
The integer logarithmic totient law also has domain $s>-1$, with the
exceptional factor at $2$ and first positive zero $\pi/\log2$
(\Cref{cor:integer-totient-amplitude}).

For the logarithmic divisor-sum laws,
\Cref{thm:sigma-amplitude} gives complex limits with finite modulus
moments of every real order. The separate temporal bounds in
\Cref{cor:sigma-temporal-moments-mas22} give convergence for
$s>-1/2$ in the odd-prime and integer cases and for $s>-1/4$ in
the full shifted-prime case, together with logarithmic moments.
Mixed polynomial moments also converge. All local factors use
one shared Haar character, so the higher-valuation phases are dependent.

These asymptotics keep the moment order fixed as $T\to\infty$;
for the shifted-prime totient modulus, finite-window tail bounds are supplied separately by
\Cref{thm:uniform-fourier-moments}.

Taking $S=\varnothing$ and $K=(\log\log T)^2$ in
\eqref{eq:uniform-fourier-tail}, and putting
$g(T):=(\log\log T)^2/\log T$, gives absolute constants $C,c>0$ such that
\begin{equation}\label{eq:typical-decay}
 \frac1T\Bigl|\Bigl\{\tau\in[0,T]:|\FT(\tau)|>g(T)\Bigr\}\Bigr|
 \le C\exp\!\bigl\{-c(\log\log T)^2\bigr\}\qquad(T\ge e^e).
\end{equation}

Since $cu^2\ge Au-A^2/(4c)$, the bound in \eqref{eq:typical-decay} is
$O_A((\log T)^{-A})$ for every $A>0$, while $g(T)=(\log T)^{-1+o(1)}$.
As $g$ decreases on $[\exp(e^2),\infty)$ and $|\FT|$ is even, the set
$E:=\{t:|t|\ge\exp(e^2),\ |\FT(t)|>g(|t|)\}$ satisfies
$E\cap[-T,T]\subseteq\{t\in[-T,T]:|\FT(t)|>g(T)\}$, hence
$|E\cap[-T,T]|\ll_AT(\log T)^{-A}$ for every $A>0$.

The mean and variance theorem and \Cref{lem:tail-moment} are proved in
\Cref{app:cesaro-moments}, using \Cref{thm:linind,lem:prime-log-stieltjes}
and the tail estimates of \Cref{lem:amplitude-tails}.
\Cref{lem:tail-moment} gives convergence of each fixed even centered tail moment
as $T\to\infty$, followed by vanishing as the prime cutoff tends to
infinity; its bounds depend on the fixed moment order.

For locally integrable $h$ write
\[
 \overline h_T:=\frac1T\int_0^Th(\tau)\,d\tau,\qquad
 \Var_T(h):=\overline{(h-\overline h_T)^2}_T.
\]

\KeepHeadingSpace{5\baselineskip}
\begin{theorem}[Mean and variance of the cosine sum $S$]\label{thm:variance}
As $T\to\infty$,
\[
 \overline S_T=2\log\log T+C_D+O(1/\log T),
 \qquad
 \Var_T(S)\longrightarrow
 \frac12\sum_{\ell\ge3}d_\ell^2.
\]
Here, with $c_{\mathrm M}$ as in \Cref{lem:prime-log-stieltjes},
\begin{equation}\label{eq:dissipation-mean-constant}
 C_D=2c_{\mathrm M}-1-2\sum_{\ell\ge3}
             \frac1{\ell(\ell-1)^2}.
\end{equation}
\end{theorem}

The following tail estimates are used for every amplitude law of this chapter and in
\Cref{app:cesaro-moments}. Part~\textup{(b)} is stated for the first-jump functions $v$ of
\Cref{ssec:global-modulus-order}; recall $v_\varphi(\ell)=\omega_\ell$ and
$v_\sigma(\ell)=\gsig(\ell)$. In part~\textup{(a)}, a bound
$\mathbb E|\xi_p|\le C'(p-1)^{-2}$ gives the hypothesis with $C=4C'$.

\begin{lemma}[Tails of sampled prime sums and products]
\label{lem:amplitude-tails}
\begin{enumerate}[label=\textup{(\alph*)},leftmargin=*]
\item Let $C\ge0$ and $T\ge3$. For the primes $p>T$ let $\xi_p$ be real random variables
with $\mathbb E|\xi_p|\le C/p^2$, and put $\phi_p(t):=\mathbb Ee^{it\xi_p}$. Then, uniformly
for $|t|\le T$ and with absolute implied constants,
\begin{equation}\label{eq:amplitude-far-tail}
 \sum_{p>T}|\phi_p(t)-1|\le CT\sum_{p>T}\frac1{p^2}\ll\frac C{\log T},
 \qquad
 0\le\sum_{p>T}\bigl(1-|\phi_p(t)|^2\bigr)\ll\frac C{\log T}.
\end{equation}
There is $T_0(C)$ such that, for $T\ge T_0(C)$ and $|t|\le T$, every $\phi_p(t)$ with $p>T$ is
nonzero, the product $\prod_{p>T}\phi_p(t)$ converges absolutely, and
\[
 \prod_{p>T}\phi_p(t)=1+O\Bigl(\frac C{\log T}\Bigr),\qquad
 \sum_{p>T}\bigl|\log|\phi_p(t)|\bigr|\ll\frac C{\log T}.
\]
\item Let $v$ satisfy \eqref{eq:first-jump-hypotheses}, let $C\ge0$, let $2\le y<x$ and
$\Lambda>0$ be real, and let $b_p$ be complex numbers with $|b_p|\le C/p$ for the primes
$y<p\le x$. Then
\begin{equation}\label{eq:amplitude-moving-tail}
 \frac1\Lambda\int_0^\Lambda\Bigl|\sum_{y<p\le x}b_pe^{itv(p)}\Bigr|^2dt
 \le\sum_{y<p\le x}|b_p|^2+\frac{6\pi C^2}{c_v}\,\frac{\pi(x)}\Lambda .
\end{equation}
In particular, for $x=\Lambda=T>y\ge3$,
\[
 \frac1T\int_0^T\Bigl|\sum_{y<p\le T}b_pe^{itv(p)}\Bigr|^2dt
 \ll_vC^2\Bigl(\frac1{y\log y}+\frac1{\log T}\Bigr).
\]
\end{enumerate}
\end{lemma}

\begin{proof}
\textup{(a)} Since $|e^{iu}-1|\le|u|$, $|\phi_p(t)-1|\le|t|\,\mathbb E|\xi_p|\le CT/p^2$ for
$|t|\le T$, and $\sum_{p>T}p^{-2}\ll1/(T\log T)$ by \eqref{eq:endpoint-tail-square}. This is
the first bound in \eqref{eq:amplitude-far-tail}; the second follows from
$0\le1-|z|^2\le2|1-z|$ for $|z|\le1$. Put $\epsilon:=CT\sum_{p>T}p^{-2}$, and choose $T_0(C)$ so
that $\epsilon\le1/2$ for $T\ge T_0(C)$. Then $|\phi_p(t)-1|\le1/2$, so $\phi_p(t)\ne0$ and the
principal logarithms satisfy $|\log\phi_p(t)|\le2|\phi_p(t)-1|$. Hence
$\sum_{p>T}|\log\phi_p(t)|\le2\epsilon\le1$, the product converges absolutely,
$|\prod_{p>T}\phi_p(t)-1|\le e^{2\epsilon}-1\le4\epsilon$, and
$|\log|\phi_p(t)||\le|\log\phi_p(t)|$.

\textup{(b)} Let $P$ be the set of primes in $(y,x]$. If $P$ has at most one element, the
left side equals $\sum_{p\in P}|b_p|^2$. Otherwise put
$\delta_p:=\min_{q\in P,\ q\ne p}|v(p)-v(q)|$. By \eqref{eq:first-jump-spacing},
$v(p)-v(q)\ge c_v(1/p-1/(p+1))$ if $q>p$ and $v(q)-v(p)\ge c_v(1/(p-1)-1/p)$ if $q<p$,
so $\delta_p\ge c_v/(p(p+1))\ge c_v/(2p^2)$. The Montgomery--Vaughan mean-value inequality
\cite[Corollary~2, (1.9)]{MV74}, in its $\exp(i\lambda_pt)$ convention with
$\lambda_p=v(p)$, gives
\[
 \frac1\Lambda\int_0^\Lambda\Bigl|\sum_{p\in P}b_pe^{itv(p)}\Bigr|^2dt
 \le\sum_{p\in P}|b_p|^2+\frac{3\pi}\Lambda\sum_{p\in P}\frac{|b_p|^2}{\delta_p}
 \le\sum_{p\in P}|b_p|^2+\frac{6\pi C^2}{c_v}\,\frac{|P|}\Lambda ,
\]
which is \eqref{eq:amplitude-moving-tail}. For $x=\Lambda=T$, Chebyshev's bound gives
$\pi(T)/T\ll1/\log T$, and $\sum_{y<p\le T}|b_p|^2\le C^2\sum_{p>y}p^{-2}\ll C^2/(y\log y)$ by
\eqref{eq:endpoint-tail-square}.
\end{proof}

\section{A common Haar amplitude for rational products}
\label{sec:rational-haar-amplitude}

The amplitude construction uses separated first jumps, with all local
phases evaluated at one Haar character on $\Q_{>0}$. All local laws
and sampling densities in this section are fixed. The Fourier convention is
$\widehat\nu(t)=\int e^{itx}\,d\nu(x)$.

\begin{definition}[Rational amplitude data]
\label{def:rational-amplitude-data}
Fix $p_0\ge2$ and a function $v$ satisfying
\eqref{eq:first-jump-hypotheses}.  Let $R_{\rm head}$ and
$(R_p)_{p>p_0}$ be independent positive rational-valued random
variables, with an arbitrary countably atomic law for $R_{\rm head}$.
For every prime $p>p_0$, fix $\epsilon_p\in\{-1,1\}$ and
$r_p\in\Q_{>0}$ such that $\log r_p=\epsilon_pv(p)$, and assume
\begin{equation}\label{eq:rational-amplitude-hypotheses}
\begin{gathered}
 a_p:=\mathbb P(R_p=1)>0,\qquad
 a_p=1-\frac1p+O(p^{-2}),\\
 b_p:=\mathbb P(R_p=r_p)=\frac1p+O(p^{-2}),\qquad
 \mathbb E|\log R_p|=O(p^{-2}).
\end{gathered}
\end{equation}
The implied constants are uniform in $p$.  Put
\[
 R=R_{\rm head}\prod_{p>p_0}R_p,\qquad
 \nu=\operatorname{Law}(\log R).
\]
The logarithmic tail is absolutely convergent almost surely, so
$0<R<\infty$ almost surely.  No moment assumption is imposed on
$R_{\rm head}$ or its logarithm.
\end{definition}

\begin{lemma}[Signed first-jump inputs]
\label{lem:signed-amplitude-inputs}
Let $v$ satisfy \eqref{eq:first-jump-hypotheses} and fix signs
$\epsilon_p\in\{-1,1\}$.
\begin{enumerate}[label=\textup{(\alph*)},leftmargin=*]
\item If $2\le y<x$, $\Lambda>0$ and $|c_p|\le C/p$, then
\begin{equation}\label{eq:signed-amplitude-moving-tail}
 \frac1\Lambda\int_0^\Lambda
 \left|\sum_{y<p\le x}c_pe^{it\epsilon_pv(p)}\right|^2dt
 \le 2\sum_{y<p\le x}|c_p|^2+
       \frac{12\pi C^2}{c_v}\frac{\pi(x)}\Lambda.
\end{equation}
\item In \Cref{thm:valuation-fourier-moments}\textup{(a)},
the hypothesis $g(p)=v(p)$ may be replaced by
$g(p)=\epsilon_pv(p)$, with all other hypotheses and the bound
\eqref{eq:valuation-moments} unchanged.  The constant can be chosen
independently of the signs.  In particular, for the data of
\Cref{def:rational-amplitude-data} there is $A\ge e$ such that
\begin{equation}\label{eq:rational-temporal-moment-bound}
 \frac1{2T}\int_{-T}^T|\widehat\nu(t)|^r\,dt
 \le\left(\frac{Ar}{\log T}\right)^r
 \qquad(r\ge1,\ T\ge3).
\end{equation}
\end{enumerate}
\end{lemma}

\begin{proof}[Proof of \Cref{lem:signed-amplitude-inputs}]
For \textup{(a)}, split the prime sum into $S_++S_-$ according to
$\epsilon_p=1$ or $-1$.  Apply
\Cref{lem:amplitude-tails}\textup{(b)} to $S_+$, setting the
unused coefficients to zero.  The conjugate of $S_-$ has positive
frequencies and coefficients $\overline c_p$, so the same lemma
applies to it.  Add the estimates and use
$|S_++S_-|^2\le2|S_+|^2+2|S_-|^2$.  In each application the
proof of that lemma bounds the error by
$(6\pi C^2/c_v)\#\{p:\epsilon_p=\pm1,\ y<p\le x\}/\Lambda$:
one uses nearest-neighbour spacing within the indicated subset.
The two cardinalities sum to at most $\pi(x)$, proving
\eqref{eq:signed-amplitude-moving-tail}.

For \textup{(b)}, write $\xi_p=g(p^{K_p})$.  The two distinct
atoms at $0$ and $\epsilon_pv(p)$ give
\[
 |\mathbb Ee^{it\xi_p}|^2
 \le\exp\{-2\mathbb P(K_p=0)\mathbb P(K_p=1)
                    (1-\cos(tv(p)))\}.
\]
Here $\cos(t\epsilon_pv(p))=\cos(tv(p))$.  Thus the
Euler-product majorant \eqref{eq:euler-majorant} holds verbatim
for these signed atoms and the same cosine sum.  To spell out the
moment step, choose a fixed $H_F\ge1$ such that
$e^{H_F}>\max(F\cup\{3\})$ and
$B_v/H_F\le\mathfrak b$, and set
$L=H_Fr$, $y_0=e^L$.  For $T>y_0$, independence of the selected
block from the remaining sum and Mertens' estimate give
\[
 |\widehat\nu(t)|^r
 \le \mathfrak C_2^r\exp\{r\mathcal T^v_{y_0,T}(t)\}
       (L/\log T)^r e^{O(r/L)}.
\]
The BMO input \eqref{eq:general-exp-moment} applies since
$rB_v\le\mathfrak b L$.  Its normalized integral is at
most $C_{\rm JN}e^{r/m_v}$.  Enlarging $A$ in terms of $v,F,C_2$
gives $(Ar/\log T)^r$.  For $3\le T\le y_0$, the same bound
follows from $|\widehat\nu|\le1$ and $A\ge H_F$.  This proves the asserted signed extension.

For the rational data, the distinct atoms $1,r_p$ have masses
$a_p,b_p$, with $a_pb_p\ge1/p-C_2/p^2$ after increasing $C_2$.
Enumerate the other rational atoms by states $k\ge2$, take
$g(p^0)=0$, $g(p)=\log r_p$, and assign their logarithms to the
remaining states.  Put $H=\log R_{\rm head}$ and
$F=\{p:p\le p_0\}$.  Tonelli's theorem and
\eqref{eq:rational-amplitude-hypotheses} show that
$\sum_{p>p_0}|\log R_p|<\infty$ almost surely.  This proves
\eqref{eq:rational-temporal-moment-bound}.
\end{proof}

Let $\mathcal X=\widehat{\Q_{>0}}$ be the compact dual of the
discrete multiplicative group $\Q_{>0}$, equipped with Haar
probability measure $m$.  A point $\chi\in\mathcal X$ is one
character on all positive rationals.  Define continuous functions
\[
 h(\chi)=\mathbb E_{R_{\rm head}}\chi(R_{\rm head}),\qquad
 A_p(\chi)=\frac{\mathbb E_{R_p}\chi(R_p)}{a_p}.
\]
For fixed $\chi$, these expectations are taken over the indicated
local laws. Their defining countable character sums converge uniformly.

\KeepHeadingSpace{6\baselineskip}
\begin{theorem}[Haar amplitudes of rational local products]
\label{thm:rational-haar-amplitude}
\index{Fourier amplitudes!rational local products}
Assume \Cref{def:rational-amplitude-data}.
\begin{enumerate}[label=\textup{(\alph*)},leftmargin=*,beginpenalty=10000]
\item The positive finite normalization is
\begin{equation}\label{eq:rational-amplitude-normalization}
\begin{aligned}
 C_0&:=\lim_{y\to\infty}(\log y)\prod_{p_0<p\le y}a_p\\
 &=e^{-\gamma}\prod_{p\le p_0}(1-1/p)^{-1}
                 \prod_{p>p_0}\frac{a_p}{1-1/p}\in(0,\infty).
\end{aligned}
\end{equation}
The products, ordered by increasing primes,
\begin{equation}\label{eq:rational-haar-product}
 P_y(\chi):=C_0h(\chi)\prod_{p_0<p\le y}A_p(\chi)
 \longrightarrow
 Z_\nu(\chi):=C_0h(\chi)\prod_{p>p_0}A_p(\chi)
\end{equation}
converge almost surely and in $L^q(m)$ for every fixed
$0<q<\infty$.  In particular
$0<\mathbb E_m|Z_\nu|^q<\infty$.
\item Let $w$ be a fixed bounded nonnegative measurable function
supported on a fixed interval $[0,B]$, $0<B<\infty$, with
$\int_0^Bw(u)\,du=1$.  Give $T_w$ density $T^{-1}w(t/T)$.
Then
\begin{equation}\label{eq:rational-weighted-amplitude-limit}
 (\log T)\widehat\nu(T_w)\Rightarrow Z_\nu.
\end{equation}
For every fixed $s>0$ and nonnegative integers $j,k$,
\begin{equation}\label{eq:rational-amplitude-moments}
\begin{aligned}
 \mathbb E| (\log T)\widehat\nu(T_w)|^s
   &\longrightarrow\mathbb E_m|Z_\nu|^s,\\
 \mathbb E\left[((\log T)\widehat\nu(T_w))^j
       \overline{((\log T)\widehat\nu(T_w))}^{\,k}\right]
   &\longrightarrow\mathbb E_m[Z_\nu^j\overline Z_\nu^{\,k}].
\end{aligned}
\end{equation}
\item For any fixed nonzero integers $k_1,\ldots,k_d$ and positive
rationals $q_1,\ldots,q_e$, the joint limit is
\begin{equation}\label{eq:rational-joint-amplitudes}
 \left(((\log T)\widehat\nu(k_iT_w))_{i=1}^d,
             (q_j^{iT_w})_{j=1}^e\right)
 \Rightarrow
 \left((Z_\nu(\chi^{k_i}))_{i=1}^d,
             (\chi(q_j))_{j=1}^e\right).
\end{equation}
Every mixed polynomial moment in these variables and their
conjugates converges.  So does every joint modulus moment
$\mathbb E\prod_{i=1}^d| (\log T)\widehat\nu(k_iT_w)|^{s_i}$
for fixed $s_i\ge0$.
\end{enumerate}
\end{theorem}

\begin{proof}[Proof of \Cref{thm:rational-haar-amplitude}]
We give the product and temporal arguments separately.

\emph{Normalization and local expansion.}
Since $a_p/(1-1/p)=1+O(p^{-2})>0$, its infinite product converges
to a positive finite value.  Mertens' product formula gives
\eqref{eq:rational-amplitude-normalization}.  The atoms $1,r_p$
are distinct, and the remaining local mass is
$1-a_p-b_p=O(p^{-2})$, nonnegative.  Consequently, uniformly on
$\mathcal X$,
\begin{equation}\label{eq:rational-local-expansion}
 A_p(\chi)=1+d_p\chi(r_p)+E_p(\chi),\qquad
 d_p:=b_p/a_p=p^{-1}+O(p^{-2}),\quad
 \|E_p\|_\infty=O(p^{-2}).
\end{equation}
Choose a fixed $p_1\ge\max(p_0,3)$ large enough that
$|A_p-1|\le1/2$ for $p>p_1$.  The principal logarithm is then
defined for these factors, and the power-series estimate for
$\log(1+z)$ gives
\begin{equation}\label{eq:rational-local-log}
 \log A_p(\chi)=\frac{\chi(r_p)}p+G_p(\chi),
 \qquad \|G_p\|_\infty\ll p^{-2}.
\end{equation}
The same estimates hold at $\chi_t(q)=q^{it}$, uniformly in real
$t$.  

\emph{Orthogonality and almost-sure convergence.}
The rationals $r_p$ are distinct and different from $1$:
opposite signs put them on opposite sides of $1$, while equal
signs use the strict monotonicity of $v(p)$.  Haar invariance
therefore gives
$\mathbb E_m\chi(r_p)=0$ and
$\mathbb E_m[\chi(r_p)\overline{\chi(r_q)}]=\mathbf1_{\{p=q\}}$.
If a rational $u\ne1$ has nonzero exponent at
a prime $\ell$, varying $\chi(\ell)$ shows that
$\int\chi(u)\,dm=0$; the prime coordinates are independent
uniform phases under Haar measure.  It follows that
$\sum_{p>p_1}\chi(r_p)/p$ converges in $L^2(m)$.

Here is the maximal argument needed for convergence almost
surely.  List the primes beyond $p_1$ increasingly as $p_n$ and
put $X_n=\chi(r_{p_n})/p_n$.  On the index block
$I_k=\{2^k,\ldots,2^{k+1}-1\}$ let
$M_k$ be the maximum modulus of its initial partial sums.
Every initial subblock is a disjoint union of at most $k+1$
dyadic subblocks.  By Cauchy--Schwarz its squared sum is at most
$k+1$ times the sum of the squared sums over those subblocks,
which is bounded by the sum over all dyadic subblocks at all
$k+1$ levels.  Orthogonality gives
\begin{equation}\label{eq:rational-orthogonal-maximal}
 \mathbb E_mM_k^2\le(k+1)^2\sum_{n\in I_k}p_n^{-2}
 \ll(k+1)^2 2^{-k},
\end{equation}
since $p_n\ge n+1$.  Hence
$\sum_k\mathbb E_mM_k<\infty$.  Tonelli gives
$\sum_kM_k<\infty$ almost surely.  The block sums then converge
absolutely, and every intermediate partial sum differs from a
block endpoint by at most $M_k\to0$.  This proves convergence
of the series in its stated prime order.  The summable uniform
remainders in \eqref{eq:rational-local-log} prove almost-sure
convergence of the tail product to a nonzero finite value.
Multiplication by the bounded function
$C_0h\prod_{p_0<p\le p_1}A_p$ gives almost-sure convergence of
$P_y$; this bounded function itself may vanish.

\emph{Uniform Haar moments from the temporal BMO input.}
Fix $q>0$.  Choose $y_0\ge p_1$ large enough that
$qB_v\le\mathfrak b\log y_0$.  For a fixed finite
$y>y_0$, finite-character averaging gives
\begin{align*}
 &\int_{\mathcal X}
 \exp\left(q\operatorname{Re}\sum_{y_0<p\le y}\frac{\chi(r_p)}p\right)dm(\chi)\\
 &\qquad=\lim_{L\to\infty}\frac1{2L}\int_{-L}^L
       \exp\left(q\sum_{y_0<p\le y}\frac{\cos(tv(p))}p\right)dt
 \le C_{\rm JN}e^{q/m_v}.
\end{align*}
For the equality, uniformly approximate the continuous function of
these finitely many phases by Laurent polynomials. A monomial
has form $\chi(u)$ with $u\in\Q_{>0}$; its temporal average
$\frac1{2L}\int_{-L}^Le^{it\log u}dt$ tends to
$\mathbf1_{\{u=1\}}$, exactly its Haar average.  The bound is
\eqref{eq:general-exp-moment} with $L\ge y$; the signs disappear
inside the cosine.  Formula \eqref{eq:rational-local-log}, its
summable uniform error, and the bounded factors through $y_0$
now imply
\begin{equation}\label{eq:rational-uniform-haar-moments}
 \sup_{y\ge p_0}\mathbb E_m|P_y|^q<\infty
 \qquad(0<q<\infty).
\end{equation}
Applying this with $2q$, Fatou and uniform integrability imply
$\mathbb E_m|P_y-Z_\nu|^q\to0$.  This works also for $q<1$,
using $|u-v|^q\le C_q(|u|^q+|v|^q)$.  At the identity character,
$h=1$ and every one of the finitely many factors through $p_1$
is positive.  Continuity makes their product nonzero on a
nonempty open subset of $\mathcal X$, which has positive Haar
measure.  The remaining product is nonzero almost surely, so
$\mathbb P_m(|Z_\nu|>0)>0$.  This proves all assertions in
\textup{(a)}, including strict positivity of its positive moments.

\emph{Fixed heads and a fixed sampling density.}
For $u\in\Q_{>0}$,
\begin{equation}\label{eq:rational-weighted-character}
 \frac1T\int_0^{BT}w(t/T)u^{it}dt
 =\int_0^Bw(s)e^{iTs\log u}ds
 \longrightarrow\mathbf1_{\{u=1\}}.
\end{equation}
For $u\ne1$, this follows first for step functions from direct
integration, then for $w$ by $L^1$ approximation; for $u=1$ it
uses $\int w=1$.  Finite products of local character sums,
including $h$, are uniform limits of character polynomials:
truncate each of the finitely many probability laws, whose
omitted mass bounds the uniform error.  A continuous test
function of a fixed collection of these bounded functions and
finitely many characters is uniformly approximable on its
compact range by polynomials in the variables and their
conjugates.  Thus \eqref{eq:rational-weighted-character} proves
the joint fixed-head convergence, in particular
\[
 P_y(\chi_{T_w})\Rightarrow P_y(\chi)
 \qquad\hbox{for each fixed }y.
\]

\emph{Moving and far tails.}
Put $D=\max(1,B)$ and $X=DT$.  For fixed $y\ge p_1$ and
$X>y$ define
\[
 V_{T,y}(t)=\sum_{y<p\le X}\log A_p(\chi_t).
\]
The signed estimate \eqref{eq:signed-amplitude-moving-tail},
with $x=\Lambda=X$ and $c_p=1/p$, and
\eqref{eq:rational-local-log} give
\begin{equation}\label{eq:rational-moving-tail-bound}
 \mathbb E|V_{T,y}(T_w)|^2
 \ll_{v,D,\|w\|_\infty}
 \sum_{p>y}p^{-2}+\frac1{\log X}
                    +\left(\sum_{p>y}p^{-2}\right)^2.
\end{equation}
Indeed a nonnegative integral with the sampling weight is at
most $D\|w\|_\infty$ times its uniform average on $[0,X]$.
Consequently $e^{V_{T,y}(T_w)}-1$ tends to zero in probability
in the successive limits $T\to\infty$, then $y\to\infty$:
use $|e^z-1|\le |z|e^{|z|}$ and Chebyshev's inequality.

Write $L_p(t)=\mathbb ER_p^{it}$.  The independent logarithmic
tail converges absolutely almost surely, and its characteristic
function is $\prod_{p>X}L_p(t)$.  By
\Cref{lem:amplitude-tails}\textup{(a)}, applied to
$\xi_p=\log R_p$, uniformly for $|t|\le X$,
\begin{equation}\label{eq:rational-far-tail-bound}
 F_X(t):=\prod_{p>X}L_p(t)=1+O(1/\log X).
\end{equation}
Thus the exact factorization on the sampled interval is
\begin{equation}\label{eq:rational-temporal-factorization}
 (\log T)\widehat\nu(t)
 =\frac{(\log T)\prod_{p_0<p\le X}a_p}{C_0}
       P_y(\chi_t)e^{V_{T,y}(t)}F_X(t).
\end{equation}
Its scalar prefactor tends to $1$ because
$\log X/\log T\to1$ and
\eqref{eq:rational-amplitude-normalization} holds.

The fixed-head convergence and
\eqref{eq:rational-uniform-haar-moments} imply, for any $q>0$,
\[
 \limsup_{T\to\infty}\mathbb P(|P_y(\chi_{T_w})|>M)
 \le C_qM^{-q},
\]
uniformly in $y$, by applying fixed-head convergence to the
continuous function $|P_y|^q$ on its bounded range.  Given
$\eta>0$, first restrict the head modulus to at most $M$ in
\eqref{eq:rational-temporal-factorization}, then use
\eqref{eq:rational-moving-tail-bound} and
\eqref{eq:rational-far-tail-bound}, and finally let $M\to\infty$.
It follows that
\[
 \lim_{y\to\infty}\limsup_{T\to\infty}
 \mathbb P\left(\left|(\log T)\widehat\nu(T_w)
                        -P_y(\chi_{T_w})\right|>\eta\right)=0.
\]
Fixed-head convergence, followed by $P_y\to Z_\nu$ in
probability, now proves \eqref{eq:rational-weighted-amplitude-limit}.
One may see this directly with bounded Lipschitz test functions:
truncate the approximation error at $\eta$, take the temporal
limit for fixed $y$, and then send $y\to\infty$ and $\eta\to0$.

\emph{Temporal moments.}
The signed prerequisite \eqref{eq:rational-temporal-moment-bound}
and $X=DT\ge T$ yield, for each fixed $r\ge1$,
\begin{equation}\label{eq:rational-weighted-uniform-moments}
 \mathbb E| (\log T)\widehat\nu(T_w)|^r
 \le D\|w\|_\infty(Ar)^r
          (\log T/\log X)^r
 \le D\|w\|_\infty(Ar)^r\quad(T\ge3).
\end{equation}
Evenness of $|\widehat\nu|$ identifies the one-sided uniform
average with the symmetric average in that prerequisite.
For $s>0$, choose fixed $r>\max(s,1)$; for the monomial
$z^j\overline z^{\,k}$ choose $r>\max(j+k,1)$.
These bounds give uniform integrability.  Truncating the
continuous test functions and removing the truncation proves
\eqref{eq:rational-amplitude-moments}.

\emph{Joint dilations and finite characters.}
For a fixed nonzero integer $k$, evaluation at $kt$ replaces
$\chi_t$ by $\chi_t^k$, the jump $\epsilon_pv(p)$ by
$\operatorname{sgn}(k)\epsilon_p|k|v(p)$, and
$\mathbb E|\log R_p|$ by $|k|\mathbb E|\log R_p|$.
The signed moving-tail proof applies to $|k|v$ and the far-tail
proof applies with a cutoff $X=\max(1,B\max_i|k_i|)T$.
Its normalization still has $\log X/\log T\to1$.
The fixed-head averaging calculation applies simultaneously to
all $P_y(\chi^{k_i})$ and $\chi(q_j)$: every monomial is still
a single rational character.  A finite union bound controls all
moving tails.  This proves \eqref{eq:rational-joint-amplitudes}.
The map $\chi\mapsto\chi^k$ preserves Haar measure, since it
raises each independent uniform prime coordinate to a nonzero
integer power.  Thus the preceding Haar moment bounds apply
to every limit coordinate.  A change of variables, or the
signed moment argument for $|k|v$, gives all fixed positive
temporal moment bounds for each coordinate.  H\"older's
inequality bounds any higher moment of a finite product of
their powers; the extra character coordinates have modulus
one.  Uniform integrability and the joint weak limit prove
the last assertions of \textup{(c)}.
\end{proof}

\begin{remark}[Specializations and normalization]
\label{rem:rational-amplitude-specializations}
The positive-moment and complex-limit portions of the subsequent
amplitude results are instances of
\Cref{thm:rational-haar-amplitude}, as follows.  
\begin{enumerate}[label=\textup{(\alph*)},leftmargin=*]
\item For the shifted-prime totient law use reciprocal ratios:
$R_{\rm head}=2$ and, at every odd prime, $R_p=p/(p-1)$ with
probability $1/(p-1)$ and $R_p=1$ otherwise.  Thus
$a_p=(p-2)/(p-1)$, $b_p=1/(p-1)$, $v=v_\varphi$,
$\mathbb E|\log R_p|=\omega_p/(p-1)\ll p^{-2}$, and
$C_0=\kappa_{\mathrm{end}}$ by
\eqref{eq:endpoint-Q-quantitative}.  The Haar product is
$\kappa_{\mathrm{end}}\chi(2)\prod_{p\ge3}
(1+\chi(p/(p-1))/(p-2))$.
\item For the integer totient law use $R_p=p/(p-1)$ with
probability $1/p$, and $R_p=1$ otherwise.  Taking $p_0=2$,
put the prime-$2$ factor into $R_{\rm head}$.  Then
$a_p=1-1/p$, $b_p=1/p$ for $p>2$,
$\mathbb E|\log R_p|\le2/p^2$, and $C_0=2e^{-\gamma}$.
Since $h(\chi)=(1+\chi(2))/2$, the full normalization becomes
$e^{-\gamma}\prod_{p\ge2}(1+\chi(p/(p-1))/(p-1))$.
In both totient cases the independent-angle representation
requires the additional arithmetic input \Cref{thm:linind}.
\item For divisor sums put $R_{p,k}=1+p^{-1}+\cdots+p^{-k}$.
The shifted-prime law at odd $p$ has probabilities
$a_p=(p-2)/(p-1)$ at $k=0$ and $p^{-k}$ at $k\ge1$.
Thus $b_p=1/p$ at $r_p=R_{p,1}$ and $v=v_\sigma$.
The integer law has probabilities $(1-1/p)p^{-k}$ at
$k\ge0$, giving $a_p=1-1/p$ and $b_p=(1-1/p)/p$.
In both cases $0\le\log R_{p,k}\le\omega_p\le2/p$, while
the total active mass is $1/(p-1)$ or $1/p$, respectively;
hence the required full-law logarithmic expectation is
$O(p^{-2})$.  The odd shifted law has head $1$ and
$C_0=\kappa_{\mathrm{end}}$; the full shifted law has head
$R_{2,K}$ with $\mathbb P(K=k)=2^{-k}$, $k\ge1$, and the
same $C_0$.  For the integer law, take the actual prime-$2$
factor as head and $C_0=2e^{-\gamma}$.  These substitutions give exactly the three
products \eqref{eq:sigma-haar-products}, with all higher
valuation phases evaluated at the same character.
\item Finite deletions and fixed progressions fit by moving every
exceptional prime into the head, retaining its actual conditional
law.  Increasing $p_0$ to $p_1$ replaces $h$ by
$h\prod_{p_0<p\le p_1}\mathbb E\chi(R_p)$ and $C_0$ by
$C_0/\prod_{p_0<p\le p_1}a_p$, so the complete Haar expression
is unchanged.  Deleting a finite set $S$ of free odd
shifted-prime factors multiplies the normalization of the
remaining free product by
$\prod_{p\in S}(p-1)/(p-2)$.
For a totient progression the deterministic head is
$A_{Q,b}^{-1}$; for a divisor-sum progression it is the
independent product of the local conditional factors at
$p\mid2Q$ in \Cref{cor:ap-sigma-law}.  Thus the normalizations
are precisely $\kappa_{F,S}$ in
\Cref{cor:deleted-amplitude,cor:ap-sigma-amplitude}.
\end{enumerate}
All these applications allow the fixed bounded sampling densities
of \Cref{thm:rational-haar-amplitude}\textup{(b)}.  Its part
\textup{(c)} supplies the positive-moment part of the joint
integer-dilation result.  If $\nu^{*m}$ is a fixed convolution
power, then $\widehat{\nu^{*m}}=\widehat\nu^{\,m}$; continuous
mapping and the higher positive moments give normalization
$(\log T)^m$ and limit $Z_\nu^m$.  
\end{remark}

\section{The normalized totient amplitude}

Factoring out the inactive probabilities $(\ell-2)/(\ell-1)$ in
\Cref{lem:factor} gives $Q(T)\sim\kappa_{\mathrm{end}}/\log T$
by \eqref{eq:def-Q} and \eqref{eq:endpoint-Q-quantitative}, explaining the
normalization and its limiting constant.

For a fixed finite set of primes, linear independence of the jumps and
Kronecker--Weyl equidistribution replace the sampled phases by independent
uniform angles as $T\to\infty$. The mean-square estimate of
\Cref{lem:amplitude-tails}\textup{(b)} then controls the infinite tail
as the prime cutoff grows.

Negative moments of the totient modulus are controlled after removal of the factor at $3$,
which has zeros. Here $\mathcal N_{\{3\}}$ is the deleted product of
\Cref{thm:uniform-fourier-moments} and $S_5$ is as in \Cref{lem:dissipation-comparisons}.

\begin{lemma}[Uniform inverse moments of the normalized totient modulus]
\label{lem:totient-inverse-moments}
For $T\ge3$ put $Y_T(\tau):=(\log T)\,|\mathcal N_{\{3\}}(\tau)|$.
\begin{enumerate}[label=\textup{(\alph*)},leftmargin=*]
\item For every fixed $b>0$,
\begin{equation}\label{eq:fourier-deleted-inverse}
 \sup_{T\ge3}\frac1T\int_0^TY_T(\tau)^{-b}\,d\tau<\infty.
\end{equation}
\item For every fixed $0<a<1$,
\begin{equation}\label{eq:fourier-inverse-moment}
 \sup_{T\ge3}\frac1T\int_0^T\bigl((\log T)|\FT(\tau)|\bigr)^{-a}\,d\tau<\infty.
\end{equation}
\end{enumerate}
\end{lemma}

\begin{proof}
\textup{(a)} By \eqref{eq:deleted-comparison},
$\log |\mathcal N_{\{3\}}(\tau)|=-S_5(\tau)/2+O(1)$ uniformly in $\tau$.
The $\ell$th factor of \eqref{eq:euler} is
$|\widehat{\nu_\ell}(\tau)|^2=1-d_\ell(1-\cos(\tau\omega_\ell))$, where $\widehat{\nu_\ell}$ is the
characteristic function of $B_\ell\omega_\ell$ and
$\mathbb EB_\ell\omega_\ell=\omega_\ell/(\ell-1)\le(\ell-1)^{-2}$. Hence, for $0\le\tau\le T$, the
tail of $S_5$ beyond $T$ is $O(1/\log T)$ by \Cref{lem:amplitude-tails}\textup{(a)}.
Since $d_\ell/2=1/\ell-1/[\ell(\ell-1)^2]$, Mertens' theorem gives
\begin{equation}\label{eq:fourier-deleted-log}
 \log Y_T(\tau)=\sum_{5\le\ell\le T}
                    \frac{\cos(\tau\omega_\ell)}\ell+O(1).
\end{equation}
With the constants $B,\mathfrak b$ of
\Cref{lem:euler-majorant-bmo}, choose a fixed $y_0=y_0(b)\ge5$ such
that $bB\le\mathfrak b\log y_0$. The sum of terms with
$5\le\ell\le y_0$ in \eqref{eq:fourier-deleted-log} is bounded.
For $T>y_0$, \eqref{eq:cosine-exp-moment}, with $y=T$ and coefficient
$-b$, gives
\[
 \frac1{2T}\int_{-T}^T
       e^{-b\mathcal T_{y_0,T}(\tau)}\,d\tau
 \le C_{\mathrm{JN}}e^b.
\]
This proves \eqref{eq:fourier-deleted-inverse} for $T>y_0$, by
\eqref{eq:fourier-deleted-log} and evenness. On bounded intervals of $\tau$,
the series of logarithms of the factors of $|\mathcal N_{\{3\}}|$ converges uniformly,
while each factor is
strictly positive. Thus $|\mathcal N_{\{3\}}|$ is positive and continuous, which covers
the remaining bounded range of $T$.

\textup{(b)} By \eqref{eq:deleted-comparison}, the prime-$3$ factor gives the exact identity
\[
 (\log T)|\FT(\tau)|=|\cos(\tau\omega_3/2)|Y_T(\tau).
\]
Take $h>1$ with $ah<1$; set $h'=h/(h-1)$.
H\"older's inequality gives
\begin{align*}
 &\frac1T\int_0^T\bigl((\log T)|\FT(\tau)|\bigr)^{-a}\,d\tau\\
 &\quad\le\left(\frac1T\int_0^T
       |\cos(\tau\omega_3/2)|^{-ah}\,d\tau\right)^{1/h}
       \left(\frac1T\int_0^TY_T(\tau)^{-ah'}\,d\tau\right)^{1/h'}
 \ll_a1\qquad(T\ge3).
\end{align*}
The first factor is uniformly bounded by periodic integrability since
$ah<1$; the second is bounded by
\textup{(a)}.
\end{proof}

Retain $U_T$ as above, and let
$(\Theta_\ell)_{\ell\ge3}$ be independent uniform angles on $[0,2\pi]$.
Put
\[
\begin{aligned}
 \varrho_\ell&:=\frac1{\ell-2},&
 \psi_\ell(\theta)&:=\log|1+\varrho_\ell e^{i\theta}|,\\
 Z_F&:=\sum_{\ell\ge3}\psi_\ell(\Theta_\ell),&
 W_F&:=\kappa_{\mathrm{end}}e^{Z_F}.
\end{aligned}
\]
For $0\le v<1$ and every real $s$, and also for $v=1$ when $s>-1$,
put
\begin{equation}\label{eq:fourier-mellin-local}
 \mathcal H_s(v):=\frac1{2\pi}\int_0^{2\pi}|1+ve^{i\theta}|^s\,d\theta
 ={}_2F_1\!\left(-\frac s2,-\frac s2;1;v^2\right).
\end{equation}
At $v=1$,
\begin{equation}\label{eq:fourier-mellin-three}
 \mathcal H_s(1)=\frac{2^s\Gamma((s+1)/2)}{\sqrt\pi\,\Gamma(1+s/2)}
 \qquad(s>-1).
\end{equation}
The defining integral at $v=1$ diverges for real $s\le-1$.

\begin{theorem}[Limiting distribution and real moments of the normalized Fourier modulus]
\label{thm:fourier-amplitude-law}
\index{Fourier amplitudes!totient law}
The series $Z_F$ converges almost surely and in $L^m$ for every
$1\le m<\infty$.
\begin{enumerate}[label=\textup{(\roman*)},leftmargin=*]
\item As $T\to\infty$,
\begin{equation}\label{eq:fourier-amplitude-law}
\begin{aligned}
 \log|\FT(U_T)|+\log\log T
 &\Rightarrow\log\kappa_{\mathrm{end}}+Z_F,\\
 (\log T)|\FT(U_T)|&\Rightarrow W_F.
\end{aligned}
\end{equation}
For every fixed nonnegative integer $m$, the $m$th moment of
$\log|\FT(U_T)|+\log\log T$ converges to that of
$\log\kappa_{\mathrm{end}}+Z_F$. In particular,
\begin{equation}\label{eq:typical-exponent}
 \frac{\log|\FT(U_T)|}{\log\log T}\xrightarrow{\mathbb P}-1.
\end{equation}
\item For every fixed real $s>-1$,
\begin{equation}\label{eq:fourier-mellin-asymp}
\begin{aligned}
 \frac1T\int_0^T|\FT(\tau)|^s\,d\tau
 &\sim\frac{M_F(s)}{(\log T)^s},\\
 M_F(s):=\mathbb EW_F^s
 &=\kappa_{\mathrm{end}}^s\prod_{\ell\ge3}\mathcal H_s(\varrho_\ell),
\end{aligned}
\end{equation}
with a convergent positive product.
\item For $s\le-1$ and $T\ge\pi/\omega_3$,
$\int_0^T|\FT(\tau)|^s\,d\tau=\infty$.
\end{enumerate}
\end{theorem}

\begin{proof}
For $\ell\ge5$, uniformly in $\theta$,
\begin{equation}\label{eq:fourier-local-log}
 \psi_\ell(\theta)
 =\sum_{j\ge1}\frac{(-1)^{j+1}\varrho_\ell^j}{j}\cos(j\theta)
 =\varrho_\ell\cos\theta+O(\varrho_\ell^2).
\end{equation}
For $\ell=3$ the series converges in $L^2[0,2\pi]$ to
$\psi_3(\theta)=\log(2|\cos(\theta/2)|)$. Consequently,
\begin{equation}\label{eq:fourier-local-log-moments}
 \mathbb E\psi_\ell(\Theta_\ell)=0,\qquad
 \mathbb E\psi_\ell(\Theta_\ell)^2
 =\frac12\sum_{j\ge1}\frac{\varrho_\ell^{2j}}{j^2}.
\end{equation}
The variances are summable, which proves almost-sure and $L^2$
convergence. Moreover, $|\psi_\ell|\ll\varrho_\ell$ for $\ell\ge5$, and
centering gives
$\mathbb E e^{b\psi_\ell(\Theta_\ell)}=1+O_b(\varrho_\ell^2)$ for
each fixed real $b$. Independence therefore bounds the positive and
negative exponential moments of every finite tail uniformly in its
endpoints. Together with almost-sure convergence, these bounds imply
convergence in each fixed $L^m$. The single term $\psi_3(\Theta_3)$ has
a finite absolute moment of every positive integer order.

Write $F(\tau):=|\FT(\tau)|$. By \Cref{lem:factor}, $\widehat{\nu_\ell}$ is the characteristic
function of $B_\ell\omega_\ell$, and $\mathbb EB_\ell\omega_\ell=\omega_\ell/(\ell-1)\le(\ell-1)^{-2}$.
For $0\le\tau\le T$ and $T$ sufficiently large, \Cref{lem:amplitude-tails}\textup{(a)}
therefore gives
\begin{equation}\label{eq:fourier-log-tail}
 \left|\sum_{\ell>T}\log|\widehat{\nu_\ell}(\tau)|\right|\ll\frac1{\log T}.
\end{equation}
Factoring the remaining local terms
exactly yields, outside their discrete zero set,
\begin{equation}\label{eq:fourier-log-truncation}
 \log F(\tau)=A_T+\sum_{3\le\ell\le T}\psi_\ell(\tau\omega_\ell)
 +O(1/\log T),\qquad
 A_T:=\sum_{3\le\ell\le T}\log\frac{\ell-2}{\ell-1}.
\end{equation}
Here $e^{A_T}=Q(T)$ in the notation of \eqref{eq:def-Q}, so
\eqref{eq:endpoint-Q-quantitative} gives
\begin{equation}\label{eq:fourier-normalization}
 e^{A_T}=Q(T)\sim\frac{\kappa_{\mathrm{end}}}{\log T}.
\end{equation}

Fix $y\ge5$. Since $\omega_\ell=v_\varphi(\ell)$ and $\varrho_\ell\le3/\ell$,
\Cref{lem:amplitude-tails}\textup{(b)}, with $v=v_\varphi$, $b_\ell=\varrho_\ell$ and
$x=\Lambda=T$, gives
\begin{equation}\label{eq:fourier-moving-tail}
 \frac1T\int_0^T
 \left|\sum_{y<\ell\le T}\varrho_\ell e^{i\tau\omega_\ell}\right|^2
 \,d\tau
 \ll\sum_{\ell>y}\varrho_\ell^2+\frac1{\log T}.
\end{equation}
With $\|\cdot\|_{2,T}$ denoting the normalized $L^2[0,T]$ norm,
\eqref{eq:fourier-local-log}--\eqref{eq:fourier-moving-tail} imply
\begin{multline}\label{eq:fourier-head-approximation}
 \limsup_{T\to\infty}
 \left\|\log F+\log\log T-\log\kappa_{\mathrm{end}}
       -\sum_{3\le\ell\le y}\psi_\ell(\tau\omega_\ell)\right\|_{2,T}
 \\
 \ll\left(\sum_{\ell>y}\varrho_\ell^2\right)^{1/2}
       +\sum_{\ell>y}\varrho_\ell^2.
\end{multline}
Values at the discrete zero set may be assigned arbitrarily in this
$L^2$ statement. Averaging the finite sum over $3\le\ell\le y$ uses the almost-periodic
argument of \cite[Lemma~2.3 and Theorem~2.9]{ANS14}.
For fixed $y$, \Cref{thm:linind} and Kronecker--Weyl equidistribution~\cite[Satz~5]{Wey16}
show that its limiting distribution is the law of
$\sum_{3\le\ell\le y}\psi_\ell(\Theta_\ell)$. To justify the
singular prime-$3$ term, replace $\psi_3$ by
$\max(\psi_3,-R)$, a continuous periodic function. For each fixed $R$,
the set of $\tau\in[0,T]$ on which this changes the sum has relative
Lebesgue measure tending, as $T\to\infty$, to
$\mathbb P(\psi_3(\Theta_3)<-R)=O(e^{-R})$. Letting first
$T\to\infty$, then $R\to\infty$, proves the claimed convergence
for each fixed $y$. Now let $y\to\infty$
in \eqref{eq:fourier-head-approximation}. This proves the logarithmic
limit in \eqref{eq:fourier-amplitude-law}; exponentiation proves the
amplitude limit. The logarithmic limit makes
$\log|\FT(U_T)|+\log\log T$ tight; division by $\log\log T\to\infty$
proves \eqref{eq:typical-exponent}.

For $s>0$, choose a fixed $r>\max\{s,1\}$. By
\Cref{thm:uniform-fourier-moments},
\[
 \sup_{T\ge3}\frac1T\int_0^T\bigl((\log T)F(\tau)\bigr)^r\,d\tau
 <\infty.
\]
Thus the order-$s$ powers are uniformly integrable. For $-1<s<0$, take $-s<a<1$. By
\Cref{lem:totient-inverse-moments}\textup{(b)}, the moments of order $-a$ of
$(\log T)F(U_T)$ are bounded uniformly in $T\ge3$. This proves uniform integrability of
the order-$s$ powers, so weak convergence gives their moment convergence throughout $s>-1$.

The binomial series and integration over the phase give
\eqref{eq:fourier-mellin-local} for $v<1$. Beta integration at $v=1$
gives \eqref{eq:fourier-mellin-three}, equivalently the endpoint value
of the hypergeometric series for $s>-1$. For $\ell\ge5$,
\[
 \mathcal H_s(\varrho_\ell)
 =1+\frac{s^2}{4}\varrho_\ell^2+O_s(\varrho_\ell^4).
\]
The bounds on positive and negative exponential moments of the random
tail justify taking limits of its moments, while the independent prime-$3$ factor has the finite
moment \eqref{eq:fourier-mellin-three}. This identifies $\mathbb E W_F^s$
with the product in \eqref{eq:fourier-mellin-asymp}. At
$\tau_0=\pi/\omega_3$,
$F(\tau)\le|\cos(\tau\omega_3/2)|\ll|\tau-\tau_0|$, by
\eqref{eq:deleted-comparison}.
Its $s$-th power is therefore nonintegrable for $s\le-1$, even when
$\tau_0$ is the right endpoint.

Finally, for every fixed $0<a<1$, \Cref{thm:uniform-fourier-moments} and
\Cref{lem:totient-inverse-moments}\textup{(b)} give
\[
 \sup_{T\ge3}\mathbb E\exp\!\left(
 a\left|\log\bigl((\log T)F(U_T)\bigr)\right|\right)<\infty.
\]
All positive integer powers of the logarithm of the normalized modulus
are consequently uniformly integrable, and their expectations converge. Equation~\eqref{eq:fourier-local-log-moments}
also shows that the limit has positive variance.
\end{proof}

\begin{definition}[Joint totient amplitude variables]\label{def:totient-amplitude-variables}
Let $\Theta_0$ and $(\Theta_\ell)_{\ell\ge3}$ be independent uniform
angles on $[0,2\pi]$, and put
\[
\begin{aligned}
 D_F&:=-\sum_{\ell\ge3}d_\ell\cos\Theta_\ell,\qquad
 L_F:=\log\kappa_{\mathrm{end}}+Z_F,\\
 Z_{F,\mathbb C}&:=\kappa_{\mathrm{end}}e^{i\Theta_0}
       \prod_{\ell\ge3}(1+\varrho_\ell e^{i\Theta_\ell}).
\end{aligned}
\]
The independent uniform phase gives $Z_{F,\mathbb C}$ a circularly symmetric law.
\end{definition}

\KeepHeadingSpace{6\baselineskip}
\begin{corollary}[Joint law of dissipation and complex Fourier amplitude]\label{cor:complex-amplitude-law}
\index{Fourier amplitudes!complex law}
With the variables of \Cref{def:totient-amplitude-variables},
the series defining $D_F$ converges almost surely and in $L^m$ for
$1\le m<\infty$, and the product converges almost surely and in $L^2$.
For $D_T:=S(U_T)-\overline S_T$, $C_T:=(\log T)\FT(U_T)$ and
$L_T:=\log|C_T|$,
\begin{equation}\label{eq:joint-fourier-dissipation}
 (D_T,L_T,C_T)\Rightarrow(D_F,L_F,Z_{F,\mathbb C}).
\end{equation}
This convergence remains joint with any fixed finite family
$(e^{iU_T\omega_\ell})$, whose limits are $(e^{i\Theta_\ell})$.
All mixed polynomial moments in $D_T,L_T,C_T,\overline{C_T}$ converge.
For nonnegative integers $j,k$ and real $s>-1$, also
\[
 \mathbb E[D_T^jL_T^k|C_T|^s]
 \longrightarrow\mathbb E[D_F^jL_F^k|Z_{F,\mathbb C}|^s].
\]
In particular,
\begin{equation}\label{eq:complex-amplitude-law}
 (\log T)\FT(U_T)\Rightarrow Z_{F,\mathbb C}.
\end{equation}
\end{corollary}

\begin{proof}
The convergence of $D_F$ follows from summable variances and
\eqref{eq:finite-cosine-moment}. The finite products over $\ell\ge5$
form a mean-one complex martingale with second moments bounded by
$\prod_{\ell\ge5}(1+\varrho_\ell^2)<\infty$.
Their almost-sure and $L^2$ limit is nonzero almost surely: the analytic
logarithms $\log(1+\varrho_\ell e^{i\Theta_\ell})$ have mean zero and
summable variances, so their real and imaginary parts converge almost
surely by the independent-series convergence criterion. The independent
prime-$3$ factor is bounded and nonzero almost surely. This identifies
$|Z_{F,\mathbb C}|=e^{L_F}$.

For $0\le t\le T$, \Cref{lem:amplitude-tails}\textup{(a)}, applied to $\xi_\ell=B_\ell\omega_\ell$
(for which $\mathbb E\xi_\ell\le(\ell-1)^{-2}$), shows that the complex product tail
$\prod_{\ell>T}\widehat{\nu_\ell}(t)$ is $1+O(1/\log T)$ uniformly. For $\ell\ge5$,
\[
 \log(1+\varrho_\ell e^{it\omega_\ell})
 =\varrho_\ell e^{it\omega_\ell}+O(\varrho_\ell^2).
\]
Consequently \Cref{lem:amplitude-tails}\textup{(b)}, with $v=v_\varphi$ and $b_\ell=\varrho_\ell$,
together with the normalization \eqref{eq:endpoint-Q-quantitative}, gives a complex-logarithmic error
$E_{T,y}$ such that, outside the discrete prime-$3$ zero set,
\[
 C_T=\kappa_{\mathrm{end}}e^{iU_T\log2}
   \prod_{3\le\ell\le y}(1+\varrho_\ell e^{iU_T\omega_\ell})
   e^{E_{T,y}(U_T)},
\]
\[
 \lim_{y\to\infty}\limsup_{T\to\infty}
 \mathbb P(|E_{T,y}(U_T)|>\eta)=0\qquad(\eta>0).
\]
The uniform positive moments in \Cref{thm:uniform-fourier-moments}
make $C_T$ tight. Multiplication by $1-e^{-E_{T,y}(U_T)}$ shows that
its difference from the displayed finite head tends to zero in
probability in these successive limits. Likewise, writing
$D_y(t):=-\sum_{3\le\ell\le y}d_\ell\cos(t\omega_\ell)$,
\Cref{lem:tail-moment} and $\overline{D_y}_T\to0$ give
\[
 \limsup_{T\to\infty}\mathbb E|D_T-D_y(U_T)|^2
 \ll\sum_{\ell>y}d_\ell^2\longrightarrow0.
\]
By \Cref{thm:linind}, Kronecker--Weyl applies to $t\log2$ and the
prime phases in these heads, jointly with any additional fixed prime
phases. Letting $y\to\infty$ proves the joint convergence of
$(D_T,C_T)$. Since $Z_{F,\mathbb C}\ne0$ almost surely, continuous
mapping gives \eqref{eq:joint-fourier-dissipation}.

H\"older's inequality, \Cref{lem:tail-moment}, \Cref{thm:uniform-fourier-moments} and
\Cref{lem:totient-inverse-moments}\textup{(b)}
give uniform integrability of all the asserted mixed moments; for
$s<0$ take the H\"older exponent attached to the negative modulus
power sufficiently close to one. Logarithmic powers are absorbed by
arbitrarily small positive and negative modulus powers, keeping every
required negative power greater than $-1$.
\end{proof}

\begin{corollary}[Circular totient amplitude and its moments]\label{cor:totient-amplitude-circular-moments}
For $Z_{F,\mathbb C}$ of \Cref{def:totient-amplitude-variables},
its modulus has law $W_F$, and its angle is uniform and independent
jointly of $(D_F,Z_F)$. Moreover,
\begin{equation}\label{eq:joint-fourier-covariance}
 \operatorname{Cov}(D_F,L_F)
 =-\frac12\sum_{\ell\ge3}d_\ell\varrho_\ell
 =-\sum_{\ell\ge3}\frac1{(\ell-1)^2}.
\end{equation}
Moreover,
\[
 \mathbb E Z_{F,\mathbb C}^{m}\overline{Z_{F,\mathbb C}}^{n}
 =\begin{cases}0,&m\ne n,\\ M_F(2m),&m=n,\end{cases}
 \qquad(m,n\in\mathbb Z_{\ge0}).
\]
\end{corollary}

\begin{proof}
Condition on the entire prime-angle family. The nonzero prime
product contributes a fixed angle, and its sum modulo $2\pi$ with the
independent uniform $\Theta_0$ is uniform. The conditional angle law is
therefore independent of that family, and hence jointly of $(D_F,Z_F)$.
The modulus is $e^{L_F}=W_F$ by the product construction and
\Cref{thm:fourier-amplitude-law}. Integrating $e^{i(m-n)\Theta_0}$
gives zero for $m\ne n$; for $m=n$ the monomial is $W_F^{2m}$ and
its expectation is $M_F(2m)$. All these moments are finite.
Finally, \eqref{eq:fourier-local-log} and its $L^2$ endpoint at $\ell=3$
give $\mathbb E[\cos\Theta_\ell\,\psi_\ell(\Theta_\ell)]
=\varrho_\ell/2$. Independence and $L^2$ convergence yield
\eqref{eq:joint-fourier-covariance}.
\end{proof}

\begin{corollary}[Fourier amplitudes after finite deletions]
\label{cor:deleted-amplitude}
\index{Fourier amplitudes!finite deletions}
For a fixed finite set $S$ of odd primes, put
\[
 \kappa_{F,S}:=\kappa_{\mathrm{end}}
        \prod_{\ell\in S}\frac{\ell-1}{\ell-2},\qquad
 W_{F,S}:=\kappa_{F,S}
        \exp\!\left(\sum_{\ell\notin S,\ \ell\ge3}
                    \psi_\ell(\Theta_\ell)\right).
\]
Then $(\log T)|\mathcal N_S(U_T)|\Rightarrow W_{F,S}$. Moreover,
\[
 \frac1T\int_0^T|\mathcal N_S(t)|^s\,dt
 \sim\frac{\mathbb EW_{F,S}^s}{(\log T)^s}
\]
for every fixed real $s$ if $3\in S$, and for $s>-1$ otherwise.
If $3\notin S$ and $s\le-1$, then $\mathbb EW_{F,S}^s=\infty$ and
$\int_0^T|\mathcal N_S(t)|^s\,dt=\infty$ for $T\ge\pi/\omega_3$.
For fixed $Q\ge1$, $\gcd(b,Q)=1$ and
$S=\{\ell\ge3:\ell\mid Q\}$, the same assertions hold for the modulus
of the Fourier transform of $\mu_{f,Q,b}=\operatorname{Law}(-\log X_{Q,b})$.
In addition,
\begin{equation}\label{eq:ap-complex-amplitude}
 (\log T)\widehat\mu_{f,Q,b}(U_T)\Rightarrow
 Z_{F,S}:=\kappa_{F,S}e^{i\Theta_0}
       \prod_{\substack{\ell\ge3\\\ell\notin S}}
             (1+\varrho_\ell e^{i\Theta_\ell}),
\end{equation}
where $\Theta_0$ is uniform on $[0,2\pi]$ and independent of the prime
angles. The limit is circularly symmetric, has modulus $W_{F,S}$,
and depends on $(Q,b)$ only through $S$.
All mixed polynomial moments in the complex variable and its conjugate
converge; for nonnegative integers $m,n$,
\[
 \mathbb E Z_{F,S}^m\overline{Z_{F,S}}^{\,n}
 =\begin{cases}
 0,&m\ne n,\\
 \mathbb EW_{F,S}^{2m},&m=n.
 \end{cases}
\]
\end{corollary}

\begin{proof}
Run the proof of \Cref{thm:fourier-amplitude-law} with
$y>\max(S\cup\{5\})$, omitting the coordinates $\ell\in S$ from the
fixed-$y$ head. The tail bounds of \Cref{lem:amplitude-tails} are unchanged, and
\eqref{eq:fourier-normalization} acquires the factor
$\prod_{\ell\in S}(\ell-1)/(\ell-2)$; this gives the modulus limit.
Positive powers are uniformly integrable by
\Cref{thm:uniform-fourier-moments} and evenness. Deleted factors have
modulus at most $1$, so $|\mathcal N_S|\ge|\FT|$ if $3\notin S$ and
$|\mathcal N_S|\ge|\mathcal N_{\{3\}}|$ if $3\in S$. For admissible $s<0$,
uniform integrability follows from \Cref{lem:totient-inverse-moments}\textup{(b)} with
$a\in(|s|,1)$ in the first case, and from \Cref{lem:totient-inverse-moments}\textup{(a)} with
$b>|s|$ in the second. Weak convergence gives the moment asymptotics.
If $3\notin S$, the simple zeros of the factor at $3$ give divergence
for $s\le-1$.

For the progression law, \eqref{eq:XQ} gives
\[
 \widehat\mu_{f,Q,b}(t)=e^{it\eta_{Q,b}}\mathcal N_S(t),\qquad
 \eta_{Q,b}=-\log A_{Q,b}
 =\log2+\sum_{\ell\in S}e_\ell(b)\omega_\ell.
\]
By \Cref{thm:linind}, $\eta_{Q,b}$ is linearly independent over $\Q$
of every finite set of the free jumps $\omega_\ell$, $\ell\notin S$.
The proofs of \Cref{cor:complex-amplitude-law,cor:totient-amplitude-circular-moments},
with these finite omissions, therefore give \eqref{eq:ap-complex-amplitude} and the
independent uniform angle. Uniform positive modulus moments imply
convergence of all mixed polynomial moments. Circular symmetry gives
the displayed formula.
\end{proof}

\begin{corollary}[Sampling densities and complex amplitude moments]
\label{cor:weighted-amplitudes-mas17}
Fix a finite set $S$ of odd primes. Let $A(t)=|\mathcal N_S(t)|$ and
$W=W_{F,S}$ as in \Cref{cor:deleted-amplitude}; the same statements
hold for the modulus in any fixed progression with that deletion set.
Let $w$ be a fixed bounded nonnegative measurable function supported
on $[0,B]$, with $B<\infty$ and $\int_0^B w(u)\,du=1$, and sample
$T_w$ with density $T^{-1}w(t/T)$. Then
\begin{equation}\label{eq:weighted-amplitude-limit-mas17}
 (\log T)A(T_w)\Rightarrow W.
\end{equation}
For complex $z$, define positive powers by $v^z=e^{z\log v}$ using
the real logarithm. For $\operatorname{Re}z>-1$ when $3\notin S$,
and for every $z\in\C$ when $3\in S$,
\begin{equation}\label{eq:weighted-complex-moments-mas17}
 \frac1T\int_0^{BT}w(t/T)\bigl((\log T)A(t)\bigr)^z\,dt
 \longrightarrow\mathbb EW^z
 =\kappa_{F,S}^z\prod_{\substack{\ell\ge3\\\ell\notin S}}
       \mathcal H_z(\varrho_\ell).
\end{equation}
Here $\mathcal H_z$ is the integral in
\eqref{eq:fourier-mellin-local} with exponent $z$ in the indicated
domain. In particular $w=\mathbf1_{[1,2]}$ gives sampling on $[T,2T]$.
For a fixed positive integer $m$, the $m$th convolution power has
modulus $A(t)^m$, normalization $(\log T)^m$, limit $W^m$, and
complex moment domain $\operatorname{Re}(mz)>-1$ if $3\notin S$.
\end{corollary}

\begin{proof}
For fixed $0<a<b$, subtract the averages on $[0,bT]$ and $[0,aT]$
in \Cref{cor:deleted-amplitude}. Since
$\log(cT)/\log T\to1$ for each fixed $c>0$, the normalized amplitude
has the same weak limit on $[aT,bT]$; intervals starting at zero use
only the first average. Finite step densities follow by linearity.
Approximate $w$ in $L^1[0,B]$ by nonnegative step functions on intervals.
For bounded test functions the error is at most their supremum norm
times this $L^1$ error, proving \eqref{eq:weighted-amplitude-limit-mas17}.

Choose $\delta>0$ so that $\operatorname{Re}z\pm\delta$ remain in
the asserted real-moment domain. The moment bounds of
\Cref{cor:deleted-amplitude} on $[0,BT]$, multiplied by $\|w\|_\infty$,
make the powers in \eqref{eq:weighted-complex-moments-mas17} uniformly
integrable, both at zero and at infinity. The weak limit therefore
proves their convergence. Independence gives the product for finite
prime heads. For the tail over primes $\ell\ge5$, write
$P_y=\prod_{5\le\ell\le y,\,\ell\notin S}
 |1+\varrho_\ell e^{i\Theta_\ell}|$.
The centered logarithms converge almost surely. Independence and
$\mathcal H_a(\varrho_\ell)=1+O_a(\varrho_\ell^2)$ give
$\sup_y\mathbb E P_y^a<\infty$ for every fixed real $a$.
For any fixed $\eta>0$, the bound at
$a=(1+\eta)\operatorname{Re}z$ makes $P_y^z$ uniformly integrable.
Its expectations therefore converge to that of its almost-sure limit.
Combining this with the independent finite head gives the stated
infinite product.
Its local factors are $1+O_z(\varrho_\ell^2)$ for
$\ell\ge5$, uniformly on compact subsets of the moment domain.
The identity $\widehat{\nu^{*m}}=(\widehat\nu)^m$ proves the last
assertion and the changed moment domain.
\end{proof}

The same sampling argument applies to the complex limits in
\Cref{cor:complex-amplitude-law,cor:deleted-amplitude,cor:integer-totient-amplitude,thm:sigma-amplitude}:
replace $U_T$ by $T_w$ and keep the normalization $\log T$ and the
same limiting variable. Mixed polynomial moments still converge,
since the available positive absolute moments give uniform
integrability. For the integer totient modulus the complex moment
domain is $\operatorname{Re}z>-1$. For divisor-sum moduli,
\Cref{cor:sigma-temporal-moments-mas22} supplies the temporal bounds
needed for the same argument: the domain is $\operatorname{Re}z>-1/2$
for the odd shifted and integer laws, and $\operatorname{Re}z>-1/4$
for the full shifted law. For their fixed $m$th convolution powers,
replace $z$ by $mz$ in these sufficient domains and use normalization
$(\log T)^m$.

A density that varies with $T$ can select exceptional phases.
Analytic continuation of the gamma or
hypergeometric expressions beyond the real-part moment domain does
not extend the moment-convergence statement.

\begin{corollary}[Integer totient amplitudes]\label{cor:integer-totient-amplitude}
\index{Fourier amplitudes!integer totient law}
Let $\nu=\mu_F^{\mathrm{int},\log}$, and let
$(\Theta_\ell)_{\ell\ge2}$ be independent uniform angles on $[0,2\pi]$.
The product
\[
 Z_{\mathrm{int}}:=e^{-\gamma}\prod_{\ell\ge2}
       \left(1+\frac{e^{i\Theta_\ell}}{\ell-1}\right)
\]
converges almost surely and in $L^2$, and is nonzero almost surely.
As $T\to\infty$,
\begin{equation}\label{eq:integer-totient-amplitude}
 (\log T)\widehat\nu(U_T)\Rightarrow Z_{\mathrm{int}}.
\end{equation}
For every fixed real $s>-1$,
\begin{equation}\label{eq:integer-totient-amplitude-moments}
 \begin{aligned}
 \frac1T\int_0^T|\widehat\nu(t)|^s\,dt
 &\sim\frac{M_{\mathrm{int}}(s)}{(\log T)^s},\\
 M_{\mathrm{int}}(s)&:=e^{-\gamma s}
       \prod_{\ell\ge2}\mathcal H_s((\ell-1)^{-1})
       =\mathbb E|Z_{\mathrm{int}}|^s.
 \end{aligned}
\end{equation}
All mixed polynomial moments in the complex variable and its conjugate
converge in \eqref{eq:integer-totient-amplitude}. Also
\[
 \log|\widehat\nu(U_T)|+\log\log T
 \Rightarrow\log|Z_{\mathrm{int}}|,
\]
with convergence of every fixed nonnegative integer moment.
For $s\le-1$ and $T\ge\pi/\log2$,
$\int_0^T|\widehat\nu(t)|^s\,dt=\infty$.
Finally, $\mathbb EZ_{\mathrm{int}}^m=e^{-\gamma m}$ for every integer
$m\ge1$, so the complex limit is not circularly symmetric.
\end{corollary}

\begin{proof}
Put $\omega_2=\log2$. By \eqref{eq:integer-totient-factorization},
\[
 \widehat\nu(t)=\prod_{\ell\ge2}a_\ell(t),\qquad
 a_\ell(t):=1-\frac1\ell+\frac{e^{it\omega_\ell}}\ell
 =(1-1/\ell)\left(1+\frac{e^{it\omega_\ell}}{\ell-1}\right).
\]
For $0\le t\le T$, the factors with $\ell>T$ have product $1+O(1/\log T)$ by
\Cref{lem:amplitude-tails}\textup{(a)}: $a_\ell$ is the characteristic function of a variable
equal to $\omega_\ell$ with probability $1/\ell$ and to $0$ otherwise, whose mean is at most
$2\ell^{-2}$. \Cref{thm:mertens} gives
$\prod_{\ell\le T}(1-1/\ell)\sim e^{-\gamma}/\log T$.
For $\ell\ge3$, the analytic logarithm of the second local factor is
$(\ell-1)^{-1}e^{it\omega_\ell}+O(\ell^{-2})$ uniformly in $t$.
\Cref{lem:amplitude-tails}\textup{(b)} applies with $v=v_\varphi$ and
$b_\ell=(\ell-1)^{-1}$.
Thus the finite-head approximation in the proof of
\Cref{cor:complex-amplitude-law} applies, without its deterministic
rotation. By \Cref{thm:linind}, the phases including $t\omega_2$
equidistribute jointly. This proves \eqref{eq:integer-totient-amplitude}.
The random finite products are mean-one martingales before multiplication
by $e^{-\gamma}$, with second moments bounded by
$\prod_{\ell\ge2}(1+(\ell-1)^{-2})<\infty$.
Their logarithms over $\ell\ge3$ have independent centered terms
and summable variances; the factor at $2$ is nonzero almost surely.
This proves the asserted product convergence and nonvanishing.

For the moment bounds put $A_{\mathrm{int}}(t):=\prod_{\ell\ge3}|a_\ell(t)|$.
The factor at $3$ lies between $1/3$ and $1$. For $\ell\ge5$, the
integer and shifted-prime local factors are bounded away from zero,
and their difference is $O(\ell^{-2})$ uniformly in $t$. Hence
\[
 A_{\mathrm{int}}(t)\asymp|\mathcal N_{\{3\}}(t)|
 \qquad(t\in\R),
\]
with absolute constants. \Cref{thm:uniform-fourier-moments} and
\eqref{eq:fourier-deleted-inverse} therefore bound every fixed positive
and negative moment of $(\log T)A_{\mathrm{int}}(U_T)$ uniformly for
$T\ge3$. Since
\[
 |\widehat\nu(t)|=|\cos(t\log2/2)|A_{\mathrm{int}}(t),
\]
the H\"older argument in the proof of \Cref{lem:totient-inverse-moments}\textup{(b)},
with $\log2$ in place of $\omega_3$ and $(\log T)A_{\mathrm{int}}$ in place of $Y_T$,
gives uniform inverse moments of
every order less than one. Uniform integrability proves all the
asserted moment convergences, including logarithmic moments; independence
and \eqref{eq:fourier-mellin-local} identify
$M_{\mathrm{int}}(s)$. At $t_0=\pi/\log2$ the cosine factor has a
simple zero, so its powers of order $s\le-1$ are not locally integrable.
Every positive integer power of a finite product has constant term one.
The uniform positive moment bounds and almost-sure product convergence
therefore give $\mathbb EZ_{\mathrm{int}}^m=e^{-\gamma m}$ for $m\ge1$.
\end{proof}

\section{Divisor-sum amplitudes}

For divisor sums, the higher-valuation phases satisfy the
multiplicative relations encoded by $\chi(ab)=\chi(a)\chi(b)$.
Independent uniform values on a multiplicative basis determine all
rational values; independence of the arithmetic prime valuations does
not imply independence of these phases. The Haar limits have finite
modulus moments of every real order, whereas temporal convergence of
negative moments is proved only in the ranges of
\Cref{cor:sigma-temporal-moments-mas22}. Positive absolute and mixed
polynomial moments converge. The totient amplitudes admit the same
character representation (\Cref{rem:totient-anchor-coefficients}).

\begin{lemma}[Inverse moments of an exceptional Haar factor]
\label{lem:exceptional-haar-inverse}
Let $\chi$ be one Haar character of $\Q_{>0}$ and put
\[
 F(\chi)=1+\sum_{k\ge1}w_k\chi(q_k),\qquad
 w_k>0,\quad \sum_{k\ge1}w_k=1,\quad q_k\in\Q_{>0}.
\]
Suppose the family $(q_k)$ contains arbitrarily large
multiplicatively independent subfamilies. Then $F\ne0$ almost surely
and $\mathbb E|F|^{-t}<\infty$ for every $t>0$.
\end{lemma}
\begin{proof}
Choose $N$ multiplicatively independent members. Their Haar values
are independent uniform phases. If $|F|<\varepsilon$, then
\[
 0\le\sum_k w_k(1+\operatorname{Re}\chi(q_k))
       =\operatorname{Re}F<\varepsilon.
\]
For a uniform angle $\Theta$,
$\mathbb P(1+\cos\Theta<u)\ll\sqrt u$. Applying this bound to
the selected phases gives
$\mathbb P(|F|<\varepsilon)\ll_N\varepsilon^{N/2}$, with a
constant depending on their weights. Thus $F\ne0$ almost surely.
For $N>2t$,
\[
 \mathbb E|F|^{-t}
 \le1+t\int_0^1\varepsilon^{-t-1}
                   \mathbb P(|F|<\varepsilon)\,d\varepsilon<\infty.
\]
\end{proof}

\begin{theorem}[Divisor-sum amplitudes and modulus moments]
\label{thm:sigma-amplitude}
\index{Fourier amplitudes!divisor-sum law}
Let $\chi$ be one Haar character of the discrete multiplicative group
$\Q_{>0}$; equivalently, its values on the primes are independent
uniform points of the unit circle. Put
\[
 R_{p,k}:=1+p^{-1}+\cdots+p^{-k},\qquad r_p:=R_{p,1},
\]
\[
 \begin{aligned}
 A_p(\chi)&:=1+\frac{p-1}{p-2}\sum_{k\ge1}p^{-k}\chi(R_{p,k})
       &&(p\ge3),\\
 B_2(\chi)&:=\sum_{k\ge1}2^{-k}\chi(R_{2,k}),&\qquad
 J_p(\chi)&:=1+\sum_{k\ge1}p^{-k}\chi(R_{p,k}).
 \end{aligned}
\]
The following products converge almost surely and in $L^2$, are nonzero
almost surely, and have finite modulus moments of every real order:
\begin{equation}\label{eq:sigma-haar-products}
 \begin{aligned}
 Z_{\sigma,\mathrm{odd}}&:=\kappa_{\mathrm{end}}\prod_{p\ge3}A_p(\chi),\\
 Z_{\sigma,\mathrm{sp}}&:=\kappa_{\mathrm{end}}B_2(\chi)
                                      \prod_{p\ge3}A_p(\chi),\\
 Z_{\sigma,\mathrm{int}}&:=e^{-\gamma}\prod_{p\ge2}J_p(\chi).
 \end{aligned}
\end{equation}
For $U_T$ uniform on $[0,T]$, with the convention
$\widehat\nu(t)=\int e^{itx}\,d\nu(x)$,
\[
 (\log T)\widehat\nu(U_T)\Rightarrow Z_\sigma,
\]
where $(\nu,Z_\sigma)$ is respectively
$(\mugodd,Z_{\sigma,\mathrm{odd}})$,
$(\mug,Z_{\sigma,\mathrm{sp}})$, or
$(\mu_\sigma^{\mathrm{int},\log},Z_{\sigma,\mathrm{int}})$.
For every fixed $s>0$,
\begin{equation}\label{eq:sigma-positive-moments}
 \frac{(\log T)^s}{T}\int_0^T|\widehat\nu(t)|^s\,dt
 \longrightarrow\mathbb E|Z_\sigma|^s\in(0,\infty).
\end{equation}
All mixed polynomial moments in $(\log T)\widehat\nu(U_T)$ and its
conjugate converge to those of $Z_\sigma$. For every integer $m\ge1$,
\[
 \mathbb EZ_{\sigma,\mathrm{odd}}^m=\kappa_{\mathrm{end}}^m,\qquad
 \mathbb EZ_{\sigma,\mathrm{sp}}^m=0,\qquad
 \mathbb EZ_{\sigma,\mathrm{int}}^m=e^{-\gamma m}.
\]
Thus the odd-prime and integer limits are not circularly symmetric.
\end{theorem}

\begin{proof}
The family $\{2\}\cup\{r_p:p\ge3\}$ is a multiplicative basis of
$\Q_{>0}$. In a finite relation, the largest odd prime $p$ occurs
only in the denominator of its own $r_p$: every prime factor of
$q+1$ for odd $q\le p$ is at most $(q+1)/2<p$.
Its exponent is therefore zero; descent proves independence.
Induction from $p=(p+1)/r_p$ proves generation, since every prime
factor of $p+1$ is smaller than $p$.
Thus $\chi(r_p)$ are independent uniform phases. This does not extend
to all valuations; for example,
\[
 R_{3,2}=\frac{13}{9}=\frac{r_3^2}{r_7r_{13}}.
\]
Uniformly over all characters,
\[
 A_p(\chi)=1+c_p\chi(r_p)+O(p^{-2}),\qquad
 c_p=\frac{p-1}{p(p-2)}=p^{-1}+O(p^{-2}),
\]
and $J_p(\chi)=1+p^{-1}\chi(r_p)+O(p^{-2})$ for odd $p$.
For $p\ge5$ the logarithms are consequently
$c_p\chi(r_p)+O(p^{-2})$ and $p^{-1}\chi(r_p)+O(p^{-2})$,
respectively. Their centered leading series have summable variances
and converge almost surely; the remainders converge absolutely.
For every fixed real $a$,
$\mathbb E\exp\{a\operatorname{Re}(c_p\chi(r_p))\}=1+O_a(p^{-2})$.
Independence and the bounded logarithmic remainder therefore bound
every real modulus moment of the nonexceptional finite products
uniformly; Fatou's lemma gives the corresponding limit moments.
For $p\ge5$, $|A_p-1|\le1/(p-2)<1$; for $p\ge3$,
$|J_p-1|\le1/(p-1)<1$.
The exceptional factors $A_3,J_2$ and
$2\chi(R_{2,1})^{-1}B_2$ have the form
$F=1+\sum_k w_k\chi(q_k)$, with $w_k>0$ and $\sum_k w_k=1$;
their rational families are, respectively,
$R_{3,k}$, $R_{2,k}$ ($k\ge1$) and
$R_{2,k}/R_{2,1}$ ($k\ge2$), with weights
$2\cdot3^{-k}$, $2^{-k}$ and $2^{1-k}$.
Each family contains arbitrarily many multiplicatively independent
members. With $p$ its base prime, recursively choose a prime $q$ exceeding $p,p-1,3$
and every prime in the supports of all previously selected rationals,
and set $k=q-2$. Then $q\mid p^{q-1}-1$ by Fermat's theorem, while
$q$ divides neither $p^{q-2}(p-1)$ nor the extra denominator factor
$3$ in the last family. Thus its valuation is nonzero in the new
rational and zero in all preceding ones. The last member of any
finite relation has exponent zero, proving independence by descent.
The corresponding Haar values are independent uniform phases.
\Cref{lem:exceptional-haar-inverse} therefore gives every inverse
modulus moment of all three exceptional factors.
H\"older's inequality combines them with the nonexceptional products,
proving the moment and nonvanishing
assertions. In particular the finite prime products, including their
exceptional factors, have uniformly bounded fourth moments. Their
almost-sure limits have finite fourth moments by Fatou's lemma.
The squared differences from the limits are therefore uniformly
integrable, and almost-sure convergence implies convergence in $L^2$.

The substitutions in \Cref{rem:rational-amplitude-specializations}
and \Cref{thm:rational-haar-amplitude}\textup{(b)} give the weak
limits, \eqref{eq:sigma-positive-moments}, and all mixed polynomial
moment limits.

In a finite prime head, every nonconstant
monomial of a positive integer power of the $A_p$ or $J_p$ factors is
$\chi(q)$ with $q>1$, whose Haar mean is zero. Its constant term is one;
a head containing $B_2$ has no constant term. The valuation series
converge uniformly for each fixed prime, and the finite heads have
uniformly bounded positive moments. Removing both truncations gives
the stated holomorphic moments. The nonzero first moments exclude
circular symmetry in the odd-prime and integer cases.
\end{proof}
\begin{corollary}[Temporal divisor-sum inverse and logarithmic moments]
\label{cor:sigma-temporal-moments-mas22}
For each of the three pairs $(\nu,Z_\sigma)$ in
\Cref{thm:sigma-amplitude}, set
\[
 a_\nu=\begin{cases}
  1/2,&\nu=\mugodd\text{ or }\mu_\sigma^{\mathrm{int},\log},\\
  1/4,&\nu=\mug,
 \end{cases}
 \qquad W_T=(\log T)|\widehat\nu(U_T)|.
\]
For every fixed real $s>-a_\nu$,
\begin{equation}\label{eq:sigma-temporal-moments-mas22}
 \mathbb E W_T^s\longrightarrow\mathbb E|Z_\sigma|^s.
\end{equation}
Moreover $\log W_T\Rightarrow\log|Z_\sigma|$, with convergence
of all fixed polynomial moments. In particular,
\begin{equation}\label{eq:sigma-log-means-mas22}
 \frac1T\int_0^T\log|\widehat\nu(t)|\,dt
 =-\log\log T+
 \begin{cases}
  \log\kappa_{\mathrm{end}},&\nu=\mugodd,\\
  -\gamma,&\nu=\mu_\sigma^{\mathrm{int},\log},\\
  \log(\kappa_{\mathrm{end}}/2),&\nu=\mug,
 \end{cases}
 +o(1).
\end{equation}
\end{corollary}

\begin{proof}
Write $A_p(t),B_2(t),J_p(t)$ for the factors in
\Cref{thm:sigma-amplitude} evaluated at the character $q\mapsto q^{it}$.
For the odd, integer and full shifted laws, respectively, factor
\[
 (\log T)|\widehat\nu(t)|=E_\nu(t)V_T(t),\qquad
 E_\nu(t)=|A_3(t)|,\quad |J_2(t)|,\quad |A_3(t)B_2(t)|,
 \qquad V_T(t)>0.
\]
Here and below it suffices to take $T\ge5$ and $0\le t\le T$.
The inactive products through $T$, namely
$\prod_{3\le p\le T}(p-2)/(p-1)$ for the odd and full shifted laws and
$\prod_{p\le T}(1-1/p)$ for the integer law, multiplied by $\log T$,
are bounded above and below by positive constants. Every remaining
normalized local
factor at a fixed prime is bounded away from zero: indeed
$|A_p-1|\le1/(p-2)<1$ for $p\ge5$ and
$|J_p-1|\le1/(p-1)<1$ for $p\ge3$.
Their logarithmic moduli, above a fixed cutoff $y$, equal
$\cos(tv_\sigma(p))/p+O(p^{-2})$ uniformly in $t$.
The original factors above $T$ have product $1+O(1/\log T)$,
by \Cref{lem:amplitude-tails}\textup{(a)}.
Consequently, for each fixed $b\in\R$, choose $y$ sufficiently
large in terms of $|b|$ in \eqref{eq:sigma-exponential} and at least $T_0(4)$ from
\Cref{lem:amplitude-tails}\textup{(a)}. The bounded
finite head and the summable $O(p^{-2})$ remainders then give
\begin{equation}\label{eq:sigma-regular-inverse-mas22}
 \sup_{T\ge5}\frac1T\int_0^T V_T(t)^b\,dt<\infty.
\end{equation}
For the compact range $5\le T\le y$, the same assertion follows
from continuity and nonvanishing of these nonexceptional factors,
and uniform convergence of their tails.

The exceptional factors require a separate estimate. If
$F(t)=1+\sum_j w_j e^{it\lambda_j}$ with $w_j>0$ and
$\sum_jw_j=1$, then
\[
 |F(t)|\ge\operatorname{Re}F(t)
 =\sum_jw_j(1+\cos(t\lambda_j))
 \ge2w_{j_0}\cos^2(t\lambda_{j_0}/2).
\]
Using the first term of each exceptional series gives explicitly
\begin{equation}\label{eq:sigma-exceptional-lower-mas22}
 \begin{aligned}
 |A_3(t)|&\ge\tfrac43\cos^2\bigl(t\log(4/3)/2\bigr),\\
 |J_2(t)|&\ge\cos^2\bigl(t\log(3/2)/2\bigr),\\
 |B_2(t)|&\ge\tfrac12\cos^2\bigl(t\log(7/6)/2\bigr).
 \end{aligned}
\end{equation}
For the last line apply the preceding observation to
$2e^{-it\log(3/2)}B_2(t)$; its remaining first weight is $1/2$.
For every fixed nonzero $c$ and $0<b<1/2$, periodicity and
integrability at the simple zeros of the cosine imply
\[
 \sup_{T\ge5}\frac1T\int_0^T|\cos(ct)|^{-2b}\,dt<\infty.
\]
If $0<a<1/2$, choose $q>1$ with $aq<1/2$ and use
H\"older with \eqref{eq:sigma-regular-inverse-mas22} and the
single exceptional factor. This bounds $\mathbb E W_T^{-a}$
uniformly for the odd and integer laws. For the full shifted law,
if $0<a<1/4$, choose $q>2$ with $aq<1/2$; H\"older exponents
$q,q,(1-2/q)^{-1}$ allocate one factor to each cosine and the
last to $V_T^{-a}$. Thus
\[
 \sup_{T\ge5}\mathbb E W_T^{-a}<\infty\qquad(0<a<a_\nu).
\]
A slightly larger inverse order proves uniform integrability for
each negative power asserted in
\eqref{eq:sigma-temporal-moments-mas22}; positive powers use
\eqref{eq:sigma-moments}. The weak limit, whose modulus is positive
almost surely, proves the moment limits. For sufficiently small
$\delta>0$, the positive and negative bounds also bound
$\mathbb E\exp(\delta|\log W_T|)$ uniformly. They imply uniform
integrability of every fixed power of $\log W_T$, proving the
logarithmic assertion.

It remains to identify the means of the limiting logarithms.
For a finite normalized factor $F=1+G$ occurring here, or its
uniformly convergent valuation series, $|G|\le1$, and every
nonconstant monomial of $G$ is a positive multiple of $\chi(q)$
with $q>1$. For $0<r<1$, the absolutely convergent logarithm series
therefore gives $\mathbb E\log|1+rG|=0$.
The identity
\[
 |1+rG|^2-r|1+G|^2=(1-r)(1-r|G|^2)\ge0
\]
and the inverse Haar moments proved in
\Cref{thm:sigma-amplitude} justify letting $r\uparrow1$ in $L^1$.
This applies to $A_3,J_2$ and to
$2\chi(3/2)^{-1}B_2$, whose other rational arguments
$R_{2,k}/(3/2)$ are all greater than one. It applies also to every
nonexceptional factor, where $|G|<1$. Hence each normalized
factor has mean logarithmic modulus zero, while
$\mathbb E\log|B_2|=-\log2$.
To pass from finite products to their limits, the uniform positive
and negative Haar moments from the same theorem make their
logarithmic moduli uniformly integrable. Their almost-sure
convergence thus preserves the zero means. The deterministic
multipliers in \eqref{eq:sigma-haar-products} now give
\eqref{eq:sigma-log-means-mas22}.
\end{proof}

\begin{corollary}[Divisor-sum amplitudes at integer dilations]
\label{cor:sigma-joint-dilations-mas22}
Fix one of the three divisor-sum logarithmic laws and fixed nonzero
integers $k_1,\ldots,k_d$. With one common Haar character $\chi$,
\begin{equation}\label{eq:sigma-joint-dilations-mas22}
 \bigl((\log T)\widehat\nu(k_jU_T)\bigr)_{j=1}^d
 \Rightarrow\bigl(Z_\sigma(\chi^{k_j})\bigr)_{j=1}^d,
 \qquad \chi^k(q):=\chi(q)^k.
\end{equation}
All mixed polynomial moments in these coordinates and their
conjugates converge. For fixed real exponents $s_1,\ldots,s_d$,
\[
 \mathbb E\prod_{j=1}^d
       |(\log T)\widehat\nu(k_jU_T)|^{s_j}
 \longrightarrow
 \mathbb E\prod_{j=1}^d|Z_\sigma(\chi^{k_j})|^{s_j}
\]
whenever
\begin{equation}\label{eq:sigma-joint-inverse-budget-mas22}
 \sum_{j=1}^d(-s_j)_+<a_\nu.
\end{equation}
In particular all positive mixed modulus moments converge.
\end{corollary}

The limiting coordinates $Z_\sigma(\chi^{k_j})$ are formed from the
one character $\chi$ and are in general dependent. For instance,
$k_2=-k_1$ makes the second limiting coordinate the complex conjugate
of the first. A zero dilation is excluded because its normalized
coordinate is $(\log T)\widehat\nu(0)=\log T$.

\begin{proof}
\Cref{thm:rational-haar-amplitude}\textup{(c)}, with the divisor-sum
substitutions in \Cref{rem:rational-amplitude-specializations}, proves
\eqref{eq:sigma-joint-dilations-mas22} and the mixed polynomial and
positive modulus moment limits.

Each nonzero power $\chi\mapsto\chi^{k_j}$ preserves Haar measure,
and the marginal temporal positive and inverse bounds follow from
\Cref{cor:sigma-temporal-moments-mas22} by the change of interval
$[0,T]\mapsto[0,|k_j|T]$. Under the strict inequality
\eqref{eq:sigma-joint-inverse-budget-mas22}, choose H\"older
exponents so that each negative coordinate uses an inverse order
strictly below $a_\nu$, leaving a positive amount of the H\"older
budget for all positive coordinates. These have moments of every
fixed order. The strict margins permit the same choice after all
$s_j$ are multiplied by some $1+\eta>1$. This bounds a moment
strictly greater than one of the product, giving uniform
integrability. The joint weak limit proves the asserted convergence.
\end{proof}

\section{Exact collision constants and concentration saturation}
\label{sec:collision-saturation-mas04}

The second moments of the amplitude laws determine exact collision
constants, including for the raw ratios. We use the Fourier average
over $[0,T]$ and the collision event $|X-X'|\le h$.
The lower part of \Cref{fig:fourier-questions} summarizes the route from
second Fourier means to collisions and saturation.

\begin{lemma}[A constant-preserving collision equivalence]
\label{lem:collision-tauberian-mas04}
Let $Y,Y'$ be independent with common law $\nu$, put $D=Y-Y'$,
$A(T)=\int_0^T|\widehat\nu(t)|^2\,dt$, and
$k(v)=(\sin(v/2)/(v/2))^2$, with $k(0)=1$. Then
\begin{equation}\label{eq:collision-kernel-exact-mas04}
 \mathbb E k(TD)=\frac2{T^2}\int_0^T A(u)\,du.
\end{equation}
For each $C\in(0,\infty)$, the following are equivalent:
\begin{equation}\label{eq:collision-equivalence-mas04}
 A(T)\sim\frac{CT}{\log^2T},\qquad
 \mathbb P(|D|\le h)\sim\frac C{\log^2(1/h)}.
\end{equation}
\end{lemma}
\begin{proof}
The triangular Fourier identity \eqref{eq:triangular-kernel}, with
$u=1/T$, gives
$\mathbb E k(TD)=2T^{-1}\int_0^T(1-t/T)|\widehat\nu(t)|^2\,dt$.
Fubini gives \eqref{eq:collision-kernel-exact-mas04}.
Write $P(h)=\mathbb P(|D|\le h)$ and $K(h)=\mathbb E k(D/h)$.
For fixed $0<\varepsilon<1$, the estimates
$k(v)=1+O(v^2)$ near zero and
$0\le k(v)\le\min(1,4/v^2)$ give
\begin{equation}\label{eq:collision-power-sandwich-mas04}
 (1-O(h^{2\varepsilon}))P(h^{1+\varepsilon})
 \le K(h)\le P(h^{1-\varepsilon})+4h^{2\varepsilon}.
\end{equation}
If $P(h)\sim C/\log^2(1/h)$, this bounds the lower and upper limits
of $\log^2(1/h)K(h)$ by $C/(1+\varepsilon)^2$ and
$C/(1-\varepsilon)^2$. Letting $\varepsilon\downarrow0$ gives
$K(h)\sim C/\log^2(1/h)$. Conversely, if this last asymptotic
holds, substitute $h^{1/(1+\varepsilon)}$ and
$h^{1/(1-\varepsilon)}$ into \eqref{eq:collision-power-sandwich-mas04}.
The resulting bounds for $\log^2(1/h)P(h)$ are
$C(1-\varepsilon)^2$ and $C(1+\varepsilon)^2$, so the same limit
follows. The power errors vanish after multiplication by
$\log^2(1/h)$ in both directions.

Set $B(T)=\int_0^T A(u)\,du$. If $A(T)\sim CT/\log^2T$, integration
gives $B(T)\sim CT^2/(2\log^2T)$: this follows, for example, by
bounding $A(u)$ above and below by $(C\pm\eta)u/\log^2u$ after
a fixed threshold and differentiating $T^2/(2\log^2T)$ to evaluate
the comparison integrals. Thus \eqref{eq:collision-kernel-exact-mas04}
gives the required asymptotic for $K(1/T)$.
In the other direction that kernel asymptotic gives
$B(T)\sim CT^2/(2\log^2T)$. Since $A$ is nondecreasing, for
$\lambda>1$,
\[
 \frac{B(T)-B(T/\lambda)}{T-T/\lambda}
 \le A(T)\le
 \frac{B(\lambda T)-B(T)}{(\lambda-1)T}.
\]
After normalization by $T/\log^2T$, the limiting lower and upper
bounds are $C(1+\lambda^{-1})/2$ and $C(1+\lambda)/2$.
Letting $\lambda\downarrow1$ proves the converse with exactly the
same constant $C$.
\end{proof}

\KeepHeadingSpace{5\baselineskip}
\begin{lemma}[Raw and logarithmic collisions]
\label{lem:raw-log-collision-mas04}
Let $X>0$ almost surely and suppose
$\mathbb E(X^r+X^{-r})<\infty$ for some $r>0$.
For independent copies $X,X'$ and $C\in(0,\infty)$,
\[
 \mathbb P(|X-X'|\le h)\sim\frac C{\log^2(1/h)}
 \quad\Longleftrightarrow\quad
 \mathbb P(|\log X-\log X'|\le h)
       \sim\frac C{\log^2(1/h)}.
\]
\end{lemma}
\begin{proof}
Fix $0<\varepsilon<1$ and restrict both copies to
$[h^\varepsilon,h^{-\varepsilon}]$. Markov's inequality bounds the
excluded probability by $O(h^{\varepsilon r})$. On this event the
mean-value theorem gives
\[
 h^\varepsilon|\log X-\log X'|
 \le |X-X'|\le h^{-\varepsilon}|\log X-\log X'|.
\]
Writing $P_X,P_{\log X}$ for the two collision probabilities, we get
\[
 P_{\log X}(h^{1+\varepsilon})-O(h^{\varepsilon r})
 \le P_X(h)\le
 P_{\log X}(h^{1-\varepsilon})+O(h^{\varepsilon r}).
\]
The displayed mean-value bounds also give the same sandwich with
$P_X$ and $P_{\log X}$ interchanged. If either collision asymptotic
holds, normalization by $\log^2(1/h)$ bounds the other between
$C/(1+\varepsilon)^2$ and $C/(1-\varepsilon)^2$.
Let $\varepsilon\downarrow0$.
\end{proof}

\begin{theorem}[Exact second Fourier means for the four ratio laws]
\label{thm:exact-collision-mas04}
In each row of the following table, $\mu$ is a raw law, $\nu$ is its
indicated logarithmic law, and $C$ is the common constant:
\[
\begin{array}{ccl}
 \mu&\nu&C\\[2pt]
 \mu_F^{\mathrm{int}}&\mu_F^{\mathrm{int},\log}
 &M_{\mathrm{int}}(2)=e^{-2\gamma}
       \displaystyle\prod_{p\ge2}(1+(p-1)^{-2})\\[3pt]
 \mu_G&\mu_f&M_F(2)=\kappa_{\mathrm{end}}^2
       \displaystyle\prod_{p\ge3}(1+(p-2)^{-2})\\[3pt]
 \mu_\sigma^{\mathrm{int}}&\mu_\sigma^{\mathrm{int},\log}
       &\mathbb E|Z_{\sigma,\mathrm{int}}|^2\\[2pt]
 \muH&\mug&\mathbb E|Z_{\sigma,\mathrm{sp}}|^2.
\end{array}
\]
For independent copies $X,X'\sim\mu$ and $Y,Y'\sim\nu$,
\begin{equation}\label{eq:four-exact-collisions-mas04}
 \mathbb P(|X-X'|\le h)\sim
 \mathbb P(|Y-Y'|\le h)\sim\frac C{\log^2(1/h)},
\end{equation}
and
\begin{equation}\label{eq:four-exact-second-means-mas04}
 \frac1T\int_0^T|\widehat\mu(t)|^2\,dt
 \sim\frac1T\int_0^T|\widehat\nu(t)|^2\,dt
 \sim\frac C{\log^2T}.
\end{equation}
The same conclusions hold for $\nu=\mugodd$ and
$\mu=\exp_*\mugodd$, with $C=\mathbb E|Z_{\sigma,\mathrm{odd}}|^2$.
\end{theorem}

\begin{proof}
The logarithmic second-mean asymptotics are
\Cref{thm:fourier-amplitude-law,cor:integer-totient-amplitude,thm:sigma-amplitude}
at moment order two. In the totient cases reflection changes neither
collisions nor the Fourier modulus; this accounts for the convention
$-\log$ in the table. Apply \Cref{lem:collision-tauberian-mas04},
then \Cref{lem:raw-log-collision-mas04}, and then
\Cref{lem:collision-tauberian-mas04} once more to the raw law.
The moment hypothesis is \Cref{lem:four-ratio-moments}; deleting the
prime-$2$ coordinate gives the same bound for the odd divisor-sum law.
The identity $\mathcal H_2(v)=1+v^2$ gives the two explicit products.
\end{proof}

\begin{remark}[Totient anchor coefficients]
\label{rem:totient-anchor-coefficients}
Let $\mathcal S$ be the family of all finite sets of odd primes,
including $\varnothing$; below $p$ is prime.
With the independent uniform angles of
\Cref{def:totient-amplitude-variables}, put
$\chi_S=e^{i(\Theta_0+\sum_{p\in S}\Theta_p)}$.
For every real $y\ge3$, \Cref{def:anchor-weights} and finite-product
expansion give
\[
 \kappa_{\mathrm{end}}e^{i\Theta_0}
 \prod_{3\le p\le y}\left(1+\frac{e^{i\Theta_p}}{p-2}\right)
 =\sum_{S\subseteq\{p:3\le p\le y\}}\kappa_S\chi_S.
\]
The characters are orthonormal by independence and uniformity.
The $L^2$ convergence in \Cref{cor:complex-amplitude-law},
monotone exhaustion of $\mathcal S$, and
$\sum_{p\ge3}(p-2)^{-2}<\infty$ therefore give, by Parseval,
\begin{equation}\label{eq:totient-anchor-parseval}
\begin{aligned}
 \sum_{S\in\mathcal S}\kappa_S^2
 &=\mathbb E|Z_{F,\mathbb C}|^2\\
 &=\kappa_{\mathrm{end}}^2\prod_{p\ge3}\bigl(1+(p-2)^{-2}\bigr)
 =M_F(2),
\end{aligned}
\end{equation}
where the last equality uses $|Z_{F,\mathbb C}|\overset d=W_F$.
Thus $\kappa_S$ is the coefficient of $\chi_S$ in the normalized limit
$Z_{F,\mathbb C}$ of $(\log T)\FT(U_T)$.
The proof of \Cref{cor:complex-amplitude-law} associates $\chi_S$ with
$e^{it(\log2+\sum_{p\in S}\omega_p)}$; its temporal frequency is
$-\log x_S=\log2+\sum_{p\in S}\omega_p$.

For fixed $S$, \Cref{thm:anchor-sharp} gives the full raw left-window
mass $G(x_S)-G(x_S-h)\sim\kappa_S/\log(1/h)$;
\eqref{eq:anchor-main-exact} concerns only its main stratum, with
$h=x_S\varepsilon$ and $\varepsilon\le1/\max S$ if $S\ne\varnothing$.
The deletion normalization in \Cref{cor:deleted-amplitude} is instead
$\kappa_{F,S}=\kappa_{\mathrm{end}}\prod_{p\in S}(p-1)/(p-2)$.
By \eqref{eq:weight-trichotomy}, the maximal weight
$\kappa_{\mathrm{end}}$ occurs exactly at $S=\varnothing$ and $S=\{3\}$.

Let $\chi$ be a Haar character of $\Q_{>0}$ as in
\Cref{thm:sigma-amplitude}. By \Cref{thm:linind}, $2$ and
$p/(p-1)$, $p\ge3$, are multiplicatively independent. Thus one may take
\[
 e^{i\Theta_0}=\chi(2),\qquad
 e^{i\Theta_p}=\chi\bigl(p/(p-1)\bigr),
\]
so that $\chi_S=\chi(1/x_S)$ and the preceding $L^2$ expansion becomes
\[
 Z_{F,\mathbb C}=\sum_{S\in\mathcal S}\kappa_S\chi(1/x_S).
\]
For the progression product in \Cref{cor:deleted-amplitude}, put
$S=\{p\ge3:p\mid Q\}$ and take
$e^{i\Theta_0}=\chi(A_{Q,b}^{-1})$.
Finite-product expansion gives coefficient $\kappa_{F,S}\mathcal K(J)$
at $\chi(A_{Q,b}^{-1}\prod_{p\in J}p/(p-1))$ for every finite set
$J$ of odd primes outside $S$.
For \Cref{cor:integer-totient-amplitude}, take
$e^{i\Theta_p}=\chi(p/(p-1))$ for all primes $p$; the coefficient at
$\chi(n/\varphi(n))$ is $e^{-\gamma}/\varphi(n)$ for squarefree $n$,
including $n=1$. Both coefficient expansions converge in $L^2$ by the
cited product limits and orthogonality.
The normalization divides each free prime factor by its inactive
probability; a deterministic head contributes its character, and a
conditioned head contributes its full local character sum.
For any of these totient laws or the three divisor-sum laws of
\Cref{thm:sigma-amplitude}, let $P_y$ be its local ratio product through
$y$, including its head and taking reciprocal ratios for totients.
The coefficient at $\chi(\rho)$, $\rho\in\Q_{>0}$, is
$\lim_{y\to\infty}(\log y)\mathbb P(P_y=\rho)$.
All occurring rationals are at least one; averaging powers of finite
products and passing to the limit using their uniform positive moments
therefore gives $\mathbb EZ^m=c_1^m$ for integers $m\ge1$, where $c_1$ is the
coefficient of the trivial character in the corresponding limit $Z$.
\end{remark}

\paragraph*{Rational anchor coefficients}
The next proposition isolates the coefficient argument from any special
choice of first-jump phases. All expectations inside a character sum
integrate the indicated rational random variables, with the character fixed.

\begin{proposition}[Aggregated rational anchor coefficients]
\label{prop:rational-anchor-parseval}
Let $H$ and $(R_p)_{p>p_0}$ be independent positive rational-valued
random variables, with $H$ countably atomic, and put
$a_p=\mathbb P(R_p=1)>0$. Suppose
\[
 A_y:=\prod_{p_0<p\le y}a_p,\qquad
 (\log y)A_y\longrightarrow C_0\in(0,\infty).
\]
On the dual of the discrete group $\Q_{>0}$ with probability Haar
measure, suppose the products
\[
 P_y(\chi):=\frac{C_0}{A_y}\,
       \mathbb E\chi\left(H\prod_{p_0<p\le y}R_p\right)
\]
converge to $Z$ in $L^2$. For $\rho\in\Q_{>0}$ define
\begin{equation}\label{eq:rational-anchor-coefficients}
 w(\rho):=\lim_{y\to\infty}\frac{C_0}{A_y}
       \mathbb P\left(H\prod_{p_0<p\le y}R_p=\rho\right).
\end{equation}
These limits exist monotonically, are finite, and are the Fourier
coefficients of $Z$ at $\chi\mapsto\chi(\rho)$. In particular,
\begin{equation}\label{eq:rational-anchor-parseval}
 Z=\sum_{\rho\in\Q_{>0}}w(\rho)\chi(\rho)\quad\hbox{in }L^2,
 \qquad C:=\mathbb E|Z|^2=\sum_\rho w(\rho)^2\in(0,\infty).
\end{equation}
The limit in \eqref{eq:rational-anchor-coefficients} also equals
$\lim_{y\to\infty}(\log y)\mathbb P(H\prod_{p_0<p\le y}R_p=\rho)$.
All configurations with the same rational product are added before
squaring their coefficient in \eqref{eq:rational-anchor-parseval}.
\end{proposition}

\begin{proof}
For a fixed cutoff, expanding the finitely many independent factors
and the head gives an absolutely and uniformly convergent character
series: the sum of the absolute coefficients is $C_0/A_y<\infty$.
Collecting every term whose rational argument is $\rho$ gives exactly
\[
 c_y(\rho)=\frac{C_0}{A_y}
       \mathbb P\left(H\prod_{p_0<p\le y}R_p=\rho\right).
\]
If the cutoff is enlarged by one prime $q$, the neutral state gives
$c_y(\rho)$ and all other states give nonnegative contributions;
explicitly,
\[
 c_{y'}(\rho)=c_y(\rho)+
       \sum_{r\ne1}\frac{\mathbb P(R_q=r)}{a_q}c_y(\rho/r).
\]
Thus $c_y(\rho)$ increases. The coefficient functional at a fixed
character has norm one on $L^2$, so convergence of $P_y$ in $L^2$
identifies its finite limit with the Fourier coefficient of $Z$.

Unique factorization identifies the character space with the countable
product of circles indexed by the primes.
A rational $\rho=\prod_q q^{n_q}$, with finitely many nonzero integer
$n_q$, gives the corresponding monomial character. Integration in a
coordinate with nonzero exponent proves orthogonality. Finite linear
combinations of these characters contain constants, are closed under
conjugation and multiplication, and separate points of the compact
character space. The Stone--Weierstrass theorem and density of
continuous functions in Haar $L^2$ therefore prove completeness.
Parseval gives both assertions in \eqref{eq:rational-anchor-parseval}.
Choose an atom $\eta$ of $H$ of mass $d>0$. The configuration with
all local factors neutral contributes $C_0d$ to $c_y(\eta)$ for every
$y$, so $C\ge(C_0d)^2>0$. Finally,
$(\log y)\mathbb P(H\prod_{p_0<p\le y}R_p=\rho)
=((\log y)A_y/C_0)c_y(\rho)$, proving the last limit.
\end{proof}

Put $R(m)=\sigma(m)/m$ and, for odd primes, set
$a_p(j)=(p-1)/(p^j(p-2))$ for $j\ge1$, with $a_p(0)=1$.
For $\rho\in\Q_{>0}$ define
\begin{equation}\label{eq:sigma-class-weights-mas04}
\begin{aligned}
 w_{\mathrm{int}}(\rho)
   &=e^{-\gamma}\sum_{\substack{m\ge1\\R(m)=\rho}}\frac1m,\\
 w_{\mathrm{odd}}(\rho)
   &=\kappa_{\mathrm{end}}
       \sum_{\substack{u\ge1\ \mathrm{odd}\\R(u)=\rho}}
          \prod_{p^j\Vert u}a_p(j),\\
 w_{\mathrm{sp}}(\rho)
   &=\kappa_{\mathrm{end}}
       \sum_{\substack{k\ge1,\ u\ge1\ \mathrm{odd}\\R(2^ku)=\rho}}
          2^{-k}\prod_{p^j\Vert u}a_p(j)
     =\frac{\kappa_{\mathrm{end}}}{2}\Omega(\rho).
\end{aligned}
\end{equation}
Empty products equal one. For one of the three corresponding
logarithmic divisor-sum laws $\nu$, write $w$ for its coefficient
function and $Z_\sigma$ for its Haar amplitude. Extend $w$ to
$\R_{>0}$ by zero outside $\Q_{>0}$.

\begin{lemma}[Abundancy coefficients and Parseval's identity]
\label{lem:sigma-abundancy-coefficients-mas04}
The values $w(\rho)$ are the Fourier coefficients of $Z_\sigma$
with respect to the characters $\chi\mapsto\chi(\rho)$. In particular,
\begin{equation}\label{eq:sigma-class-parseval-mas04}
 C_\nu:=\mathbb E|Z_\sigma|^2
       =\sum_{\rho\in\Q_{>0}}w(\rho)^2<\infty.
\end{equation}
\end{lemma}
\begin{proof}
Apply \Cref{prop:rational-anchor-parseval} to the local laws and
heads in \Cref{rem:rational-amplitude-specializations}\textup{(c)}.
For the integer law the normalized probability of valuation $j$
is $p^{-j}$; the prime-$2$ head and $C_0=2e^{-\gamma}$ give the
weight $e^{-\gamma}/m$. For the odd shifted law these probabilities
are $a_p(j)$, with normalization $\kappa_{\mathrm{end}}$.
The full shifted law adds the head weights $2^{-k}$, $k\ge1$.
Grouping by $R(m)=\rho$ gives exactly
\eqref{eq:sigma-class-weights-mas04}. The required $L^2$ convergence
is \Cref{thm:sigma-amplitude}; the proposition therefore gives
\eqref{eq:sigma-class-parseval-mas04}.
\end{proof}

\KeepHeadingSpace{5\baselineskip}
\begin{lemma}[Local logarithmic saturation]
\label{lem:local-log-saturation}
Let $\nu$ be a Borel probability measure on $\R$, and let $A$ be a
countable set of distinct real anchors with weights $w(a)\ge0$.
For $0<h<1$, put $L(h)=\log(1/h)$ and let $W_a(h)$ be nonempty Borel windows
such that
\[
 W_a(h)\subset[a-h,a+h],\qquad \operatorname{diam}W_a(h)\le h.
\]
For independent $X,X'\sim\nu$, write
$P_\nu(h)=\mathbb P(|X-X'|\le h)$. Suppose
$C=\sum_{a\in A}w(a)^2<\infty$ and
\begin{align*}
 \liminf_{h\downarrow0}L(h)\nu(W_a(h))&\ge w(a)\quad(a\in A),\\
 \limsup_{h\downarrow0}L(h)^2P_\nu(h)&\le C.
\end{align*}
Then the following hold, with all suprema over intervals $I$ of length $h$.
\begin{enumerate}[label=\textup{(\roman*)},leftmargin=*]
\item For every finite $F\subset A$, put $R_F=\sum_{a\in A\setminus F}w(a)^2$.
As $h\downarrow0$,
\begin{align*}
 L(h)\sup_{I:\ I\cap\bigcup_{a\in F}W_a(h)=\varnothing}\nu(I)
     &\le\sqrt{R_F}+o_F(1),\\
 L(h)\sup_{I:\ I\cap W_a(h)\ne\varnothing}\nu(I)
     &\le\sqrt{w(a)^2+R_F}+o_F(1)\qquad(a\in F).
\end{align*}
\item For every $a\in A$,
$\lim_{h\downarrow0}L(h)\nu(W_a(h))=w(a)$.
At every fixed $x$ outside the positive-weight anchors,
$\lim_{h\downarrow0}L(h)\nu([x,x+h])=0$.
\item If $C>0$, the weights attain a positive maximum $w_*$, and
\[
 Q_\nu(h)\sim\frac{w_*}{L(h)},\qquad w_*:=\max_{a\in A}w(a).
\]
If the maximum occurs only at $a_*$, every family $I_h$ of intervals
of length $h$ with
$\nu(I_h)/Q_\nu(h)\to1$ meets $W_{a_*}(h)$ for all sufficiently small $h$.
\end{enumerate}
\end{lemma}
\begin{proof}
For finitely many pairwise disjoint Borel sets $J_i$ of diameter at
most $h$, the events $\{X,X'\in J_i\}$ are disjoint and imply
$|X-X'|\le h$. Hence
\begin{equation}\label{eq:local-log-saturation-budget}
 \sum_i\nu(J_i)^2\le P_\nu(h).
\end{equation}
Fix a finite $F\subset A$. Attachment to distinct anchors makes its
windows disjoint for small $h$. An interval of length $h$ can then
meet at most one of them: meeting windows attached to $a,b$ would
give $|a-b|\le3h$.
If $I$ meets none, apply \eqref{eq:local-log-saturation-budget} to
$I$ and all the $F$-windows. If $I$ meets $W_a(h)$, use $I$ and
the windows indexed by $F\setminus\{a\}$. The lower bounds for
their masses and the collision upper bound give, uniformly in $I$,
\[
 L(h)^2\nu(I)^2\le R_F+o_F(1)
 \quad\hbox{or}\quad
 L(h)^2\nu(I)^2\le w(a)^2+R_F+o_F(1),
\]
respectively. Taking square roots gives (i), with additive errors
also when the displayed constants vanish.

For fixed $a\in A$, apply \eqref{eq:local-log-saturation-budget}
to the windows of any finite $F$ containing $a$. It gives
\[
 \limsup_{h\downarrow0}L(h)\nu(W_a(h))
       \le\sqrt{w(a)^2+R_F}.
\]
Exhausting $A$ by finite sets makes $R_F\to0$, and the assumed lower
bound proves the first limit in (ii), including $w(a)=0$.
If $x$ is not a positive-weight anchor, $[x,x+h]$ eventually avoids
the windows of each fixed finite set of positive-weight anchors.
Part (i), followed by exhaustion of those anchors, proves the zero
limit in (ii).

Suppose $C>0$. Square summability makes the set of weights above
any positive threshold finite, so their positive supremum $w_*$
is attained. Every window of diameter at most $h$ is contained in
an interval of length $h$; its lower bound therefore gives
$\liminf_{h\downarrow0}L(h)Q_\nu(h)\ge w_*$.
For finite $F$ containing a maximizing anchor, (i) gives
$\limsup_{h\downarrow0}L(h)Q_\nu(h)\le\sqrt{w_*^2+R_F}$.
Exhaustion proves the concentration asymptotic.
If the maximizing anchor is unique, square summability also gives
$w_{**}:=\sup_{a\ne a_*}w(a)<w_*$, taking this supremum as zero
when there are no other anchors. Choose finite $F$ containing
$a_*$ with $w_{**}^2+R_F<w_*^2$.
An interval avoiding $W_{a_*}(h)$ meets either no $F$-window or one
indexed by a different anchor. Part (i) bounds its normalized mass
by $\sqrt{w_{**}^2+R_F}+o_F(1)<w_*$ for all sufficiently small $h$.
The concentration asymptotic excludes such intervals along any
sequence of relative near-maximizers, proving (iii).
\end{proof}

\begin{corollary}[Signed rational saturation and conditional raw transfer]
\label{cor:rational-log-saturation}
Use the hypotheses and notation of \Cref{prop:rational-anchor-parseval}.
Suppose in addition that $\mathbb E|\log R_p|\ll p^{-2}$ and, for
$R=H\prod_{p>p_0}R_p$ and $\nu=\operatorname{Law}(\log R)$,
\begin{equation}\label{eq:rational-second-mean-input}
 \frac{(\log T)^2}{T}\int_0^T|\widehat\nu(t)|^2\,dt
       \longrightarrow C.
\end{equation}
The product defining $R$ is then positive and finite almost surely.
Extend $w$ by zero off $\Q_{>0}$. For $0<h<1$ set
$L(h)=\log(1/h)$ and
$W_\rho^{\log}(h)=[\log\rho-h/2,\log\rho+h/2]$.
Then
\[
 \mathbb P(|\log R-\log R'|\le h)\sim C/L(h)^2,
 \qquad L(h)\nu(W_\rho^{\log}(h))\longrightarrow w(\rho),
\]
where $R'$ is an independent copy, and the second assertion holds
for every $\rho>0$. For any finite set $\mathcal F$ of positive-weight
rational anchors, put $R_{\mathcal F}=C-\sum_{\rho\in\mathcal F}w(\rho)^2$.
Uniformly over intervals $I$ of length $h$,
\begin{align*}
 L(h)\sup_{I:\ I\cap\bigcup_{\rho\in\mathcal F}W_\rho^{\log}(h)=\varnothing}
       \nu(I)&\le\sqrt{R_{\mathcal F}}+o_{\mathcal F}(1),\\
 L(h)\sup_{I:\ I\cap W_\rho^{\log}(h)\ne\varnothing}\nu(I)
       &\le\sqrt{w(\rho)^2+R_{\mathcal F}}+o_{\mathcal F}(1)
       \quad(\rho\in\mathcal F).
\end{align*}
Consequently $Q_\nu(h)\sim w_*/L(h)$, where
$w_*:=\max_\rho w(\rho)>0$. If its maximizing class $\rho_*$ is unique,
every family of relative near-maximizers of length $h$ meets
$W_{\rho_*}^{\log}(h)$ for all sufficiently small $h$.

If also $\mathbb E(R^r+R^{-r})<\infty$ for some $r>0$, all these
conclusions hold for $\mu=\operatorname{Law}(R)$ with windows
$W_\rho^{\rm raw}(h)=[\rho-h/2,\rho+h/2]$ and the same weights and
constant. In addition,
\begin{equation}\label{eq:rational-raw-transfer}
 \mathbb P(|R-R'|\le h)\sim\frac C{L(h)^2},\qquad
 \frac1T\int_0^T|\widehat\mu(t)|^2\,dt\sim\frac C{\log^2T}.
\end{equation}
If $R_p\ge1$ almost surely for every $p>p_0$, the logarithmic
windows may instead be $[\log\rho,\log\rho+h]$; under the additional
raw moment hypothesis the raw windows may likewise be $[\rho,\rho+h]$.
\end{corollary}

\begin{proof}
The series of expectations of $|\log R_p|$ is finite, so its random
sum is finite almost surely by Tonelli. Since $H$ is a positive finite
rational almost surely, this proves the assertion about $R$.
Partial summation of $\pi(x)\ll x/\log x$ gives
\[
 \mathbb E\sum_{p>y}|\log R_p|
       \ll\sum_{p>y}p^{-2}\ll\frac1{y\log y}.
\]
Fix a rational $\rho$ and a finite collection of configurations,
each consisting of a value of $H$ and prescribed nonneutral states
at finitely many primes, whose finite rational product equals $\rho$.
Take $y$ beyond all those primes and prescribe neutrality at every
other prime in $(p_0,y]$. Different configurations give disjoint
cylinders. For a configuration with head value $\eta$ and nonneutral
states $r_p$ on a finite set $E$, its probability is
\[
 A_y\,\mathbb P(H=\eta)
           \prod_{p\in E}\frac{\mathbb P(R_p=r_p)}{a_p}.
\]
Multiplication by $\log y$ therefore tends to its normalized weight
$C_0\mathbb P(H=\eta)\prod_{p\in E}\mathbb P(R_p=r_p)/a_p$.
These are precisely the summands that exhaust $w(\rho)$ in
\eqref{eq:rational-anchor-coefficients}; every finite-cutoff character
expansion is countable and has nonnegative coefficients.

Take $y=1/h$ and let $U_y=\sum_{p>y}\log R_p$. It is independent of
all the cylinders just prescribed, and Markov's inequality gives
\[
 \mathbb P(|U_y|>h/2)\ll\frac1{\log(1/h)}.
\]
On a cylinder with $|U_y|\le h/2$, $\log R\in W_\rho^{\log}(h)$.
Sum over the fixed finite collection and then exhaust the configurations.
The result is
$\liminf_{h\downarrow0}L(h)\nu(W_\rho^{\log}(h))\ge w(\rho)$.
The assertion is trivial for a zero-weight anchor.
The second-mean assumption and \Cref{lem:collision-tauberian-mas04}
give the logarithmic collision asymptotic, with constant $C$ exactly.
Apply \Cref{lem:local-log-saturation} to the distinct anchors
$\log\Q_{>0}$, their weights, and the symmetric windows of diameter
$h$. Parseval provides the entire budget $C=\sum_\rho w(\rho)^2$.
This proves the finite-class bounds, all fixed rational window limits,
and the concentration and intersection conclusions. At a real anchor
outside this countable set, the same budget argument gives zero:
a fixed symmetric window eventually avoids every window in a fixed
finite set of positive-weight anchors, and the residual tends to zero.
This also proves the asserted fixed-window limits for irrational $\rho$.

Under the two raw moment assumptions,
\Cref{lem:raw-log-collision-mas04} transfers the logarithmic collision
asymptotic to the raw law; \Cref{lem:collision-tauberian-mas04} then
gives its second mean. To obtain the raw-window lower bounds directly,
fix $\rho>0$ and in the preceding cylinder argument require
$|U_y|\le h/(4\rho)$. Its probability again tends to one for $y=1/h$.
For sufficiently small $h$, $|U_y|\le1$ and
$|e^{U_y}-1|\le2|U_y|$, so
$|R-\rho|=\rho|e^{U_y}-1|\le h/2$.
Thus $\liminf L(h)\mu(W_\rho^{\rm raw}(h))\ge w(\rho)$.
Use the raw collision budget in \Cref{lem:local-log-saturation}, now
on the distinct real anchors $\Q_{>0}$; the same exhaustion also
gives zero at all fixed real points outside its positive-weight anchors.
All raw saturation conclusions follow with exactly the same residuals.

Finally, if every local factor is at least one then $U_y\ge0$.
The conditions $0\le U_y\le h$ and $0\le U_y\le h/(2\rho)$,
respectively, place the logarithmic and raw variables in the stated
right windows; the preceding Markov estimates still tend to one.
The budget lemma applies without changing its scale since each
window has diameter $h$.
\end{proof}

For the data of \Cref{def:rational-amplitude-data},
\Cref{thm:rational-haar-amplitude} supplies the $L^2$ product limit
and \eqref{eq:rational-second-mean-input} (take uniform sampling and
modulus order two).

\begin{remark}[Limits of the rational-product conclusions]
\label{rem:rational-amplitude-scope}
Right windows need not recover the anchor coefficient when negative
jumps are allowed. For instance, take
$H=1$ and, independently for primes $p>3$, take $R_p=p/(p+1)$ with
probability $1/p$ and $R_p=1$ otherwise. The first-jump magnitude is
$\log(1+1/p)$ and $\mathbb E|\log R_p|\ll p^{-2}$.
The coefficient at $\rho=1$ is $C_0>0$, since every nonneutral finite
product is less than one. Nevertheless $R<1$ almost surely:
$\sum_p1/p=\infty$ and independence imply infinitely many activations,
while absolute convergence of the logarithms makes the limit positive.
Thus $\nu([0,h])=0$ for every $h>0$.

Raw transfer is a separate condition. In that example replace the head
by $H=2^{-N}$, independently, where $N$ is a nonnegative integer with
$\mathbb P(N\ge n)\asymp n^{-\alpha}$, $0<\alpha<1$.
Since $0<R\le H$, two independent copies satisfy
\[
 \mathbb P(|R-R'|\le h)\ge\mathbb P(H\le h)^2
       \gg(\log(1/h))^{-2\alpha}.
\]
This excludes automatic raw collision asymptotics of order
$\log^{-2}(1/h)$ for arbitrary atomic rational heads; in particular
all positive raw moments alone do not suffice here.

Nor does such a head imply nonvanishing of its Haar factor. Fix
$0<\beta<\pi$ and let $H=2^N$, $N\in\mathbb Z$, with probabilities
\[
 d_0=\frac\beta{2\pi},\qquad
 d_n=\frac{1-\cos(n\beta)}{\pi\beta n^2}\quad(n\ne0).
\]
These are the nonnegative Fourier coefficients of the periodic tent
$g(\theta)=(1-|\theta|/\beta)_+$ on $[-\pi,\pi]$.
Their absolute summability and Fourier inversion give
$\sum_nd_n=g(0)=1$ and
$\mathbb E\chi(H)=g(\arg\chi(2))$.
This factor is zero on a set of positive Haar measure. Any $L^2$
limit of the normalized products is zero there as well, since every
finite product is zero there. Inverse modulus moments therefore do
not follow from positive or mixed moments. Likewise, cancellation of nontrivial
characters alone does not force a generic holomorphic identity:
the head factor $(\chi(2)+\chi(1/2))/2$ has zero Haar mean but its
square has Haar mean $1/2$.
\end{remark}

Return to a logarithmic divisor-sum law $\nu$, with $w,Z_\sigma,C_\nu$
as in \Cref{lem:sigma-abundancy-coefficients-mas04}.
For the following theorem and its proof, given a finite set
$F\subset\Q_{>0}$ with $w(\rho)>0$ on $F$, put
$R_F=C_\nu-\sum_{\rho\in F}w(\rho)^2$ and
$W_\rho(h)=[\log\rho,\log\rho+h]$.

\KeepHeadingSpace{8\baselineskip}
\begin{theorem}[Uniform saturation of the divisor-sum laws]
\label{thm:sigma-saturation-mas04}
For each of the three logarithmic divisor-sum laws above, the following hold.
\begin{enumerate}[label=\textup{(\roman*)},leftmargin=*,beginpenalty=10000]
\item For every finite set $F$ of positive-weight anchors and intervals
$I$ of length $h$,
\begin{align}
 \limsup_{h\downarrow0}\log(1/h)
 \sup_{I:\ I\cap\bigcup_{\rho\in F}W_\rho(h)=\varnothing}\nu(I)
     &\le\sqrt{R_F},\label{eq:sigma-away-windows-mas04}\\
 \limsup_{h\downarrow0}\log(1/h)
 \sup_{I:\ I\cap W_\rho(h)\ne\varnothing}\nu(I)
     &\le\sqrt{w(\rho)^2+R_F}\quad(\rho\in F).
       \label{eq:sigma-meeting-window-mas04}
\end{align}
\item For every fixed
$a\in\R$,
\begin{equation}\label{eq:sigma-fixed-anchor-mas04}
 \lim_{h\downarrow0}\log(1/h)\,\nu([a,a+h])=w(e^a).
\end{equation}
\item The concentration function satisfies
\begin{equation}\label{eq:sigma-saturated-concentration-mas04}
 Q_\nu(h)\sim\frac{w_*}{\log(1/h)},\qquad
 w_*:=\max_{\rho}w(\rho)>0.
\end{equation}
If exactly one class $\rho_*$ has weight $w_*$, then every family
$I_h$ of intervals of length $h$ with
$\nu(I_h)/Q_\nu(h)\to1$ meets $W_{\rho_*}(h)$ for all sufficiently
small $h$.
\end{enumerate}
\end{theorem}
\begin{proof}
Use the local laws and heads in
\Cref{rem:rational-amplitude-specializations}\textup{(c)}.
They satisfy $R_p\ge1$ and $\mathbb E\log R_p\ll p^{-2}$.
\Cref{lem:sigma-abundancy-coefficients-mas04} identifies their
coefficients with $w$, and \Cref{thm:sigma-amplitude} at $s=2$
gives \eqref{eq:rational-second-mean-input} with $C=C_\nu$.
Thus \Cref{cor:rational-log-saturation}, with right windows,
applies. Its cylinder bound is
\begin{equation}\label{eq:sigma-cylinder-lower-mas04}
 \liminf_{h\downarrow0}\log(1/h)\nu(W_\rho(h))\ge w(\rho),
\end{equation}
and its saturation conclusions give \textup{(i)}--\textup{(iii)}
with the stated residual $R_F$. The summands with $m=1$, $u=1$
and $(k,u)=(1,1)$ in \eqref{eq:sigma-class-weights-mas04} give
$C_\nu>0$ in the respective laws.
\end{proof}

\begin{remark}[Moving windows need not contain an anchor]
\label{rem:moving-anchor-windows-mas04}
The intersection conclusion in \Cref{thm:sigma-saturation-mas04}
does not say that a near-maximizing interval contains the anchor
itself. Indeed, fix an anchor $a$ of positive weight $w$, and write
$H_a(s)=\nu([a,a+s])$. The fixed-anchor asymptotic and atomlessness give
\[
 \nu([a+h^2,a+h^2+h])
 =H_a(h+h^2)-H_a(h^2)
 \sim\frac{w}{2\log(1/h)}.
\]
Thus a moving interval missing its anchor can have a nonzero leading
coefficient. More strongly, with
$\delta_h=\exp\{-(\log(1/h))^2\}$,
\[
 \nu([a+\delta_h,a+\delta_h+h])
 =H_a(h+\delta_h)-H_a(\delta_h)
 \sim\frac w{\log(1/h)}.
\]
At a maximal-weight anchor these are relative near-maximizers that
omit $a$. 
\end{remark}

\paragraph*{Divisor-sum laws in fixed progressions}
Fix $Q\ge1$ and $\gcd(b,Q)=1$, and put
\[
 S:=\{p\ge3:p\textup{ prime},\ p\mid Q\},\qquad
 m:=[2,\gcd(b-1,Q)],\qquad
 \kappa_{F,S}:=\kappa_{\mathrm{end}}
                  \prod_{p\in S}\frac{p-1}{p-2}.
\]
Let $\rho_{Q,b}$ be as in \Cref{cor:ap-endpoint-expansions}, and let
$Y_{Q,b}$ be the independent product in \Cref{cor:ap-sigma-law}, with
distribution function $H_{\sigma,Q,b}$. Write
$\mu_{g,Q,b}:=\operatorname{Law}(\log Y_{Q,b})$.
Use the single Haar character $\chi$ and the factors $R_{p,k}$,
$r_p$, $A_p(\chi)$ and $B_2(\chi)$ of \Cref{thm:sigma-amplitude},
and set $R_{p,0}:=1$. For $p\mid2Q$ and $k\ge0$, define
\[
 \lambda_{p,Q,b}(k):=
 \begin{cases}
  2^{-k}\mathbf1_{\{k\ge1\}},&p=2,\ 2\nmid Q,\\
  \mathbf1_{\{k=v_p(b-1)\}},
       &p\mid Q,\ v_p(b-1)<v_p(Q),\\
  (p-1)p^{v_p(Q)-1-k}\mathbf1_{\{k\ge v_p(Q)\}},
       &p\mid Q,\ v_p(b-1)\ge v_p(Q).
 \end{cases}
\]
Here $v_p(0)=\infty$. Define
$\Lambda_p(\chi):=\sum_{k\ge0}\lambda_{p,Q,b}(k)\chi(R_{p,k})$;
explicitly,
\[
 \Lambda_p(\chi)=\begin{cases}
 B_2(\chi),&p=2,\ 2\nmid Q,\\
 \chi(R_{p,v_p(b-1)}),&p\mid Q,\ v_p(b-1)<v_p(Q),\\
 \displaystyle(p-1)p^{v_p(Q)-1}
       \sum_{k\ge v_p(Q)}p^{-k}\chi(R_{p,k}),
       &p\mid Q,\ v_p(b-1)\ge v_p(Q).
 \end{cases}
\]
Put
\begin{equation}\label{eq:ap-sigma-haar-product}
 Z_{\sigma,Q,b}:=\kappa_{F,S}
       \prod_{p\mid2Q}\Lambda_p(\chi)
       \prod_{p\nmid2Q}A_p(\chi),
\end{equation}
where the last product is ordered by increasing primes. Let
\[
 \mathcal I_{Q,b}:=\{n\ge1:2\mid n,\
                  \gcd(n,Q)=\gcd(b-1,Q)\}.
\]
With $a_p(j)$ as defined before \eqref{eq:sigma-class-weights-mas04},
set, for $n\in\mathcal I_{Q,b}$,
\[
 \begin{aligned}
 \varpi_{Q,b}(n)&:=
  \prod_{p\mid2Q}\lambda_{p,Q,b}(v_p(n))
  \prod_{\substack{p^j\Vert n\\p\nmid2Q}}a_p(j),\\
 w_{\sigma,Q,b}(\rho)&:=\kappa_{F,S}
       \sum_{\substack{n\in\mathcal I_{Q,b}\\R(n)=\rho}}
                       \varpi_{Q,b}(n)
 \quad(\rho\in\Q_{>0}),
 \end{aligned}
\]
where $R(n)=\sigma(n)/n$. Extend $w_{\sigma,Q,b}$ by zero to
$\R_{>0}\setminus\Q_{>0}$.

\KeepHeadingSpace{5\baselineskip}
\begin{corollary}[Divisor-sum amplitudes in a fixed progression]
\label{cor:ap-sigma-amplitude}
\index{Arithmetic progressions!divisor-sum amplitudes}
With the fixed progression and local model defined above, the product \eqref{eq:ap-sigma-haar-product} converges almost surely
and in $L^2$, is nonzero almost surely, and satisfies
$\mathbb E|Z_{\sigma,Q,b}|^s<\infty$ for every real $s$.
For $\nu=\mu_{g,Q,b}$ and $U_T$ uniform on $[0,T]$,
\[
 \begin{gathered}
 (\log T)\widehat\nu(U_T)\Rightarrow Z_{\sigma,Q,b},\\
 \frac{(\log T)^s}{T}\int_0^T|\widehat\nu(t)|^s\,dt
       \longrightarrow\mathbb E|Z_{\sigma,Q,b}|^s\in(0,\infty)
       \quad(s>0).
 \end{gathered}
\]
All mixed polynomial moments in $(\log T)\widehat\nu(U_T)$ and its
conjugate converge to those of $Z_{\sigma,Q,b}$, and
$\mathbb EZ_{\sigma,Q,b}^j=0$ for every integer $j\ge1$.
\end{corollary}

\begin{proof}
\emph{The model.} Condition the independent shifted-prime valuations
of \eqref{eq:valuation-laws} on the cylinder in the proof of
\Cref{cor:ap-sigma-law}. Independence is preserved. If $p\nmid Q$,
the probabilities remain $(p-2)/(p-1)$ at zero and $p^{-k}$ at
$k\ge1$, including zero mass at zero for $p=2$.
If $p\mid Q$, write $a=v_p(Q)$. When $v_p(b-1)<a$, the valuation is
fixed at $v_p(b-1)$; otherwise conditioning on $V_p\ge a$ divides
$p^{-k}$ by $\sum_{j\ge a}p^{-j}=p^{1-a}/(p-1)$.
These give exactly $\lambda_{p,Q,b}$, whose sums are one.
At $p=2\mid Q$, reducedness forces $v_2(b-1)\ge1$; thus every
admissible valuation vector has a positive $2$-adic valuation.
The product $Y_{Q,b}=\prod_pR_{p,V_p}$ converges almost surely:
its finite head is finite and the remaining nonnegative logarithms
have expectations $O(p^{-2})$. Replacing $\chi(q)$ by $q^{it}$ gives
the local characteristic factor $\Lambda_p$ at $p\mid2Q$ and
$\frac{p-2}{p-1}A_p$ at every other prime.

\emph{Haar factors.} For $p\ge5$, $p\nmid Q$, put
$c_p=(p-1)/(p(p-2))$. Uniformly in $\chi$,
\[
 \log A_p(\chi)=c_p\chi(r_p)+O(p^{-2}).
\]
The values $\chi(r_p)$ are independent uniform phases by the
multiplicative basis proved in \Cref{thm:sigma-amplitude}.
The leading series therefore converges almost surely, since its
centered independent terms have summable variances; the error is
absolutely summable. For every fixed real $s$,
$\mathbb E\exp(s\operatorname{Re}(c_p\chi(r_p)))=1+O_s(p^{-2})$.
Independence and the uniformly bounded sum of logarithmic errors
give uniform bounds for all real modulus moments of these partial
products and finite moments of their nonzero limit.

A fixed-valuation head factor has modulus one. For an odd prime
$p\mid Q$ with $a=v_p(Q)\le v_p(b-1)$,
\[
 \Lambda_p=(1-1/p)\chi(R_{p,a})
       \left(1+\sum_{k>a}p^{a-k}\chi(R_{p,k}/R_{p,a})\right),
 \qquad \frac{p-2}{p}\le|\Lambda_p|\le1,
\]
because $\sum_{k>a}p^{a-k}=1/(p-1)$.
If the valuation at $2$ is not fixed, put $a=1$ when $2\nmid Q$ and
$a=v_2(Q)$ otherwise. Then
\[
 F_a:=2\chi(R_{2,a})^{-1}\Lambda_2
   =1+\sum_{k>a}2^{a-k}\chi(q_k),\qquad
 q_k:=\frac{R_{2,k}}{R_{2,a}}
     =\frac{2^a(2^{k+1}-1)}{2^k(2^{a+1}-1)}.
\]
The positive weights sum to $\sum_{k>a}2^{a-k}=1$.
This rational family contains arbitrarily many multiplicatively
independent members. Recursively choose a prime $q>a+2$ not dividing
$2^{a+1}-1$ and larger than every prime in the supports of the
previously selected rationals, and take $k=q-2>a$.
Fermat's theorem gives $q\mid2^{k+1}-1$, while $q$ divides neither
the denominator nor any previous rational. The valuation at this
fresh prime eliminates the last exponent in any finite relation;
descent proves independence. The corresponding Haar phases are
therefore independent and uniform.

\Cref{lem:exceptional-haar-inverse} gives every inverse modulus
moment of $F_a$. The same conclusion for $A_3$, when $3\nmid Q$,
was proved in \Cref{thm:sigma-amplitude}.
All these factors have finite positive moments as they are bounded.
H\"older's inequality combines the nonexceptional product, $A_3$
when present, and $F_a$ when present, together with the bounded
head factors. Each factor is taken with a correspondingly higher
moment order.
Thus the full finite prime products have uniformly bounded modulus
moments of every real order. They converge almost surely to a
nonzero limit, which has the same moment finiteness by Fatou's lemma.
Uniform fourth-moment bounds make their squared differences from
the limit uniformly integrable, proving convergence in $L^2$.

\emph{Temporal limit and moments.} The conditioned head is
$\prod_{p\mid2Q}R_{p,V_p}$ with the independent probabilities
$\lambda_{p,Q,b}$. It is countably atomic and
independent of the free valuations. For $p\nmid2Q$, the local law
is the shifted-prime divisor-sum law, with
$\mathbb E\log R_{p,V_p}\ll p^{-2}$. Removing the inactive
factors at $p\in S$ gives
\[
 (\log T)\prod_{\substack{3\le p\le T\\p\nmid Q}}
       \frac{p-2}{p-1}\longrightarrow\kappa_{F,S}.
\]
The finite-head regrouping in
\Cref{rem:rational-amplitude-specializations}\textup{(d)} therefore
identifies \eqref{eq:ap-sigma-haar-product} with the product in
\Cref{thm:rational-haar-amplitude}. Part \textup{(b)} of that theorem
gives the weak limit and all positive and mixed polynomial moment
limits.

Every character in a finite head has the form $\chi(R(n))$ with
$n\in\mathcal I_{Q,b}$. Such $n$ is even, so $R(n)\ge3/2$.
Every term of a positive integer power therefore has a rational
argument greater than one and Haar mean zero. Uniform valuation
convergence and the uniform higher positive moments permit removal
of both truncations, giving $\mathbb EZ_{\sigma,Q,b}^j=0$.
\end{proof}

\KeepHeadingSpace{5\baselineskip}
\begin{proposition}[Progression coefficients and collision constants]
\label{prop:ap-sigma-coefficients-collisions}
With the fixed progression and local model defined above, the weights $w_{\sigma,Q,b}(\rho)$ are the Fourier coefficients
of $Z_{\sigma,Q,b}$ at the characters $\chi\mapsto\chi(\rho)$, and
\begin{equation}\label{eq:ap-sigma-class-parseval}
 C_{\sigma,Q,b}:=\mathbb E|Z_{\sigma,Q,b}|^2
       =\sum_{\rho\in\Q_{>0}}w_{\sigma,Q,b}(\rho)^2\in(0,\infty).
\end{equation}
For $\mu=\operatorname{Law}(Y_{Q,b})$, $\nu=\mu_{g,Q,b}$ and
$C=C_{\sigma,Q,b}$, the raw and logarithmic collision asymptotics
\eqref{eq:four-exact-collisions-mas04} and both second-mean asymptotics
\eqref{eq:four-exact-second-means-mas04} hold.
\end{proposition}

\begin{proof}
\emph{Coefficients and collisions.} Apply
\Cref{prop:rational-anchor-parseval} with the conditioned head
$\prod_{p\mid2Q}R_{p,V_p}$ and the free shifted-prime local laws,
regrouping the finite head as in
\Cref{rem:rational-amplitude-specializations}\textup{(d)}.
The normalized valuation probabilities are
$\lambda_{p,Q,b}(j)$ in the head and $a_p(j)$ at the free primes.
Thus the coefficient at $\chi(\rho)$ in the product through
$y>P^+(2Q)$ is $\kappa_{F,S}$ times the sum of
$\varpi_{Q,b}(n)$ over $n\in\mathcal I_{Q,b}$ with $R(n)=\rho$
and all prime factors outside $2Q$ at most $y$.
The $L^2$ convergence in \Cref{cor:ap-sigma-amplitude} and the
proposition give the coefficients $w_{\sigma,Q,b}(\rho)$ and
\eqref{eq:ap-sigma-class-parseval}.
The admissible integer $m$ has positive weight, proving
$C_{\sigma,Q,b}>0$.

\Cref{cor:ap-sigma-amplitude} at $s=2$ gives the logarithmic second mean.
\Cref{lem:collision-tauberian-mas04} gives the logarithmic collision
asymptotic with the same constant. Since $Y_{Q,b}\ge1$ and
$\mathbb EY_{Q,b}^r=N_{\sigma,Q,b}(r)<\infty$ for $r>0$
by \Cref{cor:ap-sigma-law}, both positive and negative raw moments
required by \Cref{lem:raw-log-collision-mas04} are finite.
That lemma gives the raw collision asymptotic, and
\Cref{lem:collision-tauberian-mas04} applied to the raw law gives
its second mean.
\end{proof}

\KeepHeadingSpace{5\baselineskip}
\begin{corollary}[Progression divisor-sum saturation]
\label{cor:ap-sigma-saturation}
With the fixed progression and local model defined above, assertions \textup{(i)}--\textup{(iii)} of
\Cref{thm:sigma-saturation-mas04} hold for $\nu=\mu_{g,Q,b}$, with
$w=w_{\sigma,Q,b}$, $C_\nu=C_{\sigma,Q,b}$ and
$W_\rho(h)=[\log\rho,\log\rho+h]$. Here $F$ is replaced by a finite
set $\mathcal F\subset\Q_{>0}$ of positive-weight classes, and
\[
 R_{\mathcal F}:=C_{\sigma,Q,b}
       -\sum_{\rho\in\mathcal F}w_{\sigma,Q,b}(\rho)^2
\]
replaces $R_F$. Moreover,
\[
 Q_{\mu_{g,Q,b}}(h)\sim Q_{\operatorname{Law}(Y_{Q,b})}(h)
       \sim\frac{w_*}{\log(1/h)},\qquad
 w_*:=\max_\rho w_{\sigma,Q,b}(\rho)>0.
\]
\end{corollary}

\begin{proof}
Apply \Cref{cor:rational-log-saturation} to the conditioned head
and free local laws of \Cref{cor:ap-sigma-amplitude}.
The free factors satisfy $R_p\ge1$ and
$\mathbb E\log R_p\ll p^{-2}$.
\Cref{cor:ap-sigma-amplitude} gives the $L^2$ product limit, and
\Cref{prop:ap-sigma-coefficients-collisions} gives the coefficient
identity and second mean required by the corollary.
Its right-window conclusions prove the logarithmic assertions with
the stated weights and residual $R_{\mathcal F}$, including
intersection with its anchor window when the maximal class is unique.

Since $Y_{Q,b}\ge1$, the logarithmic image of the intersection of a
raw interval of length $h$ with $[1,\infty)$ is contained in an
interval of length at most $h$. Hence
$Q_{\operatorname{Law}(Y_{Q,b})}(h)\le Q_{\mu_{g,Q,b}}(h)$.
At a class $\rho_*$ of weight $w_*$, the raw interval
$[\rho_*,\rho_*+h]$ has logarithmic image
$[\log\rho_*,\log\rho_*+t_h]$, where
$t_h=\log(1+h/\rho_*)\sim h/\rho_*$. The fixed-anchor limit gives
\[
 \mathbb P(\rho_*\le Y_{Q,b}\le\rho_*+h)
       \sim\frac{w_*}{\log(1/t_h)}
       \sim\frac{w_*}{\log(1/h)}.
\]
This is the matching lower bound and proves the raw concentration asymptotic.
\end{proof}

\KeepHeadingSpace{5\baselineskip}
\begin{proposition}[The progression endpoint class and its gap]
\label{prop:ap-sigma-endpoint-class}
With the fixed progression and local model defined above, the only member of $\mathcal I_{Q,b}$ with abundancy $R(m)$ is
$m$, and
\[
 w_*\ge w_{\sigma,Q,b}(R(m))=\kappa_{F,S}\rho_{Q,b}.
\]
Let $T=\{p\mid Q:p>2,\ v_p(b-1)\ge v_p(Q)\}$. If $|T|\le1$,
the endpoint $R(m)$ is the unique maximal class. This includes every
prime-power modulus and every modulus $2^j p^k$ with $p$ odd.

More generally, freeze the coordinates in $T$ at their minima and
normalize the residual class weights by the endpoint weight. If all
nonendpoint residual classes have weight at most $g<1$, a sufficient
condition for the original endpoint to be uniquely maximal is
\[
 g+\prod_{p\in T}(1-1/p)^{-1}-1<1.
\]
One may take $g=2/3$ for a fixed dyadic valuation, $g=g_1$ for a
variable dyadic valuation of minimum $1$, and $g=34/45$ for a variable
dyadic valuation of higher minimum. Whenever the endpoint is the unique
maximal class, $w_*=\kappa_{F,S}\rho_{Q,b}$, and for all sufficiently
small $h$ the unique closed maximizing intervals for the logarithmic and
raw laws start at $\log R(m)$ and $R(m)$, respectively.
\end{proposition}

\begin{proof}
\emph{The endpoint class.} If $2\mid Q$, reducedness forces $b$ odd
and $m=\gcd(b-1,Q)$; if $2\nmid Q$, then
$m=2\gcd(b-1,Q)$. Thus $m\in\mathcal I_{Q,b}$. Every
$n\in\mathcal I_{Q,b}$ is even and satisfies
$v_p(n)\ge\min(v_p(b-1),v_p(Q))=v_p(m)$ for $p\mid Q$.
It follows that $m\mid n$. Each factor
$R(p^j)=1+p^{-1}+\cdots+p^{-j}$ strictly increases with $j$, so
$R$ strictly increases under proper divisibility. Consequently
$R(n)=R(m)$ within $\mathcal I_{Q,b}$ only when $n=m$.
In $\varpi_{Q,b}(m)$, the factor at $p=2\nmid Q$ is $1/2$, the
factor at a fixed valuation is $1$, and the factor where
$v_p(b-1)\ge v_p(Q)$ is
$(p-1)p^{v_p(Q)-1-v_p(Q)}=1-1/p$. There are no other active
coordinates. Hence
\[
 \varpi_{Q,b}(m)=2^{-\mathbf1_{\{2\nmid Q\}}}
       \prod_{\substack{p\mid Q\\v_p(b-1)\ge v_p(Q)}}(1-1/p)
       =\rho_{Q,b},
\]
which proves the stated endpoint weight.

\emph{The residual dyadic and free odd factors.}
Remove the fixed prime powers, dividing the target abundancy by their
contribution, and normalize by the endpoint weight. If the prime powers
in $T$ are also fixed at their minima, the residual law with dyadic
valuation fixed has class sums bounded by the $F$ of
\Cref{lem:abundancy-class-gap-mas05} (disallowed odd primes are deleted).
For variable dyadic valuation of minimum $a\ge1$, its normalized
weights at exponents $a+k$ are $2^{-k}$ by \eqref{eq:ap-sigma-haar-product}
and the formula for $\lambda_{2,Q,b}$. The residual class sums are at most
\[
 \Omega_a(r):=\sum_{k\ge0}2^{-k}F(r/R(2^{a+k})).
\]
Equality at the endpoint gives weight one. For $a=1$, the bound off
the endpoint is $g_1<4/5$ by \Cref{lem:abundancy-class-gap-mas05}.

For $a\ge2$, if a residual equals $1$, solitude of the dyadic prime
power bounds the off-endpoint class by $1/2$. Otherwise at most one
residual is $4/3$. If none is, the class is bounded by $8/15$;
if it occurs at $k\ge1$, the bound is $11/15$ as in the proof of
\Cref{lem:abundancy-class-gap-mas05}. It remains to consider
$r=(4/3)R(2^a)$. The residuals
$s_k=(4/3)R(2^a)/R(2^{a+k})$ for $k\ge1$ lie strictly between $1$ and
$4/3$, so no odd member is divisible by $3$. For $a=2$,
$s_k=7\cdot2^{k+2}/[3(2^{k+3}-1)]$ has reduced denominator divisible
by $3$, excluding every member. For $a\ge3$ we have
$s_k>(4/3)(1-2^{-a-1})\ge5/4$. A prime power $q^j$ with $q\ge5$
has abundancy strictly below $5/4$. If the class is nonempty its
reduced denominator therefore has a prime $q\ge5$, and removing $q$
leaves only nonendpoint odd residuals without $3$. Their bound $4/15$
times $1/(q-2)$ gives $F(s_k)\le4/45$. Hence the entire class is at
most $2/3+4/45=34/45$. These bounds persist under deletion of free
odd primes. In particular $T=\varnothing$ is covered.

\emph{One variable odd valuation.} Suppose $T=\{p\}$, with minimum
$v=v_p(Q)$. Write $D$ for the residual class sum with $p$ omitted
and all fixed factors removed, and $r$ for the correspondingly rescaled
target. A fixed dyadic factor, when present, is removed as well. The
normalized class weight is
\[
 \sum_{h\ge0}p^{-h}D(r/R(p^{v+h})).
\]
The residual endpoint has weight one. By strict increase of $R(p^{v+h})$,
at most one $h$ gives that endpoint; for a nonendpoint target it must
satisfy $h\ge1$. Thus if $D\le g$ off its endpoint, the sum is at most
$g p/(p-1)+(1-g)/p$.
For $p\ge5$ and variable dyadic minimum $a\ge2$, this is at most
$(34/45)(5/4)+(11/45)/5=149/150$. For $p=3$ with any variable
dyadic minimum, the no-$3$ odd bound $4/15$ gives $g=8/15$ after
summing the dyadic weights, allowing also the solitary dyadic powers
of off-endpoint weight at most $1/2$. The resulting bound is $43/45$.
With fixed dyadic exponent, if any odd residual is $1$, solitude leaves
a single off-endpoint term of weight at most $1/p$. Otherwise the sum
is at most $(2/3)p/(p-1)$ for $p\ge5$ or $(4/15)(3/2)$ for $p=3$,
both strictly below one.

Finally assume the minimum dyadic exponent is $a=1$ and $p\ge5$.
If there is no residual endpoint, the sum is at most $g_1 p/(p-1)<1$.
If it occurs at $h\ge1$, the residual for index zero is
\[
 R(2)\frac{R(p^{v+h})}{R(p^v)}
 =\frac32\frac{p^{v+h+1}-1}{p^h(p^{v+1}-1)} .
\]
Its reduced denominator contains $p$, whereas every integer contributing
to $D$ avoids $p$. Thus the index-zero contribution vanishes, and the
remaining sum is at most $1/(p-1)$.

\emph{Several variable odd valuations.}
The all-minimum excess vector contributes at most $g$ off the endpoint.
Any other excess vector $(h_p)_{p\in T}$ has coefficient
$\prod_{p\in T}p^{-h_p}$ and contributes at most that coefficient,
since every residual class is bounded by one. Summing gives the stated
sufficient condition. Finally, \Cref{cor:ap-sigma-law} gives atomlessness
and full support on the raw ray, and \Cref{cor:ap-sigma-saturation} gives
both hypotheses of \Cref{lem:saturated-endpoint-maximization} when the
endpoint class is uniquely maximal. The exact logarithmic and raw
concentrations are therefore the endpoint masses. In particular the
raw concentration has the full series \eqref{eq:ap-endpoint-series};
for the logarithmic concentration at width $h$ use raw width
$R(m)(e^h-1)$ in that series.
\end{proof}

\section{Density, tails, and resonance}

For the totient amplitude, isolating the prime-$3$ factor gives the
density, the leading small-amplitude tail and the Mellin transform.
The transform extends meromorphically beyond its moment domain,
although the real moments diverge for $s\le-1$.

Put $V_F:=\kappa_{\mathrm{end}}\exp(\sum_{\ell\ge5}
\psi_\ell(\Theta_\ell))$ and $C_F:=\mathbb EV_F^{-1}/\pi$.
Thus $W_F=2V_F|\cos(\Theta_3/2)|$ with independent factors.

\begin{corollary}[Density and lower tail of the limiting Fourier modulus]\label{cor:amplitude-density}
\begin{enumerate}[label=\textup{(\roman*)},leftmargin=*]
\item The law of $W_F$ is absolutely continuous with support $[0,\infty)$.
One density, defined almost everywhere, is
\[
 f_F(w)=\mathbb E\left[
 \frac{\mathbf1_{\{0<w<2V_F\}}}
 {\pi V_F\sqrt{1-w^2/(4V_F^2)}}\right].
\]
The circular amplitude $Z_{F,\mathbb C}$ has planar density
$z\mapsto f_F(|z|)/(2\pi|z|)$ for $z\ne0$.
\item For every fixed integer $N\ge0$, as $w\downarrow0$,
\[
 \mathbb P(W_F\le w)=
 \sum_{n=0}^N\frac{\binom{2n}{n}\mathbb EV_F^{-(2n+1)}}
 {\pi16^n(2n+1)}w^{2n+1}+O_N(w^{2N+3}).
\]
In particular, $\mathbb P(W_F\le e^{-u})=C_Fe^{-u}+O(e^{-3u})$
as $u\to\infty$, and
\begin{equation}\label{eq:amplitude-iterated-lower-tail}
 \lim_{w\downarrow0}\frac1w\lim_{T\to\infty}
 \mathbb P\bigl((\log T)|\FT(U_T)|\le w\bigr)=C_F.
\end{equation}
\end{enumerate}
\end{corollary}

\begin{proof}
For $\ell\ge5$, centering in \eqref{eq:fourier-local-log-moments} and
$|\psi_\ell|\ll\varrho_\ell$ give
$\mathbb E e^{b\psi_\ell(\Theta_\ell)}=1+O_b(\varrho_\ell^2)$
for every real $b$. Summability and independence therefore give every
positive and negative moment of $V_F$. Independently of $V_F$, the prime-$3$ factor is
$A=2|\cos(\Theta_3/2)|$, so $W_F=V_F A$. Its distribution function is
$(2/\pi)\arcsin(z/2)$ on $[0,2]$ and $1$ above $2$.
Conditional densities give the displayed density by Tonelli's theorem.
By \Cref{cor:totient-amplitude-circular-moments}, the angle of
$Z_{F,\mathbb C}$ is independent and uniform. Polar integration,
$dz=r\,dr\,d\theta$, therefore gives its stated planar density.
The finite sum $\sum_{5\le\ell\le y}\psi_\ell(\Theta_\ell)$ has
essential supremum $\sum_{5\le\ell\le y}\log(1+\varrho_\ell)\to\infty$.
Its independent tail is integrable and centered, hence exceeds $-1$
with positive probability. Independence therefore makes the essential
supremum of $V_F$ infinite. The conditional support $[0,2V_F]$ proves full
support of $W_F$.

Taylor's formula for the arcsine distribution function has, for each fixed $N$, a global
remainder $O_N(z^{2N+3})$ for $z\ge0$: use Taylor on $[0,1]$ and
bound the distribution function and the polynomial above $1$.
Substitute $z=w/V_F$ and take expectations. The finite inverse moment
$\mathbb EV_F^{-(2N+3)}$ controls the remainder. This proves the expansion.
Continuity of the limiting distribution function and \Cref{thm:fourier-amplitude-law}
give the inner limit in \eqref{eq:amplitude-iterated-lower-tail};
the expansion with $N=0$ gives the outer limit.
\end{proof}

\begin{proposition}[Mellin continuation and vertical decay of the Fourier modulus]\label{prop:amplitude-mellin-continuation}
The Mellin transform continues meromorphically to $\C$ by
\[
 M_F(s)=\mathbb EV_F^s\,
 \frac{2^s\Gamma((s+1)/2)}{\sqrt\pi\,\Gamma(1+s/2)}.
\]
Its poles are exactly the simple poles $s=-(2n+1)$, $n\ge0$, with residues
$\binom{2n}{n}\mathbb EV_F^{-(2n+1)}/(\pi16^n)$.
Its real zeros are simple and occur exactly at the negative even integers, and
$\lim_{s\downarrow-1}(s+1)M_F(s)=C_F$.
For real $s\le-1$, $\mathbb EW_F^s=\infty$.

In particular the continuation represents the Mellin expectation exactly
on $\operatorname{Re}s>-1$. If $E(s):=\mathbb EV_F^s$, then $E$ is entire,
positive on the real axis, and, for every compact interval $J\subset\R$
and every $D>0$, satisfies, uniformly for $\sigma\in J$ and $t\in\R$,
\begin{equation}\label{eq:amplitude-mellin-vertical}
 E(\sigma+it)\ll_{J,D}(1+|t|)^{-D}.
\end{equation}
The continued $M_F$ also has arbitrarily rapid vertical decay on such
strips away from its real poles.
\end{proposition}

\begin{proof}
For $s\in\C$ and $0\le v<1$, extend $\mathcal H_s(v)$ by the
integral in \eqref{eq:fourier-mellin-local}, with
$|1+ve^{i\theta}|^s:=\exp(s\log|1+ve^{i\theta}|)$. Then
\[
 E(s):=\kappa_{\mathrm{end}}^s\prod_{\ell\ge5}\mathcal H_s(\varrho_\ell)
       =\mathbb EV_F^s
\]
is entire: each factor is entire and
$\mathcal H_s(\varrho_\ell)=1+O_K(\varrho_\ell^2)$ on every compact
$K\subset\C$. It is positive on the real axis.
For $\operatorname{Re}s>-1$, independence gives the gamma quotient
in \eqref{eq:fourier-mellin-three} times $E(s)$. Since $E$ is entire
and positive at real arguments, the gamma quotient gives the asserted
poles, residues, and real zeros, as well as the limit at $-1$.
Because $2|\cos(\Theta_3/2)|$ has simple zeros, the expectation
$\mathbb E W_F^s$ diverges for real $s\le-1$.

To prove the vertical bounds, fix $0<v<1$ and a compact interval
$J\subset\R$. Put $g_v(\theta)=\log|1+ve^{i\theta}|$. Its derivative is
\[
 g_v'(\theta)=-\frac{v\sin\theta}{1+2v\cos\theta+v^2},
\]
so its only stationary points are $0$ and $\pi$, both nondegenerate.
For $|t|\ge1$, remove neighborhoods of these points of radius
$|t|^{-1/2}$ from the integral defining $\mathcal H_{\sigma+it}(v)$.
The removed integrals are $O_{v,J}(|t|^{-1/2})$. On their complement,
integration by parts against $e^{it g_v(\theta)}$ gives the same bound:
$|g_v'|$ is comparable to the distance from the nearest stationary
point, and the boundary terms and the integral of
$|g_v''|/|g_v'|^2$ are $O_v(|t|^{1/2})$ before division by $|t|$.
The amplitude $e^{\sigma g_v}$ and its derivative are bounded uniformly
for $\sigma\in J$. Hence
\[
 \mathcal H_{\sigma+it}(v)\ll_{v,J}(1+|t|)^{-1/2}.
\]
Apply this estimate to arbitrarily many fixed prime factors in $E(s)$.
For the remaining factors use
$|\mathcal H_{\sigma+it}(\varrho_\ell)|\le\mathcal H_\sigma(\varrho_\ell)$;
their product is bounded uniformly for $\sigma\in J$ by the real moment
product already established. This proves
\eqref{eq:amplitude-mellin-vertical} for every $D>0$.
Stirling's formula bounds the displayed gamma quotient by
$O_J(|t|^{-1/2})$ for $|t|\ge1$. Thus $M_F(s)$ has arbitrarily rapid
vertical decay on each bounded real strip, away from its real poles.
\end{proof}

\begin{corollary}[Schwartz density with zero Taylor radius]\label{cor:amplitude-schwartz-density}
The density $f_F$ of \Cref{cor:amplitude-density} has a strictly positive $C^\infty$ version on
$[0,\infty)$ whose even extension to $\R$ is a Schwartz function.
Its derivatives at zero are
\begin{equation}\label{eq:amplitude-density-derivatives}
 \frac{f_F^{(2n)}(0)}{(2n)!}
 =\frac{\binom{2n}{n}\mathbb EV_F^{-(2n+1)}}{\pi16^n},\qquad
 f_F^{(2n+1)}(0)=0\qquad(n\ge0).
\end{equation}
Nevertheless, its Taylor series at zero has radius of convergence zero.
\end{corollary}

\begin{proof}
Fix $c>-1$. The finite measure
$e^{cx}\operatorname{Law}(\log W_F)(dx)$ has Fourier transform
$M_F(c+it)$. Fourier inversion and a change of variables give the
smooth density
\[
 f_F(w)=\frac1{2\pi i}\int_{c-i\infty}^{c+i\infty}
             M_F(s)w^{-s-1}\,ds\qquad(w>0).
\]
Arbitrary differentiation under this integral is justified by
\eqref{eq:amplitude-mellin-vertical}. Moving the contour to arbitrarily
large positive real part shows that every derivative decreases faster
than every inverse power as $w\to\infty$. Moving it arbitrarily far to
the left crosses precisely the poles $s=-(2n+1)$ in \Cref{prop:amplitude-mellin-continuation}.
The horizontal contour integrals tend to zero by the uniform vertical
bounds. The residues give the even powers
$\binom{2n}{n}\mathbb EV_F^{-(2n+1)}w^{2n}/(\pi16^n)$, with a remainder
that can be differentiated any fixed number of times and made smaller
than any prescribed power of $w$ by moving the contour farther left.
This proves \eqref{eq:amplitude-density-derivatives} and the smooth even
extension, which is therefore Schwartz.
For any $w_0>0$, choose $w_0/2<a<b<\infty$ with
$\mathbb P(a\le V_F\le b)>0$, using the unbounded essential supremum
proved in \Cref{cor:amplitude-density}. In a neighborhood of $w_0$, the conditional density in
\Cref{cor:amplitude-density}\textup{(i)}, restricted to this event, is at least $1/(\pi b)$.
Thus the density has a positive uniform lower bound almost everywhere
in that neighborhood; its continuous version is positive at $w_0$.
At zero its value is $C_F>0$.

Finally, the essential infimum of $V_F$ is zero. The finite logarithmic
head has essential infimum
$\sum_{5\le\ell\le y}\log(1-\varrho_\ell)\to-\infty$; its independent
centered tail is at most $1$ with positive probability. Choosing the
head near its minimum therefore makes $V_F$ arbitrarily small with
positive probability. For each $R>0$ this gives
$\mathbb EV_F^{-(2n+1)}\ge R^{2n+1}\mathbb P(V_F^{-1}>R)$, so
\[
 \bigl(\mathbb EV_F^{-(2n+1)}\bigr)^{1/(2n)}\longrightarrow\infty.
\]
Since $\{\binom{2n}{n}/16^n\}^{1/(2n)}\to1/2$, the Taylor coefficients
in \eqref{eq:amplitude-density-derivatives} have unbounded root growth.
The Taylor series consequently has radius zero.
\end{proof}

\begin{corollary}[Density after finite prime deletions]\label{cor:deleted-amplitude-density}
For a fixed finite set $S$ of odd primes, let $W_{F,S}$ and
$Z_{F,S}$ be as in \Cref{cor:deleted-amplitude}. The modulus $W_{F,S}$
has support $[0,\infty)$ and a smooth density $f_S$ on $(0,\infty)$.
If $3\notin S$, put
\[
 V_S=\kappa_{F,S}\exp\!\left(\sum_{\substack{\ell\ge5\\\ell\notin S}}
                  \psi_\ell(\Theta_\ell)\right).
\]
Then $f_S$ has an even Schwartz extension, with
$f_S(0)=\mathbb EV_S^{-1}/\pi>0$.
If $3\in S$, the zero extension of $f_S$ is Schwartz and flat at zero.
For $z\ne0$ the planar density of $Z_{F,S}$ is
\[
 h_S(z)=\frac{f_S(|z|)}{2\pi|z|}.
\]
It extends to a Schwartz function on $\R^2$ if and only if $3\in S$.
When $3\notin S$,
\[
 h_S(z)\sim\frac{\mathbb EV_S^{-1}}{2\pi^2|z|}\quad(z\to0),
 \qquad h_S\in L^p(\R^2)\ \Longleftrightarrow\ 1\le p<2
 \quad(1\le p\le\infty).
\]
Thus for any fixed progression the planar Schwartz case is precisely
$3\mid Q$.
\end{corollary}

\begin{proof}
Finite deletion leaves infinitely many factors with
$0<\varrho_\ell<1$. Centering and summable variances imply convergence
of their logarithmic sum; its finite heads have interval supports whose
left and right endpoints tend to $-\infty$ and $+\infty$, since
$\sum_{\ell\notin S}\varrho_\ell=\infty$. Their independent centered
tails tend to zero in $L^2$. Consequently every open interval of the
real line has positive probability under this logarithmic sum: choose
a head whose support contains a prescribed smaller interval and a
tail cutoff with positive probability of lying sufficiently close to
zero. This proves support $[0,\infty)$ after exponentiation. If the
prime $3$ is retained, the independent cosine factor gives the same
support by conditioning.

Write
\[
 E_S(s)=\kappa_{F,S}^s
       \prod_{\substack{\ell\ge5\\\ell\notin S}}
                         \mathcal H_s(\varrho_\ell).
\]
The uniform compact estimates $\mathcal H_s(\varrho_\ell)=1+O_K(\varrho_\ell^2)$
prove that $E_S$ is entire and is the Mellin expectation of the
nonexceptional product. Apply the stationary-phase bound from the proof
of \Cref{prop:amplitude-mellin-continuation} to arbitrarily many
retained fixed primes, bounding the remaining product by its real
moments. This gives
$E_S(\sigma+it)\ll_{S,J,D}(1+|t|)^{-D}$ for every bounded interval
$J$ and every $D>0$. This is the finite-factor smoothing mechanism of
Jessen--Wintner~\cite[\S\S7--8]{JW35}.
If $3\notin S$, the Mellin transform is $E_S(s)\mathcal H_s(1)$.
Conditioning on $V_S$ gives the density formula of
\Cref{cor:amplitude-density} with $V_F$ replaced by $V_S$.
Mellin inversion and contour shifting as justified by the displayed
vertical bound give even residues at the negative odd integers;
the first is $\mathbb EV_S^{-1}/\pi$. Moving right gives rapid decay
of every derivative at infinity, and moving left gives a smooth even
extension with this value at zero.
If $3\in S$, the Mellin transform is the entire function $E_S$.
The same inversion crosses no poles on shifting left, giving
$f_S^{(j)}(w)=O_{S,j,N}(w^N)$ at zero for every $j,N\ge0$.
Shifting right again proves rapid decay at infinity. These estimates
prove that its zero extension is Schwartz and flat.

The independent angle in \Cref{cor:deleted-amplitude}, followed by
polar integration, proves the planar formula. When $3\in S$, each
radial derivative costs only finitely many powers of $|z|$; flatness
absorbs them at zero, and rapid decay absorbs them at infinity.
When $3\notin S$, continuity and the positive value $f_S(0)$ give
the stated $|z|^{-1}$ asymptotic, excluding a continuous, hence a
Schwartz, extension. In polar coordinates local $L^p$ integrability
is equivalent for finite $p$ to $\int_0^1r^{1-p}\,dr<\infty$,
which holds exactly when $p<2$; $p=\infty$ is excluded by the same
asymptotic. Rapid decay handles infinity. Finally
$S=\{\ell\ge3:\ell\mid Q\}$ in \eqref{eq:ap-complex-amplitude},
so $3\in S$ is equivalent to $3\mid Q$.
\end{proof}

\KeepHeadingSpace{3\baselineskip}
\begin{corollary}[High moments and upper tail of $W_F$]
\label{cor:amplitude-upper-tail}
\index{Fourier amplitudes!tails}
As real $r\to\infty$ and $w\to\infty$, respectively,
\begin{equation}\label{eq:amplitude-tail}
\begin{aligned}
 \log\mathbb EW_F^r&=r\log\log r+O(r/\log r),\\
 \log\log\frac1{\mathbb P(W_F>w)}&=w-\log w+O(1).
\end{aligned}
\end{equation}
Consequently $\mathbb E e^{aW_F}<\infty$ for every fixed $a>0$.
\end{corollary}

\begin{proof}
The normalization in \eqref{eq:endpoint-Q-quantitative} gives
\[
 \kappa_{\mathrm{end}}\prod_{3\le\ell\le r}(1+\varrho_\ell)
 =\frac{\kappa_{\mathrm{end}}}{Q(r)}
 =\log r\{1+O(1/\log r)\}.
\]
For $\ell>r$, centering in \eqref{eq:fourier-local-log-moments} and
$|\psi_\ell|\ll\varrho_\ell$ give
$\log\mathbb Ee^{r\psi_\ell}=O(r^2\varrho_\ell^2)$.
Summing gives $O(r/\log r)$ and proves the upper moment bound.
For the lower bound restrict the circular distance of
$\Theta_\ell$ from zero to at most $c\sqrt{\ell/r}$ for
$3\le\ell\le r$, with a fixed sufficiently small $c>0$.
The identity
\[
 \frac{|1+\varrho e^{i\theta}|^2}{(1+\varrho)^2}
 =1-\frac{2\varrho(1-\cos\theta)}{(1+\varrho)^2}
\]
shows that each logarithmic loss from the local maximum is $O(1/r)$,
including $\ell=3$. Raising the head to the $r$th power therefore
costs $\exp\{-O(\pi(r))\}$. The negative logarithm of the event
probability is
\[
 O\!\left(\pi(r)+\sum_{3\le\ell\le r}\log(r/\ell)\right)
 =O(r/\log r),
\]
since the sum is $\int_2^r(\pi(t)-1)\,dt/t$.
The independent centered logarithmic tail has exponential moment at
least one by Jensen's inequality. This proves the lower bound.

Write $K(r)=\log\mathbb EW_F^r$.
Markov's inequality with $\log r=w-C$, for a sufficiently large fixed
$C$, gives
$\mathbb P(W_F>w)\le\exp\{-c_1e^w/w\}$.
Indeed $K(r)-r\log w=r\log(1-C/w)+O(r/w)$.
Also
$2K(r)-K(2r)\ge-C_2r/\log r$ and
$(\tfrac12\mathbb EW_F^r)^{1/r}=\log r+O(1)$.
Paley--Zygmund applied to $W_F^r$, now with $\log r=w+C$ and $C$
sufficiently large, gives
$\mathbb P(W_F>w)\ge\exp\{-C_3e^w/w\}$.
Taking two logarithms proves the second assertion and hence the
exponential-moment conclusion.
\end{proof}

The finite-window exponential moments are uniformly bounded for all
sufficiently small $a>0$ by \eqref{eq:valuation-moment-optimality}.
For fixed $a>1$, they diverge: choose
$c\in(1/a,1]$ and $\eta>0$ with $|\FT|\ge c$ on $[0,\eta]$, using
$|\FT(0)|=1$ and continuity. Then, for $T\ge\max\{1,\eta\}$,
\[
 \frac1T\int_0^T\exp\bigl(a(\log T)|\FT(t)|\bigr)\,dt
 \ge\eta T^{ac-1}\longrightarrow\infty.
\]

\begin{proposition}[Quantitative Fourier resonance]
\label{prop:fourier-resonance}
\label{cor:sigma-integer-resonance-mas22}
\index{Fourier decay!resonant lower bounds}
Let $\nu$ be any of $\mu_f$, $\mu_F^{\mathrm{int},\log}$, $\mugodd$,
$\mug$, or $\mu_\sigma^{\mathrm{int},\log}$.
For every sufficiently large $T$ there is $t_T\in(0,T]$ such that
$\log t_T/\log T\to1$ and
\begin{equation}\label{eq:fourier-resonance-window}
 |\widehat\nu(t_T)|\ge(1-o(1))\frac{\log\log T}{\log T}
 \qquad(T\to\infty).
\end{equation}
In particular,
\begin{equation}\label{eq:fourier-endpoint-obstruction}
 \limsup_{|\tau|\to\infty}
 \frac{\log|\tau|}{\log\log|\tau|}|\widehat\nu(\tau)|\ge1.
\end{equation}
\end{proposition}

\begin{proof}
We use a positive-kernel resonator as in the resonance method
\cite{Sou08}. Put
\[
 H=\log T,\qquad L=\log H,\qquad m=\lceil L\rceil,
 \qquad y=\frac{H}{8m},\qquad
 w_T(t)=T^{-1}(1-|t|/T)_+,
\]
so that $\log y=L-\log L+O(1)$. All products in the proof are over
primes.
The Fej\'er kernel
\[
 K_m(\theta)=\frac1{m+1}\left|\sum_{j=0}^m e^{ij\theta}\right|^2
 =\sum_{|j|\le m}\left(1-\frac{|j|}{m+1}\right)e^{ij\theta}
\]
is nonnegative, is at most $m+1$, and has nonnegative Fourier
coefficients.
Direct integration gives
\[
 \int_{\R}w_T(t)e^{it\lambda}\,dt
 =\left(\frac{\sin(T\lambda/2)}{T\lambda/2}\right)^2
 \in\left[0,\min\left\{1,\frac4{T^2\lambda^2}\right\}\right]
 \quad(\lambda\ne0),\qquad \int_{\R}w_T(t)\,dt=1.
\]

Put $\omega_2=\log2$, and let $r_p$ and $R_{p,k}$ be as in
\Cref{thm:sigma-amplitude}. For the two totient laws take
\[
 R_T(t)=\prod_{p_0\le p\le y}K_m(t\omega_p),
\]
with $p_0=3$ for $\mu_f$ and $p_0=2$ for $\mu_F^{\mathrm{int},\log}$.
For all three divisor-sum laws take instead
\begin{equation}\label{eq:sigma-resonator-mas22}
 R_T(t)=K_m(t\log2)\prod_{3\le p\le y}K_m(t\log r_p).
\end{equation}
In each case let $B_T=R_T(0)$. The numbers $\omega_p$, over all
primes $p$, are linearly independent over $\Q$ by \Cref{thm:linind},
and $\{2\}\cup\{r_p:p\ge3\}$ is a multiplicative basis of $\Q_{>0}$,
as is proved at the beginning of the proof of
\Cref{thm:sigma-amplitude}. In each case the frequencies of the
factors of $R_T$ are therefore linearly independent over $\Q$, and the
constant coefficient of $R_T$ is one. Every other frequency
$\lambda$ is the logarithm of a quotient $a/b\ne1$ of positive
integers which are both at most
\[
 \exp\!\left(m\left(\log2+\sum_{p\le y}\log(p+1)\right)\right)
 \le e^{4my}=T^{1/2}
\]
for large $T$; this also bounds the totient resonators. Since
$a\ne b$, integration of $1/u$ between $a$ and $b$ gives
$|\lambda|\ge1/\max(a,b)\ge T^{-1/2}$. The Fourier coefficients of
$R_T$ are nonnegative and sum to $B_T$. The triangular kernel
therefore gives
\begin{equation}\label{eq:sigma-resonator-mass-mas22}
 1\le D_T:=\int_\R w_T(t)R_T(t)\,dt
 \le1+4B_T/T=1+o(1).
\end{equation}
Indeed the prime number theorem gives
$\log B_T\asymp H\log L/L^2=o(H)$, and
$\log B_T/L\to\infty$.

Write $L_p(t)$ for the local characteristic factors of $\nu$ and
$P_T(t)=\prod_{p\le T}|L_p(t)|^2$,
omitting $p=2$ for $\mugodd$, which has no factor there, and for
$\mu_f$, whose factor there is $e^{it\log2}$ by \eqref{eq:BC}.
Each local square has nonnegative Fourier coefficients: if its
valuation probabilities are $a_{p,k}$, its expansion is
\[
 |L_p(t)|^2=\sum_{k,l}a_{p,k}a_{p,l}
       e^{it(\log R_{p,k}-\log R_{p,l})}
\]
for divisor sums, with $R_{p,0}=1$. The sums converge absolutely.
The same positivity holds for the two totient laws, using the two
possible local values. Since the Fourier transform of $w_T$ is
nonnegative, every selection of zero-frequency terms of $P_TR_T$
is a lower bound for $\int w_TP_TR_T$.

For the two totient laws, let $a_{p,0}$ and $a_{p,1}=1-a_{p,0}$ be the
probabilities of the local values $0$ and $\omega_p$, and put
$d_p=2a_{p,0}a_{p,1}$, so that
$|L_p(t)|^2=1-d_p+d_p\cos(t\omega_p)$. Thus
$d_p=2p^{-1}(1-p^{-1})$ in the integer law, and in the shifted-prime
law $d_p=2(p-2)/(p-1)^2$ is the coefficient of \eqref{eq:def-dell};
in both, $d_p=2/p+O(p^{-2})$.
Retaining all valuation pairs for $p_0\le p\le y$, with their matching
Fej\'er coefficients, and diagonal pairs for $y<p\le T$, gives
\[
\begin{aligned}
 \int_\R w_TP_TR_T&\ge
 \prod_{p_0\le p\le y}\left(1-\frac{d_p}{m+1}\right)
 \prod_{y<p\le T}(1-d_p)\\
 & =(1+o(1))\left(\frac{\log y}{\log T}\right)^2
 =(1+o(1))(L/H)^2.
\end{aligned}
\]
The first product tends to one since
$\sum_{p\le y}d_p/m\ll\log\log y/m=o(1)$.
The second follows from $d_p=2/p+O(p^{-2})$ and Mertens' theorem.

For the divisor-sum laws, the probabilities at odd primes are
\[
 \begin{array}{c|cc}
 &a_{p,0}&a_{p,k}\ (k\ge1)\\ \hline
 \text{shifted}&(p-2)/(p-1)&p^{-k}\\
 \text{integer}&1-1/p&(1-1/p)p^{-k}.
 \end{array}
\]
At $2$ they are $(1-1/2)2^{-k}$ for $k\ge0$ in the integer law,
and $2^{-k}$ for $k\ge1$ in the full shifted law; the odd law has
no factor there. Fix a prime cutoff $x\ge3$ and then a valuation
cutoff $K$. For $p\le x$, retain every pair $0\le k,l\le K$
allowed by the law (start at $1$ at $2$ for the full shifted law).
Let
\[
 M_{x,K}:=\prod_{p\le x}\sum_{\substack{0\le k\le K\\k\ \mathrm{allowed}}}
 a_{p,k},
\]
again omitting $2$ in the odd law. This finite head has mass
$M_{x,K}\uparrow1$ as $K\to\infty$ for fixed $x$.
For $x<p\le y$, retain just the four pairs $k,l\in\{0,1\}$;
for $y<p\le T$, retain all diagonal pairs $k=l$.

Every ratio contributed by the fixed truncated head has a finite
expression in the basis $\{2\}\cup\{r_p:p\ge3\}$.
For all large $T$ its basis support is included in
\eqref{eq:sigma-resonator-mas22}, and its exponents are bounded by
a constant depending only on $x,K$. At any basis coordinate outside
this fixed support, an intermediate-prime pair has exponent $0$ or
$\pm1$, with Fej\'er weight $1$ or $m/(m+1)$.
On the fixed support, adding a head exponent changes that weight
by at most $O_{x,K}(1/m)$. Since there are finitely many such
coordinates, the coefficient matching any selected total frequency
is at least $1-O_{x,K}(1/m)$ times the product of the
intermediate-pair weights. Summing these nonnegative terms gives
\begin{align*}
 \int_\R w_TP_TR_T
 &\ge(1-O_{x,K}(1/m))M_{x,K}^2
 \prod_{x<p\le y}\left((a_{p,0}+a_{p,1})^2
             -\frac{2a_{p,0}a_{p,1}}{m+1}\right)
 \prod_{y<p\le T}\sum_{k\ge0}a_{p,k}^2.
\end{align*}
Extra rational dependencies can only add nonnegative terms.
Uniformly at odd primes,
\[
 a_{p,0}+a_{p,1}=1-O(p^{-2}),\qquad
 a_{p,0}a_{p,1}=O(p^{-1}),\qquad
 \sum_{k\ge0}a_{p,k}^2=1-2/p+O(p^{-2}).
\]
The intermediate product is therefore at least
$\exp\{-C\sum_{p>x}p^{-2}-o(1)\}$, while the last product is
$(1+o(1))(\log y/\log T)^2$. Taking the lower limit as
$T\to\infty$, then $K\to\infty$ for fixed $x$, and finally
$x\to\infty$, proves
\begin{equation}\label{eq:sigma-resonator-numerator-mas22}
 \int_\R w_TP_TR_T\ge(1-o(1))(L/H)^2.
\end{equation}

To pass to the full transform, we estimate the prime tail. At $p>T$ the local logarithmic jump has magnitude $O(p^{-1})$
and is nonzero with probability $O(p^{-1})$, so its mean is $O(p^{-2})$.
By the second bound in \eqref{eq:amplitude-far-tail},
$\sum_{p>T}(1-|L_p(t)|^2)\ll1/\log T$ for $|t|\le T$. Since
$\prod_p(1-u_p)\ge1-\sum_pu_p$ for $u_p\in[0,1]$, the tail product
$\prod_{p>T}|L_p(t)|^2$ is $1-O(1/\log T)$ uniformly there.
This applies to both totient laws as well. Hence
$|\widehat\nu(t)|^2\ge(1-o(1))P_T(t)$ for $|t|\le T$.
Finally let $b_T=T/B_T^2=T^{1-o(1)}$. The contribution of
$|t|<b_T$ to $\int w_TR_T|\widehat\nu|^2$ is at most
$2/B_T=o((L/H)^2)$, since $|\widehat\nu|\le1$ and $R_T\le B_T$.
Divide the remaining integral lower bound by
\eqref{eq:sigma-resonator-mass-mas22}. A point $t_T\in[b_T,T]$
then has modulus at least $(1-o(1))L/H$, by continuity and
evenness. Since $\log t_T/\log T\to1$ and $\log\log t_T/L\to1$, this
proves \eqref{eq:fourier-resonance-window} and
\eqref{eq:fourier-endpoint-obstruction}.
\end{proof}

\begin{proposition}[Fourier resonance for the raw ratio laws]
\label{prop:raw-resonance}
\index{Fourier decay!resonant lower bounds}
For any raw law $\mu$ of \Cref{thm:raw-ratio-normality}, put
\[
 q_y=\operatorname{lcm}(1,\ldots,\lfloor y\rfloor),\qquad
 M_y=\lceil e^{\sqrt y}\rceil.
\]
For every sufficiently large $y$ there is an integer $1\le m_y<M_y$
such that
\begin{equation}\label{eq:raw-resonance}
 \operatorname{Re}\widehat\mu(4\pi m_yq_y)
 \ge(1-o(1))\frac{\log y}{y}.
\end{equation}
Consequently
\begin{equation}\label{eq:raw-resonance-limsup}
 \limsup_{t\to\infty}\frac{\log t}{\log\log t}
                      \operatorname{Re}\widehat\mu(t)\ge1.
\end{equation}
\end{proposition}

\begin{proof}
Let $X\sim\mu$ be the corresponding random product
$\Phi$ or $\Sigma$ in \Cref{lem:four-ratio-moments}, using the
integer or shifted-prime valuation law as appropriate. Let $R_y$ be its
product over $p\le y$ and write $M=M_y$, $q=q_y$.
For totients its denominator divides $q$. For divisor sums impose
the head event
\[
 A_y=\left\{K_p\le\left\lfloor\frac{\log y}{\log p}\right\rfloor
                  \text{ for every }p\le y\right\}.
\]
On $A_y$ every denominator $p^{K_p}$ divides $q$, so $qR_y$ is an
integer. For either valuation law and every integer $b\ge1$,
$\mathbb P(K_p>b)\le2p^{-b-1}$. The primes at most $\sqrt y$ give
a total exceptional probability $O(y^{-1/2})$, since each summand
is at most $2/y$; the remaining primes give
$O(\sum_{n>\sqrt y}n^{-2})=O(y^{-1/2})$. Thus
$\mathbb P(A_y)=1-o(1)$, including the coordinate at $2$.
For totients let $A_y$ be the whole sample space. In both cases,
uniformly over the head, $R_y\le C\log y$ for an absolute $C$, by
$R_y\le\prod_{p\le y}(1-1/p)^{-1}$ and Mertens' theorem.

Set $z=qM e^{\sqrt y}$ and require $K_p=0$ for $y<p\le z$.
This gap event, denoted by $B_y$, has probability
\[
 \mathbb P(B_y)
 =\prod_{y<p\le z}(1-p^{-1}+O(p^{-2}))
 =(1+o(1))\frac{\log y}{\log z}.
\]
Indeed, the logarithm of the product is
$-\sum_{y<p\le z}p^{-1}+O(\sum_{p>y}p^{-2})$; Mertens' formula
has an $o(1)$ error at both endpoints, so its fixed constant cancels.
The absolute logarithmic tail $U_z$ above $z$ has mean
$O(1/(z\log z))$. Its event
\[
 C_y=\left\{U_z\le\frac1{2CyMq\log y}\right\}
\]
has probability $1-o(1)$, since Markov's inequality gives
\[
 \mathbb P(C_y^c)\ll\frac{yMq\log y}{z\log z}=o(1).
\]
The events $A_y,B_y,C_y$ are independent. On their intersection,
$X=R_y e^{\pm U_z}$ and
$|X-R_y|\le1/(yMq)$, using $1-e^{-u}\le u$ and
$e^u-1\le2u$ for small $u\ge0$.

The nonnegative kernel
\[
 K_{M-1}(\theta)=\frac1M\left|\sum_{j=0}^{M-1}e^{ij\theta}\right|^2
 =1+2\sum_{m=1}^{M-1}(1-m/M)\cos(m\theta)
\]
equals $M(1-o(1))$ at $\theta=4\pi qX$ on that intersection:
$4\pi qR_y\in2\pi\mathbb Z$ and
$|4\pi q(X-R_y)|\le4\pi/(yM)$.
Its nonnegativity elsewhere gives
\[
 \mathbb E K_{M-1}(4\pi qX)
 \ge(1-o(1))M\frac{\log y}{\log z}.
\]
Since $2\sum_{m=1}^{M-1}(1-m/M)=M-1$, some $1\le m_y<M$ satisfies
\[
 \operatorname{Re}\widehat\mu(4\pi m_yq)
 \ge\frac{(1-o(1))M\log y/\log z-1}{M-1}.
\]
The prime number theorem gives $\log q\sim y$, while
$\log M=O(\sqrt y)$; hence $\log z\sim y$ and
$M\log y/\log z\to\infty$. This proves
\eqref{eq:raw-resonance}. Uniformly for $1\le m<M$,
$\log(4\pi mq)\sim y$ and $\log\log(4\pi mq)\sim\log y$,
which imply \eqref{eq:raw-resonance-limsup}.
\end{proof}

\begin{corollary}[Logarithmic moments and nonintegrability]
\label{cor:fourier-log-statistics}
As $T\to\infty$,
\begin{align*}
 \frac1T\int_0^T\log|\FT(\tau)|\,d\tau
 &=-\log\log T+\log\kappa_{\mathrm{end}}+o(1),\\
 \Var_T(\log|\FT|)
 &\longrightarrow\frac12\sum_{\ell\ge3}
       \operatorname{Li}_2((\ell-2)^{-2}),
\end{align*}
where $\operatorname{Li}_2(v)=\sum_{j\ge1}v^j/j^2$.
Moreover, $\int_{\R}|\FT(\tau)|^s\,d\tau=\infty$ for every $s>0$.
\end{corollary}

\begin{proof}
The logarithmic assertions follow from moment convergence and
\eqref{eq:fourier-local-log-moments}. Nonintegrability follows from
\eqref{eq:fourier-mellin-asymp}, since
$M_F(s)T/(\log T)^s\to\infty$.
\end{proof}

\section{Dissipation fluctuations}

The cosine sum $S$ has mean $2\log\log T+O(1)$ and a non-Gaussian
centered limit. The preceding joint law relates this limit to the
logarithmic amplitude.
Symmetry permits the positive-cosine convention for its marginal law.
The fixed-order Ces\`aro estimates of \Cref{app:cesaro-moments}
control the infinite tail.

Put $X_D:=\sum_{\ell\ge3}d_\ell\cos\Theta_\ell$ and
$V_D:=\sum_{\ell\ge3}d_\ell^2$.

\begin{corollary}[Limiting law of the centered sum $S$]\leavevmode\phantomsection\label{cor:dissipation-law}
\index{Dissipation!limiting law}\ignorespaces
The series $X_D$ converges almost surely and in $L^m$ for every
$1\le m<\infty$, and
\[
 S(U_T)-\overline S_T\ \Rightarrow\ X_D,
 \qquad S(U_T)-2\log\log T\ \Rightarrow\ C_D+X_D,
\]
with convergence of the $m$th moment for every fixed nonnegative integer
$m$. The law of $X_D$ is
symmetric, atomless and non-Gaussian, and
\begin{equation}\label{eq:dissipation-subgaussian}
 \mathbb E e^{sX_D}\le e^{s^2V_D/4}\quad(s\in\R),\qquad
 \mathbb P(|X_D|\ge u)\le2e^{-u^2/V_D}\quad(u>0).
\end{equation}
The constant $V_D/4$ in the first bound cannot be decreased, and, for
every fixed $u>0$,
$\frac1T|\{\tau\in[0,T]:|S(\tau)-\overline S_T|\ge u\}|\to\mathbb P(|X_D|\ge u)$.
\end{corollary}

\begin{proof}
Since $X_D=-D_F$ has the same law as $D_F$, convergence of the series
and the first limit, including moments, follow from
\Cref{cor:complex-amplitude-law}. The mean formula in
\Cref{thm:variance} gives the second limit, including moments.
One cosine term has an atomless arcsine law and is independent of the
rest, so $X_D$ is atomless; with weak convergence this gives the limiting
proportions. By \Cref{lem:tail-moment}, the fourth cumulant of $X_D$ is
$-\frac38\sum_\ell d_\ell^4<0$, so $X_D$ is not Gaussian. Summing the bound
in \eqref{eq:finite-cosine-moment} over the series for
$\mathbb Ee^{sX_D}=\mathbb E\cosh(sX_D)$ gives the first bound in
\eqref{eq:dissipation-subgaussian}; Chernoff's inequality with $s=2u/V_D$
gives the second. Since $\Var X_D=V_D/2$, expanding both
sides at $s=0$ proves the asserted optimality.
\end{proof}

\begin{proposition}[Regularity and tails of the limiting dissipation]
\label{prop:dissipation}
The law of $X_D$ has a Schwartz density and characteristic function
\[
 \phi_D(s)=\prod_{\ell\ge3}J_0(sd_\ell),
\]
where $J_0$ is the Bessel function of the first kind of order zero.
It is not infinitely divisible. Moreover, as $u\to\infty$,
\begin{equation}\label{eq:dissipation-tail}
 \log\log\frac1{\mathbb P(X_D>u)}
 =e^{(u-C_D)/2}-\frac{u-C_D}{2}+O(1).
\end{equation}
The same formula holds with $X_D<-u$ or $|X_D|>u$.
In particular, the triple logarithm divided by $u$ tends to $1/2$.
\end{proposition}

\begin{proof}
The Bessel integral~\cite[(10.9.1)]{DLMF} and independence give the
product; $1-J_0(sd_\ell)=O_K(d_\ell^2)$ on compact $s$-sets gives local
uniform convergence. For $|s|\le\ell\le2|s|$, the numbers $|s|d_\ell$
lie in a fixed compact subinterval of $(0,\infty)$. The strict triangle
inequality in that integral gives $|J_0(v)|<1$ for $v\ne0$; the prime
number theorem therefore gives
$|\phi_D(s)|\le\exp\{-c|s|/\log|s|\}$ for large $|s|$.
For a fixed integer $m\ge0$, partition the ordered odd primes cyclically
into $m+1$ classes. Each class contains a positive proportion of the
primes in $[|s|,2|s|]$, up to bounded error, and its cosine series has
all moments by the proof of \eqref{eq:dissipation-subgaussian}.
Consequently its characteristic function has bounded derivatives of
all orders. In every term of the $m$th derivative of the product of
these $m+1$ characteristic functions, one factor is undifferentiated.
It supplies the same exponential decay, with a constant depending on
$m$. Thus $\phi_D$ is Schwartz, and Fourier inversion proves the
assertion about its density.

Since $d_3=1/2$, the first positive zero $j_{0,1}$ of $J_0$ gives
$\phi_D(2j_{0,1})=0$, excluding infinite divisibility
by \cite[Theorem~4]{LS53}.

The cosine exponential-moment bound applies to every tail. For fixed
$s$, it bounds the second moments of the exponentials of partial sums,
so their exponential moments converge. The identity
$d_\ell=2/\ell-2/[\ell(\ell-1)^2]$ and \eqref{eq:mertens-dvp} give
\[
 \sum_{3\le\ell\le s}d_\ell
 =2\log\log s+C_D+O(1/\log s),\qquad
 \sum_{\ell>s}d_\ell^2\ll\frac1{s\log s},
\]
with exactly the constant in \eqref{eq:dissipation-mean-constant}.
Splitting at $s$ therefore gives
\[
 K_D(s):=\log\mathbb Ee^{sX_D}
 \le s(2\log\log s+C_D)+O(s/\log s).
\]
For the reverse bound restrict the circular distance of $\Theta_\ell$
from zero to at most $c\sqrt{\ell/s}$ for $3\le\ell\le s$, with
fixed $0<c<1$. Each loss $sd_\ell(1-\cos\Theta_\ell)$ is $O(1)$,
and the negative logarithm of this event's probability is
\[
 O\!\left(\pi(s)+\sum_{3\le\ell\le s}\log(s/\ell)\right)
 =O(s/\log s).
\]
The independent centered tail has exponential moment at least one.
Consequently
\begin{equation}\label{eq:dissipation-cumulant}
 K_D(s)=s(2\log\log s+C_D)+O(s/\log s)
 \qquad(s\to\infty).
\end{equation}

Put $V=e^{(X_D-C_D)/2}>0$. By \eqref{eq:dissipation-cumulant},
for real $r\to\infty$,
\[
 \log\mathbb EV^r=K_D(r/2)-rC_D/2
 =r\log\log r+O(r/\log r).
\]
Markov's inequality and the Paley--Zygmund argument in the proof of
\Cref{cor:amplitude-upper-tail}, applied to these moments at $r$ and
$2r$, give
\[
 \log\log\frac1{\mathbb P(V>w)}=w-\log w+O(1)
 \qquad(w\to\infty).
\]
For $w=e^{(u-C_D)/2}$, the events $V>w$ and $X_D>u$ coincide.
This proves \eqref{eq:dissipation-tail}; symmetry and a factor of two
give the other versions.
\end{proof}

For $\mu_D=\operatorname{Law}(X_D)$ and positive integers $k,m$,
$\mu_D^{*k}$ has an $m$th iid convolution root if and only if $m\mid k$,
by the classical argument in \Cref{rem:iid-convolution-roots}.
Indeed, \eqref{eq:dissipation-subgaussian} makes $\phi_D$ entire.
Its zero at $2j_{0,1}$ is simple: $J_0$ has a simple first positive
zero~\cite[\S10.21(i)]{DLMF}, $d_3>d_\ell$ for $\ell\ge5$, and the
remaining positive factors have nonzero product since
$\sum_\ell d_\ell^2<\infty$. A putative root inherits all exponential
moments by Tonelli; comparison of zero multiplicities then gives
$m\mid k$, and the converse is immediate.

For $T\ge3$ the dissipation has a two-sided deviation bound:
\begin{equation}\label{eq:dissipation-exceptional}
 \frac1T\left|\left\{t\in[0,T]:
   |S(t)-2\log\log T|>E\right\}\right|
 \ll \exp\{-c e^{E/2}\}\qquad(E\ge0)
\end{equation}
for an absolute $c>0$. Indeed, $d_p=2/p+O(p^{-3})$ and
$1-\cos(t\omega_p)\ll T^2/p^2$ for $0\le t\le T$, $p>T$.
The tail above $T$ is $O(T^2\sum_{n>T}n^{-3})=O(1)$, so Mertens'
theorem gives, uniformly on this interval,
\[
 S(t)-2\log\log T=-2\sum_{3\le p\le T}\frac{\cos(t\omega_p)}p+O(1).
\]
For $r\ge1$ choose $y=e^{C_1r}$, with $C_1$ fixed large enough that
$2rB\le\mathfrak b\log y$. The primes up to $\min(y,T)$ contribute
in absolute value at most $2\log r+O(1)$, including the factor two.
If $T>y$, apply \eqref{eq:cosine-exp-moment} with $b=\pm2r$ to the
remaining sum. Evenness gives the same integral on the positive
half-window. In either case, for $U_T$ uniform on $[0,T]$,
\[
 \mathbb E\exp\{\pm r(S(U_T)-2\log\log T)\}\le(C_2r^2)^r.
\]
Markov's inequality with $r=e^{E/2}/\sqrt{eC_2}$ bounds each tail
by $e^{-r}$ when $r\ge1$. A larger implied constant covers smaller $E$.

\begin{corollary}[Dissipation outside a fixed sparse set]
\label{cor:dissipation-sparse-mas11}
For every fixed $\delta>0$ there are constants $C,t_0$ and a Borel
set $B_\delta\subseteq[t_0,\infty)$ such that
\begin{equation}\label{eq:dissipation-sparse-mas11}
 \begin{gathered}
 |B_\delta\cap[0,T]|\ll_{\delta,A} T/(\log T)^A\quad(A>0),\\
 |S(t)-2\log\log t|\le C+(2+\delta)\log\log\log t
 \quad(t\ge t_0,\ t\notin B_\delta).
 \end{gathered}
\end{equation}
Consequently
\begin{equation}\label{eq:dissipation-density-one-mas11}
 \lim_{\substack{t\to\infty\\t\notin B_\delta}}
       \frac{S(t)}{\log\log t}=2.
\end{equation}
\end{corollary}

\begin{proof}
Fix $C=2$, take $t_0$ sufficiently large, and define $B_\delta$
as the set of violations of the pointwise inequality in
\eqref{eq:dissipation-sparse-mas11}. Local uniform convergence of
the cosine series gives continuity of $S$ and hence Borel measurability.
For $\sqrt T\le t\le T$, the two centerings $2\log\log t$ and
$2\log\log T$ differ by at most $2\log2$, and the triple logarithms
differ by $O(1/\log\log T)$. Thus \eqref{eq:dissipation-exceptional}
bounds
\[
 |B_\delta\cap[0,T]|\ll\sqrt T+
 T\exp\{-c_\delta(\log\log T)^{1+\delta/2}\}.
\]
This proves the measure bound for every $A>0$. Division of the
pointwise bound by $\log\log t$ gives the limit on the same complement.
The unrestricted limits are discussed in
\eqref{eq:dissipation-limit-ranges-mas11}.
\end{proof}

\chapter{Prime cancellation and activation thresholds}\label{ssec:annulus}

Fourier bounds require prime cancellation and sufficient weight in
the cancellation range. For the phases $\log(u/(u-1))$,
$\log(1+1/u)$ and $1/u$, \Cref{prop:vk-prime-prefix} applies on
$64\le N\le b_V|\tau|^{1/5}$ with an absolute effective constant
$b_V>0$. The proof of \Cref{thm:vk-fourier} uses the subwindow from
a dyadic rounding of
$\exp(K(\log|\tau|)^{2/3}(\log\log|\tau|)^{1/3})$ to
$b_V|\tau|^{1/5}$, with fixed sufficiently large $K$. Its growing
reciprocal-prime mass supplies the logarithmic dissipation bound.

The theorem below gives a power saving for the same phases on
$|\tau|^a\le N\le|\tau|^{b_A}$, where
$a=1/(r+1)+\delta$ and $b_A=1-\delta$.
For fixed $a,b_A$, this annulus has reciprocal-prime mass
$\log(b_A/a)+o(1)$; the heavier weights of
\Cref{cor:heavier-activation} supply mass proportional to
$(b_A^\alpha-a^\alpha)(\log|\tau|)^\alpha$ when $\alpha>0$.
The constants and thresholds depend on $r,\delta,j,\gamma$, fixed
before $|\tau|\to\infty$, with $\gamma<\gamma_0$.

To pass from integer interval sums to primes, the proof in
\Cref{ssec:annulus-consequences} applies Vaughan's identity
\eqref{eq:vk-vaughan}. A small outer factor gives the Type~I estimate
\eqref{eq:typeI-inner-direct}; Cauchy--Schwarz gives differenced
phases and the bilinear estimate \eqref{eq:bilinear-L2}.
The scale comparisons in \Cref{cor:bilinear-weighted} yield the
power saving.

Recall $e(z)=e^{2\pi iz}$.
For real $t>1$, write $\omega_t=\log(t/(t-1))$ and
$\omega_t^+=\log(1+1/t)$; $u\sim U$ means $U<u\le2U$.

\begin{theorem}[Cancellation in dyadic prime intervals]\label{thm:annulus}
\index{Prime cancellation!dyadic intervals}
Fix an integer $r\ge3$ and $0<\delta<1/(4r(r+1))$, and put
$a=1/(r+1)+\delta$, $b=1-\delta$. Choose an integer $j\ge2$ with
$a(j+1-j/r)>1$, and set
\[
 \begin{gathered}
 \theta=\min\left\{\frac{a(j+1-j/r)-1}{2^j-2},1-b\right\},\qquad
 \eta=\frac{(r+1)^2\delta}{2(2^{2r}-2)},\\
 \gamma_0=\min\{\theta,a\eta/2,a/2\}>0.
 \end{gathered}
\]
For every $0<\gamma<\gamma_0$, there are effective constants $C,T_0>0$
such that, for every real $\tau$ with $|\tau|\ge T_0$ and every dyadic
$N$ with $|\tau|^a\le N\le|\tau|^b$,
\begin{equation}\label{eq:annulus-retained-power}
 \sup_{N\le x\le2N}
 \left|\sum_{N<p\le x}e^{i\tau\omega_p}\right|
 \le C\frac N{\log N}|\tau|^{-\gamma}.
\end{equation}
In particular, for every $A>0$, after increasing the threshold,
\begin{equation}\label{eq:annulus_bound}
 \sup_{N\le x\le2N}
 \left|\sum_{N<p\le x}e^{i\tau\omega_p}\right|
 \le C_{r,A,\delta}^{\mathrm{ann}}\frac N{(\log N)^A}.
\end{equation}
Both assertions hold also with $\omega_p$ replaced by $\omega_p^+$ or
$1/p$, with common constants for all three phases.
\end{theorem}

One admissible choice of the parameters in \Cref{thm:annulus} is
\[
 r=3,\quad\delta=1/100,\quad j=5,\quad
 a=13/50,\quad b=99/100,\quad\gamma=1/20000;
\]
it is the choice used in the proofs of
\Cref{cor:positive-activation,thm:loglog-activation-mas03}.
In \Cref{cor:infinite-weighted-ac} the derivative order is chosen
as a function of the positive coefficient-count liminf.
\Cref{fig:sum-regularity} shows how the proof of \Cref{cor:infinite-weighted-ac} combines the
resulting cancellation estimate with \Cref{lem:conditional-block-smoothing}.

\section{Derivative bounds and exponential-sum estimates}\label{ssec:annulus-phase}

For each fixed derivative order, both logarithmic phases have the
reciprocal derivative profile with a controlled remainder. This lets
one use the same interval-sum estimate for all three phases. Uniformity on subintervals is needed later because
the product restrictions change the interval with the endpoint $x$.

\begin{lemma}[Phase derivatives]\label{lem:phase_derivs}
For an integer $k\ge1$, define, for $t>1$,
\begin{equation}\label{eq:rho-series}
 \rho_k(t):=\sum_{m\ge1}\frac1{m+1}\binom{k+m}{m}t^{-m}.
\end{equation}
The series converges absolutely, and the following hold.
\begin{enumerate}[label=\textup{(\alph*)}]
\item For $t>1$,
\[
 \omega_t^{(k)}=(-1)^{k-1}(k-1)!\bigl(t^{-k}-(t-1)^{-k}\bigr)
 =(-1)^kk!t^{-k-1}(1+\rho_k(t)),
\]
and $|\omega_t^{(k)}|$ is strictly decreasing.
\item The function $\rho_k$ is completely monotone on $(1,\infty)$.
For $a=(k+1)/2$, every integer $j\ge0$, and $t>a$,
\begin{equation}\label{eq:rho-all-orders}
 0<(-1)^j\rho_k^{(j)}(t)\le\frac{j!a}{(t-a)^{j+1}}.
\end{equation}
For $t\ge2(k+1)$,
$|\rho_k^{(j)}(t)|\le c_j^\rho(k+1)t^{-j-1}$ with
$c_j^\rho=(j!/2)(4/3)^{j+1}$; in particular,
\begin{equation}\label{eq:rho-bounds}
 \rho_k(t)\le\frac{2(k+1)}{3t},\qquad
 |\rho_k'(t)|\le\frac{8(k+1)}{9t^2}.
\end{equation}
\item For $\tau>0$, $\phi(t)=\tau\omega_t/(2\pi)$, and $t\ge2(k+1)$,
\[
 \frac{k!}{2\pi}<\frac{t^{k+1}|\phi^{(k)}(t)|}{\tau}
 \le\frac{2k!}{3\pi}.
\]
Thus $|\phi^{(k)}(t)|\asymp_k\tau/N^{k+1}$ with constant sign for
$t\sim N$, $N\ge4(k+1)$.
\item The plus-log phase satisfies
\[
 (\omega^+)^{(k)}(t)=(-1)^kk!t^{-k-1}(1+\rho_k^+(t)),\qquad
 \rho_k^+(t)=\int_0^1(1+v/t)^{-k-1}\,dv-1,
\]
where $-(k+1)/(2t)\le\rho_k^+(t)<0$.
$(\omega^+)^{(k)}(t)$ has sign $(-1)^k$ and strictly decreasing absolute value.
\end{enumerate}
\end{lemma}

\begin{proof}
Differentiating $\omega_t=\log t-\log(t-1)$ gives the closed form.
The derivative of $(t-1)^{-k}-t^{-k}$ is
$-k((t-1)^{-k-1}-t^{-k-1})<0$, proving strict decrease.
Also
$t^{-k}-(t-1)^{-k}=-k\int_{t-1}^{t}u^{-k-1}\,du$, whence, with the
substitution $s=t-u$,
\[
  \omega_t^{(k)}
  \;=\;(-1)^k\,k!\int_0^1(t-s)^{-k-1}\,ds
  \;=\;\frac{(-1)^k\,k!}{t^{k+1}}\int_0^1(1-s/t)^{-k-1}\,ds.
\]
For $|s/t|<1$ the binomial series gives
$(1-s/t)^{-k-1}=\sum_{m\ge 0}\binom{k+m}{m}(s/t)^m$, and term-by-term
integration yields $\rho_k(t)=\sum_{m\ge 1}\binom{k+m}{m}t^{-m}/(m+1)$,
which is~\eqref{eq:rho-series}; every term is positive, so $\rho_k(t)>0$.

Put $b_{k,m}=\binom{k+m}{m}/(m+1)$. Since
$b_{k,m}=\prod_{i=1}^m(k+i)/(i+1)\le a^m$, termwise differentiation
of the positive Laurent series proves complete monotonicity and bounds
$(-1)^j\rho_k^{(j)}(t)$ by the corresponding derivative of
$\sum_{m\ge1}(a/t)^m=a/(t-a)$ for $t>a$. This proves
\eqref{eq:rho-all-orders}. For $t\ge4a$ it gives the stated $c_j^\rho$,
including $c_0^\rho=2/3$ and $c_1^\rho=8/9$ in \eqref{eq:rho-bounds}.
Equality holds for $k=1$, $t=4$, since $\rho_1(t)=1/(t-1)$;
hence all these uniform constants are optimal.

Finally, $\phi^{(k)}=(\tau/2\pi)\omega_t^{(k)}$, so
\[
 \frac{t^{k+1}|\phi^{(k)}(t)|}{\tau}
 =\frac{k!}{2\pi}(1+\rho_k(t)).
\]
For $t\ge2(k+1)$, \eqref{eq:rho-bounds} gives
$0<\rho_k(t)\le1/3$, proving the stated bounds and the dyadic
estimate, with sign $(-1)^k$.
Finally, differentiating $\omega_t^+=\int_0^1(t+v)^{-1}\,dv$ gives
part~\textup{(d)}. The remainder bound follows by integrating
$0\le1-(1+x)^{-k-1}\le(k+1)x$; positivity of the derivative
integrands gives the sign and strict decrease.
\end{proof}

For integers $j\ge2$ put
\begin{equation}\label{eq:GK-setup}
 (\kappa_j,\nu_j):=\Bigl(\frac1{2^j-2},\,1-\frac{j-1}{2^j-2}\Bigr),
\end{equation}
the exponent pair of the $j$-th derivative test.

\begin{lemma}[Uniform estimate for phases with derivatives close to those of $-y/x$]\label{lem:GK-input}
Let \(j\ge2\) be an integer, let \(I=[U,2U]\), where \(U\ge2\), and let
\(\phi\in C^j(I)\) be real-valued. Suppose that there are
\(y\in\mathbb R\setminus\{0\}\) and
functions \(R_k:I\to\mathbb R\) such that
\begin{equation}\label{eq:GK-profile}
 \phi^{(k)}(x)
 =(-1)^{k-1}k!\,y\,x^{-(k+1)}(1+R_k(x)),
 \qquad
 \|R_k\|_{\infty,I}\le1/130
 \quad(1\le k\le j).
\end{equation}
Set \(Y_0:=|y|/U^2\). Then, for every subinterval \(J\subseteq I\),
\begin{equation}\label{eq:GK-input}
 \left|\sum_{n\in J}e(\phi(n))\right|
 \le42\bigl(Y_0^{\kappa_j}U^{\nu_j}+Y_0^{-1}\bigr).
\end{equation}
\end{lemma}

\begin{proof}
We adapt \cite[\S\S~2.1--2.4]{GK91} (Theorems~2.1, 2.2, 2.8, 2.9 and Lemmas~2.5, 2.7), making the constants explicit. For the treatment of logarithmic phases, see the proof of \cite[Theorem~2.14, (2.5.2), p.~19]{GK91} and the remark after \cite[Theorem~2.12, p.~18]{GK91}.
Conjugation permits $y>0$. Put $\varepsilon=1/130$.
We first estimate sums over intervals $(a,b]\subseteq I$ with
$a,b\in\Z$, and take $L\ge\max\{1,b-a\}$ to bound the length here
and for each correlation sum below. Write $S=\sum_n e(g(n))$ over the integers
in the current interval.

If $g'$ is monotone and takes values in $[m+\delta,m+1-\delta]$,
where $m\in\Z$ and $0<\delta\le1/2$, then
\begin{equation}\label{eq:phase-first-derivative}
 \left|\sum_n e(g(n))\right|\le\frac2{\pi\delta}
\end{equation}
on any consecutive integer block in its domain. Indeed, the increments
$\theta_n=g(n+1)-g(n)$ are monotone in the same band. Summation by parts
with
$w_n=(1-e(\theta_n))^{-1}=\tfrac12+\tfrac i2\cot(\pi\theta_n)$
bounds the endpoints by $\csc(\pi\delta)$ and the variation by
$\cot(\pi\delta)$. Their sum is
$\cot(\pi\delta/2)\le2/(\pi\delta)$; singletons satisfy the same bound.

Put $c_A=5/2+2/\pi$ and $c_B=5+6/\pi$.
If $\lambda\le|g''|\le\alpha\lambda$, with $\alpha\ge1$, on an
interval $(a,b]\subseteq I$ with $a,b\in\Z$ and $b-a\le L$, then
\begin{equation}\label{eq:phase-second-derivative}
 \left|\sum_n e(g(n))\right|
 \le c_A\alpha L\sqrt\lambda+c_B\lambda^{-1/2}
\end{equation}
for its half-open integer block. For $\lambda<1/4$, conjugate if necessary
and put $\delta=\sqrt\lambda$. The range of $g'$ has length at most
$\alpha L\lambda$. At most $\alpha L\lambda+2$ intervals
$(m-\delta,m+\delta)$ meet that range; each inverse image contains at
most $2\delta/\lambda+1$ integers. The complement has at most
$\alpha L\lambda+3$ components. Applying
\eqref{eq:phase-first-derivative} to their integer blocks gives
\[
 |S|\le(\alpha L\lambda+2)(2\lambda^{-1/2}+1)
       +(\alpha L\lambda+3)\frac2{\pi\sqrt\lambda},
\]
which implies \eqref{eq:phase-second-derivative} since
$\lambda\le\sqrt\lambda/2$ and $2\le\lambda^{-1/2}$.
For $\lambda\ge1/4$ the trivial bound $|S|\le L$ suffices.

Under hypothesis \eqref{eq:GK-profile}, set $F=y/U$ and $L=U$.
Its first two derivatives give
\[
 \frac{(1-\varepsilon)F}{4U}\le \phi'\le
 \frac{(1+\varepsilon)F}{U},\qquad
 \lambda_2=\frac{(1-\varepsilon)F}{4U^2},\qquad
 \alpha_2=\frac{8(1+\varepsilon)}{1-\varepsilon}.
\]
If $(1+\varepsilon)F/U\le1/2$, use
\eqref{eq:phase-first-derivative}; otherwise $U/F<2(1+\varepsilon)$,
and \eqref{eq:phase-second-derivative} gives
\begin{equation}\label{eq:phase-base-estimate}
 |S|\le C_0\max\{F^{1/2},U/F\},\qquad
 C_0=\max\left\{\frac8{\pi(1-\varepsilon)},
 \frac{4(1+\varepsilon)(c_A+c_B)}{\sqrt{1-\varepsilon}}\right\}.
\end{equation}

For higher derivatives, extend the sequence of $N\le L$ summands by
zero outside its interval.
For real $1\le H\le L$, use $H_0=\lceil H\rceil$ shifts and
Cauchy--Schwarz. The shifted support has $N+H_0-1$ points, so
\begin{equation}\label{eq:phase-one-step}
 |S|^2\le\frac{N+H_0-1}{H_0}
 \left(N+2\sum_{1\le h<H_0}(1-h/H_0)|S_1(h)|\right)
 \le\frac{2L^2}{H}+\frac{4L}{H}\sum_{1\le h<H}|S_1(h)|.
\end{equation}
Here $\Delta_hg(n):=g(n+h)-g(n)$ and, for $q\ge1$, $S_q(h_1,\ldots,h_q):=\sum_n e(\Delta_{h_1}\cdots\Delta_{h_q}g(n))$, where the sum runs over the intersection of the supports of all the shifted summands.
For $q\ge1$, put
$Q=2^q$, $H_i=H^{2^i/Q}$ $(1\le i\le q)$, and
$P=\prod_iH_i=H^{2-2/Q}$. Iteration gives
\begin{equation}\label{eq:phase-differencing}
 |S|^Q\le8^{Q-1}\left\{
 \frac{L^Q}{H}+\frac{L^{Q-1}}P
 \sum_{1\le h_i<H_i}|S_q(h_1,\ldots,h_q)|\right\}.
\end{equation}
For $q=1$ this follows from \eqref{eq:phase-one-step}.
For the induction, use \eqref{eq:phase-differencing} at $\sqrt H$, so
$P_{\mathrm{old}}=H^{1-1/Q}$. Square, apply Cauchy--Schwarz to at most
$P_{\mathrm{old}}$ correlations, and apply \eqref{eq:phase-one-step} to each:
\[
\begin{aligned}
 |S|^{2Q}
 &\le2\cdot8^{2Q-2}\left\{\frac{L^{2Q}}H
       +\frac{L^{2Q-2}}{P_{\mathrm{old}}}\sum|S_q|^2\right\}\\
 &\le6\cdot8^{2Q-2}\frac{L^{2Q}}H
       +8\cdot8^{2Q-2}\frac{L^{2Q-1}}{P_{\mathrm{old}}H}
         \sum|S_{q+1}|.
\end{aligned}
\]
Both coefficients are at most $8^{2Q-1}$, and the new product is
$P_{\mathrm{old}}H$. Every correlation is taken on the intersection of
its shifted domains, with length at most $L$.
If $0<H<1$, the shift sums are empty and
$L^Q/H\ge L^Q\ge|S|^Q$, so \eqref{eq:phase-differencing} still holds.

Suppose $\lambda\le|\phi^{(q+2)}|\le\alpha\lambda$.
The second derivative of each $q$-fold difference is an iterated
integral of $\phi^{(q+2)}$ and hence has absolute value between
$\lambda h_1\cdots h_q$ and $\alpha\lambda h_1\cdots h_q$.
Its domain is the intersection of the shifted original intervals;
its half-open integer block has at most $L$ terms. Thus
\eqref{eq:phase-second-derivative} applies with the same $L$.
Using
$\sum_{1\le h<T}h^{1/2}\le T^{3/2}$ and
$\sum_{1\le h<T}h^{-1/2}\le2T^{1/2}$ in
\eqref{eq:phase-differencing} gives
\begin{equation}\label{eq:EN-unoptimized}
 |S|^Q\le8^{Q-1}\left\{
 \frac{L^Q}{H}+c_A\alpha\sqrt\lambda L^QH^{1-1/Q}
 +Qc_B\lambda^{-1/2}L^{Q-1}H^{-1+1/Q}\right\}.
\end{equation}
Write its braces as $a/H+bH^\beta+cH^{-\beta}$, where
$\beta=1-1/Q$, and take
\[
 H=\min\left\{L,\max\left\{
 (a/b)^{1/(1+\beta)},(c/b)^{1/(2\beta)}\right\}\right\}.
\]
If $H<L$, the increasing term dominates the two decreasing terms;
if $H=L$, at least one decreasing term dominates the increasing one.
Consequently the braces are at most
\[
 3\max\{a^{\beta/(1+\beta)}b^{1/(1+\beta)},
          \sqrt{bc},a/L,cL^{-\beta}\}.
\]
After taking $Q$th roots, the $a/L$ term is dominated by the
$\sqrt{bc}$ term. Therefore, with $D_q=(3\cdot8^{Q-1})^{1/Q}$,
\begin{equation}\label{eq:phase-higher-derivative}
\begin{split}
 |S|\le D_q\max\bigl\{&
 c_A^{1/(2Q-1)}L(\alpha^2\lambda)^{1/(4Q-2)},\\
 &(Qc_Ac_B\alpha)^{1/(2Q)}L^{1-1/(2Q)},\\
 &(Qc_B)^{1/Q}L^{(1-1/Q)^2}\lambda^{-1/(2Q)}\bigr\}.
\end{split}
\end{equation}

For $j\ge3$, put $q=j-2\ge1$ and $L=U$; hypothesis
\eqref{eq:GK-profile} permits
\[
 \lambda=(q+2)!(1-\varepsilon)2^{-q-3}FU^{-q-2},
 \qquad \alpha=2^{q+3}\frac{1+\varepsilon}{1-\varepsilon}.
\]
Equation~\eqref{eq:phase-higher-derivative} is
$|S|\le\max\{\mathcal C_1M_3,\mathcal C_2M_4,\mathcal C_3M_5\}$, where
\[
\begin{aligned}
 M_3&=F^{1/(4Q-2)}U^{1-(q+2)/(4Q-2)},\\
 M_4&=U^{1-1/(2Q)},\\
 M_5&=F^{-1/(2Q)}U^{1+(q-2)/(2Q)+1/Q^2},\\
 \mathcal C_1&=D_qc_A^{1/(2Q-1)}
 \left(\frac{2^{q+3}(q+2)!(1+\varepsilon)^2}{1-\varepsilon}
 \right)^{1/(4Q-2)},\\
 \mathcal C_2&=D_q\left(2^{2q+3}c_Ac_B
           \frac{1+\varepsilon}{1-\varepsilon}\right)^{1/(2Q)},\\
 \mathcal C_3&=D_q(2^qc_B)^{1/Q}
 \left((q+2)!(1-\varepsilon)2^{-q-3}\right)^{-1/(2Q)}.
\end{aligned}
\]
Put $R=(1+\varepsilon)/(1-\varepsilon)$,
$T=(1+\varepsilon)^2/(1-\varepsilon)$, and $d=\log(8/3)$.
First,
\[
 (\mathcal C_3/\mathcal C_2)^{2Q}
 =\frac{Qc_B}{c_A(q+2)!(1+\varepsilon)}<1,
\]
since $Q/(q+2)!$ decreases from $1/3$ and $c_B<3c_A$.
Next, for $b_0=\tfrac12\log((9/8)c_Ac_BR)>0$,
\[
 \log(\mathcal C_2/8)=\frac{q\log2+b_0}{Q},
\]
which strictly decreases with $q$, since its value at $q$ minus its
value at $q+1$ is $((q-1)\log2+b_0)/(2Q)>0$.
For $A_q=2\log c_A+(q+3)\log2+\log((q+2)!)+\log T$ and
$s_q=\log(2(q+3))$, one has
\[
 \log(\mathcal C_1/8)=\frac{A_q}{4Q-2}-\frac dQ,
 \qquad A_{q+1}=A_q+s_q.
\]
Here $A_1-s_1=\log(12c_A^2T)>\log108>4d$, and
$A_q-s_q$ increases because
$2s_q-s_{q+1}=\log(2(q+3)^2/(q+4))>0$. Hence
\[
 \frac{4QA_q-(4Q-2)s_q}{(4Q-2)(8Q-2)}>\frac d{2Q},
\]
so $\mathcal C_1$ also decreases with $q$.
At $q=1$,
$(\mathcal C_1/\mathcal C_2)^{12}
 =9c_A(1-\varepsilon^2)/(32c_B^3)<1$.
Thus, for every $q\ge1$,
\begin{equation}\label{eq:phase-order-constant}
 \max_i\mathcal C_i(q)\le\mathcal C_2(1)
 =(18432c_Ac_BR)^{1/4}
 <\left(\frac{22536192}{43}\right)^{1/4}<27,
\end{equation}
using $c_A<4$, $c_B<7$, and $R=131/129$.

The induction in the order incurs no further constant.
Suppose its preceding bound is
$|S|\le C\max\{M_1,M_2\}$, with
\[
 M_1=F^{1/(2Q-2)}U^{1-(q+1)/(2Q-2)},\qquad M_2=U/F,
\]
and the higher-derivative bound is
$|S|\le C\max\{M_3,M_4,M_5\}$ for the same $C$.
The exact identities
\[
 M_1^{(Q-1)/(2Q-1)}M_4^{Q/(2Q-1)}=M_3,
\]
\[
 M_1^{(Q-1)(3Q-1)/(2Q-1)^2}M_5^{Q^2/(2Q-1)^2}
 =M_3U^{-(Q-1)/(2(2Q-1)^2)}\le M_3
\]
show that, if $M_1>\max\{M_2,M_3\}$, both $M_4$ and $M_5$ are at
most $M_3$; otherwise the preceding bound suffices. Therefore
$|S|\le C\max\{M_2,M_3\}$ at every order.
The base case is \eqref{eq:phase-base-estimate}. Since $\pi>31/10$ and
$\sqrt{130/129}<259/258$, its two constants satisfy
\[
 \frac8{\pi(1-\varepsilon)}<3,
 \qquad
 \frac{4(1+\varepsilon)(c_A+c_B)}{\sqrt{1-\varepsilon}}
 <\frac{4241125}{103974}<41.
\]
Together with \eqref{eq:phase-order-constant}, this permits $C=41$
for all orders: the derivative-test constants are bounded uniformly in
$q$, and each interpolation uses powers summing to one, preserving
the same $C$. For $j\ge3$, $M_3=Y_0^{\kappa_j}U^{\nu_j}$ and $M_2=Y_0^{-1}$.
For $j=2$, \eqref{eq:phase-base-estimate} gives the desired estimate since
$F^{1/2}=Y_0^{1/2}U^{1/2}$ and $U/F=Y_0^{-1}$.

For an arbitrary subinterval $J\subseteq[U,2U]$ with at least two
integer points, let $a,b$ be its least and greatest integers and remove
the single term at $a$. The remaining sum is over $(a,b]$, with
$b-a\le U$, and has the preceding bound. Moreover $\max\{Y_0^{\kappa_j}U^{\nu_j},Y_0^{-1}\}\ge1$:
use the second term for $Y_0\le1$, and the first for $Y_0\ge1$.
Thus restoring the removed term, or treating an empty block or singleton directly,
gives \eqref{eq:GK-input}.
\end{proof}

\KeepHeadingSpace{7\baselineskip}
\section{Divisor coefficient bounds}
\label{ssec:annulus-coefficients}

Arithmetic convolutions create coefficients by collecting many
factorizations of the same integer. Recall that $\mathbf d_k(n)$ counts
ordered factorizations of $n$ into $k$ positive integers. The bilinear
estimate will depend on squared coefficient sums, so second-moment
bounds for $\mathbf d_k$ control the arithmetic cost of Cauchy--Schwarz.

For integers $r\ge2$, put
\begin{equation}\label{eq:HB-constants}
 \widetilde c_r=2^{2r+1},\qquad
 B_r=r^3 2^{2r},\qquad H_r=2^{r+1}.
\end{equation}
These constants control the block scales, number of pieces and coefficient
bounds, respectively, in the Heath--Brown reduction of
\Cref{lem:HB-typeI-II}.

\begin{lemma}[Explicit divisor bounds]\label{lem:eff-divisor-moment}
For integers $k\ge1$, $r\ge2$, $m\ge0$, $n\ge1$, and real $x\ge1$,
\begin{equation}\label{eq:binom-sq-domination}
 \binom{m+k-1}{k-1}^{2}\le\binom{m+k^2-1}{k^2-1},
\end{equation}
\begin{equation}\label{eq:taur-sq-domination}
 \mathbf d_r(n)^2\le\mathbf d_{r^2}(n),
\end{equation}
\begin{equation}\label{eq:divisor-harmonic-bound}
 \sum_{n\le x}\mathbf d_k(n)\le
 x\left(\sum_{m\le x}m^{-1}\right)^{k-1}\le x(1+\log x)^{k-1},
\end{equation}
\begin{equation}\label{eq:mardjanichvili}
 \sum_{n\le x}\mathbf d_k(n)\le x(\log x+k)^{k-1}/(k-1)!.
\end{equation}
In particular, with $q=r^2$,
\begin{equation}\label{eq:eff-div-moment-full}
 \sum_{n\le x}\mathbf d_r(n)^2\le
 x\min\{(1+\log x)^{q-1},(\log x+q)^{q-1}/(q-1)!\}.
\end{equation}
For $U\ge2$,
\begin{equation}\label{eq:eff-div-moment-dyadic}
 \sum_{U<n\le2U}\mathbf d_r(n)^2\le2^qU(1+\log U)^{q-1}.
\end{equation}
\end{lemma}
\begin{proof}
The ratio of the right to left binomials in
\eqref{eq:binom-sq-domination} starts at $1$ when $m=0$ and, for $m\ge1$,
has consecutive ratio $m(m+k^2-1)/(m+k-1)^2\ge1$: the numerator
minus the denominator is $(m-1)(k-1)^2$. Multiplicativity proves
\eqref{eq:taur-sq-domination}; cf.~\cite[(4), p.~180]{Pan94}.
Counting ordered factorizations and summing the last factor proves
\eqref{eq:divisor-harmonic-bound}. For the factorial refinement,
a weakened form of Mardzhanishvili's inequality~\cite{Mar39},
induct using $\mathbf d_k=\mathbf1*\mathbf d_{k-1}$.
For $k\ge2$, $L=\log x$ and $A=L+k-1$, the decreasing function
$g(t)=t^{-1}(\log(x/t)+k-1)^{k-2}$ gives
\[
 \sum_{d\le x}g(d)\le g(1)+\int_1^xg(t)\,dt
 =A^{k-2}+\frac{A^{k-1}-(k-1)^{k-1}}{k-1}
 \le\frac{(A+1)^{k-1}}{k-1}.
\]
The last step is the binomial theorem, proving
\eqref{eq:mardjanichvili}; cf.~\cite[p.~181]{Pan94}.
The base case is $\lfloor x\rfloor\le x$. The classical order
of the divisor-moment bound is also recorded in
\cite[equation~(1.80)]{IK04}.
The second-moment bound follows from \eqref{eq:taur-sq-domination}.
Apply its harmonic alternative at $2U$ and use
$1+\log(2U)\le2(1+\log U)$ for the dyadic bound.
\end{proof}
For every fixed integer $k\ge2$, write, initially for $\Re s>1$,
\[
 \sum_{n\ge1}\frac{\mathbf d_k(n)^2}{n^s}=\zeta(s)^{k^2}G_k(s),\qquad
 G_k(s)=\prod_p(1-p^{-s})^{k^2}
             \sum_{m\ge0}\binom{m+k-1}{k-1}^{2}p^{-ms}.
\]
The $p^{-s}$ coefficient cancels in each local factor, and the
remaining Taylor coefficients grow polynomially in their degree. Thus
$G_k$ has an absolutely convergent Dirichlet series on $\Re s>1/2$
and is analytic there; locally its factors are $1+O_k(p^{-2\Re s})$.
At $s=1$ every factor
is positive and their deviations from $1$ are summable; hence
$c_k=G_k(1)>0$. Selberg--Delange~\cite[Theorem~II.5.2, p.~281]{Ten15}
therefore gives
\[
 \sum_{n\le x}\mathbf d_k(n)^2\sim
 c_k x(\log x)^{k^2-1}/\Gamma(k^2).
\]
Thus the logarithmic exponent in \Cref{lem:eff-divisor-moment}
is optimal for fixed $k$.

\section{Type~I and Type~II estimates}\label{ssec:annulus-estimates}

In Type~I sums, fixing the short outer factor $m$ leaves a long
interval sum in $n$. Apply the derivative test to this inner sum first.
Abel summation then handles a logarithmic inner weight, and divisor
bounds control the sum over the outer coefficients.

\begin{lemma}[Uniform inner Type~I estimate]\label{lem:typeI-uniform}
Fix integers $r\ge3$, $j\ge2$, a constant $C>0$, and
$0<a\le b<1$ such that $a(j+1-j/r)>1$. Put
\[
 \kappa=(2^j-2)^{-1},\qquad
 \theta=\min\{\kappa[a(j+1-j/r)-1],1-b\}>0.
\]
For $\varphi(t)\in\{\log(t/(t-1)),\log(1+1/t),1/t\}$, every
$\tau\ge\max\{2,N_*^{1/a}\}$, where
$N_*=\max\{4,130(j+1),(2C)^{r/(r-1)}\}$, every
$\tau^a\le N\le\tau^b$, every integer $1\le m\le CN^{1/r}$,
and every interval $J\subseteq[N/m,2N/m]$ satisfy
\begin{equation}\label{eq:typeI-inner-direct}
 \left|\sum_{n\in J\cap\mathbb Z}e^{i\tau\varphi(mn)}\right|
 \le K_{j,C}\frac Nm\tau^{-\theta},\qquad
 K_{j,C}=42\{(2\pi)^{-\kappa}C^{j\kappa}+2\pi\}.
\end{equation}
\end{lemma}
\begin{proof}
Set $U=N/m$ and $\phi_m(t)=\tau\varphi(mt)/(2\pi)$.
The threshold gives $U\ge N^{1-1/r}/C\ge2$.
For $1\le k\le j$ and $U\le t\le2U$,
\[
 \phi_m^{(k)}(t)=(-1)^{k-1}k!y_m t^{-k-1}(1+R_k(t)),\qquad
 y_m=-\frac{\tau}{2\pi m}.
\]
For the minus-log phase, \Cref{lem:phase_derivs} gives
$R_k(t)=\rho_k(mt)$; since $mt\ge N\ge130(j+1)$,
\eqref{eq:rho-bounds} gives $0\le R_k(t)\le2(j+1)/(3N)<1/130$.
For the plus-log phase, $R_k(t)=\rho_k^+(mt)$ and
$|R_k(t)|\le(j+1)/(2N)<1/130$. For the reciprocal phase $R_k=0$.
Thus \Cref{lem:GK-input} applies with $Y_0=\tau m/(2\pi N^2)$.
Since $\nu_j=1-(j-1)\kappa$, it gives
\[
 \left|\sum_{n\in J\cap\mathbb Z}e^{i\tau\varphi(mn)}\right|
 \le42\frac Nm\left\{
 (2\pi)^{-\kappa}\left(\frac{\tau m^j}{N^{j+1}}\right)^\kappa
 +2\pi\frac N\tau\right\}.
\]
Now $m\le CN^{1/r}$ and $N\ge\tau^a$ bound the first expression
in braces by
\[
 (2\pi)^{-\kappa}C^{j\kappa}
 \tau^{-\kappa[a(j+1-j/r)-1]},
\]
while $N\le\tau^b$ bounds the second by
$2\pi\tau^{-(1-b)}$. This proves the assertion.
\end{proof}

\begin{corollary}[Divisor-weighted Type~I sums]\label{cor:typeI-weighted}
\index{Type I sums}
Fix $r,j,a,b$ and $\theta$ as in \Cref{lem:typeI-uniform}. Suppose
$\tau^a\le N\le\tau^b$ is dyadic and $\mathcal M,\mathcal W$ are
dyadic blocks independent of $x$, of scales $M,W$, allowing
$\mathcal M=\{1\}$ with $M=1$. Assume
$M\le\widetilde c_rN^{1/r}$ and $N/4<MW<2N$.
For any phase in \Cref{lem:typeI-uniform}, complex coefficients
independent of $x$, and $w(n)=1$ or $w(n)=\log n$, put
\begin{equation}\label{eq:typeI-weighted-form}
\begin{gathered}
 \mathcal T(x)=\sum_{m\in\mathcal M}a_m
  \sum_{\substack{n\in\mathcal W\\N<mn\le x}}
       w(n)e^{i\tau\varphi(mn)},\\
 |a_m|\le H_r\mathbf d_{2r}(m)(\log N)^{2r}.
\end{gathered}
\end{equation}
For sufficiently large $\tau$, uniformly over these data,
\begin{equation}\label{eq:typeI-retained-power}
 \sup_{N\le x\le2N}|\mathcal T(x)|
 \ll_{r,j,a,b}H_rN\tau^{-\theta}(\log N)^{4r+1},
\end{equation}
with an effective implied constant and threshold.
For every $A>0$, after increasing the threshold, also
\[
 \sup_{N\le x\le2N}|\mathcal T(x)|
 \ll_{r,A,a,b}H_rN(\log N)^{-A}.
\]
\end{corollary}
\begin{proof}
Apply \eqref{eq:typeI-inner-direct} with $C=2\widetilde c_r$.
For each $m\in\mathcal M$, one has $m\le2M$, and
$\mathcal W\cap(N/m,x/m]$ is an interval of integers inside
$[N/m,2N/m]$. Every prefix obeys the same bound.
Abel summation incurs a factor at most $\log(4N/m)\le2\log N$
for $w(n)=\log n$, and none for $w=1$.
By ordered factorization, for $X\ge1$,
\[
 \sum_{m\le X}\frac{\mathbf d_{2r}(m)}m
 \le\left(\sum_{d\le X}\frac1d\right)^{2r}
 \le(1+\log X)^{2r}.
\]
Consequently, uniformly in $x$,
\[
 |\mathcal T(x)|\ll_{r,j}
 H_rN\tau^{-\theta}(\log N)^{2r+1}(1+\log(2M))^{2r}
 \ll_{r,j}H_rN\tau^{-\theta}(\log N)^{4r+1},
\]
proving \eqref{eq:typeI-retained-power}. For the last conclusion,
choose an admissible $j$ in terms of $r,a$ alone.
Since $\tau^{-\theta}\le N^{-\theta/b}$, this power absorbs
$(\log N)^{A+4r+1}$ after an effective increase of the threshold.
\end{proof}

For any fixed $0<a\le b<1$, one can choose $j$ so that
$a(j+1-j/r)>1$. Thus the restriction $a>1/(r+1)$ in
\Cref{thm:annulus} comes from the Type~II estimate.

In Type~II sums both variables carry coefficients. Cauchy--Schwarz
separates one coefficient sequence into its squared norm and creates
correlations between two values $v_1,v_2$ of the other variable. The
diagonal is bounded directly. For the off-diagonal terms, the useful
fact is that the phase difference still has a reciprocal derivative
profile, with size depending on the separation $h=v_2-v_1$.

\begin{lemma}[Differenced phase derivatives]\label{lem:typeII-A}
Let $\tau>0$, $0<v_1<v_2$, $h=v_2-v_1$, $a_i=1/v_i$, and $u>a_1$.
For $\psi_h(u)=\tau(\omega_{uv_2}-\omega_{uv_1})/(2\pi)$, the
following hold for every integer $j\ge0$.
\begin{enumerate}[label=\textup{(\alph*)}]
\item The exact derivative is
\begin{equation}\label{eq:psi-integral}
 \psi_h^{(j)}(u)=\frac{(-1)^{j+1}\tau j!}{2\pi}
       \int_{a_2}^{a_1}(u-s)^{-j-1}\,ds.
\end{equation}
\item Writing
\begin{equation}\label{eq:psi-derivative}
\begin{gathered}
 \psi_h^{(j)}(u)=\frac{(-1)^{j+1}\tau j!h}
 {2\pi u^{j+1}v_1v_2}(1+R_j),\\
 R_j=\frac1{a_1-a_2}\int_{a_2}^{a_1}(1-s/u)^{-j-1}\,ds-1,
\end{gathered}
\end{equation}
one has $0<R_j\le(1-1/(uv_1))^{-j-1}-1$ and, for $uv_1>j+1$,
$R_j\le(j+1)/(uv_1-j-1)$. Also
\[
 \frac{j+1}{u-1/v_2}\le
 \frac{|\psi_h^{(j+1)}(u)|}{|\psi_h^{(j)}(u)|}
 \le\frac{j+1}{u-1/v_1}.
\]
\item For $j\ge1$, $u\in[U,2U]$, $v_i\in(V,2V]$, and
$UV\ge3(j+1)$,
\begin{equation}\label{eq:psi-magnitude}
 \frac{j!}{2^{j+4}\pi}\frac{\tau h}{U^{j+1}V^2}
 \le|\psi_h^{(j)}(u)|
 \le\frac{3j!}{4\pi}\frac{\tau h}{U^{j+1}V^2},
\end{equation}
and the ratio of consecutive derivative magnitudes is
$\asymp_j(j+1)/U$.
\item For $\psi_h^+(u)=\tau(\omega_{uv_2}^+-\omega_{uv_1}^+)/(2\pi)$,
\[
 (\psi_h^+)^{(j)}(u)
 =\frac{(-1)^{j+1}\tau j!}{2\pi}
       \int_{a_2}^{a_1}(u+s)^{-j-1}\,ds
 =\frac{(-1)^{j+1}\tau j!h}{2\pi u^{j+1}v_1v_2}(1+R_j^+),
\]
where $-(j+1)/(uv_1)\le R_j^+<0$. The derivative ratio lies between
$(j+1)/(u+1/v_1)$ and $(j+1)/(u+1/v_2)$.
Under the hypotheses of part~\textup{(c)}, $|(\psi_h^+)^{(j)}(u)|$ satisfies
\eqref{eq:psi-magnitude} with the lower bound multiplied by $2/3$,
and the same derivative-ratio comparison holds.
\end{enumerate}
\end{lemma}

\begin{proof}
Canceling $\log u$ gives
$\psi_h(u)=\tau[\log(u-a_1)-\log(u-a_2)]/(2\pi)$.
Integration and differentiation give \eqref{eq:psi-integral}; the
quantity $1+R_j$ is the average of $(1-s/u)^{-j-1}$ over $[a_2,a_1]$.
Positivity and monotonicity give \eqref{eq:psi-derivative}.
Bernoulli's inequality $(1-x)^{j+1}\ge1-(j+1)x$ gives the rational
remainder bound when its denominator is positive. The derivative ratio
is $(j+1)$ times a positive weighted average of $(u-s)^{-1}$, giving
its bounds. Under the assumption $UV\ge3(j+1)$, $R_j\le1/2$ and
$V^2<v_1v_2\le4V^2$; these imply \eqref{eq:psi-magnitude} and
$R_j\le3(j+1)/(2uv_1)$.
For part~\textup{(d)}, cancel $\log u$ and write
$\psi_h^+(u)=\tau[\log(u+a_2)-\log(u+a_1)]/(2\pi)$.
The same integration gives the displayed formula. Integrating
$0\le1-(1+x)^{-j-1}\le(j+1)x$ gives the remainder bound.
The derivative ratio is $(j+1)$ times a positive weighted average of
$(u+s)^{-1}$. Finally $1+R_j^+\ge2/3$ under the hypotheses of
part~\textup{(c)}, proving its magnitude bounds.
\end{proof}

\begin{lemma}[Differenced-phase exponent-pair estimate]\label{lem:typeII-B}
Let \(j\ge2\) be an integer, let \(U\ge2\), \(V>0\), and \(\tau>0\).  Let
\(v_1,v_2\in(V,2V]\) with \(v_2>v_1\), and
let \(\psi_h\) be either logarithmic phase of \Cref{lem:typeII-A}. Suppose
\[
 UV>K_j:=131(j+1).
\]
With \(h=v_2-v_1\), define
\begin{equation}\label{eq:lambda-j-def}
 \lambda_{j,0}:=\frac{\tau h}{2\pi U^{j+1}v_1v_2},\qquad
 \lambda_j(h):=\frac{\tau h}{U^{j+1}V^2}.
\end{equation}
Then, for every subinterval \(J\subseteq[U,2U]\),
\begin{equation}\label{eq:inner-u-bound-standard}
 \left|\sum_{u\in J}e(\psi_h(u))\right|
 \le42\left(\lambda_{j,0}^{\kappa_j}U
       +\frac1{\lambda_{j,0}U^{j-1}}\right).
\end{equation}
In terms of $\lambda_j(h)$,
\begin{equation}\label{eq:inner-u-bound-standard-nominal}
 \left|\sum_{u\in J}e(\psi_h(u))\right|
 \le336\pi\left(\lambda_j(h)^{\kappa_j}U
       +\frac1{\lambda_j(h)U^{j-1}}\right).
\end{equation}
\end{lemma}

\begin{proof}
Parts~\textup{(b)} and~\textup{(d)} of \Cref{lem:typeII-A}, applied at each
\(1\le k\le j\), give, with the respective signed remainder,
\[
 \psi_h^{(k)}(u)
 =(-1)^{k-1}k!\,y_\psi u^{-(k+1)}
   \bigl(1+R_k(u,v_1,v_2)\bigr),
 \qquad
 y_\psi:=\frac{\tau h}{2\pi v_1v_2},
\]
and
\[
 \max_{1\le k\le j}\|R_k\|_\infty
 \le\frac{j+1}{UV-j-1}
 <\frac1{130}.
\]
Applying \Cref{lem:GK-input} with $y=y_\psi$, so that
$Y_0=|y_\psi|/U^2=\lambda_{j,0}U^{j-1}$, and using
\((j-1)\kappa_j+\nu_j=1\), proves
\eqref{eq:inner-u-bound-standard}.  Since
\(V^2<v_1v_2\le4V^2\), one has
\[
 \frac{\lambda_j(h)}{8\pi}\le\lambda_{j,0}
 <\frac{\lambda_j(h)}{2\pi}.
\]
Thus the positive-power term in \eqref{eq:inner-u-bound-standard}
has coefficient at most $42$, and its reciprocal term,
$(\lambda_{j,0}U^{j-1})^{-1}$, has coefficient
at most $42\cdot8\pi=336\pi$ after conversion to the scale
$\lambda_j(h)$, proving \eqref{eq:inner-u-bound-standard-nominal}.
\end{proof}
\begin{proposition}[Bilinear estimate for arbitrary coefficients]
\label{prop:bilinear}
\index{Type II sums}
Let $r\ge2$ be an integer, $U\ge2$, $V\ge1$,
$UV>131(2r+1)$, and $\tau>0$.
Let $(a_u)$ and $(\beta_v)$ be arbitrary complex coefficients supported
on $(U,2U]\cap\mathbb Z$ and $(V,2V]\cap\mathbb Z$, respectively, and put
\[
 A_2=\sum_{u\sim U}|a_u|^2,\qquad
 B_2=\sum_{v\sim V}|\beta_v|^2,\qquad
 \kappa=\frac1{2^{2r}-2}.
\]
For $\varphi(t)\in\{\log(t/(t-1)),\log(1+1/t),1/t\}$, set
\[
 W(q,Q)=\sum_{u\sim U}a_u\sum_{v\sim V}\beta_v
       \mathbf1_{\{q<uv\le Q\}}e^{i\tau\varphi(uv)}.
\]
Uniformly for real $0\le q\le Q$,
\begin{equation}\label{eq:bilinear-L2}
 |W(q,Q)|^2\le\frac{672\pi}{\log2}\,A_2B_2UV
 \left\{\frac1V+
 \left(\frac{\tau}{U^{2r}UV}\right)^\kappa
 +\frac{UV}{\tau}\log(2V)\right\}.
\end{equation}
\end{proposition}

\begin{proof}
Cauchy--Schwarz in $u$ gives
\begin{equation}\label{eq:CS-truncated}
 |W(q,Q)|^2\le A_2\sum_{u\sim U}
 \left|\sum_{v\sim V}\beta_v
 \mathbf1_{\{q<uv\le Q\}}e^{i\tau\varphi(uv)}\right|^2.
\end{equation}
The diagonal in the last sum is at most $2UB_2$, since
$(U,2U]$ contains at most $U+1\le2U$ integers.
For $v_2>v_1$ put $h=v_2-v_1$ and
$\psi_h(u)=\tau(\varphi(uv_2)-\varphi(uv_1))/(2\pi)$.
The off-diagonal part is
\begin{equation}\label{eq:offdiag-truncated}
 2\operatorname{Re}\sum_{\substack{v_1,v_2\sim V\\v_2>v_1}}
 \beta_{v_2}\overline{\beta_{v_1}}
 \sum_{u\in J(v_1,v_2;q,Q)\cap\mathbb Z}e(\psi_h(u)),
\end{equation}
where the exact support intersection is
\[
 J(v_1,v_2;q,Q)=(U,2U]\cap
 (\max(q/v_1,q/v_2),\min(Q/v_1,Q/v_2)].
\]
This is an interval or empty. Take $j=2r$.
For either logarithmic phase, \Cref{lem:typeII-B} applies because
$U\ge2$ and $UV>131(j+1)$. For the reciprocal phase,
\[
 \psi_h(u)=-\frac{y_\psi}{u},\qquad
 y_\psi=\frac{\tau h}{2\pi v_1v_2}>0,
\]
so \Cref{lem:GK-input} applies with zero remainder. In all three cases,
with $\lambda_j(h)=\tau h/(U^{j+1}V^2)$, the same conversion
$\lambda_j(h)/(8\pi)\le\lambda_{j,0}<\lambda_j(h)/(2\pi)$ gives
\begin{equation}\label{eq:inner-u-applied}
 \left|\sum_{u\in J(v_1,v_2;q,Q)\cap\mathbb Z}e(\psi_h(u))\right|
 \le336\pi\left\{
 U\left(\frac{\tau h}{U^{2r+1}V^2}\right)^\kappa
 +\frac{U^2V^2}{\tau h}\right\}.
\end{equation}
Extend $\beta_v$ by zero outside its block. For every positive integer $h$,
Cauchy--Schwarz gives $\sum_v|\beta_v\beta_{v+h}|\le B_2$.
Every nonempty pair has $1\le h\le V$. Since $V\ge1$,
\[
 \sum_{1\le h\le V}h^\kappa\le V^{1+\kappa},\qquad
 \sum_{1\le h\le V}\frac1h\le1+\log V
 \le\frac{\log(2V)}{\log2}.
\]
Thus \eqref{eq:CS-truncated}--\eqref{eq:inner-u-applied} imply
\[
 |W(q,Q)|^2\le A_2B_2\left\{2U+
 672\pi UV\left(\frac{\tau}{U^{2r+1}V}\right)^\kappa
 +\frac{672\pi}{\log2}\frac{U^2V^2}{\tau}\log(2V)\right\},
\]
which proves \eqref{eq:bilinear-L2}.
\end{proof}

For integers $r\ge2$, $\delta>0$, and $\tau>1$, use the scale range
\begin{equation}\label{eq:N-range}
 \tau^{1/(r+1)+\delta}\le N\le\tau^{1-\delta}.
\end{equation}

\begin{corollary}[Divisor-bounded coefficients on dyadic blocks]
\label{cor:bilinear-weighted}
Let $r\ge2$ be an integer and $0<\delta<1/(4r(r+1))$, and put
\begin{equation}\label{eq:eta-prime}
 \eta_{\mathrm{II}}(r,\delta)=
 \frac{(r+1)^2\delta}{2(2^{2r}-2)},\qquad e_r=4r^2+4r.
\end{equation}
There are effective $C_r^{\mathrm{II}}(\delta)>0$ and
$\tau_0(r,\delta)\ge1$ with the following property.
Let $\tau\ge\tau_0(r,\delta)$, let $N$ be dyadic in \eqref{eq:N-range},
and suppose
\begin{equation}\label{eq:typeII-vaughan-range}
 \widetilde c_r^{-1}N^{1/(2r)}\le V\le U
 \le\widetilde c_rN^{1-1/(2r)},\qquad N/4<UV\le2N.
\end{equation}
If $A_{\mathrm{coef}},B_{\mathrm{coef}}\ge1$ and
\[
 |a_u|\le A_{\mathrm{coef}}\mathbf d_{2r}(u)(\log N)^{2r},\qquad
 |\beta_v|\le B_{\mathrm{coef}}\mathbf d_{2r}(v)(\log N)^{2r},
\]
then the preceding $W(q,Q)$, for any of the three phases, satisfies uniformly in
real $0\le q\le Q$
\begin{equation}\label{eq:typeII-bound}
 |W(q,Q)|\le C_r^{\mathrm{II}}(\delta)
 A_{\mathrm{coef}}B_{\mathrm{coef}}
 N^{1-\eta_{\mathrm{II}}(r,\delta)/2}(\log N)^{e_r}.
\end{equation}
In particular, suppose the blocks lie in the narrower ordered range
\begin{equation}\label{eq:bilinear-ordered-range}
 \widetilde c_r^{-1}N^{1/r}\le V\le U
 \le\widetilde c_rN^{1-1/r},\qquad N/4<UV<2N.
\end{equation}
The coefficient envelopes
$|a_u|\le H_r\mathbf d_{2r}(u)(\log N)^{2r}$ and
$|\beta_v|\le H_r\mathbf d_{2r}(v)(\log N)^{2r}$ give, for
\begin{equation}\label{eq:typeII-weighted-form}
 \mathcal T(x)=\sum_{u\sim U}a_u
 \sum_{\substack{v\sim V\\N<uv\le x}}\beta_v e^{i\tau\varphi(uv)}
 =W(N,x),
\end{equation}
$\sup_{N\le x\le2N}|\mathcal T(x)|\le
 C_r^{\mathrm{II}}(\delta)H_r^2
 N^{1-\eta_{\mathrm{II}}(r,\delta)/2}(\log N)^{e_r}$.
The blocks and coefficients in this specialization are fixed independently
of $x$.
\end{corollary}

\begin{proof}
Put $\kappa=(2^{2r}-2)^{-1}$ and increase the effective threshold
so that $N\ge e$, $V\ge2$,
$UV>131(2r+1)$, and $2U\le N$.
By \Cref{lem:eff-divisor-moment} at divisor parameter $2r$,
\[
 A_2\ll_r A_{\mathrm{coef}}^2U(\log N)^{e_r-1},\qquad
 B_2\ll_r B_{\mathrm{coef}}^2V(\log N)^{e_r-1}.
\]
Hence \eqref{eq:bilinear-L2}, $UV\le2N$, and
$\log(2V)\le\log N$ give
\begin{equation}\label{eq:T-squared-pre}
\begin{aligned}
 |W(q,Q)|^2&\ll_r
 A_{\mathrm{coef}}^2B_{\mathrm{coef}}^2(\log N)^{2e_r}\\
 &\quad\times\left\{(UV)^2\left(\frac{\tau}{U^{2r}UV}\right)^\kappa
       +\frac{(UV)^3}{\tau}+UN\right\}.
\end{aligned}
\end{equation}
Here $V\le U$ implies $U^{2r}UV>(N/4)^{r+1}$.
Write $\eta=\eta_{\mathrm{II}}(r,\delta)$.
The lower bound on $N$ and $(r+1)\delta<1$ imply
\[
 \frac\tau{N^{r+1}}
 \le N^{-(r+1)^2\delta/(1+(r+1)\delta)}
 \le N^{-\eta/\kappa}.
\]
The upper bound on $N$ and the scale bounds give
\[
 \frac{(UV)^3}{\tau}\ll N^{2-\delta/(1-\delta)},\qquad
 UN\le\widetilde c_rN^{2-1/(2r)}.
\]
On the smaller range \eqref{eq:bilinear-ordered-range}, the latter improves to
$UN\le\widetilde c_rN^{2-1/r}$.
Finally, for integer $r\ge2$,
\[
 \eta\le\frac{9\delta}{28}<\frac{\delta}{1-\delta},\qquad
 \eta<\frac{r+1}{8r(2^{2r}-2)}<\frac1{2r}.
\]
Every term in braces in \eqref{eq:T-squared-pre} is
$O_{r,\delta}(N^{2-\eta})$. Taking square roots proves
\eqref{eq:typeII-bound}; setting
$A_{\mathrm{coef}}=B_{\mathrm{coef}}=H_r$ and $q=N$, $Q=x$
gives the final specialization.
\end{proof}

\Needspace{5\baselineskip}
\section{Prime cancellation and Bernoulli series}
\label{ssec:annulus-consequences}

\leavevmode\par\nobreak
\begin{proof}[Proof of \Cref{thm:annulus}]
Conjugation reduces to $\tau>0$. Apply \eqref{eq:vk-vaughan} with
$z=N^{1/(2r)}$. Its first two terms have outer variables $m\le z^2$
and coefficients bounded by $1$ and $\log m$, respectively; only the
first has inner weight $\log n$. The estimate
\eqref{eq:typeI-inner-direct}, with $C=1$ and $m\le N^{1/r}$, gives
$O_{r,j}((N/m)\tau^{-\theta})$ uniformly on every subinterval of
$(N/m,2N/m]$. Abel summation and
$\sum_{m\le z^2}(1+\log m)/m\ll_r(\log N)^2$ therefore bound the
combined Type~I contribution by $O_{r,j}(N\tau^{-\theta}(\log N)^2)$.

The third term has variables $u,v>z$ and coefficients
$a(u)=\sum_{d\mid u,\,d>z}\mu(d)$ and $\Lambda(v)\mathbf1_{\{v>z\}}$.
Dyadic subdivision and interchange give $O((\log N)^2)$ block pairs
with $z/2\le V\le U\le2N/z$ and $N/4<UV\le2N$.
The blocks are fixed independently of $x$; the threshold indicators
remain in their coefficients. Since $|a(u)|\le\mathbf d_2(u)$ and
$\Lambda(v)\le\log(2N)$, \Cref{lem:eff-divisor-moment} gives
$A_2B_2\ll UV(\log N)^5$, also after interchange.
Equation~\eqref{eq:bilinear-L2} and the power comparisons in the proof of
\Cref{cor:bilinear-weighted} yield
\[
 |W(N,x)|^2\ll_r N^2(\log N)^6
 \left\{N^{-1/(2r)}+
 \left(\frac{\tau}{N^{r+1}}\right)^{1/(2^{2r}-2)}
 +\frac N\tau\right\}
 \ll_{r,\delta}N^{2-\eta}(\log N)^6.
\]
Consequently
\[
 \sup_{N\le x\le2N}
 \left|\sum_{N<n\le x}\Lambda(n)e^{i\tau\omega_n}\right|
 \ll_{r,\delta,j}N\tau^{-\theta}(\log N)^2
       +N^{1-\eta/2}(\log N)^5.
\]
Prime powers of exponent at least two contribute
$O(N^{1/2}\log N)$. Partial summation against $1/\log u$, uniformly on
$[N,x]$, bounds the left side of \eqref{eq:annulus-retained-power} by
\[
 \frac{CN}{\log N}
 \left\{\tau^{-\theta}(\log N)^2
       +N^{-\eta/2}(\log N)^5+N^{-1/2}\log N\right\}.
\]
Since $\tau^a\le N\le\tau^b$, the braces are
$O(\tau^{-\gamma_0}(\log\tau)^5)$.
The strict inequality $\gamma<\gamma_0$ absorbs the logarithm and proves
\eqref{eq:annulus-retained-power}; it also implies
\eqref{eq:annulus_bound}. All constants and thresholds are effective
for fixed parameters. For the numerical choice displayed after the statement,
$\theta=19/4500$, $\eta=1/775$ and $\gamma_0=13/77500$.

For the plus-log and reciprocal phases, \eqref{eq:typeI-inner-direct}
and \eqref{eq:bilinear-L2} hold with the same displayed constants.
Applying them to the same Vaughan blocks, followed by prime-power
removal and partial summation, proves both conclusions with common
constants for all three phases.
\end{proof}

The original power annulus will remain useful for conditioning on a
single active prime. For the endpoint activation asymptotic we need a
different upper edge, within a fixed logarithmic factor of $|t|$.
The next estimate retains the terms that determine that edge.

\begin{lemma}[An annulus with a logarithmic upper edge]
\label{lem:extended-annulus-mas09}
Let $\tau\ge2$, $\tau^{9/50}\le N\le\tau$, and put $\ell_N=\log N$.
For each of the three phases
$w(u)=\log(u/(u-1))$, $\log(1+1/u)$, or $1/u$, one has,
for sufficiently large $\tau$,
\begin{equation}\label{eq:extended-annulus-raw-mas09}
\begin{aligned}
 &\sup_{N\le x\le2N}\left|\sum_{N<p\le x}e^{i\tau w(p)}\right|\\
 &\quad\ll\frac{N}{\ell_N}\Bigg\{
 \ell_N^2\left[\left(\frac{\tau}{N^{29/5}}\right)^{1/62}
                      +\frac N\tau\right]\\
 &\qquad\quad+\ell_N^5\left[N^{-1/20}
       +\left(\frac{\tau}{N^6}\right)^{1/2044}
       +\left(\frac N\tau\right)^{1/2}\right]
       +N^{-1/2}\ell_N^2\Bigg\}.
\end{aligned}
\end{equation}
Consequently, for every fixed $A>0$, with $C_A=2A+12$,
\begin{equation}\label{eq:extended-annulus-saving-mas09}
 \sup_{N\le x\le2N}\left|\sum_{N<p\le x}e^{i\tau w(p)}\right|
 \ll_A\frac{N}{(\log N)^A}
 \quad\left(\tau^{9/50}\le N\le
                \tau(\log\tau)^{-C_A}\right).
\end{equation}
The constants are common to all three phases; $N$ need not be dyadic.
\end{lemma}

\begin{proof}
Use Vaughan's identity \eqref{eq:vk-vaughan} with $r=5$ and
$z=N^{1/10}$. The two Type~I terms have $m\le z^2=N^{1/5}$.
Apply \Cref{lem:GK-input} at order $j=6$ to their inner phases,
using the derivative remainders in \Cref{lem:phase_derivs}.
For every inner subinterval $J\subseteq[N/m,2N/m]$, this gives
\[
 \left|\sum_{n\in J\cap\Z}e^{i\tau w(mn)}\right|
 \ll\frac Nm\left\{
        \left(\frac{\tau m^6}{N^7}\right)^{1/62}
                       +\frac N\tau\right\}
 \le\frac{C N}{m}\left\{
        \left(\frac{\tau}{N^{29/5}}\right)^{1/62}
                       +\frac N\tau\right\}.
\]
The same derivative calculation is valid for the plus-log phase,
with its signed remainder, and for the reciprocal phase with zero
remainder. The derivative-test hypotheses hold once $N$ is sufficiently
large, so this estimate remains valid up to $N=\tau$.
Abel summation costs one factor $\log N$ for the first Type~I
term, and the second has coefficient at most $\log m$.
Summing the resulting harmonic weights bounds their total by
\[
 N\ell_N^2\left\{
       (\tau/N^{29/5})^{1/62}+N/\tau\right\}.
\]

For Type~II, dyadic subdivision and interchange give
$O(\ell_N^2)$ block pairs with
\[
 N^{1/10}/2\le V\le U\le2N^{9/10},\qquad N/4<UV\le2N.
\]
The coefficients are bounded by $\mathbf d_2$ and $\log(2N)$,
respectively. Their squared norms satisfy
$A_2B_2\ll UV\ell_N^5$, also after interchange, by
\Cref{lem:eff-divisor-moment}. In \eqref{eq:bilinear-L2}, use
$V^{-1}\ll N^{-1/10}$, $U^{10}UV\gg N^6$, and
$\log(2V)\ll\ell_N$. For every product-truncated block this yields
\[
 |W(N,x)|^2\ll N^2\ell_N^6
 \left\{N^{-1/10}+(\tau/N^6)^{1/1022}+N/\tau\right\}.
\]
Taking square roots and summing the $O(\ell_N^2)$ blocks gives
\[
 N\ell_N^5\left\{N^{-1/20}
       +(\tau/N^6)^{1/2044}+(N/\tau)^{1/2}\right\}.
\]
Prime powers of exponent at
least two contribute $O(N^{1/2}\ell_N^2)$. Abel summation against
$1/\log u$ on $[N,x]$ costs $O(1/\ell_N)$ and proves
\eqref{eq:extended-annulus-raw-mas09}.

Write $h=\log\tau$. Since $9h/50\le\ell_N\le h$, the three
power-saving terms and the prime-power error in braces are at most
$O(\tau^{-1/25550}h^5)$. Indeed,
\[
 \frac{(9/50)(29/5)-1}{62}=\frac{11}{15500},\qquad
 \frac{6(9/50)-1}{2044}=\frac1{25550},\qquad
 \frac{9/50}{20}=\frac9{1000}.
\]
When $N/\tau\le h^{-C_A}$, the two remaining terms are bounded by
$h^{2-C_A}+h^{5-C_A/2}=O(h^{-A-1})$.
The power of $\tau$ absorbs every fixed power of $h$, proving
\eqref{eq:extended-annulus-saving-mas09}.
\end{proof}

The activation results below concern the three jump families of
\Cref{thm:annulus}. The first decay proof chooses fixed
$r,\delta,j,\gamma$ for each prescribed $K,\alpha>0$ and
$0<c<K/\alpha$. The endpoint coefficient is quantified separately
in \Cref{cor:activation-pointwise-mas09}, using
\Cref{lem:extended-annulus-mas09}. The sharp critical conclusions use
\Cref{thm:activation-scale} as well as cancellation, under the stated
hypothesis that the head is a finite sum of independent
arithmetic-progression variables. For the common-sign absolute
continuity conclusion when $0<\alpha\le1$, the proof instead uses
\Cref{lem:conditional-block-smoothing} and the fixed numerical choice
following \Cref{thm:annulus}.

Fix $K>0$, $\alpha\in\R$, and $L\ge2$ such that
$p_\ell=K(\log\ell)^\alpha/\ell\in(0,1)$ for primes $\ell>L$.
Let $B_\ell$ be independent Bernoulli variables with probabilities
$p_\ell$, let $\epsilon_\ell\in\{-1,1\}$ be fixed, and let $H$ be any
independent real random variable. The series in
\[
 Y=H+\sum_{\ell>L}\epsilon_\ell B_\ell\omega_\ell
\]
converges absolutely almost surely. Write
$\nu=\operatorname{Law}(Y)$, $X_\alpha=\sum_{\ell>L}B_\ell\omega_\ell$
and $\mu_\alpha=\operatorname{Law}(X_\alpha)$.
Finitely many probability-one coordinates may be restored as a
translation of $H$.
In \Cref{cor:heavier-activation,cor:positive-activation}, $\omega_\ell$
may be replaced throughout by $\log(1+1/\ell)$ or by $1/\ell$.

\KeepHeadingSpace{6\baselineskip}
\begin{corollary}[Fourier decay and critical regularity]
\label{cor:heavier-activation}
\index{Bernoulli series!heavier activation}\leavevmode\par
\begin{enumerate}[label=\textup{(\alph*)},leftmargin=*]
\item If $\alpha>0$, then, for every fixed $0<c<K/\alpha$,
\begin{equation}\label{eq:activation-fourier-decay}
 |\widehat\nu(t)|\ll_{K,\alpha,L,c}
       \exp\{-c(\log|t|)^\alpha\}\qquad(|t|\to\infty).
\end{equation}
If $\alpha>1$, $\nu$ has a smooth bounded density with every derivative
bounded, for arbitrary signs and $H$.
If $\alpha=1$, then for every $0<\beta<K$,
\begin{equation}\label{eq:heavier-fourier-decay}
 |\widehat\nu(t)|\ll_{K,L,\beta}|t|^{-\beta},\qquad
 \dim_H\nu\ge\dim_F\nu\ge\min(1,2K).
\end{equation}
In particular, $\nu$ has an $L^2$ density if $\alpha=1$ and $K>1/2$.
\item Suppose $\alpha=1$ and
$H=z+\sum_{j=1}^m\eta_jH_j$, where $m\ge0$, $z,\eta_j\in\R$, and
the $H_j$ are independent variables supported on finite arithmetic
progressions, independent of the $B_\ell$.
Then
\[
 \dim_F\nu=\min(1,2K),\qquad
 \widehat\nu\in L^q(\R)\ \Longleftrightarrow\ qK>1\quad(q>0).
\]
For every $s\in\R$, as a tempered distribution,
$\nu\in H^s(\R)$ if and only if $s<K-\tfrac12$.
For $s\ge0$, this is equivalent to having a density in $H^s(\R)$.
\item Suppose $\alpha>0$ and $H$ is a constant plus a finite sum of
independent variables supported on finite arithmetic progressions,
independent of all $B_\ell$. Then
\begin{equation}\label{eq:heavier-log-average}
 \frac1T\int_T^{2T}-\log|\widehat\nu(t)|^2\,dt
 =\frac{2K}{\alpha}(\log T)^\alpha
       +O\bigl(1+(\log T)^{\alpha-1}\bigr),
\end{equation}
and
\[
 \liminf_{|t|\to\infty}
 \frac{-\log|\widehat\nu(t)|}{(\log|t|)^\alpha}=\frac K\alpha.
\]
\end{enumerate}
\end{corollary}

\begin{proof}
The expected absolute series is
$O_{K,\alpha}(\sum_{n>L}(\log n)^\alpha/n^2)<\infty$.
Also $\sum_\ell p_\ell^2<\infty$, and
\[
 2p_\ell(1-p_\ell)
 =\frac{2K(\log\ell)^\alpha}{\ell}
       +O_{K,\alpha}((\log\ell)^{2\alpha}/\ell^2).
\]
For part~\textup{(a)}, conjugation permits $t>0$.
Choose $r,\delta$ in \Cref{thm:annulus} so that, with
$a=1/(r+1)+\delta$, $b=1-\delta$,
$K(b^\alpha-a^\alpha)/\alpha>c$.
Let $N_1,N_2$ be the least and greatest dyadic numbers with
$t^a\le N_1$ and $2N_2\le t^b$. The weight
$w(u)=2K(\log u)^\alpha/u$ decreases eventually. Abel summation and
\eqref{eq:annulus-retained-power}, for some fixed $\gamma>0$, give
\[
 \left|\sum_{N<\ell\le2N}w(\ell)e^{it\omega_\ell}\right|
 \ll t^{-\gamma}(\log N)^{\alpha-1}
 \qquad(N_1\le N\le N_2).
\]
Summing the $O(\log t)$ blocks and applying
\Cref{lem:prime-log-stieltjes} to their nonoscillatory weights yields
\[
 \begin{split}
 S_p(t)&:=\sum_{\ell>L}2p_\ell(1-p_\ell)
                         (1-\cos(t\omega_\ell))\\
 &\ge\frac{2K}{\alpha}(b^\alpha-a^\alpha)(\log t)^\alpha
       -O(1+(\log t)^{\alpha-1})
       -O(t^{-\gamma}(\log t)^\alpha).
 \end{split}
\]
The local squared Fourier modulus is
$1-2p_\ell(1-p_\ell)(1-\cos(t\omega_\ell))$, independently of the
sign. Since $|\widehat{\operatorname{Law}(H)}|\le1$,
$|\widehat\nu(t)|\le e^{-S_p(t)/2}$, proving
\eqref{eq:activation-fourier-decay}. For $\alpha>1$ this is integrable
against every polynomial weight; Fourier inversion and differentiation
under the integral give the asserted bounded derivatives.
For $\alpha=1$, take $c=\beta$, and let $\beta\uparrow K$ in
\eqref{eq:measure-Fourier-dimension}; \Cref{lem:fourier-Hausdorff}
gives the Hausdorff bound. If $K>1/2$, choose $\beta>1/2$ and apply
\Cref{lem:l2-inversion}.

For parts~\textup{(b)} and~\textup{(c)}, increase a fixed cutoff
$L_0\ge L$ until $p_\ell\le1/4$ for $\ell>L_0$, and let $\nu_0$
be the prime-tail law.
The Bernoulli logarithm expansion gives, uniformly in real $t$,
\begin{equation}\label{eq:heavier-log-comparison}
 -\log|\widehat\nu_0(t)|^2
 =2K\sum_{\ell>L_0}\frac{(\log\ell)^\alpha}{\ell}
                (1-\cos(t\omega_\ell))+O(1).
\end{equation}
For part~\textup{(b)}, put
$\Lambda(T)=\sum_{\ell>L_0,\,\omega_\ell\ge T^{-1}}p_\ell$.
Here $\alpha=1$ and $\omega_\ell\ge T^{-1}$ is equivalent to
$\ell\le(1-e^{-1/T})^{-1}=T+O(1)$, so
\Cref{lem:prime-log-stieltjes} gives $\Lambda(T)=K\log T+O(1)$.
Put $\mathcal D_0(t)=\sum_{\ell>L_0}p_\ell
(1-\cos(t\omega_\ell))$. The dissipation estimate in the proof of
part~\textup{(a)}, with $\alpha=1$, gives
$\mathcal D_0(t)\ge\beta\log|t|-O_\beta(1)$ for every fixed
$0<\beta<K$. Consequently
$E(t):=\max\{K\log|t|-\mathcal D_0(t),0\}=o(\log|t|)$,
which gives the cancellation hypothesis in
\eqref{eq:activation-scale-hypotheses}. Thus
\Cref{thm:activation-scale} proves part~\textup{(b)} and gives
\begin{equation}\label{eq:heavier-shell-lower}
 \int_T^{2T}|\widehat\nu(t)|^q\,dt\gg_q T^{1-qK}
 \qquad(q>0).
\end{equation}

For part~\textup{(c)}, put
$\Lambda_\alpha(T)=\sum_{\ell>L_0,\,\omega_\ell\ge T^{-1}}p_\ell$.
The same cutoff and \Cref{lem:prime-log-stieltjes} give
$\Lambda_\alpha(T)=(K/\alpha)(\log T)^\alpha+O(1)$.
Partial summation also gives
\[
 \frac1T\sum_{\substack{\ell>L_0\\\omega_\ell\ge T^{-1}}}
            \frac{p_\ell}{\omega_\ell}
 +T^2\sum_{\substack{\ell>L_0\\\omega_\ell<T^{-1}}}
            p_\ell\omega_\ell^2
 \ll 1+(\log T)^{\alpha-1}.
\]
Direct cosine integration in \eqref{eq:heavier-log-comparison}
therefore gives mean $2\Lambda_\alpha(T)+O(1+(\log T)^{\alpha-1})$.
The finite-head calculation in the proof of \Cref{thm:activation-scale}
adds $O(1)$, proving \eqref{eq:heavier-log-average}.
Part~\textup{(a)} bounds the liminf below by $K/\alpha$; a larger
liminf would contradict this mean, since $\log t/\log T\to1$
uniformly on $[T,2T]$. This proves part~\textup{(c)}.

The same arguments apply to either alternative jump family,
by \Cref{thm:annulus}. Their cutoffs at jump size $T^{-1}$ are
$(e^{1/T}-1)^{-1}$ and $T$, in place of $(1-e^{-1/T})^{-1}$;
all equal $T+O(1)$. Thus the activation coefficient is unchanged.
\end{proof}

Fix $K,\alpha>0$ in the heavier-activation model, with arbitrary
fixed signs and any independent real head $H$. Write
$h=\log|t|$ and
\[
 \mathcal E_\alpha(h)
 =h^{2\alpha/3}(\log h)^{\alpha/3}
       +h^{\alpha-1}\log h+1.
\]
Choose a fixed $L_0\ge L$ such that $p_\ell\le1/4$ for every
prime $\ell>L_0$, and define
\[
 \nu_0=\operatorname{Law}\left(
             \sum_{\ell>L_0}\epsilon_\ell B_\ell w(\ell)\right),
 \qquad
 w(u)\in\{\log(u/(u-1)),\log(1+1/u),1/u\}.
\]

\begin{corollary}[Pointwise endpoint for a specified prime tail]
\label{cor:activation-pointwise-mas09}
\leavevmode\par
\begin{enumerate}[label=\textup{(\alph*)},leftmargin=*]
\item For sufficiently large $|t|$,
\begin{equation}\label{eq:activation-endpoint-lower-mas09}
 -\log|\widehat\nu(t)|^2
 \ge\frac{2K}{\alpha}h^\alpha
                       -O_{K,\alpha,L}(\mathcal E_\alpha(h)),
\end{equation}
where $-\log0=+\infty$.
\item $\widehat\nu_0(t)\ne0$ for every real $t$ and, as $|t|\to\infty$,
\begin{equation}\label{eq:activation-endpoint-twosided-mas09}
 -\log|\widehat\nu_0(t)|^2
 =\frac{2K}{\alpha}h^\alpha
                       +O_{K,\alpha,L_0}(\mathcal E_\alpha(h)).
\end{equation}
In particular,
\[
 \lim_{|t|\to\infty}
 \frac{-\log|\widehat\nu_0(t)|}{(\log|t|)^\alpha}
       =\frac K\alpha.
\]
\end{enumerate}
All conclusions hold for each of the three displayed jump families.
\end{corollary}

\begin{proof}
Put $\tau=|t|$ and $h=\log\tau$. First we calculate the weighted
dissipation
\[
 \mathcal D_\alpha(t)
 =K\sum_{\ell>L_0}\frac{(\log\ell)^\alpha}{\ell}
                          (1-\cos(tw(\ell))).
\]
Take $A=\alpha+3$ in \Cref{lem:extended-annulus-mas09}, set
$C=C_A$, and put
$G=D h^{2/3}(\log h)^{1/3}$, where the fixed $D>0$ will be chosen
below. Let $x_-$ be the least power of two at least $e^G$, and
$x_+$ the greatest power of two at most $\tau h^{-C}$. Thus
\[
 \log x_-=G+O(1),\qquad \log x_+=h-C\log h+O(1).
\]
Both exceed $L_0$ eventually. Every dyadic block $(N,2N]$ in
$(x_-,x_+]$ lies either in the range of
\Cref{prop:vk-prime-prefix} or in that of
\Cref{lem:extended-annulus-mas09}: these ranges overlap because
$\tau^{9/50}<b\tau^{1/5}$ eventually, where $b>0$ is the constant
in the former proposition. This is the reason for the fixed choice
$r=5$ in the extended annulus.

For $v(u)=(\log u)^\alpha/u$, eventually decreasing, Abel summation
gives on each such block
\[
 \left|\sum_{N<\ell\le2N}v(\ell)e^{itw(\ell)}\right|
 \le v(N)\sup_{N\le x\le2N}
              \left|\sum_{N<\ell\le x}e^{itw(\ell)}\right|.
\]
There are $O(h)$ blocks. For the blocks assigned to the Vinogradov
estimate, their total is at most
\[
 O_\alpha\!\left(
 h^{\alpha+9}e^{-cG^3/h^2}+h^{\alpha+9}e^{-G/6}\right)=O(1)
\]
once $cD^3>\alpha+10$, with $c$ as in
\eqref{eq:vk-prime-prefix}. For the remaining blocks, $\log N\asymp h$;
\eqref{eq:extended-annulus-saving-mas09} bounds their total by
$O_\alpha(h^{\alpha-A+1})=O(h^{-2})$. Consequently
\begin{equation}\label{eq:activation-weighted-cancel-mas09}
 \sum_{x_-<\ell\le x_+}
       \frac{(\log\ell)^\alpha}{\ell}e^{itw(\ell)}=O(1).
\end{equation}

Applying \Cref{lem:prime-log-stieltjes} with $g(U)=U^\alpha$ shows
\[
 \sum_{L_0<\ell\le x}\frac{(\log\ell)^\alpha}{\ell}
       =\frac{(\log x)^\alpha}{\alpha}+O_{\alpha,L_0}(1).
\]
Here the remainder is bounded because
$U^\alpha\rho_{\mathrm M}(U)$ is bounded and
$\int^\infty U^{\alpha-1}|\rho_{\mathrm M}(U)|\,dU<\infty$.
Together with \eqref{eq:activation-weighted-cancel-mas09}, this gives
\[
 K\sum_{x_-<\ell\le x_+}\frac{(\log\ell)^\alpha}{\ell}
                          (1-\cos(tw(\ell)))
 =\frac K\alpha h^\alpha
    +O_{K,\alpha,L_0}\!\left(G^\alpha
                        +h^{\alpha-1}\log h+1\right).
\]
The contribution below $x_-$ is nonnegative and at most
$O_{K,\alpha,L_0}(1+G^\alpha)$. Between $x_+$ and $\tau$ it is at most
\[
 2K\sum_{x_+<\ell\le\tau}\frac{(\log\ell)^\alpha}{\ell}
 \ll_{K,\alpha}1+h^{\alpha-1}\log h;
\]
the mean value theorem applies since $\log x_+\sim h$, also when
$0<\alpha<1$. Finally $w(\ell)\le2/\ell$ and
$1-\cos u\le u^2/2$, so the contribution above $\tau$ is bounded by
\[
 O_{K,\alpha}\!\left(
 \tau^2\sum_{\ell>\tau}\frac{(\log\ell)^\alpha}{\ell^3}\right)
 \ll_{K,\alpha}\sum_{k\ge0}2^{-2k}(h+k)^{\alpha-1}
 \ll_{K,\alpha}h^{\alpha-1}.
\]
The first inequality after the sum follows by dividing the primes
into $(2^k\tau,2^{k+1}\tau]$ and using $\pi(u)\ll u/\log u$.
Combining the four ranges gives
\begin{equation}\label{eq:activation-weighted-error-mas09}
 \mathcal D_\alpha(t)
 =\frac K\alpha h^\alpha
               +O_{K,\alpha,L_0}(\mathcal E_\alpha(h)).
\end{equation}

The local squared Fourier modulus is
$1-q_\ell(t)$, with
\[
 q_\ell(t)=2p_\ell(1-p_\ell)(1-\cos(tw(\ell)))\in[0,3/4].
\]
The signs disappear from this expression. For fixed $t$,
$\sum_\ell q_\ell(t)<\infty$ by $w(\ell)\le2/\ell$, so the product
is positive. Since $-\log(1-q)=q+O(q^2)$ uniformly on $[0,3/4]$,
$\sum_\ell p_\ell^2<\infty$ gives, uniformly in real $t$,
\[
 -\log|\widehat\nu_0(t)|^2
 =2\mathcal D_\alpha(t)+O_{K,\alpha,L_0}(1).
\]
This proves \eqref{eq:activation-endpoint-twosided-mas09}.
The omitted finite coordinates and the independent head have
Fourier modulus at most one, so
$|\widehat\nu(t)|\le|\widehat\nu_0(t)|$, proving
\eqref{eq:activation-endpoint-lower-mas09}. The cutoff $L_0$ may be
chosen in terms of $K,\alpha,L$, which gives its stated dependence.
Finally $\mathcal E_\alpha(h)=o(h^\alpha)$ for every fixed $\alpha>0$.
For $\alpha=1$, its $\log h+1$ terms are absorbed by
$h^{2/3}(\log h)^{1/3}$; division by two and exponentiation give
\eqref{eq:activation-endpoint-alphaone-mas09}.
\end{proof}

For $\alpha=1$, \eqref{eq:activation-endpoint-twosided-mas09} gives
\begin{equation}\label{eq:activation-endpoint-alphaone-mas09}
 |\widehat\nu_0(t)|
 =|t|^{-K}\exp\!\left(O_{K,L_0}
       \bigl(h^{2/3}(\log h)^{1/3}\bigr)\right).
\end{equation}

The specified prime-tail cutoff in
\Cref{cor:activation-pointwise-mas09} is essential for its two-sided
conclusion. Merely setting the independent head $H$ equal to zero
does not suffice: a retained coordinate with $p_\ell=1/2$ has
Fourier factor $(1+e^{it\epsilon_\ell w(\ell)})/2$, which vanishes
at $t=(2k+1)\pi/w(\ell)$ for every integer $k$. Such zeros are
compatible with the one-sided bound but not with
\eqref{eq:activation-endpoint-twosided-mas09}.

\begin{corollary}[Type and local regularity of activation laws]
\label{cor:positive-activation}\leavevmode\par
\begin{enumerate}[label=\textup{(\alph*)},leftmargin=*]
\item If $\alpha>0$ and all $\epsilon_\ell$ have the same sign, then
$\nu$ is absolutely continuous and $\dim_H\nu=1$, with arbitrary
independent $H$.
\item If $\alpha<0$, $\mu_\alpha$ is
countably atomic; if $\alpha=0$, it is singular continuous and has a
Borel full carrier of Hausdorff dimension zero.
\item If $0<\alpha<1$, then
\begin{equation}\label{eq:activation-endpoint}
 \mathbb P(X_\alpha\le h)\sim C_{K,\alpha,L}
 \exp\left\{-\frac K\alpha(\log(1/h))^\alpha\right\},
 \qquad h\downarrow0,
\end{equation}
where
\[
 C_{K,\alpha,L}=\lim_{x\to\infty}
 e^{(K/\alpha)(\log x)^\alpha}
          \prod_{L<\ell\le x}(1-p_\ell)\in(0,\infty).
\]
In this range $\supp\mu_\alpha=[0,\infty)$, and its density belongs to
no $L^q(I)$, $q>1$, on any nonempty open interval
$I\subset(0,\infty)$. Moreover,
\[
 \dim_F\mu_\alpha=0,\qquad
 \underline D_q(\mu_\alpha)=\overline D_q(\mu_\alpha)=0\quad(q>1).
\]
\end{enumerate}
\end{corollary}

For $0<\alpha<1$, the local exclusion and dimension conclusions in
\Cref{cor:positive-activation}\textup{(c)} also hold for every positive
finite convolution power. After convolution with an independent
law having an atom at $a$, the local exclusion holds on every nonempty
open interval $I\subset(a,\infty)$, and both dimension conclusions
remain valid. If the head is supported on $[a,\infty)$ and has an
atom at $a$, this covers the entire support interior.

\begin{proof}[Proof of \Cref{cor:positive-activation} and its extensions]
For part~\textup{(a)}, reflection and convolution permit positive signs
and $H=0$. If $\alpha>1$,
\Cref{cor:heavier-activation}\textup{(a)} applies. Suppose
$0<\alpha\le1$, and omit finitely many primes so that $p_\ell<1/2$.
For sufficiently large integers $k$, put
\[
 I_k=\{\ell:2^k<\ell\le2^{k+1}\},\qquad
 N_k=\sum_{\ell\in I_k}B_\ell,\qquad
 w_\ell=\frac{p_\ell}{1-p_\ell},\qquad
 W_k=\sum_{\ell\in I_k}w_\ell.
\]
The prime number theorem and $\sum_{I_k}p_\ell^2=o(W_k)$ give
\[
 \begin{gathered}
 W_k\sim K(\log2)^\alpha k^{\alpha-1},\\
 q_k:=\mathbb P(N_k=1)=W_k\prod_{\ell\in I_k}(1-p_\ell)
 \sim\begin{cases}
 K(\log2)^\alpha k^{\alpha-1},&0<\alpha<1,\\
 K(\log2)2^{-K},&\alpha=1.
 \end{cases}
 \end{gathered}
\]
Conditional on $N_k=1$, the block transform is
$\psi_k(t)=W_k^{-1}\sum_{\ell\in I_k}w_\ell e^{it\omega_\ell}$.
Use the concrete $a,b,\gamma$ displayed after \Cref{thm:annulus}.
The odds are eventually decreasing in $\ell$, and
$w_\ell\asymp(\log 2^k)^\alpha/2^k$ uniformly for $\ell\in I_k$,
while $W_k\asymp(\log 2^k)^{\alpha-1}$. Abel summation in
\eqref{eq:annulus-retained-power} therefore gives
\begin{equation}\label{eq:activation-singleton-power}
 |\psi_k(t)|\ll_{K,\alpha}|t|^{-\gamma}
 \quad\text{whenever}\quad |t|^a\le2^k\le|t|^b,
\end{equation}
uniformly in real $t$ and $k$.

For each sufficiently large integer $m$, put
\[
 J_m=\{k:\lceil a(m+1)\rceil\le k\le\lfloor bm\rfloor\}.
\]
Every retained block satisfies \eqref{eq:activation-singleton-power}
for every $2^m\le|t|<2^{m+1}$, and the singleton asymptotics give
$\sum_{k\in J_m}q_k\asymp m^\alpha$.
\Cref{lem:conditional-block-smoothing}\textup{(b)}, with $d=1$, $\rho=\alpha$
and $\delta=\gamma$, applied to
$Y_k=\sum_{\ell\in I_k}B_\ell\omega_\ell$, proves absolute continuity.
Restoring the finite coordinates and the arbitrary independent $H$
preserves absolute continuity, and hence gives $\dim_H\nu=1$.

For part~\textup{(b)}, if $\alpha<0$, partial summation gives
$\sum_\ell p_\ell<\infty$. Borel--Cantelli leaves only finitely many
active primes almost surely, so the set of possible sums is countable.
For $\alpha=0$, the hypotheses of
\Cref{cor:prime-series-zero-dimension} hold: the activation is $K/\ell$,
the expected jump is $O(\ell^{-2})$, and the Bernoulli entropy is
$O((\log\ell)/\ell)$. Thus there is a dimension-zero full carrier.
Since $\sum_\ell p_\ell=\infty$, \Cref{lem:levy-atomless} gives
atomlessness and hence singular continuity.

For part~\textup{(c)}, let $0<\alpha<1$ and $x=1/h$. For sufficiently
small $h$, one has $0<h<1$ and $x>L$.
By the empty-cylinder identity \eqref{eq:endpoint-cylinder}, applied to
the independent indicators $B_\ell$, $\ell>L$ (and $\xi_{\ell,1}=0$ for $\ell\le L$),
\[
 \mathbb P(X_\alpha\le h)=\prod_{L<\ell\le x}(1-p_\ell)\,
 \mathbb P\Bigl(\sum_{\ell>x}B_\ell\omega_\ell\le h\Bigr).
\]
Partial summation from the prime number
theorem gives
\[
 \mathbb E\sum_{\ell>x}B_\ell\omega_\ell
 \sim \frac{K(\log x)^{\alpha-1}}x=o(h).
\]
Markov's inequality consequently gives
\[
 \mathbb P(X_\alpha\le h)
 =\prod_{L<\ell\le x}(1-p_\ell)(1+o(1)).
\]
By \Cref{lem:prime-log-stieltjes},
$\sum_{L<\ell\le x}p_\ell=(K/\alpha)(\log x)^\alpha+C_0+o(1)$.
Summability of $p_\ell^2$ proves positivity and convergence of
$C_{K,\alpha,L}$ and proves \eqref{eq:activation-endpoint}.
The same empty-cylinder argument for $d$ independent copies gives
\eqref{eq:activation-endpoint} with constant $C_{K,\alpha,L}^{\,d}$
and exponent coefficient $dK/\alpha$.

For each integer $d\ge1$, let $\mu=\mu_\alpha^{*d}$. The finite-pattern
argument in the proof of \Cref{prop:scalar-density}\textup{(a)} applies because
$0<p_\ell<1$, $\omega_\ell\to0$ and $\sum_{\ell>L}\omega_\ell=\infty$.
Its finite-pattern anchors are dense in $[0,\infty)$; deleting a fixed
head preserves the endpoint asymptotic with a positive constant.
Thus every anchor $z$ satisfies
\[
 \mu([z,z+h])\ge c_z
 \exp\{-dK(\log(1/h))^\alpha/\alpha\}
\]
for small $h$.
These masses exceed every fixed positive power of $h$, and the same
support argument gives $\supp\mu=[0,\infty)$.
Every support-interior interval contains an anchor, so H\"older's
inequality excludes an $L^q$ density there for every $q>1$.
For any of these laws $\mu$, positive Fourier dimension would give
$|\widehat\mu(t)|\ll(1+|t|)^{-u}$ for some $0<u<1$;
Esseen's inequality \cite{Ess68} would then give
$Q_\mu(h)\ll h\int_{-1/h}^{1/h}|\widehat\mu(t)|\,dt\ll h^u$,
a contradiction. Finally \Cref{prop:renyi-collapse}\textup{(a)} gives
both dyadic dimensions zero. If the additional independent law has an
atom of mass $w>0$ at $a$, its convolution dominates $w$ times the
translate by $a$ of the positive law, or of its convolution power.
The anchor bounds therefore give the local exclusion on $(a,\infty)$;
the endpoint bound at $a$ preserves both dimension conclusions.
If the head is supported on $[a,\infty)$ and has an atom at $a$,
the convolution support is $[a,\infty)$.

For either alternative phase, the argument for part~\textup{(a)}
applies by \Cref{thm:annulus}. The cutoff at jump size $h$ is likewise
$h^{-1}+O(1)$, and the expected remaining tail is unchanged to leading
order. The cylinder product is therefore multiplied by $1+o(1)$,
so its constant is unchanged. The convergence, support and finite-pattern
arguments use only positivity and jumps asymptotic to $1/\ell$.

\end{proof}

The conclusions of \Cref{cor:heavier-activation,cor:positive-activation}
and the preceding extensions also hold with
activation probabilities $q_\ell\in[0,1]$ satisfying
$\sum_{\ell>L}|q_\ell-p_\ell|<\infty$, after absorbing probability-one
coordinates into $H$; implied constants may additionally depend on
$(q_\ell)$. For the zero-endpoint assertion, require $q_\ell<1$ and replace
$C_{K,\alpha,L}$ by
\[
 C_q=C_{K,\alpha,L}\prod_{\ell>L}\frac{1-q_\ell}{1-p_\ell}
 \in(0,\infty).
\]
Indeed, beyond a fixed cutoff both probabilities are at most $1/4$,
and the corresponding tail transforms satisfy uniformly in $t$
\[
 \bigl|\log|\widehat\nu_q^0(t)|-
             \log|\widehat\nu_p^0(t)|\bigr|
 \le4\sum_{\ell>L_0}|q_\ell-p_\ell|.
\]
The bounded finite-head averages preserve \eqref{eq:heavier-log-average}
and, with the pointwise decay, its liminf conclusion.
The activation mass and dissipation change by $O(1)$, so
\Cref{thm:activation-scale} preserves the critical thresholds.
\Cref{cor:summable-local-coupling} transfers absolute continuity and
the pure-type and dimension-zero assertions. For $0<\alpha<1$,
the change in the expected tail beyond $x$ is at most
$O(x^{-1}\sum_{\ell>x}|q_\ell-p_\ell|)=o(x^{-1})$; the cylinder argument
therefore gives $C_q$. The available jumps still have divergent sum
when $\alpha\ge0$, since the primes with $q_\ell=0$ have finite
$\sum1/\ell$. Hence the finite-pattern local exclusions persist.

In \Cref{lem:slow-jump-all-ac}, the reference probabilities are
$c_\ast/[\ell(\log\log\ell)^\beta]$ with $\beta\ge0$, and the
jumps are $(\log\ell)^{-A}$. In the following theorem the probabilities
are $K(\log\log\ell)^\beta/\ell$ with $\beta\in\R$, and the jumps
satisfy $a_\ell\sim1/\ell$. After reversing the sign of the activation exponent, the range
$-1\le\beta\le0$ gives absolute continuity for the logarithmic
jumps in that lemma, but singular continuity for the jumps
$a_\ell\sim1/\ell$ in the theorem below.

\begin{theorem}[Log-log activation trichotomy]
\label{thm:loglog-activation-mas03}
Fix $K>0$, $\beta\in\R$ and $\epsilon\in\{-1,1\}$, and choose
$L$ sufficiently large that
$p_\ell:=K(\log\log\ell)^\beta/\ell$ belongs to $(0,1/2)$ for
every prime $\ell>L$. Let $(B_\ell)_{\ell>L}$ be independent
Bernoulli variables with these probabilities. Choose one of the
three jump families
\[
 a_\ell=\log\frac\ell{\ell-1},\qquad
 a_\ell=\log(1+1/\ell),\qquad a_\ell=1/\ell,
\]
and put $S_\beta=\epsilon\sum_{\ell>L}B_\ell a_\ell$.
The series converges absolutely almost surely, and its law is
\[
 \begin{cases}
 \text{absolutely continuous},&\beta>0,\\
 \substack{\text{singular continuous with a Borel full carrier}\\
 \text{of Hausdorff dimension zero}},&-1\le\beta\le0,\\
 \text{countably atomic},&\beta<-1.
 \end{cases}
\]
The first conclusion persists on adding any independent random
variable $H$. The second and third persist on adding an independent
$H$ with countably atomic law.
\end{theorem}

\begin{proof}
Since $a_\ell\asymp1/\ell$, the series of absolute means is bounded
by a constant times
$\sum_{\ell>L}(\log\log\ell)^\beta/\ell^2<\infty$.
Tonelli's theorem proves convergence. Reflection permits
$\epsilon=1$ for the rest of the proof.

Suppose first that $\beta>0$. For large integers $k$, use the blocks
\[
 I_k=\{\ell:2^k<\ell\le2^{k+1}\},\qquad
 N_k=\sum_{I_k}B_\ell,\qquad
 Y_k=\sum_{I_k}B_\ell a_\ell,
\]
and write $w_\ell=p_\ell/(1-p_\ell)$ and $W_k=\sum_{I_k}w_\ell$.
Partial summation from the prime number theorem, and
$\sum_{I_k}p_\ell^2=o(W_k)$, give
\[
 W_k\sim K\frac{(\log k)^\beta}{k},\qquad
 q_k:=\mathbb P(N_k=1)
 =W_k\prod_{\ell\in I_k}(1-p_\ell)
 \sim K\frac{(\log k)^\beta}{k}.
\]
Here $W_k\to0$, so the product tends to one. On $N_k=1$, the
selected prime has probabilities $w_\ell/W_k$. The smooth
interpolation of $w_\ell$ is eventually decreasing, and on
$N<\ell\le2N$ it is comparable to
$K(\log\log N)^\beta/N$.
Abel summation in \eqref{eq:annulus-retained-power}, dividing by
$W_k\asymp K(\log\log 2^k)^\beta/\log 2^k$, consequently gives
\[
 |\mathbb E(e^{itY_k}\mid N_k=1)|\ll_{K,\beta}|t|^{-\gamma}
 \quad\text{if}\quad |t|^{13/50}\le2^k\le|t|^{99/100},
 \qquad \gamma=1/20000.
\]
This uses the common constants for the three families in
\Cref{thm:annulus}.

Set $T_n=2^{2^n}$ and
\[
 J_n=\left\{k:\left\lceil\tfrac{13}{25}2^n\right\rceil
             \le k\le\left\lfloor\tfrac{99}{100}2^n\right\rfloor
       \right\}.
\]
Every block in $J_n$ satisfies the preceding estimate throughout
$T_n\le|t|<T_{n+1}$. Uniformity of the singleton asymptotic and
comparison with the integral of $(\log u)^\beta/u$ give
\[
 M_n:=\sum_{k\in J_n}q_k
 \sim K(n\log2)^\beta\log\frac{99}{52}.
\]
Thus $\sum_n e^{-M_n/8}<\infty$ for every $\beta>0$.
\Cref{lem:conditional-block-smoothing}\textup{(a)} proves absolute continuity,
which is preserved by restoring the finite blocks and adding any
independent $H$.

Now suppose $\beta\le0$. For a fixed $C>0$ and all sufficiently
large primes, $p_\ell\le C/\ell$. Hence
\[
 \begin{gathered}
 \sum_{\ell>z}\mathbb E|B_\ell a_\ell|\ll z^{-1},\\
 \mathbb E I(y)=\sum_{L<\ell\le y}
   \{-p_\ell\log p_\ell-(1-p_\ell)\log(1-p_\ell)\}
 \ll\sum_{\ell\le y}\frac{\log\ell}{\ell}\ll\log y.
 \end{gathered}
\]
For the entropy bound, use that the Bernoulli entropy is increasing
on $(0,1/2)$ and bound it at $C/\ell$ after enlarging the cutoff.
\Cref{thm:dimension-zero-criterion} applies with its activation
envelope, tail exponent $1$ and information exponent $0$.
It gives a Borel full carrier of dimension zero.
Moreover, weighted Mertens summation gives
\[
 \sum_{L<\ell\le x}p_\ell
 =K\int_{\log\log L}^{\log\log x}s^\beta\,ds+O(1).
\]
The error is bounded because the Stieltjes remainder in
\Cref{lem:prime-log-stieltjes} decays exponentially in
$\sqrt{\log x}$. The activation sum diverges exactly when
$\beta\ge-1$. In that case \Cref{lem:levy-atomless} gives
atomlessness, and the dimension-zero carrier proves singularity.
When $\beta<-1$, Borel--Cantelli leaves only finitely many active
primes almost surely; their finite subsets form a countable set,
so the law is countably atomic.

Finally, convolution with a countably atomic law is a countable
mixture of translates. It preserves a dimension-zero full carrier,
preserves atomlessness when present, and preserves countable
atomicity. This proves the asserted head restrictions. They cannot
be omitted in the last two cases: an independent Gaussian head
makes either convolution absolutely continuous.
\end{proof}

\begin{corollary}[Normality for singular log-log activation]
\label{cor:loglog-singular-normality}
In \Cref{thm:loglog-activation-mas03}, suppose $-1<\beta<0$ and put
$a=1+\beta$. For any of the three jump families, any fixed signs
$\epsilon_\ell\in\{-1,1\}$, and any independent real head $H$, let
$\nu$ be the law of $H+\sum_{\ell>L}\epsilon_\ell B_\ell a_\ell$.
Then
\begin{equation}\label{eq:singular-loglog-fourier}
 \liminf_{|t|\to\infty}\frac{-\log|\widehat\nu(t)|}
  {(\log\log|t|)^a}\ge \frac K a\bigl(1-(2/3)^a\bigr).
\end{equation}
Almost every real number for $\nu$ is absolutely normal. If $H$ is
countably atomic, $\nu$ is singular continuous with a Borel full
carrier of Hausdorff dimension zero.
\end{corollary}

\begin{proof}
Put $u=\log|t|$ and use the window $(N_-,N_+]$ of the proof of
\Cref{thm:vk-fourier}, where $\log N_-\asymp u^{2/3}(\log u)^{1/3}$
and $\log N_+\asymp u$. The dyadic prefix estimates there, using
\Cref{prop:vk-prime-prefix} for the selected phase, bound the
$1/\ell$-weighted exponential sum by $O(1)$ uniformly up to every
endpoint of the window. Partial summation with the bounded decreasing
multiplier $(\log\log\ell)^\beta$ therefore gives
$\sum_{N_-<\ell\le N_+} p_\ell\cos(ta_\ell)=O(1)$.
Weighted Mertens summation as above gives
\begin{align*}
 \sum_{N_-<\ell\le N_+}p_\ell
 &=\frac K a\bigl((\log\log N_+)^a-(\log\log N_-)^a\bigr)+O(1)\\
 &=\left[\frac K a(1-(2/3)^a)+o(1)\right](\log u)^a.
\end{align*}
The local Bernoulli identity gives
$-\log|\widehat\nu(t)|\ge\sum_{N_-<\ell\le N_+}
p_\ell(1-p_\ell)(1-\cos(ta_\ell))$.
The cosines are even, the omitted factors have modulus at most one,
and $\sum p_\ell^2<\infty$. This proves \eqref{eq:singular-loglog-fourier}.
In particular $|\widehat\nu(t)|\ll(\log\log|t|)^{-2}$, so
\Cref{lem:iterlog-normality} gives almost-sure absolute normality.
The absolute-tail, information and activation estimates in the
preceding proof hold for all the signed terms. Thus
\Cref{thm:dimension-zero-criterion,lem:levy-atomless} give a
dimension-zero full carrier and atomlessness for the series.
Countably many translates retain both properties for a countably
atomic head.
\end{proof}

Let $m,n\ge0$ be integers with $m+n\ge1$, and let
\[
 Z_j=\sum_{\ell\ge3}B_{j,\ell}(\log\ell)^{-A_j}
 \quad(1\le j\le m),\qquad
 W_i=\sum_{\ell>L_i}C_{i,\ell}\omega_\ell
 \quad(1\le i\le n),
\]
where $A_j,c_j,K_i>0$, all Bernoulli variables are independent,
\[
 \begin{gathered}
 \sum_{\ell\ge3}|p_{j,\ell}-c_j/\ell|<\infty,
 \quad p_{j,\ell}=\mathbb P(B_{j,\ell}=1)\in[0,1],\\
 \sum_{\ell>L_i}|r_{i,\ell}-K_i\log\ell/\ell|<\infty,
 \quad r_{i,\ell}=\mathbb P(C_{i,\ell}=1)\in[0,1].
 \end{gathered}
\]
Let $\lambda_j,\kappa_i$ be nonzero real numbers.
Let $H$ be a real random variable independent of all
$B_{j,\ell}$ and $C_{i,\ell}$. Put
\[
 X=H+\sum_j\lambda_jZ_j+\sum_i\kappa_iW_i,
 \qquad b=\sum_j\frac{c_j}{A_j}+\sum_iK_i,
 \qquad \mu=\operatorname{Law}(X).
\]
In each $W_i$, independently of the other components, $\omega_\ell$
may be replaced throughout by $\log(1+1/\ell)$ or by $1/\ell$.
\Cref{fig:activation-scale} shows how the components $Z_j$ and $W_i$ enter
\Cref{thm:activation-scale} through the single coefficient~$b$, and which
further conclusions hold for positive coefficients.

\begin{figure}[t]
\centering
\begin{tikzpicture}[figdiagram, x=1pt, y=1pt,
  % Grid (pt).  Theorem lane x = 0.5..270, gutter 270..290,
  % absolute-continuity lane x = 290..383.5.
  % Levels: row 1 top 0; junction bar -66.2; row 2 bottom -133.8;
  % row 3 top -156.4.
  row1/.style={minimum height=47.2pt},
  row3/.style={minimum height=51.8pt, anchor=north}]
% ---- row 1: the two prime models inside the frame of the mixed sum ----
\node[finp, row1, anchor=north west] (S) at (5,0)
  {Slow components\enspace $Z_j=\sum_{\ell\ge3}B_{j,\ell}(\log\ell)^{-A_j}$\\
   with\enspace $\sum_{\ell\ge3}|p_{j,\ell}-c_j/\ell|<\infty$\\
   $(\log\ell)^{-A_j}\ge T^{-1}\iff\ell\le\exp(T^{1/A_j})$\\
   \figref{proof of \Cref{prop:slow-jump-regularity}; uses \Cref{lem:slow-jump-dissipation}}};
\node[finp, row1, anchor=north east] (K) at (379,0)
  {Critical components\enspace $W_i=\sum_{\ell>L_i}C_{i,\ell}\,\omega_\ell$\\
   with\enspace $\sum_{\ell>L_i}|r_{i,\ell}-K_i\log\ell/\ell|<\infty$\\
   $\omega_\ell\ge T^{-1}\iff\ell\le T+O(1)$\\
   \figref{proof of \Cref{cor:heavier-activation}\textup{(b)}; uses \Cref{thm:annulus}}};
\begin{scope}[on background layer]
  \node[fgroup, inner sep=0pt, fit={(0,7) (384,-52.2)}] (G) {};
\end{scope}
\node[ftag, anchor=west] (tagG) at (8,8.5)
  {Mixed sum $X=H+\sum_j\lambda_jZ_j+\sum_i\kappa_iW_i$ ($\lambda_j,\kappa_i\ne0$),\ \ $\mu=\operatorname{Law}(X)$;\ \ all variables independent};
% ---- junction: the two coefficients are needed jointly and add --------
\coordinate (bar) at (0,-66.2);
\node[fjoin] (J) at (135.25,-66.2) {};          % centred above T
\draw[ffeed] (S.south) |- (J);
\draw[ffeed] (K.south) |- (J);
\node[fhyp, anchor=east] (lS) at ($(S.south |- bar)+(-1.5,7)$) {coefficient $c_j/A_j$};
\node[fhyp, anchor=east] (lK) at ($(K.south |- bar)+(-1.5,7)$) {coefficient $K_i$};
% ---- row 2: the theorem (left lane) and conditional smoothing (right lane);
%      bottoms aligned, so that all arrows into row 3 have one length
\node[fres, minimum width=269.5pt, anchor=south west] (T) at (0.5,-133.8)
  {Regularity from activation at the jump scale\\
   activation mass\enspace $\Lambda(T)=\sum_{a_n\ge T^{-1}}p_n=b\log T+O(1)$,\\
   cancellation\enspace $\mathcal D(t)=\sum_np_n(1-\cos(ta_n))\ge b\log|t|-o(\log|t|)$\\
   \figref{\Cref{thm:activation-scale}, hypotheses~\eqref{eq:activation-scale-hypotheses}}};
\node[fmech, minimum width=93.5pt, anchor=south east] (M) at (383.5,-133.8)
  {Conditional\\ block smoothing\\ of each component\\
   \figref{\Cref{lem:conditional-block-smoothing,lem:slow-jump-all-ac};}\\
   \figref{\Cref{cor:positive-activation}\textup{(a)}}};
\draw[fuse] (J) -- node[fhyp, right=1pt, yshift=0.25pt, align=left] (lJ)
  {independent components add:\\ $b=\sum_jc_j/A_j+\sum_iK_i$} (J |- T.north);
\draw[fuse] (M.north |- G.south) -- node[flab, right=1pt, align=left] (lM)
  {each $Z_j$\\ and $W_i$} (M.north);
% ---- row 3: results; one top line, one height --------------------------
\node[fres, row3, minimum width=112.5pt] (E) at (56.75,-156.4)
  {Exact criteria ($q>0$, $s\in\R$)\\
   $\widehat\mu\in L^q(\R)\iff qb>1$\\
   $\mu\in H^s(\R)\iff s<b-\tfrac12$\\
   {\figsmall both fail at equality}\\
   \figref{\Cref{cor:mixed-activation-regularity}}};
\node[fres, row3, minimum width=146pt] (P) at (197,-156.4)
  {Positive sums\\
   globally $C^k$ density $\iff b>k+1$\\
   $\mathbb P(X\le h)\sim\mathrm{const}\cdot h^{b}$\enspace$(h\downarrow0)$\\
   {\figsmall $\Lambda(T)=b\log T+d+o(1)$; Tauberian step}\\
   \figref{\Cref{prop:activation-scale-endpoint}; \Cref{cor:positive-mixed-activation}\textup{(a)}}};
\node[fres, row3, minimum width=93.5pt] (A) at (336.75,-156.4)
  {Absolutely continuous,\\ $\dim_H\mu=1$,\\ every $b>0$\\
   \figref{\Cref{cor:mixed-activation-regularity}}};
% the head hypothesis runs to the right of its arrow, which therefore enters
% E left of centre; the positivity hypotheses stay inside the theorem lane
\draw[fuse] (33,0 |- T.south) -- node[fhyp, right=1pt, align=left] (lE)
  {$H$: constant + finite sum of independent\\ variables on finite arithmetic progressions} (33,0 |- E.north);
\draw[fuse] (195,0 |- T.south) -- node[fhyp, right=1pt, align=left] (lP)
  {$\lambda_j,\kappa_i>0$,\\ $p_{j,\ell},r_{i,\ell}<1$,\ $H=0$} (195,0 |- P.north);
\draw[fuse] (A.north |- M.south) -- node[flab, right=1pt] (lA) {convolution} (A.north);
\end{tikzpicture}
\caption{Slow and critical components of the mixed sum preceding
\Cref{cor:mixed-activation-regularity} enter \Cref{thm:activation-scale} through one
coefficient~$b$; one family may be empty, and $a_n$, $p_n$ run over the absolute jump sizes and
activation probabilities of all terms $\lambda_jZ_j$, $\kappa_iW_i$. Common activation of
several coordinates (\Cref{thm:shared-prime-anisotropic,prop:shared-prime-sobolev}) is a
different problem.}
\label{fig:activation-scale}
\end{figure}

\KeepHeadingSpace{6\baselineskip}
\begin{corollary}[Regularity of mixed prime sums]
\label{cor:mixed-activation-regularity}
\index{Activation intensity!jump-scale, mixed laws}
Suppose $H$ is a constant plus a finite sum of independent variables
supported on finite arithmetic progressions.
Then $\mu$ is absolutely continuous, $\dim_H\mu=1$, and
$\dim_F\mu=\min(1,2b)$. For $q>0$ and $s\in\R$,
\[
 \int_{\R}(1+t^2)^s|\widehat\mu(t)|^q\,dt<\infty
 \quad\Longleftrightarrow\quad qb>2s+1,
\]
and its density belongs to $H^s(\R)$ exactly when $s<b-\tfrac12$.
For every integer $k\ge0$ with $k+1<b$, its density has bounded
continuous derivatives through order $k$. Its $k$th derivative is
globally $\alpha$-H\"older continuous if $0<\alpha\le1$ and
$k+\alpha+1<b$, as in \Cref{thm:activation-scale}.
\end{corollary}

\begin{proof}
The slow components satisfy \eqref{eq:activation-scale-hypotheses}
with coefficient $c_j/A_j$, by
\Cref{prop:slow-jump-regularity}; the critical components satisfy
it with coefficient $K_i$, by the proof of
\Cref{cor:heavier-activation}\textup{(b)}. Nonzero scaling preserves
the coefficient, and independence adds the activation masses and
dissipations. Thus \Cref{thm:activation-scale} applies.
Each component is absolutely continuous by
\Cref{lem:slow-jump-all-ac} and
\Cref{cor:positive-activation}\textup{(a)}, so
convolution gives absolute continuity and Hausdorff dimension one.
\end{proof}

\begin{corollary}[Endpoint and density of positive mixed sums]
\label{cor:positive-mixed-activation}
Suppose $\lambda_j,\kappa_i>0$ and $p_{j,\ell},r_{i,\ell}<1$.
\begin{enumerate}[label=\textup{(\alph*)},leftmargin=*]
\item Let $H=0$, and put
\[
 C_j=\lim_{x\to\infty}(\log x)^{c_j}
                       \prod_{3\le\ell\le x}(1-p_{j,\ell}),\qquad
 D_i=\lim_{x\to\infty}x^{K_i}
                       \prod_{L_i<\ell\le x}(1-r_{i,\ell}).
\]
These limits are positive and finite. As $h\downarrow0$,
\begin{equation}\label{eq:mixed-activation-endpoint}
 \mathbb P(X\le h)\sim\frac{e^{-\gamma b}}{\Gamma(1+b)}
 \left(\prod_jC_j\lambda_j^{-c_j/A_j}\right)
 \left(\prod_iD_i\kappa_i^{-K_i}\right)h^b.
\end{equation}
For every integer $k\ge0$, a globally $C^k$ density exists exactly
when $b>k+1$, with bounded derivatives through order $k$.
\KeepHeadingSpace{4\baselineskip}
\item Let $H\ge z_0$ with $\mathbb P(H=z_0)>0$.
The same $C^k$ criterion holds, and $\supp\mu=[z_0,\infty)$.
If $0<b<1$, its density $g$ is essentially unbounded on every
nonempty open interval $I\subset(z_0,\infty)$, has no finite-valued
representative continuous at an interior point, and satisfies
\[
 g\notin L^p(I)\qquad\left(\frac1{1-b}\le p\le\infty\right).
\]
For $1/2<b<1$,
\[
 g\in L^p(\R)\quad\Longleftrightarrow\quad1\le p<\frac1{1-b}.
\]
These density and support conclusions remain valid after restoration
of probability-one coordinates, with their deterministic translation
absorbed into $H$ and $z_0$.
\end{enumerate}
\end{corollary}

In particular, the pure critical series with $K=3/4$ has exact global
range $1\le p<4$.

\begin{proof}
Put $b_j=c_j/A_j$ and
\[
 e_j=\sum_{\ell\ge3}(p_{j,\ell}-c_j/\ell),\qquad
 R_j=\sum_{\ell\ge3}\bigl(\log(1-p_{j,\ell})+p_{j,\ell}\bigr).
\]
The hypotheses give $p_{j,\ell}\to0$, $\sum_\ell p_{j,\ell}^2<\infty$
and finite expected jump sum. Since every $p_{j,\ell}<1$, the finitely
many large probabilities cause no divergence in $R_j$.
Mertens' theorem and the definition of $M_{\mathrm{odd}}$ give
\[
 C_j=e^{-c_jM_{\mathrm{odd}}-e_j+R_j}\in(0,\infty).
\]
The scaled cutoff is exactly
$\ell\le\exp((\lambda_jT)^{1/A_j})$, so its activation mass is
\[
 \Lambda_j(T)=b_j\log T+b_j\log\lambda_j
                  +c_jM_{\mathrm{odd}}+e_j+o(1).
\]
For the critical components define
\[
 \begin{split}
 d_i&=\lim_{x\to\infty}
       \left(\sum_{L_i<\ell\le x}r_{i,\ell}-K_i\log x\right),\\
 R_i^{\rm c}&=\sum_{\ell>L_i}
       \bigl(\log(1-r_{i,\ell})+r_{i,\ell}\bigr).
 \end{split}
\]
\Cref{lem:prime-log-stieltjes}, with $g(U)=U$ after a fixed cutoff,
and the summable probability discrepancy give finite $d_i$.
Summability of $r_{i,\ell}^2$ gives finite
$R_i^{\rm c}$, and hence $D_i=e^{-d_i+R_i^{\rm c}}\in(0,\infty)$.
For each of the three allowed phases the scaled critical cutoff is
\[
 \ell\le\kappa_iT+O(1),
\]
so this component contributes
$K_i\log T+K_i\log\kappa_i+d_i+o(1)$ to the activation mass.
Consequently $\Lambda(T)=b\log T+d+o(1)$, where
\[
 \begin{split}
 d&=\sum_j(b_j\log\lambda_j+c_jM_{\mathrm{odd}}+e_j)
       +\sum_i(K_i\log\kappa_i+d_i),\\
 R&=\sum_jR_j+\sum_iR_i^{\rm c},\qquad
 e^{-d+R}=\prod_jC_j\lambda_j^{-b_j}\prod_iD_i\kappa_i^{-K_i}.
 \end{split}
\]
\Cref{prop:activation-scale-endpoint} now proves the endpoint and
exact smoothness assertions, including the minimum-atom head
and deterministic translation clauses.

For the local density assertions in \textup{(b)}, apply the proof of
\Cref{prop:scalar-density}\textup{(a)}. The available positive jumps tend to zero
and have divergent unweighted sum, so their finite-pattern anchors are
dense in $[z_0,\infty)$. Every such pattern has positive probability.
Deleting its finite head preserves the exponent $b$ in
\eqref{eq:mixed-activation-endpoint}, giving
$\mu([z,z+h])\ge c_z h^b$ for all sufficiently small $h$.
These bounds identify the support. When $0<b<1$, they exclude a
locally bounded density. For finite $p\ge1/(1-b)$, H\"older's inequality
and absolute continuity of the integral of $|g|^p$ give the local
$L^p$ exclusion, including equality, as in the cited proof.
For $1/2<b<1$, \Cref{cor:mixed-activation-regularity} with $s=0$
gives $L^q$ integrability of the head-free transform for every $q>1/b$.
Multiplication by $\widehat{\operatorname{Law}(H)}$, of modulus at most
one, preserves this integrability. Inverse Hausdorff--Young and
interpolation with $L^1$ give the complementary global $L^p$ range.
\end{proof}

\KeepHeadingSpace{6\baselineskip}
\begin{corollary}[Density of signed mixed sums]
\label{cor:signed-mixed-activation}
In the mixed-sum model above, suppose the coefficients
$\lambda_j,\kappa_i$ include both signs and $H$ has an atom.
Then $\mu$ is absolutely continuous and $\supp\mu=\R$ for every $b>0$.
For $0<b<1$, the local density conclusions of
\Cref{cor:positive-mixed-activation}\textup{(b)} hold on every nonempty
open interval $I\subset\R$; for $1/2<b<1$, its global $L^p$ range also holds.
Probability-one coordinates may be absorbed into $H$.
\end{corollary}

\begin{proof}
Use the signed-anchor argument of \Cref{prop:scalar-density}\textup{(b)}.
After a common cutoff in every component,
\Cref{cor:positive-mixed-activation}\textup{(a)} applied to the absolute
weighted tail gives the symmetric-window mass bound.
For $1/2<b<1$, the Fourier-integrability argument in the proof of
\Cref{cor:positive-mixed-activation}\textup{(b)} gives the same global
$L^p$ range.
\end{proof}

\KeepHeadingSpace{7\baselineskip}
\section{Auxiliary Heath-Brown decomposition}
\label{ssec:annulus-decomposition}

The main proof of \Cref{thm:annulus} uses Vaughan's identity and the
wider Type~II range \eqref{eq:typeII-vaughan-range}. The auxiliary
Heath-Brown reduction below uses the narrower range
\eqref{eq:bilinear-ordered-range}. It makes the factor grouping
explicit: a single long variable yields Type~I; otherwise products of
several variables form the two Type~II variables. The blocks and
coefficients are fixed independently of the endpoint $x$, so the
resulting identities apply simultaneously to all prefixes.

Fix the parameters of that theorem and, by conjugation, take $\tau>0$.
Combining the reduction's
$O_r((\log N)^{2r})$ pieces with
\Cref{cor:typeI-weighted,cor:bilinear-weighted} bounds the von Mangoldt
sum by
\[
 \begin{gathered}
 O_{r,\delta,j}\!\left(
 N\tau^{-\theta}(\log N)^{6r+1}
 +N^{1-\eta/2}(\log N)^{e_r+2r}\right),\\
 \eta=\eta_{\mathrm{II}}(r,\delta),\qquad e_r=4r^2+4r.
 \end{gathered}
\]
Prime-power removal and partial summation then give
\eqref{eq:annulus-retained-power} for every $\gamma<\gamma_0$:
the strict gap absorbs these fixed logarithmic powers. Thus the two
routes retain the same final power, with different intermediate
logarithmic costs.

\KeepHeadingSpace{3\baselineskip}
\begin{lemma}[Heath-Brown identity {\cite[Lemma~1, equation~(6)]{HB82}; \cite[Proposition~13.3]{IK04}}]\label{lem:HB}
For every integer $K\ge 1$ and $z\ge 1$, and every integer $1\le n\le z^K$,
\[
  \Lambda(n)\;=\;\sum_{j=1}^{K}(-1)^{j-1}\binom{K}{j}\!\!
  \sum_{\substack{n=m_1\cdots m_j\,n_1\cdots n_j\\ m_1,\ldots,m_j\le z}}
  \!\!\mu(m_1)\cdots\mu(m_j)\,(\log n_1).
\]
\end{lemma}

\begin{proof}
Let $\varepsilon$ be the identity for Dirichlet convolution, put
$\mu_z(n)=\mu(n)\mathbf1_{\{n\le z\}}$, and set
$D=\mu_z*\mathbf1$, $E=\varepsilon-D$.
Since $E(n)=0$ for $n\le z$, one has
$E^{*K}(n)=0$ for $n\le z^K$. The binomial identity gives
\[
 \sum_{j=1}^K(-1)^{j-1}\binom Kj D^{*j}*\Lambda
 =(\varepsilon-E^{*K})*\Lambda.
\]
For $n\le z^K$ the right side equals $\Lambda(n)$.
Using $\mathbf1*\Lambda=\log$, one has
$D^{*j}*\Lambda=\mu_z^{*j}*\mathbf1^{*(j-1)}*\log$, with
$\mathbf1^{*0}=\varepsilon$, which gives the asserted sum.
\end{proof}

Use the constants in \eqref{eq:HB-constants}.
The sums \(\mathcal T_\sigma\) in parts~(a) and~(b) of the next lemma are called
Type~I and Type~II sums, respectively.

\begin{lemma}[Heath-Brown Type~I/II reduction]\label{lem:HB-typeI-II}
Let $r\ge3$ be an integer, let $N\ge N_0(r):=2^{2r^2/(r-1)}$ be dyadic, and let
$F:[N,2N]\to\mathbb C$.
There are a finite set $\mathcal S=\mathcal S_{\mathrm I}\sqcup
\mathcal S_{\mathrm{II}}$, with $|\mathcal S|\le B_r(\log N)^{2r}$,
and the following data, independent of $x$, such that
\[
 \sum_{N<n\le x}\Lambda(n)F(n)=\sum_{\sigma\in\mathcal S}
 \mathcal T_\sigma(x)\qquad(N\le x\le2N).
\]
\begin{enumerate}[label=\textup{(\alph*)}]
\item For $\sigma\in\mathcal S_{\mathrm I}$, there are dyadic blocks
$\mathcal M_\sigma,\mathcal W_\sigma$ of scales $M_\sigma,W_\sigma$,
allowing $\mathcal M_\sigma=\{1\}$ with $M_\sigma=1$,
with $M_\sigma\le\widetilde c_rN^{1/r}$ and
$N/4<M_\sigma W_\sigma<2N$, a weight $w_\sigma\equiv1$ or
$w_\sigma(n)=\log n$, and coefficients satisfying
\begin{equation}\label{eq:typeI-form}
\begin{gathered}
 \mathcal T_\sigma(x)=\sum_{m\in\mathcal M_\sigma}a_m^{(\sigma)}
 \sum_{\substack{n\in\mathcal W_\sigma\\N<mn\le x}}
 w_\sigma(n)F(mn),\\
 |a_m^{(\sigma)}|\le H_r\mathbf d_{2r}(m)(\log N)^{2r}.
\end{gathered}
\end{equation}
\item For $\sigma\in\mathcal S_{\mathrm{II}}$, there are dyadic blocks
$\mathcal U_\sigma,\mathcal V_\sigma$ of scales satisfying
\begin{equation}\label{eq:typeII-ordered}
 \widetilde c_r^{-1}N^{1/r}\le V_\sigma\le U_\sigma
 \le\widetilde c_rN^{1-1/r},\qquad N/4<U_\sigma V_\sigma<2N,
\end{equation}
and coefficients satisfying
\begin{equation}\label{eq:typeII-form}
 \mathcal T_\sigma(x)=\sum_{u\in\mathcal U_\sigma}a_u^{(\sigma)}
 \sum_{\substack{v\in\mathcal V_\sigma\\N<uv\le x}}
 b_v^{(\sigma)}F(uv),
\end{equation}
\[
 |a_u^{(\sigma)}|\le H_r\mathbf d_{2r}(u)(\log N)^{2r},\qquad
 |b_v^{(\sigma)}|\le H_r\mathbf d_{2r}(v)(\log N)^{2r}.
\]
\end{enumerate}
\end{lemma}

\begin{proof}
Apply \Cref{lem:HB} with \(K=r\) and \(z=(2N)^{1/r}\), and partition
each variable into
\[
 \{1\}\sqcup\bigsqcup_{k\ge0}(2^k,2^{k+1}].
\]
For a fixed term and a fixed tuple of blocks, suppress variables equal
to \(1\), write the remaining block scales as
\(A_1,\ldots,A_L\), where \(L\le2r\), and put
\(P_k=\prod_{i\le k}A_i\) after reordering when necessary.  If the
term is nonzero on \(N<n\le2N\), then
\begin{equation}\label{eq:HB-scale-product}
  \frac{N}{2^L}<P_L<2N.
\end{equation}
Since \(N\ge N_0(r)\), each variable has at most
\(2\log N\) admissible blocks. The outer sum in the identity has $r$
summands, indexed by $1\le j\le r$. A product of $k$ dyadic blocks
lies in $(P_k,2^kP_k]$ and meets at most $k$ dyadic blocks.
Each tuple therefore yields at most
$\max\{1,L-1,k(L-k)\}\le r^2$ pieces, since $L\le2r$.
Summing these bounds gives
\[
 |\mathcal S|\le r^3 2^{2r}(\log N)^{2r}.
\]

Suppose first that some \(A_{i_0}\ge N^{1-1/r}\).  This is an
unrestricted \(n\)-variable, since every restricted \(m_i\le(2N)^{1/r}\).
Merge the other variables into \(m\).  Their scale \(M_0\) satisfies
\(M_0<2N^{1/r}\), and their product ranges over at most \(\max\{1,L-1\}\)
dyadic blocks.  On each such block the resulting term has the form
\eqref{eq:typeI-form}; its other block is the original
\(A_{i_0}\)-block.  The logarithm is either retained as
\(w_\sigma(n)=\log n\) or absorbed into \(a_m^{(\sigma)}\), in which
case \(w_\sigma\equiv1\).  Iterated convolution gives
\[
 |a_m^{(\sigma)}|
 \le\binom rj\mathbf d_{2r-1}(m)\log(2N)
 \le2^{r+1}\mathbf d_{2r}(m)(\log N)^{2r}.
\]
The outer scale is at most
\(2^{L-1}M_0\le\widetilde c_rN^{1/r}\).

Now suppose that every \(A_i<N^{1-1/r}\).  Order the scales
decreasingly and let
\[
 k^\star=\min\{k:P_k\ge N^{1/r}\}.
\]
The set is nonempty by \eqref{eq:HB-scale-product} and the definition
of \(N_0(r)\).  Minimality and monotonicity give
\[
 N^{1/r}\le P_{k^\star}<N^{1-1/r}.
\]
Indeed, for \(k^\star\ge2\),
\(P_{k^\star}<N^{2/r}\le N^{1-1/r}\), while the case
\(k^\star=1\) is immediate.

Moreover \(k^\star<L\).  Since \(\log_2P_L\) and \(\log_2N\) are integers,
\eqref{eq:HB-scale-product} sharpens to \(P_L\ge2N/2^{L}\); at \(L=1\) this
reads \(P_1\ge N\), contradicting \(A_1<N^{1-1/r}\), so \(L\ge2\).  If
\(k^\star=L\), then \(P_{L-1}<N^{1/r}\) by minimality, while the decreasing
order gives \(A_L^{L-1}\le P_{L-1}\) and hence
\(P_L=P_{L-1}A_L\le P_{L-1}^{L/(L-1)}<N^{L/(r(L-1))}\).  Combined with
\(P_L\ge2N/2^{L}\) this forces
\[
 \log_2N<\frac{r(L-1)^{2}}{r(L-1)-L}
 \;\le\;\frac{(2r-1)^{2}}{2r-3}
 \;<\;\frac{2r^{2}}{r-1},
\]
the middle inequality because the left member increases in \(L\) on
\(2\le L\le2r\), and the right because
\((2r-1)^{2}(r-1)-2r^{2}(2r-3)=-2r^{2}+5r-1<0\) for \(r\ge3\).  This
contradicts \(N\ge N_0(r)\).

Merge the first \(k^\star\) variables
into \(u\) and the others into \(v\), and subdivide both product
ranges into dyadic blocks.  Index each nonzero block pair
by its own $\sigma$, interchanging \(u\) and \(v\) when
necessary so that \(V_\sigma\le U_\sigma\).  For a group of $k\le2r$ variables, the absolute convolution
coefficient is at most $\mathbf d_k(n)$, or $\mathbf d_k(n)\log(2N)$ if
the group contains the logarithmic weight. Assign the binomial factor
$\binom rj\le2^r$ to the first group. Since
$\mathbf d_k(n)\le\mathbf d_{2r}(n)$ and $\log(2N)\le2\log N$, both groups
satisfy the asserted majorant $H_r\mathbf d_{2r}(n)(\log N)^{2r}$.
Subdivision only restricts their supports.

For every retained Type~I or Type~II block pair, the support
\(N<uv\le2N\) implies
\[
 \frac N4<UV<2N.
\]
In the Type~II case the values $u,v$ themselves satisfy
\(u>P_{k^\star}\ge N^{1/r}\) and
\(u\le2^{k^\star}P_{k^\star}<2^{2r}N^{1-1/r}\), whence
\(v>N/u>2^{-2r}N^{1/r}\) and \(v\le2N/u<2N^{1-1/r}\).  Both blocks have this form, since \(k^\star<L\), so each consists of values in \((S,2S]\) for its
scale \(S\); applying this to the \(u\)- and \(v\)-blocks gives
\[
 \widetilde c_r^{-1}N^{1/r}=2^{-2r-1}N^{1/r}<V_\sigma\le U_\sigma
 <2^{2r}N^{1-1/r}<\widetilde c_rN^{1-1/r},
\]
which is \eqref{eq:typeII-ordered}.  The same support argument gives the
Type~I product relation, where the \(m\)-block may be \(\{1\}\): there
\(M_\sigma\le m\le2M_\sigma\) and \(W_\sigma<n\le2W_\sigma\) suffice.

The blocks and coefficients do not depend on the upper limit \(x\).
Only the restrictions
\(\mathbf1_{\{N<mn\le x\}}\) or
\(\mathbf1_{\{N<uv\le x\}}\) depend on it.  Summing the fixed block identities proves
the assertion simultaneously for \(N\le x\le2N\).
\end{proof}

\chapter{Determinant densities and explicit bounds}
\label{sec:explicit}

For a fixed prime $p$, the probability that a uniform matrix over
$\Fp$ has primitive-root determinant is the product of an
invertibility factor and $\varphi(p-1)/(p-1)$. This chapter derives
extremal bounds for that density and rational certificates from partial
factorizations of $p-1$. The finite-dimension estimates are uniform in
the matrix dimension; the arithmetic averages use the progression laws
of \Cref{ssec:progression-regimes}.

\Cref{ssec:certgen} applies the certificates to prime selection and
studies sampling without replacement for residue number systems. It
also treats friable predecessors and prime cofactors.
\Cref{sec:composite-modulus} then gives the exact determinant laws over
finite rings and the limiting laws obtained by varying the prime base
or extension degree of a finite field.

The final application chapter, \Cref{sec:syd21-app}, gives an
independent algebraic LWE reduction with a prescribed nonzero
determinant law, followed by determinant statistics for independent-entry
matrices over fixed fields. The lattice results of
\Cref{app:cyclotomic-smoothing} are developed separately.

\section{The determinant density}
\label{sec:det-density}
\index{Determinant densities!uniform matrices}

Write $p_k$ for the $k$th prime and use $\omega$ and $\vartheta$ as in
\eqref{eq:arithmetic-notation}. Put
\[
 N_k:=\prod_{i\le k}p_i,\qquad
 A_k:=\prod_{i\le k}(1-1/p_i).
\] Let $p^*(k)$ be the least prime
congruent to $1\pmod{N_k}$; $P$ is extended to real arguments as in
\Cref{lem:matrix-product-bounds}.

Define (see~\cite[\S~7, p.~126]{SYD21})
\[
 M_n^{\mathrm{prim}}(\Fp)
 :=\{M\in\Mn:\det(M)\text{ generates }\Fp^\times\}.
\]
Since $\det:\GL_n(\mathbb F_p)\to\mathbb F_p^\times$ is a surjective
homomorphism and $|\GL_n(\mathbb F_p)|=p^{n^2}\prod_{j=1}^n(1-p^{-j})$
(see Z.-X. Wan~\cite[Theorem~13.3.3]{MP13}),
\[
 c_n(p)=\frac{|M_n^{\mathrm{prim}}(\mathbb F_p)|}{|M_n(\mathbb F_p)|}
 =\frac{\varphi(p-1)}{p-1}\prod_{j=1}^n(1-p^{-j}),
\]
as in~\cite[Corollary~4]{SYD21}, and letting $n\to\infty$ recovers
\eqref{eq:def-P} (see~\cite[Remark~3(i)]{SYD21}); put $c_\infty(p):=c(p)$.
Prime-power fields and finite or varying $n$ are treated in
\Cref{prop:finite-ring-det,prop:finite-matrix-product-error,prop:finite-matrix-deficit,cor:finite-dimension,cor:ap-sharp-extrema}.
For $p-1=r_1^{e_1}\cdots r_k^{e_k}$, \eqref{eq:def-P} reads
\begin{equation}\label{eq:cp-nt}
 c(p)=P(p)\prod_{i=1}^k(1-1/r_i).
\end{equation}

\section{Universal bounds and sharp extremal order}
\label{sec:explicit-universal}

For a fixed number of distinct prime factors, the totient ratio is
smallest when those factors are the smallest primes. This comparison
gives a lower bound for each individual determinant density. To approach
the extremal order as $p$ varies, the complementary construction forces
many small primes to divide $p-1$ by placing $p$ in suitable progressions.
These two uses of primorials underlie the bounds below.

The factor-count bound is governed by $\omega(p-1)$, which has
normal order $\log\log p$~\cite{Erd35} and satisfies the shifted-prime
analogue~\cite[(1), p.~15, and Theorem~3, pp.~24--27]{Hal56} of the
Erd\H{o}s--Kac theorem~\cite{EK40}.

\begin{proposition}[Explicit factor-count bound]\label{prop:factor-count-bound}
\index{Determinant densities!factor-count bounds}
For every prime $p\ge3$, with $k=\omega(p-1)$,
\[
 c(p)\ge P(N_k+1)A_k\ge(1-1/N_k)A_k.
\]
In particular, $k\le10$ implies $c(p)>0.1579$.
\end{proposition}

\begin{proof}
The distinct prime factors of $p-1$ give $p-1\ge N_k$ and
$\varphi(p-1)/(p-1)\ge A_k$. By \Cref{lem:matrix-product-bounds},
$P(p)\ge P(N_k+1)\ge1-1/N_k$.
\Needspace{5\baselineskip}
Exact rational comparison for $1\le k\le10$ gives
\[
 \min_{1\le k\le10}(1-1/N_k)A_k
 =\frac{476997542387712}{3019980554856265}>\frac{1579}{10000}.
\]
\end{proof}

\begin{corollary}[An explicit bound of Rosser--Schoenfeld type]
\label{cor:rosser-schoenfeld}
For every prime $p\ge3$,
\begin{equation}\label{eq:rs-cp}
 \frac1{c(p)}
 <\left(1+\frac1{p-2}\right)
 \left(e^\gamma\log\log p+\frac{5}{2\log\log p}\right).
\end{equation}
\end{corollary}

\begin{proof}
Rosser and Schoenfeld~\cite[Theorem~15, (3.41)--(3.42), p.~72]{RS62} give
\[
 \frac n{\varphi(n)}
 <e^\gamma\log\log n+\frac{5}{2\log\log n}
 \qquad(n\ge3,\ n\ne223092870).
\]
At the exceptional integer, and hence for every integer $n\ge3$,
the bound with $2.50637/\log\log n$ in place of $5/(2\log\log n)$ holds.
The exception cannot equal $p-1$ for a prime $p$, since
$223092870+1=317\cdot703763$.
Put $U_{\mathrm{RS}}(t):=e^\gamma\log\log t+5/(2\log\log t)$. Its derivative is
\[
 U_{\mathrm{RS}}'(t)
 =\frac{e^\gamma-5/(2(\log\log t)^2)}{t\log t}>0
 \qquad(t\ge27).
\]
Thus $(p-1)/\varphi(p-1)<U_{\mathrm{RS}}(p)$ for $p\ge29$.
For $p\in\{3,5,7,11,13,17,19,23\}$, the same inequality follows from
$(p-1)/\varphi(p-1)\le3<2\sqrt{(5/2)e^\gamma}\le U_{\mathrm{RS}}(p)$.
Finally, \Cref{lem:matrix-product-bounds} gives
$P(p)\ge1-1/(p-1)$, so \eqref{eq:cp-nt} proves \eqref{eq:rs-cp}.
\end{proof}

\begin{lemma}[An unrestricted integer Rankin bound]
\label{lem:integer-rankin-mas12}
For a finite set $E$ of primes, put
\[
 F_E=\prod_{\ell\in E}(1-1/\ell),\qquad
 a_E=\frac{e^{-\gamma}}{F_E},\qquad
 V_E(n)=\prod_{\substack{\ell\mid n\\\ell\notin E}}(1-1/\ell)^{-1}.
\]
For every real $r>0$ and $X\ge1$,
\begin{equation}\label{eq:integer-rankin-mas12}
 \sum_{n\le X}V_E(n)^r
 \le X\prod_{\ell\notin E}
   \left(1+\frac{(1-1/\ell)^{-r}-1}{\ell}\right).
\end{equation}
As $r\to\infty$, the logarithm of the product is at most
\[
 r\bigl(\log\log r+\gamma+\log F_E+O_E(1/\log r)\bigr).
\]
In particular, for each $0<\delta<1$ there are $c_\delta>0$ and
$T_0=T_0(E,\delta)$ such that, for every $X\ge1$ and $T\ge T_0$,
\begin{equation}\label{eq:integer-double-tail-mas12}
 \#\{n\le X:V_E(n)\ge T\}
 \le X\exp\{-c_\delta\exp((1-\delta)a_E T)\}.
\end{equation}
\end{lemma}

\begin{proof}
Set $a_r(\ell)=(1-1/\ell)^{-r}-1\ge0$ and extend multiplicatively to
squarefree integers supported outside $E$, with $a_r(1)=1$.
Expanding over divisors and then using positivity gives
\[
 \sum_{n\le X}V_E(n)^r
 =\sum_{\substack{d\le X\text{ squarefree}\\\ell\mid d\Rightarrow\ell\notin E}}
      a_r(d)\left\lfloor\frac Xd\right\rfloor
 \le X\sum_{\substack{d\ge1\text{ squarefree}\\\ell\mid d\Rightarrow\ell\notin E}}
       \frac{a_r(d)}d.
\]
The last series is the convergent Euler product in
\eqref{eq:integer-rankin-mas12}: for fixed $r$, its local increments
are $O_r(\ell^{-2})$. For $r$ larger than every prime in $E$, the
factors with $\ell\le r$ are at most $(1-1/\ell)^{-r}$, whereas those
with $\ell>r$ have logarithms $O(r/\ell^2)$. Thus Mertens' theorem gives
\[
 \log\prod_{\ell\notin E}(1+a_r(\ell)/\ell)
 \le r\sum_{\substack{\ell\le r\\\ell\notin E}}-\log(1-1/\ell)+O(1)
 =r(\log\log r+\gamma+\log F_E)+O_E(r/\log r).
\]
Choose $r=\exp((1-\delta)a_ET)$ and apply Markov's inequality.
The logarithm of the resulting bound after division by $X$ is at most
\[
 r\{-\log T+\log\log r+\gamma+\log F_E+o_E(1)\}
 =r\{\log(1-\delta)+o_E(1)\}\le-c_\delta r
\]
for all sufficiently large $T$, proving the last assertion uniformly in $X$.
\end{proof}

\begin{corollary}[All lower ranks in a fixed progression]
\label{cor:ap-sharp-extrema}
\index{Extremal orders!arithmetic progressions}
Fix integers $m\ge1$ and $b$ with $\gcd(b,m)=1$. Put
\[
 E:=\{\ell\textup{ prime}:\ell\mid m,\ b\not\equiv1\pmod\ell\},
 \qquad F_E:=\prod_{\ell\in E}(1-1/\ell),\qquad N:=\pi(x;m,b).
\]
For arbitrary dimension arrays $n_x(p)\in\Z_{\ge1}\cup\{\infty\}$, let
$d_{k,x}$ be the $k$th smallest of the values $c_{n_x(p)}(p)$
$(p\le x,\ p\equiv b\pmod m)$, counted with multiplicity. Whenever
$N/k\to\infty$,
\begin{equation}\label{eq:all-lower-ranks-mas12}
 d_{k,x}\sim\frac{e^{-\gamma}}{F_E\log\log(N/k)}.
\end{equation}
For every $r_0(x)\to\infty$, this asymptotic is uniform over all arrays
and all $1\le k\le N/r_0(x)$. More precisely, the minimum-scale
asymptotic holds exactly for the following ranks:
\begin{equation}\label{eq:ap-sharp-extrema}
 d_{k,x}\sim\frac{e^{-\gamma}}{F_E\log\log x}
 \quad\Longleftrightarrow\quad
 \log\log(N/k)\sim\log\log x,
\end{equation}
where the assertion on the right includes $N/k\to\infty$.
Consequently, for every fixed dimension assignment $n(p)\in\Z_{\ge1}\cup\{\infty\}$,
\[
 \limsup_{\substack{p\to\infty\\p\equiv b\pmod{m}}}
 \frac1{c_{n(p)}(p)\log\log p}=e^\gamma F_E;
\]
in particular, $c_{n(p)}(p)$ has infimum and lower limit $0$ on the progression.
\end{corollary}

\begin{proof}
Write $a_E=e^{-\gamma}/F_E$ and $R=N/k$, and fix $0<\varepsilon<1/2$.
Throughout, $R\to\infty$; all estimates below are uniform once
$R\ge r_0(x)\to\infty$. On the progression no prime in $E$ divides
$p-1$, so $V_E(p-1)=(p-1)/\varphi(p-1)$.

We first show that fewer than $k$ entries lie at or below
$t_-=(1-\varepsilon)a_E/\log\log R$.
For $R\le\sqrt{\log x}$, the fixed-progression CDF estimate of
\Cref{thm:sharp-prime-convergence} and
$c(p)\le c_{n_x(p)}(p)\le\varphi(p-1)/(p-1)$ give
\[
 \frac1N\#\{p\le x:p\equiv b\pmod m,\ c_{n_x(p)}(p)\le t_-\}
 \le G_{m,b}(t_-)+O_m(1/\log x).
\]
By \eqref{eq:ap-small-tail},
\[
 G_{m,b}(t_-)
 =\exp\{-(\log R)^{1/(1-\varepsilon)+o_m(1)}\}=o(1/R),
 \qquad \frac1{\log x}\le\frac1{R^2}=o(1/R).
\]
This proves the required lower rank bound in this range.

For $R\ge\sqrt{\log x}$ choose a fixed $p_0=p_0(\varepsilon)$ such
that $P(p)\ge1-\varepsilon/4$ for $p>p_0$. The finitely many values
$c(p)$ with $p\le p_0$ are positive, and consequently eventually exceed
$t_-$, uniformly over the dimension arrays. For the remaining primes,
$c_{n_x(p)}(p)\ge P(p)\varphi(p-1)/(p-1)$ shows that an entry at most
$t_-$ requires
\[
 V_E(p-1)\ge
 T:=\frac{\log\log R}{(1-\varepsilon/2)a_E}.
\]
Apply \eqref{eq:integer-double-tail-mas12} with $X=x$ and
$\delta=\varepsilon/4$. With
$\beta=(1-\varepsilon/4)/(1-\varepsilon/2)>1$, the number of such
entries is at most
\[
 x\exp\{-c_\varepsilon(\log R)^\beta\}=o(N/R).
\]
Indeed, $N\asymp_m x/\log x$ and
$\log\log x\le2\log R$ here, so the logarithm of the ratio to $N/R$
is at most $3\log R-c_\varepsilon(\log R)^\beta+O_m(1)\to-\infty$.
Thus $d_{k,x}>t_-$ in both ranges.

For the upper bound we use quantitative Linnik in a form that counts
primes. Corollary~18.8 of \cite{IK04} gives an absolute exponent $L_0$
and a lower bound
\[
 \sum_{\substack{n\le t\\n\equiv a\pmod q}}\Lambda(n)
 \gg\frac{t}{\varphi(q)\sqrt q}\qquad(t\ge q^{L_0},\ (a,q)=1)
\]
for sufficiently large $q$. The contribution of proper prime powers
is $O(\sqrt t\log^2t)$. Enlarge the exponent to an absolute
$L>\max(L_0,4)$. Uniformly for $t\ge q^L$, this contribution is
$o(t/q^{3/2})$, hence also
$o(t/(\varphi(q)\sqrt q))$. Subtraction and division by $\log t$ give
\begin{equation}\label{eq:linnik-prime-count-mas12}
 \pi(t;q,a)\gg\frac{t}{\varphi(q)\sqrt q\log t}
 \qquad(t\ge q^L).
\end{equation}
Choose a fixed $0<c<\min(1/(4L),1/6)$ and put
\[
 y=c\log R,\qquad R_y=\prod_{\substack{\ell\le y\\\ell\nmid m}}\ell,
 \qquad q=mR_y.
\]
The conditions $p\equiv b\pmod m$ and $p\equiv1\pmod{R_y}$ determine
one reduced class $a_y\pmod q$. Since
$\log q=y+o_m(y)$, eventually $q\le R^{2c}$ and $q^L\le x$.
Consequently
\[
 \pi(x;q,a_y)\gg\frac{x}{q^{3/2}\log x}
 \ge\frac{xR^{-3c}}{\log x}\gg_m NR^{-3c}>N/R=k
\]
for all sufficiently large $R$. Every such prime has
$\ell\mid p-1$ for each $\ell\le y$ outside $E$: divisors of $m$
are forced by $b$, and all others by $R_y$. Uniformly in the dimension,
\[
 c_{n_x(p)}(p)\le\frac{\varphi(p-1)}{p-1}
 \le\prod_{\substack{\ell\le y\\\ell\notin E}}(1-1/\ell)
 =\frac{a_E+o_m(1)}{\log y}
 =\frac{a_E+o_m(1)}{\log\log R}.
\]
This proves the upper rank bound. Letting $\varepsilon\downarrow0$
proves \eqref{eq:all-lower-ranks-mas12} with the asserted uniformity.

To prove the minimum-scale criterion, observe that if $d_{k,x}\to0$, the same CDF upper bound gives
$k/N\le G_{m,b}(d_{k,x})+O_m(1/\log x)\to0$.
Thus the left side of \eqref{eq:ap-sharp-extrema} forces $R\to\infty$,
and \eqref{eq:all-lower-ranks-mas12} proves both directions of that
criterion.

Finally set $D=\prod_{\ell\in E}\ell$. Since $(D,p-1)=1$,
\[
 \frac{p-1}{\varphi(p-1)}
 =F_E\frac{D(p-1)}{\varphi(D(p-1))}.
\]
The Rosser--Schoenfeld bound~\cite[Theorem~15]{RS62} and $P(p)\to1$
give the upper reciprocal limit $e^\gamma F_E$ for every dimension
assignment. A prime attaining $d_{1,x}$ tends to infinity because
$d_{1,x}\to0$; since $\log\log p\le\log\log x$, the already proved
asymptotic for $d_{1,x}$ supplies the matching lower reciprocal limit.
The remaining assertions follow.
\end{proof}

For example, if $h(x)\to\infty$ and
$\log h(x)=o(\log\log x)$, then every
$1\le k\le x\exp\{-\log x/h(x)\}$ satisfies this criterion uniformly.
Indeed, with $R=N/k$, the bound
$\log R\ge\log x/h-\log\log x-O_m(1)$ and $R\le N$ give
$\log\log R\sim\log\log x$. 

Taking $m=k=1$ and $n_x(p)=n(p)=\infty$ gives
\[
 \min_{p\le x}c(p)\sim\frac{e^{-\gamma}}{\log\log x},\qquad
 \limsup_{p\to\infty}\frac1{c(p)\log\log p}=e^\gamma,
 \qquad \inf_pc(p)=\liminf_{p\to\infty}c(p)=0.
\]

The set $E$ consists precisely of the odd prime divisors $\ell$ of $m$ for
which the exponent $e_\ell(b)$ in \Cref{thm:ap-law}, with $Q=m$, equals zero;
for example, $F_E=2/3$ for $(b,m)=(2,3)$.

The prime-averaging laws also hold uniformly over finite or varying
matrix dimensions. We first isolate the sharp product comparison, then the
rescaled deficit, and finally their distributional and metric uses.
For arrays $\mathbf n_x=(n_x(p))_{p\le x}$ with
$n_x(p)\in\Z_{\ge1}\cup\{\infty\}$, write
\[
 F_{\mathbf n_x,x}(t):=\frac1{\pi(x)}
    \#\{p\le x:c_{n_x(p)}(p)\le t\},\qquad
 X(p):=\frac{\varphi(p-1)}{p-1}.
\]
The entries of a dimension array may depend on both $x$ and $p$.

\begin{proposition}[Sharp finite matrix-product error]
\label{prop:finite-matrix-product-error}
For every prime $p$ and integer $n\ge1$,
\begin{equation}\label{eq:finite-dimension-ratio}
 0<\frac{c_n(p)}{c(p)}-1\le C_\ast p^{-(n+1)},\qquad
 C_\ast:=\frac2{P(2)}-4=2.925493238910\ldots.
\end{equation}
The constant is optimal, with equality at $(p,n)=(2,1)$.
\end{proposition}

\begin{proof}
Put $q=1/p$, $r=n+1$ and $T_r(q)=\prod_{j\ge r}(1-q^j)^{-1}$, so that
$c_n(p)/c(p)=T_r(q)>1$. Expanding the product,
$q^{-r}(T_r(q)-1)$ is the sum of the monomials
$q^{(h-1)r+\sum_{j\ge0}jm_j}$ over nonzero finitely supported
$(m_j)_{j\ge0}\subset\Z_{\ge0}$, where $h=\sum_jm_j\ge1$.
All exponents are nonnegative. The convergent sum therefore increases
with $q\le1/2$ and decreases with $r\ge2$. Its maximum is
$4(T_2(1/2)-1)=2/P(2)-4$, attained at $q=1/2$, $r=2$.
This proves the bound and its optimality.
\end{proof}

\begin{proposition}[Rescaled finite matrix deficit]
\label{prop:finite-matrix-deficit}
For every odd prime $p$ and $n\in\Z_{\ge1}\cup\{\infty\}$,
\[
 0\le p\bigl(X(p)-c_n(p)\bigr)-X(p)
    \le\frac{X(p)}{p-1}\le\frac1{2(p-1)}.
\]
Put $T_x(p):=p(X(p)-c_{n_x(p)}(p))$. For fixed $Q\ge1$, let
\[
 H_{\mathbf n_x,x,Q,b}(t):=\frac1{\pi(x;Q,b)}
   \#\{p\le x:p\equiv b\pmod Q,\ T_x(p)\le t\}.
\]
As $x\to\infty$, uniformly over all arrays and reduced residues $b\pmod Q$,
\begin{equation}\label{eq:finite-dimension-deficit-cdf}
 \frac{\kappa_{\mathrm{end}}D(Q)+o_Q(1)}{\log x}
 \le\|H_{\mathbf n_x,x,Q,b}-G_{Q,b}\|_\infty
       \ll_Q\frac1{\log x}.
\end{equation}
The upper bound and its threshold may be ineffective.
\end{proposition}

\begin{proof}
The product formula \eqref{eq:prime-finite-product} and the complements
of \eqref{eq:matrix-product-bounds}, also after passage to the infinite
product, give
$X(p)\le T_x(p)\le X(p)(1+1/(p-1))$.
For odd $p$, the factor $2\mid p-1$ gives $X(p)\le1/2$.
Fix $Q$ and a reduced residue $b$. Deleting $p\le\sqrt x$ costs
progression proportion $O_Q(x^{-1/2})$. On the remaining primes the
displacement is between zero and $h_x=1/(2(\sqrt x-1))$.
If $F^X_{x,Q,b}$ is the empirical CDF of $X(p)$, bracketing gives
\[
 F^X_{x,Q,b}(t-h_x)-O_Q(x^{-1/2})
 \le H_{\mathbf n_x,x,Q,b}(t)\le F^X_{x,Q,b}(t).
\]
By \Cref{thm:sharp-prime-convergence} and
\eqref{eq:ap-sharp-concentration}, the empirical error and the model
increment at width $h_x$ are both $O_Q(1/\log x)$.
This proves the upper bound.

The lower bound uses an arithmetic support restriction. Put
$A=A_{Q,b}$ and $t_x=A(1-1/x)$. If an odd progression prime $p$ has
a prime factor $q\mid p-1$ with $q\nmid2Q$, then $q\le(p-1)/2$.
All factors forced by the progression already occur in $A$, so
\[
 X(p)\le A\left(1-\frac2{p-1}\right),\qquad
 T_x(p)\le A\left(1-\frac2{p-1}\right)
                 \left(1+\frac1{p-1}\right)\le t_x.
\]
Every remaining $p-1$ is supported on the prime divisors of $2Q$.
There are $O_Q((\log x)^{\omega(2Q)})$ such integers up to $x$,
including the possible prime $p=2$ as an additional exception.
Their progression proportion is $o_Q(1/\log x)$, and hence
$H_{\mathbf n_x,x,Q,b}(t_x)=1-o_Q(1/\log x)$.
The endpoint formula \eqref{eq:ap-sharp-concentration}, with $h=A/x$,
gives
$1-G_{Q,b}(t_x)=(\kappa_{\mathrm{end}}D(Q)+o_Q(1))/\log x$.
Subtracting proves the lower bound. Every estimate is uniform in
the arrays and in the finitely many reduced residues modulo $Q$.
\end{proof}

\begin{corollary}[Finite and varying matrix dimension]
\label{cor:finite-dimension}
All conclusions hold uniformly over dimension arrays.
\begin{enumerate}[(i)]
\item As $x\to\infty$,
\begin{equation}\label{eq:finite-dimension-cdf}
 \frac{\kappa_{\mathrm{end}}+o(1)}{\log x}
 \le\|F_{\mathbf n_x,x}-G\|_\infty\ll\frac1{\log x}.
\end{equation}
The fixed-progression laws of \Cref{thm:ap-law} also hold.
\item The Mellin moment and derivative estimate
\eqref{eq:prime-mellin-ap} holds with $U(p)=c_{n_x(p)}(p)$.
\item For every fixed $B>0$ and $1\le r<\infty$, the empirical laws
of $c_{n_x(p)}(p)$ and $-\log c_{n_x(p)}(p)$ over primes at most $x$
in a reduced class $b\pmod Q$ have $r$-Wasserstein distances
$O((\log x)^{-B})$ from the laws of $X_{Q,b}$ and
$-\log X_{Q,b}$, respectively.
\end{enumerate}
Parts \textup{(ii)} and \textup{(iii)} inherit, respectively, the
modulus ranges, fixed-parameter dependences and exceptional sets of
\Cref{thm:prime-mellin} and \Cref{prop:prime-wasserstein}.
The upper bounds may be ineffective. In the almost-all modulus
range the exceptional sets may depend on the requested accuracy,
compact set and derivative or transport order, but not on the
reduced residue or the dimension array.
\end{corollary}

\begin{proof}
Part (i) follows from \Cref{cor:ap-sharp-convergence}, with $Q=1$ for
\eqref{eq:finite-dimension-cdf}.
For (ii), apply \Cref{lem:mellin-perturbation} with
$r_x(p)=\prod_{k\le n_x(p)}(1-p^{-k})$; the lemma includes
$n_x(p)=\infty$ and the uniform constant $C_0=4$.
The population bound \eqref{eq:ap-polylog-population}, in precisely
the admissible ranges and outside the exceptional sets used in
\Cref{thm:prime-mellin}, gives perturbation error
$O_{K,j}(\varphi(Q)\log x/\sqrt x)$.
For polylogarithmic $Q$, fixed-base powers at most
$x^{1/3}\exp\{-(\log\log x)^3\}$, and the permitted almost-all
$Q\le x^{1/2-\delta}$, this is smaller than every fixed inverse
power of $\log x$. Combining with the totient-ratio estimate in
\Cref{thm:prime-mellin} proves (ii), with unchanged exceptional sets.
Part (iii) is the dimension-array assertion of
\Cref{prop:prime-wasserstein}; its proof controls both raw and
negative-logarithmic costs uniformly over these same products.
\end{proof}

\section{Effective additive extremal bounds}
\label{sec:explicit-linnik}

For real $y\ge2$, put
\[
 \mathcal M(y):=\prod_{\ell\le y}(1-1/\ell)^{-1},
\]
so $A_k=\mathcal M(p_k)^{-1}$.

\begin{theorem}[Effective additive bounds for $1/c(p)$]
\label{thm:effective_sharp}
\index{Extremal orders!effective bounds}
Put $T(p)=\log\log p$. There are effective absolute constants $a,K_+>0$
and an effective threshold $p_0$ such that every prime $p\ge p_0$ satisfies
\begin{equation}\label{eq:eff-universal-exp}
 \frac1{c(p)}\le e^\gamma T(p)+K_+ e^{-a\sqrt{T(p)}}.
\end{equation}
If effective constants $C_0\ge1$ and $L>0$ satisfy
$p^*(k)\le C_0N_k^L$ for every $k\ge1$, then there are
$K_->0$ and $k_0$, effective in terms of $C_0,L$, such that
\begin{equation}\label{eq:eff-sharp}
 \frac1{c(p^*(k))}
 \ge e^\gamma T(p^*(k))-e^\gamma\log L
       -K_- e^{-a\sqrt{T(p^*(k))}}
 \qquad(k\ge k_0).
\end{equation}
\end{theorem}

\begin{proof}
Throughout the proof, $b>0$ is an effective absolute constant that may
decrease to absorb polynomial factors in logarithms; its final value is $a$.
The prime number theorem with the de la Vall\'ee Poussin error term
\cite[Theorem~6.9]{MV06} gives
\[
 \vartheta(y)=y\bigl(1+O(e^{-b\sqrt{\log y}})\bigr).
\]
Inverting \eqref{eq:mertens-product-dvp} gives
\[
 \mathcal M(y)=e^\gamma\log y
       \bigl(1+O(e^{-b\sqrt{\log y}})\bigr).
\]

For $N=p-1$, let $N_j$ be the largest primorial not exceeding $N$ and
put $y=p_j$, $t=\log y$, $U=\log\log N$.
By the primorial extremal principle~\cite[(1.2)]{Axl24},
\[
 \frac N{\varphi(N)}\le\frac{N_j}{\varphi(N_j)}=\mathcal M(y),
 \qquad \vartheta(y)=\log N_j\le\log N.
\]
Taking logarithms in $\vartheta(y)\le\log N$ and using the estimate for
$\vartheta$ give $t+\log(1-Ce^{-b\sqrt t})\le U$, hence
$t\le U+O(e^{-b\sqrt U})$. The Mertens estimate gives an effective $C_1$ with
$\mathcal M(e^t)\le e^\gamma t+C_1e^{-b\sqrt t}$ for large $t$; this
majorant is eventually increasing in $t$, so
\[
 \frac N{\varphi(N)}\le e^\gamma U+O(e^{-b\sqrt U}).
\]
Since $P(p)^{-1}=1+O(1/p)$ and
$U=T(p)+O(1/(p\log p))$, this proves \eqref{eq:eff-universal-exp}.

For $y=p_k$, the bounds
$\vartheta(y)<\log p^*(k)\le\log C_0+L\vartheta(y)$ give
\begin{equation}\label{eq:eff-loglog}
 \log y+O(e^{-b\sqrt{\log y}})
 \le T(p^*(k))
 \le\log y+\log L+O_{C_0,L}(e^{-b\sqrt{\log y}}).
\end{equation}
In particular, $\log y=T(p^*(k))+O_{C_0,L}(1)$.
All primes through $y$ divide $p^*(k)-1$, so
$1/c(p^*(k))\ge\mathcal M(y)$.
The upper side of \eqref{eq:eff-loglog} and the lower Mertens estimate
prove \eqref{eq:eff-sharp}.
\end{proof}

The constants $a,K_+,p_0$ are effective through \cite[Theorem~6.9]{MV06} and
\Cref{lem:prime-log-stieltjes}.
Xylouris~\cite[Theorem~1]{Xyl18} (see also~\cite[Theorem~2.1]{Xyl11d}) supplies
the hypothesis with $L=L_X=5$ and an effective $C_0$; this gives the offset
$e^\gamma\log5\approx2.87$ in \eqref{eq:eff-sharp}.
\Cref{cor:rosser-schoenfeld,prop:factor-count-bound,thm:factoring-free-cert}
give fully explicit bounds for individual moduli.

\begin{lemma}[Explicit Mertens product upper bound]
\label{lem:explicit-mertens-product}
For every real $y\ge2\,278\,382$,
\begin{equation}\label{eq:eff-mertens-up}
 \mathcal M(y)\le e^\gamma\log y\left(1+\frac{0.2}{\log^3y}\right).
\end{equation}
\end{lemma}

We derive the bound stated in~\cite[Theorem~5.9]{Dus18}
from the corrected estimates of~\cite{BFKNW26}; the published
Dusart proof is invalid (see their introductory remark~3).

\begin{proof}
Put $y_0=2\,278\,382$, $x_0=y_0+1$, and $A=0.19991658$.
The special $\mathcal A_3(x_0)$ entry in \cite[\S~1.3]{BFKNW26},
with \cite[Lemma~30 and Theorem~32(i)]{BFKNW26}, gives
\[
 \frac{\mathcal M(x)}{e^\gamma\log x}
 \le\exp\left(\frac A{\log^3x}\right)\qquad(x\ge x_0).
\]
Since $e^u\le(1-u)^{-1}$ for $0\le u<1$ and
\[
 \frac A{1-A/(\log x_0)^3}<0.199929321<0.2,
\]
the required bound holds for $x\ge x_0$.
On $[y_0,y_0+1]$, use the rational certificate
\[
 \mathcal M(y_0)<26.074738559728<26.074738785290
 <e^\gamma\left(\log y_0+\frac{0.2}{\log^2y_0}\right).
\]
Put $D=10^{50}$; starting with $A_D=D$, replace $A_D$ by
$\lceil pA_D/(p-1)\rceil$ at each of the $168\,064$ primes $p\le y_0$.
Then $\mathcal M(y_0)\le A_D/D$.
For rational $1\le r\le2$, put $z=(r-1)/(r+1)$ and use
\[
 L_-(r):=2\sum_{j=0}^{99}\frac{z^{2j+1}}{2j+1},\qquad
 L_+(r):=L_-(r)+\frac{2z^{201}}{201(1-z^2)}.
\]
These enclose $\log r$; extend them to $r\ge1$ by
$L_\pm(2^kr)=kL_\pm(2)+L_\pm(r)$, $1\le r<2$.
With $n=10^5$, the harmonic-sum bound for $\gamma$ gives
\[
 0<g:=D^{-1}\sum_{j=1}^n\lfloor D/j\rfloor
       -L_+(n)-\frac1{2n}<\gamma.
\]
Put $g_D=\lfloor Dg\rfloor$, $t_0=D$, and
$t_j=\lfloor t_{j-1}g_D/(Dj)\rfloor$ for $1\le j\le84$.
Then $E:=D^{-1}\sum_{j=0}^{84}t_j<e^\gamma$, and exact rational
arithmetic gives
\[
 \frac{A_D}{D}<26.074738559728,\qquad
 E\left(L_-(y_0)+\frac1{5L_+(y_0)^2}\right)>26.074738785290.
\]
The same enclosures give $\log x_0>14.6389$, certifying the
coefficient comparison. Neither $y_0$ nor $y_0+1$ is prime, so
$\mathcal M(x)=\mathcal M(y_0)$ on $[y_0,y_0+1]$.
Finally,
\[
 \frac d{dx}\left(\log x+\frac{0.2}{\log^2x}\right)
 =\frac1x\left(1-\frac{0.4}{\log^3x}\right)>0\qquad(x\ge y_0),
\]
which extends the endpoint bound across this interval.
\end{proof}

\begin{corollary}[A lower bound along $p^*(k)$ via Chang's theorem]\label{cor:chang-offset}
As $k\to\infty$,
\begin{equation}\label{eq:chang-offset}
 \frac1{c(p^*(k))}
 \ge e^\gamma\log\log p^*(k)-e^\gamma\log(12/5)-o(1).
\end{equation}
\end{corollary}

\begin{proof}
Since $\log P^+(N_k)/\log N_k=\log p_k/\vartheta(p_k)\to0$,
Chang~\cite[arXiv version, Corollary~11]{Cha14} gives
$p^*(k)<N_k^{12/5+o(1)}$. Hence
$\log\log p^*(k)\le\log p_k+\log(12/5)+o(1)$.
Every prime $\ell\le p_k$ divides $p^*(k)-1$, so the Mertens estimate in
the proof of \Cref{thm:effective_sharp} gives
$1/c(p^*(k))\ge\mathcal M(p_k)=e^\gamma\log p_k+o(1)$.
\end{proof}

\section{Explicit certificates without complete factorization}
\label{sec:explicit-certificates}

Complete factorization of $p-1$ is unnecessary for a certified interval
containing $1/c(p)$. There are two sources of uncertainty. Truncating
the determinant product leaves a geometric tail; removing the prime
powers up to a cutoff leaves a cofactor whose prime divisors are unknown.
The first error is bounded directly. For the second, maximize the
reciprocal totient ratio over all integers no larger than the residual
cofactor whose prime factors exceed the cutoff. The product of the
smallest such primes that fits within this size attains that maximum,
which bounds the ratio for the actual cofactor. Combining these bounds
gives rational endpoints and makes the effect of the cutoff explicit.

\begin{lemma}[Finite-product enclosure for $P(p)$]
\label{lem:euler-product-enclosure}
For every prime $p\ge3$ and integer $J\ge1$, put
\[
 P_J^+(p):=\prod_{j=1}^J(1-p^{-j}),\qquad
 P_J^-(p):=P_J^+(p)-\frac{p^{-J}}{p-1}.
\]
Then
\[
 \frac12\le1-\frac1{p-1}\le P_J^-(p)\le P(p)\le P_J^+(p),\qquad \frac{P_J^+(p)}{P_J^-(p)}\le1+p^{-J}.
\]
\end{lemma}

\begin{proof}
Since $0<P_J^+(p)\le1$ and $\prod_{j>J}(1-p^{-j})\ge1-\sum_{j>J}p^{-j}$
(as in the proof of \Cref{lem:matrix-product-bounds}),
\[
 0\le P_J^+(p)-P(p)=P_J^+(p)\Bigl(1-\prod_{j>J}(1-p^{-j})\Bigr)
 \le\sum_{j>J}p^{-j}=\frac{p^{-J}}{p-1}.
\]
The first bound in \eqref{eq:matrix-product-bounds} with $t=p$, $n=J$
gives $P_J^-(p)\ge1-\frac{1-p^{-J}}{p-1}-\frac{p^{-J}}{p-1}=1-\frac1{p-1}\ge\frac12$.
Finally, $P_J^+/P_J^-=1+p^{-J}/((p-1)P_J^-)\le1+2p^{-J}/(p-1)\le1+p^{-J}$.
\end{proof}

\paragraph*{Evaluation procedure}\phantomsection\label{par:evaluation-procedure}
For a prime $p\ge3$ and an integer $D\ge1$, given the factorization of
$p-1$, take $J=\lceil D\log 10/\!\log p\rceil$ and multiply the interval of
\Cref{lem:euler-product-enclosure} by $\varphi(p-1)/(p-1)$. This uses
$O(\omega(p-1)+J)$ rational-arithmetic operations and yields an interval of
width at most $\tfrac14\,10^{-D}$ containing $c(p)$; correct rounding to $D$
decimal places is certified when its two endpoints round identically.

The primorial extremal principle for $n/\varphi(n)$
(see Axler~\cite[equation~(1.2)]{Axl24}) persists when all prime
factors exceed a prescribed cutoff.

\begin{lemma}[Maximum of $n/\varphi(n)$ for $n\le R$ with prime factors above $B$]\leavevmode\phantomsection\label{lem:rough-cofactor-envelope}\ignorespaces
Let $B\ge2$ and $R\ge1$ be real, list the primes greater than $B$ as
$r_1<r_2<\cdots$, and define
\[
 k(B,R):=\max\{k\ge0:r_1\cdots r_k\le R\},\qquad
 V(B,R):=\prod_{j=1}^{k(B,R)}\frac{r_j}{r_j-1}.
\]
Empty products are one.
\begin{enumerate}[\upshape(a)]
\item The value $V(B,R)$ is the maximum of $n/\varphi(n)$ over
integers $1\le n\le R$ all of whose prime factors exceed $B$.
\item There is an effective absolute constant $C_1\ge1$ such that,
with $k=k(B,R)$,
\begin{equation}\label{eq:rough-cofactor-size-envelope}
 r_{k+1}\le C_1(B+\log R),\qquad
 V(B,R)\le\frac{\mathcal M(C_1(B+\log R))}{\mathcal M(B)}.
\end{equation}
In particular, for $R\ge3$,
\begin{equation}\label{eq:tail-envelope}
 V(B,R)\le\frac{\max\{\log\log R,\log B\}+O(1)}{\log B}
 \Bigl(1+O\Bigl(\frac1{\log B}\Bigr)\Bigr),
\end{equation}
with absolute implied constants.
\item $\displaystyle\log V(B,R)\le\frac{k(B,R)}{r_1-1}\le\frac{\log R}{(r_1-1)\log B}$.
\end{enumerate}
\end{lemma}

\begin{proof}
Let $1\le n\le R$ have distinct prime factors $q_1<\cdots<q_j$, all
exceeding $B$. Then $q_i\ge r_i$, so $r_1\cdots r_j\le n\le R$ and
$j\le k(B,R)$. Since $t/(t-1)$ decreases for $t>1$,
$n/\varphi(n)\le\prod_{i\le j}r_i/(r_i-1)\le V(B,R)$, with equality
for $n=r_1\cdots r_{k(B,R)}$. This proves \textup{(a)}.

For \textup{(b)}, let $k=k(B,R)$ and $r_0:=B$. Then
$\vartheta(r_k)-\vartheta(B)=\log(r_1\cdots r_k)\le\log R$, so effective
Chebyshev bounds give $r_k\ll B+\log R$, and Bertrand's postulate gives
$r_{k+1}\le2r_k$; this is the first bound, with an effective absolute
$C_1\ge1$. Since $V(B,R)=\mathcal M(r_k)/\mathcal M(B)$ and
$r_k<r_{k+1}\le C_1(B+\log R)$, monotonicity of $\mathcal M$ gives the second.
For $R\ge3$, $\log(C_1(B+\log R))=\max\{\log\log R,\log B\}+O(1)$, while
$\mathcal M(t)=e^\gamma(\log t+O(1))$ for $t\ge2$~\cite[Theorems~7--8]{RS62}
and $\mathcal M(B)\ge2$; hence the second bound gives \eqref{eq:tail-envelope}.
Part \textup{(c)} follows from $-\log(1-1/r)\le1/(r-1)\le1/(r_1-1)$ for $r\ge r_1$ and $B^{k(B,R)}\le r_1\cdots r_{k(B,R)}\le R$.
\end{proof}

For a prime $p\ge3$ and an integer $B\ge3$, put
\[
 \begin{aligned}
 \mathsf{E}_B(p)&:=\prod_{\substack{\ell\le B\\\ell\mid p-1}}
             \left(1-\frac1\ell\right)^{-1},&
 R_B(p)&:=\frac{p-1}{\prod_{\ell\le B}\ell^{v_\ell(p-1)}},\\
 T_B(p)&:=\prod_{\substack{\ell>B\\\ell\mid p-1}}
             \left(1-\frac1\ell\right)^{-1},&
 U_B(p)&:=\frac{\mathsf{E}_B(p)V(B,R_B(p))}{P(p)}.
 \end{aligned}
\]

\begin{theorem}[Two-sided bound from partial factorization]
\label{thm:factoring-free-cert}
\index{Prime certificates!partial factorization}
Let $p\ge3$ be prime and $B\ge3$ be an integer.
\begin{enumerate}[\upshape(a)]
\item With the infinite product $P(p)$,
\begin{equation}\label{eq:ff-cert}
 \begin{aligned}
 \frac{\mathsf{E}_B(p)}{P(p)}
 &\le\frac1{c(p)}\le U_B(p)\\
 &\le\frac{\mathsf{E}_B(p)}{P(p)}
       \exp\!\left(\frac{\log R_B(p)}{B\log B}\right)
 \le\frac{\mathsf{E}_B(p)}{P(p)}
       \exp\!\left(\frac{\log p}{B\log B}\right).
 \end{aligned}
\end{equation}
\item For every integer $J\ge1$,
\begin{equation}\label{eq:rough-cofactor-certificate}
 \frac{\mathsf{E}_B(p)}{P_J^+(p)}
 \le\frac1{c(p)}
 \le\frac{\mathsf{E}_B(p)V(B,R_B(p))}{P_J^-(p)}.
\end{equation}
\end{enumerate}
\end{theorem}

\begin{proof}
By \eqref{eq:cp-nt}, $1/c(p)=\mathsf{E}_B(p)T_B(p)/P(p)$ and
$T_B(p)=R_B(p)/\varphi(R_B(p))$, so
\Cref{lem:rough-cofactor-envelope}\textup{(a)} with $R=R_B(p)$ gives
$1\le T_B(p)\le V(B,R_B(p))$. With \Cref{lem:euler-product-enclosure},
this gives \eqref{eq:rough-cofactor-certificate} and the first line of
\eqref{eq:ff-cert}. Since $B$ is an integer, $r_1-1\ge B$; as $R_B(p)<p$,
\Cref{lem:rough-cofactor-envelope}\textup{(c)} gives the second line.
\end{proof}

\begin{corollary}[Verification without complete factorization]
\label{cor:factoring-free-verification}
Fix $C>0$. For a prime $p\ge3$, an integer $3\le B\le(\log p)^C$ and a
rational $0<\varepsilon<1$, let $J$ be the least power of two with
$p^J\ge1/\varepsilon$, and put
\[
 U_{B,\varepsilon}^{-}(p)
 :=\frac{\mathsf{E}_B(p)V(B,R_B(p))}{P_J^+(p)},\qquad
 U_{B,\varepsilon}^{+}(p)
 :=\frac{\mathsf{E}_B(p)V(B,R_B(p))}{P_J^-(p)}.
\]
Then
\[
 \frac1{c(p)}\le U_B(p)\le U_{B,\varepsilon}^{+}(p)
 \le(1+\varepsilon)U_{B,\varepsilon}^{-}(p)\le(1+\varepsilon)U_B(p),
\]
and a deterministic algorithm computes $\mathsf{E}_B(p)$, $V(B,R_B(p))$ and
$U_{B,\varepsilon}^{\pm}(p)$ from $(p,B,\varepsilon)$, without completely
factoring $p-1$, in bit cost polynomial in $\log p$ and the binary
encoding length of $\varepsilon$, of degree depending only on $C$.
\end{corollary}

\begin{proof}
Sieve the primes up to $B$ and remove their full powers from $p-1$; this
gives $R_B(p)$ and $\mathsf{E}_B(p)$ exactly. Multiply consecutive primes
$r_1,r_2,\ldots>B$ until the product first exceeds $R_B(p)$; the
preceding product gives $k=k(B,R_B(p))$ and $V(B,R_B(p))$. By
\Cref{lem:rough-cofactor-envelope}\textup{(b)} with $R=R_B(p)\le p$,
$r_{k+1}\ll B+\log p$, so a sieve with a doubling endpoint stops at
$O(B+\log p)$; these prime products and the numerators and denominators
of $\mathsf{E}_B(p)$ and $V(B,R_B(p))$ have $O(\log p)$ bits.
By \Cref{lem:euler-product-enclosure},
$U_{B,\varepsilon}^-(p)\le U_B(p)\le U_{B,\varepsilon}^+(p)$ and
\[
 \frac{U_{B,\varepsilon}^+(p)}{U_{B,\varepsilon}^-(p)}
 =\frac{P_J^+(p)}{P_J^-(p)}\le1+p^{-J}\le1+\varepsilon,
\]
while $1/c(p)\le U_B(p)$ by \eqref{eq:ff-cert}. If $\varepsilon$ has
encoding length $L$, then $\varepsilon\ge2^{-L}$ and $J=O(1+L/\log p)$,
so $P_J^\pm(p)$ have rational representations with $O(J^2\log p)$ bits,
and exact rational arithmetic gives the claimed bit bound.
\end{proof}

Let $K$ consist of verified prime divisors of $p-1$, including every
such divisor at most $B$. Strip their full powers and put
$R_K=(p-1)/\prod_{\ell\in K}\ell^{v_\ell(p-1)}$ and
$\mathsf E_K=\prod_{\ell\in K}\ell/(\ell-1)$. Define $V_K$ by the
same consecutive-prime maximization with product at most $R_K$, using
primes greater than $B$ outside $K$. The proof of \Cref{lem:rough-cofactor-envelope}\textup{(a)}
then gives
\[
 \frac{\mathsf E_K}{P_J^+(p)}\le\frac1{c(p)}
 \le\frac{\mathsf E_KV_K}{P_J^-(p)};
\]
adding fully stripped primes narrows this enclosure, since every new
residual multiplied by the removed prime power was formerly admissible.

The endpoints $U_{B,\varepsilon}^{\pm}(p)$ enclose the conservative
bound $U_B(p)$. Increasing $J$ improves the approximation to $P(p)$
but leaves the factor $V(B,R_B(p))/T_B(p)$ between $U_B(p)$ and
$1/c(p)$ unchanged. A narrow enclosure of $U_B(p)$ therefore need not
give a narrow enclosure of $1/c(p)$, whose lower endpoint in
\eqref{eq:rough-cofactor-certificate} is $\mathsf{E}_B(p)/P_J^+(p)$.
In \hyperref[par:worked-search-certificate]{the worked search certificate},
additional factor information verifies that a prime meets a threshold
which the prescribed certificate fails to certify.

Pollack and Thompson~\cite[arXiv version, p.~10, before Lemma~3.1]{PT14} bound
$\prod_{\ell\mid m,\,\ell>\log N}(1-1/\ell)^{-1}$ by $1+O(1/\log\log N)$ for positive integers
$m\le3N$; with $m=p-1$ and $N=p$ this gives $T_{B(p)}(p)=1+O(1/\log\log p)$ at
$B(p)=\lfloor\log p\rfloor$, and \Cref{thm:critical-tail-cutoff}\textup{(a)}
recovers this order. The next two theorems treat general cutoffs $B(p)$.

\begin{theorem}[Normalized extremal sizes from partial factorization]
\label{thm:sharp-tail-cutoff}
\index{Prime certificates!tail cutoff}
Let $B(p)\ge3$ be integers with $B(p)\to\infty$, and put
\[
 A(p):=\max\left\{1,\frac{\log\log p}{\log B(p)}\right\}.
\]
Then
\begin{equation}\label{eq:normalized-tail-cutoff}
 \begin{aligned}
 \limsup_{p\to\infty}\frac{T_{B(p)}(p)}{A(p)}
 &=\limsup_{p\to\infty}
   \frac{V(B(p),R_{B(p)}(p))}{A(p)}\\
 &=\limsup_{p\to\infty}\frac{U_{B(p)}(p)c(p)}{A(p)}=1.
 \end{aligned}
\end{equation}
In particular, if $\log B(p)/\log\log p\to a\in(0,\infty)$,
then $T_{B(p)}(p)$, $V(B(p),R_{B(p)}(p))$ and $U_{B(p)}(p)c(p)$
have limsup $\max\{1,a^{-1}\}$. If $A(p)\to1$ (for instance, if $a\ge1$),
all three quantities tend to one.
All assertions hold with $p$ restricted to any fixed reduced residue class.
\end{theorem}

\begin{proof}
Put $u=\log\log p$ and $v=\log B$. As in the proof of
\Cref{thm:factoring-free-cert}, and since $V(B,R)$ is nondecreasing in $R$
and $R_B(p)\le p$,
\[
 1\le T_B(p)\le V(B,R_B(p))\le V(B,p).
\]
By \eqref{eq:cp-nt} and the definition of $U_B$,
\[
 1\le U_B(p)c(p)=\frac{V(B,R_B(p))}{T_B(p)}\le V(B,p).
\]
Since $v\to\infty$, \eqref{eq:tail-envelope} with $R=p$ gives
$V(B,p)\le(1+o(1))A(p)$, proving the upper limits.

Fix a reduced progression $b\pmod m$, with $m=1$ permitted, put
$M_y=\prod_{\ell\le y,\,\ell\nmid m}\ell$, and let $p_y$ be the least prime
with $p_y\equiv1\pmod{M_y}$ and $p_y\equiv b\pmod m$. Linnik's
theorem~\cite{Lin44}, with a fixed absolute exponent $L>0$, and the prime number theorem give
$M_y<p_y\ll_m M_y^L$ and $u_y:=\log\log p_y=\log y+O_m(1)$.
Put $v_y=\log B(p_y)$; eventually $B(p_y)$ exceeds every prime divisor
of $m$. If $B(p_y)\le y$, every prime in $(B(p_y),y]$ then divides
$p_y-1$, so~\cite[Theorems~7--8]{RS62} give
\begin{equation}\label{eq:tail-lower}
 T_{B(p_y)}(p_y)\ge\frac{\mathcal M(y)}{\mathcal M(B(p_y))}
 =\frac{u_y+O_m(1)}{v_y+O(1)}.
\end{equation}
If moreover $u_y\ge v_y$, then $A(p_y)=u_y/v_y$, and
$T_{B(p_y)}(p_y)/A(p_y)\ge1-o(1)$ since $u_y,v_y\to\infty$.
If $u_y<v_y$ or $B(p_y)>y$, then $A(p_y)\le1+O_m(1/\log y)$, and
$T_B\ge1$ suffices. Together with $V\ge T$ and the upper bound,
this proves the asserted limsups for $T_B$ and $V$.

For the actual overestimate, choose $h_x\to\infty$ with
$\log h_x=o(\log\log x)$, and put $L=\log x$,
$\varepsilon_x=e^{-L/h_x}$ and $y=L/(2h_x)$.
Apply \eqref{eq:ap-endpoint-transfer} at $x$ and $2x$ with the same
$\varepsilon_x$. It gives $\gg_m xh_x/L^2$ primes $p\in(x,2x]$,
$p\equiv b\pmod m$, with
$A_{m,b}(1-\varepsilon_x)<X(p)\le A_{m,b}$.
Discard the $O_m(L^{\omega(2m)})$ predecessors supported on $2m$.
For each remaining prime write $p-1=d_0r$, with $d_0$ supported on
$2m$, $(r,2m)=1$ and $r>1$. Then
$X(p)/A_{m,b}=\prod_{\ell\mid r}(1-1/\ell)>1-\varepsilon_x$,
so every prime divisor of $r$ exceeds $\varepsilon_x^{-1}$.
Eventually $B(p)>P^+(2m)$. If $B=B(p)\le y$, then $R_B(p)=r$ and
$\prod_{\ell\le y}\ell<\varepsilon_x^{-1}<r$ by the prime number theorem.
Hence
\[
 \begin{aligned}
 T_B(p)&=\frac{A_{m,b}}{X(p)}=1+O(\varepsilon_x),\\
 U_B(p)c(p)&\ge(1-O(\varepsilon_x))\frac{\mathcal M(y)}{\mathcal M(B)}.
 \end{aligned}
\]
Since $\log y=u-\log h_x+O(1)$ uniformly on $(x,2x]$, Mertens'
estimate gives $U_B(p)c(p)/A(p)\ge1-o(1)$ when $B\le y$.
If $B>y$, then $A(p)=1+o(1)$ and $U_B(p)c(p)\ge1$ suffices.
This proves the remaining limsup. The specializations follow from
$A(p)\to\max\{1,a^{-1}\}$ and the preceding bounds.
\end{proof}

\begin{theorem}[Relative error bounds from partial factorization]
\label{thm:critical-tail-cutoff}
Put $u=\log\log p$, $v=\log B$ and $d=u-v$.
\begin{enumerate}[\upshape(a)]
\item Uniformly for sufficiently large primes $p$ and integers $B\ge3$
with $0\le d\le u/2$,
\begin{equation}\label{eq:critical-envelope-upper}
 0\le T_B(p)-1\le V(B,R_B(p))-1
 \le\frac du+O\!\left(\frac1u+\frac{d^2}{u^2}\right).
\end{equation}
\item Let $B=B(p)\ge3$ be integer-valued with $v\to\infty$ and
$d\to\infty$. For each fixed integer $J\ge1$, let $Z(p)$ denote
$T_{B(p)}(p)$, $V(B(p),R_{B(p)}(p))$, $U_{B(p)}(p)c(p)$, or the endpoint ratio
$V(B(p),R_{B(p)}(p))P_J^+(p)/P_J^-(p)$ of \eqref{eq:rough-cofactor-certificate}.
Then
\begin{equation}\label{eq:critical-tail-general}
 \limsup_{p\to\infty}\frac vd\bigl(Z(p)-1\bigr)=1.
\end{equation}
If $d=o(u)$, $v/d$ may be
replaced by $u/d$.
\item For fixed $\kappa>0$ and
$B(p)=\lfloor\log p/(\log\log p)^\kappa\rfloor$, each of the
four quantities in \textup{(b)} satisfies
\[
 \limsup_{p\to\infty}
 \frac{\log\log p}{\log\log\log p}\bigl(Z(p)-1\bigr)=\kappa.
\]
\end{enumerate}
Assertions \textup{(b)} and \textup{(c)} hold with $p$ restricted to
any fixed reduced residue class.
\end{theorem}

\begin{proof}
As in the proof of \Cref{thm:sharp-tail-cutoff},
$1\le T_B(p)\le V(B,R_B(p))\le V(B,p)$, and we apply
\eqref{eq:tail-envelope} with $R=p$; there $\max\{u,v\}=u$ when $d\ge0$.

\textup{(a)} Since $v=u-d\ge u/2$,
\[
 \log V(B,p)
 \le\log\frac{u+O(1)}{u-d}+O\Bigl(\frac1u\Bigr)
 =\frac du+O\!\left(\frac1u+\frac{d^2}{u^2}\right).
\]
Exponentiation, using $d/u\le1/2$, proves \eqref{eq:critical-envelope-upper}.

\textup{(b)} Eventually $d>0$, and \eqref{eq:tail-envelope} gives
\[
 V(B,p)\le\frac uv+O\!\left(\frac1v+\frac u{v^2}\right).
\]
Since $u=v+d$, multiplication of the excess above one by $v/d$
gives upper bound $1+O(1/d+1/v)$.
For the primes $p_y$ of the proof of \Cref{thm:sharp-tail-cutoff},
$\log y=u_y+O_m(1)$, so $v_y\to\infty$ and $d_y\to\infty$ force
eventually $P^+(m)<B(p_y)<y$. Then \eqref{eq:tail-lower} gives
$T_{B(p_y)}(p_y)-1\ge(d_y+O_m(1))/(v_y+O(1))$, and multiplication by
$v_y/d_y$ gives $1-o(1)$. Together with $T\le V$ and the upper bound,
this proves the assertion for $T_B$ and $V$.
For $U_Bc$, choose $h_x$ in the endpoint construction of the preceding
proof so slowly that, in the fixed progression $b\pmod m$,
\[
 \log h_x=o\!\left(\inf_{\substack{p\ge x\\p\equiv b\pmod m}}d(p)\right).
\]
The resulting primes have $B<y$, and the lower bound there gives
\[
 U_B(p)c(p)-1\ge
 \frac{d-\log h_x+O(1)}{v+O(1)}-O(\varepsilon_x u/v).
\]
Multiplication by $v/d$ gives $1-o(1)$, since $v,d\to\infty$ and
$\varepsilon_x u/d\to0$. The upper bound follows from $U_Bc\le V$.
For the endpoint ratio $V(B,R_B(p))P_J^+(p)/P_J^-(p)$,
\Cref{lem:euler-product-enclosure} gives
$1\le P_J^+(p)/P_J^-(p)=1+O(p^{-J})$; the additional normalized error
is $O((1+v/d)p^{-J})=o(1)$, since eventually $v<u=\log\log p$.
If $d=o(u)$, then $v/u\to1$.

\textup{(c)} Here $d=\kappa\log u+O(u^\kappa/\log p)$, so $d\to\infty$,
$d=o(u)$ and $v\to\infty$; apply \textup{(b)} with $u/d$.
\end{proof}

For $\kappa=1$ in \Cref{thm:critical-tail-cutoff}\textup{(c)}, the last exponential multiplier in
\eqref{eq:ff-cert} tends to $e$, whereas $V(B(p),R_{B(p)}(p))\to1$.

\chapter{Prime selection and arithmetic sampling}\label{ssec:certgen}

Prime selection with controlled reciprocal determinant density $1/c(q)$
requires candidate verification and acceptance estimates.
\Cref{thm:factoring-free-cert,cor:factoring-free-verification} give
conservative certificates without completely factoring $q-1$;
arithmetic distribution estimates give acceptance probabilities.
\Cref{thm:typical-certificate-general-cutoff} describes the typical excess
of both certificate conventions for fixed parameters.
\Cref{thm:finite-population-clt,cor:finite-population-centered,prop:finite-population-moments}
give normal approximation and moment estimates for sampling without replacement,
which are then specialized to prime populations.

\section{Certified prime search}\label{sec:certified-prime-search}
The uniform statements below concern powers of two $N$; the
\hyperref[par:gencert-fixed-length]{fixed-length extension} at the end
of the section permits arbitrary $N\ge1$. For $\beta>0$,
continuity of $G$ gives $\mathbb P(1/X\le\beta)=1-G(1/\beta)$.

\paragraph*{Search guarantees}\phantomsection\label{par:search-guarantees}
\begin{enumerate}[label=\textup{(\roman*)},leftmargin=*]
\item \emph{Correctness.} Every accepted candidate satisfies the exact
 density bound in \Cref{thm:gencert}\textup{(i)}.
\item \emph{Cost.} Verification and the conditional expected cost of
 one iteration have effective polynomial bit bounds. The iteration
 bound holds for every admissible $x\ge64$
 (\Cref{thm:gencert}\textup{(i)--(ii)}).
\item \emph{Numerical certification.}
 \Cref{prop:gencert-numerics,prop:cdf-quantile-certification} certify
 probabilities and quantiles of the limiting model. These procedures
 terminate.
\item \emph{Search success and halting.} For fixed $N$ and rational
 $\beta>2$, \Cref{thm:gencert-effective-mas23} supplies a computable
 halting threshold and an effective acceptance bound.
 \Cref{thm:gencert}\textup{(iii)} gives a sharper asymptotic for the
 expected number of sampled candidates, with a possibly ineffective
 error term.
\end{enumerate}

For varying powers of two $N$, \Cref{cor:gencert-polylog} gives
acceptance estimates for $2N\le x^{1/3}\exp\{-(\log\log x)^3\}$
unconditionally, and for $2N\le x^{1/2-\delta}$ under GRH, with fixed
$0<\delta<1/2$; its error is absolute. For fixed rational $\beta>2$,
the acceptance and output-distribution errors are smaller than every
fixed inverse power of $\log\log\log x$.
The smaller-cutoff variant in \Cref{prop:gencert-small-cutoff} retains
the same quantitative bounds. \Cref{cor:gencert-moving-threshold} gives
a sufficient range for rational thresholds decreasing to $2$.

\Cref{fig:certified-search} separates the exact certificate of an accepted
prime from the estimates for the acceptance frequency, the output law and
the waiting time, and records the inputs of these estimates from
\Cref{sec:moments,sec:global-modulus,ssec:sharp-prime-convergence}.

% Diagram (design brief F6): From model window masses to certified prime search.
\begin{figure}[p]
\centering
\begin{tikzpicture}[figdiagram,
  % local helpers: slightly tighter vertical padding (the densest figure),
  % and one common height per row of boxes
  fbox/.append style={inner ysep=2pt},
  rowA1/.style={minimum height=1.35cm}, rowA2/.style={minimum height=1.41cm},
  rowB1/.style={minimum height=1.41cm}, rowB2/.style={minimum height=1.76cm},
  rowB3/.style={minimum height=1.81cm}, rowB4/.style={minimum height=1.43cm},
  % notes are set ragged right without hyphenation
  fnote/.append style={align=flush left, execute at begin node={\hyphenpenalty=10000\exhyphenpenalty=10000\binoppenalty=10000\relpenalty=10000}}]
% ===================== panel (a) =====================================
\node[fhead, anchor=west] (hA) at (0,0) {(a) Analytic inputs};
\draw[frule] (hA.south west) ++(0,-1.5pt) -- ++(13.46,0);
% --- row a1: the model (Chapters 6 and 7) ----------------------------
\node[fres, rowA1, anchor=north west] (conc) at (6.12,-0.38)
  {Sharp concentration:\\ $Q_G(\varepsilon)=H_{\mathrm{end}}(\varepsilon)+O(\mathcal R(\varepsilon))$;\\
   maximizers $t_\varepsilon\to\tfrac13$\\
   \figref{\Cref{thm:sharp-global-modulus}}};
\node[fres, rowA1, anchor=west] (end) at (0,0 |- conc) {Endpoint benchmark:\\ $H_{\mathrm{end}}(\varepsilon)\sim\kappa_{\mathrm{end}}/z$\\ \figref{\Cref{prop:endpoint-half}}};
\draw[fuse] (end) -- node[flab, above] (labdel) {uniform deletion,\\ prime-$3$ comparison}
  node[flab, below, text width=1.75cm] (labthree) {\upshape\Cref{lem:shifted-uniform-deletion,lem:prime-three-split}} (conc);
% support strip of mu_G = [0,1/2]: ticks at proportional positions
\def\sx{10.55}\def\sL{1.80}\def\sy{-1.40}
\draw[frule, line width=0.5pt] (\sx,\sy) -- ++(\sL,0);
\foreach \f/\t/\n in {0/$0$/tz,0.6667/$\tfrac13$/tt,1/$\tfrac12$/th}
  {\draw[frule, line width=0.5pt] (\sx+\f*\sL,\sy) -- ++(0,2pt) node[above, inner sep=1.5pt, font=\figsmall] (\n) {\t};}
\draw[frule] (\sx+0.6667*\sL,\sy) -- ++(0,-0.1);
\node[fnote, anchor=north, align=center] (lmax) at (\sx+0.6667*\sL,\sy-0.1) {limiting\\ maximizing\\ location};
\node[fnote, anchor=south east, align=left] (lend) at ($(13.46,0 |- th.north)+(0,0.06)$) {endpoint\\ benchmark};
\draw[frule] (th.north) -- (th.north |- lend.south);
% --- row a2: the primes (Chapter 8) ----------------------------------
\node[fjoin] (j2) at ($(6.55,0 |- conc.south)+(0,-0.385)$) {};
\node[fmech, rowA2, anchor=north west] (ht) at ($(0,0 |- j2)+(0,-0.57)$)
  {Finite-head comparison;\\ both remainder means\\
   \figref{\Cref{prop:head-tail-cdf};}\\ \figref{\Cref{lem:ap-finite-head}; \eqref{eq:ap-empirical-tail}, \eqref{eq:ap-model-tail}}};
\node[fres, rowA2, anchor=north east] (bridge) at (13.46,0 |- ht.north)
  {Value window, model mass $m_x$:\\ error of its probability $\le2\eta_x$;\\ relative asymptotic if $\eta_x=o(m_x)$\\
   \figref{\eqref{eq:moving-modulus-endpoint}; $Q=1$: \Cref{cor:prime-shrinking-anchor}}};
\node[fres, rowA2, anchor=north east] (cdf) at ($(bridge.north west)+(-0.62,0)$)
  {CDF error $\eta_x$ against $G_{Q,b}$:\\ permitted growing moduli:\\ $O(D(Q)L_2/L_1)$\\ \figref{\Cref{prop:ap-kolmogorov-pow2}\textup{(a),(b)}}};
\draw[fuse] (cdf) -- (bridge);
% junction: finite-head comparison and model concentration, jointly
\draw[ffeed] (ht.north) |- (j2);
\draw[ffeed] (conc.south -| j2) -- (j2);
\draw[fuse] (j2) -- (j2 |- cdf.north);
\node[flab, anchor=west, align=left] (labconc) at ($(j2)+(0.1,0)$)
  {model concentration\\ $\ll D(Q)/\log(1/h)$, {\upshape\Cref{lem:ap-uniform-concentration}}};
\node[fnote, anchor=north west, text width=13.35cm] (nsharp) at ($(ht.south west)+(0,-0.1)$) {Fixed $Q$: sharp order $1/L_1$, by the separate smooth-pattern recursion of \Cref{thm:addfn-sharp-order}.\\ Ranges and exceptions: \cref{ssec:progression-regimes}.};
% ===================== panel (b) =====================================
\node[fhead, anchor=north west] (hB) at ($(nsharp.south west)+(0,-0.15)$) {(b) Certification versus acceptance};
\draw[frule] (hB.south west) ++(0,-1.5pt) -- ++(13.46,0);
% --- exact lane: the certificate of one output -------------------------
\node[ftag, anchor=north west] (tE) at ($(hB.south west)+(0,-0.10)$) {exact lane: every output $q$ of \Cref{def:gencert-search}, cutoff $B$; no distribution estimate};
\node[finp, rowB1, anchor=north west] (data) at ($(tE.south west)+(0,-0.10)$)
  {Small factors of $q-1$:\\ $(\ell,v_\ell(q-1))$, $\ell\le B$;\\ cofactor $R_B(q)$,\\ not factored};
\node[fres, rowB1, anchor=north west] (ub) at (6.70,0 |- data.north)
  {Upper certificate:\\ $1/c(q)\le U_B(q)\le U_B^+(q)$\\ \figref{\Cref{thm:factoring-free-cert},}\\ \figref{\Cref{cor:factoring-free-verification}}};
\node[fres, rowB1, anchor=north east] (cert) at (13.46,0 |- data.north)
  {Every output $q$:\\ $U_B^+(q)\le\beta$, hence\\ $1/c(q)\le\beta$\\ \figref{\Cref{thm:gencert}\textup{(i)}}};
\draw[fuse] (data) -- node[flab, above] (labenc) {rational enclosures} node[fhyp, below] (lablem) {\Cref{lem:euler-product-enclosure,lem:rough-cofactor-envelope}} (ub);
\draw[fuse] (ub) -- (cert);
% --- sampling lane: a uniform prime of the window ------------------------
\draw[frule] ($(data.south west)+(0,-0.12)$) coordinate (rS) -- (13.46,0 |- rS);
\node[ftag, anchor=north west] (tS) at ($(rS)+(0,-0.04)$) {sampling lane: $q$ uniform in $\mathcal Q_x$, the primes $q\equiv1\ (\mathrm{mod}\ 2N)$ in $[x,2x]$};
\node[fres, rowB2, anchor=north west] (cdfw) at ($(tS.south west)+(0,-0.08)$)
  {CDF error on $\mathcal Q_x$\\ of $c(q)$ against $G$:\\ $O(L_2/L_1)$\\ \figref{\Cref{prop:ap-kolmogorov-pow2}\textup{(c)}}\\ \figref{at $x$, $2x$}};
\node[fres, rowB2, anchor=north east] (gap) at (13.46,0 |- cdfw.north)
  {Uniform threshold bound:\\ mass between outer sets\\ $\le Q_G(1-r_x^{-1})=O(L_3^{-1})$\\ \figref{\Cref{thm:sharp-global-modulus};}\\ \figref{$r_x-1\asymp L_2^{-2}$}};
\path ($(cdfw.east)!0.5!(gap.west)$) coordinate (smid);
\node[fmech, rowB2, minimum width=6.85cm, anchor=north] (sand) at (smid |- cdfw.north)
  {Certificate sandwich, $r_x\to1$:\\[1pt]
   $\{1/c(q)\le\beta/r_x\}\subseteq\underbrace{\{U_B^+(q)\le\beta\}}_{\makebox[0pt]{\figsmall accepted set $\mathcal A_x$}}\subseteq\underbrace{\{1/c(q)\le\beta\}}_{\makebox[0pt]{\figsmall target set $\mathcal T_x$}}$\\
   \figref{proof of \Cref{cor:gencert-polylog}}};
\draw[fuse] (9.28,0 |- ub.south) -- (9.28,0 |- sand.north);
\node[fhyp, anchor=west] (labff) at (9.36,0 |- tS) {\eqref{eq:ff-cert}, enclosure};
% junction: window CDF error, sandwich and model mass, jointly
\node[fjoin] (j4) at ($(6.86,0 |- cdfw.south)+(0,-0.16)$) {};
\draw[ffeed] (cdfw.south) |- (j4);
\draw[ffeed] (gap.south) |- (j4);
\draw[ffeed] (sand.south -| j4) -- (j4);
% --- acceptance and output -------------------------------------------------
\node[fres, rowB3, anchor=north] (acc) at ($(j4)+(0,-0.56)$)
  {Accepted fraction $|\mathcal A_x|/|\mathcal Q_x|$\\ $=1-G(1/\beta)+O(L_3^{-1})$, all real $\beta>0$;\\
   errors $O_{\beta,M}(L_3^{-M})$ for $\mathcal A_x$, $\mathcal T_x$,\\ fixed rational $\beta>2$\\ \figref{\Cref{cor:gencert-polylog}}};
\draw[fuse] (j4) -- (acc.north);
\node[fres, rowB3, anchor=north east] (out) at (13.46,0 |- acc.north)
  {Output law, if $\mathcal A_x\ne\varnothing$:\\ $\mathrm{Unif}(\mathcal A_x)$ exactly;\\ within $O_{\beta,M}(L_3^{-M})$\\ of $\mathrm{Unif}(\mathcal T_x)$ in $d_{\mathrm{TV}}$\\ \figref{\Cref{cor:gencert-polylog}\textup{(ii)}}};
\draw[fuse] (acc) -- (acc -| out.west);
\node[fnote, anchor=west, text width=3.35cm] (caut) at (0,0 |- acc)
  {Fixed rational threshold refinement: the \mbox{one-sided} window $(1/\beta,r_x/\beta]$ is bounded by \Cref{lem:binomial-uniform-deletion}, in place of $Q_G$. Error terms and thresholds in $x$ may be ineffective.};
% --- waiting time ------------------------------------------------------------
\node[fres, rowB4, anchor=north] (wait) at ($(acc.south)+(0,-0.56)$)
  {Halts a.s.\ for $x\ge x_0(N,\beta)$;\\ $\mathbb E[\#\text{sampled candidates}]$\\ $\sim\log x/(2(1-G(1/\beta)))$\\ \figref{\Cref{thm:gencert}\textup{(iii)}}};
\draw[fuse] (acc) -- node[fhyp, left=1pt] (labpos) {$1-G(1/\beta)>0$, \Cref{lem:gencert-positive}}
  node[fhyp, right=1pt] (labsw) {prime density $\sim2/\log x$, \Cref{thm:sw}} (wait);
\node[fres, rowB4, anchor=north west] (eff) at (0,0 |- wait.north)
  {Effective threshold:\\ halts a.s.\ for $x\ge X(N,\beta)$,\\ $\mathbb P(\text{draw accepted})\ge c_\beta/\log x$\\ \figref{\Cref{thm:gencert-effective-mas23}; $X$, $c_\beta$ computable}};
\draw[fuse] (eff) -- node[flab, above] (labeff) {$x_0$} (eff -| wait.west);
\node[fres, rowB4, anchor=north east] (mov) at (13.46,0 |- wait.north)
  {Thresholds $\beta(x)=2+d_x\to2$:\\ same sandwich and CDF error,\\ $|\mathcal A_x|/|\mathcal Q_x|\sim\kappa_{\mathrm{end}}/\log(1/d_x)$\\ \figref{\Cref{cor:gencert-moving-threshold}}};
\node[fres, anchor=north east] (side) at ($(13.46,0 |- mov.south)+(0,-0.81)$) {Endpoint asymptotic \figref{\eqref{eq:hend-asymp}}};
\draw[fuse] (side.north -| 10.6,0) -- node[flab, left=1pt] (labmovl) {in place of $Q_G$}
  node[fhyp, right=1pt, align=left] (labmov) {if $\beta(x)\in\mathbb Q$,\\ $d_x\ge3L_2^{-2}$ eventually} (mov.south -| 10.6,0);
\node[fnote, anchor=north west, text width=8.3cm] (fl) at ($(eff.south west)+(0,-0.07)$) {A window below $x_0$ can have $\mathcal A_x=\varnothing$.
  Arbitrary fixed $N\ge1$, rational $\beta>2\prod_{\ell\mid N,\,\ell>2}(1-1/\ell)^{-1}$: accepted~fraction~$\to1-G_{2N,1}(1/\beta)$ (no effective threshold given), $\mathbb E[\#]\sim\varphi(2N)\log x/(2N(1-G_{2N,1}(1/\beta)))$.};
\end{tikzpicture}
\caption{Analytic inputs to certified prime search.
\textup{(a)}~Model window masses on $\supp\mu_G=[0,\tfrac12]$ and the arithmetic CDF error $\eta_x$; relative counts
in shrinking windows follow when $\eta_x=o(m_x)$. \textup{(b)}~For a power of two $N$, rational enclosures certify
each accepted prime; arithmetic CDF estimates and model concentration control the frequency of acceptance and the
output law. Here $z=\log(1/\varepsilon)$, $\mathcal R(\varepsilon)=e^{-\frac15\sqrt{z\log z}}$, $L_1=\log x$,
$L_2=\log L_1$ and $L_3=\log L_2$; $M>0$ is arbitrary.}
\label{fig:certified-search}
\end{figure}

\Needspace{5\baselineskip}
\begin{lemma}[Computable positive acceptance bound]\label{lem:gencert-positive}
For an integer $y\ge3$, put $\overline m_y:=2(y-1)/(y-2)$.
For every such $y$ and every real $\beta>\overline m_y$,
\[
 1-G(1/\beta)
 \ge\left(1-\frac{\overline m_y}{\beta}\right)
       \prod_{\substack{3\le\ell\le y\\\ell\textup{ prime}}}
          \frac{\ell-2}{\ell-1}>0.
\]
For rational $\beta>2$, the choice
$y=3+\lceil2/(\beta-2)\rceil$ makes the final bound a computable
positive rational.
\end{lemma}

\begin{proof}
In the representation
$1/X=2\prod_{\ell\ge3}(1-1/\ell)^{-B_\ell}$ of
\Cref{lem:factor}, let $A_y=\{B_\ell=0:3\le\ell\le y\}$.
Independence, monotone convergence and the factor means
$\mathbb E[(\ell/(\ell-1))^{B_\ell}]=1+(\ell-1)^{-2}$ (proof of
\Cref{cor:ap-moments}) give
\[
 \begin{aligned}
 \mathbb P(A_y)&=\prod_{\substack{3\le\ell\le y\\\ell\textup{ prime}}}
                 \frac{\ell-2}{\ell-1},\qquad\\
 \mathbb E[1/X\mid A_y]&=2\prod_{\ell>y}\Bigl(1+\frac1{(\ell-1)^2}\Bigr)\\
 &\le2\prod_{n\ge y}\frac{n^2}{n^2-1}=\frac{2y}{y-1}<\overline m_y,
 \end{aligned}
\]
since the integers $\ell-1$ $(\ell>y)$ are distinct and at least $y$, and
the last product telescopes. Markov's inequality conditional on $A_y$ gives
$1-G(1/\beta)\ge\mathbb P(A_y)\,\mathbb P(1/X\le\beta\mid A_y)
\ge\mathbb P(A_y)(1-\overline m_y/\beta)$. For $y=3+\lceil2/(\beta-2)\rceil$,
$y-2>2/(\beta-2)$, hence $\overline m_y<\beta$.
\end{proof}

An odd prime $q$ is a \emph{fully splitting negacyclic number-theoretic
transform (NTT) prime} for $N$ if $q\equiv1\pmod{2N}$, equivalently
if $X^N+1=\Phi_{2N}(X)$ splits completely over $\mathbb F_q$;
for the finite-field transform of length $d\mid q-1$ see
Pollard~\cite[\S~1]{Pol71}, and for the negacyclic NTT and this splitting see
Lyubashevsky and Seiler~\cite[\S\S~1 and~4; Theorem~2.3]{LS18}.
By \Cref{cor:ap-moments}, $X_{2N,1}\overset d=X$; hence, for each fixed $N$,
$c(q)$ has limiting distribution function $G$ and $1/c(q)$ has limiting mean
$M(-1)$ along $q\equiv1\pmod{2N}$, by \Cref{thm:ap-law,thm:prime-mellin}.
The moment convergence is uniform for powers of two
$2N\le x^{1/3}\exp\{-(\log\log x)^3\}$, with arbitrary fixed
inverse-logarithmic accuracy and possibly ineffective constants and thresholds.

The progression estimates used below are compared in
\Cref{ssec:progression-regimes}, with proofs in
\Cref{ssec:sharp-prime-convergence}.

\begin{definition}[Certified search for fully splitting negacyclic NTT primes]\label{def:gencert-search}
\index{Certified prime search!definition}
\index{Number-theoretic transforms!fully splitting primes}
Fix a power of two $N$, a rational $\beta>2$, and an integer
$x\ge16$, given in binary, with $2N\le x$. We call such inputs
\emph{admissible}.
Put $n_x:=\lceil\log_2x\rceil$ and
$B:=n_x\lceil\log_2n_x\rceil$. For an odd prime $q$, put
$\varepsilon_q:=1/\lceil\log_2q\rceil$ and set
\[
 U_B^\pm(q):=U_{B,\varepsilon_q}^\pm(q)
 =\frac{\mathsf{E}_B(q)V(B,R_B(q))}{P_1^\mp(q)}
\]
as in \Cref{cor:factoring-free-verification}.

The search repeatedly samples an integer $q$ independently and uniformly
from $\{m\in\mathbb Z\cap[x,2x]:m\equiv1\pmod{2N}\}$, applies a fixed deterministic polynomial-time
primality test~\cite{AKS04}, and outputs $q$ precisely when $q$ is prime
and $U_B^+(q)\le\beta$.

The certificate of an output $q$ consists of the pairs
$(\ell,v_\ell(q-1))$ for primes $\ell\le B$ dividing $q-1$
and the number $k(B,R_B(q))$. All one-time initialization is charged
to the first iteration. Let $T_{N,\beta}(x)$ be the supremum, over
finite reachable histories preceding termination, including the empty
history before initialization, of the conditional expected bit cost
of the next iteration.
\end{definition}

Since $q\ge\lceil\log_2q\rceil=1/\varepsilon_q$, the truncation in
\Cref{cor:factoring-free-verification} is $J=1$.

\paragraph*{An exact search certificate}\phantomsection\label{par:worked-search-certificate}
Take $N=16$, $x=1024$, $\beta=3$ and $q=1889$ in
\Cref{def:gencert-search}. Then $n_x=10$, $B=40$,
$\varepsilon_q=1/11$ and $J=1$. The candidate lies in $[1024,2048]$
and is $1\pmod{32}$. Trial division by the primes at most
$\sqrt{1889}<44$ proves primality. Removing the prime powers at most
$40$ from $1888$ gives
\[
 (\ell,v_\ell(q-1))=(2,5),\qquad
 \mathsf{E}_B(q)=2,\qquad R_B(q)=59.
\]
Since $41\le59<41\cdot43$, the certificate has $k(B,R_B)=1$ and
$V(B,R_B)=41/40$. Thus
\[
 P_1^+(q)=\frac{1888}{1889},\qquad
 P_1^-(q)=\frac{1887}{1888}.
\]
The target interval from \eqref{eq:rough-cofactor-certificate} and the
enclosure of the conservative bound are, respectively,
\[
 \frac1{c(1889)}\in
 \left[\frac{1889}{944},\frac{19352}{9435}\right],\qquad
 U_B(1889)\in
 \left[\frac{77449}{37760},\frac{19352}{9435}\right].
\]
The ratios of their endpoints are
$18268288/17822715$ and $3564544/3564543$, respectively.
The verifier checks the interval, congruence, primality, complete
small-factor list and $k$, and accepts because $19352/9435<3$.

The conservatism of this certificate is visible from the additional
fact that $59$ is prime. Thus $T_B=59/58$ and
\[
 \frac1{c(1889)}\le\frac{111392}{54723}<\frac{41}{20}
 <\frac{77449}{37760}\le U_B(1889).
\]
At $\beta=41/20$, the prescribed $B=40$ search therefore rejects a
candidate whose true reciprocal density meets the threshold.

\paragraph*{A finite candidate population}\phantomsection\label{par:finite-search-population}
For the same $N=16$ and $x=1024$, the integer candidate set is
$\mathcal C_x=\{1025+32j:0\le j\le31\}$. Exact trial division gives
the eight primes below. The table uses $B=40$, $J=1$, and exact
rational comparison with $\beta=3$.
\begin{center}
\begin{tabular}{rrrrrl}
\toprule
$q$ & $\mathsf{E}_B(q)$ & $R_B(q)$ & $k(B,R_B)$ & $U_B^+(q)$ & Accepted \\
\midrule
$1153$ & $3$ & $1$ & $0$ & $3456/1151$ & no \\
$1217$ & $19/9$ & $1$ & $0$ & $23104/10935$ & yes \\
$1249$ & $13/4$ & $1$ & $0$ & $4056/1247$ & no \\
$1409$ & $11/5$ & $1$ & $0$ & $15488/7035$ & yes \\
$1601$ & $5/2$ & $1$ & $0$ & $4000/1599$ & yes \\
$1697$ & $2$ & $53$ & $1$ & $17384/8475$ & yes \\
$1889$ & $2$ & $59$ & $1$ & $19352/9435$ & yes \\
$2017$ & $7/2$ & $1$ & $0$ & $7056/2015$ & no \\
\bottomrule
\end{tabular}
\end{center}
Thus independent uniform draws from $\mathcal C_x$ are prime with
probability $8/32=1/4$, and a prime drawn uniformly from this window is
accepted with probability $5/8$. Each integer draw succeeds with
probability $5/32$, so the mean number of independent draws is $32/5$;
the output is uniform on the five accepted primes. At $\beta=41/20$
none of the eight primes passes the same certificate test, so this
fixed-window search never terminates.

\begin{corollary}[Distribution of $U_B^+$ and the search output]
\label{cor:gencert-polylog}
Use the cutoff $B$ and bound $U_B^+$ of
\Cref{def:gencert-search}. For a reduced class $b\pmod Q$, let $q$
be a uniform prime in that class and in $[x,2x]$.
\begin{enumerate}[\upshape(i)]
\item Uniformly over $Q=2^a\le x^{1/3}\exp\{-(\log\log x)^3\}$,
$a\in\Z_{\ge0}$, reduced $b\pmod Q$, and all real $\beta>0$,
\begin{equation}\label{eq:gencert-threshold-uniform}
 \mathbb P(U_B^+(q)\le\beta)
 =1-G(1/\beta)+O((\log\log\log x)^{-1}).
\end{equation}
\item For the search with fixed rational $\beta>2$, let $\mathcal Q_x$ be
the set of primes $q\in[x,2x]$ with $q\equiv1\pmod{2N}$, and put
\[
 \mathcal A_x:=\{q\in\mathcal Q_x:U_B^+(q)\le\beta\},\qquad
 \mathcal T_x:=\{q\in\mathcal Q_x:1/c(q)\le\beta\}.
\]
If $\mathcal A_x\ne\varnothing$, the output is uniform on $\mathcal A_x$.
Put $L_3=\log\log\log x$. For every fixed $M>0$, both fractions
$|\mathcal A_x|/|\mathcal Q_x|$ and $|\mathcal T_x|/|\mathcal Q_x|$ equal
$1-G(1/\beta)+O_{\beta,M}(L_3^{-M})$.
Both sets are eventually nonempty, and the output law is within total
variation distance $O_{\beta,M}(L_3^{-M})$ of
$\operatorname{Unif}(\mathcal T_x)$, uniformly for
$2N\le x^{1/3}\exp\{-(\log\log x)^3\}$.
\item Under GRH, for fixed $0<\delta<1/2$, \textup{(i)} holds for
$Q=2^a\le x^{1/2-\delta}$ with error
$O_{\delta}((\log\log\log x)^{-1})$; \textup{(ii)} holds for
$2N\le x^{1/2-\delta}$ with error
$O_{\delta,\beta,M}(L_3^{-M})$ for every fixed $M>0$.
\end{enumerate}
The constants and thresholds in $x$ may be ineffective.
\end{corollary}

\begin{proof}
Apply \eqref{eq:ap-kolmogorov-power-range} with base $2$ at $x$ and
$2x$. By \Cref{cor:ap-finite-head-ranges}\textup{(iii)} with base $2$, the
progression population asymptotic \eqref{eq:ap-polylog-population}
holds at $x$ and at $2x$, uniformly in the stated range, which is
increasing with $x$
for all sufficiently large $x$. Subtracting gives a window population
asymptotic to $x/(\varphi(Q)\log x)$ and Kolmogorov error
$O(\log\log x/\log x)$; a boundary prime is negligible uniformly.
The bound \eqref{eq:ff-cert} and the rational enclosure give
\[
 \frac1{c(q)}\le U_B^+(q)\le\frac{r_x}{c(q)},\qquad
 r_x:=\left(1+\frac1{\log x}\right)
       \exp\!\left(\frac{\log(2x)}{B\log B}\right)\longrightarrow1.
\]
Hence $\{1/c(q)\le\beta/r_x\}\subseteq\{U_B^+(q)\le\beta\}\subseteq\{1/c(q)\le\beta\}$ for every real $\beta>0$.
The model probabilities of these events differ by
$G(r_x/\beta)-G(1/\beta)\le Q_G(1-1/r_x)$ (take $t=r_x/\beta$ in the
definition of $Q_G$). Since $r_x-1\asymp(\log\log x)^{-2}$,
\Cref{thm:sharp-global-modulus} bounds this by
$O((\log\log\log x)^{-1})$ uniformly in $\beta>0$, which proves
\eqref{eq:gencert-threshold-uniform}.

For fixed $Q,b$ and rational $t\in(0,A_{Q,b})$, apply
\Cref{lem:binomial-uniform-deletion} to $X_{Q,b}$ with $k=1$,
$E=\{\ell>2:\ell\mid Q\}$, $A=A_{Q,b}$,
upper endpoint $t+h$, $\varepsilon=h/(t+h)$ and deletion exponents
$1/10,2/5$. A surviving anchor would satisfy
$t<a<(t+h)/(1-\varepsilon^{2/5})$ and have denominator
$O_Q(\varepsilon^{-1/10})$. Rational separation gives
$a-t\gg_{Q,t}\varepsilon^{1/10}$, exceeding the bracket width
$O_{Q,t}(\varepsilon^{2/5})$. Hence there is no survivor, and
\[
 G_{Q,b}(t+h)-G_{Q,b}(t)
 \ll_{Q,t}\mathcal R_c(h)
\]
for some $c>0$. For powers of two the model is $G$, so take $Q=1$
and $t=1/\beta$. Write $L_2=\log\log x$ and $L_3=\log L_2$.
Whenever $r_x-1=O(L_3/L_2)$, the lost window $(t,r_xt]$ is contained
in one of width $h=C_\beta L_3/L_2$, with $\log(1/h)\sim L_3$.
Its model mass is $O_{\beta,M}(L_3^{-M})$ for every fixed $M>0$;
the window Kolmogorov error above is smaller.

Independent uniform candidate draws with deterministic acceptance give
uniform output on $\mathcal A_x$ whenever $\mathcal A_x\ne\varnothing$.
The sandwich therefore gives both stated fraction estimates.
Their main term is positive by \Cref{lem:gencert-positive}, so both sets
are eventually nonempty. Since $\mathcal A_x\subseteq\mathcal T_x$ by the
second inclusion above,
\[
 d_{\mathrm{TV}}(\operatorname{Unif}(\mathcal A_x),
                 \operatorname{Unif}(\mathcal T_x))
 =1-\frac{|\mathcal A_x|}{|\mathcal T_x|}
 \ll_{\beta,M}L_3^{-M}.
\]

Under GRH, apply \Cref{thm:ap-grh-law} at $x$ and $2x$; by
\eqref{eq:ap-grh-population} the window population is
$\asymp_{\delta}x/(\varphi(Q)\log x)$, so subtraction preserves the
distribution function error of \eqref{eq:ap-grh-cdf}. For \textup{(i)},
the sandwich adds $Q_G(1-1/r_x)=O((\log\log\log x)^{-1})$.
For fixed rational $\beta$, the right-window argument and nested-set
identity give the asserted $O_{\delta,\beta,M}(L_3^{-M})$ error.
This proves \textup{(iii)}.
\end{proof}

The error in \eqref{eq:gencert-threshold-uniform} is absolute.
For shrinking thresholds, the endpoint comparison in
\Cref{cor:gencert-moving-threshold} gives relative acceptance
asymptotics and eventual halting. Under GRH, the $k=0$
case of \Cref{thm:bv-grh} at $x$ and $2x$ gives primes
$q\equiv1\pmod{2N}$ in $[x,2x]$ for all sufficiently large $x$ whenever
$2N(\log x)^2/\sqrt x\to0$, but \Cref{cor:gencert-polylog}(iii) estimates acceptance only for $2N\le x^{1/2-\delta}$.

\paragraph*{Candidate counts and model probabilities}
For any fixed admissible window, write $\mathcal C_x$ for the integer
candidate set in \Cref{def:gencert-search}, and use the prime and
accepted sets $\mathcal Q_x,\mathcal A_x$ of
\Cref{cor:gencert-polylog}. If $\mathcal Q_x\ne\varnothing$, the
success probability of one uniform candidate draw is exactly
\[
 \frac{|\mathcal A_x|}{|\mathcal C_x|}
 =\frac{|\mathcal Q_x|}{|\mathcal C_x|}
  \frac{|\mathcal A_x|}{|\mathcal Q_x|}.
\]
When $\mathcal A_x\ne\varnothing$, independent draws have mean
$|\mathcal C_x|/|\mathcal A_x|$ until success. For fixed $\beta>2$,
the two factors are asymptotic to $2/\log x$ and
$F(\beta):=1-G(1/\beta)$, uniformly over the power-of-two ranges in
\Cref{cor:gencert-polylog}. Indeed, $|\mathcal C_x|=x/(2N)+O(1)$,
and the population estimates in its proof give the first factor;
$\varphi(2N)=N$ gives the constant $2$.
For example, the model enclosure in \Cref{prop:gencert-numerics} gives
\[
 \frac1{2F(3)}\in
 \left[\frac{50000}{49149},\frac{25000}{24573}\right].
\]
This interval encloses the coefficient of $\log x$ in the asymptotic
mean number of candidate draws; the finite-window mean is given by
the exact count ratio above. 

\begin{theorem}[Certified generation of fully splitting negacyclic NTT primes]\label{thm:gencert}
\index{Certified prime search!generation guarantees}
Fix $N$ and $\beta$ as in \Cref{def:gencert-search}.
There are effectively computable constants $C,C_\beta>0$ and
a threshold $x_0=x_0(N,\beta)\ge\max(64,2N)$ such that:
\begin{enumerate}[\upshape(i)]
\item Every output $q$ is a prime $q\equiv1\pmod{2N}$ in $[x,2x]$
satisfying
\[
 1/c(q)\le U_B(q)\le U_B^+(q)\le\beta.
\]
Its certificate is verifiable in $O_\beta((\log q)^C)$ bit operations
without completely factoring $q-1$.
\item For every admissible $x\ge64$,
\[
 T_{N,\beta}(x)\le C_\beta(\log x)^C.
\]
\item For $x\ge x_0$, the search halts with probability $1$, and
\[
 \mathbb E[\#\textup{sampled candidates}]
 =(1+o(1))\frac{\log x}{2(1-G(1/\beta))}.
\]
Moreover,
\[
\begin{aligned}
 \mathbb E[\textup{bit operations}]
 &\le(1+o(1))\frac{T_{N,\beta}(x)\log x}{2(1-G(1/\beta))}\\
 &\le(1+o(1))\frac{C_\beta(\log x)^{C+1}}{2(1-G(1/\beta))}.
\end{aligned}
\]
\end{enumerate}
The threshold $x_0$ can be chosen effectively by
\Cref{thm:gencert-effective-mas23}; the $o(1)$ terms may be ineffective.
\end{theorem}

\Cref{prop:gencert-small-cutoff} gives a variant with a smaller
trial-division cutoff and the same asymptotic expected number of candidates.

\begin{proof}
\emph{(i)} By \Cref{thm:factoring-free-cert,cor:factoring-free-verification}, every output satisfies $U_B^+(q)\le\beta$ and
\[
 1/c(q)\le U_B(q)\le U_B^+(q)\le(1+\varepsilon_q)U_B^-(q)\le(1+\varepsilon_q)U_B(q).
\]
Given the admissible input $(N,\beta,x)$, the proposed output $q$ and
its certificate, the verifier recomputes $n_x,B,\varepsilon_q$ and
checks the interval, congruence and primality conditions. It reruns
the algorithm of \Cref{cor:factoring-free-verification}, which needs
no factorization of $R_B(q)$, compares its output with the certificate,
and tests $U_B^+(q)\le\beta$ exactly.

\emph{(ii)} The integers $n_x$ and $\lceil\log_2n_x\rceil$ are the
binary lengths of $x-1$ and $n_x-1$, respectively. Thus the cutoff
and the endpoints and cardinality of the candidate set are computable
by integer operations in polynomial bit cost. For $x\ge64$ and
$q\in[x,2x]$, one has $n_x\le2\log x$ and
\[
 3\le B\le n_x^2\le4(\log x)^2\le(\log q)^3.
\]
The encoding length of $\varepsilon_q$ is $O(\log\log x)$.
Thus \Cref{cor:factoring-free-verification}, with exponent $3$,
bounds all certificate computations by an effective polynomial in
$\log x$. Since $2N\le x$, the candidate set is nonempty; exact uniform
sampling by rejection uses $O(\log x)$ random bits on average, and the
primality test, integer arithmetic and comparison with the fixed rational
$\beta$ have effective polynomial bit cost. These bounds hold uniformly
over all reachable histories and all $2N\le x$, proving the estimate for
$T_{N,\beta}(x)$ with $C_\beta$ independent of $N$.
They also prove the verification bound in \textup{(i)}, after increasing
its constant to cover $16\le x<64$.

\emph{(iii)} By \Cref{thm:sw} and $\varphi(2N)=N$, the
$(1+o(1))x/(2N)$ integers of the sampled class contain
$(1+o(1))x/(N\log x)$ primes, so a uniform draw is prime with probability
$(1+o(1))2/\log x$. By \eqref{eq:gencert-threshold-uniform} with $Q=2N$
and $b=1$, a fraction $1-G(1/\beta)+o(1)$ of these primes is accepted,
and $1-G(1/\beta)>0$ by \Cref{lem:gencert-positive}. Thus each draw
succeeds with probability $(1+o(1))2(1-G(1/\beta))/\log x$.
Independent trials give almost-sure halting and the stated expected
number of draws. Summing conditional iteration costs, with
initialization charged to the first iteration, over the geometric
survival probabilities gives
$\mathbb E[\textup{bit operations}]\le T_{N,\beta}(x)\,\mathbb E[\#\textup{sampled candidates}]$,
which, with \textup{(ii)}, proves the cost inequalities.
\end{proof}

\begin{theorem}[An effective fixed-input search bound]
\label{thm:gencert-effective-mas23}
Fix a power of two $N$ and a rational $\beta>2$, and use the original
cutoff of \Cref{def:gencert-search}. There are a computable rational
$c_\beta>0$ and a computable threshold $X(N,\beta)$ such that, for every
admissible $x\ge X(N,\beta)$,
\[
 \mathbb P(\textup{one candidate is accepted})\ge\frac{c_\beta}{\log x},
 \qquad
 \mathbb E[\#\textup{sampled candidates}]\le\frac{\log x}{c_\beta}.
\]
The search halts almost surely, and its expected bit cost is at most
$C_\beta(\log x)^{C+1}/c_\beta$, with the effective constants of
\Cref{thm:gencert}\textup{(ii)}.
\end{theorem}

\begin{proof}
Write $Q=2N$ and choose a positive rational $\delta$ with
$2e^\delta<\beta$, using certified approximations to the exponential.
For an integer $y\ge3$, put
\[
 M_y=\prod_{3\le\ell\le y}\ell,\qquad
 \alpha_y=\prod_{3\le\ell\le y}\frac{\ell-2}{\ell-1},
\]
where products indexed by $\ell$ are over primes. Choose $y$ by finite
search until
\begin{equation}\label{eq:effective-head-choice-mas23}
 \frac{32}{\delta y}<\frac{\alpha_y}{4}.
\end{equation}
This search terminates: the identity
\[
 \alpha_y=\prod_{3\le\ell\le y}(1-1/\ell)
             \prod_{3\le\ell\le y}(1-1/(\ell-1)^2)
\]
and Mertens' theorem show $\alpha_y\asymp1/\log y$. All quantities
in the stopping test are rational and computable; $y$ depends only
on $\beta$.

We first count primes in $(x,2x]$, $q\equiv1\pmod Q$, for which no
odd prime at most $y$ divides $q-1$. Since $Q$ is a power of two,
$(Q,M_y)=1$. Finite inclusion--exclusion gives their exact number
\[
 H_y(x)=\sum_{d\mid M_y}\mu(d)
       \{\pi(2x;Qd,1)-\pi(x;Qd,1)\}.
\]
Theorem~1.3 of Bennett--Martin--O'Bryant--Rechnitzer
\cite[Theorem~1.3, equations~(1.10)--(1.11)]{BMOR18} gives
\[
 \left|\pi(t;m,a)-\frac{\operatorname{Li}(t)}{\varphi(m)}\right|
 \le c_0(m)\frac{t}{(\log t)^2}\quad(t\ge x_0(m),\ m\ge3),
\]
with explicit $c_0(m)$ and $x_0(m)$, including moduli with possible
exceptional zeros. For $m=2$, sum the two reduced classes modulo $4$;
use $c_0(2):=2c_0(4)$ and $x_0(2):=x_0(4)$. Thus, with the computable
finite sum $E_{Q,y}:=\sum_{d\mid M_y}c_0(Qd)$,
\[
 \left|H_y(x)-\frac{\alpha_y}{\varphi(Q)}
                    \int_x^{2x}\frac{dt}{\log t}\right|
 \le\frac{3E_{Q,y}x}{(\log x)^2}
 \qquad\left(x\ge\max_{d\mid M_y}x_0(Qd)\right).
\]
In particular, on increasing $x$ effectively so that
$\log x\ge\max\{7\log2,24\varphi(Q)E_{Q,y}/\alpha_y\}$, we have
\begin{equation}\label{eq:effective-head-count-mas23}
 H_y(x)\ge\frac{3\alpha_y}{4}\frac{x}{\varphi(Q)\log x}.
\end{equation}
Here $\int_x^{2x}dt/\log t\ge x/\log(2x)$.
If $N$ had an odd prime factor, this avoidance condition would be
incompatible with the progression once $y$ included that factor.

Put $z=\sqrt x/Q$, and increase $x$ effectively so that $z>y$
and the factor $1+O(1/\log(x/(Q\ell)))$ in
Brun--Titchmarsh~\cite[Theorem~3.9]{MV06} is at most $2$, uniformly for
$\ell\le z$. Since $\log(x/(Q\ell))\ge\tfrac12\log x$, this gives, for $y<\ell\le z$,
\[
 \#\{q\in(x,2x]:q\textup{ prime},\ q\equiv1\pmod{Q\ell}\}
 \le\frac{8x}{\varphi(Q)(\ell-1)\log x}.
\]
Write $w_\ell=\log(\ell/(\ell-1))\le1/(\ell-1)$.
Summing over all primes $q\equiv1\pmod Q$ in the window yields
\[
 \sum_q\sum_{\substack{y<\ell\le z\\\ell\mid q-1}}w_\ell
 \le\frac{16}{y}\frac{x}{\varphi(Q)\log x},
\]
since $\sum_{n>y}(n-1)^{-2}\le1/(y-1)\le2/y$.
Markov's inequality and \eqref{eq:effective-head-choice-mas23} show
that fewer than $(\alpha_y/4)x/(\varphi(Q)\log x)$ primes have this
middle-tail sum greater than $\delta/2$. The remaining tail satisfies
the pointwise bound
\[
 \sum_{\substack{\ell>z\\\ell\mid q-1}}w_\ell
 \le\frac{\log(2x)}{(z-1)\log z},
\]
because there are at most $\log(2x)/\log z$ such distinct factors.
For $x\ge Q^4$, the already imposed condition $z>y\ge3$ gives
$\log z\ge(\log x)/4$, $\log(2x)\le5(\log x)/4$ and
$z-1\ge2z/3$. Thus this is at most
$15Q/(2\sqrt x)\le8Q/\sqrt x$; requiring also
$x\ge(16Q/\delta)^2$ makes it at most $\delta/2$.
Combining these facts with
\eqref{eq:effective-head-count-mas23} leaves at least
\begin{equation}\label{eq:effective-good-count-mas23}
 \frac{\alpha_y}{2}\frac{x}{\varphi(Q)\log x}
\end{equation}
primes for which $(q-1)/\varphi(q-1)\le2e^\delta$.

It remains to control the prescribed certificate uniformly.
For each of these primes, \eqref{eq:ff-cert} and
$P_1^-(q)=(q-2)/(q-1)$ give
\[
 U_B^+(q)\le
 2e^\delta\exp\!\left(\frac{\log(2x)}{B\log B}\right)
                  \frac{x-1}{x-2}.
\]
Writing $a=\log2$, $L=\log x$ and $u=\log\log x$, the original
cutoff satisfies $B\ge Lu/a^2$ and $\log B\ge u$ for large $x$.
Consequently the last two factors are at most
\[
 \exp(2a^2/u^2)\frac{x-1}{x-2}.
\]
This explicit decreasing envelope tends to one. A computable threshold
therefore makes it less than $\beta/(2e^\delta)$ for every larger $x$.
All primes in \eqref{eq:effective-good-count-mas23} are then accepted.
Finally, for $x\ge2Q$, the integer candidate set has size at most
$x/Q+2\le2x/Q$, and $\varphi(Q)=Q/2$. Thus one may take
$c_\beta=\alpha_y/2$. Independent trials give the mean and halting
claims; summing the conditional cost bound over survival probabilities
gives the stated expected bit cost.
\end{proof}

\begin{proposition}[A smaller search cutoff]\label{prop:gencert-small-cutoff}
\index{Certified prime search!smaller cutoff}
The search of \Cref{def:gencert-search} may instead use
\[
 B=\max\{3,\lfloor n_x/\lceil\log_2n_x\rceil\rfloor\},
 \qquad n_x=\lceil\log_2x\rceil.
\]
For fixed $N$ and $\beta$, this
variant has the soundness, eventual almost-sure halting and expected
number of sampled candidates of \Cref{thm:gencert}.
All conclusions of \Cref{cor:gencert-polylog} hold for this cutoff,
with the same error orders and modulus ranges.
Trial division stops at $B$, and $V(B,R)$ is computed by a sieve to $O(\log x)$.
Verification has effective polynomial bit cost, and the conditional
expected bit cost of each iteration is bounded by an effective polynomial in
$\log x$ for every admissible $x\ge64$. The halting threshold
and asymptotic rate may be ineffective.
\end{proposition}

\begin{proof}
Since $n_x=\log x/\log2+O(1)$ and
$\lceil\log_2n_x\rceil=\log\log x/\log2+O(1)$, uniformly for primes
$q\in[x,2x]$ one has
$d_q:=\log\log q-\log B=\log\log\log x+O(1/\log\log x)$, so
$0\le d_q\le\frac12\log\log q$ for all sufficiently large $x$, and
\eqref{eq:critical-envelope-upper} gives
\[
 V(B,R_B(q))=1+O\!\left(\frac{\log\log\log x}{\log\log x}\right).
\]
The sieve range follows from \Cref{lem:rough-cofactor-envelope}(b), since
$B+\log q=O(\log x)$. Use the same $\varepsilon_q$ and $P_1^\pm$ enclosure.
By \eqref{eq:ff-cert}, $1/c(q)\le U_B(q)=V(B,R_B(q))\mathsf{E}_B(q)/P(q)\le
V(B,R_B(q))/c(q)$; as in the proof of \Cref{thm:gencert}\textup{(i)},
$U_B(q)\le U_B^+(q)\le(1+\varepsilon_q)U_B(q)$ with
$\varepsilon_q\le1/\log x$. Hence $1/c(q)\le U_B^+(q)\le r_x/c(q)$, where
\[
 r_x:=\Bigl(1+\frac1{\log x}\Bigr)\max\bigl(\{1\}\cup
   \{V(B,R_B(q)):q\in[x,2x]\textup{ prime}\}\bigr)\longrightarrow1,
\]
and the inclusions in the proof of \Cref{cor:gencert-polylog}\textup{(i)}
hold with this $r_x$. Since
$r_x-1=O(\log\log\log x/\log\log x)$,
\Cref{thm:sharp-global-modulus} gives
\[
 Q_G(1-1/r_x)\ll
 \left(\log\frac{\log\log x}{\log\log\log x}\right)^{-1}
 \ll(\log\log\log x)^{-1}.
\]
The same window estimates and probability sandwich therefore prove
\Cref{cor:gencert-polylog}\textup{(i)} for this cutoff, uniformly in
$\beta>0$. The fixed-rational right-window argument in that proof
applies to this $r_x$ and gives \textup{(ii)} with every fixed inverse
power of $\log\log\log x$. The GRH window estimate gives \textup{(iii)}.
In particular, for fixed $N$ and $\beta>2$, the
accepted fraction tends to $1-G(1/\beta)>0$. The proof
of \Cref{thm:gencert}\textup{(iii)} then gives eventual almost-sure halting
and $(1+o(1))\log x/(2(1-G(1/\beta)))$ expected sampled candidates.
The cutoff is computable in polynomial bit cost by integer division
and the binary-length operations in the proof of
\Cref{thm:gencert}\textup{(ii)}. For every admissible $x\ge64$, one has
$n_x\le2\log x$ and $\lceil\log_2n_x\rceil\ge3$, whence
\[
 3\le B\le\max\{3,n_x/3\}\le\log x\le(\log q)^3.
\]
Thus the proofs of \Cref{thm:gencert}\textup{(i)},\textup{(ii)} give
soundness, effective polynomial verification cost and an effective polynomial bound for the
conditional expected bit cost of each iteration throughout this range.
\end{proof}

\begin{theorem}[Typical certificate excess for general logarithmic cutoffs]
\label{thm:typical-certificate-general-cutoff}
\index{Prime certificates!typical excess}
Fix $N=2^m$ with $m\in\mathbb Z_{\ge0}$, a real constant $c>0$ and a real
exponent $a$. Let $B=B(x)\ge3$ be integer-valued, with
$B\sim c(\log x)(\log\log x)^a$ as $x\to\infty$, and sample $q$
uniformly from the primes $q\equiv1\pmod{2N}$ in $[x,2x]$.
For real-valued statistics $f_x(q)$ and $g_x(q)$, with $g_x(q)>0$
for these primes once $x$ is large, write $f_x(q)\sim g_x(q)$
\emph{in prime probability} if, for every fixed $\eta>0$, the fraction
of these primes with $|f_x(q)/g_x(q)-1|>\eta$ tends to zero as
$x\to\infty$; these primes form the \emph{exceptional set} for $\eta$.
Put
\begin{equation}\label{eq:typical-general-certificate-scale}
 e_x:=\frac{\log(1+(\log x)/B)}{\log B}.
\end{equation}
With the certificates of \Cref{thm:factoring-free-cert}, define the
rational upper certificate using the same $J=1$ enclosure as in
\Cref{def:gencert-search}:
\[
 U_B^+(q):=\frac{\mathsf E_B(q)V(B,R_B(q))}{P_1^-(q)}.
\]
Then
\begin{equation}\label{eq:typical-general-certificate-excess}
 U_B(q)c(q)-1\sim e_x,
 \qquad U_B^+(q)c(q)-1\sim e_x
 \quad\hbox{in prime probability}.
\end{equation}
The numerical enclosure satisfies the pointwise bound
$0\le U_B^+(q)/U_B(q)-1\le q^{-1}$.
Writing $u=\log\log x$, the exact scale in
\eqref{eq:typical-general-certificate-scale} satisfies
\begin{equation}\label{eq:typical-general-certificate-regimes}
 e_x\sim
 \begin{cases}
 c^{-1}u^{-a-1},&a>0,\\
 \log(1+1/c)/u,&a=0,\\
 (-a)\log u/u,&a<0.
 \end{cases}
\end{equation}
\end{theorem}
\begin{proof}
Put $Q=2N$, $L=\log x$, $u=\log L$ and $z=\sqrt x/Q$.
All probabilities and expectations below use the prime population in
the statement. Its size is asymptotic to $x/(\varphi(Q)L)$ by
\Cref{thm:sw}; in particular it is nonempty for large $x$.
We have
\[
 \log B=u+a\log u+\log c+o(1)\sim u,\qquad
 \frac LB\sim c^{-1}u^{-a}.
\]
These identities give \eqref{eq:typical-general-certificate-regimes}.
Consequently $e_x\to0$, and, for some fixed $K>0$ depending on $a$,
$e_x\gg_{c,a}u^{-K}$. More precisely,
\begin{equation}\label{eq:typical-general-tail-scale-check}
 Be_x\sim
 \begin{cases}
 L/u,&a>0,\\
 c\log(1+1/c)L/u,&a=0,\\
 c(-a)L u^{a-1}\log u,&a<0.
 \end{cases}
\end{equation}
Each quantity on the right tends to infinity. Thus $B^{-1}=o(e_x)$
and $x^{-1}=o(e_x)$. Also $3\le B\le L^2<z$ eventually: every fixed
power of $u$ is $o(L)$.

Define the completely stripped small-prime part by
\[
 s_B(q-1):=\prod_{\ell\le B}\ell^{v_\ell(q-1)},\qquad
 R_B(q)=(q-1)/s_B(q-1).
\]
We first prove the estimates
\begin{equation}\label{eq:typical-stripped-size-mas23}
 \mathbb E\log s_B(q-1)=O_N(\log B),\qquad
 \log R_B(q)/L\longrightarrow1\quad\hbox{in probability}.
\end{equation}
Expand $\log s_B=\sum_{\ell\le B}\log\ell\sum_{j\ge1}
\boldsymbol1_{\ell^j\mid q-1}$. For odd $\ell$ and $\ell^j\le z$,
Brun--Titchmarsh~\cite[Theorem~3.9]{MV06}, applied to modulus
$Q\ell^j\le\sqrt x$ and divided by the fixed-progression population,
gives probability $O(1/\varphi(\ell^j))$.
The prime $2$ uses modulus $\operatorname{lcm}(Q,2^j)$;
the finitely many forced powers and the remaining geometric sum give
$O_N(1)$ after multiplication by $\log2$.
Since $\sum_{j\ge1}1/\varphi(\ell^j)=\ell/(\ell-1)^2$, the small
powers contribute
\[
 O_N(1)+O\!\left(\sum_{\ell\le B}\frac{\log\ell}{\ell}\right)
 =O_N(\log B).
\]
The last bound follows by partial summation from
$\vartheta(t)\ll t$. For powers $\ell^j>z$, the ordinary-multiple
count in $[1,2x]$ gives probability
$O_N(L(\ell^{-j}+x^{-1}))$. For each $\ell$ the geometric tail is
$O(1/z)$, and there are at most $\log(2x)/\log\ell$ relevant
exponents. Summation therefore contributes at most
\[
 O_N\!\left(\frac{LB}{z}+\frac{L^2B}{x}\right)
 =O_N\!\left(\frac{L^3}{\sqrt x}+\frac{L^4}{x}\right)=o(1).
\]
This proves the expectation bound. Since
$\log(q-1)=L+O(1)$ uniformly and $\log B=O(u)=o(L)$,
Markov's inequality proves the convergence in
\eqref{eq:typical-stripped-size-mas23}.

For $w_\ell=\log(\ell/(\ell-1))\le1/(\ell-1)$, the same
Brun--Titchmarsh estimate gives
\[
 \mathbb E\sum_{\substack{B<\ell\le z\\\ell\mid q-1}}w_\ell
 \ll_N\sum_{\ell>B}\frac1{(\ell-1)^2}\ll_N B^{-1}.
\]
There are at most $\log(2x)/\log z$ distinct prime factors above $z$
in $q-1$. Thus, pointwise, their weights sum to at most
$\log(2x)/((z-1)\log z)$, proving
\begin{equation}\label{eq:typical-actual-tail-mas23}
 \mathbb E\log T_B(q)
 =O_N\!\left(\frac1B+\frac{L}{(z-1)\log z}\right).
\end{equation}
Here $\log z\sim L/2$, so the second term is $O_N(x^{-1/2})$.
Both terms are $o(e_x)$, by
\eqref{eq:typical-general-tail-scale-check} and the polynomial lower
bound for $e_x$ in $u$. Markov's inequality yields
$\log T_B(q)=o_{\mathbb P}(e_x)$.

We now estimate the maximizing cofactor envelope at relative scale
$e_x$. On an event of probability tending to one,
$\log R_B(q)=(1+o(1))L$. On this event $k=k(B,R_B(q))\ge1$:
indeed the first prime above $B$ is at most $2B$, while
$\log(2B)=O(u)=o(L)$. Let $r=r_k$. By maximality of the prime
product and \Cref{lem:rough-cofactor-envelope}\textup{(b)},
\[
 0\le\log R_B(q)-\{\vartheta(r)-\vartheta(B)\}
 <\log r_{k+1}=O(\log(B+L)),\qquad r\ll B+L.
\]
The effective prime number theorem
$\vartheta(t)=t+O(te^{-\gamma\sqrt{\log t}})$, for an absolute
$\gamma>0$~\cite[Theorem~6.9]{MV06}, now gives
\[
 r-B=\log R_B(q)
   +O\!\left(\log(B+L)+(B+L)e^{-\gamma\sqrt{\log B}}\right).
\]
After division by $L$ the deterministic error tends to zero, because
\begin{equation}\label{eq:typical-general-envelope-error}
 \begin{aligned}
 \frac{\log(B+L)}L&=O_{c,a}(u/L)=o(1),\\
 \frac{B+L}{L}e^{-\gamma\sqrt{\log B}}
 &=O_{c,a}\!\left((1+u^{\max(a,0)})e^{-\gamma'\sqrt u}\right)=o(1)
 \end{aligned}
\end{equation}
for a fixed $\gamma'>0$. In particular
$r=B+(1+o_{\mathbb P}(1))L$.

By \Cref{lem:prime-log-stieltjes} and
$\sum_{\ell>B}|w_\ell-1/\ell|=O(1/B)$,
\begin{equation}\label{eq:typical-envelope-log-mas23}
 \log V(B,R_B(q))
 =\log\frac{\log r}{\log B}
       +O(e^{-\gamma\sqrt{\log B}}+1/B),
\end{equation}
where $\gamma$ may be decreased. The error is $o(e_x)$ because its
first term decays faster than every fixed inverse power of $u$, and
$B^{-1}=o(e_x)$ by \eqref{eq:typical-general-tail-scale-check}.
To compare the main term with $e_x$, including when $L/B\to0$, put
$\rho=L/B$ and $\delta=(r-B)/L-1=o_{\mathbb P}(1)$.
For $0<\varepsilon<1$ and $|\delta|\le\varepsilon$, concavity of
$\log(1+t)$ and its value zero at $t=0$ give
\[
 (1-\varepsilon)\log(1+\rho)
 \le\log(1+(1+\delta)\rho)
 \le(1+\varepsilon)\log(1+\rho).
\]
Consequently $\log(r/B)=(1+o_{\mathbb P}(1))\log(1+L/B)$,
uniformly with respect to the possible sizes of $\rho$ in this
argument. Since $e_x\to0$,
\[
 \log\frac{\log r}{\log B}
 =\log\!\left(1+\frac{\log(r/B)}{\log B}\right)
 =(1+o_{\mathbb P}(1))e_x.
\]

The identity $U_B(q)c(q)=V(B,R_B(q))/T_B(q)$ gives
$\log(U_B(q)c(q))=(1+o_{\mathbb P}(1))e_x$.
Exponentiation proves the first equivalent in
\eqref{eq:typical-general-certificate-excess}.
Finally \Cref{lem:euler-product-enclosure} with $J=1$ gives, pointwise,
\[
 1\le\frac{U_B^+(q)}{U_B(q)}
 =\frac{P(q)}{P_1^-(q)}
 \le\frac{P_1^+(q)}{P_1^-(q)}\le1+q^{-1}.
\]
Its logarithm is $O(x^{-1})=o(e_x)$, proving the second equivalent
and the stated numerical-error bound.
\end{proof}
\begin{corollary}[Typical excess of the two search certificates]
\label{thm:typical-certificate-excess-mas23}
Fix a power of two $N$. For a uniform prime $q\equiv1\pmod{2N}$ in
$[x,2x]$, use $n_x=\lceil\log_2x\rceil$ and either search cutoff.
For the original cutoff $B=n_x\lceil\log_2n_x\rceil$,
\begin{equation}\label{eq:typical-original-certificate-mas23}
 U_B(q)c(q)-1\sim\frac{(\log2)^2}{(\log\log q)^2}
 \quad\hbox{in prime probability}.
\end{equation}
For $B=\max\{3,\lfloor n_x/\lceil\log_2n_x\rceil\rfloor\}$,
\begin{equation}\label{eq:typical-small-certificate-mas23}
 U_B(q)c(q)-1\sim\frac{\log\log\log q}{\log\log q}
 \quad\hbox{in prime probability}.
\end{equation}
Both equivalents hold with $U_B^+$ in place of $U_B$.
With $n_q=\lceil\log_2q\rceil$ used in the cutoff, for every fixed
$\eta>0$, the exceptional sets for $\eta$ defined by these ratios have
relative density zero among all primes $q\equiv1\pmod{2N}$ up to $x$.
\end{corollary}

\begin{proof}
Write $b_2=\log2$, $L=\log x$ and $u=\log L$.
Then $n_x=L/b_2+O(1)$ and
$\lceil\log_2n_x\rceil=u/b_2+O(1)$. The original cutoff therefore
satisfies
\[
 B=\frac{Lu}{b_2^2}(1+O(1/u)),
\]
and the smaller cutoff satisfies
\[
 B=\frac Lu(1+O(1/u)).
\]
The floor changes the latter by $O(1)$, and the maximum with $3$
eventually has no effect. Apply
\Cref{thm:typical-certificate-general-cutoff} with, respectively,
$(c,a)=(b_2^{-2},1)$ and $(c,a)=(1,-1)$. Its scales are
$b_2^2/u^2$ and $\log u/u$. Uniformly for $q\in[x,2x]$,
$\log\log q=u+O(1/L)$, so these are precisely the equivalents
in \eqref{eq:typical-original-certificate-mas23} and
\eqref{eq:typical-small-certificate-mas23}, including the assertion
for $U_B^+$.

The replacement of $n_x$ by $n_q$ permits the
cutoff to vary within a window. Let $B_-$ and $B_+$ be its minimum
and maximum over that window. The preceding two cutoff asymptotics
hold uniformly in $q$, so $B_-/B_+\to1$ and each has the stated
asymptotic value. The small-prime expectation in the proof of
\Cref{thm:typical-certificate-general-cutoff} is bounded using $B_+$,
and the tail expectation using $B_-$. They give
$\log R_{B(q)}(q)/L\to1$ in probability and
$\log T_{B(q)}(q)=o_{\mathbb P}(e_x)$ at the respective scale.
The deterministic estimates for $r-B(q)$ and the error in
\eqref{eq:typical-envelope-log-mas23} are uniform for
$B_-\le B(q)\le B_+$, as is the concavity comparison in that proof.
Moreover $\log(1+L/B(q))/\log B(q)$ is uniformly equivalent to the
same $e_x$. The proof therefore gives the asserted exceptional
fraction tending to zero in every dyadic window, now for the single
statistic whose cutoff is defined from $q$.

Fix a relative-error tolerance $\eta>0$. Partition primes up to $x$
into consecutive full windows $(x/2^{j+1},x/2^j]$ until the last left
endpoint lies in $[\sqrt x,2\sqrt x)$. The omitted primes are at most
$2\sqrt x$, and hence form $o(1)$ of the fixed-progression population
up to $x$, by \Cref{thm:sw}. The exceptional fraction in the union
of the full windows is at most the supremum of their exceptional
fractions. This supremum tends to zero, since all their left endpoints
tend to infinity and the dyadic assertion holds as a limit in the
left endpoint. This proves relative density zero in the full prime
population. Including or excluding a boundary prime does not affect
any limit.
\end{proof}

\begin{corollary}[Slowly shrinking search thresholds]
\label{cor:gencert-moving-threshold}
\index{Certified prime search!moving thresholds}
Let $\beta(x)>2$ be rational, with $d_x:=\beta(x)-2\to0$, and put
$\xi_x:=\log(1/d_x)$. Use either cutoff in
\Cref{def:gencert-search,prop:gencert-small-cutoff}, and set, respectively,
\[
 \vartheta_x:=(\log\log x)^{-2}
 \quad\hbox{or}\quad
 \vartheta_x:=\frac{\log\log\log x}{\log\log x}.
\]
Suppose that $d_x\ge3\vartheta_x$ eventually, and put
\[
 F_x:=1-G(1/\beta(x)),\qquad
 R_x:=\frac{\vartheta_x}{d_x\xi_x}+\xi_x^{-2}
          +\frac{\xi_x\log\log x}{\log x}.
\]
Then $R_x\to0$.
Uniformly over powers of two $N$ with
$2N\le x^{1/3}\exp\{-(\log\log x)^3\}$, the search with threshold
$\beta(x)$ eventually halts almost surely and
\[
 \begin{aligned}
 \frac{|\mathcal A_x|}{|\mathcal Q_x|}&=F_x(1+O(R_x)),\\
 d_{\mathrm{TV}}\bigl(\mathrm{Unif}(\mathcal A_x),
                       \mathrm{Unif}(\mathcal T_x)\bigr)&\ll R_x.
 \end{aligned}
\]
Here $\mathcal Q_x,\mathcal A_x,\mathcal T_x$ are the prime, accepted
and target sets of \Cref{cor:gencert-polylog}, with threshold $\beta(x)$.
Moreover,
\[
 \mathbb E[\#\textup{sampled candidates}]
 \sim\frac{\log x}{2F_x}
 \sim\frac{\log x\,\xi_x}{2\kappa_{\mathrm{end}}}.
\]
Under GRH the same conclusions hold uniformly for
$2N\le x^{1/2-\delta}$, for each fixed $0<\delta<1/2$, with constants
depending on $\delta$.
The constants and halting thresholds may be ineffective.
\end{corollary}

\begin{proof}
Write the certificate factor in the proofs of \Cref{cor:gencert-polylog,prop:gencert-small-cutoff} as $r_x=1+\eta_x$.
For the original cutoff,
$B\log B\sim(\log x)(\log\log x)^2/(\log2)^2$, so
$\eta_x/\vartheta_x\to(\log2)^2$. For the smaller cutoff,
\eqref{eq:critical-envelope-upper} and the estimate for $d_q$ in
\Cref{prop:gencert-small-cutoff} give
$\eta_x\le(1+o(1))\vartheta_x$. Thus in either case
$0\le2\eta_x/d_x\le2/3+o(1)$.
Put
\[
 s_x=\frac{d_x}{2+d_x},\qquad
 s'_x=\frac{d_x-2\eta_x}{2+d_x}>0.
\]
Then $F_x=H_{\mathrm{end}}(s_x)$ and
$1-G(r_x/\beta(x))=H_{\mathrm{end}}(s'_x)$. Their logarithmic widths
satisfy
\[
 L_x:=\log(1/s_x)=\xi_x+O(1),\qquad
 \log(1/s'_x)-L_x
 =\log\frac{d_x}{d_x-2\eta_x}=O(\vartheta_x/d_x)=O(1).
\]
By \eqref{eq:hend-asymp},
\[
 F_x\sim\frac{\kappa_{\mathrm{end}}}{\xi_x},\qquad
 0\le\frac{H_{\mathrm{end}}(s_x)-H_{\mathrm{end}}(s'_x)}{F_x}
 \ll\frac{\vartheta_x}{d_x\xi_x}+\xi_x^{-2}.
\]
The arithmetic window CDF error in the proof of
\Cref{cor:gencert-polylog} is $O(\log\log x/\log x)$, with a
$\delta$-dependent constant under GRH. The certificate sandwich and
$\xi_x=O(\log\log\log x)$ therefore give the displayed relative
acceptance error and $|\mathcal T_x|/|\mathcal Q_x|=F_x(1+O(R_x))$.
Both sets are eventually nonempty, and
$\mathcal A_x\subseteq\mathcal T_x$ gives the total-variation bound.
The same population estimates give prime density $(2+o(1))/\log x$
among the integer candidates, uniformly in the respective modulus
range. Independent trials have success probability
$(2+o(1))F_x/\log x$, proving almost-sure halting and the mean
candidate count. Exact comparison with the rational threshold has a
cost depending additionally on its encoding length.
\end{proof}

\KeepHeadingSpace{5\baselineskip}
\paragraph*{Arbitrary fixed transform length}
\phantomsection\label{par:gencert-fixed-length}
Fix any integer $N\ge1$, put
\[
 A_N=A_{2N,1}=\frac12\prod_{\substack{\ell\mid N\\\ell>2}}
                    (1-1/\ell),\qquad
 F_N(\beta)=1-G_{2N,1}(1/\beta),
\]
and fix a rational $\beta>A_N^{-1}$. The algorithm of
\Cref{def:gencert-search}, with either cutoff above and $2N\le x$,
retains its soundness and polynomial verification and per-iteration
cost bounds. It halts almost surely for all sufficiently large $x$,
and, as $x\to\infty$,
\[
 \mathbb E[\#\textup{sampled candidates}]
 \sim\frac{\varphi(2N)\log x}{2N F_N(\beta)}.
\]
Its output is uniform on the accepted set. For every fixed $M>0$, its
total variation distance from the uniform law on prime candidates
satisfying $1/c(q)\le\beta$ is
\[
 O_{N,\beta,M}((\log\log\log x)^{-M}).
\]
The expected bit cost is at most $T_{N,\beta}(x)$ times the expected
number of candidates.

\begin{proof}
An element $\zeta\in\mathbb F_q^\times$ of order $2N$ gives the $N$
distinct roots $\zeta^{2j+1}$ of $X^N+1$, $0\le j<N$.
The verification and iteration bounds in \Cref{thm:gencert} use only
$2N\le x$.

By \Cref{cor:ap-sharp-convergence}, the window CDF differs from
$G_{2N,1}$ by $O_N(1/\log x)$. Apply
\Cref{lem:binomial-uniform-deletion} with $k=1$, $A=A_N$,
$E=\{\ell>2:\ell\mid N\}$ and rational cut $t=1/\beta$, as in the
right-window argument of \Cref{cor:gencert-polylog}.
The certificate sandwich gives accepted fraction
$F_N(\beta)+O_{N,\beta,M}((\log\log\log x)^{-M})$ for either cutoff
and every fixed $M>0$. The support $[0,A_N]$ and atomlessness give
$F_N(\beta)>0$.

Siegel--Walfisz gives prime density
$(2N/\varphi(2N)+o(1))/\log x$ among the candidates. Independent trials
and the nested-set argument of \Cref{cor:gencert-polylog} give halting,
the expected candidate count and the output-distribution bound.
Multiplying the expected candidate count by the conditional
per-iteration bound $T_{N,\beta}(x)$ gives the stated bit cost.
\end{proof}

\section{Random prime bases for residue number systems}
\label{sec:rns-prime-bases}

For a residue number system (RNS), several distinct prime moduli contribute
to one simultaneous determinant condition. For fixed distinct primes
$q_1,\ldots,q_k$, the Chinese remainder theorem shows that a uniform matrix in
$M_d(\Z/(q_1\cdots q_k)\Z)$ has determinant a primitive root modulo every
$q_i$ with probability $\prod_{i=1}^k c_d(q_i)\to\prod_{i=1}^k c(q_i)$ as $d\to\infty$.
This identity motivates studying products of reciprocal densities as the
primes vary. Taking logarithms of these reciprocal products turns
multiplication into addition. The finite-dimensional products, including
deterministic varying dimensions, satisfy the same sampling conclusions
by \Cref{cor:rns-varying-dimension}.
The idealized independent reciprocal-density model uses independent
copies $X_1,X_2,\ldots$ of $X$; put
\[
 Z_i=-\log X_i,\qquad S_k=\sum_{i=1}^kZ_i,
 \qquad \Lambda_k=e^{S_k}=\prod_{i=1}^kX_i^{-1},
\]
\[
 K(t)=\log M(-t),\qquad I(a)=\sup_{t\in\R}\{ta-K(t)\}.
\]
\Cref{thm:finite-population-clt,cor:finite-population-centered,prop:finite-population-moments} give
the general sampling statements; their arithmetic specializations
\Cref{thm:rns-arithmetic-clt,prop:rns-finite-population} relate this model
to uniform sampling of distinct primes. Sampling without replacement
introduces dependence: the normal approximation acquires a finite-population
variance factor, and at a positive sampling fraction the large-deviation
rate changes. The sampling scales below keep these regimes separate.

\paragraph*{Sampling scales}\phantomsection\label{par:rns-sampling-scales}
For sampling without replacement, write $n=\pi(x;Q,b)$ and $f_n=k/n$.
For $1\le k\le n$, let $q_1,\ldots,q_k$ be sampled uniformly without
replacement from the primes $p\le x$ with $p\equiv b\pmod Q$, and put
$S_{x,Q,b,k}:=\sum_{i=1}^k-\log c(q_i)$.
The population mean and variance of $-\log c(p)$ are $m_x,v_x$; the array notation
below calls them $m_n,v_n$. The arithmetic input has two parts:
the CDF estimates in \Cref{prop:ap-kolmogorov-pow2,thm:ap-grh-law}
give weak convergence, while \Cref{thm:prime-mellin} and
\eqref{eq:ap-grh-mellin} give Mellin and derivative control.
\Cref{lem:prime-populations} combines these inputs. The growing-modulus sampling results
use the power-of-two cases of these Mellin estimates, for which
$M_{Q,b}=M$. Both inputs hold throughout the unconditional range below;
the moment estimates use the finite-head total-variation bound
\eqref{eq:ap-truncated-tv} in the range of
\Cref{cor:ap-finite-head-ranges}\textup{(iii)}.

The principal unconditional arithmetic populations below have
\[
 Q=2^a\le x^{1/3}\exp\{-(\log\log x)^3\},
\]
with integer $a\ge0$ and reduced $b\pmod Q$; fixed reduced progressions are treated in
\Cref{cor:rns-fixed-progression}, and the later general clauses apply
to families with the stated properties.
Where explicitly stated, GRH permits $Q=2^a\le x^{1/2-\delta}$
for fixed $0<\delta<1/2$. Unconditional prime-averaged constants
and thresholds retain the possible ineffectivity in the cited results.

\begin{itemize}
\item \emph{Normal approximation.}
\Cref{thm:rns-arithmetic-clt} centers at $km_x$ and uses the exact
variance $k(n-k)v_x/(n-1)$. Its empirical central limit holds whenever
$\min\{k,n-k\}\to\infty$, including samples near the whole population.
Centering at $k\mu$, where $\mu=\mathbb E(-\log X)=-M'(0)$, requires
the additional error control in \eqref{eq:rns-arithmetic-be}.
The corollary gives the sufficient range $k\to\infty$,
$k\le(\log x)^D$ for fixed $D>0$.
Under GRH, \Cref{cor:rns-grh-clt} instead gives the sufficient range
$k\to\infty$, $k\le x^\eta$ with
$0<\eta<\min\{2\delta,1/2+\delta\}$.
\item \emph{Relative product moments.}
For fixed real $s$, \Cref{prop:rns-finite-population} compares the
$s$th moment of the sampled product $\prod_{i=1}^k c(q_i)^{-1}$
with $\bigl(n^{-1}\sum_{p\le x,\ p\equiv b\pmod Q}c(p)^{-s}\bigr)^k$
when $k^2/n=o(1)$, where $q_1,\ldots,q_k$ are sampled uniformly without
replacement from the population.
Replacing this factor by $M(-s)^k$ uses the separately stated
polylogarithmic sample range in \eqref{eq:rns-polylog-moments}.
These arithmetic assertions use the unconditional populations of that
corollary. The generic \Cref{prop:finite-population-moments} gives the
first comparison for every family with a limiting loss law in the
strong sense defined below.
\item \emph{Dilute large deviations.}
For $k\to\infty$ and $k/n\to0$, \Cref{thm:rns-arithmetic-ldp}
gives the independent-model rate $I$, throughout the permitted
unconditional or GRH populations. This larger sampling range follows
from saddle estimates and sampling pressure
(\Cref{lem:population-saddle,thm:sampling-pressure}).
\item \emph{A positive sampling fraction.}
When $f_n\to f\in(0,1)$, \Cref{thm:positive-fraction-ldp}
specializes the classical finite-population large-deviation principle
\cite{DZ96,Tra02}, giving the logistic pressure and rate $I_f$. Its domain is
$[a_f^-,a_f^+]$, where $a_f^-$ and $a_f^+$ are the means of the lowest
and highest fractions $f$ of the limiting law of $-\log X$, respectively.
The displayed tail identities are for interior thresholds.
\item \emph{Near-complete sampling.}
If $\ell=n-k\to\infty$ and $\ell/n\to0$,
\hyperref[par:complement-sampling]{the complement identity} gives speed $\ell$ and rate
$z\mapsto I(\mu-z)$ for $(S_{x,Q,b,k}-km_x)/\ell$.
At $k=n$ the sum is deterministic.
\item \emph{Moderate deviations and local probabilities.}
For the same populations with $k\to\infty$ and $f_n\to f\in[0,1)$,
\Cref{thm:finite-population-mdp,thm:finite-population-interval-llt}
center at $km_n$ (their joint prime application is \Cref{thm:mdp-llt}):
moderate deviations
use $a_n\to\infty$, $a_n=o(\sqrt k)$, and the interval local limit
uses each fixed width $h>0$ and the exact finite-population variance.
For $f>0$, the full moderate-deviation range follows also from classical estimates
by the truncation argument preceding \Cref{thm:finite-population-mdp}.
Replacing $m_n$ by $\mu$ requires, respectively,
$\sqrt k\,|m_n-\mu|/a_n\to0$ and
$\sqrt k\,|m_n-\mu|\to0$.
The local-limit proof uses the mixed count-and-sum Fourier estimate
and interval support.
\end{itemize}

\Cref{fig:sampling-regimes} places the normal approximation and the
three large-deviation regimes along the sampling fraction $f_n$, with the
relative product-moment comparison at the dilute end, below the two
arithmetic inputs that \Cref{lem:prime-populations} combines; moderate
deviations and local limits (\Cref{thm:mdp-llt}) are not displayed.

% Diagram (design brief F11): Sampling distinct primes: different fractions, different limits.
\begin{figure}[t]
\centering
\begin{tikzpicture}[figdiagram,
  % local styles:
  %   fend    detached endpoint marker (a square, unlike the junction dot)
  %   fregime the three large-deviation boxes: equal size, hung from a tick
  %   fbar    a schematic range bar in the dilute inset
  fend/.style={rectangle, fill=figink, draw=none, inner sep=0pt,
    minimum size=3.6pt, outer sep=0pt},
  fregime/.style={fres, anchor=north, minimum width=4.1cm, minimum height=1.9cm},
  fbar/.style={draw=figink, line width=1.6pt, line cap=butt}]
% Grid (cm): x from 0 to 13.5; the frame contents run from x = 0.2 to 13.3.
% Regime columns (axis ticks = box centres): dilute, positive fraction, near-complete.
\def\xd{2.25}\def\xf{6.75}\def\xn{11.25}
\def\ya{-4.97}   % the axis
\def\yb{-5.22}   % common top of the three regime boxes
\def\yc{-7.60}   % horizontal run of the complement map
\def\yt{-8.06}   % top of the dilute band
% ---- tier 1: the two arithmetic inputs, joined, and the population lemma ----
\node[fres, anchor=north west, minimum width=5.9cm, minimum height=1.14cm] (A) at (0,0)
  {CDF estimates give\\ weak convergence \hyperref[par:rns-limiting-law]{\textup{(C2)}}\\
   \figref{\Cref{prop:ap-kolmogorov-pow2}; under GRH, \Cref{thm:ap-grh-law}}};
\node[fres, anchor=north east, minimum width=5.9cm, minimum height=1.14cm] (B) at (13.5,0)
  {Mellin and derivative estimates give\\ exponential-moment convergence \hyperref[par:rns-limiting-law]{\textup{(C3)}}\\
   \figref{\Cref{thm:prime-mellin}; under GRH, \eqref{eq:ap-grh-mellin}}};
\node[fjoin] (j) at ($(A.east)!0.5!(B.west)$) {};
\draw[ffeed] (A.east) -- (j);
\draw[ffeed] (B.west) -- (j);
\node[fres, anchor=north, minimum width=13.5cm] (L) at (\xf,-1.29)
  {The power-of-two populations have limiting loss law $\nu=\operatorname{Law}(-\log X)$\\
   \figref{\Cref{lem:prime-populations}}{\figsmall: $Q=2^a\le x^{1/3}\exp\{-(\log\log x)^3\}$,
   or, under GRH, $Q=2^a\le x^{1/2-\delta}$ with fixed $0<\delta<1/2$}};
\draw[fuse] (j) -- (L.north);
% ---- tier 2: the regime frame -----------------------------------------------
% normal-approximation bar: spans the axis exactly (from the dilute tick to the
% near-complete tick) and stops before the detached endpoint k = n
\node[fres, anchor=north, minimum width=\xn cm-\xd cm] (clt) at (\xf,-2.83)
  {Normal approximation: $(S-km_x)\big/\sqrt{k(n-k)v_x/(n-1)}$\\
   tends in law to $\mathcal N(0,1)$ whenever $\min\{k,n-k\}\to\infty$\\
   \figref{\Cref{cor:finite-population-centered}; \Cref{thm:rns-arithmetic-clt} via \Cref{thm:prime-mellin}}};
\node[fnote, anchor=north west] (cn) at ($(clt.south -| 0.2,0)+(0,-0.07)$)
  {Centering at $k\mu_\nu$ instead of $km_x$ requires the additional error control in
   \eqref{eq:finite-population-centered-be} and \eqref{eq:rns-arithmetic-be}.};
% schematic axis of the sampling fraction f_n = k/n; no numerals
\draw[frule, line width=0.6pt] (\xd,\ya) -- (\xn,\ya);
\foreach \x in {\xd,\xf,\xn} {\draw[frule, line width=0.6pt] (\x,\ya) ++(0,-2pt) -- ++(0,4pt);}
\node[fhyp, anchor=east] (fn) at (\xd-0.2,\ya+0.05) {$f_n=k/n$};
% regime hypotheses: end labels aligned inward on the end ticks, middle one centred
\node[fhyp, anchor=south west] (a0) at (\xd-0.1,\ya+0.07) {$k\to\infty$, $k/n\to0$};
\node[fhyp, anchor=south] (af) at (\xf,\ya+0.07) {$k/n\to f\in(0,1)$};
\node[fhyp, anchor=south east] (a1) at (\xn+0.1,\ya+0.07) {$\ell=n-k\to\infty$, $\ell/n\to0$};
% detached endpoint k = n: a square off the end of the axis, its note beside the bar
\node[fnote, anchor=east, align=center] (knl) at (13.3,0 |- clt.center)
  {$k=n$:\\ the sum is\\ deterministic};
\node[fend] (kn) at (knl.south |- 0,\ya) {};
\draw[frule] (knl.south) -- (kn);
% three large-deviation regimes, each centred under its tick
\node[fregime] (D) at (\xd,\yb)
  {Dilute large deviations\\ $S/k$: speed $k$, rate $I_\nu$\\ of the independent model\\
   \figref{\Cref{thm:rns-arithmetic-ldp}}};
\node[fregime] (F) at (\xf,\yb)
  {Positive sampling fraction\\ $S/k$: speed $k$, rate $I_f=\Lambda_f^{*}$,\\
   logistic pressure $\Lambda_f$ \eqref{eq:logistic-logmgf};\\ $I_f=\infty$ outside $[a_f^-,a_f^+]$\\
   \figref{\Cref{thm:positive-fraction-ldp}}};
\node[fregime] (N) at (\xn,\yb)
  {Near-complete sampling\\ $(S-km_x)/\ell$: speed $\ell$,\\
   rate $z\mapsto I_\nu(\mu_\nu-z)$\\
   \figref{\hyperref[par:complement-sampling]{complement identity} (p.~\pageref{par:complement-sampling})}};
\foreach \x/\b in {\xd/D,\xf/F,\xn/N} {\draw[frule] (\x,\ya) -- (\b.north);}
% complement map D -> N, routed below the three boxes
\draw[fmap] (D.south) -- (\xd,\yc) -- (\xn,\yc) -- (N.south);
\node[flab] (cl) at (\xf,\yc) {complement identity: {\upshape the omitted sample is dilute}};
% ---- tier 3: the dilute end in detail ----------------------------------------
\node[fres, anchor=north west] (PM) at (0.35,\yt-0.2)
  {Relative product moments, for fixed real $s$:\\
   $\mathbb E\,e^{sS}=\mathfrak m_x(s)^k\{1+O_s(k^2/n)\}$ when $k^2/n=o(1)$\\
   \figref{\Cref{prop:finite-population-moments}; \Cref{prop:rns-finite-population} via \Cref{thm:prime-mellin}}};
% nested ranges on a schematic k-scale: the short bar ends before sqrt n,
% the long bar before n (orders of magnitude, not positions)
\def\xs{9.85}\def\xr{10.95}\def\xq{12.4}\def\xe{12.9}
\def\yra{\yt-0.33}\def\yrb{\yt-0.71}\def\yrc{\yt-0.97}
\node[fnote, anchor=east] (b1l) at (\xs-0.15,\yra) {product moments};
\node[fnote, anchor=east] (b2l) at (\xs-0.15,\yrb) {rate $I_\nu$};
\draw[fbar] (\xs,\yra) -- (\xr-0.3,\yra);
\draw[fbar] (\xs,\yrb) -- (\xq-0.3,\yrb);
\draw[frule, ->] (\xs,\yrc) -- (\xe,\yrc);
\foreach \x in {\xr,\xq} {\draw[frule] (\x,\yrc) ++(0,-2pt) -- ++(0,4pt);}
\node[fnote, anchor=north] (t1) at (\xr,\yrc-0.06) {$\sqrt n$};
\node[fnote, anchor=north] (t2) at (\xq,\yrc-0.06) {$n$};
\node[fnote, anchor=west] (t3) at (\xe+0.02,\yrc) {$k$};
\begin{scope}[on background layer]
  \node[fband, fit={(0.2,\yt)(13.3,\yt)($(PM.south)+(0,-0.1)$)($(t1.south)+(0,-0.08)$)},
    inner sep=0pt] (band) {};
  \node[fgroup, fit={(0.2,0 |- clt.north)(13.3,0 |- band.south)($(clt.north)+(0,0.1)$)},
    inner xsep=3.5pt, inner ysep=3.2pt] (g) {};
\end{scope}
\node[ftag, anchor=west] at ($(g.north west)+(3mm,0)$) {Sampling $k$ distinct primes out of $n$};
\node[ftag, fill=none, anchor=south west] at ($(band.north west)+(4.5mm,0.6pt)$) {Dilute end in detail};
\draw[frule] (0.55,0 |- D.south) -- (0.55,0 |- band.north);
% ---- the lemma feeds every statement in the frame ----------------------------
\draw[fuse] (L.south) -- (L.south |- g.north);
\node[fhyp, anchor=west] at ($(L.south)!0.5!(L.south |- g.north)+(0.12,0)$)
  {\hyperref[par:rns-limiting-law]{\textup{(C1)--(C3)}}};
\end{tikzpicture}
\caption{The same limiting loss law $\nu$ supports different conclusions at dilute,
positive-fraction and near-complete sampling fractions $f_n=k/n$. Here $S=S_{x,Q,b,k}$ is
the sum of the sampled losses $-\log c(q_i)$, with population mean $m_x$ and variance $v_x$;
$\mathfrak m_x(s)=n^{-1}\sum_pc(p)^{-s}$; $I_\nu$ and $\mu_\nu$ are the independent-model
rate and the mean of $\nu$; $a_f^-$ and $a_f^+$ are the means of its lowest and highest
fractions $f$. The relative moment comparison is stated for $k^2/n=o(1)$, a smaller range
than $k/n\to0$. Thin grey lines attach regime boxes to the sampling scale.}
\label{fig:sampling-regimes}
\end{figure}

\paragraph*{Limiting loss laws}\phantomsection\label{par:rns-limiting-law}
The limit theorems below separate properties of the limiting loss law
from the moment convergence and population-size estimates needed for
sampling. In this section a \emph{prime population}
consists of the $n=\pi(x;Q,b)$ primes $p\le x$ in a reduced class
$b\pmod Q$, together with losses $y(p)\ge0$; unless other losses are
named, $y(p)=-\log c(p)$. Its empirical loss law is
$\nu_n=n^{-1}\sum_p\delta_{y(p)}$, and for $1\le k\le n$ its sample sum
$S_{x,Q,b,k}$ is the sum of the losses of $k$ primes drawn uniformly
without replacement from the population; for $y(p)=-\log c(p)$ this is
the sum defined in \Cref{thm:rns-arithmetic-clt}. A \emph{limiting loss
law} is a probability law $\nu$ on $[0,\infty)$ such that
\begin{enumerate}[label=\textup{(L\arabic*)},leftmargin=*]
\item $\mathfrak m_\nu(s):=\int e^{sy}\,d\nu(y)<\infty$ for every
      $s\in\R$;
\item $\nu$ is atomless;
\item $\supp\nu=[\alpha_\nu,\infty)$ for some $\alpha_\nu\ge0$.
\end{enumerate}
For such a law put
\[
\begin{aligned}
 \mu_\nu&=\int y\,d\nu(y),\qquad
 \varsigma_\nu^2=\int(y-\mu_\nu)^2d\nu(y)>0,\\
 K_\nu(s)&=\log\mathfrak m_\nu(s),\qquad
 I_\nu(a)=\sup_{s\in\R}\{sa-K_\nu(s)\}.
\end{aligned}
\]
For a limiting loss law $\nu$, a family of prime populations \emph{has
limiting loss law $\nu$} if, along every sequence of its populations
with $x\to\infty$,
\begin{enumerate}[label=\textup{(C\arabic*)},leftmargin=*]
\item $n\to\infty$;
\item $\nu_n\Rightarrow\nu$;
\item for every $T>0$ and integer $j\ge0$,
\[
 \sup_{|s|\le T}\Bigl|\int y^je^{sy}\,d\nu_n(y)
 -\mathfrak m_\nu^{(j)}(s)\Bigr|\longrightarrow0.
\]
\end{enumerate}
Along every such sequence, and for every choice of sample sizes
$1\le k<n$ with $k\to\infty$ and $k/n\to f\in[0,1)$, the hypotheses of
\Cref{lem:population-saddle} hold with $\mathfrak m=\mathfrak m_\nu$.
Requiring \textup{(C1)}--\textup{(C3)} along every sequence is the same
as requiring them uniformly over the family as $x\to\infty$: the
population sizes tend to infinity, and the Kolmogorov distance between
$\nu_n$ and $\nu$ (which metrizes \textup{(C2)}, because $\nu$ is
atomless) and the suprema in \textup{(C3)} tend to zero, uniformly over
the populations of the family at $x$. This is how uniformity in $Q$,
$b$ and in the choice of losses is carried below.
\Cref{lem:prime-populations} shows that the power-of-two populations
have limiting loss law $\operatorname{Law}(-\log X)$;
\Cref{cor:rns-fixed-progression,cor:rns-varying-dimension} treat fixed
reduced progressions and finite matrix dimensions.

\subsection{Normal approximation and finite-population moments}
\label{ssec:rns-normal-moments}

\begin{proposition}[Independent reciprocal-density model]\label{prop:rns-loss}
The function $K$ is finite and strictly convex on $\R$, with
$K'(\R)=(\log2,\infty)$. Put
$\mu=K'(0)=-M'(0)$ and
$\varsigma^2=K''(0)=M''(0)-M'(0)^2>0$.
For every real $s$, $\mathbb E\Lambda_k^s=M(-s)^k$; also
$\exp(\mathbb E\log\Lambda_k)=e^{k\mu}$.
Moreover,
\[
 \frac{S_k}{k}\longrightarrow\mu\quad\textup{almost surely},\qquad
 d_K\!\left(\operatorname{Law}\!\left(
 \frac{S_k-k\mu}{\varsigma\sqrt k}\right),\mathcal N(0,1)\right)
 \ll k^{-1/2},
\]
and
\begin{equation}\label{eq:rns-model-cramer}
 \begin{aligned}
 \lim_{k\to\infty}k^{-1}\log\mathbb P(S_k\ge ka)&=-I(a),&&a>\mu,\\
 \lim_{k\to\infty}k^{-1}\log\mathbb P(S_k\le ka)&=-I(a),&&\log2<a<\mu.
 \end{aligned}
\end{equation}
More generally, for every limiting loss law $\nu$ the function $K_\nu$
is finite and strictly convex on $\R$, with
$K_\nu'(\R)=(\alpha_\nu,\infty)$, $K_\nu'(0)=\mu_\nu$ and
$K_\nu''(0)=\varsigma_\nu^2$, and $I_\nu$ is finite and continuous on
$(\alpha_\nu,\infty)$, decreasing below $\mu_\nu$ and increasing above
it. The law of $-\log X$ is a limiting loss law, with
$\alpha_\nu=\log2$, $\mathfrak m_\nu(s)=M(-s)$, $K_\nu=K$, $I_\nu=I$,
$\mu_\nu=\mu$ and $\varsigma_\nu=\varsigma$.
\end{proposition}

\AddToHookNext{build/column/before}{\raggedbottom}
\begin{proof}
Let $\nu$ be a limiting loss law and let $Z_\nu$ have law $\nu$. By
\textup{(L1)}, $Z_\nu$ has all exponential moments, hence a finite
third absolute centered moment, and by \textup{(L3)},
$\supp\operatorname{Law}(Z_\nu)=[\alpha_\nu,\infty)$; in particular
$Z_\nu$ is nondegenerate, $\mathbb P(Z_\nu\le u)>0$ for every
$u>\alpha_\nu$, and $\mathbb P(Z_\nu\ge u)>0$ for every real $u$.
Under $d\mathbb P_t=e^{tZ_\nu-K_\nu(t)}d\mathbb P$,
$K_\nu''(t)=\Var_tZ_\nu>0$ and $K_\nu'(t)=\mathbb E_tZ_\nu>\alpha_\nu$;
at $t=0$ these are $\varsigma_\nu^2$ and $\mu_\nu$. For
$u>\alpha_\nu$, the bounds
$K_\nu(t)\ge tu+\log\mathbb P(Z_\nu\le u)$ $(t<0)$ and
$K_\nu(t)\ge tu+\log\mathbb P(Z_\nu\ge u)$ $(t>0)$ give
$\lim_{t\to-\infty}K_\nu(t)/t\le u\le\lim_{t\to+\infty}K_\nu(t)/t$, so
$K_\nu'(\R)=(\alpha_\nu,\infty)$. Thus $I_\nu$ is finite and continuous
on $(\alpha_\nu,\infty)$, decreasing below $\mu_\nu$ and increasing
above it. This proves the assertions for a general limiting loss law.
Now let $\nu$ be the law of $Z=-\log X$. Then \textup{(L1)} holds with
$\mathfrak m_\nu(s)=M(-s)$ by \Cref{thm:moments}, and
\Cref{thm:phi-dist}, under $u\mapsto-\log u$, gives \textup{(L2)} and
\textup{(L3)} with $\alpha_\nu=\log2$. The assertions on $K$, $I$,
$\mu$ and $\varsigma$ therefore follow from the general case, and $Z$
has a finite third absolute centered moment. Independence gives the moments
of $\Lambda_k$.
The strong law and Berry--Esseen theorem
\cite[Chapter~VII, \S~8, Theorem~1; Chapter~XVI, \S~5, Theorem~1]{Fel71}
give the almost-sure law and normal approximation. For $a\in(\log2,\infty)$ put $t_a=(K')^{-1}(a)$. By concavity of
$t\mapsto ta-K(t)$, $I(a)=t_aa-K(t_a)$. Since $K<\infty$ on $\R$ and $Z$ is
nondegenerate, Cram\'er's theorem in the form~\cite[Theorem~2.7.7, p.~108]{Dur19}
gives \eqref{eq:rns-model-cramer}: apply it
to $(Z_i)$ with $t_a>0$ when $a>\mu$, and to $(-Z_i)$ with $-t_a>0$ when
$\log2<a<\mu$.
\end{proof}

\paragraph*{Finite-population notation}
For a finite population $y_1,\ldots,y_n\ge0$, let $Y$ be one uniform
population value, and let $S_{x,k}$ be the sum of a uniform $k$-element
subset. The parameter $x$ indexes populations. Put
\[
\begin{gathered}
 m_x=\mathbb EY,\qquad v_x=\Var(Y),\qquad
 \rho_x=\mathbb E|Y-m_x|^3,\\
 \mathfrak m_x(s)=\frac1n\sum_{j=1}^n e^{sy_j},
 \qquad \Lambda_{x,k}=e^{S_{x,k}}.
\end{gathered}
\]
For $n\ge2$, write $\theta_{n,k}=(n-k)/(n-1)$.

\begin{theorem}[Finite-population normal approximation]
\label{thm:finite-population-clt}
\index{Finite-population sampling!normal approximation}
For every finite population in the preceding notation with $n\ge2$,
$v_x>0$ and $1\le k<n$,
\begin{equation}\label{eq:rns-empirical-be}
 d_K\!\left(\operatorname{Law}\!\left(
 \frac{S_{x,k}-km_x}{\sqrt{k\theta_{n,k}v_x}}\right),
 \mathcal N(0,1)\right)
 \le\frac{451\rho_x}{v_x^{3/2}\sqrt{k(1-k/n)}}.
\end{equation}
\end{theorem}

\begin{proof}
Enumerate the population values as $y_1,\ldots,y_n$, put
$z_j=y_j-m_x$, $f=k/n$, and form the deterministic array
\[
 a_{ij}=(\mathbf1_{i\le k}-f)z_j\qquad(1\le i,j\le n).
\]
Its row and column sums vanish. For a uniform permutation $\pi$,
$\sum_i a_{i\pi(i)}$ has the law of $S_{x,k}-km_x$. The variance
and absolute third-moment sums are
\[
 \sigma_{n,k}^2:=\frac1{n-1}\sum_{i,j}a_{ij}^2
       =k\theta_{n,k}v_x,
 \qquad
 \sum_{i,j}|a_{ij}|^3
       =n^2f(1-f)\{f^2+(1-f)^2\}\rho_x.
\]
For $1\le k<n$, apply the combinatorial Berry--Esseen inequality
of Chen and Fang~\cite[Theorem~1.1 and (1.5)]{CF15comb} to
$a_{ij}/\sigma_{n,k}$. It bounds the Kolmogorov distance by
\[
 \frac{451}{n\sigma_{n,k}^3}\sum_{i,j}|a_{ij}|^3
 =\frac{451\rho_x}{v_x^{3/2}\sqrt{k(1-k/n)}}
   \left(1-\frac1n\right)^{3/2}\{f^2+(1-f)^2\},
\]
which proves \eqref{eq:rns-empirical-be}.
\end{proof}

\begin{corollary}[Normal approximation with limiting parameters]
\label{cor:finite-population-centered}
Consider a family of finite populations in the preceding notation with
$n\ge2$ and a parameter $x\to\infty$. Suppose, uniformly over the
family, that for fixed $\mu_*$ and
$\varsigma_*>0$,
\[
 \varepsilon_x=|m_x-\mu_*|+|v_x-\varsigma_*^2|\longrightarrow0,
 \qquad \mathbb EY^3=O(1).
\]
There is a constant $C$, depending only on these uniform bounds and
$\mu_*,\varsigma_*$, such that, for sufficiently large $x$,
$v_x\ge1/C$, $\rho_x\le C$, the right-hand side of
\eqref{eq:rns-empirical-be} is at most $C/\sqrt{k(1-k/n)}$, and
\begin{equation}\label{eq:finite-population-centered-be}
\begin{split}
 &d_K\!\left(\operatorname{Law}\!\left(
 \frac{S_{x,k}-k\mu_*}{\varsigma_*\sqrt{k\theta_{n,k}}}\right),
 \mathcal N(0,1)\right)\\
 &\qquad\le C\left\{\frac1{\sqrt{k(1-k/n)}}
 +\left(1+\sqrt{\frac{k}{\theta_{n,k}}}\right)\varepsilon_x\right\}
 \qquad(1\le k<n),
\end{split}
\end{equation}
\begin{equation}\label{eq:finite-population-iid-scale-be}
\begin{split}
 &d_K\!\left(\operatorname{Law}\!\left(
 \frac{S_{x,k}-k\mu_*}{\varsigma_*\sqrt{k}}\right),
 \mathcal N(0,1)\right)\\
 &\qquad\le C\left\{k^{-1/2}+\frac kn+(1+\sqrt{k})\varepsilon_x\right\}
 \qquad(1\le k\le n/2).
\end{split}
\end{equation}
Thus empirical normalization converges to the standard normal law
whenever $\min\{k,n-k\}\to\infty$. Both normalizations centered at
$k\mu_*$ converge whenever $k\to\infty$, $k/n\to0$ and
$\sqrt{k}\varepsilon_x\to0$.
Every family with limiting loss law $\nu$ in the sense of
\hyperref[par:rns-limiting-law]{\textup{(C1)}--\textup{(C3)}} satisfies
these hypotheses with $(\mu_*,\varsigma_*)=(\mu_\nu,\varsigma_\nu)$.
\end{corollary}

\begin{proof}
By \Cref{thm:finite-population-clt}, the empirical bound holds.
Since $\varsigma_*^2>0$ and $Y\ge0$, one has
$v_x\gg1$ and $\rho_x\le4(\mathbb EY^3+m_x^3)=O(1)$ uniformly.
For \eqref{eq:finite-population-centered-be}, apply the positive affine map with
scale $r=\sqrt{v_x}/\varsigma_*$ and translation
$h=\sqrt{k/\theta_{n,k}}(m_x-\mu_*)/\varsigma_*$.
Kolmogorov distance is invariant under such a map, and
$d_K(\mathcal N(h,r^2),\mathcal N(0,1))\ll |h|+|r-1|$
for $r$ in a fixed neighborhood of one. This follows by the bounded
normal density for translations and by
$\sup_u|u|e^{-u^2/2}<\infty$ for changes of scale.
Since $|h|+|r-1|\ll(1+\sqrt{k/\theta_{n,k}})\,\varepsilon_x$, this
proves \eqref{eq:finite-population-centered-be}.
If $k\le n/2$, then $\theta_{n,k}\ge1/2$ and
$|\sqrt{\theta_{n,k}}-1|\ll k/n$. Using instead scale
$\sqrt{\theta_{n,k}v_x}/\varsigma_*$ and translation
$\sqrt{k}(m_x-\mu_*)/\varsigma_*$ proves \eqref{eq:finite-population-iid-scale-be}.
Since $k(1-k/n)\ge\tfrac12\min\{k,n-k\}$, the empirical limit follows
from \eqref{eq:rns-empirical-be}; the limits centered at $k\mu_*$
follow from the two bounds, because $\theta_{n,k}\to1$ when $k/n\to0$.
A family with limiting loss law $\nu$ satisfies the hypothesis with
$(\mu_*,\varsigma_*)=(\mu_\nu,\varsigma_\nu)$ by \textup{(C3)} at $s=0$
with $j=1,2,3$, which holds uniformly over the family; here
$\varsigma_\nu>0$ by \textup{(L3)}.
\end{proof}

\begin{proposition}[Finite-population moments]
\label{prop:finite-population-moments}
\index{Finite-population sampling!moments}
For every finite population in the preceding notation with $n\ge2$
and $1\le k\le n$,
\begin{equation}\label{eq:rns-population-variance}
 \mathbb ES_{x,k}=km_x,\qquad
 \Var(S_{x,k})=\frac{k(n-k)}{n-1}v_x.
\end{equation}
For $n=1$ the only sample sum is deterministic, as is $S_{x,n}$ for
every $n$. For every $n\ge1$, real $s$, and $1\le k\le n$ satisfying
$k^2\mathfrak m_x(2s)\le n\mathfrak m_x(s)^2$,
\begin{equation}\label{eq:finite-population-tilted-bound}
 \left|\frac{\mathbb E\Lambda_{x,k}^s}{\mathfrak m_x(s)^k}-1\right|
 \le\frac{k^2\mathfrak m_x(2s)}{n\mathfrak m_x(s)^2}.
\end{equation}
Consequently, for a family with $m_x\to\mu_*$ and $v_x=O(1)$
uniformly, $S_{x,k}/k\to\mu_*$ in $L^2$ whenever $k\to\infty$ and
$k\le n$. For any fixed real $s$, if
$\mathfrak m_x(2s)/\mathfrak m_x(s)^2=O_s(1)$ uniformly, then
\begin{equation}\label{eq:finite-population-relative-moments}
 \mathbb E\Lambda_{x,k}^s
 =\mathfrak m_x(s)^k(1+O_s(k^2/n))\qquad\text{when }k^2/n=o(1).
\end{equation}
In particular, every family with limiting loss law $\nu$ satisfies
these conclusions with $\mu_*=\mu_\nu$; the implied constants may
depend on the fixed law and family as well as on $s$.
\end{proposition}

\begin{proof}
For $n\ge2$, each sampled value has mean $m_x$ and variance $v_x$,
and two values with distinct sample indices have covariance
$-v_x/(n-1)$: the centered population values sum to zero.
Summing these covariances gives \eqref{eq:rns-population-variance}.
It also gives, for every real $\mu_*$,
\[
 \mathbb E|S_{x,k}/k-\mu_*|^2
 =\frac{n-k}{k(n-1)}v_x+(m_x-\mu_*)^2,
\]
which proves the $L^2$ assertion. The deterministic cases follow
directly from the definition.

Let $\mathcal D$ be the event of distinct indices in $k$ independent
draws. Give index $j$ the tilted probability
$e^{sy_j}/(n\mathfrak m_x(s))$. Summing over ordered distinct tuples
and comparing with uniform draws gives exactly
\[
 \frac{\mathbb E\Lambda_{x,k}^s}{\mathfrak m_x(s)^k}
 =\frac{\mathbb P_{\mathrm{tilt}}(\mathcal D)}
        {\mathbb P_{\mathrm{unif}}(\mathcal D)}.
\]
The union bound on colliding pairs gives
\[
 1-\mathbb P_{\mathrm{unif}}(\mathcal D)\le\frac{\binom{k}{2}}n,
 \qquad
 1-\mathbb P_{\mathrm{tilt}}(\mathcal D)
 \le\binom{k}{2}\frac{\mathfrak m_x(2s)}{n\mathfrak m_x(s)^2}.
\]
Since $\mathfrak m_x(s)^2\le\mathfrak m_x(2s)$, the hypothesis of
\eqref{eq:finite-population-tilted-bound} makes both probabilities
at least $1/2$. Their difference in absolute value is at most
$k^2\mathfrak m_x(2s)/(2n\mathfrak m_x(s)^2)$; division by the
uniform probability proves the bound, including $n=k=1$.
A uniformly bounded tilted second-moment ratio and $k^2/n=o(1)$
eventually imply its hypothesis, proving
\eqref{eq:finite-population-relative-moments}.
For a family with limiting loss law, \textup{(C3)} gives
$m_x\to\mu_\nu$, $v_x\to\varsigma_\nu^2$, and
$\mathfrak m_x(2s)/\mathfrak m_x(s)^2\to
\mathfrak m_\nu(2s)/\mathfrak m_\nu(s)^2<\infty$ uniformly.
This proves the final assertion.
\end{proof}

\begin{corollary}[Central limit theorem for random prime RNS bases]\label{thm:rns-arithmetic-clt}
\index{Residue number systems!central limit theorem}
Fix $B>0$. For $Q=2^a\le x^{1/3}\exp\{-(\log\log x)^3\}$,
$a\in\Z_{\ge0}$, and a reduced class $b\pmod Q$, put
$n=\pi(x;Q,b)$. For $1\le k\le n$, sample
$q_1,\ldots,q_k$ uniformly without replacement from these $n$ primes,
equivalently by conditioning $k$ independent uniform draws to be distinct.
Put
\[
 S_{x,Q,b,k}:=\sum_{i=1}^k\log\frac1{c(q_i)},\qquad
 \theta_{n,k}:=\frac{n-k}{n-1}\quad(n\ge2),
\]
and let $\mu=-M'(0)$ and $\varsigma>0$ be as in \Cref{prop:rns-loss}.
For one uniform population prime, write $Y=-\log c(p)$ and
\[
 m_x=\mathbb EY,\qquad v_x=\Var(Y),\qquad
 \rho_x=\mathbb E|Y-m_x|^3.
\]
For all sufficiently large $x$, uniformly in $Q,b$, one has
$v_x\gg1$ and $\rho_x\ll1$. For $1\le k<n$, \eqref{eq:rns-empirical-be} holds with
$S_{x,k}=S_{x,Q,b,k}$ and right-hand side at most
$C/\sqrt{k(1-k/n)}$. Moreover,
\begin{equation}\label{eq:rns-arithmetic-be}
\begin{split}
 &d_K\!\left(\operatorname{Law}\!\left(
 \frac{S_{x,Q,b,k}-k\mu}{\varsigma\sqrt{k\theta_{n,k}}}\right),
 \mathcal N(0,1)\right)\\
 &\qquad\le C_B\left\{
 \frac1{\sqrt{k(1-k/n)}}
 +\left(1+\sqrt{\frac{k}{\theta_{n,k}}}\right)(\log x)^{-B}
 \right\}.
\end{split}
\end{equation}
For $1\le k\le n/2$, the normalization without the finite-population
factor satisfies
\begin{equation}\label{eq:rns-iid-scale-be}
 d_K\!\left(\operatorname{Law}\!\left(
 \frac{S_{x,Q,b,k}-k\mu}{\varsigma\sqrt{k}}\right),\mathcal N(0,1)\right)
 \le C_B\left\{k^{-1/2}+\frac kn
                  +(1+\sqrt{k})(\log x)^{-B}\right\}.
\end{equation}
The implied constants and threshold may be ineffective.
In particular, the empirical normalization in
\eqref{eq:rns-empirical-be} converges to the standard normal law
whenever $\min\{k,n-k\}\to\infty$. Both normalizations centered at
$k\mu$ do so whenever $k\to\infty$ and $k\le(\log x)^D$ for a fixed
$D>0$, uniformly in the permitted populations.
\end{corollary}

\begin{proof}
For the power-of-two populations, the derivatives of orders $1,2,3$ at
zero in \Cref{thm:prime-mellin}\textup{(b)}, with $D_0=2$ and
$M_{Q,b}=M$, give, uniformly in $Q,b$,
\[
 m_x=\mu+O_B((\log x)^{-B}),\qquad
 v_x=\varsigma^2+O_B((\log x)^{-B}),\qquad
 \mathbb EY^3=O(1).
\]
Thus the hypothesis holds with $(\mu_*,\varsigma_*)=(\mu,\varsigma)$
and $\varepsilon_x\ll_B(\log x)^{-B}$, and \Cref{cor:finite-population-centered} gives
the bounds for $v_x$ and $\rho_x$ together with
\eqref{eq:rns-empirical-be}, \eqref{eq:rns-arithmetic-be} and
\eqref{eq:rns-iid-scale-be}.
The quantity $k(1-k/n)$ lies between
$\tfrac12\min\{k,n-k\}$ and $\min\{k,n-k\}$, proving the empirical
limit. Finally \eqref{eq:ap-polylog-population} gives
$n\gg x/(\varphi(Q)\log x)\gg x^{2/3}/\log x$, so $k/n\to0$ uniformly for
$k\le(\log x)^D$. Taking $B>D/2$ in the displayed bounds proves
the two limits centered at $k\mu$.
\end{proof}

\begin{corollary}[Product moments for random prime RNS bases]\label{prop:rns-finite-population}
\index{Finite-population sampling!moments}
Use the notation and the power-of-two populations of
\Cref{thm:rns-arithmetic-clt}.
The exact mean and variance in \eqref{eq:rns-population-variance}
hold with $S_{x,k}=S_{x,Q,b,k}$ for $n\ge2$, $1\le k\le n$.
Consequently $S_{x,Q,b,k}/k\to\mu$ in $L^2$ as $x\to\infty$ whenever
$k\to\infty$ and $k\le n$, uniformly in these populations.
For every fixed real $s$, put
\[
 \mathfrak m_x(s)=\frac1n\sum_{\substack{p\le x\\p\equiv b\pmod{Q}}}c(p)^{-s},\qquad
 \Lambda_{x,Q,b,k}=\prod_{i=1}^kc(q_i)^{-1}.
\]
Whenever $k^2/n=o(1)$, uniformly in these populations,
\begin{equation}\label{eq:rns-tilted-moments}
 \mathbb E\Lambda_{x,Q,b,k}^{s}
 =\mathfrak m_x(s)^k\bigl(1+O_s(k^2/n)\bigr).
\end{equation}
For every fixed $B,D>0$, uniformly for $1\le k\le(\log x)^D$,
\begin{equation}\label{eq:rns-polylog-moments}
 \mathbb E\Lambda_{x,Q,b,k}^{s}
 =M(-s)^k\bigl(1+O_{B,D,s}((\log x)^{-B})\bigr).
\end{equation}
The thresholds in $x$ for these prime-averaged estimates may be ineffective.
\end{corollary}

\begin{proof}
By \Cref{thm:prime-mellin}\textup{(b)} with $D_0=2$, $z=0$ and
$j=1$, and $M_{Q,b}=M$, one has $m_x=-M'(0)+o(1)=\mu+o(1)$
uniformly. By \Cref{thm:rns-arithmetic-clt}, $\rho_x=O(1)$ uniformly;
Lyapunov's inequality gives $v_x^{3/2}\le\rho_x$, hence $v_x=O(1)$.
Thus \Cref{prop:finite-population-moments} gives the mean,
variance and $L^2$ conclusion. By \Cref{thm:prime-mellin},
$\mathfrak m_x(s)\to M(-s)>0$ and
$\mathfrak m_x(2s)/\mathfrak m_x(s)^2=O_s(1)$ uniformly.
Thus \eqref{eq:finite-population-relative-moments} gives
\eqref{eq:rns-tilted-moments}.
If $k\le(\log x)^D$, the same theorem with $B+D$ in place of $B$ gives
$\mathfrak m_x(s)^k=M(-s)^k(1+O((\log x)^{-B}))$, while
$k^2/n\ll(\log x)^{2D+1}x^{-2/3}$ by
\eqref{eq:ap-polylog-population}. This proves
\eqref{eq:rns-polylog-moments}.
\end{proof}

\subsection{Sampling pressure and large deviations}
\label{ssec:rns-pressure-deviations}

The coefficient argument below uses the classical representation of
rejective sampling by independent Bernoulli variables conditioned on
their sum \cite{Haj64}. Its saddlepoint antecedents are Robinson
\cite{Rob77} and Hu--Robinson--Wang \cite[Theorem~3.1, (3.3)]{HRW07}:
the latter gives the corresponding moment-generating-function
approximation when $|s|\max_j|y_{nj}-m_n|$ is bounded.
The saddle estimates below establish the stated uniformity on compact
real tilt sets under convergence of the empirical exponential moments.

For $n\ge1$, let $y_{n1},\ldots,y_{nn}\ge0$ and put
\begin{align*}
 \nu_n&=\frac1n\sum_{j=1}^n\delta_{y_{nj}},
 &\mathfrak m_n(s)&=\int e^{sy}\,d\nu_n(y),\\
 m_n&=\int y\,d\nu_n,
 &v_n&=\int(y-m_n)^2d\nu_n.
\end{align*}
Let $S_n$ be the sum of a uniformly chosen $k$-element subset, and put
$f_n=k/n$ and $H(u)=-u\log u-(1-u)\log(1-u)$, continuously extended
to $[0,1]$.

\begin{lemma}[Constrained sampling coefficients]
\label{lem:population-coefficients}
For every $1\le k<n$ and $s\in\R$, there is a unique
$\lambda_n(s)\in\R$ satisfying
\[
 \frac1n\sum_j\frac{e^{\lambda_n(s)+sy_{nj}}}
                       {1+e^{\lambda_n(s)+sy_{nj}}}=f_n.
\]
Put
\begin{align*}
 p_{nj}(s)&=\frac{e^{\lambda_n(s)+sy_{nj}}}
                  {1+e^{\lambda_n(s)+sy_{nj}}},
 &V_n(s)&=\sum_jp_{nj}(s)(1-p_{nj}(s)),\\
 g_n(s)&=\frac1n\sum_j\log(1+e^{\lambda_n(s)+sy_{nj}})
                  -f_n\lambda_n(s).
\end{align*}
If $K_s$ is the sum of independent Bernoulli variables with
parameters $p_{nj}(s)$, then
\begin{equation}\label{eq:coefficient}
 \log\mathbb Ee^{sS_n}
 =n\{g_n(s)-g_n(0)\}
   +\log\frac{\mathbb P(K_s=k)}{\mathbb P(K_0=k)},
 \qquad g_n(0)=H(f_n).
\end{equation}
For every $1\le k\le n$,
\begin{equation}\label{eq:rns-symmetric-mgf-upper}
 \mathbb Ee^{sS_n}\le\mathfrak m_n(s)^k\qquad(s\in\R).
\end{equation}
\end{lemma}

\begin{proof}
The logistic mean increases continuously from zero to one as
$\lambda$ increases, proving uniqueness and existence of the saddle.

The elementary-symmetric-polynomial identity
\[
 \prod_j(1+ze^{sy_{nj}})
 =\sum_{\ell=0}^n e_\ell(e^{sy_{n1}},\ldots,e^{sy_{nn}})z^\ell
\]
gives
$\mathbb P(K_s=k)=e^{-ng_n(s)}e_k(e^{sy_{n1}},\ldots,e^{sy_{nn}})$.
At zero tilt $p_{nj}(0)=f_n$ and $g_n(0)=H(f_n)$; dividing by the
zero-tilt identity proves \eqref{eq:coefficient} exactly.
Maclaurin's inequality bounds the normalized elementary symmetric
mean by the $k$th power of the arithmetic mean, giving
\eqref{eq:rns-symmetric-mgf-upper}. This bound is also the exponential
specialization of Hoeffding's comparison between sampling without and
with replacement \cite[Theorem~4]{Hoe63}.
Indeed, with all other weights and the sum of two weights fixed,
$e_k$ is nondecreasing in their product. Replacing the largest and
smallest weights by their mean increases $e_k$; iteration converges
to the constant vector, and continuity proves the inequality.
\end{proof}

\begin{lemma}[Sampling saddle estimates]
\label{lem:population-saddle}
Use the preceding notation. Suppose $\nu_n\Rightarrow\nu$, where
$\nu$ is nondegenerate, and
$\mathfrak m_n(s)\to\mathfrak m(s):=\int e^{sy}d\nu(y)<\infty$
for every $s\in\R$. Let $1\le k<n$, $k\to\infty$, and
$f_n\to f\in[0,1)$. Uniformly on compact real tilt sets,
\begin{equation}\label{eq:count-local}
 V_n(s)\asymp k,\qquad
 \mathbb P(K_s=k)=\frac{1+O(k^{-1/2})}{\sqrt{2\pi V_n(s)}}.
\end{equation}
\begin{enumerate}[label=\textup{(\alph*)}]
\item \emph{Positive sampling fraction.}
If $f>0$, then $V_n(s)\asymp n$ and $\lambda_n(s)$ is bounded
on each compact real tilt set.
\item \emph{Vanishing sampling fraction.}
If $f=0$, then, uniformly on compact real tilt sets,
\begin{equation}\label{eq:dilute-saddle}
 \begin{aligned}
 e^{\lambda_n(s)}&=\frac{f_n}{\mathfrak m_n(s)}(1+O(f_n)),\\
 V_n(s)&=k(1+O(f_n)),\\
 g_n(s)-H(f_n)&=f_n\log\mathfrak m_n(s)+O(f_n^2).
 \end{aligned}
\end{equation}
\end{enumerate}
\end{lemma}

\begin{proof}
For each fixed $T>0$ and nonnegative integer $r$, a bounded
$(T+1)$-exponential moment controls the tails of $y^r e^{Ty}$:
for $y>R$ this integrand is at most
$C_r e^{-R/2}e^{(T+1)y}$. Truncation and equicontinuity therefore
give locally uniform convergence of $\mathfrak m_n$ and all its
fixed-order derivatives. In particular $\mathfrak m_n(s)$ is bounded
above and away from zero on compact tilt sets, and all the required
polynomial moments are uniformly integrable.

\textup{(a)} Fix a compact tilt set. If $f>0$, choose a bounded interval whose
$\nu_n$-mass exceeds $\max(f_n,1-f_n)$ by a fixed positive margin.
The weights $e^{sy}$ on it are bounded above and away from zero.
Its contribution to the logistic constraint bounds $\lambda_n(s)$
above; bounding the complementary contribution by its mass gives
a lower bound. On this interval $p_{nj}(s)$ is consequently bounded
away from zero and one. Thus $V_n(s)\asymp n\asymp k$ uniformly.

\textup{(b)} If $f=0$, write $z_n=e^{\lambda_n(s)}$. A bounded interval with
mass at least $\alpha>0$ and weights at least $c>0$ gives
$f_n\ge\alpha z_nc/(1+z_nc)$, hence $z_n=O(f_n)$.
For every $u\ge0$,
\[
 0\le u-\frac{u}{1+u}\le u^2,\qquad
 0\le u-\log(1+u)\le u^2/2.
\]
After averaging these inequalities, one obtains
\[
 f_n=z_n\mathfrak m_n(s)+O(z_n^2\mathfrak m_n(2s)),\qquad
 \sum_jp_{nj}(s)^2\le nz_n^2\mathfrak m_n(2s)=O(nf_n^2).
\]
The first equality gives the asserted formula for $z_n$; the second
gives $V_n=k-\sum_jp_{nj}^2=k(1+O(f_n))$.
Also
\[
 g_n(s)=f_n-f_n\log f_n+f_n\log\mathfrak m_n(s)+O(f_n^2).
\]
Since $H(f_n)=f_n-f_n\log f_n+O(f_n^2)$, this proves
\eqref{eq:dilute-saddle}. All estimates are uniform on the
chosen compact tilt set.

For the count estimate, put $q=1-p$ and fix $0<u_0<\pi$.
The continuous logarithm
$h_p(u)=\log(q+pe^{iu})-ipu$, with $h_p(0)=0$, satisfies
\[
 h_p'''(u)=-\frac{ipq e^{iu}(q-pe^{iu})}{(q+pe^{iu})^3},\qquad
 |q+pe^{iu}|\ge\cos(u_0/2)\quad(|u|\le u_0).
\]
Thus $|h_p'''(u)|\le C_{u_0}p(1-p)$ uniformly even when $p$ is
near zero or one. Since the count mean is exactly $k$, summation gives
\[
 \log\mathbb Ee^{iu(K_s-k)}
 =-V_n(s)u^2/2+O(V_n(s)|u|^3).
\]
Moreover, for $|u|\le\pi$,
\[
 \left|\mathbb Ee^{iu(K_s-k)}\right|
 \le\exp\{-2V_n(s)\sin^2(u/2)\}.
\]
In Fourier inversion put $u=w/\sqrt{V_n(s)}$; this bound is at most
$e^{-2w^2/\pi^2}$ throughout the rescaled interval. On $|u|\le u_0$,
the difference from $e^{-V_n(s)u^2/2}$ is bounded by
$C V_n(s)|u|^3e^{-cV_n(s)u^2}$, using the logarithmic expansion
and integrating the exponential along the segment between its two
exponents. Its integral is $O(V_n(s)^{-1})$; the remaining tails
are exponentially small. Fourier inversion therefore gives
$(2\pi V_n(s))^{-1/2}(1+O(V_n(s)^{-1/2}))$, uniformly whenever
$V_n(s)\to\infty$. This proves \eqref{eq:count-local}.
\end{proof}

For a positive sampling fraction, the large-deviation principle below
is a consequence of the empirical-measure results of Dembo--Zeitouni
\cite[Theorem~1.1 and Remarks~(3)]{DZ96} and Trashorras
\cite[Theorem~1]{Tra02}; the logistic pressure is the dual of their
entropy rate. Their bounded-observable conclusions extend to the
present nonnegative populations by truncation: replacing $y$ by
$y\wedge L$ changes every sample average by at most
\[
 f_n^{-1}\int(y-L)_+\,d\nu_n(y),
\]
which tends uniformly to zero as $L\to\infty$ by uniform integrability.
The direct coefficient proof below also covers the dilute regime.

\begin{theorem}[Sampling pressure and large deviations]
\label{thm:sampling-pressure}
\index{Finite-population sampling!pressure}
Under the hypotheses and notation of
\Cref{lem:population-saddle}, define
$\Lambda_0(s)=\log\mathfrak m(s)$. For $0<f<1$, let
$\lambda_f(s)$ be the unique solution of
\[
 \int p_s(y)d\nu(y)=f,\qquad
 p_s(y)=\frac{e^{\lambda_f(s)+sy}}{1+e^{\lambda_f(s)+sy}},
\]
and set
\begin{equation}\label{eq:logistic-logmgf}
 \Lambda_f(s)=\frac1f\left\{
 \int\log(1+e^{\lambda_f(s)+sy})d\nu(y)
 -f\lambda_f(s)-H(f)\right\}.
\end{equation}
Then, locally uniformly for real $s$,
\[
 \frac1k\log\mathbb Ee^{sS_n}\longrightarrow\Lambda_f(s).
\]
Each $\Lambda_f$ is finite, smooth and strictly convex on $\R$.
The variables $S_n/k$ satisfy the full large-deviation principle
with speed $k$ and good rate
\[
 I_f(a)=\sup_{s\in\R}\{sa-\Lambda_f(s)\}.
\]
For the fixed limiting law $\nu$, $\Lambda_f\to\Lambda_0$ locally
uniformly as $f\downarrow0$, and for every $0\le f<1$,
\[
 \Lambda_f'(0)=\int y\,d\nu,
 \qquad \Lambda_f''(0)=(1-f)\Var_\nu(Y).
\]
\end{theorem}

\begin{proof}
If $f>0$, the compact saddle bounds in
\Cref{lem:population-saddle}, weak convergence and uniqueness
imply $\lambda_n(s)\to\lambda_f(s)$ locally uniformly. Indeed,
any sequence of tilts in a compact set has a convergent subsequence;
boundedness of its saddles gives a further subsequence, and the
limiting logistic constraint identifies the limit uniquely.
Uniform first-moment integrability controls the logarithmic integrand,
which grows at most linearly in $y$ on these compact parameter sets.
Consequently $[g_n(s)-H(f_n)]/f_n\to\Lambda_f(s)$ locally uniformly.
For $f=0$, \eqref{eq:dilute-saddle} gives the same assertion
with limit $\log\mathfrak m(s)$. In both cases
\eqref{eq:count-local} gives, uniformly on compact tilt sets,
\[
 \log\frac{\mathbb P(K_s=k)}{\mathbb P(K_0=k)}
 =-\frac12\log\frac{V_n(s)}{V_n(0)}+O(k^{-1/2})=O(1).
\]
Here $V_n(0)=k(1-f_n)$, and for $f=0$ the last bound improves to
$O(f_n+k^{-1/2})$. Dividing \eqref{eq:coefficient} by $k$
proves the pressure limit.
Repeating the averaged inequalities used for
\eqref{eq:dilute-saddle}, now with the fixed law $\nu$ and
$f\downarrow0$, gives
$\Lambda_f(s)=\log\mathfrak m(s)+O_T(f)$ for $|s|\le T$.
This proves continuity at zero fraction.

For $f>0$, the implicit-function theorem and exponential moment
bounds justify differentiation of the constraint and pressure to
every order. With $w_s=p_s(1-p_s)$ and
$\beta_s=\int yw_s\,d\nu/\int w_s\,d\nu$, they give
\begin{align*}
 \Lambda_f'(s)&=\frac1f\int yp_s(y)d\nu(y),\\
 \Lambda_f''(s)&=\frac1f\int w_s(y)(y-\beta_s)^2d\nu(y)>0.
\end{align*}
The last inequality follows from nondegeneracy and $w_s>0$.
At zero $p_0=f$, giving the two derivative formulas.
For $f=0$, the derivatives are the mean and variance under the
exponential tilt $e^{sy}/\mathfrak m(s)$; this also proves smoothness,
strict convexity, and the formulas at zero.

Write $D=\Lambda_f'(\R)=(a_-,a_+)$, allowing an
infinite endpoint. For $a=\Lambda_f'(s)\in D$, strict convexity gives
$I_f(a)=sa-\Lambda_f(s)$. Under the tilted measure with density
$e^{sS_n}/\mathbb Ee^{sS_n}$, the upper and lower Chernoff bounds
for $S_n/k$ outside $(a-\epsilon,a+\epsilon)$ have limiting exponents
\[
 \Lambda_f(s+t)-\Lambda_f(s)-t(a+\epsilon),\qquad
 \Lambda_f(s-t)-\Lambda_f(s)+t(a-\epsilon),
\]
respectively. Both are negative for sufficiently small $t>0$.
Thus the tilted laws concentrate exponentially at $a$. Reversing
the tilt on $|S_n/k-a|<\epsilon$ gives
\[
 \liminf\frac1k\log\mathbb P\{|S_n/k-a|<\epsilon\}
 \ge-I_f(a)-|s|\epsilon.
\]
Convexity gives $\Lambda_f(s)/s\to a_\pm$ as $s\to\pm\infty$,
so $I_f$ is infinite outside $\overline D$. If a finite endpoint
has finite rate, its convex combinations with any fixed point of
$D$ approach it with convergent rates: convexity gives the upper
limit and lower semicontinuity gives the lower limit. Every open set
containing that endpoint therefore inherits the lower bound from
$D$. An endpoint with infinite rate imposes no additional lower-bound
obligation. Taking shrinking neighborhoods inside an arbitrary open
set proves the full open-set lower bound.

For a compact set $C$ and $L<\inf_C I_f$, each $a\in C$ admits a
real $s$ with $sa-\Lambda_f(s)>L$. This strict inequality persists
on a neighborhood of $a$; exponential Markov bounds on those
neighborhoods and a finite cover give the compact upper bound.
Since $S_n\ge0$, for every fixed $t>0$,
\[
 \limsup\frac1k\log\mathbb P\{S_n/k\ge R\}
 \le\Lambda_f(t)-tR.
\]
This proves exponential tightness and extends the upper bound to
every closed set. Finally $\Lambda_f(s)\le0$ for $s\le0$, so
$I_f(a)=\infty$ for $a<0$; the inequality
$I_f(a)\ge ta-\Lambda_f(t)$ for $t>0$ bounds each rate sublevel set.
Lower semicontinuity then proves goodness.
\end{proof}

\paragraph*{Complementary samples}\phantomsection\label{par:complement-sampling}
For a population of size $n$ and mean $m_n$, write $S_{n,k}$ for a
uniform $k$-subset sum. Taking complements gives
\[
 S_{n,k}\overset d=nm_n-S_{n,n-k},\qquad
 \mathbb Ee^{sS_{n,k}}=e^{snm_n}\mathbb Ee^{-sS_{n,n-k}}.
\]
Thus, with $\mu=\int y\,d\nu$ and $0<f<1$,
\[
 \begin{aligned}
 f\Lambda_f(s)&=s\mu+(1-f)\Lambda_{1-f}(-s),\\
 fI_f(a)&=(1-f)I_{1-f}\!\left(\frac{\mu-fa}{1-f}\right).
 \end{aligned}
\]
This is the complement symmetry of the finite-population rate
in~\cite{DZ96,Tra02}; the first identity follows from
\Cref{thm:sampling-pressure}, and the second by Legendre duality.
If $\ell=n-k\to\infty$ and $\ell/n\to0$, the same theorem applied to
the omitted sample shows that
$(S_{n,k}-km_n)/\ell$ satisfies the full large-deviation principle
with speed $\ell$ and rate $z\mapsto I_0(\mu-z)$. Moreover,
\[
 \frac1\ell\log\mathbb Ee^{s(S_{n,k}-km_n)}
 \longrightarrow s\mu+\Lambda_0(-s)\qquad(s\in\R).
\]
These follow from
$(S_{n,k}-km_n)/\ell\overset d=m_n-S_{n,\ell}/\ell$ and $m_n\to\mu$.
For a family of prime populations with limiting loss law $\nu$,
$\Lambda_0=K_\nu$ and $I_0=I_\nu$; for the power-of-two populations
these are $K$ and $I$.
Replacing $km_n$ by $k\mu$ requires
$k|m_n-\mu|/\ell\to0$.

\begin{lemma}[Prime population limits]
\label{lem:prime-populations}
The power-of-two populations of \Cref{thm:rns-arithmetic-clt}, with
losses $-\log c(p)$, have limiting loss law
$\nu=\operatorname{Law}(-\log X)$, for which $\alpha_\nu=\log2$ and
$\mathfrak m_\nu(s)=M(-s)$. Under GRH the same holds for
$Q=2^a\le x^{1/2-\delta}$ with fixed $0<\delta<1/2$, in every
reduced class and with integer $a\ge0$. In both cases
$m_n=m_x\to\mu$, $v_n=v_x\to\varsigma^2>0$, and the centered third
absolute moments are uniformly bounded. More precisely, for every $T>0$ and integer
$j\ge0$, uniformly in $|s|\le T$ and in the permitted populations,
\[
 \begin{aligned}
 &\int y^j e^{sy}d\nu_n(y)-(-1)^jM^{(j)}(-s)\\
 &\qquad=\begin{cases}
 O_{B,T,j}((\log x)^{-B}),&\text{unconditionally},\\
 O_{\delta,T,j}(x^{-\delta}(\log x)^{E_{T,j}}),&\text{under GRH},
 \end{cases}
 \end{aligned}
\]
for every $B>0$ in the first case and some fixed $E_{T,j}>0$ in the second.
\end{lemma}

\begin{proof}
The inequality $0<c(p)<1$ makes the losses nonnegative. The distribution estimates in
\Cref{prop:ap-kolmogorov-pow2,thm:ap-grh-law} give weak convergence;
the population sizes tend to infinity uniformly by
\eqref{eq:ap-polylog-population} and \eqref{eq:ap-grh-population}, respectively.
Apply \Cref{thm:prime-mellin}, specifically
\eqref{eq:prime-mellin-ap}, at $z=-s$, using $M_{Q,b}=M$ for powers
of two. Under GRH use \eqref{eq:ap-grh-mellin} in the same way.
These give exactly the displayed exponentially weighted derivative
estimates, throughout the two stated ranges.
\Cref{thm:moments} identifies $M(-s)=\int e^{sy}d\nu<\infty$ for
every real $s$. The moment limits at zero and
$|y-m_n|^3\le4(y^3+m_n^3)$ give the centered third-moment bound.
Finally \Cref{thm:phi-dist}, under $u\mapsto-\log u$, gives
atomlessness, the support, and the strictly positive limiting variance.
\end{proof}

\begin{corollary}[Central limit theorem for random prime RNS bases under GRH]
\label{cor:rns-grh-clt}
Assume GRH, fix $0<\delta<1/2$, and allow
$Q=2^a\le x^{1/2-\delta}$ in \Cref{thm:rns-arithmetic-clt}.
The empirical bound \eqref{eq:rns-empirical-be} holds with a constant
depending on $\delta$. For some fixed $C_0>0$, the bounds
\eqref{eq:rns-arithmetic-be} and \eqref{eq:rns-iid-scale-be} hold
with $(\log x)^{-B}$ replaced by
$x^{-\delta}(\log x)^{C_0}$ and $C_B$ by $C_\delta$.
Hence the empirical normalization converges whenever
$\min\{k,n-k\}\to\infty$. Both normalizations centered at $k\mu$
converge uniformly whenever $k\to\infty$ and $k\le x^\eta$, for a
fixed $0<\eta<\min\{2\delta,1/2+\delta\}$.
\end{corollary}

\begin{proof}
Equation~\eqref{eq:ap-grh-mellin} gives the first three moments of
$-\log c(p)$ with error
$O_\delta(x^{-\delta}(\log x)^{C_0})$ for a fixed $C_0$.
\Cref{cor:finite-population-centered}, with
$(\mu_*,\varsigma_*)=(\mu,\varsigma)$ and
$\varepsilon_x\ll_\delta x^{-\delta}(\log x)^{C_0}$, then gives all
three bounds.
Moreover \eqref{eq:ap-grh-population} implies
$n\gg_\delta x^{1/2+\delta}/\log x$.
Thus $k/n\to0$ for $\eta<1/2+\delta$, so $\theta_{n,k}\to1$;
the centering errors tend to zero for $\eta<2\delta$.
These observations prove the stated limits.
\end{proof}

\begin{corollary}[Large deviations for sampled prime RNS bases]
\label{thm:rns-arithmetic-ldp}
\index{Residue number systems!large deviations}
Consider a family of prime populations with limiting loss law $\nu$.
Along every sequence of its populations and sample sizes with
$x\to\infty$, $k\to\infty$ and $k/n\to0$, sampling without
replacement satisfies, for every fixed real $s$,
\begin{equation}\label{eq:rns-arithmetic-log-mgf}
 \frac1k\log\mathbb E\exp(sS_{x,Q,b,k})\longrightarrow K_\nu(s),
\end{equation}
and
\begin{equation}\label{eq:rns-arithmetic-cramer}
\begin{aligned}
 \lim\frac1k\log\mathbb P(S_{x,Q,b,k}\ge ka)&=-I_\nu(a),&&a>\mu_\nu,\\
 \lim\frac1k\log\mathbb P(S_{x,Q,b,k}\le ka)&=-I_\nu(a),
   &&\alpha_\nu<a<\mu_\nu.
\end{aligned}
\end{equation}
For the power-of-two populations of \Cref{lem:prime-populations},
unconditionally or under GRH, $K_\nu=K$, $I_\nu=I$, $\mu_\nu=\mu$ and
$\alpha_\nu=\log2$.
\end{corollary}

\begin{proof}
Along such a sequence the hypotheses of \Cref{lem:population-saddle}
hold with $f=0$. The case $f=0$ of \Cref{thm:sampling-pressure} gives
$\Lambda_0(s)=\log\mathfrak m_\nu(s)=K_\nu(s)$ and the full
large-deviation principle with rate $I_\nu$. By \Cref{prop:rns-loss},
$I_\nu$ is continuous on $(\alpha_\nu,\infty)$, decreasing below
$\mu_\nu$ and increasing above it.
Sandwiching each tail between its open and closed rays proves
\eqref{eq:rns-arithmetic-cramer}. 
The last assertion follows from \Cref{lem:prime-populations} and the
identifications at the end of \Cref{prop:rns-loss}.
\end{proof}

\begin{corollary}[Large deviations at a positive sampling fraction]
\label{thm:positive-fraction-ldp}
\index{Finite-population sampling!positive sampling fraction}
Consider a family of prime populations with limiting loss law $\nu$
and a sequence of its populations and sample sizes with $x\to\infty$
and $f_n=k/n\to f\in(0,1)$; write $S_n=S_{x,Q,b,k}$.
The pressure convergence and
full large-deviation principle of \Cref{thm:sampling-pressure}
hold, with $\Lambda_f$ as in \eqref{eq:logistic-logmgf} and
good rate $I_f=\Lambda_f^*$.
Writing $F_\nu^{-1}$ for the quantile function of $\nu$, put
\begin{equation}\label{eq:trimmed-support}
 a_f^- =\frac1f\int_0^f F_\nu^{-1}(u)\,du,
 \qquad
 a_f^+ =\frac1f\int_{1-f}^1 F_\nu^{-1}(u)\,du.
\end{equation}
One has $\alpha_\nu<a_f^-<\mu_\nu<a_f^+<\infty$;
$I_f$ is infinite outside $[a_f^-,a_f^+]$, strictly convex inside it,
vanishes only at $\mu_\nu$, and satisfies
\[
 I_f(a_f^-)=I_f(a_f^+)=\frac{H(f)}f,\qquad
 \Lambda_f'(0)=\mu_\nu,\qquad \Lambda_f''(0)=(1-f)\varsigma_\nu^2.
\]
In particular,
\[
 \begin{aligned}
 \lim k^{-1}\log\mathbb P(S_n\ge ka)&=-I_f(a),&&\mu_\nu<a<a_f^+,\\
 \lim k^{-1}\log\mathbb P(S_n\le ka)&=-I_f(a),&&a_f^-<a<\mu_\nu.
 \end{aligned}
\]
For the power-of-two populations of \Cref{lem:prime-populations},
unconditionally or under GRH, $\nu=\operatorname{Law}(-\log X)$,
$\alpha_\nu=\log2$, $\mu_\nu=\mu$ and $\varsigma_\nu=\varsigma$.
\end{corollary}

\begin{proof}
Along such a sequence the hypotheses of \Cref{lem:population-saddle} hold.
The pressure and large-deviation assertions follow from
\Cref{thm:sampling-pressure}, and its derivative formulas
give the identities at zero. It remains to identify the
endpoints. The pointwise identity
$\log(1+e^t)=\max_{0\le r\le1}\{tr+H(r)\}$ gives
\begin{equation}\label{eq:selection-entropy}
 f\Lambda_f(s)+H(f)
 =\sup_{\substack{0\le r\le1\\\int r\,d\nu=f}}
       \left\{s\int yr\,d\nu+\int H(r)\,d\nu\right\},
\end{equation}
with optimizer $r=p_s$. For the upper quantile cutoff
$q_+=F_\nu^{-1}(1-f)$, the upper selector $r_+=1_{\{y>q_+\}}$
has integral $f$. For any competing selector,
$\int(y-q_+)(r-r_+)d\nu\le0$, so $\int yr\,d\nu\le fa_f^+$.
The lower selector $r_-=1_{\{y<q_-\}}$, where
$q_-=F_\nu^{-1}(f)$, similarly minimizes the integral at $fa_f^-$.
Atomlessness makes these extreme selectors unique up to null sets;
both have zero entropy.

As $s\to+\infty$, the threshold $-\lambda_f(s)/s$ tends to $q_+$.
Indeed, if it exceeds $q_++\epsilon$, then
$p_s(y)\le1_{\{y>q_++\epsilon/2\}}+e^{-s\epsilon/2}$, whose
integral is eventually smaller than $f$. If it is below
$q_+-\epsilon$, the analogous lower bound has integral larger
than $f$. Both contradict the constraint, since the atomless law
has interval support. The same argument for $s\to-\infty$ gives
threshold $q_-$ and the lower selector. Thus $p_s\to r_\pm$
almost everywhere at the corresponding extremes. Dominated
convergence for $yp_s$ and $H(p_s)$ gives
\[
 \Lambda_f'(s)\longrightarrow a_f^\pm\quad(s\to\pm\infty),
 \qquad
 \Lambda_f(s)-sa_f^\pm\longrightarrow-H(f)/f.
\]
For the second limit, \eqref{eq:selection-entropy} bounds
$f\Lambda_f(s)+H(f)-sfa_f^\pm$ between zero and
$\int H(p_s)d\nu$ in the respective directions.
The selector itself also gives
$sa_f^\pm-\Lambda_f(s)\le H(f)/f$ for every real $s$.
These observations prove the effective domain and both endpoint
values of $I_f$. Positivity of $\Lambda_f''$ gives strict convexity
inside that domain; its derivative at zero gives the unique minimum
at $\mu_\nu$. Atomlessness \textup{(L2)} and interval support
\textup{(L3)} give all the strict
inequalities in \eqref{eq:trimmed-support}.
The large-deviation principle and continuity and monotonicity of
$I_f$ give the stated interior tail limits by open and closed rays.
The empirical trimmed endpoints converge to $a_f^-,a_f^+$ by
convergence of quantiles in $L^1$, following from weak convergence
and uniform integrability of the first moments. They may approach
from either side, so this convergence alone does not give fixed-threshold
tail limits at $a_f^-$ or $a_f^+$.
The last assertion follows from \Cref{lem:prime-populations} and the
identifications at the end of \Cref{prop:rns-loss}.
\end{proof}

\subsection{Moderate deviations and local limits}
\label{ssec:rns-moderate-local}

When $k/n\to f\in(0,1)$, the full moderate-deviation range below also follows from
Hu--Robinson--Wang \cite[Theorem~1.1]{HRW07} and Hoeffding's comparison
\cite[Theorem~4]{Hoe63}. For each sufficiently large fixed truncation level $L$, the former
applies to the bounded population $y_{nj}\wedge L$; its logarithmic
normal-tail error is $O_L(a_n^3/\sqrt k)=o(a_n^2)$ whenever
$a_n\to\infty$ and $a_n=o(\sqrt k)$.
To remove the truncation, put $r_L(y)=(y-L)_+$, let $S_n^{(L)}$ be the
truncated sample sum, set $m_n^{(L)}=\int(y\wedge L)d\nu_n$, and write
$D_{n,L}=S_n-S_n^{(L)}-k(m_n-m_n^{(L)})$.
For fixed $\theta\in\R$ and $t_n=\theta a_n/\sqrt k$, Hoeffding's
comparison and the exponential Taylor bound give, eventually,
\[
 \frac1{a_n^2}\log\mathbb E e^{t_nD_{n,L}}
 \le \frac{\theta^2}{2}
       \int r_L(y)^2e^{|t_n|r_L(y)}d\nu_n(y)
 \le C\theta^2e^{-L}.
\]
The last bound follows from $|t_n|\le1/2$ and the uniformly bounded
empirical exponential moment $\int e^y d\nu_n$.
Chernoff's inequality, followed by $n\to\infty$, $L\to\infty$ and
$|\theta|\to\infty$, proves that these centered truncations are
exponentially good approximations at speed $a_n^2$.
The limiting truncated population variances tend to $\varsigma_\nu^2$ as
$L\to\infty$, giving the stated quadratic rate on the entire range.
Below we give a direct proof including $f=0$, together with the fixed-width
interval local limit.

\paragraph*{Array assumptions for moderate and local limits}
\phantomsection\label{par:finite-population-mdp-llt-setup}
In the next two theorems, let $y_{n1},\ldots,y_{nn}\ge0$ be finite
populations, with empirical laws $\nu_n=n^{-1}\sum_j\delta_{y_{nj}}$,
means $m_n$ and variances $v_n$. Suppose $\nu_n\Rightarrow\nu$ for a
nondegenerate probability law on $[0,\infty)$, and that for every
$T>0$ and integer $j\ge0$,
\[
 \sup_{|s|\le T}\left|\int y^je^{sy}\,d\nu_n(y)
                    -\int y^je^{sy}\,d\nu(y)\right|\longrightarrow0,
 \qquad \int e^{Ty}\,d\nu(y)<\infty.
\]
Write $\mu_\nu=\int y\,d\nu$ and
$\varsigma_\nu^2=\int(y-\mu_\nu)^2d\nu>0$. Take $1\le k<n$ with
$k\to\infty$ and $f_n=k/n\to f\in[0,1)$, and let $S_n$ be the sum
of a uniform $k$-element subset. Its exact variance is
\[
 \sigma_n^2=\frac{k(n-k)}{n-1}v_n.
\]
These hypotheses hold along every sequence of a family with limiting
loss law in the defined strong sense \textup{(C1)}--\textup{(C3)}.

\begin{theorem}[Moderate deviations for finite populations]
\label{thm:finite-population-mdp}
\index{Finite-population sampling!moderate deviations and local limits}
Under the preceding array assumptions, for every $a_n\to\infty$
with $a_n=o(\sqrt k)$, the variables
$(S_n-km_n)/(a_n\sqrt k)$ satisfy the large-deviation principle with
speed $a_n^2$ and rate $z^2/(2(1-f)\varsigma_\nu^2)$.
The same assertion holds with $m_n$ replaced by $\mu_\nu$ if
$\sqrt k\,|m_n-\mu_\nu|/a_n\to0$.
\end{theorem}

\begin{proof}
Write $\mu=\mu_\nu$ and $\varsigma=\varsigma_\nu$. By
the array assumptions, $m_n\to\mu$, $v_n\to\varsigma^2$, and for each fixed
$s_0>0$ the weighted moments $\int y^je^{sy}\,d\nu_n(y)$,
$0\le j\le3$, $|s|\le s_0$, are bounded uniformly in $n$; these are
the moment bounds used below.
Use $p_{nj},\lambda_n,g_n,V_n,K_s$ from
\Cref{lem:population-coefficients}. The hypotheses of
\Cref{lem:population-saddle} follow from the array assumptions.
Write $w_{nj}(s)=p_{nj}(s)(1-p_{nj}(s))$ and
$\beta_n(s)=\sum_j y_{nj}w_{nj}(s)/\sum_jw_{nj}(s)$.
Implicit differentiation of the empirical constraint gives
$\lambda_n'(s)=-\beta_n(s)$ and
\begin{align*}
 g_n'(0)&=f_nm_n,&g_n''(0)&=f_n(1-f_n)v_n,\\
 g_n'''(s)&=\frac1n\sum_jw_{nj}(s)(1-2p_{nj}(s))
                    (y_{nj}-\beta_n(s))^3.
\end{align*}
Uniformly near zero, \eqref{eq:count-local} gives $V_n(s)\asymp k$.
For $f=0$, \eqref{eq:dilute-saddle} and
$w_{nj}(s)\le p_{nj}(s)\le e^{\lambda_n(s)+sy_{nj}}$ give
$\sum_jw_{nj}(s)(1+y_{nj}^3)\ll k$; for $f>0$ this follows from
$k\asymp n$ and the moment bounds. Hence $\beta_n(s)=O(1)$ and
$\sup_{|s|\le s_0}|g_n'''(s)|=O(f_n)$.
For $s_n=\theta a_n/\sqrt k$, the count correction satisfies
\[
 \log\frac{\mathbb P(K_{s_n}=k)}{\mathbb P(K_0=k)}
 =-\frac12\log\frac{V_n(s_n)}{V_n(0)}+O(k^{-1/2})=o(1).
\]
Here \eqref{eq:count-local} applies since $V_n(s)\asymp k\to\infty$,
and $|V_n'(s)|\ll k$ by the weighted moment bound and boundedness of $\beta_n$.
Thus the two logarithmic count terms cancel even when $a_n$ grows
arbitrarily slowly. Taylor's formula in
\eqref{eq:coefficient} gives, for every fixed $\theta\in\R$,
\[
 \frac1{a_n^2}\log\mathbb E
    \exp\left(\frac{\theta a_n}{\sqrt k}(S_n-km_n)\right)
 \longrightarrow \frac{\theta^2}2(1-f)\varsigma^2,
\]
since the normalized cubic error is $O(a_n/\sqrt k)$.
Put $W_n=(S_n-km_n)/(a_n\sqrt k)$, $b_n=a_n^2$, and
$q(\theta)=c\theta^2/2$, where $c=(1-f)\varsigma^2>0$. The displayed
limit is $b_n^{-1}\log\mathbb Ee^{b_n\theta W_n}\to q(\theta)$.
For every $M,\theta>0$, Chernoff bounds on $W_n\ge M$ and
$W_n\le-M$ give limiting upper exponents at most
$-\theta M+q(\theta)$ and $-\theta M+q(-\theta)$.
Letting $M\to\infty$ proves exponential tightness.

For the lower bound at any $z\in\R$, put $\theta=z/c$ and tilt by
$e^{b_n\theta W_n}/\mathbb Ee^{b_n\theta W_n}$.
For every $\eta>0$ and $t>0$, Chernoff under the tilt gives
\[
 \limsup b_n^{-1}\log\mathbb P_\theta(W_n\ge z+\eta)
 \le -t(z+\eta)+q(\theta+t)-q(\theta)
 =-t\eta+ct^2/2.
\]
The lower tail has the same bound using $-t$. Taking $0<t<2\eta/c$
shows that the tilted probability of $|W_n-z|<\eta$ tends to one
with exponentially small complement. Reversing the tilt on that
interval yields limiting lower exponent at least
$q(\theta)-\theta z-|\theta|\eta=-z^2/(2c)-|\theta|\eta$.
For an open set containing $z$, take $\eta\downarrow0$ inside it,
then take the infimum of $z^2/(2c)$ over the set.
For an upper bound on a compact set, each point $z$ has a neighborhood
on which $\theta_z w\ge\theta_z z-\epsilon$, where $\theta_z=z/c$.
Exponential Markov inequality bounds the probability of that
neighborhood with limiting upper exponent at most
$q(\theta_z)-\theta_z z+\epsilon=-z^2/(2c)+\epsilon$.
A finite subcover, then $\epsilon\downarrow0$, proves the compact-set
upper bound. Exponential tightness extends it to closed sets, giving
the full large-deviation principle.
Finally, replacing $m_n$ by $\mu_\nu$ translates $W_n$ by the
deterministic quantity $\sqrt{k}(m_n-\mu_\nu)/a_n=o(1)$.
The open/closed neighborhood bounds just proved preserve the same
continuous good rate under this translation.
\end{proof}

\begin{theorem}[Fixed-width interval local limit for finite populations]
\label{thm:finite-population-interval-llt}
Under the array assumptions in
\hyperref[par:finite-population-mdp-llt-setup]{the preceding setup},
assume additionally that $\supp\nu$ contains a nondegenerate interval.
Put $\phi(t)=(2\pi)^{-1/2}e^{-t^2/2}$.
For every fixed $h>0$,
\begin{equation}\label{eq:interval-llt}
 \sup_{z\in\R}\left|
 \sigma_n\mathbb P\{S_n-km_n\in[z,z+h]\}
       -h\phi(z/\sigma_n)\right|\longrightarrow0.
\end{equation}
The same limit holds with either or both interval endpoints omitted.
It also holds with $m_n$ replaced by $\mu_\nu$ if
$\sqrt k\,|m_n-\mu_\nu|\to0$.
\end{theorem}

\begin{proof}
Write $\mu=\mu_\nu$. The array assumptions give $v_n\to\varsigma_\nu^2>0$
and bounded centered third moments. Take independent $B_j\sim\operatorname{Bernoulli}(f_n)$
and set
\[
 z_{nj}=y_{nj}-m_n,\quad K=\sum_jB_j,\quad
 T=\sum_jB_jz_{nj},\quad
 c_n^2=nf_n(1-f_n),\quad \tau_n^2=c_n^2v_n.
\]
Conditionally on $K=k$, $T$ has the law of $S_n-km_n$, and
$c_n,\tau_n\asymp\sqrt k$.
For its mixed characteristic function
$\Psi_n(u,t)=\mathbb Ee^{i(u(K-k)+tT)}$, one has
\begin{equation}\label{eq:mixed-bound}
 |\Psi_n(u,t)|\le
 \exp\left\{-2f_n(1-f_n)\sum_j
          \sin^2((u+tz_{nj})/2)\right\}.
\end{equation}
On each fixed rectangle $[-\pi,\pi]\times[-R,R]$, the normalized
sine-square sum converges uniformly to
\[
 J(u,t)=\int\sin^2((u+t(y-\mu))/2)d\nu(y).
\]
Weak convergence gives pointwise convergence, and the bounded first
moments give equicontinuity in $(u,t)$. A zero of $J$ forces the
support of $\nu$ into $\{y:u+t(y-\mu)\in2\pi\Z\}$.
The interval contained in the support therefore leaves only $(u,t)=(0,0)$ in this
rectangle. Hence \eqref{eq:mixed-bound} is $O(e^{-ck})$
on compact sets away from the origin.

Choose a bounded interval on which the limiting truncated second
moment matrix of $(1,y-\mu)$ is positive definite, with endpoints
of zero $\nu$-mass. Weak convergence on this interval makes the
corresponding empirical matrix uniformly positive definite.
For sufficiently small fixed $|u|+|t|$, its bounded population values
then give
$\sum_j\sin^2((u+tz_{nj})/2)\ge cn(u^2+t^2)$.
This yields the envelope $e^{-ck(u^2+t^2)}$ near zero independently
of the remaining population values. On bounded $(v,w)$-sets,
\[
 \Psi_n(v/c_n,w/\tau_n)
 \longrightarrow e^{-(v^2+w^2)/2}.
\]
Indeed, if $f<1/2$, eventually $f_n\le p_*<1/2$. The centered
Bernoulli logarithm $h_p(v)=\log(1-p+pe^{iv})-ipv$ satisfies globally
\[
 \sup_v|h_p'''(v)|\le\frac{p(1-p)}{(1-2p)^3}
 \qquad(0<p\le p_*).
\]
For $f=0$ one may take $p_*=1/4$, giving the bound $8p(1-p)$.
Since $\sum_jz_{nj}=0$ and the centered third moments are bounded,
the quadratic term is $-(c_n^2u^2+\tau_n^2t^2)/2$, with remainder
$O(k(|u|+|t|)^3)$, without a restriction on the individual phases.
If $f\ge1/2$, then $k\asymp n$ and
$\max_j|z_{nj}|=O(n^{1/3})=o(\sqrt n)$; the small-phase expansion
in the proof of \Cref{lem:population-saddle} gives the same remainder.
The Gaussian envelope and the off-origin bound give, with
$D_n=[-\pi c_n,\pi c_n]\times[-R\tau_n,R\tau_n]$ of area $O(k)$,
\begin{equation}\label{eq:fourier-l1}
 \int_{\R^2}\left|
 \mathbf1_{D_n}(v,w)\Psi_n(v/c_n,w/\tau_n)
                  -e^{-(v^2+w^2)/2}\right|dv\,dw\longrightarrow0.
\end{equation}

Use the Fourier convention
$\widehat F(t)=\int_{\R}F(x)e^{-itx}dx$.
For each fixed $R>0$, the Selberg interval brackets give continuous integrable
functions $F_R^-,F_R^+$ with
\[
 F_R^-\le\mathbf1_{[0,h]}\le F_R^+,\qquad
 \supp\widehat F_R^\pm\subset[-R,R],\qquad
 \int F_R^\pm=h\pm\frac{2\pi}{R}.
\]
These are the interval majorant and minorant in
\cite[p.~2]{CEGK22}, with its bandwidth $\delta=R/(2\pi)$.
Their Fourier transforms are bounded and continuous, hence integrable
by their compact support. Pointwise Fourier inversion for the brackets,
finite-interval inversion for the integer-valued count $K$, and Fubini's
theorem (the integrands are absolutely integrable) together with
\eqref{eq:fourier-l1} therefore give, for either
bracket and uniformly in $z\in\R$,
\[
 c_n\tau_n\mathbb E\{\mathbf1_{\{K=k\}}F_R^\pm(T-z)\}
 =\frac{\int F_R^\pm}{2\pi}e^{-z^2/(2\tau_n^2)}+o(1).
\]
Indeed, the inversion integral is
$(2\pi)^{-2}\int\!\int\widehat F_R^\pm(t)e^{-itz}\Psi_n(u,t)\,du\,dt$;
after rescaling, replace the characteristic function in $L^1$ by
the Gaussian and $\widehat F_R^\pm(w/\tau_n)$ by
$\widehat F_R^\pm(0)$. The translation phase has modulus one,
so both errors are uniform in $z$.

By \eqref{eq:count-local},
$\mathbb P(K=k)\sim(\sqrt{2\pi}c_n)^{-1}$.
Division and the brackets show that the limsup of the error in
\eqref{eq:interval-llt}, with $\tau_n$ in place of $\sigma_n$,
is at most $2\pi/(R\sqrt{2\pi})$. First let $n\to\infty$ and then
$R\to\infty$. Finally
$\sigma_n^2/\tau_n^2=n/(n-1)\to1$ by
\eqref{eq:rns-population-variance}; Gaussian rescaling is uniform in
location, proving \eqref{eq:interval-llt}.
Applying its upper bound to arbitrarily short fixed intervals gives
$\sigma_n\sup_z\mathbb P(T=z\mid K=k)\to0$, proving the endpoint
assertion.
For the last centering assertion put $d_n=k(m_n-\mu_\nu)$.
The probability with limiting centering at location $z$ equals the
empirically centered probability at $z-d_n$. Since
$d_n/\sigma_n=O(\sqrt{k}|m_n-\mu_\nu|)\to0$ and $\phi$ is uniformly
continuous, its Gaussian term can be replaced uniformly by
$h\phi(z/\sigma_n)$.
\end{proof}

\begin{corollary}[Moderate deviations and local limits for prime populations]
\label{thm:mdp-llt}
\index{Residue number systems!moderate deviations and local limits}
Every family of prime populations with limiting loss law $\nu$
satisfies both \Cref{thm:finite-population-mdp} and
\Cref{thm:finite-population-interval-llt} along every population and
sample-size sequence with $x\to\infty$, $k\to\infty$ and
$k/n\to f\in[0,1)$. The centering is empirical and the variance is
$k(n-k)v_n/(n-1)$. Replacement by the limiting mean has the distinct
sufficient conditions stated in those two theorems.
For the populations of \Cref{lem:prime-populations},
$\varsigma_\nu=\varsigma$.
\end{corollary}

\begin{proof}
The definition \textup{(C1)}--\textup{(C3)} gives exactly the shared
array assumptions, along every sequence and therefore uniformly over
the family in the stated sequential sense. The support condition
\textup{(L3)} supplies the additional interval for
\Cref{thm:finite-population-interval-llt}. The last assertion is
\Cref{lem:prime-populations}.
\end{proof}

The condition $k/n\to0$ in \Cref{thm:rns-arithmetic-ldp} allows
samples much larger than $\sqrt n$. The independent-model rate
$I_\nu$ requires a restriction on the sampling fraction: at $k=n$ the
sum is deterministic, and \eqref{eq:rns-population-variance} exhibits
the finite-population correction already at positive sampling fractions.
\Cref{fig:sampling-regimes} shows the range $k/n\to0$ together with the
smaller range $k^2/n=o(1)$ of \eqref{eq:rns-tilted-moments}, the
positive-fraction regime, near-complete samples and the deterministic
case $k=n$.

\KeepHeadingSpace{5\baselineskip}
\begin{corollary}[Sampling in a fixed reduced progression]
\label{cor:rns-fixed-progression}
\index{Finite-population sampling!fixed progressions}
Fix $Q\ge1$ and $\gcd(b,Q)=1$. The law
$\nu=\nu_{Q,b}^{\log}=\operatorname{Law}(-\log X_{Q,b})$ is a limiting
loss law, with
\[
\begin{gathered}
 \alpha_\nu=-\log A_{Q,b},\qquad
 \mathfrak m_\nu(s)=M_{Q,b}(-s),\\
 \mu_\nu=-M_{Q,b}'(0),\qquad
 \varsigma_\nu^2=M_{Q,b}''(0)-M_{Q,b}'(0)^2,
\end{gathered}
\]
and the populations of primes $p\le x$, $p\equiv b\pmod Q$, with losses
$-\log c(p)$, have limiting loss law $\nu$. Moreover
$n=\pi(x;Q,b)\gg_Qx/\log x$ and, for all fixed $T,B>0$ and integers
$j\ge0$, uniformly for $|s|\le T$,
\[
 \int y^je^{sy}\,d\nu_n(y)=(-1)^jM_{Q,b}^{(j)}(-s)
   +O_{Q,b,T,B,j}((\log x)^{-B}).
\]
Consequently
\Cref{thm:rns-arithmetic-ldp,thm:positive-fraction-ldp,thm:mdp-llt}
and the \hyperref[par:complement-sampling]{complement conclusions}
apply with this $\nu$, as do
\Cref{prop:finite-population-moments} and
\Cref{cor:finite-population-centered}, the latter with
$(\mu_*,\varsigma_*)=(\mu_\nu,\varsigma_\nu)$ and
$\varepsilon_x\ll_{Q,b,B}(\log x)^{-B}$ for every fixed $B>0$.
In particular, for fixed $B,D>0$ and real $s$, uniformly for
$1\le k\le(\log x)^D$,
\[
 \mathbb E\Lambda_{x,Q,b,k}^{s}
 =M_{Q,b}(-s)^k\bigl(1+O_{Q,b,B,D,s}((\log x)^{-B})\bigr),
\]
and both normalizations centered at $k\mu_\nu$ converge to the standard
normal law whenever $k\to\infty$ and $k\le(\log x)^D$.
All constants and thresholds may depend on the fixed $Q,b$, and the
prime-averaged estimates may be ineffective.
\end{corollary}

\begin{proof}
By \Cref{thm:ap-law}, under $u\mapsto-\log u$, the law $\nu$ is
atomless with support $[-\log A_{Q,b},\infty)$ and the empirical loss
laws converge to it; by \Cref{cor:ap-moments},
$\mathfrak m_\nu(s)=\mathbb E[X_{Q,b}^{-s}]=M_{Q,b}(-s)<\infty$ for
every real $s$. Thus $\nu$ is a limiting loss law, \textup{(C2)} holds,
and differentiation at $s=0$ gives $\mu_\nu$ and $\varsigma_\nu^2$.
The losses are nonnegative because $0<c(p)<1$, and
$n\sim\operatorname{Li}(x)/\varphi(Q)$ by \Cref{thm:sw}, which gives
\textup{(C1)} and the lower bound for $n$. For each fixed $T,B>0$ and
integer $j\ge0$, \Cref{thm:prime-mellin}\textup{(a)}, applied with
$U=c$ at $z=-s$, gives the displayed estimate; this is \textup{(C3)},
at $s=0$ with $j=1,2$ it gives
$\varepsilon_x\ll_{Q,b,B}(\log x)^{-B}$, and with $j=3$ it gives
$\mathbb EY^3=O_{Q,b}(1)$. The cited results therefore apply. If $k\le(\log x)^D$, the displayed estimate with $j=0$ and
$B+D$ in place of $B$ gives
$\mathfrak m_x(s)^k=M_{Q,b}(-s)^k(1+O((\log x)^{-B}))$, while
$k^2/n\ll_Q(\log x)^{2D+1}/x$; with
\eqref{eq:finite-population-relative-moments} this
proves the moment estimate. Finally $k/n\to0$, and
$\sqrt k\,\varepsilon_x\to0$ on taking $B>D/2$, which gives the two
centered limits.
\end{proof}

\begin{corollary}[Sampling with finite and varying matrix dimensions]
\label{cor:rns-varying-dimension}
\index{Residue number systems!varying matrix dimensions}
Let $U_x(p)$ be $X(p)$ or $c_{n_x(p)}(p)$, where
$n_x(p)\in\Z_{\ge1}\cup\{\infty\}$ is a deterministic array on each
prime population. In each of the three ranges of
\Cref{lem:prime-populations,cor:rns-fixed-progression}, the populations
with losses $-\log U_x(p)$, taken over all such arrays, form a family
with the same limiting loss law as for the losses $-\log c(p)$, and the
moment estimates displayed in those two statements hold for
$y(p)=-\log U_x(p)$ uniformly over the arrays.
In each of these three population regimes, all sampling conclusions
of \Cref{sec:rns-prime-bases} for families with a limiting loss law
apply to the losses $-\log U_x(p)$ and the sampled products
$\prod_iU_x(q_i)^{-1}$, uniformly over the deterministic dimension arrays.
The general results
\Cref{thm:finite-population-clt,cor:finite-population-centered,prop:finite-population-moments}
apply as well. The quantitative bounds and sample-size ranges in
\Cref{thm:rns-arithmetic-clt,prop:rns-finite-population,cor:rns-grh-clt,cor:rns-fixed-progression}
remain valid in their respective regimes.
All empirical means, variances and centerings use the chosen losses.
\end{corollary}

\begin{proof}
The losses are nonnegative because $0<U_x(p)\le1$, and the population
sizes do not depend on the losses: they are bounded below by
\eqref{eq:ap-polylog-population}, \eqref{eq:ap-grh-population} and
\Cref{thm:sw} in the three ranges, which gives \textup{(C1)}. The CDF
estimates of \Cref{prop:ap-kolmogorov-pow2,thm:ap-grh-law} are uniform
over these arrays: \eqref{eq:ap-kolmogorov-power-range} with base $2$
(where $G_{Q,b}=G$ by \Cref{cor:ap-moments}),
\eqref{eq:ap-grh-cdf}, and \eqref{eq:ap-kolmogorov-general} for the
fixed modulus. Since the limits are continuous, this gives
\textup{(C2)} in the three ranges, uniformly over the arrays. The Mellin
derivative estimates have the same uniformity by
\Cref{cor:finite-dimension}\textup{(ii)} (items \textup{(b)} with
$D_0=2$ and \textup{(a)} of \Cref{thm:prime-mellin}) and by
\eqref{eq:ap-grh-mellin}; for $U_x=X$ use \Cref{thm:prime-mellin}
\textup{(b)} and \textup{(a)} with $U(p)=\varphi(p-1)/(p-1)$. Their limits
$(-1)^jM_{Q,b}^{(j)}(-s)$ do not depend on the choice of $U_x$, so
\textup{(C3)} holds with the same limiting loss law and the displayed
moment estimates hold with the same errors. The results for families
with a limiting loss law and the general results then apply,
with the same bounds for $\varepsilon_x$ (the moment estimates at
$s=0$) and for $n$. The two
general bounds \eqref{eq:finite-population-centered-be} and
\eqref{eq:finite-population-iid-scale-be} with these
$\varepsilon_x$ are \eqref{eq:rns-arithmetic-be},
\eqref{eq:rns-iid-scale-be} and the bounds of \Cref{cor:rns-grh-clt},
and the convergence criterion in \Cref{cor:finite-population-centered}
gives the sample-size ranges. In the
two unconditional ranges and for $k\le(\log x)^D$, the moment estimate
with $j=0$ and $B+D$ in place of $B$ gives
$\mathfrak m_x(s)^k=\mathfrak m_\nu(s)^k(1+O((\log x)^{-B}))$, while
$k^2/n$ is smaller; with
\eqref{eq:finite-population-relative-moments} this gives
\eqref{eq:rns-polylog-moments} and the moment estimate of
\Cref{cor:rns-fixed-progression}.
\end{proof}

\section{Bounds and a conditional law for primes with friable \texorpdfstring{$p-1$}{p-1}}
\label{ssec:friable-unconditional}
If every prime factor of $p-1$ is at most $y$, a finite Mertens product
bounds the reciprocal totient factor. We first combine this estimate
with the determinant correction to obtain value bounds uniform in the
matrix dimension. We then derive the conditional limiting law under
Lamzouri's weak Elliott--Halberstam hypothesis. The unconditional
distribution and sharpness questions are recorded in
\Cref{ssec:friable-open}.

For every prime $p$ and real cutoff $y\ge3$, put
\[
 \mathsf{E}_y(p)=\prod_{\substack{\ell\le y\\\ell\mid p-1}}
       \left(1-\frac1\ell\right)^{-1},\qquad
 R_y(p)=\frac{p-1}{\prod_{\ell\le y}\ell^{v_\ell(p-1)}},\qquad
 T_y(p)=\prod_{\substack{\ell>y\\\ell\mid p-1}}
       \left(1-\frac1\ell\right)^{-1}.
\]
These extend the definitions preceding \Cref{thm:factoring-free-cert};
empty products are one. Then
\[
 \frac1{c(p)}=P(p)^{-1}\mathsf{E}_y(p)T_y(p).
\]
In particular, \Cref{lem:rough-cofactor-envelope} applies with
$R=R_y(p)\le p$ and gives
\[
 T_y(p)\le V(y,R_y(p))
 \le\frac{\mathcal M(C_1(y+\log p))}{\mathcal M(y)},
\]
where $C_1$ is the absolute constant in
\eqref{eq:rough-cofactor-size-envelope}.
The same estimate as in \Cref{lem:rough-cofactor-envelope}\textup{(c)},
with real cutoff $y$, gives
\[
 1\le T_y(p)\le\exp\!\left(\frac{\log p}{(y-1)\log y}\right).
\]
Thus $T_y(p)=1+o(1)$ if $(y-1)\log y/\log p\to\infty$.
Put $\mathcal S_y=\{p\textup{ prime}:P^+(p-1)\le y\}$, with $P^+(1)=1$.
On $\mathcal S_y$ one has $T_y(p)=1$.
With $\mathcal M$ as in \Cref{lem:explicit-mertens-product}, put
$r_M(y)=\mathcal M(y)/(e^\gamma\log y)$.

\begin{proposition}[Reciprocal determinant densities for primes in $\mathcal S_y$]\label{prop:friable-value}
\index{Friable predecessors!determinant bounds}
Put $y\#=\prod_{\ell\le y}\ell$ and
\begin{equation}\label{eq:friable-E}
 E(y)=e^\gamma\max\left\{\frac{y\#}{y\#-1}r_M(y)-1,0\right\}.
\end{equation}
For every $y\ge3$, $p\in\mathcal S_y$ and $n\in\mathbb Z_{\ge1}\cup\{\infty\}$,
\begin{equation}\label{eq:friable-largep}
 2\le\frac1{c_n(p)}\le\frac1{c(p)}
 <\frac{y\#}{y\#-1}\mathcal M(y)
 \le(e^\gamma+E(y))\log y.
\end{equation}
Equality in the first inequality holds exactly at $(p,n)=(2,1)$;
$1/c_n(p)<1/c(p)$ for finite $n$.
The function $E$ is finite and tends to zero. It is computable to
arbitrary precision for exact rational cutoffs $y$, and satisfies
\[
 E(y)\le\frac2{y\#-1}+\frac{0.4}{\log^3y}
 \qquad(y\ge2\,278\,382).
\]
The determinant correction $1/(y\#-1)$ equals
$\exp(-(1+o(1))y)$.
\end{proposition}

\begin{proof}
The identity \eqref{eq:cp-nt} gives $1/c(p)=P(p)^{-1}\mathsf{E}_y(p)$ on $\mathcal S_y$.
For $p\ge3$, the product inequality used in
\Cref{lem:matrix-product-bounds} gives the strict bound
\[
 1<P(p)^{-1}<\frac{p-1}{p-2},
\]
since $P(p)>1-\sum_{j\ge1}p^{-j}=(p-2)/(p-1)$; strictness follows
from two positive factors. If $5\le p\le y+2$, a prime $q\mid p-2$
is at most $y$ and does not divide $p-1$. Its missing factor compensates
for $P(p)^{-1}$:
\begin{equation}\label{eq:friable-smallp}
 \frac1{c(p)}\le\mathcal M(y)(1-1/q)\frac{p-1}{p-2}
 \le\mathcal M(y)\left(1-\frac1{(p-2)^2}\right)<\mathcal M(y).
\end{equation}
If $p>y+2$ and some prime $q\le y$ does not divide $p-1$, then
$q<p-1$ and $(1-1/q)(p-1)/(p-2)<1$, giving the same conclusion. If every prime
$q\le y$ divides $p-1$, then $p-1\ge y\#$, so
$P(p)^{-1}<(p-1)/(p-2)\le y\#/(y\#-1)$.
For $p=2,3$, the bounds $P(2)\ge9/32$ and $P(3)>5/9$ give
$1/c(p)<18/5$. The upper bound $y\#\mathcal M(y)/(y\#-1)$ is $18/5$
for $3\le y<5$ and exceeds
$\mathcal M(y)\ge15/4$ for $y\ge5$.

For odd $p$, $\varphi(p-1)/(p-1)\le1/2$ and every determinant product
is below $1$, so $1/c_n(p)>2$. At $p=2$, the reciprocal product equals
$2$ for $n=1$ and exceeds $2$ thereafter, including $n=\infty$.
For finite $n$, its positive tail product is below $1$, giving
$c_n(p)>c(p)$.
The definition of $E$ proves the final inequality in
\eqref{eq:friable-largep}. By \Cref{lem:explicit-mertens-product},
$r_M(y)\le1+0.2/\log^3y\le1.0001$ above the stated threshold; hence
\[
 E(y)\le e^\gamma\left(\frac{1.0001}{y\#-1}
                  +\frac{0.2}{\log^3y}\right)
 \le\frac2{y\#-1}+\frac{0.4}{\log^3y}.
\]
For exact rational $y$, the finite prime products are exactly
computable, and effective approximations to $e^\gamma$ and $\log y$
give $E(y)$ to arbitrary precision. \Cref{thm:mertens} gives $E(y)\to0$,
and $\log(y\#)=\vartheta(y)\sim y$ gives the determinant correction.
\end{proof}
Direct evaluation of \eqref{eq:friable-E} gives
$E(50)<0.06186$, $E(200)<0.035568$ and $E(1000)<0.006915$.

\begin{remark}[Sharpness]\label{rem:friable-sharp}
We do not know whether the upper bound in \eqref{eq:friable-largep}
is attained in limsup, that is, whether
\[
 \limsup_{y\to\infty}\frac1{\log y}\sup_{p\in\mathcal S_y}\frac1{c(p)}=e^\gamma
\]
holds. Infinitely many primorial primes $N_k+1$ (computations:~\cite{CG02}) would imply it, since
$1/c(N_k+1)=P(N_k+1)^{-1}\mathcal M(p_k)\sim e^\gamma\log p_k$.
A sequence attaining this constant must have
$\sum_{\ell\le y,\,\ell\nmid p-1}1/\ell=o(1)$.
The proof of \Cref{thm:effective_sharp} does not establish that its
witnesses $p^*(k)$ have $p_k$-friable predecessors.
More generally, if the displayed limsup is $L>0$, then for each
$A>e^\gamma/L$ there are infinitely many primes $p$ with
$P^+(p-1)\le (\log p)^A$. Indeed, choose $y\to\infty$ and
$p\in\mathcal S_y$ with $1/c(p)\ge(L-o(1))\log y$. Then $p\to\infty$,
and the unrestricted extremal upper bound
$1/c(p)\le(e^\gamma+o(1))\log\log p$ from
\Cref{cor:rosser-schoenfeld} gives the claim. Thus if $L=e^\gamma$,
these witnesses satisfy $P^+(p-1)\le(\log p)^{1+o(1)}$.
\end{remark}

\begin{proposition}[A conditional law for friable shifted primes]
\label{prop:friable-joint-limit}
\index{Friable predecessors!conditional limiting law}
Let $\rho_{\mathrm D}$ denote the Dickman--de Bruijn function. Assume
Lamzouri's weak Elliott--Halberstam hypothesis
\cite[Conjecture~1.2]{Lam07}: for every fixed $\varepsilon>0$,
\[
 \sum_{d\le x^{1-\varepsilon}}
 \left|\pi(x;d,1)-\frac{\pi(x)}{\varphi(d)}\right|=o(\pi(x)).
\]
Then, for every fixed $u>1$ and continuous $H:[2,\infty)\to\R$
of at most polynomial growth,
\begin{equation}\label{eq:friable-joint-limit}
 \frac1{\pi(x)}
 \sum_{\substack{p\le x\\P^+(p-1)\le x^{1/u}}}H(1/c(p))
 \longrightarrow\rho_{\mathrm D}(u)\,\mathbb E H(1/X),
 \qquad X\sim\mu_G.
\end{equation}
Let $\nu_{x,u}$ be the empirical law of $1/c(p)$ conditioned on
$p\le x$ and $P^+(p-1)\le x^{1/u}$. For every fixed $1\le r<\infty$
and $s\in\R$,
\[
 \mathcal W_r(\nu_{x,u},\operatorname{Law}(1/X))\longrightarrow0,
 \qquad \int z^s\,d\nu_{x,u}(z)\longrightarrow M(-s).
\]
\end{proposition}

\begin{proof}
Fix a finite set $A$ of odd primes, put
$M_A=\prod_{\ell\in A}\ell$ and $y=x^{1/u}$, and let
$\Psi(t,y)$ count positive integers at most $t$ with no prime divisor
exceeding $y$. The allowed prime set $\{\ell\le y\}\setminus A$
has excluded reciprocal sum
$\sum_{\ell\in A}1/\ell+\sum_{y<\ell\le x}1/\ell=O_{A,u}(1)$
for all sufficiently large $x$. Thus \cite[Theorem~1.3]{Lam07}
applies. Inclusion--exclusion gives its integer count as
\[
 \sum_{d\mid M_A}\mu(d)\Psi(x/d,y)
 \sim x\rho_{\mathrm D}(u)\prod_{\ell\in A}(1-1/\ell).
\]
Lamzouri's correction at each excluded odd prime is
$1-1/(\ell-1)^2$; its product over $\ell>y$ tends to one.
Since $(1-1/\ell)(1-1/(\ell-1)^2)=(\ell-2)/(\ell-1)$, this yields
\[
 \frac{\#\{p\le x:P^+(p-1)\le y,\ (p-1,M_A)=1\}}{\pi(x)}
 \longrightarrow
 \rho_{\mathrm D}(u)\prod_{\ell\in A}\frac{\ell-2}{\ell-1}.
\]
Another inclusion--exclusion gives every finite Bernoulli pattern in
\Cref{lem:factor}. The logarithmic tail is nonnegative, and
\eqref{eq:ap-tail} with $Q=1$ gives, for each fixed cutoff $B\ge3$,
\[
 \frac1{\pi(x)}
 \sum_{\substack{p\le x\\P^+(p-1)\le y}}
 \sum_{\substack{\ell>B\\\ell\mid p-1}}\omega_\ell
 \ll B^{-1}+o(1).
\]
The model tail has the same bound. Taking first $x\to\infty$ and
then $B\to\infty$ proves logarithmic weak convergence. Since
$P(p)\to1$, the determinant correction vanishes, and exponentiation
gives \eqref{eq:friable-joint-limit} for bounded continuous $H$.
Since $\rho_{\mathrm D}(u)>0$, division by the counting marginal
$H=1$ gives the conditional weak limit. For every fixed $v>0$,
\Cref{thm:prime-mellin} bounds the unconditional averages of $c(p)^{-v}$,
and division by this positive limiting proportion gives
$\sup_{x\ge x_0(u)}\int z^v\,d\nu_{x,u}(z)<\infty$.
Truncation therefore proves \eqref{eq:friable-joint-limit} for every
continuous $H$ of polynomial growth and all the stated Mellin limits;
negative powers are bounded on $[2,\infty)$.
Couple $\nu_{x,u}$ and $\operatorname{Law}(1/X)$ by their quantiles
at one uniform variable. Weak convergence gives almost-sure convergence
of this coupling, and the uniform $(r+1)$st moments make its $r$th
power differences uniformly integrable. Their expectations tend to
zero, proving the Wasserstein assertion.
\end{proof}

\section{Endpoint counts and prime cofactors}\label{ssec:prime-cofactor-detectors}
A small positive deficit of the totient ratio from $1/2$ excludes small
odd prime factors of $p-1$. The first proof below combines this exclusion with the
size of $p-1$ to limit the number of factors, counted with multiplicity.
It then compares the totient deficit with the determinant deficit;
their distinction matters at boundary scales and for finite-range
exceptions.

For an odd prime $p$, write $p-1=2^am$ with $m$ odd, let $\Omega(m)$ be the
number of prime factors of $m$ counted with multiplicity, and put
\[
 \Delta(p):=\tfrac12-\frac{\varphi(p-1)}{p-1},\qquad
 \Delta_n(p):=\tfrac12-c_n(p)
 \quad(n\in\mathbb Z_{\ge1}\cup\{\infty\}).
\]

\Needspace{4\baselineskip}
\begin{proposition}[Prime and almost-prime odd parts of $p-1$]
\label{prop:safe-prime-obstruction}
\index{Prime cofactors!endpoint criterion}\leavevmode\par
\begin{enumerate}[label=\textup{(\roman*)}]
\item For every $x\ge2$, integer $a\ge1$, and
$0<\varepsilon\le2^{(a-3)/2}x^{-1/2}$, the primes $x<p\le2x$ with
$v_2(p-1)=a$ and $0<\Delta(p)<\varepsilon$ are exactly those with $m$
prime and
\begin{equation}\label{eq:endpoint-exact-clip}
 p>1+\frac{2^{a-1}}{\varepsilon}.
\end{equation}
\item Fix $a\ge1$, $C>0$ and $1/2<\theta<1$. For all sufficiently large
$x$, the primes $x<p\le2x$ with $v_2(p-1)=a$ and
$0<\Delta(p)<Cx^{-\theta}$ are exactly those with $m$ prime.
For $\theta=1/2$ the same holds if $C\le2^{(a-3)/2}$, or if $a$ is odd
and $C<2^{(a-1)/2}$. These identities hold with $\Delta_n$ in place of
$\Delta$ for every $n\in\mathbb Z_{\ge1}\cup\{\infty\}$, with a threshold
independent of $n$.
\item Fix $a\ge1$, $C>0$ and an integer $k\ge1$. If $\theta>1/(k+1)$, then every
prime $x<p\le2x$ with $v_2(p-1)=a$ and $\Delta(p)<Cx^{-\theta}$ has
$\Omega(m)\le k$ for all sufficiently large $x$. At $\theta=1/j$,
with an integer $j\ge2$, the condition $C\le2^{(a-1)/j-1}$ implies $\Omega(m)\le j-1$
for every $x\ge2$.
\end{enumerate}
\end{proposition}

\begin{proof}
The totient product gives
\begin{equation}\label{eq:endpoint-detector}
 \Delta(p)=\frac12\left(1-\prod_{q\mid m}(1-1/q)\right)
 \ge\frac1{2q}\qquad(q\mid m,\ q\textup{ prime}),
\end{equation}
so $\Delta(p)=0$ only for $m=1$, and $\Delta(p)=1/(2m)=2^{a-1}/(p-1)$
for prime $m$. If $\Delta(p)<\eta$, every prime factor of $m$ exceeds
$1/(2\eta)$; as $m=(p-1)/2^a<2^{1-a}x$, $\Omega(m)\ge h$ then forces
$(2\eta)^{-h}<2^{1-a}x$. For $\eta=Cx^{-\theta}$ this reads
$x^{h\theta-1}<2^{1-a}(2C)^h$, which fails for all large $x$ if
$h=k+1$ and $\theta>1/(k+1)$, and for every $x\ge2$ if $h=j$,
$\theta=1/j$ and $C\le2^{(a-1)/j-1}$. This proves \textup{(iii)}. Under
\textup{(i)}, $(2\varepsilon)^{-2}\ge2^{1-a}x$, so the case
$\eta=\varepsilon$, $h=2$ gives $\Omega(m)\le1$; as $\Delta(p)>0$, $m$ is
prime, and $\Delta(p)=2^{a-1}/(p-1)<\varepsilon$ is equivalent to
\eqref{eq:endpoint-exact-clip}.

For \textup{(ii)}, the determinant product formula,
\eqref{eq:matrix-product-bounds} and $\varphi(p-1)/(p-1)\le1/2$ give
$\Delta(p)\le\Delta_n(p)\le\Delta(p)+1/(2(p-1))$ for every
$n\in\mathbb Z_{\ge1}\cup\{\infty\}$. For prime $m$ this yields
$0<\Delta(p)\le\Delta_n(p)\le(2^{a-1}+1/2)/(p-1)=O_a(x^{-1})<Cx^{-\theta}$
for large $x$, since $\theta<1$. Conversely, once $x\ge2^a+1$ we have
$m\ne1$, so $0<\Delta_n(p)<Cx^{-\theta}$ implies $0<\Delta(p)<Cx^{-\theta}$,
and it suffices to show that the latter forces $m$ prime. If
$\theta>1/2$, or $\theta=1/2$ and $C\le2^{(a-3)/2}$, this is
\textup{(iii)} with $k=1$, resp.\ $j=2$. Let $a$ be odd, $\theta=1/2$
and $C<2^{(a-1)/2}$; \textup{(iii)} with $k=2$ gives $\Omega(m)\le2$
eventually. If $m=q^2$ and $q\ne3$, odd $a$ gives
$2^aq^2+1\equiv0\pmod3$; $q=3$ yields only one fixed candidate. For
distinct primes $q,r$ with $m=qr$,
\[
 \Delta(p)=\frac{q+r-1}{2m}\ge\frac1{\sqrt m}-\frac1{2m},\qquad
 \sqrt x\,\Delta(p)\ge2^{(a-1)/2}-O_a(x^{-1/2})>C
\]
for large $x$.
\end{proof}

\begin{remark}[Endpoints and finite-range exceptions]
\label{rem:endpoint-detector-scope}
At $a=1$, prime cofactors $m$ are Sophie Germain primes and $p=2m+1$ is a
safe prime.
For even $a$ and $2^{(a-3)/2}<C<2^{(a-1)/2}$, a square cofactor is
possible: if $q$ is an odd prime and $p=2^aq^2+1$ is prime, then
$\Delta(p)=1/(2q)$, so $x<p\le2x$ and $0<\Delta(p)<C/\sqrt x$ hold exactly when
$\max\{\sqrt{(x-1)/2^a},\sqrt x/(2C)\}<q\le\sqrt{(2x-1)/2^a}$;
since $C>2^{(a-3)/2}$, this interval is nonempty for all sufficiently large $x$.
The qualifier ``for all sufficiently large $x$'' in part \textup{(iii)} cannot be dropped:
$p=19$, $x=10$, $a=1$, $C=1$ and $\theta=0.6$ give
$\Delta(p)=1/6<10^{-0.6}$ although $\Omega(9)=2$.

The finite-range cutoff does not transfer to the determinant deficits
$\Delta_n(p)$. At
$p=3$, $x=2$, $a=1$ and $\varepsilon=1/4$, one has $\Delta(3)=0$,
$\Delta_1(3)=1/6$, and $0<\Delta_\infty(3)\le2/9<1/4$, since
$P(3)\ge(1-1/3)(1-\sum_{j\ge2}3^{-j})=5/9$.
For prime $m$ the deficits separate at the scale $\theta=1$:
$\Delta(p)=2^{a-1}/(p-1)$, while $\Delta_n(p)=(2^{a-1}+1/2)/p+O(p^{-2})$
for every $n\in\mathbb Z_{\ge1}\cup\{\infty\}$; the implied constant is absolute,
and the error term vanishes when $n=1$, since
$0\le1-\prod_{2\le j\le n}(1-p^{-j})\le1/(p(p-1))$.
\end{remark}

Summing over all $2$-adic classes gives a common endpoint count;
a sieve estimate treats the boundary exponent $1/2$.

\begin{proposition}[A common count for power-sized endpoint windows]
\label{prop:endpoint-common-count}
Fix $C>0$ and $1/2\le\theta<1$. Let $U_x(p)$ be $X(p)$ or
$c_{n_x(p)}(p)$, with arbitrary $n_x(p)\in\mathbb Z_{\ge1}\cup\{\infty\}$
on the sampled primes. Write $m(p)=(p-1)/2^{v_2(p-1)}$ and put
\[
\begin{aligned}
 N_{\theta,C}^{U}(x)
 &:=\#\{3\le p\le x:p\textup{ prime},\ 0<\tfrac12-U_x(p)<Cx^{-\theta}\},\\
 S(x)&:=\#\{3\le p\le x:p\textup{ prime},\ m(p)\textup{ prime}\}.
\end{aligned}
\]
Uniformly over the dimension arrays,
\begin{equation}\label{eq:endpoint-common-count}
  N_{\theta,C}^{U}(x)=S(x)+
 \begin{cases}
 O_C(x/(\log x)^3),&\theta=1/2,\\
 O_{C,\theta}(x^\theta),&1/2<\theta<1.
 \end{cases}
\end{equation}
The same formula holds with both counts restricted to $x<p\le2x$,
keeping width $Cx^{-\theta}$. Consequently the limiting-law prediction
\[
 N_{\theta,C}^{U}(x)\sim
 \frac{\kappa_{\mathrm{end}}}{\theta}\frac{x}{(\log x)^2}
\]
can hold for at most one $\theta\in[1/2,1)$, even if different exponents
use different fixed $C$ or different dimension arrays.
\end{proposition}

\begin{proof}
Put $\Delta_U(p)=1/2-U_x(p)$. The determinant-product bound in the proof of
\Cref{prop:safe-prime-obstruction} gives
\[
 \Delta(p)\le\Delta_U(p)\le\Delta(p)+\frac1{2(p-1)}.
\]
If $m=m(p)>1$, \eqref{eq:endpoint-detector} shows that admission to
either window forces every prime divisor $q\mid m$ to exceed
$x^\theta/(2C)$. If $\theta>1/2$, a composite $m$ would have two
prime factors counted with multiplicity, and hence
$m>x^{2\theta}/(4C^2)>x$ for large $x$, whereas $p\le2x$ gives
$m\le(p-1)/2<x$. Thus $m$ is prime in this case.

For $\theta=1/2$, put $z=\sqrt{x}/(2C)$. An admitted composite
$m$ has exactly two prime factors $q,r>z$, counted with multiplicity,
since $z^3>x$ for large $x$. Writing $p-1=2^aqr$ gives
$2^a<8C^2$ and $z<q\le(2x-1)/(2^az)$. Apply
\eqref{eq:prime-model-sieve} with population bound $2x$ and modulus
$2^aq\le(2x-1)/z$. The complementary cofactor is $r>z$, so the
number of admitted composite cofactors is at most
\[
 \ll\frac{x}{\log^2z}
 \sum_{\substack{a\ge1\\2^a<8C^2}}2^{1-a}
 \sum_{\substack{z<q\le(2x-1)/(2^az)\\q\textup{ prime}}}\frac1{q-1}
 \ll_C\frac{x}{(\log x)^3}.
\]
Indeed, there are $O_C(1)$ outer terms, and each inner interval has
endpoints comparable to $\sqrt{x}$ with bounded ratio; Chebyshev's
bound gives a reciprocal-prime sum $O_C(1/\log x)$.
The case $m=1$ gives only
$O(\log x)$ points; their totient deficit is zero, but their determinant
deficit may be positive.

Conversely, if $p-1=2^aq$ with $q$ an odd prime, then
\[
 0<\Delta_U(p)\le\frac1{2q}+\frac1{2\cdot2^aq}\le\frac3{4q}.
\]
Thus every such point with $q>3x^\theta/(4C)$ is admitted. The number
of possibly missing points in either sampling range is at most
\[
 \frac{\log(2x)}{\log2}\,
 \pi\!\left(\frac{3x^\theta}{4C}\right)
 \ll_{C,\theta}x^\theta
\]
by Chebyshev's bound. This and the power-of-two exceptions prove both
forms of \eqref{eq:endpoint-common-count}, uniformly over the arrays.

\Cref{prop:endpoint-half} gives
$H_{\mathrm{end}}(2Cx^{-\theta})\sim\kappa_{\mathrm{end}}/(\theta\log x)$,
which, multiplied by $\pi(x)$, is the stated limiting-law prediction.
The error in either case of \eqref{eq:endpoint-common-count} is
$o(x/(\log x)^2)$, so that prediction would force
$S(x)\sim(\kappa_{\mathrm{end}}/\theta)x/(\log x)^2$.
Two distinct exponents prescribe different positive constants for the
same $S(x)$, proving the last assertion.
\end{proof}

For fixed $1/2<\theta<1$, the same count at any width
$h(x)=x^{-\theta+o(1)}$ equals $S(x)+o(x/\log^2x)$, uniformly over
the dimension arrays, by sandwiching between widths
$x^{-(\theta+\delta)}$ and $x^{-(\theta-\delta)}$ with
$1/2<\theta-\delta<\theta+\delta<1$.

\begin{theorem}[Linear-sieve bounds for endpoint counts]
\label{thm:endpoint-sieve}
\index{Prime cofactors!sieve estimates}
Retain the count $N^U_{\theta,C}(x)$ of
\Cref{prop:endpoint-common-count}, now with fixed $C>0$ and
$0<\theta<1/2$. Let $f_{\mathrm{sieve}},F_{\mathrm{sieve}}$ be the
dimension-one linear-sieve functions. Then, uniformly over the dimension arrays,
\begin{equation}\label{eq:endpoint-sieve}
\begin{split}
 f_{\mathrm{sieve}}\!\left(\frac4{7\theta}\right)
 &\le\liminf_{x\to\infty}
       \frac{\theta(\log x)^2N^U_{\theta,C}(x)}{\kappa_{\mathrm{end}}x}\\
 &\le\limsup_{x\to\infty}
       \frac{\theta(\log x)^2N^U_{\theta,C}(x)}{\kappa_{\mathrm{end}}x}
 \le F_{\mathrm{sieve}}\!\left(\frac4{7\theta}\right).
\end{split}
\end{equation}
In particular, for $0<\theta<2/7$,
$N^U_{\theta,C}(x)\asymp_{C,\theta}x/(\log x)^2$.
Both assertions hold also for counts restricted to $(x,2x]$,
with the same width $Cx^{-\theta}$ and normalization.
\end{theorem}

\begin{proof}
Write $R(x,z)$ for the count of odd primes $p\le x$ whose odd part
$m(p)$ has no prime divisor below $z$, temporarily including $m=1$.
The detector \eqref{eq:endpoint-detector} implies that admission to
the endpoint window requires roughness at $z=x^\theta/(2C)$.
Conversely, if every odd prime divisor exceeds $Kx^\theta$, there
are at most $2/\theta$ such divisors eventually, whence
\[
 \frac12-X(p)\le\frac12\sum_{q\mid m(p)}\frac1q
                 \le\frac{x^{-\theta}}{K\theta}.
\]
Choose $K>2/(C\theta)$. Discarding primes $p\le x^\beta$ with
$\theta<\beta<1$ makes the determinant correction
$1/(2(p-1))$ smaller than half the window and costs
$o(x/(\log x)^2)$. The cases $m=1$ cost $O(\log x)$.
Thus two fixed-multiple roughness counts sandwich the desired count
with negligible errors.

Sieve the full sequence of odd shifted primes $p-1$, using only
odd sieve primes. For odd squarefree $d$ its density is
$g(d)=1/\varphi(d)$ and the residue is the fixed integer $1$.
Furthermore
\[
 V(z)=\prod_{3\le q<z}\left(1-\frac1{q-1}\right)
 \sim\frac{\kappa_{\mathrm{end}}}{\log z},\qquad
 \kappa_{\mathrm{end}}=\Smer e^{-\gamma}.
\]
This also verifies the dimension-one product-ratio hypothesis.
The linear-sieve functions are normalized by $f_{\mathrm{sieve}}(s)=0$
for $s\le2$, $F_{\mathrm{sieve}}(s)=2e^\gamma/s$ for $1\le s\le3$,
and the usual delay equations
$(sf_{\mathrm{sieve}}(s))'=F_{\mathrm{sieve}}(s-1)$ for $s>2$ and
$(sF_{\mathrm{sieve}}(s))'=f_{\mathrm{sieve}}(s-1)$ for $s>3$.
In particular $f_{\mathrm{sieve}}(s)>0$ exactly when $s>2$.

Restoring the prime $2$ before sieving and removing it afterward costs
at most one point, so we use $\pi(x)$ as the total mass.
The factorable linear sieve~\cite[Theorem~2.10 and (2.4)]{Lic24}
gives upper and lower main terms $F_{\mathrm{sieve}}(s)+O(\epsilon)$ and
$f_{\mathrm{sieve}}(s)+O(\epsilon)$, where $s=\log E/\log z$ and $E$ is the
construction level. Fix $0<\epsilon<1/4$ sufficiently small
in terms of $\theta$. We first work with $2\le s<\epsilon^{-1}$
and justify $1\le s<2$ below. Put $\tau=\epsilon^9$
and $u=E^{\epsilon^2}$. Use boxes $(E_i,E_i^{1+\tau}]$ on the grid
$E^{\epsilon^2(1+\tau)^j}$, $j\ge0$, restricted to odd primes below $z$.
For weakly decreasing vectors $E_1\ge\cdots\ge E_r$ with
$E_1\le\sqrt E$, put $C_i=E_1\cdots E_{i-1}E_i^3$.
Let $\rho_\pm$ be the upper and lower Rosser weights of level
$E^\epsilon$ on odd primes at most $u$. The two box families and their
small-prime weights are
\[
\begin{array}{c|c|c|c|c}
 \text{bound}&r&\text{tested }i&\text{condition}&\rho\\ \hline
 \text{upper}&\text{even}&\text{odd}&C_i<E&\rho_+\\
 \text{upper}&\text{odd}&\text{odd}&C_i<E^{1/(1+\tau)}&\rho_-\\
 \text{lower}&\text{odd}&\text{even}&C_i<E&\rho_+\\
 \text{lower}&\text{even}&\text{even}&C_i<E^{1/(1+\tau)}&\rho_-
\end{array}
\]
Include the empty vector. Prime choices are distinct and run over
labelled slots, with repeated-box multiplicity divided out once,
as in \cite[(5.10)--(5.11)]{Lic24}. With small primes first removed
exactly, order each selected prime set decreasingly and put
$C_i(p)=p_1\cdots p_{i-1}p_i^3$. The inequalities
$C_i<C_i(p)\le C_i^{1+\tau}$ compare these terms with the
Rosser inequalities: the outer tests admit additional positive upper
and negative lower terms, whereas the inner tests omit negative upper
and positive lower terms. Applying the heads in the table therefore
preserves both inequalities. There are $O_\epsilon(1)$ components,
each with $r\le\epsilon^{-2}$ slots and $|\rho|\le1$, supported on
squarefree $b\le E^\epsilon$ with prime factors at most $u$.
Both families of decreasing box vectors $(E_1,\ldots,E_r)$ satisfy
\[
 E_1\cdots E_{i-1}E_i^2\le E\qquad(1\le i\le r).
\]
Indeed, the cubic restriction gives this directly at its prescribed
parity; at the other parity use the preceding index and
$E_i\le E_{i-1}$, with $E_1^2\le E$ covering the first lower-sieve
index.

Distinct boxes are disjoint. Write $I_j$
for the distinct boxes, $k_j$ for their multiplicities, and $h_j$
for the indicator of primes in $I_j$, and put
$\gamma_{\mathrm{box}}=\prod_j k_j!$. Each squarefree product in this
profile has exactly $\gamma_{\mathrm{box}}$ representations in the
labelled tuple sum: the $k_j$ distinct primes in box $j$ may be
permuted among its $k_j$ slots. Its external factor
$1/\gamma_{\mathrm{box}}$ therefore gives the squarefree coefficient
$\rho(b)$, with sign $(-1)^r$.

These coefficients retain the two main terms of
\cite[(25)--(28)]{Iwa80}. Put
$V_u=\prod_{3\le q\le u}(1-g(q))$ and, over all grid boxes $I$,
$t_I=\sum_{q\in I}g(q)$.
The fundamental lemma gives
$\sum_b\rho_\pm(b)g(b)=V_u\{1+O(e^{-1/\epsilon})+o_E(1)\}$.
Distinct-box profiles agree with the classical assembly. On repeated
profiles the multiplicity discrepancy is at most
$\gamma_{\mathrm{box}}$; since $k!e_k\le t_I^k$ for the squarefree
$k$-selection weight $e_k$ in box $I$, their total discrepancy is
\[
\begin{aligned}
 \ll V_u\left\{\prod_I(1-t_I)^{-1}-\prod_I(1+t_I)\right\}
 &\le V_u\prod_I(1-t_I)^{-1}\sum_It_I^2\\
 &\ll V(z)\epsilon^{-4}\tau\log(1/\epsilon).
\end{aligned}
\]
Here $\max_It_I\ll\tau$, $\sum_It_I\ll\log(1/\epsilon)$ and
$\prod_I(1-t_I)^{-1}\ll\epsilon^{-2}$ follow from the
dimension-one product estimate, as does $V_u/V(z)\ll\epsilon^{-2}$.
Endpoint conventions contribute $o_E(V(z))$. Thus the density sums
$A_\pm$ of the normalized coefficients satisfy
\[
\begin{aligned}
 A_+&\le V(z)\{F_{\mathrm{sieve}}(s)+O(\epsilon)+o_E(1)\},\\
 A_-&\ge V(z)\{f_{\mathrm{sieve}}(s)-O(\epsilon)-o_E(1)\}.
\end{aligned}
\]
Before choosing any factorization level, define the fixed extension
\[
 H=\rho*\mathop{*}_{j}
                 \frac{h_j^{*k_j}}{k_j!}.
\]
Here $*$ denotes Dirichlet convolution, and an empty convolution is
the unit mass at $1$. Thus $(-1)^rH$ extends the signed tuple
component after its multiplicity has been divided out. For an integer
in the prescribed box profile, uniquely written as
$n=b\prod_{p>u}p^{e_p}$, this fixed extension has coefficient
\[
 H(n)=\frac{\rho(b)}{\prod_{p>u}e_p!},\qquad
 \sum_{p\in I_j}e_p=k_j.
\]
It therefore agrees with the indicator component on squarefree integers,
is $1$-bounded, and every nonsquarefree integer in its support is
divisible by $q^2$ for a prime $q>u$.

We verify well-factorability at the enlarged level
$L=E^{1+\tau+\epsilon}$ in the sense of
\cite[Definition~1]{MayWF20}. Given any real split $L=MN$ with
$1\le M\le N$, assign the small-prime factor to the $N$ side.
The remaining capacities $M$ and $N/E^\epsilon$ are at least one
and have product $E^{1+\tau}$, since
$N\ge\sqrt L\ge E^\epsilon$. Allocate the prime slots in order,
with upper bounds $a_i=E_i^{1+\tau}$, to either capacity whenever
they fit. If $a_i$ fitted neither side, the products $P,Q$ already
allocated would satisfy
\[
 Pa_i>M,\qquad Qa_i>N/E^\epsilon,
 \qquad
 PQa_i^2=(E_1\cdots E_{i-1}E_i^2)^{1+\tau}
          \le E^{1+\tau},
\]
a contradiction. Thus the allocation succeeds, including empty
parts. If $r_j$ copies of box $j$ go to the $M$ side, set
\[
 U=\mathop{*}_{j}\frac{h_j^{*r_j}}{r_j!},\qquad
 V=\rho*\mathop{*}_{j}
                   \frac{h_j^{*(k_j-r_j)}}{(k_j-r_j)!}.
\]
The same coefficient formula makes $U,V$ $1$-bounded, with supports
in $[1,M]$ and $[1,N]$, and
\[
 H=c(U*V),\qquad
 c=\prod_j\binom{k_j}{r_j}^{-1}\le1.
\]
Absorb $c$ and the sign $(-1)^r$ into one factor; both have
modulus at most one. Thus $H$ and its signed component have the
required factorizations for every split,
including $r=0$. Oddness and the condition that every prime factor is below
$z$ are completely multiplicative masks; multiplying both factors
by either mask preserves their convolution identity and support.

The Bombieri--Friedlander--Iwaniec theorem in
\cite[Theorem~A]{MayWF20} now controls each masked extension when
$L\le x^{4/7-\delta}$, with arbitrary logarithmic savings. Replacing
the squarefree remainder by its extension costs at most
\[
 \ll_\epsilon\sum_{q>u}\sum_{\substack{d\le L\\q^2\mid d}}
       \left(\frac xd+1+\frac{\pi(x)}{\varphi(d)}\right)
 \ll_\epsilon\frac{x\log(2L)+L}{u}.
\]
For the last bound, use $\sum_{q>u}q^{-2}\ll u^{-1}$,
$\varphi(q^2n)\ge\varphi(q^2)\varphi(n)$, and
$\sum_{n\le T}1/\varphi(n)\ll\log(2T)$; the latter follows by
expanding $n/\varphi(n)=\sum_{d\mid n}\mu^2(d)/\varphi(d)$.

Take
\[
 L=x^{4/7-\delta},\qquad
 E=L^{1/(1+\epsilon+\tau)},\qquad u=E^{\epsilon^2},
\]
and either fixed multiple $z\asymp x^\theta$ above. Choose the
fixed losses small enough that $E>z$, $u<z$, and
$\log E/\log z<\epsilon^{-1}$ eventually. Then
\[
 \frac{\log E}{\log z}\longrightarrow
 \frac{4/7-\delta}{(1+\epsilon+\tau)\theta},\qquad
 \pi(x)V(z)\sim\frac{\kappa_{\mathrm{end}}x}{\theta\log^2x}.
\]
If $1\le s<2$, apply the upper sieve instead at $z_0=E^{1/2}$:
$R(x,z)\le R(x,z_0)$, and $F_{\mathrm{sieve}}(s)=2e^\gamma/s$ gives
$V(z_0)F_{\mathrm{sieve}}(2)\sim V(z)F_{\mathrm{sieve}}(s)$;
its $O(\epsilon)$ main-term error
remains $O(\epsilon)V(z)$ since $\log z/\log z_0=2/s\le2$.
The lower bound is then $f_{\mathrm{sieve}}(s)=0$, by nonnegativity.
The extension error is smaller than every fixed logarithmic saving.
First let $x\to\infty$ at fixed $\delta,\epsilon$, and then let
these losses decrease to zero. Continuity of $f_{\mathrm{sieve}},F_{\mathrm{sieve}}$ proves
\eqref{eq:endpoint-sieve}; positivity gives $\theta<2/7$.
For the dyadic assertion apply the sieve directly to the interval
sequence: its remainder is the difference of the progression errors
at $2x$ and $x$, to which the same distribution theorem applies,
and $\pi(2x)-\pi(x)\sim x/\log x$ gives the same normalization.
\end{proof}

\begin{remark}[The sharp Kolmogorov coefficient]
\label{rem:kolmogorov-conjecture}
The assertion $(\log x)D_x^U\to\kappa_{\mathrm{end}}$ remains
conjectural. It would imply
\[
 \kappa_{\mathrm{end}}\le
 \liminf\frac{(\log x)^2S(x)}x
 \le\limsup\frac{(\log x)^2S(x)}x
 \le2\kappa_{\mathrm{end}}.
\]
Indeed, at the threshold $1/2-Cx^{-\theta}$ with fixed
$1/2\le\theta<1$, \Cref{prop:endpoint-common-count} and the
endpoint expansion would give
$|\kappa_{\mathrm{end}}/\theta-(\log x)^2S(x)/x|
\le\kappa_{\mathrm{end}}+o(1)$.
Taking $\theta=1/2$ gives the lower bound, and letting $\theta\uparrow1$
gives the upper bound. The count $S(x)$
includes every prime $p\le x$ of the form $p=2^a q+1$, $a\ge1$,
with $q$ an odd prime. The raw-statistic coefficient conjecture in the
class $3\pmod4$ and its Sophie Germain consequence are stated below.
Even the weaker condition $L:=\limsup_{x\to\infty}(\log x)D_x^U
<2\kappa_{\mathrm{end}}$ forces a positive limsup for the normalized
prime-pair count for some fixed exponent $a$. Indeed, the same comparison at
$\theta=1/2$ gives $\liminf (\log x)^2 S(x)/x\ge 2\kappa_{\mathrm{end}}-L$.
For $z=x^{1/4}$ the pairs $p=2^a q+1$ with $q\le z$ contribute
$O(z\log x)$ in total. A pair with $q>z$ has $2^a\le(x-1)/z$;
\eqref{eq:prime-model-sieve} with divisor $2^a$ bounds the number for
this $a$ by $O(x/(\varphi(2^a)\log^2 z))$. Summing for $a\ge A$
costs $O(2^{-A}x/\log^2 x)$ with an absolute constant. After choosing
$A$ sufficiently large, the finitely many $1\le a<A$ therefore
supply $\gg_L x/\log^2 x$ pairs. For at least one fixed $a$, along
a sequence $x\to\infty$ this class supplies that order. In terms
of $y=(x-1)/2^a$, it follows that
\[
 \limsup_{y\to\infty}\frac{(\log y)^2}{y}
       \#\{q\le y:q\ \textup{and}\ 2^a q+1\ \textup{prime}\}>0.
\]
The transition to rough composite cofactors is discussed
\hyperref[par:endpoint-buchstab]{below}.
\end{remark}

\paragraph*{A progression coefficient and Sophie Germain primes}
For the raw statistic $X(p)=\varphi(p-1)/(p-1)$, put
\[
 D_{x;4,3}:=\|F^X_{x,4,3}-G\|_\infty.
\]
Here $G_{4,3}=G$ by \Cref{thm:ap-law}: the factor at $2$ is $1/2$
and all odd-prime coordinates remain free. The progression analogue of
\Cref{rem:kolmogorov-conjecture} is the conjecture
\begin{equation}\label{eq:progression-coefficient-conjecture}
 (\log x)D_{x;4,3}\longrightarrow\kappa_{\mathrm{end}}.
\end{equation}

\begin{proposition}[A CDF bound and fixed prime pairs]
\label{prop:progression-coefficient-sophie-germain}
\index{Sophie Germain primes}
\index{Kolmogorov distance!prime-pair consequences}
Fix $a\ge1$, put $Q_a=2^{a+1}$, $b_a=1+2^a$, and let
\[
 \operatorname{SG}_a(y):=\#\{q\le y:q\textup{ and }2^aq+1
                                   \textup{ are prime}\}.
\]
Let $U_x(p)=X(p)$ or $c_{n_x(p)}(p)$, with arbitrary
$n_x(p)\in\Z_{\ge1}\cup\{\infty\}$, and put
$D^U_{x,a}=\|F^{U_x}_{x,Q_a,b_a}-G\|_\infty$.
If $\limsup_{x\to\infty}(\log x)D^U_{x,a}\le L<\infty$, then
\[
 \begin{aligned}
 \liminf_{y\to\infty}\frac{\log^2y}{y}\operatorname{SG}_a(y)
 &\ge\max\{0,2\kappa_{\mathrm{end}}-L\},\\
 \limsup_{y\to\infty}\frac{\log^2y}{y}\operatorname{SG}_a(y)
 &\le\kappa_{\mathrm{end}}+L.
 \end{aligned}
\]
For $a=1$ and $U_x=X$, \eqref{eq:progression-coefficient-conjecture}
gives lower and upper Sophie Germain coefficients
$\kappa_{\mathrm{end}}$ and $2\kappa_{\mathrm{end}}$.
For each fixed $a$, already $L<2\kappa_{\mathrm{end}}$ implies
$\operatorname{SG}_a(y)\gg_{a,L}y/\log^2y$.
\end{proposition}

\begin{proof}
Fix $C>0$ and $1/2<\theta<1$. Let $N(x)$ count primes $p\le x$,
$p\equiv b_a\pmod{Q_a}$, with
$0<1/2-U_x(p)<Cx^{-\theta}$, and put $y=(x-1)/2^a$.
Here $p-1=2^am$ with $m$ odd. Apart from the single possible $m=1$,
\eqref{eq:endpoint-detector} forces every prime divisor of an admitted
$m$ to exceed $x^\theta/(2C)$; two factors counted with multiplicity
would exceed $y$ for all sufficiently large $x$.
Conversely, for $m=q$ prime, the bound in the proof of
\Cref{prop:safe-prime-obstruction} gives, uniformly in the dimensions,
\[
 0<\frac12-U_x(p)\le\frac1{2q}+\frac1{2^{a+1}q}\le\frac3{4q}.
\]
Thus every $q>3x^\theta/(4C)$ is admitted, and
\[
 N(x)=\operatorname{SG}_a(y)+O_{a,C,\theta}(x^\theta/\log x+1).
\]
The exceptional cofactor $q=2$ and the possible zero-deficit point
cost $O(1)$. At $t_x=1/2-Cx^{-\theta}$ it follows that
\[
 1-F^{U_x}_{x,Q_a,b_a}(t_x)
 =\frac{N(x)+O(1)}{\pi(x;Q_a,b_a)}.
\]
By \Cref{thm:ap-law}, $G_{Q_a,b_a}=G$, and
\Cref{prop:endpoint-half} gives
\[
 1-G(t_x)=H_{\mathrm{end}}(2Cx^{-\theta})
 =\frac{\kappa_{\mathrm{end}}}{\theta\log x}
    +O_{C,\theta}((\log x)^{-2}).
\]
Comparison costs $D^U_{x,a}$, while
$\pi(x;Q_a,b_a)\sim x/(2^a\log x)\sim y/\log y$.
The finite bound $L$ first gives
$\operatorname{SG}_a(y)=O_{a,L}(y/\log^2y)$ and then lower and upper
normalized coefficients $\kappa_{\mathrm{end}}/\theta-L$ and
$\kappa_{\mathrm{end}}/\theta+L$. Let $\theta\downarrow1/2$ in the
lower bound and $\theta\uparrow1$ in the upper bound; nonnegativity
completes the proof.
\end{proof}

\paragraph*{A Buchstab transition in a fixed $2$-adic class}\phantomsection\label{par:endpoint-buchstab}
\index{Buchstab transition}
\index{Shrinking windows!prime-pair conjectures}
Let $\omega_{\mathrm B}$ be the continuous Buchstab function, defined by
\[
 \omega_{\mathrm B}(u)=1/u\quad(1\le u\le2),\qquad
 (u\omega_{\mathrm B}(u))'=\omega_{\mathrm B}(u-1)\quad(u>2).
\]
For a fixed integer $a\ge1$, $C>0$ and $0<\theta<1$, put
\[
 N_{a,\theta,C}(x):=
 \#\{p\le x:v_2(p-1)=a,\ 0<1/2-X(p)<Cx^{-\theta}\}.
\]
A conjectural transition law is
\begin{equation}\label{eq:endpoint-buchstab-conjecture}
 N_{a,\theta,C}(x)\sim\frac{\Smer}{2^a}
       \frac{\omega_{\mathrm B}(1/\theta)}{\theta}
       \frac{x}{\log^2x}.
\end{equation}

An equivalent formulation uses rough odd cofactors.
With $M=(x-1)/2^a$, define
\[
 \begin{aligned}
 \Phi(M,z)&:=\#\{1<m\le M:P^-(m)>z\},\\
 R_a(x,z)&:=\#\{1<m\le M:P^-(m)>z,\ 2^am+1\textup{ prime}\}.
 \end{aligned}
\]
Here $P^-(m)$ is the least prime factor of $m$; for $z>2$ every counted
$m$ is odd, so the valuation is exactly $a$. The correlation formulation is
\begin{equation}\label{eq:rough-prime-correlation-conjecture}
 R_a(x,\lambda x^\theta)\sim
       \frac{\Smer}{\log x}\Phi(M,\lambda x^\theta)
 \qquad(\lambda>0\textup{ fixed}).
\end{equation}
Indeed, the integer rough-number asymptotic
\cite[\S~6.1, Lemma~6.1]{Lic24} gives
\[
 \Phi(M,\lambda x^\theta)\sim
 \frac{\omega_{\mathrm B}(1/\theta)}{2^a\theta}
 \frac{x}{\log x}.
\]
For an odd prime $\ell$, conditioning $m\not\equiv0\pmod\ell$
leaves one forbidden class for $2^am+1$ among $\ell-1$ possibilities.
The correction to the ordinary prime density is therefore
\[
 \frac{1-1/(\ell-1)}{1-1/\ell}
 =1-\frac1{(\ell-1)^2}.
\]
The factor at $2$ is $2$, giving the product $\Smer$ in
\eqref{eq:rough-prime-correlation-conjecture}.

For each fixed $a,\theta$ as above and $C,\lambda>0$ one has
unconditionally
\begin{equation}\label{eq:buchstab-count-comparison}
 N_{a,\theta,C}(x)=R_a(x,\lambda x^\theta)
             +O_{a,\theta,C,\lambda}(x/\log^3 x).
\end{equation}
To prove this, first compare the roughness thresholds. For odd $m>1$ and prime $p=2^am+1$,
\[
 \frac1{2P^-(m)}\le\frac12-X(p)
 =\frac12\left(1-\prod_{q\mid m}(1-1/q)\right)
 \le\frac12\sum_{q\mid m}\frac1q,
\]
where the product and sum use distinct prime divisors.
Choose an integer $K>1/\theta$ and $B>K/(2C)$.
If $P^-(m)>Bx^\theta$, then
$\Omega(m)\le\log M/\log(Bx^\theta)<K$ eventually, and the last
bound is less than $Cx^{-\theta}$. The first bound gives the reverse
inclusion at the smaller cutoff, so for all sufficiently large $x$,
\begin{equation}\label{eq:endpoint-roughness-sandwich}
 R_a(x,Bx^\theta)\le N_{a,\theta,C}(x)
       \le R_a(x,x^\theta/(2C)).
\end{equation}
Put $y=x^\theta$. For any fixed $0<c_1<c_2$, a cofactor in the
difference $R_a(x,c_1y)-R_a(x,c_2y)$ has the form $m=qk$ with
$q=P^-(m)\in(c_1y,c_2y]$. The case $k=1$ contributes $O(y)$.
Otherwise $P^-(k)\ge q>c_1y$ and $2^a q\le(x-1)/(c_1y)$, so
\eqref{eq:prime-model-sieve} bounds the remaining difference by
\[
 \ll \frac{x}{\log^2(c_1y)}
      \sum_{\substack{c_1y<q\le c_2y\\q\ \mathrm{prime}}}
              \frac1{\varphi(2^a q)}
 \ll_{a,c_1,c_2,\theta} \frac{x}{\log^3 x}.
\]
Here Chebyshev's bound gives a reciprocal-prime sum $O_{c_1,c_2}(1/\log y)$.
As $y=o(x/\log^3 x)$, this proves \eqref{eq:buchstab-count-comparison}
by \eqref{eq:endpoint-roughness-sandwich}. The integer rough-number
asymptotic above now makes \eqref{eq:endpoint-buchstab-conjecture} for
one fixed $C$ equivalent to \eqref{eq:rough-prime-correlation-conjecture}
for every fixed $\lambda$, and conversely.

The same comparison \eqref{eq:buchstab-count-comparison} holds with
$X(p)$ replaced by $c_{n_x(p)}(p)$, uniformly over the dimension arrays.
For fixed $\theta<\eta<1$, discarding $p\le x^\eta$ costs
$o(x/\log^3 x)$. On the remaining primes
$0\le X(p)-c_{n_x(p)}(p)\le1/(2(p-1))=o(x^{-\theta})$;
apart from the one possible raw zero-deficit point, the determinant
count is bracketed by the raw counts at widths $(C/2)x^{-\theta}$
and $Cx^{-\theta}$. Both have the same comparison by
\eqref{eq:buchstab-count-comparison}.
As a family over fixed $0<\theta<1$, the conjecture also gives the
same main term for widths $h(x)=x^{-\theta+o(1)}$, by sandwiching
between $x^{-(\theta+\delta)}$ and $x^{-(\theta-\delta)}$ and then
letting $\delta\downarrow0$, using continuity of
$\omega_{\mathrm B}(1/\theta)/\theta$.

The proposed profile has the explicit regimes
\[
 \frac{\omega_{\mathrm B}(1/\theta)}{\theta}
 =\begin{cases}
 1,&1/2\le\theta<1,\\
 1+\log(1/\theta-1),&1/3\le\theta\le1/2,
 \end{cases}
\]
Moreover, $\omega_{\mathrm B}(1/\theta)/\theta\sim e^{-\gamma}/\theta$
as $\theta\downarrow0$, after the fixed-$\theta$ limit in $x$.
On the plateau, prime cofactors give the Bateman--Horn prediction
\cite{BatH62} for the pair $t,2^at+1$: their product has one root
modulo $2$ and two modulo every odd prime, hence singular series
$\Smer$. At $\theta=1/2$, square and semiprime cofactors must also be considered when $C$ is arbitrary. The sieve argument in the proof of
\Cref{prop:endpoint-common-count} bounds their number by
\[
 \ll_{a,C}\frac{x}{\log^2x}
 \sum_{\substack{q\asymp_{a,C}\sqrt x\\q\textup{ prime}}}\frac1{q-1}
 \ll_{a,C}\frac{x}{\log^3x}.
\]
For $\theta<1/2$, ordinary
Bateman--Horn for fixed polynomial pairs supplies no uniformity for
the growing almost-prime cofactors in
\eqref{eq:rough-prime-correlation-conjecture}.

The exact class is $1+2^a\pmod{2^{a+1}}$, with prime count
asymptotic to $x/(2^a\log x)$ and model law $G$. Its model endpoint count is
therefore asymptotic to
$\kappa_{\mathrm{end}}x/(2^a\theta\log^2x)$.
Under \eqref{eq:endpoint-buchstab-conjecture}, the ratio of the
actual count to this model count tends to
$e^\gamma\omega_{\mathrm B}(1/\theta)$; on $[1/2,1)$ they agree
only at $\theta=e^{-\gamma}$.

\chapter{Determinant laws over finite rings and fields}\label{sec:composite-modulus}

The determinant probabilities computed at a fixed prime lead to a further
question: what changes when the coefficient ring or extension degree
changes? We first sample uniform matrices in a fixed finite ring, using
reduction to residue fields and equal determinant fibers to count
prescribed unit determinants. We then regard the primitive-determinant
density of a finite field as an arithmetic statistic and sample the
prime base or the extension degree. These are separate sources of
randomness. In the field statements, $n=\infty$ denotes the
infinite-product density defined below.

\section{Exact determinant laws over finite rings}
\label{ssec:finite-ring-identities}

For prime powers $q$ and $n\in\mathbb Z_{\ge1}\cup\{\infty\}$, extend the notation by
\[
 c_n(q)=\frac{\varphi(q-1)}{q-1}\prod_{j=1}^n(1-q^{-j}).
\]
For an arbitrary integer $Q\ge2$ and $R=\Z/Q\Z$, the full determinant law is classical
\cite[Corollary~2.2 and the following paragraph]{BM87};
\cite[arXiv version, Theorems~3.1, 3.5, 3.8, 3.11 and~4.1]{CJU17} treats finite
commutative principal ideal rings. For finite $n\ge1$ and uniform
$M\in M_n(\Z/p^e\Z)$, put $g_s=[z^s]\prod_{j=1}^n(1-p^{-j})/(1-p^{-j}z)$. For nonzero $a$
of valuation $s<e$,
\begin{equation}\label{eq:prime-power-det-point}
 w_{p,e}(a):=\mathbb P(\det M=a)=\frac{g_s}{\varphi(p^{e-s})},\qquad
 w_{p,e}(0)=1-\sum_{s=0}^{e-1}g_s.
\end{equation}
Here $g_s=\mathbb P(v_p(\det M)=s)$ for $0\le s<e$: the first column has
minimal valuation $t<e$ with mass $(1-p^{-n})p^{-nt}$, after which row
operations leave a uniform $(n-1)$-square block modulo $p^{e-t}$; unit row
scaling spreads $g_s$ equally over the $\varphi(p^{e-s})$ residues of
valuation $s$. By the Chinese remainder theorem, for $Q=\prod_{p\mid Q}p^{e_p}$,
uniform $M\in M_n(\Z/Q\Z)$ and $A\subseteq\Z/Q\Z$,
\begin{equation}\label{eq:composite-arbitrary-det}
 \mathbb P(\det M\in A)=\sum_{a\in A}\prod_{p\mid Q}
 w_{p,e_p}(a\bmod p^{e_p}).
\end{equation}

\begin{proposition}[Finite commutative rings]\label{prop:finite-ring-det}
\index{Finite rings!determinant laws}
Let $R$ be a nonzero finite commutative ring with identity, with distinct
maximal ideals $\mathfrak m_1,\ldots,\mathfrak m_r$, and let $n\ge1$. Put
$q_i=|R/\mathfrak m_i|$, $J=\bigcap_i\mathfrak m_i$ and
$G_i=\prod_{j=1}^n(1-q_i^{-j})$. For uniform $M\in M_n(R)$ and
$A\subseteq R^\times$,
\begin{equation}\label{eq:finite-ring-det}
 \mathbb P(\det M\in A)=\frac{|A|}{|R^\times|}\prod_{i=1}^rG_i.
\end{equation}
For every $B\subseteq R/J\cong\prod_i\mathbb F_{q_i}$,
\begin{equation}\label{eq:radical-det-predicate}
 \mathbb P(\det M\bmod J\in B)=\sum_{b\in B}\prod_iw_i(b_i),\qquad
 w_i(u)=\begin{cases}1-G_i,&u=0,\\G_i/(q_i-1),&u\ne0.\end{cases}
\end{equation}
In particular, for every prime power $q$, the fraction of matrices in
$M_n(\mathbb F_q)$ whose determinant generates $\mathbb F_q^\times$ is $c_n(q)$.
\end{proposition}

\begin{proof}
A matrix is invertible precisely when its reduction modulo $J$ is
invertible. The Chinese remainder theorem and the equal $|J|^{n^2}$ lift counts give
$|\GL_n(R)|/|R|^{n^2}=\prod_iG_i$. The determinant map
$\GL_n(R)\to R^\times$ is surjective, by diagonal matrices, and has equal
fibers, proving \eqref{eq:finite-ring-det}. The field reductions are
independent and uniform; their determinant point masses are $w_i$.
This proves \eqref{eq:radical-det-predicate}. Finally,
$\mathbb F_q^\times$ is cyclic with $\varphi(q-1)$ generators.
\end{proof}

For a cyclic quotient $\pi:R^\times\twoheadrightarrow C$,
\eqref{eq:finite-ring-det} with $A=\pi^{-1}\{u\in C:\langle u\rangle=C\}$
gives conditional density $\varphi(|C|)/|C|$ among invertible matrices.
For $R=\Z/Q\Z$, $Q\ge2$, one may take $C=R^\times$ precisely when
$Q=2,4,p^k$ or $2p^k$ ($p$ odd prime, $k\ge1$); for $Q=2^k$, $k\ge3$, the
cyclic factor of order $2^{k-2}$ gives density $1/2$.
For $R=\Z/Q\Z$, the determinants of maximal unit order $\lambda(Q)$,
Carmichael's primitive $\lambda$-roots, have conditional density
$\prod_{\ell\mid\lambda(Q)}(1-\ell^{-r_\ell})$, where $r_\ell$ counts
the cyclic factors of $R^\times$ whose $\ell$-primary component has
maximal exponent~\cite[Theorem~2.1]{Li98}.
The same primary-component count applies to every finite $R^\times$,
with $\lambda(Q)$ replaced by its exponent; multiplying by
$\prod_iG_i$ gives the unconditional density in
\eqref{eq:finite-ring-det}, with conditional density $3/4$ for $Q=8$.

\begin{proposition}[Rectangular matrices and row-scaling-invariant laws]
\label{prop:finite-ring-rectangular}
\index{Finite rings!rectangular matrices}
For $R,q_i$ as above, a uniform $m\times n$ matrix, $n\ge m\ge1$, has a
right inverse with probability
\[
 \prod_i\prod_{j=0}^{m-1}(1-q_i^{j-n}).
\]
If a law on $M_n(R)$ is invariant under multiplying the first row by every
unit, then, for $A\subseteq R^\times$,
\[
 \mathbb P(\det M\in A)=\mathbb P(M\textup{ invertible})
                       \frac{|A|}{|R^\times|}.
\]
\end{proposition}

\begin{proof}
A right inverse modulo $J$ lifts to $B$ with $MB\equiv I_m\pmod J$;
then $MB$ is invertible and $B(MB)^{-1}$ is a right inverse over $R$.
Equal lift counts and successive independent rows over each residue
field give the product. Unit row scaling equates all unit determinant
masses, proving the second formula.
\end{proof}
Left $\mathrm{SL}_n(R)$-invariance alone does not suffice: the uniform law
on $\mathrm{SL}_2(\mathbb F_3)$ has determinant identically $1$.

\paragraph*{The field ensembles}\phantomsection\label{par:field-ensembles}
The ring identities average over matrices in a fixed ring. We now
average determinant densities over three families of fields.
The three families are set side by side in \Cref{fig:field-ensembles}, which records in each case
the sampled and the fixed parameter, the local structure or comparison, and the limiting model.

% Diagram (design brief F10): Finite-field ensembles: what varies and what is independent?
\begin{figure}[!ht]
\centering
\begin{tikzpicture}[figdiagram]
% Grid (pt): three columns with common left/right edges, 20pt gutters, 384pt in all.
% Every lane hangs its boxes from one top line (\ya, \yb, \yc); the row arrows run
% 17pt (lane 3: 18.35pt) below it, through the centres of the Sample and Limit boxes.
\def\xa{0pt}\def\xaa{100pt}       % column 1: Sample
\def\xb{120pt}\def\xbb{237.5pt}   % column 2: Local structure
\def\xc{257.5pt}\def\xr{384pt}    % column 3: Limit
\def\wa{93pt}\def\wb{110.5pt}\def\wc{119.5pt}   % text widths (box = text + 7pt)
\def\xj{270pt}                    % vertical feed from R1 and junction dot
\def\ya{-41.5pt}\def\yb{-114.5pt}\def\yc{-193.9pt}   % lane tops
% ---- entry note and column headings
\node[fnote, anchor=north west] (n0) at (0,0) {$c_n(q)=\frac{\varphi(q-1)}{q-1}\prod_{j=1}^{n}(1-q^{-j})$: fraction of $M_n(\mathbb F_q)$ with determinant generating $\mathbb F_q^\times$ (\Cref{prop:finite-ring-det})};
\node[fhead, anchor=base] (h1) at (50pt,-22.5pt) {Sample};
\node[fhead, anchor=base] (h2) at (178.75pt,-22.5pt) {Local structure};
\node[fhead, anchor=base] (h3) at (320.75pt,-22.5pt) {Limit};
\draw[frule] (0,-26.5pt) -- (\xr,-26.5pt);
% ---- lane 1: fixed extension degree, varying prime base
\node[ftag, anchor=base west] (t1) at (0,\ya+6.5pt) {Fixed extension degree, varying prime base};
\node[fnote, anchor=base east] (pf) at (\xr,\ya+6.5pt) {\Cref{fig:information-reuse}};
\node[fnote, anchor=base east] (pi) at ($(pf.base west)+(-20pt,0)$) {information bounds for $X_m$: \Cref{cor:fixed-degree-information-quantization}};
\draw[fptr] (pi.east) -- (pf.west);
\node[finp, anchor=north west, text width=\wa, minimum height=35.5pt] (S1) at (\xa,\ya) {sampled: primes $p\le x$\\ fixed: degree $m\ge1$\\ $c_{n_x(p)}(p^m)$};
\node[fmech, anchor=north west, text width=\wb] (L1) at (\xb,\ya) {$\gcd(m,\ell-1)$ roots of $u^m=1$\\ in $(\Z/\ell\Z)^*$ for each prime $\ell$;\\ independent local indicators,\\ $\mathbb P(B_{\ell,m}=1)=\frac{\gcd(m,\ell-1)}{\ell-1}$\\ \figref{finite sets: CRT, \Cref{thm:sw}}};
\node[fres, anchor=north west, text width=\wc, minimum height=35.5pt] (R1) at (\xc,\ya) {$X_m=A_m\prod_{\ell\notin\mathcal E_m}(1-1/\ell)^{B_{\ell,m}}$\\[1.5pt] independent field-prime factors\\ \figref{\Cref{cor:fixed-degree-fields}}};
\draw[fuse] (\xaa,\ya-17.75pt) -- (\xb,\ya-17.75pt);
\draw[fuse] (\xbb,\ya-17.75pt) -- (\xc,\ya-17.75pt);
\node[fnote, anchor=north east, align=right] (na) at ($(R1.south east)+(0,-1.7pt)$) {probability one iff $\ell-1\mid m$:\\ finite set $\mathcal E_m$, product $A_m$};
\draw[frule] (0,\yb+15pt) -- (\xr,\yb+15pt);
% ---- lane 2: sparse prime divisors in the degree
\node[ftag, anchor=base west] (t2) at (0,\yb+6.5pt) {Sparse prime divisors in the degree};
\node[fnote, anchor=base west] (nr) at ($(t2.base east)+(9pt,0)$) {no upper restriction on $r(x)$};
\node[finp, anchor=north west, text width=\wa, minimum height=47.6pt] (S2) at (\xa,\yb) {sampled: primes $p\le x$\\ fixed: $d\ge1$; $r(x)$ varies\\ $\sum_{\substack{s\mid r(x)\\s\ {\rm prime}}}\frac{1+\log\log(3s)}{s}\to0$\\ $c_{n_x(p)}(p^{d\,r(x)})$};
\node[fmech, anchor=north west, text width=\wb, minimum height=47.6pt] (L2) at (\xb,\yb) {same-prime coupling:\\ mean logarithmic difference\\ $\frac1{\pi(x)}\sum_{p\le x}\log\frac{T_d(p)}{T_{dr}(p)}\to0$\\ \figref{\eqref{eq:sparse-degree-comparison}}};
\node[fnote, anchor=north] (nt) at ($(L2.south)+(0,-1.5pt)$) {$T_j(p)=\varphi(p^j-1)/(p^j-1)$};
% joint input: the coupling bound and Theorem 16.3 at m=d
\node[fjoin] (j) at (\xj,\yb-17pt) {};
\node[fres, anchor=north west, text width=\wc] (R2) at (\xc,\yb-35pt) {the same limit $X_d$\\ \figref{\Cref{cor:growing-prime-degree-fields}}};
\draw[fuse] (\xaa,\yb-17pt) -- (\xb,\yb-17pt);
\draw[ffeed] (\xbb,\yb-17pt) -- (j.center);
\draw[ffeed] (R1.south -| j) -- (j.center);
\draw[fuse] (j) -- (j |- R2.north);
\node[fhyp, anchor=west] (hj) at (\xj+2pt,\yb-5pt) {at $m=d$};
\draw[frule] (0,\yc+15pt) -- (\xr,\yc+15pt);
% ---- lane 3: fixed field, sampled degrees (no connector to lanes 1-2)
\node[ftag, anchor=base west] (t3) at (0,\yc+6.5pt) {Fixed field, sampled degrees};
\node[finp, anchor=north west, text width=\wa, minimum height=36.7pt] (S3) at (\xa,\yc) {sampled: degrees $d\le D$\\ fixed: prime power $q$\\ $c_{n_D(d)}(q^d)$};
\node[fmech, anchor=north west, text width=\wb] (L3) at (\xb,\yc) {$o_\ell=\operatorname{ord}_\ell(q)$, $\ell\nmid q$:\\ $g_q(m)=\sum_{o_\ell=m}\log\frac{\ell}{\ell-1}$\\ $\log\frac{q^d-1}{\varphi(q^d-1)}=\sum_{m\mid d}g_q(m)$\\ \figref{growth bound: \Cref{lem:fixed-q-order-weights};}\\ \figref{\Cref{thm:profinite-divisor-law}}};
\node[fres, anchor=north west, text width=\wc, minimum height=36.7pt] (R3) at (\xc,\yc) {$\mathcal X_q=\exp\bigl(-\sum_{m\ge1}g_q(m)\mathbf1_{m\mid U}\bigr)$\\ $U$ Haar distributed on $\widehat{\mathbb Z}$\\ \figref{\eqref{eq:degree-law}; \Cref{thm:fixed-q-degree}}};
\draw[fuse] (\xaa,\yc-18.35pt) -- (\xb,\yc-18.35pt);
\draw[fuse] (\xbb,\yc-18.35pt) -- (\xc,\yc-18.35pt);
% ---- closing note: the central distinction of lane 3
\node[fnote, font=\figfont, text=black, anchor=north west] (n3) at ($(\xa,0 |- L3.south)+(0,-5pt)$) {$\mathbb P(u\mid U,\ v\mid U)=1/\operatorname{lcm}(u,v)$: order indicators dependent; prime valuations $v_r(U)$ independent};
\end{tikzpicture}
\caption{Three arithmetic ensembles for primitive-determinant densities. A varying prime base
at fixed degree gives independent prime activations; a degree factor $r(x)$ with sparse prime
divisors preserves the fixed-degree limit $X_d$ by the
comparison~\eqref{eq:sparse-degree-comparison}. Sampling degrees over a fixed field gives a
profinite divisor model with dependent order indicators. The dimension arrays
$n_x(p),n_D(d)\in\Z_{\ge1}\cup\{\infty\}$ ($n=\infty$: infinite-product density) are arbitrary
and each convergence in distribution is uniform in them; the same limits hold for the totient
ratios $\varphi(Q-1)/(Q-1)$, $Q=p^m,\,p^{d\,r(x)},\,q^d$.}
\label{fig:field-ensembles}
\end{figure}

\begin{itemize}
\item \emph{Fixed extension degree, varying prime base.}
In \Cref{cor:fixed-degree-fields}, $m\ge1$ is fixed and primes $p\le x$
are averaged. The limit $X_m$ has independent field-prime factors.
The logarithmic empirical laws converge in every finite Wasserstein
order, and every fixed real Mellin moment of the ratios converges.
Its raw law has a Liouville carrier and Hausdorff dimension zero.
The laws of $-\log X_m$ and of every fixed power $X_m^s$, $s\ne0$,
are Rajchman and give full mass to $\mathfrak N$, with Fourier transforms
$O_{m,s}(\rho^{\tau(m)})$ (with no $s$ in the logarithmic case).
The periodic estimates obtained from \Cref{thm:sw} are ineffective.
Further geometric, information, joint-law and endpoint results are
in \Cref{ssec:fields-fixed-degree}.
\item \emph{Sparse prime divisors in the degree.}
In \Cref{cor:growing-prime-degree-fields}, $d\ge1$ is fixed,
$r=r(x)$ satisfies \eqref{eq:sparse-degree-condition}, and the average
is over primes $p\le x$ at degree $dr(x)$. There is no upper restriction
on $r(x)$; arbitrary powers of a growing prime are included, and the
limit is $X_d$.
\item \emph{Fixed field, sampled degrees.}
In \Cref{thm:fixed-q-degree}, the prime power $q$ is fixed
and degrees $1\le d\le D$ are averaged. The limit is the Haar model
$\mathcal X_q=e^{-\mathcal S_q}$, with dependent order indicators.
The general framework for fixed nonnegative weights satisfying
\eqref{eq:profinite-weight-growth} is given in
\Cref{thm:profinite-divisor-law,prop:profinite-divisor-mellin};
\Cref{thm:fixed-q-degree} proves continuity and support for this model.
Its empirical Mellin transforms and all fixed derivatives converge
at the rate \eqref{eq:degree-empirical-mellin}, uniformly in the
matrix-dimension arrays (\Cref{prop:fixed-q-degree-moments}).
\Cref{prop:fixed-q-degree-concentration} gives endpoint estimates,
including \eqref{eq:degree-endpoint}. Constants and thresholds may
depend on $q$.
\end{itemize}

\KeepHeadingSpace{8\baselineskip}
\section{Fixed extension degree and varying prime base}
\label{ssec:fields-fixed-degree}

\begin{theorem}[Fields of fixed extension degree]
\label{cor:fixed-degree-fields}
\index{Finite fields!fixed extension degree}
Fix an integer $m\ge1$, and put
\[
 \mathcal E_m:=\{\ell\text{ prime}:\ell-1\mid m\},\qquad
 A_m:=\prod_{\ell\in\mathcal E_m}(1-1/\ell),\qquad
 q_{\ell,m}:=\frac{\gcd(m,\ell-1)}{\ell-1}.
\]
Let $(B_{\ell,m})_{\ell\notin\mathcal E_m}$ be independent Bernoulli
variables of probabilities $q_{\ell,m}$, and define
\begin{equation}\label{eq:fixed-degree-field-law}
 X_m:=A_m\prod_{\ell\notin\mathcal E_m}(1-1/\ell)^{B_{\ell,m}}.
\end{equation}
\begin{enumerate}[label=\textup{(\alph*)},leftmargin=*]
\item \emph{Convergence.}
Uniformly in real thresholds and dimension arrays
$n_x(p)\in\mathbb Z_{\ge1}\cup\{\infty\}$, the empirical distribution
functions of $c_{n_x(p)}(p^m)$ over primes $p\le x$ converge to that of
$X_m$. The same limit holds for $\varphi(p^m-1)/(p^m-1)$.
\item \emph{Raw law.}
The law of $X_m$ is singular continuous, has support $[0,A_m]$ and
Hausdorff dimension zero, and gives full mass to $\mathcal L$.
\item \emph{Logarithmic law.}
The law of $-\log X_m$ has Hausdorff dimension zero, is Rajchman, gives
full mass to $\mathfrak N$, and has Fourier transform
$O_m(\rho(\tau)^{\tau(m)})$. Constants and thresholds may depend on
$m$; effectivity is not asserted for this refinement.
\item \emph{Rational approximation.}
For the actual reduced prefixes through the active primes outside
$\mathcal E_m$, including $A_m$, the approximation dichotomy
\eqref{eq:scaled-exclusion}--\eqref{eq:prefix-liminf} holds with
$\kappa=\tau(m)$ for every function allowed in \Cref{thm:intensity}.
In particular, for fixed $C>0$ and $\beta\in\R$, the boundary in
\Cref{cor:boundaries} is
$\psi(q)=C(\log\log q)^{1/\tau(m)}
(\log\log\log q)^\beta$, with divergence exactly when
$\tau(m)\beta\le1$.
\end{enumerate}
\end{theorem}

\begin{proof}
The cyclic group $(\Z/\ell\Z)^*$ contains exactly $\gcd(m,\ell-1)$
roots of $u^m=1$. For each fixed finite prime set, the Chinese remainder
theorem and \Cref{thm:sw} therefore give the independent local indicators
in~\eqref{eq:fixed-degree-field-law}; the finitely many prime bases in
that set do not affect the limit. The probability-one factors are
exactly those in the finite set $\mathcal E_m$.

For $y<\ell\le\sqrt x$, Brun--Titchmarsh in these reduced root classes
gives
\[
 \frac1{\pi(x)}\sum_{p\le x}
 \sum_{\substack{y<\ell\le\sqrt x\\\ell\mid p^m-1}}\omega_\ell
 \ll_m\sum_{\ell>y}\frac{\omega_\ell}{\ell-1}\ll_m y^{-1}.
\]
Since $p^m-1<x^m$, there are at most $2m$ distinct prime divisors
exceeding $\sqrt x$, and their total weight is $O_m(x^{-1/2})$.
The model has expected logarithmic tail $O_m(y^{-1})$ as well.
Truncation and the finite-coordinate limits prove convergence in law
to $-\log X_m$, and hence ratio convergence.

For every sufficiently large prime,
$1/(\ell-1)\le q_{\ell,m}\le m/(\ell-1)<1/2$.
The sum of these probabilities diverges, so \Cref{lem:levy-atomless}
gives atomlessness. Both local states have positive probability outside
$\mathcal E_m$, the jumps tend to zero, and
$\sum_{\ell\notin\mathcal E_m}\omega_\ell=\infty$.
Thus \Cref{lem:phi-even-dense,lem:supp_conv} gives support $[0,\infty)$ for the free
logarithmic sum, and support $[0,A_m]$ for $X_m$.

For the Diophantine conclusion choose $\ell_0>\max\mathcal E_m$ and
put all earlier factors into the independent positive rational head.
The elementary divisor identity and partial summation in
\Cref{thm:sw}, for each fixed divisor $d\mid m$, give
\[
\begin{aligned}
 \sum_{\ell_0<\ell\le y}\frac{\gcd(m,\ell-1)}\ell
 &=\sum_{d\mid m}\varphi(d)
   \sum_{\substack{\ell_0<\ell\le y\\\ell\equiv1\pmod d}}\frac1\ell\\
 &=\tau(m)\log\log y+C_m+o(1),
\end{aligned}
\]
where $\tau(m)$ is the number of positive divisors of $m$ and $C_m$ is
a constant. Since
$q_{\ell,m}=\gcd(m,\ell-1)/\ell+O_m(\ell^{-2})$, this yields
\[
 \sum_{\ell_0<\ell\le y}-\log(1-q_{\ell,m})
 =\tau(m)\log\log y+C'_m+o(1).
\]
The expected denominator logarithm of a local factor is
$q_{\ell,m}\log\ell$, whose sum up to $y$ is $O_m(\log y)$, and
the expected absolute logarithmic tail is $O_m(z^{-1})$.
Hence \Cref{lem:rational-blocks} applies with
$c_\ast=\tau(m)$ and $\beta=1$. Atomlessness and
\Cref{lem:joint-rational-carrier} give full mass to $\mathcal L$ and
dimension zero for both the ratio and logarithmic laws.
The same hazard asymptotic verifies \eqref{eq:cumulative-intensity}.
Apply \Cref{thm:intensity,cor:boundaries} with $s=-1$ and
$\kappa=\tau(m)$, absorbing the finitely many earlier factors into the
rational head. Its finite-alteration assertion preserves the actual
prefixes through the active primes outside $\mathcal E_m$.

For the Fourier assertion, omit the finite set of primes dividing $m$
and the forced primes $\mathcal E_m$. On the remaining primes put
$a_m(\ell)=\gcd(m,\ell-1)$, viewed as a function of the unit class
of $\ell$ modulo $m$. The divisor identity and character orthogonality
give
\begin{equation}\label{eq:fixed-degree-character-identity}
 \gcd(m,\ell-1)
 =\sum_{d\mid m}\varphi(d)\mathbf1_{\ell\equiv1\pmod d}
 =\sum_{d\mid m}\sum_{\chi\bmod d}\chi(\ell).
\end{equation}
The reduction map $(\mathbb Z/m\mathbb Z)^\times\to
(\mathbb Z/d\mathbb Z)^\times$ is surjective for $d\mid m$:
this follows prime by prime and then from the Chinese remainder theorem.
Thus each of its fibers has size $\varphi(m)/\varphi(d)$, and the
mean of $a_m$ on the unit classes modulo $m$ is $\tau(m)$.
The convention for $m=1$ is the one-element unit group and $a_1=1$.

For each retained prime, the squared modulus of the Bernoulli factor is
\[
 1-2q_{\ell,m}(1-q_{\ell,m})
                 (1-\cos(\tau\omega_\ell)),\qquad
 2q_{\ell,m}(1-q_{\ell,m})
       =\frac{2a_m(\ell)}\ell+O_m(\ell^{-2}).
\]
Set $\lambda_\ell=2q_{\ell,m}(1-q_{\ell,m})$ on these primes and
zero on the finite omitted set. These nonnegative weights satisfy
\Cref{thm:periodic-weight-dissipation} with $D=m$ and
$\kappa=\tau(m)$. The inequality $1-u\le e^{-u}$, independence,
and convergence of the logarithmic sum give
\[
 \left|\mathbb E e^{i\tau(-\log X_m)}\right|^2
 \le\exp\left(-\sum_\ell\lambda_\ell
                             (1-\cos(\tau\omega_\ell))\right).
\]
The finite independent head has characteristic function of modulus at
most one. Hence the Fourier transform is $O_m(\rho(\tau)^{\tau(m)})$.
Apply \Cref{lem:iterlog-normality}.
Finally, uniformly in $n\in\mathbb Z_{\ge1}\cup\{\infty\}$,
\[
 P(p^m)\le
 \frac{c_n(p^m)}{\varphi(p^m-1)/(p^m-1)}\le1,
 \qquad 1-P(p^m)=O(p^{-m}).
\]
Discarding primes below a growing cutoff and using continuity of the
limit distribution, the distribution-function comparison in the proof of
\Cref{prop:head-tail-cdf} gives convergence uniformly in thresholds
and dimension arrays.
\end{proof}

\begin{corollary}[Singularity at distinct field degrees]
\label{cor:fixed-degree-field-singularity}
\index{Finite fields!fixed-degree mutual singularity}
The laws of $X_m,X_n$ in \Cref{cor:fixed-degree-fields} are mutually
singular whenever $m\ne n$, including degrees with $\tau(m)=\tau(n)$.
Their images under either map $u\mapsto\log u$ or $u\mapsto-\log u$
are also mutually singular.
\end{corollary}

\begin{proof}
Restore the forced coordinates by
putting $B_{p,m}=1$ at odd $p\in\mathcal E_m$. Since $2\in\mathcal E_m$,
\eqref{eq:fixed-degree-field-law} becomes
\[
 X_m=\frac12\prod_{p\ge3}(1-1/p)^{B_{p,m}},
 \qquad \mathbb P(B_{p,m}=1)=\frac{\gcd(m,p-1)}{p-1}.
\]
The hazard estimate in the proof of \Cref{cor:fixed-degree-fields}
verifies \eqref{eq:cumulative-intensity} on a tail; all earlier
probabilities, including the forced coordinates, are allowed by
\Cref{prop:canonical-decoder-mas08}. Thus the same map
$\mathcal D_{1/2,-1}$, with $p_0=3$, recovers their coordinates for
every fixed $m$.
For $L=\operatorname{lcm}(m,n)$ put
\[
 A_L(y)=\sum_{\substack{3\le p\le y\\p\equiv1\pmod L}}\frac1{p-1}.
\]
Partial summation in \Cref{thm:sw} gives
$A_L(y)\sim\varphi(L)^{-1}\log\log y\to\infty$.
For degree $m$, the count
$N_m(y)=\sum_{3\le p\le y,\ p\equiv1\pmod L}B_{p,m}$ has mean
$mA_L(y)$ and variance at most $mA_L(y)$. Choose the first cutoff
$y_k$ at which $A_L(y_k)\ge k^2$; its bounded jumps give
$A_L(y_k)=k^2+O(1)$. Chebyshev and Borel--Cantelli yield
$N_m(y_k)/A_L(y_k)\to m$ almost surely. Monotonicity of both counts
and $A_L$, and $A_L(y_{k+1})/A_L(y_k)\to1$, extend this to every
$y\to\infty$. Under degree $n$ the same argument gives limit $n$.
The Borel set on which the decoded coordinates have normalized count
limit $m$ therefore has full $X_m$-measure and zero $X_n$-measure.
This proves mutual singularity for all $m\ne n$, and injectivity of
both logarithmic maps gives their stated images.
\end{proof}

\begin{corollary}[Fourier decay of the raw fixed-degree law]
\label{cor:fixed-degree-raw-fourier}
Fix an integer $m\ge1$ and a real $s\ne0$, and let $X_m$ be as in
\eqref{eq:fixed-degree-field-law}. Then
\begin{equation}\label{eq:fixed-degree-raw-fourier}
 \left|\mathbb E e^{itX_m^s}\right|\ll_{m,s}\rho(t)^{\tau(m)}
 \qquad(|t|\to\infty).
\end{equation}
In particular the law of $X_m^s$ is Rajchman and gives full mass to
$\mathfrak N$, and $\mathbb P(X_m\in\mathfrak N\cap\mathcal L)=1$.
The constant and threshold may depend on $m,s$; effectivity is not asserted.
\end{corollary}

\begin{proof}
Let $p_0$ exceed $2m$ and every prime in $\mathcal E_m$ and dividing
$m$. All factors at primes at most $p_0$, including the deterministic
factor $A_m$, form a positive finite head $H_0$, independent of the
remaining coordinates and bounded above and away from zero. For
$\ell>p_0$ take $\xi_\ell=B_{\ell,m}\in\{0,1\}$ and
$\vartheta_1(z)=1-z$. Put $D=m$ and
$a_1(b)=\gcd(m,b-1)$ on the unit classes modulo $m$, extended by zero
off them. For $m=1$ the unit group has one element and $a_1=1$.
The exact probabilities and odds are
\[
 \mathbb P(\xi_\ell=1)=q_{\ell,m}
       =\frac{a_1(\ell)}\ell+O_m(\ell^{-2}),\qquad
 \mathbb P(\xi_\ell=0)=1-q_{\ell,m}\ge1-2m/\ell,
\]
\[
 \frac{q_{\ell,m}}{1-q_{\ell,m}}
       =\frac{a_1(\ell)}{\ell-1-a_1(\ell)}
       =\frac{a_1(\ell)}\ell+O_m(\ell^{-2}).
\]
The averaging calculation following
\eqref{eq:fixed-degree-character-identity} gives mean $\tau(m)$ for $a_1$.

The product has a positive almost-sure limit because
\[
 \sum_{\ell\notin\mathcal E_m}
       \mathbb E(B_{\ell,m}\omega_\ell)
 \le\sum_\ell\frac{m\omega_\ell}{\ell-1}<\infty.
\]
Furthermore $0<X_m\le A_m$, and for every fixed $u>0$ independence
and monotone convergence give
\[
 \mathbb EX_m^{-u}=A_m^{-u}
  \prod_{\ell\notin\mathcal E_m}
    \{1+q_{\ell,m}[(1-1/\ell)^{-u}-1]\}<\infty.
\]
Indeed $q_{\ell,m}\le m/(\ell-1)$ and
$0\le(1-1/\ell)^{-u}-1\le u2^{u+1}/\ell$, so the logarithms
of the product factors are summable. Thus
$\mathbb E(X_m^s+X_m^{-s})<\infty$ for every fixed $s\ne0$.
The function $1-z$ is positive and analytic on a neighbourhood of
$[0,1/p_0]$, with derivative $-1$ at zero. All hypotheses of
\Cref{thm:periodic-raw-product-fourier} now hold with one distinguished
state and $\kappa=\tau(m)$, proving \eqref{eq:fixed-degree-raw-fourier}
and normality. The Liouville assertion for $X_m$ is that of
\Cref{cor:fixed-degree-fields}.
\end{proof}

\begin{remark}[Constant-weight Fourier bounds]
Put $g=\gcd(m,2)$. For all sufficiently large odd $\ell$,
$\gcd(m,\ell-1)\ge g$ and
$2q_{\ell,m}(1-q_{\ell,m})\ge g d_\ell-O_m(\ell^{-2})$.
The local characteristic factors, $1-u\le e^{-u}$ and
\eqref{eq:vk-dissipation} therefore give
\[
 \left|\mathbb E e^{i\tau(-\log X_m)}\right|
 \le C_m e^{-gS(\tau)/2}\ll_m\rho(\tau)^g,
\]
with the summable error and finitely many omitted primes absorbed in $C_m$.

For the raw law, use the cutoff $p_0$ of the preceding proof. For
$\ell>p_0$, realize $B_{\ell,m}=\mathbf1_{\{U_\ell^m=1\}}$ with independent
uniform $U_\ell\in\mathbb F_\ell^\times$. The $g$ roots of $u^g=1$
are roots of $u^m=1$; designate them as state $1$, nonroots as state $0$,
and the remaining roots as an additional state with factor $1-1/\ell$.
The distinguished probability is $g/(\ell-1)=g/\ell+O_m(\ell^{-2})$,
and the zero-state probability is at least $1-2m/\ell$.
With the finite head, convergence and moment checks above,
\Cref{thm:raw-product-fourier} gives
$|\mathbb E e^{itX_m^s}|\ll_{m,s}\rho(t)^g$ for every fixed $s\ne0$.
For $s=1$, the constant and a sufficient threshold in this bound are
effective for each fixed $m$: all local parameters are effective and
\[
 \mathbb E(X_m+X_m^{-1})\le1+A_m^{-1}e^m,
\]
since $2\in\mathcal E_m$ and
$\sum_{\ell\ge3}(\ell-1)^{-2}<1$.
\end{remark}

Fix $m\ge1$ and keep the notation $X_m$, $A_m$, $\mathcal E_m$, $q_{\ell,m}$,
$B_{\ell,m}$ from the beginning of this section. Write $Z_m=-\log X_m$, and
choose $L$ so large that every forced prime in $\mathcal E_m$, including $2$,
is at most $L$ and $q_{\ell,m}<1/2$ for $\ell>L$. Thus
\[
 Z_m=H_m+\sum_{\ell>L}B_{\ell,m}\omega_\ell,
 \qquad
 H_m=-\log A_m+
       \sum_{\substack{\ell\le L\\\ell\notin\mathcal E_m}}
          B_{\ell,m}\omega_\ell.
\]
The finite head $H_m$ is atomic and independent of the remaining coordinates.
For these coordinates the bounds and cumulative hazard established in the
proof of \Cref{cor:fixed-degree-fields} give
\begin{equation}\label{eq:fixed-degree-geometry-inputs-mas19}
\begin{gathered}
 \frac1{\ell-1}\le q_{\ell,m}\le\frac m{\ell-1},\\
 \Theta_m(y):=\sum_{L<\ell\le y}-\log(1-q_{\ell,m})
   =\tau(m)\log\log y+C_{m,L}+o(1),\\
 \sum_{\ell>y}\mathbb E(B_{\ell,m}\omega_\ell)\ll_m y^{-1}.
\end{gathered}
\end{equation}
The coefficient $\tau(m)$ belongs to the summed hazard; the individual
probabilities need not equal $\tau(m)/\ell+O_m(\ell^{-2})$.

\Cref{lem:periodic-prime-mass} gives the fixed-progression estimate
\eqref{eq:fixed-degree-progression-mertens} with its strong remainder.

\begin{proposition}[Local, packing and R\'enyi dimensions at fixed degree]
\label{prop:fixed-degree-geometry-mas19}
Fix $m\ge1$ and let $\lambda$ be the law of either $X_m$ or
$-\log X_m$ from \Cref{cor:fixed-degree-fields}. Then
\[
 (\underline d_\lambda(x),\overline d_\lambda(x))=(0,1)
 \quad\text{for }\lambda\text{-almost every }x,
 \qquad
 \dim_P E=1\quad\text{whenever }\lambda(E)>0
\]
for every Borel set $E$. Moreover,
\begin{equation}\label{eq:fixed-degree-renyi-mas19}
 \underline D_\theta(\lambda)=\overline D_\theta(\lambda)=
 \begin{cases}
 1,&0<\theta<1,\\
 0,&\theta>1.
 \end{cases}
\end{equation}
\end{proposition}

\begin{proof}
Let $Z_m$, $L$, $H_m$ and $\Theta_m$ be as introduced before the proposition,
in the paragraph containing \eqref{eq:fixed-degree-geometry-inputs-mas19}.
The Bernoulli entropies of the coordinates $B_{\ell,m}$, $\ell>L$, satisfy
\[
 -q_{\ell,m}\log q_{\ell,m}
 -(1-q_{\ell,m})\log(1-q_{\ell,m})
 \ll_m\frac{\log\ell}{\ell}.
\]
Summing gives expected coordinate information $O_m(\log y)$.
Consequently \Cref{thm:dimension-zero-criterion} applies with
$c_\ast=\tau(m)$, $\alpha=0$ and $\beta=1$ after deletion of the
finite head. Restoring that head preserves a Hausdorff-zero full
carrier.

We next construct the diffuse submeasures needed for the other
dimensions. Fix $0<s<1$ and choose the exponents in
\eqref{eq:selected-prime-exponents}. Apply \Cref{lem:selected-prime-blocks} to
$\xi_\ell=B_{\ell,m}$, taking $h$ small enough that all block primes
exceed $L$. Here $z_\ell=1-q_{\ell,m}>1/2$, and the pointwise bound in
\eqref{eq:fixed-degree-geometry-inputs-mas19} gives the selected-state odds
$u_{\ell,m}:=q_{\ell,m}/(1-q_{\ell,m})=O_m(\ell^{-1})$.
Moreover,
$u_{\ell,m}=q_{\ell,m}+O_m(\ell^{-2})$ and
$-\log(1-q_{\ell,m})=q_{\ell,m}+O_m(\ell^{-2})$.
Differencing the summed hazard in
\eqref{eq:fixed-degree-geometry-inputs-mas19} at $h^{-b_j}$ and
$h^{-a_j}$ gives its block limit $\tau(m)\log(b_j/a_j)$.
Since $\sum_{\ell>h^{-a_j}}\ell^{-2}=o(1)$, the block limits and
event probability in \Cref{lem:selected-prime-blocks} are therefore
\[
\begin{aligned}
 W_{j,h}:=\sum_{\ell\in\mathcal I_{j,h}}u_{\ell,m}
   &\longrightarrow \tau(m)\log(b_j/a_j),\\
 \prod_{\ell\in\mathcal I_{j,h}}(1-q_{\ell,m})
   &\longrightarrow(a_j/b_j)^{\tau(m)},\\
 \mathbb P(E_h)=\prod_{j=1}^k
      \left(W_{j,h}\prod_{\ell\in\mathcal I_{j,h}}(1-q_{\ell,m})\right)
   &\longrightarrow
   c_{m,s}:=\prod_{j=1}^k
       \tau(m)\log(b_j/a_j)(a_j/b_j)^{\tau(m)}>0.
\end{aligned}
\]
These odds and block limits verify every selected-block hypothesis of
\Cref{cor:prime-block-diffuse}, with $\Xi=H_m$, $H=H_m$ and $D=\Omega$.
Apply it with $V=0$ and $F(u,v)=u$, respectively $F(u,v)=e^{-u}$.
It gives submeasures $\sigma_{s,h}$ and $\widetilde\sigma_{s,h}$ for
$Z_m$ and $X_m=e^{-Z_m}$ satisfying
\eqref{eq:dominated-diffuse-submeasures}. For the exponential image,
that corollary restricts to a compact interval, where the absolute
derivative of $u\mapsto e^{-u}$ is bounded away from zero.

For each fixed $t>0$,
\[
 \mathbb E e^{tB_{\ell,m}\omega_\ell}
 =1+q_{\ell,m}(e^{t\omega_\ell}-1)=1+O_{m,t}(\ell^{-2}).
\]
Independence and the bounded finite head give
$\mathbb E e^{tZ_m}<\infty$. In particular, for every
$0<\theta<1$ the logarithmic law satisfies the required unit-cell
summability, since
\[
 \mathbb P(Z_m\in[j,j+1))^\theta
 \le (\mathbb E e^{tZ_m})^\theta e^{-t\theta j}
 \quad(j\ge0).
\]
The raw law is supported in $[0,A_m]$, so its unit-cell sum is finite
as well. The diffuse construction and
\Cref{prop:dominated-diffuse-submeasures}\textup{(a)} prove
\eqref{eq:fixed-degree-renyi-mas19} for $0<\theta<1$ for both laws.
For $\theta>1$, the head $H_m$ has a positive-mass atom, the tail
has the power bound in \eqref{eq:fixed-degree-geometry-inputs-mas19},
and $\Theta_m(y)=o(\log y)$. Apply the general part of
\Cref{prop:renyi-collapse}\textup{(b)} with $T(u)=u$ and
$T(u)=e^{-u}$, both Lipschitz near that head atom. This proves
both zero dimensions.

Finally, \Cref{prop:dominated-diffuse-submeasures}\textup{(c)} makes
the essential supremum of the upper local dimension of the logarithmic
law equal to one. The independent atomic-series zero--one statement
\Cref{lem:atomic-series-dimensions} makes that dimension constant
almost everywhere and makes every positive-mass Borel set have packing
dimension one. Its Hausdorff-zero full carrier gives lower local
dimension zero by the same lemma. The exponential map is bi-Lipschitz
on every $[0,K]$ and $Z_m<\infty$ almost surely; hence it preserves
the local exponents. If a raw Borel set has positive mass, its inverse
image has a positive-mass intersection with some $[0,K]$, and that
intersection has packing dimension one. Bi-Lipschitz invariance on
this interval gives the same conclusion for the raw set.

\end{proof}

\begin{corollary}[Information and quantization at fixed degree]
\label{cor:fixed-degree-information-quantization}
Fix $m\ge1$ and let $\lambda$ be the law of either $X_m$ or $-\log X_m$
from \Cref{cor:fixed-degree-fields}.
For every fixed $r>0$, $\lambda$ satisfies \eqref{eq:quantization} and
\eqref{eq:rate-distortion}. All mesh entropies of $\lambda$ are finite,
the laws of $X_m$ and $-\log X_m$ have the same lower and the same upper
information dimension, and
\begin{equation}\label{eq:fixed-degree-information-mas19}
 \begin{gathered}
 1-2^{-\tau(m)}\le c^{\mathrm{low}}_{\tau(m)}
 \le\underline d_1(\lambda)\le\overline d_1(\lambda)\le d_{\tau(m)},\\
 d_{\tau(m)}<1-\frac{e^{-\gamma\tau(m)-1}}{\tau(m)!}.
 \end{gathered}
\end{equation}
Here $c^{\mathrm{low}}_{\tau(m)}$ is the functional
\eqref{eq:convolution-lower-functional} and $d_{\tau(m)}$ is the coding constant
of \eqref{eq:coding-functional}, both at the integer $\tau(m)$; for $m=1$
they are $c_L$ and $d_*$. In this scalar case the lower bound
improves to $c_L^+$ of \eqref{eq:scalar-strengthened-lower}.
\end{corollary}

\begin{proof}
Keep $Z_m=-\log X_m$, the cutoff $L$, the independent finite-valued
head $H_m$ and the tail probabilities $q_{\ell,m}$ from
\eqref{eq:fixed-degree-geometry-inputs-mas19}.
The submeasures $\sigma_{s,h}$ and $\widetilde\sigma_{s,h}$ constructed
in the proof of \Cref{prop:fixed-degree-geometry-mas19} verify \eqref{eq:dominated-diffuse-submeasures} for the logarithmic
and raw laws, respectively. The logarithmic law has every positive
moment, since $Z_m\ge0$ and $\mathbb E e^{tZ_m}<\infty$ for every $t>0$;
the raw law is supported in $[0,A_m]$.
\Cref{prop:dominated-diffuse-submeasures}\textup{(b)} therefore gives
\eqref{eq:quantization} for both laws and every $r>0$.
Since $\mathbb EZ_m<\infty$, \Cref{lem:transfers}\textup{(i)} gives finite
mesh entropies and
$|H_h(X_m)-H_h(Z_m)|=O_m(1)$ for $0<h\le1$, by applying its exponential
comparison to $-Z_m$ and using the reflection bound. Thus both the lower
and upper information dimensions agree between the two laws.
For either law $\lambda$, finite mesh entropy allows us to apply
\Cref{lem:mesh-entropy-rate-distortion}, giving
\[
 H_h(\lambda)-C_r\le R_{\lambda,r}(h^r)\le H_h(\lambda),
\]
where $C_r$ depends only on $r$. Dividing by $\log(1/h)$ and taking the
lower and upper limits separately proves \eqref{eq:rate-distortion}.

For the information bounds, note that
\[
 \gcd(m,\ell-1)
 =\sum_{d\mid m}\varphi(d)\mathbf1_{\ell\equiv1\pmod d},
 \qquad
 0\le q_{\ell,m}-\frac{\gcd(m,\ell-1)}\ell
 \le\frac{m}{\ell(\ell-1)}.
\]
The latter correction is summable, with tail $O_m(y^{-1})$.
With $c'_m:=\min_{d\mid m}c_d>0$, summing \eqref{eq:fixed-degree-progression-mertens} over
the fixed divisors of $m$ therefore gives a constant $\mathfrak b_{m,L}$
such that
\[
 \sum_{L<\ell\le y}q_{\ell,m}
 =\tau(m)\log\log y+\mathfrak b_{m,L}
   +O_m\!\left(\sqrt{\log y}\,e^{-c'_m\sqrt{\log y}}+y^{-1}\right).
\]
The remainder is $o(1/\log y)$, so
\eqref{eq:real-intensity-mertens-mas13} holds with intensity $\tau(m)$.
The envelope \eqref{eq:real-intensity-envelope-mas13} holds with constant
$2m$ by \eqref{eq:fixed-degree-geometry-inputs-mas19}.
Apply \Cref{thm:real-intensity-mas13} to one copy, with coefficient $1$,
branch $\omega_\ell=\log(\ell/(\ell-1))$, cutoff $L$, probabilities
$q_{\ell,m}$ and the finite-valued independent head $H_m$.
Its head has finite entropy and finite absolute mean, and its tail
probabilities are less than $1/2$ by the choice of $L$.
Here $Z=Z_m$ and $e^{-Z}=X_m$. The theorem gives the inner bounds in
\eqref{eq:fixed-degree-information-mas19}; its single-copy lower
functional at intensity $\tau(m)$ is $c^{\mathrm{low}}_{\tau(m)}$.
The pair, cluster and triple terms are nonnegative, as in the proof of
\Cref{cor:convolution-information}, giving the lower outer inequality.
For $m=1$, $X_1$ is the shifted-prime totient ratio and
\Cref{thm:shannon-dimension} gives the strengthened scalar bound
for $X_1$ and its logarithmic law.
Equation~\eqref{eq:coding-deficit}, with
$\Gamma(\tau(m)+1)=\tau(m)!$, gives the strict upper inequality.
\end{proof}

The envelope \eqref{eq:real-intensity-envelope-mas13} neither fixes the
intensity nor implies \eqref{eq:real-intensity-mertens-mas13}.
Together, these two conditions are incomparable with the summable
perturbation hypothesis of \Cref{cor:general-bias-fourier}.
For $O(p^{-1})$ probabilities, after a finite cutoff ensuring they are
at most $1/2$, the cumulative hazard and probability sum differ by a
constant plus $O(x^{-1})$. The remainders $O(1)$ in
\eqref{eq:cumulative-intensity}, constant-plus-$o(1)$ in
\eqref{eq:fixed-degree-geometry-inputs-mas19}, and
constant-plus-$o(1/\log x)$ in \eqref{eq:real-intensity-mertens-mas13}
are therefore successively stronger.
The preceding proof verifies the last for fixed degree, although
$q_{\ell,m}=\tau(m)/\ell+O_m(\ell^{-2})$ holds only for $m\le2$.
\Cref{fig:information-reuse} places the application of
\Cref{thm:real-intensity-mas13} in the preceding proof beside those in
\Cref{sec:information-dimensions}, to independent sums and through them
to the basic laws.

\paragraph*{Cyclotomic specialization}
In \Cref{def:polynomial-root-model}, take $F_j(T)=T^{m_j}-1$.
Its distinct irreducible factors are $\Phi_a$ for $a\mid m_j$, so
$r_j=\tau(m_j)$ and the factor union is indexed by
$\mathcal D_{\boldsymbol m}=\bigcup_j\{a:a\mid m_j\}$.
Consequently the denominator in \Cref{cor:polynomial-minimal-head-corners}
is
$\prod_{a\in\mathcal D_{\boldsymbol m}}
       \max_{j:a\mid m_j}\log(1/\varepsilon_j)$.
The field normalization can restore a jointly admissible endpoint:
at each prime choose a generator of $\mathbb F_\ell^\times$.
It activates coordinate $j$ exactly when $\ell-1\mid m_j$.
All other head states are coordinatewise at least this forced vector,
so this vector is the unique minimal head atom, of mass
\[
 w_b=\prod_{\ell\le b}\alpha_{\ell,\boldsymbol m}>0,
 \quad
 \alpha_{\ell,\boldsymbol m}
 =P\{B_{\ell,j}=0\text{ for every }j
                       \text{ with }\ell-1\nmid m_j\}.
\]
For $b$ beyond the exceptional and deterministic primes, its coordinates
are $-\log A_{m_j}$, and the exact product definitions give
$w_bK_b=\kappa_{\boldsymbol m}$ of
\Cref{cor:joint-degree-endpoint}. This identifies both the denominator
and the constant after head restoration, for arbitrarily disparate
widths.

The next corollary gives the cyclic-subgroup intersection probabilities,
the uniform matrix-dimension correction and uniform joint
distribution-function convergence. The following endpoint results give the explicit scalar
product \eqref{eq:kappa-degree} and joint formula
\eqref{eq:joint-degree-endpoint}.

\begin{corollary}[Joint laws at fixed extension degrees]
\label{cor:joint-fixed-degree-fields}
\index{Finite fields!joint extension degrees}
Fix $m_1,\ldots,m_d\ge1$. Independently for each prime $\ell$, let
$U_\ell$ be uniform on $\mathbb F_\ell^\times$, and couple the scalar
limits of \Cref{cor:fixed-degree-fields} by
\[
 B_{\ell,j}=\mathbf1_{U_\ell^{m_j}=1},\qquad
 X_{m_j}=\prod_\ell(1-1/\ell)^{B_{\ell,j}}.
\]
For every nonempty $S\subseteq\{1,\ldots,d\}$,
\begin{equation}\label{eq:joint-degree-local}
 \mathbb P(B_{\ell,j}=1\text{ for every }j\in S)
 =\frac{\gcd(\ell-1,\gcd_{j\in S}m_j)}{\ell-1}.
\end{equation}
As $x\to\infty$, the empirical joint distribution functions over
primes $p\le x$ of
\[
 \bigl(c_{n_{j,x}(p)}(p^{m_j})\bigr)_{j=1}^d
\]
converge uniformly on $\R^d$ to that of $(X_{m_j})_{j=1}^d$,
uniformly over all arrays
$n_{j,x}(p)\in\mathbb Z_{\ge1}\cup\{\infty\}$.
The same assertion holds with the determinant densities replaced by
$\varphi(p^{m_j}-1)/(p^{m_j}-1)$. In particular, $m_i\mid m_j$
implies $X_{m_j}\le X_{m_i}$ in this coupling.
\end{corollary}

\begin{proof}
Intersections of root subgroups in the cyclic group
$\mathbb F_\ell^\times$ give \eqref{eq:joint-degree-local}.
The Chinese remainder theorem and \Cref{thm:sw} give the corresponding
joint patterns at every finite set of primes. Summing the coordinate
tail estimates in the proof of \Cref{cor:fixed-degree-fields} bounds
the empirical omitted logarithmic sum by
$O_{\boldsymbol m}(y^{-1}+x^{-1/2})$ and its model counterpart by
$O_{\boldsymbol m}(y^{-1})$. Truncation therefore proves joint weak
convergence of the ratios. Uniformly in $n\ge1$, including $n=\infty$,
\[
 0\le1-\frac{c_n(p^{m_j})}{\varphi(p^{m_j}-1)/(p^{m_j}-1)}
 \le\frac1{p^{m_j}-1}.
\]
Discarding a growing initial segment of negligible relative size
transfers convergence uniformly over the dimension arrays. Each
limiting marginal is continuous by \Cref{cor:fixed-degree-fields};
a finite grid then gives uniform convergence of the joint distribution
functions. Finally, $m_i\mid m_j$ implies
$B_{\ell,i}\le B_{\ell,j}$ at every prime, proving the last assertion.
\end{proof}

\begin{proposition}[Endpoint at fixed extension degree]
\label{prop:fixed-degree-endpoint}
For each fixed $m\ge1$, let $X_m,A_m,\mathcal E_m,q_{\ell,m}$ be as in
\Cref{cor:fixed-degree-fields}, and put $r_m=\tau(m)$. The limit
\[
 \kappa_m=\lim_{y\to\infty}(\log y)^{r_m}
       \prod_{\substack{\ell\le y\\\ell\notin\mathcal E_m}}
                      (1-q_{\ell,m})
\]
exists in $(0,\infty)$. As $\varepsilon\downarrow0$,
\begin{equation}\label{eq:fixed-degree-endpoint}
 \mathbb P\{X_m>A_m(1-\varepsilon)\}
 =\frac{\kappa_m}{(\log(1/\varepsilon))^{\tau(m)}}
 \left(1+O_m\left(\frac1{\log(1/\varepsilon)}\right)\right).
\end{equation}
The same formula holds with a non-strict inequality. More explicitly,
if
\[
 C_d=\lim_{y\to\infty}(\log y)^{1/\varphi(d)}
       \prod_{\substack{\ell\le y\\\ell\equiv1\pmod d}}(1-1/\ell),
 \qquad h_{\ell,m}=\gcd(m,\ell-1),
\]
then
\begin{equation}\label{eq:kappa-degree}
 \kappa_m=
 \prod_{d\mid m}C_d^{\varphi(d)}
 \prod_{\ell\in\mathcal E_m}(1-1/\ell)^{-(\ell-1)}
 \prod_{\ell\notin\mathcal E_m}
       \frac{1-h_{\ell,m}/(\ell-1)}{(1-1/\ell)^{h_{\ell,m}}}.
\end{equation}
The last product converges absolutely.
\end{proposition}

\begin{proof}
The progression estimate \eqref{eq:fixed-degree-progression-mertens}
proves existence and positivity of $C_d$ with a relative
$O_d(1/\log y)$ remainder. The identity
$h_{\ell,m}=\sum_{d\mid m}\varphi(d)\mathbf1_{\ell\equiv1\pmod d}$
and $\log(1-q_{\ell,m})=-h_{\ell,m}/\ell+O_m(\ell^{-2})$
give
\[
 Q_m(y):=\prod_{\substack{\ell\le y\\\ell\notin\mathcal E_m}}
                   (1-q_{\ell,m})
 =\kappa_m(\log y)^{-\tau(m)}(1+O_m(1/\log y)).
\]
Factoring the corresponding products of $(1-1/\ell)^{h_{\ell,m}}$
proves~\eqref{eq:kappa-degree}. On $\mathcal E_m$ one has
$h_{\ell,m}=\ell-1$; every probability-one factor is therefore removed.
The remaining logarithmic corrections are $O_m(\ell^{-2})$.

Write $W_m=-\log(X_m/A_m)$ and, for $0<t<1/\max\mathcal E_m$, put
$y=1/t$.
Apply \Cref{lem:empty-cylinder} with $d=1$ and $\xi_{\ell,1}=B_{\ell,m}$
for $\ell\notin\mathcal E_m$, $\xi_{\ell,1}=0$ for $\ell\in\mathcal E_m$,
so that the indicators remain independent across primes and
$\mathbb P(\xi_{\ell,1}=1)\le m/(\ell-1)\le2m/\ell$. For either
$\diamond\in\{<,\le\}$, parts~\textup{(a)} and~\textup{(b)} give
\[
 \mathbb P(W_m\diamond t)
 =Q_m(y)\,\mathbb P\Bigl(\sum_{\ell>y}B_{\ell,m}\omega_\ell\diamond t\Bigr)
 =Q_m(y)\{1+O_m(1/\log y)\}.
\]
Substitution
$t=-\log(1-\varepsilon)=\varepsilon+O(\varepsilon^2)$ proves the result.
For example the normalization gives
\[
 \kappa_1=2e^{-\gamma}\prod_{\ell\ge3}(1-(\ell-1)^{-2}),\qquad
 \kappa_2=9e^{-2\gamma}\prod_{\ell\ge5}
                         \frac{\ell^2(\ell-3)}{(\ell-1)^3}.
\]
\end{proof}
\begin{corollary}[Joint endpoints at fixed extension degrees]
\label{cor:joint-degree-endpoint}
\index{Finite fields!joint endpoints}
For the coupled variables of \Cref{cor:joint-fixed-degree-fields}, put
\[
 \mathcal D_{\boldsymbol m}:=\bigcup_{j=1}^d\{a\ge1:a\mid m_j\},
 \qquad r_{\boldsymbol m}:=|\mathcal D_{\boldsymbol m}|,
\]
and define
\[
 \alpha_{\ell,\boldsymbol m}:=
 \mathbb P\{B_{\ell,j}=0\text{ for every }j
                         \text{ with }\ell\notin\mathcal E_{m_j}\}.
\]
Then
$\kappa_{\boldsymbol m}:=\lim_{y\to\infty}(\log y)^{r_{\boldsymbol m}}
\prod_{\ell\le y}\alpha_{\ell,\boldsymbol m}$ exists in $(0,\infty)$.
For $0<\varepsilon_j<1$, put
\[
 L_j:=\log(1/\varepsilon_j),\qquad
 L_a^*:=\max_{j:\,a\mid m_j}L_j
 \quad(a\in\mathcal D_{\boldsymbol m}).
\]
Uniformly as $\min_jL_j\to\infty$,
\begin{equation}\label{eq:joint-degree-endpoint}
\begin{split}
 &\mathbb P\{X_{m_j}>A_{m_j}(1-\varepsilon_j)\text{ for every }j\}\\
 &\qquad=\frac{\kappa_{\boldsymbol m}}
                {\prod_{a\in\mathcal D_{\boldsymbol m}}L_a^*}
 \left(1+O_{\boldsymbol m}\left(\sum_{j=1}^d\frac1{L_j}\right)\right).
\end{split}
\end{equation}
The same formula holds with non-strict inequalities. Equal widths give
exponent $r_{\boldsymbol m}$. For degrees $2,3$ the denominator is
$L_1L_2\max(L_1,L_2)$, giving exponent $3$ at equal widths, whereas
independent marginals have exponent $\tau(2)+\tau(3)=4$.
\end{corollary}

\begin{proof}
At each prime, the excluded root subgroups are proper subgroups of
the cyclic group $\mathbb F_\ell^\times$. A generator belongs to none
of them, so $\alpha_{\ell,\boldsymbol m}>0$. For all sufficiently large
$\ell$, no coordinate is deterministic, and counting elements by order gives
\[
 1-\alpha_{\ell,\boldsymbol m}
 =\frac1{\ell-1}
   \sum_{\substack{a\in\mathcal D_{\boldsymbol m}\\a\mid\ell-1}}\varphi(a).
\]
The progression Mertens estimate \eqref{eq:fixed-degree-progression-mertens}
for each modulus $a\in\mathcal D_{\boldsymbol m}$, and the remaining summable
$O_{\boldsymbol m}(\ell^{-2})$ logarithmic corrections prove the
assertion about $\kappa_{\boldsymbol m}$.

Put $W_j=-\log(X_{m_j}/A_{m_j})$, $t_j=-\log(1-\varepsilon_j)$,
$y_j=1/t_j$ and $Y_a=\max_{j:\,a\mid m_j}y_j$.
Let $\mathcal A$ be the event that $B_{\ell,j}=0$ whenever $\ell\le y_j$
and $\ell\notin\mathcal E_{m_j}$. It is the event
$\mathcal A_{\boldsymbol t}$ of \Cref{lem:empty-cylinder} for
$\boldsymbol t=(t_1,\ldots,t_d)$ and the indicators
$\xi_{\ell,j}=B_{\ell,j}\mathbf1_{\{\ell\notin\mathcal E_{m_j}\}}$,
whose weighted sums are the $W_j$. These vectors are independent across
primes because the $U_\ell$ are independent. Take all $t_j\in(0,1)$
and let $\min_jy_j$ exceed all deterministic primes. At every sufficiently
large prime the probability excluded by $\mathcal A$ is
\[
 \frac1{\ell-1}
 \sum_{\substack{a\in\mathcal D_{\boldsymbol m}\\
                  a\mid\ell-1,\ \ell\le Y_a}}\varphi(a).
\]
Compare the factors of $\mathbb P(\mathcal A)$ with the products of
$(1-1/\ell)^{\varphi(a)}$ over $\ell\equiv1\pmod a$, $\ell\le Y_a$.
Their ratios are $1+O_{\boldsymbol m}(\ell^{-2})$ and coincide with
those defining $\kappa_{\boldsymbol m}$ below $\min_jy_j$, including
the finite deterministic-prime corrections. Mertens' estimates therefore give
\[
 \mathbb P(\mathcal A)
 =\frac{\kappa_{\boldsymbol m}}{\prod_{a\in\mathcal D_{\boldsymbol m}}\log Y_a}
 \left(1+O_{\boldsymbol m}\left(\sum_j\frac1{\log y_j}\right)\right).
\]
Since $\mathbb P(\xi_{\ell,j}=1)\le m_j/(\ell-1)\le2m_j/\ell$,
\Cref{lem:empty-cylinder}\textup{(b)} gives, for either
$\diamond\in\{<,\le\}$,
\[
 \mathbb P(W_j\diamond t_j\text{ for every }j)
 =\mathbb P(\mathcal A)
  \left(1+O_{\boldsymbol m}\left(\sum_j\frac1{\log y_j}\right)\right).
\]
Since $\log y_j=L_j+O(\varepsilon_j)$, this proves
\eqref{eq:joint-degree-endpoint}, also with non-strict inequalities.
\end{proof}

For $m=2$, one has $A_2=1/3$ and
$X_2\overset d=\tfrac12e^{-U-V}$ for
$(U,V)\sim\mu_{\mathrm{joint}}$, expressing the square-field law through the joint model of
\Cref{thm:diff-dist}. The totient-ratio distribution and mean for every
fixed $m$ already follow from
Aivazidis--Sofos~\cite[Theorems~1.1(i) and~1.3(i)]{AS15}, with parameter
$n=m$ and their normalization multiplied by $m$, since nonprime bases
have relative density zero in their prime-power ensemble.

\section{Sparse prime divisors in the extension degree}
\label{ssec:fields-growing-prime-degree}

\begin{corollary}[Sparse prime divisors in the extension degree]
\label{cor:growing-prime-degree-fields}
\index{Finite fields!growing prime degree}
Fix an integer $d\ge1$ and positive integers $r=r(x)$ satisfying
\begin{equation}\label{eq:sparse-degree-condition}
 \sum_{\substack{s\mid r(x)\\s\textup{ prime}}}\frac{1+\log\log(3s)}{s}
 \longrightarrow0.
\end{equation}
There is no upper restriction on $r(x)$. Uniformly in thresholds and
arrays $n_x(p)\in\mathbb Z_{\ge1}\cup\{\infty\}$, the empirical laws over
primes $p\le x$ of
\[
 \frac{\varphi(p^{dr}-1)}{p^{dr}-1}
 \quad\hbox{and}\quad c_{n_x(p)}(p^{dr})
\]
converge to the law of $X_d$ in \eqref{eq:fixed-degree-field-law}.
This includes $r(x)=s(x)^{k(x)}$ for primes $s(x)\to\infty$ and arbitrary
positive integers $k(x)$, and permits $r(x)=1$ on a subsequence.
It also includes $r=\prod_{t<s\le2t}s$, over primes, for any
$t=t(x)\to\infty$: the sum in \eqref{eq:sparse-degree-condition} is
$O((1+\log\log t)/\log t)$, whereas
$\sum_{t<s\le2t}(\log s)/s\to\log 2$ by the prime number theorem.
\end{corollary}

\begin{proof}
Put $T_j(p)=\varphi(p^j-1)/(p^j-1)$. Condition
\eqref{eq:sparse-degree-condition} implies $(d,r)=1$ eventually.
Every prime $\ell\mid p^{dr}-1$, $\ell\nmid p^d-1$, then has order
$\operatorname{ord}_\ell(p)=ae$ for unique $a\mid d$ and $e\mid r$,
$e>1$. Let $\mathcal G_e(p)$ be the group with this $e$.
Its primes satisfy $\ell\equiv1\pmod e$ and $\ell\mid p^{de}-1$.
There are at most $de$ reduced roots of $u^{de}=1$ modulo $\ell$.
Brun--Titchmarsh~\cite[Theorem~3.9]{MV06}, uniformly for
$\ell\le\sqrt x$, therefore gives
\[
 \frac1{\pi(x)}\sum_{p\le x}
 \sum_{\substack{\ell\in\mathcal G_e(p)\\\ell\le\sqrt x}}\omega_\ell
 \ll de\sum_{\substack{\ell\equiv1\ (e)}}(\ell-1)^{-2}
 \ll d/e.
\]
At most $2de$ primes in $\mathcal G_e(p)$ exceed $\sqrt x$, since
their product divides $p^{de}-1<x^{de}$.
Those above $e^2$ have total weight $O_d(1/e)$, and the primes
$\ell\equiv1\pmod e$ at most $2e$ also have total weight $O(1/e)$.
On $2^je<\ell\le 2^{j+1}e$, $j\ge1$, Brun--Titchmarsh gives
\[
 \sum_{\substack{2^je<\ell\le 2^{j+1}e\\\ell\equiv1\pmod e}}
       \omega_\ell\ll \frac1{j\varphi(e)}.
\]
Summing these blocks up to $e^2$, and including the small-divisor
estimate, gives
\begin{equation}\label{eq:sparse-degree-comparison}
 \frac1{\pi(x)}\sum_{p\le x}\log\frac{T_d(p)}{T_{dr}(p)}
 \ll_d\sum_{\substack{e\mid r\\e>1}}
                  \frac{1+\log\log(3e)}{\varphi(e)}.
\end{equation}
The right side tends to zero uniformly in the exponents of $r$.
Indeed $\mathcal A_r:=\sum_{e\mid r}1/\varphi(e)
\le\exp(C\sum_{s\mid r,\ s\ {\rm prime}}1/s)=1+o(1)$, by the Euler
product and $\sum_{j\ge1}1/\varphi(s^j)=s/(s-1)^2$.
For $e>1$,
$1+\log\log(3e)\ll1+\log(1+\log e)$ and
$\log(1+\log e)\le\sum_{s^j\parallel e}\log(1+j\log s)$, where
$s$ ranges over primes. The contribution of the constant term is
$O(\mathcal A_r-1)$. The rest is at most a constant times
\[
 \mathcal A_r\sum_{s\mid r,\ s\ {\rm prime}}
        \sum_{j\ge1}\frac{\log(1+j\log s)}{\varphi(s^j)}
 \ll\mathcal A_r\sum_{s\mid r,\ s\ {\rm prime}}
                     \frac{1+\log\log(3s)}{s}=o(1).
\]
The inner estimate follows from $\log(1+j\log s)\le\log j+
\log(1+\log s)$ and geometric summation. Empty sums cover $r=1$.

For prime $r$, the predecessors of the at most $2dr$ large new prime
divisors are distinct multiples of $r$
and exceed $\sqrt x/2$. Integral comparison bounds their total weight by
\[
 \frac2{\sqrt x}+\frac1r\log\left(1+\frac{4dr^2}{\sqrt x}\right)
 \le\frac2{\sqrt x}+2\sqrt d\,x^{-1/4}.
\]
Together with the small-divisor estimate this gives
\[
 \frac1{\pi(x)}\sum_{p\le x}\log\frac{T_d(p)}{T_{dr}(p)}
 \ll \frac d r+\sqrt d\,x^{-1/4}.
\]

Since $0\le T_d-T_{dr}\le\log(T_d/T_{dr})$, Markov's inequality and
\Cref{cor:fixed-degree-fields} give convergence in law to $X_d$.
Moreover, uniformly in $n\in\mathbb Z_{\ge1}\cup\{\infty\}$ and $r\ge1$,
\[
 0\le1-\frac{c_n(p^{dr})}{T_{dr}(p)}
 \le1-P(p^{dr})\le\sum_{j\ge1}p^{-drj}
 =\frac1{p^{dr}-1}\le\frac1{p^d-1}.
\]
The prime average of the last bound tends to zero, by discarding a
fixed initial segment and then letting its cutoff tend to infinity.
Markov's inequality therefore transfers convergence uniformly over
all dimension arrays. Continuity of the law of $X_d$, supplied by
\Cref{cor:fixed-degree-fields}, and the distribution-function comparison
in the proof of \Cref{prop:head-tail-cdf} give uniform convergence in thresholds.
\end{proof}

At each fixed degree $d$, \Cref{thm:polynomial-root-law} applied to
$T^d-1$ gives convergence in every finite Wasserstein order for the
empirical negative logarithms and convergence of all real Mellin
moments for the ratios. These conclusions hold also for the determinant
densities, uniformly over the dimension arrays: the added logarithmic
term is nonnegative, bounded and $O_d(p^{-d})$, hence tends to zero in
every fixed mean power; its exponential is uniformly bounded at every
fixed order. Uniform integrability thus transfers the moments.

For the sparse varying degrees of \Cref{cor:growing-prime-degree-fields},
\eqref{eq:sparse-degree-comparison} gives the cost of the same-prime
coupling in $\mathcal W_1$ for the logarithms, and the matrix
corrections again have mean $o(1)$. In both degree regimes the raw
laws converge in every fixed $\mathcal W_r$, $1\le r<\infty$, since
their support lies in $[0,1]$.

\section{Divisor sums on the profinite integers}
\label{ssec:profinite-divisor-sums}

The probabilistic part of the fixed-field argument has the following
form. Fix a function $g:\mathbb N\to[0,\infty)$ and constants
$C\in\mathbb R$, $a\ge0$ such that, for every integer $M\ge1$,
\begin{equation}\label{eq:profinite-weight-growth}
 G(M):=\sum_{m\le M}g(m)\le C+a\log\log(M+2).
\end{equation}
Increasing $C$ when necessary, we may take $C\ge1$. Put $G(0)=0$,
and let $U$ have Haar probability measure on $\widehat{\mathbb Z}$.
Define
\begin{equation}\label{eq:profinite-divisor-model}
 S_g=\sum_{m\ge1}g(m)\mathbf1_{m\mid U},\qquad
 X_g=e^{-S_g},\qquad s_g(n)=\sum_{m\mid n}g(m).
\end{equation}
For uniform $N_D\in\{1,\ldots,D\}$ write $S_{g,D}=s_g(N_D)$
and $X_{g,D}=e^{-S_{g,D}}$.
For a finite nonempty set $F\subset\mathbb N$,
\begin{equation}\label{eq:profinite-divisor-dependence}
 \mathbb P(m\mid U\text{ for every }m\in F)
 =\frac1{\operatorname{lcm}(F)}.
\end{equation}
In particular the divisor indicators are generally dependent; it is
the prime valuations $V_r=v_r(U)$ that are independent, with
$\mathbb P(V_r\ge k)=r^{-k}$. These assertions follow from uniform
reduction modulo every positive integer and the Chinese remainder
theorem.

\begin{theorem}[Profinite divisor-sum laws]
\label{thm:profinite-divisor-law}
\index{Profinite integers!divisor-sum laws}
Under \eqref{eq:profinite-weight-growth}, $S_g$ is almost surely finite.
For every $t>0$,
\begin{equation}\label{eq:profinite-exponential-moments}
 \sup_{D\ge1}\mathbb E e^{tS_{g,D}}
 \le\mathbb E e^{tS_g}<\infty.
\end{equation}
As $D\to\infty$, $S_{g,D}$ converges weakly to $S_g$ and $X_{g,D}$
converges weakly to $X_g$. Both convergences hold in every fixed
$\mathcal W_r$, $1\le r<\infty$. Each limiting law gives full mass
to a Borel set of Hausdorff dimension zero. If its limiting
distribution function is continuous, the corresponding empirical
distribution functions converge uniformly in the threshold.
\end{theorem}

\begin{proof}
First put $T_g(M)=\sum_{m>M}g(m)/m$. Summation by parts, or its
dyadic form, gives
\begin{equation}\label{eq:profinite-weight-tail}
 T_g(M)\ll_{C,a}\frac{1+\log\log(M+3)}M\quad(M\ge1).
\end{equation}
Indeed, the contribution of $2^kM<m\le2^{k+1}M$ is at most
$G(\lfloor2^{k+1}M\rfloor)/(2^kM)$; the growth bound and
$\log\log(2^{k+1}M+2)\ll1+\log\log(M+3)+\log(k+2)$
make the sum converge with the stated bound. Thus
$\mathbb ES_g=\sum_m g(m)/m<\infty$, by Tonelli, proving almost
sure finiteness.

For the exponential moments let $b_m(t)=e^{tg(m)}-1\ge0$.
Expanding a finite product and using
\eqref{eq:profinite-divisor-dependence} yields
\[
 \mathbb E\exp\left(t\sum_{m\le M}g(m)\mathbf1_{m\mid U}\right)
 =\sum_{F\subset\{1,\ldots,M\}}
       \frac{\prod_{m\in F}b_m(t)}{\operatorname{lcm}(F)}.
\]
The empty-set term is one. For nonempty $F$,
$\operatorname{lcm}(F)\ge\max F$, whereas
$1/\max F$ is exactly the joint probability of
$\{V\le1/m:m\in F\}$ for a single uniform $V\in(0,1)$.
Coefficientwise comparison with these nested indicators, followed
by monotone convergence, proves
\begin{equation}\label{eq:profinite-nested-moment-bound}
 \mathbb E e^{tS_g}
 \le\int_0^1 e^{tG(\lfloor1/v\rfloor)}\,dv
 \ll_{C,a,t}\int_0^1(1+\log(1/v))^{at}\,dv<\infty.
\end{equation}
For the empirical model the corresponding coefficient is
$D^{-1}\lfloor D/\operatorname{lcm}(F)\rfloor\le
1/\operatorname{lcm}(F)$; no $m>D$ divides $N_D$.
The same positive expansion proves
\eqref{eq:profinite-exponential-moments}.

For a fixed integer $M$, the pattern of divisibility by $m\le M$
depends only on the residue modulo
$L_M=\operatorname{lcm}(1,\ldots,M)$. Each residue has empirical
probability differing from $1/L_M$ by at most $1/D$. Thus the
truncated empirical sum converges to its Haar counterpart.
Both omitted nonnegative means are bounded by $T_g(M)$, since
$D^{-1}\lfloor D/m\rfloor\le1/m$. For any bounded
$1$-Lipschitz test function the difference of the full-law
expectations is at most $2T_g(M)$ plus the finite-pattern error.
Letting $D\to\infty$ and then $M\to\infty$ proves weak convergence.
The map $u\mapsto e^{-u}$ is continuous on $[0,\infty)$, giving
the raw weak limit.

For the transport limits, fix $r\ge1$. Either tail $Z$ beyond $M$
(empirical or Haar) satisfies $\mathbb EZ\le T_g(M)$. By Cauchy--Schwarz,
\[
 \mathbb EZ^r\le(\mathbb EZ)^{1/2}
                     (\mathbb EZ^{2r-1})^{1/2}
 \ll_{C,a,r}T_g(M)^{1/2},
\]
because $Z$ is bounded by its full sum and
\eqref{eq:profinite-exponential-moments} bounds every positive
power uniformly in $D$. A maximal coupling of the two residue
patterns has disagreement probability at most $L_M/(2D)$;
the corresponding sums lie in $[0,G(M)]$. Their $r$th-power
transport cost is at most $G(M)^rL_M/(2D)$.
The triangle inequality for $\mathcal W_r$, first $D\to\infty$
and then $M\to\infty$, proves transport convergence for the full
sums. Since $e^{-u}$ is $1$-Lipschitz, it proves the raw assertion
as well. If a limiting CDF is continuous, choose a finite grid
with small limiting mass between successive points and small
outer tails. Convergence at the grid points and monotonicity
then bound the uniform CDF error.

We construct the small-dimensional carrier using only independent
prime valuations. Fix an integer $B>10$, put
$y_j=\exp(2^{j^2})$, and let $D_y=\prod_{r\le y}r^{V_r}$,
where here and below $r$ ranges over primes. Prime partial summation
and \Cref{thm:mertens} give
\[
 \mathbb E\log D_y
 =\sum_{r\le y}\frac{\log r}{r-1}\sim\log y.
\]
For all sufficiently large $j$, $y_{j-1}^B<y_j$.
The old sum through $y_{j-1}^B$ has mean
$O_B(\log y_{j-1})$, and
$\log y_{j-1}/\log y_j=2^{-2j+1}$ is summable.
Markov's inequality and Borel--Cantelli therefore make this old
sum at most $\log y_j$ eventually, almost surely.
The fresh sum over $y_{j-1}^B<r\le y_j$ has mean at most
$2\log y_j$ for large $j$, so it is at most $4\log y_j$
with probability at least $1/2$. Independently of that fresh sum,
the gap $y_j<r\le y_j^B$ is empty with probability
\[
 \prod_{y_j<r\le y_j^B}(1-1/r)\longrightarrow\frac1B.
\]
The events just involving the fresh sum and the empty gap use
disjoint prime blocks for different $j$. Independent
Borel--Cantelli implies that almost surely infinitely many $j$
satisfy
\begin{equation}\label{eq:degree-good-block}
 D_{y_j}\le y_j^5,\qquad V_r=0\quad(y_j<r\le y_j^B).
\end{equation}

Condition on the valuations through $y$ and on the empty gap.
A surviving divisor $m>y^B$ must have the form $m=ab$, where
$a\mid D_y$ and every prime factor of $b$ exceeds $y^B$.
Its conditional activation probability is $1/b=a/m\le D_y/m$;
all other $m$ have conditional probability zero. Hence, on
$D_y\le y^5$, the conditional expected tail beyond $y^B$ is at most
\[
 y^5T_g(\lfloor y^B\rfloor)
 \ll_{C,a,B}y^{5-B}(1+\log\log(y+3)).
\]
Conditional Markov followed by averaging shows that the probability
of \eqref{eq:degree-good-block} together with a tail exceeding
$\varepsilon_j=y_j^{5-B/2}$ is
$O_{C,a,B}(y_j^{-B/2}(1+\log\log(y_j+3)))$.
These probabilities are summable. Almost surely, therefore,
infinitely many good blocks have tail at most $\varepsilon_j$.
On a good block all active divisors at most $y_j^B$ are exactly
the divisors of $D_{y_j}$: the gap is empty and
$D_{y_j}\le y_j^5<y_j^B$. Consequently
\[
 K_B=\limsup_{j\to\infty}
 \bigcup_{1\le d\le\lfloor y_j^5\rfloor}
       [s_g(d),s_g(d)+\varepsilon_j]
\]
is a deterministic Borel set of full $S_g$-measure. For
$s>10/(B-10)$,
\[
 \sum_j\lfloor y_j^5\rfloor\varepsilon_j^s
 \le\sum_j y_j^{5+s(5-B/2)}<\infty.
\]
The tails of this cover show that $\mathcal H^s(K_B)=0$.
Intersecting $K_B$ over all integers $B>10$ gives a full-measure
Borel set $K\subset[0,\infty)$ with $\dim_H K=0$.
The exponential image is Borel, since $u\mapsto e^{-u}$ is a
homeomorphism from $[0,\infty)$ onto $(0,1]$, and is also a
Hausdorff-zero carrier by its Lipschitz property.
\end{proof}

\begin{proposition}[Transforms of profinite divisor sums]
\label{prop:profinite-divisor-mellin}
Under \eqref{eq:profinite-weight-growth},
$M_g(s)=\mathbb E X_g^s=\mathbb E e^{-sS_g}$ is entire, and
$X_g$ has finite negative moments of every order. For every fixed
$R\ge0$ and integer $j\ge0$,
\begin{equation}\label{eq:profinite-empirical-mellin}
 \begin{split}
 &\sup_{|s|\le R}\left|
 \frac1D\sum_{n\le D} e^{-s s_g(n)}(-s_g(n))^j
       -\mathbb E[e^{-sS_g}(-S_g)^j]\right|\\
 &\qquad\ll_{C,a,R,j}
 \frac{\log^{aR}(D+2)[1+\log\log(D+3)]^j}{D}.
 \end{split}
\end{equation}
The powers in the Mellin transform use the real logarithm of $X_g$;
the derivative convention at $j=0$ is $(-u)^0=1$, including $u=0$.
\end{proposition}

\begin{proof}
The exponential moment bound in
\Cref{thm:profinite-divisor-law} dominates $e^{-sS_g}$ and each
of its derivatives on every compact disk (enlarge the exponential
moment parameter to absorb any power of $S_g$). This proves
entire holomorphy and the derivative formula
$M_g^{(j)}(s)=\mathbb E[e^{-sS_g}(-S_g)^j]$.
It also proves all negative moments.

We give the coefficient and tail estimates without enlarging $R$,
so that the exponent $aR$ in the rate is retained, including at
$R=0$. For a finite set $F\subset\mathbb N$, put
\[
 L_F=\operatorname{lcm}(F),\qquad
 a_F(s)=\prod_{m\in F}(e^{-sg(m)}-1),\qquad
 L_\varnothing=a_\varnothing(s)=1.
\]
Let $b_m(R)=e^{Rg(m)}-1$. The exponential power series gives, for
every integer $k\ge0$ and $|s|\le R$,
\[
 |\partial_s^k(e^{-sg(m)}-1)|\le\partial_R^k b_m(R).
\]
For $k=0$ this uses the series with zero constant coefficient;
for $k\ge1$ the right side is $g(m)^ke^{Rg(m)}$.
All coefficients and derivatives on the right are nonnegative,
even at $R=0$. Leibniz's rule therefore gives
\[
 \sum_{F\subset\{1,\ldots,M\}}|a_F^{(j)}(s)|
 \le\partial_R^j\prod_{m\le M}(1+b_m(R))
 =G(M)^je^{RG(M)}=:H_j(M),
\]
where $H_0(0)=1$ and $H_j(0)=0$ for $j\ge1$.
More precisely, differentiating the positive sum of products with
maximum index $m$ gives
\begin{equation}\label{eq:profinite-max-index-bound}
 \sum_{\max F=m}|a_F^{(j)}(s)|
 \le H_j(m)-H_j(m-1)=:\Delta H_j(m)\ge0.
\end{equation}
Write
\[
 B_{R,j}(M)=\log^{aR}(M+2)[1+\log\log(M+3)]^j.
\]
The weight hypothesis gives $H_j(M)\ll_{C,a,R,j}B_{R,j}(M)$.
For fixed $R,j$,
\begin{equation}\label{eq:profinite-abel-tail}
 \begin{split}
 \sum_{m>D}\frac{\Delta H_j(m)}m
 &=-\frac{H_j(D)}{D+1}
       +\sum_{m>D}\frac{H_j(m)}{m(m+1)}\\
 &\ll_{C,a,R,j}\frac{B_{R,j}(D)}D.
 \end{split}
\end{equation}
To justify the identity, sum first to $N$ and then let $N\to\infty$;
$H_j(N)/N\to0$. For the bound, dyadic intervals and
\[
 B_{R,j}(2^{k+1}D)
 \ll_{a,R,j}B_{R,j}(D)(k+2)^{aR}[1+\log(k+2)]^j
\]
reduce it to a convergent series with an additional factor $2^{-k}$.

Since $L_F\ge\max F$, \eqref{eq:profinite-max-index-bound}
and \eqref{eq:profinite-abel-tail} prove absolute convergence of
$\sum_F a_F^{(j)}(s)/L_F$, uniformly on every compact disk.
Finite-product expansion, dominated convergence of the random
products, and these coefficient bounds give
\begin{equation}\label{eq:profinite-transform-expansions}
 \begin{split}
 M_g(s)&=\sum_{F\subset\mathbb N\text{ finite}}\frac{a_F(s)}{L_F},\\
 \frac1D\sum_{n\le D}e^{-s s_g(n)}
 &=\sum_{F\subset\{1,\ldots,D\}}
                  a_F(s)\frac{\lfloor D/L_F\rfloor}D.
 \end{split}
\end{equation}
Every fixed derivative may be taken term by term.
For $\max F\le D$ the two multipliers differ by at most $1/D$,
also when $L_F>D$; for $\max F>D$ the empirical multiplier is zero.
Thus the absolute $j$th derivative error is at most
\[
 \frac{H_j(D)}D+\sum_{m>D}\frac{\Delta H_j(m)}m,
\]
which proves \eqref{eq:profinite-empirical-mellin}.
\end{proof}

The growth bound alone supplies neither continuity nor interval support:
$g(2)=1$ and $g(m)=0$ for $m\ne2$ give $S_g\in\{0,1\}$ with equal
probabilities, although the empirical distribution functions converge
uniformly. More generally, finite-support weights give finite laws.
For nonnegative divisor sums, Engberg--Pollack~\cite[Theorem~5.4 and the following variant]{EP21}
prove that positivity of $g(r^e)$ for every $e\ge1$ and every prime $r$
above a fixed cutoff implies continuity whenever the limiting
distribution exists. The field application below uses arithmetic
arguments for continuity and support.

\section{Fixed field and sampled extension degrees}
\label{ssec:fields-fixed-base}

We next fix the field size and average over extension degrees. For a fixed
prime power $q\ge2$, put
\[
 \begin{gathered}
 o_\ell=\operatorname{ord}_\ell(q)\quad(\ell\nmid q),\qquad
 g_q(m)=\sum_{o_\ell=m}\log\frac\ell{\ell-1},\\
 S_q(d)=\sum_{m\mid d}g_q(m)=\log\frac{q^d-1}{\varphi(q^d-1)}.
 \end{gathered}
\]
The existence of the degree-sampling distribution follows already from
Aivazidis--Sofos~\cite[Theorem~1.3(iii)]{AS15}: their $d p_d(q)$ is
$\varphi(q^d-1)/(q^d-1)$. For $q=2$, Luca's discussion~\cite[p.~458]{Luc03}, referring to
\cite{Luc05}, already records existence of the distribution of these
ratios; their density in $[0,1]$ is proved in
\cite[Theorem~(ii)]{Luc03}. Existence is distinct from continuity:
Engberg--Pollack~\cite[\S1, footnote~2]{EP21} note that the proofs
in \cite[Theorem~3(2)]{Luc05} establish existence but leave continuity
unaddressed. Deshouillers--Luca
\cite[\S5]{DL11} state that their method for the Fibonacci totient-ratio
law also applies to Mersenne numbers, giving the $q=2$ full-support
conclusion as a published extension with details omitted.

Atomlessness for every fixed $q$ also follows from the general
continuity criterion of Engberg--Pollack~\cite[Theorem~5.4 and its
following variant]{EP21}, applied to $S_q(d)=\sum_{m\mid d}g_q(m)$.
Indeed, $g_q(r^e)>0$ for every prime $r>q$ and $e\ge1$: a prime
factor $\ell$ of $\Phi_{r^e}(q)$ has order $r^e$ unless $\ell=r$,
and $\Phi_{r^e}(q)\equiv1\pmod r$ excludes that exception.
The following results give direct proofs of these qualitative assertions,
along with a Hausdorff-zero carrier and quantitative moment and
concentration bounds.

Let $U$ be Haar distributed on $\widehat{\mathbb Z}$ and define
\begin{equation}\label{eq:degree-law}
 \mathcal S_q=\sum_{m\ge1}g_q(m)\mathbf1_{m\mid U},\qquad
 \mathcal X_q=e^{-\mathcal S_q},\qquad
 A_q=\frac{\varphi(q-1)}{q-1}.
\end{equation}
For each $D$, let $U_D(d)$, $1\le d\le D$, denote any one of
\[
 \frac{\varphi(q^d-1)}{q^d-1},\qquad
 \frac{\varphi(q^d-1)}{q^d},\qquad c_{n_D(d)}(q^d)
\]
where $n_D(d)\in\mathbb Z_{\ge1}\cup\{\infty\}$ is an arbitrary
dimension array.

For positive integers $u,v$,
\[
 \mathbb P(u\mid U,\ v\mid U)=\frac1{\operatorname{lcm}(u,v)}.
\]
Thus the order indicators are dependent.
The degree-prime valuations $V_r=v_r(U)$ are independent with
$\mathbb P(V_r\ge k)=r^{-k}$.
In \Cref{fig:field-ensembles} this structure is set beside the independent
field-prime factors of the fixed-degree limits.
Put $G_q(M)=\sum_{m\le M}g_q(m)$ and
$T_q(M)=\sum_{m>M}g_q(m)/m$.

\begin{lemma}[Order weights at a fixed field]
\label{lem:fixed-q-order-weights}
For $M\ge2$,
\begin{equation}\label{eq:order-tails}
 G_q(M)\le C_q+\log\log(M+2),\qquad
 T_q(M)\ll_q\frac{\log\log(M+2)}M.
\end{equation}
The series defining $\mathcal S_q$ has finite expectation and is almost
surely finite.
\end{lemma}

\begin{proof}
The prime divisors of
$N_M=\prod_{m\le M}(q^m-1)$ are exactly the field primes of order at
most $M$, and $\log N_M\ll_q M^2$. For $M\ge2$, the contribution to
$G_q(M)$ from primes at most $M^2$ is at most
$\log\log M+O(1)$, by \Cref{thm:mertens}. There are
$O_q(M^2/\log M)$ larger prime divisors, each of weight at most
$2/M^2$. Partial summation gives \eqref{eq:order-tails}.
Since $\mathbb P(m\mid U)=1/m$, \eqref{eq:order-tails} gives finite
expectation for the series in \eqref{eq:degree-law}.
\end{proof}

\begin{proposition}[Concentration and endpoints at a fixed field]
\label{prop:fixed-q-degree-concentration}
For all sufficiently small $h>0$,
\begin{equation}\label{eq:degree-concentration}
 Q_{\mathcal X_q}(h):=\sup_{a\in\mathbb R}
       \mathbb P\{\mathcal X_q\in[a,a+h]\}
 \le\frac{C_q}{\sqrt{\log\log\log(1/h)}}.
\end{equation}
Moreover,
\begin{equation}\label{eq:degree-endpoint}
 \liminf_{h\downarrow0}\log(1/h)\,
 \mathbb P\{\mathcal X_q\in[A_q-h,A_q]\}\ge e^{-\gamma}.
\end{equation}
The endpoint also satisfies
\begin{equation}\label{eq:degree-endpoint-upper-mas19}
 \mathbb P\{\mathcal X_q\in[A_q-h,A_q]\}
 \ll_q\frac1{\log\log(1/h)}\qquad(h\downarrow0).
\end{equation}
All constants and thresholds may depend on the fixed $q$.
\end{proposition}

\begin{proof}
Represent $V_r=B_rH_r$, where
$B_r$ is Bernoulli of parameter $1/r$, and $H_r$ has the conditional
law of $V_r$ given $V_r\ge1$, independently of $B_r$ and of the other
coordinates. For every prime $r>q$, a prime factor $\ell$ of
$1+q+\cdots+q^{r-1}$ has order $r$: its order divides $r$, while
order one would force $\ell\mid q-1$ and $\ell\mid r$, which is
impossible. Thus
\begin{equation}\label{eq:prime-order-jump}
 g_q(r)>q^{-r}.
\end{equation}
Condition on the $H_r$ and all $B_r$ outside a finite set $\mathcal R$
of primes exceeding $q$. Almost surely every Boolean state in
$\mathcal R$ gives a finite sum, since each state has positive
conditional probability. Increasing one of its coordinates increases
$\mathcal S_q$ by at least $g_q(r)$; all other changes are nonnegative.
If $g_q(r)>u$ for every $r\in\mathcal R$, the inverse image of an
interval of length $u$ is therefore an antichain. The product-measure
antichain inequality of Yehuda--Yehudayoff
\cite[Theorems~1 and~4]{YY25} bounds its conditional probability by
\[
 C\left(\sum_{r\in\mathcal R}\frac1r\left(1-\frac1r\right)\right)^{-1/2}.
\]
Take $\mathcal R=\{r:q<r\le R\}$ with
$R=\log(1/u)/(2\log q)$. By \eqref{eq:prime-order-jump},
\[
 Q_{\mathcal S_q}(u)\ll_q(\log\log R)^{-1/2}.
\]
In particular the logarithmic and raw laws are atomless. On
$\{\mathcal X_q\ge\sqrt h\}$, a raw interval of length $h$ maps
under $-\log$ into an interval of length at most $\sqrt h$; its
complement has probability at most
$2\mathbb E\mathcal S_q/\log(1/h)$. This proves the upper bound in
\eqref{eq:degree-concentration}.

Let $E_y=\{V_r=0:r\le y\}$. Its probability is
$\sim e^{-\gamma}/\log y$, by \Cref{thm:mertens}. Given $E_y$, only orders
with all prime factors greater than $y$ can occur, apart from $1$,
and their conditional activation probabilities remain $1/m$.
Consequently
\[
 \mathbb E[\mathcal S_q-g_q(1)\mid E_y]\le T_q(y).
\]
Take $y=h^{-1}\log(1/h)$. Then \eqref{eq:order-tails} gives
$T_q(y)=o_q(h)$, and Markov's inequality puts a proportion $1-o_q(1)$
of this cylinder in $[g_q(1),g_q(1)+h]$.
Since $e^{-g_q(1)}=A_q$ and $u\mapsto e^{-u}$ is $1$-Lipschitz on
$[0,\infty)$, its image lies in $[A_q-h,A_q]$.
Now $\log y\sim\log(1/h)$ proves \eqref{eq:degree-endpoint}.

For the upper bound, let $0<h<A_q$ be sufficiently small and put
\[
 h'=-\log(1-h/A_q),\qquad R=\frac{\log(1/h')}{\log q}.
\]
The event in \eqref{eq:degree-endpoint-upper-mas19} is exactly
$\{\mathcal S_q-g_q(1)\le h'\}$. If a degree prime $r$ satisfies
$q<r\le R$ and $V_r\ge1$, then the order-$r$ term is active and
\eqref{eq:prime-order-jump} gives
$\mathcal S_q-g_q(1)\ge g_q(r)>q^{-r}\ge h'$.
Thus the endpoint event forces $V_r=0$ for all these degree primes.
Their valuations are independent, so Mertens' theorem gives
\[
 \mathbb P\{\mathcal S_q-g_q(1)\le h'\}
 \le\prod_{q<r\le R}(1-1/r)
 \ll_q\frac1{\log R}
 \ll_q\frac1{\log\log(1/h)}.
\]
Here the products run over primes, and $h'\sim h/A_q$ with $q$
fixed.
\end{proof}

\begin{theorem}[A fixed field and varying extension degree]
\label{thm:fixed-q-degree}
\index{Finite fields!sampled extension degrees}
As $D\to\infty$, the empirical distribution functions of $U_D(d)$ over
$1\le d\le D$ converge uniformly in the threshold to that of
$\mathcal X_q$, uniformly over the dimension arrays in the determinant
case. The law of $\mathcal X_q$ is atomless, has support $[0,A_q]$, and
gives full mass to a Borel set of Hausdorff dimension zero. The same
dimension statement holds for $\mathcal S_q$.
\end{theorem}

\begin{proof}
By \Cref{lem:fixed-q-order-weights}, the nonnegative function $g_q$
satisfies \eqref{eq:profinite-weight-growth} with $a=1$; increasing
$C_q$ includes $M=1$. Its divisor sum is exactly $S_q(d)$ and its
Haar sum is $\mathcal S_q$. Thus \Cref{thm:profinite-divisor-law}
gives weak convergence of $S_q(d)$ and of
$\varphi(q^d-1)/(q^d-1)$, and Hausdorff-zero Borel carriers for
$\mathcal S_q$ and $\mathcal X_q$.
Furthermore, uniformly in $n\ge1$, including $n=\infty$,
\[
 0\le1-\frac{c_n(q^d)}{\varphi(q^d-1)/(q^d-1)}
 \le\sum_{j\ge1}q^{-dj}\le2q^{-d}.
\]
The primitive-element ratio has multiplier $1-q^{-d}$. Discarding
any fixed initial set of degrees transfers weak convergence to both
statistics, uniformly over the dimension arrays.
By \Cref{prop:fixed-q-degree-concentration}, the limiting law is
atomless, so weak convergence of the uncorrected ratio gives uniform
CDF convergence. For the corrected ratios the distribution-function
sandwich at thresholds displaced by any fixed $\varepsilon>0$
has an error bounded uniformly in the arrays by the proportion of
degrees with $2q^{-d}>\varepsilon$. This proportion tends to zero.
Uniform continuity of the limiting CDF and then
$\varepsilon\downarrow0$ prove uniformity in both thresholds and
dimension arrays.

It remains to prove full support. First every finite anchor $S_q(d)$
belongs to $\operatorname{supp}(\mathcal S_q)$. For $y$ exceeding all
prime factors of $d$, condition $V_r=v_r(d)$ for every $r\le y$.
This cylinder has positive probability. Its orders composed entirely
of these primes are exactly the divisors of $d$. Any extra surviving
order $m$ has conditional probability at most $d/m$ and exceeds $y$.
Therefore
\begin{equation}\label{eq:degree-cylinder}
 0\le\mathbb E[\mathcal S_q-S_q(d)\mid
       V_r=v_r(d)\ (r\le y)]\le dT_q(y)\longrightarrow0.
\end{equation}
Markov's inequality proves the assertion for each neighborhood of
$S_q(d)$.

We now construct arbitrarily large anchors with small successive
increments. Write $q=p^f$. Choose a fixed reduced progression
$\ell\equiv b_0\pmod{M_0}$ in which
$\ell\equiv3\pmod4$, $q$ is a quadratic residue modulo $\ell$,
and no odd prime divisor of $M_0$ divides $\ell-1$.
For $p=2$, take $b_0=7$, $M_0=8$. For odd $p$, take $M_0=4p$
and impose $\ell\equiv b\pmod p$ with $b\ne0,1$.
If $f$ is even any such $b$ works. If $f$ is odd, choose
\[
 \left(\frac bp\right)=(-1)^{(p-1)/2}.
\]
Such a $b$ exists: for $p=3$ use $b=2$; for $p\equiv1\pmod4$
there is a nonidentity quadratic residue, and for
$p\equiv3\pmod4$ there is a nonresidue. Quadratic reciprocity
then gives $\left(\frac q\ell\right)=1$.

For large $t$ put $y=\log\log t$, and in this progression exclude
$\ell\equiv1\pmod r$ for every odd prime $r\le y$ not dividing
$M_0$. The combined modulus is
$M_0\prod_{3\le r\le y,\ r\nmid M_0}r\ll_q(\log t)^2$.
There are $\prod(r-2)$ allowed reduced classes, whose total relative
density is
\[
 \frac1{\varphi(M_0)}
 \prod_{\substack{3\le r\le y\\r\nmid M_0}}\frac{r-2}{r-1}
 \asymp_q\frac1{\log y}.
\]
The last comparison follows from Mertens' product, since the ratio
of each factor to $1-1/r$ is $1-1/(r-1)^2$.
Summing \Cref{thm:sw} over at most $O_q((\log t)^2)$ classes gives
at least $c_qt/(\log t\log y)$ field primes. Remove those for which
$r^2\mid\ell-1$ for some odd prime $r>y$. Brun--Titchmarsh
\cite[Theorem~3.9]{MV06} bounds the removal with $r\le t^{1/4}$ by
\[
 \ll\frac t{\log t}\sum_{r>y}\frac1{r(r-1)}
 \ll\frac t{y\log t}.
\]
For $t^{1/4}<r\le\sqrt t$, the integer count gives at most
\[
 \sum_{t^{1/4}<r\le\sqrt t}\left(\frac t{r^2}+1\right)
 \ll t^{3/4}+\sqrt t.
\]
Both errors are $o(t/(\log t\log y))$, and $4\nmid\ell-1$.
Thus the fixed set $\mathcal P_q$ of field primes satisfying the
chosen progression and
\[
 \ell-1\text{ squarefree},\qquad
 P^-((\ell-1)/2)>\log\log\ell,
 \qquad o_\ell\mid(\ell-1)/2
\]
has counting function $\gg_q t/(\log t\log\log\log t)$.
Here $P^-(1)=\infty$; the final divisibility follows from the
quadratic-residue condition. The construction at $t$ imposes
roughness stronger than the displayed condition for every $\ell\le t$.
Partial summation now gives
$\sum_{\ell\in\mathcal P_q}1/\ell=\infty$.

Given any large $Y$, omit finitely many field primes so that their
orders are squarefree and all their prime factors exceed $Y$.
A finite selection then has arbitrarily large total weight
$\sum\log(\ell/(\ell-1))$. Its order least common multiple $D$
is squarefree and $Y$-rough, and $S_q(D)$ is at least that weight.
List the prime factors of $D$ increasingly, and form the chain of
their successive products starting at $1$.
Every field prime of order $m\ge2$ is $1\pmod m$, and their product
divides $q^m-1$. Their number is $O_q(m)$, whence
\[
 g_q(m)\le\frac1m\sum_{j\le C_qm}\frac1j
 \ll_q\frac{\log(m+1)}m.
\]
If $r$ is the next chain prime and $d$ the preceding product, the
new divisors are $ru$, $u\mid d$. Consequently
\begin{equation}\label{eq:degree-increment}
 \begin{split}
 S_q(dr)-S_q(d)
 &\ll_q\frac1r\left(\log r\sum_{u\mid d}\frac1u
                  +\sum_{u\mid d}\frac{\log u}{u}\right)\\
 &\ll_q\frac{(\log r)^2}{r}.
 \end{split}
\end{equation}
Indeed $\sum_{u\mid d}1/u=\prod_{s\mid d}(1+1/s)\ll\log r$, and
\[
 \sum_{u\mid d}\frac{\log u}{u}
 =\prod_{s\mid d}(1+1/s)\sum_{s\mid d}\frac{\log s}{s+1}
 \ll(\log r)^2,
\]
because every prime $s\mid d$ is smaller than $r$.
The chains start at $g_q(1)$, end arbitrarily high, and have maximum
increment tending to zero as $Y\to\infty$. They therefore cross
every target in $[g_q(1),\infty)$ with arbitrarily small overshoot.
The anchors are dense there, and \eqref{eq:degree-cylinder}
proves $\operatorname{supp}(\mathcal S_q)=[g_q(1),\infty)$.
Since $e^{-g_q(1)}=A_q$, exponentiation and closure give
$\operatorname{supp}(\mathcal X_q)=[0,A_q]$.
\end{proof}

\KeepHeadingSpace{6\baselineskip}
\begin{proposition}[Mellin moments at a fixed field]
\label{prop:fixed-q-degree-moments}
The Mellin transform
$\mathbb E\mathcal X_q^s$ is entire, with all negative moments finite.
For every fixed $R\ge0$ and integer $j\ge0$,
\begin{equation}\label{eq:degree-empirical-mellin}
\begin{split}
 &\sup_{|s|\le R}\left|
 \frac1D\sum_{d\le D}U_D(d)^s(\log U_D(d))^j
 -\mathbb E\bigl[\mathcal X_q^s(\log\mathcal X_q)^j\bigr]\right|\\
 &\qquad\ll_{q,R,j}
 \frac{\log^R(D+2)[1+\log\log(D+3)]^j}{D}.
\end{split}
\end{equation}
The estimate is uniform over the dimension arrays, and complex powers
use the real logarithm.
\end{proposition}

For related moment-method existence results on recurrence sequences,
see Luca--Shparlinski~\cite{LS07}, as used for the Mersenne
prime-divisor and divisor-sum laws in Engberg--Pollack~\cite[\S1]{EP21}.
The elementary comparison, with $v_d=(q^d-1)/(q-1)$, is
\[
 \frac{q^d-1}{\varphi(q^d-1)}
 \le\frac{q-1}{\varphi(q-1)}\frac{v_d}{\varphi(v_d)}
 \le\frac{q-1}{\varphi(q-1)}\zeta(2)\frac{\sigma(v_d)}{v_d}.
\]

\begin{proof}
The application $g=g_q$, $a=1$ of
\Cref{prop:profinite-divisor-mellin}, justified by
\Cref{lem:fixed-q-order-weights}, gives entire holomorphy, all
negative moments, and \eqref{eq:degree-empirical-mellin} for the
uncorrected ratios $T_d=e^{-S_q(d)}$. In this proof write $G=G_q$.
In particular, the positive-expansion comparison
\eqref{eq:profinite-nested-moment-bound} gives
\begin{equation}\label{eq:degree-mellin}
 \mathbb E e^{t\mathcal S_q}
 \le\int_0^1e^{tG_q(1/v)}\,dv
 \ll_{q,t}\int_0^1(1+\log(1/v))^t\,dv<\infty
 \qquad(t>0),
\end{equation}
where $G_q$ is extended to real arguments by its defining sum.

For the determinant statistic write $c_n(q^d)=T_de^{-v_{d,n}}$.
Uniformly in $n\in\mathbb Z_{\ge1}\cup\{\infty\}$,
\[
 0\le v_{d,n}\le\sum_{k\ge1}-\log(1-q^{-dk})
 \le\frac{q^{-d}}{(1-q^{-d})^2}\le4q^{-d}.
\]
The primitive-element ratio has correction
$-\log(1-q^{-d})\le2q^{-d}$. Since $S_q(d)\le G(d)$, the
mean-value theorem applied to $e^{-su}(-u)^j$ bounds the sum of either
correction by
\[
 \ll_{q,R,j}\sum_{d\ge1}q^{-d}(1+G(d))^j e^{RG(d)}<\infty.
\]
After division by $D$ this is $O_{q,R,j}(D^{-1})$, completing
\eqref{eq:degree-empirical-mellin} uniformly in the arrays.
\end{proof}

For fixed $q$, \Cref{thm:fixed-q-degree,prop:fixed-q-degree-moments}
give convergence of the three empirical laws in every fixed
$\mathcal W_r$, $1\le r<\infty$, both for the
raw statistics and their negative logarithms, uniformly over the
dimension arrays. Weak convergence and the moment bounds in
\eqref{eq:degree-empirical-mellin} give uniform integrability of each
required power, hence this transport convergence.

\section{Extrema over degrees and prime powers}
\label{ssec:field-extrema}
\index{Finite fields!extremal determinant densities}

The least-common-multiple degrees in \Cref{rem:finite-field-extrema}
force determinant densities toward zero at each fixed base prime.

\begin{remark}[Finite-field extrema]\label{rem:finite-field-extrema}
For every fixed prime $r$ and every dimension assignment $n(a)\in\mathbb Z_{\ge1}\cup\{\infty\}$,
\[
 \inf_{a\ge1}c_{n(a)}(r^a)=\liminf_{a\to\infty}c_{n(a)}(r^a)=0.
\]
Over all prime powers, uniformly in dimension arrays
$n_x(q)\in\mathbb Z_{\ge1}\cup\{\infty\}$,
\[
 \min_{\substack{q\le x\\q\textup{ prime power}}}c_{n_x(q)}(q)
 \sim\frac{e^{-\gamma}}{\log\log x}.
\]
For base primes $p\le x$ and integer degrees $1\le d_x(p)\le D(x)$
with $\log D(x)=o(\log\log x)$ one also has
\begin{equation}\label{eq:variable-degree-base-prime-minimum}
 \min_{p\le x,\ p\ \mathrm{prime}}
       c_{n_x(p)}(p^{d_x(p)})\sim\frac{e^{-\gamma}}{\log\log x},
\end{equation}
uniformly over the permitted degree and matrix-dimension choices
for each such envelope $D(x)$.
For every fixed dimension assignment
$n(q)\in\mathbb Z_{\ge1}\cup\{\infty\}$,
\[
 \limsup_{\substack{q\to\infty\\q\textup{ prime power}}}
 \frac1{c_{n(q)}(q)\log\log q}=e^\gamma.
\]

Indeed, put $A_y=\operatorname{lcm}(1,\ldots,\lfloor y\rfloor)$. Each prime
$\ell\le y$, $\ell\ne r$, divides $r^{A_y}-1$, so uniformly in dimension
\[
 c_{n(A_y)}(r^{A_y})\le\prod_{\substack{\ell\le y\\\ell\ne r}}(1-1/\ell)
 \longrightarrow0\qquad(y\to\infty).
\]
The Rosser--Schoenfeld bound for $(q-1)/\varphi(q-1)$ used in
\Cref{cor:rosser-schoenfeld}, $c_n(q)\ge c_\infty(q)$ and $P(q)\to1$ give
$c_n(q)\ge(e^{-\gamma}+o(1))/\log\log q$ uniformly in $n$, and
$c_n(q)\ge c_\infty(q)>0$ for each of the finitely many small $q$.
Restricted to primes, \Cref{cor:ap-sharp-extrema} with $m=k=1$ gives the
matching upper bound for the minimum and lower bound for the reciprocal
limsup. For \eqref{eq:variable-degree-base-prime-minimum}, use the same
lower bound at $q=p^{d_x(p)}$, since
$\log\log q\le\log\log x+\log D(x)=(1+o(1))\log\log x$.
For the opposite bound $p-1\mid p^{d_x(p)}-1$ implies
$c_{n_x(p)}(p^{d_x(p)})\le\varphi(p-1)/(p-1)$.
The prime witnesses for the raw ratio in \Cref{cor:ap-sharp-extrema}
therefore work for every allowed degree assignment.
\end{remark}

\chapter{Matrix models and determinant-constrained LWE}\label{sec:syd21-app}

The principal result is a reduction to matrix secrets uniform conditional
on a prescribed unit determinant, over finite fields and integer residue
rings. Its proof combines a column hybrid with the action of the special
linear group on completable frames. After the reduction and its sampling
hypotheses, we study independent-entry matrices under a fixed-field
Fourier-gap hypothesis (\Cref{thm:finite-field-entry-determinant}) and
its prime-field largest-atom specialization (\Cref{thm:chi-det}).
The determinant comparisons in \Cref{cor:entry-determinant-comparisons}
control the determinant laws of these matrices after conditioning on
invertibility, and hidden special-linear randomization in
\Cref{cor:determinant-fiber-rerandomization} then turns a distinguisher
for secrets uniform within determinant fibers into one for the
conditioned secret laws, a one-way advantage transfer. The final section
applies the explicit
density bounds to the prescribed moduli of an earlier set-system
construction.

\section{Determinant-constrained LWE}\label{sec:determinant-lwe-reduction}

\paragraph*{LWE models}\phantomsection\label{par:lwe-models}
For a prime $p$, an integer $n\ge1$ and a law $\chi$ on $\Fp$,
an LWE sample~\cite{Reg09} with vector secret
$s\in\Fp^n$ is $(a,\langle a,s\rangle+e)$, where $a$ is uniform on
$\Fp^n$ and $e\sim\chi$ is independent. Ordinary decision LWE takes
$s$ uniform on $\Fp^n$. A square matrix-LWE sample with secret $S\in\Mn$ is
$(A,B)=(A,AS+E)$, where $A$ is uniform on $\Mn$ and $E$ has independent
$\chi$-distributed entries, independently of $A,S$. The uniform-secret and
PRIM-LWE models take $S$ uniform on $\Mn$ and $M_n^{\mathrm{prim}}(\Fp)$,
respectively. A common-secret transcript uses one secret in all samples, with
independent public matrices and errors; independent instances use independent
secrets. In both cases the null law consists of independent uniform pairs
$(A,B)$, and the advantage of a decision algorithm is the absolute difference
between its acceptance probabilities under the model law and under the null law.
These are the vector and square-matrix models of
\cite[Definition~17 and Theorem~10]{SYD21}.

Over a represented finite field $\mathbb F_q$, field-valued vector LWE
in dimension $k$ has samples $(a,\langle a,s\rangle+e)$ sharing one
secret $s\in\mathbb F_q^k$ drawn from a specified law. The secret,
all public vectors, and all errors are mutually independent, with $a$
uniform on $\mathbb F_q^k$ and $e\sim\chi$ in $\mathbb F_q$.
Its null law consists of independent uniform pairs in
$\mathbb F_q^k\times\mathbb F_q$. The matrix models and their null
laws use the same formulas over $\mathbb F_q$.
This is a field-valued source problem; passage to prime-field coordinates
does not in general give scalar LWE with the same error law.
For $R=\Z/Q\Z$, the same formulas define $R$-valued vector and matrix
LWE, with secrets, public entries and errors in $R$.

\begin{remark}[Determinant-fiber sampling and public affine maps]
\label{rem:direct-sampling}
The fibers of $\det:\GL_n(\Fp)\to\Fp^\times$ are the cosets of
$\mathrm{SL}_n(\Fp)$. Hence, if $g$ is uniform among the primitive
roots and $U$ is independently uniform on $\mathrm{SL}_n(\Fp)$, then
$U\operatorname{diag}(g,1,\ldots,1)$ is uniform on
$M_n^{\mathrm{prim}}(\Fp)$. The setup conditions following
\Cref{thm:determinant-lwe} give an exact sampler for $g$ with
polynomial expected running time.

For square matrix-LWE samples, the randomized map
\[
 (A,B)\longmapsto(AM^{-1},BN+AM^{-1}R),
 \qquad M,N\in\GL_n(\Fp),\quad R\in\Mn,
\]
with $(M,N,R)$ public and chosen independently of the samples, sends $S$
to $MSN+R$ and $E$ to $EN$; it preserves the prescribed i.i.d.\ error law
only when right multiplication by $N$ is a symmetry of that law. Each
fixed $(M,N,R)$ acts on $\Mn$ by an affine bijection, so a uniform secret
remains uniform and has primitive-root determinant with probability
$c_n(p)$, or $\varphi(p-1)/(p-1)$ conditional on invertibility; this
gives no acceptance test for a hidden secret.
\end{remark}

\begin{proposition}[Common-secret masking]\label{prop:common-secret-masking}
\index{Learning with errors!common-secret masking}
Let $t\ge1$, and consider the $t$-sample common-secret matrix models
over a finite field $\mathbb F_q$ and their joint uniform null law.
Let $R$ be uniform on $M_n(\mathbb F_q)$, hidden from the
distinguisher, and independent of $S$, all public matrices, all errors, and
all pairs under the null law.
Then
\[
 (A_j,B_j)_{j=1}^t\longmapsto(A_j,B_j+A_jR)_{j=1}^t
\]
sends the PRIM-LWE law to the uniform-secret matrix-LWE law and
preserves the joint null law. It gives a reduction from decision
PRIM-LWE to decision uniform-secret matrix LWE with the same advantage.
\end{proposition}

This map is the matrix form of the secret re-randomization in the proof
of~\cite[Lemma~4.1, p.~34:33]{Reg09}.

\begin{proof}
We have $B_j+A_jR=A_j(S+R)+E_j$, and conditional on $S,(A_j,E_j)_{j=1}^t$
the secret $S+R$ is uniform on $M_n(\mathbb F_q)$; hence it is uniform and independent of
the public matrices and errors, whose joint law is unchanged. Under the null
law, for fixed $(A_j)_{j=1}^t$ and $R$, the map
$(B_j)_{j=1}^t\mapsto(B_j+A_jR)_{j=1}^t$ is a translation of $M_n(\mathbb F_q)^t$ and
preserves its uniform law. Composing a distinguisher with the map therefore
preserves its advantage.
\end{proof}

The masking map runs in the opposite direction to the following
reduction. We combine the matrix-column hybrid of~\cite[Lemma~2.9]{Mic18}
with a change of basis on completable frames. Hidden invertible changes
of basis for vector secrets occur in~\cite[\S~3.4, Lemma~5]{BGMRR16}.

\KeepHeadingSpace{6\baselineskip}
\paragraph*{Determinant-law model}
Let $R$ be either a finite field with a public polynomial-basis
representation $\mathbb F_\ell[T]/(f)$, where $\ell$ is prime and $f$
is supplied monic irreducible, or $\Z/Q\Z$ with $Q\ge2$. Put $q=|R|$.
Arithmetic in either representation has polynomial bit cost.
Let $n\ge2$ and $t\ge1$ be integers, and let $\chi,\pi$ be laws
on $R,R^\times$, respectively. Write $L$ for the
bit length of the public setup, including the representation of $R$,
shared by the source and target problems.
Define $\mathcal S_\pi$ by drawing $d\sim\pi$ and then drawing $S$
uniformly from $\{S\in M_n(R):\det S=d\}$. The target transcript is
\[
 (A_j,A_jS+E_j)_{j=1}^t,\qquad S\sim\mathcal S_\pi,
\]
where the $A_j$ are independent uniform matrices and all entries of
the $E_j$ are independent with law $\chi$, independently of $S$ and
the $A_j$. The null transcript is jointly uniform. All distributional
statements below are conditional on each fixed public setup.

\begin{theorem}[LWE with a prescribed determinant law]
\label{thm:determinant-lwe}
\index{Learning with errors!determinant-constrained reduction}
\index{Determinant fibers!uniform sampling}
In the determinant-law model above, assume exact samplers for $\chi$
and $\pi$ with expected time polynomial in $n,t,\log q,L$.
A target distinguisher $D$ of advantage $\varepsilon$ gives a randomized
algorithm using $nt$ $R$-valued vector-LWE samples in dimension $n-1$,
with error law $\chi$ and one call to $D$, whose advantage is exactly
$\varepsilon/n$. Its expected overhead is polynomial in $n,t,\log q,L$,
including sampler costs. The source may have any common-secret law on
$R^{n-1}$ independent of the public vectors and errors;
no sampler for this law is required, and the guarantees are uniform
over these laws. For $R=\Z/Q\Z$, no factorization of $Q$ is required.
\end{theorem}

\begin{proof}
For $r<n$, a completable $r$-frame consists of the first $r$ columns
of an invertible matrix over $R$. Scaling a complementary column by
the inverse determinant gives a special-linear completion. Thus
$\mathrm{SL}_n(R)$ acts transitively on these frames. Its action on
each unit determinant fiber is also transitive, since $GF^{-1}$ has
determinant one when $\det G=\det F\in R^\times$.
Consequently a uniform special-linear matrix sends
each fixed frame, or fixed matrix in a determinant fiber, to the
uniform law on its orbit.

For $0\le r<n$, let $H_r$ be the transcript obtained by choosing a
uniform completable $r$-frame $F$, using the same $F$ in all samples,
and setting the first $r$ columns of $B_j$ to $A_jF$ plus independent
$\chi$-errors. Fill the remaining columns independently and uniformly.
The public matrices are independent uniform matrices, independently
of the frame and errors. Then $H_0$ is the null law; let $H_n$ be
the target law in the statement.

Choose $r$ uniformly from $\{1,\ldots,n\}$ and choose a hidden
uniform $M\in\mathrm{SL}_n(R)$, once for the whole transcript.
Group the $nt$ challenge pairs into $t$ blocks of $n$ rows. Write
a challenge row as $(a_i,b_i)$, with $a_i\in R^{n-1}$, append an
independent uniform $u_i\in R$, and let $C_j$ be the resulting
$n\times n$ matrix in block $j$. Set $z=1$ for $r<n$; for $r=n$,
draw $z\sim\pi$ once. In every block form $B_j$ by taking its
column $k<r$ to be $C_je_k$ plus independent $\chi$-errors, its
column $r$ to have entries $b_i+zu_i$, and its columns $k>r$ to be
independent uniform vectors. Output
\[
 (C_jM^{-1},B_j)_{j=1}^t.
\]
Neither $M$ nor $r$ is supplied to $D$.

On a real source transcript with any fixed secret $s\in R^{n-1}$,
the encrypted columns before the change of basis are
\[
 F_0=[e_1,\ldots,e_{r-1},w],\qquad
 w=(s_1,\ldots,s_{n-1},z)^{\mathsf T}.
\]
For $r<n$, appending $e_r,\ldots,e_{n-1}$ to $F_0$ gives determinant
$\pm1$, since the last coordinate of $w$ is one. Thus $F_0$ is
completable and $MF_0$ is a uniform $r$-frame. For $r=n$, $\det F_0=z\in R^\times$,
so conditional on $z$, $MF_0$ is uniform on its determinant fiber.
Conditional on $s,M,z$ and all error variables, the $C_j$ are
independent uniform matrices. Right multiplication by $M^{-1}$
is a bijection, and this conditional law is constant. Consequently
the public matrices $C_jM^{-1}$ are independent of $s,M,z$ and all
errors, and in particular of the resulting secret frame. The
challenge errors occupy column $r$ unchanged; the other encrypted
columns have independently sampled errors of the same law. Thus the
real output is exactly $H_r$ for every fixed $s$ and hence for every
source-secret law.

On a null source transcript, the $b_i$ are independent uniform
scalars conditional on all $a_i,u_i,M,z$ and freshly sampled errors.
Their translates $b_i+zu_i$ therefore form an independent uniform
column. The surviving encrypted frame
$[Me_1,\ldots,Me_{r-1}]$ is uniform, with the same public-matrix
independence. The output law is exactly $H_{r-1}$. Writing
$q_r=\mathbb P(D(H_r)=1)$, the signed acceptance gap is
\[
 \frac1n\sum_{r=1}^n(q_r-q_{r-1})=\frac{q_n-q_0}{n}.
\]
Taking absolute values gives $\varepsilon/n$.

Draw uniform $U\in M_n(R)$ until $\det U\in R^\times$ and
return $U\operatorname{diag}((\det U)^{-1},1,\ldots,1)$.
Each special-linear matrix has $|R^\times|$ preimages. Over a field,
the acceptance probability is at least $\prod_{i\ge1}(1-2^{-i})>1/4$.
For $R=\Z/Q\Z$, \Cref{prop:finite-ring-det} gives
\[
 \mathbb P(U\in\GL_n(R))
 =\prod_{p\mid Q}\prod_{j=1}^n(1-p^{-j})
 \ge\frac{\varphi(Q)}{4Q}
 \ge\frac1{4(1+\log_2Q)}.
\]
Indeed, $\prod_{j\ge2}(1-p^{-j})\ge1-1/(p(p-1))$; the factor for
$p=2$ is at least $1/2$, and the remaining product is at least
$1-\sum_{m\ge3}1/(m(m-1))=1/2$. Ordering the distinct prime divisors
as $p_i\ge i+1$ gives $Q/\varphi(Q)\le1+\log_2Q$.
Compute determinants and adjugates by the division-free Berkowitz
algorithm~\cite{Ber84}, using polynomially many ring operations.
Modulo $Q$, test the determinant by its gcd with $Q$ and invert it
by extended Euclid only after acceptance. Reduction modulo $Q$ at each
step gives polynomial bit cost without factoring $Q$; the adjugate gives
$M^{-1}$ since $\det M=1$.
Uniform residues, including field coefficients, are sampled by binary
rejection with success probability at least $1/2$. These operations
and the assumed samplers for $\chi,\pi$ give the asserted expected overhead.
\end{proof}

\begin{corollary}[Exact-secret search reduction]
\label{cor:determinant-lwe-search-mas21}
\index{Learning with errors!exact-secret search}
Under the ring, distribution and exact-sampling hypotheses of
\Cref{thm:determinant-lwe}, suppose an oracle, given $t$ common-secret
determinant-law samples, returns the actual planted matrix secret
$S\sim\mathcal S_\pi$ with probability at least $\varepsilon$.
There is a randomized algorithm that recovers the actual vector secret
from $nt$ common-secret $R$-valued vector-LWE samples in dimension $n-1$
with probability at least $\varepsilon$, using one oracle call and
expected overhead polynomial in $n,t,\log q,L$.
The guarantee holds for every source-secret law independent of the
public vectors and errors, without a sampler for that law; for
$R=\Z/Q\Z$, it requires no factorization of $Q$.
\end{corollary}

\begin{proof}
Use the simulator of \Cref{thm:determinant-lwe} with the embedded
column fixed at $r=n$. Explicitly, group the source pairs
$(a_i,b_i)$ into $t$ blocks of $n$ rows, independently append a
uniform $u_i\in R$ to each $a_i\in R^{n-1}$, and call the resulting
matrices $C_j$. Draw one $z\sim\pi$ and one independent uniform
$M\in\mathrm{SL}_n(R)$ for the whole transcript. For $k<n$, set
column $k$ of $B_j$ to $C_je_k$ plus fresh independent
$\chi$-errors; set its last column to the entries $b_i+zu_i$.
Give the oracle $(C_jM^{-1},B_j)_{j=1}^t$.

Conditional on any source secret $s\in R^{n-1}$, put
\[
 w=(s_1,\ldots,s_{n-1},z)^{\mathsf T},\qquad
 F_0=[e_1,\ldots,e_{n-1},w],\qquad S=MF_0.
\]
Since $\det F_0=z\in R^\times$, for every fixed $s,z$ the map
$M\mapsto MF_0$ is a bijection from $\mathrm{SL}_n(R)$ onto
the determinant-$z$ fiber: the inverse sends $G$ to $GF_0^{-1}$.
Thus $S$ is exactly uniform on that fiber conditional on every
$s,z$, and its unconditional law is $\mathcal S_\pi$.
The last-column errors are precisely the original vector-LWE errors;
all other entries of the error matrices are independently sampled
from $\chi$. Conditional on $s,M,z$ and every error variable, the
$C_j$ are independent uniform matrices. The bijections
$C_j\mapsto C_jM^{-1}$ leave this conditional law constant, so the
public matrices are independent uniform and independent of $S$ and
all errors. Consequently the oracle receives the exact target
joint law, with the actual planted secret $S$.

If the oracle returns a matrix $\widehat S$, return the first $n-1$
coordinates of $M^{-1}\widehat S e_n$; otherwise declare failure.
On the event $\widehat S=S$, the whole column is
$M^{-1}Se_n=w=(s,z)^{\mathsf T}$, and those first coordinates
equal $s$. This event has probability at least $\varepsilon$.
The samplers and ring operations are those already justified in
\Cref{thm:determinant-lwe}, proving the expected-overhead assertion.
\end{proof}

The search reduction assumes recovery of the planted matrix, not merely
some secret consistent with the observed noisy equations, and fixes the
embedded column at $r=n$. The decision reduction in
\Cref{thm:determinant-lwe} instead chooses $r$ uniformly and obtains
advantage $\varepsilon/n$ by telescoping the signed acceptance gaps.

\Needspace{10\baselineskip}
\begin{corollary}[Bounded-time reduction]
\label{cor:determinant-lwe-cutoff}
Under the hypotheses of \Cref{thm:determinant-lwe}, let $\lambda$ be a
security parameter and suppose $n,t,\log q,L$ are polynomially bounded
in $\lambda$. Assume that the samplers for $\chi$ and $\pi$ have known
polynomial cutoffs with negligible total cutoff probability in the
simulation. Then the reduction has polynomially bounded overhead and
advantage at least $\varepsilon/n-\operatorname{negl}(\lambda)$.
More precisely, total cutoff probability at most $\delta$ in each
source experiment gives advantage at least $\varepsilon/n-2\delta$,
uniformly over the source-secret laws.
\end{corollary}

\begin{proof}
The uniform residue and finite-set samplers have geometric tails.
For either $R$, the special-linear sampler has success probability at
least $1/[4(1+\log_2q)]$ per matrix trial; after
$\lceil4(1+\log_2q)\lambda^2\rceil$ trials its failure probability is
at most $e^{-\lambda^2}$. These cutoffs, including those for every
residue draw, and the assumed cutoffs for $\chi,\pi$ give polynomially
bounded overhead with negligible total cutoff probability.
Coupling the exact and truncated executions bounds their total
variation distance by the cutoff probability in each experiment.
The two acceptance probabilities therefore change by at most
$\delta$ each.
\end{proof}

Taking $\pi$ to be the point mass at any prescribed $d\in R^\times$ gives
the theorem for a uniform determinant-$d$ secret, in particular for
a uniform secret in $\mathrm{SL}_n(R)$. For $R=\mathbb F_q$, taking $\pi$ uniform on
the primitive roots gives PRIM-LWE. For the latter specialization,
a supplied certified primitive root $g_0$ permits exact sampling by
drawing $u$ uniformly modulo $q-1$ until $\gcd(u,q-1)=1$ and returning
$g_0^u$. Alternatively, a complete certified factorization of $q-1$
permits testing uniform nonzero candidates $g$ by
$g^{(q-1)/v}\ne1$ for every prime $v\mid q-1$. Each sampler uses
$(q-1)/\varphi(q-1)=O(1+\log\log q)$ expected trials and has
geometric tails; see the bound in the proof of
\Cref{cor:rosser-schoenfeld}. This setup is also supplied to the
source distinguisher.

\KeepHeadingSpace{6\baselineskip}
\begin{corollary}[Same-dimension reduction]
\label{cor:determinant-lwe-same-dimension}
\index{Learning with errors!same-dimension reduction}
Under the hypotheses of \Cref{thm:determinant-lwe} with $R=\mathbb F_q$, a target
distinguisher of advantage $\varepsilon$ gives a dimension-$n$
vector-LWE distinguisher using $nt$ common-secret samples, one target
call, and advantage exactly $\varepsilon/(nq)$ under exact sampling.
The source-secret law on $\mathbb F_q^n$ is arbitrary and independent
of the public vectors and errors. The cutoff qualification of
\Cref{cor:determinant-lwe-cutoff} applies uniformly over these laws. For $n,t,q$ polynomial in a
security parameter, this gives a polynomial loss; in the primitive
case, no supplied factorization or generator is needed in this regime.
\end{corollary}

\begin{proof}
Write the source secret as $(s',s_n)$ and its samples as
$((a_i',a_{i,n}),b_i)$. Draw one uniform $v\in\mathbb F_q$ for the whole
transcript and pass $(a_i',b_i-va_{i,n})$ to the reduction of
\Cref{thm:determinant-lwe}, discarding $a_{i,n}$. With probability
$1/q$, $v=s_n$; this event preserves the law of the source secret,
and the result is dimension-$(n-1)$ LWE with the marginal secret law
of $s'$. Otherwise, conditional on $s',s_n,v$, the discarded
independent uniform $a_{i,n}$ make
$a_{i,n}(s_n-v)$ independent uniform scalars, so the whole transcript
has the null law. Null input remains null. Hence the signed
acceptance gap is multiplied by $1/q$, proving the assertion.
For polynomial $q$, trial division factors $q-1$ in polynomial time
in the security parameter, after which the primitive-root sampler
above applies.
\end{proof}

\begin{corollary}[Randomization within determinant fibers]
\label{cor:determinant-fiber-rerandomization}
\index{Learning with errors!secret randomization}
For either $R$ in \Cref{thm:determinant-lwe}, let
$n\ge2$, $t\ge1$ and let $S\sim\eta$ be supported on
$\GL_n(R)$, independently of the public matrices and errors.
Put $\gamma=\operatorname{Law}(\det S)$.
One hidden, independent uniform $U\in\mathrm{SL}_n(R)$ gives
\[
 (A_j,B_j)_{j=1}^t\longmapsto(A_jU^{-1},B_j)_{j=1}^t.
\]
This sends the $\eta$-secret transcript exactly to the
$\mathcal S_\gamma$ transcript, preserves the error law and joint
uniform null law, and has expected polynomial overhead.
Consequently a distinguisher of advantage $\varepsilon$ for
$\mathcal S_\pi$ gives advantage at least $\varepsilon-\delta$, where
$\delta=d_{\mathrm{TV}}(\gamma,\pi)$.
A cutoff probability at most $\delta_{\mathrm{cut}}$ in each experiment
changes this lower bound to $\varepsilon-\delta-2\delta_{\mathrm{cut}}$.
No sampler for the secret or determinant laws is required.
\end{corollary}

\begin{proof}
This is the determinant-fiber analogue of~\cite[Lemma~5]{BGMRR16}.
The orbit argument in \Cref{thm:determinant-lwe} makes $US$ uniform
conditional on $\det S$. Conditional on $S,U$ and all errors,
right multiplication by $U^{-1}$ leaves the public matrices independent
uniform, with a conditional law independent of those variables.
The identity $B_j=(A_jU^{-1})(US)+E_j$ proves the real-law assertion;
the same bijection preserves the null law. Since
$d_{\mathrm{TV}}(\mathcal S_\gamma,\mathcal S_\pi)
=d_{\mathrm{TV}}(\gamma,\pi)$, data processing gives the advantage bound.
The special-linear sampler in \Cref{thm:determinant-lwe} and the
coupling argument in \Cref{cor:determinant-lwe-cutoff} give the
computational assertions.
\end{proof}

\Needspace{22\baselineskip}
\paragraph*{Reduction interfaces}\phantomsection\label{par:lwe-reductions}
Each row gives the source advantage obtained from a distinguisher
for the problem in the middle column, over $R$ as above. The first and
third rows require $R=\mathbb F_q$.
Each transcript has one common secret, and the prescribed error law is
$\chi$ throughout. The stated advantages use exact sampling.
\begin{center}
\begin{tabular}{@{}p{0.34\linewidth}p{0.40\linewidth}p{0.16\linewidth}@{}}
\toprule
Source problem and samples & Distinguisher used & \shortstack[l]{Source\\advantage} \\
\midrule
PRIM-LWE; $t$ matrix samples & Uniform-secret matrix LWE; $t$ samples
 & $\varepsilon$ \\
Vector LWE, arbitrary secret law; dimension $n-1$, $nt$ samples
 & Determinant-law matrix LWE; one call on $t$ samples
 & $\varepsilon/n$ \\
Vector LWE over $\mathbb F_q$, arbitrary secret law; dimension $n$, $nt$ samples
 & The same determinant-law target; one call on $t$ samples
 & $\varepsilon/(nq)$ \\
Matrix LWE, secret law $\eta$ on $\GL_n(R)$; $t$ samples
 & Determinant-law matrix LWE; one call on $t$ samples
 & $\ge\varepsilon-\delta$ \\
\bottomrule
\end{tabular}
\end{center}
The rows are \Cref{prop:common-secret-masking},
\Cref{thm:determinant-lwe}, \Cref{cor:determinant-lwe-same-dimension}
and \Cref{cor:determinant-fiber-rerandomization}, respectively.
In the last three, the target has unit determinant law $\pi$ and is
uniform conditional on that determinant, for each fixed public setup.
In the last row $\delta=d_{\mathrm{TV}}(\operatorname{Law}(\det S),\pi)$;
only uniform special-linear sampling is used.
Rows two and three have the expected-time and sampling guarantees of
\Cref{thm:determinant-lwe}. Under the hypotheses of
\Cref{cor:determinant-lwe-cutoff}, cutoff probability at most
$\delta_{\mathrm{cut}}$ in each experiment costs at most
$2\delta_{\mathrm{cut}}$ in advantage. These guarantees are uniform over
source-secret laws independent of the public vectors and errors;
no source-secret sampler is required. For primitive-root sampling,
the public setup is as above. When $n,t,q$ are polynomial in the
security parameter, trial division supplies the factorization and
the same-dimension loss is polynomial; polynomial $\log q$ alone
is insufficient.

The factor-$n$ reduction has one dimension of overhead. For prime
$q=p$, the same-dimension loss $np$ is $O_p(n)$, but does not give a
uniform $O(n^2)$ bound when $p$ grows without a restriction on $p/n$.
The $O(n^2)$ bound in~\cite[Theorem~10]{SYD21} uses the
$p$-dependent constant of its Proposition~5; its proof does not
establish uniformity in this regime.
Both results concern target secrets uniform within determinant fibers;
they do not supply such a law for the bounded-norm encoding
of~\cite[\S~8.2]{SYD21} or establish security of that protocol.

\begin{remark}[Entropy deficit of the determinant condition]\label{rem:entropy}
Here $H_\alpha$ denotes R\'enyi entropy of order $\alpha$ (Shannon
entropy when $\alpha=1$), and $D_{\mathrm{KL}}$ denotes
Kullback--Leibler divergence; both use base-$2$ logarithms.
For the law $\mathcal S_\pi$ over $\mathbb F_q$, every nonzero determinant fiber
has size $F=|\mathrm{SL}_n(\mathbb F_q)|$, so
\[
 \begin{aligned}
 H_\alpha(\mathcal S_\pi)&=\log_2F+H_\alpha(\pi)
                         \quad(0\le\alpha\le\infty),\\
 D_{\mathrm{KL}}(\mathcal S_\pi\Vert\mathrm{Unif}(M_n(\mathbb F_q)))
 &=n^2\log_2q-H_1(\mathcal S_\pi).
 \end{aligned}
\]
Let $\mathsf P$ and $\mathsf Q$ be the uniform laws on $\Mn$ and on
$M_n^{\mathrm{prim}}(\Fp)$, and put $c=c_n(p)$. Since $\mathsf Q$ is uniform
on $cp^{n^2}$ points, every R\'enyi entropy in base $2$ satisfies
\[
 H_\alpha(\mathsf Q)=n^2\log_2p-\log_2(1/c)
 \quad(0\le\alpha\le\infty),\qquad
 D_{\mathrm{KL}}(\mathsf Q\Vert\mathsf P)=\log_2(1/c).
\]
As $c_n(p)>c(p)$, the deficit is
at most $\log_2(1/c(p))=O(\log\log\log p)$ bits.
Its mean in the limiting law over primes is
$-M'(0)/\log2\approx1.459$ bits, below the Jensen bound
$\log_2\mathbb E[1/X]\approx1.499$ bits (\Cref{cor:avg-loss}).

Entropic LWE reductions carry specific noise and dimension hypotheses:
\cite[Theorem~4]{GKPV10} treats binary secrets, and \cite[Theorem~4.1]{BD20}
dependent vector coordinates with average conditional min-entropy given an
observation of the secret plus Gaussian noise. For $B=AS+E$, with
$\operatorname{vec}$ stacking matrix columns, the public matrix in
$\operatorname{vec}(AS)=(I_n\otimes A)\operatorname{vec}(S)$ is structured,
so applying them to the joint transcript requires a further argument for the
conditional secret laws and noise and dimension parameters.
\Cref{thm:determinant-lwe} uses the uniform laws on completable frames
rather than these entropy identities.
\end{remark}

\section{Determinant statistics for secrets with independent entries}\label{ssec:chi-secret}

We first work over a fixed finite field $\mathbb F_q$, $q=p^f$,
under a uniform gap for additive Fourier coefficients. The prime-field
largest-atom hypothesis will then give a specialization.
The entry laws may differ between positions and dimensions.
Write $\psi_q(x)=\exp(2\pi i\operatorname{Tr}_{\mathbb F_q/\mathbb F_p}(x)/p)$,
$P_n(q)=\prod_{j=1}^n(1-q^{-j})$, $P(q)=\prod_{j\ge1}(1-q^{-j})$,
$c_n(q)=\varphi(q-1)P_n(q)/(q-1)$ and
$c(q)=\varphi(q-1)P(q)/(q-1)$.

\paragraph*{Determinant laws and hidden randomization}\phantomsection\label{par:determinant-matrix-laws}
The secret-entry laws are distinct from the LWE error law $\chi$ in
the \hyperref[par:lwe-models]{LWE models paragraph}. The bounds in
\Cref{cor:entry-determinant-comparisons}\textup{(iii)} compare
determinant-valued random variables and their conditional laws,
with the field and Fourier gap fixed as $n\to\infty$.
After conditioning on a nonzero determinant, hidden randomization in
\Cref{cor:determinant-fiber-rerandomization} turns a distinguisher for
secrets uniform within determinant fibers into one for the conditioned
entry law, losing at most the total variation distance between their
determinant laws.

For i.i.d.\ entries, the asymptotic probability of each nonzero
determinant value is stated in~\cite[Theorem~1.2]{Map13}; its proof uses
Proposition~2.3, which is incorrect~\cite[Remark~A.10]{NP20}.
For i.i.d.\ entries drawn from a fixed distribution on a fixed prime
field whose support contains at least two elements, the asymptotic in \eqref{eq:maples-fixed} also follows from Lvov's
joint cokernel--determinant universality
\cite[Theorems~1.1\textup{(B)} and~2.2]{Lvo26}.
The uniformity over entry laws under the stated largest-atom bound
comes from the direct proof below.

\begin{lemma}[Product rows in a finite-field quotient]
\label{lem:finite-field-entry-quotient}
Fix $q$ and $0<\lambda<1$. Let $X_1,\ldots,X_n\in\mathbb F_q^n$
have mutually independent coordinates, with
$|\mathbb E\psi_q(a(X_i)_j)|\le\lambda$ for every $a\ne0$.
Put $\beta=(1+(q-1)\lambda)/q<1$ and $b=-\log\beta$.
There is $0<\varepsilon<1/4$, depending only on $q,\lambda$, such that,
for $r=\lfloor\varepsilon n\rfloor\ge1$ and $m=n-r$, the following holds. Except on an event of probability
at most
\[
 a_n:=\frac{\beta^{r+1}}{1-\beta}+e^{-bn/2},
\]
the space $W=\langle X_1,\ldots,X_m\rangle$ has dimension $m$,
and, conditional on the exposed first $m$ rows, the joint law of
the remaining $r$ rows modulo $W$ is within
$r e^{-\kappa n}/2$ in total variation of $r$ independent uniform
elements of $\mathbb F_q^n/W$, for some $\kappa>0$ depending only
on $q,\lambda$.
\end{lemma}

\begin{proof}
Fourier inversion gives every entry atom the upper bound
$q^{-1}(1+(q-1)\lambda)=\beta$.
A product row lies in a fixed codimension-$d$ subspace with
probability at most $\beta^d$: choose $d$ pivot coordinates in
defining linear equations and condition on all other coordinates.
The pivot coordinates are uniquely specified and remain independent.
This is the product-measure form of Odlyzko's
lemma~\cite{Odl88}; see \cite[Lemma~2.2]{Map13}.
Sequential exposure therefore gives
\[
 \mathbb P(\dim\langle X_1,\ldots,X_m\rangle<m)
 \le\sum_{k=0}^{m-1}\beta^{n-k}
 \le\frac{\beta^{r+1}}{1-\beta}.
\]
Let $H(u)=-u\log u-(1-u)\log(1-u)$. Choose $0<\delta<1/2$ so that
$h:=H(\delta)+\delta\log(q-1)<b/4$, and then choose
$0<\varepsilon<1/4$ so small that
$\kappa:=-\delta\log\lambda-\varepsilon\log q>0$.
For each nonzero $v\in\mathbb F_q^n$, the pivot bound gives
$\mathbb P(v\cdot X_i=0)\le\beta$.
With $z=\delta/((1-\delta)(q-1))<1$,
\[
 \sum_{k\le\delta n}\binom nk(q-1)^k
 \le z^{-\delta n}(1+(q-1)z)^n=e^{nh}.
\]
Independence of the first $m$ rows and a union bound show that
\[
 \mathbb P\bigl(\exists v\ne0:\ |\supp v|\le\delta n,
                  \ v\cdot X_i=0\ (1\le i\le m)\bigr)
 \le e^{nh}\beta^m\le e^{-bn/2},
\]
because $m\ge3n/4$. Outside the two exceptional events, every
nonzero vector in $W^\perp$ has support greater than $\delta n$.

The additive characters of $\mathbb F_q^n/W$ are
$x+W\mapsto\psi_q(v\cdot x)$ for $v\in W^\perp$.
Indeed, the trace pairing is nondegenerate, and its annihilator
of an $\mathbb F_q$-linear space is its ordinary dot-product
orthogonal complement: if $v\cdot w\ne0$, some scalar multiple
of $w$ has nonzero trace pairing with $v$.
For a remaining row $\xi$, independence of coordinates bounds
each nontrivial character expectation by $\lambda^{\delta n}$.
Fourier inversion on the quotient, which has $q^r$ elements, gives
\[
 d_{\mathrm{TV}}(\operatorname{Law}(\xi+W),
                    \mathrm{Unif}(\mathbb F_q^n/W))
 \le\frac{q^r-1}{2}\lambda^{\delta n}
 \le\frac12 e^{-\kappa n}.
\]
This is the code argument of
\cite[Definition~2 and Lemma~2.1, p.~388]{Woo19}
(cf.~\cite[Definition~4.4 and Lemma~4.5]{NW22}).
Conditioning on the first $m$ rows leaves the remaining rows
independent. Replacing their projected laws one at a time and
using the triangle inequality for product measures gives the
claimed joint total variation bound.
\end{proof}

\begin{theorem}[Independent-entry determinants over a fixed finite field]
\label{thm:finite-field-entry-determinant}
\index{Determinant densities!extension fields}
Fix $q=p^f$ and $0<\lambda<1$. For each $n\ge1$, let
$S_n\in M_n(\mathbb F_q)$ have independent entries satisfying
\begin{equation}\label{eq:finite-field-entry-gap}
 \sup_{n,i,j}\max_{a\in\mathbb F_q^\times}
       |\mathbb E\psi_q(a(S_n)_{ij})|\le\lambda.
\end{equation}
There are $C_{q,\lambda},c_{q,\lambda}>0$, depending only on
$q,\lambda$, such that, uniformly over all these ensembles,
all $n\ge1$ and $t\in\mathbb F_q^\times$,
\begin{equation}\label{eq:finite-field-entry-determinant}
 \left|\mathbb P(\det S_n=t)-\frac{P(q)}{q-1}\right|
 \le C_{q,\lambda}e^{-c_{q,\lambda}n}.
\end{equation}
If $U_n$ is uniform on $M_n(\mathbb F_q)$, then also
\[
 d_{\mathrm{TV}}(\operatorname{Law}(n-\operatorname{rank}S_n),
                 \operatorname{Law}(n-\operatorname{rank}U_n))
 \ll_{q,\lambda}e^{-c'n}
\]
for some $c'=c'(q,\lambda)>0$.
\end{theorem}

\begin{proof}
Use the rows, parameters and good event in
\Cref{lem:finite-field-entry-quotient}. On the good event complete
the first $m$ rows to an invertible matrix $B$. Then
\[
 S_nB^{-1}=\begin{pmatrix}I_m&0\\ *&Y\end{pmatrix},\qquad
 \det S_n=(\det B)\det Y,
\]
where the rows of $Y$ are the remaining rows in quotient coordinates.
For a uniform $r\times r$ matrix, every nonzero determinant value
has probability $P_r(q)/(q-1)$: its nonsingularity probability is
$P_r(q)$ by successive independent row counts, and scaling one
row by a nonzero element permutes determinant fibers transitively.
Multiplication by $\det B\ne0$ preserves this common mass.
Thus averaging the conditional total variation bound, including
the bad event, and using the product bound
$0\le P_r(q)-P(q)\le\sum_{j>r}q^{-j}=q^{-r}/(q-1)$, gives
\[
 \left|\mathbb P(\det S_n=t)-\frac{P(q)}{q-1}\right|
 \le\frac{\beta^{r+1}}{1-\beta}+e^{-bn/2}
       +\frac r2e^{-\kappa n}+\frac{q^{-r}}{(q-1)^2}.
\]
Since $r=\lfloor\varepsilon n\rfloor$, this is exponentially small
with constants depending only on $q,\lambda$ whenever $r\ge1$.
The left side is at most one, so enlarging the constant covers
the finitely many smaller $n$.

On the same good event, $n-\operatorname{rank}S_n$ is
$r-\operatorname{rank}Y$. Its law is within $a_n+r e^{-\kappa n}/2$
of the corank law of a uniform $r\times r$ matrix.
For $U_n$, its first $m$ rows fail to be independent with
probability at most $q^{-(r+1)}/(1-q^{-1})$, by the same pivot
argument with atom bound $q^{-1}$. Conditional on independent
first rows, its remaining quotient rows are exactly independent
uniform, so its corank has that same $r\times r$ law. The triangle
inequality proves the asserted corank estimate, again enlarging
the constant to cover small $n$.
\end{proof}

For established rank and cokernel universality results, see
\cite[Theorem~1.1]{Map13}, \cite[Appendix~A and Remark~A.10]{NP20},
\cite[Theorem~1.2]{KN21} and \cite[Corollary~3.5]{Woo19}.
A common atom-mass bound alone does not suffice: entries uniform
on a proper subfield have every determinant in that subfield.
Thus nonzero values outside it have probability zero, and primitive
determinants never occur. The trace-pairing annihilator of this
proper additive subgroup supplies a nontrivial Fourier coefficient
of modulus one, so such laws do not satisfy
\eqref{eq:finite-field-entry-gap}.

\begin{corollary}[Determinant events, acceptance and conditional comparisons]
\label{cor:entry-determinant-comparisons}
Fix a finite field $\mathbb F_q$. Suppose a matrix ensemble satisfies
\[
 |\mathbb P(\det S_n=t)-P(q)/(q-1)|\le Ce^{-cn}
 \quad(n\ge1,\ t\ne0)
\]
for fixed $C,c>0$, as in \Cref{thm:finite-field-entry-determinant}.
Define together
\[
 \begin{gathered}
 w:=\frac{P(q)}{q-1},\qquad \eta_n:=Ce^{-cn},\qquad
 E_n:=\varphi(q-1)\eta_n+2^{-n},\\
 n_0(q;C,c):=\min\{n\ge1:E_n\le c(q)/2\},\qquad
 \rho_n:=\mathbb P(\det S_n\text{ generates }\mathbb F_q^\times),
 \end{gathered}
\]
and let $U_n$ be uniform on $M_n(\mathbb F_q)$.
\begin{enumerate}[\upshape(i)]
\item For $A\subseteq\mathbb F_q^\times$,
$|\mathbb P(\det S_n\in A)-|A|w|\le|A|\eta_n$.
In particular $|\rho_n-c(q)|\le\varphi(q-1)\eta_n$ and
$|\rho_n-c_n(q)|\le E_n$.
\item For every $n\ge n_0(q;C,c)$, $\rho_n\ge c_n(q)/2>0$.
\item For every $n\ge1$,
\[
 d_{\mathrm{TV}}(\operatorname{Law}(\det S_n),
                 \operatorname{Law}(\det U_n))
 \le(q-1)\eta_n+\frac{q^{-n}}{q-1}.
\]
If $\eta_n<w$, then for every nonempty $A\subseteq\mathbb F_q^\times$,
\[
 d_{\mathrm{TV}}(\operatorname{Law}(\det S_n\mid\det S_n\in A),
                 \mathrm{Unif}(A))
 \le\frac{\eta_n}{w-\eta_n}.
\]
\end{enumerate}
\end{corollary}

\begin{proof}
Summing the assumed estimate over $A$ proves the first assertion.
For the $\varphi(q-1)$ primitive elements the reference mass is $c(q)$.
Moreover,
\[
 0\le c_n(q)-c(q)\le P_n(q)-P(q)
 \le\frac{q^{-n}}{q-1}\le2^{-n},
\]
which proves (i). The positive sequence $E_n$ decreases to zero,
so $n_0$ exists. For $n\ge n_0$, (i) gives
$\rho_n\ge c_n(q)-E_n\ge c_n(q)-c(q)/2\ge c_n(q)/2$.
For (iii), write $\mathbb P(\det S_n=t)=w+e_t$ for $t\ne0$,
where $|e_t|\le\eta_n$. The uniform matrix has nonzero determinant
mass $P_n(q)/(q-1)$ at each such $t$. The discrepancy at zero is
the negative sum of the nonzero discrepancies. Half the sum of
all absolute discrepancies is consequently at most
$(q-1)\eta_n+P_n(q)-P(q)$, proving the first bound.
For nonempty $A$ let $E_A=\sum_{t\in A}e_t$. The event probability
is $|A|w+E_A\ge|A|(w-\eta_n)>0$, and its conditional distance is
\[
 \frac{\sum_{t\in A}|e_t-E_A/|A||}{2(|A|w+E_A)}
 \le\frac{|A|\eta_n}{|A|(w-\eta_n)}.
\]
This proves the final bound.
\end{proof}

\begin{corollary}[Determinant probabilities over a fixed prime field]
\label{thm:chi-det}
\index{Determinant densities!independent entries}
Fix a prime $p$ and $p^{-1}\le\beta<1$. For each $n\ge1$, let
$S_n\in M_n(\Fp)$ have independent entries, all of whose atom
masses are at most $\beta$. There are $C_{p,\beta},c_{p,\beta}>0$
depending only on $p,\beta$ such that, uniformly over these ensembles,
all $n\ge1$ and $t\in\Fp^\times$,
\begin{equation}\label{eq:maples-fixed}
 \left|\mathbb P(\det S_n=t)-\frac{P(p)}{p-1}\right|
 \le C_{p,\beta}e^{-c_{p,\beta}n}.
\end{equation}
Fix such a pair and abbreviate
$n_0(p,\beta)=n_0(p;C_{p,\beta},c_{p,\beta})$.
All three parts of \Cref{cor:entry-determinant-comparisons} apply
with $q=p$, $C=C_{p,\beta}$ and $c=c_{p,\beta}$, as does the
corank conclusion of \Cref{thm:finite-field-entry-determinant}.
For i.i.d.\ entries from a fixed non-point-mass law
$\chi_{\mathrm{sec}}$, one may take
$\beta=\max_x\chi_{\mathrm{sec}}(x)$ and write
$\rho_n(\chi_{\mathrm{sec}})=\rho_n$.
\end{corollary}

\begin{proof}
For any entry law $\mu$ and $a\in\Fp^\times$,
\[
 \begin{split}
 1-|\widehat\mu(a)|^2
 &=\sum_{x,y}\mu(x)\mu(y)(1-\cos(2\pi a(x-y)/p))\\
 &\ge(1-\cos(2\pi/p))(1-\sum_x\mu(x)^2)\\
 &\ge(1-\cos(2\pi/p))(1-\beta)>0.
 \end{split}
\]
Here $\widehat\mu(a)=\sum_x\mu(x)e^{2\pi iax/p}$.
The expression $d=(1-\cos(2\pi/p))(1-\beta)$ lies in $(0,1]$;
choose $\lambda=(1+\sqrt{1-d})/2\in(0,1)$.
Every nontrivial entry Fourier coefficient is bounded by $\lambda$.
The finite-field theorem and its comparison corollary now apply,
with constants depending only on $p,\beta$.
\end{proof}

\begin{corollary}[Rejection sampling from independent-entry matrices]
\label{cor:chi-rejection-sampling}
\index{Rejection sampling!determinant constraints}
With $p,\beta,S_n$ and the pair $(C_{p,\beta},c_{p,\beta})$ of
\Cref{thm:chi-det}, let $n\ge n_0(p,\beta)$ and assume exact samplers
for all entry laws. Rejection sampling from independent copies of $S_n$
has a geometric number of attempts with parameter $\rho_n$.
Its output has law
$\operatorname{Law}(S_n\mid\det S_n\textup{ generates }\Fp^\times)$, and
\[
 \left|\mathbb E[\#\textup{attempts}]-\frac1{c_n(p)}\right|
 \le\frac{2E_n}{c(p)^2}.
\]
For $m_p=\lceil1/c(p)\rceil$, the probability of at least one acceptance
among $m_p$ independent draws is at least
$1-\exp(-1+m_pE_n)=1-e^{-1}+o(1)$ as $n\to\infty$.
\end{corollary}

\begin{proof}
By \Cref{cor:entry-determinant-comparisons}\textup{(i),(ii)},
$\rho_n\ge c_n(p)/2\ge c(p)/2>0$ and
$|\rho_n-c_n(p)|\le E_n$. Independent trials give the geometric
law and the conditional law of the accepted matrix, and
\[
 \left|\frac1{\rho_n}-\frac1{c_n(p)}\right|
 =\frac{|\rho_n-c_n(p)|}{\rho_n c_n(p)}
 \le\frac{2E_n}{c(p)^2}.
\]
Finally, $(1-\rho_n)^{m_p}\le e^{-m_p\rho_n}$,
$\rho_n\ge c(p)-E_n$ and $m_pc(p)\ge1$ give the probability bound.
\end{proof}

The sampler in \Cref{cor:chi-rejection-sampling} tests determinants of
known matrices and returns the displayed conditioned product law.
Its entries may be dependent and its determinant fibers nonuniform.
If the exact entry samplers have
expected costs uniformly polynomial in $n$ and their setup length,
then, for fixed $p,\beta$, the expected total cost is polynomial:
each trial samples $n^2$ entries and tests its determinant in the fixed
field.

Let $n\ge2$, $\eta_n<w$ and $\varnothing\ne A\subseteq\Fp^\times$.
Consider a common-secret transcript whose secret has law
$\operatorname{Law}(S_n\mid\det S_n\in A)$ and is independent of
the public matrices and errors. By
\Cref{cor:entry-determinant-comparisons}\textup{(iii)} and
\Cref{cor:determinant-fiber-rerandomization}, a distinguisher of advantage
$\varepsilon$ for the uniform law on $\{S:\det S\in A\}$ gives advantage
at least $\varepsilon-\eta_n/(w-\eta_n)-2\delta_{\mathrm{cut}}$ for this
conditioned prior. Taking $A$ to be the primitive roots gives PRIM-LWE.
The next corollary gives the fixed-extension-field version.

\begin{corollary}[Sampling and conditioning with an additive Fourier gap]
\label{cor:chi-det-extension-field}
Under \eqref{eq:finite-field-entry-gap}, use
$C=C_{q,\lambda}$, $c=c_{q,\lambda}$ in
\Cref{cor:entry-determinant-comparisons}. Its event, acceptance,
determinant-TV and conditional-TV conclusions hold uniformly over
these ensembles. With exact entry samplers and
$n\ge n_0(q;C,c)$, all conclusions of
\Cref{cor:chi-rejection-sampling} hold with $p$ replaced by $q$.
For $n\ge2$, $\eta_n<w$ and nonempty $A\subseteq\mathbb F_q^\times$,
the conditioned prior $\operatorname{Law}(S_n\mid\det S_n\in A)$
in the common-secret setting of
\Cref{cor:determinant-fiber-rerandomization}, independent of the public
matrices and errors, also has the one-way advantage transfer just
described, with loss
at most $\eta_n/(w-\eta_n)+2\delta_{\mathrm{cut}}$.
The constants and dimension thresholds depend only on the fixed
$q,\lambda$ (and the chosen admissible pair $C,c$).
\end{corollary}

\begin{proof}
Theorem~\ref{thm:finite-field-entry-determinant} supplies the premise
of \Cref{cor:entry-determinant-comparisons}, including
$\rho_n\ge c_n(q)/2\ge c(q)/2$ and $|\rho_n-c_n(q)|\le E_n$.
For independent trials the first acceptance has the original matrix
law conditioned on the primitive-determinant event, and its attempt
count is geometric with mean $1/\rho_n$. Consequently
$|1/\rho_n-1/c_n(q)|\le2E_n/c(q)^2$.
For $m_q=\lceil1/c(q)\rceil$,
$(1-\rho_n)^{m_q}\le\exp(-m_q\rho_n)
\le\exp(-1+m_qE_n)$, proving all sampling conclusions.
The conditioned prior on $A$ is supported on invertible matrices.
Its determinant law is within $\eta_n/(w-\eta_n)$ of
$\mathrm{Unif}(A)$ by \Cref{cor:entry-determinant-comparisons}\textup{(iii)}. Apply
\Cref{cor:determinant-fiber-rerandomization} with this prior and
target determinant law $\mathrm{Unif}(A)$ to obtain the advantage
bound, including the stated cutoff loss.
\end{proof}

\begin{remark}[A dimension threshold is necessary]\label{rem:chi-witness}
Take $p=5$, and let $\chi_{\mathrm{sec}}$ be uniform on
$\{1,4\}=\{\pm1\}\subset\mathbb F_5^*$.
For $n=3$, subtracting the first row from each later row shows that
the integer determinant is divisible by $2^{n-1}=4$.
Hadamard's inequality gives $|\det S_3|\le3^{3/2}<8$.
Hence $\det S_3\in\{0,\pm4\}\equiv\{0,1,4\}\pmod5$, so
$\rho_3(\chi_{\mathrm{sec}})=0$ whereas $c_3(5)>0$.
Thus $n_0(5,1/2)>3$, so the dimension threshold in
\Cref{cor:entry-determinant-comparisons}\textup{(ii)} and
\Cref{cor:chi-rejection-sampling} cannot be removed.
\end{remark}

\begin{remark}[Permutation and public affine sampling obstructions]\label{rem:permutation-mixture}
Let $\mu$ be a probability on a finite set $\Omega$ and $E$ an event with
$\mu(E)>0$; put $S=\supp\mu\cap E$ and $t=|S|$, and order the atom masses as
$\mu_{(1)}\ge\mu_{(2)}\ge\cdots$. Let $\nu$ be the law of $\sigma(\omega)$,
where $\omega\sim\mu$ and $\sigma$ is a random permutation of $\Omega$
independent of $\omega$. Since $|\sigma^{-1}(S)|=t$, we have
$\nu(S)\le\sum_{j\le t}\mu_{(j)}$, while $\mu(S\mid E)=1$; hence
\[
 d_{\mathrm{TV}}(\nu,\mu(\,\cdot\mid E))
 \ge1-\sum_{j=1}^t\mu_{(j)}.
\]
If $\mu$ is uniform on its support, the right side equals $1-\mu(E)$,
and $\sigma=\mathrm{id}$ attains it.

For matrices, let $n\ge2$ and let $\chi_{\mathrm{sec}}$ be a law on
$\Fp$ with $\rho_n(\chi_{\mathrm{sec}})>0$. Every map
$S\mapsto MSN+R$, with $M,N\in\GL_n(\Fp)$ and $R\in\Mn$, permutes
$\Mn$. Thus the bound above, with
$\mu=\chi_{\mathrm{sec}}^{\otimes n\times n}$ and the
primitive-determinant event, applies to every public random choice of
such a map independent of the input.
The lower bound is strictly positive: for any
$x\in\supp\chi_{\mathrm{sec}}$, the matrix with every entry equal to
$x$ has positive probability and determinant zero.
For uniform matrix input the exact minimum is $1-c_n(p)$.
\end{remark}

\section{Prescribed moduli in a set-system construction}
\label{ssec:lwe-prescribed-moduli}

The proof of~\cite[Theorem~10]{SYD21} uses $m_p:=\lceil1/c(p)\rceil$
independent matrix-LWE instances, the bound
$1-(1-c(p))^{m_p}\ge1-e^{-1}$ and the claimed advantage factor
$m_pn^2/(1-e^{-1})$. It specifies no procedure extracting a PRIM-LWE
instance from the hidden secrets. Subject to the scope limitations in
\hyperref[par:lwe-reductions]{Reduction interfaces}, we bound its
reciprocal-density factor for the prescribed moduli.

Fix integers $l,r\ge2$ and distinct odd primes
$l<\ell_1<\cdots<\ell_r$, and put $m=\prod_{i=1}^r\ell_i$.
Theorem~1 of~\cite{SYD21} supplies constants
$c_{\mathrm{TS}}=c_{\mathrm{TS}}(r,m,l)>0$ and
$h_0=h_0(r,m,l)\ge lm$ such that, for every integer $h\ge h_0$ and
$t\ge2$, there is an explicitly constructible set system
$\mathcal H\subseteq2^{[h]}$ with $t$-wise restricted intersections
modulo $m$ in the sense of~\cite[Definition~2]{SYD21} and
\begin{equation}\label{eq:setsystem-size}
 |\mathcal H|\ge
 \exp\!\left(\frac{c_{\mathrm{TS}}l(\log h)^r}
                        {(\log\log h)^{r-1}}\right)
 +l\exp\!\left(\frac{c_{\mathrm{TS}}(\log h)^r}
                        {(\log\log h)^{r-1}}\right).
\end{equation}
For fixed $r,m,l$, this lower bound is superpolynomial in $h$.
The detailed verifiable secret sharing construction specializes to
$l=2$, $\ell_1=3$ and
$p=2m+1$ prime~\cite[\S~8.2]{SYD21}.

\begin{proposition}[Reciprocal density of the prescribed moduli]
\label{prop:setsystem-loss}
Assume that $p=2m+1$ is prime. Let $q_1<\cdots<q_r$ be the first
$r$ primes greater than $l$, and put $A=2\prod_{i=1}^rq_i+1$.
\begin{equation}\label{eq:loss-bound}
 \frac1{c(p)}\le\frac2{P(A)}\prod_{i=1}^r\frac{q_i}{q_i-1},
\end{equation}
with equality if and only if $p=A$.
\end{proposition}

\begin{proof}
By \eqref{eq:cp-nt}, $1/c(p)=(2/P(p))\prod_{i=1}^r\ell_i/(\ell_i-1)$.
Since $\ell_i\ge q_i$ for all $i$, so that $p\ge A$, \eqref{eq:loss-bound}
follows because $P$ is strictly increasing (\Cref{lem:matrix-product-bounds})
and $t/(t-1)$ is strictly decreasing on $(1,\infty)$; equality holds if and only if $p=A$, since $p=A$ forces $\ell_i=q_i$ for all $i$.
\end{proof}

The proof applies verbatim, with the same equality case, when $p-1$ has
exactly $r$ distinct odd prime factors, all greater than $l$, with arbitrary
positive exponents, since then still $p\ge A$. For $l=2$, \eqref{eq:loss-bound}
is \Cref{prop:factor-count-bound} with $k=r+1$.

Along the least primes $p^*(r+1)\equiv1\pmod{N_{r+1}}$,
\Cref{thm:effective_sharp} with $k=r+1$ and \eqref{eq:eff-loglog} give
$1/c(p^*(r+1))\sim e^\gamma\log p_{r+1}\sim e^\gamma\log\log p^*(r+1)$ as
$r\to\infty$.

\paragraph*{When $2m+1$ is composite}
Let $p$ be the least prime congruent to $1$ modulo $2m$; by
\cite[Theorem~1]{Xyl18}, $p\le C_0(2m)^{L_X}$ for an effective absolute
constant $C_0\ge1$. As $p-1$ may have further prime factors,
\eqref{eq:loss-bound} need not apply; instead $1/c(p)$ is bounded by
\Cref{prop:factor-count-bound} with $k=\omega(p-1)$ or, without factoring
$p-1$, by \eqref{eq:rs-cp}.

\begin{table}[htbp]
\centering
\caption{Reciprocal determinant densities for the prescribed moduli
($2\le r\le4$) and the least primes $p^*(r+1)\equiv1\pmod{N_{r+1}}$
($5\le r\le7$). Here $\ell_i=p_{i+1}$, $m=\prod_i\ell_i$, and $p=2m+1$ when
prime, while otherwise $p=p^*(r+1)$. The values were computed from
the evaluation procedure in \Cref{sec:explicit-certificates}; $c(p)$ and $1/c(p)$ are rounded to six and two
decimal places, respectively.}
\label{tab:setsystem-frontier}
\[
\renewcommand{\arraystretch}{1.15}
\begin{array}{ccccccc}
\toprule
r & \{\ell_i\} & m & p & \omega(p-1) & c(p) & 1/c(p) \\
\midrule
2 & 3,5 & 15 & 31 & 3 & 0.257787 & 3.88 \\
3 & 3,5,7 & 105 & 211 & 4 & 0.227483 & 4.40 \\
4 & 3,5,7,11 & 1155 & 2311 & 5 & 0.207702 & 4.81 \\
5 & 3,5,7,11,13 & 15015 & 120121 & 6 & 0.191807 & 5.21 \\
6 & 3,5,7,11,13,17 & 255255 & 4084081 & 7 & 0.180525 & 5.54 \\
7 & 3,5,7,11,13,17,19 & 4849845 & 106696591 & 8 & 0.171024 & 5.85 \\
\bottomrule
\end{array}
\]
\end{table}

In \Cref{tab:setsystem-frontier}, $2m+1$ is composite for
$r\in\{5,6,7\}$. In every row the
distinct prime factors of $p-1$ are exactly $p_1,\ldots,p_{r+1}$, so
$c(p)=P(p)A_{r+1}$; for example, $120120=2^3\cdot3\cdot5\cdot7\cdot11\cdot13$
when $r=5$. Without the constraint $p=2m+1$, the reciprocal density has mean
$M(-1)\approx2.83$ over the primes (\Cref{thm:prime-mellin,cor:avg-loss}),
and its limiting law has a doubly exponential upper tail
(\Cref{cor:small-x-tail}).

For $p=p(\lambda)$ polynomial, respectively exponential, in a security
parameter $\lambda$, \Cref{cor:rosser-schoenfeld} gives
$m_{p(\lambda)}=O(\log\log\lambda)$, respectively $O(\log\lambda)$; hence, in
the claimed advantage factor $m_pn^2/(1-e^{-1})$ of \cite[Theorem~10]{SYD21},
$m_{p(\lambda)}=o(n(\lambda)^2)$ whenever
$n(\lambda)^2/\log\log\lambda\to\infty$, respectively
$n(\lambda)^2/\log\lambda\to\infty$.

\chapter{Concluding remarks and open questions}\label{sec:conclusion}
The prime-coordinate models separate carrier geometry, information at
finite resolution and Fourier regularity. The basic laws have
dimension-zero carriers, positive information dimensions and quantitative
Fourier decay; suitable infinite weighted sums have smooth densities.
Open questions concern exact information dimensions, the location of the
freely sliding totient maximizer, optimal uniform Fourier decay and the
window sizes and modulus ranges for arithmetic approximation.

\paragraph*{Geometry and information}
The four basic laws and their logarithmic images satisfy
$0.55043653<c_L^+\le\underline d_1\le\overline d_1\le d_*<0.570489495$.
For each arithmetic function, integer and shifted-prime sampling have
the same lower information dimension and the same upper information
dimension (\Cref{thm:shannon-dimension}). It remains open whether these
two dimensions coincide. \Cref{prop:scalar-extra-triples} improves the
constructed lower bound $c_L$ by $8/138915+1/57600$. The exact dimensions,
possible equality with the upper coding bound $d_*$, and equality between
the totient and divisor-sum dimensions remain open.

Finite nonzero linear combinations of independent samples from the
basic laws and their logarithmic images retain dimension-zero carriers
(\Cref{cor:independent-sum-geometry}). For independent basic logarithmic
samples and any fixed sequence of nonzero weights, both information
dimensions of the partial sums increase to one
(\Cref{cor:convolution-information}). Information dimension can thus
approach one while every finite sum retains a dimension-zero carrier.
For a locally Lipschitz scalar image of finitely many independent basic
logarithmic samples, both R\'enyi dimensions vanish at every $q>1$ after
conditioning on any positive-probability event in the joint valuation
sigma-field (\Cref{prop:conditioned-renyi-images}).

\paragraph*{Infinite sums and sparse coefficients}
For independent basic logarithmic samples, infinite sums with infinitely
many nonzero coefficients have information dimension one under absolute
summability, or square summability after centering
(\Cref{thm:infinite-weighted-sums}). Under these hypotheses, the
coefficient-count conditions \eqref{eq:infinite-sum-smoothing-criterion}
and \eqref{eq:infinite-sum-ac-criterion-mas01} give, respectively,
Schwartz densities and absolute continuity with Hausdorff dimension one.
Geometric and polynomial weights in the corresponding summability
regimes satisfy the stronger condition. The convergence, Fourier bounds
and Schwartz criterion extend to independent raw-ratio summands
(\Cref{rem:infinite-sum-raw-laws}); the weaker absolute-continuity
criterion is stated for logarithmic summands.

The weaker criterion also applies to $a_i=\exp\{-\exp(\exp(ci))\}$ for
each fixed $c>0$, by conditioning on prime activations. At the other extreme,
a residual set of coefficients in real $\ell^1$, or in real $\ell^2$ for centered
summands, gives Hausdorff dimension zero and information dimension one.
Coefficients giving smooth densities are dense in both norm topologies
(\Cref{cor:generic-coefficient-carriers}).

\paragraph*{Digit statistics}
Each basic ratio law gives full mass to absolutely normal Liouville
numbers (\Cref{thm:liouville-carriers,thm:raw-ratio-normality}). Normality
concerns asymptotic block frequencies in individual expansions; entropy
measures the distribution of finite prefixes. For $X\sim\mu_G$, the most
likely $B$-bit prefix of $2X$ is eventually unique. For odd $B$ it is
the all-ones prefix. For even $B$ it is the cell of index
$\lfloor 2^{B+1}/3\rfloor$, with indexing from zero. Min-entropy is
$\log_2 B+O(1)$, while Shannon entropy in bits lies between
$(c_L^+-o(1))B$ and $(d_*+o(1))B$ (\Cref{cor:quantized-randomness}).
Whether the average information per digit has a limit remains open.

\paragraph*{Fourier behavior and slow increments}
Normalized complex Fourier amplitudes distinguish the logarithmic
totient laws: the shifted-prime limit is circular, whereas the integer
limit is noncircular
(\Cref{cor:totient-amplitude-circular-moments,cor:integer-totient-amplitude}).
These limits concern amplitudes at frequencies sampled uniformly from
growing intervals.

For increments $(\log\ell)^{-A}$, the integer and shifted-prime laws
are absolutely continuous for every $A>0$
(\Cref{lem:slow-jump-all-ac}), disproving the Erd\H{o}s--K\'atai
singularity conjecture for $A>2$. The scalar and joint criteria for
Fourier integrability and smoothness are given in
\Cref{ssec:slow-jump-ac}. For fixed nonproportional affine tuples,
\eqref{eq:affine-fourier-sobolev-thresholds} gives the exact thresholds
even when exponents coincide. In the sparse activation model with
$-1<\beta<0$, \Cref{cor:loglog-singular-normality} proves absolute
normality for all three jump families.

\paragraph*{Divisor-sum concentration}
For the integer, odd shifted-prime and full shifted-prime divisor-sum laws,
\Cref{thm:sigma-concentration-constants-mas05} gives the exact maximizing
interval at every sufficiently small width in both raw and logarithmic
coordinates. For the integer law, the leading constant already follows
from Toulmonde's analysis \cite[\S~10]{Tou06}. For the two shifted laws,
the full endpoint expansions give the corresponding concentration expansions.
Relative near-maximizers may omit the endpoint.
In a fixed reduced progression, \Cref{cor:ap-sigma-saturation}
identifies the leading coefficient as the largest class weight $w_*$.
The exact endpoint conclusion holds whenever the endpoint class is uniquely
maximal. This includes every progression with at most one saturated odd
prime, irrespective of the dyadic head, and further progressions covered by
the strict inequality in \Cref{prop:ap-sigma-endpoint-class}. Determining
the maximizing class in the remaining progressions is open.

\section{Lower bounds for \texorpdfstring{$S(\tau)$}{S(tau)}}\label{ssec:open_effective}
\index{Fourier decay!open questions}
The amplitude law in \Cref{thm:fourier-amplitude-law} describes a uniformly
sampled frequency on a growing interval. A sharp uniform decay bound
must also control exceptional frequencies, including those in
\Cref{prop:fourier-resonance}. The remaining pointwise question can be
expressed through the cosine sum. Put
\[
 c_{\mathrm{diss}}=\liminf_{|\tau|\to\infty}
                  \frac{S(\tau)}{\log\log|\tau|}.
\]
Then $2/3\le c_{\mathrm{diss}}\le2$, whereas the global upper limit
of $S(\tau)/\log\log|\tau|$ lies in $[2,10/3]$, by
\Cref{prop:dissipation-upper-mas11}. For each $\delta>0$,
\Cref{cor:dissipation-sparse-mas11} gives a single sparse set $B_\delta$
of positive frequencies. By evenness, whenever $|\tau|\notin B_\delta$,
the discrepancy $\bigl|S(\tau)-2\log\log|\tau|\bigr|$ is at most
$(2+\delta)\log\log\log|\tau|+O_\delta(1)$ as $|\tau|\to\infty$.
The exact value of $c_{\mathrm{diss}}$ remains open, as does removal of
the factor $(\log\log|\tau|)^{1/3}$ from
$|\FT(\tau)|\ll(\log\log|\tau|/\log|\tau|)^{1/3}$.

Put $H=\log|\tau|$. The cancellation estimates in the proof of
\Cref{thm:vk-fourier} require the starting scale
$\exp\{K H^{2/3}(\log H)^{1/3}\}$ to absorb logarithmic losses.
An $O(1)$ bound for the weighted prime cosine sum from
$\exp\{K H^{2/3}\}$ to $b|\tau|^{1/5}$, with fixed $K,b>0$, would
remove the log-log loss: Mertens' theorem gives mass
$\frac23\log H+O(1)$, and \eqref{eq:forward} gives
$|\FT(\tau)|\ll H^{-1/3}$.

A stronger open target is
\[
 S(\tau)\ge2\log\log|\tau|-2\log\log\log|\tau|-O(1)
 \qquad(|\tau|\to\infty).
\]
By \eqref{eq:forward}, this would give
$|\FT(\tau)|\le e^{-S(\tau)/2}\ll\log\log|\tau|/\log|\tau|$.
Equation~\eqref{eq:fourier-endpoint-obstruction} gives
$\limsup_{\tau\to\infty}(\log\tau/\log\log\tau)|\FT(\tau)|\ge1$,
so this decay order would be sharp. The equality $c_{\mathrm{diss}}=2$
alone does not control the second logarithmic term.

For every fixed $0<\delta<1$, \Cref{prop:log-scale-obstruction}
excludes a uniform bound of order $N/(\log N)^A$, $A>1$, for
$\sum_{N<p\le2N}e^{i\tau\omega_p}$ throughout dyadic
$N\in[\log\tau,\tau^{1-\delta}]$.

\section{Modulus of continuity and growing moduli}\label{ssec:open_modulus}
\index{Concentration functions!open questions}
\paragraph*{Maximizing intervals and pointwise regularity}
For the sharp relative and additive moduli in
\Cref{thm:sharp-global-modulus,cor:sharp-additive-modulus},
uniqueness and rationality of the maximizing upper endpoint and its
possible equality to $1/3$ for every sufficiently small window remain
open. For a relative-width maximizer,
$0\le t_\varepsilon-1/3\ll\varepsilon z^2e^{-\frac15\sqrt{z\log z}}$,
where $z=\log(1/\varepsilon)$; the additive-width maximizer satisfies
the analogous bound with $\varepsilon,z$ replaced by $h,\log(1/h)$.
In particular, $t_\varepsilon-1/3=o(\varepsilon z^{-A})$ for every fixed
$A>0$. A sharp displacement asymptotic remains open.

For every fixed nonempty finite set $S$ of odd primes,
\Cref{cor:anchor-residual} gives two-sided bounds for
$q_\varepsilon(x_S)-\mathcal K(S)H_{\mathrm{end}}(\varepsilon)$;
a sharp asymptotic remains open. At a prescribed non-anchor
$t\in(0,1/2)$, a sharp asymptotic for
$G(t)-G(t(1-\varepsilon))$ also remains open in general;
\Cref{cor:non-anchor-decay} gives rational and Diophantine upper bounds.

Although $G'=0$ Lebesgue-almost everywhere by
\Cref{rem:erdos-partial}, the existence of a non-anchor point with a
finite positive right-hand or two-sided derivative remains open.
The same remark records the integer analogue of the two-sided question.

\paragraph*{Arithmetic approximation with varying modulus}
The fixed-progression laws retain dimension zero and quantitative
Fourier decay for their negative logarithms
(\Cref{cor:ap-liouville,cor:ap-rajchman}).
\Cref{prop:ap-kolmogorov-pow2} gives quantitative approximation for
all polylogarithmic moduli, almost all moduli below $x^{1/2-\delta}$,
and every $Q=B^a\le x^{1/3}\exp\{-(\log\log x)^3\}$ with fixed
integer $B\ge2$. The last range uses Elliott's theorem~\cite{Elliott07}
and has CDF error $O_B(\log\log x/\log x)$ against $G_{Q,b}$.
Under GRH, \Cref{thm:ap-grh-law} covers every
$Q\le x^{1/2-\delta}$, for fixed $0<\delta<1/2$.

The concentration loss
$D(Q)=\prod_{\ell\mid Q,\,\ell>2}(\ell-1)/(\ell-2)$ reflects deletion
of free prime coordinates. A concentration constant independent of $Q$ is impossible:
if $Q$ contains every prime through $y$, the remaining logarithmic sum
has mean $O((y\log y)^{-1})$, so its concentration at width $y^{-1/2}$
tends to one. The order $1/\log x$ is sharp in every fixed reduced
progression (\Cref{cor:ap-sharp-convergence}). The sharper coefficient
assertions for unrestricted primes and for the raw statistic in the
class $3\pmod4$ remain conjectural; see
\Cref{rem:kolmogorov-conjecture,prop:progression-coefficient-sophie-germain}
for their distinct prime-cofactor consequences.

Unconditional CDF bounds uniform over arbitrary individual moduli
beyond the polylogarithmic range remain open; the fixed-base power
families above have a stronger available range, with constants depending
on the base. In the admissible CDF regimes, \eqref{eq:moving-modulus-endpoint}
gives the moving-modulus endpoint coefficient $\kappa_{\mathrm{end}}D_{\le y}(Q)$
at relative width $1/y$, with a relative empirical asymptotic for
$y\to\infty$, $\log y=o(\log x/\log\log x)$ and $Q\le x$.
Optimal ranges remain open.

\paragraph*{Transport rates}
For every fixed $r\ge1$, \Cref{prop:wasserstein-bounds} gives
Wasserstein lower bounds of scale $x^{-1/2}(\log x)^{-2/r}$ throughout
the admissible progression-population regimes, with the stated endpoint
and deletion factors. Under Dirichlet GRH,
\Cref{cor:ap-grh-wasserstein} gives stretched-exponential upper bounds.
Unconditionally, \Cref{prop:prime-wasserstein} gives arbitrary
inverse-logarithmic savings in the polylogarithmic, fixed-base power
and almost-all modulus ranges of \Cref{thm:prime-mellin}, with
exceptional sets depending on the requested accuracy and order.
Even at $Q=1$, these bounds do not determine the true transport rate.

\section{Shrinking endpoint windows}\label{ssec:shrinking-endpoint-windows}
For unrestricted primes, \Cref{cor:prime-shrinking-anchor} gives
relative endpoint and fixed-anchor asymptotics whenever
$\varepsilon=\varepsilon(x)\downarrow0$ and
$\log(1/\varepsilon)=o(\log x)$. For a moving anchor $x_S$, put
$z=\log(1/\varepsilon)\to\infty$. Formula~\eqref{eq:prime-moving-anchor}
gives a relative asymptotic if $D_S\le\exp(b\sqrt{z\log z})$ for some
fixed $0<b<1/\sqrt{20}$ and $z/(\kappa_S\log x)\to0$.
For every fixed reduced class $b\pmod Q$, \Cref{cor:ap-sharp-convergence} and
\eqref{eq:ap-sharp-concentration} give the same endpoint range:
the model mass of $(A_{Q,b}(1-\varepsilon),A_{Q,b}]$ is asymptotic to
$\kappa_{\mathrm{end}}D(Q)/\log(1/\varepsilon)$. The empirical window
probability differs from this model mass by a relative error
$O_Q(\log(1/\varepsilon)/\log x)=o(1)$.
\Cref{cor:prime-shrinking-anchor}\textup{(ii)} also gives uniform
relative counts on compact subintervals of $(0,1/2]$ for
$\varepsilon=(\log x)^{-a}$, $0<a<1$.

At power widths, endpoint membership can force the odd part of $p-1$
to be prime (\Cref{prop:safe-prime-obstruction}). The coefficient
conjecture \eqref{eq:progression-coefficient-conjecture} is constrained
by this prime-pair problem. At fixed $2$-adic valuation, the Buchstab
transition \eqref{eq:endpoint-buchstab-conjecture} is equivalent to the
rough-cofactor asymptotic by the unconditional comparison
\eqref{eq:buchstab-count-comparison}. The common main term remains conjectural.

\paragraph*{A progression coefficient and Sophie Germain primes}
The coefficient conjecture \eqref{eq:progression-coefficient-conjecture}
for the raw statistic in the class $3\pmod4$ remains open.
\Cref{prop:progression-coefficient-sophie-germain} gives its necessary
consequences for Sophie Germain primes: the lower and upper asymptotic
coefficients lie between $\kappa_{\mathrm{end}}$ and
$2\kappa_{\mathrm{end}}$. Already
$\limsup_{x\to\infty}(\log x)D_{x;4,3}<2\kappa_{\mathrm{end}}$ would imply
$\operatorname{SG}_1(y)\gg y/\log^2y$. For the unrestricted discrepancy,
a limsup coefficient strictly below $2\kappa_{\mathrm{end}}$ forces a
positive normalized prime-pair limsup in some fixed $2$-adic exponent
class (\Cref{rem:kolmogorov-conjecture}).

For unrestricted primes, \Cref{prop:endpoint-common-count} allows the
limiting-model endpoint constant at power widths for at most one
$\theta\in[1/2,1)$. The two-sided order $x/(\log x)^2$ for
$0<\theta<2/7$ follows from \Cref{thm:endpoint-sieve}; the relative
asymptotic and its leading constant remain open. It also remains to
determine the sharp power-width ranges at nonempty anchors and the optimal
ranges for relative counts in varying progressions.
\section{Primes with friable \texorpdfstring{$p-1$}{p-1}}\label{ssec:friable-open}
\index{Friable predecessors!open questions}
The bounds in \Cref{ssec:friable-unconditional} control determinant
densities at primes whose predecessors have no large prime factor.
We do not know whether the upper bound in \eqref{eq:friable-largep}
is attained in limsup
(\Cref{rem:friable-sharp}). A positive limsup $L$ would force infinitely
many primes with $P^+(p-1)\le(\log p)^A$ for each $A>e^\gamma/L$.
A distributional question requires counts with prescribed finite
divisibility patterns as well as friability.
Unconditionally, for fixed $u>1$, determine the weak limit,
if it exists, of
\[
 \frac{
   \displaystyle\sum_{\substack{p\le x,\ p\textup{ prime}\\
                    P^+(p-1)\le x^{1/u}}}\delta_{1/c(p)}
 }{
   \displaystyle\#\{p\le x:p\textup{ prime},\ P^+(p-1)\le x^{1/u}\}
 }
 \qquad (x\to\infty).
\]
\begingroup\brokenpenalty=10000
Earlier quantitative bounds for shifted primes without large prime
factors were obtained by Baker and Harman~\cite{BH98}.
For fixed $1<u<32\sqrt e/15=3.5172\ldots$, Lichtman's unconditional
dyadic estimate~\cite[Theorem~1.1]{Lic22} gives
$\#\{p\le x:P^+(p-1)\le x^{1/u}\}\gg x/(\log x)^{C(u)}$.
Erd\H{o}s conjectured that, for every $\varepsilon>0$, infinitely
many primes satisfy $P^+(p-1)\le p^\varepsilon$
(\cite[\S~1]{Lic22}).

The predicted asymptotic
$\#\{p\le x:P^+(p-1)\le x^{1/u}\}\sim\rho_{\mathrm D}(u)\pi(x)$, with $\rho_{\mathrm D}$
the Dickman--de Bruijn function, is the fixed-$u$ specialization of
Pomerance's heuristic~\cite[p.~86]{Pom80}; for every fixed $u>1$ it
follows from Lamzouri's weak Elliott--Halberstam hypothesis
\cite[Conjecture~1.2 and Lemma~2.1]{Lam07}.
A preprint dated September 2026 claims the full Poisson--Dirichlet law
for the ranked logarithms of prime factors of $p-1$, normalized by
$\log(p-1)$, and hence this counting asymptotic
\cite[Theorem~1.1]{OAI26PD}.

Banks, Harcharras and Shparlinski~\cite[Theorem~2]{BHS04} prove
$\#\{p\le x:p\equiv b\pmod q,\ P^+(p-1)\le y\}
\ll_{\varepsilon,A}u\rho_{\mathrm D}(u)\pi(x)/\varphi(q)$, where
$u=\log x/\log y$, uniformly for $\gcd(b,q)=1$,
$q\le(\log y)^A$ and
$\exp\bigl(\sqrt{\log x\,(\log\log x)^{1+\varepsilon}}\bigr)\le y\le x$,
with fixed $\varepsilon,A>0$ and sufficiently large $x$.

Under Lamzouri's hypothesis, \Cref{prop:friable-joint-limit} identifies
the conditional limit as the law of $1/X$ for each fixed $u>1$.
Convergence holds in every finite Wasserstein order and for all real
Mellin moments.
The ranked large-factor marginal alone does not determine this joint
limit; the finite divisibility patterns are also required.
\par\endgroup

For fixed $y\ge3$, even the infinitude of $\mathcal S_y$ is open; $y=2$ is the
Fermat-prime problem. With
$\mathcal E_y(p)=\{\ell\le y:\ell\textup{ odd prime},\ell\mid p-1\}$,
convergence $1/c(p)\to2$ along $p\in\mathcal S_y$, $p\to\infty$, forces
$\mathcal E_y(p)=\varnothing$ eventually, hence $p-1=2^a$; only
finitely many such primes are conjectured to exist. By contrast, the
support of $1/X$ in \Cref{thm:phi-dist,lem:factor} contains
$2\ell/(\ell-1)\to2$ as the odd prime $\ell\to\infty$, arising from
$B_\ell=1$ and all other odd-prime indicators zero. For fixed $y$, this
selected prime eventually exceeds $y$.

\appendix

\chapter{Divisibility patterns of \texorpdfstring{$p-1$}{p-1}: Bombieri--Vinogradov and GRH estimates}
\label{app:bv-grh}

\paragraph*{Arithmetic inputs}
A prime-coordinate pattern specifies which odd primes divide $p-1$
and which do not. For their product $L$ and a reduced class $a\pmod m$
with $\gcd(m,L)=1$, inclusion--exclusion gives
\eqref{eq:crt-inclusion-exclusion} and the prime-counting error bound
\eqref{eq:pattern-error-sum}. \Cref{thm:bv-bivariate} gives an
unconditional average over its stated range of $m$.
The individual-modulus estimate in \Cref{thm:bv-grh} requires $mL\le x^2$
and GRH for every primitive Dirichlet $L$-function whose conductor
divides $mL$. It includes ordinary progression counts through the
empty-pattern convention $k=0$, $L=1$, and is used in
\Cref{thm:ap-grh-law,cor:wasserstein-grh-global,cor:ap-grh-wasserstein}.

The notation $\pi_S(x;m,a)$ and $c_S$ is that of
\Cref{prop:ap-bivariate}.
Let $k\ge0$ and $m\ge1$ be integers and let $a\in\Z$. Put
$L=\ell_1\cdots\ell_k$ for distinct odd primes $\ell_j$, and suppose
$\gcd(m,L)=\gcd(a,m)=1$. Given
$S\subseteq\{1,\ldots,k\}$, put $S^c=\{1,\ldots,k\}\setminus S$.
For $T\subseteq S^c$, define
$M_{m,T}=m\prod_{j\in S\cup T}\ell_j$, and let $a^*_{m,T}$ be the
unique reduced residue class modulo $M_{m,T}$ satisfying
$a^*_{m,T}\equiv a\pmod m$ and
$a^*_{m,T}\equiv1\pmod{\ell_j}$ for $j\in S\cup T$.
Inclusion--exclusion and the Chinese remainder theorem give, for $x\ge2$,
\begin{equation}\label{eq:crt-inclusion-exclusion}
\begin{split}
\pi_S(x;m,a)-c_S\frac{\operatorname{Li}(x)}{\varphi(m)}
={}&\sum_{T\subseteq S^c}(-1)^{|T|}
 \left[\pi(x;M_{m,T},a^*_{m,T})
 -\frac{\operatorname{Li}(x)}{\varphi(M_{m,T})}\right].
\end{split}
\end{equation}
Put
\[
 E_x(M):=\max_{\gcd(b,M)=1}\left|\pi(x;M,b)-\frac{\operatorname{Li}(x)}{\varphi(M)}\right|.
\]
By \eqref{eq:crt-inclusion-exclusion}, for $x\ge2$,
\begin{equation}\label{eq:pattern-error-sum}
 \left|\pi_S(x;m,a)-c_S\frac{\operatorname{Li}(x)}{\varphi(m)}\right|
 \le\sum_{T\subseteq S^c}E_x(M_{m,T}),\qquad M_{m,T}\le mL.
\end{equation}
For $k=0$ we use $L=1$, $S=\varnothing$, $c_S=1$ and
$\pi_\varnothing(x;m,a)=\pi(x;m,a)$.

\begin{theorem}[Bombieri--Vinogradov theorem for divisibility patterns of $p-1$]\label{thm:bv-bivariate}
\index{Bombieri--Vinogradov theorem!divisibility patterns}
For every $A>0$ there are $B(A),C(A)>0$ such that, for all $x\ge2$,
$k\ge1$, odd primes $\ell_1<\dots<\ell_k$ with product $L$, and
$S\subseteq\{1,\dots,k\}$,
\[
\sum_{\substack{m\le x^{1/2}/((\log x)^{B(A)}L)\\
 \gcd(m,L)=1}}
 \max_{\gcd(a,m)=1}
 \left|\pi_S(x;m,a)-c_S\frac{\operatorname{Li}(x)}{\varphi(m)}\right|
 \le C(A)\frac{x}{(\log x)^A}.
\]
\end{theorem}

\begin{proof}
By \eqref{eq:pattern-error-sum}, the left side is at most
\[
 \sum_m\sum_{T\subseteq S^c}E_x(M_{m,T}).
\]
Since $\gcd(m,L)=1$, the $\ell_j$-adic valuations of $M_{m,T}$ determine
$S\cup T$, hence $T$ and $m$. Thus $(m,T)\mapsto M_{m,T}$ is injective,
with values at most $x^{1/2}(\log x)^{-B(A)}$.
The Bombieri--Vinogradov theorem~\cite{Bom65,Vin65} in the $\pi$-form
$\sum_{M\le x^{1/2}(\log x)^{-B(A)}}E_x(M)\le C(A)x(\log x)^{-A}$
of \cite[Lemma~4]{Kat68} completes the proof. Enlarging $C(A)$, if necessary,
extends the asymptotic estimate to every $x\ge2$.
\end{proof}

\begin{theorem}[Divisibility patterns of $p-1$ at individual moduli under GRH]\leavevmode\phantomsection\label{thm:bv-grh}
\index{Generalized Riemann hypothesis!divisibility patterns}\ignorespaces
There is an absolute, effectively computable constant $A_0$ with the
following property. Let $x\ge2$, let $k,m,a,\ell_1,\dots,\ell_k,L,S$ be as at the start of \Cref{app:bv-grh} with $mL\le x^2$, and assume GRH for every primitive Dirichlet $L$-function whose conductor divides $mL$. Then
\[
 \pi_S(x;m,a)=c_S\frac{\operatorname{Li}(x)}{\varphi(m)}+R_{m,a,S}(x),
 \qquad
 |R_{m,a,S}(x)|\le A_0\,2^{k-|S|}\sqrt{x}\log x.
\]
For $x\ge4$,
\[
 \left|\frac{\pi_S(x;m,a)}{c_S\operatorname{Li}(x)/\varphi(m)}-1\right|
 \le\frac{8A_0}{3}\,\frac{mL(\log x)^2}{\sqrt{x}}.
\]
\end{theorem}

\begin{proof}
By \eqref{eq:pattern-error-sum}, it suffices to show
$E_x(M)\ll\sqrt x\log x$, with an absolute effective constant, for each of
the $2^{k-|S|}$ moduli $M=M_{m,T}\le mL\le x^2$.
Every character modulo $M$ is induced by a primitive character of conductor
dividing $mL$, so the assumed GRH applies.

For $M\le\sqrt x$, the required estimate is
\cite[Theorem~13.7 and Corollary~13.8]{MV06} (their $\operatorname{li}$ is the
$\operatorname{Li}$ of \Cref{sec:prime-counting-inputs}).
For $\sqrt x<M\le x^2$, the bounds $\pi(x;M,b)\le1+x/M$ and
$M/\varphi(M)\ll\log\log M$ give the same estimate.
The bounded range of $x$ is absorbed into the constant.

For $x\ge4$, use $\operatorname{Li}(x)\ge x/(2\log x)$ and
$\varphi(m)\le m$. The local factors satisfy
\[
 \frac{2^{k-|S|}}{c_SL}
 =\prod_{j\in S}\frac{\ell_j-1}{\ell_j}
  \prod_{j\notin S}\frac{2(\ell_j-1)}{\ell_j(\ell_j-2)}
 \le\frac43,
\]
since only the factor at $\ell=3$ can exceed one. Dividing the
absolute-error estimate by its main term proves the relative bound.
\end{proof}
\chapter{Certified distribution functions and quantiles}
\label{app:gencert-numerics}

For $\beta>0$, put $F(\beta):=1-G(1/\beta)=\mathbb P(1/X\le\beta)$. For the search
in \Cref{def:gencert-search} with fixed $N$, this is the limiting fraction
of primes $q\in[x,2x]$, $q\equiv1\pmod{2N}$, that satisfy
$U_B^+(q)\le\beta$, by \Cref{cor:gencert-polylog}.

\paragraph*{Enclosures and quantiles}
The finite Bernoulli head, outward rounding and tail bounds in the first
proof certify the four values in \Cref{prop:gencert-numerics}.
The separate \Cref{prop:cdf-quantile-certification} uses finite rational
truncations and continuity to give terminating procedures for rational
arguments $\beta>2$ and rational quantile levels $q\in(0,1)$, with any
prescribed positive rational accuracy. The same procedure applies to the
four basic laws and their fixed-progression variants, with a valuation
cutoff for divisor sums.

\begin{proposition}[Certified distribution function enclosures]\leavevmode\phantomsection\label{prop:gencert-numerics}
\index{Distribution functions!certified enclosures}\ignorespaces
The values of $F$ satisfy
\[
\begin{array}{cc}
\toprule
\beta & \textup{certified enclosure for }F(\beta)\\
\midrule
3      &[0.49146,0.49149]\\
4      &[0.95378,0.95385]\\
5      &[0.99922,0.99924]\\
3.0001 &[0.5559,0.5634]\\
\bottomrule
\end{array}
\]
In particular, the median of $1/X$ lies in $(3,3.0001)$.
\end{proposition}

\begin{proof}
By the definition of $X$,
$1/X=2\exp\bigl(\sum_{\ell\ge3}B_\ell\omega_\ell\bigr)$. Writing
$R=\sum_{\ell\ge5}B_\ell\omega_\ell$ and $c=\log(\beta/2)$, and
conditioning on $B_3$, which equals $1$ with probability $1/2$, gives
\[
 F(\beta)=\tfrac12\mathbb P(R\le c)
          +\tfrac12\mathbb P(R\le c-\omega_3).
\]
At $\beta=3$, where $c=\omega_3$, the second term vanishes: since $\omega_\ell>0$,
\eqref{eq:def-Q} gives $\mathbb P(R\le0)=\mathbb P(B_\ell=0\ \forall\ell\ge5)=\lim_{y\to\infty}2Q(y)=0$.

Put $L:=50{,}000$, $R_H:=\sum_{5\le\ell\le L}B_\ell\omega_\ell$ and
$R_T:=\sum_{\ell>L}B_\ell\omega_\ell$, so that $R=R_H+R_T$. Each
logarithm needed for a summand or a threshold is enclosed in a rational
interval of width below $2^{-220}$ by truncating
$\log r=2\sum_{j\ge0}z^{2j+1}/(2j+1)$, $z=(r-1)/(r+1)$, after the term
$j=n$; for $r>1$ the remainder is positive and less than
$2z^{2n+3}/((2n+3)(1-z^2))$. Rounding these enclosures outward gives
integers $s_\ell^-\le2^{20}\omega_\ell\le s_\ell^+$, and
$R_H^\pm:=2^{-20}\sum_{5\le\ell\le L}B_\ell s_\ell^\pm$ satisfy
$R_H^-\le R_H\le R_H^+$, so that
$\mathbb P(R_H^+\le\theta)\le\mathbb P(R_H\le\theta)\le\mathbb P(R_H^-\le\theta)$
for every $\theta$. The probabilities $p_j$ of $R_H^\pm$ at the grid points
$2^{-20}j$ are computed by adjoining the primes $5\le\ell\le L$ one at a
time; adjoining $\ell$ maps $(p_j)$ to
\[
 p'_j=\Bigl(1-\frac1{\ell-1}\Bigr)p_j+\frac1{\ell-1}p_{j-s_\ell^\pm}
 \qquad(p_j:=0\text{ for }j<0).
\]
In exact arithmetic each update is an $\ell^1$-contraction on nonnegative
vectors. In IEEE-754 binary64 arithmetic with unit roundoff $u=2^{-53}$,
the two coefficient operations, two products and one addition contribute
less than $128u$ in $\ell^1$ per prime; over the $5{,}131$ primes in
$[5,L]$ this is below $7.4\cdot10^{-11}$, and underflow contributes less
than $10^{-290}$. Each $s_\ell^+$ rounds up an upper bound for
$2^{20}\omega_\ell$ of excess below $2^{-200}$, so
$s_\ell^+<2^{20}\omega_\ell+1+2^{-200}$. Since $\omega_\ell<(\ell-1)^{-1}$ and
exact rational summation gives $\sum_{5\le\ell\le L}(\ell-1)^{-1}<1.917$,
the largest support index $J_+:=\sum_{5\le\ell\le L}s_\ell^+$ satisfies
$J_+<1.917\cdot2^{20}+5{,}132<2^{21}-1$, and $\sum_\ell s_\ell^-\le J_+$.
Hence each computed vector has fewer than $2^{21}$ entries, each
cumulative probability is a partial sum of fewer than $2^{21}$
nonnegative binary64 numbers, and sequential summation has relative
error at most $\gamma_{2^{21}}:=2^{21}u/(1-2^{21}u)<2.4\cdot10^{-10}$.
The total floating-point error is therefore below $3.2\cdot10^{-10}$, and
subtracting $10^{-9}$ from each computed lower bound for the distribution
function of $R_H$ and adding $10^{-9}$ to each computed upper bound
absorbs it.

Since $\omega_\ell<1/(\ell-1)$, with $M:=80{,}000$,
\[
 \mathbb E R_T
 \le\sum_{L<\ell\le M}\frac1{(\ell-1)^2}+\frac1{M-1}\le1.32\cdot10^{-5}.
\]
Fix $s:=0.00015>\mathbb ER_T$. For every real $\theta$,
\[
 \mathbb P(R_H\le\theta-s)-\mathbb P(R_T>s)
 \le\mathbb P(R\le\theta)
 \le\mathbb P(R_H\le\theta),
\]
where Bennett's inequality~\cite{Bennett62}, with
$\mu:=\mathbb ER_T$, $a:=s-\mu>0$,
$v:=\sum_{\ell>L}\omega_\ell^2/(\ell-1)$ and
$b:=\max_{\ell>L}\omega_\ell$, gives
\[
 \mathbb P(R_T>s)\le
 \exp\!\left[-\frac v{b^2}
 \left\{\left(1+\frac{ba}{v}\right)
 \log\left(1+\frac{ba}{v}\right)-\frac{ba}{v}\right\}\right].
\]
Indeed, the centered summands are independent, bounded above by $b$,
and have total variance at most $v<\infty$. Apply the finite-sum
inequality to truncations of $R_T$, whose means are at most $\mu$,
and pass to the increasing limit. As in the bound for $\mathbb ER_T$
above, $\omega_\ell<1/(\ell-1)$ and comparison of the sums over
$\ell>M$ with integrals give $\mu\le\bar\mu$, $v\le\bar v$ and
$b\le\bar b$, where
\[
 \begin{gathered}
 \bar\mu:=10^{-60}\sum_{L<\ell\le M}
     \Bigl\lceil\frac{10^{60}}{(\ell-1)^2}\Bigr\rceil+\frac1{M-1},\\
 \bar v:=10^{-60}\sum_{L<\ell\le M}
     \Bigl\lceil\frac{10^{60}}{(\ell-1)^3}\Bigr\rceil+\frac1{2(M-1)^2},
 \qquad\bar b:=\frac1L.
 \end{gathered}
\]
These rational bounds give a valid upper bound in Bennett's inequality;
the bound increases with each of these parameters while $s>\mu$.
At $(\mu,v,b)=(\bar\mu,\bar v,\bar b)$ the Bennett bound displayed above
is less than $2.3375\cdot10^{-8}$. To certify this, reduce
the logarithm to arguments in $[1,2]$ and retain $140$ terms of its
positive $\operatorname{arctanh}$ series. If $A$ is the negative logarithm
of this bound, expand $\exp(A/32)$ through degree $100$.
Geometric remainder bounds and outward rounding to multiples of
$10^{-115}$ enclose both series; exact powers and reciprocals give
an enclosure for $e^{-A}$. The verifier cited below evaluates the same
bound at these inputs in $80$-digit decimal arithmetic and adds
$10^{-40}$. Direct rational comparison with the enclosure verifies that
the result is an upper bound for these inputs:
\[
 \begin{aligned}
 e^{-A}&<2.3374882469032218255158869712147156\cdot10^{-8},\\
 \text{verifier value}&>2.3374882469032218255158869712147255\cdot10^{-8}.
 \end{aligned}
\]
At $\beta=3.0001$,
put $t:=c-\omega_3=\log(\beta/3)>0$. The lower bound for
$\mathbb P(R\le t)$ is sharpened as in the proof of
\Cref{lem:gencert-positive}: on the event that $B_\ell=0$ for every
$\ell\ge5$ with $\omega_\ell>t$, $R$ equals the sum over the remaining
$\ell\ge5$, which is independent of this event, so Markov's inequality gives
\[
 \mathbb P(R\le t)\ge
 \prod_{\substack{\ell\ge5\\\omega_\ell>t}}\Bigl(1-\frac1{\ell-1}\Bigr)
 \Bigl(1-\frac1t\sum_{\substack{\ell\ge5\\\omega_\ell\le t}}\frac{\omega_\ell}{\ell-1}\Bigr).
\]
The product is bounded below by exact fixed-point multiplication, and the
sum above by an exact fixed-point sum over primes $\ell\le M_\ast:=2{,}000{,}000$
plus $\sum_{\ell>M_\ast}(\ell-1)^{-2}\le(M_\ast-1)^{-1}$.
Rounding these bounds to the displayed precision and then widening each
rounded endpoint outward by one unit in its last decimal place gives the
displayed intervals.
\end{proof}

\paragraph*{Code availability}
The deterministic CDF verifier for \Cref{prop:gencert-numerics} is archived in
\href{https://doi.org/10.5281/zenodo.24288792}{Zenodo record~24288792}.

\begin{proposition}[Computable probabilities and quantiles]
\label{prop:cdf-quantile-certification}
\index{Quantiles!numerical certification}
Let $\eta>0$ be rational. For rational $\beta>2$, a terminating procedure
computes rational lower and upper bounds for $F(\beta)$ differing by less
than $\eta$. For rational $q\in(0,1)$, a terminating procedure computes
rationals $2<a<\beta_q<b$ with $b-a<\eta$, where $\beta_q$ is the unique
solution of $F(\beta_q)=q$.
The same conclusions hold for each of the four basic ratio laws,
the fixed-progression laws of \Cref{thm:ap-law,cor:ap-sigma-law}, and
their positive logarithmic images: their CDFs at rational arguments
and their quantiles at rational levels in $(0,1)$ admit terminating
rational enclosures of any prescribed positive rational width.
\end{proposition}

\begin{proof}
For integers $n\ge2$ put $L=n^2$ and
$V_L=2\prod_{3\le\ell\le L}(\ell/(\ell-1))^{B_\ell}$, and let $F_L$
be its distribution function. Enumerating the finitely many assignments
to $(B_\ell)_{3\le\ell\le L}$ computes $F_L$ exactly at every rational
argument, since all atoms and their probabilities are rational.
Writing $1/X=V_Le^{T_L}$ with $T_L\ge0$, one has
$\mathbb ET_L<\sum_{m>L}(m-1)^{-2}\le1/(L-1)$.
Markov's inequality at level $\log(1+1/n)\ge1/(n+1)$ gives
\[
 F_L\!\left(\frac{n\beta}{n+1}\right)-\frac1{n-1}
 \le F(\beta)\le F_L(\beta).
\]
Increase $n$ until the width of these rational bounds is below $\eta$.
Since $V_L\uparrow1/X$, one has $F_L(\beta)\downarrow F(\beta)$ and
$F_L(n\beta/(n+1))\ge F(n\beta/(n+1))$, so continuity of $F$ makes the
width tend to zero and the procedure terminates.
For a quantile, dovetail the procedure above over all rational pairs
$2<a<b$ with $b-a<\eta$, refining every pair indefinitely, and stop when
the bounds certify $F(a)<q<F(b)$, which forces $a<\beta_q<b$. Such a pair
exists because $F$ is continuous and strictly increasing on $(2,\infty)$.

For the other laws, let $R$ be the reciprocal totient ratio or the
divisor-sum ratio. Retain primes at most an integer $y$, beyond all
progression primes, and replace every retained valuation $V_\ell$ by
$\min(V_\ell,J)$. The resulting $R_{y,J}$ has finitely many rational
atoms and rational probabilities. Its logarithmic remainder satisfies
\[
 0\le\mathbb E\log(R/R_{y,J})
 \le B_{y,J}:=\frac2{y-1}
       +\sum_{\ell\le y}\frac{2\mathbb P(V_\ell>J)}{\ell-1}.
\]
Indeed, every positive logarithmic local factor is at most
$\log(\ell/(\ell-1))\le1/(\ell-1)$, and the unconditioned tail
activation probability is at most $1/(\ell-1)$.
The retained valuation tails are explicit rational geometric tails,
also in a fixed progression. Choose increasing $y_n,J_n\ge n$ with
$B_{y_n,J_n}\le n^{-3}$, and denote the finite CDF by $F_n$.
The preceding argument gives, for rational $t>0$,
\[
 F_n\!\left(\frac{nt}{n+1}\right)-(n+1)B_{y_n,J_n}
 \le\mathbb P(R\le t)\le F_n(t).
\]
Continuity makes the gap tend to zero. Reciprocal transformation gives
the totient CDFs; rational enclosures of exponentials give the
logarithmic CDFs by monotone squeezing. For the integer divisor-sum
law, the interval support follows from
\Cref{lem:supp_conv,cor:regularity}. The laws are continuous and
strictly increasing on their interval supports, so the same rational
bracket search computes their quantiles.
\end{proof}

\chapter{Ces\`aro moments of the cosine sum \texorpdfstring{$S$}{S}}\label{app:cesaro-moments}
\index{Cesaro moments@Ces\`aro moments}

\paragraph*{Moment estimates}
The finite cosine expansion \eqref{eq:finite-cosine-moment} and
\Cref{lem:tail-moment} control the centered tails of $S$ in every fixed
even moment norm. The latter takes $T\to\infty$ at fixed prime cutoff
$y$, then $y\to\infty$, with bounds depending on the moment order.
The argument uses \Cref{thm:linind,lem:amplitude-tails}; the mean
calculation also uses \Cref{lem:prime-log-stieltjes}.
These estimates give the mean and variance in \Cref{thm:variance},
the joint limit and mixed moments in \Cref{cor:complex-amplitude-law},
and the non-Gaussian and exponential-moment conclusions in
\Cref{cor:dissipation-law}.

For locally integrable $h$, write $\overline h_T=T^{-1}\int_0^T h(\tau)\,d\tau$.
Let $(\Theta_\ell)_{\ell\ge3}$, indexed by odd primes, be independent
uniform phases on $[0,2\pi]$.
Recall $X_D=\sum_{\ell\ge3}d_\ell\cos\Theta_\ell$ and
$V_D=\sum_{\ell\ge3}d_\ell^2$.
For a finite set $A$ of odd primes put
$X_A:=\sum_{\ell\in A}d_\ell\cos\Theta_\ell$ and
$V_A:=\sum_{\ell\in A}d_\ell^2$. Since odd cosine moments vanish and
$\mathbb E\cos^{2m}\Theta=\binom{2m}{m}/4^m$, the multinomial theorem and
$(m!)^{-2}\le(m!)^{-1}$ give, for every integer $q\ge1$,
\begin{equation}\label{eq:finite-cosine-moment}
 \mathbb E X_A^{2q}
 =\frac{(2q)!}{4^q}
  \sum_{\substack{m_\ell\in\mathbb Z_{\ge0}\ (\ell\in A)\\\sum_{\ell\in A}m_\ell=q}}
  \prod_{\ell\in A}\frac{d_\ell^{2m_\ell}}{(m_\ell!)^2}
 \le\frac{(2q)!}{4^q q!}\,V_A^q.
\end{equation}
Since $\sum_\ell d_\ell^2<\infty$, applying this bound to differences of
partial sums makes $\sum_\ell d_\ell\cos\Theta_\ell$ Cauchy in every fixed
$L^{2q}$. Completeness and monotone convergence of the nonnegative
coefficients extend the identity and bound to the infinite series and
each of its tails.

\begin{lemma}[Centered Ces\`aro moments of the remainder $R_y$]\label{lem:tail-moment}
For every fixed integer $q\ge1$ and real $y\ge2$, with
$R_y(\tau):=\sum_{\ell>y}d_\ell(1-\cos(\tau\omega_\ell))$, so that $R_2=S$,
\[
 \lim_{T\to\infty}\overline{|R_y-\overline{R_y}_T|^{2q}}_T
 =\mathbb E\left|\sum_{\ell>y}d_\ell\cos\Theta_\ell\right|^{2q}
 \ll_q\left(\sum_{\ell>y}d_\ell^2\right)^q
 \xrightarrow[y\to\infty]{}0.
\]
In particular, for every integer $n\ge1$,
\[
 \mathfrak M_{2n}:=\lim_{T\to\infty}\overline{(S-\overline S_T)^{2n}}_T
 =\mathbb EX_D^{2n}\le\frac{(2n)!}{4^nn!}V_D^n,
\]
with $\mathbb EX_D^{2n}$ given by the infinite-series extension of
\eqref{eq:finite-cosine-moment}; moreover,
$\mathfrak M_4=\tfrac34V_D^2-\tfrac38\sum_\ell d_\ell^4$.
\end{lemma}

\begin{proof}
Split the primes at $P:=T^{1/(4q)}$ and $T$. The $\ell$th factor of \eqref{eq:euler} is
$|\widehat{\nu_\ell}(\tau)|^2=1-d_\ell(1-\cos(\tau\omega_\ell))$, so the second bound in
\eqref{eq:amplitude-far-tail}, applied to $\xi_\ell=B_\ell\omega_\ell$ (whose mean is at most
$(\ell-1)^{-2}$), shows that the primes $\ell>T$ contribute $O(1/\log T)$, uniformly in
$0\le\tau\le T$. For $P<\ell\le T$ put
\[
 Z_{P,T}(\tau):=\sum_{P<\ell\le T}d_\ell e^{i\tau\omega_\ell}.
\]
\Cref{lem:amplitude-tails}\textup{(b)}, with $v=v_\varphi$, $b_\ell=d_\ell\le2/\ell$,
$y=P$ and $x=\Lambda=T$, gives
\[
 \overline{|Z_{P,T}|^2}_T
 \ll\sum_{\ell>P}d_\ell^2+\frac1{\log T}
 \ll\frac1{P\log P}+\frac1{\log T}=o(1).
\]
After subtracting its average over $[0,T]$, the contribution from
$P<\ell\le T$ is $-\operatorname{Re}(Z_{P,T}-\overline{Z_{P,T}}_T)$, whose
mean square over $[0,T]$ is at most $\overline{|Z_{P,T}|^2}_T=o(1)$. Its
$L^\infty$ norm is $O_q(1)$ by Mertens' theorem, so interpolation makes
its centered $L^{2q}$ norm $o(1)$.

For $y<\ell\le P$, the centering terms are $O(P/T)=o(1)$ because
$\sum_{\ell\le P}d_\ell/(T\omega_\ell)\ll P/T$.  Expand the $2q$-th power.
A nonzero frequency is
$\log(m/n)$ with $m,n\le P^{2q}$, hence has modulus at least $P^{-2q}$;
the total contribution of the nonzero frequencies is
\[
 O_q\!\left((\log\log T)^{2q}\frac{P^{2q}}T\right)=o(1).
\]
By \Cref{thm:linind}, the zero-frequency terms are exactly those in which
each prime occurs equally often with each sign, and their total
contribution is
\[
 \mathbb E\left|\sum_{y<\ell\le P}d_\ell\cos\Theta_\ell\right|^{2q}.
\]
As $T\to\infty$, this tends to $\mathbb E|\sum_{\ell>y}d_\ell\cos\Theta_\ell|^{2q}$,
whose stated bound follows by taking limits in \eqref{eq:finite-cosine-moment};
Minkowski's inequality combines the three ranges.
Taking $y=2$, so that $R_2=S$, gives the asserted even central moments.
The infinite-series extension of \eqref{eq:finite-cosine-moment}
gives the bound and, for $q=2$, the fourth-moment identity.
\end{proof}

\begin{proof}[Proof of \Cref{thm:variance}]
For the mean, Tonelli gives
\[
 \overline S_T=\sum_\ell d_\ell
 \left(1-\frac{\sin(T\omega_\ell)}{T\omega_\ell}\right).
\]
Since $d_\ell/\omega_\ell\ll1$, the primes $\ell\le T$ contribute
$\sum_{3\le\ell\le T}d_\ell+O(\pi(T)/T)$. The primes $\ell>T$ contribute
the Ces\`aro mean over $[0,T]$ of the tail of $S$ beyond $T$, which is
$O(1/\log T)$ by the second bound in \eqref{eq:amplitude-far-tail}, applied to
$\xi_\ell=B_\ell\omega_\ell$: by \eqref{eq:euler} the $\ell$th term of that tail is
$1-|\widehat{\nu_\ell}(\tau)|^2$. Hence
$\overline S_T-\sum_{3\le\ell\le T}d_\ell=O(1/\log T)$.
The identity $d_\ell=2/\ell-2/[\ell(\ell-1)^2]$ and
\eqref{eq:mertens-dvp} give the stated $C_D$; the prime $2$ contributes
$-1$, and the convergent correction has an $O(T^{-2})$ tail.

Finally, \Cref{lem:tail-moment} with $y=2$ and $q=1$ gives the
variance limit $V_D/2$.
\end{proof}

\chapter{Type-\texorpdfstring{$A$}{A} tensor products and cyclotomic codifferents}
\label{app:cyclotomic-smoothing}

This appendix treats Gaussian smoothing of cyclotomic integer lattices
through the shells of their codifferents. Its connection with the
determinant applications of \Cref{sec:explicit} and the prime search in
\Cref{ssec:certgen} concerns the negacyclic rings in
\Cref{def:gencert-search}: for $N\ge2$, $m=2N$ is a power of two and
the codifferent is a scaled $\Z^N$.
In this case \eqref{eq:cyclotomic-block-identity} reduces to
$\eta_\epsilon(\Z^N)=\eta_{(1+\epsilon)^{1/N}-1}(\Z)$, which appears in the proof
of~\cite[Corollary~12]{GLWT25}.
For arbitrary fixed $N$ in
\hyperref[par:gencert-fixed-length]{the fixed-length search}, the polynomial
$X^N+1$ may have several cyclotomic factors. The analysis below treats
individual conductors, including those with odd prime factors.

The shell classification and codifferent decomposition provide the
geometric inputs for the smoothing estimates in
\Cref{sec:lattice-uniform-gap}.
\Cref{fig:tensor-smoothing} indicates how these inputs are used.

\section{Tensor products of type-\texorpdfstring{$A$}{A} root lattices}
\label{sec:A-tensor-shells}

For $n\ge1$, write
\[
 A_n=\left\{x\in\Z^{n+1}:\sum_{j=1}^{n+1}x_j=0\right\}.
\]
In the basis $e_j-e_{n+1}$, $1\le j\le n$, its Gram matrix is
$I_n+J_n$, where $J_n$ is the all-ones matrix.  Let $k\ge1$, let
$n_1,\ldots,n_k\ge1$, and let
$X\in\Z^{n_1\times\cdots\times n_k}$.  For $S\subseteq[k]$, define
\[
 \operatorname{marg}_S(X)_{(i_s)_{s\in S}}
 =\sum_{(j_t)_{t\notin S}}X_{j_1,\ldots,j_k},
 \qquad
 \Sigma_k(X)=\sum_{S\subseteq[k]}
       \lVert\operatorname{marg}_S(X)\rVert_2^2,
\]
where $j_s=i_s$ for $s\in S$. Here $\operatorname{vec}(X)$ lists
entries in lexicographic order of $(i_1,\ldots,i_k)$, with $i_k$
varying fastest. Expanding the Kronecker product shows that
\[
 \operatorname{vec}(X)^{\!T}
 \left[\bigotimes_{i=1}^k(I_{n_i}+J_{n_i})\right]
 \operatorname{vec}(X)=\Sigma_k(X),
\]
so $\Sigma_k$ is the squared-norm form on
$A_{n_1}\otimes\cdots\otimes A_{n_k}$.

Let $k\ge1$ and $n_1,\ldots,n_k\ge1$, and put
$r=\#\{i:n_i\ge2\}$. If $r=1$, write $n_*$ for the rank of the unique
non-$A_1$ factor.
For $r\ge2$, an alternating signed six-cycle in axes $i,j$, in the
ambient coordinates of $A_{n_i}\subset\Z^{n_i+1}$ and
$A_{n_j}\subset\Z^{n_j+1}$, has six entries $\pm1$ alternating around
a cycle on three positions in each axis, and zero elsewhere.

The minimum and minimal-vector classification below follows from
Kitaoka's $E$-type results \cite[Proposition~1 and Theorem~2]{Kit77}:
each $A_n$ has minimum two and positive integral sublattice discriminants,
and $E$-type is preserved under tensor products.
We include a self-contained proof in the present normalization.

\begin{lemma}[Minimum of type-$A$ tensor products]
\label{lem:A-tensor-minimum}
\index{Root lattices!tensor minimum}
The squared minimum of $A_{n_1}\otimes\cdots\otimes A_{n_k}$ is $2^k$,
and its minimal vectors are exactly the pure tensors of roots of the factors.
The first-shell multiplicity is
\[
 N_1=2^{1-k}\prod_{i=1}^k n_i(n_i+1).
\]
\end{lemma}

\begin{proof}
Realize the tensor product as the integer arrays of shape
$(n_1+1)\times\cdots\times(n_k+1)$ whose sum on each coordinate line is zero.
Indeed, the tensor basis $\bigotimes_i(e_{j_i}-e_{n_i+1})$, $j_i\le n_i$,
has these line sums, and the entries with every $j_i\le n_i$ determine
all remaining entries by the line-sum conditions. The squared norm in
this realization is the sum of the squares of all entries.

For $k=1$, a nonzero integer vector with coordinate sum zero has squared
norm at least two, with equality precisely at the roots $e_a-e_b$,
$a\ne b$. Inductively, slice along the first axis. Every nonzero slice
belongs to the tensor product of the remaining factors and has squared
norm at least $2^{k-1}$. The slices sum to zero, so there are at least two
nonzero slices. Hence the squared norm is at least $2^k$.
Equality forces exactly two opposite slices $Y,-Y$, each minimal.
Induction then gives precisely $(e_a-e_b)\otimes Y$, a pure tensor of
roots. Conversely every such tensor has squared norm $2^k$.

There are $n_i(n_i+1)$ roots in the $i$th factor. Equality of two nonzero
pure tensors forces their corresponding factors to be proportional;
for roots of equal length the proportionality factors are signs.
Exactly the $2^{k-1}$ sign choices with product one represent the same
tensor, giving the stated count. In particular all $A_1$ factors give
a rank-one lattice with minimum $2^k$ and two shortest vectors.
\end{proof}

\KeepHeadingSpace{6\baselineskip}
\begin{theorem}[Type-$A$ tensor shells below twice the squared minimum]
\label{thm:A-tensor-second-shell}
\index{Root lattices!tensor shells}
The distinct nonzero squared norms strictly below $2^{k+1}$,
normalized by $2^k$, form the set
\[
 \begin{cases}
  \{2-2^{1-j}:1\le j\le r\},&r\ge1,\\
  \{1\},&r=0.
 \end{cases}
\]
For $r\ge1$, every vector in this range is a sum of two minimal
pure tensors.
For $3\le j\le r$, the shell of squared norm $2^k(2-2^{1-j})$ has
multiplicity
\[
 N_j=2^{j-1}N_1
       \sum_{\substack{S\subseteq[k]\\|S|=j}}\prod_{i\in S}(n_i-1),
\]
and each of its vectors has a unique unordered decomposition into two
minimal pure tensors.

For $r\ge2$, the second shell consists exactly of these six-cycles
tensored with roots in the remaining factors; its vectors are non-pure, and its
multiplicity is
\begin{equation}\label{eq:tensor-second-count}
 N_2=\frac23N_1\sum_{i<j}(n_i-1)(n_j-1).
\end{equation}
Equivalently, for $r\ge2$, a second-shell vector is the sum of two
minimal vectors with inner product $-2^{k-2}$; each such vector has
exactly three unordered decompositions of this form.

\end{theorem}

\begin{proof}
We first establish the elementary lower gap needed in the classification.
Strip off the $A_1$ factors: tensoring with their generators multiplies
all squared norms by $2^{k-r}$ and changes neither ratios nor multiplicities.
For $r=0$ the lattice before rescaling is $\Z$, whose only nonzero squared
norm below two is one. For $r=1$ every squared norm is even, since
$\sum x_a^2\equiv\sum x_a=0\pmod2$, and hence every nonminimal squared
norm is at least four.

For the lower-gap induction suppose now that all $n_i\ge2$ and $k\ge2$.
In the coefficient coordinates of $\Sigma_k$, slice $X$ as
$Y_1,\ldots,Y_{n_1}$ and put $Z=\sum_iY_i$. The marginal definition gives
\begin{equation}\label{eq:A-slice-recursion}
 \Sigma_k(X)=\sum_{i=1}^{n_1}\Sigma_{k-1}(Y_i)+\Sigma_{k-1}(Z).
\end{equation}
Assume inductively that a nonzero $(k-1)$-tensor has squared norm
$2^{k-1}$ or at least $3\cdot2^{k-2}$; the preceding parity argument
supplies the base $k-1=1$. Let $t$ count the nonzero coefficient slices.
If $t\ge3$, the first sum in~\eqref{eq:A-slice-recursion} is at least
$3\cdot2^{k-1}$. If $t=1$, the total is twice the squared norm of its
nonzero slice. If $t=2$ and $Z=0$, the same doubling holds. In either
doubling case the value is $2^k$ or at least $3\cdot2^{k-1}$.
If $t=2$ and $Z\ne0$, the three nonzero terms in
\eqref{eq:A-slice-recursion} have total at least $3\cdot2^{k-1}$.
Thus every nonminimal vector has squared norm at least $3\cdot2^{k-1}$.

Return to the ambient line-sum-zero realization in the proof of
\Cref{lem:A-tensor-minimum}. Its slices, unlike the coefficient slices
in the preceding recursion, sum to zero and contribute their squared
norms without an additional term.

Assume $r\ge2$. At squared norm $3\cdot2^{r-1}$ there are
either two opposite nonzero slices, reducing the assertion to $r-1$
axes, or three minimal slices summing to zero. The first alternative
is impossible when $r=2$, since a single $A_n$ has no squared norm three;
it therefore terminates in the second alternative after induction. In a relation
$u+v+w=0$ among three nonzero pure tensors, the factors can vary
projectively in at most one axis. Indeed, if the factors of $u,v$ are
independent in one axis, flattening against that axis forces the
remaining tensor factors of $u,v$ to be proportional; otherwise
$u+v$ has matrix rank two and cannot equal $-w$. Proportional roots of
equal length differ by a sign. There must be a varying axis, because three signed copies of one
minimal tensor cannot sum to zero. The three roots in the varying
axis thus form a triangle: their oriented coordinate differences
cancel successively. Together with the slicing axis they give a signed
six-cycle. Conversely each such cycle has the asserted norm. It is
non-pure since its flattenings in its two core axes have rank two;
all other flattenings have rank one. Thus the core pair is unique.
A six-cycle is a sum of two root rectangles sharing one edge with
opposite signs; after tensoring with the other roots their inner
product is $-2^{k-2}$. Conversely a sum with this inner product has
the second-shell norm and hence the asserted form.
For a pair of minimal tensors with inner product $-2^{k-2}$, no root
inner product is zero. In exactly two axes its absolute value is one,
and in all remaining axes it is two. These two varying axes are the
core axes, as their flattenings have rank two. The cancelling edge of the two
rectangles is one of the three absent matching edges of the core;
each choice determines one unordered decomposition.

For core axes $i,j$, choose three positions in each. There are six
choices of the omitted perfect matching in the resulting $3$ by $3$
array and two alternating signs. Multiplying by the root counts in
the other factors and dividing by $2^{k-2}$ for sign
redistributions gives
\[
 \frac{12}{2^{k-2}}\binom{n_i+1}{3}\binom{n_j+1}{3}
       \prod_{h\ne i,j}n_h(n_h+1).
\]
Summation over the unique core pair proves
\eqref{eq:tensor-second-count}.

For all ratios below two, strip the $A_1$ factors and induct on $r$.
For $r=0,1$ only ratio one occurs, by the rank-one case and parity.
For $r=1$, split each root as
$e_a-e_b=(e_a-e_c)+(e_c-e_b)$ with $c\notin\{a,b\}$.
For $r\ge2$, the slices sum to zero and each nonzero slice has squared
norm at least $2^{r-1}$. A nonzero tensor of squared norm below
$2^{r+1}$ therefore has two or three nonzero slices.
Two are opposite and preserve a slice's
normalized squared norm; its inductive two-term decomposition tensors
with a root in the slicing axis. With three slices, at least two are minimal:
two nonminimal slices would give total squared norm at least
$(3/2+3/2+1)2^{r-1}=2^{r+1}$. If these minimal slices are $U,V$ at
positions $a,b$, the third, at $c$, is $-U-V$, and the tensor is
\[
 (e_a-e_c)\otimes U+(e_b-e_c)\otimes V.
\]
If the remaining slice has ratio $q$,
then $q<2$ and the full ratio is $1+q/2$. Induction restricts the
ratios to $2-2^{1-j}$, $1\le j\le r$.
For each such $j\ge2$, choose root pairs with inner products of absolute
value one in $j$ axes and identical roots elsewhere, with signs making
the product negative. The resulting minimal tensors have inner product
$-2^{r-j}$, so their sum has squared norm $2^{r+1}-2^{r-j+1}$.
The minimum supplies $j=1$, so this realizes every stated ratio.
Restoring the $A_1$ factors rescales norms and changes no multiplicity
or ratio, and preserves the two-term representation for $r\ge1$.

For $3\le j\le r$, the two minimal summands $P,Q$ of a vector of
normalized squared norm $2-2^{1-j}$ have inner product $-2^{k-j}$.
Their root factors have nonzero inner products; since these have
absolute value one or two, the factors are nonproportional in exactly $j$ axes. Group the axes into
three modes with an independent pair in each, writing the vector as
$a\otimes b\otimes c+x\otimes y\otimes z$.
The third-mode flattening has image
$\operatorname{span}\{a\otimes b,x\otimes y\}$.
In the independent bases $(a,x)$ and $(b,y)$ this space consists of
diagonal $2$ by $2$ matrices, so its only rank-one lines are the two
displayed lines. Any other two-term pure decomposition has independent
third factors and must use these same lines. Their third coefficients
are then determined, proving uniqueness up to interchange.

Fix $P$ and a $j$-element set $S$ of varying axes. A root of $A_{n_i}$
has $4(n_i-1)$ roots with inner-product magnitude one and two
proportional roots. Halving the factor choices for the negative sign
and dividing by $2^{k-1}$ for signed representations gives
$2^j\prod_{i\in S}(n_i-1)$ partners $Q$.
Sum over $S$, multiply by $N_1$, and divide by two for ordering;
uniqueness gives the stated multiplicity. Factors $A_1$ contribute
$n_i-1=0$ and cannot be varying axes.
\end{proof}

\begin{corollary}[Second and third squared-norm gaps of type-$A$ tensors]
\label{cor:A-tensor-gap-table}
\index{Root lattices!tensor gaps}
The ratio of the second
distinct nonzero squared norm to $2^k$ is
\[
 \begin{cases}
  \tfrac32,&r\ge2,\\[2pt]
  3,&r=1,\ n_*=2,\\[2pt]
  2,&r=1,\ n_*\ge3,\\[2pt]
  4,&r=0.
 \end{cases}
\]
For $r\ge2$, its multiplicity is given by \eqref{eq:tensor-second-count}.
If $r=0$, then $N_2=2$; if $r=1$, then $N_2=6$ for
$n_*=2$, and $N_2=6\binom{n_*+1}{4}$ for $n_*\ge3$.

The ratio $q_3$ of the third distinct nonzero squared norm to the
minimum is determined by the non-$A_1$ factors as follows:
\begin{equation}\label{eq:tensor-third-gap}
\begin{array}{c|c}
 \text{non-$A_1$ factors}&q_3\\ \hline
 \text{none}&9\\
 A_2&4\\
 A_n\ (n\ge3)&3\\
 A_2\otimes A_2&5/2\\
 A_a\otimes A_b\ (a,b\ge2,\ \max(a,b)\ge3)&2\\
 \text{at least three factors}&7/4
\end{array}
\end{equation}
\end{corollary}

\begin{proof}
For $r\ge2$, \Cref{thm:A-tensor-second-shell} gives second ratio $3/2$,
its six-cycle multiplicity, and, when $r\ge3$, third ratio $7/4$.
If $r=0$, stripping the $A_1$ factors leaves $\Z$, with successive
distinct squared norms $1,4,9$ and multiplicity two in every shell.
For a single $A_n$, squared norms are even. If $n\ge3$, a vector with
two entries one and two entries minus one attains squared norm four.
For $A_2$, writing a vector as $(x,y,-x-y)$ gives squared norm
$2(x^2+xy+y^2)$, and $x^2+xy+y^2\equiv(x-y)^2\pmod3$ cannot be two.
The vector $(2,-1,-1)$ attains squared norm six.
A single $A_n$, $n\ge3$,
has squared-norm-four vectors with exactly two entries $1$ and two
entries $-1$, giving $6\binom{n+1}{4}$. For $A_2$ the three permutations of $(2,-1,-1)$ and their
negatives give its second shell.

For the third gaps with at most two non-$A_1$ factors, first consider a single factor. In $A_n$,
$n\ge3$, squared norms are even, and $(2,-1,-1,0,\ldots,0)$
attains six after the norms two and four. In $A_2$, the next value
after two and six is eight, attained by twice a root; the rank-one
case is $1,4,9$ before rescaling. For two factors every squared norm
is even, since $\sum x_j^2\equiv\sum x_j=0\pmod2$.
Thus there is no value between six and eight. If one rank is at least
three, its squared-norm-four vector tensored with a root attains
eight. In a balanced $3$ by $3$ array, an entry of absolute value at
least two contributes at least four; the other entries in its row
and column contribute at least two each, and the complementary
$2$ by $2$ block has sum of absolute value at least two and contributes
at least two. The squared norm is therefore at least ten.
If all entries are $0,\pm1$, each nonzero row has exactly two
nonzero entries, so the squared norm is at most six. This excludes
eight for $A_2\otimes A_2$; the rows
$(2,-1,-1),(-1,1,0),(-1,0,1)$ attain ten.

Restoring the $A_1$ factors rescales all squared norms and does not change
any of these ratios or multiplicities.
\end{proof}

\section{Cyclotomic codifferents}
\label{sec:cyclotomic-gap}

Let $m\ge3$. Write $\Lambda_m$ for the canonical embedding of $\Z[\zeta_m]$ and
$\Lambda_m^*$ for its dual with respect to the trace form
$\operatorname{Tr}_{\Q(\zeta_m)/\Q}(x\bar y)$, the embedded codifferent.  Factor
\[
 m=2^{a_0}\prod_{i=1}^k \ell_i^{a_i},
 \qquad
 D=2^{\max(a_0-1,0)}\prod_{i=1}^k \ell_i^{a_i-1},
 \qquad
 R_m=\prod_{i=1}^k\ell_i,
\]
where $a_0\ge0$, $a_i\ge1$ and the $\ell_i$ are distinct odd primes; empty
products are $1$ and the empty tensor product is $\Z$.

\begin{proposition}[Cyclotomic codifferent structure]
\label{prop:cyclotomic-structure}
\index{Cyclotomic codifferents!tensor decomposition}
There is an isometry
\begin{equation}\label{eq:cyclotomic-lattice-structure}
 \Lambda_m^*\cong(DR_m)^{-1/2}
 \left(A_{\ell_1-1}\otimes\cdots\otimes A_{\ell_k-1}\right)^{\oplus D}.
\end{equation}
Moreover, $\lambda_1(\Lambda_m^*)^2=2^k/(DR_m)$, and the number $K_m$ of
shortest vectors of $\Lambda_m^*$ is
\[
 K_m=D\,2^{1-k}\prod_{i=1}^k\ell_i(\ell_i-1).
\]
\end{proposition}

\begin{proof}
For an odd prime power $p^a$, grouping the power-basis indices modulo
$p^{a-1}$ gives the trace Gram matrix
$p^{a-1}I_{p^{a-1}}\otimes(pI_{p-1}-J_{p-1})$; since
$(pI_{p-1}-J_{p-1})(I_{p-1}+J_{p-1})=pI_{p-1}$, its inverse is
$p^{-a}I_{p^{a-1}}\otimes(I_{p-1}+J_{p-1})$. For $2^a$, $a\ge1$, the
inverse trace Gram is $2^{-(a-1)}I_{2^{a-1}}$. The integral tensor
decomposition over the coprime prime-power components is isometric for the
trace forms \cite[\S~2.3, pp.~44--45]{LPR13}.

In a trace-dual tensor basis, therefore,
\[
 \operatorname{Gram}(\Lambda_m^*)
 =\frac1{DR_m}I_D\otimes
   \bigotimes_{i=1}^k(I_{\ell_i-1}+J_{\ell_i-1}),
\]
which is \eqref{eq:cyclotomic-lattice-structure}.
The case $k=0$ is immediate.  For $k\ge1$,
\Cref{lem:A-tensor-minimum} gives the
minimum and minimal-vector classification.  Since $A_{\ell_i-1}$ has $\ell_i(\ell_i-1)$ roots and each pure tensor has
$2^{k-1}$ representations by signed factors, multiplication by $D$ gives
the stated value of $K_m$.
\end{proof}

Tensor products of the root bases, followed by the orthogonal sum and
scaling in \eqref{eq:cyclotomic-lattice-structure}, give a basis of
shortest vectors, so every ordinary successive minimum of $\Lambda_m^*$
equals $\lambda_1(\Lambda_m^*)$.

For a lattice $\Gamma$, let $\mu_2(\Gamma)$ be its second distinct nonzero vector
length, and put $g_m:=\mu_2(\Lambda_m^*)/\lambda_1(\Lambda_m^*)$. Write
$B:=\bigotimes_{p\mid R_m}A_{p-1}$, so that $\Lambda_m^*\cong(DR_m)^{-1/2}B^{\oplus D}$ by
\eqref{eq:cyclotomic-lattice-structure}, and $\rho_B:=\mu_2(B)^2/\lambda_1(B)^2$.

\begin{theorem}[Ratio of the first two lengths in a cyclotomic codifferent]
\label{thm:cyclotomic-gap}
\index{Cyclotomic codifferents!successive norms}
For every $m\ge3$,
\[
 g_m^2=\begin{cases}
  \tfrac32,&\omega_{\mathrm{odd}}(m)\ge2,\\[2pt]
  3,&m\in\{3,6\},\\[2pt]
  2,&\text{otherwise},
 \end{cases}
\]
where $\omega_{\mathrm{odd}}(m)=k$ is the number of distinct odd prime
divisors of $m$.
Let $H_m$ count the vectors in the second shell, and let $q_{3,m}$
be the third distinct nonzero squared norm divided by
$\lambda_1(\Lambda_m^*)^2$.
Then
\begin{equation}\label{eq:cyclotomic-second-count}
 H_m=\begin{cases}
 \displaystyle\frac23K_m\sum_{p<q,\ p,q\mid R_m}(p-2)(q-2),&k\ge2,\\[4pt]
 \displaystyle6D\binom p4+\binom D2[p(p-1)]^2,&R_m=p\ge5,\\[4pt]
 36\binom D2,&R_m=3,\ D\ge2,\\[2pt]
 6,&m=3,6,\\[2pt]
 4\binom D2,&k=0,
 \end{cases}
\end{equation}
and
\begin{equation}\label{eq:cyclotomic-third-gap}
 q_{3,m}=\begin{cases}
 7/4,&k\ge3,\\
 2,&k=2,\\
 4,&m\in\{3,4,6\},\\
 3,&\text{otherwise}.
 \end{cases}
\end{equation}
\end{theorem}

\begin{proof}
If $\Gamma$ has first two distinct nonzero squared norms $\lambda^2<\mu^2$, then
those of $\Gamma^{\oplus d}$, $d\ge2$, are $\lambda^2$ and $\min(\mu^2,2\lambda^2)$: squared
norms add, and a vector of squared norm above $\lambda^2$ has a component of squared
norm at least $\mu^2$ or two nonzero components. Hence $g_m^2=\rho_B$ if $D=1$ and
$g_m^2=\min(\rho_B,2)$ if $D\ge2$. Each $A_{\ell_i-1}$ has rank at least $2$, so
\Cref{cor:A-tensor-gap-table} gives $\rho_B=3/2$ if $k\ge2$, $\rho_B=3$ if $R_m=3$,
and $\rho_B=2$ if $R_m\ge5$ is prime. If $k=0$, then $B=\Z$, $\rho_B=4$, and $D\ge2$
because $m\ge4$ is a power of two. Since $D=1$ and $R_m=3$ exactly when $m\in\{3,6\}$,
the stated cases follow.

For $k\ge2$ the second ratio is less than two, so a second-shell
vector occupies one block. Equation~\eqref{eq:tensor-second-count}
therefore gives the first case of
\eqref{eq:cyclotomic-second-count}. When $R_m=p\ge5$, the
second ratio two is obtained either within one $A_{p-1}$ block or
from shortest vectors in two different blocks. These give,
respectively, $6D\binom p4$ and
$\binom D2[p(p-1)]^2$. For $R_m=3,D\ge2$ only the latter
possibility occurs, giving $36\binom D2$; for $D=1$ the second
shell of $A_2$ has six vectors. When $k=0$, the lattice before
rescaling is $\Z^D$, with $D\ge2$, giving $4\binom D2$.

Distinct odd primes exclude the core $A_2\otimes A_2$. Thus
\eqref{eq:tensor-third-gap} gives $q_{3,m}=7/4$ for $k\ge3$
and two for $k=2$; vectors using several blocks cannot precede
ratio two. For $R_m=p\ge5$, the single-block ratios two and three
are present and all block squared norms are integral multiples of
the minimum. For $R_m=3,D\ge2$, the ratios two and three are
present (two shortest blocks, and a second-shell vector in one
block); for $D=1$ the third ratio is four. Finally, for $k=0$,
$D=2$ gives the ratios $1,2,4$, whereas $D\ge4$ admits three
nonzero unit coordinates and gives $1,2,3$. Here $D$ is a power of
two, and $D=2$ is exactly $m=4$. This proves all the exceptions in
\eqref{eq:cyclotomic-third-gap}.
\end{proof}

\Cref{fig:tensor-smoothing} illustrates, from
\Cref{sec:A-tensor-shells,sec:cyclotomic-gap}, the first two shells of a
type-$A$ tensor product, in the example $A_2\otimes A_2$, and the passage
from one block to its orthogonal copies in $\Lambda_m^*$.
Panel~(c) indicates how these results enter the smoothing estimates of
\Cref{sec:lattice-uniform-gap}.

% Diagram (design brief F12): Tensor shells, codifferent blocks and smoothing.
\begin{figure}[p]
\centering
\begin{tikzpicture}[figdiagram,
    % local helpers: the two columns of panel (c) share widths and heights
    lcol/.style={fres, anchor=west, minimum width=4.85cm, minimum height=1.2cm},
    rcol/.style={fres, anchor=west, minimum width=6.2cm, minimum height=1.2cm},
    % entries of the small grid inside the structure box of panel (b)
    cell/.style={fnote, inner sep=0pt, anchor=base}]
% ================= (a) two exact arrays ==============================
\node[fhead, anchor=west] (hA) at (0,0) {(a) The first two shells of $A_2\otimes A_2$: integer arrays with zero row and column sums};
\draw[frule] (hA.south west) ++(0,-1.5pt) -- ++(13.5,0);
% both arrays have three columns of the width of "-1"; both text cells have one width
\node[fres, anchor=north west] (a1) at (0,-0.5) {%
  \begin{tabular}{@{}c@{\hspace{6pt}}c@{}}
  $\left[\begin{array}{@{}r@{\;\,}r@{\;\,}r@{}}1&-1&\hphantom{-}0\\-1&1&0\\0&0&0\end{array}\right]$ &
  \makebox[28.3mm]{\begin{tabular}{@{}c@{}}Minimal vectors:\\ pure tensors of roots,\\ squared norm $4$\\ \figref{\Cref{lem:A-tensor-minimum}}\end{tabular}}
  \end{tabular}};
\node[fres, anchor=north east] (a2) at (13.5,-0.5) {%
  \begin{tabular}{@{}c@{\hspace{6pt}}c@{}}
  $\left[\begin{array}{@{}r@{\;\,}r@{\;\,}r@{}}1&-1&0\\0&1&-1\\-1&0&1\end{array}\right]$ &
  \makebox[28.3mm]{\begin{tabular}{@{}c@{}}Second shell:\\ alternating six-cycles,\\ squared norm $6$\\ \figref{\Cref{thm:A-tensor-second-shell}}\end{tabular}}
  \end{tabular}};
\draw[fmap, <->] (a1.east) -- node[flab, above=1.5pt] (acl) {squared-norm ratio $\tfrac32$}
  node[fnote, below=2pt, align=center] (acr) {\figref{\Cref{cor:A-tensor-gap-table}}} (a2.west);
% ================= (b) conductor data, block structure, gap table ====
\node[fhead, anchor=west] (hB) at (0,-2.61) {(b) The codifferent $\Lambda_m^*$: orthogonal copies of one block $B$};
\draw[frule] (hB.south west) ++(0,-1.5pt) -- ++(13.5,0);
\node[finp, anchor=north west, minimum width=13.5cm] (bin) at (0,-3.11) {%
  conductor $m=2^{a_0}\prod_{i=1}^{k}\ell_i^{a_i}\ge3$;\quad $a_0\ge0$, $a_i\ge1$ ($1\le i\le k$);\\
  distinct odd primes $\ell_i$;\quad odd radical $R_m=\prod_i\ell_i$;\quad block multiplicity $D=2^{\max(a_0-1,0)}\prod_i\ell_i^{a_i-1}$;\\
  block $B=A_{\ell_1-1}\otimes\cdots\otimes A_{\ell_k-1}$ \ ($B=\Z$ if $k=0$)};
% table of the first two shells (right); the arrow enters at its title row
\node[anchor=north east, inner sep=0pt, font=\figfont] (btab) at (13.5,-4.76) {%
  \begin{tabular}{@{}l@{\hspace{1.1em}}c@{\hspace{1.1em}}c@{\hspace{1.1em}}l@{}}
  \multicolumn{4}{@{}l@{}}{$g_m^2=\rho_B$ if $D=1$;\quad $g_m^2=\min(\rho_B,2)$ if $D\ge2$}\\[1pt]
  \toprule
  & & \multicolumn{2}{c@{}}{$g_m^2$}\\[-1pt]
  \cmidrule(l{0.25em}){3-4}
  block $B$ & $\rho_B$ & $D=1$ & $D\ge2$\\
  \midrule
  $k\ge2$ & $3/2$ & $3/2$ & $3/2$, by (i)\\
  $A_{p-1}$, $R_m=p\ge5$ & $2$ & $2$ & $2$, by (i), (ii)\\
  $A_2$, $R_m=3$ & $3$ & $3$ & $2$, by (ii)\\
  $\Z$, $R_m=1$ & $4$ & --- & $2$, by (ii)\\
  \bottomrule
  \multicolumn{4}{@{}l@{}}{\rule{0pt}{9pt}\figref{\Cref{thm:cyclotomic-gap}; $\rho_B$ for $k\ge1$: \Cref{cor:A-tensor-gap-table}}}
  \end{tabular}};
% structure box (left): its frame spans the height of the table
\coordinate (bTL) at ($(0,0 |- btab.north)+(0,3.6pt)$);
\coordinate (bBR) at ($(6.3,0 |- btab.south)+(0,-2.4pt)$);
\node[fres, fit=(bTL)(bBR), inner sep=0pt] (bstr) {};
\draw[fuse] (bin.south -| bstr.north) -- (bstr.north);
% title on the baseline of the table title, reference on the baseline of the table footer
\node[anchor=base, inner sep=0pt] (bstrT) at ($(bstr.center |- btab.north)+(0,-6.72pt)$) {$\Lambda_m^*\cong(DR_m)^{-1/2}\,B^{\oplus D}$};
\node[anchor=base, inner sep=0pt] (bstrR) at ($(bstr.center |- btab.south)+(0,2.88pt)$) {\figref{\Cref{prop:cyclotomic-structure}; rows (i), (ii): \Cref{thm:cyclotomic-gap}}};
\draw[fuse] ($(bstr.east |- bstrT.base)+(0,2.2pt)$) -- ($(btab.west |- bstrT.base)+(-1.5mm,2.2pt)$);
% grid inside the structure box: block row, then rows (i) and (ii)
\coordinate (bMid) at ($(bstrT.south)!0.5!(bstrR.north)$);
\begin{scope}[shift={($(bstr.west |- bMid)+(0,0.36)$)}]
  \def\cA{1.07}\def\cB{2.07}\def\cC{3.77}\def\cR{4.32}
  \foreach \i/\x in {1/\cA,2/\cB,3/\cC}
    {\node[finp, minimum width=5mm, minimum height=4.4mm, inner sep=0pt] (B\i) at (\x,0) {$B$};}
  \node[font=\figfont, inner sep=0pt] (op1) at ($(B1)!0.5!(B2)$) {$\oplus$};
  \node[font=\figfont, inner sep=0pt] (op2) at ($(B2)!0.5!(B3)$) {$\oplus\cdots\oplus$};
  \draw[decorate, decoration={brace, amplitude=2.2pt}, frule] ($(B1.north west)+(0,1.5pt)$) -- ($(B3.north east)+(0,1.5pt)$)
    node[midway, above=2.5pt, fnote, inner sep=0.5pt] (bcop) {$D$ orthogonal copies};
  \node[cell, anchor=mid west] (ksr) at (\cR,0) {squared ratio};
  % row (i): one component of length mu_2
  \node[cell, anchor=base east] (ki) at (0.72,-0.80) {(i)};
  \node[cell] (ki1) at (\cA,-0.80) {$\mu_2$};
  \node[cell] (ki2) at (\cB,-0.80) {$0$};
  \node[cell] (kid) at ($(B2)!0.5!(B3)+(0,-0.80)$) {$\cdots$};
  \node[cell] (ki3) at (\cC,-0.80) {$0$};
  \node[cell, anchor=base west] (kit) at (\cR,-0.80) {$\rho_B$};
  % row (ii): two components of length lambda_1
  \node[cell, anchor=base east] (kii) at (0.72,-1.28) {(ii)};
  \node[cell] (kii1) at (\cA,-1.28) {$\lambda_1$};
  \node[cell] (kii2) at (\cB,-1.28) {$\lambda_1$};
  \node[cell] (kiid) at ($(B2)!0.5!(B3)+(0,-1.28)$) {$\cdots$};
  \node[cell] (kii3) at (\cC,-1.28) {$0$};
  \node[cell, anchor=base west] (kiit) at (\cR,-1.28) {$2$\ \ ($D\ge2$)};
\end{scope}
% ================= (c) the ladder of smoothing estimates ==============
\coordinate (cH) at ($(0,0 |- bstr.south)+(0,-0.56)$);
\node[fhead, anchor=west] (hC) at (cH) {(c) Smoothing estimates: shell data, and exact reduction to one block};
\draw[frule] (hC.south west) ++(0,-1.5pt) -- ++(13.5,0);
% left column: inputs and general tools; centres 1.5 cm apart
\coordinate (c0) at ($(cH)+(0,-1.1)$);
\node[lcol] (SH) at ($(c0)+(0,0)$)    {$\Lambda_m^*$: $K_m$ shortest vectors, gap $g_m$\\ \figref{\Cref{prop:cyclotomic-structure,thm:cyclotomic-gap}}};
\node[lcol] (T9) at ($(c0)+(0,-1.48)$) {Explicit bounds, every full-rank $\Lambda$\\ with $\mu_2(\Lambda^*)\ge g\,\lambda_1(\Lambda^*)$, $g\in(1,2]$\\ \figref{\Cref{thm:lattice-gap-uniform}}};
\node[lcol] (BI) at ($(c0)+(0,-2.96)$) {Block identity, $\delta=(1+\epsilon)^{1/D}-1$:\\ $\eta_\epsilon(\Lambda_m)=\sqrt{DR_m}\,\eta_\delta(B^*)$\\ \figref{\eqref{eq:cyclotomic-block-identity} in \Cref{prop:cyclotomic-block-smoothing}}};
\node[lcol] (T7) at ($(c0)+(0,-4.44)$) {Fixed lattice $\Lambda$, $\epsilon\downarrow0$: second-shell\\ correction; constants depend on $\Lambda$\\ \figref{\Cref{prop:fixed-lattice-second-shell}}};
% stub heights: T9 and BI emit two stubs, 2 mm above and below their centres
\coordinate (T9u) at ($(T9.east)+(0,0.2)$);  \coordinate (T9l) at ($(T9.east)+(0,-0.2)$);
\coordinate (BIu) at ($(BI.east)+(0,0.2)$);  \coordinate (BIl) at ($(BI.east)+(0,-0.2)$);
% junction dots on one vertical line x = 6.65; each result box is centred on its dot
\def\xch{6.65}
\node[fjoin] (J1) at (\xch,0 |- SH.east) {};
\node[fjoin] (J3) at (\xch,0 |- T9l) {};
\node[fjoin] (J2) at (\xch,0 |- T7.east) {};
\node[rcol] (C10) at ($(J1)+(0.65,0)$) {All $m\ge3$ with $1<g\le g_m$, at $\epsilon=2^{-c\varphi(m)}$:\\ explicit bound if $c>\log_2(1+2g)/(g^2-1)$\\ \figref{\Cref{cor:cyclotomic-smoothing}}};
\node[rcol] (P17) at ($(J3)+(0.65,0)$) {$m=2^a\ge4$, $n=m/2=D$, all $0<\epsilon<1$:\\ $0\le\pi\eta_\epsilon(\Lambda_m)^2/n-\log(2/\delta)\le\log(1+5\delta^3/8)$\\ \figref{\Cref{prop:power-two-smoothing-mas15}}};
\node[rcol] (P16) at ($(J2)+(0.65,0)$) {Fixed odd radical $R_m$, $D\to\infty$: expansions\\ uniform in $0<\epsilon<1$, with $O_{R_m}$ remainders\\ \figref{\Cref{prop:cyclotomic-fixed-radical}}};
% joint inputs: feeds without tips into the dot, one arrow out
\draw[ffeed] (SH.east) -- (J1);
\draw[ffeed] (T9u) -| (J1);
\draw[fuse] (J1) -- (C10);
\draw[ffeed] (T9l) -- (J3);
\draw[ffeed] (BIu) -| (J3);
\draw[fuse] (J3) -- (P17);
\draw[ffeed] (BIl) -| (J2);
\draw[ffeed] (T7.east) -- (J2);
\draw[fuse] (J2) -- (P16);
% hypotheses, all left-aligned 1 mm right of the left column
\node[fhyp, anchor=south west] (h1) at ($(SH.east)+(0.1,0)$) {$g\le g_m$};
\node[fhyp, anchor=south west] (h2) at ($(T9u)+(0.1,0)$) {part (ii)};
\node[fhyp, anchor=north west] (h3) at ($(T9l)+(0.1,0)$) {part (i)};
\node[fhyp, anchor=south west] (h4) at ($(BIu)+(0.1,0)$) {$R_m=1$};
\node[fhyp, anchor=north west] (h5) at ($(BIl)+(0.1,0)$) {$R_m$ fixed};
\node[fhyp, anchor=south west] (h6) at ($(T7.east)+(0.1,0)$) {on $B^*$ at $\delta$};
% shared parameter setting (dotted, no tip), hanging under the power-of-two result
\node[finp, anchor=north west, minimum width=6.2cm] (NEG) at ($(P17.south west)+(0,-0.62)$) {negacyclic rings, $m=2N\ge4$: \figref{\Cref{def:gencert-search}}};
\draw[flink] ($(P17.south west)+(0.5,0)$) -- node[flab, right=1pt] (shl) {shared parameter setting} ($(NEG.north west)+(0.5,0)$);
\end{tikzpicture}
\caption{\textup{(a)} Squared norms are sums of squared entries.
\textup{(b)} Tensor geometry determines the squared-norm ratio $\rho_B=\mu_2(B)^2/\lambda_1(B)^2$ of
the first two shells of the block $B$, while orthogonal repetition introduces vectors supported in
two blocks; $\lambda_1<\mu_2$ are the first two distinct nonzero lengths of a lattice, and
$g_m=\mu_2(\Lambda_m^*)/\lambda_1(\Lambda_m^*)$. \textup{(c)} These data give gap-based smoothing
estimates. The block decomposition also gives an exact reduction to one block; fixed-lattice
asymptotics and bounds uniform over conductors are distinct routes.}
\label{fig:tensor-smoothing}
\end{figure}

\section{Uniform bounds for the smoothing parameter}
\label{sec:lattice-uniform-gap}

\Cref{thm:lattice-gap-uniform} gives explicit bounds in terms of the
dimension, the number of shortest dual vectors, and a lower bound on the
ratio of the second distinct nonzero dual length to the first; they hold
for every full-rank lattice and have no remainder term.
The fixed-lattice theta-series approach is that of~\cite[\S~6.1, p.~84]{EWY23}.
For lattices similar to rational lattices, the revision
\cite[Propositions~18 and~21]{EWY26} gives second-shell asymptotics and
asymptotic bounds under its additional hypotheses. By homogeneity, the
factor $\lambda_1(\Lambda^*)^{-2}$ in the coefficient displayed
in~\cite[Proposition~18]{EWY26} should be omitted.
\Cref{thm:lattice-gap-uniform} uses instead \Cref{lem:lattice-packing-count};
at $\epsilon=2^{-cn}$, for $c$ above an explicit threshold depending on the
gap, its relative error is exponentially small in $n$
(\Cref{thm:lattice-gap-uniform}\textup{(ii)}).

\begin{definition}[Smoothing parameter]\label{def:lattice-smoothing}
\index{Smoothing parameter!definition}
For a full-rank lattice $\Lambda\subset\R^n$ and $s>0$, set
\[
 \Phi_\Lambda(s)=\sum_{0\ne w\in\Lambda^*}
       e^{-\pi s^2\lVert w\rVert_2^2}.
\]
For $0<\epsilon<1$, define
\[
 \eta_\epsilon(\Lambda)=\inf\{s>0:\Phi_\Lambda(s)\le\epsilon\}.
\]
Let $\lambda=\lambda_1(\Lambda^*)$, let $K$ be the number of dual vectors
of length $\lambda$, and put
\[
 s_{\mathrm{sh}}(\Lambda,\epsilon)
 =\frac1\lambda\sqrt{\frac{\log(K/\epsilon)}{\pi}}.
\]
\end{definition}

At $s=s_{\mathrm{sh}}(\Lambda,\epsilon)$, the $K$ shortest dual vectors
contribute exactly $\epsilon$ to $\Phi_\Lambda(s)$. All remaining terms
are positive, so
\begin{equation}\label{eq:shortest-shell-bound}
 \eta_\epsilon(\Lambda)\ge s_{\mathrm{sh}}(\Lambda,\epsilon).
\end{equation}
This sharpens~\cite[Claim~2.13]{Reg09}, which has $\log(1/\epsilon)$ in place of
$\log(K/\epsilon)$; for $\Lambda=\Z$, where $K=2$, it is~\cite[Lemma~7]{GLWT25}.
Since $\eta_\epsilon$ and $s_{\mathrm{sh}}(\cdot,\epsilon)$ scale linearly under dilation of $\Lambda$, while $K$ and
the ratios of dual lengths are unchanged, the arguments below assume $\lambda_1(\Lambda^*)=1$ and use
\[
 F(t):=\sum_{0\ne w\in\Lambda^*}e^{-t\lVert w\rVert_2^2}=\Phi_\Lambda(\sqrt{t/\pi})\qquad(t>0).
\]
Every nonzero dual vector then has squared norm at least $1$, so
\begin{equation}\label{eq:min-length-shift}
 F(t+\Delta)\le e^{-\Delta}F(t)\qquad(t>0,\ \Delta\ge0).
\end{equation}

\paragraph*{Selecting a smoothing bound}
The shell classifications in
\Cref{cor:A-tensor-gap-table,thm:cyclotomic-gap} distinguish,
respectively, the $A_1$, single-factor and $A_2\otimes A_2$ cases,
and the small-conductor exceptions $m=3,4,6$.
The following bounds concern the
ratio $\eta_\epsilon/s_{\mathrm{sh}}$, with $L=\log(K/\epsilon)$ in the
general lattice setting.
\begin{enumerate}[label=\textup{(\roman*)},leftmargin=*]
\item \emph{Fixed lattice, $\epsilon\downarrow0$.}
\Cref{prop:fixed-lattice-second-shell} gives the second-shell correction.
Its remainder constants depend on the fixed lattice $\Lambda$.
\item \emph{A prescribed gap, finite dimension and error.}
For $1<g\le2$ with $\mu_2(\Lambda^*)\ge g\lambda_1(\Lambda^*)$,
\Cref{thm:lattice-gap-uniform}\textup{(i)} gives an explicit bound when
$L\ge2n/[g(1+2g)]$, with its alternative bound available for
$L>n/[g(1+2g)]$. At $\epsilon=2^{-cn}$, part~\textup{(ii)} requires
$c>\log_2(1+2g)/(g^2-1)$.
\item \emph{Varying cyclotomic conductor and error.}
Put $n=\varphi(m)$, $k=\omega_{\mathrm{odd}}(m)$,
$\beta=2^k\log(2n+1)$ and $E=\log(1/\epsilon)$.
For $m\notin\{3,6\}$, \Cref{prop:cyclotomic-second-shell} gives the
uniform second-shell expansion \eqref{eq:uniform-second-shell}, and
\Cref{prop:cyclotomic-second-bracket-mas15} gives an explicit finite-error
bracket. The relative remainder in \eqref{eq:uniform-second-shell} tends
to zero if $E-7\beta\to\infty$ for $k\ge2$, or $E-3\beta\to\infty$ for
$k\le1$; in particular this holds at $\epsilon=\exp(-cn^\alpha)$ for
fixed $c,\alpha>0$ as $n\to\infty$.
At $\epsilon=2^{-cn}$, \Cref{cor:cyclotomic-smoothing} uses the preceding
explicit gap threshold, whereas \eqref{eq:all-conductor-smoothing} gives relative
error $O_c(2^{-cn/2}/n)$ for every fixed $c>0$.
\item \emph{Fixed odd radical and growing block multiplicity.}
With $R_m$ fixed and $D\to\infty$, \eqref{eq:fixed-radical-smoothing}
gives relative error $O_{R_m}(1/\log(D/\epsilon))$, uniformly for
$0<\epsilon<1$. The finer expansions in
\Cref{prop:cyclotomic-fixed-radical} have the same
uniformity, with constants and thresholds depending only on $R_m$;
they include exponentially decreasing $\epsilon$.
For $\epsilon=2^{-cn}$, \Cref{prop:cyclotomic-block-smoothing}\textup{(iii)}
is uniform in $R_m$ under
$\theta=c/2-\log_2(1+\sqrt6)/D>0$. At fixed $\epsilon$, allowing the odd
radical to grow can destroy convergence, as
\Cref{rem:growing-radical-smoothing} shows.
\end{enumerate}

\Needspace{6\baselineskip}
\paragraph*{Fixed-lattice asymptotics}\phantomsection\label{par:fixed-lattice-asymptotics}

\begin{proposition}[Correction to the shortest-vector lower bound for a fixed lattice]
\label{prop:fixed-lattice-second-shell}
Fix a full-rank lattice $\Lambda$, with $\lambda$ and $K$ as in
\Cref{def:lattice-smoothing}. Let $1<q<q_3$ be its second and third
distinct nonzero dual squared norms divided by $\lambda^2$, and let $H$ be
the number of dual vectors of squared norm $q\lambda^2$. With $L=\log(K/\epsilon)$ and
$\rho=\min(2(q-1),q_3-1)$, as $\epsilon\downarrow0$,
\[
 \pi\lambda^2\eta_\epsilon(\Lambda)^2
 =L+HK^{-q}\epsilon^{q-1}+O_\Lambda(\epsilon^\rho),
\]
\[
 \frac{\eta_\epsilon(\Lambda)}{s_{\mathrm{sh}}(\Lambda,\epsilon)}
 =1+\frac{HK^{-q}\epsilon^{q-1}}{2L}
   +O_\Lambda(\epsilon^\rho/L).
\]
\end{proposition}

\begin{proof}
Isolating the dual vectors of squared norms $1$ and $q$, and
bounding the sum over the remaining dual vectors as in
\eqref{eq:min-length-shift}, gives $F(t)=Ke^{-t}+He^{-qt}+O_\Lambda(e^{-q_3t})$
for $t\ge1$. The continuous strictly decreasing function $F$ takes the value
$\epsilon$ at $t=\pi\eta_\epsilon(\Lambda)^2$.
Writing $t=L+h$, \eqref{eq:shortest-shell-bound} and \eqref{eq:min-length-shift} give
$0\le h\le\log(F(L)/\epsilon)=O_\Lambda(\epsilon^{q-1})$.
Consequently
\[
 1=e^{-h}+HK^{-q}\epsilon^{q-1}e^{-qh}
       +O_\Lambda(\epsilon^{q_3-1}),
\]
so $h=HK^{-q}\epsilon^{q-1}+O_\Lambda(\epsilon^\rho)$.
The square-root expansion gives the second formula.
\end{proof}

\paragraph*{Uniform bounds from dual-shell gaps}\phantomsection\label{par:uniform-dual-shell-gaps}

\begin{lemma}[Counting lattice points in a ball]\label{lem:lattice-packing-count}
If $\Gamma\subset\R^n$ is a full-rank lattice with $\lambda_1(\Gamma)=1$,
then, for $u\ge0$,
\[
 \#\{w\in\Gamma:\lVert w\rVert_2\le u\}\le(1+2u)^n.
\]
\end{lemma}

\begin{proof}
The open balls of radius $1/2$ centered at these lattice points are
disjoint and lie in the ball of radius $u+1/2$.  Comparison of volumes
gives the claim.
\end{proof}

\begin{theorem}[Uniform smoothing bounds from the first two dual lengths]\label{thm:lattice-gap-uniform}
\index{Smoothing parameter!uniform bounds}
Let $\Lambda\subset\R^n$, $n\ge1$, be a full-rank lattice with $K$ as in
\Cref{def:lattice-smoothing}, and, with $\mu_2$ as in \Cref{sec:cyclotomic-gap}, suppose
that $\mu_2(\Lambda^*)\ge g\,\lambda_1(\Lambda^*)$ for some $g\in(1,2]$. For $0<\epsilon<1$ put
\[
 \mathfrak a:=\frac{n}{g(1+2g)},\qquad L:=\log(K/\epsilon),\qquad
 A:=2(1+2g)^nK^{-g^2}\epsilon^{g^2-1}.
\]
\begin{enumerate}[\upshape(i)]
\item If $L\ge2\mathfrak a$, then
\begin{equation}\label{eq:finite-gapped-smoothing}
 1\le\frac{\eta_\epsilon(\Lambda)}{s_{\mathrm{sh}}(\Lambda,\epsilon)}
 \le\sqrt{1+\frac{\log(1+A)}L}\le1+\frac A{2L}.
\end{equation}
If $L>\mathfrak a$, the same inequalities hold with
$A_*:=(1-\mathfrak a/L)^{-1}A/2$ in place of $A$.
\item Let $c>0$ and $\epsilon=2^{-cn}$. If
$\kappa:=c(g^2-1)-\log_2(1+2g)>0$, then
\begin{equation}\label{eq:uniform-gapped-smoothing}
 s_{\mathrm{sh}}(\Lambda,\epsilon)\le\eta_\epsilon(\Lambda)
 \le\left(1+\frac{K^{-g^2}2^{-\kappa n}}L\right)
 s_{\mathrm{sh}}(\Lambda,\epsilon).
\end{equation}
\end{enumerate}
\end{theorem}

\begin{proof}
Let $F$ be as above and write $F(t)=Ke^{-t}+T(t)$; by hypothesis, $T(t)$ comes
from dual vectors of norm at least $g$.  The identity
$e^{-tr^2}=2t\int_r^\infty u e^{-tu^2}\,du$, Tonelli's theorem, and
\Cref{lem:lattice-packing-count} give
\[
 T(t)\le2t\int_g^\infty u(1+2u)^n e^{-tu^2}\,du.
\]
Since $\tfrac{d}{d(u^2)}\log(1+2u)^n=n/(u(1+2u))\le\mathfrak a$ for $u\ge g$,
$(1+2u)^n\le(1+2g)^n e^{\mathfrak a(u^2-g^2)}$, and integration gives
\begin{equation}\label{eq:gapped-tail}
 T(t)\le\frac{t}{t-\mathfrak a}(1+2g)^n e^{-tg^2}
 \qquad(t>\mathfrak a).
\end{equation}
At $t=L\ge2\mathfrak a$, this yields $T(L)/\epsilon\le A$.
Under the weaker condition $L>\mathfrak a$, it yields the same assertion
with $A_*$ in place of $A$.
Since $Ke^{-L}=\epsilon$, $F(L)\le\epsilon(1+A)$, and \eqref{eq:min-length-shift}
with $\Delta=\log(1+A)$ gives $F(L+\Delta)\le\epsilon$, that is,
$\pi\eta_\epsilon(\Lambda)^2\le L+\log(1+A)$, where $L=\pi s_{\mathrm{sh}}(\Lambda,\epsilon)^2$.
With \eqref{eq:shortest-shell-bound} this gives the first two inequalities of
\eqref{eq:finite-gapped-smoothing}; the last follows from $\log(1+A)\le A$ and
$\sqrt{1+u}\le1+u/2$ for $u\ge0$.

For $\epsilon=2^{-cn}$, $\kappa>0$ gives
$L\ge cn\log2>n\log(1+2g)/(g^2-1)>2n/(g(1+2g))=2\mathfrak a$, since
$\log(1+2g)>1$ and $2(g^2-1)<g(1+2g)$. Also $A=2K^{-g^2}2^{-\kappa n}$, so (i)
gives \eqref{eq:uniform-gapped-smoothing}.
\end{proof}

For $1<g\le\sqrt3$ put
\[
 c_g:=\frac{\log_2(1+2g)}{g^2-1},\qquad
 \kappa_g:=c(g^2-1)-\log_2(1+2g)\quad(c>c_g).
\]

\begin{corollary}[Cyclotomic smoothing with a prescribed gap]
\label{cor:cyclotomic-smoothing}
\index{Smoothing parameter!cyclotomic lattices}
Let $1<g\le\sqrt3$ and $c>c_g$.
For every $m\ge3$ with $g_m\ge g$ and $\epsilon=2^{-c\varphi(m)}$,
\[
 s_{\mathrm{sh}}(\Lambda_m,\epsilon)\le\eta_\epsilon(\Lambda_m)
 \le\Bigl(1+\frac{K_m^{-g^2}2^{-\kappa_g\varphi(m)}}{\log K_m+c\varphi(m)\log2}\Bigr)
 s_{\mathrm{sh}}(\Lambda_m,\epsilon).
\]
\end{corollary}

\begin{proof}
Apply \Cref{thm:lattice-gap-uniform}\textup{(ii)} with
$\Lambda=\Lambda_m$, $n=\varphi(m)$ and $K=K_m$.
\end{proof}

By \Cref{thm:cyclotomic-gap}, $g^2=3/2$ is admissible for every
$m\ge3$, with $c_g=2\log_2(1+\sqrt6)$ and
$\kappa_g=c/2-\log_2(1+\sqrt6)$.
If $\omega_{\mathrm{odd}}(m)\le1$ and $m\notin\{3,6\}$, one may use
$g^2=2$, with $c_g=\log_2(1+2\sqrt2)\approx1.937$.
For $m\in\{3,6\}$, one may use $g^2=3$, with
$c_g=\tfrac12\log_2(1+2\sqrt3)\approx1.079$.

\paragraph*{Cyclotomic second-shell asymptotics}\phantomsection\label{par:cyclotomic-second-shell-asymptotics}

For the second-shell estimate, let $m\ge3$, $m\notin\{3,6\}$, and put
$n=\varphi(m)$, $k=\omega_{\mathrm{odd}}(m)$,
$\beta=2^k\log(2n+1)$, $E=\log(1/\epsilon)$ and
$L_m=\log(K_m/\epsilon)$, where $0<\epsilon<1$. Set
\[
 q=\begin{cases}3/2,&k\ge2,\\2,&k\le1,\end{cases}
 \qquad
 \mathcal E_m=\begin{cases}
 e^{(7\beta-E)/4},&k\ge2,\\
 e^{3\beta-E},&k\le1.
 \end{cases}
\]

\KeepHeadingSpace{6\baselineskip}
\begin{proposition}[Uniform cyclotomic second-shell correction]
\label{prop:cyclotomic-second-shell}
With the preceding notation, suppose $E\ge7\beta$ for $k\ge2$, or
$E\ge3\beta$ for $k\le1$. Uniformly over conductors and errors,
\begin{equation}\label{eq:uniform-second-shell}
 \frac{\eta_\epsilon(\Lambda_m)}
 {s_{\mathrm{sh}}(\Lambda_m,\epsilon)}-1
 =\frac{H_m\epsilon^{q-1}}{2K_m^q L_m}
   \bigl\{1+O(\epsilon^{q-1}+\mathcal E_m)\bigr\}.
\end{equation}
The relative remainder tends to zero when $E-7\beta\to\infty$ for
$k\ge2$, or $E-3\beta\to\infty$ for $k\le1$.
\end{proposition}

\begin{proof}
The Gram matrix $I_j+J_j$ of $A_j$
is at least $I_j$. After normalizing the dual minimum to one,
\eqref{eq:cyclotomic-lattice-structure} therefore gives a basis
with Gram matrix at least $2^{-k}I_n$, also for $k=0$.
An integer coefficient vector $z$ of a point of norm at most $u$
satisfies $\lVert z\rVert_1\le\lVert z\rVert_2^2\le2^ku^2$.
It is representable by a word of length $\lfloor2^ku^2\rfloor$
in $\{0,\pm e_1,\ldots,\pm e_n\}$. Hence the normalized counting
function satisfies
\begin{equation}\label{eq:cyclotomic-counting}
 N(u)\le e^{\beta u^2},\qquad \beta=2^k\log(2n+1)=n^{o(1)}.
\end{equation}
Indeed, ordering the odd prime divisors as $p_1<\cdots<p_k$ gives
$p_j\ge2j+1$ and $n=D\prod_j(p_j-1)\ge2^k k!$.
Stirling's formula yields $2^k=n^{o(1)}$ uniformly as $n\to\infty$;
bounded $k$ is immediate. Thus $E/\beta\to\infty$ at every fixed
$\epsilon=\exp(-cn^\alpha)$, $c,\alpha>0$.
If $T_Q(t)$ is the Gaussian sum over any subset of
vectors of squared norm at least $Q$, Tonelli's theorem gives
\begin{equation}\label{eq:cyclotomic-tail}
 T_Q(t)\le t\int_Q^\infty N(\sqrt v)e^{-tv}\,dv
 \le\frac{t}{t-\beta}e^{-Q(t-\beta)}\qquad(t>\beta).
\end{equation}

The exact first-shell count gives
\begin{equation}\label{eq:cyclotomic-first-count-polynomial-mas15}
 \frac{K_m}{n^2}
 =\frac{2^{1-k}}D\prod_{p\mid R_m}\frac p{p-1}\le2.
\end{equation}
Indeed, each odd prime has $p/(p-1)\le3/2$, so the middle expression
is at most $2(3/4)^k/D\le2$, including $k=0$.
Consequently $L_m=E+O(\log n)$. If $k\ge2$, set $a_m=H_mK_m^{-3/2}$.
Since $\sum_{p<q}(p-2)(q-2)\ge3$, one has $H_m\ge2K_m$.
For each pair in this sum use $p-2\le\sqrt{p(p-1)}$ and
$p(p-1)\ge6$ for each remaining prime. Equation
\eqref{eq:cyclotomic-second-count} yields
\begin{equation}\label{eq:second-coefficient-bounds}
 \frac{\sqrt2}{n}\le a_m\le
 \frac{2\sqrt2}{3\sqrt D}\binom k2 3^{-(k-2)/2}
 \le\frac{4\sqrt2}{3\sqrt D}.
\end{equation}
Indeed, the ratio of consecutive values of
$\binom k2 3^{-(k-2)/2}$ is $(k+1)/((k-1)\sqrt3)$;
the maximum is its value two at $k=4$.

Separate the first two shells in the normalized theta sum:
$F(t)=K_me^{-t}+H_me^{-qt}+T_3(t)$.
For $k\ge2$, \eqref{eq:cyclotomic-third-gap} gives
$Q=7/4$ as a lower bound for the remaining squared norms.
At $t=L_m$, \eqref{eq:cyclotomic-tail} and $H_m\ge2K_m$ give
\[
 \frac{T_3(L_m)}{\epsilon a_m\sqrt\epsilon}
 \le\frac{L_m}{L_m-\beta}
       \frac{K_m^{-7/4}}{a_m}e^{(7\beta-E)/4}
 \ll e^{(7\beta-E)/4}=\mathcal E_m,
\]
since $K_m^{-7/4}/a_m\le K_m^{-5/4}/2$ and
$L_m/(L_m-\beta)\le7/6$. All constants are absolute.

For $k\le1$, excluding $m=3,6$, put $a_m=H_m/K_m^2$ and $q=2$.
For $R_m=p\ge5$, direct substitution gives
\[
 a_m=\frac{D-1}{2D}+
       \frac{(p-2)(p-3)}{4Dp(p-1)}\in[3/40,1/2].
\]
The second summand at $D=1$ is increasing for $p\ge5$ and equals
$3/40$ at $p=5$; for $D\ge2$ the first summand is at least $1/4$,
and the upper bound follows since the fraction with denominator
$4p(p-1)$ is less than $1/2$. If $R_m=3,D\ge2$, or $k=0$,
$a_m=(D-1)/(2D)\in[1/4,1/2]$.
Now $\beta\le2\log(2n+1)$ and the remaining squared norms are
at least three, including the exceptional conductor $m=4$.
Thus \eqref{eq:cyclotomic-tail} gives
\[
 \frac{T_3(L_m)}{\epsilon a_m\epsilon}
 \le\frac{L_m}{L_m-\beta}\frac{K_m^{-3}}{a_m}e^{3\beta-E}
 \ll e^{3\beta-E}=\mathcal E_m,
\]
since $L_m/(L_m-\beta)\le3/2$.

In both classes write
$\pi\lambda_1(\Lambda_m^*)^2\eta_\epsilon(\Lambda_m)^2=L_m+h$
and $b=a_m\epsilon^{q-1}$. The minimum-length shift gives
\[
 0\le h\le\log\left(1+b+T_3(L_m)/\epsilon\right)=O(b),
\]
since $\mathcal E_m\le1$. At the smoothing solution,
\[
 1=e^{-h}+b e^{-qh}+T_3(L_m+h)/\epsilon.
\]
Expansion and the tail bounds give
$h=b+O(b^2+b\mathcal E_m)$.
Since $a_m\ll1$ and $L_m\ge1$, taking the square root yields
\eqref{eq:uniform-second-shell}.
\end{proof}

Since $\beta=n^{o(1)}$, the relative remainder in
\eqref{eq:uniform-second-shell} tends to zero uniformly as $n\to\infty$
for $\epsilon=\exp(-cn^\alpha)$ with fixed $c,\alpha>0$.

\begin{proposition}[Explicit second-shell bracket]
\label{prop:cyclotomic-second-bracket-mas15}
Let $m\ge3$, $m\notin\{3,6\}$, $0<\epsilon<1$, and retain
$K_m,H_m,q,\beta,L_m$ from \Cref{prop:cyclotomic-second-shell}.
Suppose $L_m>\beta$ and put
\[
 b=H_mK_m^{-q}\epsilon^{q-1},\qquad
 Q=\begin{cases}7/4,&k\ge2,\\3,&k\le1,\end{cases}
 \qquad
 R=\frac{L_m}{L_m-\beta}K_m^{-Q}\epsilon^{Q-1}e^{Q\beta}.
\]
Then
\begin{equation}\label{eq:cyclotomic-second-bracket-mas15}
 \sqrt{1+\frac{b}{L_m(1+qb)}}-1
 \le \frac{\eta_\epsilon(\Lambda_m)}
 {s_{\mathrm{sh}}(\Lambda_m,\epsilon)}-1
 \le \sqrt{1+\frac{\log(1+b+R)}{L_m}}-1.
\end{equation}
\end{proposition}

\begin{proof}
Normalize the dual minimum to one and write
$F(t)=K_me^{-t}+H_me^{-qt}+T_Q(t)$.
By \eqref{eq:cyclotomic-third-gap}, all terms in $T_Q$ have
squared norm at least $Q$, including the case $m=4$.
At $t=L_m$, \eqref{eq:cyclotomic-tail} gives
$T_Q(L_m)/\epsilon\le R$.
Writing $\pi\lambda_1(\Lambda_m^*)^2\eta_\epsilon(\Lambda_m)^2=L_m+h$,
the minimum-length shift gives $0\le h\le\log(1+b+R)$.
At the smoothing solution,
\[
 1=e^{-h}+be^{-qh}+T_Q(L_m+h)/\epsilon.
\]
Since $1-e^{-h}\le h$ and the tail is nonnegative,
$h\ge be^{-qh}\ge b(1-qh)$, hence $h\ge b/(1+qb)$.
Taking square roots proves both bounds.
\end{proof}

\begin{proposition}[All cyclotomic shells below squared ratio two]
\label{prop:cyclotomic-multishell-mas15}
Let $m\ge3$ have $k\ge2$ distinct odd prime divisors and let
$0<\epsilon<1$. Retain $L_m=\log(K_m/\epsilon)$ and
$\beta=2^k\log(2\varphi(m)+1)$. For $2\le j\le k$,
put $q_j=2-2^{1-j}$ and let $H_j$ count dual vectors of squared norm
$q_j\lambda_1(\Lambda_m^*)^2$. Thus $H_2=H_m$, and for $3\le j\le k$,
\[
 H_j=2^{j-1}K_m
 \sum_{\substack{S\subseteq\{p:p\mid R_m\}\\|S|=j}}
       \prod_{p\in S}(p-2).
\]
For fixed $m$, as $\epsilon\downarrow0$,
\begin{equation}\label{eq:cyclotomic-multishell-expansion-mas15}
 \pi\lambda_1(\Lambda_m^*)^2\eta_\epsilon(\Lambda_m)^2
 =L_m+\sum_{j=2}^k H_jK_m^{-q_j}\epsilon^{q_j-1}+O_m(\epsilon).
\end{equation}
For a uniform statement with $L_m>\beta$, set
\[
 S=\sum_{j=2}^k H_jK_m^{-q_j}\epsilon^{q_j-1},\qquad
 R_2=\frac{L_m}{L_m-\beta}K_m^{-2}\epsilon e^{2\beta},\qquad
 h=\pi\lambda_1(\Lambda_m^*)^2\eta_\epsilon(\Lambda_m)^2-L_m.
\]
Then the explicit residual bracket is
\begin{equation}\label{eq:cyclotomic-multishell-residual-mas15}
 -2S^2\le h-S\le R_2.
\end{equation}
\end{proposition}

\begin{proof}
A vector with squared ratio less than two occupies one orthogonal
block. \Cref{thm:A-tensor-second-shell} therefore lists all such
ratios and gives the multiplicities above after multiplication by $D$.
Writing $b_j=H_jK_m^{-q_j}\epsilon^{q_j-1}$, the normalized smoothing
equation is
\[
 1=e^{-h}+\sum_{j=2}^k b_je^{-q_jh}+T_2(L_m+h)/\epsilon,
\]
where $T_2$ consists of squared ratios at least two. By
\eqref{eq:cyclotomic-tail}, $T_2(L_m)/\epsilon\le R_2$, so
$h\le\log(1+S+R_2)\le S+R_2$.
Conversely, $q_j<2$, positivity and $1-e^{-h}\le h$ give
\[
 h\ge\sum_{j=2}^k b_je^{-q_jh}
 \ge Se^{-2h}\ge S(1-2h).
\]
Thus $h\ge S/(1+2S)\ge S-2S^2$, proving the residual bracket.
For fixed $m$, $S=O_m(\sqrt\epsilon)$ and $R_2=O_m(\epsilon)$,
which proves \eqref{eq:cyclotomic-multishell-expansion-mas15}.
The quadratic interactions can already be of order $\epsilon$.
\end{proof}

\KeepHeadingSpace{6\baselineskip}
\begin{corollary}[Cyclotomic smoothing at exponential error]
\label{cor:cyclotomic-exponential}
Fix $c>0$. For $m\ge3$, put $n=\varphi(m)$,
$k=\omega_{\mathrm{odd}}(m)$ and $\epsilon=2^{-cn}$.
Uniformly over all $m\ge3$,
\begin{equation}\label{eq:all-conductor-smoothing}
 \frac{\eta_{2^{-c\varphi(m)}}(\Lambda_m)}
 {s_{\mathrm{sh}}(\Lambda_m,2^{-c\varphi(m)})}
 =1+O_c\!\left(\frac{2^{-c\varphi(m)/2}}{\varphi(m)}\right).
\end{equation}
For sufficiently large $n$, with positive constants depending only on $c$,
\[
 C_c n^{-2}2^{-cn/2}\le
 \frac{\eta_\epsilon(\Lambda_m)}{s_{\mathrm{sh}}(\Lambda_m,\epsilon)}-1
 \le C'_c\frac{2^{-cn/2}}{n\sqrt D}\qquad(k\ge2),
\]
and the excess is $\Theta_c(2^{-cn}/n)$ when $k\le1$.
Along every sequence of conductors with $n\to\infty$ and $k\ge2$
\textup{(}respectively $k\le1$\textup{)}, the excess is not
$O(2^{-\kappa n}/n)$ for any $\kappa>c/2$
\textup{(}respectively $\kappa>c$\textup{)}.
\end{corollary}

\begin{proof}
For $\epsilon=2^{-cn}$, one has $L_m\asymp_c n$ and
$\mathcal E_m=o_c(1)$, so \Cref{prop:cyclotomic-second-shell} and the coefficient bounds
in its proof give all stated upper and lower estimates. The finitely many conductors with bounded
$\varphi(m)$, including $m=3,6$, are absorbed in the constant of
\eqref{eq:all-conductor-smoothing}. Dividing the lower estimates by
$2^{-\kappa n}/n$ proves the claimed failure at larger exponential rates.
\end{proof}

For fixed odd radical $R_m$ with $k\ge2$ and $D\to\infty$, write
$K_m=DK_B$, $H_m=DH_B$ and $c_B=H_B/K_B^{3/2}$, and put
\[
 h=\pi\lambda_1(\Lambda_m^*)^2\eta_\epsilon(\Lambda_m)^2-L_m,
 \qquad b=c_B\sqrt{\epsilon/D}.
\]
The relative second-shell expansion $h\sim b$ holds if and only if
$\epsilon D\to0$. The
$\binom D2K_B^2$ vectors formed from shortest vectors in two different
blocks contribute
$\frac{D-1}{2D}\epsilon e^{-2h}$ to $T_3(L_m+h)/\epsilon$.
An expansion $h\sim b$ forces this contribution to be $o(b)$, hence
$\epsilon D\to0$. Conversely, \eqref{eq:fixed-radical-second} gives
$h\sim b$: with its notation, $u\sim\epsilon$,
$C=O(\epsilon)=o(\sqrt{\epsilon/D})$ and $\rho>1/2$.
For varying radicals with $D\to\infty$, the two-block contribution
requires $\epsilon D/c_B^2\to0$; the coefficient bound makes
$\epsilon D\to0$ necessary there as well.

\paragraph*{Orthogonal blocks and fixed odd radical}\phantomsection\label{par:orthogonal-blocks-fixed-radical}

For the block estimates, let $D,R_m,B,\rho_B$ be as in
\Cref{sec:cyclotomic-gap}, and put $r:=\varphi(R_m)$,
$n:=Dr=\varphi(m)$, $k:=\omega_{\mathrm{odd}}(m)$,
$K:=\#\{b\in B:\lVert b\rVert_2=\lambda_1(B)\}$,
$g_B:=\sqrt{\rho_B}$ and $a_B:=1+2g_B$. Explicitly,
\[
 \rho_B=\begin{cases}
 4,&R_m=1\ (B=\Z),\\
 3,&R_m=3,\\
 2,&R_m\ge5\text{ prime},\\
 3/2,&k\ge2.
 \end{cases}
\]
For $0<\epsilon<1$ set $\delta:=(1+\epsilon)^{1/D}-1$ and
$L:=\log(DK/\epsilon)$.
\Cref{fig:tensor-smoothing}\textup{(c)} indicates how the identity
\eqref{eq:cyclotomic-block-identity} below is combined with
\Cref{thm:lattice-gap-uniform,prop:fixed-lattice-second-shell}.

\begin{proposition}[Smoothing bounds from the orthogonal decomposition]
\label{prop:cyclotomic-block-smoothing}
\index{Smoothing parameter!block decomposition}
With the preceding notation, the following hold.
\begin{enumerate}[\upshape(i)]
\item The smoothing parameters satisfy
\begin{equation}\label{eq:cyclotomic-block-identity}
 \eta_\epsilon(\Lambda_m)=\sqrt{DR_m}\,\eta_\delta(B^*).
\end{equation}
\item If $\log(K/\delta)\ge2r/(g_Ba_B)$, then
\begin{equation}\label{eq:cyclotomic-block-bound}
 0\le\frac{\eta_\epsilon(\Lambda_m)}{s_{\mathrm{sh}}(\Lambda_m,\epsilon)}-1
 \le\frac{\epsilon/2+2a_B^rK^{-\rho_B}(\epsilon/D)^{\rho_B-1}}{2L}.
\end{equation}
Moreover, for fixed $R_m$, as $D\to\infty$,
\begin{equation}\label{eq:fixed-radical-smoothing}
 \frac{\eta_\epsilon(\Lambda_m)}{s_{\mathrm{sh}}(\Lambda_m,\epsilon)}
 =1+O_{R_m}\!\left(\frac1{\log(D/\epsilon)}\right)
 \quad(0<\epsilon<1),
\end{equation}
uniformly in $\epsilon$, without the condition on $\log(K/\delta)$.
\item Let $c>0$, $\epsilon=2^{-cn}$ and $\theta:=c/2-\log_2(1+\sqrt6)/D$.
If $\theta>0$, then
\[
 0\le\frac{\eta_\epsilon(\Lambda_m)}{s_{\mathrm{sh}}(\Lambda_m,\epsilon)}-1
 \le\frac{2^{1-\theta n}}{cn}.
\]
\end{enumerate}
\end{proposition}

\begin{proof}
Since Gaussian sums factor over orthogonal direct sums (as in
\cite[proof of Lemma~4]{EWY23}), \eqref{eq:cyclotomic-lattice-structure} gives
\[
 1+\Phi_{\Lambda_m}(s)
 =\bigl(1+\Phi_{B^*}(s/\sqrt{DR_m})\bigr)^D,
\]
whence \eqref{eq:cyclotomic-block-identity}, as $x\mapsto(1+x)^D$ is increasing. Put
$\Delta:=\log(\epsilon/(D\delta))$. Concavity and $e^u\ge1+u$ give
\[
 \log(1+\epsilon)\le D\delta\le\epsilon,
 \qquad
 0\le\Delta\le\log\frac{\epsilon}{\log(1+\epsilon)}
 \le\epsilon/2,
\]
where the last step uses
$\log(1+\epsilon)\ge2\epsilon/(2+\epsilon)$.

The listed values of $\rho_B$ follow from
\Cref{cor:A-tensor-gap-table}, with $\mu_2(\Z)=2$ for $B=\Z$.
Thus \Cref{thm:lattice-gap-uniform}\textup{(i)} applies to $B^*$
in dimension $r$ with its true gap $g_B$, error $\delta$, and
$\mathfrak a=r/(g_Ba_B)$.
Since $\lambda_1(\Lambda_m^*)^2=\lambda_1(B)^2/(DR_m)$ and $K_m=DK$,
\[
 \left(\frac{\eta_\epsilon(\Lambda_m)}
 {s_{\mathrm{sh}}(\Lambda_m,\epsilon)}\right)^2
 \le1+\frac{\Delta+
  \log(1+2a_B^rK^{-\rho_B}\delta^{\rho_B-1})}{L}.
\]
Use $\delta\le\epsilon/D$, $\log(1+u)\le u$, and
$\sqrt{1+u}\le1+u/2$ to obtain \eqref{eq:cyclotomic-block-bound}.
For fixed $R_m$, this applies for all sufficiently large $D$, uniformly
over $0<\epsilon<1$, since $\log(K/\delta)\ge\log(KD/\epsilon)$.
Since $\rho_B>1$ and $\epsilon/D<1$, the numerator in
\eqref{eq:cyclotomic-block-bound} is $O_{R_m}(1)$, and
$L\ge\log(D/\epsilon)$ because $K\ge1$; this gives
\eqref{eq:fixed-radical-smoothing}.
For the universal consequence \textup{(iii)}, use the same argument
with the valid lower gap $g_0=\sqrt{3/2}$ and $a_0=1+\sqrt6$,
regardless of the larger true block gap. Let $\epsilon=2^{-cn}$ and
$\theta=c/2-\log_2(a_0)/D>0$. Then
\[
 \log(K/\delta)\ge L\ge cn\log2
 =cDr\log2>2r\log a_0>2r/(g_0a_0),
\]
so the coarse version of the preceding bound applies. Since
$a_0^r=2^{n\log_2(a_0)/D}$, $K,D\ge1$, and $\theta<c$,
its numerator is at most
$2^{-cn-1}+2\cdot2^{-\theta n}\le\tfrac52\,2^{-\theta n}$.
Finally $2L\ge2cn\log2$ and $5/(4\log2)<2$, proving \textup{(iii)}.

\end{proof}

\KeepHeadingSpace{6\baselineskip}
\begin{proposition}[Smoothing with fixed odd radical]
\label{prop:cyclotomic-fixed-radical}
Retain the block notation, fix $R_m$, and put
$u:=\log(1+\epsilon)$, $C:=\log(\epsilon/u)$ and
$\beta:=\min(1,\rho_B-1)$.
Uniformly for $0<\epsilon<1$, as $D\to\infty$,
\[
 \frac{\eta_\epsilon(\Lambda_m)}{s_{\mathrm{sh}}(\Lambda_m,\epsilon)}
 =1+\frac{C}{2L}
 +O_{R_m}\!\left(\frac{C^2}{L^2}+\frac{(u/D)^\beta}{L}\right).
\]
If $k\ge2$, let $H_B$ count the second shell of $B$, and put
\[
 c_B=H_BK^{-3/2},\qquad
 \rho=\begin{cases}1,&k=2,\\3/4,&k\ge3.\end{cases}
\]
Then
\begin{equation}\label{eq:fixed-radical-second}
 \left(\frac{\eta_\epsilon(\Lambda_m)}
 {s_{\mathrm{sh}}(\Lambda_m,\epsilon)}\right)^2
 =1+\frac{C+c_B\sqrt{u/D}
             +O_{R_m}((u/D)^\rho)}{L}.
\end{equation}
All implied constants and the threshold for $D$ depend only on $R_m$.
\end{proposition}

\begin{proof}
For fixed $R_m$, the lattice $B$ is fixed. Since
$0<\delta\le2^{1/D}-1$, the estimates in
\Cref{prop:fixed-lattice-second-shell} apply for every $0<\epsilon<1$
once $D$ exceeds a threshold depending only on $R_m$. Put $x=u/D$.
Uniformly for $0<x\le\log2$,
\[
 \delta=x(1+O(x)),\qquad
 \log(K/\delta)=L+C+O(x),\qquad
 \sqrt\delta=\sqrt x+O(x^{3/2}),
\]
with absolute implied constants. By \eqref{eq:cyclotomic-block-identity},
the fixed-lattice expansion gives
\[
 \frac{\pi\lambda_1(B)^2}{DR_m}\eta_\epsilon(\Lambda_m)^2
 =\log(K/\delta)+O_{R_m}(\delta^{\rho_B-1})
 =L+C+O_{R_m}(x^\beta).
\]
Since $0\le C\le\epsilon/2$, division by $L$ and Taylor expansion
at one give the first expansion; the quadratic remainder is
$O_{R_m}((C^2+x^{2\beta})/L^2)$.

If $k\ge2$, the fixed block has second ratio $3/2$ and third
ratio two for $k=2$, or $7/4$ for $k\ge3$. Applying
\Cref{prop:fixed-lattice-second-shell} at $\delta=e^x-1$ gives
\[
 \frac{\pi\lambda_1(B)^2}{DR_m}\eta_\epsilon(\Lambda_m)^2
 =\log(K/\delta)+c_B\delta^{1/2}+O_{R_m}(\delta^\rho).
\]
The preceding uniform estimates and $\rho\le1$ yield
\eqref{eq:fixed-radical-second} after division by $L$.
Taylor expansion about $1+C/L\ge1$, retaining its square root,
gives the equivalent square-root form below: the additional quadratic error is
$O_{R_m}(x/L^2)$ and is absorbed in $O_{R_m}(x^\rho/L)$.
\end{proof}

For $k\ge2$, the equivalent unsquared expansion is
\[
\begin{aligned}
 \frac{\eta_\epsilon(\Lambda_m)}{s_{\mathrm{sh}}(\Lambda_m,\epsilon)}
 &=\sqrt{1+C/L}+
 \frac{c_B\sqrt{u/D}}{2L\sqrt{1+C/L}}\\
 &\quad+O_{R_m}\!\left(\frac{(u/D)^\rho}{L}\right),
\end{aligned}
\]
with the same uniformity and dependence on $R_m$.

To compare the universal-gap bounds, put $a_0=1+\sqrt6$,
$\epsilon=2^{-cn}$ and $c>2\log_2a_0$.
The coarse specialization $g_0^2=3/2$ of the block argument above gives
the upper bound
\[
 U_{\mathrm{block}}=
 \frac{\epsilon/2+2a_0^{n/D}K^{-3/2}(\epsilon/D)^{1/2}}{2L}
\]
for the relative smoothing excess. The corresponding global-gap bound
in \Cref{cor:cyclotomic-smoothing} is
$U_{\mathrm{global}}=K_m^{-3/2}2^{-cn/2}a_0^n/L$.
Since $K_m=DK$, their quotient is
\[
 \frac{U_{\mathrm{block}}}{U_{\mathrm{global}}}
 =Da_0^{-(1-1/D)n}
   +\frac{K_m^{3/2}}4\,2^{-(c/2+\log_2a_0)n}.
\]
By \eqref{eq:cyclotomic-first-count-polynomial-mas15},
$K_m\le2n^2$; also $D\le n$. For $D\ge2$ the quotient is therefore
at most
\[
 na_0^{-n/2}+\frac{(2n^2)^{3/2}}4\,
 2^{-(c/2+\log_2a_0)n},
\]
which is exponentially small in $n$. For $D=1$ the displayed
quotient exceeds one. Equation~\eqref{eq:cyclotomic-block-bound}
additionally uses the actual block gap when it is larger.

\Cref{prop:cyclotomic-block-smoothing}\textup{(iii)} gives an explicit
bound for every $c>0$ when $D$ is sufficiently large; \eqref{eq:all-conductor-smoothing} holds for all $D$.

\begin{proposition}[Power-of-two smoothing bracket]
\label{prop:power-two-smoothing-mas15}
For $n\ge1$, $0<\epsilon<1$ and $\delta=(1+\epsilon)^{1/n}-1$,
\begin{equation}\label{eq:integer-power-smoothing-mas15}
 \log(2/\delta)\le\pi\eta_\epsilon(\Z^n)^2
 \le\log(2/\delta)+\log(1+5\delta^3/8).
\end{equation}
For a cyclotomic conductor $m=2^a$ with $a\ge2$ and $n=m/2$,
$\Lambda_m$ is isometric to $\sqrt n\,\Z^n$, so equivalently
\begin{equation}\label{eq:power-two-smoothing-mas15}
 \log(2/\delta)\le\frac{\pi\eta_\epsilon(\Lambda_m)^2}{n}
 \le\log(2/\delta)+\log(1+5\delta^3/8).
\end{equation}
The true one-dimensional logarithmic excess satisfies
\begin{equation}\label{eq:integer-smoothing-excess-mas15}
 \pi\eta_\delta(\Z)^2-\log(2/\delta)
 =\delta^3/8+O(\delta^6)\qquad(\delta\downarrow0).
\end{equation}
\end{proposition}

\begin{proof}
Factorization of the Gaussian sum gives
$\eta_\epsilon(\Z^n)=\eta_\delta(\Z)$, also for $n=1$.
Apply \Cref{thm:lattice-gap-uniform}\textup{(i)} to $\Z$ with
$K=2$, $g=2$ and error $\delta$. Its threshold is
$\log(2/\delta)\ge1/5$, which holds since $0<\delta<1$;
its tail constant is $A=2\cdot5\cdot2^{-4}\delta^3=5\delta^3/8$.
This proves \eqref{eq:integer-power-smoothing-mas15}.
The cyclotomic assertion follows from
\eqref{eq:cyclotomic-lattice-structure} with $R_m=1$ and $D=n$.

For the expansion put $L_0=\log(2/\delta)$ and
$h_0=\pi\eta_\delta(\Z)^2-L_0$. The bracket gives $h_0=O(\delta^3)$.
The squared norms in $\Z$ are $1,4,9,\ldots$, each of multiplicity two,
so its smoothing equation is
\[
 1=e^{-h_0}+\frac{\delta^3}{8}e^{-4h_0}+O(\delta^8).
\]
The tail estimate here follows by summing
$2\sum_{j\ge3}(\delta/2)^{j^2}/\delta=O(\delta^8)$;
$h_0\ge0$ can only reduce that tail.
Expanding the first two exponentials gives
$h_0=\delta^3/8+O(\delta^6)$, as claimed.
\end{proof}

The upper bound \eqref{eq:integer-power-smoothing-mas15} is strictly
smaller than the bound of~\cite[Theorem~15]{GLWT25} throughout its
nonempty range
\[
 \sqrt{\frac{24n+1}{n^2-1}}\le\epsilon\le\sqrt2-1,
 \qquad n>1.
\]
Here that theorem's squared bound, multiplied by $\pi$, is
$\log((n-1)(1+2/\epsilon))$.
To verify the comparison analytically, write $u=\log(1+\epsilon)$.
For $0\le x\le1/2$, differentiation gives
\[
 \frac{d}{dx}\left(\log(1+x)-\frac{x}{1+x/2-x^2/16}\right)
 =\frac{x^2(x^2-32x+16)}
 {256(1+x)(1+x/2-x^2/16)^2}\ge0.
\]
Consequently $1/u\le1/\epsilon+1/2-\epsilon/16$.
Since $\delta=e^{u/n}-1\ge u/n$ and $\delta\le\epsilon$,
\[
 \frac2\delta+\frac54\delta^2
 \le\frac{2n}{\epsilon}+n-\frac{n\epsilon}{8}
       +\frac54\epsilon^2.
\]
The lower endpoint of the cited range implies $n\epsilon^2>24$.
Subtracting the right side from $(n-1)(1+2/\epsilon)$ gives
\[
 \frac{n\epsilon}{8}-1-\frac2\epsilon-\frac54\epsilon^2
 >\frac1\epsilon-1-\frac54\epsilon^2
 \ge\frac{11}{16}>0,
\]
because $\epsilon\le\sqrt2-1<1/2$. Taking logarithms proves the
strict comparison.
The bracket also applies for $\epsilon=2^{-cn}$ for every $c>0$,
which lies outside the cited range for all sufficiently large $n$.

\begin{remark}[Dependence on the odd prime divisors at fixed error]\leavevmode\phantomsection\label{rem:growing-radical-smoothing}\ignorespaces
Part~\textup{(iii)} of \Cref{prop:cyclotomic-block-smoothing}, for
$\epsilon=2^{-cn}$, is uniform in $R_m$. At fixed $0<\epsilon<1$ the
convergence in \eqref{eq:fixed-radical-smoothing} is not uniform in $R_m$.
Put $R_k:=N_{k+1}/2$, the product of the first $k$ odd primes, and
$m_k:=2^kN_{k+1}$. Then $D=2^k$ and $R_m=R_k$. Let $B,r,n,K,L,\delta$ be
as in \Cref{prop:cyclotomic-block-smoothing} for $m=m_k$. Then, with $\det(B)$
the covolume of $B$,
\[
 \lambda_1(B)^2=2^k,\qquad
 \det(B)^{2/r}=\prod_{p\mid R_k}p^{1/(p-1)}\le3^{k/2}.
\]
Poisson summation for the Schwartz Gaussian on the full-rank lattice $B$ gives
\[
 1+\Phi_{B^*}(s)
 =\frac{s^{-r}}{\det(B)}\sum_{w\in B^*}e^{-\pi\lVert w\rVert_2^2/s^2}
 \ge\frac{s^{-r}}{\det(B)}.
\]
At $s=\eta_\delta(B^*)$,
$1+\Phi_{B^*}(s)=1+\delta$, so
$s^2\ge\det(B)^{-2/r}(1+\delta)^{-2/r}$.
Using \eqref{eq:cyclotomic-block-identity},
\[
 \left(\frac{\eta_\epsilon(\Lambda_{m_k})}
 {s_{\mathrm{sh}}(\Lambda_{m_k},\epsilon)}\right)^2
 =\frac{\pi\lambda_1(B)^2s^2}{L}
 \ge\frac{\pi(2/\sqrt3)^k}{L}(1+\epsilon)^{-2/n}.
\]
Since $K_{m_k}=2^kK$, \Cref{prop:cyclotomic-structure} gives $K=2^{1-k}\prod_{p\mid R_k}p(p-1)$,
and Bertrand's bound $p_{j+1}\le2^{j+1}$ for the $j$th odd prime gives
$L=O_\epsilon(k^2)$, so the ratio diverges.
\end{remark}

\chapter*{AI assistance}
The author used GPT-6 Astra (OpenAI) to assist with drafting and editorial
revision, and with developing, reviewing and refining mathematical arguments
in this monograph. The author takes full responsibility for its content,
including all mathematical statements and proofs, the exposition, and the
references.

\clearpage
\begingroup
\let\cleardoublepage\clearpage
\printindex
\endgroup

\end{document}